\documentclass[11pt,a4paper,english,makeidx,figuresright]{book}
\usepackage[T1]{fontenc}
\usepackage[latin9]{inputenc}
\usepackage{fancyhdr}
\usepackage{array}
\usepackage{longtable}
\usepackage{varioref}
\usepackage{float}
\usepackage{rotfloat}
\usepackage{url}
\usepackage{amsthm}
\usepackage{amsmath}
\usepackage{amssymb}
\usepackage{makeidx}
\makeindex
\usepackage{graphicx}
\usepackage{setspace}
\usepackage[authoryear]{natbib}
\makeatletter

\pdfpageheight\paperheight
\pdfpagewidth\paperwidth

\providecommand{\LyX}{L\kern-.1667em\lower.25em\hbox{Y}\kern-.125emX\@}
\providecommand{\tabularnewline}{\\}
\newcommand{\lyxdot}{.}

 \numberwithin{section}{chapter}
 \theoremstyle{plain}    
 
 \numberwithin{figure}{chapter} 
 \numberwithin{table}{chapter}  

\newcommand{\setpagesize}{%
   \setlength{\oddsidemargin}  {1.5cm}
   \addtolength{\oddsidemargin}{-1in}
   \setlength{\evensidemargin}{\oddsidemargin}
   \setlength{\evensidemargin} {-0.5cm}    
   \setlength{\marginparwidth} {40\p@}
   \setlength{\marginparsep}   {10\p@}     
   \setlength{\topmargin}      {-0.6cm}    
   \setlength{\headheight}     {15\p@}     
   \setlength{\headsep}        {0.5cm}
   \setlength{\textwidth}      {14.1cm}    
   \setlength\@tempdima{\paperheight}
   \addtolength\@tempdima{-35mm}  
   \addtolength\@tempdima{-25mm}  
   \divide\@tempdima\baselineskip
   \@tempcnta=\@tempdima
   \setlength\textheight{\@tempcnta\baselineskip}
   \addtolength\textheight{\topskip}
   \setlength{\topskip}{1\topskip \@plus 1\baselineskip}
   \setlength{\parskip}{\bigskipamount}
   \raggedbottom
}

\renewcommand{\cleardoublepage}{\clearpage\if@twoside \ifodd\c@page\else
  \thispagestyle{empty}
  \hbox{}\newpage\if@twocolumn\hbox{}\newpage\fi\fi\fi}

\renewcommand{\thefigure}{\thechapter.\@arabic\c@figure}
\renewcommand{\thetable}{\thechapter.\@arabic\c@table}
\renewcommand{\theequation}{\thechapter.\@arabic\c@equation}
\renewcommand \thepart {\@Roman\c@part}
\renewcommand \thechapter {\@arabic\c@chapter}
\renewcommand \thesection {\thechapter-\@arabic\c@section}
\renewcommand\thesubsection   {\thesection.\@arabic\c@subsection}
\renewcommand\thesubsubsection{\thesubsection .\@arabic\c@subsubsection}
\renewcommand\theparagraph    {\thesubsubsection.\@arabic\c@paragraph}
\renewcommand\thesubparagraph {\theparagraph.\@arabic\c@subparagraph}

\renewcommand{\@footnotetext}[1]{\insert\footins{%
   \linespread{1}\normalfont\footnotesize%
   \interlinepenalty\interfootnotelinepenalty
   \splittopskip\footnotesep
   \splitmaxdepth \dp\strutbox \floatingpenalty \@MM
   \hsize\columnwidth \@parboxrestore
   \protected@edef\@currentlabel{%
     \csname p@footnote\endcsname\@thefnmark}%
   \color@begingroup
     \@makefntext{%
       \rule\z@\footnotesep\ignorespaces#1\@finalstrut\strutbox}%
   \color@endgroup}}

\renewcommand{\@mpfootnotetext}[1]{%
 \global\setbox\@mpfootins\vbox{%
   \unvbox \@mpfootins
   \linespread{1}\normalfont\footnotesize%
   \hsize\columnwidth
   \@parboxrestore
   \protected@edef\@currentlabel{%
     \csname p@mpfootnote\endcsname\@thefnmark}%
   \color@begingroup
     \@makefntext{%
      \rule\z@\footnotesep\ignorespaces#1\@finalstrut\strutbox}%
  \color@endgroup}}

\renewcommand{\bibname}{References}

\renewcommand{\title}[1]{\gdef\@title{#1}}
\renewcommand{\author}[1]{\gdef\@author{#1}}
\newcommand{\school}[1]{\gdef\@school{#1}}
\newcommand{\degreetype}[1]{\gdef\@degreetype{#1}}
\newcommand{\submitdate}[1]{\gdef\@submitdate{#1}}

\renewcommand{\@title}{}
\renewcommand{\@author}{}
\newcommand{\@school}{School of Physics}
\newcommand{\@degreetype}{Doctor of Philosophy}
\newcommand{\@university}{The University of New South Wales}
\newcommand{\@submitdate}{\ifcase\the\month\or
 January\or February\or March\or April\or May\or June\or
 July\or August\or September\or October\or November\or December\fi
 \space \number\the\year}
\ifnum\month=12
   \@tempcnta=\year \advance\@tempcnta by 1
   \edef\@copyrightyear{\number\the\@tempcnta}
\else
   \newcommand{\@copyrightyear}{\number\the\year}
\fi

\newcommand{\titlep}{%
       \thispagestyle{empty}%
       \begin{titlepage}
       \null\vskip2.5cm%
       \begin{center}
               {\rmfamily\huge\textbf{\expandafter{\@title}}}
       \end{center}
       \vskip1cm
       \begin{center}
               {\rmfamily\Large by \\ \vskip0.5cm \huge \expandafter{\@author}}
       \end{center}
       \vfill
       \begin{center}
               \textsc{\large A thesis submitted for the degree of \\
               \Large \expandafter{\@degreetype}}
       \end{center}
       \vfill
       \vskip1cm
\begin{center}
\includegraphics[width=4cm]{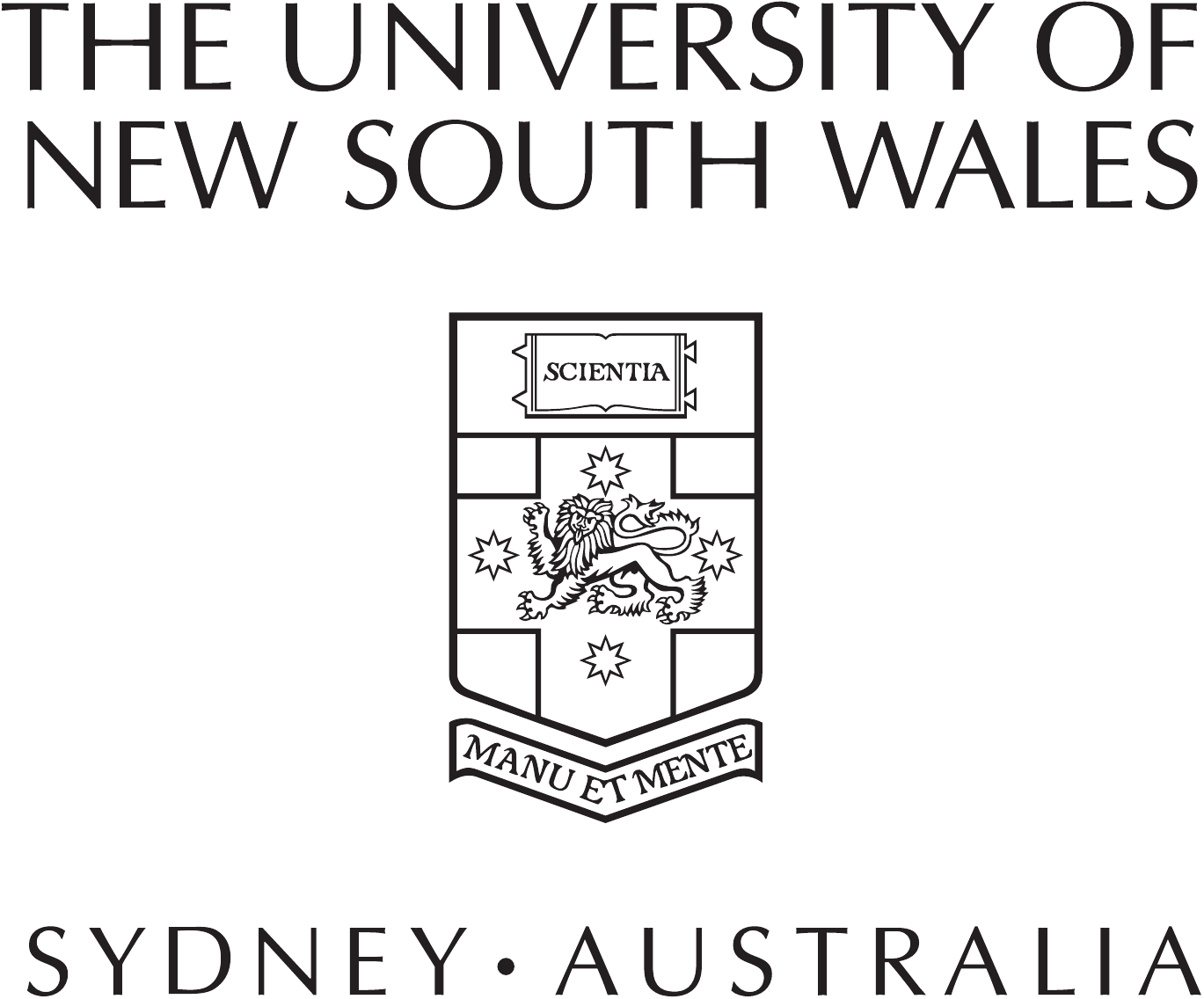}
\end{center}
       \vskip1cm
       \begin{center}
               {\rmfamily\Large \expandafter{\@school},\\
               \expandafter{\@university}.}
               \vskip1cm
               {\rmfamily\LARGE 9 December 2010}
\afterpage{\thispagestyle{empty}}
       \end{center}
       \vskip1cm
       \end{titlepage}
       \newpage}

\renewcommand{\maketitle}{%
       \thispagestyle{empty}%
       \newpage%
       \titlep
       \raggedbottom
   }
\usepackage{graphicx}

\usepackage{lmodern}
\usepackage{pdflscape}
\usepackage{url}
\usepackage{afterpage}
\usepackage{listings}

\usepackage[bf]{caption}
 \usepackage[colorlinks=true, bookmarks, bookmarksnumbered,  
 linkcolor=blue, citecolor=red, urlcolor=blue, filecolor=blue,
 pdfpagelayout=OneColumn, pdfnewwindow=true,
 pdfstartview=XYZ, plainpages=false, pdfpagelabels,
 pdfauthor={Julian King}, pdftex,
 pdftitle={PhD thesis, 2010}]{hyperref}

\usepackage{geometry}
\usepackage{lotsmorefloats}

\renewcommand{\chaptermark}[1]%
{\markboth{\thechapter.\ #1}{}}
\renewcommand{\sectionmark}[1]%
{\markright{\thesection.\ #1}}
\usepackage{array}
\newcolumntype{:}{>{\global\let\currentrowstyle\relax}}
\newcolumntype{;}{>{\currentrowstyle}}
\newcommand{\rowstyle}[1]{\gdef\currentrowstyle{#1}%
  #1\ignorespaces
}

\makeindex

\makeatother

\usepackage{babel}
\begin{document}

\title{Searching for variations in the fine-structure constant and the proton-to-electron
mass ratio using quasar absorption lines}

\author{Julian A. King}

\maketitle
\textbf{\pagenumbering{roman}}

\newpage{}


\newpage
\thispagestyle{empty}
\null\vfill
\begin{center}
\begin{minipage}{13cm}
\setlength{\parindent}{0cm}
\setlength{\parskip}{2ex}
{\underline{Copyright and DAI Statement}}
\par
{\rmfamily\normalsize
'I hereby grant the University of New South Wales or its agents the 
right to archive and to make available my thesis or dissertation in  
whole or part in the University libraries in all forms of media, now 
or here after known, subject to the provisions of the Copyright Act 1968. 
I retain all proprietary rights, such as patent rights. I also retain the 
right to use in future works (such as articles or books) all or part of 
this thesis or dissertation. 

I also authorise University Microfilms to use the 350 word abstract of 
my thesis in Dissertation Abstracts International (this is applicable to 
doctoral theses only).        

I have either used no substantial portions of copyright material in my 
thesis or I have obtained permission to use copyright material; where 
permission has not been granted I have applied/will apply for a 
partial restriction of the digital copy of my thesis or dissertation.'
}
\par
{\rmfamily\normalsize
Signed:\hrulefill\hspace{1cm}Date:\hrulefill
}
\par
\vspace{1.5cm}
{\underline{Authenticity Statement}}
\par
{\rmfamily\normalsize
'I certify that the Library deposit digital copy is a direct equivalent of 
the final officially approved version of my thesis. No emendation of 
content has occurred and if there are any minor variations in  
formatting, they are the result of the conversion to digital format.'
}
\par
{\rmfamily\normalsize
Signed:\hrulefill\hspace{1cm}Date:\hrulefill
}
\end{minipage}
\end{center}
\vfill\null


\newpage
\thispagestyle{empty}
\null\vfill
\begin{center}
\begin{minipage}{13cm}
\setlength{\parindent}{0cm}
\setlength{\parskip}{2ex}
{\underline{Originality Statement}}
\par
{\rmfamily\normalsize
'I hereby declare that this submission is my own work and to the best of my  
knowledge it contains no materials previously published or written by 
another person, or substantial proportions of material which have been 
accepted for the award of any other degree or diploma at UNSW or any other 
educational institution, except where due acknowledgement is made in the thesis. 
Any contribution made to the research by others, with whom I have 
worked at UNSW or elsewhere, is explicitly acknowledged in the thesis. I 
also declare that the intellectual content of this thesis is the product of my 
own work, except to the extent that assistance from others in the project's 
design and conception or in style, presentation and linguistic expression is 
acknowledged.'
}
\par
\vspace{1.5cm}
{\rmfamily\normalsize \hspace{2cm}Signed:\hrulefill}
\end{minipage}
\end{center}
\vfill\null

\newpage{}

\textbf{\thispagestyle{plain}\setcounter{page}{1}}\thispagestyle{plain} \phantomsection \addcontentsline{toc}{section}{Abstract}\textbf{\begin{center}Abstract\end{center}}

Quasar absorption lines provide a precise test of the assumed constancy
of the fundamental constants of physics over cosmological times and
distances. We have used quasar absorption lines to investigate potential
changes in the fine-structure constant, $\alpha\equiv e^{2}/(4\pi\epsilon_{0}\hbar c)$,
and the proton-to-electron mass ratio, $\mu\equiv m_{p}/m_{e}$.

The many-multiplet method allows one to use optical fine-structure
transitions to constrain $\Delta\alpha/\alpha$ at better than the
$10^{-5}$ level. We present a new analysis of 154 quasar absorbers
with $0.2<z<3.7$ in VLT/UVES spectra. From these absorbers we find
$2.2\sigma$ evidence for angular variations in $\alpha$ under a
dipole+monopole model. Combined with previous Keck/HIRES observations,
we find $4.1\sigma$ evidence for angular (and therefore spatial)
variations in $\alpha$, with maximal increase of $\alpha$ occurring
in the direction $\mathrm{RA=(17.3\pm1.0)\,\mathrm{hr}}$, $\mathrm{dec.}=(-61\pm10)^{\circ}$.
Under a model where the observed effect is proportional to the lookback-time
distance the significance increases to $4.2\sigma$. Importantly,
dipole models fitted to the VLT and Keck samples independently yield
consistent estimates of the dipole direction, which suggests that
the effect is not caused by telescope systematics. Similarly, dipole
models fitted to $z<1.6$ and $z>1.6$ sub-samples also point in a
consistent direction. The observed dipole effect is stable under iterative
trimming of potentially outlying $\Delta\alpha/\alpha$ values, implying
that the result is not being generated by a subset of the data. We
consider a number of systematic effects, including potential wavelength
scale distortions and evolution in the abundance of Mg isotopes, and
show that they are unable to explain the observed dipole effect. If
these results are correct, they directly demonstrate the incompleteness
of the Standard Model and violation of the Einstein Equivalence Principle.

Optical spectra of $z\gtrsim2$ molecular hydrogen absorbers can probe
evolution in $\mu$. We have used spectra of the quasars Q0405$-$443,
Q0347$-$383 and Q0528$-$250 from VLT/UVES to investigate the absorbers
at $z=2.595$, 3.025 and 2.811 in these spectra respectively. We find
that $\Delta\mu/\mu=(10.1\pm6.6)\times10^{-6}$, $(8.2\pm7.5)\times10^{-6}$
and $(-1.4\pm3.9)\times10^{-6}$ in these absorbers respectively.
A second spectrum of Q0528$-$250 provides an additional constraint
of $\Delta\mu/\mu=(0.2\pm3.2_{\mathrm{stat}}\pm1.9_{\mathrm{sys}})\times10^{-6}$.
The weighted mean of these values yields $\Delta\mu/\mu=(1.7\pm2.4)\times10^{-6}$,
the most precise constraint on evolution in $\mu$ at $z>1$.

We also demonstrate the application of Markov Chain Monte Carlo methods
to determining $\Delta\alpha/\alpha$ from quasar spectra.

\newpage{}\newcommand{\isc}{\textsc{i}}
\newcommand{\ii}{\textsc{ii}}
\newcommand{\iii}{\textsc{iii}}
\newcommand{\iv}{\textsc{iv}}
\newcommand{\iscs}{\textsc{i} }
\newcommand{\iis}{\textsc{ii} }
\newcommand{\iiis}{\textsc{iii} }
\newcommand{\ivs}{\textsc{iv} }
\def \aap{A\&A} \def \aaps{A\&AS} \def \al{Astron. Lett.} \def \aj{AJ} \def \ap{Appl. Phys.} \def \apj{ApJ} \def \apjl{ApJ} \def \apjs{ApJS} \def \asp{Astron. Soc. Pac.} \def \apss{AP\&SS} \def \ca{Comments on Astrophys.}  \def \josb{J. Opt. Soc. Am. B} \def \jpcrd{J. Phys. Chem. Ref. Data} \def \jqsrt{J. Quant. Spectrosc. Radiat. Transfer} \def \met{Metrologia} \def \mnras{MNRAS} \def \nat{Nature} \def \npb{Nucl. Phys. B} \def \nsrds{Natl. Stand. Rel. Data Ser.} \def \pasp{PASP} \def \plb{Phys. Lett. B} \def \pr{Phys. Rev.} \def \pra{Phys. Rev. A} \def \prc{Phys. Rev. C} \def \prd{Phys. Rev. D} \def \prep{in preparation} \def \prl{Phys. Rev. Lett.} \def \psc{Phys. Scr.} \def \rmp{Rev. Mod. Phys.} \def \zpa{Z. Phys. A} \def \physrep{Phys. Rep.} \def \araa{Ann. Rev. Astron. Astrophys.} \def \memsai{Mem.~Soc.~Astron.~Italiana}
\renewcommand{\bibname}{References}

\singlespacing \cleardoublepage \phantomsection\addcontentsline{toc}{section}{Contents}\tableofcontents

\listoffigures
\addcontentsline{toc}{section}{List of Figures}

\listoftables
\addcontentsline{toc}{section}{List of Tables}\onehalfspacing\newpage{}

\noindent \begin{center}
\thispagestyle{plain}\textbf{\large Acknowledgments}
\par\end{center}{\large \par}

\phantomsection \addcontentsline{toc}{section}{Acknowledgements}Firstly,
I must thank my supervisor, John Webb. Although astrophysics was already
a passion of mine, I think it's unlikely that I would have abandoned
law (at least, so rapidly) to undertake my PhD without such an intriguing
and important topic. His support and encouragement have made this
work extremely enjoyable. I must also thank Michael Murphy, for not
only doing much of the spectral reduction legwork (something I am
exceedingly glad was already done) but also for many in-depth conversations
about rather technical issues, as well as prompt/1am replies to my
(usually double-barrelled) emails begging for help. I am similarly
grateful for the existence of his thesis%
\footnote{\citet{Murphy:PhD}%
}, which made my life far easier than it probably should have been.
I am also grateful to my co-supervisor, Victor Flambaum, for the theoretical
power he has brought to bear on this subject and friendly advice,
as well as occasional tersely worded emails urging us to publish.
Thanks must go to Bob Carswell, both for insightful discussions and
the enormous amount of effort he has invested in maintaining \textsc{vpfit}.
Thanks also to Julian Berengut, for lively discussions, rants, and
beers.

I am grateful to UNSW and the Australian Government for the support
of an Australian Postgraduate Award. Also, UNSW kindly provided me
within funding under the Postgraduate Research Support Scheme to attend
the IAU XXVII General Assembly in Rio de Janeiro, to present some
of my work. Similarly, I thank Wim Ubachs for providing funding to
travel to Amsterdam and collaborate with him and others at the Vrije
Universiteit. Thankyou to the Centre for Astrophysics \& Supercomputing,
at Swinburne University, for time on their supercomputer (the Green
Machine).\afterpage{\thispagestyle{plain}}

To all the crazy astrophysics students I have met from time to time,
both at UNSW and the week of binge drinking that is HWWS/ASA: thanks
for lots of laughs, and teaching me things along the way. Respect
to the denizens of room 169, who without question spent more time
there than I did (CJ, Mike W, Steve P and Vicki L in particular; Colin
B is an honorary member, and would have been there if he knew what
was good for him).

I have made extensive use of NASA's Astrophysics Data System (ADS).
This thesis was written in \LyX{} --- I am extremely grateful to the
creators for the interface. I can't imagine hundreds of pages of raw
\LaTeX. Plots have helpfully/frustratingly been made with \textsc{pgplot}.
I am grateful to the authors of Numerical Recipes \citep{NumericalRecipes:92,NumericalRecipes:2007}
for their computational tome. Thanks to Linux, for mostly working.

To my colleagues at Pottinger --- thanks for tolerance of my physics
rants, and my sometimes-sporadic attendence at work. I am particularly
grateful to Cassandra and Nigel, who have been extremely flexible
in helping me to blend ``work work'' with ``uni work'', who have always
been a good source of advice and who have always supported me.

To all my friends: you have kept me (mostly) sane throughout the last
few years, although at differing levels of sobriety. These include
(in a far from exhaustive fashion): Amber, Angela, Annabelle (vexer
\#1), Bec~M, Bronwyn, Cameron, Emily~F, Hannah~M, Hannah~T, Helen~D,
Georgie, Glenn, Iris, Mike~J, Kamala, Kate~D, John~V, Jon W, Josh~W,
Katrina, Kimberley, Laura ``loose/loud'' S, Marisel, Marshy, Marty,
Nat~K, Nathan, Nina~F, Renee~L, Rodd, Seb, Smurf, Tanner, Tanya,
Tia (vexer \#2), Tim~A, Tim~F, Tommy (Tom-E), Zoe\ldots{} (I'm
sure I've missed people, but figured this was as good a place as any
to name drop my friends ={]} ). In an separate but equally important
category are my friends-at-arms, Chris, James and Steve, for helping
me combine booze and guns (although never at the same time) --- WBAF
\& penguins all round.

I am indebted to the late Michael Bishop and Andrew Haines at Sydney
Grammar School, along with many other teachers there, who nurtured
my interest in physics and set me on the path that led to this thesis.

To Michelle: thanks for putting up with my inane behaviours (particularly
our disparate views on the value of navels), for keeping me fed and
mostly clean, and for lots of love.

Finally, I cannot thank my parents enough, for their support of my
academic endeavours, for tolerating me, and for everything in general. 

\newpage{}

\noindent \begin{center}
\thispagestyle{plain}\textbf{\large Preface}
\par\end{center}{\large \par}

\phantomsection \addcontentsline{toc}{section}{Preface}A significant
part of the content of this thesis was (or is to be) published. In
particular, the following papers are relevant:
\begin{itemize}
\item The analysis of molecular hydrogen data in section \ref{sec:mu:results}
of chapter \ref{cha:mu} to search for variations in $\mu$ was published
in Physical Review Letters as \citet{King:08}. The second part, relating
to an analysis of a new spectrum of Q0528$-$250 (section \ref{sub:Q0528-250 revisited}),
has not yet been submitted for publication but will be in the near
future.
\item The analysis of the fine-structure constant, $\alpha$, in chapter
\ref{cha:alpha}, was submitted in brief to Physical Review Letters
in \citet{King:10a}, and will be submitted in full to Monthly Notices
of the Royal Astronomical Society (MNRAS).
\item The analysis of potential systematics in relation to $\Delta\alpha/\alpha$
(chapter \ref{cha:da systematic errors}) will be submitted in full
to MNRAS.
\item The discussion in chapter \ref{cha:Discussion} concerning the variation
of certain ratios of fundamental constants mirrors that found in \citet{Berengut:10b}.
The analysis of $\Delta\mu/\mu$ in that paper was developed by Julian
Berengut, however I had developed a similar analysis independently,
and that is presented here. The discussion of the dimensionless constants
$F$ and $G$ in that paper was written by Julian Berengut, but was
based upon my initial analysis. The analysis of those constants is
given in chapter \ref{cha:Discussion} in more detail in any event.
\item The primary results of the Markov Chain Monte Carlo analysis in chapter
\ref{cha:MCMC} were published in \citet{King:09} (an unrefereed
conference proceedings), although we provide considerably more detail
here.\afterpage{\thispagestyle{plain}}
\end{itemize}
I have retained the use of the first person plural throughout this
work, although the work is all mine except where noted. Of course,
all the work has been developed in close collaboration with my supervisor,
John Webb. 

This thesis was proof-read by John Webb, and in part by Julian Berengut,
Michelle Crisp, Adrian Malec, Daniel Mortlock and Michael Murphy.
John Webb, Michael Murphy and Daniel Mortlock are professional astronomers.
Except where noted, the contributions of these people are limited
to linguistic and stylistic suggestions, and suggestions for further
work.

Most of this thesis derives results from the fitting of Voigt profiles
to quasar spectra using the program \textsc{vpfit}. Various people
have contributed to \textsc{vpfit} since its creation two decades
ago. The primary authors are R. F. Carswell, J. K. Webb, M. J. Irwin
and A. J. Cooke, although large number of people have contributed
bits and pieces%
\footnote{Full credits are available at \url{http://www.ast.cam.ac.uk/~rfc/vpfit.html}.%
}.

Velocity plots of $\Delta\alpha/\alpha$ absorbers have been produced
with \textsc{plabsys}, a program written by Michael Murphy specifically
for this purpose (although subsequently modified in a relatively minor
fashion by me). Where I have clipped data from spectra, I have used
the program \textsc{uves popler}, also written by Michael Murphy for
this purpose. The molecular hydrogen plots in the appendices were
modelled on those in \citet{Malec:10}, however the code is mine.
All the code used to analyse the $\Delta\mu/\mu$ results in chapter
\ref{cha:mu} and the $\Delta\alpha/\alpha$ results in chapter \ref{cha:alpha}
is mine. The Markov Chain Monte Carlo (MCMC) code in chapter \ref{cha:MCMC}
is added on to \textsc{vpfit}, however the actual code is mine. 

I have adopted a referencing style which includes the titles of articles
as, I think, this is more useful than the typical astrophysical journal
style, which omits the title.

Contributions made by others in specific chapters are noted where
they occur, but for clarity the following contributions are noted
here:
\begin{itemize}
\item \textbf{Chapter \ref{cha:mu}}: The spectral reductions were undertaken
by Helene Ménager and Michael Murphy. Manipulation of the Q0528:B2
spectrum to emulate systematics was performed by Michael Murphy.
\item \textbf{Chapter \ref{cha:alpha}}: The spectral reductions were undertaken
by a wide variety of people as part of the UVES SQUAD (UVES Spectroscopic
QUasar Absorption Database) project. My particular thanks go to Michael
Murphy, who has coordinated this project, and worked on a particularly
large number of reductions. The following people have also contributed
to the spectral reductions: Matthew Bainbridge, Ruth Buning, Huw Campbell,
Robert Carswell, Ankur Chaudhary, Glenn Kacprzak, Ronan McSwiney,
Helene Ménager, Daniel Mountford, Jon Ouellet, Tang Wei, Berkeley
Zych. The Voigt profile fits to the absorbers (see appendix \ref{cha:MM fits})
were reviewed by John Webb; he provided various suggestions for improvement. 
\item \textbf{Chapter \ref{cha:da systematic errors}}: The $\Delta v$
test was developed in collaboration with John Webb, F.~Elliot Koch
and Robert Carswell. The $\Delta v$ values used to investigate potential
inter-telescope systematics were generated by Matthew Bainbridge.
The modelling of the $\Delta v$ values, and the analysis of the impact
they have on $\Delta\alpha/\alpha$ is my work. The $\Delta v$ function
used to emulate the intra-order distortions found by \citet{Whitmore:10}
was created by F. Elliot Koch. The discussion about the potential
impact of the isotopic abundance of different Mg isotopes was developed
in close collaboration with John Webb and Michael Murphy.
\item \textbf{Chapter \ref{cha:MCMC}}: Although the MCMC code is mine,
the initial development occurred with suggestions from Daniel Mortlock,
to whom I am grateful.\afterpage{\pagenumbering{arabic}}
\end{itemize}
\newpage{}

\chapter{Introduction\label{cha:Introduction}}

\setcounter{page}{1}This chapter provides an introduction to the
field of varying constants. We give an overview of the early history
of the field, as well as some interesting recent developments. Our
primary work is a two-pronged analysis of the proton-to-electron mass
ratio, $\mu$, and the fine-structure constant, $\alpha$, using quasar
spectra. In this chapter, we introduce these constants, and review
constraints on changes in these constants derived from methods which
do not utilise quasar spectra. We consider constraints derived from
quasars in chapters \ref{cha:mu} and \ref{cha:alpha} respectively.
We give more details of the structure of this work in section \ref{sec:Structure of work}.
Chapters \ref{cha:mu} through \ref{cha:MCMC}, all involve the analysis
of quasar absorption lines, and so we describe our methodology where
it is common to these chapters in chapter \ref{cha:Common methodology}.

\section{What are fundamental constants?}

Any formulation of physics is inextricably linked with a system of
units. All numerical quantities must ultimately be compared with some
standard. Many units in any system are superfluous, and can be reduced
to combinations of some set of base units. The SI system lists seven
units, namely: the kilogram, the second, the metre, the ampere, the
kelvin, the mole, and the candela. These in turn relate to underlying
dimensions, namely mass, length and time, electric charge and temperature.
Amongst these dimensions, some consider only mass, length and time
as fundamental. The use of the metre, the kilogram and the second
as base units of length, mass and time respectively are not unique
even within modern physics. \citet{Uzan:03} noted that the SI system
is only useful for measurements that are ``of human size''. One
can construct unit systems that are more appropriate for other regimes.
A common basis for high-energy physics is to construct a unit mass
$m_{e}$, a unit length $4\pi\epsilon_{0}h^{2}/m_{e}e^{2}$ and a
unit time $2\epsilon_{0}h^{3}/\pi m_{e}e^{4}$ \citep{Uzan:03}.

Although the SI system is convenient, the various units in it are
not fundamental. A more rational system for fundamental science associates
units with certain physical constants, which seem to be intrinsic
properties of our universe. \citet{FlowersPetley:01} considered the
following as fundamental: the electron charge $e$, the proton mass
$m_{p}$, the reduced Planck constant $\hbar$, the velocity of light
in a vacuum $c$, the Avogadro constant $N_{A}$, the Boltzmann constant
$k_{B}$, the Newtonian gravitational constant $G$, and the permittivity
and permeability of free space $\epsilon_{0}$ and $\mu_{0}$. Clearly
some of these quantities are not independent, as for example $\epsilon_{0}\mu_{0}=1/c^{2}$,
and in theory Avogadro's constant can be derived from a sufficient
ability to weigh and count atoms. \citet{Okun:91} considered that
only three fundamental quantities are necessary: the metre, the second
and the kilogram. 

This work is concerned primarily with two constants of fundamental
importance. The first is the proton-to-electron mass ratio, $\mu\equiv m_{p}/m_{e}$,
which is simply the ratio of the proton mass to the electron mass.
The second is the fine-structure constant, $\alpha\equiv e^{2}/(4\pi\epsilon_{0}\hbar c)$.
The importance of these two constants is that they, along with an
energy scale, completely define the gross structure of atomic and
molecular systems \citep{Born:35}. Besides the obvious physical implications,
this also means that much of chemistry ultimately hinges on these
two numbers. Thus, these two numbers ultimately have an enormous impact
on the universe.

The constants that appear in our theories determine the proportionality
between different quantities. As our knowledge increases, some constants
are deprecated by our new-found ability to relate different quantities.
The clearest historical example of this is gravitation, where prior
to Newton it was widely believed that the acceleration due to gravity,
$g$, was a universal quantity. Newton's inverse square law of gravitation
yielded the force between two masses as $F=Gm_{1}m_{2}/r^{2}$, which
replaced one constant with another, but yielded a relationship of
much broader generality. $G$ still retains the status of a fundamental
constant today, the value of which we are unable to predict from other
quantities. In fact, a reasonable definition of a fundamental physical
constant at present is any proportionality constant of a fundamental
theory which cannot be predicted. 

The standard model\index{standard model} of physics, together with
gravity, requires 22 unknown constants: the Newtonian constant, six
Yukawa couplings for the quarks and three for the leptons, the mass
and vacuum expectation value of the Higgs field, four parameters for
the Cabibbo-Kobayashi-Maskawa (CKM) matrix, three coupling constants,
a UV cutoff including the speed of light and Planck's constant \citep{Hogan:00a,Uzan:09a}.
We require three constants to define a system of units, leaving 19
unexplained dimensionless parameters. Indeed, the problem becomes
worse with the discovery that neutrinos must be massive \citep[see for instance][]{ParticleDataGroup:08}.
This implies at least seven more parameters in the standard model
(three Yukawa couplings and four CKM parameters) \citep{Uzan:09a}.
The existence of a large number of free parameters in our fundamental
theories is almost a \emph{prima facie} suggestion that these theories
are incomplete; one would hope that an all-encompassing fundamental
theory would have far fewer free parameters (and perhaps none). On
the other hand, it remains to be seen whether any of the fundamental
constants can truly be predicted from theory --- some or all of them
may turn out to be independent properties of the universe, set seemingly
at random and with no relationship to the laws which describe the
dynamics of particles.

There is a strong (although oft-forgotten) assumption within physics
that constants are just that: constant. In fact, we already know that
our constants are not the same under all regimes: the coupling constants
of the three forces of the standard model ``run'' with energy, but
the high-energy values can nevertheless be expressed in terms of their
low-energy values. This fact aside, the accepted position is that
our constants are invariant throughout time and space. 

How does one look for a change in constants? A naive approach is to
search for variations in a convenient constant, such as the speed
of light, in different times and places. Although such a variation
might be found, the interpretation is severely hampered. A variation
in $c$ could mean any or all of the following: \emph{i)} the physics
underlying the propagation of light is changing; \emph{ii)} the length
of the metre is changing, \emph{iii)} the length of the second is
changing. These possibilities cannot be disentangled. \citet{Dicke:62a}
notes the solution to this problem: to only search for variation in
dimensionless quantities. Detection of variation in a dimensionless
quantity guarantees that it is the quantity under consideration which
is changing, and not any aspect of the unit system. Changes in dimensionful
quantities can be measured, but this necessarily entails an explicit
statement about which units are assumed to be held fixed. Measured
change in a dimensionless quantity would therefore unambiguously imply
that physics is changing.

\section{History of varying constants}

\subsection{Early considerations}

The first investigations into whether the fundamental constants vary
were due to Milne \citep{Milne:35a,Milne:37a} and Dirac\index{Dirac, Paul}
\citep{Dirac:37a}, who suggested that $G$ might vary with cosmological
time. Dirac's work in particular was a response to several observed
apparent coincidences between large numbers (this subsequently became
known as the Large Number Hypothesis\index{Large Number Hypothesis (LNH)},
or LNH). In particular, Dirac noted that in atomic units of time the
age of the universe is $\sim10^{40}$, whilst the number of protons
in the observable universe is $\sim10^{80}$. $10^{40}$ is of the
same order of magnitude of the ratio of the strength of the electrical
forces between a proton and electron to the strength of the gravitational
force between them. This led Dirac to speculate that perhaps these
quantities were fundamentally interrelated, and that, perhaps, $G\propto1/t$
and $M\propto t^{2}$ (where $M$ is the amount of mass in the universe).
\citet{Teller:48} objected that the implications of this cosmology
were inconsistent with paleontological data. However, \citet{Gamow:67a}
showed that interpreting the LNH as allowing for time variation of
$e$ rather than $G$ side-stepped Teller's objections. Although the
suggestions of Milne and Dirac were based on numerological rather
than physical grounds, the discussion around them serves the purpose
of showing that people have seriously considered the variation of
fundamental constants for quite some time.

\citet{BransDicke:61} placed the variation of $G$ on a more rigorous
footing by developing a self-consistent scalar-tensor extension of
Einstein's General Relativity (GR), where the tensor component describes
the classical GR behaviour, whilst the scalar part describes the propagation
of a scalar field, which itself is a source of space-time curvature.
So-called Brans-Dicke theories\index{Brans-Dicke theory} (and their
extensions) are still considered as objects of interest, although
constraints on them have become increasingly stringent. The theory
predicts that the post-Newtonian $\gamma$ parameter will deviate
from the standard GR value of 1, instead giving $\gamma=(\omega+1)/(\omega+2)$
\citep{Weinberg:72} where $\omega$ is the dimensionless coupling
constant of the scalar field. Precision measurements of the Cassini
spacecraft require $|\omega|\gtrsim40,000$ \citep{Bertotti:03a},
which is ``uncomfortably large'' \citep{Moffat:10a}.\index{Brans-Dicke theory}

\subsection{Motivations for varying constants}

Varying constants in the 21st century? A crazy thought, some might
think. Modern physical theories have enormous predictive power, and
some might be content with the status quo (although certainly not
Sir Karl Popper). Yet, it is well known that General Relativity (which
describes gravity) and the Standard Model (which describes electromagnetism,
and the strong and weak nuclear forces) are incompatible. We do not
have a good theory of quantum gravity at present. The incompleteness
of our theories is reason enough to try to subject them to every test
we can imagine. 

Fundamentally, it must be noted that it is known experimentally that
the contants ``run'' with energy (that is, they take on different
values at high energy scales), and so variation of the physical constants
under different local energy regimes has already been shown. There
is no known law or symmetry principle -- other than an assumption
for the sake of simplicity -- which prevents the constants of nature
from varying in space and time. Thus, it is necessary to check this
assumption experimentally. 

Notwithstanding the desire to try to falsify some aspect of our current
understanding of physics, there are lines of argument which suggest
that a variation of the fundamental constants might be possible, or
even desirable. It has been explicitly shown that cosmological variation
in the constants may proceed differently in different places and times
\citep{Forgacs:79a,Barrow:87a,Damour:94a,Li:98a}. Additionally, in
any model of the universe with extra dimensions then the constants
of nature must vary, although the magnitude of the variation is not
constrained by theory \citep{kaluza:21,Klein:26a,Forgacs:79a,Barrow:87a,Li:98a}. 

Two interesting considerations exist which are worth presenting. We
consider here the well-known argument about the triple-$\alpha$ process,
which is of significant historical interest, and also a line of argument
which has emerged from an apparent cosmological deficit of $^{7}$Li
relative to theoretical predictions (the ``lithium problem'').

\subsubsection{The anthropic principle, the triple-$\alpha$ process and fine-tuning\label{sub:triple-alpha process}}

An extremely well-known prediction made by Fred Hoyle concerns the
existence of the triple $\alpha$ resonance\index{triple-alpha
 process@triple-$\alpha$ process}, the reaction through which $^{12}$C is produced in our sun. The
production of $^{12}$C depends crucially on the carbon energy level
at 7.65 MeV, which is only 0.3 MeV higher than the sum of the masses
of three $\alpha$ particles \citep{Okun:96a}; the relatively small
difference enhances the cross section of the reaction $3\alpha\rightarrow\,^{12}\mathrm{C}$.
$^{8}$Be is unstable, and therefore $^{12}$C cannot be produced
in sufficient quantities through the reaction $\alpha+\,^{8}\mathrm{Be}\rightarrow\,^{12}C$.
Additionally, without the resonance carbon would disappear through
the reaction $\alpha+{}^{12}C\rightarrow\,^{16}O$. Hoyle deduced
that a resonance must exist before it was discovered experimentally,
noting that without such a resonance we would not see the observed
quantities of $^{12}$C in the universe \citep{Dunbar:53a,Hoyle:54a}.
That is, without this resonance, humans would not exist! The requisite
excited state of $^{12}$C has become known as the Hoyle level \citep{Ekstrom:09a}.

The reaction rate is extremely sensitive to the energy of the resonance.
If $Q_{\alpha\alpha\alpha}$ is the energy of the resonance, then
the sensitivity of the reaction rate to a variation of $Q_{\alpha\alpha\alpha}$
is 
\[
s=\frac{\mathrm{d\,\ln}\lambda_{3\alpha}}{\mathrm{d\,\ln Q_{\alpha\alpha\alpha}}}=-\frac{Q_{\alpha\alpha\alpha}}{k_{b}T}\sim\left(\frac{-4.4}{T^{9}}\right)
\]
\citep{Ekstrom:09a}. If the energy level of the Hoyle level were
to increase, the amount of observed $^{12}$C would be reduced on
account of rapid processing to $^{16}$O (thereby increasing the quantity
of $^{16}$O), with the converse effect for a reduced energy of the
level. It has been estimated that carbon or oxygen production would
be suppressed by a factor of between 30 and 1000 if the fine-structure
constant differed by more than about four percent (or if the strong
force was different in strength by more than about half a percent)
\citep{Oberhummer:00a,Oberhummer:03,Csoto:01,Schlattl:04a}. Indeed,
our universe is effectively on-resonance. Thus, some argue that our
universe appears to be uniquely fine-tuned\index{fine-tuning} for
the existence of life\ldots{} or at least, life as we understand
it \citep[see for instance][]{Davies:03a}. 

Hoyle's argument from the existence of humans for the existence of
the resonance is perhaps the only known good prediction made using
the anthropic principle. Although one might take this to argue in
favour of a deity or designer of some form, another argument via the
weak anthropic principle is that there exists a statistical ensemble
of universes, in which different values of fundamental constants are
realised. The weak anthropic principle then yields that we simply
find ourselves in one of the most life-friendly universes \citep{Okun:96a}. 

What are the actual requirements for life to form? Given our lack
of understanding of the origin of life, these clearly remain unknown.
Nevertheless, it is clear that the laws of physics must allow for
the creation of complex structures. \citet{Rees:00a} considered six
dimensionless constants to be important in creating a universe which
is amenable to complicated structure (and thus life): the ratio of
the strength of electromagnetism to that of gravity, the strength
of the nucleon binding force, the relative importance of gravity and
expansion energy in the universe, the cosmological constant ($\Lambda$),
the ratio of the gravitational energy required to unbind a galaxy
to its mass energy equivalent and the number of spatial dimensions.
Other viewpoints are possible. 

However, whether our universe is fine-tuned or not remains a point
of contention. For instance, \citet{Stenger:00} considered two numbers
of interest, $N_{1}$ and $N_{2}$. $N_{1}$ is the ratio of the strength
of the electromagnetic force to the gravitational force between two
electrons, and is given by $N_{1}\approx10^{39}$. $N_{2}$ is the
ratio of a typical stellar lifetime to the time for light to traverse
the radius of a proton. \citet{Dirac:37a} noted that $N_{1}\sim N_{2}$.
\citet{Dicke:61a} noted that $N_{2}$ must be large in a universe
with life, so that stars live long enough to generate heavy elements.
He also noted that $N_{1}$ must be of similar magnitude in order
for the universe to have elements heavier than lithium. \citeauthor{Stenger:00}
simulated different universes in which fundamental parameters differ.
In particular, he varied $\alpha$ (the fine-structure constant),
$\alpha_{s}$ (the strong nuclear interaction strength at low energy),
$m_{e}$ (the electron mass) and $m_{p}$ (the proton mass). If one
defines the dimensionless gravitational strength as $\alpha_{G}=Gm_{p}^{2}(\hbar c)^{-1}$,
then $N_{1}=(\alpha/\alpha_{G})\mu$ and $N_{2}=\alpha\alpha_{s}\mu N_{1}$
\citep[see][and references therein]{Stenger:00}. In 100 toy universes,
\citeauthor{Stenger:00} generated each of the four parameters above
from a range of four orders of magnitude below their values in our
universe to four orders of magnitude above. For this range of parameters,
$N_{1}>10^{33}$ and $N_{2}>10^{20}$ in most cases. He noted that
although $N_{1}\sim N_{2}$ does not occur in most cases, nevertheless
an approximate coincidence between these two quantities is not rare
either. He concluded that a rather wide variation in the fundamental
constants still produces universes in which complex matter can form,
and thus perhaps life. 

It must be said that Stenger's arguments are themselves not without
criticism. \citet{Barnes:10a} gave a wide range of criticisms of
Stenger's analysis. He noted that one only needs to find a single
instance of fine-tuning for the universe to be fine-tuned. Conversely,
showing that simple toy universes are amenable to life does not prove
that the universe is fine-tuned. \citeauthor{Barnes:10a} suggested
that Stenger ignores the following requirements in his model universes:
\emph{i) }the stability of atoms; \emph{ii) }the need to be able to
form complex structures (showing that stars can exist does not show
that the complex chemical structures necessary for life are stable);
\emph{iii)} the need for suitable stars; \emph{iv)} the need for large
planets (if gravity is too strong, planets which support life will
be too small to support ecosystems), and; \emph{iv) }other constraints
on the masses of fundamental particles. \citeauthor{Barnes:10a} also
noted that the choice of priors on the parameters necessarily leads
to the conclusion of long-lived stars in half of the universes considered.
Another objection is simply that the analysis is too simplistic for
the strength of the claim made; the detailed work of \citet[and other references above]{Oberhummer:00a}
seems to clearly show the constraints placed on the strength of electromagnetism
and the strong force. \citet{Hogan:00a} considered constraints on
fundamental particles necessary for complex structure, and noted that
the difference between the mass of the up and down quarks appears
quite finely tuned.

Ultimately, whether our universe is fine-tuned for life is difficult
to resolve. However, if one accepts that our universe is fine-tuned
for life, then arguments such as that of \citet{Okun:96a} provide
an escape from invoking the existence of a creator; variations of
the fundamental constants can help defuse what seems otherwise to
be a rather special situation that we find ourselves in. Of course,
perhaps if we understood physics better (and thus understood the true
origin of the physical constants, if such an understanding is possible)
then this problem might be solved in any event. We return to consideration
of the triple $\alpha$ process in light of the results of chapter
\ref{cha:alpha} in section \ref{sub:size of habitable universe}.

\subsubsection{The ``lithium problem''\label{sub:lithium_problem}}

A review of this problem has been recently presented by \citet{Berengut:10a},
so we present the issues in summary here.\index{lithium problem, the}

Big Bang nucleosynthesis (BBN) theory attempts to predict the observed
abundances of elements from fundamental physics. BBN theory, coupled
with precise measurements of the neutron half-life and the WMAP measurements
of the baryon-to-photon ratio, $\eta$, have made the theory essentially
parameter free \citep{Berengut:10a,ParticleDataGroup:08,Cyburt:08}.
There is excellent agreement between the predicted abundances of deuterium
and $^{4}$He, however BBN overpredicts the abundance of $^{7}$Li
\citep{ParticleDataGroup:08}. This is known as the ``lithium problem''.
BBN overpredicts the amount of $^{7}$Li produced by a factor of between
$2.4$ and $4.3$ compared with observation \citep{Cyburt:08}. This
difference is significant at the $4$ to $5\sigma$ level. The abundance
of $^{7}$Li is determined from metal-poor population II stars in
our galaxy \citep{Asplund:06a,Bonifacio:07a,Hosford:09a}. It is noted
observationally that the lithium abundance does not vary particularly
over many orders of metallicity in the stars considered (an effect
known as the Spite plateau) \citep{Spite:82a}.

The rates of reactions which produce $^{7}$Li are sensitive to the
value of certain fundamental constants or derivatives thereof. In
particular, the predictions of BBN are sensitive to the deuterium
binding energy, $B_{d}$, as this determines the temperature at which
deuterium is subject to photo-disintegration and therefore the time
at which nucleosynthesis begins \citep{Dmitriev:04}. \citet{Dmitriev:04}
varied $B_{d}$ to minimise the $^{7}$Li discrepancy, and found that
$\Delta B_{d}/B_{d}=(-0.019\pm0.005)$ --- possible evidence for variation
of $B_{d}$. 

\citet{Flambaum:07b} and \citet{Berengut:08} considered the effect
of variation of $X_{q}\equiv m_{q}/\Lambda_{\mathrm{QCD}}$, where
$m_{q}$ is the light quark mass and $\Lambda_{\mathrm{QCD}}$ is
the pole in the running strong-coupling constant. They parameterise
$\Delta X_{q}/X_{q}=\Delta m_{q}/m_{q}$ (this does not assume that
$\Lambda_{\mathrm{QCD}}$ is constant but instead assumes that all
dimensions are in units of $\Lambda_{\mathrm{QCD}}$). They found
that allowing for variation of $m_{q}$ can resolve the discrepancy
between predicted and observed $^{7}$Li abundances, but only if one
ignores the shift in the resonances resonances for certain reactions.
In particular, they examined the effect of a variation in $X_{q}$
on the reactions $^{3}\mathrm{He}(d,p)\leftrightarrow\mathrm{^{4}He}$
and $t(d,n)\leftrightarrow\mathrm{^{4}He}$, where the reaction cross-section
is dominated by a narrow resonance. Including these resonances leads
to the conclusions that variations in $X_{q}$ may not be able to
explain the lithium problem. However, they noted that they have not
considered the effect of the $^{5}$He$^{*}$ and $^{5}$Li$^{*}$
resonances, which may be very sensitive to $\Delta X_{q}/X_{q}$,
and leave this consideration for future work.

Thus, although it seems like variation of fundamental constants might
be able to resolve the discrepancy between BBN theory and the observed
$^{7}$Li abundance, more work is clearly needed.

\section{How to find variation in a constant}

Most methods of searching for variations in a dimensionless constant
share the same fundamental derivation. For a dimensionless constant,
$P$, and observable quantity, $O$, one attempts to derive a change
in the observed quantity as a function of a change in a relevant dimensionless
ratio
\begin{equation}
\Delta O=k\frac{\Delta P}{P}+\mathcal{O}\left(\frac{\Delta^{2}P}{P^{2}}\right),
\end{equation}
where $k$ determines the sensitivity to the effect; for a particular
circumstance $k$ is referred to as the ``sensitivity coefficient''\index{sensitivity coefficient}.
The second order term can be neglected in almost all circumstances
as the variations in the fundamental constants, if they occur, are
small in all regimes in which can be currently be probed, although
clearly if exact expressions are available they should be used (i.e.\ $\Delta O=k\times f[\Delta P/P]$).
In many circumstances, multiple dimensionless constants are relevant
to the problem, in which case this becomes a sum over the constants
of interest,
\begin{equation}
\Delta O=\sum_{i}k_{i}\frac{\Delta P_{i}}{P_{i}}.
\end{equation}
One then compares the observations of $O$ at different time periods
to probe temporal evolution in the various $P_{i}$, or at different
places to probe spatial variation in $P_{i}$. Some care is required
in disentangling the effects, as observations to large distances necessarily
entail observations to the deep past due to the finite speed of light. 

As we search for a variation in $\mu$ and $\alpha$, we define the
quantity
\begin{equation}
\frac{\Delta\mu}{\mu}\equiv\frac{\mu_{z}-\mu_{0}}{\mu_{0}},
\end{equation}
where $\mu_{z}$ is the measurement of $\mu$ at some redshift $z$,
and $\mu_{0}$ is the laboratory value. Similarly, we define
\begin{equation}
\frac{\Delta\alpha}{\alpha}\equiv\frac{\alpha_{z}-\alpha_{0}}{\alpha_{0}}.
\end{equation}

No variation of a fundamental constant has been conclusively accepted
at present, and thus the goal of experimentation is to obtain higher
accuracy and precision. For temporal evolution, there are two primary
paths available. One is to use long temporal base-lines, with the
hope that small changes will become magnified. This path leads directly
to astrophysical observations, which can probe effectively the entire
age of the universe by looking to sufficiently high redshift. Certain
aspects of the solar system also carry the integrated history of the
physics they have been subjected to, which allows the probing of about
$\sim5$ billion years into the past. Although legitimate, astrophysical
methods suffer from the fact that one can only observe the past, not
experiment on it, and therefore controlling systematic errors may
be difficult. The other path is to perform experiments over human
time scales, but attempt to obtain extreme precision (usually through
application of modern technology and human ingenuity in laboratories). 

Probing spatial variation directly is rather more difficult. This
relates to the fact that humans are confined to the solar system,
and the velocity of the solar system is small in any reasonable reference
frame (particularly the cosmic microwave background (CMB) rest frame).
Present-day tests within the solar system (which make use of the Earth's
orbit around the sun) do not probe large amounts of space relative
to the observable universe, and therefore detection of spatial variation
is difficult. This problem does not hold true for astrophysical observations
to high redshift, which can not only probe most of the temporal history
of the universe but also most of the spatial volume of the observable
universe.

\section{Theories for variation of fundamental constants}

The current investigations into whether the fundamental constants
of nature vary are limited by experiment. Because the variation of
any fundamental constant has not been conclusively demonstrated, a
cornucopia of theories and models which generate variations in the
fundamental constants have been created; there is no space to detail
most of these here. However, as experimental constraints on both present-day
and past variation of the constants have improved, the parameter space
into which these theories can fit is being steadily compressed. Unfortunately,
many theories suffer from the need to introduce parameters which translate
into the magnitude of the variation of different constants%
\footnote{This seems to replace one constant with another, however discovery
of such a mechanism might yield further insights toward fundamental
theories%
}. There is often no natural magnitude for these parameters, and therefore
these theories may not be easily falsifiable (or may not be falsifiable
at all) --- experimental constraints simply keep diminishing the magnitude
of the parameters. 

Nevertheless, there are two important conclusions to draw from the
theoretical approaches to generating a variation in the fundamental
constants. Firstly, it is possible to construct theories as extensions
to existing physics which allows for the variation of some or all
of the fundamental constants. This is important, because it lends
plausibility to the idea that the constants might vary, therefore
lending weight to experimental searches. Secondly, some theories make
falsifiable predictions. This is particularly important for the case
of string theories (and their cohorts), which have held out hope as
a post-standard model framework, even if they have not yet yielded
the revolution that has been hoped for. In particular, many string-type
models make predictions as to how the fundamental constants should
vary in relation to each other. Thus, constraints on several different
constants might be used to constrain or falsify string theories. Given
the intellectual effort expended on and general lack of accessible
laboratory tests for string theories, the potential for investigations
into the fundamental constants to constrain string theories should
be taken seriously.

This work is an experimental one, and the proliferation of theoretical
frameworks for varying constants continues to grow. \citet{Murphy:PhD}
provides a brief overview. \citet{Uzan:03} and \citet{Uzan:10a}
provide a more wide-ranging treatment of different approaches which
might be taken. We therefore present here only a brief history of
the theoretical treatment of the variation of fundamental constants.

\subsection{Modern viewpoints}

Although people have attempted to construct theories to contrive a
variation in the fundamental constants, it seems that general efforts
towards unification of the four fundamental forces of nature often
naturally produce variation of the constants. \citet{Murphy:PhD}
notes that historical attempts are loosely divided into multidimensional
unification theories (of which the now well-known string theories
fall) and scalar field theories.

Kaluza-Klein theory\index{Kaluza-Klein theory} \citep{kaluza:21,Klein:26a}
derives from the fact that the solution of a 5-dimensional extension
to GR in fact looks like the standard 4-dimensional GR plus Maxwell's
equations. This observation has motivated a large interest in attempts
to unify the fundamental constants of nature through the construction
of additional dimensions. The extra spatial dimension is proposed
to be ``compactified'' on a microscopic scale, therefore explaining
why it is not directly observed. More generally, for $N$-dimensional
extensions, the 3D gauge couplings vary as the inverse square of the
mean scale of the extra dimensions. Evolution in the scale of the
extra dimensions therefore leads to variability of the observed coupling
constants in Kaluza-Klein theories, and in string theories more generally. 

\citet{Bekenstein:82a} \index{Beckenstein}proposed a self-consistent
scalar field theory incorporating a varying $\alpha$. In the limit
of constant $\alpha$, the theory reduces to Maxwell's equations.
The theory describes the evolution of a scalar field, where space-time
evolution of the scalar field produces a change in $\alpha$. Although
the impact on gravity was originally neglected, it has subsequently
been included as a modification of the theory \citep{Barrow:00a,Magueijo:00a}.
\citet{Dent:08a} considers constraints placed on coupling between
a scalar field and particular constants in light of recent data.

\subsection{Relationships between variation in different constants\label{sub:mu_alpha_relation}}

As noted earlier, within the framework of Grand Unified Theories (GUTs)\index{Grand Unified Theories},
and also string theories\index{string theory}, one can derive approximate
relationships between changes in different constants. For instance,
one obtains
\begin{equation}
\frac{\Delta\mu}{\mu}=R\frac{\Delta\alpha}{\alpha}.\label{eq:mu propto alpha relationship}
\end{equation}
The sign of $R$ may differ depending on the derivation, but researchers
typically report $|R|$ of between 30 and 40 \citep{Calmet:Fritsch:2001-1,Calmet:2002-1,Langacker:2002-1,Dent:08a}
for both GUTs and string theories. \citet{Dent:08a} noted that quite
a wide range of proportionality constants can be obtained, however.
In the circumstance where no variation in $\mu$ or $\alpha$ has
been seen, then this relationship is of no practical use. However,
in the event that variation is seen in either $\mu$ or $\alpha$,
one can then use this relationship to potentially falsify an apparently
quite wide class of theories. Nevertheless, it also seems possible
to generate smaller values of $R$, although this requires fine-tuning
of the unification model that many consider to be unnatural \citep{Dine:03a}.
We return to this proportionality later.

\subsection{Mach's principle}

An interesting argument has recently emerged based on Mach's principle\index{Mach's principle}.
Mach's principle asserts that the local laws of physics are somehow
due to non-local interaction with all the other matter in the universe.
This has been postulated to explain why a unique frame exists with
zero angular momentum (``if something is rotating, what can it be
rotating with respect to other than the rest of the universe?'').
\citet{Gogberashvili:10a} considered a simple Machian model in which
they estimate the gravitational energy of baryons and the electromagnetic
energy of radiation. They identified the total Machian energy of all
particles with that of dark energy, and concluded that the fine-structure
can be defined in terms of cosmological parameters by
\begin{equation}
\alpha=4\pi\Omega_{\Lambda}^{2}\frac{\Omega_{r}}{\Omega_{b}},
\end{equation}
where $\Omega_{\Lambda}$ is the dark energy density, $\Omega_{b}$
is the baryonic density and $\Omega_{r}$ is the radiation density,
all expressed as ratios of the critical density. Using cosmological
data and associated errors, they concluded that $\alpha\approx(7.5\pm0.4)\times10^{-3}$,
which is $\alpha^{-1}\approx133\pm7$ --- surprisingly close to the
present value of $\alpha^{-1}\approx137$. At the time of writing,
this work was unrefereed and so we are unsure of its import. Nevertheless,
Mach's principle has remained appealing to many, even if it is rather
loosely defined, and the serious return of Machian arguments would
be an interesting turn for physics.

\section{Structure of this work\label{sec:Structure of work}}

The goal of this work has been to investigate the variation of two
important fundamental constants using quasar absorption lines. This
work is divided into four primary sections:
\begin{enumerate}
\item In chapter \ref{cha:mu}, we investigate possible changes in the proton-to-electron
mass ratio, $\mu\equiv m_{p}/m_{e}$, using UV molecular hydrogen
transitions at high redshift. 
\item In chapter \ref{cha:alpha}, we use redshifted metal line absorption
in quasar spectra to investigate the possibility that the fine-structure
constant, $\alpha\equiv e^{2}/(4\pi\epsilon_{0}\hbar c)$, has changed.
Both of these chapters make use of data obtained with the Ultraviolet
and Visual Echelle Spectrograph (UVES), mounted on the Very Large
Telescope (VLT), in Chile. In chapter \ref{cha:da systematic errors},
we consider potential systematic errors for our analysis of $\Delta\alpha/\alpha$.
\item In chapter \ref{cha:Discussion}, we consider the $\mu$ and $\alpha$
results in the context of each other, and in the context of constraints
from other methods.
\item A critical concern when modelling quasar absorption lines is whether
the optimisation algorithm used to fit the models to the spectral
data has converged, and whether it returns sensible errors. Although
the reliability of \textsc{vpfit} (the program we use) has been confirmed
through simulations, recent publications draw attention to the need
to ensure that error estimates are accurate. Moreover, it would be
useful to confirm in specific cases that \textsc{vpfit} produces appropriate
parameter estimates and uncertainties, rather than relying on ensemble
results from synthetic spectra. Thus, in chapter \ref{cha:MCMC} we
apply Markov Chain Monte Carlo (MCMC) methods to confirm both that
VPFIT does indeed converge and that the uncertainty estimates it provides
are reasonable. 
\end{enumerate}
Some of the methods and methodology are common to the analysis of
both $\mu$ and $\alpha$. We discuss these in chapter \ref{cha:Common methodology}.

Finally, in chapter \ref{cha:Conclusions} we present our conclusions.
Below, we give non-quasar constraints on $\Delta\mu/\mu$ and $\Delta\alpha/\alpha$.
We give quasar constraints on $\Delta\mu/\mu$ and $\Delta\alpha/\alpha$
in chapters \ref{cha:mu} and \ref{cha:alpha} respectively.

\section{Non-quasar constraints on $\Delta\mu/\mu$ and $\Delta\alpha/\alpha$}

The most precise present-day bounds on variation of $\mu$ and $\alpha$
derive from atomic clocks. This method relies on the fact that different
transitions have different sensitivities to a variation in $\mu$
and $\alpha$. By comparing two clocks which use transitions with
significantly different sensitivities to a change in $\mu$ or $\alpha$,
one can derive strong bounds on the present-day rate of change of
these constants. Improved precision is obtained by utilising transitions
with sensitivity coefficients of greater difference, by running the
experiment for longer, or by building more precise atomic clocks. 

Atomic clock measurements only constrain the present time rate-of-change
of fundamental constants. To fully investigate the universe, we must
turn to observations of the Solar System and elsewhere. The cosmic
microwave background (CMB) allows us to derive constraints on $\Delta\mu/\mu$
and $\Delta\alpha/\alpha$ at $z\sim1100$. Big Bang nucleosynthesis,
noted earlier, allows us to probe the first minutes after the Big
Bang. Ultimately all of these avenues are of interest, because they
allow us to probe most of the history of the universe, albeit with
differing sensitivities. Very loosely, we can probe the temporal evolution
of certain constants extremely well at the present day (i.e.\ a fractional
change at the $\sim10^{-14}$--$10^{-17}$ per year level at $z=0$),
reasonably well through to redshifts of a few (i.e.\ at the $\sim10^{-6}$
level) and at the $10^{-2}$ level at the CMB era ($z\sim1100$).

\subsection{The proton-to-electron mass ratio, $\mu$}

\subsubsection{Atomic clocks}

\index{atomic clocks}A strong present-day direct bound on variation
of $\mu$ is obtained through comparison of the molecular transitions
in SF$_{6}$ to the Cs standard, yielding $\dot{\mu}/\mu=(-3.8\pm5.6)\times10^{-14}\,\mathrm{yr}^{-1}$
\citep{Shelkovnikov:08a}. The combination of a series of atomic clock
experiments from Sr, Hg$^{+}$ , Yb$^{+}$ and a H maser yields $\Delta\mu/\mu=(-1.6\pm1.7)\times10^{-15}$
per year \citep{Blatt:08a}%
\footnote{This work uses $\mu\equiv m_{e}/m_{p}$ and therefore a sign reversal
is required.%
}. \citet{Blatt:08a} similarly conclude that there is no coupling
of $\alpha$, $\mu$ and the light quark mass to the gravitational
potential at the present level of accuracy. \citet{Salumbides:PhD}
noted that use of Sr$_{2}$ transitions and the inversion transitions
of NH$_{3}$ may be able to probe variation in $\mu$ at the level
of $10^{-15}$ per year in the near future.

\citet{Shaw:10a} used the fact that the ratio of optical to Cs frequencies
are sensitive to changes in $\mu$, although there is a degeneracy
with $\alpha$. They combine the Yb$^{+}$ measurements of \citet{Peik:04a},
and other data to conclude that the gravitational coupling between
$\mu$ and gravity is $k_{\mu}=(3.9\pm3.1)\times10^{-6}$.

\subsubsection{Galactic ammonia}

\citet{Molaro:09a} investigated potential variation of $\mu$ within
the Milky Way by searching for radial velocity offsets between the
inversion transitions of NH$_{3}$\index{ammonia} --- which are sensitive
to a change in $\mu$ --- and control molecules CCS and N$_{2}$H$^{+}$,
and concluded that $|\Delta\mu/\mu|<10^{-7}$. However, they noted
positive velocity shifts between the line centres of NH$_{3}$ and
the two other molecules and noted that if this is was due to a change
in $\mu$ it would imply a spatial variation of $\Delta\mu/\mu\sim4\times10^{-8}$.
They also noted that this would conflict with atomic clock experiments
by about five orders of magnitude, thereby requiring chameleon-type
theories, in which the values of the constants have a dependence on
the local matter density. However, we note that the observations are
of emission lines. In emission, there are significant optical depth
effects, where the transitions may arise from significantly different
places both in the radial direction as well as in the spatial direction.
The beam size for the Green Bank Telescope (GBT) for their observations
corresponds to 0.04pc at the distance of the Perseus cloud observed.
Due to the significant potential systematics intrinsic to emission
observations, we would be extremely cautious about interpreting the
results of \citeauthor{Molaro:09a} as evidence for spatial variation
in $\mu$.

\subsection{The fine-structure constant, $\alpha$}

\subsubsection{Atomic clocks}

The combination of a series of atomic clock\index{atomic clocks}
experiments from Sr, Hg$^{+}$, Yb$^{+}$ and a H maser yields $\dot{\alpha}/\alpha=(-3.3\pm3.0)\times10^{-16}\,\mathrm{yr}^{-1}$
\citep{Blatt:08a}; \citeauthor{Blatt:08a} also concluded that $\alpha$,
$\mu$ and the light quark mass do not couple to the local gravitational
potential at the current experimental limit. The experiment of \citet{Rosenband:08}
compared the ratio of single-ion Al$^{+}$ and Hg$^{+}$ optical clocks
to conclude that $\dot{\alpha}/\alpha=(-5.3\pm7.9)\times10^{-17}\,\mathrm{yr}^{-1}$
--- an extremely precise constraint.

Dysprosium displays two nearly degenerate energy levels of differing
sensitivity to $\Delta\alpha/\alpha$; the resonance enhances the
sensitivity coefficients of the transitions. \citet{Cingoz:07a} utilised
these transitions to find that $\Delta\alpha/\alpha=(-2.7\pm2.6)\times10^{-15}$
per year.

\citet{Shaw:10a} searched for annual variation in the results of
\citet{Rosenband:08} to examine the coupling constant between $\alpha$
and gravity, $k_{\alpha}$, and concluded that $k_{\alpha}=(-5.4\pm5.1)\times10^{-8}$.

\subsubsection{Direct solar system observations}

\citet{Iorio:10} considers the effect of a varying speed of light%
\footnote{It is assumed that $e$ and $h$ are held constant.%
} on the precession of the perihelion of the orbits of various inner
solar system planets, and concludes that $\dot{c}/c=(0.5\pm2)\times10^{-7}\,\mathrm{yr}^{-1}$
over the past century based on astronomical observations.

In synchrotron accelerators, when electrons scatter off a laser beam
whilst in flight they emit a spectrum of radiation. The lower edge
of the spectrum, the Compton Edge (CE), depends on the velocity of
light. \citet{Gurzadyan:10a} used measurements of the CE in the GRAAL
beam-line at the European Synchrotron Radiation Facility (ESRF) in
Grenoble to constrain velocity anisotropy in the speed of light, $\Delta c/c$.
Using data from 2008, they constrain isotropy in the velocity of light
to $\Delta c/c\lesssim10^{-14}$. If one assumes constancy of $e$
and $\hbar$ this implies%
\footnote{$\Delta\alpha/\alpha=-\Delta c/c$ if $e$ and $\hbar$ are assumed
to be fixed.%
} that $\Delta\alpha/\alpha\lesssim10^{-14}$.

\subsection{The weak equivalence principle}

The weak equivalence principle (WEP)\index{weak equivalence principle (WEP)}
states that trajectory of a free-falling body under gravity is independent
of its composition \citep{Dent:08a}. This is equivalent to requiring
that inertial and gravitational masses are identical. The Einstein
equivalence principle\index{Einstein equivalence principle (EEP)}
(EEP) is a stronger statement than the WEP. The EEP requires: \emph{i)
}that the WEP holds; \emph{ii) }that the outcome of any non-gravitational
experiment conducted in free-fall is independent of the velocity of
the experiment (local Lorentz invariance\index{local Lorentz invariance (LLI)},
or LLI); and \emph{iii) }the outcome of any non-gravitational experiment
conducted in free-fall is independent of the location and time of
the experiment (local position invariance\index{local position invariance (LPI)},
or LPI) \citep{Dent:08a}. Variation of the fundamental constants
would imply a change in the composition of the object; the mass of
nucleons is in substantial part due to coupling constants to fundamental
forces, and therefore variation of fundamental constants would change
the mass of an object, thereby causing a violation of the both the
second and third points above, and thus the EEP \citep{Salumbides:06}.
The strong equivalence principle\index{strong equivalence principle}
(SEP) says that the outcome of any experiment (gravitational or not)
in a free-falling reference frame is independent of the position in
space-time. Violation of the strong equivalence principle would manifest
as a fifth force \citep{Dent:08a}.

The WEP has been stringently tested by both Eötvös-type torsion balance
experiments \citep{Schlamminger:08a} and the measurement of free-fall
of the Moon via lunar laser ranging experiments \citep{Williams:04a}.
These measure the Eötvös parameter 
\begin{equation}
\eta=2\frac{r_{a}-r_{b}}{r_{a}+r_{b}},
\end{equation}
where $r_{a}$ and $r_{b}$ are the ratios of the gravitational mass
to the inertial mass of particles $a$ and $b$ respectively. Both
experiments given above yield constraints on violation of the equivalence
principle at the $10^{-13}$ level. \citet{Schlamminger:08a} calculated
that space-fixed differential accelerations in any direction are limited
to less than $8.8\times10^{-15}\,\mathrm{ms^{-2}}$ at the 95\% confidence
level. \citet{Tobar:10a} compared various hydrogen masers to a cryogenic
sapphire oscillator for sidereal and annual modulations of the oscillator
frequency, and constrain both (and thus LLI and LPI violation) at
the few parts in 10$^{8}$ level.

\subsection{The Oklo natural nuclear reactor\label{sub:Oklo}}

\index{Oklo natural nuclear reactor}It was discovered in the 1970s
that a uranium deposit at Oklo, in Gabon, showed depletion of $^{235}$U
relative to the natural abundance, as well as anomalies in the abundance
of isotopes of other elements. The observed abundances are explained
by the operation of a water-moderated natural nuclear fission reactor
about 1.8 billion years ago \citep{Naudet:74a,Maurette:76}. This
effect was made possible by the relatively higher isotopic abundance
of $^{235}$U then (about 3.7\%) compared to today (about 0.72\%). 

The production of $^{149}$Sm by neutron capture depends strongly
on a $\approx0.1$eV resonance. Thus, the $^{149}$Sm/$^{150}$Sm
abundance ratio today constraints variation in fundamental constants
at the time of operation of the reactor. \citet{Shlyakhter:76} estimated
the shift of the resonance due to variation in $\alpha$. \citet{Damour:96a}
claimed that the measured abundance ratio leads to the constraint
$-0.9\times10^{-7}<\Delta\alpha/\alpha<1.2\times10^{-7}$. \citet{Fujii:00a}
used new samples from Oklo to find $\Delta\alpha/\alpha=(-0.04\pm0.15)\times10^{-7}$.
\citet{Gould:06a} claim $-0.11\times10^{-7}<\Delta\alpha/\alpha<0.24\times10^{-7}$.
\citet{Petrov:06a} give $-0.56\times10^{-7}<\Delta\alpha/\alpha<0.66\times10^{-7}$. 

However, \citet{Flambaum:09a} noted that the shift of the resonance
induced is
\begin{equation}
\Delta E\approx10\left(\frac{\Delta X_{q}}{X_{q}}-0.1\frac{\Delta\alpha}{\alpha}\right)\,\,\mathrm{MeV},
\end{equation}
where $X_{q}\equiv m_{q}/\Lambda_{\mathrm{QCD}}$ and $m_{q}$ is
the light quark mass. As such, the shift in the resonance is dominated
by the first term, and so the Oklo reactor measurements cannot give
any constraint on $\Delta\alpha/\alpha$ without the wholly unjustified
assumption that $\Delta X_{q}/X_{q}=0$ \citep{Flambaum:09b}. \citet{Flambaum:09b}
used the findings that $|\Delta E|<0.1\mathrm{eV}$ \citep{Fujii:00a,Gould:06a,Petrov:06a}
to give the constraint
\begin{equation}
\left|\frac{\Delta X_{q}}{X_{q}}-0.1\frac{\Delta\alpha}{\alpha}\right|\lesssim4\times10^{-9}.
\end{equation}
If one assumes linear temporal variation this leads to 
\begin{equation}
\left|\frac{\dot{X}_{q}}{X_{q}}\right|<2.2\times10^{-18}\,\mathrm{yr}^{-1}.
\end{equation}

\subsection{Cosmic microwave background (CMB)}

The CMB\index{cosmic microwave background (CMB)} provides an investigation
of the variability of the fundamental constants at very high redshift,
$z\sim1100$. This gives the longest practical baseline over which
the fundamental constants can be examined via electromagnetic radiation
at present, as the universe is opaque to light at higher redshifts.
Neutrinos can in principle be used to probe higher redshifts, but
this remains well beyond practical examination at present. The CMB
is sensitive to the variation of $\alpha$, as the strength of the
electromagnetic force affects the Thomson scattering cross-section
and the ionisation fraction \citep{Salumbides:06,Uzan:03}. Increasing
$\alpha$, for instance, increases the amount of power at small scales
in the CMB power spectrum \citep{Kaplinghat:99a,Hannestad:99a}. The
constraints are unfortunately only at the $10^{-2}$ level, although
this may improve with time. The situation is not assisted by strong
degeneracies between different parameters.

\citet{Landau:10} used 7 year WMAP data and a model-free%
\footnote{Here, model-free means that no specific model for variation of $\alpha$
or $m_{e}$ is considered.%
} approach to find $\Delta\alpha/\alpha=-0.014\pm0.007$ when only
$\alpha$ variation was considered, $\Delta\alpha/\alpha=-0.014\pm0.009$
and $\Delta m_{e}/m_{e}=-0.001\pm0.035$ when both $\alpha$ and $m_{e}$
were allowed to vary and $\Delta m_{e}/m_{e}=-0.036\pm0.025$ when
only $m_{e}$ was allowed to vary ($m_{p}$ was held constant for
these purposes due to a strong degeneracy with the baryon mass density
and number density). \citet{Nakashima:10a} allowed variation of $m_{p}$
and assumed that the variation in different coupling constants is
driven by a single scalar field (the dilaton), and obtained $-8.28\times10^{-3}<\Delta\alpha/\alpha<1.81\times10^{-3}$
(95\% confidence) and $0.52<\Delta\mu/\mu<0.17$ (95\% confidence)
in an analysis where both $\alpha$ and $\mu$ could vary. The substantial
increase in the error bar on $\Delta\mu/\mu$ (compared to $\Delta m_{e}/m_{e}$)
as a result of allowing $m_{p}$ to vary is clearly seen.

\subsection{Other}

We note with amusement the April Fool's Day spoof article on arXiv
claiming to detect a temporal variation in $\pi$ through examination
of historical calculated values \citep{Scherrer:09a}, and thank the
author for a good laugh. This ``result'' was widely circulated on
the Internet through popular science websites (e.g.\ \emph{New Scientist})
and blogs. Some commentators did not seem to realise the nature of
the paper. This demonstrates both that a fairly wide readership is
interested in the variation of fundamental constants, and also that
citing the results of papers without reading them can lead to much
embarrassment for those involved. At the time of writing, NASA's ADS
records no refereed citations to this article, and therefore we kindly
supply Robert Scherrer with one through this work.

\section{Quasar absorption lines\label{sec:quasar absorption lines}}

\index{quasars!absorption lines}The discovery of quasars \citep{Schmidt:63}
--- first observed as star-like radio-loud%
\footnote{It is now known that not all quasars are radio-loud.%
} objects --- rapidly led to a intense study of the absorption spectra
they generate. \citet{Schmidt:63} observed that 3C 273 (which has
an apparent magnitude of about 13, but an absolute magnitude of about
$-27$) exhibited a redshift of $0.16$, implying recession at $\sim47,000$
km/s. The mechanism through which quasars generate power for the observed
luminosity and redshift was initially unknown. In particular, the
light curves of quasars were initially observed to vary on the timescale
of years, implying that the power source must be contained within
parsec-sized regions \citep{Greenstein:64}. Known power generation
mechanisms were insufficient to explain the observed luminosity unless
the objects had lifetimes of $\lesssim10^{3}$ years \citep{Greenstein:64}.
However, \citet{Hawkins:10a} recently claimed that quasar light curves
do not show the expected time dilation, and therefore that intrinsic
variability may be due to other factors, such as microlensing. The
microlensing explanation seems difficult to support, as the required
population of compact galactic halo objects is incompatible with the
results from the MACHO project \citep{Alcock:97a,Hawkins:10a}.

Resolution of the power source conundrum came with the finding in
the 1970s that black hole accretion disks could generate sufficient
amounts of power to match observed luminosities \citep{Shakura:73}.
Although it was unknown originally how disk viscosities could be sufficiently
high to generate the requisite angular momentum transfer, it is now
clear that magnetohydrodynamical stresses are crucial \citep{Blaes:07a,Kuncic:07}.
A fit to 60 observed quasar and active galactic nuclei (AGN) spectra
indicated that the observed power law continuum is well modelled by
a geometrically thin, optically thick black hole accretion disk \citep{Sun:89}.
For the purposes of our work, the mechanics of power generation are
not relevant. Instead, we utilise the fact that quasars are the brightest
continuous sources known in the universe. Their extreme luminosities
allows observations at high redshifts, which can probe more than 90
percent of the time back to the Big Bang.

\citet{Gunn:65} and \citet{Bahcall:65} suggested that absorption
along the line of sight to high redshift objects could be detected
by optical observation of redshifted UV absorption lines caused by
intergalactic H \isc. \citet{Lynds:71} suggested that the ``forest''
of absorption lines almost exclusively blueward of the quasar Lyman-$\alpha$
emission line was due to Lyman-$\alpha$ absorption by intervening
H $\isc$\index{Lyman-alpha
 forest@Lyman-$\alpha$ forest}; this has since become known as the Lyman-$\alpha$ forest. \citet{Becker:01a}
claimed detection of a complete Gunn-Peterson trough, where zero flux
is observed, in observations of a $z=6.28$ quasar. See also \citet{Djorgovski:01,Fan:03,Fan:06}.
\citet[and references therein]{Murphy:PhD} noted that the lower column
density forest lines probably arise from ``the large-scale filamentary
and sheet-like structures in which galaxies are embedded''. He also
noted that the higher column density forest lines probably arise from
galaxy halos, or galaxies themselves. 

\begin{figure}[tbph]
\includegraphics[clip,width=1\textwidth]{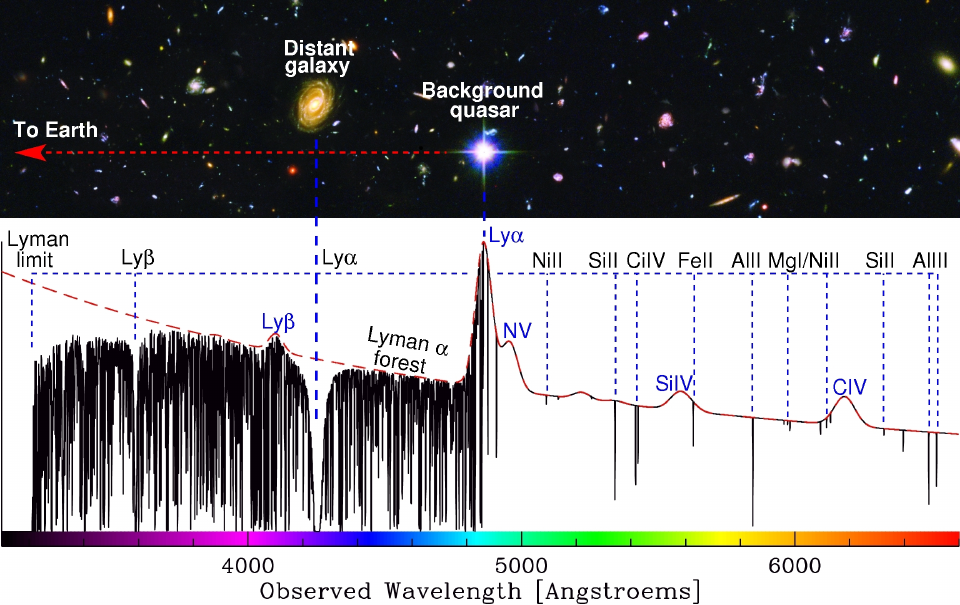}

\caption[Schematic overview of a quasar spectrum]{Schematic overview of a quasar spectrum. The emission line marked ``Ly$\alpha$'' at $\lambda \sim 4950 \AA$ is due to Lyman-$\alpha$ emission by the quasar. Bluewards of the Lyman-$\alpha$ emission peak is a dense series of absorption lines --- the Lyman-$\alpha$ forest --- caused by absorption by intervening H~\textsc{i} along the line of sight to the quasar. Clouds with sufficiently high H~\textsc{i} column density display damped wings, and are known as Damped Lyman-$\alpha$ absorbers if the H \textsc{i} column density is greater than $2\times 10^{20}$ cm$^{-2}$. Other H \textsc{i} absorbers --- Lyman limit systems (LLSs) --- still have sufficiently high column densities, of $N$(H \textsc{i}) $\gtrsim 2\times 10^{17}$ cm$^{-2}$, to cause a substantial drop in the transmitted quasar flux below the Lyman limit (at $\sim 911.8\AA$ in the rest frame of the absorber). A LLS is indicated in this system by the ``distant galaxy'' and absorption at $\lambda \sim 4250\AA$. Metal lines are often observed redwards of the Ly-$\alpha$ emission peak, indicated here by the narrow absorption lines corresponding to Ni \ii, Si \ii, C \iv, Fe \ii, Al \iis and Al \iiis (all in black text). These are due to metal line absorption along the line of sight to the quasar. Metal lines also fall in the Ly-$\alpha$ forest, but are often observed out of the forest simply because some transitions possess rest wavelengths significantly longer than the 1216$\AA$ Lyman-$\alpha$ line. These metal lines prove useful to search for a change in $\alpha$ (chapter \ref{cha:alpha}). The redshifted transitions of molecular hydrogen, which can be used to search for a change in $\mu$ (chapter \ref{cha:mu}), all possess rest wavelengths shorter than $1216\AA$, and therefore are observed only in the Lyman-$\alpha$ forest. All absorption and emission is observed with respect to an underlying power law spectrum, indicated by the dashed red line. Diagram by Michael Murphy, used with permission. \label{Flo:Schematic quasar absorption}}
\end{figure}

Quasar spectra also display metal-line absorption \citep[e.g.][]{Burbidge:66a,Stockton:66a},
which may be due to clouds either associated with the quasar host
galaxy itself or at some other (cosmological) distance along the line
of sight. Investigation of the metal absorption complexes at high
resolving powers reveals dense and complicated velocity structures.
Metals in this context refer to any element more massive than helium.
We show in figure \ref{Flo:Schematic quasar absorption} a schematic
representation of a quasar spectrum, and highlight the characteristics
of metal absorption and Lyman-$\alpha$ absorption. Although the high
redshifts of these absorbers imply they are at cosmological distances,
it is reassuring that in many cases the galaxies with which the absorbers
are associated can be identified through direct imaging \citep[see for instance][]{Zych:07a}. 

\subsection{Quasar absorption lines \& fundamental constants}

As the absorption lines displayed in the spectra of quasars occur
as a result of gas clouds at cosmological distances, they can be used
as a sensitive probe of physics at the time of absorption of the light.
Certain transitions are more sensitive to variation in one or more
fundamental constants, and it is these transitions which have been
actively targeted. A single transition cannot be used to search for
a variation in fundamental constants, because the redshift of the
absorbing gas cloud is unknown. However, the use of two or more transitions
with a differing sensitivity to a change in the constant of interest
can yield a constraint on the constants involved, as the redshift
is then no longer degenerate with a variation in the constants considered.
Metal transitions can be used to search for a change in $\alpha$,
whereas molecular transitions (and in particular, molecular hydrogen)
can be used to search for a change in $\mu$. Importantly, as will
be seen in chapters \ref{cha:mu} and \ref{cha:alpha}, the way in
which various transitions would vary if $\mu$ or $\alpha$ were different
at the time of absorption is a relatively unique fingerprint, which
is difficult to confuse with or be mimiced by some other effect.

\chapter{Common methods \& methodology\label{cha:Common methodology}}

The results of chapters \ref{cha:mu}, \ref{cha:alpha}, \ref{cha:da systematic errors}
and \ref{cha:MCMC} share much in common --- they all derive constraints
on fundamental parameters through the application of Voigt profile
fitting to quasar absorbers. Therefore, we outline here methods \&
methodology common to these chapters.

\section{General comments on Voigt profile fitting }

\subsection{Voigt profiles and VPFIT\label{sub:meth:VPFIT}}

\index{VPFIT}To fit Voigt profiles to the quasar spectra, we have
used the non-linear least squares Voigt profile fitting program \textsc{vpfit}%
\footnote{Available at \url{http://www.ast.cam.ac.uk/~rfc/vpfit.html}.%
} \citep{Webb:PhD:87}, which was specifically designed for this purpose.
A Voigt profile\index{Voigt profile} describes the observed profile
of an absorption line where the line is broadened through both Doppler
(Gaussian) and Lorentzian broadening mechanisms \citep{Armstrong:67}.
In the case of quasar spectra, the former mechanism is due to the
a combination of turbulent motions of the gas and the non-zero gas
temperature, whilst the latter is due to the finite lifetime of excited
states. Each Voigt profile for a particular transition is described
by three numbers: the redshift of the transition, $z$, the column
density, $N$, and the velocity width, $b$ (also known as the $b$-parameter).
The column density is the number of atoms per $\mathrm{cm}^{-2}$,
integrated along the line of sight. The $b$-parameter defines the
observed width of the transition (where $b=\sqrt{2}\sigma$), and
is usually specified in km/s.

\textsc{vpfit }attempts to minimise $\chi^{2}$\index{chi squared (chi^{2}
)@chi squared ($\chi^{2}$)}, where 
\begin{equation}
\chi^{2}=\sum_{i=1}^{N}\frac{\left[f(\mathbf{x})_{i}-y_{i}\right]^{2}}{\sigma_{i}^{2}}.
\end{equation}
$f(\mathbf{x})_{i}$ is the model prediction for the $i$th flux pixel
for a set of parameters $\mathbf{x}$, $y_{i}$ is the normalised
flux of the $i$th pixel and $\sigma_{i}$ is the $1\sigma$ statistical
uncertainty associated with that flux pixel. The model consists of
a series of Voigt profiles. The user must supply the number of profiles
to fit, as well as reasonable starting guesses for each of them. Clearly,
each transition must be appropriately identified, which requires identification
of the ground state, the wavelength of the transition and the atomic
mass of the species from which the transition originates. 

The optimisation proceeds iteratively until the fractional change
in $\chi^{2}$ is below some user-defined cutoff. One desires that
the change in $\chi^{2}$ should be much less than unity near the
optimisation solution \citep{NumericalRecipes:92}. We have chosen
this stopping criterion as $\Delta\chi^{2}<10^{-6}$, which fulfils
this requirement even for many thousands of degrees of freedom.  

The optimisation algorithm used by \textsc{vpfit} is described in
greater detail in section \ref{sub:Optimisation-theory}, where we
consider not only the mechanics of the algorithm but potential points
of failure. Also of interest for determining whether the model is
a good fit to the data is the normalised $\chi^{2}$, or $\chi^{2}$
per degree of freedom $\nu$, defined as $\chi_{\nu}^{2}\equiv\chi^{2}/\nu$
(see below for more on model selection). The Voigt function is non-analytic,
and therefore must be evaluated through numerical methods. A good
review of different algorithms is given by \citet{Murphy:PhD}.

As a result of the optimisation, \textsc{vpfit} provides parameter
estimates on all free parameters, as well as statistical uncertainties,
which are given by the square root of the diagonal terms of the covariance
matrix at the purported solution, multiplied by $\sqrt{\chi_{\nu}^{2}}$
for the fit. The multiplication by $\sqrt{\chi_{\nu}^{2}}$ is a first-order
correction to account for dispersion of the spectral data about the
model which is greater or less than the expected $\chi_{\nu}^{2}=1$
\citep{NumericalRecipes:92}. 

\textsc{vpfit} allows the user to link parameters which are physically
related. In particular, this means that the redshifts of components
can be tied together if they are assumed to originate from the same
location. Additionally, the $b$-parameters of transitions can be
related. The relationship imposed relates to the choice of broadening
mechanism\index{Voigt profile!line broadening mechanism}. One can
impose turbulent broadening ($b^{2}=b_{\mathrm{turb}}^{2}$), thermal
broadening ($b^{2}=2kT/M$, where $T$ is the temperature of the cloud
and $M$ is the atomic mass of the species in question) or a combination
of the two effects ($b^{2}=b_{\mathrm{turb}}^{2}+b_{\mathrm{therm}}^{2}$).
If two species of different atomic mass are fitted simultaneously,
\textsc{vpfit} can explicitly decompose the $b$-parameter into turbulent
and thermal contributions. However, in almost all cases the two contributions
are highly degenerate, leading both to very large uncertainties on
the individual contributions and poor performance of the optimisation
algorithm. For our $\alpha$ fits, we work only with the turbulent
and thermal limiting cases.

\subsection{Model selection \label{sub:Model-selection}}

\index{model selection}In fitting the quasar spectra, the objective
is to produce a model which provides a physically realistic, statistically
acceptable model of the observed absorption features. Almost all absorption
features display departures from that expected for a single Voigt
profile, thus necessitating the use of multiple Voigt profiles (``velocity
components'') to achieve a statistically acceptable fit. Unfortunately,
there is no way of knowing \emph{a priori} how many components are
required to obtain a statistically acceptable fit. The process of
modelling the observed structure amounts to adding components until
a physically realistic, statistically acceptable fit is achieved.

We have three criteria for a statistically acceptable fit:
\begin{enumerate}
\item $\chi_{\nu}^{2}\sim1$. For a statistically acceptable fit, $\chi_{\nu}^{2}$
should be of order unity. This follows from the fact that the $\chi^{2}$
distribution with $\nu$ degrees of freedom has mean $\nu$. However
this criterion is not the only one which must be used. Adding components
until $\chi_{\nu}^{2}\leq1$ only suggests that the dispersion of
the data points about the model is what one would expect for a reasonable
model. \citet{Murphy:08} demonstrate through simulations that, at
least for one synthetic spectrum considered, ``underfitting'' of
spectra may lead to significant bias in estimated values of $\Delta\alpha/\alpha$,
whereas ``overfitting'' does not seem to induce bias of the same
magnitude. We are therefore particularly cautious about underfitting
spectra. 
\item \emph{Best fit possible}. Fitting components until $\chi_{\nu}^{2}\lesssim1$
does not mean that the considered model is the best one, only that
it might be a reasonable one. $\chi^{2}$ fitting is a maximum likelihood
method, and under the maximum likelihood method one must choose whichever
model best explains the data. This means that if one can find a model
which reduces $\chi^{2}$ more than would be expected by chance, this
model should be preferred. \\
\\
A rigorous way to proceed in this fashion is to perform a statistical
significance test on every component added (for example, the $F$-test).
This process is not only laborious, but does not allow the comparison
of multiple models simultaneously. To remedy this, certain heuristics
are available which tend to lead to reasonable choices. A primary
method utilised by many practitioners is to try to find the model
which minimises $\chi_{\nu}^{2}.$ If one adds a component, and $\chi_{\nu}^{2}$
increases, this suggests that the extra component is not supported
by the data. In model selection, parsimony is valued --- one should
attempt to choose whichever model best explains the data, in the simplest
fashion.\\
\\
Other methods are available which penalise free parameters more or
less strongly. We have chosen to use the Akaike Information Criterion
(AIC) \citep{Akaike:74}, defined as $\mathrm{AIC}=\chi^{2}+2p$ where
$p$ is the number of free parameters. When comparing two models,
whichever model has the lower AIC should be preferred. The AIC is
derived by approximately minimising the Kullback-Leibler entropy\index{Kullback-Leibler entropy}
\citep{Kullback:1951}, which measures the difference between the
true distribution and the model distribution. In fact, the AIC is
only correct in the limit of large $N/p$ (where $N$ is the number
of data points fitted), which is generally not true for our fits.
Thus, we use the AIC\index{Akaike information criterion (AIC)} corrected
for finite sample sizes \citep{Sigiura:1978}, defined as
\begin{equation}
\mathrm{AICC}=\chi^{2}+2p+\frac{2p(p+1)}{(n-p-1)}.\label{eq:AICc}
\end{equation}
A significant advantage of the AICC is that it allows the comparison
of multiple models simultaneously, or two models which are not nested.
If several competing models are being considered, one chooses the
model which has the lowest AICC. The actual value of the AICC is not
important; only relative differences matter. The AICC is interpreted
according to the Jeffreys' scale \citep{Jeffreys:1961,Liddle:07}
where $\Delta\mathrm{AICC>5}$ is considered strong evidence and $\Delta\mathrm{AICC>10}$
is considered very strong evidence (this corresponds to odds ratios
of approximately 13:1 and 150:1 against the weaker model).\\
\\
Another commonly used information criterion is the Bayesian Information
Criterion (BIC)\index{Bayesian information criterion (BIC)}, introduced
by \citet{Schwarz:78}, defined as
\begin{equation}
\mathrm{BIC}=\chi^{2}+p\ln N.
\end{equation}
The BIC is obtained by approximating the Bayes factor \citep{Jeffreys:1961},
which gives the ratio of the posterior odds of one model compared
to another. For $N>8$ (i.e.\ in all practical circumstances), the
BIC penalises free parameters more strongly than the AICC. \citet{Liddle:07}
provides a good summary of the AIC, BIC and other information criteria.
Unfortunately, there is no easy decision as to which criteria is better.
\citet{Burnham:02} prefer the AIC, but note that the BIC is justified
whenever the complexity of the model does not increase with the size
of the data set. This is\emph{ not} true in the case of quasar absorption
line fitting --- although one can increase the statistical precision
of the data through longer observations, a combination of seeing and
the light collecting ability of the telescope limits the practical
resolving power. This means that the number of pixels which sample
an absorption feature of interest is limited. Moreover, although the
density of fitted components varies somewhat depending on the situation
under consideration, in general the model complexity scales roughly
with the amount of spectral data fitted. For these reasons, we use
the $\mathrm{AICC}$.
\item \emph{No long range correlation of residuals. }When fitting, one must
consider the degree of correlation of the normalised (standardised)
residuals\index{residuals, statistical}, $r_{i}$, of the fit (where
$r_{i}=[\mathrm{data}-\mathrm{model}]/\mathrm{error}$). It is clearly
possible to achieve $\chi_{\nu}^{2}\sim1$ and yet have long range
correlations in the residuals (i.e.\ a situation where many pixels
systematically deviate from $r=0$, over the range of a few to tens
of pixels). Despite the fact that $\chi_{\nu}^{2}\sim1$, this indicates
that the fit is unlikely to be adequate. An explicit calculation of
the chance probability can be made using the well-known Wald-Wolfowitz
runs test\index{runs test, Wald-Wolfowitz}%
\footnote{This is often known as just the ``runs test''.%
} \citep{Wald:40a}, although this is unnecessary in most cases. In
general, adding components which removes significant correlations
of the residuals also reduces the AICC, and we accept the addition
of components which decreases the AICC.
\end{enumerate}
Thus, in fitting the observed absorption profiles, we attempt to obtain
a fit which has $\chi_{\nu}^{2}\sim1$ \emph{and} the minimum AICC
possible \emph{and }no substantial correlations of the residuals.
However, we treat with caution any fitted component which seems to
improve the AICC significantly but seems physically implausible. This
is possible where unremoved spikes exist in the data (for instance,
as a result of uncleaned cosmic rays). When we fit metal lines to
search for $\Delta\alpha/\alpha$, the use of many transitions of
differing optical depths allows one, in most cases, to reliably fit
narrow lines. However, problems can emerge when fitting forest data
along with molecular hydrogen data to investigate $\Delta\mu/\mu$.
As will be seen in chapter \ref{cha:mu}, we only use the H \iscs
$\lambda1215.7\mathrm{\AA}$ transition to fit the observed structure
in the forest. By using only a single transition to fit the forest
data, it is possible to fit uncleaned noise spikes. The fact that
a noise feature has been fitted can generally be determined \emph{a
posteriori} as the component required to fit the noise has a velocity
width much smaller than the instrumental resolution (generally $b<1\,\mathrm{km\, s^{-1}}$).
Additionally, the errors on these parameters are very large (for the
$b$-parameter, many times larger than the value of $b$). 

In assessing whether any particular region of the spectrum is adequate,
there are two rough considerations: \emph{i) }are the magnitudes of
the residuals too large or too small? (this relates to the $\chi^{2}$
test); and \emph{ii) }are there long range correlations in the residuals?
(this relates to the runs test). If the RMS of the normalised residuals
is $\sim1$ and there are no long range correlations, the model is
likely to be adequate (though not necessarily optimal).

Unfortunately, there is a degree of subjectivity to Voigt profile
fitting, especially in equivocal cases where the signal-to-noise ratio
is low and/or the line widths are close to the instrumental resolution.
This is difficult to avoid simply because the Voigt profile decomposition
is not unique. 

In the case of H$_{2}$, for $\Delta\mu/\mu$, and given the large
number of transitions used to determine the H$_{2}$ structure, the
quantity of data is sufficiently high that it is extremely unlikely
one can subjectively bias $\Delta\mu/\mu$ through choice of the Voigt
profile model. In the case of $\Delta\alpha/\alpha$, one may theoretically
be able to introduce some bias into the value of $\Delta\alpha/\alpha$
for a particular absorber, although we regard this as extremely difficult
to do in practice. The response of $\Delta\alpha/\alpha$ to the addition
of components is not obvious except in the simplest of cases, and
therefore any attempt to systematically bias $\Delta\alpha/\alpha$
would not only require detailed calculations in each case, but would
probably be unable to be supported by the data in any event. The consequence
of this, and the fact that the absorption profiles differ from absorber
to absorber, means that any error introduced through a failure to
select the correct model will be random from absorber to absorber,
and therefore will average out when considering the results of a statistical
ensemble of absorbers. The only way to significantly bias $\Delta\alpha/\alpha$
over an ensemble of observers through the model selection process
is by using the numerical value of $\Delta\alpha/\alpha$ to inform
the model selection process --- clearly a very dangerous way to proceed.
We do not use the value of $\Delta\alpha/\alpha$ to guide our choice
of model, and therefore no bias should be introduced as a result of
our model selection methodology.

\section{Data pipeline problems\label{sub:Data pipeline problems}}

The \textsc{midas\index{MIDAS@\textsc{MIDAS}}} extraction routine
(part of the UVES pipeline) appears to incorrectly estimate the errors
associated with the flux data points in the base of saturated lines.
In particular, the dispersion of the flux data points is too large
to be accounted for by the statistical error. Fitting a straight line
through the base of saturated lines typically produces $\chi_{\nu}^{2}\gtrsim2$.
The problem is somewhat more noticeable in the blue end of the spectra.
Although it is difficult to determine precisely what happens in regions
of low, but non-zero flux, we believe that the errors there are also
underestimated. The effect of this is to give falsely high precision
on any quantity derived from these data points (including $\Delta\mu/\mu$
or $\Delta\alpha/\alpha$). Additionally, one cannot fit plausible
models to data involving regions of low or negligible flux; to achieve
a reasonable $\chi_{\nu}^{2}$ in these regions one would need to
fit very large numbers of unphysical components.

When fitting the H$_{2}$ spectra initially we adopted one approach
to adjusting for this problem (section \ref{sub:Correcting error arrays through an estimated functional form}
below), but for our second of Q0528$-$250 in section \ref{sub:Q0528-250 revisited}
and for $\Delta\alpha/\alpha$ we adopted a more automatic approach
(section \ref{sub:Correcting error arrays through consistency checks}
below).

\subsection{Correcting error arrays through an approximate functional form\label{sub:Correcting error arrays through an estimated functional form}}

One way to attempt to correct the problem in the base of saturated
lines is to try to approximate the functional form of the problem.
The errors in the continuum are acceptable, whereas those in the base
of lines are not, so presumably there is some monotonically increasing
function from a normalised flux of 1 to 0 which describes this behaviour.
If one knew the functional form, one could increase the error estimates,
thereby removing the problem. Investigation of the problem suggests
(R. F. Carswell, priv.\ communication) that the functional form
\begin{equation}
e_{i}\rightarrow e_{i}\left[a+b\left(1-\max\left(0,\min\left(1,\frac{d_{i}}{c_{i}}\right)\right)\right)^{s}\right]^{1/t}\label{eq:flux error corr formula}
\end{equation}
might be useful in correcting the problem, where $e_{i}$ is the error
on the $i$th normalised flux pixel, $d_{i}$ is the normalised flux
value at that point, $c_{i}$ is the value of the continuum at that
point and $a$, $b$, $s$ and $t$ are user-defined constants to
emulate the desired behaviour. R. F. Carswell suggested using $s=t=4$.
Clearly this form will be incorrect, but in the absence of any other
information a guess of this sort is all that is possible. One then
chooses $a=1$ to leave errors in the continuum unchanged and $b$
such that $(a+b)^{1/t}=f$ where $f$ is the factor by which errors
should be increased in the base of saturated lines.

\subsection{Correcting error arrays through consistency checks with the input
spectra \label{sub:Correcting error arrays through consistency checks}}

Another method of correcting this problem, and other problems arising
from inconsistencies between combined spectra, is by adjusting the
error arrays to account for the degree of inconsistency of the spectral
combination. When individual exposures are co-added to create a combined
spectrum using \textsc{uves\_popler}\index{UVES_POPLER@UVES\_POPLER},
\textsc{uves\_popler} provides a check on the concordance of the different
spectra, by calculating a value of $\chi_{\nu}^{2}$ for each flux
pixel in the combined spectrum by considering the dispersion of the
corresponding pixels in the contributing spectra about their weighted
mean. For each spectral data point in the combined spectrum, we take
a region of five pixels centred on that point, and take the median
of the $\chi_{\nu}^{2}$ values associated with those five points.
We then multiply the error estimate for that spectral point by the
square root of that median value (that is, $\sigma_{i}\rightarrow\sigma_{i}\times\sqrt{\mathrm{med}[\chi_{\nu}^{2}]}$).
This is a first-order correction to the error estimate to ensure that
the individual exposures are consistent with the weighted mean \citep{NumericalRecipes:92}.
Thus, this algorithm provides protection against under-estimation
of the errors in the base of saturated lines. 

Additionally, this algorithm also provides some protection against
other data combination problems (such as weak sky emission that differs
between exposures or improperly removed cosmic rays). However some
of these effects have non-zero expectation value (that is, they cannot
be averaged out with large numbers of exposures), and so data affected
by these processes should not be utilised. In particular, cosmic rays
always contribute excess flux, and therefore the impact of including
data affected by these cosmic rays would be lessened by our algorithm,
but the results which would be biased. In the case of $\Delta\mu$
and $\Delta\alpha/\alpha$, although this effect is random from transition
to transition and absorber to absorber (and therefore cannot systematically
bias $\Delta\mu/\mu$ or $\Delta\alpha/\alpha$ over a larger number
of systems), it is an extra source of uncertainty, which would make
our final error estimates larger than might otherwise be needed.

\chapter{$\mu$ --- the proton-to-electron mass ratio\label{cha:mu}}

\section{Introduction}

The proton-to-electron mass ratio\index{proton-to-electron mass ratio},
$\mu$, is defined simply as the proton mass divided by the electron
mass i.e.\ $\mu\equiv m_{p}/m_{e}$. Some works define $\mu\equiv m_{e}/m_{p}$,
and therefore caution is warranted in reading the literature. The
current 2006 CODATA recommended value is $\mu=1836.152\,672\,47(80)$
\citep{Mohr:08}, derived from two experiments using Penning ion traps.

\subsection{The importance of $\Delta\mu/\mu$}

In the Standard Model, the proton mass is proportional to $\Lambda_{\mathrm{QCD}}$,
where $\Lambda_{\mathrm{QCD}}$ is the value of the Landau pole in
the logarithm of the running strong coupling constant i.e.\ $\alpha_{s}\sim1/\ln(\Lambda_{\mathrm{QCD}}r/\hbar c)$,
if the direct $\sim10\%$ contribution from the quark masses is ignored
\citep{Berengut:10b}. The electron mass, $m_{e}$, is proportional
to the Higgs vacuum expectation value (vev)\index{Higgs vacuum expectation value (vev)},
$v_{H}$, if one assumes the Higgs mass mechanism \citep{Coc:07a}.
The Higgs vev determines the electroweak unification scale. Therefore
$\mu\equiv m_{p}/m_{e}$ depends on the ratio $\Lambda_{\mathrm{QCD}}/v_{H}$.
As a result, $\Delta\mu/\mu$ probes evolution in the strong force
relative to the electroweak scale. This contrasts with the fine-structure
constant, $\alpha$, which probes the strength of the electromagnetic
force.

\section{Quasar constraints}

Almost all direct%
\footnote{Constraints on $\Delta\mu/\mu$ may be obtained through other dimensionless
ratios, which are a combination of fundamental constants, typically
including $\Delta\alpha/\alpha$ and $\Delta g_{p}/g_{p}$, where
$g_{p}$ is the proton gyromagnetic ratio. However, determination
of $\Delta\mu/\mu$ then requires disentangling these combinations
of constants. We discuss these combinations of fundamental constants
in section \ref{sub:Combinations_of_constants}.%
} quasar constraints on $\Delta\mu/\mu$ rely on the examination of
molecular hydrogen transitions, although the inversion transitions
of ammonia now provides a strong test at moderate ($z<1$) redshifts.

\subsection{Molecular hydrogen (H$_{2}$) }

\index{molecular hydrogen}Most known baryonic matter in the universe
is hydrogen, found in either atomic or molecular form \citep{Combes:00a}.
Molecular hydrogen transitions fall in the far ultraviolet, and therefore
cannot be observed from the ground due to the UV cutoff caused by
atmospheric ozone. The first astrophysical observation of molecular
hydrogen was made in 1970, using a rocket-launched spectrometer, in
the spectrum of the star $\xi$ Persei \citep{Carruthers:70a}. The
column density ratio of H$_{2}$ to atomic hydrogen was found to be
approximately 1:3. The Far Ultraviolet Spectroscopic Explorer (FUSE)
satellite \citep{Moos:00a} made large numbers of observations of
molecular hydrogen routine \citep[see for example][]{Shull:00a,Rachford:02a,Tumlinson:02a,Richter:03a,Rachford:09a}.
The FUSE mission was concluded in 2007 after fine control over telescope
pointing was lost. 

The possibility of observing redshifted molecular hydrogen transitions
from the ground has been known for quite some time. \citet{Carlson:74a}
ascribed features in the spectrum of quasar 4C 05.34 to molecular
hydrogen, at a redshift of $z=2.64$. \citet{Aaronson:74a} conducted
a search for molecular hydrogen in quasar spectra, and tentatively
identified molecular hydrogen at $z=2.31$ in the spectrum of PHL
957. \citet{Levshakov:85a} tentatively identified molecular hydrogen
at $z=2.811$ toward Q0528$-$250. This identification was correct,
and Q0528$-$250 forms part of the analysis of this chapter.

\subsubsection{The sensitivity of molecular hydrogen to a change in $\mu$\label{sub:mu:H2 sensitivity}}

\citet{Thompson:1975} noted that molecular absorption by gas clouds
at high redshift along the line of sight to quasar sources might reveal
variation in $\mu$ over time, and identified molecular hydrogen (H$_{2}$)
as a possible tool. Unfortunately, serious examination of this idea
had to wait some time for the robust detection of H$_{2}$ at high
redshift. Due to the UV atmospheric cutoff, one needs to identify
H$_{2}$ absorbers at $z\gtrsim2$ in order to obtain a sufficient
number of lines in the optical region to make ground-based observations
practically useful. Indeed, only about a dozen absorbers are presently
known which contain the requisite redshifted H$_{2}$ lines in their
spectrum, and only several of these have yielded strong constraints
on $\Delta\mu/\mu$. The reason that the number of absorbers known
is small relates to the way in which H$_{2}$ is produced. H$_{2}$
is formed in cold clouds, typically via adhesion onto dust grains
\citep{Bechtold:99a}. The low temperature of the clouds means that
the clouds must be small, and so the chance of obtaining intersections
with the line-of-sight to the quasar is much smaller than for DLAs.

\citet{Foltz:88a} used the fact that the vibrational component of
the energy of a transition increases with increasing excited state
vibrational quantum number to obtain $|\Delta\mu/\mu|<2\times10^{-4}$
from the H$_{2}$ transitions in the $z=2.811$ absorber toward Q0528$-$250,
using a spectrum with resolving power $R\sim5,000$. However, \citet{Varshalovich:93a}
noted that different ro-vibrational transitions have a different dependence
on the reduced mass of the particular molecule. This led to the currently
used definition of the sensitivity coefficients, presented below.

For a molecular spectrum, there are three primary contributions to
the observed structure, all of which scale with the Rydberg energy,
but only two of which depend on $\mu$ \citep{Thompson:1975}. Firstly,
the electronic energy has no dependence on $\mu$. The vibrational
energy structure scales as $E_{\mathrm{vib}}\propto\mu^{-1/2}$, similar
to a harmonic oscillator. The rotational term scales as $E_{\mathrm{rot}}\propto\mu^{-1}$,
similar to a simple rotor. As such, the energy of a particular level
of the H$_{2}$ molecule is given by 
\begin{equation}
E\backsimeq E_{I}\left(c_{\text{elec}}+\frac{c_{\text{vib}}}{\sqrt{\mu}}+\frac{c_{\text{rot}}}{\mu}\right)
\end{equation}
for certain constants $c$ in the Born-Oppenheimer approximation (BOA)
\citep{Salumbides:PhD}. One can derive the sensitivity to a change
in $\mu$ for differing transitions using either \emph{ab initio}
methods, or within a semi-empirical approximation, yielding
\begin{equation}
\frac{\Delta\lambda_{i}}{\lambda_{i}^{0}}=K_{i}\frac{\Delta\mu}{\mu}+\mathcal{O}\left(\frac{\Delta\mu^{2}}{\mu^{2}}\right)\approx K_{i}\frac{\Delta\mu}{\mu},\label{eq:mu wl series expansion}
\end{equation}
where $\lambda_{i}$ is the wavelength of a transition under consideration,
$\lambda_{i}^{0}$ is the unperturbed value and $K_{i}$ is a sensitivity
coefficient\index{sensitivity coefficient} which determines the magnitude
and sign of the effect. The values of $K_{i}$ are defined in terms
of the derivatives of the energy or wavelength of the transition with
respect to $\mu$, as \citep{Reinhold:06-1,Salumbides:PhD} 
\begin{equation}
K_{i}=\frac{d\ln\lambda_{i}}{d\ln\mu}=\frac{\mu}{\lambda_{i}}\frac{d\lambda_{i}}{d\mu}=-\frac{\mu}{\nu_{i}}\frac{d\nu_{i}}{d\mu}=-\frac{\mu}{E_{e}-E_{g}}\left(\frac{dE_{e}}{d\mu}-\frac{dE_{g}}{d\mu}\right),
\end{equation}
where $E_{e}$ and $E_{g}$ are the energies of the excited and ground
states respectively. For useful H$_{2}$ transitions $K_{i}$ is typically
in the range $-0.02\lesssim K_{i}\lesssim0.05$. Early observations
established that $\left|\Delta\mu/\mu\right|\ll1$, thereby allowing
the use of only the first order term of equation \ref{eq:mu wl series expansion}
with good accuracy. 

We use the $K_{i}$ values from \citet{Ubachs:2007-01}, who use a
semi-empirical treatment based on the Dunham expansion of the energy
levels of the H$_{2}$ molecule. They include an adiabatic correction
to account for the contribution to the electronic energy of each state
from the nuclear mass, which scales as $\sim1/\mu$. They also account
for post-BOA effects by accounting for the term in the Hamiltonian
which relates to the interaction between the nuclear and electronic
motion, and give careful attention to the effect this has on H$_{2}$
level crossings. We show in figure \ref{Flo:mu:Ki vs lambda} the
values of $K_{i}$ for a variety of Lyman and Werner series H$_{2}$
transitions. 

\citet{Meshkov:06a} have derived $K_{i}$ values based on \emph{ab
initio} calculations of the H$_{2}$ molecule. \citet{Ubachs:2007-01}
compare their values of $K_{i}$ to those of \citeauthor{Meshkov:06a},
and note that the deviations, $\Delta K_{i}$, lie between $-2\times10^{-4}$
and $4\times10^{-4}$. Given the totally independent method of derivation,
this implies that the absolute accuracy of the $K_{i}$ values is
better than $5\times10^{-4}$. 

\begin{figure}[tbph]
\noindent \begin{centering}
\includegraphics[bb=50bp 111bp 463bp 766bp,clip,angle=-90,width=1\textwidth]{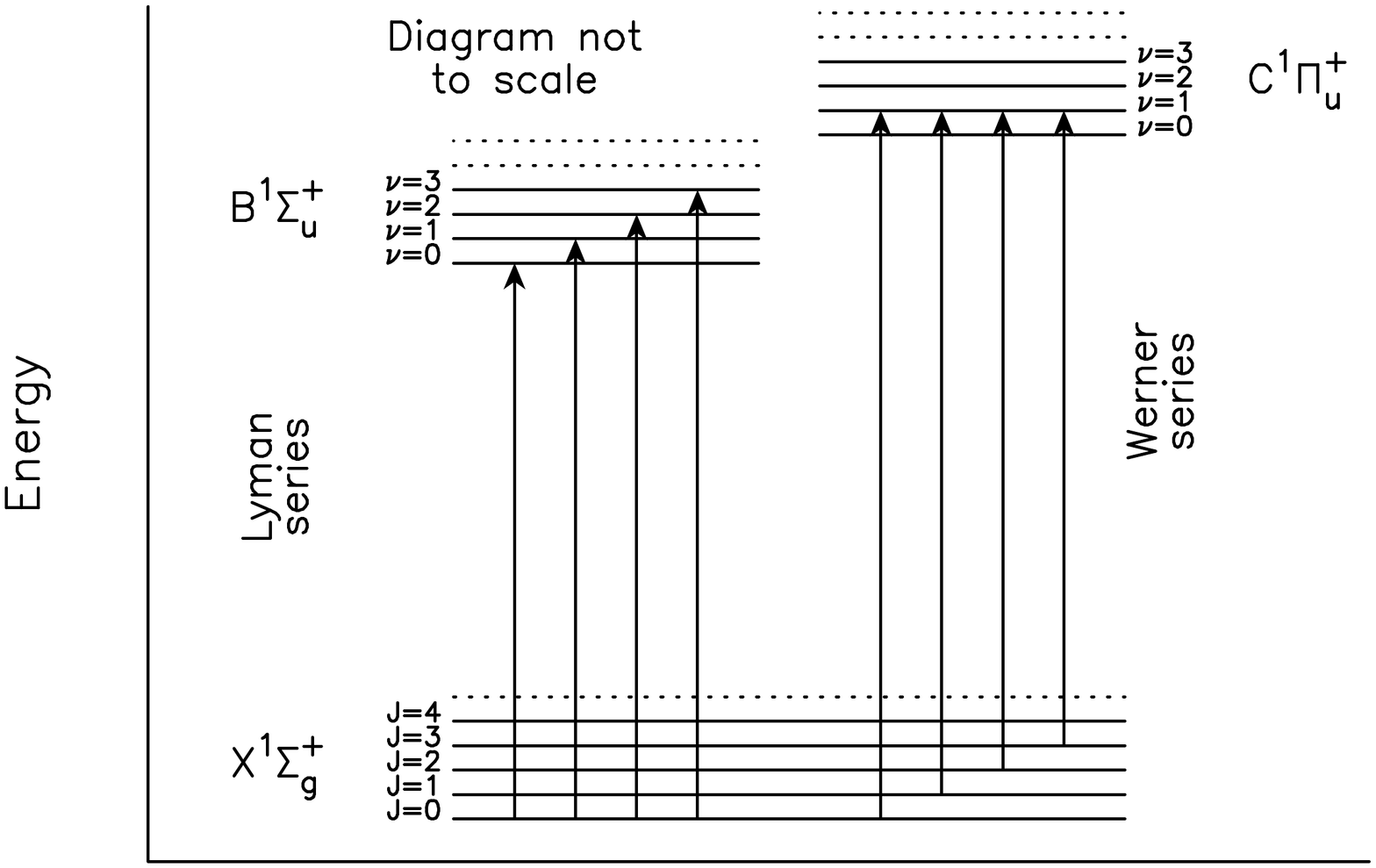}
\par\end{centering}

\caption[Schematic representation of energy levels of the H$_2$ molecule]{Schematic representation of the energy levels of the H$_2$ molecule. The left side shows the Lyman series, whilst the right side shows the Werner series. The arrows on the left side show transitions from the $J=0$ level of the ground state to various $\nu$ levels in the Lyman band. The arrows on the right side show transitions from various $J$-levels of the ground state to the $\nu=1$ vibrational level of the Werner band. Each vibrational ($\nu$) level is subdivided into states with different angular momentum (not shown). \label{Flo:mu:H2_energy_diagram}}
\end{figure}

\begin{figure}[tbph]
\noindent \begin{centering}
\includegraphics[bb=116bp 50bp 558bp 739bp,clip,angle=-90,width=1\textwidth]{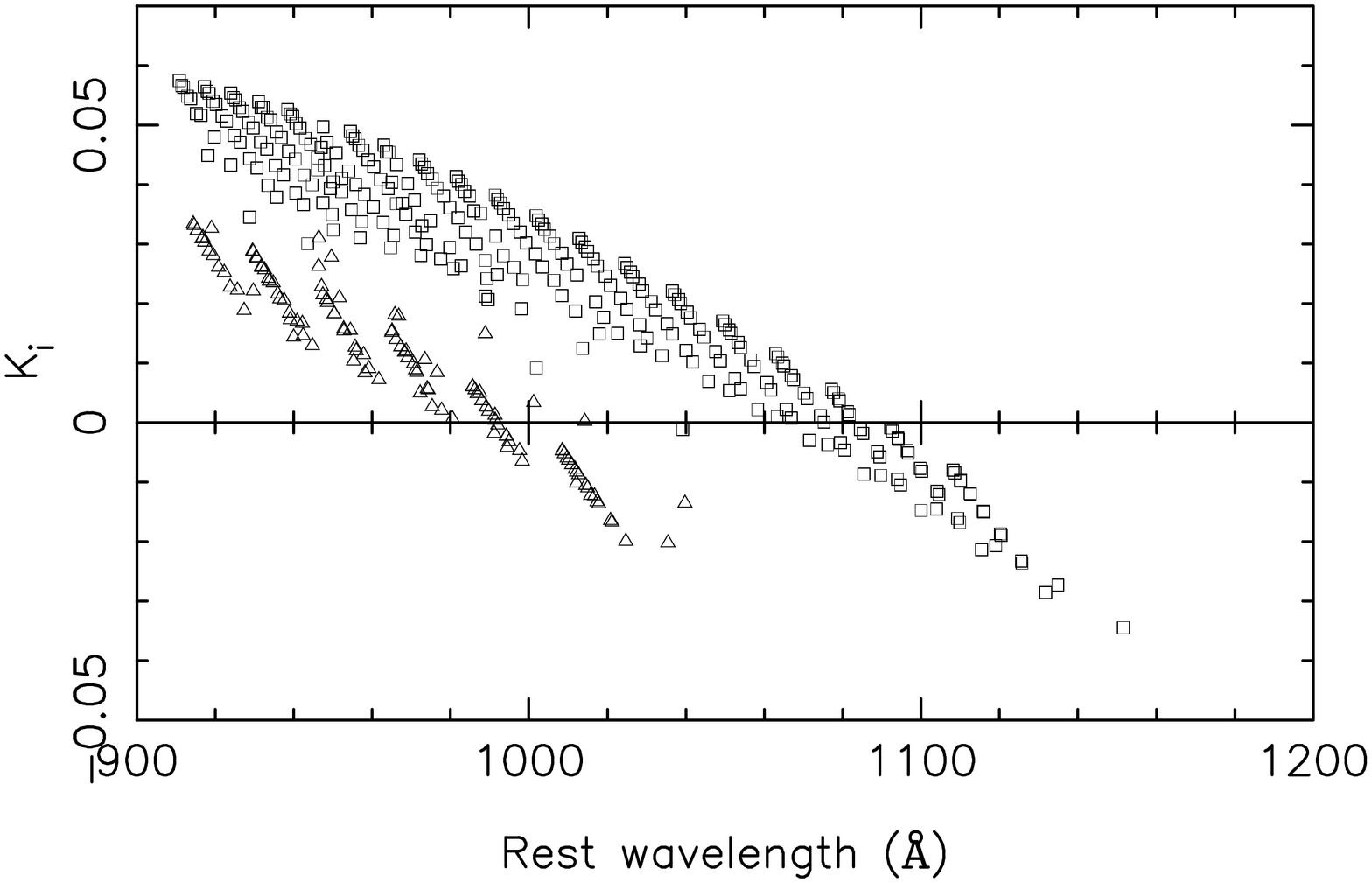}
\par\end{centering}

\caption[The $K_i$ values of certain Lyman and Werner series H$_2$ transitions]{The $K_i$ values of certain Lyman and Werner series H$_2$ transitions. Lyman series transitions are plotted as squares, whilst Werner series transitions are plotted as triangles. Note that there is a reasonable correlation of wavelength with $K_i$ when considering the Lyman series or Werner series individually. The importance of this is described in section \ref{sub:mu:Importance Werner series}. \label{Flo:mu:Ki vs lambda}}
\end{figure}

\begin{figure}[tbph]
\noindent \begin{centering}
\includegraphics[bb=89bp 40bp 568bp 757bp,clip,angle=-90,width=1\textwidth]{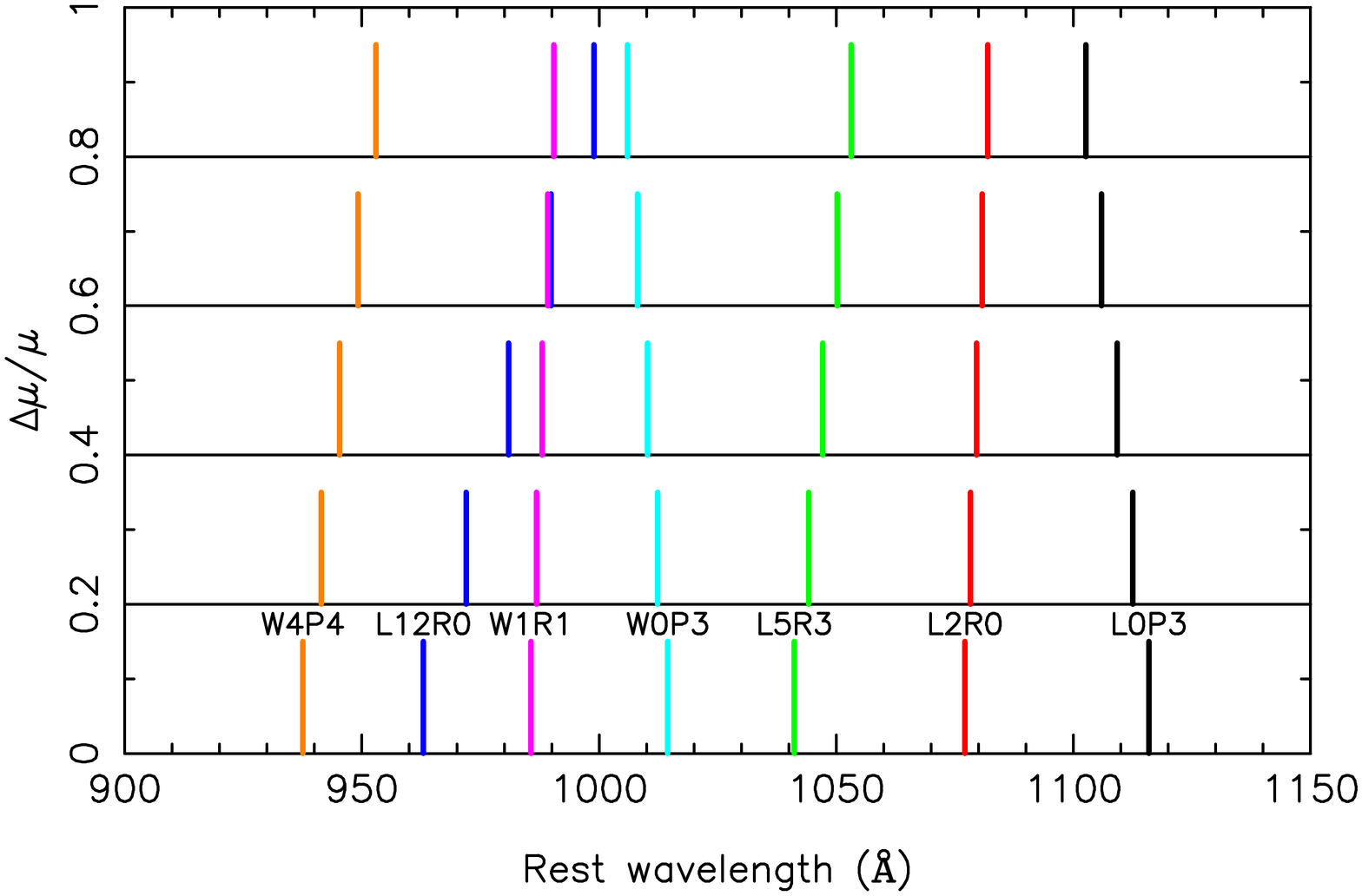}
\par\end{centering}

\caption[Exaggerated effect of $\Delta\mu/\mu$ on the position of certain H$_2$ transitions]{The exaggerated effect of $\Delta\mu/\mu$ on certain H$_2$ transitions assuming that $\lambda_i = \lambda_i^0(1 + K_i \Delta\mu/\mu)$. Although it is clear that $|\Delta\mu/\mu| \ll 1$ over most of observable time, this plot shows the effect of $\Delta\mu/\mu \sim 1$. The interpretation of the label for each transition is given in the text of section \ref{sub:mu:H2 sensitivity}. Note the presence of certain transitions which are relatively insensitive to $\mu$ variation (e.g.\ L2R0), as well as transitions which shift both to longer (W4P4, L12R0, L5R3) and shorter (W0P3, L0P3) wavelengths. Crucially, this ``fingerprint'' is rather unique, which therefore makes the measurement resistant to a wide range of systematic effects. \label{Flo:mu:mu shift plot}}
\end{figure}

Each H$_{2}$ transition is described by quantum numbers $\nu$ and
$J$, which describe the excited state vibrational quantum number
and the angular momentum of the ground state. The Lyman series is
described by the $\mathrm{X}{}^{1}\Sigma_{g}^{+}\rightarrow\mathrm{B}^{1}\Sigma_{u}^{+}$
transitions, and the Werner series by the $\mathrm{X}{}^{1}\Sigma_{g}^{+}\rightarrow\mathrm{C}^{1}\Pi_{u}$
transitions. An additional letter, P, Q or R denotes the quantity
$\Delta J=J'-J$ as $-1$, 0 and 1 respectively (where $J'$ is the
angular momentum of the excited state). A useful shorthand notation
to describe particular lines is therefore $A\nu BJ$, where A is L
or W for Lyman or Werner, B is either P, Q or R and $\nu$ and $J$
are as described previously. In figure \ref{Flo:mu:H2_energy_diagram}\index{molecular hydrogen!energy diagram}
we show a schematic representation of the Lyman and Werner series
in the H$_{2}$ molecule. The Lyman state is a $\Sigma$ state and
therefore has a total orbital angular momentum of zero, and the Werner
state is a $\Pi$ state and therefore has a total orbital angular
momentum of unity. The selection rules thus impose the following constraints
on transitions: \emph{i)} for the Lyman series, $\Delta J=\pm1$,
leading to P and R branches but no Q branch; \emph{ii) }for the Lyman
series, transitions from $J=0$ to $J'=-1$ are not possible, and
so there are no L$\nu$P0 transitions, and; \emph{iii) }for the Werner
series, there are no $J'=0$ levels, so the lowest transition in the
P branch is W$\nu$P2. We show in figure \ref{Flo:mu:mu shift plot}\index{molecular hydrogen!effect of Deltamu/mu
@effect of $\Delta\mu/\mu$} how the rest-frame wavelengths of a selection of H$_{2}$ transitions
would differ under variation of $\mu$.

The accuracy of the laboratory data for the H$_{2}$ transitions historically
meant that the laboratory errors were non-negligible. Significant
recent work has rectified this situation, such that the error budget
is now wholly dominated by non-laboratory factors. The current best
wavelengths are given in \citet{Bailly:09a} and \citet{Ubachs:2007-01},
and have been collated in \citet{Malec:10} with $K_{i}$ values,
oscillator strengths and damping coefficients.

\subsubsection{General comments on measuring $\Delta\mu/\mu$ with H$_{2}$\label{sub:mu:General H2 comments}}

It is widely acknowledged that because the H$_{2}$ transitions fall
in the Lyman-$\alpha$ forest\index{Lyman-alpha
 forest@Lyman-$\alpha$ forest} it is difficult to model the spectra. Traditionally, researchers
have discarded transitions which appear to be heavily blended with
the forest, and utilised only weakly blended transitions. An obvious
questions is: how does one decide what is weakly blended? Clearly
proceeding in this fashion introduces an element of subjectivity into
the analysis. A more appropriate way to proceed is to model the forest
explicitly, thereby allowing the uncertainty in determining the forest
structure to propagate into the uncertainty in determining $\Delta\mu/\mu$.

In order to account for the effect of the forest (which provides a
background continuum against which the H$_{2}$ absorption occurs),
in previous analyses researchers have generally fitted a low order
polynomial across the H$_{2}$ transitions to estimate the optical
depth due to the forest\index{continuum, quasar flux}. One can then
divide the flux spectrum by this polynomial continuum estimate to
obtain a H$_{2}$ profile to fit. This is particularly obvious in
\citet{Ivanchik:05a}, where many H$_{2}$ profiles are displayed
\emph{after }division by the polynomial, which masks the presence
of the forest.

There are two problems with this method:
\begin{enumerate}
\item Firstly, it appears in the literature that the uncertainty in accounting
for the local continuum does not propagate into the error in determining
$\Delta\mu/\mu$. As the forest structure is unknown, this uncertainty
should be accounted for. Any method which does not attempt to account
for the uncertainty in determining the forest structure \emph{must
}under-estimate the required uncertainty on $\Delta\mu/\mu$. Unless
one models the forest structure appropriately, one cannot tell by
how much the uncertainty is under-estimated.
\item Where it is clear that absorption is due to other gas clouds%
\footnote{Usually Lyman-$\alpha$, although metal lines are found in the forest.%
}, one is not making use of the physics that generates the absorption.
That is, one should model the absorption with a series of Voigt profiles
in order to obtain a realistic model. A polynomial continuum across
the observed H$_{2}$ profile is not constrained to any physical situation,
and therefore in principle the estimated local continuum for the H$_{2}$
transitions will be incorrect. Conversely, it must be noted that it
is often difficult to differentiate several closely-spaced forest
absorption features from an error in determining the local continuum,
and therefore there is necessarily some error introduced by an incorrect
model.\\
\\
Over a large number of molecular hydrogen transitions, one does not
expect that the use of a polynomial continuum to estimate the optical
depth of the forest in the vicinity of the H$_{2}$ transitions will
introduce a significant error into the determination of $\Delta\mu/\mu$;
the random nature of the forest structure with respect to the H$_{2}$
line profiles means that errors which bias $\Delta\mu/\mu$ to more
positive values should occur as often as those which bias $\Delta\mu/\mu$
to negative values. However, without accurately modelling the forest,
one cannot tell how legitimate this argument is or not, and what the
associated error introduced by proceeding in a more simplistic fashion
is. 
\end{enumerate}
It is for these reasons that we have modelled the Lyman-$\alpha$
forest concurrently with the H$_{2}$ transitions.

\begin{sidewaysfigure}
\noindent \begin{centering}
\includegraphics[bb=0bp 0bp 451bp 318bp,clip,width=0.65\textwidth]{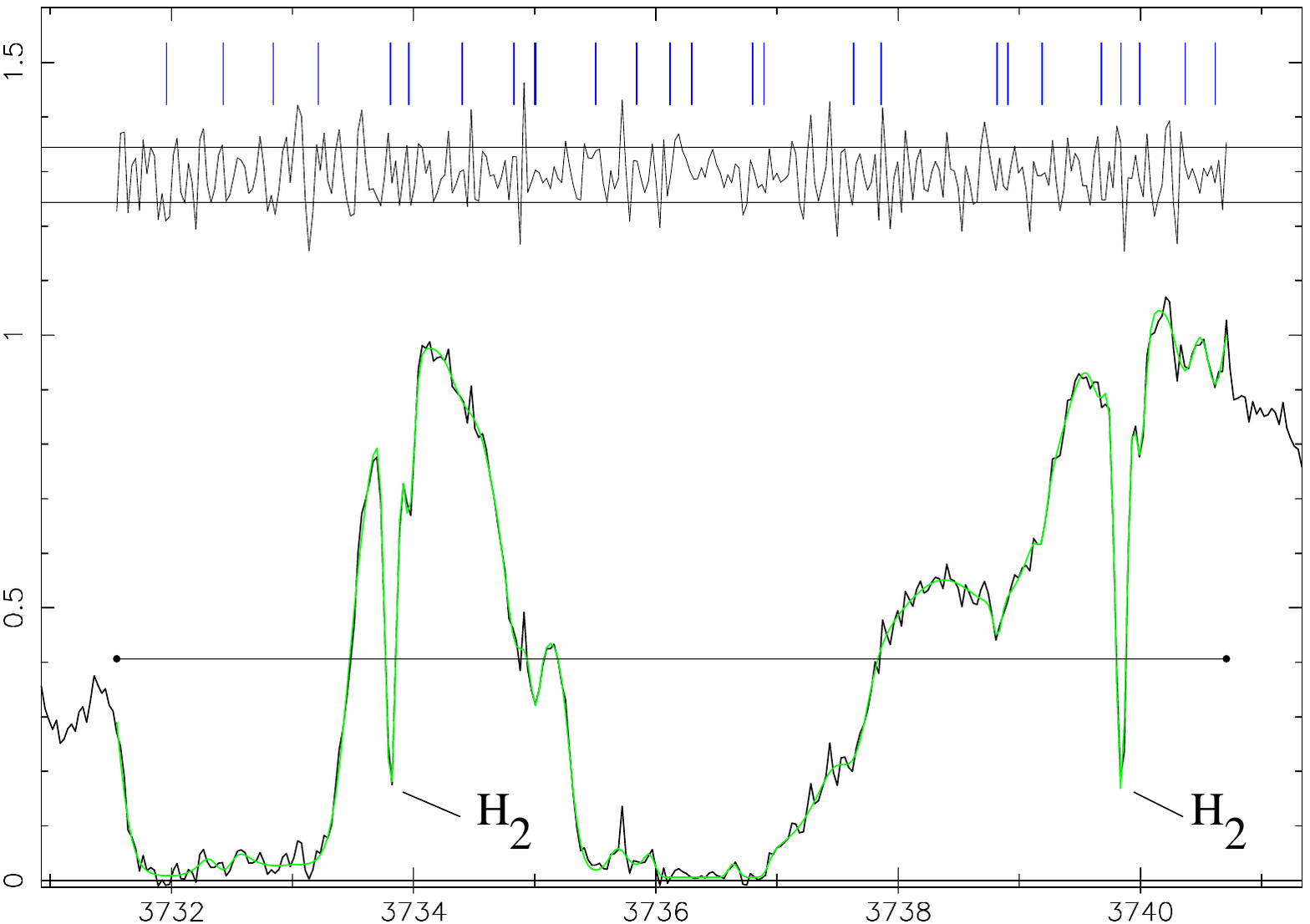}
\par\end{centering}

\caption[Example H$_2$ transitions and Ly-$\alpha$ forest structure from Q0405$-$443]{Example H$_2$ transitions and Lyman-$\alpha$ forest structure from Q0405$-$443. The horizontal axis shows wavelength in \AA, and the vertical axis shows normalised flux. The model fit to the data (green) is plotted on top of the data (black). The blue tick marks at the top indicate the position of fitted components. The normalised residuals ([data - model]/error) are drawn just below the tick marks, with the two horizontal lines indicating the $\pm 1\sigma$ range. Note that the sharp H$_2$ transitions are seen interspersed with the complicated background forest structure. In order to accurately determine the line centroids of the H$_2$ transitions, which yield $\Delta\mu/\mu$, it is necessary to simultaneously model the observed forest structure. It is unlikely that simple polynomial approximations to the local continuum for the H$_2$ transitions will produce an accurate characterisation of the absorption due to forest transitions. \label{Flo:mu:example H2 trans 1}}
\end{sidewaysfigure}

We show example H$_{2}$ transitions from Q0405$-$443 in figure \ref{Flo:mu:example H2 trans 1}.
This figure clearly shows the complexity of the forest, and the necessity
in modelling the forest simultaneously with the Lyman-$\alpha$ transitions
if one wants to ensure that the H$_{2}$ line positions are accurately
determined. 

Another concern for the measurement of $\Delta\mu/\mu$ using molecular
hydrogen is simply the paucity of known sources. In table \ref{Flo:H2 sources},
we give a list of currently known H$_{2}$ sources at sufficiently
high redshift that investigation of $\Delta\mu/\mu$ is potentially
feasible, with various references which may be of interest to the
reader. The relative lack of H$_{2}$ sources is problematic because
one cannot then use the consistency of many results to check whether
the uncertainties in individual measurements are correct. If one has
many $\Delta\mu/\mu$ results, one can use the $\chi^{2}$ test (under
some model) to determine whether the results are statistically consistent.
Inconsistency between the results is indicative of either the wrong
model or under-estimated uncertainties for the individual measurements.
More importantly, however, is that if $\mu$ varies with time and
space then many different measurements of $\Delta\mu/\mu$ in different
times and places are needed to map out the evolution of $\mu$. 

\begin{sidewaystable}
\caption[List of H$_2$ absorbers observable from the ground]{List of H$_2$ absorbers which may be observed from the ground, which are potentially useful for $\Delta\mu/\mu$. Unfortunately, some of these systems have few lines or are of low column density, reducing their utility. The column density for the absorber is given, as well as a selection of references which may be of interest to the reader. This reference list is not intended to be exhaustive.\index{molecular hydrogen!list of known absorbers at $z\gtrsim 2$}}

\label{Flo:H2 sources}

\centering{}%
\begin{longtable}{clll}
\hline 
\noalign{\vskip\doublerulesep}
Quasar name & $z_{abs}$ & $\log_{10}\left[N(\mathrm{H}_{2})/\mathrm{cm^{-2}}\right]$ & References\tabularnewline[\doublerulesep]
\hline 
\endfirsthead
\hline 
\noalign{\vskip\doublerulesep}
Quasar name & $z_{abs}$ & $\log_{10}\left[N(\mathrm{H}_{2})/\mathrm{cm^{-2}}\right]$ & References\tabularnewline[\doublerulesep]
\hline 
\endhead
Q 0000$-$2621 & 3.39 & $13.94\pm0.06$ \citep{Levshakov:01a} & \citet{Molaro:00a,Levshakov:01a}\tabularnewline
Q 0013$-$004 & 1.973 & $18.90\pm1.10$ \citep{Petitjean:02a} & \citet{Petitjean:02a}\tabularnewline
HE 0027$-$184 & 2.402 & $17.3\pm0.05$ \citep{Noterdaeme:08a} & \citet{Noterdaeme:07b}\tabularnewline
Q 0347$-$383 & 3.025 & $14.55\pm0.09$ \citep{Ledoux:03} & \citet{Ivanchik:02a,Ivanchik:05a,Levshakov:02a}\tabularnewline
 &  &  & \citet{Ledoux:03,Reinhold:06-1,King:08}\tabularnewline
Q 0405$-$443 & 2.595 & $18.16_{-0.06}^{+0.21}$ \citep{Ledoux:03} & \citet{Ivanchik:02a,Ivanchik:05a,Levshakov:02a}\tabularnewline
 &  &  & \citet{Ledoux:03,Reinhold:06-1,King:08}\tabularnewline
Q 0528$-$250 & 2.881 & $18.22_{-0.16}^{+0.23}$ \citep{Ledoux:03} & \citet{Morton:80a,Foltz:88a}\tabularnewline
 &  &  & \citet{Levshakov:85a,Cowie:Songaila:1995}\tabularnewline
 &  &  & \citet{Potekhin:98a,Ledoux:03}\tabularnewline
 &  &  & \citet{Srianand:2005,King:08}\tabularnewline
Q 0551$-$366 & 1.962 & $17.42_{-0.90}^{+0.63}$ \citep{Ledoux:02a} & \citet{Ledoux:02a}\tabularnewline
Q 0642$-$506 & 2.66 & $18.41_{-0.04}^{+0.03}$ \citep{Noterdaeme:07a} & \citet{Noterdaeme:08a}\tabularnewline
FJ 0812+32 & 2.6265 & $19.88\pm0.2$ \citep{Tumlinson:10a} & \citet{Prochaska:03a,Jorgenson:09a,Tumlinson:10a}\tabularnewline
Q 1232+082 & 2.338 & $19.67\pm0.09$ \citep{Ivanchik:10a} & \citet{Srianand:00a,Ge:01a}\tabularnewline
 &  &  & \citet{Varshalovich:01a,Ivanchik:10a}\tabularnewline
Q 1337+315 & 3.174 & $14.09\pm0.03$ \citep{Srianand:10a} & \citet{Srianand:10a}\tabularnewline
Q 1439+113 & 2.418 & $19.38\pm0.10$ \citep{Noterdaeme:08b} & \citet{Noterdaeme:08b}\tabularnewline
Q 1441+272 & 4.224 & $\approx17.6$ \citep{Ledoux:06a} & \citet{Ledoux:06a}\tabularnewline
Q 1444+014 & 2.087 & $18.3\pm0.37$ \citep{Ledoux:03} & \citet{Ledoux:03}\tabularnewline
J2123$-$0050 & 2.059 & $17.57\pm0.04$ \citep{Malec:10} & \citet{Malec:10,Milutinovic:10a,Tumlinson:10a}\tabularnewline
Q 2318-111 & 1.989 & $15.49\pm0.03$ \citep{Noterdaeme:07b} & \citet{Noterdaeme:07b}\tabularnewline
Q 2343+125 & 2.431 & $13.69\pm0.09$ \citep{Petitjean:06a} & \citet{Sargent:88a,Petitjean:06a}\tabularnewline
Q 2348$-$011 & 2.42 & $18.45_{-0.26}^{+0.27}$ \citep{Petitjean:06a} & \citet{Petitjean:06a,Noterdaeme:07a}\tabularnewline[\doublerulesep]
\hline 
\end{longtable}
\end{sidewaystable}

\subsubsection{How to measure $\Delta\mu/\mu$\label{sub:mu:how_measure_dmu}}

For a gas cloud at redshift $z$, one can relate the observed wavelength
to the laboratory wavelength as
\begin{equation}
\lambda_{i}=\lambda_{i}^{0}\left(1+z\right)\left(1+K_{i}\frac{\Delta\mu}{\mu}\right).\label{eq:mu obs wavelength eqn}
\end{equation}
The redshift of the cloud must determined simultaneously with $\Delta\mu/\mu$.
$\Delta\mu/\mu$ is not degenerate with redshift provided that at
least two transitions of differing $K_{i}$ are used. In principle,
accurate knowledge of the observed wavelengths of the different H$_{2}$
transitions are all that is needed to determine $\Delta\mu/\mu$.
However, there are different approaches one can take to arrive at
a value of $\Delta\mu/\mu$. There have been two methods used in the
literature in recent times. These are the \emph{reduced redshift method}
(RRM) and the \emph{direct $\chi^{2}$ minimisation method} (DCMM),
described below.

\emph{Reduced redshift method (RRM)} \citep[see][]{Ivanchik:02a,Reinhold:06-1}.
\index{reduced redshift method (RRM)}The RRM defines for each transition
the quantity
\begin{equation}
\zeta_{i}=\frac{(z_{i}-z_{0})}{1+z_{0}}=\frac{\Delta v}{c}=K_{i}\frac{\Delta\mu}{\mu},
\end{equation}
where $z_{i}$ is the observed redshift of the transition and $z_{0}$
is the redshift of a transition for which $K_{i}=0$. This quantity
is just the velocity difference from the unperturbed value. The individual
$z_{i}$ values can be obtained by independent Voigt profile fits
to each molecular hydrogen transition. From this relationship, a graph
of $\zeta_{i}$ vs $K_{i}$ will thus have gradient $\Delta\mu/\mu$.
$\Delta\mu/\mu$ can then be determined through standard $\chi^{2}$
minimisation of a straight line. This method is advantageous in that
one obtains a visual relationship between $\zeta_{i}$ and $K_{i}$
--- this allows one to check whether outliers exist, facilitating
either their removal or re-examination of the fit to investigate the
reason for the discrepancy. 

Unfortunately, this method is not easily applied in the situation
where the H$_{2}$ absorption displays more than one velocity component
(where the components overlap). In this case, one can generate the
$\zeta_{i}$ values, however the $\zeta_{i}$ values of the components
of each transition will be correlated. This makes it difficult to
analyse a graph of $\zeta_{i}$ vs $K_{i}$ with standard $\chi^{2}$
minimisation, as $\chi^{2}$ minimisation assumes that all data points
are independent. In principle, one can use Generalised Least Squares
--- which allows for correlated errors --- to analyse this situation,
but this has not been applied in the literature. The RRM is also a
summary method, assuming that a table of redshifts and associated
uncertainties contain all the information needed. Although this makes
calculation easy, it does not operate directly on the spectral data;
ideally one would prefer to work directly with the spectral data rather
than intermediate quantities.

\emph{Direct $\chi^{2}$ minimisation method (DCMM)} \citep{King:08,Malec:10}.
\index{direct chi^{2}
 minimisation method (DCMM)@direct $\chi^{2}$ minimisation method (DCMM)}In the DCMM, one assumes that all transitions arise from the same
cloud and therefore the same redshift. In the case of multiple components,
one assumes that corresponding components in each transition arise
from the same redshift. One then perturbs the rest wavelengths as
$\lambda_{i}^{0}\rightarrow\lambda_{i}^{0}\left[1+K_{i}(\Delta\mu/\mu)\right]$,
and then finds the value of $\Delta\mu/\mu$ which minimises $\chi^{2}$.
The value of $\Delta\mu/\mu$ which minimises $\chi^{2}$ therefore
gives the best-fitting value of $\Delta\mu/\mu$. One can model $\Delta\mu/\mu$
as an external parameter, in which case one plots $\chi^{2}$ vs $\Delta\mu/\mu$.
This graph will be approximately parabolic near the $\chi^{2}$ minimum,
with the location of the $\chi^{2}$ minimum giving the best fit value
of $\Delta\mu/\mu$. In this case, errors can be obtained by finding
$\sigma_{\Delta\mu/\mu}$ such that 
\begin{equation}
\chi^{2}\left(\Delta\mu/\mu_{\mathrm{best\, fit}}+\sigma_{\Delta\mu/\mu}\right)-\chi^{2}\left(\Delta\mu/\mu_{\mathrm{best\, fit}}\right)=1
\end{equation}
\citep{NumericalRecipes:92}. Alternatively, $\Delta\mu/\mu$ can
be included as a free parameter in the fit. The inclusion of $\Delta\mu/\mu$
as a free parameter in the fit has the advantage of being significantly
faster, as for any value of $\Delta\mu/\mu$ the first and second
derivatives of $\chi^{2}$ with respect to $\Delta\mu/\mu$ are used
to search for the minimum value of $\chi^{2}$ \citep{Murphy:PhD}.
Moreover, this method significantly reduces the number of free parameters
by imposing the physical constraint that the transitions should arise
from the same location, and therefore redshift. The reduction in the
number of free parameters should improve reliability as well as allowing
tighter confidence limits on $\Delta\mu/\mu$.  The corollary of this
is that one loses any explicit check on whether an individual transition
is consistent with the overall trend (i.e.\ whether the reduced redshift
differs greatly from the trend of $\zeta_{i}$ with $K_{i}$ given
by the other transitions). 

\emph{Comparison of the two methods. }Although the RRM is appealing
because of the simpler numerical methods required, the reduction in
the number of free transitions with the DCMM can be substantial. In
particular, the DCMM requires $n_{v}(n_{t}-1)$ fewer free parameters,
where $n_{v}$ is the number of H$_{2}$ velocity components and $n_{t}$
is the number of H$_{2}$ transitions used. The reduction in the number
of free parameters under the DCMM acts to improve the stability of
the fitting process. In particular, individual transitions may have
very poorly constrained line parameters, despite the fact that these
parameters may be well constrained in a joint fit to many transitions.
In the RRM method, this can cause certain transitions or, particularly,
components to be removed during the $\chi^{2}$ minimisation process,
rendering those transitions unsuitable for inclusion in the fit. \textsc{vpfit}
will automatically remove components during the $\chi^{2}$ minimisation
if their parameters move outside certain user-defined boundaries.
The two important ones for this scenario are that the column densities
of transitions must be greater than $10^{8}\,\mathrm{cm^{-2}}$ and
the $b$-parameters must be greater than $0.05\,\mathrm{km\, s^{-1}}$.
With the DCMM, the tying of components helps to prevent these transitions/components
from being removed, allowing for the inclusion of a greater number
of transitions.

Another assumption in the RRM is that the errors on the line redshifts
are Gaussian. In the event where a transition is blended on one side
with a forest line, the uncertainty on the redshift for the H$_{2}$
transition will almost certainly be asymmetric. This means that the
errors on the reduced redshifts, $\zeta_{i}$, will also be asymmetric
(and not Gaussian). $\chi^{2}$ minimisation of a linear fit to $\zeta_{i}$
vs $K_{i}$ assumes that the errors \emph{are} Gaussian (or at least
symmetric), and therefore the use of the standard errors from the
spectral fitting in a fit of $\zeta_{i}$ vs $K_{i}$ will only be
approximately valid. This problem should not affect the DCMM, however.
This is because the determination of $\Delta\mu/\mu$ is derived from
the (potential) velocity shifts from many transitions. Because of
the central limit theorem, $\Delta\mu/\mu$ should be approximately
Gaussian. Because $\Delta\mu/\mu$ is determined simultaneously with
all the redshift parameters, any asymmetry in the uncertainty of individual
line redshifts will be accounted for when searching for the best-fitting
value of $\Delta\mu/\mu$. Similarly, the uncertainty on $\Delta\mu/\mu$
is determined directly from the curvature of $\chi^{2}$ at the purported
best-fit, meaning that it should be robust. Thus, the estimate of
$\Delta\mu/\mu$ derived from the DCMM is more likely to be accurate
than one derived from the RRM.

Ultimately, we prefer the DCMM, as it is both faster and more reliable,
and works directly with the spectral data rather than on intermediate
quantities, although we use the RRM as a check where possible.

\subsubsection{The importance of the Werner series\label{sub:mu:Importance Werner series}}

Although a good constraint on $\Delta\mu/\mu$ is possible using only
the Lyman series, it is clear from figure \ref{Flo:mu:Ki vs lambda}
that $K_{i}$ is well correlated with rest wavelength for the Lyman
series. This implies that a simple stretching or compression of the
wavelength scale would mimic variation in $\mu$. The use of the Werner
series\index{molecular hydrogen!Werner series} helps to break this
degeneracy to some degree, as for rest wavelengths $\lambda\lesssim1020\AA$,
where the Werner series exists, the Lyman transitions move in a significantly
different fashion to the Werner transitions. Importantly, for rest
wavelengths $990\AA\lesssim\lambda\lesssim1020\AA$ the Werner series
transitions move in the opposite direction to the Lyman series transitions
if $\Delta\mu/\mu\neq0$. It will be seen in chapter \ref{cha:alpha}
that the different magnitudes and signs of the $q$ coefficients play
a similar role in providing robustness against a similar stretching
or compression of the spectrum when searching for $\Delta\alpha/\alpha$
(the $q$ coefficients are the sensitivity coefficients used, and
are analogous to the $K_{i}$ coefficients for $\mu$).

\subsubsection{Previous constraints}

\citet{Varshalovich:93a} analysed the spectrum of \citet{Foltz:88a}
of the $z=2.811$ absorber toward Q0528$-$250 to obtain $|\Delta\mu/\mu|<0.005$.
\citet{Varshalovich:95a} reanalysed the same spectrum to conclude
that $|\Delta\mu/\mu|<0.002$. \citet{Potekhin:98a} used new observations
of the same system at higher resolving power ($R\sim14,000$) to obtain
$\Delta\mu/\mu=(-10\pm8)\times10^{-5}$ using the laboratory wavelengths
for H$_{2}$ of \citet{Abgrall:93a,Abgrall:93b}.

\citet{Cowie:Songaila:1995} used a $R=36,000$ Keck observation of
the $z=2.811$ absorber toward Q0528$-$250 to produce $\Delta\mu/\mu\in[-7,5.5]\times10^{-4}$
(95 percent confidence limits). This result was the first to be obtained
with the 8--10m class optical telescopes, which supersede the previous
$\sim4\mathrm{m}$ class telescopes. The extra collecting area allows
spectra to be taken with significantly higher $R$ in a reasonable
amount of time. The precision with which $\Delta\mu/\mu$ can be determined
increases with the resolving power of the spectrum%
\footnote{Assuming that SNR is held constant%
}. Additionally, higher resolving powers are important in attempting
to determine the velocity structure of the H$_{2}$ absorbers. The
H$_{2}$ clouds are cold, leading to line widths of only a few km/s.
With low-$R$ spectra, it is extremely difficult to determine any
velocity structure present in the absorbers, as it is below the instrumental
resolution. A spectrum with $R=36,000$ corresponds to an instrumental
resolution of $\sim8.3\,\mathrm{km\, s^{-1}}$. With this $R$, it
is possible to clearly distinguish the H$_{2}$ lines from the surrounding
Lyman-$\alpha$ forest, and to start to resolve detailed velocity
structures. Higher resolving powers obviously lead to better results. 

\citet{Ivanchik:02a} analysed much higher quality ($R\sim43,000$,
$\mathrm{SNR}\sim10$ to $40$ per pixel) VLT/UVES spectra of the
$z=3.025$ system toward Q0347$-$383 and the $z=2.338$ system toward
Q1232+0815 with the RRM. Using the $K_{i}$ values of \citet{Varshalovich:95a}
they found that $\Delta\mu/\mu=(5.8\pm3.4)\times10^{-5}$ and $\Delta\mu/\mu=(14.4\pm11.4)\times10^{-5}$
for two systems respectively. A combined regression analysis gave
$\Delta\mu/\mu=(5.7\pm3.8)\times10^{-5}$, where the H$_{2}$ wavelengths
of \citet{Abgrall:93a,Abgrall:93b} are used, or $\Delta\mu/\mu=(12.5\pm4.5)\times10^{-5}$
if the wavelengths of \citet{Morton:76a} are used. 

A significant potential source of systematic error in all of the above
results arises from uncertainties in the laboratory measurements of
the H$_{2}$ wavelengths. \citet{Ivanchik:02a} state the measurement
errors in the wavelengths of \citet{Abgrall:93a,Abgrall:93b} to be
$\sim1.5\mathrm{m\AA}$. The fractional error in the wavelengths is
thus of order $\Delta\lambda/\lambda\sim1.5\,\mathrm{m\AA}/1000\mathrm{\AA}=1.5\times10^{-6}$
(the H$_{2}$ transitions have $900\mathrm{\AA}\lesssim\lambda\lesssim1150\mathrm{\AA}$).
With $\Delta\lambda/\lambda=\Delta v/c$ this implies a velocity uncertainty
of $\sim450\,\mathrm{ms^{-1}}$. This can be converted into an implied
systematic with $\Delta v/c\sim|\Delta K_{i}|(\Delta\mu/\mu)$, with
$|\Delta K_{i}|$ being the range of $K_{i}$ values used. $\Delta K_{i}$
is typically $\approx0.05$, thus implying a systematic error term
of $\Delta\mu/\mu\sim3\times10^{-5}$. However, the difference between
their two results indicates that the systematic error is larger \citep{Ivanchik:02a}.
On account of this, there has been considerable laboratory work in
recent years to generate laboratory wavelengths of sufficient accuracy
that they do not contribute appreciably to the total error budget.
\citet{CanJChem_H2} used a narrow band XUV laser source to provide
a substantially improved (although incomplete) line list, where the
errors for the highest energy level states are an order of magnitude
smaller than those from \citet{Abgrall:93a} and \citet{Abgrall:93b}.
\citet{Hollenstein:06-1} performed a similar experiment, completing
the line list of \citet{CanJChem_H2}. The fractional accuracies for
these wavelength measurements are of order $\sim5\times10^{-8}$. 

\citet{UbachsReinhold:04} used the wavelengths of \citet{CanJChem_H2}
to analyse the absorbers in the spectra of Q0528$-$250, Q0347$-$383
and Q1232+082. For the combined data, they found that $\Delta\mu/\mu=(-0.5\pm1.8)\times10^{-5}$
using the RRM. Omitting the Q0528$-$250 data, which is of poorer
quality, they obtained $\Delta\mu/\mu=(1.9\pm1.5)\times10^{-5}$.

\citet{Ivanchik:05a} analysed the Q0347$-$383 absorber, as well
as a new one toward Q0405$-$433 (from a spectrum obtained using VLT/UVES)
using the wavelengths of \citet{CanJChem_H2} to obtain $\Delta\mu/\mu=(1.47\pm0.83)\times10^{-5}$
for the system towards Q0347$-$383 and $\Delta\mu/\mu=(2.11\pm1.39)\times10^{-5}$
for the system towards Q0405$-$443 using the RRM.

\citet{Reinhold:06-1} used the wavelength data of \citet{CanJChem_H2}
and \citet{Hollenstein:06-1} to examine the spectra of \citet{Ivanchik:05a}
(Q0347$-$383 and Q0405$-$443). They recalculated the $K_{i}$ values
in a significantly more accurate fashion, as described earlier. They
noted that the $K_{i}$ values for highly excited states changed significantly
as a result of the post-BOA corrections, and that all $K_{i}$ values
experienced a systematic shift due to the adiabatic correction. They
found that $\Delta\mu/\mu=(2.06\pm0.79)\times10^{-5}$ for Q0347$-$383
and $\Delta\mu/\mu=(2.78\pm0.88)\times10^{-5}$ for Q0405$-$443 using
the RRM. It is worth noting that their points with $K_{i}<0$ demonstrate
an unusually small scatter, and indeed they conceded that their result
differs from previous works primarily as a result of the addition
of new laboratory wavelengths for the $\nu=0$ and $\nu=1$ Lyman
bands, which correspond to these $K_{i}$ values. A combined weighted
fit yielded $\Delta\mu/\mu=(2.4\pm0.6)\times10^{-5}$ using the RRM,
although $\chi_{\nu}^{2}=2.1$ for the $\zeta_{i}$ values about the
linear model for $\zeta_{i}$ vs $K_{i}$, suggesting that unmodelled
errors exist. An unweighted fit gave $\Delta\mu/\mu=(2.0\pm0.6)\times10^{-5}$.
This result seems to suggest that $\mu$ was larger in the past at
the $>3.5\sigma$ confidence level. The result of this paper formed
the motivation for the analysis of this chapter. The analysis of \citeauthor{Reinhold:06-1}
was explained in considerably more detail in \citet{Ubachs:2007-01}.

A potential systematic effect in the analysis of molecular hydrogen
concerns spatial segregation of the different $J$-levels of the ground
state. \citet{Jenkins:97a} noted that there appeared to be small
velocity shifts between $J=0$ and $J=3$ transitions of H$_{2}$
observed towards $\zeta$ Orionis A, with $\Delta v\approx0.8\,\mathrm{km\, s^{-1}}$.
However, this effect was not observed toward other stars \citep{Jenkins:00a}.
\citet{Levshakov:02a} claim to detect a gradual shift in $z_{\mathrm{abs}}$
with increasing $J$ in their analysis of the $z=3.025$ absorber
toward Q0347$-$383. \citet{Murphy:PhD} notes that similar shifts
of similar magnitude to those seen in \citet{Jenkins:97a} would lead
to a systematic error in $\Delta\mu/\mu$ of $\sim4\times10^{-5}$.
However, \citet{Reinhold:06-1} addressed these concerns by showing
no significant correlation exists between $\zeta_{i}$ and $J$ or
$\zeta_{i}$ and $\lambda_{i}^{0}$.

\citet{Wendt:08a}, \citet{Thompson:09a} and \citet{Wendt:10a} all
investigated Q0347$-$383 and Q0405$-$443 to examine the results
of \citet{Reinhold:06-1}. We defer discussion of these results to
section \ref{sub:Further-investigations-Reinhold} so that they can
be interpreted in the context of this work, which was reported first
in \citet{King:08}.

The results of \citet{King:08} (this work) are given in tables \ref{Tab:mu:DCMM results 3 quasars}
and \ref{Tab:mu:RRM results 2 quasars}.

\citet{Malec:10} analysed the $z=2.059$ molecular hydrogen system
toward J2123$-$0050 using 86 H$_{2}$ transitions from Keck observations.
They also use 7 HD (deuterated molecular hydrogen) transitions in
their analysis --- the first constraint on $\mu$ variation to utilise
HD. They found that $\Delta\mu/\mu=(+5.6\pm5.5_{\mathrm{statistical}}\pm2.9_{\mathrm{systematic}})\times10^{-6}$.
The systematic error contribution arises predominantly from wavelength
calibration uncertainties, however the estimate is model dependent.
Like this work, \citeauthor{Malec:10} applied the DCMM to reduce
the number of free parameters in the fit, and improve the robustness
of the result. They also modelled the Lyman-$\alpha$ forest in a
similar fashion to this work.

\subsection{$\Delta\mu/\mu$ from ammonia}

\index{ammonia}The inversion transitions of ammonia (NH$_{3}$),
which result from the situation where the nitrogen atom tunnels from
one side of the molecule through the potential barrier due to the
hydrogen atoms to the other side, are strongly sensitive to a change
in $\mu$, with $K\sim4.2$ \citep{Flambaum:07a}. \citet{Murphy:Flambaum:08}
and \citet{Henkel:09} compared the inversion transitions of ammonia
with rotational molecules to determine very stringent limits on $\Delta\mu/\mu$
at $z<1$. \citet{Murphy:Flambaum:08} used B0218+357 to find that
$|\Delta\mu/\mu|<1.8\times10^{-6}$ (95\% confidence) at $z=0.68$,
whilst \citet{Henkel:09} concluded that $|\Delta\mu/\mu|<0.9\times10^{-6}$
($2\sigma$ confidence) from PKS1830$-$211 at $z=0.89$. 

The ammonia method is theoretically preferable to the analysis of
molecular hydrogen, as the sensitivity coefficient is larger by a
factor of $\sim100$, and the transitions are not blended with the
Lyman-$\alpha$ forest. However, there are some drawbacks. In particular,
quasars are point sources in the optical but are manifestly extended
sources in the radio. This implies that the clouds from which the
ammonia transitions arise may not be spatially co-located with the
rotational transitions. Spatial offsets in the radial direction will
lead to velocity differences, which would mimic a change in $\mu$.
Molecular hydrogen is much less prone to this problem, because one
is comparing transitions which arise from the same molecule (albeit
from different $J$-levels of the ground state). If one only compares
transitions which arise from the same $J$-level, then the concern
of spatial segregation is eliminated. Perhaps more importantly, there
are few sources known which possess the necessary ammonia transitions,
and none at high ($z\gtrsim1$) redshift.

\section{Methods \& methodology}

Our goal was to re-analyse the work of \citet{Reinhold:06-1} and
confirm or dispute the apparent evidence for a change in $\mu$. Our
methodology differs from that of \citeauthor{Reinhold:06-1} in four
significant ways: 
\begin{enumerate}
\item We model the Lyman-$\alpha$ forest in the vicinity of the H$_{2}$
transitions using Voigt profiles (not polynomials);
\item We use the DCMM rather than the RRM (although we retain the RRM as
a check on our results); 
\item Our spectra have been re-reduced using a new thorium-argon wavelength
calibration algorithm, which yields substantially improved wavelength
calibration, and; 
\item We correct for under-estimation of the flux uncertainties in regions
of low flux in VLT/UVES spectra.
\end{enumerate}

\subsection{Spectral data}

The first stage of our analysis examined the absorbers in Q0347$-$383,
Q0405$-$443 and Q0528$-$250. We are grateful to H. Ménager and M.
Murphy, who reduced the exposures from 2D format to 1D format, and
then co-added the 1D exposures within \textsc{uves\_popler}. They
also cleaned the spectrum to remove the effect of data problems, including
removing cosmic rays which are not removed by the automatic algorithm
within \textsc{uves\_popler}, ghosts caused by reflections within
the UVES enclosure and other inconsistencies between the contributing
exposures. The analysis of the absorbers in each of these systems
leads to the results in section \ref{sec:mu:results}.

The exposures used by \citet{Ivanchik:05a} which contribute to the
spectra for Q0347$-$383 and Q0405$-$443 were obtained on VLT/UVES
in January 2002 and 2003. The exposures contributing to their spectrum
of Q0347$-$383 were obtained under program IDs 68.A-0106(A) and 68.B-0115(A),
whilst those contributing to Q0405$-$443 were obtained under program
ID 70.A-0017(A). For each object, nine exposures of 1.5 hours each
were taken with a slit width of 0.8 arcseconds, yielding a resolution
of $R\sim53,000$ and a SNR of between $\sim30$ and $\sim70$ over
the wavelength range $3290$ to $4515\mathrm{\AA}$. Prevailing seeing
was sub-arcsecond. The ThAr calibration spectra were taken before
and after the science exposures, and so the wavelength calibration
should be good. \citet{Ivanchik:05a} note that the temperature drift
at UVES is sufficiently small that the uncertainty introduced into
wavelength calibration as a result of temperature drift is negligible.
Further details can be found in \citet{Ivanchik:05a}.

Besides the exposures noted above, for Q0347$-$383 we incorporated
exposures from program ID 60.A-9022(A), although these contribute
only an additional 2.6 hours. For Q0405$-$443, we also made use of
additional exposures under program IDs 68.A-0361(A), 68.A-0600(A)
and 68.A-0361(A). 

The exposures which contribute to the spectrum for Q0528$-$250 were
obtained with VLT/UVES between 2001 and 2002 under program IDs 66.A-0594,
68.A-0600 and 68.A-0106, with a total exposure time of 21.9 hours.
A slit width of 1.0 arcseconds was used for all exposures. Seeing
was generally sub-arcsecond. 

However, Q0528$-$250 was re-observed in late 2008/early 2009 under
program ID 82.A-0087, with exposures totalling approximately 8.2 hours,
after our analysis of the previous spectrum of Q0528$-$250 was complete.
We have re-analysed the absorber in Q0528$-$250 using these new exposures
to provide an additional constraint on $\Delta\mu/\mu$. For clarity,
we refer to spectrum created from exposures under under program IDs
66.A-0594, 68.A-0600 and 68.A-0106 as Q0528:A, and give the results
for this analysis in section \ref{sec:mu:results}. We refer to the
spectrum generated from program ID 82.A-0087 as Q0528:B2, and discuss
this particular spectrum in section \ref{sub:Q0528-250 revisited}.

For our analysis of Q0405$-$443, Q0347$-$383 and Q0528$-$250 (Q0528:A),
we have used $K_{i}$ values and laboratory wavelength values from
\citet{Ubachs:2007-01}. However, for our second analysis of Q0528$-$250
(Q0528:B2) we used the data from table 1 of \citet{Malec:10}, which
includes the work of \citet{Ubachs:2007-01} but also includes newer
measurements from \citet{Bailly:09a}.

\subsection{Wavelength calibration\label{sub:mu wavelength calibration}}

When searching for variations in $\mu$ at the $10^{-5}$ to $10^{-6}$
level, the spectra must be accurately calibrated at the $\mathrm{m\AA}$
level. The wavelength scale of the science echelle exposure is calibrated
through a secondary calibration exposure, usually using a thorium-argon
(ThAr) lamp, which produces a large number of well-measured transitions
across the total wavelength coverage of an optical telescope. \citet{Murphy:07b}
considered the line list used by the UVES pipeline in detail, and
considered different factors which may introduce errors into the wavelength
calibration process. One potential problem is the use of blended ThAr
lines which are unresolved in typical UVES spectra ($R\sim30,000$
to $70,000$). Use of such lines will cause bias in the ThAr line
centroid measurement and therefore in the wavelength calibration.
Another factor considered is the use of weak lines, which may cause
false identification by the UVES pipeline. Similarly, they also consider
the fact that the existing ThAr line lists contain inaccuracies, and
therefore they reject ThAr lines which have large residuals. The result
of this is a new ThAr line list for which the wavelength calibration
residuals (RMS $\sim35$ ms$^{-1}$) are a factor of three better
than those achieved using the ESO line list or the line list of \citet{deCuyper:98a}
(RMS $\sim130$ ms$^{-1}$). \citeauthor{Murphy:07b} note that not
only are the random calibration errors significantly improved through
the use of this line list, but the existence of long-range variations
with peak-to-peak amplitudes of up to $\sim75$ ms$^{-1}$ are reduced.
Our spectra have been wavelength calibrated using the calibration
algorithm of \citet{Murphy:07b}, and therefore our spectra should
have significantly better calibration than the spectra used in previous
analyses.

\subsection{Correction for underestimated flux errors\label{sub:mu:flux error corr}}

As described in section \ref{sub:Data pipeline problems}, the uncertainty
estimates on flux values in the base of saturated lines appear to
be too low. To correct for this in the spectra for Q0405$-$443, Q0347$-$383
and our first analysis of Q0528$-$250 (Q0528:A) we applied the heuristic
correction described in section \ref{sub:Correcting error arrays through an estimated functional form}.
We show the results of a number of measurements of the ratio of the
RMS of pixels in the base of saturated lines to the average of the
RMS array in table \ref{tab:mu:saturated flux data} (the RMS array
is a modified version of the flux error array produced by \textsc{uves\_popler}
which attempts to account for inter-pixel correlations). It is clear
that the error estimates are too small by a factor of approximately
$2$. For these three quasar spectra, we modify the error arrays using
the functional form in equation \ref{eq:flux error corr formula},
choosing $b$ such that $(1+b)^{1/4}=\bar{R}$ as given in table \ref{tab:mu:saturated flux data}.

\begin{table}[tbph]
\caption[Evidence for flux uncertainty under-estimation]{Evidence for understimated uncertainties on flux values in the base of saturated lines. For each quasar spectrum, $n$ shows the number of measurements taken in the base of saturated lines. The quantity $\bar{R}=\sigma_f / \bar{\sigma}$ is the ratio of the RMS of the flux array to the average of the RMS array (a modified version of the error array which attempts to account for inter-pixel correlations). This quantity should be $\sim 1$ if the error arrays correctly reflect the noise in the spectral data, but will be larger than 1 if the error arrays are underestimated. $\sigma_{\bar{R}}$ indicates the standard error on this quantity. The final column indicates the deviation of $\bar{R}$ from the expected value of unity, expressed as a multiple of the standard error. These data clearly indicate that the flux uncertainty estimates in the base of saturated lines are too small by a factor of approximately 2. \label{tab:mu:saturated flux data}}

\noindent \centering{}%
\begin{tabular}{ccccc}
\hline 
Quasar spectrum & $n$ & $\bar{R}=\sigma_{f}/\bar{\sigma}$ & $\sigma_{\bar{R}}$ & $(\bar{R}-1)$\tabularnewline
\hline 
Q0405$-$443 & 27 & 2.08 & 0.10 & $10.8\sigma$\tabularnewline
Q0347$-$383 & 23 & 2.20 & 0.086 & $14.0\sigma$\tabularnewline
Q0528$-$250 & 7 & 2.19 & 0.079 & $15.1\sigma$\tabularnewline
\hline 
\end{tabular}
\end{table}

\begin{figure}[tbph]
\noindent \begin{centering}
\includegraphics[bb=132bp 54bp 556bp 722bp,clip,angle=-90,width=1\textwidth]{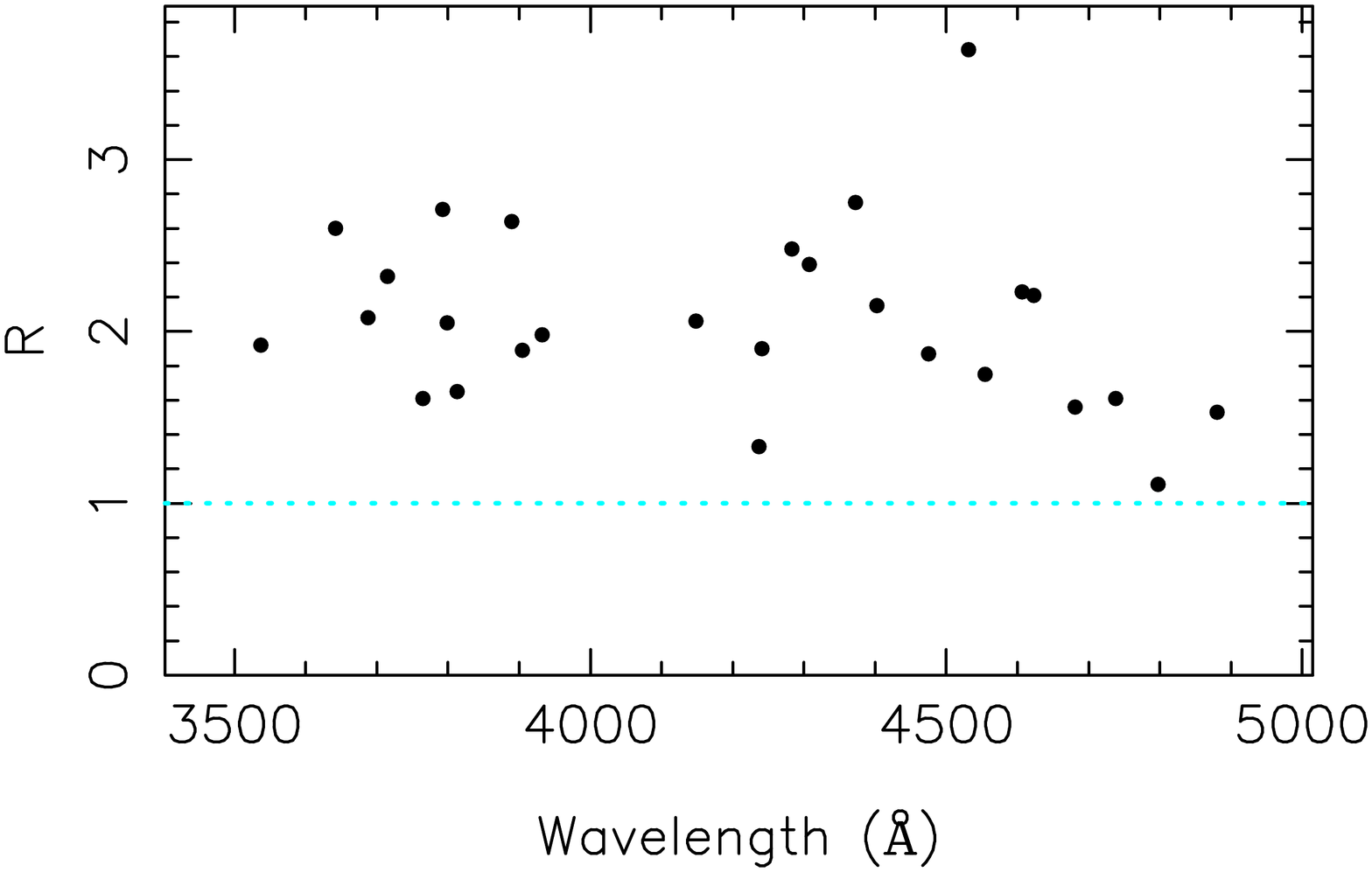}
\par\end{centering}

\caption[Factor by which errors in the base of saturated lines are underestimated for Q0405$-$443]{The factor by which errors in the base of a selection of saturated lines are underestimated in the spectrum of Q0405$-$443. The quantity $\bar{R}=\sigma_f / \bar{\sigma}$ is the ratio of the RMS of the flux array to the average of the RMS array (a modified version of the error array which attempts to account for inter-pixel correlations). The dotted blue line shows the expected value of 1 if the error arrays correctly account for interpixel dispersion. Note that there is significant scatter between individual measurements, and also that there may be a correlation of the effect with wavelength.  \label{Flo:mu:Q0405 flux uncertainty prob}}
\end{figure}

In figure \ref{Flo:mu:Q0405 flux uncertainty prob} we show the measurements
taken in the base of saturated lines for Q0405$-$443. It is clear
that there is significant scatter between the individual measurements,
and therefore the functional form of equation \ref{eq:flux error corr formula}
will only be approximately correct. Additionally, there appears to
be a weak wavelength dependency, with the problem worse in the blue
end of the spectrum. We have not investigated whether this is a true
wavelength dependency, or whether it is simply a function of SNR (which
is correlated with wavelength because the spectrograph throughput
is worse in the blue end of the spectrum). Equation \ref{eq:flux error corr formula}
can be modified to account for a wavelength dependence of the observed
problem, but we did not do this.

\subsection{Free parameters \& physical assumptions}

The observed transitions of molecular hydrogen consist of transitions
from the ground states to upper excited states for the Lyman and Werner
bands. By ``ground states'', we refer to the subdivision of the
lowest energy level into levels with different angular momentum $J$.
The different $J$-levels of the ground state have different relative
populations, which depend on the temperature of the gas cloud but
also on the influence of non-equilibrium processes (e.g.\ collisions).
The non-equilibrium processes simply cause the relative populations
in the different $J$-levels to be different from the Boltzmann distribution
\citep{Spitzer:73a,Levshakov:85a}. In particular, transitions with
high $J$ display apparent overpopulation relative to low-$J$ transitions.
Clearly, transitions which arise from the same ground state must have
the same $b$-parameter and the same column density. 

We make further physical assumptions which reduce the number of free
parameters in the fit. The most important of these, noted earlier,
is that all transitions arise from the same location, and therefore
have the same $z$. For H$_{2}$ absorbers with multiple velocity
components, this means that corresponding components in all transitions
have the same $z$. We also explore whether we can impose the requirement
that all transitions have the same $b$-parameter, irrespective of
$J$ (for H$_{2}$ absorbers with multiple velocity components, this
means that corresponding components in all transitions have the same
$b$). By minimising the number of free parameters in the fit the
optimisation process should be more robust. Similarly, by imposing
physical constraints on the problem it is more likely that our derived
value of $\Delta\mu/\mu$ will be accurate.

In order to address the concerns in section \ref{sub:mu:General H2 comments}
relating to continuum fitting, in regions where the local continuum
is uncertain we allow for a linear continuum which is determined simultaneously
with all other parameters. The uncertainty in determining the local
continuum therefore propagates in to the uncertainty on $\Delta\mu/\mu$.

We note that in addition to the under-estimation of flux uncertainty
in regions of low flux, there appears to be residual flux in the base
of many saturated lines. The typical magnitude of this effect is about
2\% of the local continuum. Whilst weak sky emission should be subtracted
as part of the flux extraction, it appears that the \textsc{midas}
pipeline systematically underestimates the subtraction required, leading
to the observed effect. A similar problem has been noted previously
by \citet{Malec:10}, albeit in relation to a Keck/HIRES spectrum
of J2123$-$0050. We attempt to correct for this problem by allowing
the zero level to vary in any region which includes absorption lines
which are saturated, or nearly saturated. As for the continuum, the
uncertainty in determining the zero level propagates into the uncertainty
on $\Delta\mu/\mu$.

\subsection{Modelling the Lyman-$\alpha$ forest with molecular hydrogen}

\index{Voigt profile!model construction}The structure of the Lyman$-\alpha$
forest\index{Lyman-alpha
 forest@Lyman-$\alpha$ forest} is unknown \emph{a priori}, and therefore must be modelled from the
observed flux profile. Our model of the molecular hydrogen transitions
with the forest was built up iteratively. With knowledge of the redshift
of the molecular hydrogen absorbers, in each spectra we searched for
molecular hydrogen transitions which we considered to be potentially
usable. We consider potentially usable transitions to be those for
which the molecular hydrogen transition can be visually distinguished
from its surrounds. This necessarily precludes the use of H$_{2}$
transitions in regions of near zero flux, but in any event these transitions
would contribute no meaningful constraint on $\Delta\mu/\mu$. 

From a list of potentially usable transitions, we then selected a
buffer region around the H$_{2}$ transition, where the region should
be large enough to include any absorption feature which might overlap
with the H$_{2}$ transition. In general, we attempted to ensure that
the fitting region was sufficiently large so as to return to the local
continuum, although this was not always possible. In each of the fitting
regions, we modelled the molecular hydrogen transition and then modelled
all surrounding features as H \isc. To do this, we added and removed
H \iscs components to attempt to achieve a statistically satisfactory
model, using the criteria set out in section \ref{sub:Model-selection}.
Note that although most transitions observed in the forest are indeed
due to H~\isc, there are also metal transitions from other absorbers
along the line of sight (including galactic and atmospheric lines).
The identification of the origin of these transitions is not necessary
if they do not overlap with the H$_{2}$ transitions; we simply modelled
them as H~\iscs in order to have a physical model for them. We describe
the treatment of metal lines which overlap with H$_{2}$ lines below.
For all transitions assumed to be H~\iscs (which we refer to hereafter
as just H~\iscs transitions), we use only the $\lambda1215.7$ transition
rather than the whole Lyman series, to prevent line misidentification
spuriously impacting regions blueward of that transition. Where Lyman-$\beta$
transitions exist in the blue region of the spectrum, we simply modelled
them with additional H \iscs components.

We then combined models from the regions fitted individually into
a model where the regions are fitted simultaneously. As the line parameters
for the individual H$_{2}$ transitions were independent when the
regions are fitted independently, at this stage we imposed physical
restrictions on the transitions by tying certain parameters together.
The H$_{2}$ absorbers in Q0347$-$383 and Q0405$-$443 appear to
be well modelled by a single component. For these absorbers, we required
that the redshifts of all of the transitions are the same and therefore
tie them together within \textsc{vpfit}. We also required that the
$b$-parameters be the same. Although the line strengths can be in
principle determined from the oscillator strengths and a single column
density, we allowed the column densities for each transition to be
determined independently (effectively fitting the oscillator strengths
as free parameters).

The absorber in Q0528$-$250 requires more than one component to model
the structure correctly. We describe how we determined the velocity
structure below in section \ref{sub:mu H2 velocity structure}. For
this absorber, we required that the redshifts of corresponding components
be the same. As above, we fitted the column densities for each transition
as free parameters. However, we wished to ensure that a physical consistency
is maintained, in that the ratios of the line strengths between different
components should be the same for transitions arising from the same
ground state. We therefore imposed the requirement that the ratio
of the column densities between the different components was the same
for transitions arising from the same $J$-level. In this way, the
total column density (effectively, oscillator strength) for each transition
was a free parameter, but the ratios of the individual column densities
within each transition were constrained. 

We then iteratively refined the fit by alternately allowing \textsc{vpfit}
to minimise $\chi^{2}$ for a particular model, then attempting to
improve that model through the addition and deletion of H \iscs components
to obtain a robust model according to the criteria in section \ref{sub:Model-selection}.

During the iterative process, it can become clear that a molecular
transition is blended with another line (presumed H \isc) when it
was not thought to be from a fit to just that region. This is because
the information from the other molecular hydrogen transitions imposes
a strong constraint on the $b$-parameter(s) and redshift(s) of that
transition, thus uncovering apparently hidden blends. These blends
necessitate the addition of H \iscs components that overlap with
the H$_{2}$ transition in question. With the addition of extra H
\iscs transitions, an acceptable fit can generally be achieved. This
demonstrates the utility of fitting all transitions simultaneously:
otherwise inconspicuous blends are generally revealed. In a few instances,
the transitions which had to be included to achieve a statistically
acceptable fit had extremely narrow $b$ parameters ($b\lesssim5\,\mathrm{km\, s^{-1}}$).
In this case, it is likely that the blend is a metal line from an
unknown absorber along the line of sight. As a result, we rejected
the transition. The reason for not accepting transitions affected
by narrow-$b$ interlopers is that any inaccuracy in modelling the
interloping transitions could lead to a significant bias in measuring
the H$_{2}$ line position --- the narrow $b$-parameter(s) of the
interloping transitions means that the absorption they cause varies
rapidly across the H$_{2}$ line profile. Ultimately, the joint fit
of all the molecular hydrogen transitions allows the detection and
rejection of transitions which are likely to be contaminated by metal
lines. Rejecting transitions which are suspected to be contaminated
cannot bias $\Delta\mu/\mu$ away from zero. Moreover, this should
not bias $\Delta\mu/\mu$ significantly. If the suspicion of contamination
in particular lines was in fact due to $\Delta\mu/\mu\neq0$, we would
expect to see this problem more frequently, and more obviously, for
transitions with larger $|K_{i}|$. The number of transitions rejected
was small, and did not appear to be correlated with $|K_{i}|$, and
hence it is unlikely that we are biasing $\Delta\mu/\mu$ towards
zero.

It is possible to add too many H~\iscs components to a particular
region, leading to ``over-fitting''. Over-fitting is undesirable
for several reasons. The primary reason is that it means that another,
simpler model can explain the data better than the over-fitted model.
Parsimony should be strongly valued in model selection, as noted in
section \ref{sub:Model-selection}. Perhaps more importantly, it means
that the performance of the optimisation algorithm can be substantially
impaired. With significant over-fitting, convergence to the $\chi^{2}$
minimum can be excessively slow. In extreme cases, convergence may
not occur at all. Over-fitting can be detected through two means:
\begin{enumerate}
\item The addition of components which increase the AICC suggests that the
components are not supported by the data. If the AICC significantly
decreases upon removal of the components, this suggests that the model
was over-fitted.
\item Over-fitting causes the uncertainty estimates on the parameters of
the components in question to be excessively large \citep[this point was discussed by][]{GMW:86}.
In fact, this is often a good way to directly identify components
which are potentially unnecessary; the AICC relates to the model as
a whole and therefore cannot suggest which components may be unnecessary.
In particular, H \iscs transitions with $\sigma_{\log_{10}N}\gtrsim1.0$
or $\sigma_{b}/b\gtrsim1$ are certainly suspicious. In regions with
substantial over-fitting, errors can easily be substantially larger
than this. The numerical cause of these large errors is strong relative
degeneracies between parameters. That is, $\chi^{2}$ is almost flat
in some direction in the parameter space relating to the offending
transitions. It is this flatness in $\chi^{2}$ which is the cause
of poor convergence.

\begin{enumerate}
\item Nevertheless, the presence of large errors on some components does
not mean that they are unnecessary. In particular, the column densities
for transitions which are saturated can be very poorly determined.
This necessarily means that saturated H~\iscs transitions will have
large errors on the column density. 
\end{enumerate}
\end{enumerate}
Because of the impact of over-fitting on the convergence of \textsc{vpfit},
we spent considerable effort trying to identify cases of over-fitting,
and removing H \iscs components as necessary to minimise the problem. 

Our final fits were obtained where we were not able to obtain any
statistically appreciable improvement.

In practice, it is not important that the structure of the forest
be modelled with total accuracy in all regions. The goal is simply
to fit all observable structure with a plausible model, so that a
plausible background flux model exists against which the molecular
hydrogen model is constructed. Although the uncertainty which propagates
into the determination of $\Delta\mu/\mu$ is likely to be somewhat
incorrect, proceeding in this fashion at least attempts to account
for the uncertainty in determining the forest structure. 

The process of fitting the spectra constitutes almost all of the effort
in obtaining $\Delta\mu/\mu$. The speed of the optimisation algorithm
unfortunately degrades rapidly with increased numbers of parameters.
For instance, consider the effort required to calculate the partial
derivatives of $\chi^{2}$ with respect to each of the parameters%
\footnote{Please see section \ref{sub:Optimisation-theory} for more details
on the theory behind the optimisation.%
}. For $n$ Voigt profiles, one needs $3n$ parameters, and therefore
there are $3n$ first-order partial derivatives%
\footnote{Ignoring for the moment the parameters which describe the linear continuum
fits and the zero-level determination.%
}. For each of these derivatives, $n$ Voigt profiles must be generated.
Similarly, the spectral density of lines is approximately constant
with wavelength, which implies that the number of pixels at which
the profile must be evaluated scales as $\mathcal{O}(n)$. Thus, the
time required to evaluate the partial derivatives at each iteration
scales as $\mathcal{O}(n^{3})$. Even after parallelisation, the time
required for one step of the iteration process for our model for Q0528$-$250
on a quad-core Intel 3.2GHz i7 processor is about 10 minutes. Many
iterations (typically $\sim$10 to 30) are needed to make the model
relatively close to optimal, with potentially many more needed if
strong degeneracies exist. Once the model is relatively optimal, human
interaction is then required to look for parameter degeneracies, areas
of poor fitting and the appropriateness of various parameters. Adjustments
are made to the model, and the optimisation restarted. One can easily
see how this process becomes extremely time-consuming. It is regrettable
that the time required to obtain a final, satisfactory model for a
particular spectrum is of the order of months. We discuss future avenues
of improvement in this regard later.

\subsection{Other details}

The Voigt profile model must be convolved with a model for the instrumental
profile in order to obtain a model which can be compared with the
observed spectrum. In the case where all exposures which contribute
to a spectrum are taken with the same slit width, and the quasar image
fills the spectrograph slit uniformly, then the instrumental profile
will be well-described by a Gaussian. The velocity width of this Gaussian
can be determined from the ThAr spectrum. However, in the case where
exposures are taken with different slit widths, or where the seeing
fluctuates such that in some exposures the seeing is significantly
better than the slit width, then the instrumental profile is difficult
to model accurately. 

We have assumed that the instrumental profile is Gaussian, with a
velocity FWHM of $6\,\mathrm{km\, s^{-1}}$ for our initial analysis
of Q0405$-$443, Q0347$-$383 and Q0528$-$250 (Q0528:A). For our
analysis of Q0528:B2, we have used an instrumental FWHM of $5.45\,\mathrm{km\, s^{-1}}$,
which appears to better reflect the observed profile in that spectrum.
Small errors made in determining the instrumental resolution will
necessarily lead to inaccuracy in modelling the spectrum. However,
because the Voigt profile is symmetric, these errors should not significantly
bias $\Delta\mu/\mu$ if a sufficiently large number of molecular
hydrogen transitions are used.

\subsection{Comments on VPFIT}

We have used a modified version of \textsc{vpfit} v9.5\index{VPFIT}
to perform our analysis of the molecular hydrogen data. Early investigations
suggested to us that \textsc{vpfit} was not adequately converging
for the full fits, which contain thousands of free parameters. We
modified \textsc{vpfit} to augment the existing Gauss-Newton optimisation
algorithm with the Levenberg-Marquardt algorithm, and found that this
produced reliable convergence. We describe this further in section
\ref{sub:Optimisation-theory} in the context of the results of that
chapter. We are grateful to R. Carswell for merging our algorithm
into the release version of \textsc{vpfit}. 

\citet{Malec:10} have investigated the convergence of \textsc{vpfit}
when determining $\Delta\mu/\mu$ from the $z=2.059$ absorber toward
J2123$-$0050 using Monte Carlo methods applied to synthetically generated
spectra. They find under 420 different realisations of a noisy spectrum
that \textsc{vpfit} returns the correct value of $\Delta\mu/\mu$
with an appropriate statistical uncertainty. The noise was generated
such that $\sigma$ was $0.8$ times the error array at each pixel.
The use of 0.8 rather than 1.0 was to ensure that marginally required
Lyman-$\alpha$ blends were always required in the simulated spectra.
We therefore believe that the parameter estimates and uncertainties
produced by \textsc{vpfit} here are likely to be reasonable.

\section{Results\label{sec:mu:results}}

\subsection{Description of the absorbers}

The absorber at $z=3.025$ toward Q0347$-$383\index{Q0347$-$383}
appears to be well modelled by a single H$_{2}$ velocity component.
The absorption system at $z=2.595$ toward Q0405$-$443\index{Q0405$-$443}
contains one main velocity component, with another weaker component.
The two components are separated by $\approx13$ km~s$^{-1}$ in
velocity space. However, many of its transitions are weak or heavily
blended, and so we have not utilised the second component. Where the
weak component is observed, we modelled it as H \iscs in order to
ensure that the observed spectral features are accounted for. This
also has the advantage of placing our analysis of this absorber on
a comparable basis to that of \citet{Reinhold:06-1}, who also analysed
only the strong component. The structure of the absorber toward Q0528$-$250
is described below.

\subsubsection{Velocity structure of H$_{2}$ in Q0528$-$250 \label{sub:mu H2 velocity structure}}

\begin{figure}[tbph]
\noindent \begin{centering}
\includegraphics[bb=78bp 59bp 533bp 750bp,clip,angle=-90,width=1\textwidth]{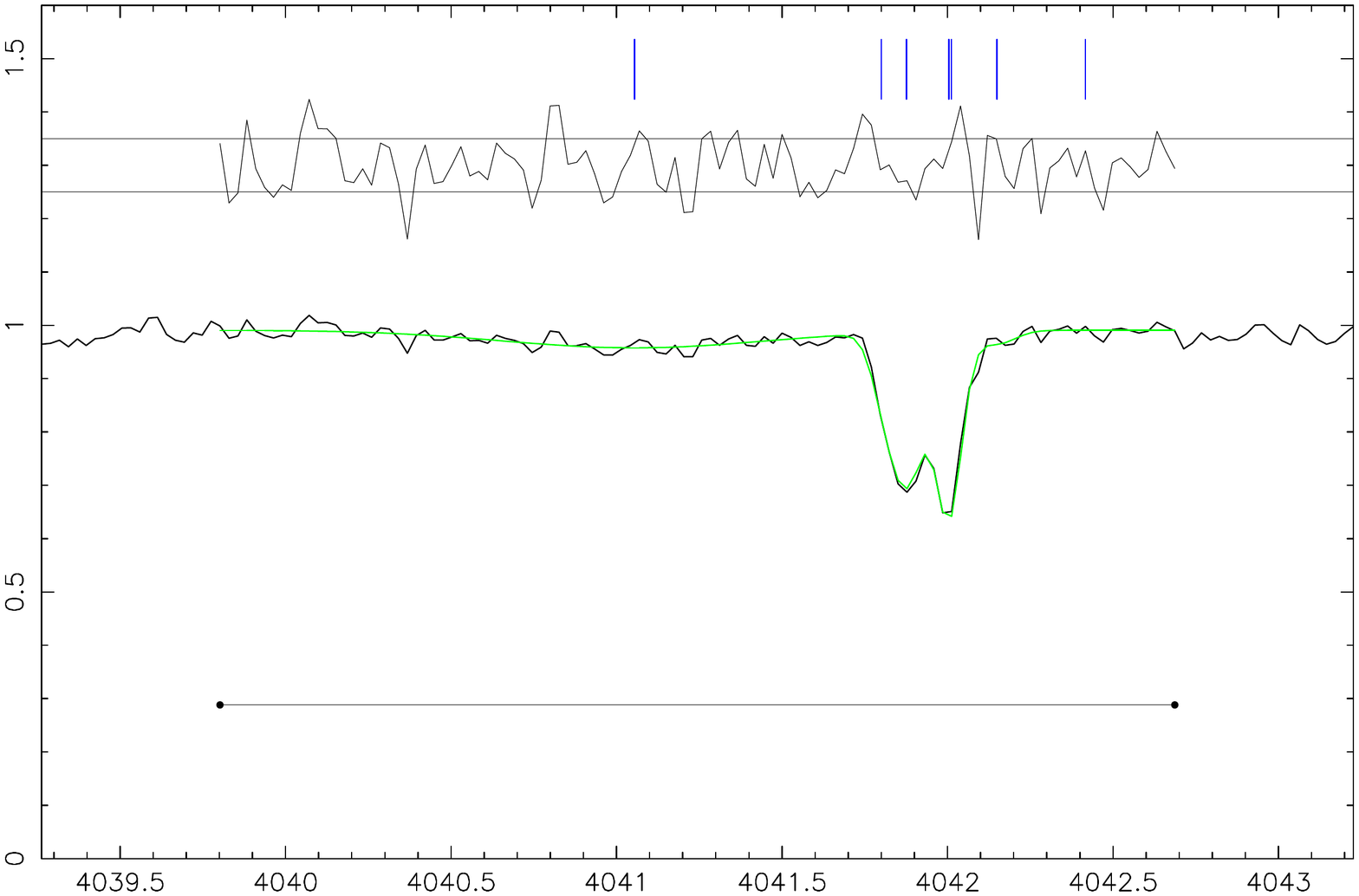}
\par\end{centering}

\caption[Demonstration of velocity structure in Q0528$-$250]{Demonstration of velocity structure in Q0528$-$250. The transition at $\sim4042\AA$ is a molecular hydrogen transition. The presence of at least two components is clearly demonstrated visually. The existence of more components must be determined through appropriate statistical techniques. The green line is a model fitted to the data (in black). \label{Flo:mu:Q0528 structure demonstration}}
\end{figure}

The system toward Q0528$-$250\index{Q0528$-$250} presents with complex
structure. We show an exemplary $J=4$ molecular hydrogen transition
in figure \ref{Flo:mu:Q0528 structure demonstration}. \citet{Ledoux:03}
reported the detection of multiple velocity components, and \citet{Srianand:2005}
modelled the absorber with two components. Two components are plainly
visible as a substantial asymmetry in every line (see figure \ref{Flo:mu:Q0528 structure demonstration}).
We have tried modelling the absorber toward Q0528$-$250 with 2, 3
and 4 velocity components. 

\begin{figure}[tbph]
\noindent \begin{centering}
\includegraphics[bb=50bp 51bp 558bp 766bp,clip,angle=-90,width=1\textwidth]{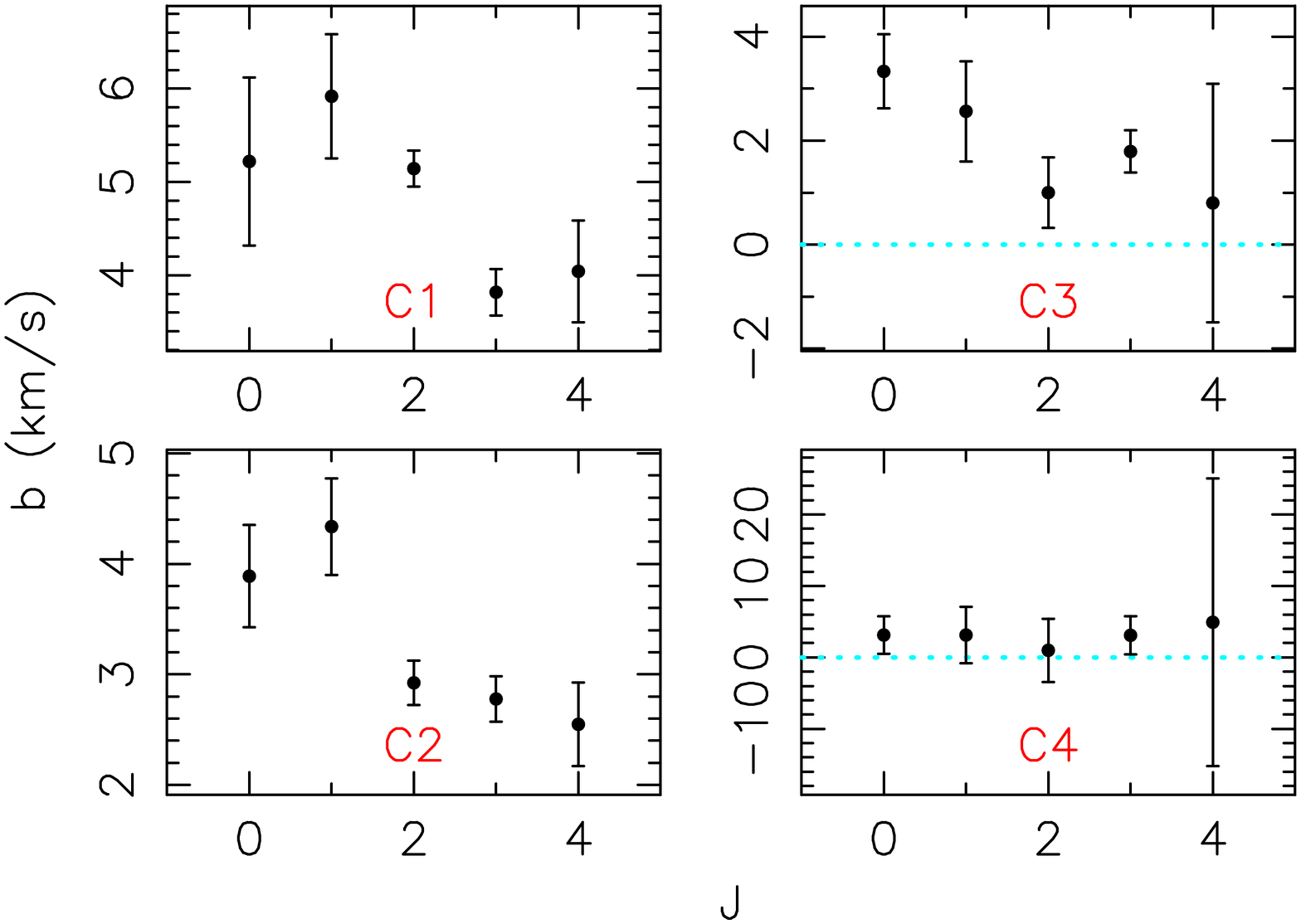}
\par\end{centering}

\caption[Relationship of $b$ with $J$ for the $z=2.811$ absorber toward Q0528$-$250]{Relationship of $b$ with $J$ for the four components of the H$_2$ fit for the $z=2.811$ absorber toward Q0528$-$250. The panels C1 through C4 show the relationship for the four components of the fit in a model where transitions with different $J$ are allowed to have different $b$ parameters. The statistical uncertainties given are derived from the covariance matrix of the fit. $b$ cannot be Gaussian if $b/\sigma_b \lesssim 1$, and so the fact that some error bars overlap with $b=0$ is of no actual consequence. A more thorough investigation would explore the actual confidence region, but that is not necessary for these purposes. The panel C4, corresponding to the weakest (4th) component reveals no obvious trend of $b$ with $J$, but this is expected due to the low column density (that is, the statistical errors are large). However, in panels C1 through C3 a trend is noted that for transitions with higher $J$ the $b$-parameters are smaller. This is clearly seen in the spectra, where the velocity structure is more obvious for higher $J$ transitions. The results here imply that forcing all transitions for a particular component to have the same $b$ parameter may not be physically realistic. Such an assumption could lead to errors in determining $\Delta\mu/\mu$. \label{Flo:mu:Q0528 b vs J}}
\end{figure}

To determine whether more than two velocity components were required,
we firstly considered the AICC\index{Q0528$-$250!velocity structure}.
A particular model with three H$_{2}$ components compared to a model
with two has $\Delta\mathrm{AICC}=-120.0$. That is, the three component
model is very strongly preferred over the two component model. A similar
model with four H$_{2}$ components compared to a model with three
has $\Delta\mathrm{AICC=-20.9}$, which again indicates that the model
with four components is strongly preferred. Using the $F$-test, the
probability that the reduction in $\chi^{2}$ from using three components
instead of two is due to chance is $p=4\times10^{-18}$. Comparing
a model with four velocity components to one with three gives $p=1.8\times10^{-8}$.
This statistical evidence suggests that a model with four velocity
components is appropriate. The use of a five component model produced
a fit that was highly unstable, by which we mean that some of the
H$_{2}$ components were rejected by \textsc{vpfit} as being statistically
unnecessary. As a result, we used the four component model as our
primary model. 

Notwithstanding the significant statistical evidence for four components,
it is interesting to consider the ``per-transition'' $\Delta\mathrm{AICC}$.
With $n=64$ transitions, $\Delta\mathrm{AICC}_{2\rightarrow3}/n=-1.88$
and $\Delta\mathrm{AICC}_{3\rightarrow4}/n=-0.33$. Using the Jeffreys'
scale \citep{Jeffreys:1961}, then only $\sim6$ average transitions
are necessary to conclude that there is very strong evidence ($\Delta\mathrm{AICC<-10})$
for three components. However to conclude that there is very strong
evidence for four components, one needs $\sim30$ average transitions.
Thus it is clear that a large amount of spectral data is required
to detect the fourth component. 

We note that the strength of the statistical evidence for 3 and 4
components depends on a number of factors, including the correctness
of the flux uncertainties and the choice of the correct instrumental
resolution. Therefore, the true statistical evidence after considering
unmodelled uncertainties is necessarily smaller. 

We also noted for this absorber that transitions of increasing $J$
appear to have smaller $b$ parameters; we show this effect in figure
\ref{Flo:mu:Q0528 b vs J}. In the spectrum, this has the effect of
making the velocity structure more pronounced for transitions with
higher $J$. For this reason, in deriving our estimates on $\Delta\mu/\mu$
we allow for transitions of different $J$ to have different $b$
parameters.

\subsection{Transitions used \& fits}

\begin{figure}[tbph]
\noindent \begin{centering}
\includegraphics[bb=50bp 82bp 478bp 789bp,clip,angle=-90,width=1\textwidth]{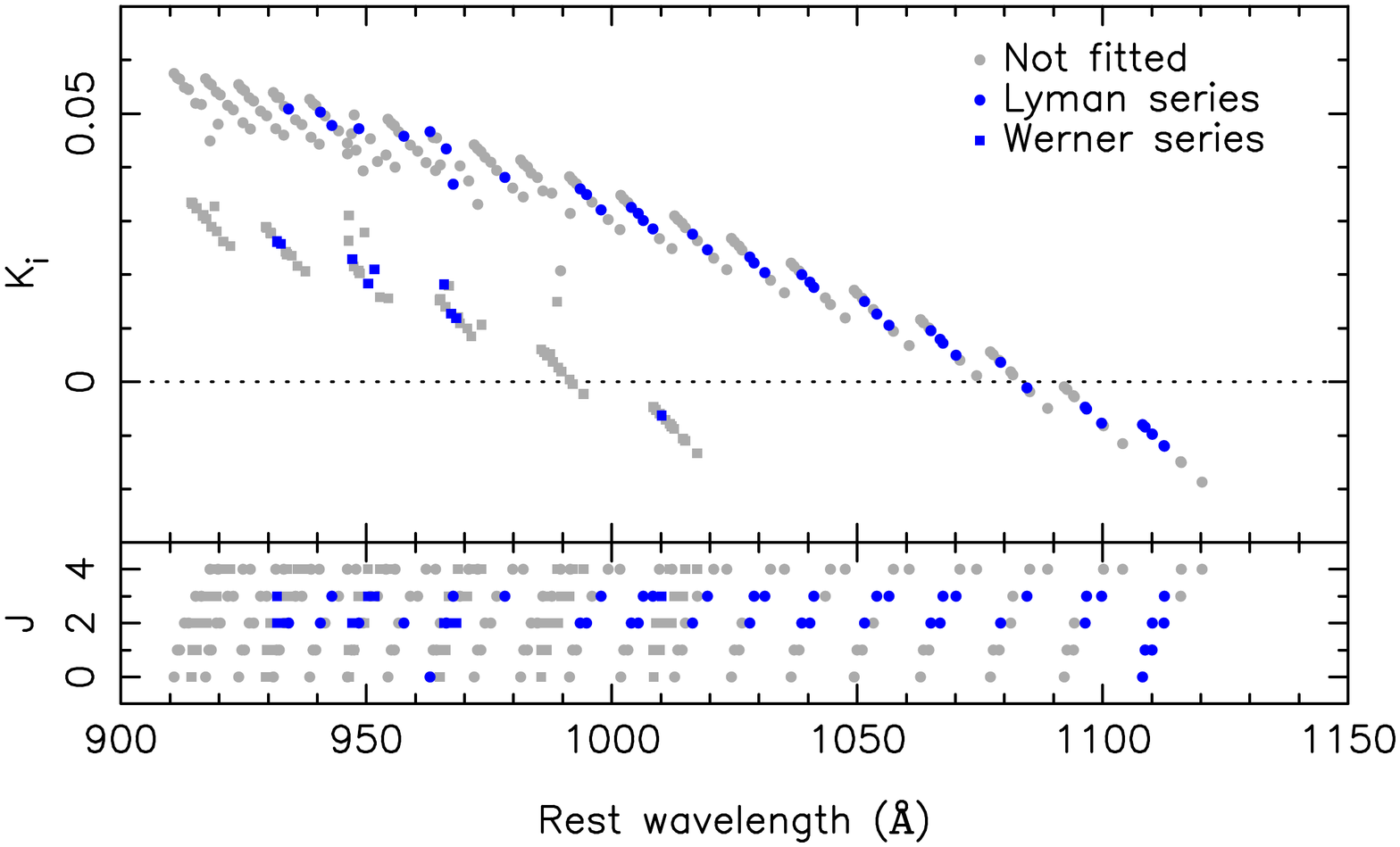}
\par\end{centering}

\caption[Relationship of $K_i$ and $J$ with $\lambda_0$ for the transitions used in Q0405$-$443]{Relationship of $K_i$ and $J$ with $\lambda_0$ for the transitions used in Q0405$-$443. \emph{Upper panel}: the sensitivity coefficients, $K_i$, for the transitions used in our analysis of Q0405$-$443 (dark blue points) and not detected or not fitted (grey points). \emph{Lower panel}: the distribution of transitions with wavelength according to their $J$-level. \label{Flo:mu:Q0405_lambdaki}}
\end{figure}

\begin{figure}[tbph]
\noindent \begin{centering}
\includegraphics[bb=50bp 82bp 478bp 789bp,clip,angle=-90,width=1\textwidth]{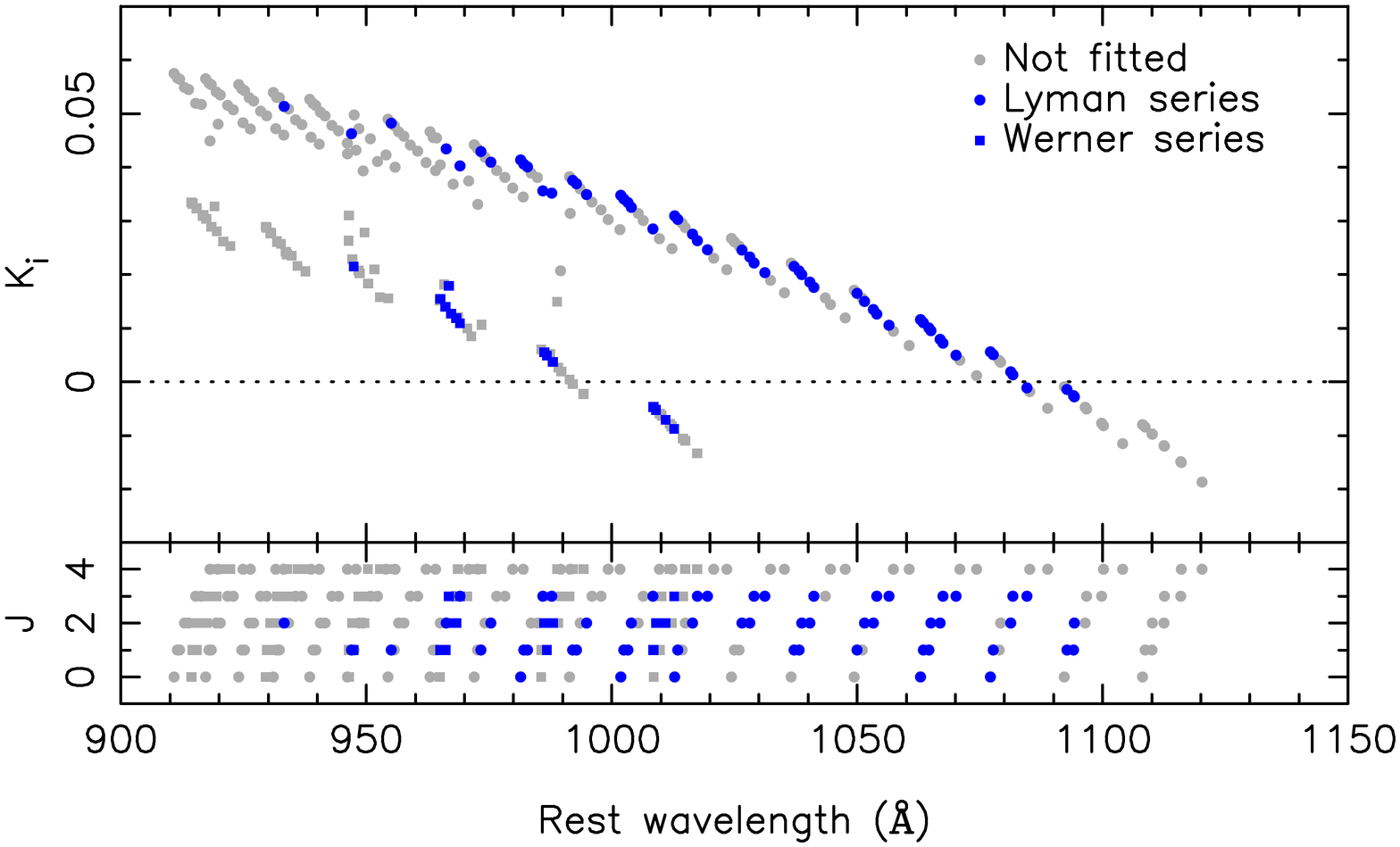}
\par\end{centering}

\caption[Relationship of $K_i$ and $J$ with $\lambda_0$ for the transitions used in Q0347$-$383]{Relationship of $K_i$ and $J$ with $\lambda_0$ for the transitions used in Q0347$-$383. \emph{Upper panel}: the sensitivity coefficients, $K_i$, for the transitions used in our analysis of Q0347$-$383 (dark blue points) and not detected or not fitted (grey points). \emph{Lower panel}: the distribution of transitions with wavelength according to their $J$-level. \label{Flo:mu:Q0347_lambdaki}}
\end{figure}

\begin{figure}[tbph]
\noindent \begin{centering}
\includegraphics[bb=50bp 82bp 478bp 789bp,clip,angle=-90,width=1\textwidth]{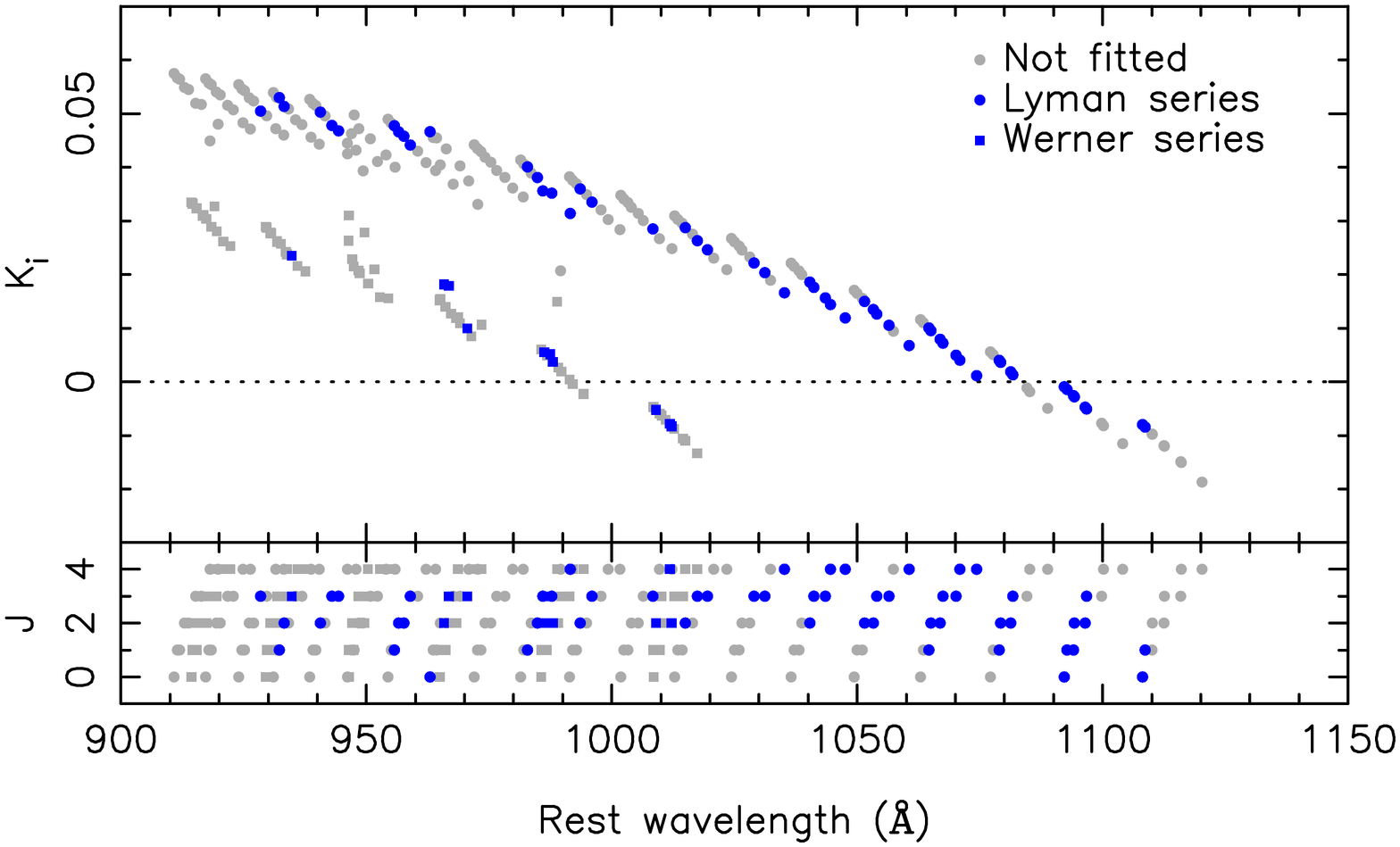}
\par\end{centering}

\caption[Relationship of $K_i$ and $J$ with $\lambda_0$ for the transitions used in Q0528$-$250(A)]{Relationship of $K_i$ and $J$ with $\lambda_0$ for the transitions used in Q0528$-$250(A). \emph{Upper panel}: the sensitivity coefficients, $K_i$, for the transitions used in our analysis of Q0528$-$250(A) (dark blue points) and not detected or not fitted (grey points). \emph{Lower panel}: the distribution of transitions with wavelength according to their $J$-level. \label{Flo:mu:Q0528_lambdaki}}
\end{figure}

We present in table \ref{Tab:mu:3 quasars transitions used} a list
of the transitions used in each of the quasar fits. In figures \ref{Flo:mu:Q0405_lambdaki},
\ref{Flo:mu:Q0347_lambdaki} and \ref{Flo:mu:Q0528_lambdaki} we show
the distribution of the $K_{i}$ values and $J$-levels with rest
wavelength for the transitions used in our analysis of Q0405$-$443,
Q0347$-$383 and Q0528$-$250 respectively. The Voigt profile fits
to Q0405$-$443, Q0347$-$383 and Q0528$-$250 may be found in appendices
\ref{cha:mu fits:Q0405}, \ref{cha:mu fits:Q0347} and \ref{cha:mu fits:Q0528}
respectively. 

\begin{table}[tbph]
\caption[Transitions used in fit for Q0405$-$443, Q0347$-$383 and Q0528$-$250]{Transitions used in our fits for Q0405$-$443, Q0347$-$383 and Q0528$-$250. $n$ gives the number of transitions used.\label{Tab:mu:3 quasars transitions used}}

\noindent \centering{}%
\begin{tabular}{ccc>{\centering}p{0.6\textwidth}}
\hline 
Quasar spectrum & $z_{\mathrm{abs}}$ & $n$ & Transitions used\tabularnewline
\hline 
\noalign{\vskip\doublerulesep}
Q0405$-$443 & 2.595 & 52 & L0P1, L0P2, L0R0, L0R1, L0R2, L0R3, L1P2, L1P3, L1R3, L2P3, L2R2,
L3P2, L3P3, L3R2, L3R3, L4P3, L4R2, L4R3, L5P2, L5R2, L5R3, L6P2,
L6P3, L6R3, L7P2, L7P3, L8P2, L8P3, L8R2, L8R3, L9P2, L9P3, L9R2,
L11P3, L12P2, L12R0, L12R3, L13P2, L14R2, L15R2, L15R3, L16P2, W0R3,
W2P2, W2R2, W2Q2, W3R2, W3Q3, W4P2, W4R3, W4Q2\tabularnewline
\noalign{\vskip\doublerulesep}
Q0347$-$383 & 3.025 & 68 & L1P2, L1R1, L1R2, L2P2, L2P3, L2R0, L2R1, L2R3, L3P1, L3P2, L3P3,
L3R0, L3R1, L3R2, L3R3, L4P2, L4P3, L4R1, L4R2, L4R3, L5P1, L5P2,
L5R1, L5R3, L6P2, L6P3, L6R2, L6R3, L7P2, L7P3, L7R0, L7R1, L7R3,
L8P1, L8P3, L8R0, L8R1, L8R2, L9P1, L9R1, L9P2, L10P1, L10P3, L10R0,
L10R1, L10R3, L11P1, L11P2, L12P2, L12P3, L13R1, L14R1, L16R2, W0R1,
W0R2, W0Q2, W0Q3, W1R2, W1Q1, W1Q2, W2P2, W2R1, W2R3, W2Q2, W2Q1,
W2Q3, W3Q1\tabularnewline
\noalign{\vskip\doublerulesep}
Q0528$-$250 & 2.811 & 64 & L0R0, L0R1, L1P1, L1P2, L1R0, L1R1, L1R2, L1R3, L2P1, L2P2, L2R2,
L2R3, L3P1, L3P2, L3P3, L3P4, L3R2, L3R3, L3R4, L4P2, L4P3, L4P4,
L4R2, L4R3, L5P2, L5P3, L5P4, L5R3, L5R4, L6P3, L6P4, L6R3, L7P3,
L7R2, L7R3, L8P3, L9R2, L9R3, L10P1, L10P2, L10P3, L10P4, L10R3, L12R0,
L13P1, L13P2, L13R2, L13R3, L15P3, L15R2, L15R3, L16P1, L16R2, L17R3,
W0P2, W0R2, W0R4, W1R3, W1Q2, W2P3, W2R2, W2R3, W4P3\tabularnewline
\hline 
\noalign{\vskip\doublerulesep}
\end{tabular}
\end{table}

\subsection{Results for Q0405$-$443, Q0347$-$383, and Q0528$-$250\index{Q0405$-$443}\index{Q0347$-$383}\index{Q0528$-$250}\label{sub:mu:first results}}

We present in tables \ref{Tab:mu:DCMM results 3 quasars} and \ref{Tab:mu:RRM results 2 quasars}
the results of the DCMM and RRM respectively applied to the absorbers
in the spectra of Q0405$-$443, Q0347$-$383, and Q0528$-$250 (Q0528:A). 

We note that the use of the DCMM results in a substantial reduction
in the number of free parameters compared to the RRM. For Q0405$-$443,
the DCMM yields 51 fewer parameters and for Q0347$-$383 it yields
67 fewer parameters. Our preferred result is that from a weighted
mean of the DCMM results, which yields $\Delta\mu/\mu=(2.6\pm3.0)\times10^{-6}$,
compared with $\Delta\mu/\mu=(24\pm6)\times10^{-6}$ from \citet{Reinhold:06-1}.
We prefer the results from the DCMM over the RRM for the reasons given
in section \ref{sub:mu:how_measure_dmu}. We therefore find that our
results are inconsistent with those from \citeauthor{Reinhold:06-1},
and are unable to produce a non-zero result. It is difficult to determine
whether the three data points are consistent about the weighted mean,
as the $\chi^{2}$ test has low statistical power to reject consistency
for small $\nu$. Nevertheless, $\chi_{\nu}^{2}=1.47$ for the three
DCMM data points about the weighted mean. A value of $\chi_{\nu}^{2}$
this large or larger has a probability $p=0.48$ of occurring by chance.
Thus, we can say that our results appear to be consistent --- at least
under a weighted mean model --- with no evidence for excess scatter
due to unmodelled systematic effects. 

\begin{sidewaysfigure}
\noindent \begin{centering}
\includegraphics[bb=89bp 42bp 571bp 739bp,clip,angle=-90,width=0.75\textwidth]{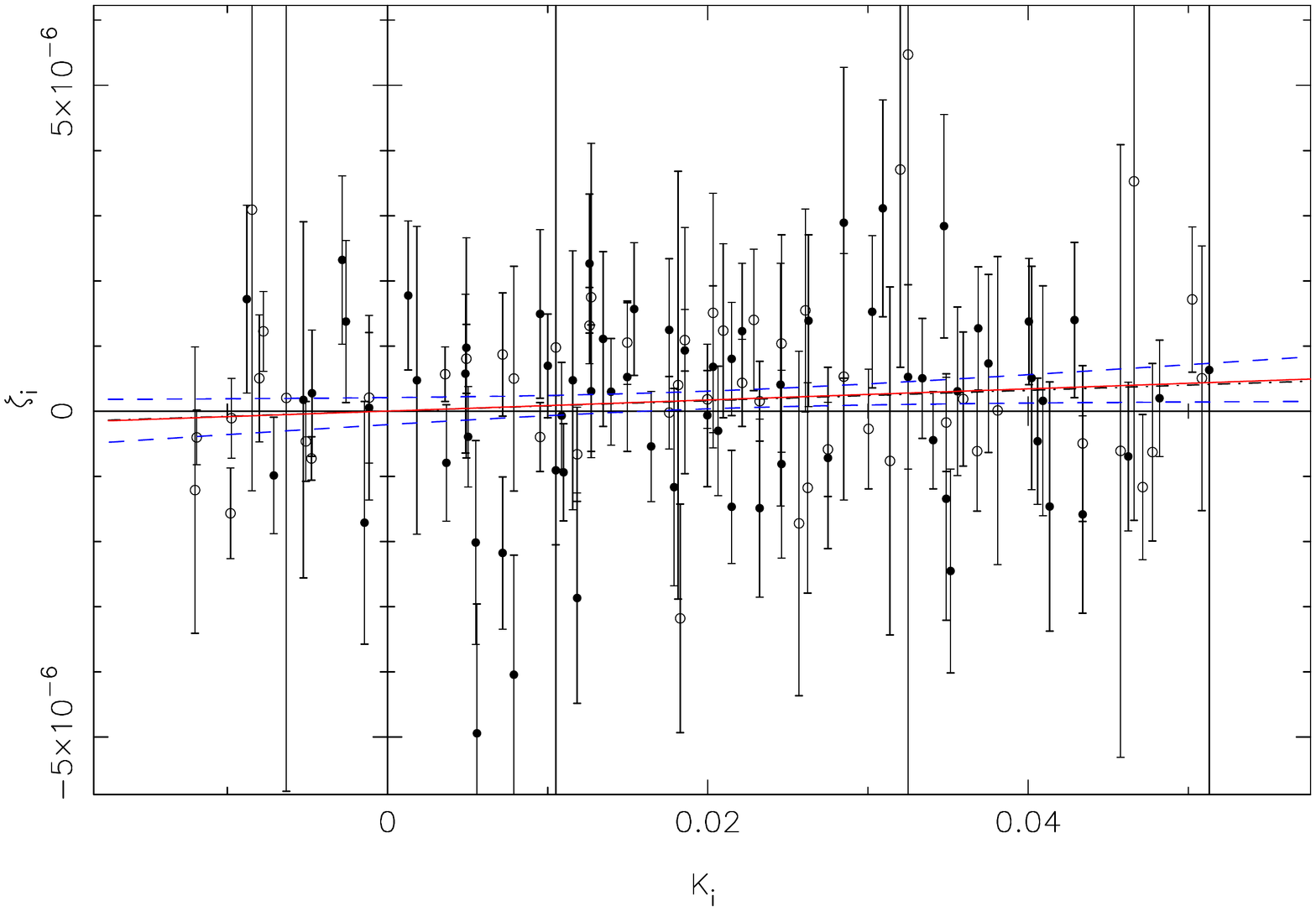}
\par\end{centering}

\caption[Reduced redshift plot for Q0405$-$443 and Q0347$-$383]{Reduced redshift plot ($\zeta_i$ vs $K_i$ for Q0405$-$443 and Q0347$-$383, with gradient $\Delta\mu/\mu=(8.5 \pm 5.7)\times10^{-6}$ (red line). Q0405$-$443 is represented by open circles, whilst Q0347$-$383 is represented by closed circles. The unweighted fit, a dot-dashed line, is obscured by the weighted fit. The dashed, blue lines give the $1\sigma$ confidence limit on the regression line. Note that this is not our preferred result because it does not include Q0528$-$250 in the fit. This graph may be directly compared with figure 2 of \citet{Reinhold:06-1}. \label{Flo:mu:reduced redshift plot}} 
\end{sidewaysfigure}

We show in figure \vref{Flo:mu:reduced redshift plot} a reduced redshift
plot for Q0405$-$443 and Q0347$-$383, which has gradient $\Delta\mu/\mu=(8.5\pm5.7)\times10^{-6}$. 

\begin{table}[tbph]
\caption[DCMM results for Q0405$-$443, Q0347$-$383 and Q0528$-$250]{Direct $\chi^2$ minimisation method (DCMM) constraints on $\Delta\mu/\mu$ for Q0405$-$443, Q0347$-$383 and Q0528$-$250. $n$ is the number of transitions. $\chi^2_\nu$ is the reduced $\chi^2$ of the spectral data about the Voigt profile model. The weighted mean is also given. \label{Tab:mu:DCMM results 3 quasars}}

\noindent \centering{}%
\begin{tabular}{ccccc}
\hline 
Quasar spectrum & $\Delta\mu/\mu$ --- DCMM & $\chi_{\nu}^{2}$ & $z_{\mathrm{abs}}$ & n\tabularnewline
\hline 
\noalign{\vskip\doublerulesep}
Q0405$-$443 & $(10.1\pm6.6)\times10^{-6}$ & 1.42 & 2.595 & 52\tabularnewline
\noalign{\vskip\doublerulesep}
Q0347$-$383 & $(8.2\pm7.5)\times10^{-6}$ & 1.28 & 3.025 & 68\tabularnewline
\noalign{\vskip\doublerulesep}
Q0528$-$250(A) & $(-1.4\pm3.9)\times10^{-6}$ & 1.22 & 2.811 & 64\tabularnewline
\noalign{\vskip\doublerulesep}
Weighted mean & $(2.6\pm3.0)\times10^{-6}$ & n/a & 2.81 & n/a\tabularnewline
\hline 
\noalign{\vskip\doublerulesep}
\end{tabular}
\end{table}

\begin{table}[tbph]
\caption[RRM results for Q0405$-$443 and Q0347$-$383]{Reduced redshift method (RRM) constraints on $\Delta\mu/\mu$ for Q0405$-$443 and Q0347$-$383 and Q0528$-$250. $n$ is the number of transitions. The RRM cannot be applied to Q0528$-$250 for reasons set out in the text. $\chi^2_\nu$ gives the reduced $\chi^2$ about the linear fit. \label{Tab:mu:RRM results 2 quasars}}

\noindent \centering{}%
\begin{tabular}{ccccc}
\hline 
Quasar spectrum & $\Delta\mu/\mu$ --- RRM & $\chi_{\nu}^{2}$ & $z_{\mathrm{abs}}$ & n\tabularnewline
\hline 
\noalign{\vskip\doublerulesep}
Q0405$-$443 & $(10.9\pm7.1)\times10^{-6}$ & 1.01 & 2.595 & 52\tabularnewline
\noalign{\vskip\doublerulesep}
Q0347$-$383 & $(6.4\pm10.3)\times10^{-6}$ & 1.13 & 3.025 & 68\tabularnewline
\noalign{\vskip\doublerulesep}
Q0405 + Q0347 & $(8.5\pm5.7)\times10^{-6}$ & 1.06 & 2.811 & 120\tabularnewline
\noalign{\vskip\doublerulesep}
Q0528$-$250(A) & n/a & n/a & n/a & n/a\tabularnewline
\hline 
\noalign{\vskip\doublerulesep}
\end{tabular}
\end{table}

We note that our in our RRM linear fit to $\zeta_{i}$ against $K_{i}$,
the $\zeta_{i}$ values demonstrate good consistency with the linear
model, with $\chi_{\nu}^{2}=1.01$ and 1.13 for Q0405$-$443 and Q0347$-$383
respectively, and $\chi_{\nu}^{2}=1.06$ for the two data sets fitted
together. This contrasts with the results of \citet{Reinhold:06-1},
where they found that $\chi_{\nu}^{2}=2.1$. The smaller $\chi_{\nu}^{2}$
in our case is likely to arise from a combination of: \emph{i) }better
wavelength calibration in our spectra compared to the spectra of \citeauthor{Reinhold:06-1};
\emph{ii) }the simultaneous fitting of the forest with the H$_{2}$
transitions, which must increase the uncertainties on the redshifts
of each H$_{2}$ transition; and, \emph{iii) }the fact that the flux
uncertainties in regions of low flux are under-estimated, and not
corrected for, in the data of \citeauthor{Reinhold:06-1}, which means
that the statistical uncertainties on redshifts will be under-estimated.
It is reassuring that $\chi_{\nu}^{2}\approx1$ for our RRM fits.

\subsubsection{Bootstrap verification}

\begin{figure}[tbph]
\noindent \begin{centering}
\includegraphics[bb=106bp 47bp 566bp 722bp,clip,angle=-90,width=1\textwidth]{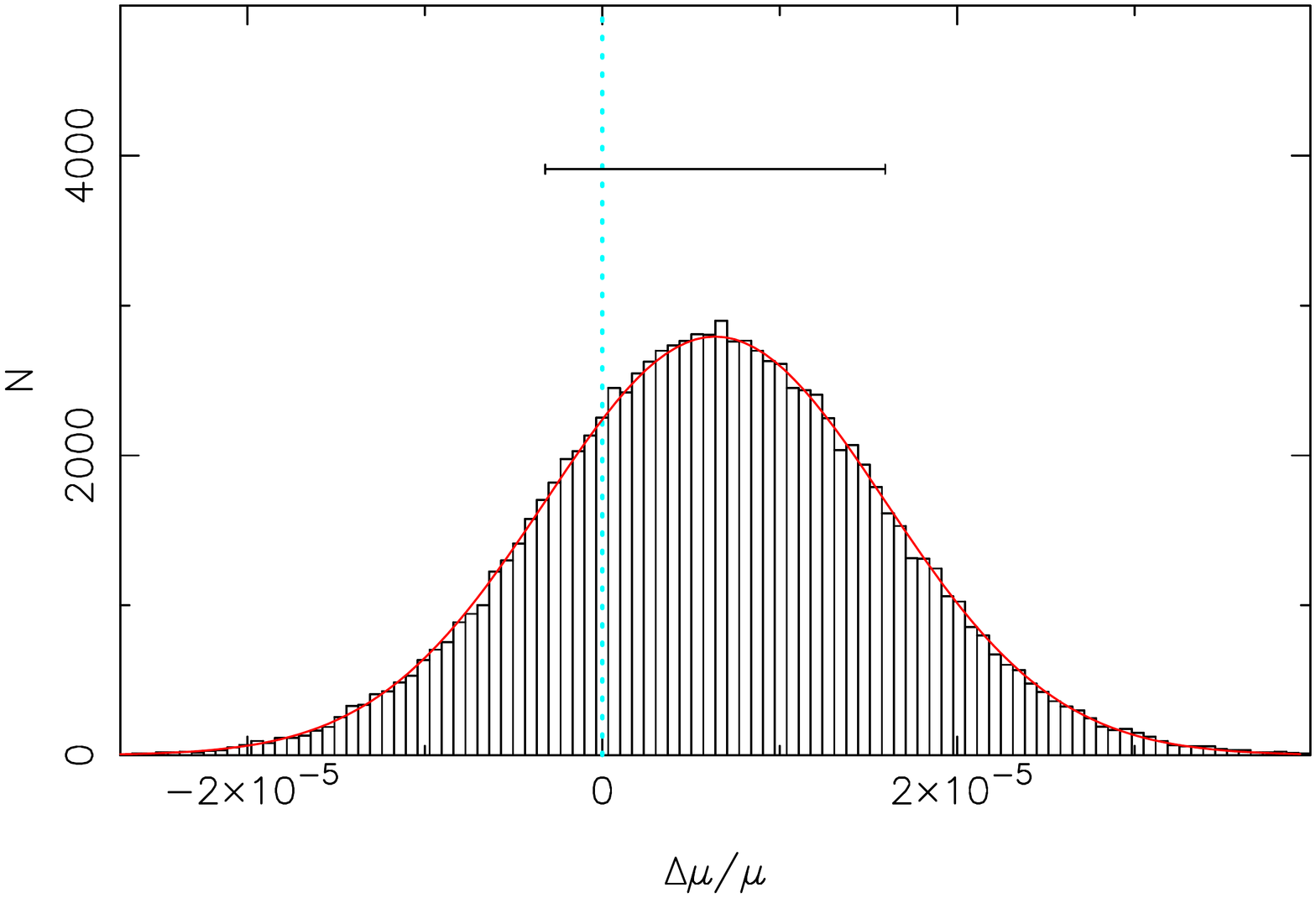}
\par\end{centering}

\caption[Bootstraped $\Delta\mu/\mu$ for RRM for the $z=3.025$ absorber toward Q0347$-$383]{Bootstrapped distribution of $\Delta\mu/\mu$ for the reduced redshift method (RRM) for the $z=3.025$ H$_2$ absorber toward Q0347$-$383. The histogram shows the distribution of values of $\Delta\mu/\mu$ from $10^5$ resamplings. The red curve shows a theoretical Gaussian distribution with the same mean and standard deviation as the bootstrapped samples. The horizontal bar indicates the $1\sigma$ confidence interval from the Gaussian. The good agreement between the curve and the histogram shows that the results are well described by a Gaussian with $\Delta\mu/\mu = (6.4\pm9.6)\times 10^{-6}$. \label{Flo:mu:Q0347 bootstrap}}
\end{figure}

\begin{figure}[tbph]
\noindent \begin{centering}
\includegraphics[bb=106bp 47bp 566bp 748bp,clip,angle=-90,width=1\textwidth]{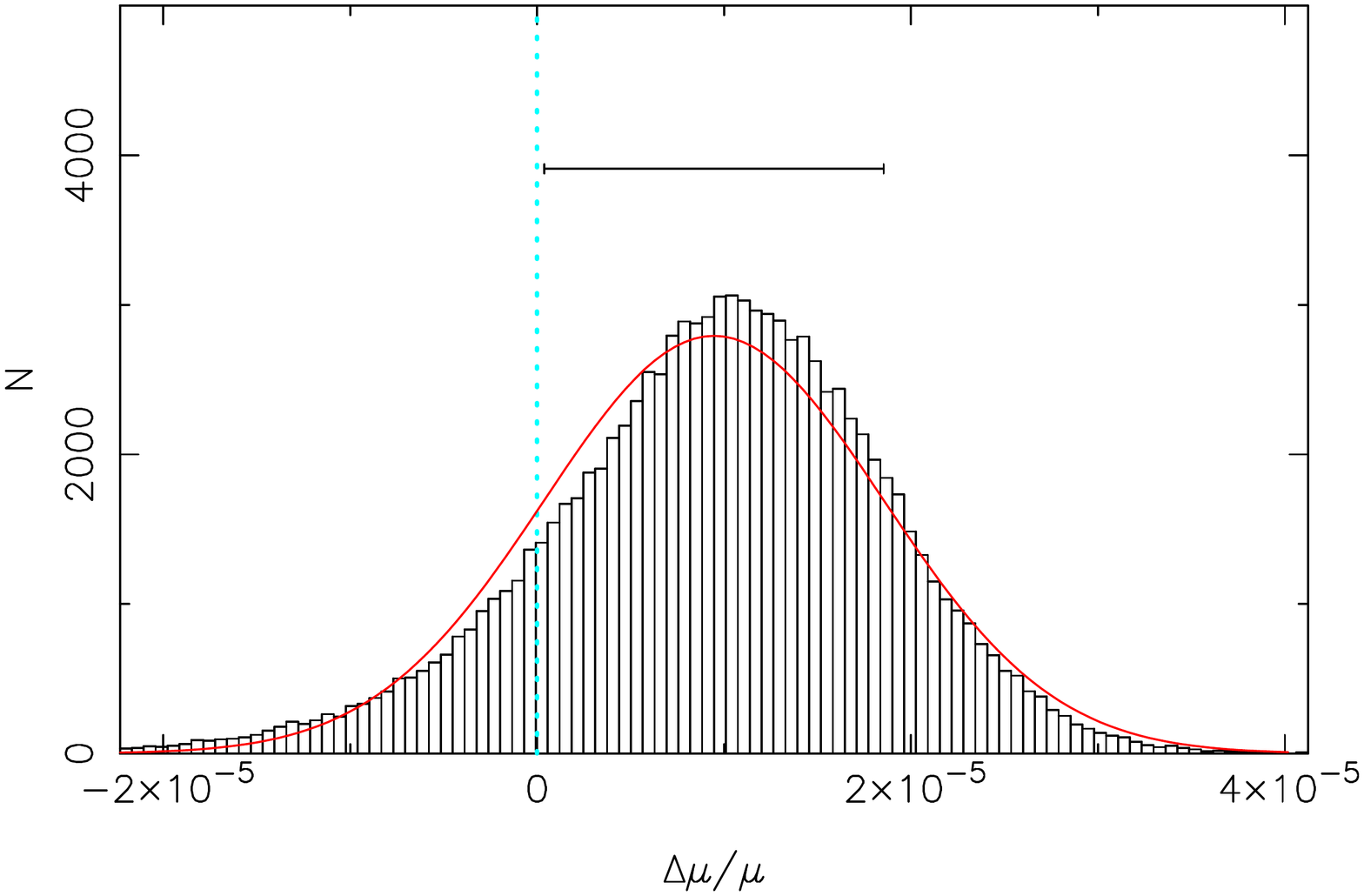}
\par\end{centering}

\caption[Bootstraped $\Delta\mu/\mu$ for RRM for the $z=2.595$ absorber toward Q0405$-$443]{Bootstrapped distribution of $\Delta\mu/\mu$ for the reduced redshift method (RRM) for the $z=2.595$ H$_2$ absorber toward Q0405$-$443. The histogram shows the distribution of values of $\Delta\mu/\mu$ from $10^5$ resamplings. The red curve shows a theoretical Gaussian distribution with the same mean and standard deviation as the bootstrapped samples. The horizontal bar indicates the $1\sigma$ confidence interval from the Gaussian. The bootstrapped distribution is mildly left-skewed, with excess skewness $\sim -0.4$. The Gaussian is described by $\Delta\mu/\mu = (9.5\pm9.1)\times 10^{-5}$; the actual $1\sigma$ confidence limits from the boostrap is $\Delta\mu/\mu \in [0.7,18.2]\times 10^{-6}$. The mode of the distribution coincides approximately with the fittted value given in table \ref{Tab:mu:RRM results 2 quasars}. Further investigation shows that the skewness arises from the presence of a few high statistical precision points at long wavelengths. Given the good agreement beetween empirical confidence limits from the bootstrap and the analytic uncertainty values on $\Delta\mu/\mu$, the mild skewness seen here is of no practical consequence.\label{Flo:mu:Q0405 bootstrap}}
\end{figure}

We have assessed the results of the reduced redshift plot using a
resampling bootstrap method \citep{NumericalRecipes:92}\index{bootstrap!resampling}.
The resampling bootstrap method proceeds as follows: \emph{i) }from
the set $\zeta_{i}$ and $K_{i}$ values for a particular absorber
(with $n$ $\zeta_{i}/K_{i}$ pairs), generate a new set of $\zeta_{i}$
and $K_{i}$ values by drawing $\zeta_{i}$/$K_{i}$ pairs with replacement
from the original set, such that the new set of $\zeta_{i}$/$K_{i}$
pairs has $n$ points; \emph{ii) }calculate $\Delta\mu/\mu$ by fitting
a linear model to $\zeta_{i}$ vs $K_{i}$, and keep this value of
$\Delta\mu/\mu$; \emph{iii)} repeat this process $10^{5}$ times
to obtain $10^{5}$ values of $\Delta\mu/\mu$. The mean and standard
deviation of this ensemble of $\Delta\mu/\mu$ values should be consistent
with the results from the original RRM fit to the $\zeta_{i}$ vs
$K_{i}$ values for that absorber. We show the results of this in
figures \ref{Flo:mu:Q0347 bootstrap} and \ref{Flo:mu:Q0405 bootstrap},
and note good agreement with the fitted values from table \ref{Tab:mu:RRM results 2 quasars}.
The good agreement seen is unsurprising given the reasonable number
of transitions used; the probability distribution of the fitted parameters
should be approximately Gaussian because of the central limit theorem,
combined with the fact that there is not a large range in the magnitudes
of the uncertainty estimates for the redshifts for different transitions.

\subsection{Comparison with the results of Reinhold et al.}

To directly compare our results with that of \citet{Reinhold:06-1}\index{Reinhold et al!comparison with},
we performed an analysis where we utilised the same transitions used
in that work. For Q0405$-$443, this removed 16 transitions and adds
3, the latter of which we initially decided were badly contaminated
and excluded from our main analysis. For Q0347$-$383, we removed
35 transitions and added 4. We used the RRM so as to yield a like-with-like
comparison. The results of this are set out in table \ref{Tab:mu:like with like dmu Reinhold comparison}.
It is difficult to compare our results directly with those of \citeauthor{Reinhold:06-1}
in a statistical fashion, because the results are derived from the
same spectra, and therefore the data are not independent. However,
in both cases we see a shift toward $\Delta\mu/\mu=0$. Although the
inclusion of the result from Q0528$-$250 shifts a combined Q0405$-$443
+ Q0347$-$383 result toward zero, our result based on the same transitions
used by \citeauthor{Reinhold:06-1} is null. 

\begin{table}[tbph]
\caption[Like with like comparison of results for Q0405$-$443 and Q0347$-$383 with Reinhold et al.]{Reduced redshift method (RRM) constraints on $\Delta\mu/\mu$ for Q0405$-$443 and Q0347$-$383 based on the same transitions utilised by \citet{Reinhold:06-1}. The use of the same transitions therefore indicates whether the differences between our results and that of \citeauthor{Reinhold:06-1} are due to the differing transition list or due to some other factor. We see that $\Delta\mu/\mu$ from our fits is significantly different to the results of \citeauthor{Reinhold:06-1} on this basis, and is consistent with our results presented in table \ref{Tab:mu:DCMM results 3 quasars}. Therefore, the difference between our results and that of \citeauthor{Reinhold:06-1} is not due to the transitions used. \label{Tab:mu:like with like dmu Reinhold comparison}}

\noindent \centering{}%
\begin{tabular}{cccc}
\hline 
Quasar spectrum & $z_{\mathrm{abs}}$ & $\Delta\mu/\mu$ --- our analysis & $\Delta\mu/\mu$ --- \citeauthor{Reinhold:06-1}\tabularnewline
\hline 
\noalign{\vskip\doublerulesep}
Q0405$-$443 & 2.595 & $(10.2\pm8.9)\times10^{-6}$ & $(27.8\pm8.8)\times10^{-6}$\tabularnewline
\noalign{\vskip\doublerulesep}
Q0347$-$383 & 3.025 & $(12.0\pm14.0)\times10^{-6}$ & $(20.6\pm7.9)\times10^{-6}$\tabularnewline
\noalign{\vskip\doublerulesep}
Combined result & 2.81 & $(10.7\pm7.5)\times10^{-6}$ & $(23.9\pm5.9)\times10^{-6}$\tabularnewline
\hline 
\noalign{\vskip\doublerulesep}
\end{tabular}
\end{table}

As noted in section \ref{sub:mu:General H2 comments}, the modelling
of the Lyman-$\alpha$ forest should not cause appreciable deviations
from a more simplistic treatment of the background continuum over
a large sample of transitions. We experimented with this by performing
simple, linear fits to the background Lyman-$\alpha$ flux in the
vicinity of certain molecular hydrogen transitions and found this
to be true. Therefore, we do not believe that the deviation of the
result of \citeauthor{Reinhold:06-1} from ours is due to inadequate
modelling of the forest. Instead, we ascribe the difference between
our results to a combination of the better wavelength calibration
of our spectral data and the fact that the \citeauthor{Reinhold:06-1}
result is dominated by a few points with $K_{i}<0$ which have particularly
small error bars.

\section{Discussion of results}

\subsection{Further investigations of \citet{Reinhold:06-1}.\label{sub:Further-investigations-Reinhold}}

Other researchers have attempted to replicate the findings of \citeauthor{Reinhold:06-1}\index{Reinhold et al!other comparisons with}:
\begin{itemize}
\item \citet{Wendt:08a} examined the Q0347$-$383 and Q0405$-$443 absorbers
to investigate the results of \citeauthor{Reinhold:06-1}, and found
that $|\Delta\mu/\mu|<4.9\times10^{-5}$ (95 percent confidence).
\item \citet{Thompson:09a} also analysed the Q0347$-$383 and Q0405$-$443
absorbers to investigate the results of \citeauthor{Reinhold:06-1}
and those of \citet{King:08} (the results presented above). They
found that $\Delta\mu/\mu=(-7\pm8)\times10^{-6}$, which is inconsistent
with the results of \citeauthor{Reinhold:06-1}
\item \citet{Wendt:10a} re-investigated the Q0347$-$383 absorber, using
additional data from program ID 68.B-0115(A), taken in 2002. Rather
than co-adding spectra, as is traditionally done, they fitted each
of the exposures simultaneously. For each exposure, they determined
a velocity offset with respect to the other exposures by maximising
spectral cross-correlation. The exposures were then shifted onto a
common wavelength scale. They noted an average inter-spectrum wavelength
deviation of $2.3\mathrm{m\AA}$ or $170\,\mathrm{ms^{-1}}$ at $4000\mathrm{\AA}$.
The Ly-$\alpha$ absorption against which the H$_{2}$ transitions
is seen was fitted as a polynomial. From an initial set of 52 lines,
they examined the effective wavelength difference between the wavelengths
of each transition between the two data sets. They noted that only
36 lines have a difference between the two sets of exposures of less
than $3\sigma$; some lines deviate by more than $5\sigma$. They
rejected lines which differ by more than $3\sigma$, leaving 36 lines
for analysis. They concluded that $\Delta\mu/\mu=(15\pm9_{\mathrm{stat}}\pm6_{\mathrm{sys}})\times10^{-6}$,
where the estimate of the systematic comes from increasing the error
bars to make the fit statistically consistent. They suggest that the
excess scatter seen in a plot of $\zeta_{i}$ vs $K_{i}$ may be due
to the way in which they have approximated the Ly-$\alpha$ flux with
a polynomial, which accords well with our earlier arguments that not
modelling the forest structure appropriately will necessarily cause
under-estimation of the statistical error on $\Delta\mu/\mu$.
\end{itemize}
All of these results are consistent with the analysis of Q0405$-$443
and Q0347$-$383 presented in section \ref{sec:mu:results}, which
is reassuring. The somewhat tighter confidence limits we obtain compared
to these works relate to the fact that these works have attempted
to be more conservative in analysing the spectra. However given that
the $\chi_{\nu}^{2}=1.06$ for our simultaneous RRM fit of Q0405$-$443
and Q0347$-$383 we do not think that our confidence limits on $\Delta\mu/\mu$
are too small.

\subsection{Convergence for Q0528$-$250}

The Q0528$-$250\index{Q0528$-$250!convergence of fit} absorber has
a sufficiently high optical depth that the low-$J$ lines are commonly
saturated. That is, the transitions fall on the flat part of the curve
of growth. The curve of growth describes the change in the equivalent
width of the transition with increasing column density \citep[see for example][and refereinces therein]{Vardavas:93a}.
In this regime, $\chi^{2}$ is relatively insensitive to changes in
the column density, which makes accurate determination of the column
density difficult from a single transition. With many transitions
of different oscillator strengths, one can sample the curve of growth
in many different places and, in principle, obtain a good constraint
on the column density. However, we have left the total column density
(effectively, the oscillator strength) free for each H$_{2}$ transition. 

Because $\chi^{2}$ is relatively flat with respect to $N$ for saturated
transitions, convergence to the final estimate of the total column
density for these transitions can be slow. Thus, for a fit to Q0528$-$250
we generally see relatively fast convergence of most parameters, but
then many tens of iterations for which the change in $\chi^{2}$ is
only a few times larger than the stopping criteria. In these iterations,
$\Delta\mu/\mu$ changes only very slightly, with $\delta(\Delta\mu/\mu)\lesssim0.1\sigma$.
Thus, although convergence for the values of $N$ for some parameters
is slow, the convergence of $\Delta\mu/\mu$ is not affected. This
relates to the fact that it is the line centroiding which determines
$\Delta\mu/\mu$, and the accuracy of the line centroiding should
not be markedly affected by reasonable changes in $N$.

Nevertheless, it would be preferable if the actual oscillator strengths\index{oscillator strength}
could be used, as this would speed convergence. The final result should
also be more reliable because of the use of fewer free parameters.
We tried to modify our existing model for Q0528$-$250 to utilise
the oscillator strengths, and found that in many transitions the model
was a reasonable match to the data. However, in some transitions we
found that the amount of absorption was over-predicted by the model
relative to the data. In general, this means that the local continuum
against which the H$_{2}$ Voigt profile is calculated has been set
too low. Determination of the true continuum is difficult in the forest,
especially in the blue end of the forest, due to the fact that there
are few regions of spectra with no apparent absorption. Given the
relative unimportance of accurate determination of the column density
on the value of $\Delta\mu/\mu$, we therefore simply retained our
model where the total column density for each transition remained
a free parameter, and did not pursue models where the column densities
were constrained by the oscillator strengths. We leave this avenue
to future research.

\section{Q0528$-$250 revisited\label{sub:Q0528-250 revisited}}

The Q0528$-$250\index{Q0528$-$250} constraint presented in section
\ref{sec:mu:results} is extremely precise. However, there are good
reasons to revisit this absorber. Some of the exposures which contribute
to the Q0528:A spectrum are not well-calibrated. By well-calibrated,
we mean that the exposures do not have ThAr calibration spectra taken
in the same observation block. The design of VLT/UVES is such that
the position of the spectrograph grating is reset between different
observation blocks. Although the specification is such that the placement
of the grating should be good to within 0.1 pixels \citep{dOdorico:00a},
the use of spectra for which the ThAr spectra were taken in different
observation blocks necessarily introduces wavelength calibration uncertainties
into the result. Additionally, the velocity structure of the absorber
is clearly complicated; further observations should allow better determination
of the velocity structure. 

A more subtle concern for the earlier observations under program IDs
P66.A-0594, P68.A-0600 and P68.A-0106 is that they suffer from the
fact that the slit width used for the observations was often significantly
larger than the prevailing seeing conditions. For these observations,
a 1 arcsecond slit was used in both the blue and red arms. The average
ratio of the slit width to seeing, where seeing is quantified by the
output of the DIMM%
\footnote{Differential Image Motion Monitor%
} and the seeing values were weighted by the duration of the exposure,
was 0.78. In one exposure of 1.6 hours, the ratio was as low as 0.48.
In the case where the seeing is significantly smaller than the slit
width, the instrumental profile will be non-Gaussian, which complicates
the analysis. Although \textsc{vpfit} allows for a numerically-provided
instrumental profile with which the Voigt profile model is convolved,
we have assumed that the instrumental profile is Gaussian. 

As noted earlier, Q0528$-$250 was re-observed in late 2008/early
2009 under program ID 82.A-0087, with exposures totalling approximately
8.2 hours. All science exposures were taken with ThAr calibration
exposures in the same observation block, meaning that we consider
these exposures to be well-calibrated. The duration-weighted average
of the ratio of the slit width to the DIMM average seeing for these
exposures is 1.09; in these exposures, the slit width is more appropriately
matched to the average seeing conditions. The exposures from P82.A-0087
were kindly reduced, co-added with the previous exposures and cleaned
using \textsc{uves\_popler} by M.\ Murphy in a similar fashion to
the previous spectra.. The co-added spectrum was then split into the
new exposures (the combination of those from P82.A-0087) and the old
exposures. The advantage of doing this is that the use of more data
provides a better constraint on the quasar flux continuum. We refer
to the new spectrum generated from the older exposures as Q0528:B1,
and the spectrum generated from the exposures from P82.A-0087 as Q0528:B2. 

On account of the considerations relating to the underestimation of
flux uncertainties in regions of low flux given in section \ref{sub:mu:flux error corr},
instead of applying a heuristic model to correct the error arrays
as was done previously, we applied the method described in section
\ref{sub:Correcting error arrays through consistency checks}, which
considers the actual inconsistency of the contributing spectra to
each flux pixel about their weighted mean. 

Although an analysis of Q0528:B1 is effectively a re-analysis of Q0528:A,
as the contributing exposures are the same, the use of a different
correction for flux uncertainty under-estimation appreciably affects
the final spectrum. Thus it is useful to compare an analysis of Q0528:B1
to Q0528:A to check that consistent estimates of $\Delta\mu/\mu$
are obtained.

The additional exposures provide significant extra information with
which the velocity structure of molecular hydrogen can be constrained,
and additionally more information to constrain the structure of the
forest. Therefore we have adopted the following approach to fitting
the spectra. Firstly, we fitted the regions of the Q0528:B2 spectrum
containing molecular hydrogen afresh. Then, we searched the region
of the spectrum above the forest for metal line absorbers from any
redshift. We modelled these absorbers if they contained atomic species
which would generate transitions in the forest, and then looked to
see if transitions from these absorbers were blended with the molecular
hydrogen transitions. If they were, we rejected these molecular hydrogen
transitions from our analysis. This caused us to reject a small number
of transitions which were in our previous fit (in particular: L1R3
and L2R3). Once a satisfactory model was achieved, we then applied
the same model to the Q0528:B1 spectrum, in order to attempt to achieve
a like-with-like comparison. We then independently refined the models
for the Q0528:B1 and Q0528:B2 spectra. The models for the two spectra
differ somewhat, due to the different SNR in different spectral regions.

Additionally, HD (deuterated molecular hydrogen) was detected in this
absorber with a column density of $\log_{10}N=13.27\pm0.08$. HD is
sensitive to a change in $\mu$, and therefore here we include HD
in our analysis of $\Delta\mu/\mu$. Although HD should display a
similar velocity structure to H$_{2}$, the low optical depth and
the small number of transitions observed means that any such structure
is unresolved. We therefore model the HD absorption with only a single
velocity component. That is, the constraint on $\Delta\mu/\mu$ from
HD is derived only by considering potential velocity shifts of the
HD lines with respect to each other, and not with respect to molecular
hydrogen. \citet{Malec:10} have collated oscillator strengths, laboratory
wavelength values and $K_{i}$ values for HD; wavelength values are
from \citet{Hollenstein:06-1} and \citet{Ivanov:08a}, $K_{i}$ values
are from \citet{Ivanov:08a} and oscillator strengths were calculated
by \citet{Malec:10} from Einstein $A$ coefficients given in \citet{Abgrall:06a}.

\subsection{Transitions used}

\begin{figure}[tbph]
\noindent \begin{centering}
\includegraphics[bb=50bp 82bp 560bp 789bp,clip,angle=-90,width=1\textwidth]{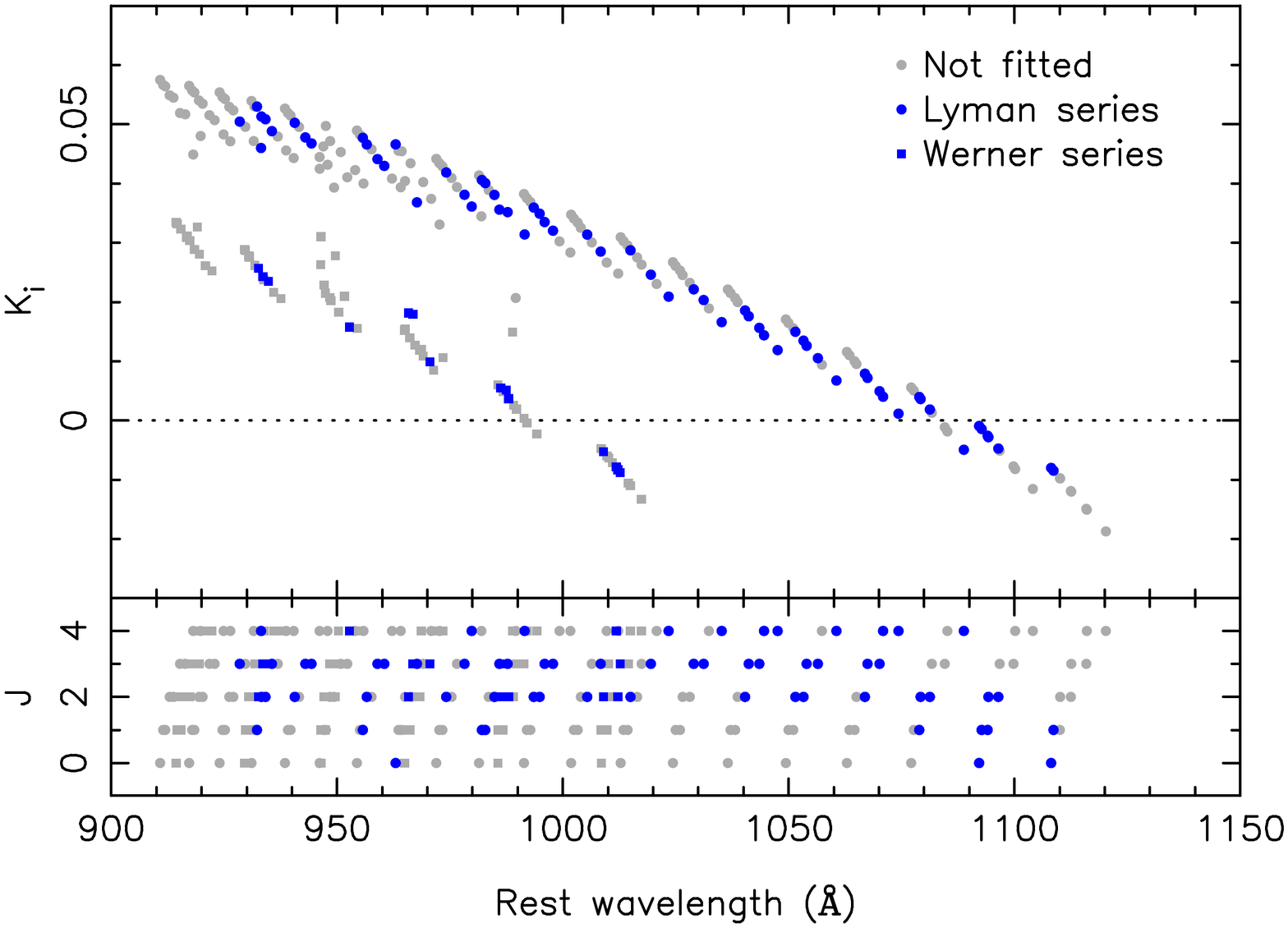}
\par\end{centering}

\caption[Relationship of $K_i$ and $J$ with $\lambda_0$ for the transitions used in Q0528:B2]{Relationship of $K_i$ and $J$ with $\lambda_0$ for the transitions used in Q0528:B2. \emph{Upper panel}: the sensitivity coefficients, $K_i$, for the transitions used in our analysis of Q0528:B2 (dark blue points) and not detected or not fitted (grey points). \emph{Lower panel}: the distribution of transitions with wavelength according to their $J$-level. \label{Flo:mu:Q0528_B2_lambdaki}}
\end{figure}

The transitions used in our re-analysis of Q0528$-$250 differ somewhat
from our earlier analysis. The transitions used are set out in table
\ref{Tab:mu:Q0528:reanalysis transitions}. In figure \ref{Flo:mu:Q0528_B2_lambdaki}
we give the relationship of $K_{i}$ and $J$ with $\lambda_{0}$
for the transitions used in our analysis.

\begin{table}[tbph]
\caption[Transitions used in our re-analysis of Q0528$-$250]{Transitions used in our re-analysis of Q0528$-$250. $n$ gives the number of transitions used. \label{Tab:mu:Q0528:reanalysis transitions}}

\noindent \centering{}%
\begin{tabular}{ccc>{\centering}p{0.6\textwidth}}
\hline 
Quasar spectrum & $z_{\mathrm{abs}}$ & $n$ & Transitions used\tabularnewline
\hline 
\noalign{\vskip\doublerulesep}
Q0528$-$250:B2 & 2.811 & 76 & L0R0, L0R1, L1P1, L1P2, L1R0, L1R1, L1R2, L2P1, L2P2, L2P4, L2R2,
L3P2, L3P3, L3P4, L3R3, L3R4, L4P2, L4P3, L4P4, L4R2, L4R3, L5P2,
L5P3, L5P4, L5R3, L5R4, L6P3, L6P4, L6R3, L7P3, L7P4, L7R2, L8P2,
L8P3, L9R2, L9P2, L9P3, L9R3, L10P1, L10P2, L10P3, L10P4, L10R1, L10R3,
L11P3, L11R2, L11R4, L12R0, L12R3, L13P1, L13P3, L13R2, L13R3, L16R3,
L15P3, L15R2, L15R3, L16P1, L16P2, L16R2, L17P4, L17R3, W0P2, W0Q3,W0R2,
W0R4, W1R2, W1R3, W1Q2, W2P3, W2R2, W2R3, W3Q4, W4P2, W4P3, W4Q3, \tabularnewline
\hline 
\noalign{\vskip\doublerulesep}
\end{tabular}
\end{table}

\subsection{Velocity structure \& $\Delta\mu/\mu$ results}

We re-examined the question of the velocity structure using the spectrum
Q0528:B2\index{Q0528$-$250!velocity structure} under different scenarios.
To do this, we considered three and four velocity components. In each
case, we applied a model in which corresponding components in transitions
with the same $J$ had the same $b$-parameter ($b=F[J]$), and also
the scenario in which corresponding transitions had the same $b$-parameter
regardless of $J$ ($b\neq F[J])$. These models are not nested, and
so we consider only the AICC rather than using the $F$-test. We give
the AICC for these scenarios, and the resulting values of $\Delta\mu/\mu$,
in table \ref{Tab:mu:Q0528B2:num components}. We note that we were
unable to obtain a stable fit for a 4-component model where $b$-parameters
for components were forced to be the same for all $J$-levels. In
this model, the column density of one of the components in the $J=1$
transitions was driven down below the detection threshold of $N=10^{8}\,\mathrm{cm}^{-2}$.
This component was the second strongest component in the $J=2$ and
$J=3$ transitions. Although we could have omitted this component,
the substantial differences in relative strength between the different
components in the different $J$-levels means that this model is very
unlikely to be a good model of the physical situation, and therefore
that the value of $\Delta\mu/\mu$ derived is unlikely to be accurate. 

\begin{table}[tbph]
\caption[Analysis of the velocity structure of Q0528$-$250 using the spectrum Q0528:B2]{Analysis of the velocity structure of Q0528$-$250 using the spectrum Q0528:B2. $n$ gives the number of components. The second column defines whether the $b$-parameter for different components is fixed or is different for transitions with different $J$. The column $\Delta AICC$ shows the difference of the AICC with respect to the best-fitting model. For the 4-component, $b\neq F(J)$ model, the column density of one component of the $J=1$ transitions was rejected, where this component was strongly detected in other $J$-levels. This implies that the model is not physically realistic, and so we label it as unstable.  \label{Tab:mu:Q0528B2:num components}}

\noindent \centering{}%
\begin{tabular}{cccccc}
\hline 
$n$ & Relationship between $b$ and $J$ & AICC & $\Delta\mathrm{AICC}$ & $\chi_{\nu}^{2}$ & $\Delta\mu/\mu$ $(10^{-6})$\tabularnewline
\hline 
\noalign{\vskip\doublerulesep}
3 & $b=F(J)$ & 11488.2 & 0 & 1.115 & $0.2\pm3.2$\tabularnewline
\noalign{\vskip\doublerulesep}
3 & $b\neq F(J)$ & 11653.8 & 165.6 & 1.141 & $0.4\pm3.2$\tabularnewline
\noalign{\vskip\doublerulesep}
4 & $b=F(J)$ & 11510.4 & 22.2 & 1.117 & $3.4\pm3.7$\tabularnewline
\noalign{\vskip\doublerulesep}
4 & $b\neq F(J)$ & Unstable fit & n/a & n/a & n/a\tabularnewline
\hline 
\noalign{\vskip\doublerulesep}
\end{tabular}
\end{table}

We immediately note the sign change from the results in section \ref{sub:mu:first results},
although in $n=3$ case the result is only marginally different from
zero. It is clear from the $n=3$ results that a model where different
$J$-levels have different $b$ parameters is preferred very strongly
over a model with the same $b$-parameter for each $J$-level. This
accords well with our findings from the earlier analysis. 

We note that in this case the 3-component model is preferred to the
4-component model, at about the same statistical significance as the
4-component model was preferred to the 3-component model in our analysis
of Q0528:A. There are several points to note here:
\begin{enumerate}
\item The seeing conditions for the earlier spectra were quite variable,
and may have induced a significantly non-Gaussian instrumental profile%
\footnote{Although the instrumental profile in this case might be non-Gaussian,
the profile should remain symmetric.%
}. The requirement for 4 components earlier may be a reflection of
the non-Gaussian profile, rather than the absorber itself. Because
the components are unresolved at these resolving powers, identification
of the correct number of components is difficult.
\item Some of the exposures contributing to the earlier spectrum were poorly
calibrated. It is conceivable that wavelength miscalibrations could
cause a model with more complexity to be favoured.
\item Although we have attempted to apply the same forest model in analysing
the 3- and 4-component model (although where in each case $\chi^{2}$
is obviously minimised with respect to all the parameters), the construction
of the forest model itself depends on the choice of the H$_{2}$ model.
Strong H$_{2}$ components will obviously have little effect on the
forest model because they are clearly distinguished from the forest.
However, the 4th component of the model is weak and unresolved visually.
We created the 3-component fit by removing the weakest component from
the 4-component fit. In most regions, the resulting fit was reasonable,
however in a small number of regions we found that we had to add weak
forest components to account for the removal of the 4th H$_{2}$ component.
To achieve a like-with-like comparison, we included these extra forest
components in the 4-component fit. This means that any test for the
statistical significance of the number of components depends somewhat
on the choice of forest model near the H$_{2}$ lines. 
\item We noted whilst we were iteratively refining the model in the 3- and
4-component cases that the model for the 4-component model was preferred
for most of the refining process, with $\Delta\mathrm{AICC\approx10}$
in favour of the 4-component model for much of the time. It was only
in the last few rounds of refining the model that the 3-component
model became preferred as a result of changes made to a small number
of regions. Therefore, the choice of the correct number of components
can be sensitive to decisions made about the forest model in a small
number number of regions.
\end{enumerate}
We conclude from this that although the Jeffreys' scale suggests that
there is very significant evidence for the 3-component model over
the 4-component model, in light of the fact that this evidence is
conditioned on the correct choice of forest model and instrumental
resolution, the actual preference for the 3-component model over the
4-component is rather weaker. These arguments apply similarly to our
earlier preference for a 4-component model over a 3-component model.
The issue of whether there are 3 or 4 components is simply very difficult
to resolve given the actual SNR and resolution of the spectra available. 

On the basis of the statistical results in table \ref{Tab:mu:Q0528B2:num components},
we choose $\Delta\mu/\mu=(0.2\pm3.2)\times10^{-6}$ as our preferred
statistical result for the analysis of Q0528:B2. This is obviously
consistent with no change in $\mu$.

\subsection{Consistency checks}

We can relax some of the assumptions made on our analysis of this
absorber to explore whether they have a meaningful impact on the result.
In particular, we explore here whether the result we obtain is significantly
affected by our assumptions about the different $J$-levels. We follow
a similar procedure to that used by \citet{Malec:10}.

\subsubsection{Different $\Delta\mu/\mu$ from different $J$-levels}

\begin{figure}[tbph]
\noindent \begin{centering}
\includegraphics[bb=50bp 92bp 516bp 799bp,clip,angle=-90,width=1\textwidth]{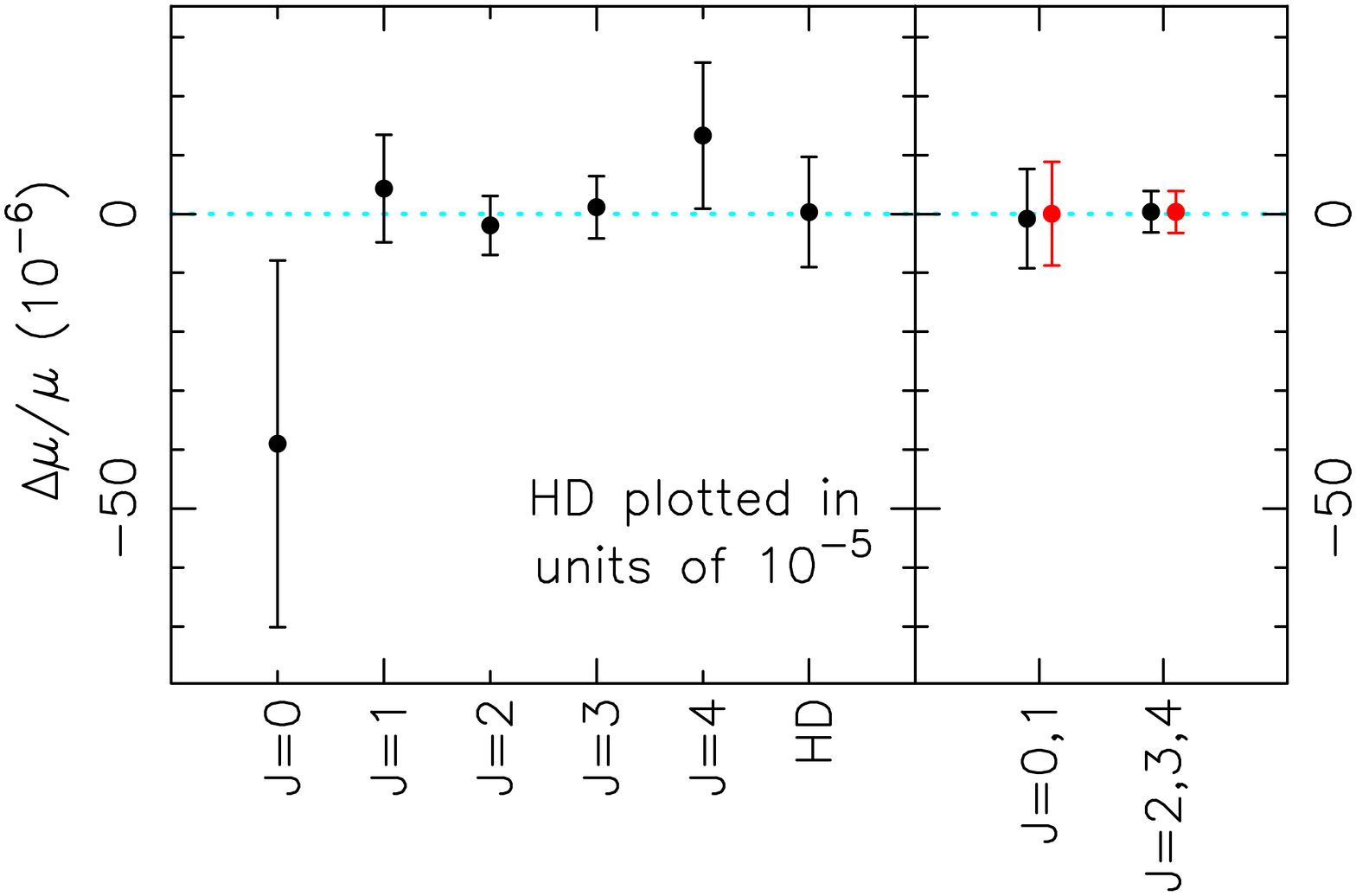}
\par\end{centering}

\caption[Relative contributions toward $\Delta\mu/\mu$ from different H$_2$ $J$-levels and HD for Q0528:B2]{\emph{Left panel:} $\Delta\mu/\mu$ for each H$_2$ $J$-level and HD, assuming that each $J$-level has the same 3-component velocity structure. The $J=0$ level has a substantially larger uncertainty than the $J \in [1,4]$ transitions because only 3 $J=0$ transitions are used in the fit. HD has been plotted in units of $10^{-5}$ to increase clarity for the H$_2$ results. \emph{Right panel:} $\Delta\mu/\mu$ for two groups of transitions, H$_2$ $J\in [0,1]$ (``cold transitions'') and $J\in[2,4]$ (``warm transitions''). The black data points are the results where the redshifts for corresponding components in the cold and warm transitions are assumed to be the same, and the red points are where the redshifts are allowed to differ between the cold and warm transitions. \label{Flo:mu:Q0528 diff J contrib}}
\end{figure}

Rather than allowing transitions from all $J$-levels to contribute
to a single value of $\Delta\mu/\mu$, within \textsc{vpfit} we can
calculate a value of $\Delta\mu/\mu$ for each $J$-level (and one
for HD separately)\index{molecular hydrogen!J-levels}. Strictly,
the values of $\Delta\mu/\mu$ obtained are not independent because:
\emph{i) }they assume that each $J$-level has the same number of
components; and, \emph{ii)} the redshifts are tied between corresponding
components in transitions arising from different $J$-levels. Nevertheless,
this is useful for quantifying the contribution that each $J$-level
makes to the final result. \citet{Ubachs:2007-01} noted that, on
account of the para-ortho distribution of H$_{2}$%
\footnote{In ortho-hydrogen the proton spins are parallel, whilst in para-hydrogen
the spins are anti-parallel.%
} the $J=1$ state is significantly populated even at low temperatures.
They suggested dividing the states into a $J\in[0,1]$ set (cold states)
and $J\geq2$ (warm states) to examine the impact of temperature.
We examine both of these cases in figure \ref{Flo:mu:Q0528 diff J contrib}.
We see that there is no clear evidence for a difference of $\Delta\mu/\mu$
obtained using transitions arising from different $J$-levels.

\subsubsection{Other consistency checks}

We checked that our derived value of $\Delta\mu/\mu$ was not unduly
affected by the starting guess that $\Delta\mu/\mu=0$ by starting
the optimisation with $\Delta\mu/\mu=10^{-5}$. The final value of
$\Delta\mu/\mu$ under this circumstance was $0.210\times10^{-6}$,
compared to $0.187\times10^{-6}$ when started from $\Delta\mu/\mu=0$.
These two numbers differ by $\mbox{\ensuremath{\approx}}0.007\sigma$,
which is entirely negligible. This demonstrates both that the final
result is insensitive to reasonable choices of the starting guess
for $\Delta\mu/\mu$ and that our optimiser is functioning adequately.

Other checks are possible. One could divide the fit into several pieces
(such as fitting odd- and even-numbered regions separately). Estimates
of $\Delta\mu/\mu$ derived from these sub-fits should be consistent
with each other. However, unless one divides the fit into a very large
number of sub-fits (each of which has only a small number of transitions
fitted) it is unlikely that significant deviations will be detected
on account of the central limit theorem. The limiting case of this
method is where one considers the impact of each transition on $\Delta\mu/\mu$.
In the RRM this is easily handled by inspecting the residuals of the
$\zeta_{i}$ values about the linear model for $\zeta_{i}$ vs $K_{i}$.
We cannot apply the RRM to Q0528$-$250 because of the complicated
velocity structure. However, one could use a jack-knife approach.
In this method, if $n$ transitions are fitted, then one generates
$n$ new fits, where in the $i$th fit one removes the $i$th H$_{2}$
transition. One can then inspect the distribution of the $n$ $\Delta\mu/\mu$
values obtained in this way to search for values which deviate strongly
from the average; such deviation implies that the fit is being strongly
affected by the transition involved. This could mean either that the
transition is yielding a very precise constraint on the line centroids
or that the transition is an outlier; further work is required to
determine which of these possibilities is relevant. However, this
requires a substantial amount of computing time, and so we did not
implement this check.

\subsection{Systematic errors}

Whilst for the results given in section \ref{sub:mu:first results}
we provided only statistical errors, given the high statistical precision
it is appropriate to attempt to estimate the impact of systematic
errors\index{molecular hydrogen!systematic errors}. \citet{Malec:10}
noted possible contributions to the systematic error budget, which
include: \emph{i) }known wavelength calibration errors due to uncertainties
in the ThAr wavelength calibration; \emph{ii) }intra-order wavelength
distortions of unknown origin; \emph{iii)} the effect of velocity
structure decisions, and; \emph{iv) }the effect of re-dispersion of
the spectra. We consider each of these in turn.

\subsubsection{Known wavelength calibration errors due to uncertainties in the ThAr
calibration}

The calibration of the ThAr wavelength scale is not perfect; each
of the ThAr transitions displays a residual velocity offset about
the best-fit polynomial solution. The RMS of the residuals is $\sim70\,\mathrm{m\, s^{-1}}$
in the blue arm and $\sim55\,\mathrm{m\, s^{-1}}$ in the red arm
(M. Murphy, priv.\ communication). However, these fluctuations are
random, and therefore will average out if a large number of H$_{2}$
transitions are used. Only systematic deviations from the true wavelength
solution should appreciably affect the best estimate of $\Delta\mu/\mu$.
There are fewer good ThAr lines in the blue end of the spectrum than
in the red end, and therefore larger deviations of the wavelength
solution from the true solution are possible. The systematic deviation
in the blue end of the spectrum relative to the red end of the forest
has an upper limit of $\sim20\,\mathrm{m\, s^{-1}}$ (M. Murphy, priv.\ communication).
The maximum $K_{i}$ value used in the fit is 0.053, whilst the minimum
is $-0.009$. This implies that the maximum possible systematic due
to this effect is given by $\delta(\Delta\mu/\mu)=(\Delta v/c)/\Delta K_{i}$,
which is $1.1\times10^{-6}$. In reality, the effect is likely to
be smaller than this as positive deviations should tend to cancel
somewhat with negative deviations. However, how to reduce the effect
is unclear; it may not simply scale as $1/\sqrt{N}$. Therefore, we
retain this estimate as the maximum possible systematic effect due
to this cause.

\subsubsection{Intra-order distortions of unknown origin\label{sub:mu:Intraorder_distortions}}

\index{wavelength distortions!intra-order}The path that the quasar
light takes through the telescope is similar but not identical to
that from the ThAr calibration lamp --- the ThAr light fills the slit
nearly uniformly, whilst the quasar light does not. Due to the different
light paths, the wavelength scale of the quasar light may be different
to that of the ThAr light; the differences between them appear as
an apparent distortion of the wavelength scale. Both long range and
short-range distortions are possible.

\citet{Griest:09} identified a pattern of distortion within echelle
orders in Keck/HIRES spectra, such that the wavelength scale at the
centre of echelle orders is distorted with respect to that at the
echelle order edges. The peak-to-peak velocity distortion is $\sim500\,\mathrm{m\, s^{-1}}$
at $\sim5,600\mathrm{\AA}$. The distortion was identified by comparing
the calibration of a spectrum obtained using a ThAr exposure to that
obtained using an I$_{2}$ absorption cell. The iodine cell is placed
in the quasar light path, and the characteristic absorption spectrum
is imprinted on the quasar spectrum. The use of an iodine cell therefore
obviates the concern about optical path differences when using a ThAr
lamp. Unfortunately, an I$_{2}$ cell is not useful for calibration
of general quasar observations, because the iodine transitions cover
only a relatively narrow part of the optical range, and because of
the loss of flux from the quasar as a result of the use of the cell.
The observed distortion pattern appears to be dependent on wavelength,
and the distortion may be larger at longer wavelengths. The precise
origin of the distortions is unknown, and similarly it is unknown
to what extent the distortions remain constant in time, and how they
depend on extrinsic factors (e.g.\ telescope orientation, temperature,
pressure and accuracy of quasar centering in the spectrograph slit).
Therefore, it is not possible at present to adequately remove these
distortions of the wavelength scale from observations.

\citet{Whitmore:10} identified a similar effect in VLT/UVES spectra,
with a peak-to-peak velocity distortion of $\sim200\,\mathrm{m\, s^{-1}}$.
The distortion appears to be much less consistent between echelle
orders than that seen by \citeauthor{Griest:09}, however. Further
observations have shown that the observed wavelength distortion is
definitely not constant over long periods of time i.e.\ more than
several nights (M. Murphy, priv.\ communication), which makes removal
of the distortion extremely difficult.

Similar to \citet{Malec:10}, we attempted to estimate the magnitude
of the error introduced into a determination of $\Delta\mu/\mu$ as
a result of the observed velocity distortions. To do this, we used
a triangular-shaped distortion, where the wavelengths of pixels at
the centre of echelle orders were displaced by $+200\,\mathrm{m\, s^{-1}}$
with respect to those at the echelle order edges. The modification
to the spectra was kindly implemented by M. Murphy within \textsc{uves\_popler}.
The shift in $\Delta\mu/\mu$ after modifying the spectrum was $-0.3\times10^{-6}$.
Clearly this value is model-dependent --- if the distortion has a
different amplitude or form, then the impact on $\Delta\mu/\mu$ may
be different. However, this estimate of the systematic is likely to
be of the correct magnitude. We therefore adopt a Gaussian with $\sigma=0.3\times10^{-6}$
as an estimate of the systematic effect due to distortions of this
type.

\subsubsection{Velocity structure \& spatial segregation}

As noted earlier, it is possible that transitions arising from different
$J$-levels might be spatially segregated \citep{Jenkins:97a,Levshakov:02a}\index{molecular hydrogen!spatial segregation}.
Assuming that all $J$-levels arise from the same redshift in this
event could spuriously produce $\Delta\mu/\mu\neq0$. Although in
all of our analyses the results are statistically consistent with
zero, it is of course possible that a non-zero $\Delta\mu/\mu$ could
be pushed \emph{towards} zero by this sort of systematic effect. Similar
to \citet{Malec:10}, we relaxed our assumption that corresponding
components in all $J$-levels arise from the same redshift. In particular,
we divided the data set into ``cold'' transitions, $J\in[0,1]$,
and ``warm'' transitions, $J\in[2,4]$, as was done earlier, but
only tie the redshifts of the different $J$-levels within these two
groups. If there is spatial segregation, this should be seen as a
statistically significant difference between the redshifts of corresponding
components between the two groups, and also as a substantial shift
in the values of $\Delta\mu/\mu$ derived from the two groups compared
to what was obtained earlier.

\begin{figure}[tbph]
\noindent \begin{centering}
\includegraphics[bb=50bp 92bp 535bp 776bp,clip,angle=-90,width=1\textwidth]{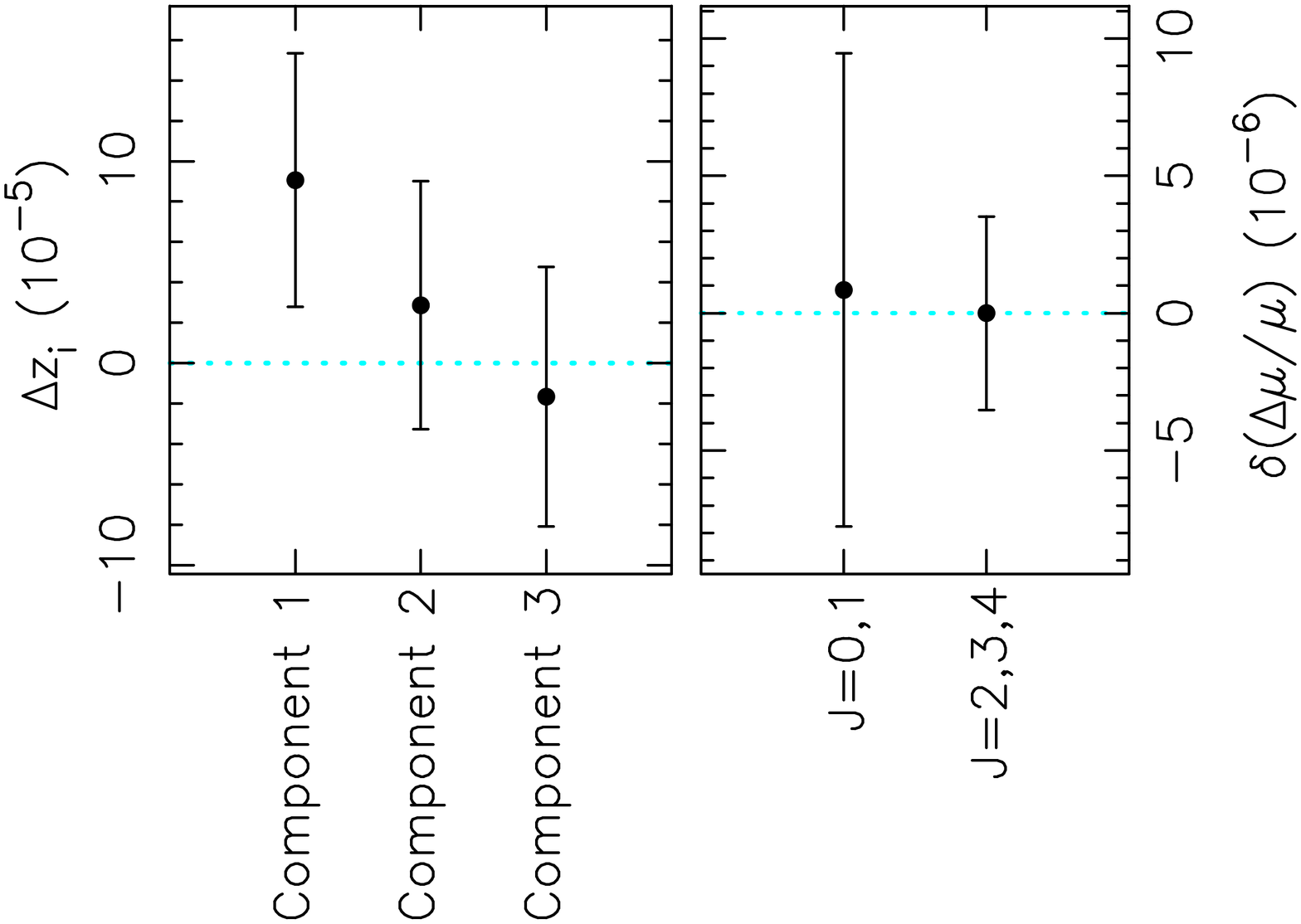}
\par\end{centering}

\caption[Effect of allowing the velocity structure to differ between cold and warm components on $\Delta\mu/\mu$ for Q0528:B2]{\emph{Left panel:} Change in the redshifts of components for the H$_2$ fit for the 3-component model to Q0528:B2 when the velocity structure was allowed to vary between the ``cold'' and ``warm'' components. The difference for each component is defined as $z(J\in[0,1])-z(J\in[2,4])$. \emph{Right panel:} The difference in  $\Delta\mu/\mu$ when the velocity structure was allowed to vary, defined as $\Delta\mu/\mu(\mathrm{structure\ allowed\ to\ vary})-\Delta\mu/\mu(\mathrm{structure\ not\ allowed\ to\ vary})$. The error estimate is calculated as the mean of the error estimates in the two cases considered; the errors do not differ appreciably between the two cases.\label{Flo:mu:Q0528 veldiff}}
\end{figure}

In the right panel of figure \ref{Flo:mu:Q0528 diff J contrib} we
directly compare $\Delta\mu/\mu$ in the case where the velocity structure
was allowed to vary between cold and warm components, and note that
there is no appreciable shift. In figure \ref{Flo:mu:Q0528 veldiff}
we show these considerations more directly by examining the differences
in the redshifts of the three components, and also the explicit difference
between $\Delta\mu/\mu$ in the two cases considered. We see that
there is no statistically significant difference between the redshifts
in any of the three components. Similarly, we see that the shift in
$\Delta\mu/\mu$ is $\lesssim0.1\sigma$, which we consider to be
effectively negligible. Thus, we conclude that there is no evidence
for a systematic shift in $\Delta\mu/\mu$ as a result of segregation
of the cold and warm $J$-levels. 

Nevertheless, to quantify the possible error introduced by our assumptions
regarding velocity structure, we examine the actual shifts in $\Delta\mu/\mu$.
The shift in $\Delta\mu/\mu$ for the $J\in[0,1]$ levels is $0.8\times10^{-6}$,
and for the $J\in[2,4]$ levels is $0.2\times10^{-6}$. We thus take
$0.8\times10^{-6}$ as an estimate of the potential error introduced
into our analysis due to assumptions about the velocity structure
in order to be conservative.

\subsubsection{Re-dispersion of spectra}

The spectrum used is the result of the co-addition of exposures taken
with different echelle grating settings. During the co-addition, the
spectra are placed on a common wavelength grid. Because the spectra
are re-binned, the choice of wavelength grid introduces correlations
between neighbouring pixels. More importantly, the choice of the wavelength
grid has the potential to affect $\Delta\mu/\mu$. To investigate
this, we examined the effect of shifting the wavelength grid by $-0.2$,
$-0.1$, 0.1 and 0.2 pixels. The modification of the spectra was kindly
implemented by M.\ Murphy within \textsc{uves\_popler}. The shifts
this caused on $\Delta\mu/\mu$ respectively are $-1.3\times10^{-6}$,
$-2.1\times10^{-6}$, $+1.0\times10^{-6}$ and $+0.2\times10^{-6}$
respectively. The standard deviation of these values is $\approx1.4\times10^{-6}$,
and so we adopt $1.4\times10^{-6}$ as an estimate of the potential
error in $\Delta\mu/\mu$ on account of the re-dispersion of the contributing
exposures.

\subsection{Result including systematic errors}

In table \ref{Tab:mu:Q0528B2:error analysis}, we accumulate the potential
systematic errors from our discussion above and give our final estimate
of $\Delta\mu/\mu$ including the systematic component. Although the
distribution of systematic errors is likely to be Gaussian in many
cases, the impact on $\Delta\mu/\mu$ arising from the distortion
in the wavelength scale (for instance) is an upper limit. The probability
distribution of the sum of random variables is given by the convolution
of their individual probability density functions. Thus, to estimate
our final uncertainty, we convolve the distributions assumed for each
of the sources of uncertainty, and give the standard deviation of
the resultant distribution as our uncertainty estimate. 

This yields our final estimate of $\Delta\mu/\mu$ for Q0528:B2, as
\begin{equation}
\frac{\Delta\mu}{\mu}=(0.2\pm3.2_{\mathrm{stat}}\pm1.9\mathrm{_{\mathrm{sys}})\times10^{-6}}=(0.2\pm3.7)\times10^{-6}.
\end{equation}
\begin{sidewaystable}
\caption[Error budget for the analysis of Q0528:B2]{Error budget for the analysis of Q0528:B2 from the sources described. The third column gives the magnitude of the uncertainty estimate. The final error estimate is calculated as the standard deviation of the convolution of the assumed distributions of the individual error estimates; the assumed distributions are given in the fourth column.  \label{Tab:mu:Q0528B2:error analysis}}

\noindent \centering{}%
\begin{tabular}{cccc}
\hline 
Source of error & $\Delta\mu/\mu$ ($10^{-6}$) & $\delta(\Delta\mu/\mu)$ ($10^{-6}$) & Assumed distribution\tabularnewline
\hline 
\noalign{\vskip\doublerulesep}
Statistical & $0.2$ & $\pm3.2$ & Gaussian\tabularnewline
\noalign{\vskip\doublerulesep}
Systematic distortion of ThAr wavelength scale &  & $\pm1.1$ & Uniform\tabularnewline
\noalign{\vskip\doublerulesep}
Intra-order wavelength scale distortions &  & $\pm0.3$ & Gaussian\tabularnewline
\noalign{\vskip\doublerulesep}
Velocity structure \& spatial segregation &  & $\pm0.8$ & Gaussian\tabularnewline
\noalign{\vskip\doublerulesep}
Re-dispersion of spectra &  & $\pm1.4$ & Gaussian\tabularnewline
\noalign{\vskip\doublerulesep}
\hline 
Final estimate & $0.2$ & $\pm3.7$ & \tabularnewline
\hline 
\noalign{\vskip\doublerulesep}
\end{tabular}
\end{sidewaystable}

\subsection{Q0528:B1}

At the time of writing, our analysis of Q0528:B1 is incomplete. However,
preliminary results suggest that the value of $\Delta\mu/\mu$ is
likely to be very similar to that obtained with the earlier spectrum
(Q0528:A).

\section{Discussion and Summary of results}

\subsection{Summary of results}

In this chapter, we presented analyses of high quality spectra of
the highly-studied quasars Q0347$-$383, Q0405$-$443 and Q0528$-$250
(Q0528:A). In our initial investigation of these absorbers, we applied
the DCMM to all three absorbers, and the RRM to the absorbers towards
Q0347$-$383 and Q0405$-$443 (the RRM cannot be applied to the absorber
towards Q0528$-$250 because of the overlapping velocity components).
Our preferred results for the absorbers towards Q0347$-$383, Q0405$-$443
and Q0528$-$250 from the analysis of these spectra are derived from
the DCMM and are $\Delta\mu/\mu=(8.2\pm7.5)\times10^{-6}$, $(10.1\pm6.6)\times10^{-6}$
and $(-1.4\pm3.9)\times10^{-6}$ respectively. 

The spectra for these quasars used were obtained from the VLT/UVES
archive. We discussed potential problems with the exposures contributing
to Q0528$-$250 in section \ref{sub:Q0528-250 revisited}, and why
analysis of a new spectrum would be advantageous. We performed this
analysis on a new spectrum generated from exposures taken specifically
for the purpose of measuring $\Delta\mu/\mu$ in 2008 and 2009. From
this spectrum we obtained the constraint $\Delta\mu/\mu=(0.2\pm3.2_{\mathrm{stat}}\pm1.9_{\mathrm{sys}})\times10^{-6}=(0.2\pm3.7)\times10^{-6}$
(Q0528:B2), where the systematic error estimate is dominated by systematic
distortions of the ThAr wavelength scale and the process whereby individual
exposures are co-added onto a common wavelength grid. 

The value of $\Delta\mu/\mu$ derived from Q0528:B2 is likely to be
more accurate than that derived from Q0528:A because the exposures
were taken with the specific purpose of searching for potential variation
in $\mu$, and should have superior wavelength calibration. Nevertheless,
the values of $\Delta\mu/\mu$ derived from Q0528:A and Q0528:B2 are
statistically consistent, and so it is difficult to demonstrate any
marked inaccuracy in the value from Q0528:A.

All of these results are consistent with each other, and with no cosmological
variation in $\mu$. Taken together, these results are the best $z>1$
constraints on cosmological evolution in $\mu$. A weighted mean of
these results yields $(\Delta\mu/\mu)_{w}=(1.7\pm2.4)\times10^{-6}$.

\subsection{Comparison with other results}

\begin{figure}[tbph]
\noindent \begin{centering}
\includegraphics[bb=50bp 103bp 542bp 766bp,clip,angle=-90,width=1\textwidth]{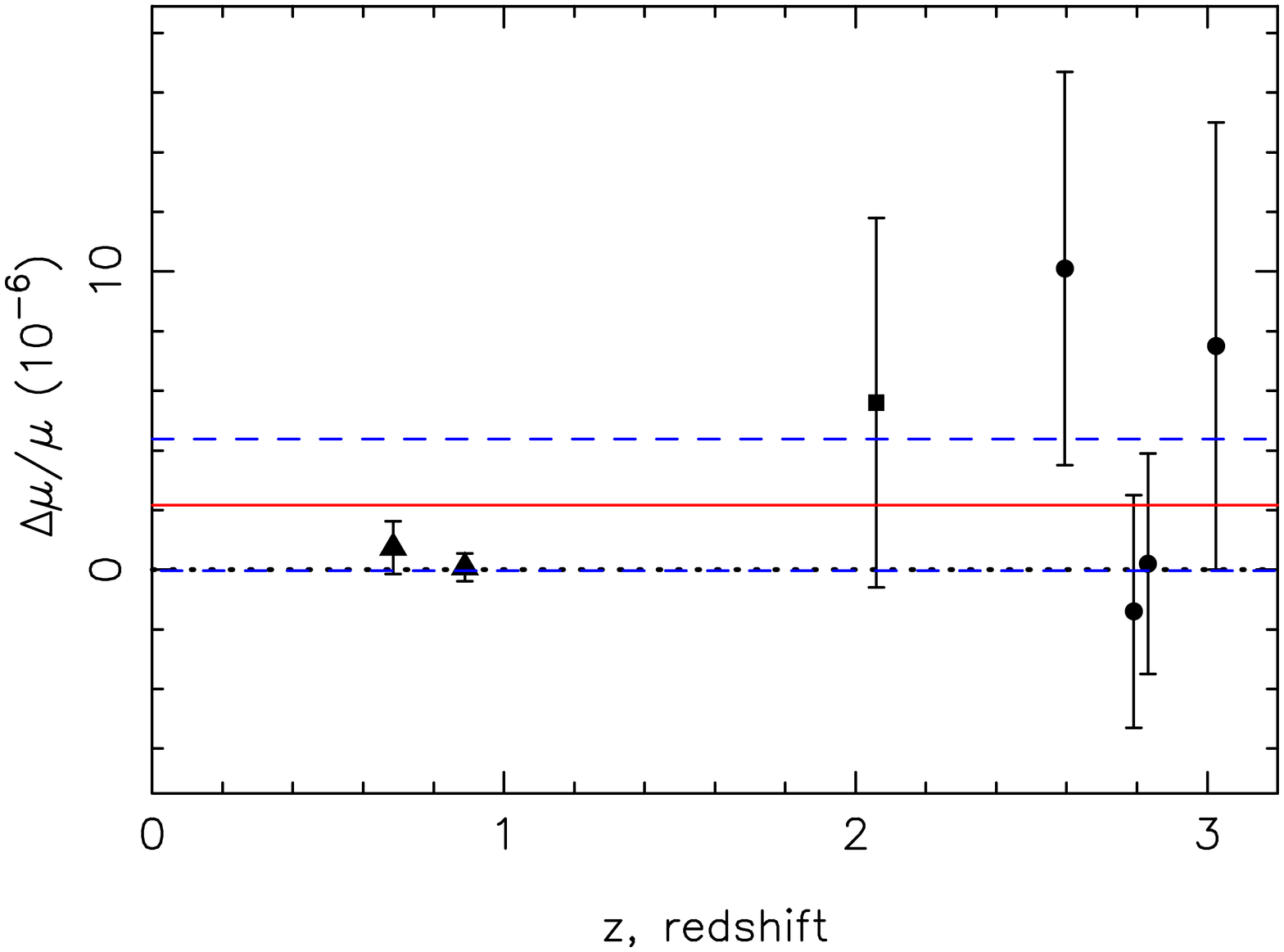}
\par\end{centering}

\caption[Current best extragalactic constraints on $\Delta\mu/\mu$]{Current best extragalactic constraints on $\Delta\mu/\mu$. The results of this work are shown as circles, the J2123$-$0050 constraint of \citet{Malec:10} is shown as a square, and the two ammonia results of \citet{Murphy:Flambaum:08} and \citet{Henkel:09} are shown as triangles. The two Q0528$-$250 points at $z=2.811$ have been slightly offset for clarity. The red line shows the weighted mean of the high-redshift constraints on $\Delta\mu/\mu$, and the blue dashed regions show the $1\sigma$ confidence limit on the weighted mean. \label{Flo:mu:all extragalactic results}}
\end{figure}

In figure \ref{Flo:mu:all extragalactic results} we show all current
extragalactic constraints on $\Delta\mu/\mu$%
\footnote{We have ignored here the recent analyses of \citet{Wendt:08a}, \citet{Thompson:09a}
and \citet{Wendt:10a}, as the data used by these studies is not independent
of the results of this chapter due to the use of common spectra.%
}. Assuming that all of the high-redshift $\Delta\mu/\mu$ points are
well-described by a single value of $\Delta\mu/\mu$, one can calculate
the weighted mean of these points, which gives $(\Delta\mu/\mu)_{w}=(2.2\pm2.2)\times10^{-6}$.
In calculating this weighted mean, we have added statistical and systematic
error estimates in quadrature where they are available. $\chi_{\nu}^{2}=0.88$
about this weighted mean, giving no evidence of excess scatter in
the data (and therefore unmodelled systematic trends).

We return to the question of whether a weighted mean model is appropriate
in chapter \ref{cha:Discussion}.

\subsection{Q0528$-$250}

The investigation of the $z=2.811$ absorber toward Q0528$-$250 is
extremely challenging. The additional spectrum of Q0528:B2 is important
in that it was specifically taken for the purposes of investigating
$\Delta\mu/\mu$. That is, ThAr exposures were taken so as to maximise
wavelength accuracy. Nevertheless, the different components of the
molecular hydrogen transitions are poorly resolved, making accurate
determination of the structure difficult. Although investigations
at higher SNR will incrementally aid this process, a more fruitful
approach would be to obtain a spectra at significantly higher resolving
powers ($R\gtrsim100,000$). With VLT/UVES, this would require slit
widths of $\lesssim0.3$ arcseconds \citep{dOdorico:00a}. This would
result in an unacceptable loss of flux, unless an image slicer is
used. The use of an image slicer makes the profile of the quasar light
in the spatial direction more complicated on the spectrograph CCD,
which makes extraction of the spectrum more complicated. However,
for a fixed-aperture telescope, increased $R$ comes at the expense
of reduced SNR per pixel. Ultimately, routine observations at such
high resolving powers may have to await the next generation of large
telescopes, for which suggested spectral resolutions may be $\sim150,000$
\citep{Pasquini:08a}. Until the individual velocity components can
be resolved, it may be difficult to ascertain the precise nature of
the uncertainty due to velocity structure mis-specification. On the
other hand, if the inter-component velocity spacing is comparable
to the intrinsic velocity widths of the transitions, then observations
with higher resolving powers will not yield such large gains. Nevertheless,
the higher SNR per pixel and higher resolving powers that will be
available with future telescopes should make the analysis of complicated
H$_{2}$ absorbers more reliable than it is at present.

\subsection{Future work}

With the addition of the $z=2.059$ molecular hydrogen system toward
J2123$-$0050 \citep{Malec:10} there are now five high-quality, independent
measurements of $\Delta\mu/\mu$ derived from H$_{2}$. An analysis
of the $z=2.059$ absorber J2123$-$0050 using VLT data is being undertaken
by Freek van Weerdenburg (Vrije Universiteit), and should be available
shortly. However, to constrain $\Delta\mu/\mu$ adequately at high
redshift, it will be imperative to increase the number of absorption
systems utilised along different sightlines, in order to confidently
map out $\Delta\mu/\mu$ in different locations and earlier times. 

As noted earlier, there is a clear lack of molecular hydrogen absorbers
at high redshift which can be used to constrain $\Delta\mu/\mu$.
\citet{Malec:10} noted that despite the large number ($>1000$) of
damped Lyman-$\alpha$ systems known \citep{Prochaska:09a}, the number
known to contain molecular hydrogen is less than 20. Ultimately, the
paucity of absorbers may mean that constraining $\Delta\mu/\mu$ may
be done more rapidly through methods which constrain combinations
of fundamental constants (see chapter \ref{cha:Discussion}).

The process of extracting values of $\Delta\mu/\mu$ from the spectra
given the necessity of modelling the Lyman-$\alpha$ forest is extremely
time-consuming when done manually owing to the large amount of time
required to optimise parameter values after each trial modification
of the fit. The continual progress in computing speeds has rendered
this process substantially easier even since this work was started.
Nevertheless, it would be ideal for the number of $\Delta\mu/\mu$
measurements from H$_{2}$ to be increased by at least an order of
magnitude. This will likely require many PhD students and much patience,
but ultimately a move to automated methods is likely. Given the time
required for the optimisation, manual inspection and alteration of
the fit at each step easily dominates any simple automated method
at present. Nevertheless, more clever approaches to automated fitting
and better computing power may yield progress in this respect sooner
rather than later.

\chapter{$\alpha$ --- the fine-structure constant\label{cha:alpha}}

\section{Introduction}

The fine-structure constant\index{fine-structure constant}, $\alpha$,
is an extremely important fundamental constant with a rich history.
In S.I. units, 
\[
\alpha\equiv\frac{e^{2}}{4\pi\epsilon_{0}\hbar c},
\]
where $e$ is the electron charge, $\epsilon_{0}$ is the electric
permittivity of free space, $\hbar$ is Planck's constant divided
by $2\pi$, and $c$ is the speed of light. In cgs units, $\alpha\equiv e^{2}/\hbar c$.
The CODATA 2006 recommended value of $\alpha$ is $1/137.035\,999\,679(94)$
\citep{Mohr:08}. $\alpha$ may be measured in many different ways,
but the most precise derive from measurements of the magnetic anomaly
of the electron and muon, where the magnetic anomaly is defined as
$a=(g-2)/2$, and $g$ is the spin $g$-factor of the particle in
question, combined with quantum electrodynamics (QED)\index{quantum electrodynamics (QED)}
calculations. The CODATA value is unfortunately significantly affected
by a significant error in the QED calculations of \citet{Gabrielse:06a},
who contributed the most precise point in the CODATA analysis. \citet{Gabrielse:07a}
updated their result to $\alpha=1/137.035\,999\,070(98)$ after correcting
for this error. A more recent CODATA value is not yet available. The
current most precise bound on $\alpha$ derives from \citet{Hanneke:08a},
who give $\alpha=1/137.035\,999\,084(51)$ based on the electron magnetic
anomaly. $\alpha$ may also be determined from the recoil effects
of cold ensembles of Rb \citep{Clade:06,Cadoret:08} and Cs \citep{Gerginov:06}
atoms; these experiments are less precise by an order of magnitude
but give results which are effectively independent of QED calculations
\citep{Salumbides:PhD}.

In quantum electrodynamics, $\alpha$ represents the strength of the
coupling between the electron and the photon, and therefore determines
the effective strength of the electromagnetic force. The fact that
QED is practically useful derives from the fact that $\alpha<1$,
which makes a perturbative expansion of the effect of QED in powers
of $\alpha$ possible. In fact, the value of $\alpha$ given above
is the low-energy value. $\alpha$ is a running coupling constant,
the value of which changes depending upon the energy scale being probed;
at the mass scale of the Z boson, $\alpha$ increases to $\alpha\approx1/129$
\citep{Okun:96a}. When we discuss evolution in $\alpha$, we are
referring to evolution in the low-energy value of $\alpha$.

\subsection{Sensitivity of transitions to a change in $\alpha$}

For an alkali doublet (AD), the separation between the two fine-structure
transitions scales as $\alpha^{2}$ \citep{Bethe:77a}. For a small
change in $\alpha$, $\Delta\alpha/\alpha\equiv(\alpha_{z}-\alpha_{0})/\alpha_{0}$,
(where $|\Delta\alpha/\alpha|\ll1$), the change in the doublet separation
is given by
\begin{equation}
\frac{\Delta\alpha}{\alpha}=\frac{c}{2}\left[\frac{(\Delta\lambda)_{z}}{(\Delta\lambda)_{0}}-1\right],
\end{equation}
 where $(\Delta\lambda)_{z}$ and $(\Delta\lambda)_{0}$ are the relative
doublet separations in the cloud rest-frame, at redshift $z$, and
in the laboratory \citep{Varshalovich:00a,Murphy:PhD}. The constant
$c$ is different for different doublets, and accounts for higher
order relativistic effects. For the Si~\iv $\lambda\lambda1393,1042$
doublet, $c\approx$1 \citep{Murphy:PhD}. Considering AD-type transitions
in quasar spectra leads to the \emph{alkali-doublet method} (AD method)\index{alkali-doublet (AD) method}.
Effectively, one compares the observed relative spacings in quasar
spectra with laboratory spectra to determine $\Delta\alpha/\alpha$.
Because two transitions are being used, $\Delta\alpha/\alpha$ is
not degenerate with the determination of the redshift. 

However, the AD method does not make use of all available information
in the quasar spectra. In particular, different atomic transitions
display significantly larger sensitivities to a variation in $\alpha$
than the Si \ivs transitions. If one considers a many-electron atom
or ion, then the correction to the energy of an external electron
due to relativistic effects can be written as 
\begin{equation}
\Delta\propto(Z_{n}\alpha)^{2}|E|^{3/2}\left[\frac{1}{j+1/2}-C(j,l)\right],\label{eq:alpha:relativistic energy correction}
\end{equation}
where $Z_{n}$ is the nuclear charge, $E$ is the electron energy
($E<0$, $|E|$ is the ionisation potential) and $j$ and $l$ are
the total and orbital electron angular momenta \citep{Murphy:PhD}.
The quantity $C(j,l)$ determines the contribution to the correction
from many-body effects. Equation \ref{eq:alpha:relativistic energy correction}
immediately suggests two strategies for probing $\alpha$ variation.
Firstly, as the effect scales with $Z_{n}^{2}$, a comparison of transitions
from light ions with those from heavy ions should lead to a large
difference in the relativistic correction, which is therefore sensitive
to a change in $\alpha$. Secondly, the term $C(j,l)$ begins to dominate
with increasing $j$. As a result of this, the correction to an $s$-$p$
transition in a heavy ion will be of the opposite sign to that for
a $d$-$p$ transition \citep{Murphy:PhD}. Thus the comparison of
different types of transitions can yield substantial differences in
the relativistic corrections. Thus, comparing many different transitions
from light and heavy atomic species simultaneously can substantially
increase the sensitivity to a variation in $\alpha$. This leads to
the \emph{many-multiplet method} (MM method)\index{many-multiplet (MM) method}.
The MM method is described in considerable detail in \citet{Webb:99,Dzuba:99:01},
and so we present the salient features here.

For experimental purposes, one can generally describe how the energy
level of a given transition varies if $\alpha$ changes, for any multiplet
and species. This yields
\begin{equation}
\omega_{z}=\omega_{0}+q_{1}x_{z}+q_{2}y_{z},
\end{equation}
 where $\omega_{z}$ is the wavenumber of the transition in the rest-frame
of the cloud at redshift $z$ \citep{Dzuba:99:01,Dzuba:99:02,Dzuba:01,Dzuba:02}.
$x_{z}$ and $y_{z}$ are related to $\Delta\alpha/\alpha$, with
\begin{equation}
x_{z}=\left(\frac{\alpha_{z}}{\alpha_{0}}\right)^{2}-1\quad\mathrm{and}\quad y_{z}=\left(\frac{\alpha_{z}}{\alpha_{0}}\right)^{4}-1.
\end{equation}
 The $q_{1}$ and $q_{2}$ coefficients\index{q@$q$ coefficients}
account for the relativistic corrections to the energy for a particular
transition. As we consider only $|\Delta\alpha/\alpha|\ll1$, we can
amalgamate $q_{1}$ and $q_{2}$ into $q=q_{1}+2q_{2}$, yielding
\begin{equation}
\omega_{z}=\omega_{0}+qx_{z}.\label{eq:omega_xz_shift}
\end{equation}
The sign and magnitude of $q$ differs significantly depending on
the species and transition under consideration. Ultimately it is not
the actual value of $q$ that constrains $\Delta\alpha/\alpha$, due
to the need to simultaneously determine the redshift of the absorber,
but the differences in the $q$ values between different transitions
used. Note that because of the functional form of equation \ref{eq:omega_xz_shift},
if $\Delta\alpha/\alpha=0$ then errors in the $q$ coefficients can
not manufacture an observed $\Delta\alpha/\alpha\neq0$ (in the absence
of systematic effects). 

Note that 
\begin{equation}
\left(\frac{\alpha_{z}}{\alpha_{0}}\right)^{2}-1=\left(\frac{\Delta\alpha}{\alpha}\right)^{2}+2\left(\frac{\Delta\alpha}{\alpha}\right)\approx2\left(\frac{\Delta\alpha}{\alpha}\right),
\end{equation}
where the approximation is valid for $|\Delta\alpha/\alpha\ll1|$.
From equation \ref{eq:omega_xz_shift}, the velocity shift%
\footnote{$\Delta v/c\approx\Delta\lambda/\lambda=-\Delta\omega/\omega$.%
}\index{many-multiplet (MM) method!velocity shifts}, $\Delta v$,
for a given transition is thus given by 
\begin{equation}
\Delta v\approx-\frac{2cq_{i}}{\omega_{0}}\left(\frac{\Delta\alpha}{\alpha}\right)=-2cq_{i}\lambda_{0}\left(\frac{\Delta\alpha}{\alpha}\right),
\end{equation}
where $q_{i}$ is the $q$ coefficient for the transition and again
$|\Delta\alpha/\alpha\ll1|$. 

\begin{figure}[tbph]
\noindent \begin{centering}
\includegraphics[bb=89bp 40bp 568bp 757bp,clip,angle=-90,width=1\textwidth]{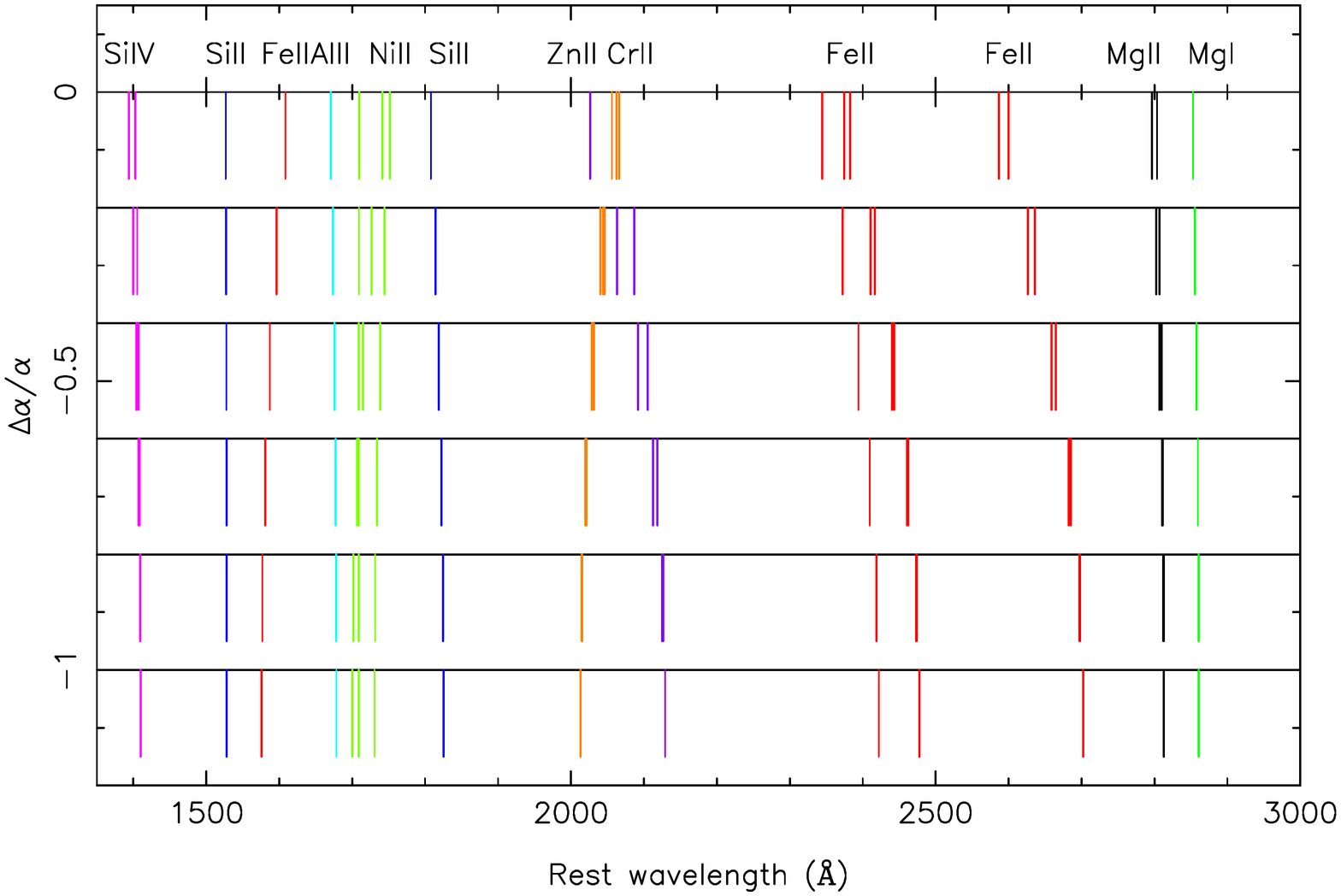}
\par\end{centering}

\caption[Exaggerated effect of $\Delta\alpha/\alpha$ on certain MM transitions]{The exaggerated effect of $\Delta\alpha/\alpha$ on the wavelengths of certain MM transitions. Note that the case $\Delta\alpha/\alpha = -1$ corresponds to the non-relativistic case where $c=\infty$. Note also the presence of transitions which are relatively insensitive to $\alpha$ variation (e.g.\ Mg \ii, Mg \isc, Si \ii \ $\lambda 1526$ and Al \ii \ $\lambda 1670$), transitions which are strongly sensitive and move to longer wavelengths (e.g.\ Fe \ii \ excluding $\lambda 1608$, and  Zn \ii) and transitions which are strongly sensitive and move to shorter wavelengths (e.g.\ Fe \ii \ $\lambda 1608$, Cr \ii \ and Ni \ii \ $\lambda\lambda 1741, 1751$).  \label{Flo:alpha:alpha shift plot}}
\end{figure}

We show the effect of $\alpha$ variation on the wavelengths of different
MM transitions in figure \ref{Flo:alpha:alpha shift plot}, and show
the relationship of the $q$ coefficients with wavelength in figure
\ref{Flo:alpha:q_vs_wl}. We give explicit values for the $q$ coefficients
in table \ref{tab:atomdata}. 

In figure \ref{Flo:alpha:alpha shift plot}, one can broadly see the
presence of three types of transitions: \emph{i)} transitions with
large, positive $q$, which shift to shorter wavelengths with increasing
$\Delta\alpha/\alpha$ (e.g.\ the Fe~\iis transitions at $\sim2400$
and $\sim2600\mathrm{\AA}$, and the Zn~\iis transitions); \emph{ii)
}transitions with large, negative $q$, which shift to longer wavelengths
with increasing $\Delta\alpha/\alpha$ (e.g.\ the Fe~\iis $\lambda1608$
transition and the Cr~\iis transitions), and; \emph{iii)} those
which are relatively insensitive to $\alpha$ variation (e.g.\ the
Mg~\iscs and Mg~\iis transitions). 

The Fe~\iis transitions are very commonly fitted in the MM method,
and it is worth considering the Fe~\iis $\lambda2382$ transition
(which has the highest oscillator strength of the Fe~\iis transitions)
as an example. For $\Delta\alpha/\alpha=+10^{-5}$, the induced velocity
shift in the Fe~\iis $\lambda2382$ transition (which has $q=1460\,\mathrm{cm^{-1}}$)
is $\approx-209\,\mathrm{m\, s^{-1}}$. The instrumental resolution
for VLT/UVES is typically $\sim6\,\mathrm{km\, s^{-1}}$, and the
pixel width is typically $\sim2\,\mathrm{km\, s^{-1}}$. This immediately
makes it obvious that searching for variations in $\alpha$ at the
$10^{-5}$ to $10^{-6}$ level requires very high quality spectra,
with good wavelength calibration.

The MM method has significant advantages over the AD method \citep{Murphy:PhD}.
In particular, the sensitivity gain (with a maximal $\Delta q$ of
$\sim4000$) is potentially much larger than for the AD method (for
which $\Delta q\sim500$ for Si \iv), yielding an order of magnitude
sensitivity improvement in the best case. Additionally, the use of
all transitions provides a statistical advantage simply through the
use of more data. Furthermore, by using many different transitions
one can better constrain the velocity structure of the absorber, and
therefore the model is more likely to be a good representation of
the physical generative processes, and thus the values of $\Delta\alpha/\alpha$
should be more accurate. Finally, the use of transitions with both
positive and negative $q$ helps to minimise systematic effects (we
saw in chapter \ref{cha:mu} how the use of both the Lyman and Werner
series played a similar role for the analysis of $\Delta\mu/\mu$
using H$_{2}$). 

A key assumption exists in the application of the MM method\index{many-multiplet (MM) method!assumptions}
that is not required in the AD method, namely that the transitions
arise from the same location. In the AD method, the transitions considered
in any particular analysis arise from the same ground state and therefore
from the same location. However, in the MM method, it is possible
that transitions from the ground states of different atoms/ions arise
from different locations. This may occur if the gas clouds containing
the relevant species are inhomogeneous. For a particular absorber,
spatial segregation of the relevant species would lead to a relative
shift in line positions, which would lead to a spurious detection
of a variation in $\alpha$ for that system if the spatial segregation
was sufficiently large. However, over an ensemble of absorbers this
could only produce a variation in $\alpha$ if the centres of mass
of some species were on average located closer to us than the centres
of mass of other species along the line of sight to the absorbers.
This would be an extreme violation of the Copernican principle, and
therefore we consider this possibility not to be relevant. Although
spatial segregation cannot bias the results of $\Delta\alpha/\alpha$
over a sufficiently large ensemble of absorbers, it will manifest
as excess scatter of the $\Delta\alpha/\alpha$ points about a fitted
model, and therefore this effect constitutes a potential random error
with expectation value zero but non-zero variance.

\subsection{Application of the alkali-doublet method}

\index{alkali-doublet (AD) method!applications}The strength of the
constraint that can be placed on variation in $\alpha$ directly relates
to the line widths of the observed transitions used; narrower lines
will yield higher precision. \citet{Savedoff:56a} investigated AD
separations in a Seyfert galaxy emission spectra. \citet{Bahcall:67a}
first applied the AD method to absorption lines (which are generally
narrower than emission lines) which seemed to be intrinsic to the
quasar 3C 191, giving $\Delta\alpha/\alpha=(-2\pm5)\times10^{-2}$
at $z\approx1.95$. \citet{Wolfe:76a} analysed Mg \textsc{ii} doublets
from a DLA at $z=0.524$. \citet{Potekhin:94a} investigated C \textsc{iv},
N \textsc{v}, O \textsc{vi}, Al \textsc{iii} and Si \textsc{iv} doublets
to find $|(1/\alpha)(\Delta\alpha/\Delta z)|<5.6\times10^{-4}$. \citet{Cowie:Songaila:1995}
investigated the Si \textsc{iv} doublet to find $|\Delta\alpha/\alpha|<3.5\times10^{-4}$. 

\citet{Varshalovich:00a} applied the AD method to 16 absorption systems
toward 6 quasars using the Si \textsc{iv} $\lambda\lambda1393$ and
$1402$ doublet, obtaining a weighted mean of $\Delta\alpha/\alpha=(-4.6\pm4.3)\times10^{-5}$
(statistical error only). They suggested that an additional error
of $\pm1.4\times10^{-5}$ is required due to uncertainties in the
laboratory doublet separation $(\Delta\lambda)_{0}$. However, \citet{Murphy:PhD}
noted that this seems optimistic given statements about the laboratory
accuracy in \citet{Ivanchik:99a}; \citet{Murphy:PhD} considered
that a systematic error of order $\sim5\times10^{-5}$ is needed.

\citet{Murphy:01a} analysed 21 Si \ivs doublets using Keck/HIRES
and found that $\Delta\alpha/\alpha=(-0.5\pm1.3)\times10^{-5}$. The
dispersion of their data about the weighted mean yields $\chi_{\nu}^{2}=0.95$,
which suggests that the statistical errors are correctly estimated.

\subsection{Application of the many-multiplet method}

\index{many-multiplet (MM) method!applications}The MM method was
first applied in \citet{Webb:99} to 30 Keck/HIRES quasar spectra
in the redshift range $0.5<z_{\mathrm{abs}}<1.6$, where they fitted
the Mg~\textsc{i} $\lambda2852$ line, the Mg~\textsc{ii} $\lambda2796,2803$
doublet and the five strongest Fe~\textsc{ii} transitions with $q>0$
($\lambda\lambda\lambda\lambda\lambda2383$, 2600, 2344, 2586 and
2374). This yielded tentative evidence that $\alpha$ was smaller
in the absorption clouds than in the laboratory, with $\Delta\alpha/\alpha=(-1.09\pm0.36)\times10^{-5}$
(a $3\sigma$ detection). This result was dominated by the 14 systems
above $z_{\mathrm{abs}}=1$, for which $\Delta\alpha/\alpha=(-1.88\pm0.53)\times10^{-5}$.
The $z_{\mathrm{abs}}<1$ systems yielded $\Delta\alpha/\alpha=(-0.17\pm0.39)\times10^{-5}$,
which is consistent with no variation. 

The addition of more absorbers by \citet{Webb:01} and \citet{Murphy:01b}
increased the significance of the detection, yielding $\Delta\alpha/\alpha=(-0.72\pm0.18)\times10^{-5}$
(a $4\sigma$ detection). Importantly, these works made use of the
Ni~\textsc{ii}/Cr~\textsc{ii}/Zn~\textsc{ii} transitions, which
display a significantly different relationship between $\lambda$
and $q$. The Fe~\textsc{ii}/Mg~\textsc{ii} subset of those works
yielded $\Delta\alpha/\alpha=(-0.70\pm0.23)\times10^{-5}$, whilst
the Ni~\textsc{ii}/Cr~\textsc{ii}/Zn~\textsc{ii} subset yielded
$\Delta\alpha/\alpha=(-0.76\pm0.28)\times10^{-5}$. The good concordance
of these results despite the quite different relationship between
$\lambda$ and $q$ for these transitions suggests that the apparent
result is not due to some simple systematic effect.

A third quasar sample augmented the results above to produce a sample
with 128 quasar absorbers \citep{Webb:03a,Murphy:03,Murphy:03a}.
This analysis found $\Delta\alpha/\alpha=(-0.57\pm0.10)\times10^{-5}$.
\citet{Murphy:04:LNP} added an additional 15 absorbers to find\index{Keck Deltaalpha/alpha
 results@Keck $\Delta\alpha/\alpha$ results}
\begin{equation}
\Delta\alpha/\alpha=(-0.573\pm0.113)\times10^{-5},\label{eq:Murphy 2004 LNP result}
\end{equation}
which we think represents the previous best constraint on variation
in $\alpha$ from quasar spectra. \citet{Murphy:01c,Murphy:03a,Murphy:04:LNP}
considered systematic effects in detail and found that the results
presented cannot be ascribed to any known systematic effect.

Other groups have applied the MM method to search for variation in
$\alpha$. Unfortunately, almost all of these works present the results
of single quasar absorbers. Without a statistical sample, it is difficult
to ascertain whether the statistical uncertainties are a good reflection
of the true uncertainty, in that one cannot compare a set of $\Delta\alpha/\alpha$
results to check for over-dispersion about a fitted model. Additionally,
most of these analyses focus on a single absorber at $z\approx1.151$
toward the bright quasar HE~0515$-$4414. \citet{Quast:04a} reported
from an analysis of that absorber using a VLT/UVES spectrum that $\Delta\alpha/\alpha=(-0.04\pm0.19\pm0.27)\times10^{-5}$.
\citet{Levshakov:05} also analysed this absorber using a VLT/UVES
spectrum to find that $\Delta\alpha/\alpha=(0.04\pm0.15)\times10^{-5}$.
A later analysis by \citet{Chand:06a} reported $\Delta\alpha/\alpha=(0.10\pm0.22)\times10^{-5}$
from VLT/UVES observations, and $\Delta\alpha/\alpha=(0.05\pm0.24)\times10^{-5}$
from HARPS observations of the same absorber. \citet{Levshakov:06}
reported $\Delta\alpha/\alpha=(-0.007\pm0.084)\times10^{-5}$ from
VLT/UVES observations of the same absorber. The other highly studied
absorber is the $z\approx1.839$ absorber toward Q1101$-$264, for
which \citet{Levshakov:05} reported $\Delta\alpha/\alpha=(0.24\pm0.38)\times10^{-5}$,
updated to $\Delta\alpha/\alpha=(0.54\pm0.25)\times10^{-5}$ in \citet{Levshakov:07a}. 

Some of these works \citep[e.g.][]{Levshakov:07a} state that they
are using a method termed the Single Ion Differential $\alpha$ Method
(SIDAM)\index{single ion differential alpha
 method (SIDAM)@single ion differential $\alpha$ method (SIDAM)}, which measures shifts between transitions of only one species (Fe
\textsc{ii}). Whilst this does obviate the potential concern of spatial
segregation between different species, it also affords a lack of sensitivity
relative to the full MM method. SIDAM is just the multiplet analogy
of the AD method, and is simply a restricted case of the MM method.
In the ideal case, the use of SIDAM should cause no problems (other
than a potential loss of sensitivity), but in practice there are three
potential concerns, two of them related:
\begin{enumerate}
\item Consider the differences in the $q$ coefficients for the Fe \textsc{ii}
transitions used. All of the transitions display similar $q\sim1500$
except the $\lambda1608$ transition ($q\sim-1300$). Although the
Fe \textsc{ii} transitions other than $\lambda1608$ do have small
differences in $q$, the effect of this configuration is that almost
all of any potential signal arises from the use of the $\lambda1608$
transition. This can lead to two potential problems:

\begin{enumerate}
\item Because of the relative differences in the $q$ coefficients, any
error in the $\lambda1608$ laboratory wavelength or $q$ coefficient
will significantly affect the measured value of $\Delta\alpha/\alpha$.
\item For the same reason, if the $\lambda1608$ transition is contaminated
in some fashion {[}either with interloping transitions (see section
\ref{sub:Interlopers}) or through other data problems, such as cosmic
rays{]}, the bias introduced into the value of $\Delta\alpha/\alpha$
measured could be significant. \\
\\
The MM method helps to obviate both of these problems by reducing
the effect of a problem in any one transition. For instance, if the
SIDAM transitions were augmented with Mg \textsc{ii}, then the impact
of any contamination of the Fe \textsc{ii} $\lambda1608$ transition
would be reduced. Further reduction in the impact of any problems
with the Fe \textsc{ii} $\lambda1608$ transition would come from
using additional MM transitions.
\end{enumerate}
\item The arrangement of $q$ coefficients with wavelength when using SIDAM
(see figure \ref{Flo:alpha:q_vs_wl}) is such that $q$ is strongly
correlated with wavelength. This means that SIDAM is sensitive to
long-range wavelength miscalibrations in a way which the MM method
is not (providing that the full range of transitions can be used).
\end{enumerate}
It is for these reasons that we recommend the adoption of the full
MM method --- it confers significant resistance to problems such as
those described here by making full use of the data, and yields a
significant gain in sensitivity to $\Delta\alpha/\alpha$.

The only other statistical sample is that from \citet{Chand:2004}\index{Chand et al},
who reported $\Delta\alpha/\alpha=(-0.06\pm0.06)$ from an analysis
of 23 absorbers from VLT/UVES. The statistical precision stated is
surprising, given the relatively small sample size compared to that
of \citet{Murphy:04:LNP}. The VLT/UVES spectra are generally of higher
signal-to-noise, but not sufficiently so to explain the quoted statistical
precision. \citet{Murphy:07-2} and \citet{Murphy:08} analysed the
results of \citet{Chand:2004} and showed that the stated precision
is far in excess of the maximum theoretical precision allowed by the
data. The same criticisms apply to \citet{Chand:06a} and \citet{Levshakov:06}.
\citet{Murphy:08} applied the models of \citet{Chand:2004} to the
same data (however with different flux error arrays) using \textsc{vpfit},
and ultimately derived a more appropriate value of $\Delta\alpha/\alpha=(-0.64\pm0.36)\times10^{-5}$.
However, \citeauthor{Murphy:08} caution that the models used probably
under-fit the spectra, and therefore this result should not be considered
an optimal treatment of these absorbers. \citet{Murphy:08} also commented
on the optimisation algorithm used by \citet{Chand:2004}. In particular,
curves of $\chi^{2}$ vs $\Delta\alpha/\alpha$ given by \citet{Chand:2004}
display point-to-point fluctuations which are not substantially smaller
than unity. This implies that the optimisation algorithm used by \citet{Chand:2004}
was terminating prematurely, and therefore that the actual values
of $\Delta\alpha/\alpha$ given in that work are meaningless. We comment
further on the design of non-linear least squares optimisation algorithms
and their points of failure in chapter \ref{cha:MCMC}, and demonstrate
through the application of Markov Chain Monte Carlo methods that the
parameter estimates and uncertainties produced by \textsc{vpfit} are
robust.

\subsection{Objective of this chapter}

The purpose of this chapter is to use publicly available data from
VLT/UVES to generate a MM sample similar in size to the Keck sample.
If $\Delta\alpha/\alpha=0$ and we could achieve a statistical precision
comparable to or better than the Keck sample, we might be able to
demonstrate inconsistency of the Keck $\Delta\alpha/\alpha$ values
with the VLT $\Delta\alpha/\alpha$ values at the $\sim3.5\sigma$
level. On the other hand, concordance of the VLT result with the Keck
result would be a significant step in verifying the work of \citet{Murphy:04:LNP}.
We set out our analysis of the VLT data below, and compare our results
with the Keck sample.

\section{Atomic data}

In table \ref{tab:atomdata} we show the atomic data and $q$-coefficients
which were used in our analysis.\index{many-multiplet (MM) method!atomic data}

\begin{landscape}

\newcommand\oldtabcolsep{\tabcolsep}
\setlength{\tabcolsep}{0.5em}
\begin{center}
\vspace{-0.5em}
{\footnotesize
\begin{longtable}{:l;l;c;l;c;c;l;l;l;c;c;c;c}
\caption[Atomic data for use in many-multiplet analyses]{Atomic data for transitions usable in many-multiplet or
    alkali-doublet analyses, i.e.~transitions with precise laboratory
    wavelengths. Information for isotopic and hyperfine components is
    given in italics. Columns 1 and 2 show the common names used for
    the transitions. Column 3 shows the mass number for each ionic
    species. The derivation of the laboratory wavenumbers, $\omega_0$,
    is summarized by the value of $X$ as follows: 0 -- Measured
    wavenumber; 1 -- Inferred from measured component wavenumbers; 2
    -- Inferred from measured composite wavenumber and measured
    component splitting; 3 -- Inferred from measured composite
    wavenumber and calculated component splitting.  Column 6 gives the
    reference(s) for the wavenumber measurement and/or calculations
    (specified below the table). Vacuum laboratory wavelengths,
    $\lambda_0$, are derived from the wavenumbers. Columns 8 and 9
    show the lower and upper/excited state electronic configurations.
    The ID letters in column 10 offer a simple shorthand for labelling
    transitions used to fit absorption systems. Column 11 shows the
    ionization potential for the relevant ion, IP$^+$, and for the ion
    with a unit lower charge, IP$^-$.  Column 12 shows the oscillator
    strengths, $f$, taken from \citet{Morton:03} or the relative
    strengths of the hyperfine or isotopic components. The latter are
    taken from \citet{Rosman:98}. The $q$ coefficients and their
    uncertainties are from \citet{Dzuba:99:01,Dzuba:99:02,Dzuba:01,Dzuba:02} and \citet{Berengut:04a,Berengut:04b}. Note that
    uncertainties in the $q$ coefficients are representative not
    statistical. Wavenumbers are on the \citet{Whaling:95} Ar{\sc
      \,ii} calibration scale; the Fe{\sc \,ii} $\lambda$1608/1611 and
    Ni{\sc \,ii} wavenumbers have been scaled from their original
    values to account for the calibration difference between the
    Ar{\sc \,ii} scales of \citet{Norlen:73} and
    \citet{Whaling:95}. The exceptions to this are the Mg{\sc
      \,i}/{\sc ii} wavenumbers which are on a highly accurate
    absolute scale generated using a frequency-comb calibration
    system. The \citet{Whaling:95} scale best agrees with this
    absolute scale. Table modified from a version compiled by Michael Murphy.\label{tab:atomdata}}\\
\hline 
\multicolumn{1}{c}{Ion}&
\multicolumn{1}{c}{Tran.}&
\multicolumn{1}{c}{$A$}&
\multicolumn{1}{c}{$\omega_0$ [cm$^{-1}$]}&
\multicolumn{1}{c}{X}&
\multicolumn{1}{c}{Ref.}&
\multicolumn{1}{c}{$\lambda_0$ [\AA]}&
\multicolumn{1}{c}{Lower state}&
\multicolumn{1}{c}{Upper state}&
\multicolumn{1}{c}{ID}&
\multicolumn{1}{c}{IP$^-$, IP$^+$ [eV]}&
\multicolumn{1}{c}{$f$ or {\it \%}}&
\multicolumn{1}{c}{$q$ [cm$^{-1}$]} \\
\hline \endfirsthead

\multicolumn{13}{c}{{Continued from previous page.}} \\
\hline 
\multicolumn{1}{c}{Ion}&
\multicolumn{1}{c}{Tran.}&
\multicolumn{1}{c}{$A$}&
\multicolumn{1}{c}{$\omega_0$ [cm$^{-1}$]}&
\multicolumn{1}{c}{X}&
\multicolumn{1}{c}{Ref.}&
\multicolumn{1}{c}{$\lambda_0$ [\AA]}&
\multicolumn{1}{c}{Lower state}&
\multicolumn{1}{c}{Upper state}&
\multicolumn{1}{c}{ID}&
\multicolumn{1}{c}{IP$^-$, IP$^+$ [eV]}&
\multicolumn{1}{c}{$f$ or {\it \%}}&
\multicolumn{1}{c}{$q$ [cm$^{-1}$]} \\
\hline \endhead

\hline \multicolumn{13}{c}{{Continued on next page.}} \\ 
\endfoot

\hline
\endlastfoot
  
Mg{\sc \,i  } & 2026   & 24.31 & 49346.772611(36) & 1 & $       $ & 2026.4749792(15) & $3\rm{s}^2~^1\rm{S}_0                      $ & $3\rm{s}4\rm{p}~^1\rm{P}_1^{\rm{o}}                  $ & $a_1$ & ---\,,~7.65  & 0.113   & $87(7)     $ \\
\rowstyle{\itshape}   && 26    & 49346.854173(40) & 0 & $a      $ & 2026.4716298(16) & $                                          $ & $                                                    $ & $   $ &              & 11.0    & $          $ \\
\rowstyle{\itshape}   && 25    & 49346.807724(40) & 0 & $a      $ & 2026.4735372(16) & $                                          $ & $                                                    $ & $   $ &              & 10.0    & $          $ \\
\rowstyle{\itshape}   && 24    & 49346.756809(35) & 0 & $a      $ & 2026.4756281(14) & $                                          $ & $                                                    $ & $   $ &              & 79.0    & $          $ \\
  & 2852   & 24.31 & 35051.28076(19)  & 1 & $       $ & 2852.962797(15)  & $                                          $ & $3\rm{s}3\rm{p}~^1\rm{P}_1^{\rm{o}}                  $ & $a_2$ &              & 1.83    & $86(10)    $ \\
\rowstyle{\itshape}   && 26    & 35051.32015(25)  & 0 & $b      $ & 2852.959591(20)  & $                                          $ & $                                                    $ & $   $ &              & 11.0    & $          $ \\
\rowstyle{\itshape}   && 25    & 35051.29784(25)  & 0 & $b      $ & 2852.961407(20)  & $                                          $ & $                                                    $ & $   $ &              & 10.0    & $          $ \\
\rowstyle{\itshape}   && 24    & 35051.27311(17)  & 0 & $b      $ & 2852.963420(14)  & $                                          $ & $                                                    $ & $   $ &              & 79.0    & $          $ \\
\end{longtable}}
{\footnotesize $^a$\citet{Hannemann:06}; $^b$\citet{Salumbides:06}; $^c$\citet{Batteiger:09}; $^d$\citet{Griesmann:00}; $^e$\citet{Berengut:03}; $^f1.4\times({\rm Mass~shift})$; $^g$\citet{Aldenius:06}; $^h$\citet{Berengut:08}; $^i$\citet{Blackwell-Whitehead:05}; $^j$\citet{Nave:10a}; $^k$\citet{Porsev:09}; $^l$S.~Johansson~(priv.~comm.); $^m$\citet{Pickering:00}; $^n$\citet{Matsubara:03a}; $^o$\citet{Dixit:08a}; $^p$\citet{Matsubara:03b}.}
\end{center}

\setlength{\tabcolsep}{\oldtabcolsep}

\end{landscape}

\subsection{The $q$ coefficients and the effect of redshift}

\begin{figure}[tbph]
\noindent \begin{centering}
\includegraphics[bb=50bp 103bp 544bp 793bp,clip,angle=-90,width=1\textwidth]{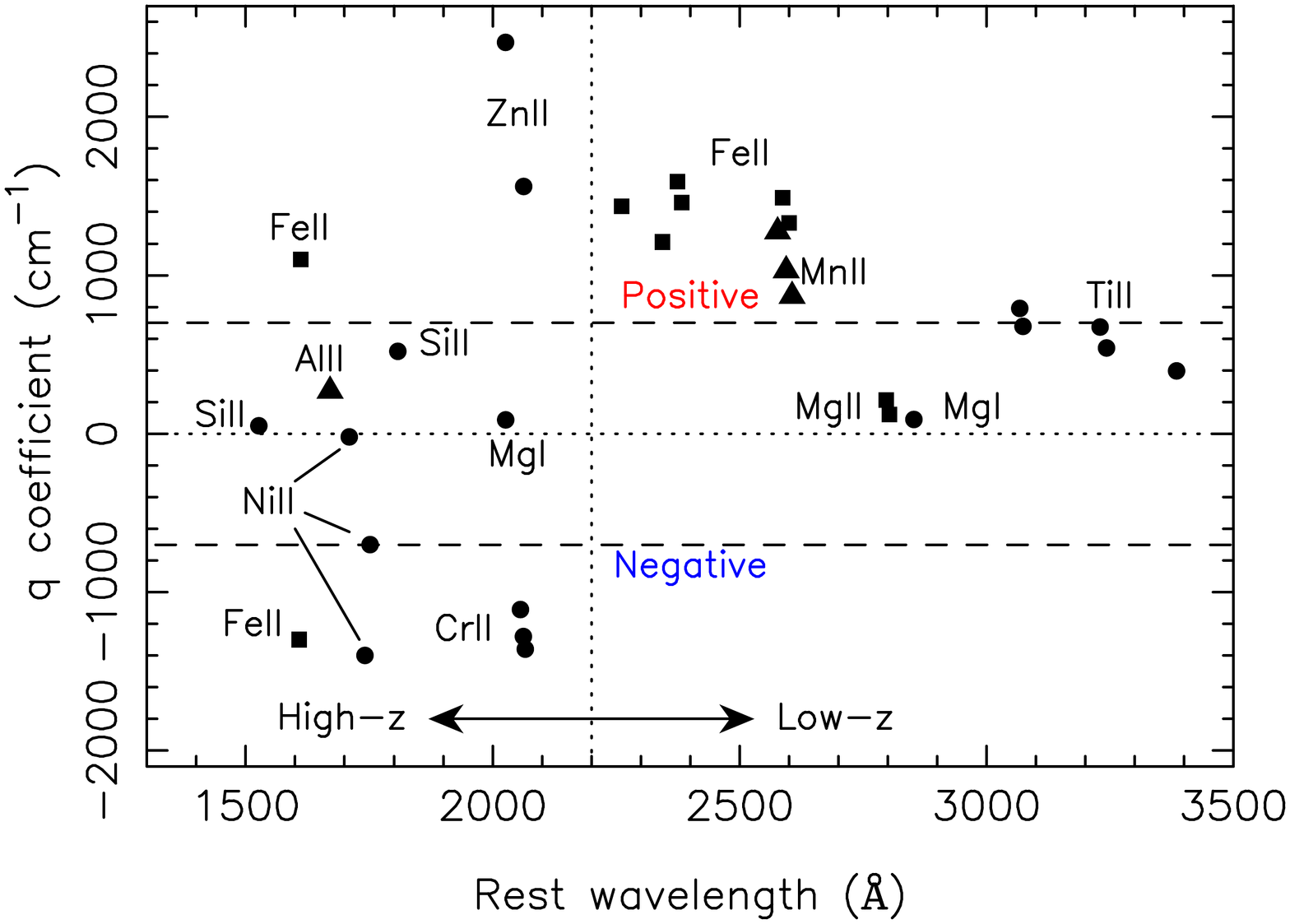}
\par\end{centering}

\caption[Relationship of $q$ coefficients with rest wavelength for the MM transitions]{Relationship of $q$ coefficients with rest wavelength. Points for different transitions have been given different shapes for clarity. In line with \citet{Murphy:PhD}, we define transitions with $q \gtrsim 700\,\mathrm{cm}^{-1}$ as ``positive shifters'' and transitions with $q \lesssim -700\,\mathrm{cm}^{-1}$ as ``negative shifters''. Similarly, we define transitions with $|q| \lesssim 300\,\mathrm{cm}^{-1}$ as ``anchor transitions'', and transitions with $300 \lesssim |q| \lesssim 700\,\mathrm{cm}^{-1}$ as ``mediocre shifters''. We have not shown Al \iiis here as we treat the Al \iiis transitions differently to previous analyses. For the low-$z$ Fe \ii/Mg \iis combination, the $q$ coefficients are strongly anti-correlated with wavelength, which makes this combination susceptible to low-order wavelength scale distortions. However, the arrangement of the $q$ coefficients with wavelength is much more complicated for the high redshift systems; this arrangement confers significant resistance to systematics. \label{Flo:alpha:q_vs_wl}}
\end{figure}

It is instructive to consider the relationship of the $q$ coefficients\index{q@$q$ coefficients}
with wavelength, and in turn with the redshifts of the absorbers.
We show the relationship of the $q$ coefficients with rest wavelength
in figure \ref{Flo:alpha:q_vs_wl}. 

At low redshifts ($z\lesssim1.3$), absorbers predominantly consist
of the Fe \ii/Mg \iis combination (with Mg \iscs sometimes included).
This combination gives a reasonable sensitivity to $\Delta\alpha/\alpha$,
with $\Delta q\sim1500$. However, it is worth noting that the arrangement
of the $q$ coefficients is such that $q$ is significantly anticorrelated
with wavelength. This means that any effect which stretches or compresses
the spectrum will mimic variation in $\alpha$. Similarly if the spectral
data regions which contain Mg and Fe are obtained at different time
periods, and there is a wavelength calibration offset between the
spectra taken at different times, then a spurious value of $\Delta\alpha/\alpha$
will emerge. The latter circumstance is possible for the earlier Keck
data, as all the data were obtained when HIRES only had a single CCD
chip; multiple exposures were required to obtain full coverage of
the optical region. Differences in quasar slit centering between the
exposures could produce significant wavelength miscalibrations between
the exposures. 

For absorbers of moderate redshift ($1.3\lesssim z\lesssim2.2$),
more transitions become useful. For high column density systems, Zn~\iis
and Cr~\iis may be observed. The Zn~\ii/Cr~\iis combination
is extremely important: the Zn~\iis transitions display strongly
positive $q$ and the Cr~\iis transitions display strongly negative
$q$. The difference between them provides $\Delta q\sim2900$ if
only the Zn~\iis $\lambda2062$ transition is used from the Zn~\iis
pair, and $\sim3800$ if the Zn~\iis $\lambda2026$ transition can
be used. Additionally, the Zn~\iis and Cr~\iis transitions are
interleaved with each other, with different transitions shifting in
different directions. This produces a unique signature if $\alpha$
varies which is difficult to mimic through systematic effects. If
Zn~\iis and Cr~\iis are observed then Ni~\iis is also generally
seen. In some cases, the Si~\iis $\lambda\lambda1526,1808$ and
Al~\iis $\lambda1670$ transitions are available. If these are fitted
together with Fe~\iis and Mg~\iis then the anticorrelation of
$q$ with wavelength disappears, creating a combination which is much
more resistant to systematics. At these redshifts, Fe~\iis $\lambda1608$
also appears for high column density systems. This provides a $\Delta q$
of $\sim2500$ between the Fe~\iis transitions. Above redshifts
of $\sim2$, it becomes difficult to use Mg~\iis because it either
falls in regions affected by sky emission or absorption, or because
the transitions are located out of the red end of the spectral coverage. 

For high redshifts ($z\gtrsim2.2$), the Mg transitions are no longer
useful. Instead, the predominant combination is some combination of
Si~\ii, Al~\iis and the Fe~\iis transitions. At even higher
redshifts the positive-$q$ Fe~\iis transitions also start to use
utility, leaving Fe~\iis $\lambda1608$ and $\lambda1611$ as the
only useful Fe~\iis transitions. For lower column density systems,
the Si~\ii/Al~\ii/Fe~\iis $\lambda1608$ combination is prevalent,
whereas for higher column density systems we see these transitions
and the Cr~\ii/Zn~\ii/Ni~\iis combination. 

Mn~\iis is seen at low to high redshifts, but the Mn~\iis transitions
have $q\sim1000$ and are situated at wavelengths between the Fe~\iis
positive-$q$ transitions and the Mg transitions. Thus, they add statistical
sensitivity but do not help break the anticorrelation of $q$ and
wavelength if used only in combination with the positive-$q$ Fe~\iis
transitions and the Mg~\iis transitions.

Ti~\iis is only seen in high-column density systems at low redshifts,
of which we have few in our sample. For $z\gtrsim1.5$ the Ti~\iis
transitions are strongly affected by sky emission/absorption and so
are difficult to use; for $z\gtrsim2$ they generally fall out of
the red end of the spectral coverage. The easiest way to search for
Ti~\iis is to search near the redshifts of DLAs, but DLAs can only
be quickly identified from the ground for $z\gtrsim1.5$, where the
$\lambda1216$ H~\textsc{i} absorption falls above the atmospheric
UV cutoff at $\sim3000\mathrm{\AA}$. Effectively, the long rest-wavelength
of Ti~\iis relative to the other transitions used is why it does
not feature prominently in our analysis. It is worth noting, however,
that Juliet Pickering and Matthew Ruffoni at Imperial College have
recently measured the wavelengths of the Ti~\iis $\lambda\lambda1910.60,1910.94$
lines to the requisite precision for use in a MM analysis (J.~Webb,
priv.\ communication). The oscillator strength of the $\lambda1610.60$
line ($f\approx0.20$) is comparable to the second strongest Ti~\iis
line in the existing MM set, the $\lambda3243$ line ($f\approx0.18$).
The $\lambda1910.60$ and $\lambda1910.94$ lines have $q$ coefficients
of $-1564$ and $-1783$ respectively \citep{Berengut:04a}. These
new measurements will help increase the number of absorbers which
are fitted with transitions having negative $q$, which will both
increase sensitivity to $\Delta\alpha/\alpha$ and help constrain
systematics.

\section{Spectral data}

Our spectral data are drawn from the archive of UVES, on the VLT.
A collaboration of researchers, coordinated by Michael Murphy, has
attempted to reduce all the publicly available quasar observations
which might be used for determining $\Delta\alpha/\alpha$ into wavelength-calibrated,
cleaned 1D normalised spectra. We are grateful for the efforts of
the following people: Matthew Bainbridge, Ruth Buning, Huw Campbell,
Robert Carswell, Ankur Chaudhary, Glenn Kacprzak, Ronan McSwiney,
Helene Ménager, Daniel Mountford, Michael Murphy, Jon Ouellet, Tang
Wei, Berkeley Zych.

Data were converted from 2D echelle spectra to 1D form using the \textsc{midas}
pipeline, provided by ESO. To calibrate the wavelength scale of the
science exposures, the \textsc{midas} routine uses a thorium-argon
(ThAr) list combined with a particular ThAr exposure. Unfortunately,
the default algorithm and line list used by the \textsc{midas} pipeline
is suboptimal for the reasons set out in section \ref{sub:mu wavelength calibration}.
As for the analysis of chapter \ref{cha:mu}, all our spectra are
calibrated with the algorithm and line list of \citet{Murphy:07b}.
The program \textsc{uves\_popler}%
\footnote{Available at \url{http://astronomy.swin.edu.au/~mmurphy/UVES_popler}.%
}\index{UVES_POPLER@UVES\_POPLER}, by Michael Murphy, was used to
combine multiple exposures into a single, 1D, normalised spectrum.
\textsc{uves\_popler} was specifically written for this purpose. All
this work was done by the aforementioned people.

\section{Methods \& methodology}

\subsection{Instrumental profile}

The exposures for most absorbers were taken over many nights, often
by different observers under significantly varying observing conditions.
In these circumstances, defining an instrumental profile is difficult.
In all cases we assumed a Gaussian instrumental profile with a velocity
FWHM of 6 km/s. Although this choice may cause inaccuracies for any
given absorber, particularly in the choice of the number of velocity
components (see section \ref{sub:alpha:modelling_velocity_structure}
below), because the nature of the error made will be random from absorber
to absorber it will average out over a sufficiently large ensemble
of absorbers.

\subsection{Modelling the velocity structure\label{sub:alpha:modelling_velocity_structure}}

\begin{figure}[tbph]
\noindent \begin{centering}
\includegraphics[bb=34bp 58bp 554bp 727bp,clip,angle=-90,width=0.9\textwidth]{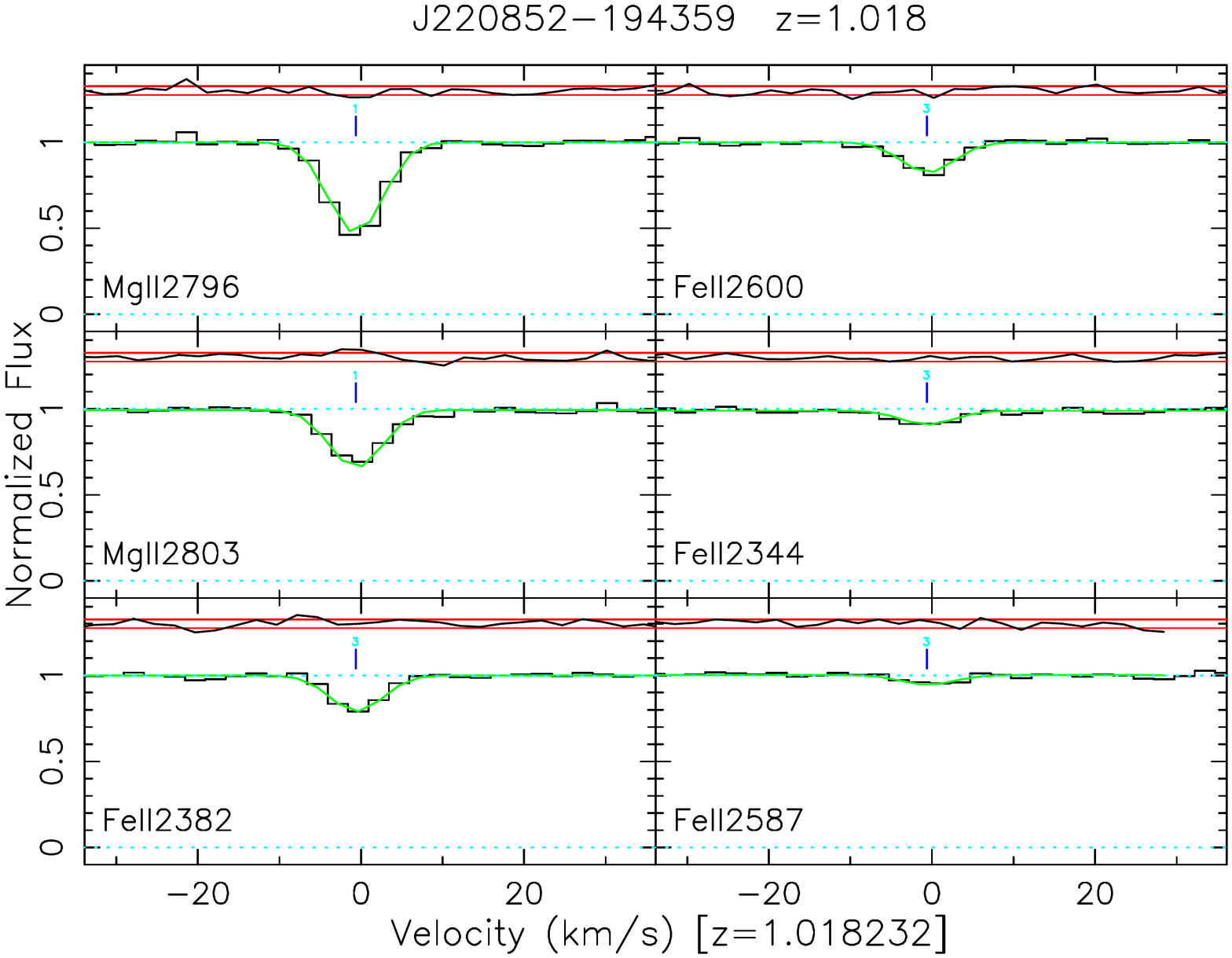}
\par\end{centering}

\caption[Fit of the $z_\mathrm{abs}=1.018$ toward J220852$-$1934359]{Our MM fit to the absorber at $z_\mathrm{abs}=1.018$ toward J220852$-$1934359, which is apparently well fitted by a single component. The horizontal scale indicates the velocity difference from the arbitrary redshift stated at the bottom for the given data points. The black line indicates the observed normalised flux, with the green line indicating our best fit solution. At the top of each box, the black line indicates the standardised residuals (that is, [data - model]/error), with the red lines indicating $\pm 1\sigma$. The position of the blue tick marks indicates the fitted position of the single component \label{J220852-194359-absorber}}
\end{figure}

\index{Voigt profile!model construction}Although a few absorbers
can be well modelled by a single Voigt profile (such as the absorber
shown in figure \ref{J220852-194359-absorber}), most absorbers display
complicated structure. The structure arises as a result of different
clouds of gas located along the line of sight, separated by non-cosmological
distances; we noted in section \ref{sec:quasar absorption lines}
that the absorbers are likely to be associated with galaxy disks and
halos. The typical velocity separation of the different components
of the absorption is typically tens to a few hundreds of kilometres
per second, which is typical of the velocity of galaxy rotation curves.
In general, the observed absorption profile can usually be adequately
modelled by adding Voigt components until a statistically acceptable
fit is achieved. 

The process of building up the Voigt profile model usually requires
synthesising the information available from the observable transitions
of different atomic species. Because the $\chi^{2}$ minimisation
process is non-linear, at each stage of building the Voigt profile
model one must supply initial guesses for the parameters. Choosing
poor parameter guesses can cause convergence to physically implausible
models, or non-convergence. Therefore the process of building the
model starts from regions which contain the most information available
to constrain the parameters of the model (that is, regions where the
structure of the model is most clearly visible ``by eye''). The
ideal region has transitions of high optical depth, but which are
not saturated. Regions of spectra where the optical depth is very
low or very high do not show the velocity structure of the absorber
clearly, and therefore contribute weaker constraints on both the velocity
structure and $\Delta\alpha/\alpha$. The model building process thus
starts with regions where the velocity structure is relatively clear,
and then proceeds to regions with less information to constrain the
parameter set. Therefore, the process of building up the Voigt profile
model proceeds as follows:
\begin{enumerate}
\item The fitting starts with the strongest unsaturated transition. For
low redshift systems, this is typically the Mg \iis $\lambda2796$
or $\lambda2803$ transition, although for the high column density
systems at low redshift this may be the Fe \iis $\lambda2383$ or
$\lambda2600$ transition. For high redshift systems, this may be
any of the Al \iis $\lambda1670$, Si \iis $\lambda1526$ or Fe
\iis $\lambda1608$ transitions. For intermediate redshift systems,
a wide variety of transitions are often visible, and any of the Fe
\ii, Mg \ii, Al \iis or Si \iis transitions were generally used.
Transitions which were clearly affected by sky absorption were not
used at this stage. Similarly, the initial fitting was generally not
done with the Mg \iis transitions where they fall at $\lambda>8000\mathrm{\AA}$
due to the possibility of contamination with sky absorption or emission.
\item For the initial transition selected, Voigt components were added until
a statistically acceptable fit was achieved (see section \ref{sub:Model-selection}
for the definition of ``statistically acceptable'').
\item This model was then applied to other transitions of the same atomic
species, if these transitions were available. This is almost always
possible for Fe \ii, for which the $\lambda\lambda\lambda\lambda2383$,
2600, 2344 and 2586 transitions are often observed simultaneously.
The Voigt profile model was then refined, adding or removing transitions
as necessary. By applying the model to transitions from the same atomic
species (i.e.\ same ground state), one is guaranteed that the same
model must be valid for all the transitions. Deviation of the data
from the model in one transition without a corresponding deviation
in the other transition therefore is a likely signal of problems with
the data or contamination by other species. 
\item The deviation of the data from the model just described is generally
due either to cosmic rays (which cause excess flux), absorption by
an interloping species, or sky emission or absorption. Where evidence
exists for pixels affected by cosmic rays, we clip out the affected
pixels so that they do not contribute to the calculation of $\chi^{2}$.
We explain the treatment of interlopers below in section \ref{sub:Interlopers}.
The \textsc{\small midas} pipeline attempts to subtract sky emission
as part of the spectral extraction process. However, we have noticed
that where sky emission is strong, the extraction appears to be imperfect,
leaving sharp spikes and dips in the spectrum. Where we find evidence
for such artifacts, we clip out the affected pixels. Where the spectra
are affected by absorption from sky lines, we either clip out the
affected pixels or do not use the affected transition. 
\item We then applied the model from the single atomic species to other
observable species. We generally first applied the model to transitions
of lower optical depth. This is because the fact that the transitions
are not saturated generally allows rapid convergence of the applied
model to a good fit. Because the line centres are generally identifiable
in the data, the parameters $N$ and $b$ for each line are relatively
uncorrelated with $z$, which makes convergence more likely to occur,
and more rapid. We then adjusted the model, adding extra components
where this decreased the AICC. 

\begin{enumerate}
\item In some cases, weak transitions were rejected by \textsc{vpfit}, because
their column density was driven below a user-adjustable cutoff (by
default $N<10^{8}\,\mathrm{cm}^{-2}$). This simply means that the
model is statistically preferred without these components. There is
no way to force the inclusion of these components in a statistically
justifiable sense. There \emph{may} be some bias introduced into the
value of $\Delta\alpha/\alpha$, but given that the components are
weak this bias should be small. Moreover, because this effect should
equally bias $\Delta\alpha/\alpha$ in a positive direction as much
as in a negative direction between different absorbers, it will average
out over many absorbers if it does exist. \citet{Murphy:PhD} investigated
the effect of fixing these dropped components at the column density
immediately before they were rejected, and found that ``{[}n{]}o
cases were found where the values of $\Delta\alpha/\alpha$ from the
different runs differed significantly''.
\end{enumerate}
\item We then applied the model to the transitions of higher optical depth.
Where saturation is present, there is a relative degeneracy between
$N$, $b$ and $z$ (although constraints on $b$ and $z$ come from
other transitions), and therefore we generally had to be more cautious
about our initial guesses for the values for $N$ for each component
of the model for those transitions. In general, we would manually
adjust the column densities for the different components to obtain
a reasonable starting fit by-eye, and then allow \textsc{\small vpfit}
to minimise $\chi^{2}$.
\item For the transitions fitted in point 6, it is sometimes necessary to
add more components in the optically thin regions of the profile.
Because the optical depths for the transitions in point 6 were higher
than those in point 3, this means that these components were not required
in the regions in point 3. These components were then introduced to
the other regions in an attempt to see whether they could be retained.
In some regions these components were retained, but, as in point 5(a),
these components were sometimes rejected in the weaker transitions.
\end{enumerate}
All of this process is guided by the model selection criteria set
out in section \ref{sub:Model-selection}.

\begin{figure}[tbph]
\noindent \begin{centering}
\includegraphics[bb=34bp 58bp 554bp 727bp,clip,angle=-90,width=1\textwidth]{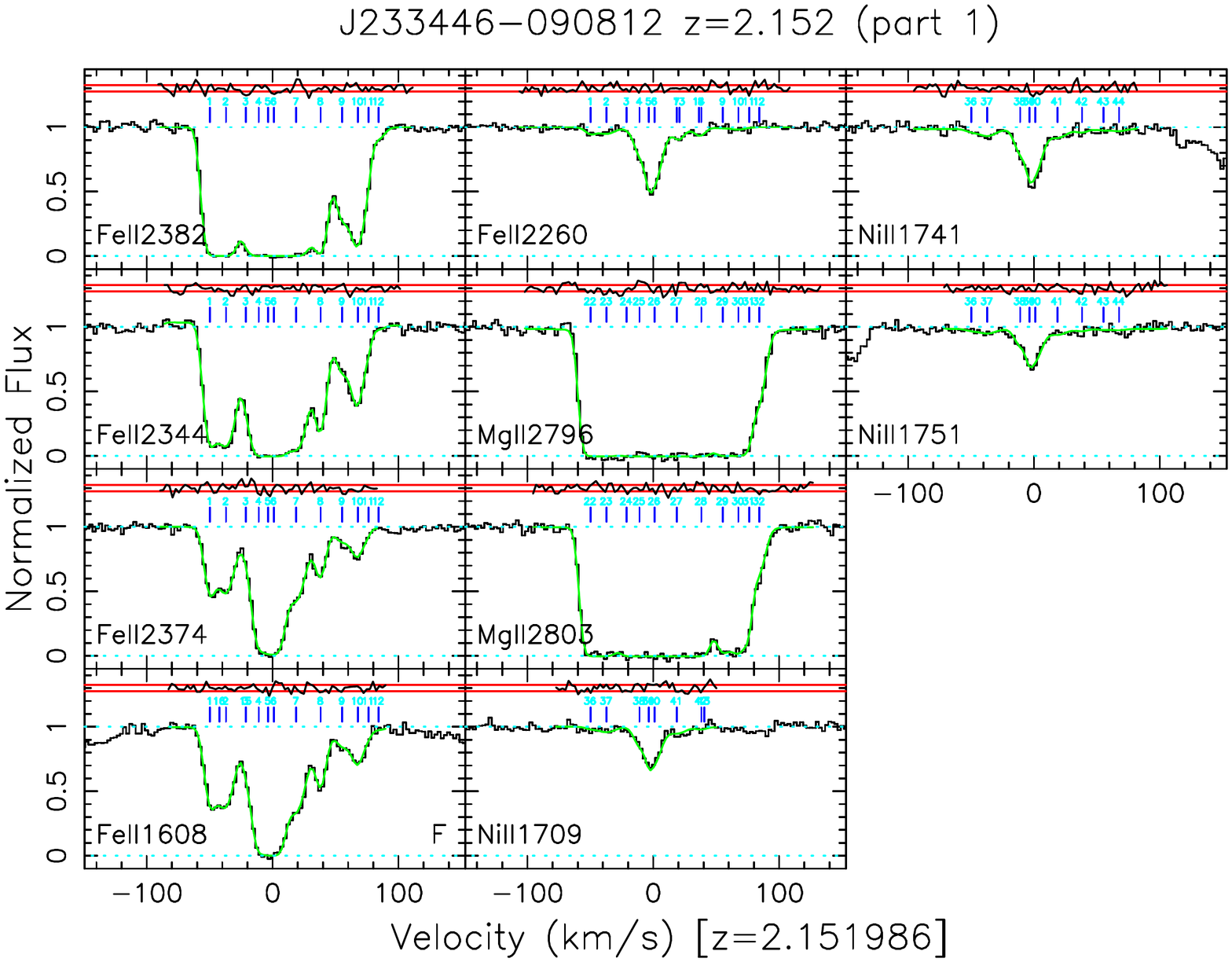}
\par\end{centering}

\caption[Part of the fit to the $z_\mathrm{abs}=2.152$ absorber toward J233446$-$090812]{Part of our MM fit to the $z_\mathrm{abs}=2.152$ absorber toward J233446$-$090812. This is a complex absorption system, requiring many components in order to achieve a statistically acceptable ($\chi^2_\nu \sim 1)$ fit. The horizontal scale indicates the velocity difference from the redshift stated at the bottom for the given data points. The black line indicates the observed normalised flux, with the green line indicating our best fit solution. At the top of each box, the black line indicates the standardised residuals (that is, [data - model]/error), with the red lines indicating $\pm 1\sigma$. The position of the blue tick marks indicates the fitted position of each component. We draw the reader's attention to the presence of a wide range of transitions, some with relatively small magnitude $q$ coefficients (Ni\,\textsc{ii} $\lambda 1709$ and the two Mg\,\textsc{ii} transitions), some with large magnitude, positive $q$ coefficients (Fe\textsc{\,ii} $\lambda\lambda\lambda 2382,2344,2374,2260$) and some with large magnitude, negative $q$ coefficients (Fe\textsc{\,ii} $\lambda 1608$, Ni\textsc{\,ii} $\lambda 1741,1751$). Also included in the fit, but not shown, are the transitions Cr\,\textsc{ii} $\lambda2056,2062,2066$, Zn\,\textsc{ii} $\lambda 2026,2062$ and Mn\,\textsc{ii} $\lambda 2576$. Note that for the stronger species, such as the Fe\,\iis $\lambda\lambda\lambda 2382$ and Mg\,\iis transitions, the centre regions of the profile are saturated, and thus a constraint on $\Delta\alpha/\alpha$ only comes from the optically thin wings. Conversely, for the weaker species (for example, Fe\,\textsc{ii} $\lambda 1608,2260$ and the Ni\,\textsc{ii} transitions) most or all of the profile is optically thin, and thus a constraint on $\Delta\alpha/\alpha$ is derived across the whole profile. Importantly, a single velocity structure model provides a good model to all the observed MM transitions. This serves to validate an assumption underlying the MM method, namely that spatial segregation of the different species --- if present --- must be relatively small. We note the presence of two weak interlopers in Fe\,\iis $\lambda2260$, and one in Fe\,\iis $\lambda1608$, yielding an additional three tick marks. \label{J223446-090812-absorber}}
\end{figure}

We show an example of a complicated Voigt profile fit in figure \ref{J223446-090812-absorber}
to illustrate the process described above.

At any stage of the above process, evidence may emerge for contamination
by cosmic rays or interlopers. The decision as to whether such contamination
exists is effectively based on consistency between the models used
for and observed flux data in different spectral regions. For instance,
with only two transitions (and no other information), one cannot determine
whether contamination is present in a particular transitions. However,
with three transitions (or with the presence of additional information),
the concordance between two transitions allows one to infer that the
third transition is problematic. As the model is progressively constructed,
the confidence with which one can infer the presence of an interloper
increases --- concordance in corresponding regions of the model between
many transitions and significant excess absorption in the corresponding
region in another transition constitutes strong evidence for an interloper. 

Of course, in some cases the interloping transition can be identified
on account of other transitions from the same atomic species. This
is particularly true for the C~\ivs and S~\ivs doublets. Nevertheless,
the process of building up a good Voigt profile model depends on the
synthesis of information from all the transitions present.

\subsubsection{Gravitational lenses}

\index{gravitational lenses}We discovered a small number of absorbers
for which the velocity structure appeared to be the same for transitions
arising from the same ground state, but where the line intensities
differed substantially. A particular example of this is the $z=0.82$
absorber along the line of sight to J081331+254503. Further investigation
demonstrated that this quasar is known to be gravitationally lensed.
The complex line-of-sight geometry causes the effect described. We
discarded any system for which this problem appeared and for which
we could identify the quasar as a known gravitationally lensed system.

\subsection{Random and systematic errors\label{sub:alpha:rand_sys_errs}}

\index{statistical errors!random and systematic}Unfortunately, the
Voigt profile decomposition is not unique. Errors in modelling the
velocity structure may impact the fitted value of $\Delta\alpha/\alpha$.
We thus distinguish between three different types of errors which
may affect $\Delta\alpha/\alpha$:
\begin{enumerate}
\item \emph{Statistical errors}. These errors are simply the errors on $\Delta\alpha/\alpha$
which derive from the propagation of uncertainty from the flux error
array via the Voigt profile model. These errors are the errors produced
by \textsc{vpfit} from the covariance matrix at the best-fitting solution.
\item \emph{Random errors}. Random errors are any effects which might cause
$\Delta\alpha/\alpha$ to be measured inaccurately when considering
a single absorber. Significant errors made in determining the correct
velocity structure could cause an error of this type. \citet{Murphy:08}
demonstrated that an under-fitted spectrum (i.e\ one with a deficient
model for the velocity structure) gives erroneous values of $\Delta\alpha/\alpha$.
Other potential causes of random errors include: \emph{i)} Spatial
segregation of different elements (which cannot be preferentially
biased along the radial sightline over a large number of absorbers);
\emph{ii)} random blends with other transitions; \emph{iii)} random
departures of the wavelength calibration solution from the true wavelength
scale; \emph{iv)} cosmic rays and other uncleaned data glitches; \emph{v)}
incorrect determination of the broadening mechanism (turbulent or
thermal) for any component. Importantly, the effect of random errors
will average to zero when considering an ensemble of absorbers. This
is because these errors will displace $\Delta\alpha/\alpha$ to be
more positive as often as they will displace it to be more negative.
When considering only a single absorber, this type of effect must
be considered a systematic error. However, when considering an ensemble
of absorbers, this impact of this type of effect is random, and merely
adds extra scatter into the data.
\item \emph{Systematic errors}. This is any error which systematically affects
the value of $\Delta\alpha/\alpha$. Such effects would include: \emph{i)}
inaccuracies in the laboratory wavelengths; \emph{ii)} a different
heavy isotope abundance for Mg in the clouds relative to terrestrial
values; \emph{iii)} systematic blends with other lines; \emph{iv)}
time-invariant differential light paths through the telescope for
different wavelengths; \emph{v)} atmospheric dispersion for spectra
taken without an image rotator; \emph{vi)} differential isotopic saturation,
and; \emph{vii)} wavelength miscalibration due to thorium-argon line
list inaccuracies. \citet{Murphy:03} considered many potential systematic
effects in detail. Note that these effects will not necessarily produce
the same spurious shift in $\Delta\alpha/\alpha$ in every absorber.
For example, if some transitions have inaccurate laboratory wavelengths,
the effect on $\Delta\alpha/\alpha$ will depend on which other transitions
are fitted in the same absorber. Nevertheless, the above effects are
considered systematic errors because they will cause similar or correlated
shifts in certain subsets of absorbers.
\end{enumerate}
There is a crucial distinction between these effects: some effects
may be considered systematics in single absorbers, but are not systematics
in an ensemble of absorbers. For this reason, we are cautious against
placing too much emphasis on the interpretation of the value of $\Delta\alpha/\alpha$
from any individual absorber. We demonstrate later that the impact
of random effects is non-negligible. We explain our treatment of random
effects in section \ref{sub:LTS method}.

\subsection{Interlopers\label{sub:Interlopers}}

\index{interlopers}Some transitions display excess absorption beyond
what is predicted from other transitions from the same atomic species.
This excess absorption is caused by another absorber located along
the line of sight to the quasar, which is usually extragalactic. Even
where a prediction cannot be made from another transition of the same
atomic species, interlopers can still be detected when many transitions
are fitted together, as the redshifts and $b$-parameters of each
component are constrained. Although the contaminated sections of spectrum
can be discarded, it is often possible to adequately model the contamination,
thereby maximising use of the spectral data.

We distinguish two types of interlopers: identifiable and unidentifiable.

\subsubsection{Identifiable interlopers}

In some cases, the interlopers can be identified and modelled simultaneously
with the MM transitions. By ``identified'' we mean that the redshift,
atomic species and wavelength of the transition which causes the excess
absorption can be determined. In principle, one needs accurate rest
wavelengths and $q$ coefficients for the interloping transitions.
This means that the interloping transition can be modelled if it is
an MM transition from an absorber at a different redshift, or if it
is Si\,\textsc{iv} $\lambda\lambda1393$ or $1402$. If the interloper
is from the C\,\textsc{iv} doublet, the contamination can also be
modelled despite the fact that the rest wavelengths for this doublet
are relatively poorly known. This is done by allowing the C\,\textsc{iv}
transitions to have a separate value of $\Delta\alpha/\alpha$, which
is then discarded. This extra parameter effectively absorbs any error
introduced through inaccurate knowledge of the rest wavelengths.

\subsubsection{Unidentifiable interlopers}

\begin{figure}[tbph]
\noindent \begin{centering}
\includegraphics[bb=34bp 58bp 554bp 727bp,clip,angle=-90,width=1\textwidth]{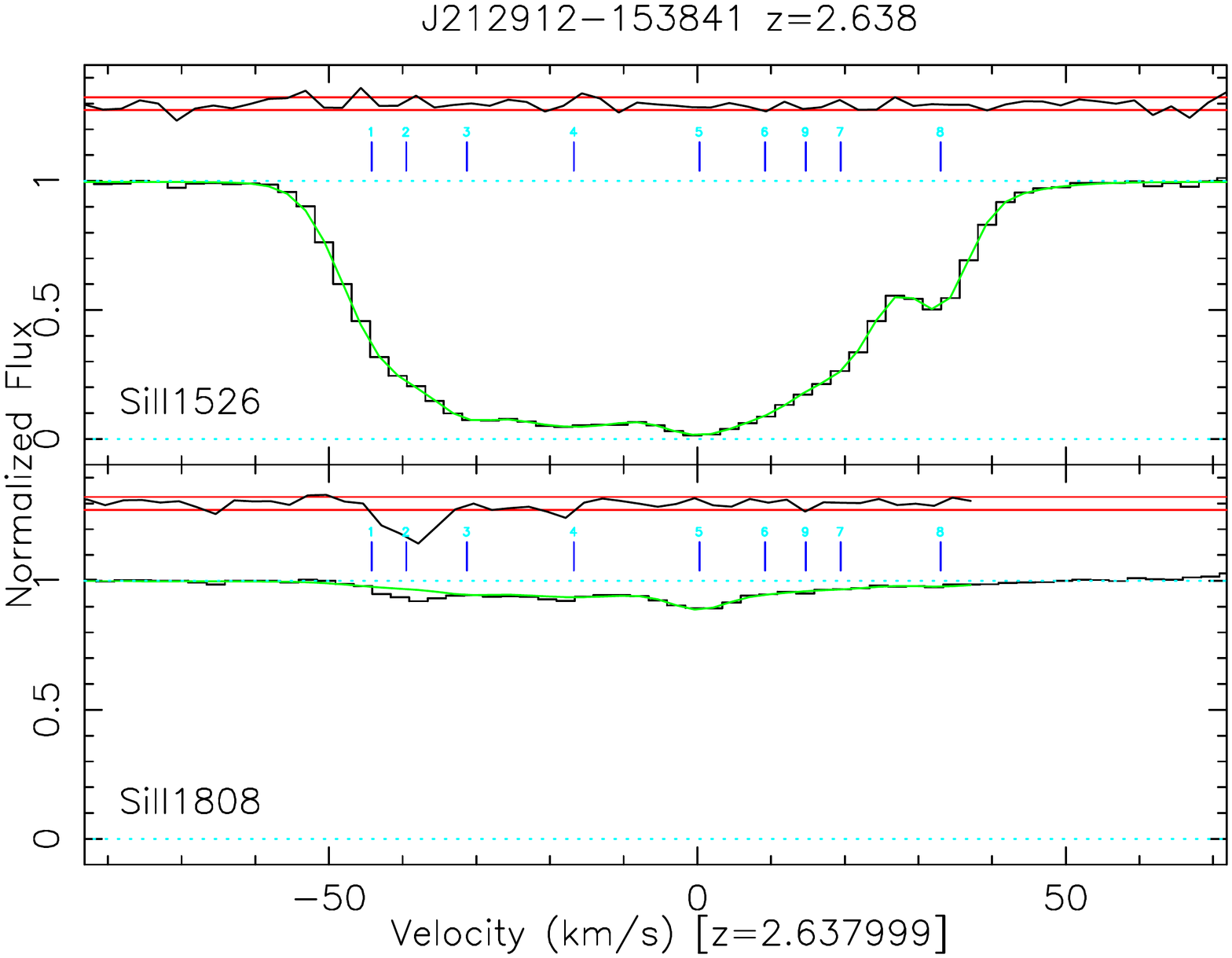}
\par\end{centering}

\caption[Example fit where the use of an interloping component provides a statistically acceptable fit ($z_\mathrm{abs}=2.638$ toward J212912$-$153848)]{The Si\,\iis transitions from our MM fit to the absorber at $z_\mathrm{abs}=2.638$ toward J212912$-$153848, shown without an included interloper. The horizontal scale indicates the velocity difference from the arbitrary redshift stated at the bottom for the given data points. The black line indicates the observed normalised flux, with the green line indicating our best fit solution. At the top of each box, the black line indicates the standardised residuals (that is, [data - model]/error), with the red lines indicating $\pm 1\sigma$. The position of the blue tick marks indicates the fitted position of the single component. The strength of the Si\,\iis $\lambda1808$ transition can be predicted from the model for Si\,\iis$\lambda 1526$. Si\,\iis $\lambda1808$ shows excess absorption at $v \approx -40\,\mathrm{km\,s^{-1}}$ which cannot be explained by Si\,\iis$\lambda 1526$. To account for this, we include a single, unconstrained interloper. After including the interloper, the fit is statistically acceptable. Our fit including the interloper may be found in figure \ref{appfig:J212912-153841-z2.638}.\label{J212912-153841-absorber}}
\end{figure}

In many cases, however, the interloping transition can not be identified.
In this case, our decision as to how to proceed depends on the degree
of contamination. If the degree of contamination is small, and confined
to a small area of the observed profile, we can include unknown interloping
transitions where the residuals of the fit ({[}data - model{]}/error)
are bad until a statistically acceptable fit is achieved. Doing this
provides a statistically acceptable model of the contamination. Note
that the contribution to $\Delta\alpha/\alpha$ of the affected MM
transition will be reduced as a result of this, as the interlopers
included are unconstrained by other spectral regions. We show an example
of this in figure \ref{J212912-153841-absorber}.

\begin{figure}[tbph]
\noindent \begin{centering}
\includegraphics[bb=34bp 58bp 554bp 727bp,clip,angle=-90,width=1\textwidth]{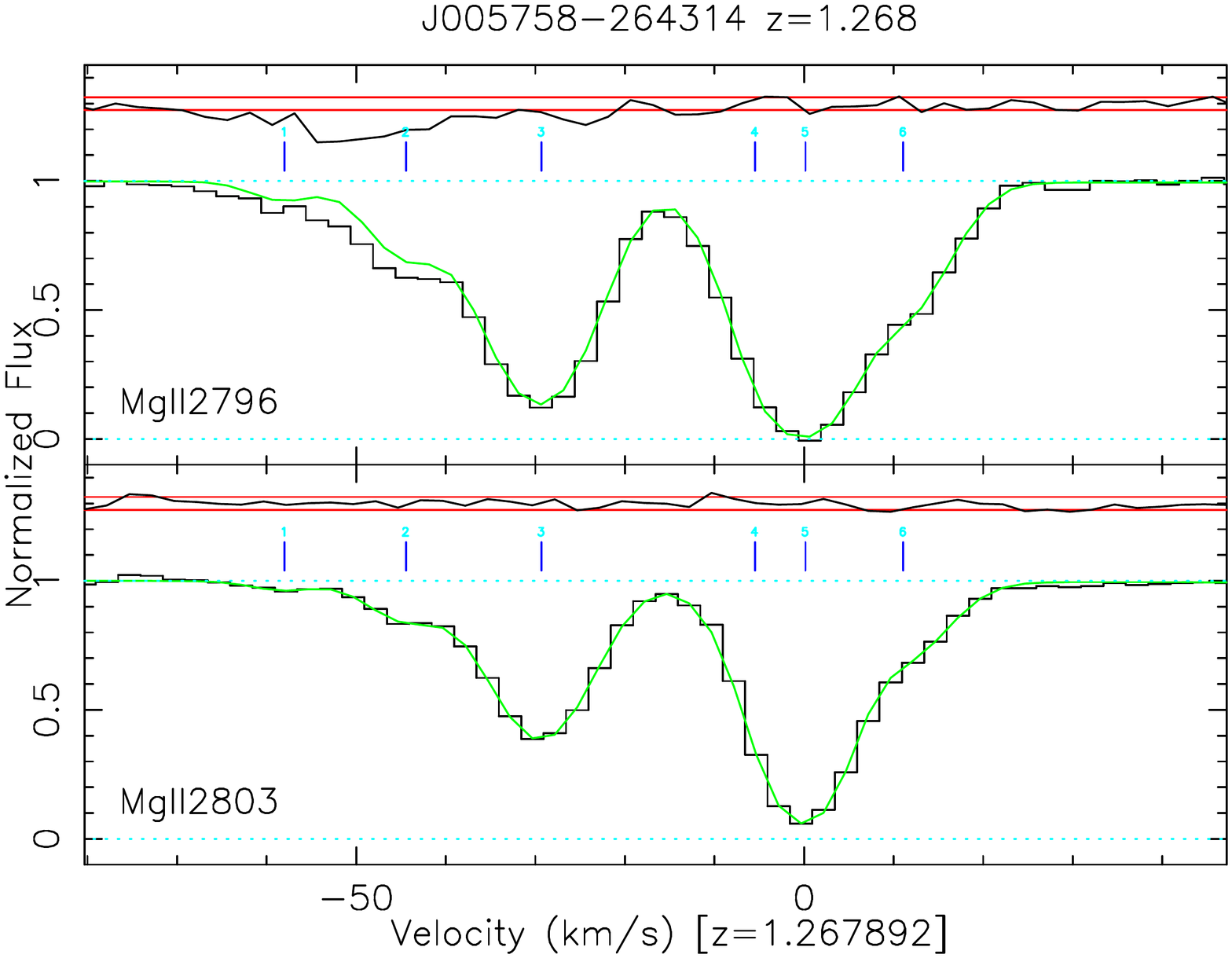}
\par\end{centering}

\caption[Example fit where we clip pixels to remove the effect of interlopers ($z_\mathrm{abs}=1.268$ toward J005758$-$264314)]{Part of our MM fit to the absorber at $z_\mathrm{abs}=1.268$ toward J005758$-$264314. The horizontal scale indicates the velocity difference from the arbitrary redshift stated at the bottom for the given data points. The black line indicates the observed normalised flux, with the green line indicating our best fit solution. At the top of each box, the black line indicates the standardised residuals (that is, [data - model]/error), with the red lines indicating $\pm 1\sigma$. The position of the blue tick marks indicates the fitted position of the single component. We show here the model from Mg\,\iis $\lambda2803$ plotted over the spectral region for Mg\,\iis $\lambda2796$. One can see that there is significant excess absorption in Mg\,\iis $\lambda2796$ for $v\lesssim -40\,\mathrm{km\,s^{-1}}$. There also appears to be excess absorption in the range $-40\,\mathrm{km\,s^{-1}} \lesssim v \lesssim -20\,\mathrm{km\,s^{-1}}$. Due to the wide range in velocity over which the absorption occurs, instead of attempting to model the absorption we clip away all pixels for $v \lesssim -20\,\mathrm{km\,s^{-1}}$ in the Mg\,\iis $\lambda2796$ region for our actual fit (which may be found in figure \ref{appfig:J005758-264314-z1.268}).\label{J005758-264314-absorber}}
\end{figure}

As the degree of contamination begins to increase, the potential error
introduced into $\Delta\alpha/\alpha$ may grow. Our treatment of
transitions affected by significant contamination depends on whether
there are other transitions from the same species available which
can be used to constrain the velocity structure for that transition.
In the case of Fe~\ii, a wide variety of transitions are often available.
Effectively, a fit to several other transitions of the same species
may allow the structure of the contamination to be determined, particularly
if the SNR is high. However, in some cases, there may be no other
transitions which can be used to obtain the velocity structure. This
may occur as a result of all of the transitions suffering from contamination
from different absorbers, or because the spectra only includes one
of the transitions of that species (due to gaps in the spectral coverage),
or because the species has no other transitions which could be used
for that purpose. The last case is particularly problematic for Al~\ii,
for which we only use the Al~\iis $\lambda1670$ transition ---
there are no other Al~\iis transitions which can be used to directly
constrain the Al~\iis structure. This issue also occurs for Si~\iis
--- although in theory both the Si~\iis $\lambda\lambda1526,1808$
transitions can be used to constrain the velocity structure of Si~\ii,
the oscillator strength for the $\lambda1526$ transition ($f\approx0.13$)
is much larger than for the $\lambda1808$ transition ($f\approx0.002$).
For many systems observed to have Si~\iis $\lambda1526$ absorption,
the column density of Si~\iis is not large enough to detect the
$\lambda1808$ transition. Even if the $\lambda1808$ transition is
detected, it may be too weak to provide a meaningful constraint on
the Si~\iis structure. 

In cases where we are unable to obtain a good constraint on the velocity
structure of a particular species from other unaffected transitions
of the same species, and the degree of contamination is not small,
we are very cautious about inserting interlopers, due to the potential
bias this could introduce into $\Delta\alpha/\alpha$ for this absorber.
Note that any bias introduced here is a \emph{random} effect, and
therefore will average out when considering an ensemble of absorbers.
Nevertheless, we wish to avoid introducing extra scatter into the
$\Delta\alpha/\alpha$ values where possible. In these cases, we clip
out the pixels which appear to be affected by contamination, leaving
a wide buffer on either side. Note that we can only do this where
another transition from the same species exists. Otherwise, components
situated in the middle of the clipped pixels might have very little,
if any spectral data to constrain them, and thus their column densities
could take on values which would not be consistent with the general
model used for the absorber. If there is no other transition for the
species in this case, we simply do not use the transition. In figure
\ref{J005758-264314-absorber} we show an example of where we have
clipped out pixels because of contamination by an interloper.

Fitting contamination has the potential to introduce a \emph{random}
bias for individual absorbers, however this may nevertheless reduce
\emph{systematic} effects. To see this consider a hypothetical absorber
where the Mg~\iis $\lambda\lambda2796,2803$, Al~\iis $\lambda1670$
and the Fe~\iis $\lambda\lambda\lambda$2383, 2600, 2344 transitions
are available, but where the Al~\iis $\lambda1670$ transition suffers
from some minor contamination in part of the observed profile, and
the absorber of the contaminating transition cannot be identified.
One could simply ignore the Al~\iis $\lambda1670$ transition and
fit the Mg~\iis and Fe~\iis transitions. Deriving $\Delta\alpha/\alpha$
from just the Mg~\ii/Fe~\iis combination is not very robust to
a simple stretching or compression of the wavelength scale \citep{Murphy:03}.
The Mg~\ii/Fe~\ii/Al~\iis combination just described is much
more robust against this effect, as the Mg~\iis and Al~\iis anchor
transitions are positioned on other side of the high-$q$ Fe~\iis
transitions. Modelling the contamination of the Al~\iis profile
here may introduce a random error, but this can be averaged out in
the context of many absorbers. The potential introduction of this
random error is easily justified by the increased resistance to systematic
effects (such as wavelength scale distortions, which in principle
could be common to many absorbers or spectra). In the situation just
described, the only choices are to discard the Al~\iis transition,
or to model the contamination. We choose the latter option where the
degree of contamination is minor, and the former where the contamination
is severe.

\subsubsection{Transitions in the Lyman-$\alpha$ forest}

It also happens that some transitions fall in the Lyman-$\alpha$
forest (``the forest''), a dense series of absorption lines blueward
of the quasar Lyman-$\alpha$ emission line. These transitions are
caused by H~\isc absorption along the line of sight to the quasar.
We are cautious about using MM transitions which fall in the forest,
due to the uncertainties in determining the structure of the forest.
Nevertheless, the use of MM transitions in the forest may afford significantly
better constraints on $\Delta\alpha/\alpha$. Although this often
occurs with low-$z$ Mg~\ii/Fe~\iis absorbers, with some Fe~\iis
transitions falling in the forest, it also occurs in high-$z$ systems,
where the Si~\iis $\lambda1526$/Al~\iis $\lambda1670$/Fe~\iis
$\lambda1608$ combination is common. Where the SNR is high for the
transitions which fall in the forest, we model the forest structure
with H~\isc absorption. If the SNR ratio is low, determination of
the forest structure can be difficult, and therefore we do not utilise
the contaminated transitions. Again, we emphasise that although this
has the potential to introduce bias for a single absorber, because
the contamination is random from absorber to absorber, it must average
out over a large number of systems. We have used transitions which
fall in the forest in 27 of the 1142 spectral fitting regions in the
VLT sample (2.4 percent).

\subsection{Cr II $\lambda2062$, Zn II and Mg I $\lambda2026$}

The Cr~\iis $\lambda2062$ $(2062.24\mathrm{\AA})$ and Zn~\iis
$\lambda2062$ $(2062.66\mathrm{\AA})$ lines are relatively closely
spaced, being separated by $\approx62\,\mathrm{km\, s^{-1}}$. For
narrow absorption systems, one can distinguish between these transitions
as they do not overlap. However, these transitions are most commonly
associated with DLAs, where the velocity structure is generally complicated,
and the system displays absorption over tens to hundreds of km/s.
In this case, the Cr~\iis $\lambda2062$ and Zn~\iis $\lambda2062$
transitions often overlap. The velocity structure for these transitions
can be determined by simultaneously modelling these transitions with
the Cr~\iis $\lambda2052,2056,2066$ and Zn~\iis $\lambda2062$
transitions. 

There is one point of caution here. For high column density systems,
a potential blend exists with Mg~\iscs $\lambda2026$. Mg~\iscs
$\lambda2026$ is weak, with oscillator strength $f=0.113$, and is
rarely seen. One can in principle use the Mg~\iscs $\lambda2852$
($f=1.83$) information to constrain the Mg~\iscs structure. In
this case, a joint fit of Zn~\iis $\lambda2026$ and Mg~\iscs
$\lambda\lambda2026,2852$ will ensure that the Zn~\iis $\lambda2026$
results are not biased by any absorption due to Mg~\iscs $\lambda2026$.
However, the absorbers for which Mg~\iscs $\lambda2026$ might be
detected are often at high redshift, in which case Mg~\iscs $\lambda2852$
is often unusable, either due to heavy contamination by sky emission
or absorption, or because it is located out of the red end of the
spectral coverage. In this particular circumstance, we are generally
cautious about fitting Zn~\iis $\lambda2026$. Where we consider
that the Zn~\iis $\lambda2026$ transition might be affected by
Mg~\iscs $\lambda2026$, and we are unable to utilise Mg~\iscs
$\lambda2852$, we do not include the Zn~\iis $\lambda2026$ transition.

\subsection{Physical constraints\label{sub:alpha:physical_constraints}}

As in the previous Keck analyses, we required that the $b$-parameter
of each modelled component for a particular species in the fit is
related to the corresponding components for other species. The two
extreme cases are wholly thermal broadening and wholly turbulent broadening.
In general, there will be contributions from the two mechanisms ($b^{2}=b_{\mathrm{therm}}^{2}+b_{\mathrm{turb}}^{2}$),
however we have found that most systems are generally well-fitted
with turbulent broadening. As noted in section \ref{sub:meth:VPFIT}
\citep[and by][]{Murphy:PhD} it is possible to explicitly determine
the degree of thermal and turbulent broadening, however in this circumstance
the $b$-parameters are generally poorly determined, which makes the
optimisation difficult. 

It turns out that the turbulent fit is preferred on the basis of the
AICC in 71 percent of the absorbers, and the thermal fit in 29 percent
of the absorbers. However, it should be noted that the fits were initially
constructed with turbulent broadening and then converted to thermal
broadening. It may be that if the fits were constructed thermally
and then converted to turbulent that these figures might change significantly.
We emphasise that mistakes made in choosing turbulent or thermal fitting
may bias $\Delta\alpha/\alpha$ for a single absorber, but these effects
must average to zero over a large number of absorbers due to the random
nature of the bias from absorber to absorber. 

The previous analyses of the Keck results required that $\Delta\alpha/\alpha$
calculated using both thermal and turbulent fits differed by no more
than $1\sigma$ for that absorber to be included in their ensemble,
where the difference is considered only in terms of the statistical
error. However, the generally higher SNR of the VLT data (compared
to the Keck data) often leads to very precise statistical bounds on
$\Delta\alpha/\alpha$. This makes the $1\sigma$-difference criterion
difficult to fulfil in a significant number of cases. We describe
below how we resolve any potential inconsistency between $\Delta\alpha/\alpha$
values from the thermal and turbulent fits.

In determining how to resolve any potential inconsistency, there are
three cases to consider. \emph{i)} Where the difference between the
fits is substantial, as measured by the AICC, one wants to take the
statistically preferred fit. \emph{ii)} Where the quality of the fits
is similar ($\mathrm{AICC}_{\mathrm{turbulent}}\approx\mathrm{AICC}_{\mathrm{thermal}}$),
and the values of $\Delta\alpha/\alpha$ are the same, then it does
not matter which fit is used. \emph{iii)} If the values of the AICC
for the thermal and turbulent fits are similar, but the values of
$\Delta\alpha/\alpha$ produced by those fits differ significantly,
then the statistical precision accorded to $\Delta\alpha/\alpha$
should be reduced to account for the conflicting evidence, and value
of $\Delta\alpha/\alpha$ should be somewhere between the two cases.

To resolve this problem, we use a method-of-moments estimator\index{method-of-moments estimator}
which takes into account the relative differences in the AICC and
the agreement, or otherwise, of the values of $\Delta\alpha/\alpha$.
We estimate the underlying probability distribution of $\Delta\alpha/\alpha$
for the absorber in question as the weighted sum of two Gaussian distributions
(one for the thermal result, one for the turbulent), with centroids
given by the best fit value of $\Delta\alpha/\alpha$ for each fit,
and $\sigma$ equal to $\sigma_{\Delta\alpha/\alpha}$ for each fit.
We weight the sum by the penalised likelihood of the fits, via the
AICC \citep[see][]{Liddle:07}. That is, if
\begin{align}
k & =\exp(-\mathrm{AICC}_{\mathrm{turbulent}}/2)+\exp(-\mathrm{AICC}_{\mathrm{thermal}}/2),\\
j_{1} & =\exp(-\mathrm{AICC}_{\mathrm{turbulent}}/2)/k,\\
j_{2} & =\exp(-\mathrm{AICC}_{\mathrm{thermal}}/2)/k,\\
a_{1} & =\Delta\alpha/\alpha_{\mathrm{turbulent}},\\
a_{2} & =\Delta\alpha/\alpha_{\mathrm{thermal}},\\
s_{1} & =\sigma(\Delta\alpha/\alpha_{\mathrm{turbulent}}),\quad\mathrm{and}\\
s_{2} & =\sigma(\Delta\alpha/\alpha_{\mathrm{thermal}})
\end{align}
then matching the first two moments of our weighted sum of distributions
with a Gaussian yields
\begin{align}
m & =\Delta\alpha/\alpha=j_{1}a_{1}+j_{2}a_{2},\quad\mathrm{and}\\
\sigma_{\Delta\alpha/\alpha} & =\sqrt{j_{1}s_{1}^{2}+j_{2}s_{2}^{2}+j_{1}a_{1}^{2}+j_{2}a_{2}^{2}-m^{2}}\label{eq:alpha:mom_s}
\end{align}
This covers all the cases described above. In particular, where the
AICC is similar but $\Delta\alpha/\alpha$ differs significantly between
the turbulent and thermal fits, the estimated error increases with
the difference between them, providing resistance to incorrectly determining
the line broadening mechanism. To see this, note that with $j_{1}+j_{2}=1$,
equation \ref{eq:alpha:mom_s} reduces to 
\begin{equation}
\sigma_{\Delta\alpha/\alpha}=\sqrt{j_{1}s_{1}^{2}+(1-j_{1})s_{2}^{2}+j_{1}(1-j_{1})(a_{1}-a_{2})^{2}}
\end{equation}
Thus, errors only ever increase from our smallest error estimate,
and therefore this method could be considered conservative. In the
event where one broadening mechanism is significantly preferred, then
our result will be effectively the same as if only that broadening
mechanism was considered. For the case where the fits are statistically
indistinguishable ($j_{1}=j_{2}$), $\Delta\alpha/\alpha$ is given
by the simple mean of the two values of $\Delta\alpha/\alpha$, and
the variance is the simple mean of the individual variances plus $0.25(a_{1}-a_{2})^{2}$.

\subsubsection{Al{\small{} III}}

In principle the Al~\iiis transitions can be included in a MM fit,
however its ionisation potential is somewhat different to the other
MM transitions described. Due to variations in the incident radiation
field, the Al~\iiis transitions may therefore not arise from the
same location, and therefore velocity, as the other MM transitions.
If the Al~\iiis transitions arise at significantly different velocities
to the other MM transitions then an error would be introduced into
$\Delta\alpha/\alpha$ for a system with Al~\iiis included (although
this effect must average to zero over a large number of absorbers,
as there is no reason for a systematic bias in the centroid of the
Al~\iiis transitions with respect to the other MM transitions along
a line of sight to Earth). 

Generally, the profiles for different transitions for the other MM
transitions used correlate well with each other. By this, we mean
that the relative column densities between corresponding velocity
components are similar for different MM transitions. However, we have
noticed that the absorption profiles for some Al~\iiis transitions
in some absorbers differ significantly in the relative line strengths
between components, when compared to other MM transitions. Importantly,
we found some absorbers where it was difficult to apply the same velocity
structure model to Al~\iiis transitions and the other MM transitions
simultaneously. For this reason, we are therefore cautious in fitting
Al~\iiis together with the other MM transitions.

Therefore, we include and model Al~\iiis if both the transitions
are available, and allow the spectral data to contribute to $\Delta\alpha/\alpha$
derived from the other MM transitions for that absorber, but do not
constrain the modelled structure with the velocity structure from
other MM transitions. Given the small difference in the $q$ coefficients
between the two Al~\iiis transitions ($\Delta q\sim250$) the statistical
contribution of Al~\iiis to $\Delta\alpha/\alpha$ is low, however
given that the exposures have already been obtained it is prudent
to try to maximise our use of the existing data.

As an example: if Al~\iiis is not utilised in the $z=1.857$ absorber
towards J013105$-$213446, the turbulent fit value of $\Delta\alpha/\alpha$
changes from $(0.30\pm1.43)\times10^{-5}$ to $(0.43\pm1.44)\times10^{-5}$.
Similarly, if Al~\iiis is not utilised in the $z=1.71$ absorber
towards J014333$-$391700, the turbulent fit value of $\Delta\alpha/\alpha$
changes from $(-2.20\pm2.67)\times10^{-5}$ to $(-2.32\pm2.68)\times10^{-5}$.

We note that previous works have included the Al~\iiis transitions
as part of the MM analysis, and it was demonstrated by \citep{Murphy:03}
that the inclusion of these transitions did not significantly alter
the Keck results. Nevertheless, the approach we have adopted is conservative. 

We have also observed less substantial relative line strength differences
between the Mg~\iscs transitions and other MM transitions, but in
no case did we find a system where we could not apply the same velocity
structure model to the Mg~\iscs and MM transitions, and so we include
the Mg~\iscs transitions in the full MM analysis.

\subsection{Aggregation of $\Delta\alpha/\alpha$ values from many absorbers}

\subsubsection{Weighted mean}

If one assumes that all the $\Delta\alpha/\alpha$ values are described
by a constant offset from the laboratory values, one can combine the
$\Delta\alpha/\alpha$ values together using a weighted mean. This
process is valid provided that the $\Delta\alpha/\alpha$ values support
a constant value of $\Delta\alpha/\alpha$. If $\Delta\alpha/\alpha\neq0$,
then this implies that there must be a transition at some point from
the present day ($\Delta\alpha/\alpha\equiv0$), and therefore the
$\Delta\alpha/\alpha$ values should be inspected to see if a transition
point can be identified. A further implication is that one must examine
the residuals about the fit for a weighted mean, plotted against various
parameters of interest (primarily redshift and sky position) to determine
if unmodelled trends exist.

\subsubsection{Dipole fit}

A dipole+monopole model constitutes the first two terms of the spherical
harmonic expansion. The simplest dipole model\index{dipole model!angular}
is of the form 
\begin{equation}
\Delta\alpha/\alpha=A\cos(\Theta)+m\label{eq:alpha:dipole eq}
\end{equation}
where $\Theta$ is the angle between the pole of the dipole and the
sky position under consideration. $A$ is an angular amplitude and
$m$ (the monopole) represents a possible offset of $\Delta\alpha/\alpha$
from the laboratory value. An equivalent (and more computationally
convenient) form is 
\begin{equation}
\Delta\alpha/\alpha=\mathbf{c}\cdot\mathbf{\hat{x}}+m,
\end{equation}
where $\mathbf{\hat{x}}$ is a unit vector pointing towards the direction
under consideration and $\mathbf{c}$ contains the amplitude and direction
information of the dipole. The components of $\mathbf{c}$, ($c_{x}$,
$c_{y}$ and $c_{z}$) are easily related to the right ascension (RA)
and declination (dec) of the direction of the dipole. $|\mathbf{c}|$
gives the magnitude of the dipole. In this form, $\Delta\alpha/\alpha$
is linear in the $c_{i}$ and so the $c_{i}$ can be determined through
weighted linear least squares. 

Although naively we might expect that $m=0$, some theories contemplate
otherwise. This could be possible if $\Delta\alpha/\alpha$ depends
on the local gravitational potential or density \citep{Khoury:04a,Mota:07a,Olive:08a}
--- laboratory conditions differ quite significantly in this regard
to the conditions in the quasar absorbers. Note that by including
the $m$ term one obtains an explicit test of this idea. Additionally,
in the presence of temporal evolution of $\alpha$, $m$ amounts to
the average effect of temporal evolution. In any particular redshift
slice, $m$ therefore represents the angle-independent value of $\Delta\alpha/\alpha$.

Clearly a model of this form is unphysical --- it makes no account
for any redshift dependence. Clearly, $A=A(z)$. Nevertheless, a model
which includes only angular dependence is useful because it provides
a method of detecting spatial variations in $\alpha$ which does not
require the specification of a functional form for $A(z)$. Use of
this model is valid under several possible circumstances. One is where
any variation in $\alpha$ with redshift in the sample along a particular
direction is small compared to variation in $\alpha$ in the opposite
direction. This might be possible in our sample, depending on how
$\alpha$ might vary, as we typically probe lookback times of greater
than 5 gigayears. Another is where $\alpha$ does vary significantly
with redshift in our sample and the distribution of absorber redshifts
does not vary greatly with sky position. With enough data, one could
simply take redshift slices and apply this model to each redshift
slice, thereby building up the functional form of $A(z)$ in model-independent
manner. However, given that we have only $\sim300$ absorbers between
the Keck and VLT samples, we simply cannot slice the data enough to
do this for more than $\sim$two redshift bins. Another consideration
is the effect of choosing a particular form of $A(z)$. An incorrect
choice of $A(z)$ may reduce sensitivity to detect an effect, and
could lead to the wrong conclusion if the choice is sufficiently bad.
As a result, we explore an angle-dependence model initially, and later
consider explicitly including distance dependence. 

Uncertainty estimates on dipole locations are derived by transforming
the covariance matrix from our fit in rectilinear coordinates, $(c_{x},c_{y},c_{z})$
to spherical coordinates $(r,\phi,\theta)$ using the standard Jacobian
matrix. That is, if $\mathbf{J}$ is the Jacobian matrix of the transformation
from rectilinear to spherical coordinates, and $\mathbf{C}$ is the
covariance matrix calculated from the fit, then $\mathbf{C}'=\mathbf{J\cdot C\cdot J^{T}}$
gives the approximate covariance matrix in spherical coordinates.
The radial component, $r$, corresponds to the amplitude of the dipole,
whereas $\phi$ and $\theta$ can be converted to the RA and dec of
the pole of the dipole. Our errors on RA and dec are thus linearised
approximations based on the covariance matrix at the best-fitting
solution, and should be regarded only as approximate. These error
estimates will be inaccurate if they subtend a large fraction of the
sky.

Note that, by virtue of the fact that $r\geq0$ in spherical coordinates,
the dipole amplitude, $A$, is not Gaussian. Thus, we perform a resampling
bootstrap analysis \citep{NumericalRecipes:92} to derive an uncertainty
for a dipole amplitude. Similarly, one cannot use a $t$-test to determine
if $A$ is significantly different from zero. Thus, we calculate the
statistical significance, $1-p$, of the dipole model over the monopole
model by using a bootstrap method where we randomise $\Delta\alpha/\alpha$
values over sightlines, and from the observed distribution of $\chi^{2}$
over many iterations determine the probability that a value of $\chi^{2}$
as good or better than that given by our observed dipole fit would
occur by chance. One can also use analytic methods \citep{Cooke:09}
if desired. These methods should yield similar answers for large sample
sizes. However, for small samples sizes, the results may differ somewhat
(especially if the statistical uncertainties vary significantly in
magnitude between the $\Delta\alpha/\alpha$ values). As the dipole+monopole
model will always improve the fit over a monopole model, the statistical
test is one-tailed, and so when we state the $\sigma$-equivalence
of a statistical significance, this is calculated as $|\mathrm{probit}(p/2)|$
in order to accord with conventional usage. $\mathrm{probit}$ is
the inverse normal cumulative distribution function. 

In principle, one can use penalised likelihood methods to determine
which model is preferred, however these information critera (e.g.
the AICC) are only heuristics, and have some drawbacks \citep{Liddle:07}.
The bootstrapping approach described above yields a direct estimation
of the preference for a dipole+monopole model over a monopole-only
model, and so we use that method here.

Unless otherwise mentioned, we multiply uncertainty estimates on monopole
values and sky coordinates by $\sqrt{\chi_{\nu}^{2}}$ as a first-order
correction for over- or under-dispersion about the fitted model \citep{NumericalRecipes:92}. 

We have checked that our optimisation code is performing adequately
by rotating the $\Delta\alpha/\alpha$ data sets into different coordinate
frames. Although clearly this will change the error estimates on the
angular position, the statistical significance tests and the value
of the dipole amplitude should not be affected, and this was found
to be the case.

\subsection{Estimating random errors \& the Least Trimmed Squares (LTS) method}

\subsubsection{Error bar inflation \& over-dispersion}

A common problem with the analysis of observational data is that the
observed scatter about the model is too high to be accounted for by
the model. This can either be caused by an incorrect model or, if
the model is a good approximation to the true underlying process,
random and systematic errors. For data with Gaussian statistical errors,
this effect is revealed by $\chi_{\nu}^{2}>1$ for large $\nu$. Indeed,
when modelling $\Delta\alpha/\alpha$ as a function of time and space,
we expect over-dispersion about any simplistic model, as the true
functional form of the variation is unknown. Therefore, over-dispersion
about a particular model will reflect not only unmodelled systematic
effects in the observations, but also some element of model mis-specification.
This does not render the modelling useless --- detection of effects
in reasonable models at high enough statistical significance is still
a demonstration of an underlying deviation from known physics. However,
it is a reminder that all the models presented here must be considered
approximations at best.

One solution to this problem is to use an unweighted model, thereby
allowing the dispersion of the model to determine the implied model
errors. Whilst this is valid in a systematic-dominated regime, typically
one operates somewhere between being statistically dominated and systematic-dominated.
In this case, an unweighted model is inappropriate, as it ignores
legitimate statistical information. The ideal solution is to model
the influence of the systematic error, however this is not always
possible, particularly in Voigt profile analysis of quasar absorption
lines, where certain systematics can be difficult to quantify \emph{a
priori}. 

Suppose that measurement values $y_{i}$ arise from true values $Y_{i}$
on account of observational statistical scatter. The probability of
observing $y_{i}$ given $Y_{i}$ is then
\begin{equation}
Pr(y_{i}|Y_{i})=\frac{1}{\sigma_{i}\sqrt{2\pi}}\exp\left(\frac{-[y_{i}-Y_{i}]^{2}}{2\sigma_{i}^{2}}\right).
\end{equation}
Now suppose some unknown random effect with uniform size $\sigma_{\mathrm{rand}}$
causes extra scatter beyond that caused by statistical errors. If
the model prediction is $f(\mathbf{x})_{i}$, then the probability
that a true value $Y_{i}$ arises from scatter about the observed
model is then
\begin{equation}
Pr(Y_{i}|f(\mathbf{x})_{i})=\frac{1}{\sigma_{\mathrm{rand}}\sqrt{2\pi}}\exp\left(\frac{-[Y_{i}-f(\mathbf{x})_{i}]^{2}}{2\sigma_{\mathrm{rand}}^{2}}\right).
\end{equation}
In this case, the probability of measuring $y_{i}$ given the model
is then \citep{Cooke:09}
\begin{align}
Pr(y_{i}|f(\mathbf{x})_{i}) & =\int_{-\infty}^{\infty}Pr(y_{i}|Y_{i})\, Pr(Y_{i}|f(\mathbf{x}))\,\mathrm{d}Y_{i}\\
 & =\frac{1}{\sqrt{2\pi(\sigma_{i}^{2}+\sigma_{\mathrm{rand}}^{2})}}\exp\left(\frac{-[y_{i}-f(\mathbf{x})_{i}]^{2}}{2(\sigma_{i}^{2}+\sigma_{\mathrm{rand}}^{2})}\right).
\end{align}
This then yields the log-likelihood, $\ln(L)$, which leads to $\chi^{2}$
as
\begin{equation}
\ln(L)=\ln\left[\prod_{i}Pr(y_{i}|f(\mathbf{x})_{i})\right]\Rightarrow\chi^{2}=\sum_{i}\frac{\left(f(\mathbf{x})_{i}-y_{i}\right)^{2}}{\sigma_{i}^{2}+\sigma_{\mathrm{rand}}^{2}}\label{eq:chisq with systematic error}
\end{equation}

For large $\nu$, the likelihood maximum of $\chi^{2}$ occurs for
$\chi_{\nu}^{2}=1$. Although one can in principle determine the maximum
of equation \ref{eq:chisq with systematic error} directly, this requires
non-linear methods even for linear models. A more practical option
is to slowly add a term $\sigma_{\mathrm{rand}}$ in quadrature with
the observational uncertainties on $\Delta\alpha/\alpha$ values,
finding the $\chi^{2}$ minimum of the model under consideration at
each iteration, until $\chi_{\nu}^{2}=1$ about the fitted model (i.e.\ $\sigma_{\mathrm{tot}}^{2}=\sigma_{\mathrm{stat}}^{2}+\sigma_{\mathrm{rand}}^{2}$).
This method has been used previously to attempt to estimate the size
of a random error, or aggregation of random errors, responsible for
any extra scatter observed data \citep{Murphy:PhD,Murphy:03,Murphy:04:LNP}.
Note that this assumes that all data points are equally affected by
the same random errors, which is unlikely to be true in practice.
Therefore, it is prudent to attempt to identify subsamples which are
affected by different random errors and correct them independently.
It is also worth noting that because of the propagation of uncertainty
for Gaussian errors, then $\sigma$ is the aggregation of any series
of Gaussian random errors with zero expectation value ($\sigma=\sqrt{\sum_{j}\sigma_{j}^{2}}$)
providing they are uncorrelated.

\subsubsection{Other robust methods\label{sub:alpha:other robust methods}}

The method described in the previous section works well provided that
one truly believes that the random errors which affect all $\Delta\alpha/\alpha$
values have the same underlying process. For quasar absorption line
Voigt profile fitting, this is unlikely to be true. We fit profiles
to a wide range of systems of varying species, some with substantial
ranges in optical depth. Additionally, different parts of the spectrum
are affected by different issues. In particular, the red end of the
spectrum displays significant sky absorption and emission, the presence
of which cannot be unequivocally excluded in certain cases, particularly
where the spectral region is of relatively low SNR. Additionally,
we cannot assume that certain processes are described by a Gaussian.
Some events are binary (e.g.\ sky emission is either present or it
is not, although clearly the magnitude of the impact could vary substantially).
Furthermore, certain events may occur with low probability most of
the time but have significant impact (e.g.\ uncleaned cosmic rays).
All of these considerations lead to the possibility of outliers in
the sample (that is, values of $\Delta\alpha/\alpha$ which do not
match the trend shown by other absorbers). 

Outliers cause two problems. Firstly, they bias parameter estimates
away from underlying values. As $\chi^{2}$ minimisation weights points
by the square of the weighted normalised residuals ($r_{i}=[\Delta\alpha/\alpha-\mathrm{model\, prediction}]/\mathrm{\sigma_{tot}}$),
even a few large-residual points can cause substantial bias in parameter
estimates. Secondly, any estimate of the average random error from
growing the error bars in quadrature with some $\sigma_{\mathrm{rand}}$
will not be a good estimate of the average random error affecting
the good points; it will over-estimate the random error affecting
most points, whilst under-estimate the systematic affecting the outliers.
A traditional solution to the second problem has been to manually
remove outliers from the sample, typically by discarding points with
$|r_{i}|>3$. However, because the outliers are included in the fit,
they tend to bias the fit towards them. This tends to mask the presence
of other outliers, and may lead to the rejection of good data. Additionally,
because one has to estimate $\sigma_{\mathrm{rand}}$ before calculating
the residuals, the overly-large estimate of $\sigma_{\mathrm{rand}}$
will tend to mask outliers. This can lead to both false positives
and false negatives.

Although there is no perfect solution to this problem, these considerations
have led to the development of robust statistical methods \citep[see][for a review of many of the basic approaches]{Rousseeuw:87}.
Although these methods mildly underperform standard least squares
methods in the presence of no contamination, for data sets with contamination
of even a few percent robust statistical methods can lead to dramatic
outperformance \citep{Rousseeuw:87}. 

A common method is to use a so-called $M$-estimator\index{robust statistics!$M$-estimator},
which minimises a maximum-likelihood type estimate of the residuals,
$\sum_{i}\rho(r_{i})$. For standard least squares, $\rho(x)=x^{2}$.
Choosing $\rho(x)=|x|$ leads to the L1-norm method of minimising
the mean absolute deviation of residuals \citep{Rousseeuw:87,NumericalRecipes:92}.
This corresponds to a probability distribution where the residuals
are distributed as a double exponential, namely
\begin{equation}
Pr(y_{i}-f(\mathbf{x})_{i})\sim\exp\left(-\left|\frac{y_{i}-f(\mathbf{x})_{i}}{\sigma_{i}}\right|\right)
\end{equation}
\citep{NumericalRecipes:92}. Other functions do not correspond to
traditional probability distributions, but instead are heuristic functions
derived to have robustness against outliers whilst still maintaining
good statistical efficiency. A widely used choice is Tukey's biweight
\citep{Rousseeuw:84}, 
\begin{equation}
\rho(x)=\left\{ \begin{array}{cc}
\frac{x^{2}}{2}-\frac{x^{4}}{2c^{2}}+\frac{x^{6}}{6c^{4}}, & |x|\leq c\\
\frac{c^{2}}{6} & |x|>c
\end{array}\right.
\end{equation}
 As $x\rightarrow0$, $\rho(x)\sim x^{2}$ and so this approximates
standard least squares fitting. On the other hand, for $|x|\rightarrow c$,
$\rho(x)\rightarrow c^{2}/6$. Thus the effect of outliers is bounded.
The effect of outliers on parameter estimates relates to the function
$\Psi(x)=\rho'(x)$. For Tukey's biweight, $\Psi(x)=0$ for $|x|>c$
and therefore these outliers have no influence on parameter determination,
which is desirable. Unfortunately, application of $M$-estimators
requires that the expected scatter of data about the model be known.
If random errors are significant, then naive application of a $M$-estimate
to statistically weighted data will simply discard many points which
are not necessarily outliers when considered in the context of the
observed scatter of the points about the model.

One solution to this is to use an $S$-estimator \citep{Rousseeuw:84b}\index{robust statistics!$S$-estimator},
which attempts to obtain a robust estimate of scale. To do this, one
defines a robust estimate of scale, $s$, as the solution of
\begin{equation}
K=\frac{1}{n}\sum_{i=1}^{n}\rho\left(\frac{r_{i}}{s}\right)\label{eq:Robust K definition}
\end{equation}
for some $K$. $K$ is typically set to be equal to the expected value
of $\rho(x)$ under a Gaussian distribution, that is
\begin{equation}
K=\int_{-\infty}^{\infty}\rho(x)\frac{1}{\sqrt{2\pi}}e^{-x^{2}/2}\,\mathrm{d}x.
\end{equation}
For $\chi^{2}$ minimisation, $\rho(x)=x^{2}$ and $K\sim1$.

There are two problems with this in a scientific context. Firstly,
$s$ enters reciprocally in the functional form of K in equation \ref{eq:Robust K definition},
implying that $s$ simply scales errors by a specific amount, which
is undesirable. Certainly, the statistical errors impose an absolute
lower bound on the precision available from each data point. The solution
to this is to solve the implicit equation
\begin{equation}
K=\frac{1}{n}\sum_{i=1}^{n}\rho\left(\frac{f(\mathbf{x})_{i}-y_{i}}{\sqrt{\sigma_{i}^{2}+\sigma^{2}}}\right)
\end{equation}
for $\sigma$ and $\mathbf{x}_{i}$. Unfortunately this improvement
leads to the second problem. One can rapidly determine through experimentation
that even a few outliers of arbitrarily large magnitude can substantially
influence the estimate of $\sigma$, because for $|r_{i}|>c$, $\rho(r_{i})$
may not change significantly upon increasing $\sigma$ by reasonable
amounts. We thus found that the application of $S$-estimators was
not appropriate for our purposes.

\subsubsection{Bayesian methodology\label{sub:Bayesian regression}}

Another approach is to use Bayesian methods, which can formally account
for an uncertainty in the error estimates, $\sigma_{i}$. Suppose
that the statistical error bar, $\sigma_{i}$, is taken as a lower
bound on the true error bar. One way of incorporating this approach
is to take a prior PDF for the true error bar, $\sigma_{i,t}$ as
\begin{equation}
\mathrm{Pr}(\sigma_{i,t}|\sigma_{i},M)=\frac{\sigma_{i}}{\sigma_{i,t}^{2}}\label{eq:robust sigma prior}
\end{equation}
where $M$ is the model used. \citep{BayesianTutorial}. A more correct
approach is to assign a Jeffreys' prior, as 
\begin{equation}
\mathrm{Pr}(\sigma_{i,t}|\sigma_{i},c,M)=\frac{1}{\ln(c/\sigma_{i})}\frac{1}{\sigma_{i,t}},
\end{equation}
however this requires specification of a finite upper bound, $c$,
to make the prior normalisable. The choice of equation \ref{eq:robust sigma prior}
should not substantially alter the conclusions of this analysis \citep{BayesianTutorial}.
Suppose we consider a single datum, $x_{i}$, then the marginal likelihood
for the data, $D$, with the unknown $\sigma_{i,t}$ integrated out
is
\[
\mathrm{Pr}(D|F,\sigma_{i},M)=\int_{\sigma_{i}}^{\infty}\mathrm{Pr}(D|F,\sigma_{i,t},M)\,\mathrm{Pr}(\sigma_{i,t}|\sigma_{i},M)\,\mathrm{d}\sigma_{i,t},
\]
where $F$ is the function being estimated. If we assume a Gaussian
PDF for $\mathrm{Pr}(D|F,\sigma_{i,t},M)$ then we obtain
\[
\mathrm{Pr}(D|F,\sigma_{i},I)=\frac{1}{\sigma_{0}\sqrt{2\pi}}\left[\frac{1-e^{-r_{i}^{2}/2}}{r_{i}^{2}}\right],
\]
where $r_{i}$ is the residual about the model, $(\mathrm{data}-\mathrm{model})/\sigma_{i}$.
Extending this analysis to a set of data of size $N$, and assigning
uniform prior PDFs to the parameters, the log posterior probability
is
\begin{equation}
L=\ln[\mathrm{Pr}(\mathbf{X}|\mathbf{D},I)]=\mathrm{constant}+\sum_{k=1}^{N}\ln\left[\frac{1-e^{-r_{i}^{2}/2}}{r_{i}^{2}}\right]\label{eq:SBLR L defn}
\end{equation}
\citep{BayesianTutorial}. Maximising $L$ (or minimising $-L$) thus
yields a robust estimate of the parameters under the assumptions described
above. For brevity, we refer to this methodology as skeptical Bayesian
regression\index{robust statistics!skeptical Bayesian regression}.
In the context of a linear fit, we call this skeptical Bayesian linear
regression (SBLR). Unfortunately, the rather broad assumption about
the validity of the $\{\sigma_{i}\}$ values leads to a loss of statistical
precision for the resultant uncertainties on model parameters in the
event that the residuals are Gaussian. On the other hand, in the event
that the residuals are not Gaussian, then the standard least squares
assumption that the residuals are Gaussian will mean that confidence
limits on model parameters are too small. In this case, the approach
described here naturally gives a highly robust method of determining
parameters whilst making use of all the data. We return to this method
later.

\subsubsection{The LTS method\label{sub:LTS method}}

\index{Least Trimmed Squares (LTS) method}\index{robust statistics!LTS method}Ideally,
what one would like to do is identify outliers and remove them from
the sample of consideration. This not only means that they cannot
bias parameter estimates, but that the estimate of $\sigma$ is likely
to be much more reasonable. For this, we have found that the Least
Trimmed Squares (LTS) method \citep{Rousseeuw:84} works well. Instead
of fitting all $n$ data points, the LTS method traditionally only
fits $k=(n+p+1)/2$ points (where $p$ is the number of free parameters)
using standard least squares, and then searches for the combination
of $k$ data points and fitted model that yields the lowest sum of
squared residuals. In our case, we modify this to include statistical
weightings on the data points. We thus wish to find the combination
of $k$ data points and model which minimises $\chi^{2}$. Essentially,
the method only fits the inner $k/n$ fraction of the distribution
of the residuals. Where a few outliers exist, they will be ignored
by this method provided that they are in the excluded fraction. 

Calculation of the LTS is computationally intensive because the target
function is highly non-linear on account of the inclusion/exclusion
of data, as well as the need to sort the residuals. To directly explore
all the $\binom{n}{k}$ possible combinations is unfeasible for the
datasets we consider. Original methods attempted to sample from this
space using a forward search algorithm \citep{Atkinson:94}, however
a newer algorithm --- Fast-LTS \citep{Rousseeuw:02} --- demonstrates
good results for hundreds to thousands of data points. We implement
the Fast-LTS algorithm.

Although in the limit $n\rightarrow\infty$ the use of $k=(n+p+1)/2$
will produce a very robust fit, for small $n$ (e.g.\ $n\lesssim20$)
we are wary of finding combinations of $(n+p+1)/2$ points by chance
that do not reflect the true trend. However, we still wish to obtain
robustness against outliers. As such, we operate with $k=0.85$, which
provides robustness against up to 15\% of the data being outliers.

To allow for the inclusion of $\sigma_{\mathrm{rand}}$, we propose
a variant of the LTS method which proceeds as follows. First, we define
a robust scatter measure as 
\begin{equation}
\chi_{\nu}^{2}(k)=\frac{1}{k-p}\sum_{i=1}^{k}\frac{(f(\mathbf{x})_{i}-y_{i})^{2}}{\sigma_{i}^{2}+\sigma_{\mathrm{rand}}^{2}},
\end{equation}
 where the sum is taken over only the smallest $k$ residuals of the
fit. We then slowly increase $\sigma_{\mathrm{rand}}$ from 0 until
$\chi_{\nu}^{2}(k)$ is what we would expect for a Gaussian distribution
with large $\nu$, refitting and recalculating $\chi_{\nu}^{2}(k)$
after each increment in $\sigma$. For $k=n$ then this yields $\langle\chi_{\nu}^{2}(n)\rangle=1$,
for but for $k<n$ we obtain
\begin{equation}
\langle\chi_{\nu}^{2}(k)\rangle=\int_{-a}^{a}x^{2}\cdot\frac{1}{\sqrt{2\pi}}e^{-x^{2}/2}\,\mathrm{d}x,
\end{equation}
 where $a=\mathrm{probit}[(1+f)/2]$ and $f=k/n$. $\mathrm{probit}$
is the inverse normal cumulative distribution function. We take the
value of $\sigma_{\mathrm{rand}}$ derived in this way as our estimate
of the additional random error for the data given the model. In this
way, if the data are contaminated by a few outliers these will not
impact the estimate of the random error which affects most points.

After applying the LTS method to estimate the random error term, we
then discard all points with $\lvert r_{i}\rvert>3$ about the LTS
fit, but only if we are applying the method to a full sample (i.e.\ the
whole VLT or Keck sample, or a combination of the two). This is because
in small-$n$ fits one does not have much data, and so it is not clear
whether outliers would become inliers with more data. 

If we remove outliers, we then reapply the LTS method to check that
no more outliers are unmasked, and to re-estimate $\sigma_{\mathrm{rand}}$.
The LTS fit is statistically inefficient because it ignores some good
data (15 percent for $k=0.85n$ if all the remaining points are inliers).
Therefore, after we discard high residual points, we apply a normal
weighted least squares fit to the remaining data to estimate the parameters
and achieve the best possible confidence limits on our modelled parameters. 

The benefits of the LTS method can be summarised as follows.
\begin{enumerate}
\item \emph{Robust estimate of $\sigma_{\mathrm{rand}}$}. If we calculate
$\sigma_{\mathrm{rand}}$ by increasing it until $\chi_{\nu}^{2}=1$,
then even a single, arbitrarily large outlier can increase $\sigma_{\mathrm{rand}}$
without bound. This is much less likely with the LTS method. A more
appropriate estimate of $\sigma_{\mathrm{rand}}$ means that false
negatives are less likely. 
\item \emph{Robust detection of outliers}. In a standard $\chi^{2}$ minimisation
fit, residuals with larger magnitude $|r_{i}|$ are weighted as $r_{i}^{2}$,
which distorts the fit towards them. This tends to mask outliers.
By distorting the fit, one might incorrectly decide that some good
points are in fact outliers. Similarly, the existence of one outlier
tends to conceal the existence of additional outliers (a masking effect). 
\item \emph{Objectivity}. Manual outlier rejection is often characterised
as subjective. The rule provided here provides an objective method
of classifying data points as outliers, thereby removing this objection.
\item \emph{More robust parameter estimates}. Even a few outliers can substantially
distort the fit. This biases parameter estimates away from their underlying
values. We are interested in the underlying values, not the values
given by a blind least squares fit. The rate of false positives should
also be decreased, as false positives can be caused by outliers. 
\end{enumerate}

\section{Many-multiplet VLT results}

\begin{figure}[tbph]
\noindent \begin{centering}
\includegraphics[bb=79bp 88bp 540bp 766bp,clip,width=0.9\textwidth]{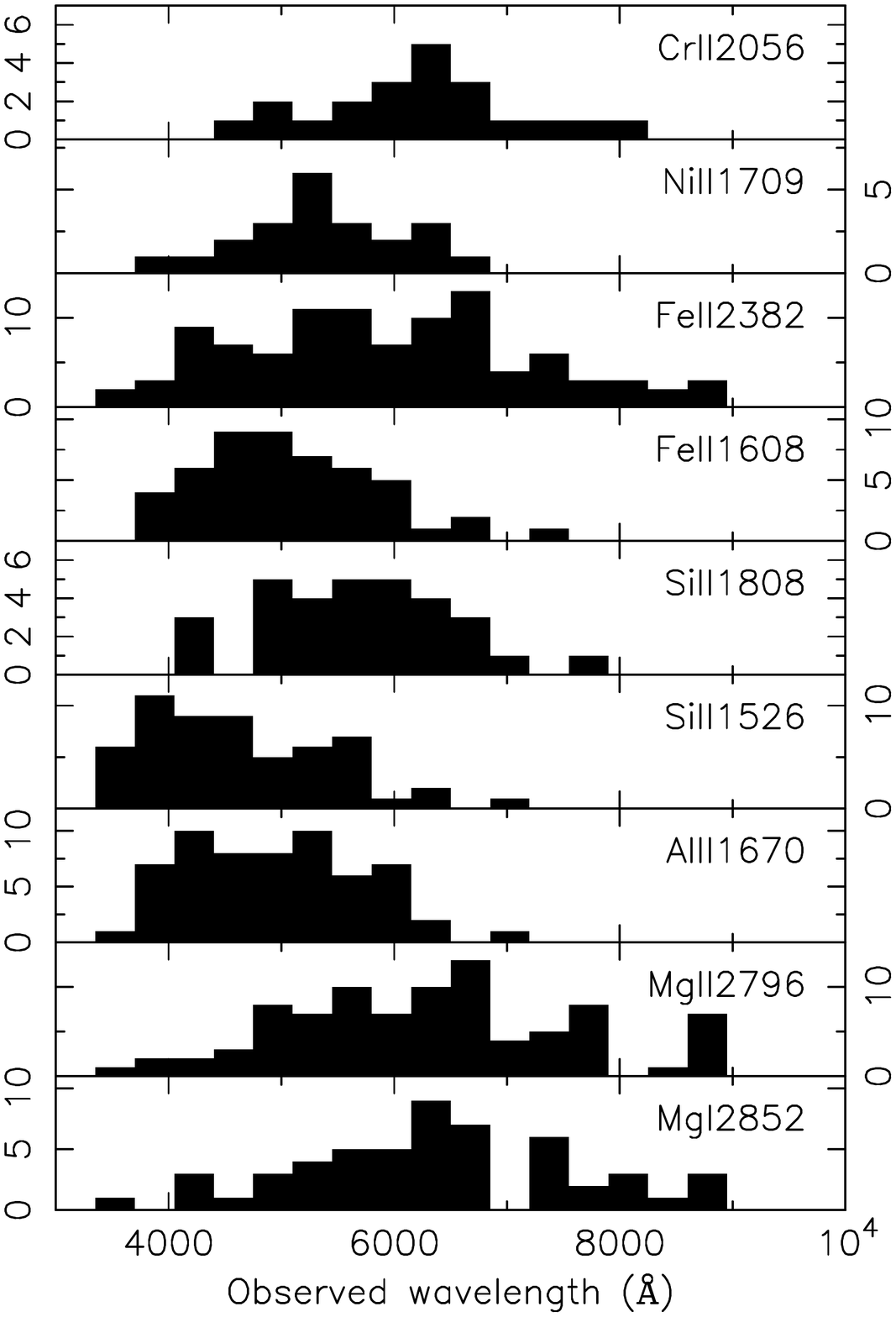}
\par\end{centering}

\caption[Distribution of certain transitions in the VLT MM sample]{Distribution of certain transitions in the VLT MM sample. The transitions shown are representative of those used at low redshifts (the Mg \isc/Mg \ii/Fe \iis combination, excluding the Fe \iis $\lambda\lambda 1608$ and $1611$ transitions), moderate redshifts (most transitions), and high redshifts (the Si \ii/Fe \iis $\lambda 1611,1608$/Al \iis combination, with the additional use of Zn \ii, Cr \iis and Ni \ii). The vertical scales alternate. \label{Flo:alpha:transition_distribution}}
\end{figure}

\begin{figure}[tbph]
\noindent \begin{centering}
\includegraphics[bb=50bp 92bp 556bp 777bp,clip,angle=-90,width=1\textwidth]{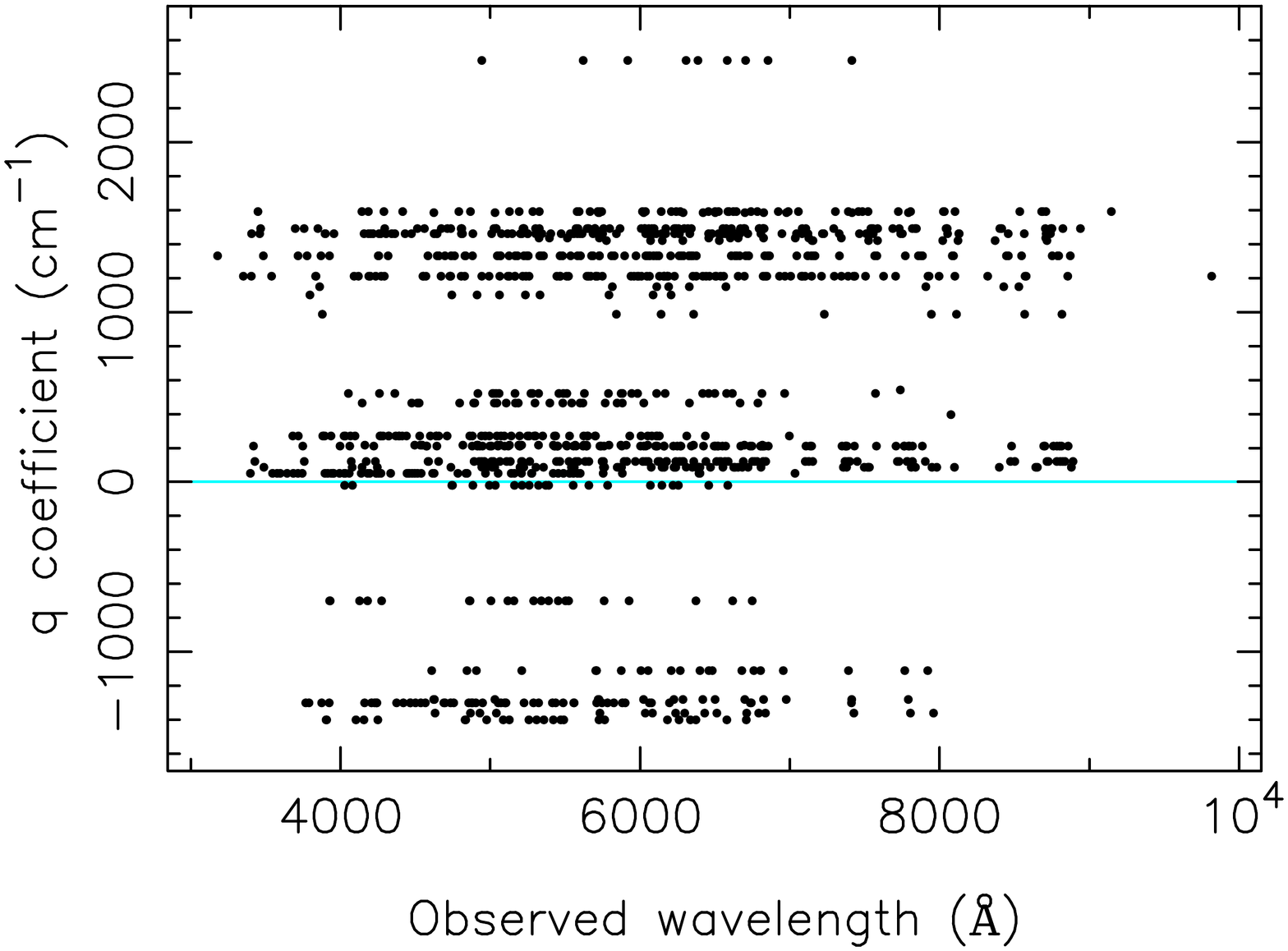}
\par\end{centering}

\caption[Relationship between $q$ coefficients with observed wavelength for the VLT sample]{Relationship between $q$ coefficients and observed wavelength for all utilised transitions in all absorbers in the VLT sample. Although the low-$z$ Fe~\ii/Mg~\iis combination is sensitive to low-order wavelength distortions because the $q$ coefficients for this combination are correlated with wavelength (see figure \ref{Flo:alpha:q_vs_wl}), one can see that for the full sample there is little correlation between observed wavelength and $q$, making the MM method resistant to systematics when many absorbers at different redshifts are used. \label{fig:alpha:wavelength_vs_q}}
\end{figure}

\begin{figure}[tbph]
\noindent \begin{centering}
\includegraphics[bb=78bp 79bp 540bp 727bp,clip,width=0.85\textwidth]{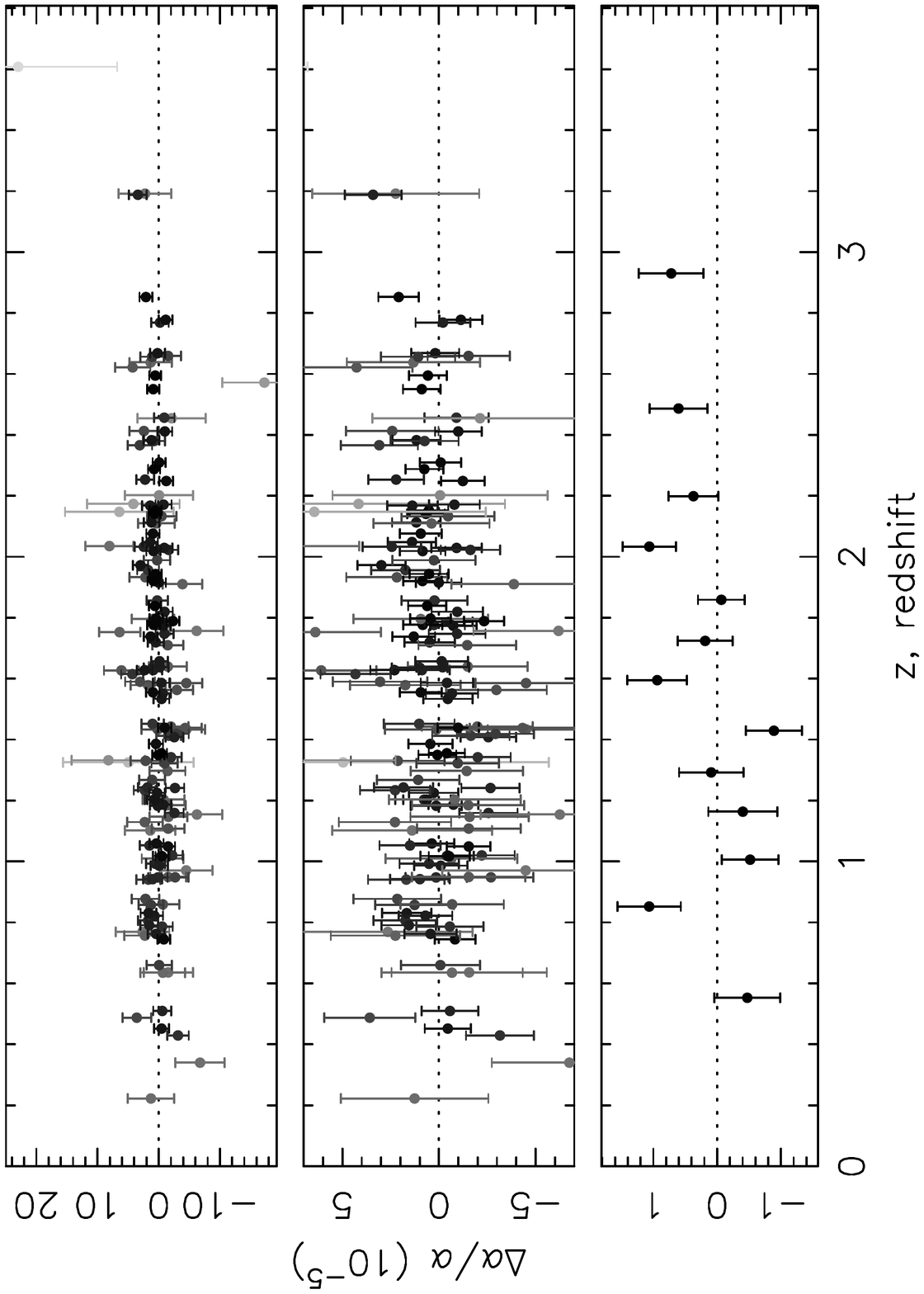}
\par\end{centering}

\caption[Values of $\Delta\alpha/\alpha$ for the VLT sample]{Values of $\Delta\alpha/\alpha$ for the VLT sample. The top panel shows all values of $\Delta\alpha/\alpha$ with error bars increased in quadrature with $\sigma_\mathrm{rand} = 0.905 \times 10^{-5}$. The middle panel shows the same data as the top panel, with with the vertical range restricted for better viewing of the higher statistical weight points. Both of the panels have been shaded according to a greyscale as the logarithm of the uncertainty estimate, with lower uncertainty points being darker. The bottom panel shows binned values of $\Delta\alpha/\alpha$ where approximately 12 points contribute to each bin. The bottom panel appears to demonstrate a change of sign for $\Delta\alpha/\alpha$ with increasing $z$. In particular, for $z<1.5$, 3 of the 5 binned points fall in the region $\Delta\alpha/\alpha < 0$. For $z>1.5$, 6 of 7 points in the binned plot fall in the region $\Delta\alpha/\alpha > 0$. This trend with redshift is different to that seen in fig.\ 6 of \citet{Murphy:04:LNP}, for which all binned points fall in the region $\Delta\alpha/\alpha < 0$. This suggests that a weighted mean model is not a good description of the data. \label{fig:alpha:zstack_VLT}}
\end{figure}

We give our many-multiplet VLT results in table \ref{tab:alpha:VLT_daoa_results}\index{many-multiplet (MM) method!VLT results}.
The values of $\Delta\alpha/\alpha$ for the VLT sample are shown
in figure \ref{fig:alpha:zstack_VLT}. The distribution of observed
wavelengths for certain representative transitions occur can be found
in table \ref{Flo:alpha:transition_distribution}. The frequency with
which all utilised transitions are fitted is given in table \ref{tab:alpha:freqtransitions}.
In figure \ref{fig:alpha:wavelength_vs_q} we show the relationship
between the $q$ coefficients and observed wavelength for all the
transitions fitted in the VLT sample. A summary of the parameters
for various models for $\Delta\alpha/\alpha$ fitted to the VLT and
Keck data may be found in table \ref{tab:alpha:combineresults_dipole}.
Plots of the fit to each absorber may be found in Appendix \vref{cha:MM fits}.

\begin{landscape}\begin{center}

{\footnotesize  
\begin{longtable}{@{}llll@{}rrcrrcr@{}}

\caption[VLT many-multiplet results for $\Delta\alpha/\alpha$]{Results for $\Delta\alpha/\alpha$ derived from MM absorbers. Errors given are purely statistical (see definition in section \ref{sub:alpha:rand_sys_errs}). The emission redshift of the quasar and absorption redshift are given by $z_\mathrm{em}$ and $z_\mathrm{abs}$ respectively. $z_\mathrm{abs}$ is given as the redshift of the strongest component in the fit. Three values of $\Delta\alpha/\alpha$ are given: the turbulent fit, the thermal fit, and the value derived from the method-of-moments estimator from section \ref{sub:alpha:physical_constraints}. The key for the transition labels, in column 4, is given in table \ref{tab:alpha:freqtransitions}. $\chi^2_\nu$ and $\nu$, the number of degrees of freedom, are given for both the turbulent and thermal fits. The last column gives the value of $\Delta\alpha/\alpha$ derived using our method-of-moments estimator. The absorber marked with an asterisk (*) has been identified as an outlier and removed from the analysis.\label{tab:alpha:VLT_daoa_results}} \\
  
\hline
 Quasar name & $z_\mathrm{em}$ & $z_\mathrm{abs}$ & Transitions & $\Delta\alpha/\alpha_\mathrm{turb}\,(10^{-5})\,$ & $\nu_\mathrm{turb}$ 
& $\chi^2_{\nu,\mathrm{turb}}$ & $\Delta\alpha/\alpha_\mathrm{therm}\,(10^{-5})$ & $\nu_\mathrm{therm}$ & $\chi^2_{\nu,\mathrm{therm}}$ &  $\Delta\alpha/\alpha_\mathrm{MoM}\,(10^{-5}) $  \\
\hline
\endfirsthead

\hline
 Quasar name & $z_\mathrm{em}$ & $z_\mathrm{abs}$ & Transitions & $\Delta\alpha/\alpha_\mathrm{turb}\,(10^{-5})\,$ & $\nu_\mathrm{turb}$ 
& $\chi^2_{\nu,\mathrm{turb}}$ & $\Delta\alpha/\alpha_\mathrm{therm}\,(10^{-5})$ & $\nu_\mathrm{therm}$ & $\chi^2_{\nu,\mathrm{therm}}$ &  $\Delta\alpha/\alpha_\mathrm{MoM}\,(10^{-5}) $  \\
\hline
\endhead

\hline
\multicolumn{5}{c}{Continued on next page.}\\
\endfoot

\hline
\endlastfoot

J000344$-$232355  & 2.28  &   0.4521  &  $  a_2 b_1 b_2 j_4 j_5 j_6 j_7 j_8  $  & $-0.498 \pm 0.758$  & 533  & 0.7653 & $-0.459 \pm 0.787$ & 533 & 0.7466 &  $-0.459 \pm 0.787$   \\ 
J000344$-$232355  & 2.28  &   0.9491  &  $  a_2 b_1 b_2 j_4 j_6 j_8  $  & $-1.390 \pm 2.700$  & 530  & 0.7766 & $-4.960 \pm 2.650$ & 531 & 0.7916 &  $-1.534 \pm 2.788$   \\ 
J000344$-$232355  & 2.28  &   1.5864  &  $  a_2 j_1 j_4 j_5 j_6 j_7 j_8 c_1 e_1  $  & $-1.630 \pm 1.040$  & 380  & 0.7061 & $-0.389 \pm 0.989$ & 380 & 0.6847 &  $-0.410 \pm 1.003$   \\ 
J000448$-$415728*  & 2.76  &   1.5419  &  $  b_1 b_2 j_4 j_5 j_6 j_7 j_8  $  & $-5.270 \pm 0.906$  & 372  & 0.7682 & $-1.920 \pm 0.741$ & 372 & 1.0146 &  $-5.270 \pm 0.906$   \\ 
J000448$-$415728  & 2.76  &   1.9886  &  $  j_4 j_6 j_7 j_8 c_1 d_1 d_2  $  & $-0.952 \pm 1.510$  & 254  & 0.8991 & $0.934 \pm 1.830$ & 254 & 0.8943 &  $0.266 \pm 1.945$   \\ 
J000448$-$415728  & 2.76  &   2.1679  &  $  b_1 j_1 j_4 j_5 j_6 c_1 e_1  $  & $2.320 \pm 0.942$  & 309  & 0.7517 & $1.370 \pm 0.938$ & 309 & 0.7229 &  $1.381 \pm 0.944$   \\ 
J001210$-$012207  & 2.00  &   1.2030  &  $  b_1 b_2 j_4 j_6 j_7 j_8 c_1  $  & $0.723 \pm 1.290$  & 505  & 0.9464 & $0.772 \pm 1.190$ & 505 & 0.8959 &  $0.772 \pm 1.190$   \\ 
J001602$-$001225  & 2.09  &   0.6351  &  $  b_1 b_2 j_4 j_6 j_8  $  & $-0.719 \pm 3.580$  & 307  & 1.1152 & $-0.089 \pm 3.000$ & 307 & 1.1318 &  $-0.673 \pm 3.545$   \\
\end{longtable}
}
\end{center}

\end{landscape}

\begin{table}[tbph]

\begin{center} \caption[The frequency of occurrence for each MM transition in our VLT fits]{The frequency of occurrence for each MM transition in our VLT fits. Note that the Mg\,\textsc{i} $\lambda 2026$ transition is fairly weak compared to the Mg\,\iscs $\lambda 2852$ transition, and so is included in few fits. Where we have included Mg\,\textsc{i} $\lambda2052$ in our fit, and Zn\,\textsc{ii} $\lambda 2026$, is included in our fit, Mg\,\textsc{i} $\lambda2026$ will also be modelled, although the contribution may be extremely minor. Nevertheless, we count this as an occurrence of Mg\,\textsc{i} $\lambda2026$, as that transition is included in our model. The transition key provides a convenient, short-hand way of referring to a particular transition. This key is used in table \ref{tab:alpha:VLT_daoa_results}. $q$-coefficients given are those used in our analysis. \label{tab:alpha:freqtransitions}}

\begin{tabular}{lccc} \hline \multicolumn{1}{c}{Transition}& $q$ (cm$^{-1}$) & Key & Frequency of occurrence \\\hline
Mg{\sc \,i} $\lambda$2026     & 87                 &  $a_1$  &  3 \\
Mg{\sc \,i} $\lambda$2852     & 86                 &  $a_2$ &   53 \vspace{0.2cm}\\ 
Mg{\sc \,ii} $\lambda$2796    & 211                  &  $b_1$ &  88  \\
Mg{\sc \,ii} $\lambda$2803    & 120                   &  $b_2$ &   86  \vspace{0.2cm}\\
Al{\sc \,ii} $\lambda$1670    & 270                  &  $c_1$ &  60  \vspace{0.2cm}\\
Al{\sc \,iii} $\lambda$1854   & 464                  &  $d_1$ &  25  \\
Al{\sc \,iii} $\lambda$1862   & 216                 &  $d_2$ &   25  \vspace{0.2cm}\\
Si{\sc \,ii} $\lambda$1526    & 50                  &  $e_1$ &   57 \\ 
Si{\sc \,ii} $\lambda$1808    & 520                  &  $e_2$ &   31 \vspace{0.2cm}\\
Cr{\sc \,ii} $\lambda$2056    & -1110                 &  $h_1$ &  21  \\
Cr{\sc \,ii} $\lambda$2062    & -1280                 &  $h_2$ &   15 \\
Cr{\sc \,ii} $\lambda$2066    & -1360                  &  $h_3$ &  17   \vspace{0.2cm}\\
Fe{\sc \,ii} $\lambda$1608    & -1300                  &  $j_1$ &   50\\ 
Fe{\sc \,ii} $\lambda$1611    & 1100                  &  $j_2$ &   9 \\ 
Fe{\sc \,ii} $\lambda$2260    & 1435                  &  $j_3$ &   12 \\
Fe{\sc \,ii} $\lambda$2344    & 1210                  &  $j_4$ &   97 \\ 
Fe{\sc \,ii} $\lambda$2374    & 1590                  &  $j_5$ &   51\\
Fe{\sc \,ii} $\lambda$2382    & 1460                   &  $j_6$ &   100 \\
Fe{\sc \,ii} $\lambda$2587    & 1490                  &  $j_7$ &    74\\
Fe{\sc \,ii} $\lambda$2600    & 1330                   &  $j_8$ &   97 \vspace{0.2cm}\\ 
Mn{\sc \,ii} $\lambda$2576    & 1420               & $i_1$ &    13 \\ 
Mn{\sc \,ii} $\lambda$2594    & 1148                 &  $i_2$ &  9  \\ 
Mn{\sc \,ii} $\lambda$2606    & 986               &  $i_3$ &    9 \vspace{0.2cm}\\ 
Ni{\sc \,ii} $\lambda$1709    & -20                 &  $k_1$ &  22 \\
Ni{\sc \,ii} $\lambda$1741    & -1400                  &  $k_2$   &24  \\
Ni{\sc \,ii} $\lambda$1751    & -700                  &  $k_3$ &  21 \vspace{0.2cm}\\ 
Ti{\sc \,ii} $\lambda$3067    & 791                 &  $g_1$ &  0\\ 
Ti{\sc \,ii} $\lambda$3073    & 677                 &  $g_2$ &  0\\ 
Ti{\sc \,ii} $\lambda$3230    & 673                 &  $g_3$ &  0\\ 
Ti{\sc \,ii} $\lambda$3342    & 541                 &  $g_4$ &  1\\ 
Ti{\sc \,ii} $\lambda$3384    & 396                 &  $g_5$ &  1\vspace{0.2cm}\\ 
Zn{\sc \,ii} $\lambda$2026    & 2479                  &  $l_1$ &  9\\ 
Zn{\sc \,ii} $\lambda$2062    & 1584                  &  $l_2$ &  13 \\\hline 
\end{tabular} \end{center} 

\end{table}

\subsection{Weighted mean for the VLT data}

We initially fit a weighted mean to our VLT points. The LTS method
indicates that the $z=1.542$ absorber toward J000448$-$415728 is
an outlier, with a residual of $4.2\sigma$ about the LTS fit, and
so we remove this point. If we do not remove this point, the weighted
mean after increasing errors is $\Delta\alpha/\alpha=(0.154\pm0.132)\times10^{-5}$,
with $\chi_{\nu}^{2}=1.17$. 

After removing this point, a weighted mean fit with our raw statistical
errors yields $\Delta\alpha/\alpha=(0.229\pm0.095)\times10^{-5}$,
with $\chi_{\nu}^{2}=1.78$. Applying the LTS method to this data
set yields a random error estimate of $\sigma_{\mathrm{rand}}=0.905\times10^{-5}$. 

After accounting for this extra random error, the weighted mean becomes
$\Delta\alpha/\alpha=(0.208\pm0.124)\times10^{-5}$, with $\chi_{\nu}^{2}=0.99$.
This result differs from that of \citet{Murphy:04:LNP} at the $\sim4.7\sigma$
level. Although this appears to be a gross inconsistency, as will
be seen below it is more likely that this reflects the fact that a
weighted mean model is not a good description of the data set.

\subsubsection{Distribution of $\Delta\alpha/\alpha$ values with redshift and validity
of a weighted mean model}

In the bottom panel of figure \ref{fig:alpha:zstack_VLT} we show
binned values of $\Delta\alpha/\alpha$ plotted against redshift for
the VLT sample. For $z<1.5$, 3 of the 5 binned points fall in the
region $\Delta\alpha/\alpha<0$. For $z>1.5$, 6 of 7 points in the
binned plot fall in the region $\Delta\alpha/\alpha>0$. This trend
with redshift is different to that seen in fig.\ 6 of \citet{Murphy:04:LNP}
for $z<1.6$, all 7 points fall in the region $\Delta\alpha/\alpha<0$,
whereas for $z>1.5$ all 6 points also fall in the region $\Delta\alpha/\alpha<0$.
The apparent change in sign of $\Delta\alpha/\alpha$ with $z$ in
the VLT sample suggests that a weighted mean model is not a good description
of the VLT data.

\subsection{Dipole fit for the VLT data\label{sub:alpha:VLT_dipole}}

In this section, we fit the dipole model of equation \ref{eq:alpha:dipole eq}
to the new VLT data. 

Inspection of the residuals about the fit, plotted as a function of
redshift, reveals no obvious trend for higher scatter at higher redshifts
and therefore we treat all absorbers the same in attempting to estimate
$\sigma_{\mathrm{rand}}$. We again identify the $z=1.542$ system
toward J000448$-$415728 as an outlier, with a residual of $4.6\sigma$
about the LTS fit, even after increasing the error bars. Thus, we
remove this system from our sample, and re-estimate $\sigma_{\mathrm{rand}}=0.905\times10^{-5}$.
We call this sample ``VLT-dipole''.

Our dipole fit parameters after adding $\sigma_{\mathrm{rand}}=0.905\times10^{-5}$
in quadrature to all error bars are: $m=(-0.109\pm0.180)\times10^{-5}$,
$A=1.18\times10^{-5}$ ($1\sigma$ confidence limits $[0.80,1.66]\times10^{-5}$),
$\mathrm{RA}=(18.3\pm1.2)\,\mathrm{hr}$ and $\mathrm{dec}=(-62\pm13)^{\circ}$.
For this fit, $\chi^{2}=141.8$ and $\chi_{\nu}^{2}=0.95$.

\begin{figure}[tbph]
\noindent \begin{centering}
\includegraphics[bb=77bp 92bp 556bp 742bp,clip,angle=-90,width=1\textwidth]{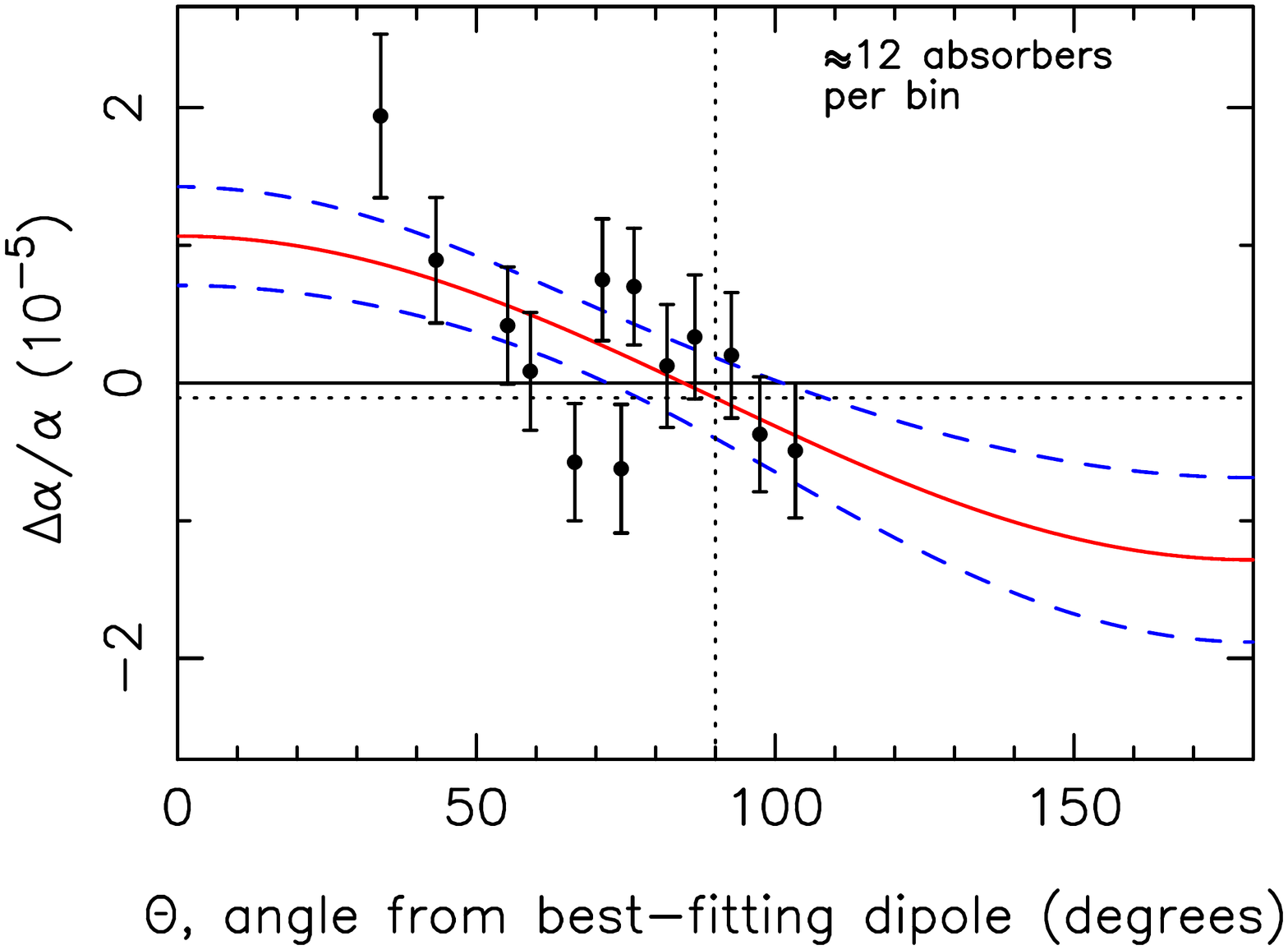}
\par\end{centering}

\caption[$\Delta\alpha/\alpha$ vs $\cos\Theta$ for the VLT sample]{Binned values of $\Delta\alpha/\alpha$ plotted against angle to the best-fitting dipole for the VLT sample. The red, solid line is the model $\Delta\alpha/\alpha = A\cos(\Theta) + m$, and the dashed, blue lines indicate the $1\sigma$ uncertainty on the dipole fit. Statistical errors have been increased prior to binning as described in the text. The dipole+monopole model is preferred over monopole-only model at the $2.2\sigma$ level. The parameters for this fit are: $m = (-0.109 \pm 0.180) \times 10^{-5}$, $A = 1.18 \times 10^{-5}$ ($1\sigma$ confidence limits $[0.80, 1.66] \times 10^{-5}$), $\mathrm{RA}= (18.3 \pm 1.2)\, \mathrm{hr}$ and $\mathrm{dec.}= (-62 \pm 13) ^\circ$.  \label{fig:alpha:VLT_dipole}}
\end{figure}

To assess the dipole fit compared to a monopole-only (weighted mean)
fit, we compare a weighted mean fit with errors adjusted according
to \emph{the same} $\sigma_{\mathrm{rand}}$ as used for the dipole
fit, in order to ensure consistency of the data points used. As the
weighted mean fit has $\chi^{2}=149.8$, the dipole fit yields a reduction
in $\chi^{2}$ of $7.9$ for an extra 3 degrees of freedom, when a
reduction of $\sim3$ would be expected by chance. Our bootstrap method
yields a significance for the dipole+monopole model over the monopole-only
model at the 97.1 percent confidence level ($2.19\sigma$), indicating
marginal evidence for the existence of a dipole when considering only
the VLT data. We demonstrate this fit in figure \ref{fig:alpha:VLT_dipole}.

We also give the parameters for a dipole-only (no monopole) fit in
table \ref{tab:alpha:combineresults_dipole}. 

\subsubsection{Effect of the choice of the method-of-moments estimator}

In section \ref{sub:alpha:physical_constraints} we suggested that
a method-of-moments estimator\index{method-of-moments estimator}
was preferable in attempting to reconcile $\Delta\alpha/\alpha$ values
from turbulent and thermal fits. It is legitimate to ask whether our
results differ if we simply choose that fit (turbulent or thermal)
which has the lowest $\chi_{\nu}^{2}$, instead of applying our method-of-moments
estimator. The results for a VLT dipole model if we do this are: $\sigma_{\mathrm{rand}}=0.928\times10^{-5}$,
$m=(-0.112\pm0.184)\times10^{-5}$, $A=1.15\times10^{-5}$ ($1\sigma$
confidence limits $[0.76,1.63]\times10^{-5}$), $\mathrm{RA}=(18.2\pm1.2)\,\mathrm{hr}$,
$\mathrm{dec}=(-62\pm14)^{\circ}$. The dipole model is preferred
over the monopole model at the 96 percent level ($2.1\sigma$). Thus
our choice of the method-of-moments estimator does not change the
results significantly, although $\sigma_{\mathrm{rand}}$ is mildly
larger if we simply choose those fits which have the lowest $\chi_{\nu}^{2}$.

\subsection{Summary of VLT results}

In this section, we have described the analysis of 154 new MM absorbers.
The VLT sample appear to display a different trend of $\Delta\alpha/\alpha$
with redshift to that seen in \citet{Murphy:04:LNP}: in the VLT sample,
$\Delta\alpha/\alpha$ appears to grow more positive with increasing
redshift, whereas fig.\ 6 of \citeauthor{Murphy:04:LNP} seems to
suggest that $\Delta\alpha/\alpha$ becomes more negative with increasing
redshift. We showed that, in the VLT sample, an angular dipole model
is preferred over a weighted mean model at the $2.2\sigma$ level,
which seems to suggest angular (and therefore spatial) variations
in $\alpha$. The direction of maximal increase in $\alpha$ is found
to be $\mathrm{RA}=(18.3\pm1.2)\,\mathrm{hr}$ and $\mathrm{dec.}=(-62\pm13)^{\circ}$
under a simple dipole model. We therefore explore the consistency
of our $\Delta\alpha/\alpha$ values and model parameters with those
derived from the same models applied to the Keck sample, and a Keck
+ VLT sample, in the next section.

We have shown that the VLT $\Delta\alpha/\alpha$ values display excess
scatter ($\chi_{\nu}^{2}>1$) about the simple models described. This
is likely due to both model mis-specification (from the use of simple
weighted mean and angular dipole models) as well as unmodelled uncertainties.
We described in section \ref{sub:alpha:rand_sys_errs} a number of
potential random effects which could give rise to excess scatter in
the data, even if our model for $\Delta\alpha/\alpha$ were correct.
It is difficult to determine the contribution of each of these effects
to the error budget, and so we have assumed that all absorbers are
affected by the same processes, and therefore increased our error
bars conservatively in quadrature with a $\sigma_{\mathrm{rand}}$
term. If the extra scatter in the $\Delta\alpha/\alpha$ values is
due to inaccuracies in modelling the velocity structure of the absorbers,
it may be that observations at higher signal-to-noise ratios and higher
resolving powers might help reduce the scatter. On the other hand,
if the inter-component spacing is comparable to the intrinsic line
widths then this may not be the case. 

We consider the specific effect of wavelength scale distortions on
the VLT sample in sections \ref{sec:asys:dv_test} and \ref{sec:asys:intraorder_distortions},
and show there how such distortions can give rise to extra scatter
in the data.

\section{Combination and comparison with previous Keck results}

In table \ref{tab:alpha:combineresults_dipole} we give the estimates
of parameters and their associated uncertainties under various models
fitted to the VLT, Keck and VLT+Keck $\Delta\alpha/\alpha$ samples.
The particular models and results are described in more detail in
the following sections.

\begin{sidewaystable}
\centering
  \caption[Constraints from different models for $\Delta\alpha/\alpha$ from Keck and VLT spectra]{\emph{First table:} Constraints from different models for $\Delta\alpha/\alpha$ from from Keck and VLT spectra. $I$ gives a number to identify the sample + model. The samples are as described in the text. $N_\mathrm{abs}$ gives the number of absorption systems used for the fit. $m$ gives the monopole component of the dipole fit for dipole models, and the weighted mean of the $\Delta\alpha/\alpha$ values for a weighted mean model. The RA and dec columns specify the right ascension and declination of the fitted pole. The $A$ column gives the amplitude of the dipole, and the $\delta(A)$ column gives the $1\sigma$ confidence limits on the dipole amplitude. $A$ is unitless, except for the $r$-dipole where $A$ has units of $\mathrm{GLyr}^{-1}$. The column labelled ``significance'' gives the significance of the dipole model over the monopole model, as both a probability and its $\sigma$ equivalent, assessed using a bootstrap method. However, for models with no monopole, the significance is given with respect to the null model ($\Delta\alpha/\alpha = 0$). \emph{Second table:} The values of $\sigma_\mathrm{rand}$ used for each model. For the Keck $\Delta\alpha/\alpha$ values, LC and HC refer to low contrast and high contrast respectively. \label{tab:alpha:combineresults_dipole}}
   {\small
\begin{tabular}{cllrccccc}
  \hline
   $I$  & Sample + model           & $N_\mathrm{abs}$ & $m\ (10^{-5})$ & RA (hr)           & dec ($^\circ$) & $A\ (10^{-5})$         & $\delta A\ (10^{-5})$       & significance \\
  \hline
   1           & Keck04-dipole           &    $140$   & $-0.465\pm0.145$     & $16.0\pm 2.7$ &  $-47\pm29$   &  $0.41$   &    $[0.29, 0.78]$    &  $36$ percent ($0.5\sigma$)\\
   2           & \# 1 with no monopole   &    $140$   &  N/A                 & $16.4\pm 1.2$ &  $-56\pm12$   &  $1.06$   &    $[0.82, 1.34]$    &  $72$ percent ($1.1\sigma$)\\ 
   3           & VLT-weighted mean     &    $153$     &   $0.208\pm0.124$  &   N/A            & N/A                  &  N/A        &  N/A               &  N/A \\
   4           & Combined weighted mean &   $293$     &   $-0.216 \pm 0.086$ &  N/A           & N/A                  &  N/A        &  N/A               & N/A\\
   5           & VLT-dipole            &    $153$     &   $-0.109\pm0.180$  &   $18.3\pm1.2$  &  $-62\pm13$  & 1.18  & $[0.80,1.66]$  &   97.1 percent ($2.19\sigma$)\\
   6           & \#5 with no monopole  &    $153$     &   N/A                &  $18.4\pm1.3$  &  $-58\pm15$  & 0.99  & $[0.70,1.37]$  &   98.4 percent ($2.39\sigma$)\\
   7           & Combined dipole       &    $293$     &   $-0.178\pm0.084$  & $17.3\pm1.0$    &   $-61\pm10$   &   $0.97$     &  $[0.77,1.19]$ &  $99.995$ percent ($4.06\sigma$)\\
   8           & \#7 with no monopole &    $293$     &   N/A             &  $17.4 \pm 0.9$   &   $-58 \pm 9$  &   $1.02$     &  $[0.83, 1.24]$ & $99.996$ percent ($4.14\sigma$)\\
   9           & Combined $r$-dipole   &    $293$     &   $-0.187\pm0.084$  & $17.5\pm1.0$     &   $-62\pm10$  &  $0.11$    & $[0.09,0.13]$   &  $99.997$ percent ($4.15\sigma$)\\
   10           & \#9 with no monopole   &    $293$     &   N/A  & $17.5\pm0.9$     &   $-58\pm9$  &  $0.11$    & $[0.09,0.14]$   &  $99.998$ percent ($4.22\sigma$)\\
   11           & $z^\beta$ dipole, $\beta = 0.46\pm0.49$   &    $293$     &   $-0.184 \pm 0.085$  & $17.5\pm1.1$     &   $-62\pm10$  &  $0.81$    & $[0.55,1.09]$   &  $99.99$ percent ($3.9\sigma$)\\
  \hline
 \\
  \end{tabular}
  \begin{tabular}{cllll}
  \hline
  $I$ & Sample + model            & $\sigma_\mathrm{rand}$(VLT) $(10^{-5})$ & $\sigma_\mathrm{rand}$(Keck LC) $(10^{-5})$ & $\sigma_\mathrm{rand}$(Keck HC) $(10^{-5})$\\
  \hline
  1   & Keck04-dipole             & N/A                         & 0            & 1.630  \\
  2   & \#1 with no monopole      & N/A                         & 0            & 1.668  \\
  3   & VLT-weighted mean         & 0.905                       & N/A          & N/A \\
  4   & Combined weighted mean    & 0.905                       & 0            & 1.743\\
  5   & VLT-dipole                & 0.905                       & N/A          & N/A \\
  6   & \#5 with no monopole      & 0.882                       & N/A          & N/A \\
  7   & Combined dipole           & 0.905                       & 0            & 1.630 \\
  8   & \#7 with no monopole      & 0.882                       & 0            & 1.668 \\
  9   & Combined $r$-dipole       & 0.858                       & 0            & 1.630\\
  10   & \#9 with no monopole      & 0.858                       & 0            & 1.630\\
  11  & $z^\beta$ dipole          & 0.812                       & 0            & 1.592\\
  \hline
  \end{tabular}}
  
\end{sidewaystable}

\subsection{Previous Keck results}

\index{Keck Deltaalpha/alpha
 results@Keck $\Delta\alpha/\alpha$ results}Although the data of \citet{Murphy:04:LNP} do not demonstrate a statistically
significant dipole, one can nevertheless calculate the location of
a (non-significant) dipole in the data. 

We note briefly that \citeauthor{Murphy:04:LNP} noticed significant
wavelength calibration problems in the spectrum of Q2206$-$1958 (J220852$-$194359)
from sample 3 of \citet{Murphy:03} for $\lambda\gtrsim5000\mathrm{\AA}$
of the order of $\sim5\,\mathrm{kms}^{-1}$ at the time of that analysis.
The two absorbers contributed by this spectrum were erroneously included
in that paper, and so we remove them from the sample. 

\citet{Murphy:04:LNP} divide their sample into two portions, a high-contrast
sample and a low-contrast sample. The high-contrast sample was defined
by 27 absorbers where there were significant differences between the
optical depth in the transitions used. \citet{Murphy:03} give arguments
as to why this might be expected to generate extra scatter in the
$\Delta\alpha/\alpha$ values. Due to the fact that many of the high
redshift ($z>1.8$) absorbers considered in \citet{Murphy:04:LNP}
are associated with damped Lyman-$\alpha$ systems, this effect manifests
itself as extra scatter in the $\Delta\alpha/\alpha$ values about
a weighted mean at high redshifts. For the VLT sample, we note that
there is no evidence for excess scatter at higher redshifts compared
to lower redshifts. 

We can examine the differences between the Keck and VLT samples in
terms of the prevalence of weak species as follows. Firstly, define
the following transitions as weak: Mg~\textsc{\small i} $\lambda2026$,
Si~\iis $\lambda1808$, the Cr~\iis transitions, Fe \iis $\lambda\lambda\lambda1608,1611,2260$,
the Mn~\iis transitions, the Ni~\iis transitions, the Ti~\iis
transitions and the Zn~\iis transitions. From the table of frequency
of occurrence in \citet{Murphy:03}, at $z<1.8$ these transitions
constitute about 3 percent of the total number of transitions used.
On the other hand, in the VLT sample these transitions constitute
about 13 percent of the sample used. The significantly greater prevalence
of these weak transitions at low redshifts in the VLT sample may explain
the lack of evidence for differential scatter between high and low
redshifts. Effectively, the greater prevalence of weak species in
the low-$z$ VLT sample may increase the scatter at low redshifts
in that sample, making any low-$z$/high-$z$ difference appear smaller.
We retain the high/low contrast distinction when analysing the Keck
sample.

\subsubsection{LTS method applied to the \citet{Murphy:04:LNP} results\label{sub:alpha:results_Keck04_LTS}}

If we apply the LTS method to the high-contrast sample to estimate
the extra error needed about a dipole model, we find that an extra
error term of $\sigma_{\mathrm{rand}}=1.630\times10^{-5}$ is needed.
With this extra term, $\chi_{\nu}^{2}=1.13$, indicating that the
distribution is mildly leptokurtic (fat-tailed). The low-contrast
sample data are already consistent under a dipole model with the LTS
method ($\sigma_{\mathrm{rand}}=0$). 

We then combine the high-contrast $\Delta\alpha/\alpha$ values (with
error bars increased) with the low-contrast $\Delta\alpha/\alpha$
values to form a new sample under a dipole model (equation \ref{eq:alpha:dipole eq}).
The LTS method applied to this set reveals that the $\Delta\alpha/\alpha$
values are consistent about dipole model. Additionally, $\chi_{\nu}^{2}=1.04$.
Nevertheless, we identify one possible outlier from this set: the
absorber with $z\approx2.84$ towards Q1946$+$7658, with $\Delta\alpha/\alpha=(-4.959\pm1.334)\times10^{-5}$,
and remove this absorber from the sample. This point has a residual
of $-3.6\sigma$ about the LTS fit. We refer to this sample as ``Keck04-dipole''. 

A dipole fitted to this sample yields $\mathrm{RA}=(16.0\pm2.7)\,\mathrm{hr}$,
$\mathrm{dec}=(-47\pm29)^{\circ}$, and $A=0.41\times10^{-5}$. $1\sigma$
confidence limits on $A$ are $[0.29,0.78]\times10^{-5}$. The monopole
is $m=(-0.465\pm0.145)\times10^{-5}$. This fit has $\chi_{\nu}^{2}=0.96$.
The dipole model is preferred over the weighted mean model at the
36 percent confidence level ($0.47\sigma$). 

The monopole offset appears to be significant at the $3.2\sigma$
confidence level, but this is related to the fact that the Keck results
alone do not clearly support a dipole interpretation.

For dipole model with no monopole ($\Delta\alpha/\alpha=A\cos\Theta$),
the fitted parameters are $A=1.06\times10^{-5}$ ($1\sigma$ confidence
limits $[0.82,1.34]\times10^{-5}$), $\mathrm{RA}=(-16.4\pm1.2)\,\mathrm{hr}$,
$\mathrm{dec.}=(-56\pm12)^{\circ}$. This model is significant at
the 72 percent confidence level ($1.1\sigma$).

\subsection{Combined weighted mean\label{sub:alpha:combine_wmean}}

We create a combined weighted mean fit by combining the VLT-dipole
sample with the Keck04-dipole sample. The VLT sample has had errors
increased in quadrature with $\sigma_{\mathrm{rand}}=0.905\times10^{-5}$,
whereas the Keck high-contrast sample has had errors increased in
quadrature with $\sigma_{\mathrm{rand}}=1.743\times10^{-5}$. The
same points identified as outliers have been removed.

This leads to a weighted mean of $(\Delta\alpha/\alpha)_{w}=(-0.216\pm0.086)\times10^{-5}$,
with $\chi_{\nu}^{2}=1.03$. However, a weighted mean model does not
appear to adequately capture all the information in the data (see
figure \ref{fig:alpha:zstack}). Comparing the weighted mean of the
$z>1.6$ points for both samples yields a simple demonstration of
the north/south difference. For the VLT sample, $\Delta\alpha/\alpha_{w}(z>1.6)=(0.533\pm0.172)\times10^{-5}$,
whereas for the Keck sample $\Delta\alpha/\alpha_{w}(z>1.6)=(-0.603\pm0.224)\times10^{-5}$.
The difference between these weighted means is $4\sigma$.

\subsection{Combined dipole fit}

\index{many-multiplet (MM) method!VLT + Keck results}To create our
combined dipole fit, we combine the VLT-dipole sample with the Keck04-dipole
sample to create the ``combined dipole'' sample, our main sample.
This sample consists of 293 MM absorbers. Importantly, both of these
sets exhibit no $\lvert r_{i}\rvert\geq3$ residuals, and thus a combined
fit is unlikely to exhibit any large residuals provided that both
data sets are well described by the same model. If the data sets are
inconsistent, one might expect large-residual points to emerge. 

For an angular dipole fit to these $\Delta\alpha/\alpha$ values ($\Delta\alpha/\alpha=A\cos\Theta+m$),
we find that $m=(-0.178\pm0.084)\times10^{-5}$, $A=0.97\times10^{-5}$,
$\mathrm{RA}=(17.3\pm1.0)\,\mathrm{hr}$, $\mathrm{dec}=(-61\pm10)^{\circ}$,
with $\chi^{2}=280.6$ and $\chi_{\nu}^{2}=0.97$. $1\sigma$ confidence
limits on $A$ are $[0.77,1.19]\times10^{-5}$. A weighted mean fit
to the same $\Delta\alpha/\alpha$ values and uncertainties yields
$\chi^{2}=303.8$, and so a dipole model yields a reduction in $\chi^{2}$
of $23.2$ for an extra 3 free parameters. With our bootstrap method,
we find that the dipole model is preferred over the weighted mean
fit at the 99.995 percent confidence level ($4.06\sigma$), thus yielding
significant evidence for the existence of angular variations in $\alpha$.
Using the method of \citet{Cooke:09}, the significance of the dipole
is found to be $4.07\sigma$. 

Importantly, the combination of the Keck04-dipole $\Delta\alpha/\alpha$
values with the VLT-dipole $\Delta\alpha/\alpha$ values yields $\chi_{\nu}^{2}\sim1$
about a dipole model. If inter-telescope systematics were present,
we would expect the combination of the Keck and VLT data to yield
a $\chi_{\nu}^{2}$ that is significantly greater than unity under
the dipole model, despite $\chi_{\nu}^{2}$ being $\sim1$ when that
model is fitted to the samples individually. Thus, there is no significant
evidence based on $\chi^{2}$ that inter-telescope systematics are
present. 

\begin{figure}[tbph]
\noindent \begin{centering}
\includegraphics[bb=78bp 40bp 526bp 751bp,clip,width=0.825\textwidth]{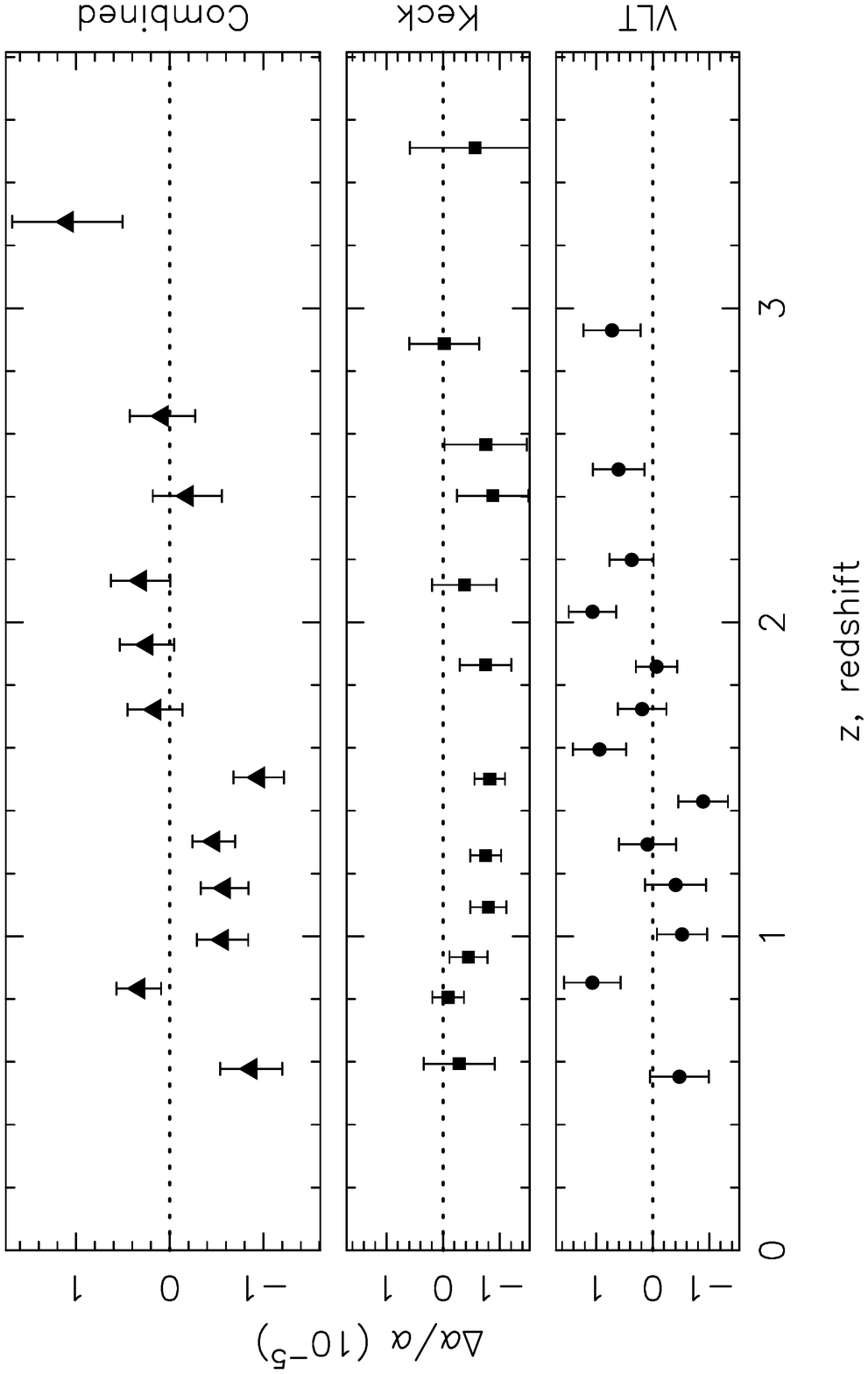}
\par\end{centering}

\caption[Binned VLT and Keck results for $\Delta\alpha/\alpha$ vs redshift]{Binned values of $\Delta\alpha/\alpha$ by redshift in the VLT-dipole sample (bottom panel, circles), the Keck04-dipole sample (middle panel, squares) and the combination of the two (top panel, triangles). The value of $\Delta\alpha/\alpha$ for each bin is calculated as the weighted mean of the values of $\Delta\alpha/\alpha$ from the contributing absorbers. The statistical errors for certain points have been increased prior to binning, as described in the text. Note that for $z \gtrsim 1.5$, the Keck data generally indicate $\Delta\alpha/\alpha < 0$, whereas the VLT data indicate $\Delta\alpha/\alpha > 0$. As Keck is located in the northern hemisphere, and VLT is in the south, this is a rough visual demonstration of the dipole effect. However, given the overlap between the samples, the proper procedure is to directly fit a dipole (see figure \ref{fig:alpha:combinedresults}). Interestingly, both the VLT and Keck data seem to to support $\Delta\alpha/\alpha < 0$ for $z \lesssim 1.5$. This effect is considered in more detail in section \ref{sub:alpha:s_monopole}. \label{fig:alpha:zstack}}
\end{figure}

\begin{figure}[tbph]
\noindent \begin{centering}
\includegraphics[bb=44bp 88bp 517bp 727bp,clip,width=0.85\textwidth]{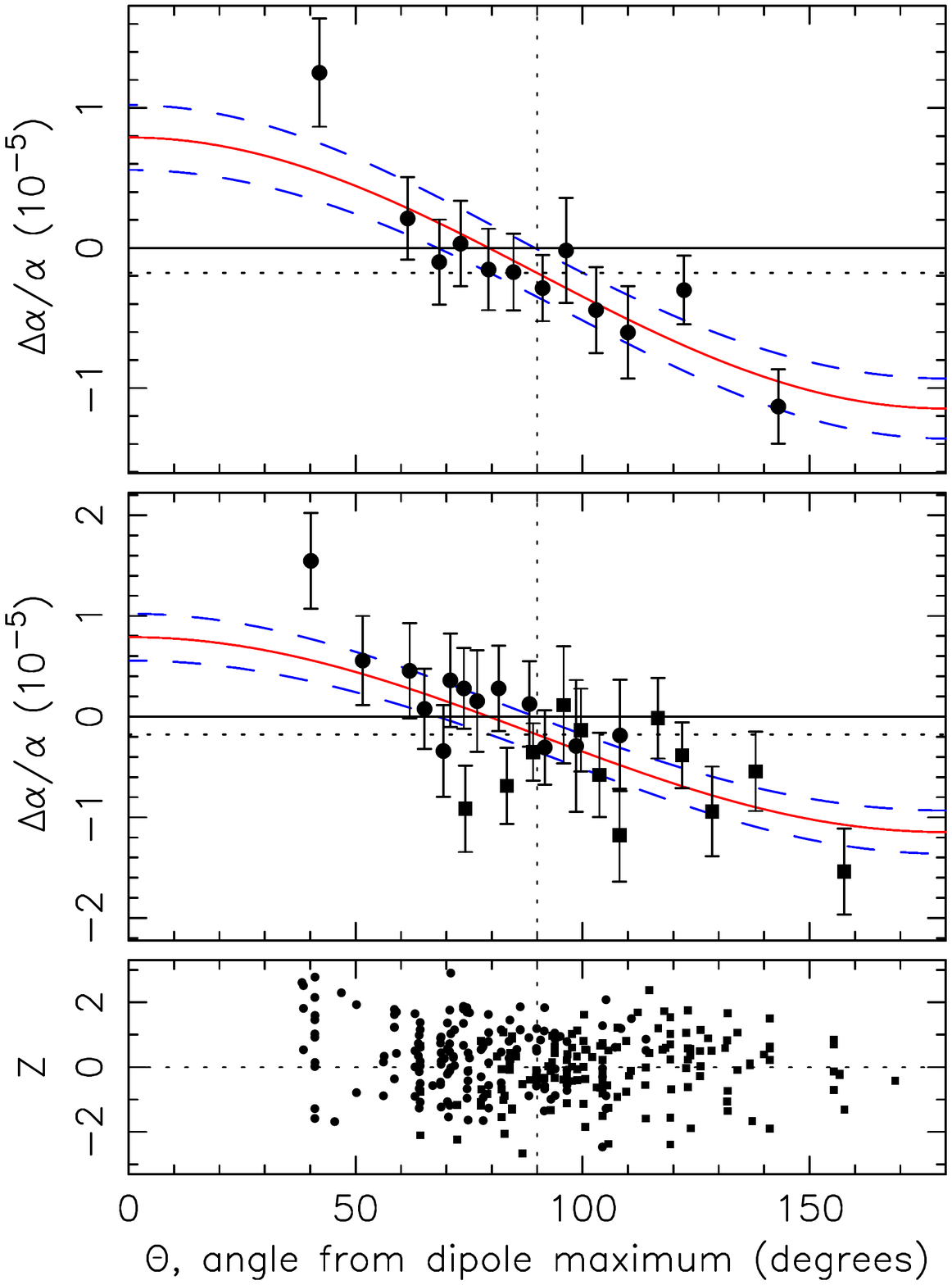}
\par\end{centering}

\caption[Binned VLT and Keck results for $\Delta\alpha/\alpha$ vs angle from the best-fitting dipole]{The top panel shows the combined results for the Keck and VLT samples, plotting $\Delta\alpha/\alpha$ against angle from the fitted dipole location for the combination of the Keck and VLT $\Delta\alpha/\alpha$ values, binned together. The middle panel shows the data from different telescopes, with Keck as squares and VLT as circles. Points in the top panel contain approximately 25 absorbers per bin, whereas points in the middle panel contain approximately 12. The model shown (red, solid line) is $\Delta\alpha/\alpha = A\cos(\Theta) + m$. The parameters for this model are: $m = (-0.178 \pm 0.084) \times 10^{-5}$, $A = 0.97 \times 10^{-5}$ ($1\sigma$ confidence limits $[0.77, 1.19] \times 10^{-5}$), $\mathrm{RA} = (17.3 \pm 1.0)\,\mathrm{hr}$, $\mathrm{dec.} = (-61 \pm 10)^\circ$. The dashed, blue lines indicate the $1\sigma$ uncertainty on the fit, including the uncertainty in determining the position of the dipole. In both the top and middle panels, the dotted horizontal line indicates the monopole value. The vertical dotted line shows $90^\circ$. The bottom panel indicates the standardised residuals ($r_i = [\mathrm{data} - \mathrm{model}]/\mathrm{error}$) about the best fit. The presence of no points with $\lvert r_i \rvert > 3$ indicates that the fit is not being dominated by a small number of large residual points. \label{fig:alpha:combinedresults}}
\end{figure}

\begin{figure}[tbph]
\noindent \begin{centering}
\includegraphics[bb=78bp 79bp 542bp 727bp,clip,width=0.85\textwidth]{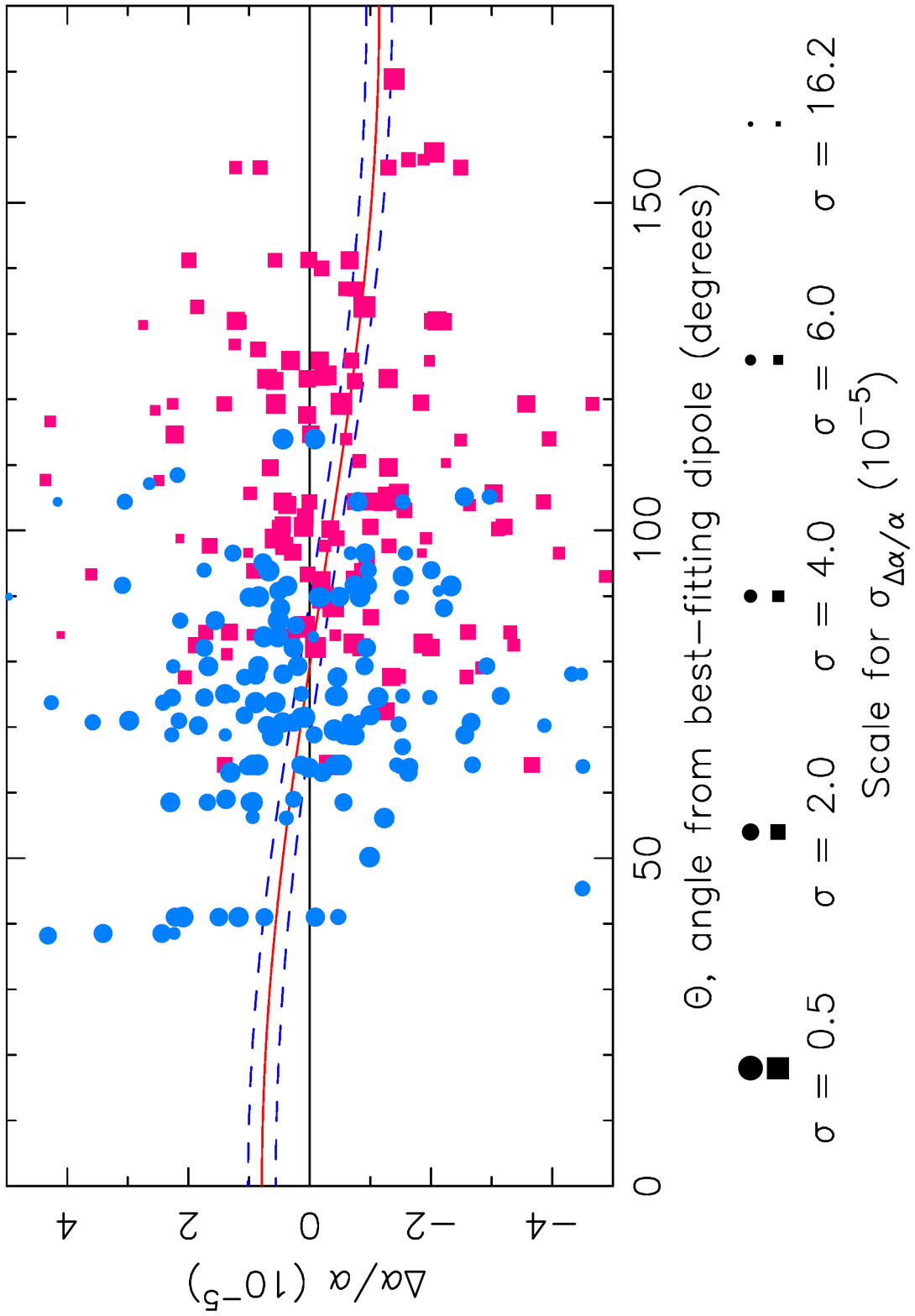}
\par\end{centering}

\caption[Unbinned VLT and Keck results for $\Delta\alpha/\alpha$ vs angle from the best-fitting dipole]{$\Delta\alpha/\alpha$ against the angle from the fitted dipole location under the model $\Delta\alpha/\alpha = A\cos(\Theta) + m$ for the VLT and Keck $\Delta\alpha/\alpha$ values. In contrast to figure \ref{fig:alpha:combinedresults}, these data are not binned. Blue circles are VLT absorbers and pink squares are Keck absorbers. Error bars have been omitted; instead, larger symbols indicate $\Delta\alpha/\alpha$ values with greater statistical weight, according to the key provided. The precision includes the effect of $\sigma_\mathrm{rand}$. The dipole trend is visible as the presence of more and larger points in the upper left and lower right quadrants. The visual cluster of points at $\Theta < 47^\circ$ is due to 4 quasars which contribute 14 values of $\Delta\alpha/\alpha$ (2 points not shown because they lie beyond the vertical range of the plot). One can investigate the consistency of the VLT and Keck $\Delta\alpha/\alpha$ values in the region near the dipole equator (defined here as $80^\circ < \Theta < 100^\circ$) by comparing the weighted mean of the $\Delta\alpha/\alpha$ values. In this case, $\Delta\alpha/\alpha_w(\mathrm{VLT}) - \Delta\alpha/\alpha_w(\mathrm{Keck}) = (0.32 \pm 0.19) \times 10^{-5}$, giving no significant evidence for a difference between the two samples. In this region, the VLT sample contributes 39 points and the Keck sample contributes 43 points. The difference here is calculated so as to include the effect of $\sigma_\mathrm{rand}$.\label{fig:alpha:anglefromdipole_size}}
\end{figure}

We show in figure \ref{fig:alpha:combinedresults} the values of $\Delta\alpha/\alpha$
for both Keck and VLT against the best-fitting dipole model. We give
binned values there, which yields a visual demonstration of the dipole
effect. We also give there a plot of the standardised residuals about
the fit, which demonstrates that the fit is statistically reasonable.
We also show binned values of $\Delta\alpha/\alpha$ for the Keck,
VLT and combined samples in figure \ref{fig:alpha:zstack}. We show
an unbinned version of these data for $|\Delta\alpha/\alpha|<5\times10^{-5}$
in figure \ref{fig:alpha:anglefromdipole_size}.

For a model with no monopole ($\Delta\alpha/\alpha=A\cos\Theta$),
the fitted parameters are $A=1.02\times10^{-5}$ ($1\sigma$ confidence
limits $[0.83,1.24]\times10^{-5}$), $\mathrm{RA}=(17.4\pm0.9)\times10^{-5}$,
$\mathrm{dec.}=(-58\pm9)^{\circ}$. This model is significant at the
99.996 percent level ($4.14\sigma$).

\begin{sidewaysfigure}
\noindent \begin{centering}
\includegraphics[bb=41bp 65bp 468bp 728bp,clip,angle=-90,width=0.75\textwidth]{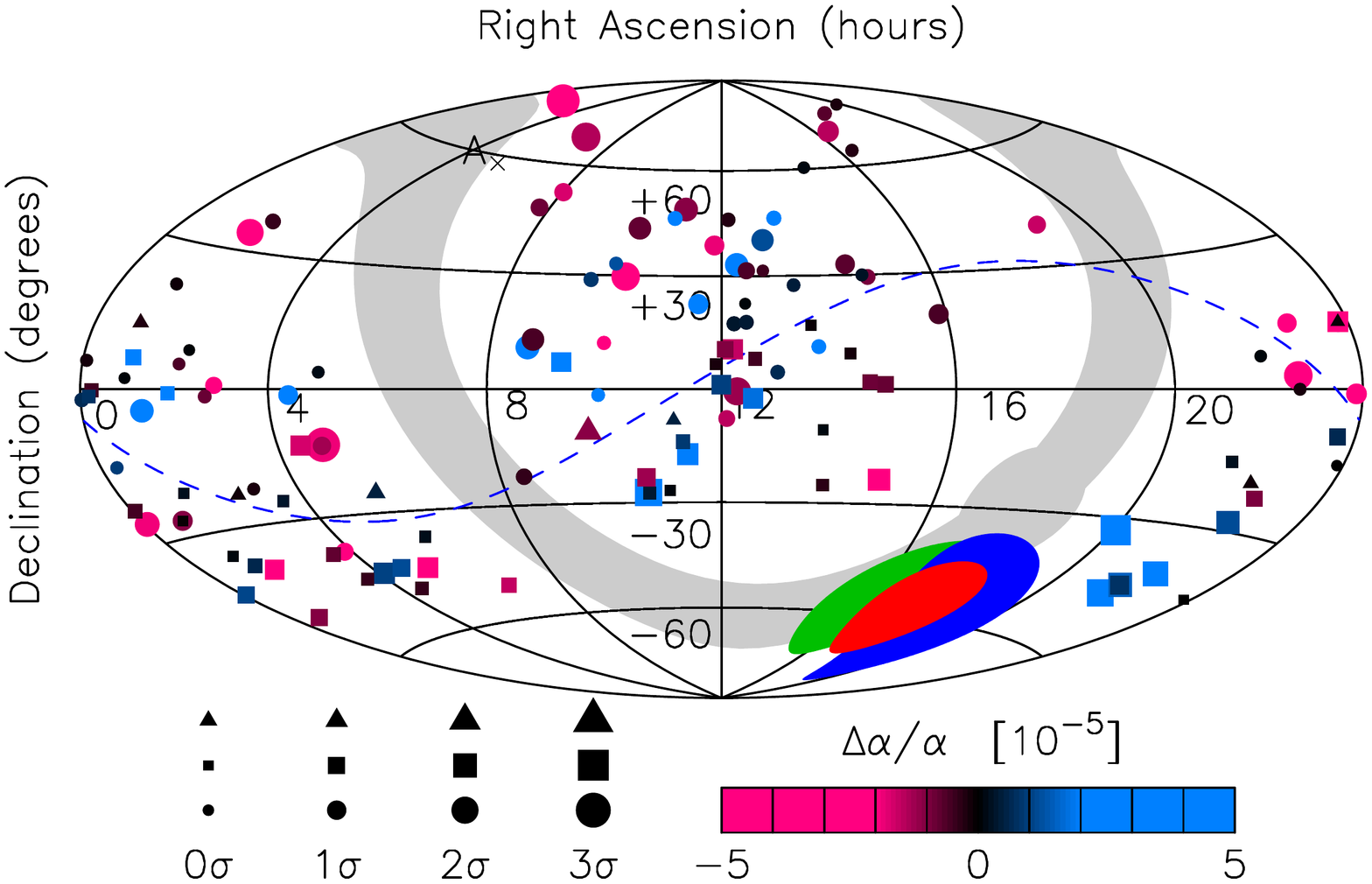}
\par\end{centering}

\caption[Whole sky map showing aggregated data for $\Delta\alpha/\alpha$]{Whole sky map showing aggregated data for $\Delta\alpha/\alpha$. The green, blue and red regions show the $1\sigma$ confidence regions on the location of the Keck, VLT and Keck+VLT dipole models respectively. In all cases, the model is $\Delta\alpha/\alpha = A\cos(\Theta)$. The antipole is marked with an ``A''. The dashed, blue line shows the equatorial region of the dipole. The data points show the weighted mean of $\Delta\alpha/\alpha$ along each quasar sightline. Circles indicate Keck quasars, squares indicate VLT quasars, and triangles show quasars which are common to both samples. The size of the symbols indicates the residual about the null model (i.e.\  [$\Delta\alpha/\alpha]/\sigma_i$). The colour scale gives the value of the difference between the value of $\Delta\alpha/\alpha$ and $\Delta\alpha/\alpha=0$. The grey region schematically indicates the galactic plane, with the bulge representing the galactic centre. The dipolar trend can be seen visually with more and larger blue points towards the pole, and more and larger pink points towards the antipole. \label{Flo:alpha:compressed_sightlines}}
\end{sidewaysfigure}

\begin{sidewaysfigure}
\noindent \begin{centering}
\includegraphics[bb=41bp 65bp 468bp 728bp,clip,angle=-90,width=0.75\textwidth]{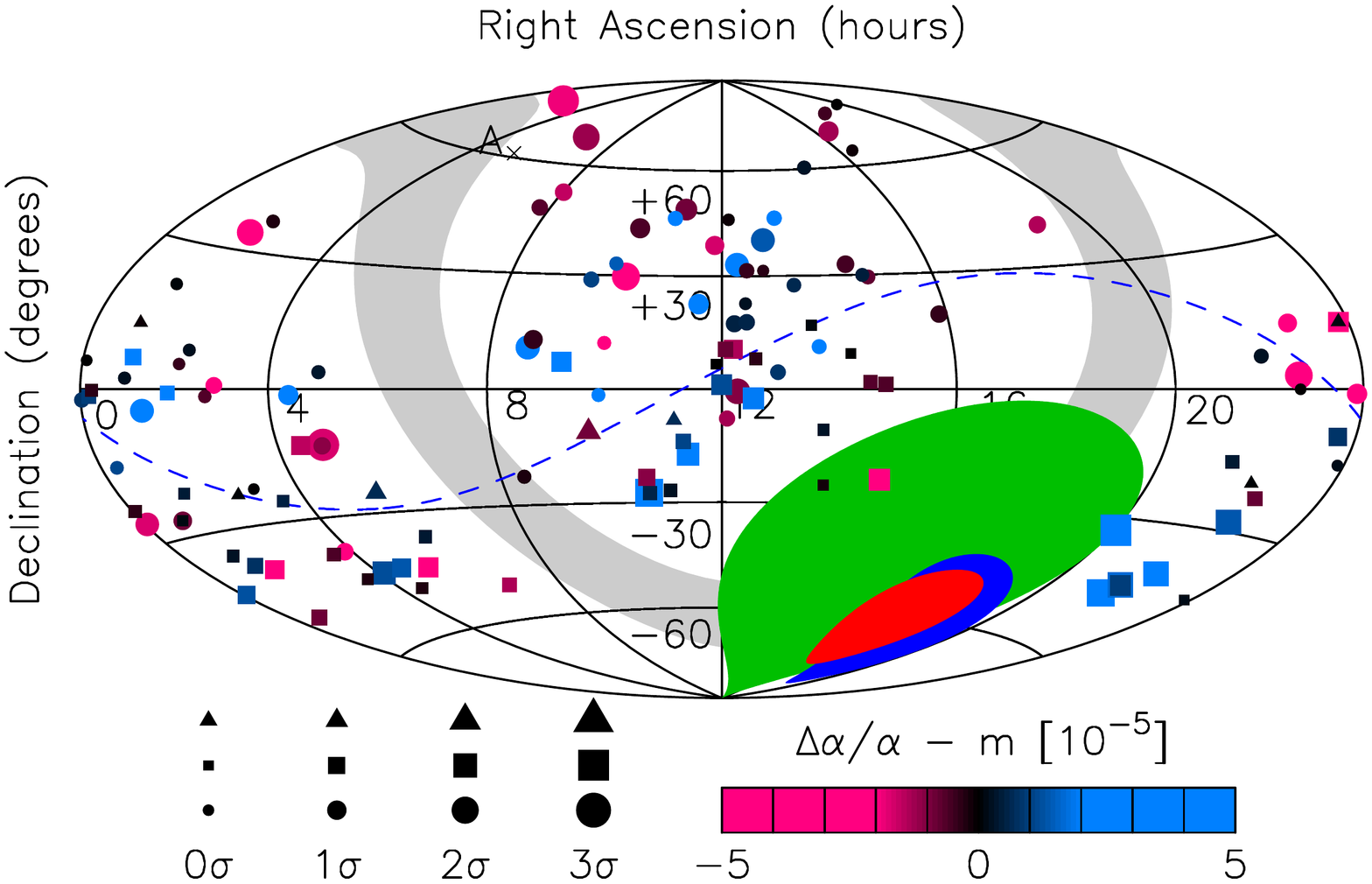}
\par\end{centering}

\caption[Whole sky map showing aggregated data for $\Delta\alpha/\alpha$]{Whole sky map showing aggregated data for $\Delta\alpha/\alpha$. The green, blue and red regions show the $1\sigma$ confidence regions on the location of the Keck, VLT and Keck+VLT dipole models respectively. In all cases, the model is $\Delta\alpha/\alpha = A\cos(\Theta)+m$. The antipole is marked with an ``A''. The dashed, blue line shows the equatorial region of the dipole. The data points show the weighted mean of $\Delta\alpha/\alpha$ along each quasar sightline. Circles indicate Keck quasars, squares indicate VLT quasars, and triangles show quasars which are common to both samples. The size of the symbols indicates the residual about the monopole (i.e.\  [$\Delta\alpha/\alpha-m]/\sigma_i$). The colour scale gives the value of the difference between the value of $\Delta\alpha/\alpha$ and the monopole. The grey region schematically indicates the galactic plane, with the bulge representing the galactic centre. The dipolar trend can be seen visually with more and larger blue points towards the pole, and more and larger pink points towards the antipole. \label{Flo:alpha:compressed_sightlines_monopole}}
\end{sidewaysfigure}

In figures \ref{Flo:alpha:compressed_sightlines} and \ref{Flo:alpha:compressed_sightlines_monopole},
we show the confidence limits on the dipole location for the Keck,
VLT and combined samples. The individual symbols illustrate the weighted
mean of $\Delta\alpha/\alpha$ along each sightline under the models
$\Delta\alpha/\alpha=A\cos\Theta$ and $\Delta\alpha/\alpha=A\cos\Theta+m$
respectively. 

There are several significant points to consider from these results:
\begin{enumerate}
\item \emph{The dipole is statistically significant.} Even after accounting
for random errors in a conservative fashion, the statistical significance
of the dipole is greater than $4\sigma$. This is strong statistical
evidence for angular and therefore spatial variation in $\alpha$.
\item \emph{Dipole models fitted to the Keck and VLT $\Delta\alpha/\alpha$
values yield consistent estimates for the pole direction.} This is
important, and would be very surprising if one assumes that a dipole
effect is not present. If two different systematic effects were operating
in each telescope so as to produce a trend in $\Delta\alpha/\alpha$,
then: \emph{a)} it is unlikely that these effects would be correlated
with sky position, and \emph{b)} even if systematic effects existed
in both telescopes which were correlated with sky position, it is
very unlikely that such effects would occur in such a way as to yield
very consistent estimates of the dipole position between the two telescopes,
with a similar amplitude, particularly when the two telescopes are
independently constructed and separated by $\sim45^{\circ}$ in latitude.
Any attempt to ascribe the observed variation in $\alpha$ to systematics
must account for the good alignment of the dipole vectors from dipole
models fitted independently to the Keck and VLT samples. Note that
telescope or instrumental systematics which depend only on wavelength
\emph{cannot} produce observed angular variation in $\alpha$ for
a sufficiently large sample of absorbers. 
\item \emph{The VLT and Keck $\Delta\alpha/\alpha$ values appear consistent
near the equatorial region of the dipole.} From the middle panel of
figure \ref{fig:alpha:combinedresults}, both the VLT and Keck results
show large variation from $\Delta\alpha/\alpha=0$ near the pole ($\Theta=0^{\circ}$)
and anti-pole ($\Theta=180^{\circ}$) of the dipole, but show much
less variation in the equatorial region ($\Theta=90^{\circ}$). So,
at least visually, the Keck and VLT points are not inconsistent in
the region where they overlap. This issue is addressed quantitatively
in the caption to figure \ref{fig:alpha:anglefromdipole_size}
\item \emph{The dipole effect is not being caused by large residual points.}
The bottom panel of figure \ref{fig:alpha:combinedresults} clearly
shows that there are no $|r_{i}|>3\sigma$ points present. 
\end{enumerate}

\subsubsection{Bayesian evidence}

In the Bayesian paradigm, a quantity of fundamental interest for model
selection is the Bayesian evidence\index{Bayesian evidence}. For
some model $M_{i}$, data set $\mathbf{D}$ and vector of parameters
$\mathbf{x}_{i}$ (of dimension $p_{i}$) the evidence is given by
\begin{equation}
\mathrm{Pr}(\mathbf{D}|M_{i})=\int\mathrm{Pr}(\mathbf{D}|\mathbf{x}_{i})\,\mathrm{Pr}(\mathbf{x}_{i})\,\mathrm{d}\mathbf{x}_{i}.
\end{equation}
Suppose a competing model for the same data set has parameters $\mathbf{x}_{i}$.
For the evidence in favour of a dipole + monopole model (described
by the 4 parameters $\mathbf{x}_{d}$) against a monopole-only model
(described by the parameter $\mathbf{x}_{m}$), the Bayes factor determines
the evidence in favour of one model over the other, namely
\begin{equation}
B=\frac{\int\mathrm{Pr}(\mathbf{D}|\mathbf{x}_{d})\,\mathrm{Pr}(\mathbf{x}_{d})\,\mathrm{d}\mathbf{x}_{d}}{\int\mathrm{Pr}(\mathbf{D}|\mathbf{x}_{m})\,\mathrm{Pr}(\mathbf{x}_{m})\,\mathrm{d}\mathbf{x}_{m}}.
\end{equation}
The evidence is computationally difficult to evaluate, especially
for high numbers of dimensions --- the integration must generally
be carried out through Monte Carlo means, and naive Monte Carlo integration
degrades exponentially with increasing dimensionality. One option
is to assume that the posterior PDF is approximately Gaussian. For
our data, this should be at least approximately true on account of
the central limit theorem, as we have 293 points in our main sample.
Although we noted that $(A,\mathrm{RA},\mathrm{dec})$ are not normally
distributed, $(c_{x},c_{y},c_{z})$ should be. 

By approximating the posterior probability as a Gaussian, one obtains
\begin{equation}
\mathrm{Pr}(\mathbf{x_{i}}|\mathbf{D})\propto\mathrm{Pr}(\hat{\mathbf{x}_{i}}|\mathbf{D})\exp\left[-\frac{1}{2}(\mathbf{x}_{i}-\hat{\mathbf{x}}_{i})^{\mathrm{T}}\mathbf{C}_{i}^{-1}(\mathbf{x}_{i}-\hat{\mathbf{x}}_{i})\right],
\end{equation}
where $\hat{\mathbf{x}}_{i}$ is the best estimate of the parameters
and $\mathbf{C}_{i}$ is the covariance matrix at the best-fitting
solution. This leads \citep{Hobson:02a} to the approximation
\begin{equation}
\mathrm{Pr}(\mathbf{D}|M_{i})\approx(2\pi)^{p_{i}/2}\lvert\mathbf{C}_{i}\rvert^{1/2}\,\mathrm{Pr}(\hat{\mathbf{x}}_{i})\,\mathrm{Pr}(\mathbf{D}|\hat{\mathbf{x}}_{i},M_{i}),
\end{equation}
where $p_{i}$ is the number of data points. This is known as the
Laplace approximation. This expression requires that $\mathrm{Pr}(\mathbf{x}_{i})$,
the prior for the parameters, and $\mathrm{Pr}(\mathbf{D}|\hat{\mathbf{x}}_{i},M_{i})$,
the likelihood function for the fit, are appropriately normalised,
such that $\int\mathrm{Pr}(\mathbf{x}_{i})\,\mathrm{d}\mathbf{x}_{i}=1$
and $\int\mathrm{Pr}(\mathbf{D}|\mathbf{x}_{i})\,\mathrm{d\mathbf{D}}=1$.

The likelihood for the $N$ data points is given by
\begin{equation}
\mathrm{Pr}(\mathbf{D}|\mathbf{x}_{i},M_{i})=\prod_{j=1}^{N}\frac{1}{\sigma_{j}\sqrt{2\pi}}\exp\left[-\frac{\left(y_{j}-f_{i}(\mathbf{x}_{i})_{j}\right)^{2}}{2\sigma_{j}^{2}}\right],
\end{equation}
where $y_{j}$ is the $j$th value of $\Delta\alpha/\alpha$, $\sigma_{j}$
is the associated uncertainty and $f_{i}(\mathbf{x}_{i})_{j}$ is
the model prediction under the $i$th model. We can drop certain terms
in here, because when comparing two models we consider the same data
set (i.e.\ the $\sigma_{j}$ are common). Thus we can use instead
\begin{equation}
\mathrm{Pr}'(\mathbf{D}|\mathbf{x}_{i},M_{i})=\prod_{j=1}^{N}\exp\left[-\frac{\left(y_{j}-f_{i}(\mathbf{x}_{i})_{j}\right)^{2}}{2\sigma_{j}^{2}}\right]=\exp\left[-\frac{\chi_{i}^{2}(\mathbf{x}_{i})}{2}\right],
\end{equation}
where we write $\chi_{i}^{2}(\mathbf{x}_{i})$ to indicate that the
model is evaluated at some value of the parameters, not necessarily
the maximum likelihood estimate.

The only issue left is to evaluate the prior, $\mathrm{Pr}(\hat{\mathbf{x}}_{i})$.
Unfortunately, the estimation of the evidence is sensitive to the
choice of priors. Firstly, note that we are comparing a dipole + monopole
model to a monopole-only model. If we assume the same uniform prior
for the monopole in both samples, it will be a common factor in the
evidence for both models and therefore will cancel. Thus we need not
choose any particular range for the prior on the monopole. However,
we must choose a prior on the dipole components. Here, it is more
convenient to work in spherical coordinates, where the dipole is naturally
expressed. Firstly, we assume a separable prior, so we can write
\begin{equation}
\mathrm{Pr}(A,\phi,\theta)=\mathrm{Pr}(A)\mathrm{Pr}(\phi,\theta).
\end{equation}
The obvious choice for $\mathrm{Pr}(\phi,\theta)$ is one which gives
no preference to any particular angle, so that there is no preference
specified for the dipole direction. The necessary prior can be derived
from the symmetry argument that the probability of a point being in
a particular region is proportional to the region's angular area.
That is, 
\begin{equation}
d\mathrm{Pr}=\frac{d\Omega}{4\pi}=\frac{\sin\theta}{4\pi}\,\mathrm{d}\theta\,\mathrm{d}\phi,
\end{equation}
where the factor of $4\pi$ is chosen to give the correct normalisation.
Thus, the prior is simply
\begin{equation}
\mathrm{Pr}(\phi,\theta)=\frac{\sin\theta}{4\pi}.
\end{equation}
We have to choose a realistic prior for $\mathrm{Pr}(A)$; an unreasonably
broad choice of prior will cause a model with more parameters to always
be disfavoured \citep{BayesianTutorial}. An ideal choice would be
$\mathrm{Pr}(A)\propto1/A$ (the Jeffreys' prior), however this prior
cannot be normalised. A heuristic choice is
\begin{equation}
\mathrm{Pr}(A)=Ce^{-A/k},
\end{equation}
for some initial scale estimate $k$ \citep{BayesianTutorial}. This
gives a preference to small amplitudes, which is what we naturally
expect. Thus, the prior required is
\begin{equation}
\mathrm{Pr}(A,\phi,\theta)=\frac{C}{4\pi}e^{-A/k}\sin\theta.
\end{equation}
To ensure that $\mathrm{Pr}(A,\phi,\theta)$ is properly normalised,
we need
\begin{equation}
\int_{A=0}^{\infty}\int_{\phi=0}^{2\pi}\int_{\theta=0}^{\pi}\frac{Ce^{-A/k}}{4\pi}\, A^{2}\sin\theta\,\mathrm{d}\theta\mathrm{\, d}\phi\,\mathrm{d}A=1,
\end{equation}
which means that
\begin{equation}
C=\frac{1}{2k^{3}}.
\end{equation}
If we define
\begin{equation}
F(h)=\int_{A=0}^{h}\int_{\phi=0}^{2\pi}\int_{\theta=0}^{\pi}\frac{e^{-A/k}\sin\theta}{8\pi k^{3}}\, A^{2}\,\mathrm{d}\theta\,\mathrm{d}\phi\,\mathrm{d}A,
\end{equation}
then note that 
\begin{eqnarray*}
F(k) & \approx & 0.08,\\
F(2k) & \approx & 0.32,\quad\mathrm{and}\\
F(3k) & \approx & 0.58.
\end{eqnarray*}
Thus, most of the probability volume is located at $A>k$. We think
that a choice of $k=1\times10^{-5}$ for the dipole amplitude as a
prior is not too controversial. This means that there is a 99.7 percent
chance that the dipole amplitude is less than $10^{-4}$, with other
probabilities as given above. In our case, $A=0.97\times10^{-5}$
and $\theta=151^{\circ}$. Thus, $\mathrm{Pr}(\hat{\mathbf{x}}_{D})=\exp(-0.97)\sin(151^{\circ})/[8\pi(10^{-5})^{3}]$,
again neglecting the monopole prior because it is common to the evidence
for the monopole. 

With these assumptions, we calculate $B=49.7$. That is, the dipole
+ monopole model is preferred to the monopole model at the 98 percent
level. Converting this to the Jeffreys' scale \citep{Jeffreys:1961}
requires us to consider $2\ln B=7.8$. On the Jeffreys' scale \citep{Jeffreys:1961},
this is considered strong evidence in favour of the dipole + monopole
model over the monopole-only model.

\subsection{Potential effect of differences in atomic data and $q$ coefficients}

If the atomic data or $q$ coefficients we used were significantly
different to those used by \citet{Murphy:04:LNP}, this could spuriously
create differences in $\Delta\alpha/\alpha$ between VLT and Keck.
This has the potential to mimic spatial variation in $\alpha$. To
check the influence of this, we re-fit the VLT spectra using the same
atomic data used by \citeauthor{Murphy:04:LNP}, and then combine
the $\Delta\alpha/\alpha$ values with the Keck values. Where we use
transitions that were not available to \citeauthor{Murphy:04:LNP}
(e.g.\ Mn~\iis and Ti~\ii) we make no modification to the atomic
data or $q$ coefficients. The frequency of occurrence of these transitions
in the sample is small and therefore this is of little consequence.
When we proceed in this way, the parameters for the model $\Delta\alpha/\alpha=A\cos\Theta+m$
are: $A=0.97\times10^{-5}$, $\mathrm{RA}=(17.5\pm1.0)\,\mathrm{hr}$,
$\mathrm{dec.}=(-60\pm10)^{\circ}$ and $m=(-0.168\pm0.084)\times10^{-5}$.
The significance of the dipole+monopole model with respect to the
monopole-only model is $4.15\sigma$. We conclude that the impact
of any variations between atomic data or the $q$ coefficients used
for our fits and those used by \citet{Murphy:04:LNP} is negligible.

\subsection{Alignment by chance between Keck and VLT\label{sub:alpha:alignment_by_chance}}

One can pose the question: ``Given the distribution of sightlines
and values of $\Delta\alpha/\alpha$ in each sample, what is the probability
of observing alignment as good or better than that observed between
the Keck and VLT samples by chance?'' To assess this, we undertake
a bootstrap analysis, where at each bootstrap iteration we randomly
reassign the values of $\Delta\alpha/\alpha$ in both the Keck and
VLT samples to different sightlines within those samples, keeping
the redshifts of the absorbers fixed. That is, we do not mix the two
samples. We then calculate the best-fitting dipole vectors for each
sample, and calculate the angle between them. We then assess over
many iterations in what percentage of cases is the fitted angle smaller
than the angle for our actual data. 

\begin{figure}[tbph]
\noindent \begin{centering}
\includegraphics[bb=101bp 54bp 556bp 696bp,clip,angle=-90,width=0.7\textwidth]{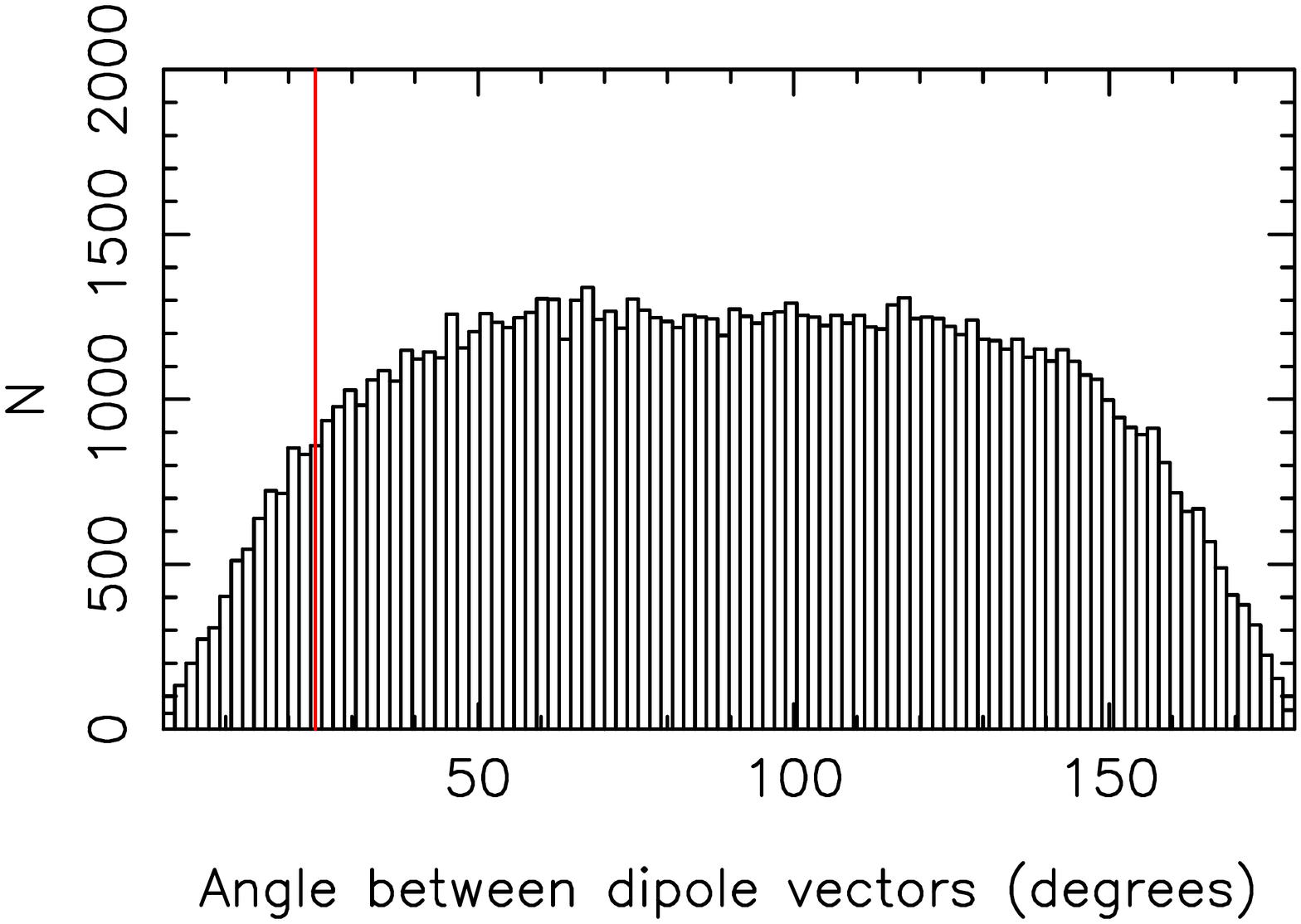}
\par\end{centering}

\caption[Bootstrap analysis to determine the probability of obtaining alignment between VLT and Keck dipole at least as good as is seen by chance]{Results of the bootstrap analysis described in section \ref{sub:alpha:alignment_by_chance} to assess the probability of obtaining alignment of the dipole vectors between the VLT and Keck samples as good as we have observed by chance. The vertical red line shows the angle between the Keck and VLT dipole vectors (24 degrees). The area to the left of the red line indicates the probability of interest, namely 6 percent. \label{fig:alpha:alignmentbychance}}
\end{figure}

For our actual Keck and VLT samples, the angle between the fitted
dipole vectors is 24 degrees, and the chance probability is $\approx6$
percent. We show the results of this bootstrap analysis in figure
\ref{fig:alpha:alignmentbychance}. Thus, it seems unlikely that inter-telescope
systematics are responsible for the observed effect. The good consistency
between the results also qualitatively supports the notion that the
measured effect is real. 

\subsection{Low-$z$ vs high-$z$ sample cuts\label{sub:alpha:lowz_vs_highz}}

We divide our sample into low-$z$ and high-$z$ absorbers to examine
the contribution of the different redshifts to the dipole detection.
Although there is no clear delineation between which transitions are
fitted for a given redshift, we can generally say that the low-$z$
sample is dominated by the Mg/Fe combination, that intermediate redshifts
display a wide range of transitions, and that high redshift systems
are dominated by the Si \ii/Al \ii/Fe~\iis $\lambda1608$ combination
with Cr~\ii/Zn~\ii/Ni~\ii. In particular, Mg~\ii, Mg~\iscs
$\lambda2852$ and the Fe~\iis transitions with $\lambda_{0}\gtrsim2200$
are not generally used when fitting absorbers at high $z$ because
they are either beyond the red cut-off in the observed spectral range,
or the transitions are affected by sky absorption or emission.

If the observed dipole effect was caused by chance or by a systematic
effect which affects some combination of transitions, then we would
not expect dipole fits to absorbers from high and low redshift to
yield the same location on the sky. Conversely, if dipole models fitted
to high and low redshift samples point in a similar direction, this
lends support to the dipole interpretation of the data. 

\begin{figure}[tbph]
\noindent \begin{centering}
\includegraphics[bb=77bp 78bp 455bp 727bp,clip,angle=-90,width=0.8\textwidth]{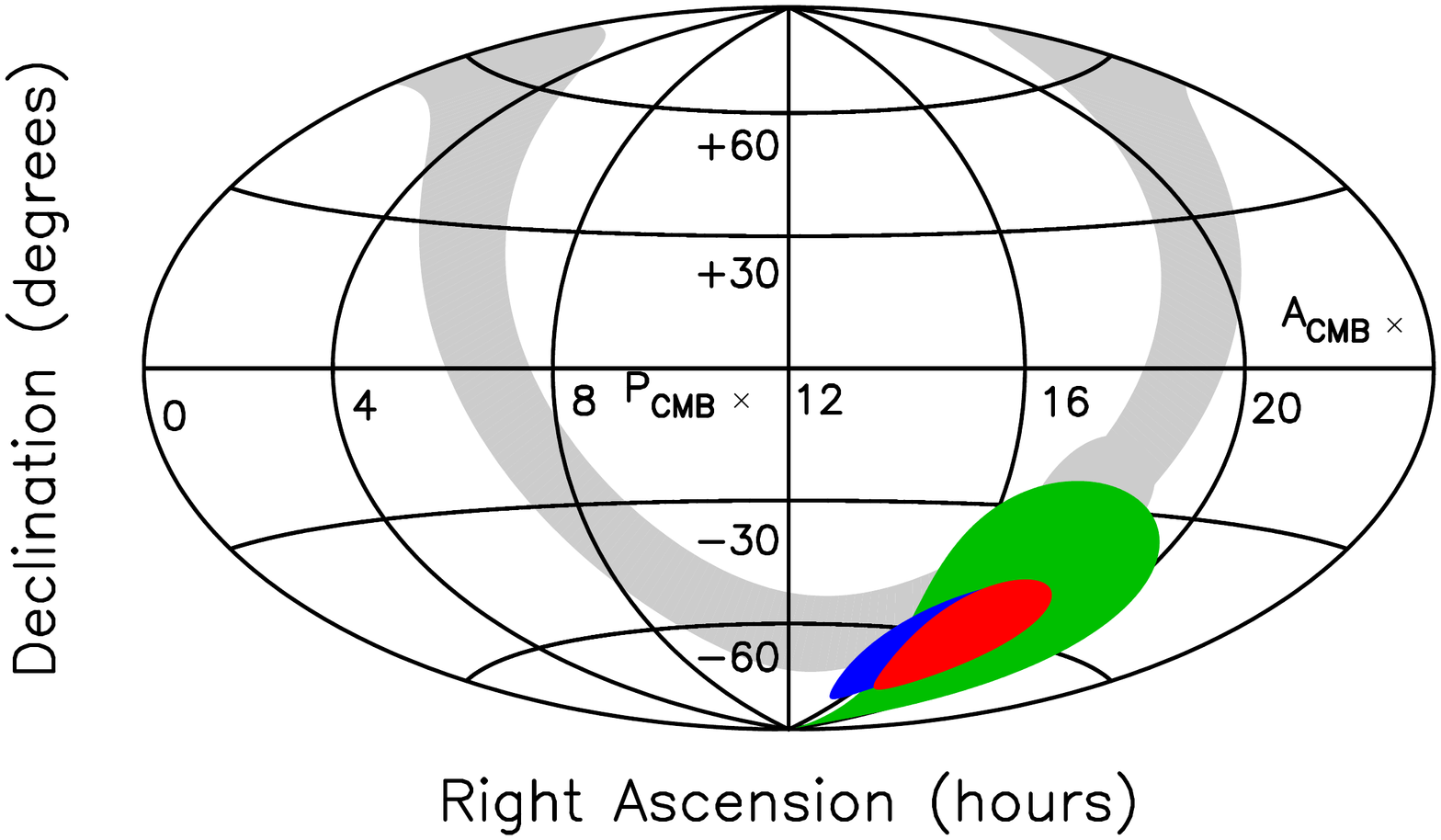}
\par\end{centering}

\caption[Sky map of Low-$z$ vs high-$z$ sample cuts]{Sky map in equatorial (J2000) coordinates showing the 68.3 percent ($1\sigma$ equivalent) confidence limits of the location of the pole of the dipole fitted to the $z<1.6$ combined sample (green region), $z > 1.6$ combined sample (blue region) and combined sample (red region) under the model $\Delta\alpha/\alpha = A\cos(\Theta) + m$. The location of the CMB dipole and antipole are marked as P$_\mathrm{CMB}$ and A$_\mathrm{CMB}$ respectively for comparison \citep{Lineweaver:97}. This figure demonstrates that the low-$z$ and high-$z$ absorbers produce consistent estimates of the dipole location, despite generally using significantly different combinations of transitions. The dipole vectors for the $z<1.6$ and $z>1.6$ sample are separated by 13 degrees. The probability of getting alignment this good or better by chance is 2 percent. \label{fig:alpha:skymap_lowvshigh}}
\end{figure}

We cut the data into a $z<1.6$ sample (low-$z$) and a $z>1.6$ sample
(high-$z$). This divides the data approximately in half, with $148$
points in the low-$z$ sample and $145$ in the high-$z$ sample.
We show in figure \ref{fig:alpha:skymap_lowvshigh} the confidence
limits on the dipole directions from separate fits to the low-$z$
and high-$z$ samples, and demonstrate that they yield consistent
estimates of the dipole location. We give the parameters to the model
$\Delta\alpha/\alpha=A\cos(\Theta)+m$ in table \ref{tab:alpha:zsplit}.
In particular, the dipole vectors are separated by 13 degrees on the
sky. 

Given the distribution of $\Delta\alpha/\alpha$ values and sightlines
in each sample, the probability of obtaining alignment this good or
better by chance is 2 percent. Given that the transitions used at
low and high redshift are significantly different (and the relationship
between the $q$ coefficients and wavelength is significantly different
for the transitions used at low and high redshift), this consistency
further supports the dipole interpretation of the data. It is also
clear that the dipole signal is significantly larger at high redshift,
although the low redshift sample contributes. 

There is no significant evidence for a high-$z$ monopole, but the
low-$z$ monopole is significant at the $3.6\sigma$ level\index{monopole, low-$z$}.
We discuss the significance of the low-$z$ monopole in section \ref{sub:alpha:s_monopole}. 

\begin{table}[tbph]
\caption[$\Delta\alpha/\alpha$ results for $z<1.6$ and $z>1.6$ cuts of the combined Keck+VLT sample]{Parameters for the model $\Delta\alpha/\alpha = A\cos(\Theta) + m$ for $z<1.6$ and $z>1.6$ samples. The column ``$\delta A$'' gives $1 \sigma$ confidence limits on $A$. The column labelled ``sig'' gives the significance of the dipole model with respect to the monopole model. Although it is clear that most of the significance comes from the $z>1.6$ sample, the $z<1.6$ sample also contributes. Additionally, a dipole model for the $z<1.6$ sample points in a similar direction to that of the $z>1.6$ sample. \label{tab:alpha:zsplit}}

\centering{}%
\begin{tabular}{ccccccc}
\hline 
Sample  & $A$ ($10^{-5}$)  & $\delta A$ ($10^{-5}$)  & RA (hr)  & dec ($^{\circ}$)  & $m$ ($10^{-5}$)  & sig\tabularnewline
\hline 
$z<1.6$  & 0.56  & $[0.38,0.85]$  & $(18.1\pm1.8)$  & $(-57\pm22)$  & $(-0.390\pm0.108)$  & $1.4\sigma$ \tabularnewline
$z>1.6$  & 1.38  & $[1.12,1.74]$  & $(16.5\pm1.4)$  & $(-63\pm11)$  & $(0.097\pm0.138)$  & $3.5\sigma$ \tabularnewline
\hline 
\end{tabular}
\end{table}

\subsection{Joint probability\label{sub:alpha:Joint-probability}}

The probability of obtaining alignment between the dipole vectors
from dipole models fitted to the Keck and VLT samples separately as
good or better than is seen by chance is about 6 percent. The chance
probability of obtaining alignment between the dipole vectors from
dipole models fitted to the low- and high-redshift samples is about
2 percent. Through a bootstrap method we have calculated the joint
probability of obtaining alignment that is at least as good as seen
for both of these conditions by chance, and it is $\approx0.1$ percent.

It is possible to conjecture that the Keck results are somehow erroneous,
with $\Delta\alpha/\alpha$ values shifted to be more negative on
average through some unknown systematic. The VLT results then show
no overall statistically significant monopole variation, and only
a marginal ($\approx2.2\sigma$) angular variation. In this case,
it would then appear that there is no statistically significant variation
of $\alpha$. However, in this case one is still left with the $\approx0.1$
percent chance probability above, which is equivalent to $\approx3.3\sigma$.
This would be a large and intriguing coincidence, but we agree that
$3.3\sigma$ is not overwhelmingly large. Ultimately, we cannot exclude
the possibility that the results presented here which seem to indicate
spatial variation of $\alpha$ are due to chance (with or without
the influence of an unknown systematic) but the joint chance probability
of 0.1 percent described here seems to suggest that this is unlikely.
We discuss potential systematic errors in chapter \ref{cha:da systematic errors}.

\subsection{Significance of the monopole\label{sub:alpha:s_monopole}}

In section \ref{sub:alpha:lowz_vs_highz} we noted that the low-$z$
sample shows evidence for a statistically significant monopole at
the $3.6\sigma$ level\index{monopole, low-$z$}. In figure \ref{fig:alpha:zstack},
the existence of the monopole in both samples can be seen at low $z$.
Note in particular the top panel, where the trend of $\Delta\alpha/\alpha$
is toward negative $\Delta\alpha/\alpha$ for $z<1.6$. 

An obvious question is whether the monopole arises from one of the
Keck or VLT samples. For a model $\Delta\alpha/\alpha=A\cos(\Theta)+m$,
the Keck sample yields a $z<1.6$ monopole of $m=(-0.404\pm0.171)\times10^{-5}$,
which differs from zero at the $2.4\sigma$ level. However, the same
model fitted to the VLT $z<1.6$ $\Delta\alpha/\alpha$ values yields
$m=(-0.373\pm0.295)\times10^{-5}$. This differs from zero at the
$1.3\sigma$ level. There are three important considerations from
these values: \emph{i)} Both data sets yield very consistent monopole
values for $\Delta\alpha/\alpha$ at low redshift; the monopole values
differ at the $0.09\sigma$ level. Therefore, whatever is generating
the monopole appears to affect both the Keck and VLT samples. \emph{ii)}
Because there is no significant difference between the monopole values
in the Keck and VLT samples, the monopole cannot be responsible for
mimicing angular variation in $\alpha$. \emph{iii)} Additionally,
most of the dipole signal originates at $z>1.6$ (where the significance
of the dipole+monopole model over the monopole-only model is $3.5\sigma$).
As such, the presence of a low-$z$ monopole does not affect the redshifts
where most of the dipole significance originates. 

There are several possible explanations for this, and we discuss each
of them in turn:
\begin{enumerate}
\item \emph{Errors in the laboratory wavelengths}. Errors in the laboratory
wavelengths of transitions which feature predominantly at low redshifts
could cause a statistically significant monopole at low redshifts.
However, this seems particularly unlikely. The Mg~\iscs/\iis wavelengths
have been accurately measured on an absolute scale generated using
a frequency-comb calibration system. The Fe~\iis wavelengths used
at $z<1.6$ have also been precisely measured (the $\lambda1608,1611$
transitions are more difficult to measure accurately, but these transitions
are used infrequently at low redshifts due to their short rest wavelengths).
For instance, the absolute velocity uncertainty in the Fe~\iis $\lambda2382$
transition is $\approx14\,\mathrm{m\, s^{-1}}$, which is significantly
smaller than the $\approx82\,\mathrm{m\, s^{-1}}$ which would be
needed to generate a monopole value of $-0.39\times10^{-5}$. This
implies a systematic error some six times larger than the existing
error budget, which seems unlikely. Additionally, the relative wavelength
scales of the different experiments which measured the transitions
used at lower redshifts are likely to be significantly better than
this.
\item \emph{Time evolution of $\alpha$}. The functional form for variation
of $\alpha$ (if $\alpha$ varies) is unknown. Recent observations
confirm the apparent acceleration of the universe at late times \citep{Astier:06},
for which dark energy is posited as an explanation. For $z\lesssim0.5$,
dark energy dominates over matter and radiation \citep{Riess:04}.
If $\alpha$ couples to dark energy, then late-time evolution of $\alpha$
might be possible. Monotonic evolution of $\alpha$ cannot by itself
be an explanation for a low-$z$ monopole, because this would imply
that $\Delta\alpha/\alpha$ should approach zero for $z\rightarrow0$,
with the greatest divergence of $\Delta\alpha/\alpha$ from zero at
high redshift. If $\alpha$ oscillates with time then a pattern such
as is seen could arise. However, this would require the period of
oscillations to be $\sim$ twice the age of the universe, with the
present day at a node of the oscillation, in order to obtain $\langle\Delta\alpha/\alpha_{z>1.6}\rangle\sim0$,
$\langle\Delta\alpha/\alpha_{0.2<z<1.6}\rangle\sim-0.4\times10^{-5}$
and $\Delta\alpha/\alpha_{z=0}=0$. This may be possible, but this
case seems rather contrived.
\item \emph{Dependence of $\alpha$ on the local environment}. If the value
of $\alpha$ depends on the local environment (e.g.\ matter density,
gravitational potential, or gradient of the gravitational potential)
then this could produce an offset between the value of $\alpha$ measured
in the quasar absorbers and the value measured in the laboratory,
even as $z\rightarrow0$. If this was the case, we would expect a
similar magnitude monopole to also be present at high redshift, which
is not seen. 
\item \emph{Telescope systematics}. Wavelength-dependent telescope systematics
seem difficult to support given the inter-telescope consistency.
\item \emph{Significantly different abundances of isotopes in the absorbers}.
The isotopic splitting scales as $\Delta\omega_{i}\propto\omega_{0}/m_{i}^{2}$,
where $m_{i}$ is the mass of the species under consideration. Mg
is the lightest atom used in the MM method, and therefore the isotopic
splitting for the Mg transitions is relatively large. If the abundance
of the three Mg isotopes differs significantly in the quasar absorbers
to terrestrial values, this would mimic a change in $\alpha$. The
low-$z$ sample is dominated by the Mg~\ii/Fe~\iis combination,
which is particularly sensitive to the effect of differences in the
abundance of the Mg isotopes \citep{Murphy:01c}. 
\end{enumerate}
It is possible that a combination of the time evolution of $\alpha$
\emph{and} dependence of $\alpha$ on the local environment could
explain the low-$z$ monopole, but this requires two different mechanisms.
Additionally, in this circumstance the magnitude of the environmental
dependence must be very similar to the magnitude of the time evolution
from $z\sim4$ to $z=0$ in order to obtain the observed distribution
of $\Delta\alpha/\alpha$ with $z$, requiring significant fine-tuning. 

On balance, evolution in the abundance of the Mg isotopes seems like
the most likely of these explanations. We explore the effect of differences
in the relative abundances for Mg isotopes between terrestrial values
and those in the quasar absorbers in section~\ref{sec:asys:isotopic_abundances}. 

The lack of a clear explanation for the low-$z$ monopole is a weakness
of the results presented here. Specifically targeted future observations
at sufficiently high resolving powers and signal-to-noise ratios may
be able to resolve the isotopic shifts for the magnesium lines (or
otherwise), thus directly determining whether the above explanation
is correct. It would be particularly interesting to map out the angle-independent
variation in $\alpha$ as a function of redshift; this would require
many $\Delta\alpha/\alpha$ measurements at all angles, binned into
redshift slices. Similarly, it would be useful to demonstrate whether
or not a $\Delta\alpha/\alpha$ monopole was present at low redshifts
by using transitions other than magnesium --- discovery of a low-$z$
monopole in this case would suggest evolution in $\alpha$ (or perhaps
some other systematic), whilst failure to detect the monopole would
imply that evolution in the abundance of the magnesium isotopes was
responsible.

\subsection{Iterative clipping of potentially outlying $\Delta\alpha/\alpha$
values\label{sub:sigma_clipping}}

\index{data clipping}We have attempted to be conservative in presenting
our results when accounting for extra scatter in the $\Delta\alpha/\alpha$
values about a model by adding a term, $\sigma_{\mathrm{rand}}$,
in quadrature with the error bars. This effectively functions as an
interpolation between a $\chi^{2}$ fit where the error bars are believed
to be correct and an unweighted fit, where the error bars are unknown. 

However, another option is to assume that the statistical error bars
for most $\Delta\alpha/\alpha$ values are a good representation of
the total uncertainty for those absorbers, and then remove points
one-by-one (``clipping'') until $\chi_{\nu}^{2}=1$. In our sample
it is difficult to determine to what extent different random processes
affect different absorbers, and therefore to determine to what extent
clipping is justified. Adding some $\sigma_{\mathrm{rand}}$ in quadrature
with all $\Delta\alpha/\alpha$ values, as we have done, is a conservative
option. Nevertheless, we explore the effect of data clipping here
to investigate the robustness of our results to the removal of $\Delta\alpha/\alpha$
values.

Traditionally, data clipping involves iteratively removing the point
with the largest residual and then re-fitting. However, for the reasons
given in section \ref{sub:LTS method}, this has the potential to
incorrectly remove points. Therefore, we use a modified method. At
each iteration, we calculate the LTS fit using the model $\Delta\alpha/\alpha=A\cos\Theta+m$
to the $\Delta\alpha/\alpha$ values with their raw statistical errors,
and then remove the point with the largest residual. However, we choose
$k=n-1$ in this case. Effectively, at each stage, we want to identify
only one point to remove, and therefore it makes sense to calculate
a fit to $n-1$ points. We then calculate a weighted fit, using only
the statistical error bars, and calculate the significance of the
dipole model. For efficiency of calculation, we avoid bootstrapping,
and so we use the method of \citet{Cooke:09} to calculate the significance
of the dipole+monopole fit with respect to the monopole fit. We then
repeat the process. At any iteration, if $\chi_{\nu}^{2}>1$, we multiply
all entries of the covariance matrix by $\chi_{\nu}^{2}$ in order
to account for excess scatter about the model. If $\chi_{\nu}^{2}<1$,
we do not adjust the covariance matrix.

Initially, one expects the significance of the fit to improve, as
one discards a few points which are not consistent with the general
trend of the fit. Eventually, one will remove enough $\Delta\alpha/\alpha$
values that the significance must decline. If the significance declines
rapidly, this implies that the dipole effect is dominated by a few
points. Conversely, if the significance of the fit is sustained or
improved for the removal of small fractions of the data (e.g.\ $\sim10$
percent), this qualitatively implies robustness of the result.

\begin{figure}[tbph]
\noindent \begin{centering}
\includegraphics[bb=44bp 23bp 556bp 805bp,clip,angle=-90,width=1\textwidth]{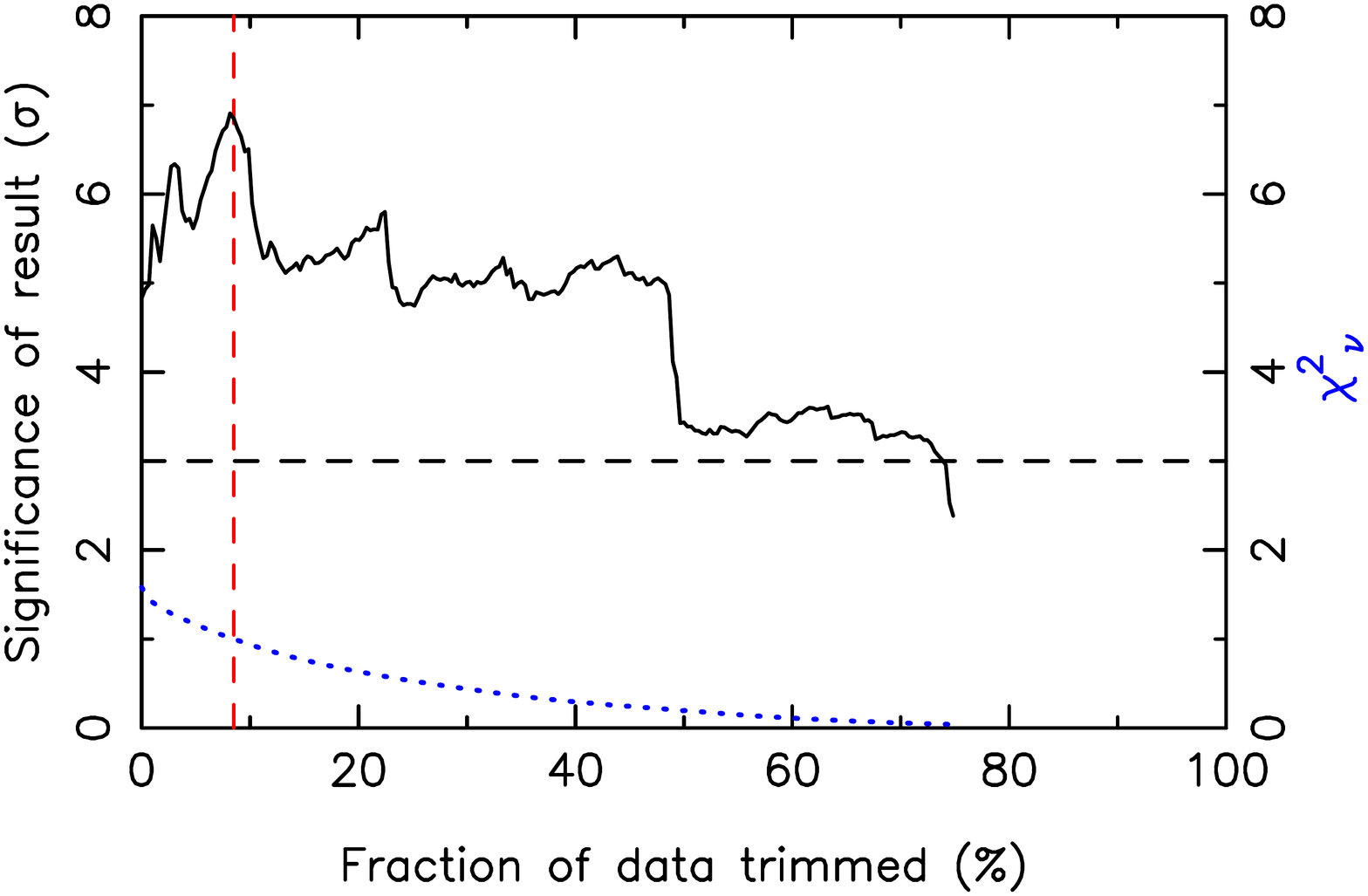}
\par\end{centering}

\caption[Effect on the significance of the dipole model for VLT + Keck of iteratively trimming data]{Effect of iteratively clipping the data on the statistical significance of the dipole model for the combined sample, as described in section \ref{sub:sigma_clipping}. The vertical axis shows the statistical significance of the dipole as determined by the method of \citet{Cooke:09} given in terms of $\sigma$ (solid line) and $\chi^2_\nu$ at that point (blue, dotted line). A dashed horizontal line is drawn at $3\sigma$ for reference. The vertical red (dashed) line indicates the point at which our clipping method reduces $\chi^2_\nu$ to below unity. We note that we have to remove more than 40 percent of data before the significance of the detection drops to about $3\sigma$. As there is no good reason to remove so much data, this implies that our result is robust. The actual significance given here is probably overstated compared to the ``true'' significance, given that no attempt has been made to account for systematic errors.  \label{fig:alpha:sigma_clipping}}
\end{figure}

We show in figure \ref{fig:alpha:sigma_clipping} the results of this
process. We find that we must remove large numbers of absorbers to
destroy the significance of the dipole. In particular, the significance
does not decrease rapidly with the number of $\Delta\alpha/\alpha$
values clipped initially, suggesting that the observed dipole effect
is not being caused by a few outlying points. If we clip until $\chi_{\nu}^{2}=1$,
the significance of the dipole is almost $7\sigma$. 

\begin{figure}[tbph]
\noindent \begin{centering}
\includegraphics[bb=50bp 23bp 556bp 750bp,clip,angle=-90,width=1\textwidth]{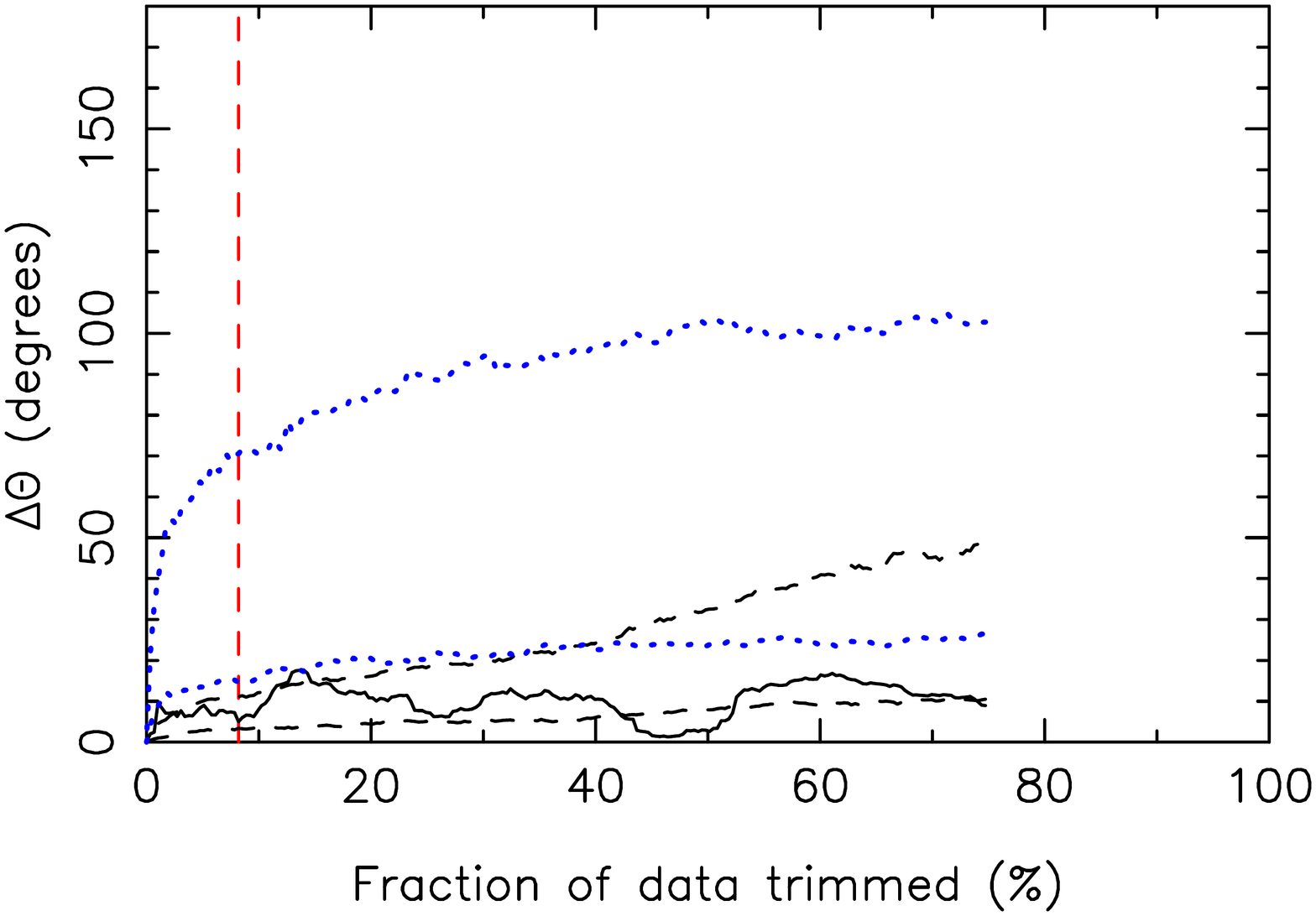}
\par\end{centering}

\caption[Effect on the location of the dipole vector for the dipole model for VLT + Keck of iteratively trimming data]{Effect of iteratively clipping the $\Delta\alpha/\alpha$ values on the location of the dipole, as described in section \ref{sub:sigma_clipping}. The vertical axis shows the deviation of the fitted angle from the untrimmed model ($\Delta \Theta$) as a function of the percentage of absorbers removed. This figure compares the results of trimming on our actual $\Delta\alpha/\alpha$ values, bootstrapped samples designed to emulate our data, and random samples. \emph{Actual data:} The solid black line shows the results of trimming for our $\Delta\alpha/\alpha$ values. If the fit is stable and not due to the presence of a small number of highly significant points, we expect to see that $\Delta\Theta$ should not grow rapidly with the amount of data removed. This is what is seen. \emph{Bootstrapped samples:} The dashed lines show the $1\sigma$ range for 300 bootstrap samples (generated as described in the text). This shows the typical range of variation at fraction of absorbers removed given distribution of sightlines, values of $\Delta\alpha/\alpha$, statistical errors and distribution of residuals in the sample. The region given reflects the $1\sigma$ range for the bootstrapped samples at each point; each individual sample may wander substantially more than is indicated by this range, and so the deviation of the path for our actual sample outside the region is not indicative of any problem with our $\Delta\alpha/\alpha$ values. \emph{Random samples:} The blue, dotted lines show the $1\sigma$ range for 300 samples where we have randomised $\Delta\alpha/\alpha$ over the sightlines. We see that $\Delta \Theta$ in this case grows rapidly with increased trimming for these samples. Our real sample does not do this, which suggests that our real sample is significantly dissimilar from a random sample.  \label{fig:alpha:sigmaclipping_dtheta_bootstrap}}
\end{figure}

In figure \ref{fig:alpha:sigmaclipping_dtheta_bootstrap} we show
the effect of clipping $\Delta\alpha/\alpha$ values on the location
of the dipole. One expects that if the dipole effect is real, then
the position of the dipole should not change dramatically with the
removal of small amounts of data (that is, $\Delta\Theta=\Theta_{i}-\Theta_{0}$
should be small). To assess how likely it is that this seemingly restricted
path is typical for our distribution of data, we apply a bootstrap
method to generate and iteratively trim 300 new samples, and examine
the distribution of $\Delta\Theta$ at each point. We cannot use a
traditional bootstrap, which resamples the data with replacement,
because how the data is trimmed depends crucially on the distribution
of residuals. Therefore, we resample the residuals of the fit to generate
new samples. To do this, we use the following process to generate
one sample: \emph{i)} calculate the model prediction for each absorber
given the model, $p_{i}=A\cos(\Theta)+m$; \emph{ii)} calculate the
residuals about the fit for each absorber, $r_{i}=(\Delta\alpha/\alpha_{i}-p_{i})/\sigma_{i}$;
\emph{iii)} randomly reassign the calculated $r_{i}$ to different
absorbers, generating $r'_{j}$; \emph{iv)} generate a new set of
$\Delta\alpha/\alpha$ values as $\Delta\alpha/\alpha'_{j}=p_{j}+r'_{j}\sigma_{j}$.
In this way, we generate new values of $\Delta\alpha/\alpha$ which
represent different possible realisations of our sample where the
actual distribution of residuals is preserved. This is demonstrated
in figure \ref{fig:alpha:sigmaclipping_dtheta_bootstrap}. We see
that the bootstrapped samples do not wander very far even when much
of the data is removed ($\Delta\Theta\lesssim20^{\circ}$).

To contrast this with the effect on a random sample, we also show
in figure \ref{fig:alpha:sigmaclipping_dtheta_bootstrap} the effect
of trimming random samples. To do this, we generate 300 new samples
by randomly reassigning values of $\Delta\alpha/\alpha$ to different
sightlines, and iteratively trimming under the model $\Delta\alpha/\alpha=A\cos(\Theta)+m$.
We see here that our actual sample is not typical of the random samples,
therefore suggesting that the actual sample is significantly dissimilar
to random samples. 

\subsection{Removal of spectra}

A further question one might ask is how sensitive our results are
to the inclusion of particular spectra. We would like to know whether
the dipole result could be dominated by a small number of spectra
which, if removed, would destroy the result. 

\begin{figure}[tbph]
\noindent \begin{centering}
\includegraphics[bb=61bp 23bp 556bp 795bp,clip,angle=-90,width=0.8\textwidth]{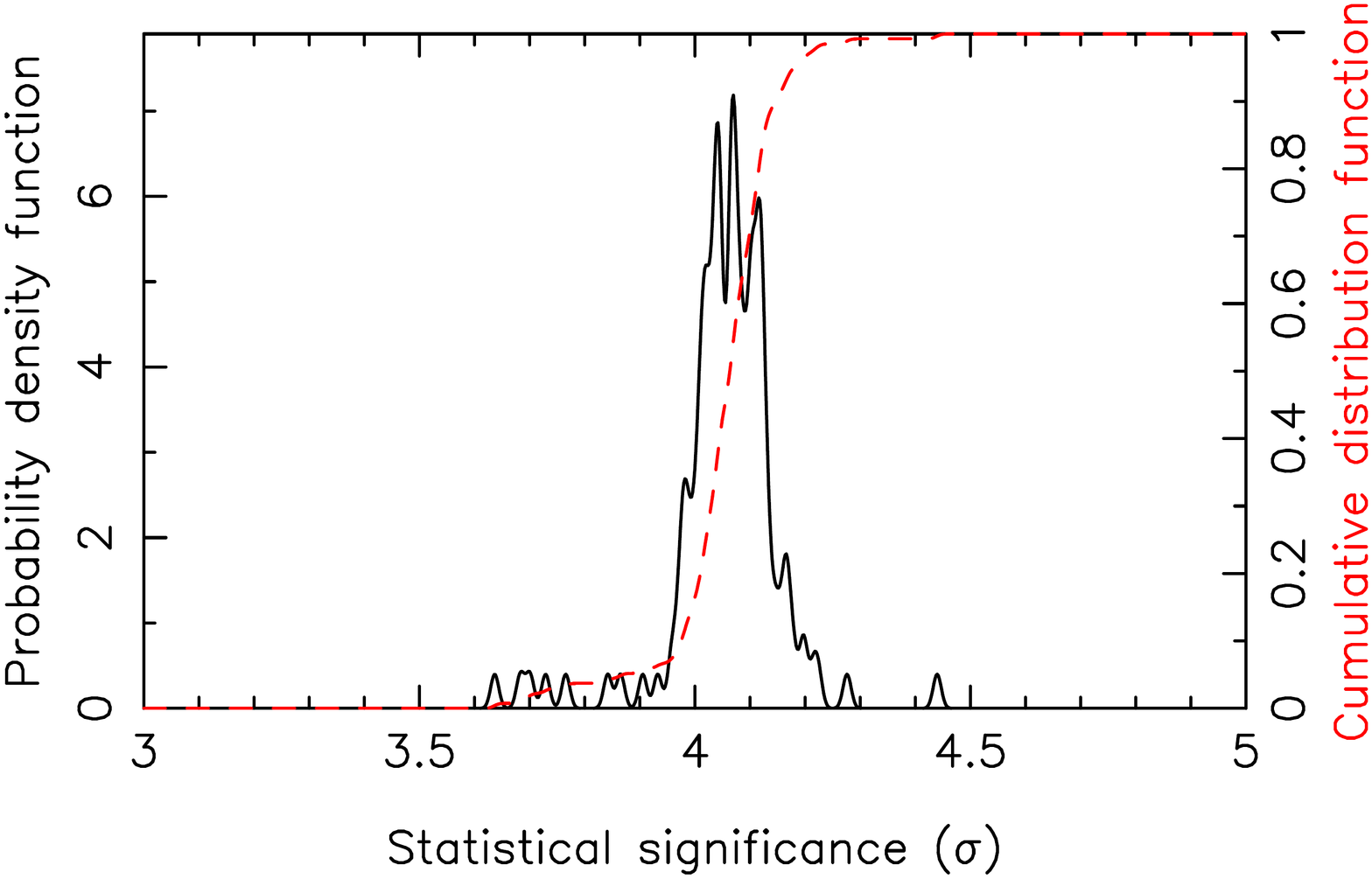}
\par\end{centering}

\caption[Effect on the significance of the dipole model of removal of single spectra]{Effect of removing quasar spectra on the statistical significance of the dipole, as assessed through a jack-knife method. Each spectrum is removed one at a time, and the value of the statistical significance of the model $\Delta\alpha/\alpha = A\cos(\Theta) + m$ is calculated with respect to the monopole model, using the bootstrap method. We use a Gaussian kernel density estimator to construct the approximate probability density function of the effect of quasar spectrum removal, where the width of the Gaussian basis functions has been chosen to be the inverse of the number of spectra. The cumulative distribution function is plotted as a dashed, red line. This demonstrates that the angular dipole effect is not due to a single spectrum.\label{fig:alpha:quasar_jackknife}}
\end{figure}

\begin{figure}[tbph]
\noindent \begin{centering}
\includegraphics[bb=61bp 23bp 556bp 795bp,clip,angle=-90,width=0.8\textwidth]{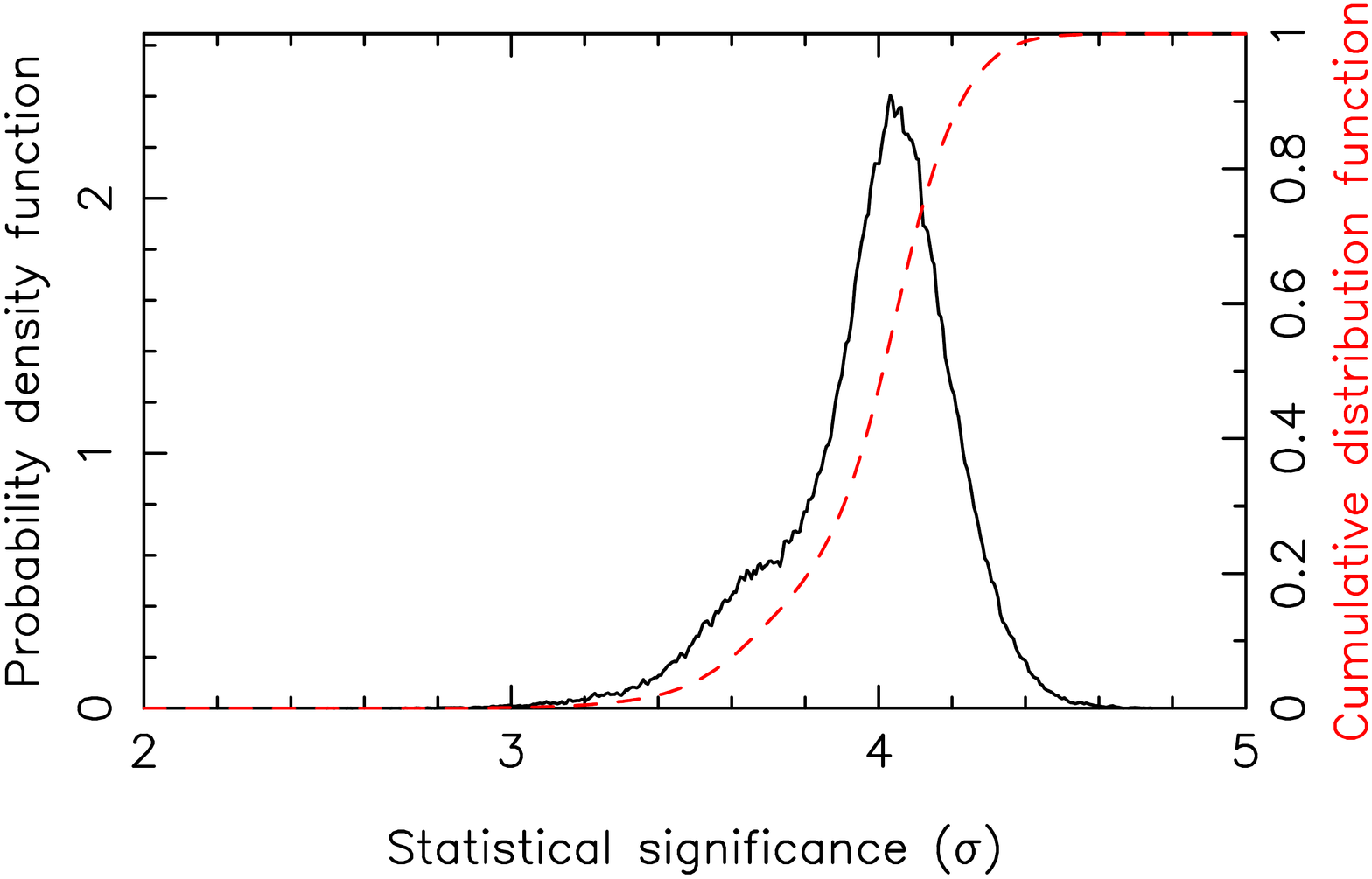}
\par\end{centering}

\caption[Effect on the significance of the dipole model of the random removal of five spectra]{Effect of removing quasar spectra on the statistical significance of the dipole, as assessed through a sampling method. For 100,000 samples, 5 quasar spectra are randomly removed from the combined Keck + VLT sample, and the statistical significance of the dipole model $\Delta\alpha/\alpha = A\cos(\Theta) + m$ is calculated through the method of \citet{Cooke:09}. Gaussian basis functions with width $\approx 1/316$ (=$10^{5/2}$) are used. This graph demonstrates that, in the absence of particular knowledge about problematic spectra, the chance of obtaining a dipole model where the statistical significance of the dipole is less than $3\sigma$ is small as a result of randomly removing 5 spectra. \label{fig:alpha:quasar_jackknife_n5}}
\end{figure}

We therefore explore this question through a jack-knife method, where
we remove one quasar at a time and recalculate the statistical significance
of the fit. We show the results of this exploration in figure \ref{fig:alpha:quasar_jackknife}.
The figure clearly demonstrates that, unsurprisingly, our result is
not due to a single quasar spectrum. We extend this in figure \ref{fig:alpha:quasar_jackknife_n5}
to show the effect of removing 5 spectra at random. We chose the number
5 in order to potentially include the cluster of 5 quasars at $\mathrm{RA}\sim22\mathrm{hr}$,
$\mathrm{dec}\sim-45^{\circ}$, where all of these sightlines demonstrate
$\Delta\alpha/\alpha>0$. Under this circumstance, the probability
of obtaining a dipole result which is insignificant ($<3\sigma$)
is small. This suggests that the dipole effect is not being created
by a small number of spectra. 

\subsection{Comment on the removal of outliers in the Keck and VLT samples}

In each of the VLT and Keck samples we have removed one putative outlier,
which in each sample represents less than one percent of the $\Delta\alpha/\alpha$
values. It is possible to calculate dipole significances and parameter
values with these points included, but it is not clear what intepretation
to place on these numbers on account of the arguments in section \ref{sub:LTS method}.
In particular, such a fit is immediately called into question on the
basis of the fact that it contains outliers. Nevertheless, we tried
such a fit and the dipole significance is not substantially altered.

\section{Translation from an angular variation model to a physical model including
a distance measure}

We now explore simple phenomenological parameterisations of the dipole
effect which attempt to account for distance dependence. In all of
these models, the same $\Delta\alpha/\alpha$ values identified as
outliers previously have been removed from considerations.

\subsection{$z^{\beta}$ dipole}

To model potential distance dependence directly with the observable
quantity, $z$, we fit a power-law relationship of the form\index{dipole model!z^{beta}
@$z^{\beta}$} 
\begin{equation}
\Delta\alpha/\alpha=Cz^{\beta}\cos(\Theta)+m\label{eq:dipole_z_beta}
\end{equation}
for some $\beta$ and amplitude $C$. For a fit to the combined Keck
+ VLT samples this gives the ``$z^{\beta}$ dipole'' sample.

\begin{figure}[tbph]
\noindent \begin{centering}
\includegraphics[bb=77bp 92bp 556bp 727bp,clip,angle=-90,width=1\textwidth]{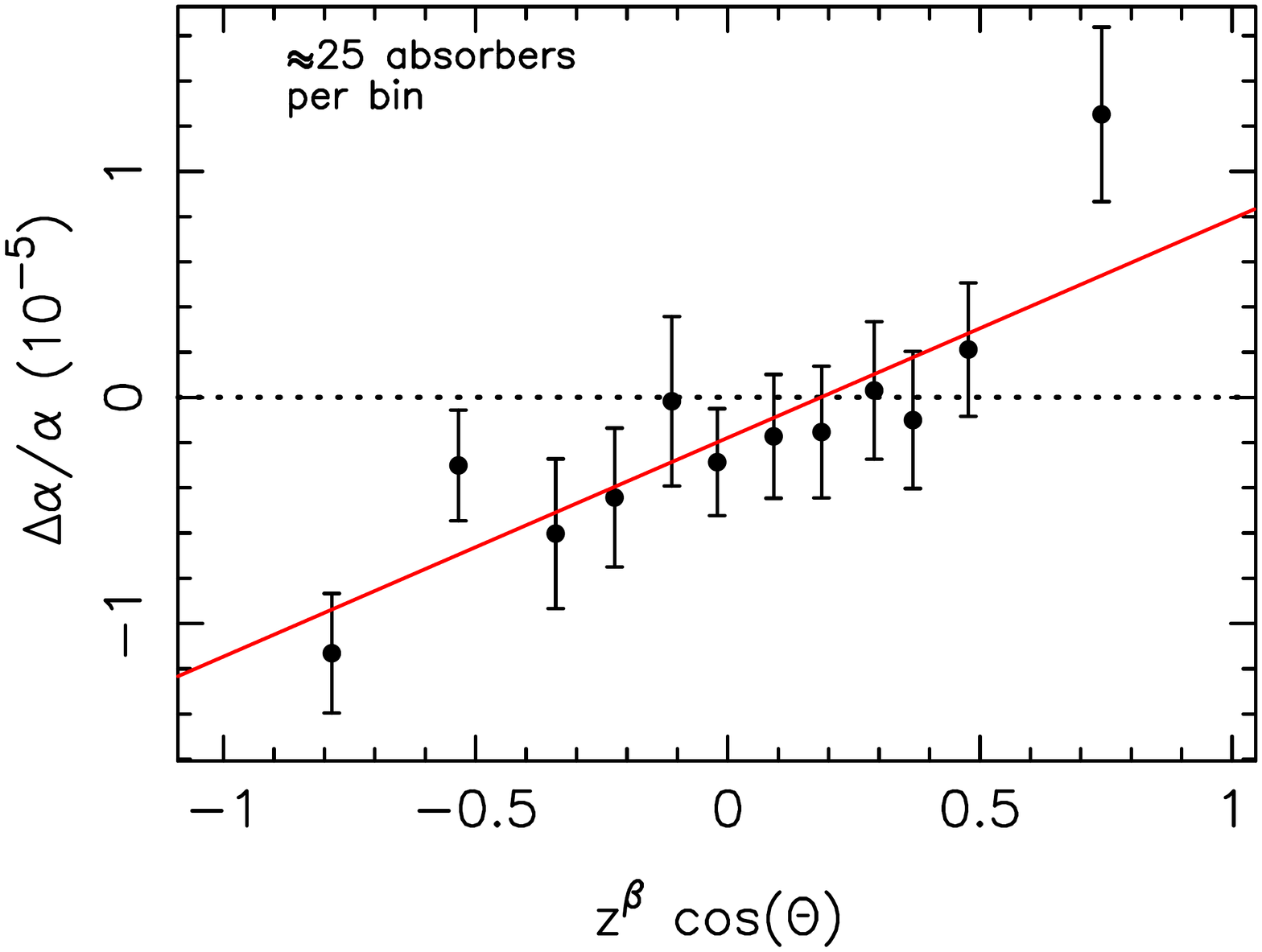}
\par\end{centering}

\caption[$\Delta\alpha/\alpha$ vs $z^\beta\cos(\Theta)$]{Binned values of $\Delta\alpha/\alpha$ plotted against $z^\beta \cos \Theta$, for $\beta = 0.41$. The $z^\beta$ dipole+monopole model is preferred over the monopole-only model at the $3.9\sigma$ level. Importantly, this plot only covers $\lvert z^\beta \cos(\Theta) \rvert \lesssim 1$. Given that it is possible to probe up to redshift $z \lesssim 4$ with the MM method, judicious choice of observational targets close to the dipole axis might be able to extend this horizontal range of this graph up to $\sim \pm 2$, thereby potentially increasing sensitivity to the effect substantially, if the effect is real. \label{fig:alpha:da_vs_zbetacostheta}}
\end{figure}

We use the Levenberg-Marquardt algorithm \citep{NumericalRecipes:92}
to fit equation \ref{eq:dipole_z_beta} to the combined sample. This
fit yields $\mathrm{RA}=(17.5\pm1.0)\,\mathrm{hr}$, $\mathrm{dec}=(-62\pm10)^{\circ}$,
$C=0.81$ ($1\sigma$ confidence limits $[0.55,1.09]\times10^{-5}$),
$m=(-0.184\pm0.085)\times10^{-5}$ and $\beta=0.46\pm0.49$. The fact
that the amplitude grows as a low power of $z$, and the fact that
it is statistically consistent with zero, is the reason that the approximation
$A\sim Cz^{0}$ yields reasonable results earlier. We show the results
of this fit in figure \ref{fig:alpha:da_vs_zbetacostheta}. This dipole
+ monopole model is statistically preferred over the monopole-only
model at the 99.99 percent confidence level ($3.9\sigma$). The reduction
in significance relative the angular dipole model occurs as a result
of the uncertainty in determining $\beta$, but is relatively small.

Note that the standard practice of fitting for $\Delta\alpha/\alpha$
as a function of redshift ($\Delta\alpha/\alpha=az+m$) is subsumed
within this analysis, which directly determines the scaling relationship
of $\Delta\alpha/\alpha$ with redshift (and whether it is statistically
compatible with linearity).

\subsection{$r$-dipole\label{sub:alpha:rdipole}}

Another plausible alternative is to try to relate the amplitude of
the dipole to some explicit distance metric. For simplicity, we use
the ``lookback-time distance''. This is defined by $r=ct$, where
$c$ is the speed of light and $t$ is the lookback time to the absorber.
Thus, we try a fit of the form\index{dipole model!$r$-dipole} 
\begin{equation}
\Delta\alpha/\alpha=Br\cos(\Theta)+m.\label{eq:alpha:ct_dipole}
\end{equation}
To calculate lookback times, we use the standard $\Lambda$CDM ($\Lambda$
Cold Dark Matter) model, with parameters given by the 5-year WMAP
(Wilkinson Microwave Anisotropy Probe) results \citep{WMAP5yr}. We
note that this calculation is derived from the FLRW (Friedmann-Lemaître-Robertson-Walker)
metric, which assumes isotropy of the universe. Our model implies
anisotropy of the universe, and therefore use of the FLRW metric is
strictly incorrect. Nevertheless, as $\Delta\alpha/\alpha\ll1$ we
assume that the FRLW metric is a good approximation to the actual
metric, and therefore that our lookback times are approximately correct.
The $\Lambda$CDM parameters used are: ($H_{0}$, $\Omega_{M}$, $\Omega_{\Lambda}$)
= (70.5, 0.2736, 0.726).

\begin{figure}[tbph]
\noindent \begin{centering}
\includegraphics[bb=78bp 79bp 488bp 727bp,clip,width=0.825\textwidth]{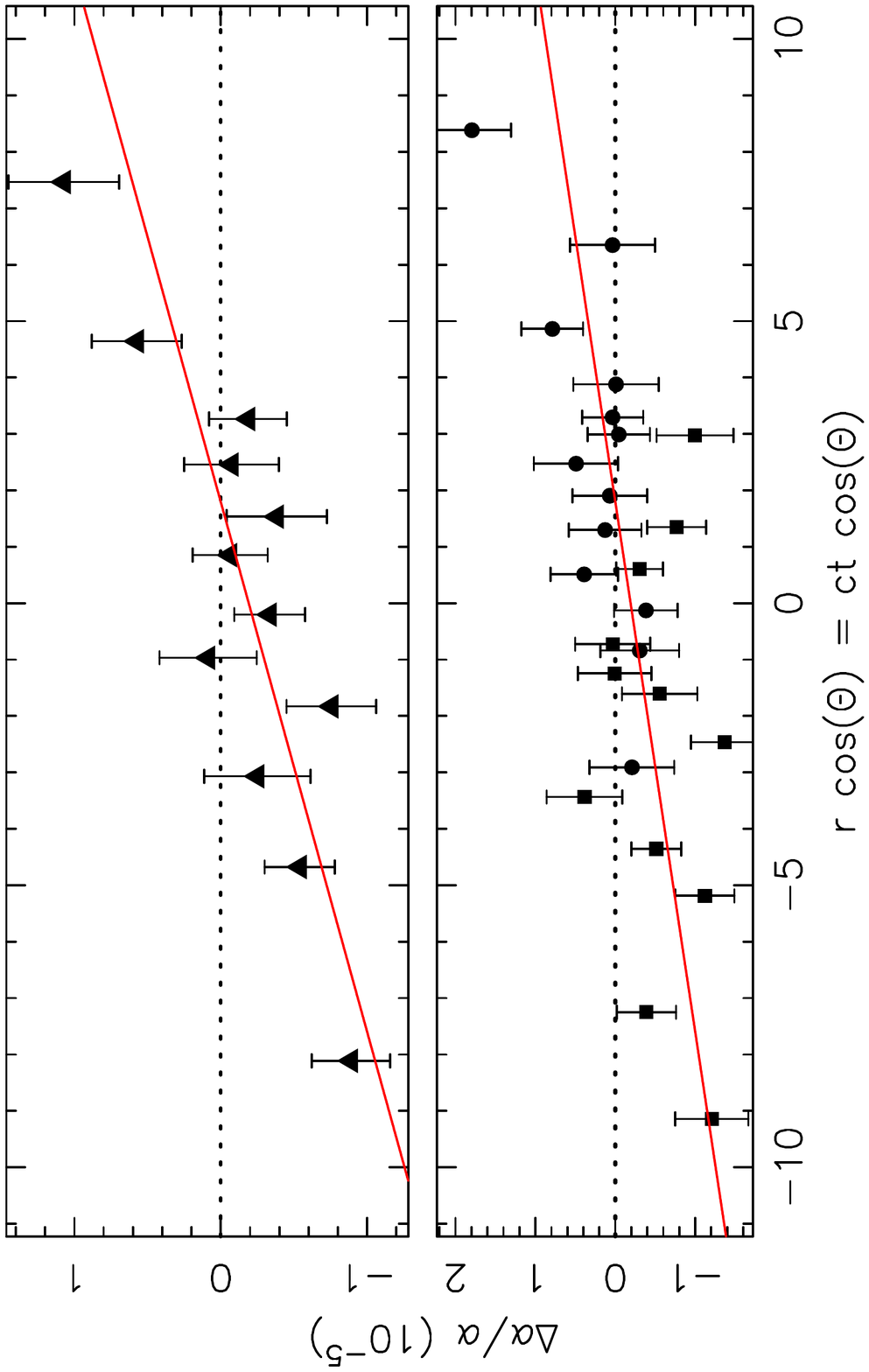}
\par\end{centering}

\caption[$\Delta\alpha/\alpha$ vs $ct\cos(\Theta)$]{Binned values of $\Delta\alpha/\alpha$ plotted against $r\cos(\Theta) \equiv ct\cos \Theta$, where $t$ is the look-back time to a redshift $z$. $\Lambda$CDM parameters are from \citet{WMAP5yr}. The top panel (triangles) shows the VLT + Keck sample, binned with approximately 25 absorbers per bin. The bottom panel shows VLT (circles) and Keck (squares) $\Delta\alpha/\alpha$ values, binned with approximately 12 absorbers per bin. The red (solid) line in both cases shows the model, $\Delta\alpha/\alpha = Br\cos(\Theta) + m$. The parameters for the fit are:  $B = 1.1\times 10^{-6}\,\mathrm{GLyr}^{-1}$ ($1\sigma$ confidence limits $[0.9, 1.3] \times 10^{-6} \,\mathrm{GLyr}^{-1}$), $\mathrm{RA} = (17.5 \pm 1.0)\,\mathrm{hr}$, $\mathrm{dec.} = (-62 \pm 10)^\circ$ and $m = (-0.187 \pm 0.084) \times 10^{-5}$.  It is interesting that this simple model is a reasonable representation of the data.\label{fig:alpha:da_vs_ctcostheta}}
\end{figure}

\begin{figure}[tbph]
\noindent \begin{centering}
\includegraphics[bb=77bp 78bp 455bp 727bp,clip,angle=-90,width=0.8\textwidth]{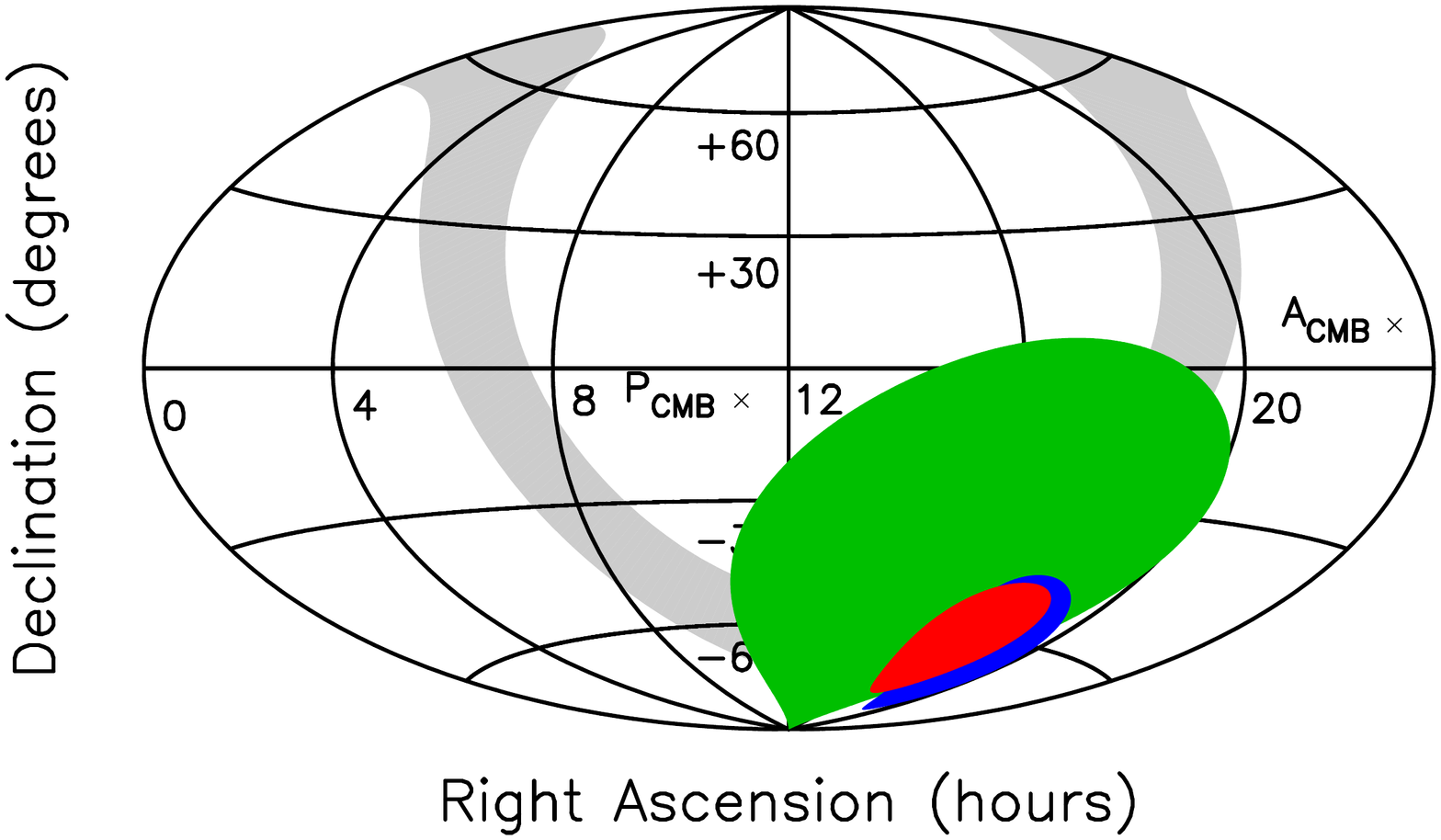}
\par\end{centering}

\noindent \begin{centering}
\includegraphics[bb=77bp 78bp 455bp 727bp,clip,angle=-90,width=0.8\textwidth]{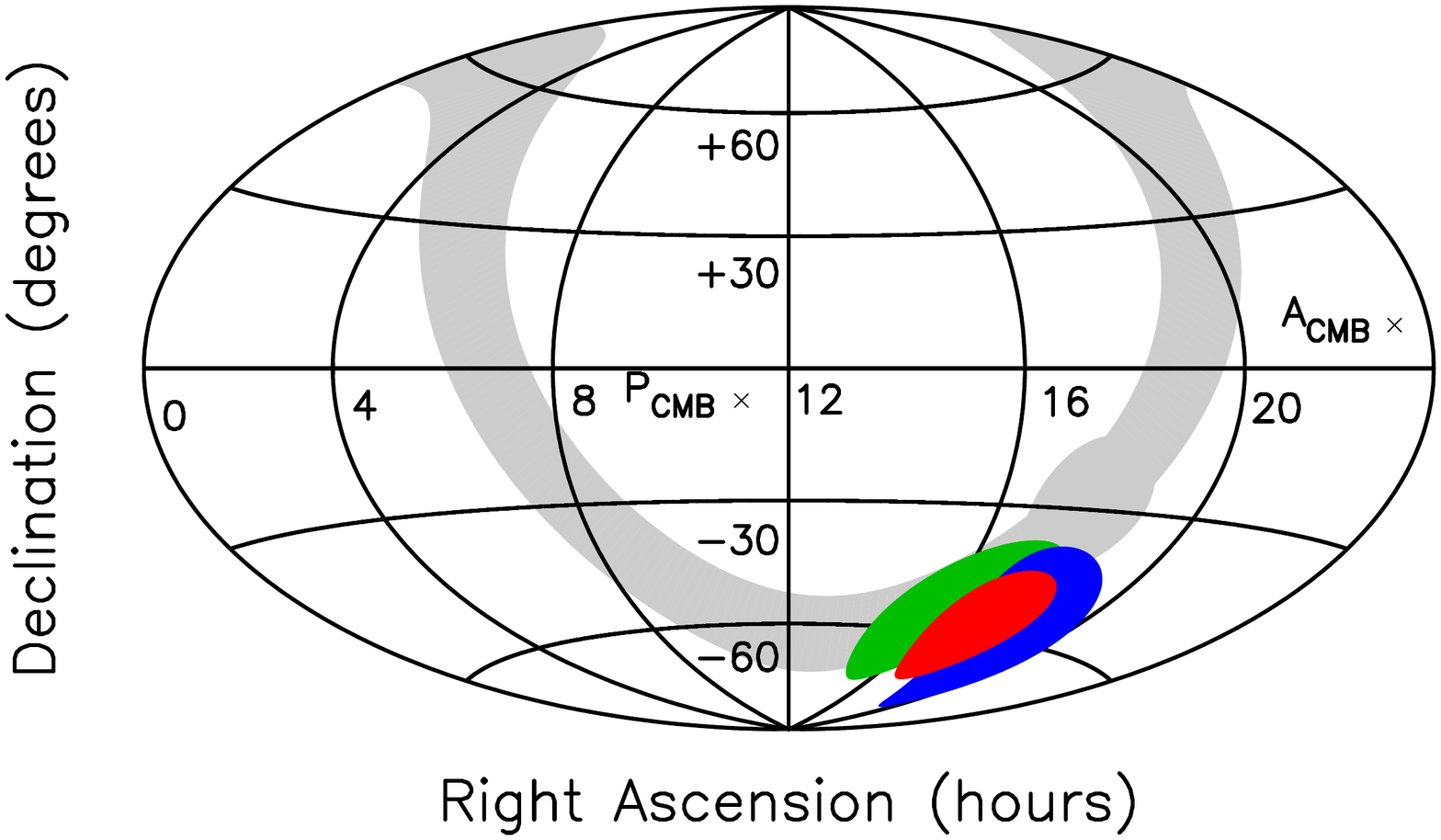}
\par\end{centering}

\caption[Sky map showing a $r$-dipole model for the VLT+Keck data, with and without a monopole]{Sky map in equatorial coordinates showing the 68.3 percent ($1\sigma$ equivalent) confidence limits of the location of the pole of the dipole for a fit to the Keck $\Delta\alpha/\alpha$ values (green region), VLT $\Delta\alpha/\alpha$ values (blue region) and combined $\Delta\alpha/\alpha$ values (red region), for a fit of $\Delta\alpha/\alpha = Br\cos(\Theta) + m$ (\emph{top figure}) and $\Delta\alpha/\alpha = Br\cos(\Theta)$ (\emph{bottom figure}), where $r=ct$ and $t$ is the lookback time to the absorber. The pole and antipole of the CMB dipole are marked as $P_\mathrm{CMB}$ and $A_\mathrm{CMB}$ respectively. In the model which includes a monopole (top figure), the Keck confidence region is large due to a relative degeneracy with the monopole; the region is much smaller in the bottom figure on account of no monopole term being included.\label{fig:alpha:fig_skymap_ct}}
\end{figure}

We show in figure \ref{fig:alpha:da_vs_ctcostheta} the fit of $\Delta\alpha/\alpha$
to a combined VLT + Keck sample (``combined $r$-dipole sample'')
against $r\cos(\Theta)=ct\cos(\Theta)$. The parameters for this fit
are: $B=1.1\times10^{-6}\,\mathrm{GLyr}^{-1}$ ($1\sigma$ confidence
limits $[0.9,1.3]\times10^{-6}\,\mathrm{GLyr}^{-1}$), $\mathrm{RA}=(17.5\pm1.0)\,\mathrm{hr}$,
$\mathrm{dec}=(-62\pm10)^{\circ}$ and $m=(-0.187\pm0.084)\times10^{-5}$.
Using the bootstrap method we assess the statistical significance
of this fit with respect to the monopole-only fit as $4.15\sigma$.
In figure \ref{fig:alpha:fig_skymap_ct}, we show the confidence regions
on the dipole location for the VLT, Keck and combined samples on the
sky. 

In galactic coordinates, the pole of this fit is at approximately
$(l,b)=(330^{\circ},-15^{\circ})$. The fact that the pole and antipole
are close to the Galactic Plane explains the relative lack of absorbers
near to the pole and antipole in both the Keck and VLT samples, a
fact made obvious in figure \ref{Flo:alpha:compressed_sightlines}
earlier. 

If we adopt a dipole-only model, 
\begin{equation}
\Delta\alpha/\alpha=Br\cos(\Theta),
\end{equation}
we derive $B=1.1\times10^{-6}\,\mathrm{GLyr}^{-1}$($1\sigma$ confidence
limits $[0.9,1.3]\times10^{-5}$), $\mathrm{RA}=(17.4\pm0.9)\,\mathrm{hr}$,
$\mathrm{dec}=(-58\pm9)^{\circ}$. The statistical significance of
the dipole model is 99.998 percent ($4.22\sigma$). The confidence
limits on the dipole location for this fit for the VLT, Keck and combined
samples are shown in figure \ref{fig:alpha:fig_skymap_ct}.

\section{Summary}

In this chapter we have presented 154 new many-multiplet constraints
on $\Delta\alpha/\alpha$ derived from spectra obtained using VLT/UVES.
A simple weighted mean analysis shows that these values of $\Delta\alpha/\alpha$
appear inconsistent with the Keck results of \citet{Murphy:04:LNP}.
However, if we consider that angular (and therefore spatial) variations
in $\alpha$ are possible, then the two data sets are rendered consistent
with each other. The combination of the two data sets yields statistically
significant evidence for angular variations in $\alpha$ at the $4.1\sigma$
level, with the best-fitting dipole having an angular amplitude of
$0.97_{-0.20}^{+0.22}\times10^{-5}$, and pointing in the direction
$\mathrm{RA}=(17.3\pm1.0)\,\mathrm{hr}$, $\mathrm{dec=(-61\pm10)^{\circ}}$.
If we consider a simple model for distance using the lookback-time
distance, we find that the statistical significance of the dipole
increases to $4.2\sigma$. In this case the dipole has an amplitude
of $(1.1\pm0.2)\times10^{-6}\,\mathrm{Glyr^{-1}}$, and points in
a similar direction.

The data display a remarkable consistency. Dipole fits to low ($z<1.6$)
and high ($z>1.6$) cuts of the sample point in a similar direction,
as do dipole fits to the Keck and VLT data separately. Similarly,
the significance of the dipole is robust under removal of data at
random. If we take a less conservative approach to treating the data,
the significance of the dipole approaches $7\sigma$. 

A weighted mean of the VLT and Keck $\Delta\alpha/\alpha$ values
(a whole-sample monopole) yields $(\Delta\alpha/\alpha)_{w}=(-0.216\pm0.086)\times10^{-5}$.
However, this value should be interpreted with caution, given the
fact that there appears to be significant angular dependence for $\alpha$
and the fact that the monopole takes on significantly different values
at low ($z<1.6$) and high ($z>1.6$) redshift. 

The cause of the difference between the monopole at low and high redshifts
is unknown, and is a weakness of the results presented here. We argued
that the most likely explanation for this is due to evolution in the
abundance of magnesium isotopes, and discussed other possible explanations.
Due to the fact that most of the significance for the dipole originates
at high redshifts, where the monopole is not present, and because
of the consistency between low- and high-redshift samples, and between
the Keck and VLT results, we do not think that this significantly
affects the evidence for spatial variation of $\alpha$. 

The results of this chapter therefore yield significant statistical
evidence for spatial variation in the fine-structure constant.

It is possible that the results presented here are the result of some
unknown systematic effect, or combination of systematic effects. We
discuss potential systematic effects in the next chapter.

\chapter{Systematic errors for $\Delta\alpha/\alpha$\label{cha:da systematic errors}}

\section{Introduction}

It is easy to conceive of a large number of possible systematic effects\index{systematic errors}
which could, if present, spuriously generate a specific form for a
non-zero $\Delta\alpha/\alpha$. In particular, if one assumes that
$\Delta\alpha/\alpha$ is well described by a weighted mean, or one
considers the monopole term of our dipole fit, there are a large number
of effects which could push either of these values away from zero.

To generate angular variation in $\alpha$ from a systematic effect
is, however, rather harder than producing an offset from $\Delta\alpha/\alpha=0$.
Any such effect --- if it exists --- must be well correlated with
sky position \emph{or} must be a combination of systematics that by
coincidence mimics angular variation in $\alpha$. On the whole, we
argue that a detection of a angular variation in $\alpha$ is relatively
robust to potential systematic effects. Nevertheless, in this chapter
we explore the potential impact of a number of systematic effects.

\citet{Murphy:01c,Murphy:03} considered a wide range of potential
systematic effects in relation to the Keck results, including: ``potential
kinematic effects, line blending, wavelength miscalibration, spectrograph
temperature variations, atmospheric dispersion and isotopic/hyperfine-structure
effects''. They concluded that only the latter two effects are potentially
large enough to be of significance, and that neither of these can
explain the Keck results. 

Some of these potential systematic errors are common to the VLT sample
because we observe the same types of absorbers as are in the Keck
sample, and the impact of many of them in the VLT should be similar
to the Keck sample because the statistical constraints on $\Delta\alpha/\alpha$
from individual absorbers in the VLT sample is of the same order of
magnitude as that from absorbers in the Keck sample. Certainly, the
same considerations regarding potential kinematic effects and line
blending apply, and so these effects should also not significantly
affect our results. UVES has operated with an image rotator since
observations commenced, and so the concern about atmospheric dispersion
that is present for some of the Keck sample does not affect the VLT
sample. Spectrograph temperature variations should also be small,
as the UVES enclosure is thermally isolated, and the VLT enclosure
is air-conditioned to minimise thermal variation \citep{dOdorico:00a}.
Given that spectrograph temperature variations are unable to explain
the Keck result, the design of UVES in this respect should ensure
that such effects are negligible in our sample.

It is conceivable that telescope flexure could induce some systematic
effect into the $\Delta\alpha/\alpha$ results. In the most obvious
case this would make $\Delta\alpha/\alpha$ correlated with the zenith
angle of the observations. \citet{Murphy:03} explicitly considered
the possibility that $\Delta\alpha/\alpha$ could be correlated with
zenith angle, and found no evidence for significant correlation, which
seems to rule out this problem in the Keck sample. We have not explicitly
addressed this concern here given the findings of \citeauthor{Murphy:03},
and note that any systematic which mimics angular variation in $\alpha$
must not only be well correlated with sky position, but must do so
in a way which is consistent between the two telescopes. A systematic
which is correlated with zenith angle is not sufficient to produce
the observed dipole effect; such an effect should produce a variation
in $\alpha$ that is approximately symmetric about the latitudes of
the telescopes projected onto the sky (i.e.~dec.$\sim20^{\circ}$
for Keck and dec. $\sim-25^{\circ}$ for VLT), which is not what is
seen. Importantly, such an effect is unable to produce the consistency
observed between the dipole locations.

Now that we are utilising data from two telescopes, the obvious question
arises as to whether some difference between the telescopes could
manufacture or alter a dipole signal. The fact that a $2.2\sigma$
dipole is seen in the VLT data alone (section \ref{sub:alpha:VLT_dipole})
and that there is good alignment between dipoles fitted to the Keck
and VLT samples (section \ref{sub:alpha:alignment_by_chance}) suggests
that inter-telescope differences are not responsible for the observed
effect. One way of trying to determine the impact of differences between
the telescopes would be to attempt to calculate any such differences
from first principles. However, any potential systematics are likely
to be extremely subtle, and depend on a variety of factors relating
to the telescopes and instruments. A more direct approach is to compare
spectra of the same objects taken by both telescopes. Absorption features
in these spectra should appear at the same wavelengths in spectra
from both telescopes%
\footnote{This is not strictly true: if the dynamical timescale of the absorption
process is comparable to the time difference between exposures, then
evolution in the absorption features is possible. We include changes
in the position of the gas clouds in the definition of the dynamical
timescale, as proper motion of the clouds could produce changes in
the observed column density. For transitions with multiple velocity
components, this will produce apparent shifts in line centroids. %
}. Any difference constitutes a relative distortion of the wavelength
scale between the two telescopes. This technique is powerful, and
does not require \emph{a priori} knowledge of how the wavelength scale
distortions are generated. Because of the importance of this technique,
we present it first as the $\Delta v$ test in section \ref{sec:asys:dv_test}.

The issue of wavelength calibration is potentially tricky. Since the
work of \citet{Murphy:03}, wavelength scale distortions have been
identified within echelle orders in both Keck/HIRES \citep{Griest:09}
and VLT/UVES \citep{Whitmore:10}. We discussed the potential origin
of these in section \ref{sub:mu:Intraorder_distortions} in the context
of $\Delta\mu/\mu$. As in that section, the fact that we use transitions
across the whole optical range combined with the non-monotonic nature
of the distortions means that any bias introduced into $\Delta\alpha/\alpha$
by these distortions should average out over a large enough sample
of absorbers. \citet{Murphy:09} explored the impact of distortions
of this type on the Keck results and found that the impact on the
weighted mean was effectively negligible. Nevertheless, it is worth
exploring this effect further, and we do this in section \ref{sec:asys:intraorder_distortions}. 

We explore the potential impact of the fact that UVES is a dual-armed
spectrograph in section \ref{sec:asys:UVES_dual_arm}.

As was done by \citet{Murphy:03}, we explore the effect of a different
heavy Mg isotope fraction in the quasar absorbers relative to terrestrial
values in section \ref{sec:asys:isotopic_abundances}.

\section{Inter-telescope systematics and the $\Delta v$ test\label{sec:asys:dv_test}}

\index{systematic errors!Delta v
 test@$\Delta v$ test}\index{Delta v
 test@$\Delta v$ test}Suppose that some systematic effect existed which was intrinsic to
the telescope which created a distortion of the wavelength scale.
Two possible types of wavelength distortions exist: stationary and
non-stationary. Stationary (i.e.\ time-invariant) distortions could
be produced due to some intrinsic aspect of the telescope or instrument.
Non-stationary distortions could be produced by a wide number of phenomena,
including atmospheric effects and the method through which the telescope
tracks the quasar source (i.e.\ the accuracy of slit centering).
All of the Keck spectra used in the analysis in this paper were acquired
whilst HIRES had only one CCD chip. In this configuration, multiple
exposures are needed to yield full wavelength coverage. If the quasar
image is not precisely centred in the spectrograph slit for every
exposure, velocity offsets between spectral segments obtained at different
times are possible. This issue should be substantially mitigated at
VLT, as UVES can acquire almost the entire spectral range in a single
observation. The effect could be exacerbated in conditions of good
seeing and could include an additional small effect due to the seeing
profile decreasing slightly towards the red end of the spectrum.

It so happens that the VLT and Keck samples have 7 quasars in common.
We give a list of the quasars common to the VLT and Keck samples in
table \ref{tab:asys:qlist}. The use of common sources allows one
to search for problems with wavelength calibration; absorption features
should be found at the same barycentric vacuum wavelength between
different exposures. This inspires a method of searching for distortions
of the wavelength scale in both the Keck and VLT spectra. In the simplest
sense, one aims to cross-correlate particular patches of spectra and
try to verify whether absorption features really do occur at the same
wavelength, or whether some correction is required to achieve a good
match. Note that the number of absorption lines which can be used
for this purpose is much larger than is used for analysing $\Delta\alpha/\alpha$.
Whilst for $\Delta\alpha/\alpha$ many absorption lines are needed
to yield a single measurement of $\Delta\alpha/\alpha$, in principle
each absorption line in the spectrum yields one constraint on potential
wavelength distortion.

\begin{table}[tbph]
\caption[Quasars common to the Keck and VLT samples]{List of quasars common to the Keck and VLT samples. Keck names are given in B1950 format, whilst VLT names are given in J2000 format.\label{tab:asys:qlist}}

\centering{}%
\begin{tabular}{cc}
\hline 
Keck sample name & VLT sample name\tabularnewline
\hline 
0216+0803 & J021857+081727\tabularnewline
0237$-$233 & J024008$-$230915\tabularnewline
0940$-$1050 & J094253$-$110426\tabularnewline
1202$-$0725 & J120523$-$074232\tabularnewline
0528$-$250 & J053007$-$250329\tabularnewline
1337+1121 & J134002+110630\tabularnewline
2206$-$1958 & J220852$-$194359\tabularnewline
\hline 
\end{tabular}
\end{table}

One possibility is to use direct cross-correlation methods, however
this suffers from the fact that the spectral resolutions of VLT and
Keck spectra are different, and so direct cross-correlation requires
rebinning of the spectra onto a common wavelength scale. A more inspired
approach is to actually model the quasar absorbers directly. By imposing
an assumption about the nature of the observed profiles (namely that
they are Voigt profiles), one can obtain substantially tighter constraints
on any wavelength distortion.

To explore potential wavelength scale distortions, we use a method
which we refer to as the $\Delta v$ test. The method proceeds as
follows: \emph{i)} for each common quasar, visually identify regions
of non-terrestrial absorption, typically having width of a few $\mathrm{\AA}$;
\emph{ii)} for each of these regions, perform a Voigt profile fit
to the VLT spectral data (identification of the transition responsible
is unimportant); \emph{iii)} fit corresponding spectral regions of
the Keck and VLT simultaneously, but with an extra free parameter,
$\Delta v$, which allows for a velocity shift between the two spectral
regions. R.~F.~Carswell has kindly modified \textsc{vpfit} to be
able to estimate $\Delta v$. The VLT spectral data for these regions
were kindly fitted by M. Bainbridge using an automated Voigt profile
fitting routine designed to fit regions of the forest automatically,
and he has provided us with $\Delta v$ values derived from the joint
fits to the Keck and VLT data. $\Delta v$ is defined hereafter as
the velocity difference $\Delta v=v(\mathrm{VLT})-v(\mathrm{Keck})$
which must be applied to minimise $\chi^{2}$ between two comparable
spectral regions. In particular, this means for a particular transition
that
\begin{equation}
z_{\mathrm{VLT}}=(1+z_{\mathrm{Keck}})\left(1+\frac{\Delta v}{c}\right)-1.
\end{equation}
Each value of $\Delta v$ provides an estimate of the velocity offset
between the two telescopes at that observed wavelength, giving $\Delta v(\lambda)$.
One can therefore examine the functional form of $\Delta v(\lambda)_{i}$,
where $i$ refers to the $i$th quasar pair under consideration. For
each set of $\Delta v$ values from a spectral pair, we use the LTS
method to calculate the weighted mean of that set of $\Delta v$ values,
which we then subtract from the $\Delta v$ values for that spectral
pair. This is to remove any constant offset resulting from mis-centering
of the quasar within the slit. We use $k=0.95n$ for the LTS fit (see
section \ref{sub:LTS method}). 

Any relative wavelength scale distortion can in principle be removed
by applying an inverse function based on the observed $\Delta v$
data. To see this, consider the form of the distortion. For an absorption
line with rest wavelength $\lambda_{0}$, observed wavelength $\lambda_{i}$,
and velocity distortion $\Delta v$ then
\begin{equation}
\lambda_{i}=\lambda_{0}(1+z)\left(1+\frac{\Delta v}{c}\right)
\end{equation}
where we have assumed that $\Delta v$ is constant over the absorption
profile under consideration. The effect of $\Delta\alpha/\alpha$
can be ignored --- whatever transition is being examined is the same
in both spectra, and so any effect due to a change in constants will
be absorbed into the determination of $z$. There are two options
to attempt to remove the wavelength scale distortion given some function
$\Delta v(\lambda_{\mathrm{obs}})$. One could modify the spectral
data, changing the observed wavelengths as 
\begin{equation}
\lambda_{\mathrm{obs}}\rightarrow\frac{\lambda_{\mathrm{obs}}}{1+\Delta v(\lambda_{\mathrm{obs}})/c}.
\end{equation}
When one fits a particular transition, the other possibility is to
perturb the rest wavelength of the transition fitted, as 
\begin{equation}
\lambda_{0}\rightarrow\lambda_{0}\left(1+\frac{\Delta v}{c}\right)\label{eq:dv rest wl modify}
\end{equation}
We use the second option for ease of implementation within \textsc{vpfit}.
Doing this means that the value of $\Delta\alpha/\alpha$ derived
from the fit will be the same as if the wavelength scale from the
other telescope in the spectral pair had been used, thereby removing
any inter-telescope differences (provided that $\Delta v$ is correctly
specified). 

In all our analysis in this section we have removed those absorbers
which were previously flagged as outliers from consideration in the
statistical analysis.

\subsection{The $\Delta v$ data}

\begin{figure}[tbph]
\noindent \begin{centering}
\includegraphics[bb=52bp 44bp 556bp 750bp,clip,width=1\textwidth]{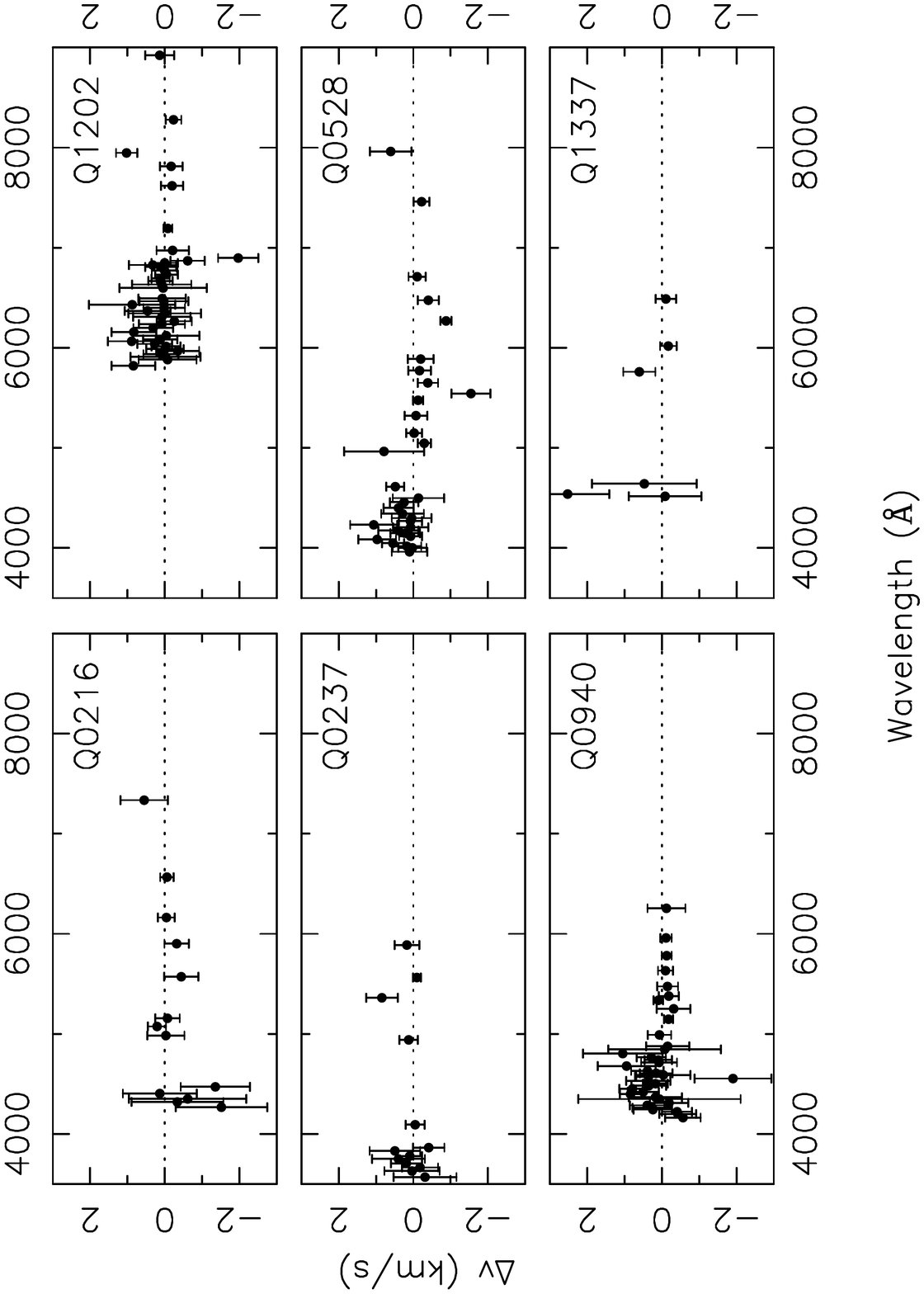}
\par\end{centering}

\caption[Binned plot of $\Delta v$ values for the six core quasar pairs]{Binned plot of $\Delta v$ values for the six core quasar pairs with 5 points per bin. $\Delta v$ values are calculated as the weighted mean of $\Delta v$ values contributing to the bin. Each set of $\Delta v$ values are normalised to $\langle \Delta v \rangle = 0$ as described in section \ref{sub:dv_analysis:core pairs}. Note that the different quasar pairs sample different regions of the wavelength space, and that some pairs provide substantially more points than others. \label{Flo:dv_analysis:six core pairs}}
\end{figure}

We show the $\Delta v$ data for 6 of the quasar spectral pairs (``core
pairs''), which appear similar to each other, in figure \ref{Flo:dv_analysis:six core pairs}.
We analyse the $\Delta v$ data from these quasars in the following
section. We noticed a problem with the 7th pair, 2206$-$1958/J220852$-$194359,
which displays variations of $\Delta v$ with wavelength which are
grossly different from the other six pairs. A systematic trend in
$\Delta v$ is seen, with a maximum difference in $\Delta v$ of $\sim2.5\,\mathrm{km\, s^{-1}}$
over the range $4000\lesssim\lambda\lesssim6000\mathrm{\AA}$. In
section \ref{sub:asys:Q2206}, we apply an inverse function derived
from the $\Delta v$ data seen in this spectral pair to all the VLT
spectra and show that a distortion of this type cannot affect all
the data. We consider the joint impact of the $\Delta v$ functions
from the 6 core quasars and from 2206$-$1958/J220852$-$194359 in
section \ref{sub:asys:overall_effect}. 

We note the presence of significant outliers within the $\Delta v$
data from the six core pairs. Therefore we rely wholly on robust statistical
methods to estimate parameters for phenomenological models of $\Delta v(\lambda)$.

\subsection{Core pairs\label{sub:dv_analysis:core pairs}}

In figure \ref{Flo:dv_analysis:six core pairs} we show binned values
of $\Delta v(\lambda)$ for the six core quasar pairs\index{Delta v
 test@$\Delta v$ test!six core pairs}. The trend in each spectral pair is different, but no common trend
is seen. For instance, it appears (by eye) from 0216/J021857 that
$\Delta v$ increases with increasing wavelength. It is difficult
to conclude what the functional form of $\Delta v(\lambda)$ is from
1337/J134002 and 0237/J024008 due to a paucity of data, although 0237/J024008
suggests no significant trend. 0528/J053007 seems to suggest that
$\Delta v$ decreases markedly with increasing wavelength. The conclusion
from 1202/J120523 is unclear, and the interpretation from 0940/J094253
is complicated by non-linear behaviour. Importantly, the functional
form of $\Delta v$ appears to differ in both magnitude and sign between
quasars. This suggests that any relative wavelength distortion is
likely to average out over a large number of absorbers. Additionally,
the wavelength coverage of the $\Delta v$ data for most spectra is
significantly smaller than the wavelength range within which MM absorbers
are fitted. This means that from each spectral pair it is impossible
to tell what the wavelength distortion might be over large amounts
of the spectral range.

\subsubsection{Linear fit}

\begin{figure}[tbph]
\noindent \begin{centering}
\includegraphics[bb=50bp 79bp 532bp 789bp,clip,angle=-90,width=1\textwidth]{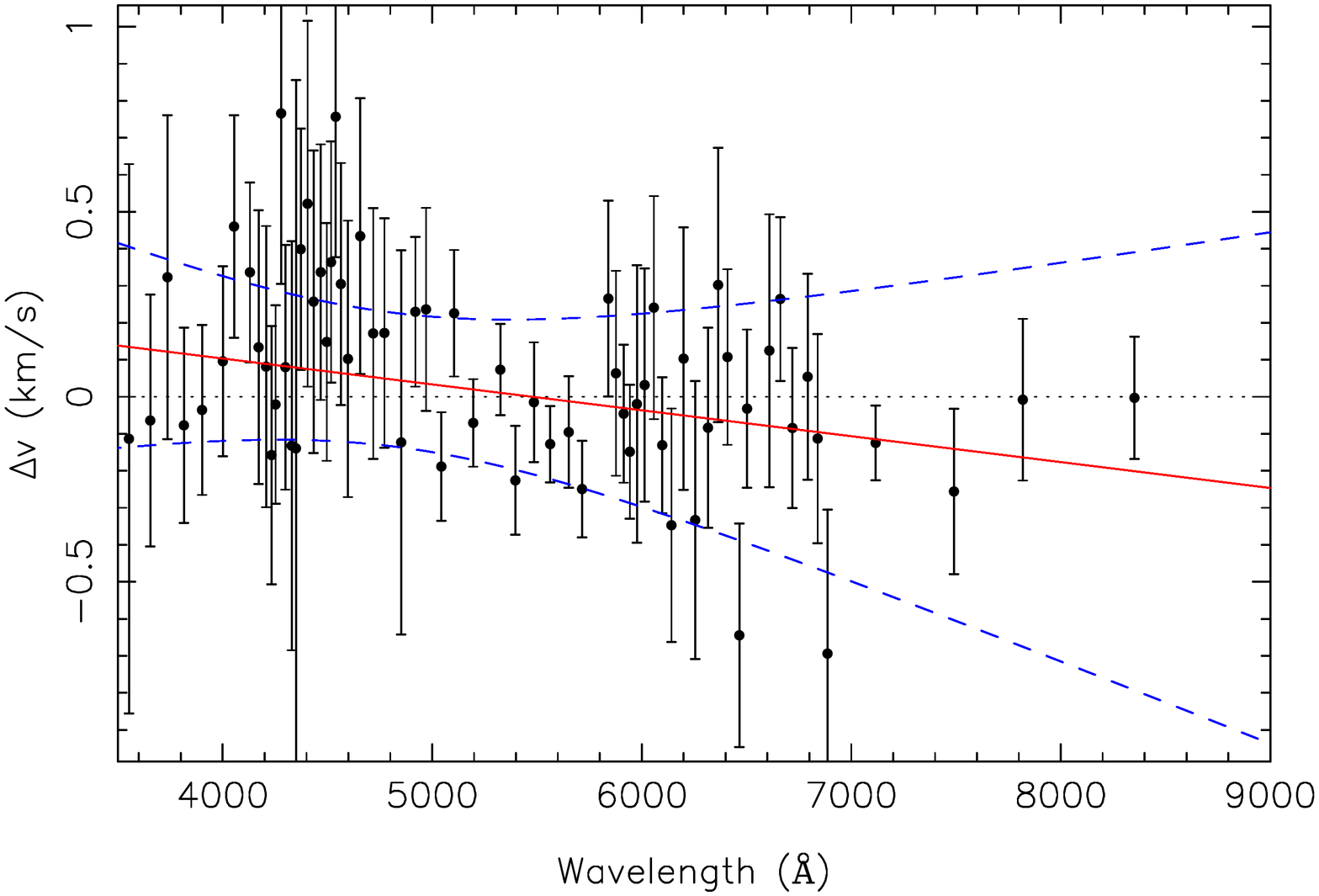}
\par\end{centering}

\caption[LTS linear fit of $\Delta v$ vs $\lambda$ for the six core quasar pairs]{Binned plot of $\Delta v$ values for the six core quasar pairs with a LTS linear fit. Only points which contribute to the fit are shown (that is, the worst 5 percent of the data has been excluded). Most transitions used in the MM analysis fall in the range $4000 \lesssim \lambda \lesssim 7000\mathrm{\AA}$; from this graph there is no significant evidence for a significant wavelength distortion in this region. \label{Flo:dv_analysis:sixcore_linear}}
\end{figure}

Due to fact that the $\Delta v$ values from each spectral pair do
not densely span the whole spectroscopic wavelength range, we combine
the $\Delta v$ values from each of the six core pairs together in
order to estimate a common function which spans the full wavelength
range. The functional form of this is unknown, however a high-order
polynomial cannot be statistically supported. We use a linear function
as a first approximation. We fit the linear function with the LTS
method, using $k=0.95n$. We show this linear fit in figure \ref{Flo:dv_analysis:sixcore_linear}.
For the form 
\begin{equation}
\Delta v=a\lambda+b,\label{eq:asys:dv_linfunc}
\end{equation}
$a=(-7\pm14)\times10^{-5}\,\mathrm{km\, s^{-1}\,\AA^{-1}}$ and $b=0.38\pm0.71\,\mathrm{km\, s^{-1}}$.
Note firstly that $a$ is statistically consistent with zero. Therefore,
it is difficult to conclude that a common linear systematic exists
in the $\Delta v$ data. Nevertheless, in section \ref{sub:asys:applytoVLT}
we apply an inverse function of this form to the VLT spectral data
to determine the effect that a wavelength distortion of this type
and magnitude would have on the dipole in section \ref{sub:asys:applytoVLT}. 

\subsubsection{Reasonableness of $k=0.95n$}

\begin{figure}[tbph]
\noindent \begin{centering}
\includegraphics[bb=96bp 40bp 559bp 734bp,clip,angle=-90,width=1\textwidth]{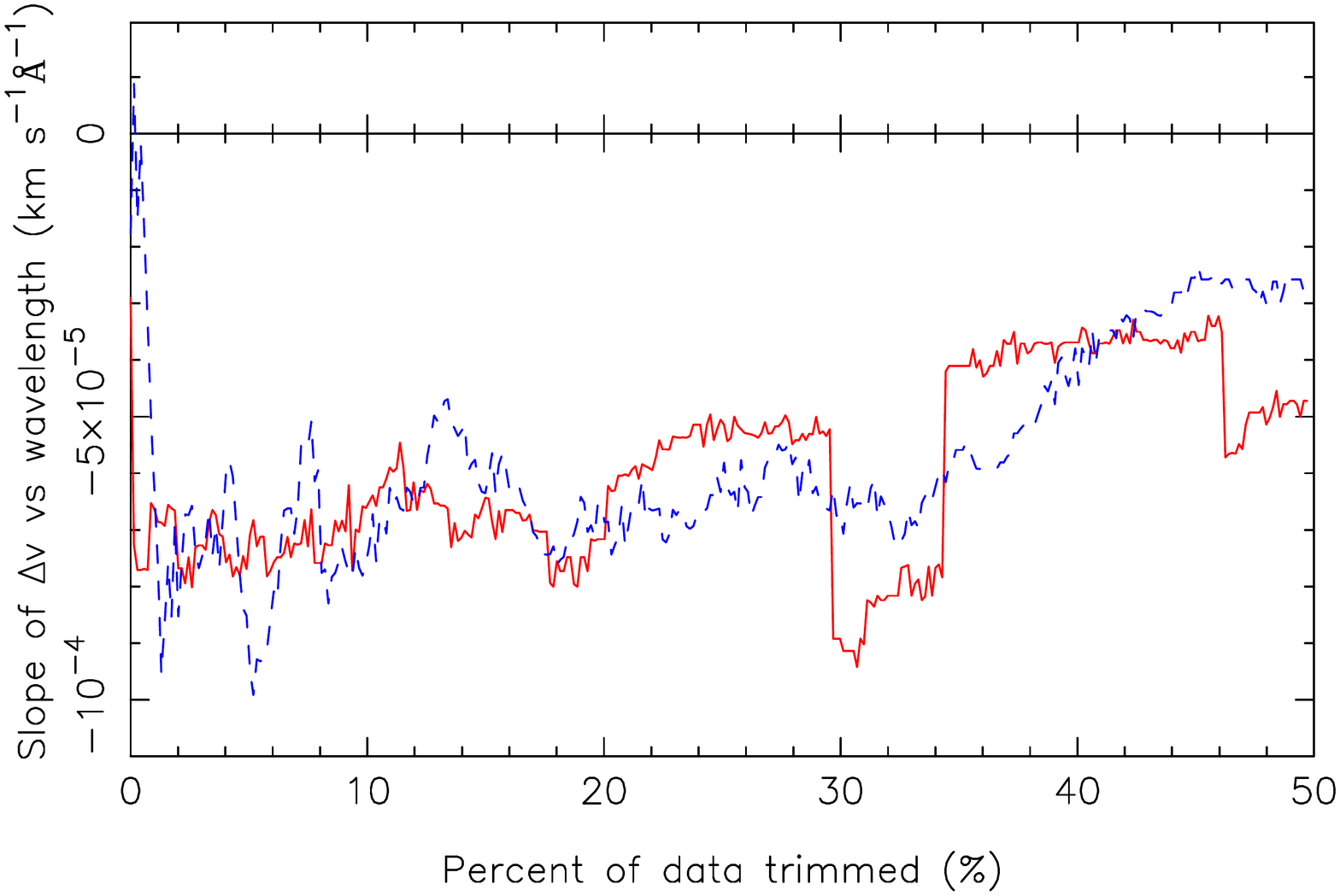}
\par\end{centering}

\caption[Effect of trimming fraction on the slope of a LTS linear fit of $\Delta v$ vs $\lambda$ for the six core quasar pairs]{Effect of trimming fraction on the slope of a LTS linear fit of $\Delta v$ vs $\lambda$ for the six core quasar pairs. The solid, red line corresponds to a weighted LTS fit, whilst the dashed, blue line corresponds to an unweighted fit. We have modelled the distortion with amplitude $\Delta v = -7 \times 10^{-5} \, \mathrm{km\, s^{-1} \AA^{-1}}$, which is as large or larger than the weighted fit over most of the range considered. As a general trend, as the trimming fraction increases, the amplitude decreases, which therefore suggests that our chosen magnitude is a reasonable estimate of the maximum distortion allowed by the data. \label{Flo:dv_analysis:sixcore_linear_trimfrac}}
\end{figure}

A legitimate question to ask is whether the choice of $k=0.95n$ is
reasonable. We show the effect of difference choices of $k$ in figure
\ref{Flo:dv_analysis:sixcore_linear_trimfrac}. Our estimate of the
slope is not overly sensitive to a choice of $k$. With the exception
of a small region around $k=0.7n$, the general trend is for the slope
to decrease with decreasing $k$. The fact that the slope decreases
with increased trimming implies that the underlying trend may be less
than what we have estimated.

\subsubsection{Application to the VLT sample\label{sub:asys:applytoVLT}}

To investigate the effect of the potential wavelength distortions
from the 6 core pairs, we apply an inverse $\Delta v$ function (equation
\ref{eq:asys:dv_linfunc}) to all the VLT absorbers by perturbing
the rest wavelengths of the transitions fitted in each absorber, as
described in section \ref{sec:asys:dv_test}. We apply the same linear
function in every VLT spectrum fitted. This therefore puts the VLT
and Keck spectral data on a common wavelength scale. Any observed
angular variation in $\alpha$ which survives the inverse function
can not be due to stable inter-telescope wavelength calibration differences. 

To investigate the effect of the potential wavelength distortions
from the 6 core pairs, we apply an inverse $\Delta v$ function (equation
) to the VLT data, which therefore puts the VLT and Keck data on a
common wavelength scale. Any observed angular variation in $\alpha$
which survives the inverse function can not be due to stable inter-telescope
wavelength calibration differences.

Because we apply the inverse function by perturbing the rest wavelengths
of transitions in absorbers fitted, we can only do this where each
fitted transition occurs in only one spectral region in a particular
fit. There are two pairs of absorbers where we have fitted both absorbers
in the pair simultaneously, because a transition from one absorber
in the pair overlaps with a transition from the other absorber (at
a different redshift) in the pair. In this case, a particular transition
can be fitted twice (in two absorbers, in two widely separated spectral
regions). However, we cannot apply two different perturbations to
a single rest wavelength. Therefore, we remove these two pairs of
absorbers to form a ``VLT reference set''. Thus, we compare the effect
of the VLT set of $\Delta\alpha/\alpha$ values where the $\Delta v$
inverse function has been applied with the $\Delta\alpha/\alpha$
values from a VLT reference set. The two pairs of absorbers which
are removed are the $z\sim2.253$ and $z\sim2.380$ absorbers associated
with J214225$-$442018 and the $z\sim1.154$ and $z\sim0.987$ absorbers
in the same spectrum (i.e.\ 4 absorbers are removed to form the VLT
reference set). This means that the VLT reference set contains 149
absorbers.

\begin{figure}[tbph]
\noindent \begin{centering}
\includegraphics[bb=77bp 79bp 455bp 727bp,clip,angle=-90,width=0.8\textwidth]{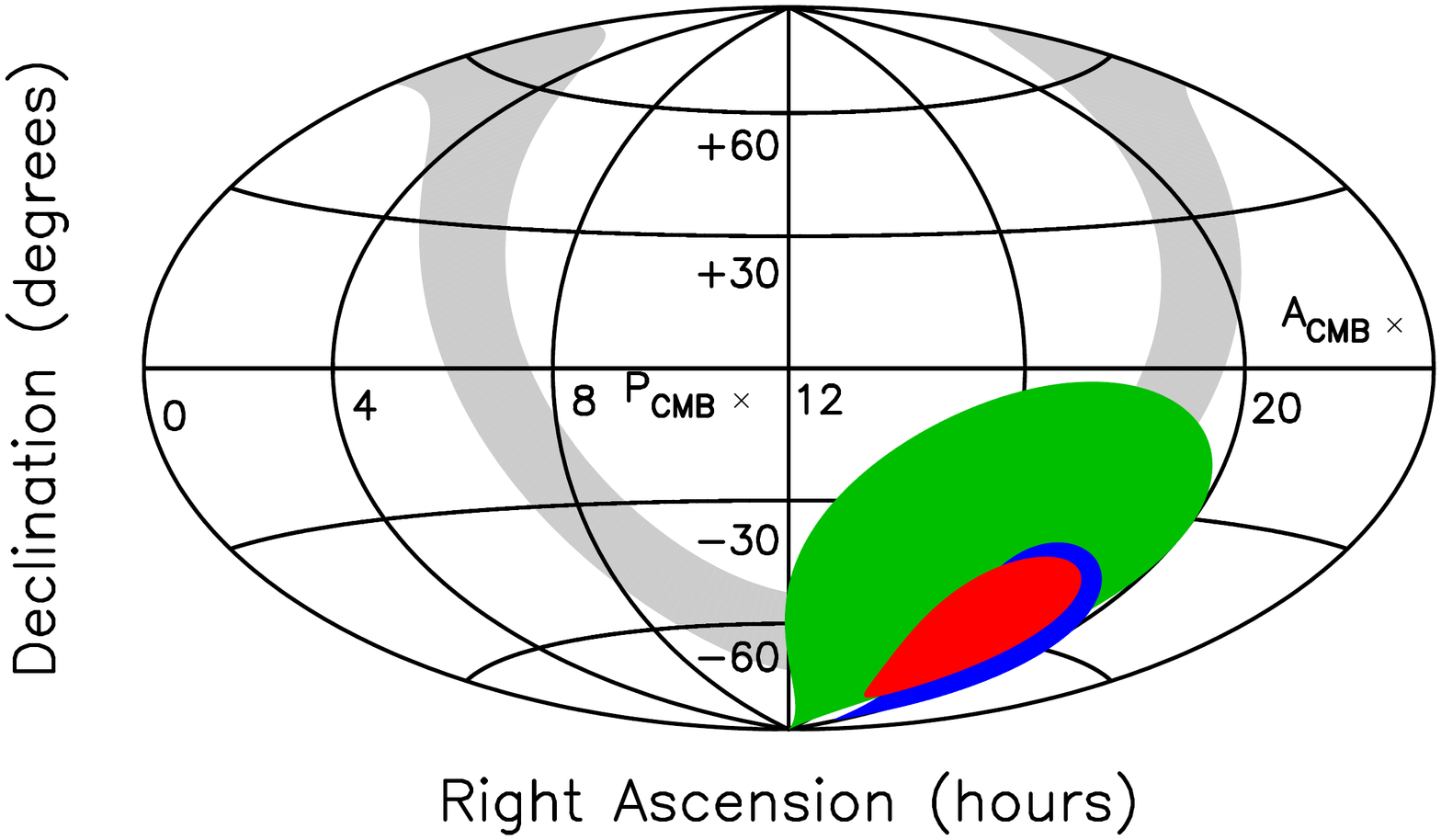}
\par\end{centering}

\caption[Confidence limits on dipole location for VLT, Keck and combined samples when the linear $\Delta v$ function from the 6 core spectral pairs is applied]{$1\sigma$ confidence regions for the Keck (green), VLT (blue) and combined (red) dipoles, where the linear inverse function derived from the $\Delta v$ data from the 6 core spectral pairs has been applied to the VLT set, under the model $\Delta\alpha/\alpha = A\cos\Theta+m$. Although the statistical significance of the dipole decreases from $3.9\sigma$ to $3.1\sigma$, the positions of the VLT and combined dipole are effectively unchanged. \label{Flo:dv_analysis:sixcore_linfit_skymap}}
\end{figure}

In table \ref{tab:asys:dv_6quasars_results}, we give the results
of applying the $\Delta v$ inverse function above those absorbers
in the VLT set. The effect is generally to push $\Delta\alpha/\alpha$
to more negative values. We show an updated plot of the confidence
regions of the Keck, VLT and combined dipole locations in figure \ref{Flo:dv_analysis:sixcore_linfit_skymap}.
Although the statistical significance of the dipole decreases from
$3.9\sigma$ (reference set) to $3.1\sigma$, the position of the
VLT (and therefore combined) dipole is effectively unchanged. This
accords well with our earlier argument that because the detection
of a dipole is a differential effect, it is difficult to emulate through
any simple systematic. The Keck and VLT dipoles in this case are separated
by $25^{\circ}$, which has a chance probability of 7 percent (see
section \ref{sub:alpha:alignment_by_chance}). Also note that introducing
this modification to the wavelength scale of the VLT spectra does
not significantly change the good alignment between the $z<1.6$ and
$z>1.6$ samples. The dipole directions in this case are separated
by $13{}^{\circ}$, which has a chance probability of 2 percent.

\begin{sidewaystable}
\centering{}\caption[Results of applying the inverse $\Delta v$ distortion from figure \ref{Flo:dv_analysis:sixcore_linear} to the VLT absorbers]{Results of applying the inverse $\Delta v$ distortion from figure \ref{Flo:dv_analysis:sixcore_linear} to the VLT absorbers under the model $\Delta\alpha/\alpha = A\cos\Theta + m$. The column ``$\delta A$'' gives $1 \sigma$ confidence limits on $A$. The column labelled ``sig'' gives the significance of the dipole+monopole model with respect to the monopole model. The origin of the reference VLT set is described in section \ref{sec:asys:dv_test}. The impact of the (non-significant) linear distortion modelled from the 6 core quasar pairs overall is a reduction in the statistical significance of the dipole from $4.1\sigma$ to $3.1\sigma$.  \label{tab:asys:dv_6quasars_results}}%
\begin{tabular}{clcccccc}
\hline 
Sample  & $\sigma_{\mathrm{rand}}$ $(10^{-5})$  & $A$ ($10^{-5}$)  & $\delta A$ ($10^{-5}$)  & RA (hr)  & dec ($^{\circ}$)  & $m$ ($10^{-5}$)  & sig\tabularnewline
\hline 
VLT reference  & 0.88 & 1.21  & $[0.80,1.72]$  & $18.3\pm1.1$  & $-61\pm13$  & $-0.110\pm0.179$  & $2.2\sigma$ \tabularnewline
VLT with $\Delta v$ function applied  & 0.95  & 1.02  & $[0.65,1.52]$  & $18.6\pm1.3$  & $-61\pm16$  & $-0.262\pm0.183$  & $1.8\sigma$ \tabularnewline
VLT reference + Keck & As above + Keck HC=1.63 & 0.97  & $[0.57,1.39]$  & $17.4\pm1.0$  & $-61\pm10$  & $-0.177\pm0.085$  & $3.9\sigma$ \tabularnewline
Keck + VLT & As above + Keck HC=1.63 & 0.789 & $[0.59,1.07]$  & $17.4\pm1.2$  & $-60\pm12$  & $-0.273\pm0.085$  & $3.1\sigma$ \tabularnewline
\hline 
\end{tabular}
\end{sidewaystable}

e would expect that any good model for a wavelength-dependent systematic
should quantitatively improve the fit of the dipole model to the $\Delta\alpha/\alpha$
values. To see if the $\Delta v$ model significantly improves the
fit, we compare the AICC of the model $\Delta\alpha/\alpha=A\cos\Theta+m$
fitted to different sets of $\Delta\alpha/\alpha$ values: \emph{i)}
$\mathrm{AICC}_{\mathrm{VLT},\Delta v}$, the AICC of the angular
dipole model fitted to the VLT absorbers in the VLT reference set
with the linear $\Delta v$ function described above applied; \emph{ii)}
$\mathrm{AICC}_{\mathrm{VLT,ref}}$, the AICC of the angular dipole
model fitted to the VLT absorbers in the VLT reference set; \emph{iii)}
$\mathrm{AICC}_{\mathrm{VLT},\Delta v+\mathrm{Keck}}$, the AICC for
the angular dipole model fitted to the absorbers in set (i) combined
with the Keck absorbers in the Keck04-dipole set, and; \emph{iv)}
$\mathrm{AICC}_{\mathrm{VLT,ref}+\mathrm{Keck}}$, the AICC for the
angular dipole model fitted to the absorbers in set (ii) combined
with the Keck absorbers in the Keck04-dipole set. In all cases we
apply the same values of $\sigma_{\mathrm{rand}}$ to the VLT $\Delta\alpha/\alpha$
values in order to compare points on a like-with-like basis. We find
that $\mathrm{AICC}_{\mathrm{VLT},\Delta v}-\mathrm{AICC}_{\mathrm{VLT},\mathrm{ref}}\approx-0.8$,
indicating that the set of VLT absorbers with the $\Delta v$ inverse
function applied is preferred, but not significantly. Comparing the
VLT+Keck set to the equivalent reference set, we find that $\mathrm{AICC}_{\mathrm{VLT},\Delta v+\mathrm{Keck}}-\mathrm{AICC}_{\mathrm{VLT,ref}+\mathrm{Keck}}\approx-3.2$,
indicating that the VLT+Keck set where the $\Delta v$ inverse function
has been applied to the VLT absorbers is weakly preferred. However,
when comparing the reference sets and the $\Delta v$ sets, the AICC
does not account for the extra two parameters for the linear model
of $\Delta v$ vs $\lambda$. Thus, with a two-parameter model for
the $\Delta v$ function there is no significant preference for the
$\Delta v$ results, and thus there is no strong evidence in the $\Delta\alpha/\alpha$
values themselves for a wavelength distortion of this type. 

\begin{figure}[tbph]
\noindent \begin{centering}
\includegraphics[bb=50bp 59bp 556bp 766bp,clip,angle=-90,width=0.8\textwidth]{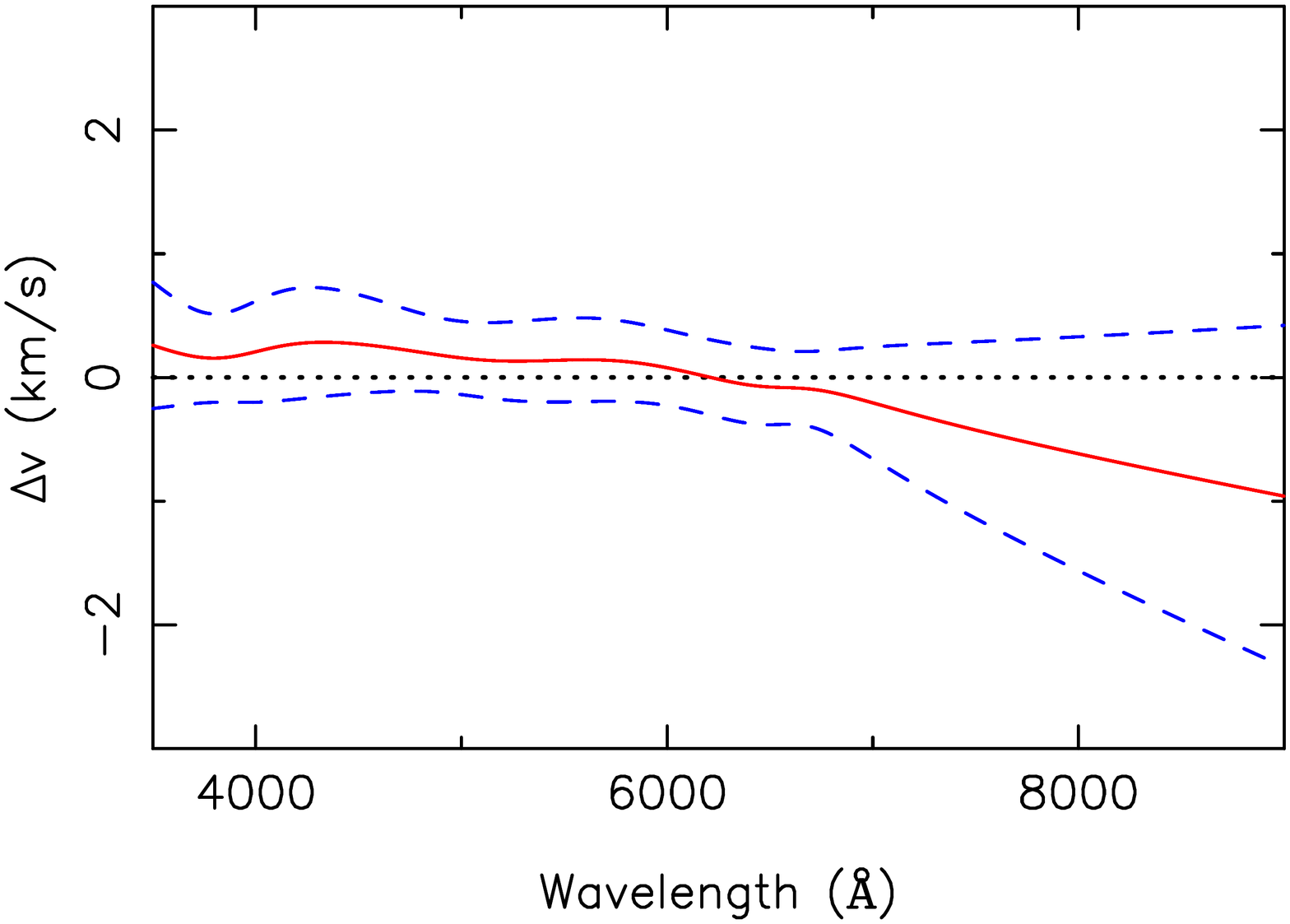}
\par\end{centering}

\caption[Joint estimate for $\Delta v(\lambda)$ from the 6 core quasar pairs made without combining the data into a single linear fit]{$\Delta v_w(\lambda)$, a joint estimate for $\Delta v(\lambda)$ from the 6 core quasar pairs made without combining the data into a single linear fit. The red (solid) line shows the estimate for $\Delta v$, and the dashed (blue) lines show the 95 percent confidence interval on the estimate. One can see that over the range where most of our $\Delta v$ data are obtained ($4000\mathrm{\AA} \lesssim \lambda \lesssim 7000\mathrm{\AA}$) that $\Delta v$ is relatively flat. $\Delta v$ diverges from zero for $\lambda \lesssim 3500\mathrm{\AA}$ (not shown) and for $\lambda \gtrsim 7000\mathrm{\AA}$, however there are few $\Delta v$ measurements to constrain $\Delta v$ in these regions. At no wavelength is $\Delta v$ significantly different from zero. Because the linear model used for each quasar fit is unlikely to be a true description of any underlying wavelength distortion, there is also uncertainty due to model specification, which is naturally not included in the confidence region shown above. As such, the confidence region shown is under-estimated. These considerations show that there is no statistically significant evidence for a common systematic from consideration of the 6 core quasar pairs.  \label{Flo:dv_analysis:sixcore_jointpred}}
\end{figure}

In deriving the results above, we have assumed that $\Delta v$ values
from different spectral pairs may be legitimately combined in order
to estimate a common systematic. This may not be a good assumption,
given the differences in the signal-to-noise of the spectral data,
the spectral range which the $\Delta v$ values cover, the potential
functional form of $\Delta v(\lambda)_{i}$ and number of exposures.
We then proceed as follows: \emph{i)} fit a linear model to the $\Delta v$
values from each spectral pair using the LTS method; \emph{ii)} from
each model, estimate $\Delta v(\lambda)$ along with an uncertainty
on the estimate; \emph{iii)} for the six estimates of $\Delta v$
at each $\lambda$, form a weighted mean of the estimates, $\Delta v_{w}(\lambda)$,
and calculate the associated uncertainty, and; \emph{iv)} plot $\Delta v_{w}(\lambda)$
as a function of wavelength. We show the result of this in figure
\ref{Flo:dv_analysis:sixcore_jointpred}. Importantly, under this
model we can find no wavelength where $\Delta v$ is statistically
different from zero. 

\subsubsection{Skeptical Bayesian Linear Regression\label{sub:asys:SBLR}}

Given the large range in the statistical error bars on the $\Delta v$
data, it is possible that by discarding even a few percent of the
data we are also discarding the data with the highest statistical
precision. Clearly, if these data are strongly inconsistent with the
general trend given by the majority of the data then their relative
influence should be downweighted. To investigate how all the data
might be used without needing to decide what fraction of the data
should be trimmed, we apply the SBLR\index{robust statistics!skeptical Bayesian linear regression (SBLR)}
method of section \ref{sub:Bayesian regression}. That is, we use
a Bayesian method where we regard the statistical errors as lower
bounds on the true error.

\begin{figure}[tbph]
\noindent \begin{centering}
\includegraphics[bb=50bp 79bp 559bp 766bp,clip,angle=-90,width=1\textwidth]{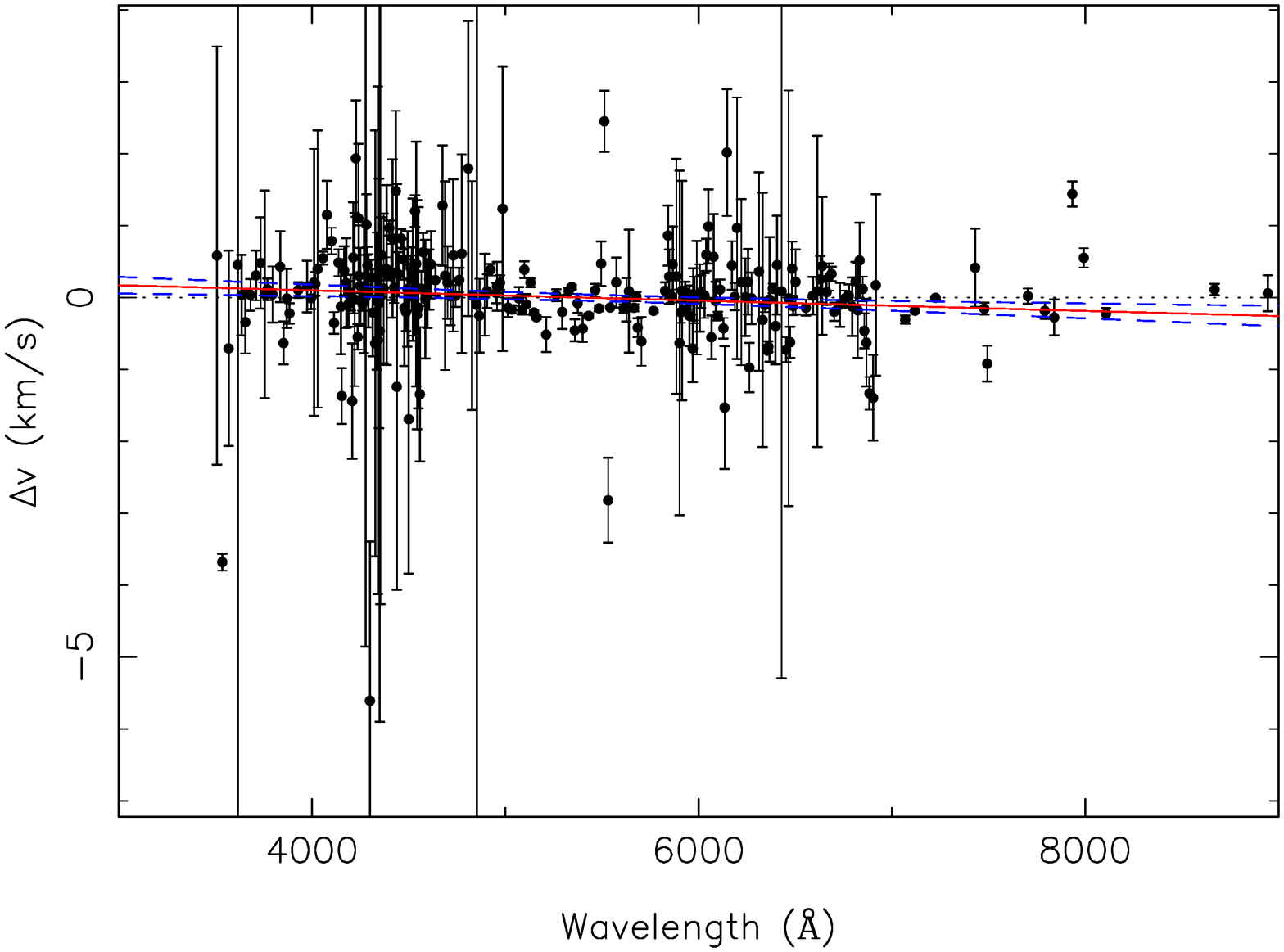}
\par\end{centering}

\caption[Skeptical Bayesian linear regression applied to the six core quasar pairs]{Skeptical Bayesian linear regression (SBLR) applied to the $\Delta v$ results for the six core quasar pairs. Data here are shown unbinned, so the full effect of the outliers can be seen. Note that although outliers are still present in the fit, they do not appear to bias the result. In particular, the point at approximately $3500\AA$ with $\Delta v\approx -3.7 \ \mathrm{km\,s^{-1}}$ would otherwise be highly damaging to the fit. The slope of this fit is $-7.2\times 10^{-5} \ \mathrm{km\,s^{-1}\AA^{-1}}$. Analytic $1\sigma$ confidence limits on the fit are shown as blue, dashed lines, but for reasons given in the text these are inappropriate. \label{Flo:dv_analysis:sixcore_bayes}}
\end{figure}

To maximise the likelihood, $L$, in equation \ref{eq:SBLR L defn}
we use a simplex algorithm from \citet{NumericalRecipes:92}, which
does not require knowledge of the derivatives of $L$ with respect
to the parameters. Because of the functional form of $L$, the possibility
for multiple likelihood maxima arises \citep{BayesianTutorial}. For
parameter estimates, one is interested in the global likelihood maximum.
To avoid this potential trap, we choose a wide range of plausible
starting values for the slope and intercept of the linear function,
and run the simplex algorithm 10,000 times. We keep the parameters
from whichever of those iterations produces the maximal $L$. Application
of SBLR to the data yields a slope of $a=-7.2\times10^{-5}\,\mathrm{km\, s^{-1}\AA^{-1}}$,
which accords well with that found through the LTS method above. We
show the result of this fit in figure \ref{Flo:dv_analysis:sixcore_bayes}.

Estimation of the uncertainty on the slope must be done carefully.
A simple approach is to estimate the Hessian matrix (the matrix of
second order partial derivatives) at the purported solution using
finite difference derivatives, take the inverse to obtain the covariance
matrix, and read off the square roots of the appropriate diagonal
entries \citep{BayesianTutorial}. In the large-data limit this approach
will be valid on account of the central limit theorem. However, because
of the functional form of equation \ref{eq:SBLR L defn}: \emph{i)}
there can be multiple likelihood maxima for small sample sizes, and
\emph{ii) }the likelihood function will have fatter tails than a Gaussian.
Therefore the formal covariance matrix at the best-fitting solution
is likely to under-estimate the true uncertainty. The formal covariance
matrix of the fit for figure \ref{Flo:dv_analysis:sixcore_bayes}
gives the error as $4.0\times10^{-5}\,\mathrm{km\, s^{-1}\AA^{-1}}$,
and so the slope differs from zero at the $1.8\sigma$ level (bearing
in mind that this must not be converted to a probability value using
Gaussian statistics unless one believes that the error is Gaussian).

\begin{figure}[tbph]
\noindent \begin{centering}
\includegraphics[bb=84bp 51bp 558bp 746bp,clip,angle=-90,width=0.85\textwidth]{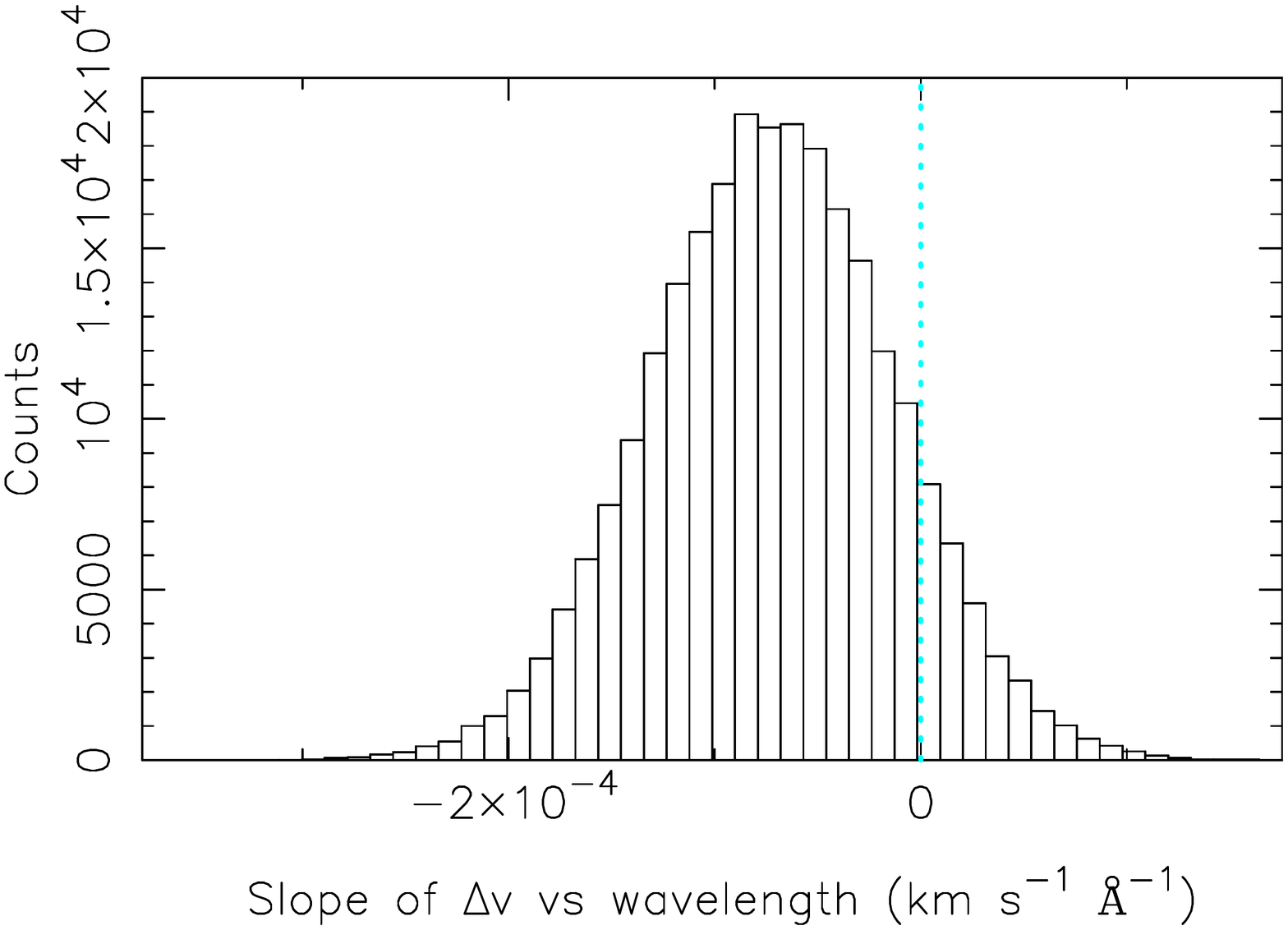}
\par\end{centering}

\caption[Probability distribution for the slope of $\Delta v$ vs wavelength for the six core pairs under SBLR]{Probability distribution for the slope of $\Delta v$ vs wavelength under SBLR for the 6 core quasar pairs. The slope is statistically consistent with zero, and therefore there is no evidence for a common linear wavelength miscalibration in the 6 core quasar pairs. \label{Flo:dv_analysis:noQ2206P_bayes_conf}}
\end{figure}

An alternative method is to explore the likelihood function directly
to obtain confidence limits on the slope which are not subject to
the Gaussian approximation. To do this, we utilise the Markov Chain
Monte Carlo (MCMC) machinery of chapter \ref{cha:MCMC}. We defer
a full explanation of the mechanics of this method until that chapter.
With $250,000$ samples of the likelihood function, we find that $a=(-7.1\pm5.9)\times10^{-5}\,\mathrm{km\, s^{-1}\AA^{-1}}$
at the 68.3 percent confidence level. That is, the 68.3 percent confidence
level is some $\approx50$ percent larger than implied by the formal
covariance matrix. At the 95 percent confidence level, the error is
$11.5\times10^{-5}\,\mathrm{km\, s^{-1}\AA^{-1}}$, and so we can
conclude that the slope is statistically consistent with zero. We
show the probability distribution of $a$ in figure \ref{Flo:dv_analysis:noQ2206P_bayes_conf}.

It is additionally worth exploring what the value of the slope is
in individual systems i.e.\ where we do not combine the data from
different quasar pairs. We give the results of this in table \ref{tab:dv:SBLR 6 core pairs}.
Both the magnitudes and signs of the slope vary between the different
quasar pairs. Importantly, all of the slopes are reasonably consistent
with zero with the exception of 0528/J053007, which nevertheless does
not deviate strongly from zero. 

\begin{table}[tbph]
\caption[SBLR applied to the 6 core quasar pairs]{Skeptical Bayesian Linear Regression (SBLR) applied to the 6 core quasar pairs. Uncertainties are derived from Markov Chain Monte Carlo methods, and are given at the 68.3 percent level ($1\sigma$-equivalent). The central estimate is taken here as the mean of the upper and lower bounds of the confidence region. Although this does not necessarily coincide with the probability mode, the difference in all cases is not significant. The only system for which the statistical significance of the slope of $\Delta v$ vs wavelength is significantly larger than ``$1\sigma$'' is 0528$-$250/J053007$-$250329. \label{tab:dv:SBLR 6 core pairs}}\medskip

\noindent \centering{}%
\begin{tabular}{cc}
\hline 
\noalign{\vskip\doublerulesep}
Quasar pair & Slope of $\Delta v$ vs wavelength ($10^{-5}\,\mathrm{km\, s^{-1}\,\mathrm{\AA}^{-1}}$)\tabularnewline[\doublerulesep]
\hline 
0216+0803/J021857+081727 & $7.1\pm33.7$\tabularnewline
\noalign{\vskip\doublerulesep}
0237$-$233/J024008$-$230915 & $0.3\pm22.6$\tabularnewline
\noalign{\vskip\doublerulesep}
0940$-$1050/J094254$-$110426 & $-26.5\pm21.3$\tabularnewline
\noalign{\vskip\doublerulesep}
1202$-$0725/J120523$-$074232 & $-4.2\pm18.7$\tabularnewline
\noalign{\vskip\doublerulesep}
0528$-$250/J053007$-$250329 & $-23.8\pm13.4$\tabularnewline
\noalign{\vskip\doublerulesep}
1337+1121/J134002+110630 & $-61.9\pm84.8$\tabularnewline
\hline 
\noalign{\vskip\doublerulesep}
\end{tabular}
\end{table}

In summary: we are unable to detect a statistically significant linear
wavelength distortion common to the 6 core spectral pairs. Applying
to the entire VLT spectral sample a simple linear model for $\Delta v(\lambda)$
from the six core pairs reduces the statistical significance of the
dipole, but the statistical significance still remains high enough
to be of interest. The systematic applied here does not destroy the
good alignment between the fitted Keck and VLT dipole vectors. We
are therefore unable to remove the dipole effect from the combined
Keck and VLT sample.

\subsection{2206$-$1958/J220852$-$194359\label{sub:asys:Q2206}}

\begin{figure}[tbph]
\noindent \begin{centering}
\includegraphics[bb=50bp 59bp 556bp 766bp,clip,angle=-90,width=1\textwidth]{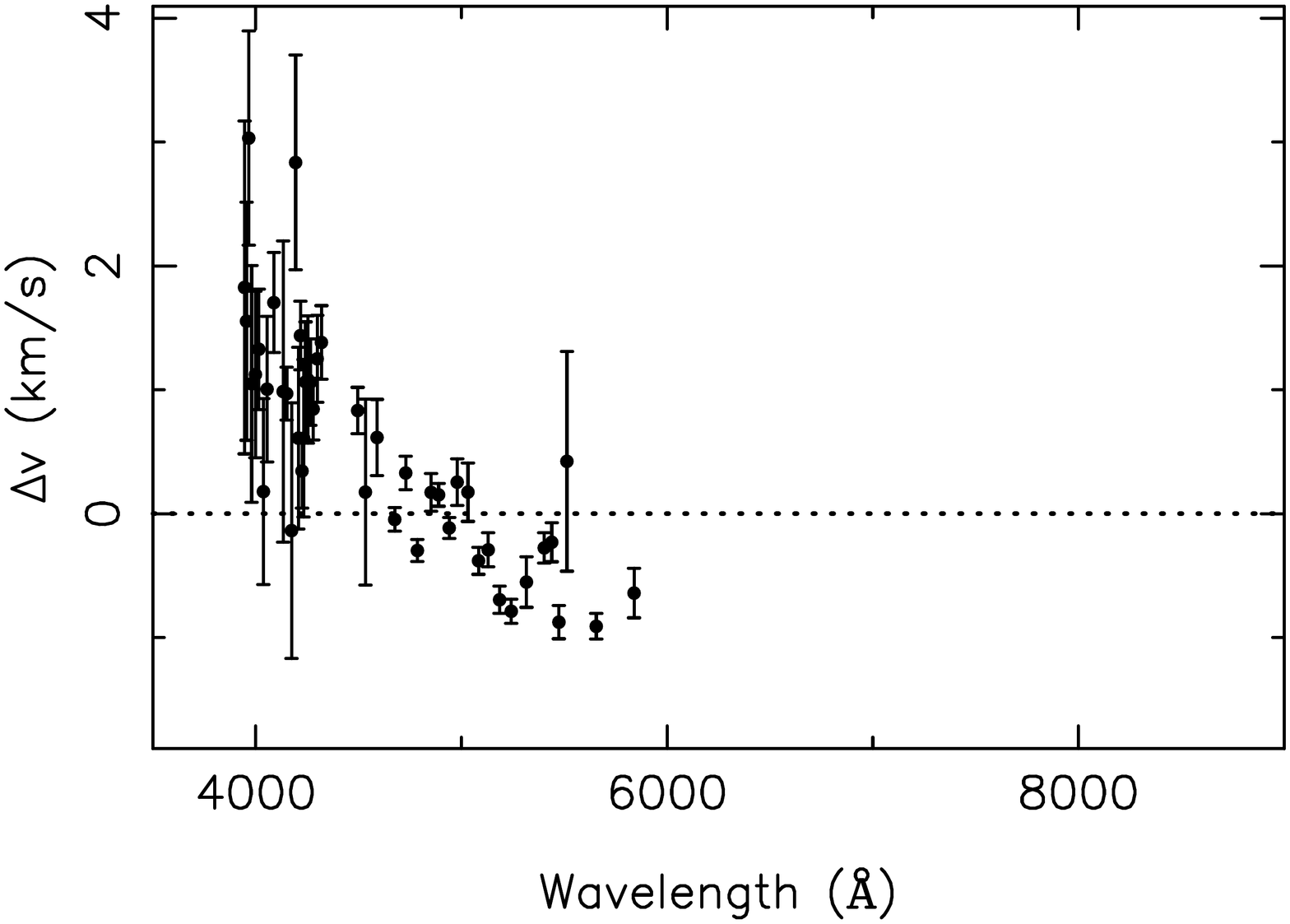}
\par\end{centering}

\caption[Binned $\Delta v$ results for the 2206$-$1958/J220852$-$194359 spectral pair]{Binned $\Delta v$ results for the 2206$-$1958/J220852$-$194359 spectral pair, with 3 points per bin.  Note the substantial slope over a relatively small wavelength range, with a range in $\Delta v$ of about $2 \ \mathrm{km\,s^{-1}}$. \label{Flo:dv_analysis:Q2206P_bin3}}
\end{figure}

\index{Q2206$-$1958/J220852$-$194359}\index{Delta v
 test@$\Delta v$ test!Q2206$-$1958/J220852$-$194359}In figure \ref{Flo:dv_analysis:Q2206P_bin3} we show the $\Delta v$
data for the 2206$-$1958/J220852$-$194359 pair. Two things are immediately
obvious. Firstly, there is a clear anticorrelation of $\Delta v$
with wavelength. Secondly, the magnitude of the effect is extremely
large --- of the order of $|\delta(\Delta v)|\approx2.5\,\mathrm{km\, s^{-1}}$
over the range considered. A wavelength distortion of this magnitude
will have a substantial impact on determining $\Delta\alpha/\alpha$
in any spectra affected by it. 

The $\Delta v$ test only examines calibration differences between
Keck and VLT, and so we cannot tell whether Keck or VLT is responsible
for significant trend in $\Delta v$ for this spectral pair.

\subsubsection{Arctangent fit}

\begin{figure}[tbph]
\noindent \begin{centering}
\includegraphics[bb=50bp 59bp 556bp 766bp,clip,angle=-90,width=1\textwidth]{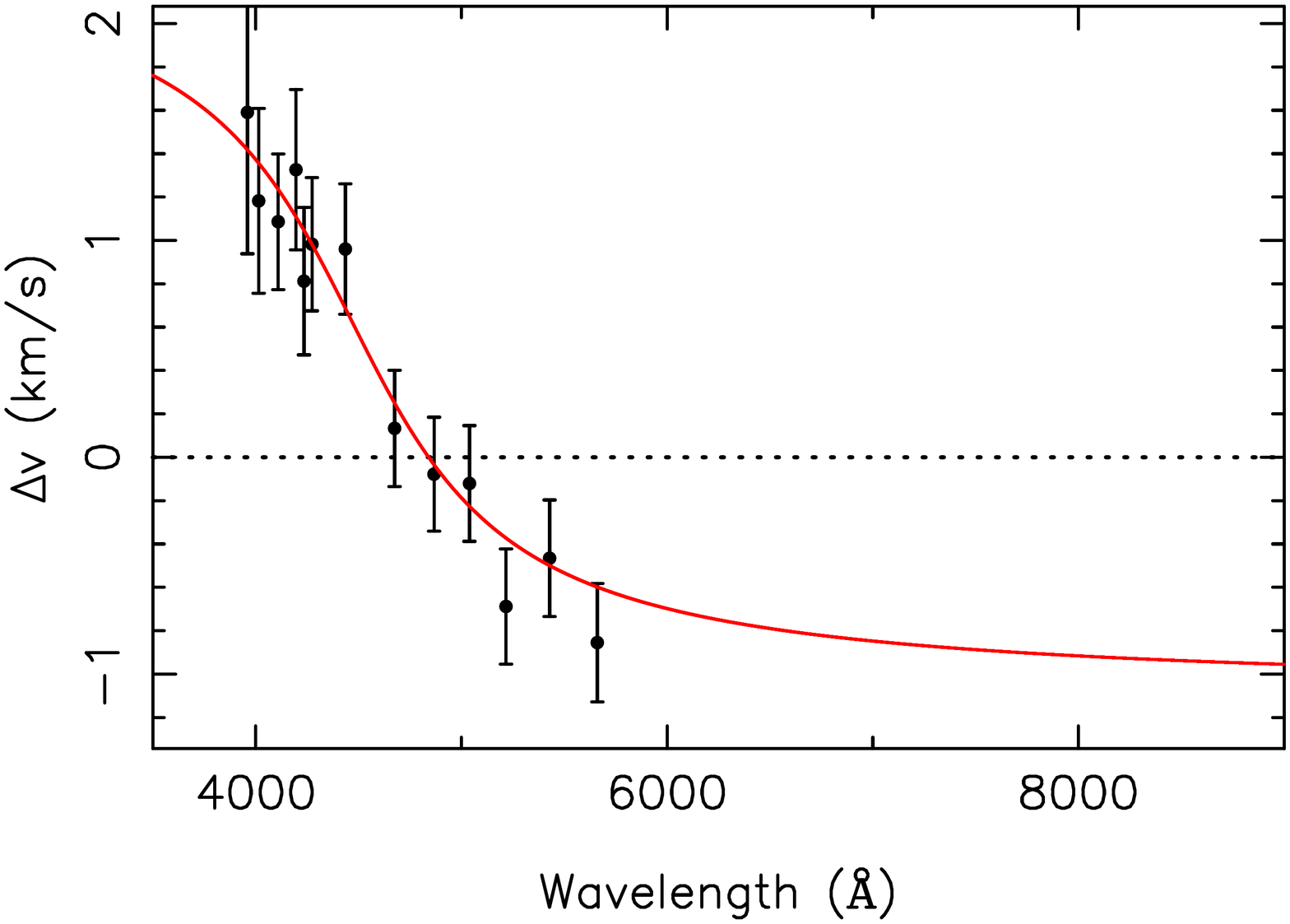}
\par\end{centering}

\caption[LTS arctan fit applied to binned $\Delta v$ results for the 2206$-$1958/J220852$-$194359 spectral pair]{LTS arctangent fit applied to $\Delta v$ results for the 2206$-$1958/J220852$-$194359 spectral pair. Results are shown binned, with 5 points per bin. 5 percent of the data is ignored in the LTS fit, and only those points which contribute to the fit contribute to the bins shown. \label{Flo:dv_analysis:Q2206P_arctan}}
\end{figure}

The limited spectral range ($4000\lesssim\lambda\lesssim6000\mathrm{\AA}$)
of the $\Delta v$ data means that we simply do not know what the
functional form of $\Delta v$ is for this spectral pair at $\lambda\gtrsim6000\mathrm{\AA}$.
In order to estimate the potential impact of the distortion present
on $\Delta\alpha/\alpha$ values in the whole sample, we need knowledge
of $\Delta v(\lambda)$ at all observed wavelengths. One possibility
is to assume that the relationship is linear, but then the extrapolation
over the whole spectral range results in a total change in $\Delta v$
of $\sim5\,\mathrm{km\, s^{-1}}$, which is comparable to the velocity
width of the spectrograph slit; this seems too extreme. Additionally,
the $\Delta v$ values in figure \ref{Flo:dv_analysis:Q2206P_bin3}
do not seem to be linearly related with wavelength. We therefore try
a phenomenologically motivated  arctangent model, 
\begin{equation}
\Delta v=A\tan^{-1}\left[k(\lambda-\lambda_{c})\right]+b.\label{eq:asys:dv_arctanfunc}
\end{equation}
Applying the LTS method to this fit, with $k=0.95n$, yields: $A=(-0.98\pm1.03)\,\mathrm{km\, s^{-1}\,\AA^{-1}}$,
$k=(1.9\pm2.6)\times10^{-3}\,\mathrm{\AA^{-1}}$, $\lambda_{c}=4547\pm526\,\mathrm{\AA}$
and $b=0.48\pm0.74\,\mathrm{km\, s^{-1}}$, where errors are derived
from the diagonal terms of the covariance matrix at the best fit.
Each $\Delta v$ uncertainty has been increased in quadrature with
$\approx0.26\,\mathrm{km\, s^{-1}}$ to account for over-dispersion
about the LTS fit. We show the results of this fit, along with an
extrapolation over the useful wavelength range, in figure \ref{Flo:dv_analysis:Q2206P_arctan}.

\subsubsection{Application to the VLT sample}

The impact of this wavelength perturbation on the values of $\Delta\alpha/\alpha$
when this function is applied to the VLT absorbers is severe. In particular,
large numbers of points are scattered away from $\Delta\alpha/\alpha=0$,
inducing a highly significant detection of $\Delta\alpha/\alpha$
at $z\sim0.8$ and $z\sim1.5$. For instance, a formal weighted mean
of all the $1.3<z<1.8$ points yields $\Delta\alpha/\alpha=(-3.96\pm0.12)\times10^{-5}$
--- a $33\sigma$ ``detection''. If one multiplies the error by $\sqrt{\chi_{\nu}^{2}}$
to account for $\chi_{\nu}^{2}=14.7$ about the weighted mean, then
one obtains $\Delta\alpha/\alpha=(-3.96\pm0.46)\times10^{-5}$, still
an $8.5\sigma$ ``detection''. Such a signal is seen in neither the
Keck or VLT samples, which immediately implies that this particular
relative wavelength distortion can not possibly apply to all of either
the VLT or Keck spectra. That is, the distortions seen in this particular
Keck/VLT spectral pair appear not to be representative of a significant
fraction of the entire sample. However, the fact that we have identified
this distortion demonstrates the power and utility of the quasar pair
analysis in identifying systematic errors, even when their actual
origin remains unknown.

\subsection{Overall effect of wavelength systematics using the $\Delta v$ test\label{sub:asys:overall_effect}}

We now investigate whether a diluted form of the above effect (i.e.\ the
effect from the 2206$-$1958/J220852$-$194359 pair) could exist in
the spectral data in combination with the (non-significant and much
smaller) effect observed from the 6 core quasars. To do this, we use
a Monte Carlo approach where at each iteration we apply the inverse
function derived in section \ref{sub:dv_analysis:core pairs} (equation
\ref{eq:asys:dv_linfunc}) to $6/7$ of the quasar spectra selected
at random in the VLT sample, and the arctan function of equation \ref{eq:asys:dv_arctanfunc}
to the remaining $1/7$ of the spectra. We then apply the LTS method
to estimate a new $\sigma_{\mathrm{rand}}$. We then add the Keck
sample to this new VLT sample. At each iteration, we calculate the
statistical significance of the dipole using the bootstrap method.
The mode of the distribution obtained is $\sim2.2\sigma$, with quite
substantial variation between iterations. 

\begin{figure}[tbph]
\noindent \begin{centering}
\includegraphics[bb=132bp 53bp 550bp 696bp,clip,angle=-90,width=0.85\textwidth]{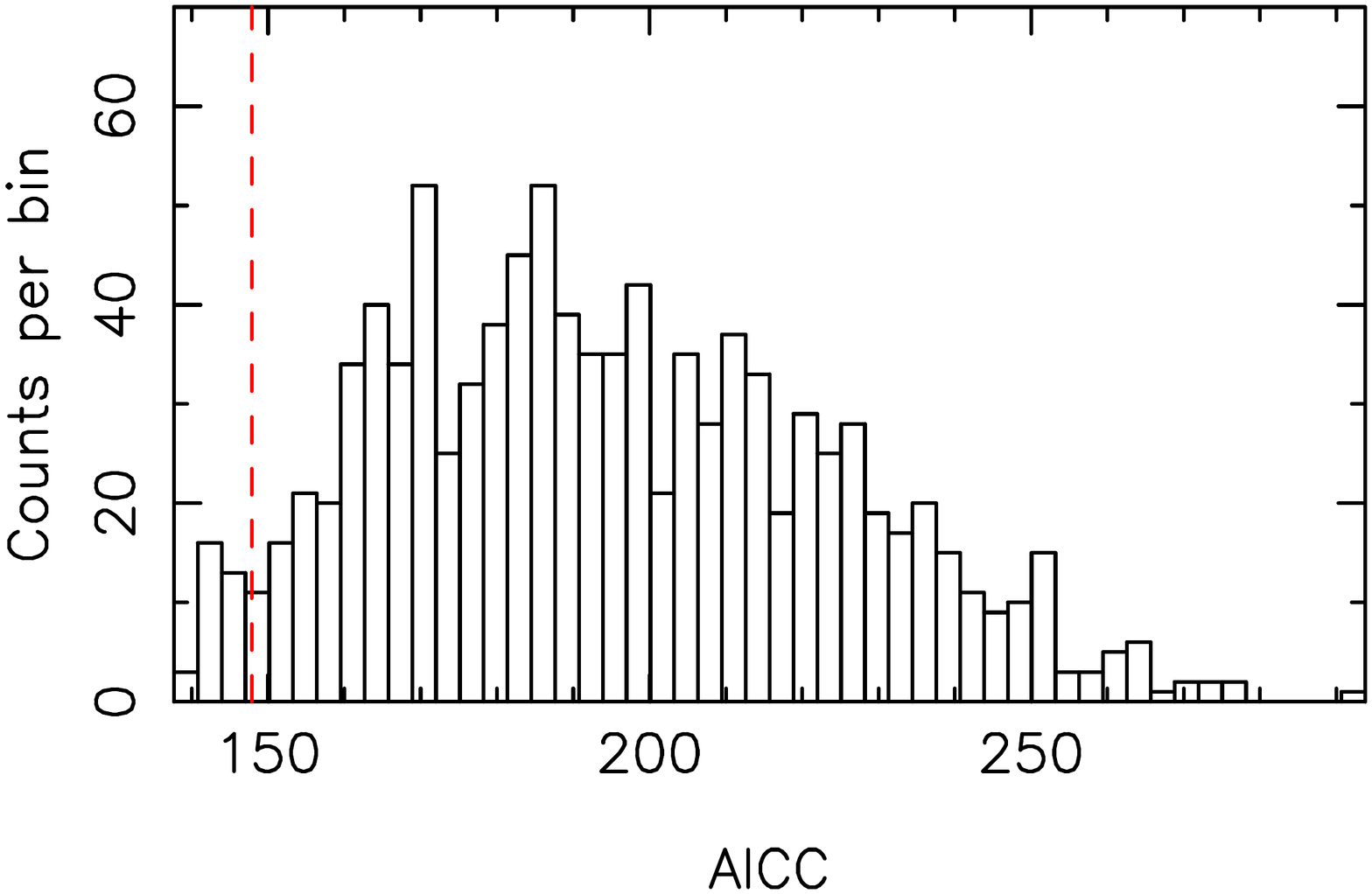}
\par\end{centering}

\caption[AICC under a Monte Carlo simulation where the $\Delta v$ function from the 6 core spectral pairs is applied to 6/7 of the VLT data, and where the $\Delta v$ function from the 2206$-$1958/J220852$-$194359 pair is applied to the remaining 1/7]{AICC for 1000 iterations of a Monte Carlo simulation where the $\Delta v$ function from the 6 core pairs (equation \ref{eq:asys:dv_linfunc}) is applied to 6/7 of the VLT quasars at random, and the $\Delta v$ function from the 2206$-$1958/J220852$-$194359 pair (equation \ref{eq:asys:dv_arctanfunc}) is applied to the remaining 1/7 of the quasars. The AICC here is calculated for the $\Delta\alpha/\alpha$ data with respect to the model $\Delta\alpha/\alpha = A\cos(\Theta)+m$ for the VLT points only. The dashed red line shows the actual AICC for the VLT reference set (see section \ref{sub:asys:applytoVLT}). Only in 3.5 percent of the iterations is the AICC lower than for the reference set, which suggests that it is unlikely that a distortion of this type is present in the VLT data. The median value of the AICC here is 191.4, corresponding to a median $\Delta \mathrm{AICC} = 43.5$. For the median case, the odds against a wavelength distortion of this type being present in the data are $\approx 3\times 10^{9}:1$. \label{Flo:dv_analysis:6q+2206_sim_AICC}}
\end{figure}

To determine whether the $\Delta v$ function significantly improves
the goodness-of-fit in the VLT sample, we compare the AICC at each
iteration in the Monte Carlo simulation for a dipole model ($\Delta\alpha/\alpha=A\cos\Theta+m$)
fitted to the VLT $\Delta\alpha/\alpha$ values in that iteration
with the AICC from a dipole model fitted to the $\Delta\alpha/\alpha$
values in the VLT reference set, where in each iteration we use $\sigma_{\mathrm{rand}}=0.88\times10^{-5}$
in order to compare $\Delta\alpha/\alpha$ values on a like-with-like
basis. We show this distribution in figure \ref{Flo:dv_analysis:6q+2206_sim_AICC}.
In only 3.5 percent of iterations is the AICC lower than in the reference
set. This implies that it is unlikely that a wavelength distortion
of this type is present in our data set. However, in almost all of
the iterations the AICC is much larger than the AICC from the VLT
reference set; the median $\Delta\mathrm{AICC}=43.5$. Importantly,
in \emph{no case} is the $\Delta\mathrm{AICC}>10$. Thus in no case
can we say that there is very strong evidence in favour of the model
with the $\Delta v$ function applied. Additionally, the AICC does
not account for the 6 parameters used in deriving $\Delta v$ model
-- we should expect a significant reduction in the AICC if the $\Delta v$
function is a good model. From this argument, we thus conclude that
a wavelength distortion of this type is unlikely to be present in
the VLT spectral data.

\section{Comment on \citet{Griest:09} I$_{2}$ and ThAr measurements on Keck/HIRES}

\index{wavelength distortions!long ranged}In section \ref{sub:mu:Intraorder_distortions}
we discussed how wavelength distortions may arise on account of the
different path that the quasar light may take through the spectrograph
compared to the thorium-argon (ThAr) light. These distortions may
be long or short ranged. \citet{Griest:09} detected long and short
ranged calibration differences by comparing the calibration using
an I$_{2}$ absorption cell and the standard ThAr calibration. In
particular, they reported drifts between the I$_{2}$ and ThAr calibration
scales of up to $2000\,\mathrm{m\, s^{-1}}$ over several nights,
and claimed that that ``this level of systematic uncertainty may make
it difficult to use Keck HIRES data to constrain the change in the
fine-structure constant''. 

The $\Delta v$ test above explicitly includes the effect of any drifts
in the wavelength calibration both within a single night and between
observation nights. From figure \ref{Flo:dv_analysis:sixcore_linear},
the RMS of the 66 binned points about the fit is $\approx250\,\mathrm{m\, s^{-1}}$.
The mean wavelength separation between these points is comparable
to the echelle order width. This RMS can therefore be compared directly
to the spread in $v_{\mathrm{shift}}$ seen in figure 5 of \citet{Griest:09}.
In contrast to their spread of $\approx2000\,\mathrm{m\, s^{-1}}$,
we see see typical wavelength distortions between VLT and Keck which
are some 8 times smaller. We have directly quantified the impact of
this on our measurements of $\Delta\alpha/\alpha$ in section \ref{sub:asys:applytoVLT}.
Our results demonstrate that it is possible to reliably use Keck/HIRES
data to constrain the fine-structure constant from quasar observations.
We deal with the intra-order distortions described in the next section.

\section{Intra-order wavelength distortions\label{sec:asys:intraorder_distortions}}

\index{wavelength distortions!intra-order}In section \ref{sub:mu:Intraorder_distortions}
we noted the presence of intra-order wavelength distortions in both
Keck/HIRES spectra \citep{Griest:09} and VLT/UVES spectra \citep{Whitmore:10}.
In this section we attempt to estimate the impact of the extra scatter
that has already been introduced into the VLT $\Delta\alpha/\alpha$
as a result of the intra-order distortions reported by \citet{Whitmore:10}.

As the distribution of MM transitions is random with respect to the
location of the echelle orders, the effect of these distortions will
be random from absorber to absorber. A distortion of this type, with
no monotonic long-range component, constitutes a random effect (see
section \ref{sub:alpha:rand_sys_errs}). \citet{Murphy:09} applied
a model of the distortion found by \citet{Griest:09} to the 2004
Keck results, and found that the impact on the weighted mean of the
$\Delta\alpha/\alpha$ values was effectively negligible. It is also
worth noting that systems which utilise a large number of MM transitions
will be less sensitive to an effect of this type. This is because
with many transitions, the distortion is sampled in many locations;
if the distortion does not have a long-range component, the average
distortion must tend to zero. It is important to note that because
the VLT spectra used here are the result of the co-addition of many
exposures, taken with different echelle grating settings and over
many nights, it is expected that any distortions of the wavelength
scale due to light path differences should be reduced in magnitude.
Thus, we consider the possible estimate of the impact of the effect
we present to be an upper limit.

\begin{figure}[tbph]
\noindent \begin{centering}
\includegraphics[bb=62bp 53bp 447bp 752bp,clip,angle=-90,width=0.9\textwidth]{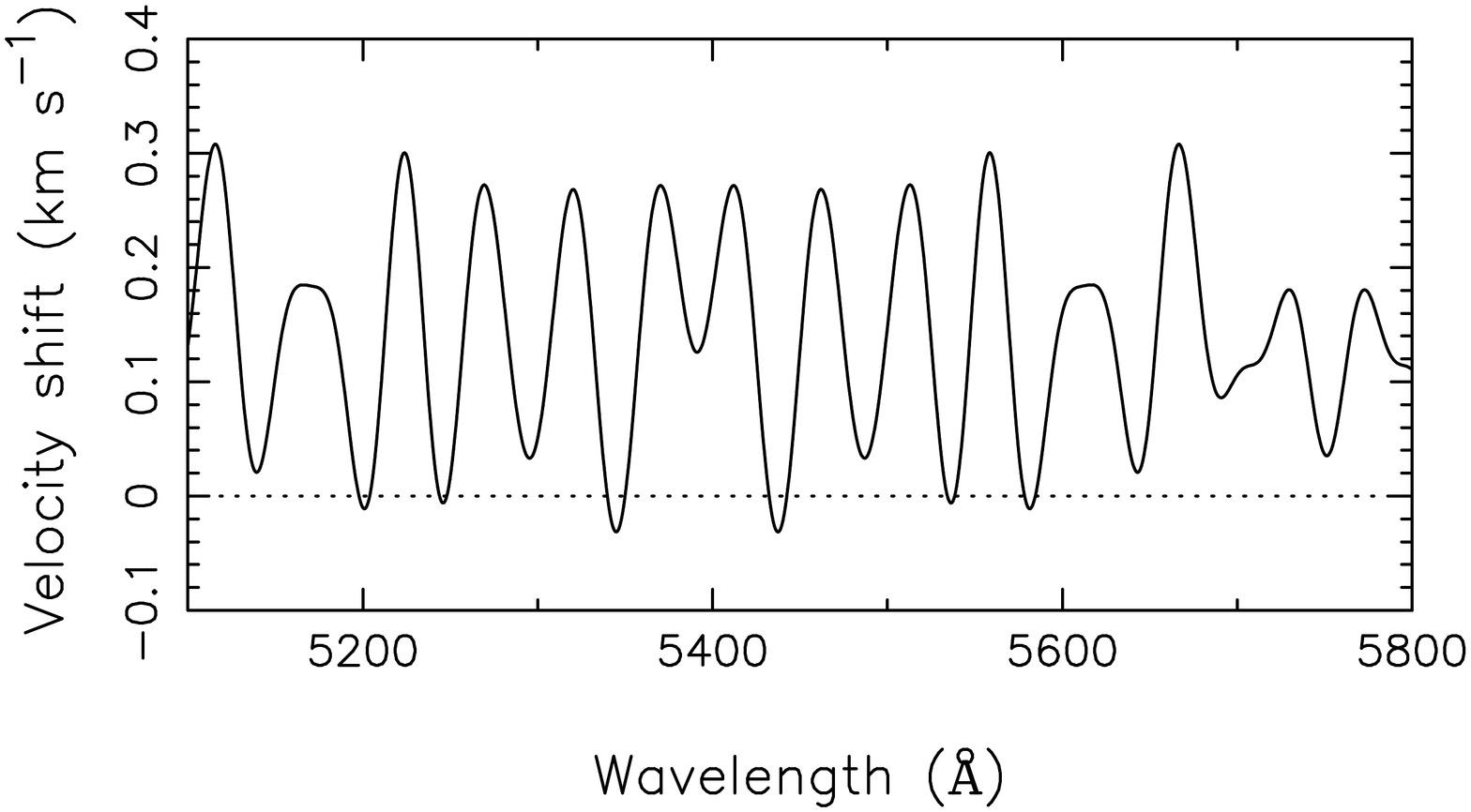}
\par\end{centering}

\caption[$\Delta v$ function to estimate the impact of intra-order distortions, from Whitmore et al.]{$\Delta v$ function used for the investigating the wavelength distortions found by \citeauthor{Whitmore:10} based on a Fourier analysis of their data. This function is repeated to longer and shorter wavelengths. This function was kindly provided by F.~E.~Koch.\label{Flo:Whitmore_plot}}
\end{figure}

To investigate the effect of the \citet{Whitmore:10} distortions
on the VLT absorbers, we used a model constructed from a Fourier analysis
of the velocity shift data presented in that paper (kindly provided
by F.~E.~Koch). The iodine cell absorption lines used to establish
the intra-order distortion results only occur over the wavelength
range $\sim$ 5000--6200$\mathrm{AA}$. We are therefore forced to
assume that our model of these distortions applies to much bluer and
redder wavelengths as well. Clearly, the important part of the model
is the amplitude rather than the period of the distortions; our model
has a maximum peak-to-peak distortion of $\sim300\,\mathrm{m\, s^{-1}}$.
We show this model in figure \ref{Flo:Whitmore_plot}.

\begin{sidewaystable}
\centering{}\caption[Results of applying the inverse $\Delta v$ function from figure \ref{Flo:Whitmore_plot} to the VLT absorbers]{Results of applying the inverse $\Delta v$ function from figure \ref{Flo:Whitmore_plot} to the VLT absorbers. The column ``$\delta A$'' gives $1 \sigma$ confidence limits on $A$. The column labelled ``sig'' gives the significance of the dipole model with respect to the monopole model. The origin of the reference VLT set is described in section \ref{sub:dv_analysis:core pairs}. For $I=2$, we remove the absorber at $z=1.6574$ toward J024008$-$230915, as it is identified as an outlier with the LTS method. Note that the estimates of the dipole location and monopole value do not differ greatly between samples 2 and 3. For $\sigma_\mathrm{rand}$, HC refers to the Keck high-contrast sample. Application of the Whitmore et.\ al. $\Delta v$ function causes a reduction in significance of the dipole model of $\approx 0.8\sigma$. This reflects the maximal amount by which the significance has already been reduced; the $3.3\sigma$ figure is not the significance after applying a correction for a systematic. \label{tab:asys:Whitmore_results}}%
\begin{tabular}{ccccccccc}
\hline 
$I$  & Sample  & $\sigma_{\mathrm{rand}}$ ($10^{-5}$) & $A$ ($10^{-5}$)  & $\delta A$ ($10^{-5}$)  & RA (hr)  & dec ($^{\circ}$)  & $m$ ($10^{-5}$)  & sig\tabularnewline
\hline 
1  & VLT with Whitmore $\Delta v$ function z  & 1.090  & 0.87  & $[0.51,1.43]$  & $19.0\pm1.6$  & $-54\pm23$  & $-0.025\pm0.215$  & $1.2\sigma$ \tabularnewline
2  & \#1 with $z=1.6574$ absorber removed  & 1.067  & 1.03  & $[0.63,1.56]$  & $18.4\pm1.1$  & $-51\pm19$  & $-0.090\pm0.204$  & $1.6\sigma$ \tabularnewline
3  & VLT reference  & 0.882  & 1.21  & $[0.80,1.72]$  & $18.3\pm1.1$  & $-61\pm13$  & $-0.110\pm0.179$  & $2.2\sigma$ \tabularnewline
4  & \#2 + Keck sample  & Keck HC=1.630  & 0.88  & $[0.68,1.12]$  & $17.4\pm1.0$  & $-56\pm11$  & $-0.217\pm0.090$  & $3.3\sigma$ \tabularnewline
\noalign{\vskip\doublerulesep}
5 & VLT reference + Keck & Keck HC=1.630 & 0.97 & $[0.57,1.39]$ & $17.4\pm1.0$ & $-61\pm10$ & $-0.177\pm0.085$ & $3.9\sigma$\tabularnewline
\hline 
\end{tabular}
\end{sidewaystable}

In table \ref{tab:asys:Whitmore_results}, we show the result of applying
the function shown in figure \ref{Flo:Whitmore_plot} to the VLT absorbers
using equation\textbf{ }\ref{eq:dv rest wl modify}. The impact on
the location of the dipole and the value of the monopole is minimal,
as expected. However, we note that the $\sigma_{\mathrm{rand}}$ required
is somewhat larger, which means that this model of the wavelength
distortion has introduced extra scatter into the $\Delta\alpha/\alpha$
values. Any good model of the systematic should reduce, not increase,
the scatter. The extra scatter reduces the significance of the dipole,
but does not destroy the good alignment between Keck and VLT, nor
between low and high redshift samples. In particular, the chance probability
of alignment for the Keck and VLT samples (where the VLT sample has
been altered with this $\Delta v$ model is 6 percent, the chance
probability of alignment between low and high redshift samples is
4 percent, and the joint chance probability for these two factors
is 0.3 percent. 

The presence of intra-order wavelength distortions would serve to
increase the scatter of the $\Delta\alpha/\alpha$ values about the
true values. These distortions can only randomise but not bias $\Delta\alpha/\alpha$
values. They can not manufacture a dipole or monopole. Were we able
to make the same quasar observations without the presence of any wavelength
scale distortions, the scatter in the $\Delta\alpha/\alpha$ values
about the model should be smaller (and so $\sigma_{\mathrm{rand}}$
would be smaller). We would therefore expect that this would increase
the significance of the dipole model. Our analysis in this section
suggests that the maximal reduction in statistical significance of
the dipole which may have occurred as a result of intra-order wavelength
distortions present is $\sim0.6\sigma$.

\section{UVES, a dual-arm spectrograph\label{sec:asys:UVES_dual_arm}}

UVES\index{UVES!spectrograph slit misalignment} is a dual-arm spectrograph,
where the incoming light is split into a red arm and a blue arm using
a dichroic mirror. In principle, misalignment of the slit in the blue
arm relative to that in the red arm would produce a distortion of
the wavelength scale between the two arms, which could mimic a change
in $\alpha$ if transitions are fitted simultaneously from spectral
data from both arms. \citet{Molaro:08} investigated the possibility
that such misalignment might cause velocity shifts between the blue
and red arms, using measurements of asteroids, and argued that the
two arms do not show separation by more than $30\,\mathrm{m\, s}^{-1}$
in the situation where the science exposures are bracketed by the
ThAr exposures. A shift of this magnitude is equivalent to $\Delta\alpha/\alpha\approx0.14\times10^{-5}$
for the Fe~\iis $\lambda2382$ transition, which is negligible in
the context of our sample.

However, we note that \citeauthor{Molaro:08} used a slit with of
$0.5"$, which is rather different to the $\sim0.7"$ to $\sim1.0"$
typical of the quasar exposures. The UVES archive indicates that,
for the observations of \citeauthor{Molaro:08}, the seeing was always
poorer than the slit width. If the slits for the blue and red arms
are misaligned, one would expect the induced effect on wavelength
calibration to depend on slit size. In the seeing-limited regime,
the slit is relatively uniformly illuminated, and therefore the observed
science wavelengths should be well calibrated through the ThAr exposure.
On the other hand, when the seeing is much better than the slit, one
might expect to see larger differences, if such differences exist.

\section{The effect of isotopic abundances\label{sec:asys:isotopic_abundances}}

\index{Mg isotope abundance}Most of the atomic species we use have
a number of stable isotopes, and each of these isotopes exhibits a
slightly different rest wavelength for a given transition. The isotopic
spacing depends on the transition and species under consideration,
but scales according to the inverse square of the mass. That is, $\Delta\omega_{i}\propto\omega_{0}/m_{i}^{2}$.
The isotopic shifts of Mg, as the lightest of the elements species
under consideration, are relatively significant. The terrestrial abundance
of the Mg isotopes is $^{24}$Mg:$^{25}$Mg:$^{26}$Mg = 79:10:11
\citep{Rosman:98}. We define the heavy isotope fraction as $\Gamma=({}^{25}\mathrm{Mg}+{}^{26}\mathrm{Mg})/\mathrm{Mg}$,
which has a terrestrial value of $\Gamma_{t}=0.21$.

We have assumed for our final fits that the quasar absorber isotopic
abundances are the same as the terrestrial abundances. However, if
the abundances in the absorbers differs from the terrestrial abundances,
this will introduce a small but potentially significant shift in the
quasar absorption lines compared to laboratory measurements. Mg will
be most affected by this, due to its low atomic mass compared to the
other species. Previous work \citep{Murphy:03} noted that the effect
could be particularly significant for low-$z$ absorbers, as these
predominantly consist of the Fe/Mg combination. High-$z$ absorbers
are less likely to be affected due to the use of more massive anchors
(Si and Al), for which this effect is less relevant. Additionally,
the use of many transitions with differing $q$ coefficients at high
redshift will tend to reduce the importance of this effect \citep{Murphy:04:LNP}. 

Both observations \citep{Gay:2000} and theoretical estimates \citep{Timmes:1995}
of stellar abundances for Mg suggest that the heavy isotope abundance
of Mg (i.e.\ the $^{25}$Mg and $^{26}$Mg isotopes) decreases with
decreasing metallicity. \citet{Murphy:03} noted that the low-$z$
Mg/Fe systems considered in the Keck sample have relative metal abundances,
{[}Fe/H{]}, in the range $-2.5$ to $0.0$, whereas the high-$z$
DLA systems have relative metal abundances of about $-1.0$. Therefore,
the quasar absorbers we consider may also have sub-solar metallicities.
However, observations of some low metallicity red giants show significant
enrichment of the heavy Mg isotopes. \citet{Ashenfelter:04a,Ashenfelter:04b}
considered a ``modest'' enhancement of the stellar initial mass function
(IMF) for intermediate mass stars ($M\approx5M_{\odot}$), and showed
that this could produce $\Gamma\sim0.4$ for {[}Fe/H{]} $\sim-1.5$.
\citet{Fenner:05a} argued that such an IMF would substantially overproduce
nitrogen relative to observations, and therefore that this mechanism
of creating $\Gamma>\Gamma_{t}$ does not seem possible. 

However, it appears that the link between stellar evolution and the
likely nitrogen abundance in quasar absorbers is not fully understood.
\citet{Centurion:03a} described observations of extremely low relative
abundances of nitrogen in DLAs, and thus argued that nitrogen production
cannot be dominated by massive stars. In a detailed study, \citet{Dessauges-Zavadsky:07a}
argued that ``no single star formation history explains the diverse
sets of abundance patterns in DLAs''. \citet{Melendez:07a} claimed
(in contrast to previous analyses) that heavy Mg isotope enrichment
due to AGB stars in the Galaxy halo does not occur until {[}Fe/H{]}
$\gtrsim-1.5$. \citet{Levshakov:09a} examined 11 metal-rich, high-redshift
($1.5<z<2.9$) quasar absorbers and argued that the nitrogen abundance
is uncorrelated with the metallicity, which implies that nitrogen
enrichment has several sources. They also claimed to observe shifts
in the Mg\textasciitilde{}\iis $\lambda2796,2803$ lines which they
ascribe to enrichment of the heavy isotopes relative to terrestrial
abundances. 

From the arguments above, it appears that the observational situation
concerning $\Gamma$ at high redshift is uncertain. There are no stringent,
independent observations which constrain $\Gamma$ in our sample.
We therefore treat $\Gamma$ as unknown and explore what happens if
we vary it.

\begin{figure}[tbph]
\noindent \begin{centering}
\includegraphics[bb=77bp 78bp 455bp 727bp,clip,angle=-90,width=0.8\textwidth]{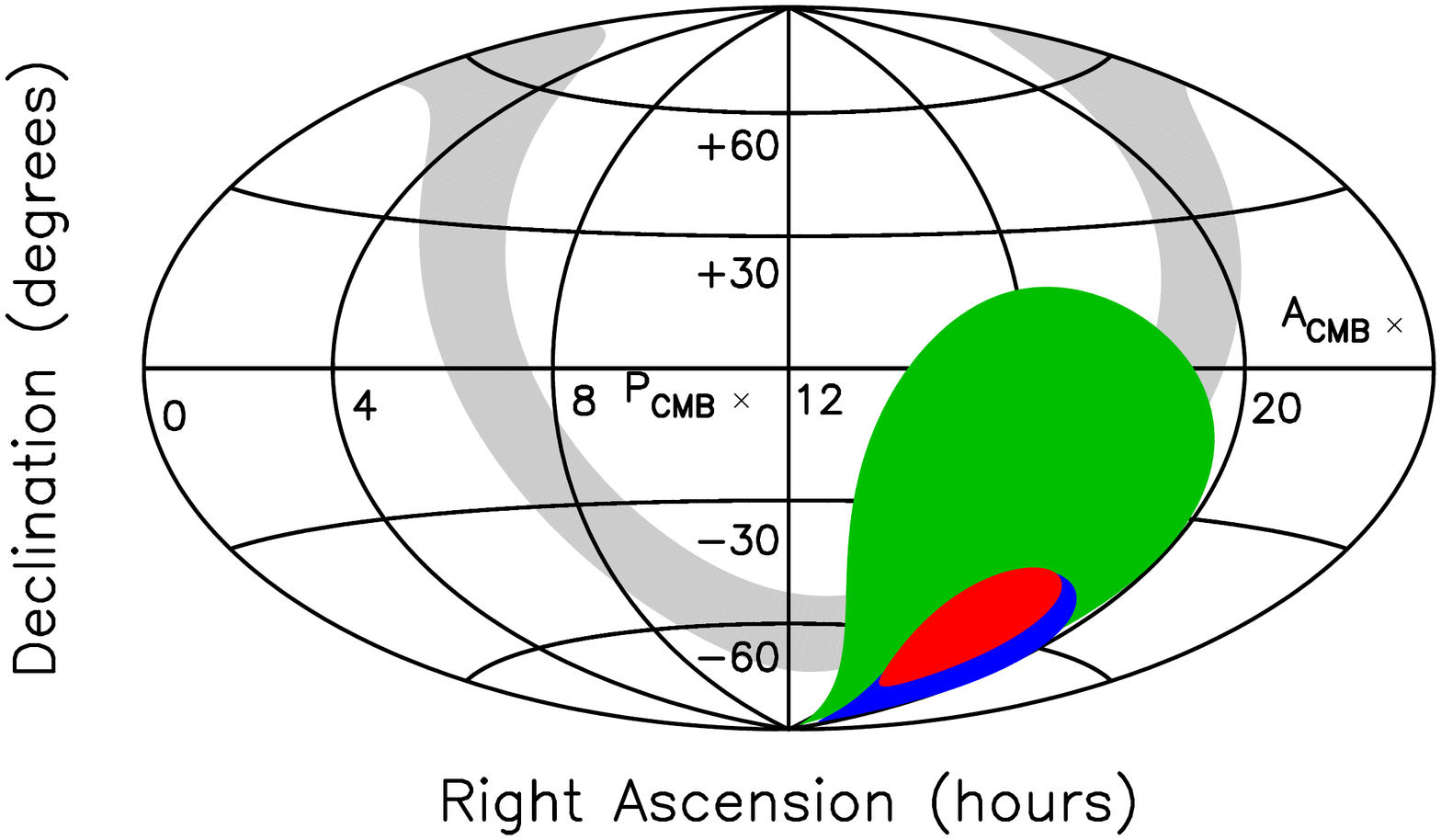}
\par\end{centering}

\caption[Sky map showing the impact of fitting with no heavy Mg isotope fraction on the VLT+Keck data]{$1\sigma$ confidence regions for the Keck (green), VLT (blue) and combined (red) dipoles in the circumstance where absorbers containing Mg are fitted with no $^{25}$Mg or $^{26}$Mg, to mimic the maximum possible effect of a lower heavy isotope fraction in the quasar absorbers compared to terrestrial values. Although the confidence regions are enlarged as a result of extra scatter introduced into the data, the reasonable alignment between the samples is still maintained. The separation between the Keck and VLT dipole vectors increases to $32^\circ$, which has a chance probability of 11 percent. \label{fig:alpha:skymap_noheavymg}}
\end{figure}

We first consider $\Gamma<\Gamma_{t}$. To place an upper limit on
the effect of $\Gamma<\Gamma_{t}$, we refit all the VLT absorbers
with no $^{25}$Mg or $^{26}$Mg, and similarly re-fit the absorbers
in \citet{Murphy:04:LNP} using no $^{25}$Mg or $^{26}$Mg. We give
the parameters for the fits to the Keck, VLT and combined samples
in this situation in table \ref{tab:asys:noheavymg}. The confidence
regions on the dipole location are shown in figure \ref{fig:alpha:skymap_noheavymg}.
Importantly, the dipole model remains statistically significant at
the $3.5\sigma$ level. The reduction in significance from $4.1\sigma$
is primarily due to extra scatter introduced into the $\Delta\alpha/\alpha$
values about the model. The extra scatter implies that the $\Gamma=0$
model is not a good model for the absorbers. Additionally, the monopole
becomes statistically significant at the $5.7\sigma$ level. Thus,
a lower heavy isotope abundance in the quasar absorbers is unable
to explain the dipole effect, and additionally increases the significance
of the monopole term. The increase in significance of the monopole
term mirrors the result in \citet{Murphy:03}. 

\begin{table}[tbph]
\centering{}\caption[Effect of removing $^{25}$Mg and $^{26}$Mg isotopes on the model $\Delta\alpha/\alpha = A\cos(\Theta) + m$]{Effect of removing $^{25}$Mg and $^{26}$Mg isotopes on the model $\Delta\alpha/\alpha = A\cos(\Theta) + m$ for the VLT, Keck and combined samples. Generally speaking the effect is to push $\Delta\alpha/\alpha$ to more negative values. The column ``$\delta A$'' gives $1 \sigma$ confidence limits on $A$. The column labelled ``significance'' gives the significance of the dipole+monopole model with respect to the monopole model. This also introduces extra scatter into the $\Delta\alpha/\alpha$ values about the dipole model, which implies (unsurprisingly) that fits with no heavy Mg isotopes are not a good representation of the absorbers. Despite the extra scatter, the dipole model is still significant at the $3.5\sigma$ level. Additionally, the monopole term here becomes significant at the $5.7\sigma$ level. $\sigma_\mathrm{rand}$ is given for the different samples; HC refers to the Keck high-contrast sample. \label{tab:asys:noheavymg}}\footnotesize%
\begin{tabular}{clcccccc}
\hline 
Sample  & $\sigma_{\mathrm{rand}}$ $(10^{-5})$  & $A$ ($10^{-5}$)  & $\delta A$ ($10^{-5}$)  & RA (hr)  & dec ($^{\circ}$)  & $m$ ($10^{-5}$)  & sig\tabularnewline
\hline 
VLT  & 1.04  & 1.20  & $[0.78,1.75]$  & $18.1\pm1.4$  & $-65\pm14$  & $-0.439\pm0.197$  & $1.9\sigma$ \tabularnewline
Keck  & 1.63 for HC & 0.42  & $[0.32,0.88]$  & $16.6\pm2.2$  & $-35\pm35$  & $-0.835\pm0.156$  & $0.4\sigma$ \tabularnewline
Keck+VLT & As above  & 0.98  & $[0.77,1.23]$  & $17.3\pm1.0$  & $-59\pm10$  & $-0.528\pm0.092$  & $3.5\sigma$ \tabularnewline
\hline 
\end{tabular}
\end{table}

We now consider the impact of increasing the Mg heavy isotope fraction
($\Gamma>\Gamma_{t}$). In section \ref{sub:alpha:s_monopole}, we
discussed the presence of a low-$z$ monopole in both samples, where
the difference between the two samples is remarkably small. Explaining
this result via alterations to the Mg isotope abundance would require
enrichment of the heavy isotope fraction relative to terrestrial values.
If we assume that all of the $z<1.6$ monopole is due to relative
enrichment of the heavy Mg isotopes, we can extrapolate from the $\Gamma=\Gamma_{t}$
and $\Gamma=0$ cases to estimate $\langle\Gamma_{z<1.6}\rangle$
using a simple linear model. A linear model may be used as the response
of $\Delta\alpha/\alpha$ to changes in $\Gamma$ is linear \citep{Murphy:04:LNP}.
This model assumes that the ratio of $^{25}$Mg/$^{26}$Mg is fixed.
For $z<1.6$, $m=(-0.390\pm0.108)\times10^{-5}$ for $\Gamma=\Gamma_{t}$,
and $m=(-0.884\pm0.115)\times10^{-5}$ for the case $\Gamma=0$. Under
our linear model, $\langle\Gamma_{z<1.6}\rangle\approx0.32$ in order
to make $m=0$. If we take $\sigma_{m}=0.108\times10^{-5}$ as a representative
error, this yields $\langle\Gamma_{z<1.6}\rangle=0.32\pm0.03$. 

In summary, variations in the magnesium heavy isotope fraction have
the potential to significantly impact the monopole component of the
angular dipole + monopole model, but cannot explain angular variations
in $\alpha$.

\section{Summary}

In this chapter, we have explored potential systematic effects to
determine whether they are able to cause the angular variation in
$\alpha$ described in chapter \ref{cha:alpha}. 

In section \ref{sec:asys:dv_test}, we used VLT and Keck spectra of
7 quasars to investigate whether inter-telescope wavelength-dependent
systematics exist which could manufacture the dipole effect. Although
we were unable to find a statistically significant common trend from
six spectral pairs, we applied an estimate of the possible wavelength
distortion to the VLT sample, and found that this reduced the statistical
significance of the VLT+Keck dipole from $3.9\sigma$%
\footnote{Calculated using a reference set.%
} to $3.1\sigma$. Importantly, this does not destroy the good alignment
between the VLT and Keck dipole directions, nor does it significantly
affect the alignment between dipole models fitted to $z<1.6$ and
$z>1.6$ sample cuts. We also investigated the significant distortion
present in the 2206$-$1958/J220852$-$194359 spectral pair. We showed
that a distortion of this type cannot apply to the whole sample. From
a combined analysis of all 7 quasars, we conclude that it is unlikely
that the combination of these two distortions in the appropriate proportions
is present in the data.

In section \ref{sec:asys:intraorder_distortions}, we examined the
potential impact of the intra-order wavelength distortions found by
\citet{Whitmore:10}, and concluded that they are unable to explain
the variation in $\alpha$ observed.

In section \ref{sec:asys:isotopic_abundances} we explored the effect
of variations in the Mg heavy isotope fraction, and showed that these
are unable to explain the observed dipole effect, but could explain
the apparent $z<1.6$ monopole in both the Keck and VLT data if the
quasar absorbers display an enriched heavy Mg isotope fraction relative
to terrestrial values.

We are thus unable to find any systematic effect which can explain
the observed angular variation in $\alpha$.

We cannot conclusively exclude the possibility that the detected angular
variation in $\alpha$ is the result of some unknown combination of
systematic effects. In section \ref{sub:alpha:Joint-probability}
we showed that the chance probability of getting as good alignment
as seen between the dipole vectors in both low- and high-redshift
sample cuts and between the Keck and VLT samples is $\approx0.1$
percent ($\approx3.3\sigma$). Thus, even if it is supposed that the
Keck $\Delta\alpha/\alpha$ results are systematically shifted to
more negative values through some unknown effect, and the VLT sample
shows no statistically significant variation in $\alpha$, one is
still left with a significant coincidence, or a conspiracy of subtle
systematic effects, or some unknown systematic effect in both telescopes
which is significantly correlated with sky position. Future observations
using a different telescope will help to rule out telescope-dependent
systematic effects, although if a systematic effect which knows about
declination (rather than zenith angle) exists which is common to multiple
telescopes it will be difficult to discover. We are unaware of any
mechanism which would cause $\Delta\alpha/\alpha$ to be specifically
correlated with declination in the same way in both telescopes.

\chapter{Further discussion on $\mu$ and $\alpha$\label{cha:Discussion}}

In this chapter we draw the work on $\mu$ and $\alpha$ from chapters
\ref{cha:mu} and \ref{cha:alpha} together, to discuss implications
which arise from the joint consideration of both sets of results.
We also discuss measurements of other dimensionless ratios.

\section{$\mu$ and $\alpha$ --- what have we learned?}

At first glance, chapter \ref{cha:mu} seems to suggest that $\Delta\mu/\mu=0$.
Certainly the low-$z$ ammonia results are extremely consistent with
with no change in $\mu$. However, chapter \ref{cha:alpha} seems
to reveal significant evidence for spatial variations in $\alpha$.
There are several ways of interpreting these results together:
\begin{enumerate}
\item The $\mu$ results are correct, and the $\alpha$ results are instead
the result of some unknown systematic. Assuming that the $\mu$ results
are correctly described by a weighted mean, then $|\Delta\mu/\mu|\lesssim3\times10^{-6}$
and $|\Delta\alpha/\alpha|\lesssim10^{-5}$. The MM method is relatively
resistant to systematic effects. The thorough investigation into potential
systematic effects by \citet{Murphy:03} was unable to find any systematic
which could explain the Keck results. Similarly, the investigations
in chapter \ref{cha:da systematic errors} are also unable to eliminate
variation in $\alpha$. This explanation is possible, but unlikely.
\item The $\mu$ results are incorrect, and the $\alpha$ results are correct.
It seems unlikely, although possible, that the ammonia results are
incorrect given the good wavelength calibration in the radio regime.
Similarly, the H$_{2}$ results are relatively resistant to systematics
on account of the large number of transitions used. An explicit analysis
of potential systematic errors for Q0528$-$250 shows that they are
relatively small. On the whole, this possibility also seems unlikely.
\item The $\mu$ results are correct and the $\alpha$ results are correct.
We note immediately that this possibility implies that $|\Delta\mu/\mu|\lesssim|\Delta\alpha/\alpha|$.
This is in conflict with the predictions given in section \ref{sub:mu_alpha_relation}
(which predict that $|\Delta\mu/\mu|\sim35|\Delta\alpha/\alpha|$
from many types of theories). Theory, of course, must be guided by
the data. We note that it is not known what the correct model for
grand unification is (or even if one exists), and so theories which
predict relationships between $\Delta\mu/\mu$ and $\Delta\alpha/\alpha$
must currently be characterised as speculative. Thus, any conflict
between theories and experiments at present tends to argue against
those theories rather than against the experiments. If both sets of
results are correct, this immediately suggests that we should be concerned
with the possible spatial variation of $\mu$.
\end{enumerate}

\subsection{Is there spatial variation in $\mu$?}

\index{proton-to-electron mass ratio!spatial variation in}If $\alpha$
varies spatially, then it seems natural to allow for spatial variation
of $\mu$ as well. It would seem natural that the spatial variation
of $\alpha$ should be tied to the spatial variation of $\mu$, although
this may not be the case. In this circumstance, we can try to look
for a dipole in the data for $\Delta\mu/\mu$, although clearly the
small number of results makes this difficult. 

\begin{figure}[tbph]
\noindent \begin{centering}
\includegraphics[bb=50bp 114bp 532bp 766bp,clip,angle=-90,width=1\textwidth]{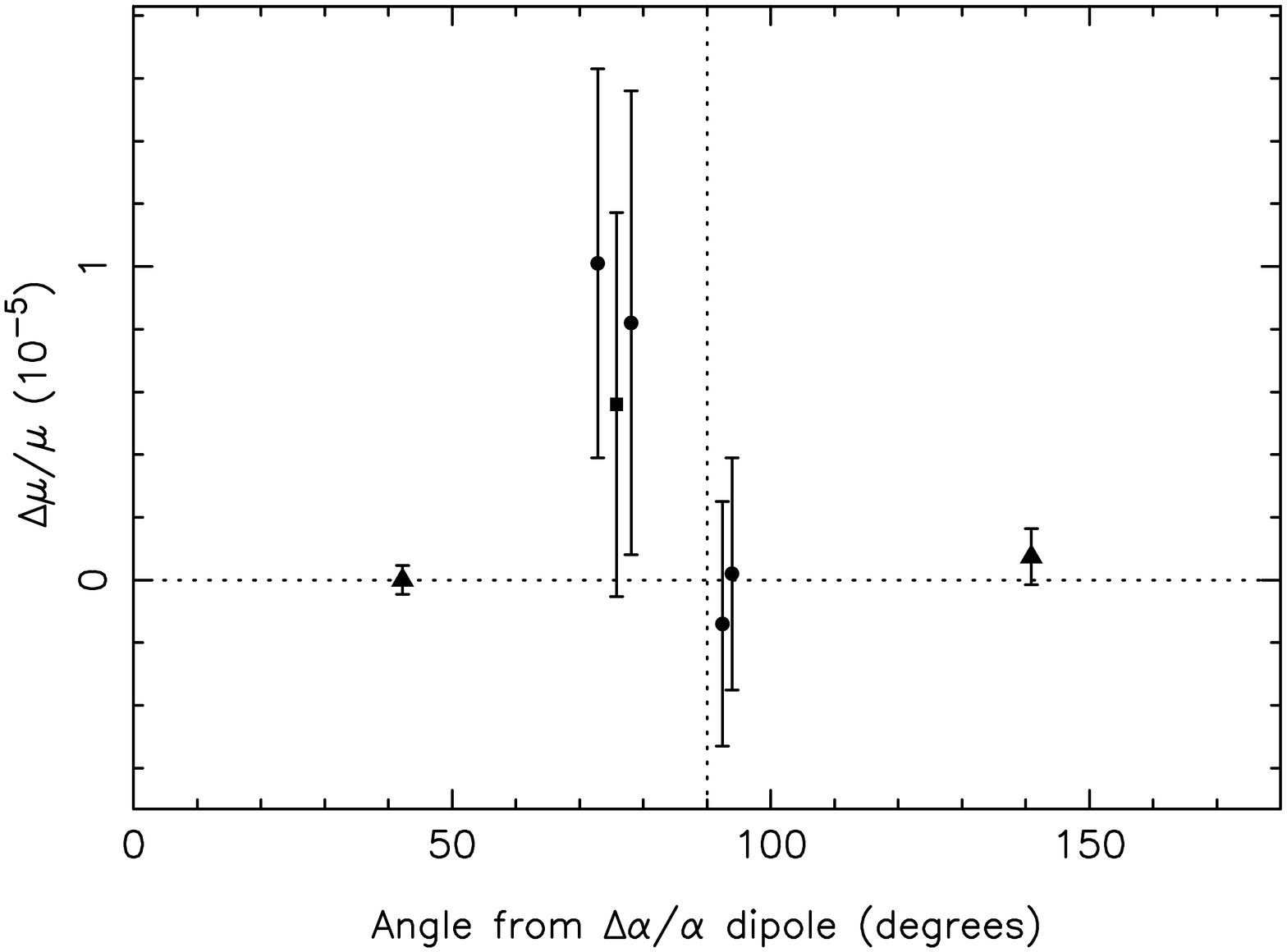}
\par\end{centering}

\caption[Extragalactic values of $\Delta\mu/\mu$ vs angle from the $\Delta\alpha/\alpha$ dipole]{Extragalactic values of $\Delta\mu/\mu$ vs angle from the $\Delta\alpha/\alpha$ dipole. The circles show the four H$_2$ results from chapter \ref{cha:mu}, the square shows the H$_2$ result from J2123$-$0050 produced by \citet{Malec:10} and the two triangles show the $z<1$ results obtained from the inversion transitions of NH$_3$ by \citet{Murphy:Flambaum:08} and \citet{Henkel:09}. The two Q0528$-$250 points are displayed slightly offset with respect to each other for clarity.\label{Flo:discuss:mu_vs_alphadipole}}
\end{figure}

In figure \ref{Flo:discuss:mu_vs_alphadipole} we show the extragalactic
constraints on $\Delta\mu/\mu$ from figure \ref{Flo:mu:all extragalactic results},
but instead plotted against angle from the $\Delta\alpha/\alpha$
$r$-dipole model from section \ref{sub:alpha:rdipole}. The two ammonia
constraints at $z<1$, which are located closer to the $\Delta\alpha/\alpha$
dipole poles than the equator ($\Theta=90^{\circ})$, seem to suggest
that there is no angular variation in $\mu$. However, the dipole
in $\Delta\alpha/\alpha$ manifests mostly at higher redshifts, and
therefore we would naturally expect ammonia results to show much less
variation than the H$_{2}$ results. 

From figure \ref{Flo:discuss:mu_vs_alphadipole} it is immediately
clear that the H$_{2}$ results are clustered near the $\Delta\alpha/\alpha$
dipole equator, and therefore sensitivity to any variation in $\mu$
(if it obeys a similar dipole relationship) should be reduced. Nevertheless,
the H$_{2}$ data could be consistent with a dipole having the same
direction as the $\Delta\alpha/\alpha$ dipole. The Q0528$-$250 points,
which are numerically closest to zero, also lie very close to the
equatorial region of the $\Delta\alpha/\alpha$ dipole, where no variation
would be expected. 

\begin{figure}[tbph]
\noindent \begin{centering}
\includegraphics[bb=71bp 92bp 556bp 742bp,clip,angle=-90,width=1\textwidth]{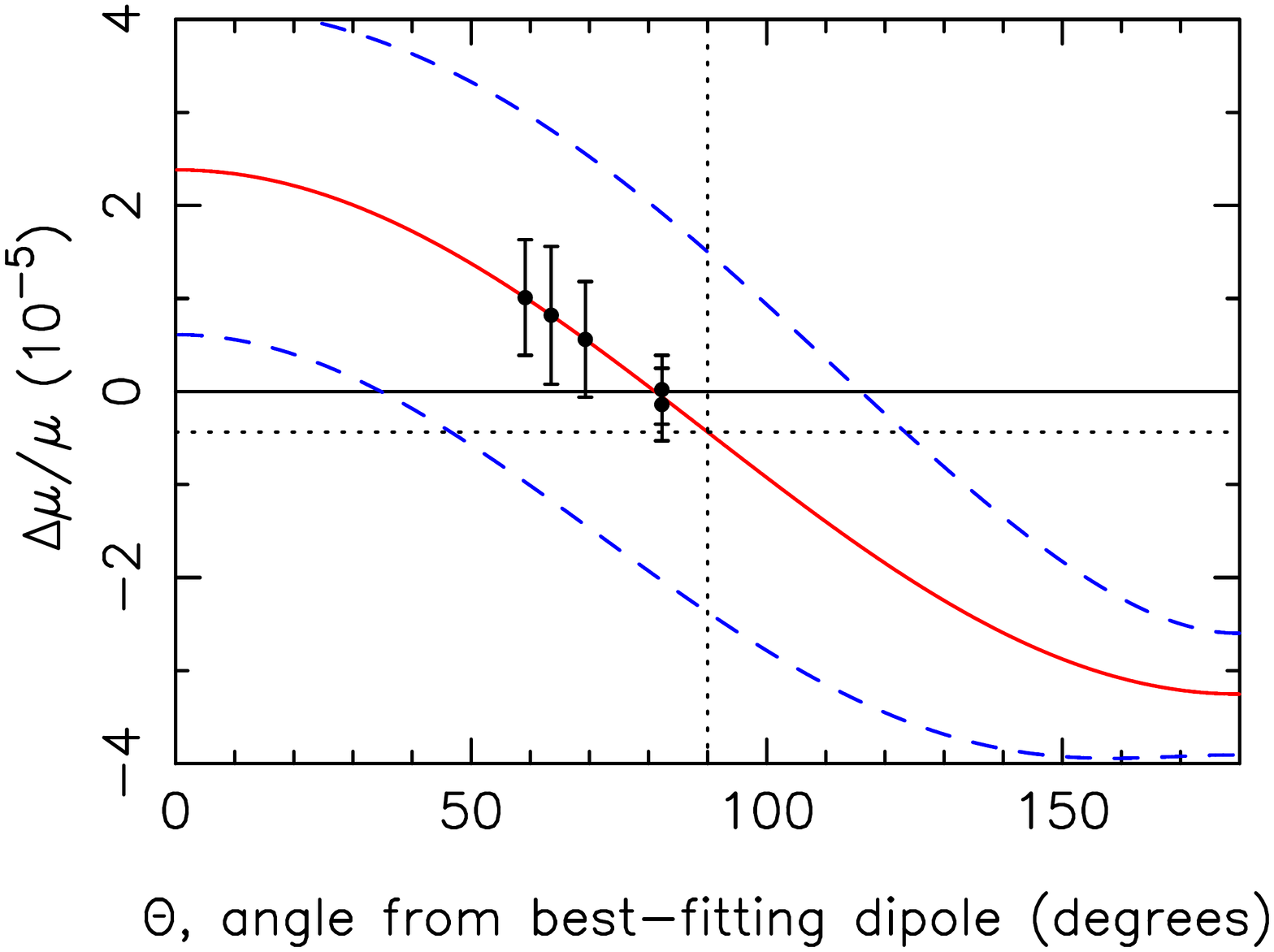}
\par\end{centering}

\caption[$\Delta\mu/\mu$ dipole model fitted to H$_2$ data]{$\Delta\mu/\mu$ dipole model fitted to the H$_2$ data, where $\Delta\mu/\mu = A\cos(\Theta)+m$. The horizontal dotted line shows the value of the monopole. The confidence limits (blue, dashed lines) are derived from the covariance matrix at the optimisation solution, and should be regarded as \emph{extremely} approximate given that the fit only has a single degree of freedom. \label{Flo:discuss:mu_dipolefit}}
\end{figure}

To investigate a $\mu$-dipole model explicitly, we apply a dipole
model to the H$_{2}$ data. For an angle-only model ($\Delta\mu/\mu=A\cos\Theta$+m),
we obtain: $A=(2.8\pm1.0)\times10^{-5}$, $\mathrm{RA=(20.0\pm4.6)\,\mathrm{hr}}$,
$\mathrm{dec=(-69\pm8)^{\circ}}$ and $m=(-0.43\pm0.90)\times10^{-5}$,
with $\chi_{\nu}^{2}=0.09$. Here, we give the error on $A$ simply
as the analytic standard error given the small sample size. This difference
between this dipole vector and that from the same model fitted to
the $\alpha$ data is $18^{\circ}$. For a distance-dependent model
($\Delta\mu/\mu=Br\cos\Theta+m$, where $r$ is the lookback time
distance to the absorbers), we obtain: $B=(2.6\pm1.0)\times10^{-6}\,\mathrm{GLyr^{-1}}$,
$\mathrm{RA=(18.7\pm5.7)\,\mathrm{hr}}$, $\mathrm{dec=(-68\pm5})^{\circ}$
and $m=(-0.27\pm0.77)\times10^{-5}$. Clearly the interpretation of
these results is hampered by the small sample size; with a 4-parameter
model, the fit only has a single degree of freedom. Nevertheless,
it is certainly intriguing that the fitted dipole points in a very
similar direction to the $\Delta\alpha/\alpha$ dipole. We show the
results of the fit to $\Delta\mu/\mu=A\cos(\Theta)+m$ in figure \ref{Flo:discuss:mu_dipolefit}.
One can calculate the statistical significance of the dipole model
over the monopole-only model using the method of \citet{Cooke:09},
to obtain $\approx2\sigma$. However, given the small sample size
we believe that the interpretation of this value is extremely limited.

Including the two ammonia constraints for a $r$-dipole model yields
the results $B=(1.5\pm0.5)\times10^{-6}\,\mathrm{GLyr^{-1}}$, $\mathrm{RA=(23.8\pm0.3)}\,\mathrm{hr}$,
$\mathrm{dec}=(-20\pm4)^{\circ}$ and $m=(-0.31\pm0.13)\times10^{-5}$,
with $\chi_{\nu}^{2}=0.36$. The two ammonia constraints place very
tight restrictions on the location of a dipole model of this form,
giving the very precise constraints on the dipole location. Obviously
these values are conditional on the correct specification of the model,
which in this case is far from certain. As for the H$_{2}$-only results,
the limited sample size impairs interpretation of these numbers. Nevertheless,
inclusion of the ammonia results destroys the good alignment seen
between the fitted $\Delta\mu/\mu$ dipole and the $\Delta\alpha/\alpha$
dipole.

The H$_{2}$-only results here are suggestive, but far from conclusive.
To determine whether spatial variation exists in $\mu$, a much larger
sample of measurements of $\Delta\mu/\mu$ will be needed at high
redshifts, preferably at $z>2$. If spatial variation in $\mu$ does
exist, and it occurs in tandem with variation in $\alpha$ such that
the relationship $\Delta\mu/\mu=R(\Delta\alpha/\alpha)$ holds, then
the combination of the H$_{2}$ $\mu$ data and the $\alpha$ data
would suggest that $R\sim3$, which contradicts the $R\sim30$ to
40  predictions made under various GUT and string-type models. If
all these data are correct, then we have ruled out an apparently quite
large range of unification theories. 

Our results for Q0528$-$250 yielded a statistical precision for $\Delta\mu/\mu$
of $\approx4\times10^{-6}$. If a dipole exists in $\Delta\mu/\mu$,
and it has an amplitude of $\approx3$ at redshifts of $2\lesssim z\lesssim3$,
then a H$_{2}$ absorber yielding the same precision on $\Delta\mu/\mu$
as Q0528$-$250 near the dipole axis might detect deviation from $\Delta\mu/\mu=0$
at the $\sim8\sigma$ level. Although H$_{2}$ absorbers are hard
to detect, this line of argument strongly suggests that future searches
for H$_{2}$ absorbers should preferentially target DLAs near the
$\Delta\alpha/\alpha$ pole and antipole.

\subsection{Is spatial variation of $\alpha$ consistent with experimental constraints?\label{sub:are_results_consistent}}

An immediate concern of the results of chapter \ref{cha:alpha} is
whether the results are consistent with other experimental constraints
on variation of $\alpha$. For instance, \citet{Murphy:03} found
for the Keck results that, if one assumes that the rate of change
of $\alpha$ is constant with time, then $\dot{\alpha}/\alpha=(6.40\pm1.35)\times10^{-16}\,\mathrm{yr^{-1}}$,
which is in conflict with the atomic clock constraints of \citet{Rosenband:08}
{[}$\dot{\alpha}/\alpha=(-5.3\pm7.9)\times10^{-17}\,\mathrm{yr}^{-1}${]}
by an order of magnitude. The conflict between these two results requires
that: \emph{i) }the Keck results are wrong, or; \emph{ii)} the atomic
clock results are wrong (which seems extremely unlikely), or; \emph{iii)
}the variation with time is not linear, or; \emph{iv) }the Keck results
are, at least in part, explained by spatial variation. We would note
that there is no known model which predicts that the variation in
$\alpha$ should be linear with time in the redshift range encompassing
both the quasar-derived results and the present-day constraints, and
so the importance of this apparent conflict is relatively low. The
results of chapter \ref{cha:alpha} also point to spatial variation
as the path to resolving this apparent conflict. \citet{Berengut:10c}
have compared the results of that chapter to existing evidence, and
found that our $\Delta\alpha/\alpha$ results are consistent with
all other experiments. We summarise their analysis here briefly. 

Atomic clock constraints should be able to detect spatial variation
of $\alpha$ given sufficient precision. \citeauthor{Berengut:10c}
considered the motion of the Earth with respect to the dipole axis;
because Earth's motion is not orthogonal to the dipole axis, atomic
clocks should in principle be able to detect the spatial variation
described in chapter \ref{cha:alpha} as a slow drift in $\alpha$
as a result of the Solar System's motion with respect to the dipole
axis, with an annual modulation due to the Earth's orbit around the
Sun. The drift would be seen as $\dot{\alpha}/\alpha_{\mathrm{lab}}=1.35\times10^{-18}\cos\psi\,\mathrm{yr}^{-1}$,
where $\psi$ defines the angle of the Solar System's motion with
respect to the dipole axis. From the results in chapter \ref{cha:alpha},
they calculated that $\psi=0.07$. They also calculate that the annual
modulation will have an amplitude $\delta\alpha/\alpha\approx1.4\times10^{-20}$.
Given the current best constraint on $\Delta\alpha/\alpha$ from atomic
clocks by \citet{Rosenband:08} at the $10^{-17}\,\mathrm{yr}^{-1}$
level, this implies that atomic clocks will need to improve by at
least two orders of magnitude to detect a spatial variation of this
sort. \citeauthor{Berengut:10c} noted the rapid improvement in the
precision of atomic clocks, and suggested that this precision may
be achievable.

We noted in section \ref{sub:Oklo} that the Oklo natural nuclear
reactor is sensitive to changes in $\alpha$, although changes in
$X_{q}$ dominate the change in the resonance level. One can assume
that $\Delta X_{q}/X_{q}=0$ to obtain the maximal possible constraints
on $\Delta\alpha/\alpha$ for comparison with our result. \citeauthor{Berengut:10c}
estimated the distance travelled by the Milky Way since the operation
of the Oklo reactor, about 1.8 billion years ago, to be $\sim3\times10^{6}$
light years. They noted that, based on the results of chapter \ref{cha:alpha},
this implies $\Delta\alpha/\alpha\sim10^{-9}$ across this distance.
However, \citet{Gould:06a} (for instance) claimed $-1.1\times10^{-8}<\Delta\alpha/\alpha<2.4\times10^{-8}$
under the assumption that $\Delta X_{q}/X_{q}=0$, which is not sensitive
enough by an order of magnitude. When $X_{q}$ is allowed the vary,
the constraint obviously worsens, and thus we can conclude that our
$\Delta\alpha/\alpha$ results are consistent with constraints from
Oklo.

\citeauthor{Berengut:10c} also considered constraints from the $\beta$-decay
of $^{187}\mathrm{Re}$ to $^{187}\mathrm{Os}$ obtained from meteorites\index{meteorites}.
One can translate measurements of the abundance of these species into
a constraint on the variation of $\alpha$. This requires the assumption
that the weak coupling constant $\alpha_{w}$ does not vary, however
\citep{Murphy:PhD,Uzan:03}. The measurements of the past decay of
$^{187}\mathrm{Re}\rightarrow{}^{187}\mathrm{Os}$ constrain the average
decay rate over the time since the meteorites were formed, 
\begin{equation}
\bar{\lambda}=\frac{1}{\Delta t}\int_{\mathrm{now}}^{\Delta t}\lambda(t)\,\mathrm{d}t.
\end{equation}
\citealt{Berengut:10c} concluded from measurements of $\bar{\lambda}$
\citep{Smoliar:96a} and $\lambda_{\mathrm{now}}$ \citep{Galeazzi:01a}
that the current experimental constraints on $(\bar{\lambda}-\lambda_{\mathrm{now}})/\lambda_{\mathrm{now}}$
are at the $10^{-2}$ level, whilst the results of chapter \ref{cha:alpha}
imply variation at the $\mbox{\ensuremath{\sim}}10^{-5}$ level. Thus,
our results are consistent with the meteorite results.

\subsection{Other observational tests for spatial variation in $\alpha$}

Spatial variation of $\alpha$ should in principle leave an imprint
on the CMB; \citet{Sigurdson:03a} consider this explicitly. Not only
is the mean power spectrum modified, but spatial variations in $\alpha$
induce ``higher order (non-Gaussian) temperature and polarization
correlations in the CMB'' \citep{Sigurdson:03a}. Unfortunately,
CMB constraints on $\Delta\alpha/\alpha$ are currently only at the
$\sim$ percent level. Depending on the mechanism of $\alpha$ variation
and how it scales with distance, the level of precision in CMB measurements
required to confirm spatial variation in $\alpha$ suggested by our
distance models in chapter \ref{cha:alpha} is unclear.

\subsection{The size of the habitable universe\label{sub:size of habitable universe}}

\index{habitable universe!size of}Traditionally, asking questions
about what might lie beyond the observable universe has been considered
metaphysics, as much of the discussion which follows is inevitably
non-falsifiable. Nonetheless, the 7-year Wilkinson Microwave Anisotropy
Probe (WMAP)\index{Wilkinson Microwave Anisotropy Probe (WMAP)}\index{cosmic microwave background (CMB)}
results give the spatial curvature, $\Omega_{k}$, as $\Omega_{k}=-0.080_{-0.093}^{+0.071}$,
where the slight preference for a closed model results from a degeneracy
with the Hubble constant \citep{Larson:10a}. Imposing further constraints
on $H_{0}$ through local distance scale measurements and adding in
the baryon acoustic oscillation (BAO) data yields $\Omega_{k}=-0.0023_{-0.0056}^{+0.0054}$
\citep{Komatsu:10a}, which is extremely consistent with a flat and
by implication infinite universe, unless the universe has non-trivial
topology (e.g.\ dodecahedral, see \citet{Luminet:03a}; c.f. \citet{Cornish:04a}).

Recall the discussion of the triple-$\alpha$ process from section
\ref{sub:triple-alpha process}, which suggests that the fine-structure
constant cannot vary by more than a few percent if we are to produce
appreciable quantities of $^{12}C$ or $^{16}O$. Under the extremely
strong assumption that something like local abundances of carbon and
oxygen are required for life (or at least, for carbon based life),
we can ask the question: how big is the habitable universe? If we
cannot observe spatial gradients in the fundamental constants, this
question is difficult to answer. However, observations of a spatial
gradient in the fine-structure constant would allow one to start to
speculate on an answer to this question (we agree that this is probably
non-falsifiable, but is extremely interesting nonetheless). In the
presence of a spatial dipole, one can make the strong assumption that
the dipole amplitude grows linearly along the dipole axis and extrapolate
until one is off the triple-$\alpha$ resonance. Obviously, in the
directions orthogonal to the dipole axis there is no constraint via
this mechanism, but one can easily take the size obtained from the
extrapolation as a lower limit boundary in all directions. We neglect
other effects induced by the variation of $\alpha$ for this simple
example.

In chapter \ref{cha:alpha} we presented evidence for a dipole in
$\alpha$ that is larger at larger distances, with an amplitude of
$A=1.1\times10^{-6}\,\mathrm{GLyr}^{-1}$ under the assumption that
the effect grows linearly with lookback time. In any event, the quasar
data probe most of the size of the observable universe, so to a first
approximation $\alpha$ changes by about 1 part in $10^{5}$ along
the radius of the observable universe toward the pole of the dipole.
If we assume a conservative figure that changing $\alpha$ more than
one percent makes carbon-based, oxygen-respiring life much less probable,
then under the argument outlined the radius of the habitable universe
is about 1000 times the radius of the observable universe. Taking
a sphere of this size as a lower limit under the argument in the previous
paragraph gives that there are about $\sim10^{9}$ observable universe
volumes in the habitable universe --- an extremely large number. 

Although this estimate is extremely rough, it is a demonstration of
how the variation of fundamental constants might suggest something
to us about the region beyond our observable universe, which otherwise
is\ldots{} unobservable. For those who are concerned that our observable
universe does not give enough room for life other than humans to emerge
by chance (a notion that we do not subscribe to), the extra nine orders
of magnitude from this calculation might give pause to reconsider
the possibility that there is life out there, somewhere.

\subsection{Implications for physics}

The dipolar variation in $\alpha$ presented in chapter \ref{cha:alpha},
if confirmed, would be a demonstration of new physics at the most
fundamental level. Importantly, it would directly demonstrate the
incompleteness of the Standard Model, which makes no allowance for
spatial variation in the fundamental constants. Additionally, it would
demonstrate that the Einstein Equivalence Principle is violated. The
combined impact on the Standard Model and General Relativity may assist
in attempts to unify these two pillars of twentieth century physics;
this unification of these theories is a problem for which a definitive
solution has proved elusive over the last few decades. 

Confirmed spatial variation of the fundamental constants would demonstrate
the existence of a preferred frame in the universe, which has significant
implications for cosmology. We explore some other claims for cosmological
anisotropy in the next section.

\section{Other evidence for dipoles \& a preferred cosmological direction}

The existence of a dipole in $\alpha$ would constitute a preferred
direction in the universe. The natural question to ask is: can such
an effect be seen in other data? The $r\cos\Theta$ dipole described
in chapter \ref{cha:alpha} points in the direction (RA, dec) $\sim(17.5\mathrm{hr},-62^{\circ})$
which is approximately $(l,b)=(330^{\circ},-15^{\circ})$ in galactic
coordinates. Below we consider other searches for preferred axes in
the literature.

\subsection{Bulk flows}

There have recently been claims for large scale bulk motions in the
universe\index{anisotropy!bulk flows}. \citet{Kashlinsky:09a} \citep[c.f.][]{Kashlinsky:09b}
presented the results of such an analysis. They used the Sunyaev-Zel'dovich
effect (SZ effect) \citep{Sunyaev:80a,Birkinshaw:99a} to measure
the line-of-sight peculiar velocity of clusters of galaxies in their
own frame of reference; the kinematic SZ effect is independent of
redshift \citep{Kashlinsky:09b}. For single cluster measurements
uncertainties are large --- of the order $\gtrsim1000\,\mathrm{km\, s^{-1}}$
per cluster \citep{Kashlinsky:09b}. However, with a sufficient number
of clusters and modern CMB measurements it is in principle possible
to determine whether a bulk flow exists. \citeauthor{Kashlinsky:09a}
analysed $\sim700$ X-ray clusters out to redshift $z\sim0.3$ and
the three-year WMAP data and found evidence for a bulk flow with amplitude
of $\gtrsim600\,\mathrm{km\, s^{-1}}$in the direction $(l,b)=(283^{\circ},11^{\circ})\pm14^{\circ}$
\citep{Kashlinsky:09a,Kashlinsky:09b}, or (RA,dec) = $(10.9\mathrm{h},-47^{\circ}$).
This is approximately $50^{\circ}$ from the dipole found in chapter
\ref{cha:alpha} on the sky.

The statistical significance of this result has been challenged by
\citet{Keisler:09a}, who argued that the result is due to correlations
between the CMB WMAP channels, and that the statistical significance
is more properly characterised at $0.7\sigma$. However, \citet{Atrio-Barandela:10a}
considered the error budget for \citet{Kashlinsky:09b} in detail
and claimed that the statistical significance is in fact $\gtrsim3$
to $3.5\sigma$. They note that the methods used to compute uncertainty
estimates have biases which cause errors to be over-predicted, and
also that if the bulk flow measurement is indeed caused by a systematic
error, it must ``have a dipole pattern, correlate with X-ray luminosity
and be present only at cluster positions''.

A less contentious measurement relates to the so-called Great Attractor\index{Great Attractor}
\citep{Lynden-Bell:88a}; there appears to be motion of the Local
Group of galaxies towards a gravitic source of extremely high mass
($M\sim5.4\times10^{16}M_{\mathrm{sun}}$) in the direction of the
Hydra/Centaurus constellations at $(l,b)=(307^{\circ},9^{\circ})$.
Further study attributed this to a significant overdensity of clusters
in that direction \citep{Raychaudhury:89a,Scaramella:89a}. More recent
study attributes 44\% of the motion of the Local Group to the Great
Attractor, with much of the remainder being attributed to the Shapley
Supercluster at about 700 Mpc in that direction \citep{Kocevski:06a}.

\subsection{Supernovae type Ia}

Measurements of type Ia supernovae (SN Ia)\index{anisotropy!supernovae type Ia}
have revealed that the expansion of the universe is accelerating \citep{Riess:98a,Perlmutter:99a,Astier:06}.
The current interpretation of this phenomenon is the existence of
a positive vacuum energy with negative pressure, which at present
is presumed to be the cosmological constant $\Lambda$. The existence
of anisotropy in the SN Ia data would imply anisotropic acceleration
of the universe, yielding a clearly preferred cosmological axis. \citet{Cooke:09}
used the Union compilation of SN Ia data to search for a dipolar anisotropy.
They found a (non-significant) $14\%\pm12\%$ increase in the acceleration
toward $(l,b)=(309^{\circ},43^{\circ})$, corresponding to (RA, dec)
= ($13.2\mathrm{h},-20^{\circ}$). This is approximately $60^{\circ}$
from the dipole described in chapter \ref{cha:alpha}. They note that
this is $31^{\circ}$ from the CMB dipole as seen from the sun, and
only $17^{\circ}$ as seen from the CMB frame of elliptical galaxies
with $v<2000\,\mathrm{km\, s^{-1}}$, where the CMB dipole is in the
direction $(l,b)=(311^{\circ},26^{\circ})$.

\subsection{CMB rings}

\citet{Kovetz:10a} examined the CMB temperature map and looked for
an axis around which ``giant rings''\index{anisotropy!CMB rings}
exist, and find such an axis in the direction $(l,b)=(276^{\circ},-1^{\circ})$
at a significance of $2.8\sigma$. This corresponds to (RA, dec) =
$(9.6\mathrm{h},-53^{\circ})$.

\subsection{Primordial deuterium abundance}

We noted in section \ref{sub:lithium_problem} that the abundance
of $^{7}\mathrm{Li}$ is sensitive to variation of the fundamental
constants. In fact, the abundance of all elements are sensitive to
variation in the fundamental constants, with differing degrees of
sensitivity. However $^{3}\mathrm{He}$ is measured only at $z\lesssim0.01$,
and $^{7}\mathrm{Li}$ only within our galaxy; only the deuterium
abundance has been measured at sufficiently high redshifts that spatial
variation in the fundamental constants might be probed \citep{Berengut:10b}. 

\citet{Berengut:10b} investigated\index{anisotropy!deuterium abundance}
the 7 constraints on the high-redshift deuterium abundance presented
in \citet{Pettini:08a} to see whether evidence for a dipole can be
found. They conclude that the data do not support a dipole model over
a monopole model on the basis of $\chi_{\nu}^{2}$, but note that
if one fits a dipole that the direction, $\mathrm{RA=(15.5\pm1.6)\,\mathrm{hr}}$,
$\mathrm{dec=(-14\pm51)^{\circ}}$, is consistent with the results
of chapter \ref{cha:alpha}. In galactic coordinates this is $(l,b)=(351^{\circ},34^{\circ})$.

\subsection{Combined analysis}

\citet{Antoniou:10a} reviewed different results which search for
a cosmologically preferred axis, and select six different types of
observations: Sn Ia data (from the Union 2 set), the CMB dipole, large
scale velocity flows (from various techniques), the anomalous alignment
of the CMB dipole\index{anisotropy!CMB low-order moments}, quadrupole
and octopole moments%
\footnote{This is the so-called ``axis of evil'' \citep{Land:05a,Land:07a}.%
} and large scale alignment in quasar optical polarisation data. They
gave the mean direction of the six axes considered as $(l,b)=(277^{\circ}\pm26^{\circ},44^{\circ}\pm27^{\circ})$,
which corresponds to (RA, dec) = $(11.6\mathrm{h},-15^{\circ})$.
Under simulations they argue that the probability of obtaining alignment
this good or better by chance is about 0.8\%. Excluding the CMB measurements,
the chance probability rises to about 7\%.

\subsection{What does this mean?}

\begin{table}[tbph]
\caption[Summary of other claims for cosmological anisotropy or preferred directions]{Summary of some claims for cosmological anisotropy or preferred directions, given in galactic coordinates $(l,b)$.}\medskip\label{tab:anisotropy_measurements}

\noindent \centering{}%
\begin{tabular}{lcc}
\hline 
Description & $l$ (degrees) & $b$ (degrees)\tabularnewline
\hline 
$r$-dipole from chapter \ref{cha:alpha} & 330 & -15\tabularnewline
\citet{Kashlinsky:09a} bulk flow measurements & 283 & 11\tabularnewline
Great Attractor & 307 & 9\tabularnewline
Supernovae type Ia & 309 & 43\tabularnewline
Primordial deuterium abundance & 351 & 34\tabularnewline
CMB rings & 276 & -1\tabularnewline
CMB dipole\footnotemark[1] & 264 & 48\tabularnewline
CMB quadrupole\footnotemark[1] & 240 & 63\tabularnewline
CMB octopole\footnotemark[1] & 308 & 63\tabularnewline
\hline 
\noalign{\vskip\doublerulesep}
\multicolumn{3}{l}{$^{1}$ {\small See references in \citet{Antoniou:10a}.}}\tabularnewline
\end{tabular}
\end{table}

We give a summary of the results described above in table \ref{tab:anisotropy_measurements}.
From these results, it seems reasonable to conclude that the alignment
between these various measures of anisotropy is suspicious, but far
from conclusive. The alignment between these phenomena may be due
to chance, or there may be some common cause. Another possibility
is that common systematics exist. Certainly, several of these phenomena
rely on CMB measurements, and thus common-mode systematics here would
be unsurprising, but this fails to explain, for instance, the reasonable
alignment with the (non-significant) SN Ia dipole. Ultimately, further
investigation is required to determine the importance of these phenomena,
and whether they they are related to the $\alpha$ dipole.

\section{Other dimensionless ratios}

\subsection{$\Delta G_{N}/G_{N}$}

Here, $G_{N}$ is the Newtonian gravitational constant. Measuring
$G_{N}$ is difficult due to the fact that gravity is weak compared
to the other three known forces. Nevertheless, it is certainly possible
to probe $\Delta G_{N}/G_{N}$ to better precision than our knowledge
of $G_{N}$. Clearly $G_{N}$ is not dimensionless; with $[G_{N}]=\mathrm{kg}^{-1}\,\mathrm{m}^{3}\,\mathrm{s^{-2}}$.
A more appropriate quantity to investigate is a gravitational fine-structure
constant, $\alpha_{g}\equiv G_{N}m_{p}^{2}/(\hbar c)$ \citep{Moss:10a}.
Note that $\alpha_{g}\approx6\times10^{-39}$, emphasising the weakness
of gravity relative to electromagnetism. The results given here are
in terms of $G_{N}$, which therefore must be interpreted as assuming
constancy of $\hbar$, $c$ and $m_{p}$. Note that $\hbar$, $c$
and $m_{p}$ can be used together to define a unit system with a unit
mass of $m_{p}$, a unit length of $\hbar/(m_{p}c)$ and a unit time
interval of $\hbar/(m_{p}c^{2})$.

Big Bang nucleosynthesis yields the constraint that $|\Delta G_{N}/G_{N}|\lesssim0.1$
from one second after the Big Bang \citep{Bambi:05a}. \citet{Kaspi:94a}
used biweekly timing observations of pulsar B1855+09 over a 7 year
period to obtain $\dot{G}_{N}/G_{N}=(-9\pm18)\times10^{-12}\,\mathrm{yr}^{-1}$.
\citet{Williams:04a} used the Lunar Laser Ranging experiment over
a 30 year baseline to achieve $\dot{G}_{N}/G_{N}=(4\pm9)\times10^{-13}\,\mathrm{yr}^{-1}$.

\citet{Bambi:08a} assumed that it may be possible to create a stable
strange star from a neutron star progenitor. The transition from hadronic
to quark matter should release extreme amounts of gamma ray energy
in a short timescale. If $G_{N}$ or $\Lambda_{\mathrm{QCD}}$ vary,
then a sufficiently large variation would cause some neutron stars
to transition to strange stars, thereby causing a gamma ray burst
(GRB). Under two strong assumptions that strange or hybrid stars exist
(and that not all compact stars are strange or hybrid stars), and
that the transition from hadronic matter to quark matter is a first
order transition, coupled with several auxiliary assumptions, they
concluded from the observed rate of long GRBs that $\dot{G}_{N}/G_{N}\lesssim10^{-17}\,\mathrm{yr}^{-1}$.

\subsection{Combinations of constants\label{sub:Combinations_of_constants}}

By comparing transitions with a totally different mechanism of generation,
one can constrain various combinations of fundamental constants. For
instance, comparison of millimetre transitions in CO and optical fine-structure
transitions constrains the quantity $F=\alpha^{2}/\mu$. These dimensionless
ratios often include $g_{p}$, which is the proton gyromagnetic ratio.
Here we present a selection of constraints on these combinations of
fundamental constants. The precision which can be obtained with these
combination of constants can be considerable, particularly given the
high precision and accuracy of wavelength measurements in the radio
domain. The downside for these measurements is that they require good
absolute wavelength calibration over a potentially very large wavelength
range; the many-multiplet method and the measurement of $\mu$ from
H$_{2}$ transitions requires only good relative wavelength calibrations. 

Detection of variation in one of these dimensionless ratios would
be extremely interesting, but the interpretation would require multiple
dimensionless ratios in order to relate the variation directly to
variation in $\mu$ or $\alpha$. The most appropriate way of investigating
these ratios would be to fit all of them simultaneously, thereby breaking
the degeneracy between the various fundamental constants. It seems
rational to require that the directional dependence of the different
constants should be the same, although other scenarios might be possible.
We leave this task to future work.

\subsubsection{$\Delta x/x$, $x\equiv\alpha^{2}g_{p}/\mu$}

Velocity differences between H \textsc{i }21cm absorption and optical
transitions constrain $x=\alpha^{2}g_{p}/\mu$. \citet{Srianand:10a}
used the recently detected absorption of 21cm and metal line absorption
at $z\approx3.174$ toward J133724+315254 to derive $\Delta x/x=(-1.7\pm1.5_{\mathrm{statistical}}\pm0.6_{\mathrm{systematic}})\times10^{-6}$.
However, the ThAr calibration exposures were not taken immediately
after the science exposures, and so there may be additional uncertainty
introduced due to wavelength miscalibration.

\citet{Kanekar:10a} also analysed H~\textsc{i} 21cm and C\textsc{~i}
absorption at $z\sim1.36$ and $z\sim1.56$ along the lines of sight
to Q2237$-$011 and Q0458$-$020 respectively. They found that $\Delta x/x=(+6.8\pm1.0_{\mathrm{stat}}\pm7.7_{\mathrm{max\, systematic}})\times10^{-6}$.
One can translate this constraint into a prediction for $\Delta\alpha/\alpha$
only if one has information about $\Delta\mu/\mu$ and $\Delta g_{p}/g_{p}$,
as $\Delta x/x=2(\Delta\alpha/\alpha)+(\Delta g_{p}/g_{p})-(\Delta\mu/\mu)$.
They use the results of \citet{King:08} (section \ref{sec:mu:results})
for $\Delta\mu/\mu$, and concluded that the Keck $\Delta\alpha/\alpha$
results are inconsistent with their findings and constraints on $\Delta\mu/\mu$
unless fractional changes in $g_{p}$ are larger than those in $\alpha$
and $\mu$.

However, we note that both of these absorbers lie close to the equatorial
region of the $\alpha$ dipole reported here ($\sim85^{\circ}$ and
$\sim115^{\circ}$ respectively). To compare our dipole model of $\alpha$
with the results of \citet{Kanekar:10a}, we need a prediction for
$\Delta\mu/\mu$ in the directions of Q2237$-$011 and Q0458$-$020,
along with the unjustified assumption that $\Delta g_{p}/g_{p}$.
We are reluctant to apply a dipole model for $\Delta\mu/\mu$ given
the small number of measurements available, and instead leave this
question to be answered when a larger sample of $\Delta\mu/\mu$ results
becomes available. Similarly, we are reluctant to calculate a constraint
on $\Delta g_{p}/g_{p}$ under the assumption that $\Delta\mu/\mu$
has no angular variation, given that we have detected apparent variation
in $\alpha$; if $\alpha$ varies across the sky, we cannot assume
that $\mu$ does not.

\subsubsection{$\Delta y/y$, $y\equiv\alpha^{2}g_{p}$}

The ratio of 21cm absorption to molecular rotational absorption constrains
the quantity $y\equiv\alpha^{2}g_{p}$. \citet{Murphy:01d} analysed
the $z=0.2467$ and $z=0.6847$ absorbers toward PKS 1413+135 and
TXS 0218+357 respectively. They gave $\Delta y/y=(-0.20\pm0.44)\times10^{-5}$
and $(-0.16\pm0.54)\times10^{-5}$ for the two systems respectively.

\subsubsection{$\Delta F/F$, \textmd{\textup{$F\equiv\alpha^{2}/\mu$}}}

\citet{Levshakov:10a} consider radial velocity differences between
galactic sub-mm- and mm-wave transitions in $^{13}$CO and the fine-structure
transitions in C \textsc{i} toward a variety of molecular clouds at
different galactocentric distances, namely TMC-1, L183, Ceph B, Orion
A/B and Cas A. They used existing radio data to constrain $|\Delta v|<110\,\mathrm{ms}^{-1}$,
which leads to $|\Delta F/F|<3.7\times10^{-7}$, with $F=\alpha^{2}/\mu$.
However, they noted that their results derive from statistical measurements
which fit single-component Gaussians to profiles which, in some absorbers,
display significant asymmetry. Although they give an $M$-estimate%
\footnote{See section \ref{sub:alpha:other robust methods} for the definition
of a $M$-estimate.%
} as a robust figure, the relatively low sample size (25 absorbers)
means that there is likely to be residual bias in their estimate,
and so their constraint is probably weaker by a moderate though unknown
amount. Although \citeauthor{Levshakov:10a} did not fit an angular
model to their results, they concluded that there is no spatial variation
in $F$; their definition of spatial variation is variation from terrestrial
values. We note that their results for Orion A/B have 9 negative values
of $\Delta v$ and 3 positive values. The chance of obtaining this
many negative values if $\Delta v=0$ based on a binomial estimate
is about 7\%. Thus, there may be residual systematics associated with
the clouds, or perhaps galactic variations in $F$ are very weakly
indicated. A more robust method would be to fit an angular variation
model to their results and determine whether angular variations in
$F$ exist.

\subsubsection{$\Delta J/J$, $J\equiv g_{p}(\alpha^{2}/\mu)^{1.57}$}

\citet{Kanekar:05a}%
\footnote{Note that in their paper the quantity $J$ is labelled $F$. We have
renamed it to avoid conflict with $F$ defined above.%
} compared the OH 18cm and H \iscs 21cm lines from the $z\sim0.765$
gravitational lens toward PMN J0134$-$0931, which constrain variation
in $J\equiv g_{p}(\alpha^{2}/\mu)^{1.57}$. They report that $\Delta J/J=(0.44\pm0.36_{\mathrm{stat}}\pm1.0_{\mathrm{sys}})\times10^{-5}$,
which is consistent with no change in $J$. We note that this absorber
is $\sim95^{\circ}$ from the pole of our dipole, and therefore minimal
variation in $\alpha$ would be expected from our results.

\subsubsection{$\Delta G/G$, $G\equiv g_{p}(\alpha^{2}\mu)^{1.85}$}

\citet{Kanekar:10b} have reported an observation of the satellite
OH 18cm lines at $z\sim0.247$ toward PKS 1413+135 (lookback time
$\sim2.9\,\mathrm{Gyrs}$). By combining results from the Westerbork
Synthesis Radio Telescope, and the Aricebo Telescope, they found that
$\Delta G/G=(-1.18\pm0.46)\times10^{-5}$ --- a $2.6\sigma$ detection.
They noted that the conjugate nature of the absorption and emission
lines provides a check on systematics, and by looking at the difference
between the optical depth in absorption and emission found no evidence
for systematic effects. We note that this absorber lies at $\sim85^{\circ}$
from our dipole pole, and therefore should be expected to show minimal
variation in $\alpha$ if our results are correct.

\section{Future avenues of research}

\subsection{$^{229}$Th nucleus optical transition}

\index{thorium nuclear optical clock}The $^{229}$Th nucleus has
the lowest known excited state of any nuclear transition --- a meagre
$7.6\pm0.5\,\mathrm{eV}$ above the ground state \citep{Beck:07a}.
The transition has not been measured directly, but instead determined
from differences of many $\gamma$-transitions to the ground level
and first excited state. The width of the level is estimated at $10^{-4}$
Hz \citep{Tkalya:00a}, explaining lack of a direct detection. As
the transition is extremely narrow, it in principle can be used as
the standard for a high accuracy clock \citep{Flambaum:09b}. The
transition is in the UV spectrum, and therefore can be excited in
principle with conventional lasers, although the experimental difficulties
in exciting a nuclear transition are considerable. 

This transition appears to be extremely sensitive to a change in fundamental
constants. A rough estimate by \citet{Flambaum:09b} gives 
\[
\frac{\Delta\omega}{\omega}\approx10^{5}\left(\frac{\Delta X_{q}}{X_{q}}+0.1\frac{\Delta\alpha}{\alpha}\right),
\]
which implies that $\Delta\omega\sim3\times10^{20}\times(\Delta X_{q}/X_{q})\,\mathrm{Hz}$.
With a width of $\sim10^{-4}\,\mathrm{Hz}$, this implies that one
could achieve sensitivities to $\Delta X_{q}/X_{q}$ of about 1 part
in $10^{24}$ per year, which is about ten orders of magnitude better
than the current constraints on the variation of $X_{q}$ \citep{Flambaum:09b}.
If such a clock could be built, the precision is several orders of
magnitude better than needed to detect the spatial variation of $\alpha$
implied by the results of chapter \ref{cha:alpha}; verification or
refutation of these results would be extremely rapid. 

\citet{Rellergert:09a} noted that, as the nucleus is well isolated
from the general environment, a thorium nuclear clock might be constructed
in the solid state (crystal) environment. Based on an analysis of
the crystal environment, they conclude that one second of photon collection
may yield a (systematic-limited) accuracy of $\Delta f/f\sim2\times10^{-16}$,
which is comparable with the precision available from present atomic
clock experiments over the course of a year.

\chapter{Markov Chain Monte Carlo methods applied to $\Delta\alpha/\alpha$\label{cha:MCMC}}

In this chapter, we set out to verify whether the Voigt profile fitting
program \textsc{\large vpfit} produces correct parameter estimates
and uncertainties for particular models. We briefly present the theory
behind \textsc{\large vpfit} in order to demonstrate both how parameter
values are estimated, and how uncertainties are derived. This also
serves to demonstrate the multiple potential points of failure for
an optimisation algorithm of this type. We then demonstrate the application
of Markov Chain Monte Carlo (MCMC) methods to show that, in the context
of simple $\Delta\alpha/\alpha$ fits, the estimates of $\Delta\alpha/\alpha$
produced by \textsc{vpfit} are good, and also that the associated
uncertainties on $\Delta\alpha/\alpha$ are reasonable.

\section{Introduction}

\subsection{Motivation}

\citet{Chand:2004} analysed 23 absorbers using VLT/UVES data, and
reported $\Delta\alpha/\alpha=(-0.06\pm0.06)\times10^{-5}$, which
appears to contradict the results of \citet{Murphy:04:LNP}, with
143 absorbers. Although the VLT/UVES data are generally of higher
signal-to-noise than the Keck data used by \citet{Murphy:04:LNP},
the statistical precision reported by \citet{Chand:2004} seems to
be too good when considering the differences in sample size. 

\citet{Chand:2004} modelled $\Delta\alpha/\alpha$ as an external
parameter to each fit, rather than including it as a free parameter
in each fit as we have done. In this approach, one steps through values
of $\Delta\alpha/\alpha$ and determines that value which minimises
$\chi^{2}$. A significant disadvantage of this method is reduced
speed, as one is not using gradient and curvature information of $\chi^{2}$
with respect to $\Delta\alpha/\alpha$ at a given point to locate
the $\chi^{2}$ minimum. Nevertheless, a plot of $\chi^{2}$ vs $\Delta\alpha/\alpha$
is instructive. Sufficiently near the $\chi^{2}$ minimum, the functional
form of $\chi^{2}$ implies that a plot of $\chi^{2}$ vs $\Delta\alpha/\alpha$
should be approximately parabolic. The uncertainty on $\Delta\alpha/\alpha$
can be determined by solving
\begin{equation}
\chi^{2}(\Delta\alpha/\alpha_{\mathrm{min}}+\sigma_{\Delta\alpha/\alpha})-\chi^{2}(\Delta\alpha/\alpha_{\mathrm{\min}})=1
\end{equation}
for $\sigma_{\Delta\alpha/\alpha}$ \citep{NumericalRecipes:92},
where $\Delta\alpha/\alpha_{\mathrm{min}}$ is the value of $\Delta\alpha/\alpha$
which gives the minimum $\chi^{2}$. In their paper, \citet{Chand:2004}
show plots of $\chi^{2}$ vs $\Delta\alpha/\alpha$, which demonstrate
fluctuations near the purported $\chi^{2}$ minimum that are much
larger than unity. This implies not only that the $\chi^{2}$ minimisation
algorithm is unlikely to have reached the true minimum, but also that
the uncertainty on $\Delta\alpha/\alpha$ has not been correctly determined.
\citet{Murphy:07-2} and \citet{Murphy:08} considered these issues
in more detail. Additionally, \citet{Murphy:08} demonstrated that
the statistical precisions quoted by \citet{Chand:2004} exceed the
theoretical maximum allowed by the spectral data and associated errors.
This suggests that the results of \citet{Chand:2004} are unreliable. 

\citet{Murphy:PhD} demonstrated that, over an ensemble of simulated
spectra, \textsc{vpfit} recovers the input value of $\Delta\alpha/\alpha$
on average, and that the mean $1\sigma$ uncertainty matches that
expected from the introduced noise. This strongly suggests that \textsc{vpfit}
is working correctly. However, this does not demonstrate that for
a \emph{particular} spectrum that \textsc{vpfit} produces good parameter
estimates and uncertainties. Although the results of \citet{Murphy:PhD}
seem robust, the results of \citet{Chand:2004} motivate us to attempt
a direct demonstration that \textsc{vpfit} is working as intended.
Markov Chain Monte Carlo methods, described below, allow direct exploration
of the likelihood function and parameter space considered in a reasonable
amount of time, therefore allowing us to verify whether or not the
output of \textsc{vpfit} is good.

\subsection{Optimisation theory\label{sub:Optimisation-theory}}

\index{VPFIT!theory}\index{optimisation!theory}When fitting a model
to data, $\chi^{2}$ minimisation techniques are widely used, which
minimise the quantity
\begin{equation}
\chi^{2}=\sum_{i=1}^{N}\frac{\left[f(\mathbf{x})_{i}-y_{i}\right]^{2}}{\sigma_{i}^{2}}\label{eq:chisq definition}
\end{equation}
where $f(\mathbf{x})_{i}$ is the prediction for the model at the
$i$th data point, $y_{i}$ is the observed value of the $i$th point,
$\sigma_{i}$ is the associated statistical uncertainty of that data
point and the sum is over $N$ points. The $\chi^{2}$ statistic is
a sufficient quantity to determine maximum likelihood parameter values
and uncertainties in the case of Gaussian likelihood. A wide variety
of methods are available to undertake this minimisation process. For
linear functions, explicit solutions exist, but otherwise an iterative
method must be applied from some starting guess at the parameters
$\mathbf{x}$. 

The most common methods utilised are Newton-type methods, which utilise
a local parabolic approximation to $\chi^{2}$ and then search towards
the projected minimum by some prescription. Sufficiently near a minimum,
we expect that $\chi^{2}$ is reasonably well approximated by a parabola.
Using the Taylor series expansion at some point $\mathbf{x}$ yields
\begin{equation}
\chi^{2}(\mathbf{x}+\mathbf{p})\sim\chi^{2}(\mathbf{x})+\mathbf{g}^{\mathbf{T}}(\mathbf{x})\mathbf{p}+\frac{1}{2}\mathbf{p}^{\mathbf{T}}\mathbf{G}(\mathbf{x})\mathbf{p}\label{eq:chisq taylor expansion}
\end{equation}
where $\mathbf{g}$ is the vector of first order partial derivatives
of $\chi^{2}$ with respect to the parameters (the gradient vector)
and $\mathbf{G}$ is the matrix of second order partial derivatives
(the Hessian matrix). If the parabolic approximation is exact, one
can find the exact minimum by finding $\mathbf{p}$ which minimises
\begin{equation}
\phi=\mathbf{g}^{\text{\textbf{T}}}(\mathbf{x})\mathbf{p}+\frac{1}{2}\mathbf{p}^{\text{\textbf{T}}}\mathbf{G}(\mathbf{x})\mathbf{p}.\label{eq:chisqmin phi equation}
\end{equation}
The stationary point (minimum) of this function can be obtained by
solving the set of linear equations
\begin{equation}
\mathbf{G}(\mathbf{x})\mathbf{p}=-\mathbf{g}(\mathbf{x}).\label{eq:chisqmin phi solution}
\end{equation}
In practice, the parabolic approximation is unlikely to be exact,
and thus one can search along the direction $\mathbf{p}$ for a lower
value of $\chi^{2}$ (the Gauss-Newton method\index{optimisation!Gauss-Newton method})
\citep{GMW:86}. Another possibility is to heuristically modify the
Hessian matrix, $\mathbf{G}$, to account for the imperfection of
the approximation by scaling the diagonal of the Hessian by a factor
$(1+\lambda)$ for different trial values of $\lambda$. The $\lambda=0$
case gives the case of equation \ref{eq:chisqmin phi solution}, whilst
the limit $\lambda\rightarrow\infty$ gives the steepest descent method,
which simply searches down the local gradient. As $\lambda\rightarrow\infty$,
the implied step size tends to zero \citep{GMW:86}. As the search
vector tends towards the local gradient descent direction, and the
step size tends to zero, a lower point will always be found unless
$\mathbf{x}$ is the solution or $\lambda$ is poorly chosen. This
method is the Levenberg-Marquardt method\index{optimisation!Levenberg-Marquardt method}
\citep{Marquardt:63,NumericalRecipes:92}. One then iterates the chosen
method until one cannot find a significantly lower point. For \textsc{vpfit},
this leads to a stopping criteria that the fractional change in $\chi^{2}$
between iterations must be smaller than some user-specified tolerance.
We have used $\Delta\chi_{\mathrm{tol}}^{2}=10^{-6}$. 

Previously, \textsc{vpfit} implemented the Gauss-Newton method, which
in most circumstances works very well. However, we found that for
the complicated molecular hydrogen fits, with thousands of free parameters,
full convergence did not seem to occur. Investigations showed this
to be due to a combination of the large number of parameters and the
presence of many parameters that were only moderately well or poorly
determined. We have modified \textsc{vpfit} to run a dual optimisation
process to overcome this. At each iteration, \textsc{vpfit} attempts
to take both a Gauss-Newton step and a Levenberg-Marquardt step, and
takes whichever of the steps produces a greater reduction in $\chi^{2}$.
We have found this algorithm to be very successful in producing apparent
convergence even for thousands of free parameters.

To implement either of the methods above, one requires knowledge of
the first and second order partial derivatives of $\chi^{2}$ with
respect to all the parameters. For the estimation of parameters from
quasar spectra, the model consists of a series of Voigt profiles.
Unfortunately, not only is the Voigt function non-analytic, but some
of the derivatives are also non-analytic. As a result, the Voigt function
is evaluated through numerical approximations, whilst derivatives
are calculated through the use of finite differencing methods on function
values. The first order partial derivatives with respect to a parameter
$x_{j}$ are given by
\begin{equation}
\frac{\partial\chi^{2}}{\partial x_{j}}=-2\sum_{i=1}^{N}\frac{\left[f(\mathbf{x})_{i}-y_{i}\right]}{\sigma_{i}^{2}}\frac{\partial f(\mathbf{x})_{i}}{\partial x_{j}},\label{eq:chisq_firstpartials}
\end{equation}
 whilst the second order partial derivatives are given by
\begin{equation}
\frac{\partial\chi^{2}}{\partial x_{j}\partial x_{k}}=2\sum_{i=1}^{N}\frac{1}{\sigma_{i}^{2}}\left[\frac{\partial f(\mathbf{x})_{i}}{\partial x_{j}}\frac{\partial f(\mathbf{x})_{i}}{\partial x_{k}}-\left[f(\mathbf{x})_{i}-y_{i}\right]\frac{\partial^{2}f(\mathbf{x}_{i})}{\partial x_{j}\partial x_{k}}\right].\label{eq:chisq_secondpartials}
\end{equation}

For even moderate numbers of fitted components, the computational
effort required to calculate the second order partial derivatives
becomes severe. However, the second term in equation \ref{eq:chisq_secondpartials}
contains the term $[f(\mathbf{x})_{i}-y_{i}]$. For a well-fitting
model, with large numbers of degrees of freedom, we expect that this
term has zero expectation value, as the model should predict $f(\mathbf{x})_{i}>y_{i}$
about as much as $f(\mathbf{x})_{i}<y_{i}$. Thus, when summed over
a large number of data points, the second order partial derivatives
of $\chi^{2}$ should be dominated by the first term in equation \ref{eq:chisq_secondpartials}.
Approximating the second order partial derivatives with only the first
term is known as the Gauss-Newton approximation, giving
\begin{equation}
\frac{\partial\chi^{2}}{\partial x_{j}\partial x_{k}}\approx2\sum_{i=1}^{N}\frac{1}{\sigma_{i}^{2}}\frac{\partial f(\mathbf{x})_{i}}{\partial x_{j}}\frac{\partial f(\mathbf{x})_{i}}{\partial x_{k}}.\label{eq:Hessian approximation}
\end{equation}
This approximation is much faster, as the second order partial derivatives
can be approximated through a combination of first order partials.
This approximation tends to be good in practice, and is used by \textsc{vpfit}.
Note that the Gauss-Newton approximation ensures that the Hessian
matrix is positive definite. This is useful --- if the Hessian matrix
is positive definite, then the Gauss-Newton direction is guaranteed
to be a descent direction (in the absence of numerical issues) \citep{NumericalRecipes:92}.
However, numerical issues may serve to push $\mathbf{p}$ sufficiently
far away from the true Gauss-Newton direction that it is no longer
a descent vector \citep{GMW:86}. Alternatively, inadequate searching
along the direction $\mathbf{p}$ may mean that a lower point is not
found even when one exists. 

Once a purported solution is reached, 
\begin{equation}
\mathbf{C}=\mathbf{G}^{-1}.
\end{equation}
gives the covariance matrix \citep{NumericalRecipes:92}. The square
roots of the diagonal terms of the covariance matrix correspond to
the estimated uncertainty on the various parameters \citep{Fisher:58}.

Note that, depending on the spectra being fitted and number of modelled
components, there may be many local minima for $\chi^{2}$ which do
not represent ideals fit to the data, or for which the fit is reasonable
but the values of some parameters are unlikely to be physically plausible.
This problem increases with increasing model complexity. Although
in the future automated methods may supplant human interaction, at
present good results are generally achieved faster through user-supplied
starting guesses.

\subsection{What can go wrong?}

Although the minimisation theory given above seems clear cut, in practice
the implementation may be severely affected by certain issues. These
include: general programming errors (bugs); errors in the calculation
of the partial derivatives, and; ill-conditioning due to model mis-specification.

\subsubsection{Programming errors}

Although it is never a topic one likes to consider, the possibility
remains that programming errors may cause the failure of an optimisation
algorithm to converge. To the extent that such errors existed in \textsc{vpfit},
our MCMC algorithm would detect them (see in particular section \ref{sub:MCMC convergence}).
However, our MCMC code has been created as an add-on to \textsc{vpfit},
and therefore to the extent that they use the same Voigt profile code
generation and other housekeeping routines, such errors will be common
to our analysis. Thus, our MCMC code allows us to demonstrate whether
or not failure of the optimisation algorithm has occurred, but does
not allow us to confirm the absolute correctness of the solution.
Nevertheless, \citet{Murphy:PhD} generated spectra using independent
Voigt profile code and found that \textsc{vpfit} recovered the input
values, suggesting that much of the ``back end'' of \textsc{vpfit}
works correctly and that our Voigt profile generator is reliable.

\subsubsection{Partial derivatives of $\chi^{2}$}

A key point of consideration is the calculation of the partial derivatives
of $\chi^{2}$ with respect to various parameters. Consider a quasar
which presents an unabsorbed continuum intensity $I_{0}(v)$. If an
absorbing cloud exists along the line of sight to Earth, the observed
intensity, $I(\nu$), is given by the convolution of the intrinsic
spectrum and the instrumental profile (IP) of the observing instrument
as
\[
I(\nu)=\Phi(\delta\nu)\otimes I_{0}(\nu)e^{-\tau(\nu)},
\]
where $\Phi(\delta\nu)$ is the instrumental profile and $\tau(\nu)$
is the optical depth of absorption due to the intervening cloud. $\chi^{2}$
is determined by the observed profile, and therefore the partial derivatives
of $\chi^{2}$ with respect to the parameters must account not only
for the non-analytic nature of the Voigt function but also the instrumental
profile. At present, the best way to obtain the partial derivatives
of $\chi^{2}$ with respect to the various parameters is using finite
differencing methods. 

With finite differencing methods, one must choose parameter step sizes
$h_{j}$ to evaluate the approximation
\[
\frac{\partial\chi^{2}(\mathbf{x})}{\partial x_{j}}\approx\frac{\chi^{2}(\mathbf{x}+h_{j}\hat{\mathbf{x}}_{j})-\chi^{2}(\mathbf{x})}{h_{j}}.
\]

If $h$ is too large, then the approximation will be poor because
the step size is too large. If $h$ is too small, the use of finite
precision calculations will result in substantial amounts of cancellation,
which will also render the approximation inaccurate. Poor choices
of $h$ will result not only in poor convergence but also in incorrect
uncertainty estimates through propagation into the covariance matrix.
If the accuracy to which $\chi^{2}$ is computed is $\epsilon_{f}$,
it can be shown that the optimal choice of $h_{j}$ is 
\[
h_{j}\sim\sqrt{\frac{\epsilon_{f}\chi^{2}(\mathbf{x})}{\left[\chi^{2}(\mathbf{x})\right]''}},
\]
where $[\chi^{2}(\mathbf{x})]''$ is the curvature scale of the function
\citep{NumericalRecipes:92}. To apply this formula, one would need
knowledge of the derivatives of the convolved function at each pixel,
which depends on the column density $N$, the dispersion parameter
$b$, the distance from the line centre and the instrumental resolution.
Rather than attempting to thoroughly investigate this 4D parameter
space, it turns out that fixed parameter step sizes work reasonably
well in most cases, provided that they are adequately chosen. \textsc{vpfit}
uses fixed parameter step sizes which have been chosen through experimentation
to yield good results, in that convergence seems to be reached and
the solution does not seem to be unduly affected by reasonable perturbations
to the starting guesses for parameters. In particular, we use $h_{z}=10^{-6}$,
$h_{\log_{10}N}=0.005$ (where $N$ is in atoms per cm$^{2}$) and
$h_{b}=0.1$ km/s. Our experience indicated in particular that values
of $h_{b}$ much smaller than this seemed to affect convergence. For
$\Delta\alpha/\alpha$ we choose $h_{\Delta\alpha/\alpha}=10^{-6}$,
which is always smaller than the statistical uncertainties we generate.
It is extremely different to optimise an arbitrary function in general,
and knowledge of the likely scale of the solution often yields insights
into how to construct a successful optimiser; the various values of
$h$ above relate to the typical scale of the uncertainties on parameters
in our Voigt profile models.

\subsubsection{Ill-conditioning}

\index{optimisation!ill-conditioning}Another possibility is that
the model is overspecified (``over-fitting''). For an over-fitted
model, the data will not discriminate adequately between some of the
parameters, leading to $\chi^{2}$ space being relatively flat in
relation to these parameters. This implies not only that strong covariances
are likely between these parameters, but also that the parameter uncertainties
will be large. This leads to two effects. Firstly, uncertainty estimates
on some parameters may be larger than they need otherwise be \citep{GMW:86}.
Perhaps more importantly, this can seriously affect the convergence
of the optimisation algorithm. This problem presents as ill-conditioning
of the Hessian matrix (and therefore the equation \ref{eq:chisqmin phi solution}).
The inverse of the Hessian matrix, can be written as $\mathbf{G}^{-1}=\mathbf{V\cdot}[\text{diag}(1/w_{j})]\cdot\mathbf{U}^{T}$,
where $\mathbf{U}$ and $\mathbf{V}$ are orthogonal square matrices
\citep{NumericalRecipes:92}; this is the singular value decomposition
(SVD) of $\mathbf{G}^{-1}$. The condition number of the Hessian matrix
is defined as the ratio of the largest to smallest $w_{j}$. As the
condition number increases, small perturbations in the inputs to equation
\ref{eq:chisqmin phi solution} will lead to large variations in the
solution. This tends to render the optimisation algorithm unstable.
A useful heuristic is that the number of significant figures lost
is equivalent to the base-10 logarithm of the condition number \citep{NumericalRecipes:92};
for double precision (with approximately 16 significant figures),
a condition number of $\sim10^{16}$ implies that no digits in the
solution of equation \ref{eq:chisqmin phi solution} are correct.
We observe that even for moderately simple problems, condition numbers
of $10^{3}$ to $10^{5}$ are common. For substantially overfitted
problems, condition numbers can reach $10^{10}$ or higher. Ill-conditioning
can push the search direction for the Gauss-Newton method arbitrarily
far from the true solution \citep{GMW:86}. This can lead to an inability
to find a lower search direction, causing premature termination of
the algorithm. The Levenberg-Marquardt algorithm is less susceptible
to this problem --- in the event of ill-conditioning, scaling the
diagonal of the Hessian reduces the condition number by forcing the
matrix to be diagonal-dominant. 

Note that, provided that $\mathbf{G}$ remains positive definite,
an optimisation algorithm \emph{should} continue to head downhill
until a solution is reached, even if substantial alterations to $\mathbf{G}$
are made \citep{NumericalRecipes:92}. This implies that parameter
uncertainties are much more likely to be affected by numerical problems
than the parameter estimates themselves, although numerical instabilities
of sufficient magnitude will also prevent convergence.

\subsection{Verification of the solution}

To address all the concerns above, one would like an independent method
of verifying not only that the purported solution is a local minimum
of $\chi^{2}$, but also that parameter uncertainties are realistic.
In principle, one could explicitly map out $\chi^{2}$ with respect
to all the model parameters, either exhaustively or through traditional
Monte Carlo methods to verify both the location of the minimum and
the curvature. Unfortunately, this problem becomes exponentially difficult
with the number of parameters --- the so-called ``curse of dimensionality''.
When modelling metal absorbers to investigate $\Delta\alpha/\alpha$,
one typically needs a few to tens of components, with several different
species, leading to a few to hundreds of parameters. When modelling
$\Delta\mu/\mu$ with H$_{2}$, models can easily reach thousands
of parameters. The number of parameters (high dimensionality) and
the time taken to evaluate the Voigt function conspire to render traditional
Monte Carlo methods useless. However, Markov Chain Monte Carlo methods
can be applied successfully to problems of moderate dimensionality
where traditional Monte Carlo methods cannot.

\section{Overview of the Markov Chain Monte Carlo method}

\index{Markov Chain Monte Carlo}A Markov chain is a series of points
for which the next point can be generated only with knowledge of the
current point. That is, any series of points which satisfies
\begin{equation}
\mathbf{x}_{i+1}=f(\mathbf{x}_{i})\label{eq:Markov property}
\end{equation}
satisfies the Markov property \citep{LiuMCMCBook}. The essential
idea of Markov Chain Monte Carlo (MCMC) methods is not to uniformly
sample some volume within which the target probability distribution,
$\mathrm{Pr}(\mathbf{x})$, is contained, but instead to construct
a Markov chain such that the stationary distribution of the chain
is the target distribution $\mathrm{Pr}(\mathbf{x})$. Each iteration
of the chain yields a sample from $\mathrm{Pr}(\mathbf{x})$. Typically,
one specifies a transition rule $T(\mathbf{x},\mathbf{x}'$) which
proposes a new point, $\mathbf{x}'$, from the current point, $\mathbf{x}$.
It turns out that the combination of the Markov property and the detailed
balance condition is sufficient to generate a chain whose stationary
distribution is the target distribution \citep{Metropolis1953,LiuMCMCBook}.
The detailed balance condition requires that the probability of jumping
from point \textbf{$\mathbf{a}$ }to point $\mathbf{b}$ is the same
as jumping from point $\mathbf{b}$ to point $\mathbf{a}$, or 
\begin{equation}
\mathrm{Pr}(\mathbf{a})T(\mathbf{a},\mathbf{b})=\mathrm{Pr}(\mathbf{b})T(\mathbf{b},\mathbf{a}).\label{eq:Detailed balance}
\end{equation}
Any Markov chain which is irreducible (that is, there is a non-zero
probability to move between any two points in the state space in a
finite number of steps), aperiodic and possesses an invariant distribution
will converge to the invariant distribution, $\pi$. For the Metropolis
algorithm (see below in section \ref{sub:Metropolis-algorithm}),
this is almost surely true \citep{Tierney:94}. Thus, even if the
algorithm is started in a region of low likelihood, it will eventually
converge to the desired distribution. We describe below in section
\ref{sub:MCMC convergence} why we think our algorithm should correctly
sample from the target distribution from the first iteration. 

Whilst naive (e.g.\ uniform) sampling of $\mathrm{Pr}(\mathbf{x})$
degrades exponentially with the number of parameters, one can construct
MCMC algorithms which degrade only polynomially with the number of
parameters%
\footnote{See section \ref{sub:Tuning g} for a justification of this.%
} --- the primary advantage of MCMC methods. However, because each
step in the chain depends on the previous step, successive steps will
be correlated. The degree of correlation is difficult to predict \emph{a
priori}, as it depends on the number of parameters, the target distribution,
the proposal distribution, and the degree to which the transition
rule $T(\mathbf{x},\mathbf{x}')$ is well tuned to the target distribution.
If the correlation is high, large numbers of steps will be required
to obtain the equivalent of one independent sample. Therefore, the
running time of MCMC algorithms is unknown at the start, and can only
be determined by examining the chain as the algorithm progresses.

\subsection{Applications of MCMC methods}

MCMC methods have found wide application in a number of fields. From
an astrophysical perspective, they have been applied to determine
posterior confidence regions from CMB data (for example, CosmoMC ---
\citet{Lewis:CosmoMC:02}, \citet{Slosar:03,Dunkley:05,Destri:08a,Destri:08b}),
for investigating CMB systematics \citep{Gold:10a}, for CMB model
selection \citep[see review by][]{Trotta:08a}, in exoplanet searches
\citep{Ford:07a,Balan:09a,Hrudkov:10a}, for analysis of exoplanet
atmospheres \citep{Madhusudhan:10a}, for investigation of dark energy
models \citep{Bozek:08,Wang:10a}, for investigating galaxy formation
models \citep{Henriques:09a,Lu:10a}, for investigating post-general
relativity models and testing general relativity \citep{Daniel:10a,Lombriser:10a},
for analysing Supernova type 1a data \citep{Gong:10a}, for analysis
of potential gravity wave data from the Laser Interferometer Gravitational-Wave
Observatory (LIGO) and other gravity wave observatories \citep{Robinson:08a,vanderSluys:09,Raymond:09a},
and for analysis of gamma ray bursts (GRBs) \citep{Gou:07}. In applications
which are directly relevant to the context of this work, \citet{Nakashima:08a}
applied MCMC methods to the 5-year WMAP data to constrain $\Delta\alpha/\alpha$
as $-0.028<\Delta\alpha/\alpha<0.026$ (using data from the Hubble
Space Telescope as a prior). Similarly, \citet{Wu:10a} applied MCMC
methods to the 5-year WMAP CMB data to constrain the change in the
gravitational constant to be $-0.083<\Delta G_{N}/G_{N}<0.095$. They
also place constraints on Brans-Dicke theories from the same data.
Clearly the utility of MCMC methods is high, and research into improving
the method is active.

\subsection{Aim of MCMC work}

Our aim is to verify that the purported solution of \textsc{vpfit}
is good and that parameter uncertainties are reasonable by sampling
from the likelihood function of our supplied model. Note that $\chi^{2}=-2\ln(L(\mathbf{x}))$,
where $L(\mathbf{x})$ is the likelihood function, up to an additive
constant which can be neglected, as we only ever consider differences
in $\chi^{2}$ for finding parameters. As such, we define the likelihood
function here as 
\begin{equation}
L=e^{-\chi^{2}/2}.\label{eq:Likelihood definition}
\end{equation}
As we typically have hundreds of degrees of freedom, a naive calculation
of $L$ will underflow in IEEE 754 floating point implementations.
To remedy this, $L$ can be calculated as $L=e^{-\chi^{2}/2+\chi_{\mathrm{min}}^{2}/2}$,
where $\chi_{\mathrm{min}}^{2}$ is the smallest $\chi^{2}$ value
under consideration. As only ratios of likelihoods or ratios of sums
of likelihoods are being considered, this extra factor will always
cancel. Another option is simply to work in $-2\ln L$, which avoids
this problem.

\subsection{Metropolis algorithm\label{sub:Metropolis-algorithm}}

\index{Markov Chain Monte Carlo!Metropolis algorithm}The Metropolis
algorithm \citep{Metropolis1953} is perhaps the simplest MCMC algorithm.
The Metropolis algorithm proposes a new position in parameter space,
$\mathbf{x}'$, based on the current position, $\mathbf{x}$, according
to some proposal function $T(\mathbf{x},\mathbf{x}')$. The only requirement
imposed is that 
\begin{equation}
T(\mathbf{x},\mathbf{x}')=T(\mathbf{x'},\mathbf{x})\label{eq:Metropolis reversibility}
\end{equation}
(that is, the proposal distribution is symmetric) \citep{LiuMCMCBook}. 

In principle, there area large number of possible proposal functions,
$T$, although in practice the most common choice is a multidimensional
Gaussian centred on the current point \citep{LiuMCMCBook}, such that
\begin{equation}
\mathbf{x}'=\mathbf{x}+gN(\mathbf{0},\mathbf{\Sigma}),\label{eq:Ndim Gaussian trial}
\end{equation}
where $\mathbf{\Sigma}$ is the covariance matrix obtained from the
optimisation algorithm at the purported best-fit solution, and $g$
is a scalar tuning factor. Note that the choice of $T$ influences
only the efficiency of the algorithm, not the formal correctness of
the solution \citep{Metropolis1953,Tierney:94}. To the extent that
the estimated covariance matrix is not a good approximation to the
true covariance matrix, the algorithm's performance will degrade.

The Metropolis algorithm generates a sequence of points, $\{\mathbf{x}^{t}\}$,
according to a two step prescription. First, from the current point,
$\mathbf{x}$, propose a new point, $\mathbf{x}'$, via $T(\mathbf{x},\mathbf{x}')$.
Then, calculate the ratio $q=L(\mathbf{x}')/L(\mathbf{x})$. Second,
with probability $\min(q,1)$ move to the new point (i.e.\ set $\mathbf{x}^{t+1}=\mathbf{x}'$).
Otherwise, retain the current point i.e.\ $\mathbf{x}^{t+1}=\mathbf{x}^{t}$.
In this fashion, proposed moves to a point which is more likely than
the existing point are always accepted, whereas moves to a point which
is less likely than the existing point are sometimes accepted, depending
on the ratio of the likelihoods. For a sufficiently large number of
iterations, the distribution of $\{\mathbf{x}^{t}\}$ will sample
from the underlying probability distribution up to a normalisation
constant. The probability distribution of each parameter is approximated
by the distribution of that parameter in the chain, $\{\mathbf{x}_{i}^{t}\}$.
The algorithm will spend most of its time in regions of high likelihood,
and little time in regions of low likelihood. It is for this reason
that MCMC outperforms traditional Monte Carlo methods in high-dimensional
parameters spaces --- for high-dimensional parameters spaces, most
of the hypervolume is located away from the region of interest, and
therefore uniformly sampling from a region will generally sample in
regions of low likelihood, whereas for MCMC samples are necessarily
concentrated near regions of high likelihood. The algorithm must be
tuned to ensure reasonable running times --- this is described below
in section \ref{MCMC:Speeding convergence}. Note that moving to the
new point only if $q\geq1$ turns the Metropolis algorithm into a
stochastic optimiser.

For reasons described below in section \ref{sub:Multiple Try Metropolis},
we implement a variant of the Metropolis algorithm known as the Multiple-Try
Metropolis method. 

The Metropolis-Hastings algorithm \citep{Hastings:70} generalises
the Metropolis algorithm to allow non-symmetric proposal functions.
However, in most cases it is not clear why a non-symmetric proposal
function should outperform a symmetric one. In any event, we expect
that the our likelihood function should be approximately symmetric
near the likelihood maximum, in which case a symmetric distribution
function seems reasonable.

\subsection{Convergence \& sampling efficiency\label{sub:MCMC convergence}}

There are two key concerns for MCMC algorithms --- converging to the
target distribution from the initialisation point (reaching stationarity),
and obtaining sufficient numbers of samples when stationary. 

In our case, the first concern relates to the fact that the algorithm
must start near the likelihood maximum. If the algorithm is started
away from the likelihood maximum, it will eventually converge to the
stationary distribution, but the time required for this is unknown.
Stationarity can be determined by inspecting the chain of samples
to see if the long term average of parameters differs significantly
from their starting values. Standard practice is to discard a certain
number of samples from the start of the chain to allow for ``burn-in''
(the exact number must be determined from the observed behaviour of
the chain). However, we start the algorithm with parameters set to
those which give a purported optimal solution from the \textsc{vpfit}
optimisation. In this case, the parameters should already be at or
near the maximum likelihood position, and so burn-in should be unnecessary.
This assumption can be verified by inspection of the chain, and we
find that our parameter values do not wander appreciably from their
starting values. 

The second concern above relates to the fact that successive samples
are correlated. For a traditional Monte Carlo estimator, the precision
of the mean of a set of samples is trivially given by $\sigma_{\bar{x}_{i}}=\sigma_{i}/\sqrt{N}$,
where $\sigma_{i}$ is the standard deviation of the samples for the
parameter of interest. If we define $\rho_{j}$ as the lag-$j$ autocorrelation
of the MCMC chain for some parameter (i.e.\ $\rho=\text{corr}[\{\mathbf{x}_{i}^{1}\},\{\mathbf{x}_{i}^{j+1}\}]$),
then \citet{LiuMCMCBook} gives 
\begin{equation}
\sigma_{\bar{x_{i}}}\approx\frac{\sigma_{i}}{\sqrt{N}}\sqrt{1+2\sum_{j=1}^{N}\rho_{j}}.
\end{equation}
If we define the integrated autocorrelation time as 
\begin{equation}
\tau=\frac{1}{2}+\sum_{j=1}^{N}\rho_{j},
\end{equation}
then 
\begin{equation}
\sigma_{\bar{x}_{i}}=\sigma\sqrt{\frac{2\tau}{N}}.
\end{equation}
The quantity $N/2\tau$ is commonly known as the effective sample
size \citep{LiuMCMCBook}. Typically, $\tau$ for even simple cases
we consider may be of order $\sim10^{2}$, meaning that many samples
are required to obtain the equivalent of a single independent sample.
Equivalently, one needs $\eta=2\tau$ samples to obtain the equivalent
of one independent sample. Although the presence of autocorrelation
increases the running time substantially, this problem is generally
outweighed by the ability to adequately sample the likelihood region
of interest.

By calculating $\tau$, one obtains a quantitative measure of the
convergence of the chain. Proposal functions which are poorly tuned
to the target distribution will eventually generate sufficient numbers
of samples from the target distribution, but this can take a prohibitively
long time. We deem a final run acceptable if $\tau$ is much smaller
than the chain length. The ideal situation is $\tau\sim1$, in which
case the chain will look like noise. The ideal circumstance is where
the chain appears stochastically invariant under random reordering
of the chain values. We describe a chain where $\tau\ll N$ as well-mixed.

\subsection{Speeding convergence \& reducing run-time\label{MCMC:Speeding convergence}}

\subsubsection{Acceptance rate\label{MCMC:Acceptance rate}}

Let us define the acceptance rate for a large number of steps as the
ratio of the number of accepted steps to the number of steps attempted.
If the algorithm takes steps which are generally much larger than
the scale of the target distribution, then the acceptance rate will
be $\sim0\%$, and the parameters will rarely change, leading to high
autocorrelations and therefore low sampling efficiency. On the other
hand, if the algorithm takes steps which are generally much smaller
than the scale of the target distribution, then the acceptance rate
will be $\sim100\%$, but it will take a long time to fully explore
the distribution. In this case, the parameters display random-walk
behaviour. It turns out that if both the target and proposal distributions
are Gaussian then the ideal acceptance rate in 1 dimension is 44\%
\citep{GRG95}, in the sense that this acceptance rate produces the
smallest autocorrelation time for the chain. It appears that if the
acceptance rate is slightly too low then efficiency is not too adversely
affected, whilst slight increases in the acceptance rate seem to confer
much worse performance penalties \citep{LiuMCMCBook}. Although naively
we might expect that an acceptance rate of about 50\% is ideal also
in higher dimensions, it surprisingly turns out that the optimal rate
for $k$ dimensions as $k\rightarrow\infty$ is 23.4\% \citep{Roberts:97a}.
Understanding the long-run acceptance rate requires large numbers
of samples. For our algorithm to be tuned such that the acceptance
rate was close to $23\%$ requires large amounts of time. As such,
we attempt to tune our algorithm such that the acceptance rate is
between 15\% and 40\%, and find that this rule works well in our cases.
The actual tuning is achieved by modifying $g$.

\subsubsection{Tuning $g$\label{sub:Tuning g}}

Consider a $k$-dimensional Gaussian proposal function (equation \ref{eq:Ndim Gaussian trial})
with a diagonal covariance matrix where all entries on the diagonal
are unity, and a target distribution consisting of a Gaussian with
some set of parameters and the same covariance matrix. The probability
of moving a radial distance $r$ is related to the $\chi$ distribution
and is given by 
\begin{equation}
P(r)\propto r^{k-1}e^{-r^{2}/2}.
\end{equation}
The $r^{k-1}$ term arises from the fact that the volume element $\mathrm{d}^{k}\mathbf{x}$
has a radial term of $r^{k-1}$ in hyperspherical coordinates of appropriate
dimension. This distribution is peaked at $r=\sqrt{k-1}$, and so
the most common step proposed will have length $\sqrt{k-1}$. However,
the target distribution only has a typical width of $\sim1$ along
any radial slice. This means that for large $k$ most steps will land
far from the likelihood maximum, meaning that almost all steps will
be rejected. This implies that we must scale the covariance matrix
of the trial distribution by $\sim1/\sqrt{k}$ in order to obtain
reasonable acceptance rates. If the target distribution is Gaussian,
and the proposal distribution is Gaussian, then the ideal acceptance
rate is achieved by setting $g=2.38/\sqrt{k}$ \citep{Roberts:97a}.
Thus, we initialised our algorithm with $g$ as $2.38/\sqrt{k}$ as
a first guess in order to hope to start with approximately good scaling.

To ensure that the tuning of $g$ is relatively optimal, before commencing
a large MCMC run, we conducted small runs of 250 iterations and then
compared the acceptance rate to the target rate range. If the acceptance
rate is too high, we increased $g$, and if the acceptance rate is
too low, we decreased $g$. The adjustment of $g$ was done automatically
by our algorithm. This process was then repeated until we a reasonable
acceptance rate was achieved. Samples obtained in this way were discarded.

\citet{Roberts:97a} also noted that the efficiency of the Metropolis
algorithm, compared to independent samples from the target distribution,
is approximately $0.3/k$. As we add more components, the computational
effort required to calculate the Voigt profiles increases as a low
power of the number of parameters. Certainly, one needs $n$ Voigt
profiles, but typically profiles with more components occupy larger
spectral regions, requiring evaluation of the Voigt function at $\mathcal{O}(k)$
points. Each Voigt profile has 3 parameters (i.e.\ $k\sim3n$), although
this may be reduced through parameter tying. In any event, this suggests
that the time required to evaluate the likelihood function scales
as $\mathcal{O}(k^{2})$. Combined with the result of \citet{Roberts:97a},
this implies that the approximate running time of our MCMC algorithm
scales as $\mathcal{O}(k^{3})$, thus justifying the earlier statement
that our algorithm degrades only polynomially with increased dimensionality.
A naive uniform Monte Carlo sampler, on the other hand, would have
running times that scale as $\mathcal{O}(c^{k})$ for some $c$.

\subsubsection{Covariance matrix re-estimation}

In order to try to ensure that the covariance matrix of our proposal
distribution is well suited to the target distribution, we ran our
MCMC algorithm multiple times (typically five to ten) with several
hundred thousand iterations per stage. After each stage, we re-estimated
the covariance matrix from the chain. With sufficient numbers of stages,
the covariance matrix should eventually converge on one which adequately
samples the target distribution. Although, as noted, the formal correctness
of the solution does not depend on $\mathbf{\Sigma}$, in practice
if $\mathbf{\Sigma}$ is badly tuned then the running time can become
unacceptably large. Performing this re-estimation process drastically
increases the chance that the final MCMC run will produce a good approximation
of the underlying probability distribution.

\subsubsection{Proposal distribution}

As a starting point, we assume that the distribution of parameters
is likely to be well approximated by a multidimensional Gaussian near
the likelihood maximum for large numbers of degrees of freedom. It
is well known, however, that the Voigt profile decomposition is not
unique. This means that, away from a particular local likelihood maximum,
we may discover multiple likelihood maxima (only one of which is a
global maximum), as well as likelihood ``shelves'', and other interesting
features. If any of these features occurs sufficiently close to the
solution returned by \textsc{vpfit}, they should be observable in
the MCMC chain. Note that the MCMC algorithm means that all of these
features will be eventually reached, but if areas of relatively high
likelihood are separated by large regions of low likelihood, the chance
of discovering other likelihood maxima in a reasonable time is exceedingly
small. 

If the target distribution is Gaussian, a Gaussian proposal distribution
will yield good performance, provided that the proposal Gaussian is
well tuned. However, if the tuning is bad initially, then sampling
may be slow. To remedy this, we use a heuristic radial proposal distribution
which has
\begin{equation}
P(r)=g\left(\frac{2}{3}r^{2}e^{-r^{2}/2}+\frac{1}{3}e^{-r}\right).\label{eq:MCMC proposal distribution}
\end{equation}
This is an admixture of the radial component of a 2D Gaussian and
an exponential distribution, similar to that used by CosmoMC%
\footnote{See http://cosmologist.info/notes/CosmoMC.pdf for more details on
this proposal function.%
} \citep{Lewis:CosmoMC:02}. The rationale for this is that if the
covariance matrix is poorly tuned, then large steps are rarely taken
on account of the $e^{-r^{2}/2}$ term. An obvious potential problem
is that initial guess for $\mathbf{\Sigma}$ is badly matched to the
true covariance matrix of the target distribution. Certainly, given
that we are trying to verify whether the parameter estimates provided
by \textsc{vpfit} are correct, we cannot assume that $\mathbf{\Sigma}$
is good. One typical problem is that one parameter uncertainty estimate
is bad (that is, the error estimate seems implausibly large), perhaps
because the data are overfitted, or because the evidence for that
component is weak. Suppose that an initial covariance matrix is supplied
where the uncertainty on one parameter is much larger than the true
uncertainty. In this event, the tuning factor $g$ will shrink until
this parameter is being reasonably well sampled, in order to achieve
a reasonable acceptance rate. However, this means that exploration
of other parameters will be very slow. As a result, re-estimation
of the covariance matrix is less likely to obtain a useful estimate
of $\mathbf{\Sigma}$, as the exploration of the other parameters
will not have occurred in a reasonable time. The exponential factor
above helps to remedy this, by occasionally taking large steps. Additionally,
for a $k$-dimensional Gaussian (where $k\geq2$) the probability
of taking small steps is minimal. This, again, means that the proposal
distribution must be well-tuned in order to achieve good exploration
of the parameter space. The admixture of the exponential yields a
non-negligible probability of taking small steps. The proposal function
given in equation \ref{eq:MCMC proposal distribution} will demonstrate
suboptimal performance if the target distribution is Gaussian and
the scaling of $\mathbf{\Sigma}$ is good. On the other hand, where
we have supplied covariance matrices that are poorly tuned, or where
the target distribution is non-Gaussian, this method appears to increase
the likelihood of the final MCMC run being useful. 

To generate trial steps, we generate a vector of perturbations $\mathbf{p}$
from a spherically symmetric distribution with radial probability
density $P(r)$ and then left multiply by $\mathbf{L}$ (where $\mathbf{L}\mathbf{L^{T}=\Sigma})$
so that the proposal has the correct covariance structure \citep{NumericalRecipes:2007}.
A new test point is thus given by $\mathbf{x}'=\mathbf{x}+\mathbf{Lp}$.

\subsection{Chain thinning}

For large problems, storing the entire chain of values can be problematic,
as this requires a matrix of size $Nk$, where $N$ is the number
of iterations and $k$ is the number of parameters. If the file is
stored as human-readable (i.e.\ a text file), the file can become
quite large for even moderate values of $k$. One solution to this
is to ``thin'' the chain. That is, one only retains only every $i$th
iteration of the chain. Provided that $i\ll\tau$, then one is is
not throwing away large amounts of useful information. Indeed, thinning
the chain has the effect of reducing the autocorrelation time as $\tau\rightarrow\sim\tau/i$.

\subsection{Multiple-Try Metropolis\label{sub:Multiple Try Metropolis}}

\index{Markov Chain Monte Carlo!Multiple-Try Metropolis}As noted
in section \ref{MCMC:Acceptance rate}, even for an optimal $\mathbf{\Sigma}$,
convergence slows down due to the need to take $g\sim1/\sqrt{k}$.
Although the acceptance rate will be reasonable, each step will typically
take a step of only $\sim\sigma_{i}/\sqrt{k}$ in each parameter,
requiring long running times to adequately explore the parameter space.
To attempt to combat this problem, we implement the Multiple-Try Metropolis
method (MTM) \citep{Liu2000,LiuMCMCBook}. Rather than attempting
a single step at each iteration, the MTM method tries many different
steps in order to try to better explore the local parameter space.
This is done in such a way as to maintain the detailed balance requirement
(equation \ref{eq:Metropolis reversibility}). The MTM proceeds as
follows. Firstly, draw $m$ independent trial proposals $\mathbf{y}_{1},\ldots,\mathbf{y}_{m}$
from a symmetric proposal function $T(\mathbf{x,\cdot})$. Compute
for each trial value $w_{j}=L(\mathbf{y}_{j})$. Now, from the trial
set $\mathbf{y}_{1},\ldots,\mathbf{y}_{m}$ select $\mathbf{y}$ with
probability proportional to $w_{m}$. Then, produce a reference set
by drawing $\mathbf{x}_{1}^{*},\ldots\mathbf{x}_{k-1}^{*}$ from the
distribution \textbf{$T(\mathbf{y},\cdot)$}. Let $\mathbf{x}_{k}^{*}=\mathbf{x}$
(which preserves the detailed balance requirement). Also, create weights
$w_{j}^{*}=L(\mathbf{x}_{j}^{*})$. Now accept \textbf{$\mathbf{y}$
}with probability 
\begin{equation}
r_{g}=\min\left(1,\frac{\sum_{j=1}^{m}w_{j}}{\sum_{j=1}^{m}w_{j}^{*}}\right).
\end{equation}
We refer to $m$ as the cloud size, as the algorithm generates a cloud
of points around the current point. Effectively, this algorithm allows
larger potential step sizes whilst still maintaining a reasonable
acceptance rate \citep{LiuMCMCBook}.

Our experience is that this method significantly reduces the required
running time by taking larger steps and by reducing the autocorrelation
length of the chain. Although at each point one must generate the
likelihood for $2m-1$ new points at any iteration (as opposed to
just 1 for the standard Metropolis algorithm), we find that this extra
computational burden appears to be more than offset in running time
by the use of the MTM algorithm \citep[similar to][]{Liu2000}. With
experimentation, we have found that $m=10$ worked well, in that the
chain autocorrelation time was much smaller. We noted that for $m\gg10$,
the autocorrelation time did not seem to decrease faster than the
computing time increased. Thus, $m=10$ appears in our case to be
a reasonable trade-off between exploring the parameter space around
the current point (local investigation) and the need to take large
numbers of steps to explore the parameter space fully (global investigation).
We note that because the trial points $\mathbf{y}_{1},\ldots,\mathbf{y}_{m}$
and then the points $\mathbf{x}_{1}^{*},\ldots\mathbf{x}_{k-1}^{*}$(after
$\mathbf{y}$ has been selected) are independent, the MTM method is
partially parallelisable. We have implemented this parallelisation
using the OpenMP extensions to \textsc{\large gfortran}.

\subsection{MCMC as a Bayesian sampler}

The MCMC method can be used to explore any probability distribution.
However, MCMC can also be used to directly estimate posterior probabilities
in the Bayesian framework with the appropriate choice of prior. The
likelihood ratio then becomes $L(\mathbf{x})\rightarrow L(\mathbf{x})\pi(\mathbf{x})$
where \textbf{$\pi(\mathbf{x})$ }is the Bayesian prior for a particular
set of parameters. For the column densities and redshifts of transitions,
we utilise improper flat priors. This method does not require normalised
priors. Given that we are interested for these purposes in parameter
estimation, we do not require normalised priors. We also use a flat
prior on $\Delta\alpha/\alpha$ as an agnostic position.

Our experience has been that for the $b$ parameters of transitions
with small $b$ (typically less than a few km/s) the algorithm tends
to propose many movements to $b<0$, which must be rejected (lines
with $b<0$ are unphysical). This is equivalent to imposing the improper
prior $\pi(b)=0$ for $b\leq0$ and $\pi(b)=1$ for $b>0$. In any
event, lines with $b/\sigma_{b}\lesssim1$ (where $\sigma_{b}$ is
the standard deviation of the $b$ parameter estimated from the covariance
matrix) cannot be Gaussian given that $b>0$. To remedy this, we heuristically
chose a flat prior for the logarithm of $b$, which tends to suppress
movement to small $b$. This is implemented by simulating $\log_{10}b$
rather than $b$, and then using a uniform prior in $\log_{10}b$.
We have found that this prior provides significantly better running
times. 

In principle, one could use observed distributions of parameters as
priors for the parameter estimates, however in most cases the statistical
constraints on our line parameters are good. In this event, the prior
is relatively flat across the region of interest of the likelihood
function, which means that parameter estimates will only marginally
be affected by this choice of prior. It is also for this reason that
our choice of a logarithmic prior $b$ does not lead to unnaturally
large estimates of $b$. The correct specification of priors is not
particularly important in our case because the data quality is high;
in this case, the likelihood is sufficiently restrictive to obtain
a normalisable posterior.

\subsection{Hard limits\label{sub:MCMC hard limits}}

Although we have implemented a uniform prior on the logarithm of the
$b$ parameters, this did not solve the problem associated with narrow
lines entirely. Whereas when fitting in $b$ the range of allowable
values is $b\in(0,\infty)$, when using $\log_{10}b$ the allowable
range is $\log_{10}b\in(-\infty,\infty)$. For transitions with $b$
much smaller than the instrumental resolution ($\sim$6 km/s for the
VLT data), there is effectively no change in $\chi^{2}$ for small
changes in $b$ on account of the convolution. This means that, for
example, $b\sim0.1$ km/s is effectively indistinguishable from $b=0.01$
km/s%
\footnote{$b$ cannot be arbitrarily small because of kinematic considerations,
but \textsc{vpfit} has no prior knowledge of cloud kinematics, and
thus in principle arbitrarily small $b$s may arise from the fitting
process. \textsc{vpfit} implements a user-adjustable lower limit to
$b$ to prevent it becoming too small.%
}. Two problems can arise from this. Firstly, our Voigt profile model
for each flux pixel was computed using a sub-binned profile for each
pixel, with some binning size. We have used $n_{\mathrm{bin}}=21$.
That is, the flux value for each pixel is calculated as the average
of the model evaluated at 21 points straddling the pixel such that
the bins are uniformly distributed between $\pm1/2$ a pixel. If $b$
becomes too small, however, this sub-binning will start to miss significant
amounts of flux, thereby rendering the profile generation incorrect.
Secondly, even if we had arbitrarily large numbers of bins, we would
observe that the distribution of $\log_{10}b$ is not normal because
of the convolution. That is, the likelihood is effectively flat for
$b_{i}\ll b_{\mathrm{IP}})$ (where $b_{\mathrm{IP}}$ is the width
of the instrumental profile), or $\log_{10}(b_{i})\lesssim0$ in this
case. Thus, for very narrow lines, $\log_{10}b$ is bounded above
but unbounded below, and $\log_{10}b$ will execute a random walk
toward $-\infty$. In finite precision algebra, this will eventually
cause zero underflow, rendering all subsequent iterations meaningless.
The only solution to this is to implement a hard boundary on $\log_{10}b$
at some point. We chose the boundary through experimentation to be
$b_{\mathrm{lim}}=1$ km/s. We trialed $b_{\mathrm{lim}}=0.1$ km/s,
but found that this significantly degraded the performance of the
algorithm. In future works, one can easily choose a smaller $b_{\mathrm{lim}}$
than we have, but we note simply that our choice of $b_{\mathrm{lim}}$
does not affect our conclusions.

\section{Application to quasar absorbers}

In this section, we present the application of our MCMC algorithm
to three quasar absorbers for the purpose of determining whether the
estimates of $\Delta\alpha/\alpha$ produced by \textsc{vpfit} are
reasonable. Redshifts given below all refer to the absorption redshift
of the system. In all three cases, we find good agreement between
the \textsc{vpfit} result and that produced by our MCMC code, although
the statistical uncertainties produced by our MCMC code are mildly
smaller than those produced by \textsc{vpfit}, indicating that \textsc{vpfit}
may be conservative. Interestingly, we note in the case of Q~0051$-$366
(section \ref{sub:MCMC:Q0551-366}) that even though some of the parameters
are clearly not Gaussian, $\Delta\alpha/\alpha$ is, and that the
\textsc{vpfit} result for $\Delta\alpha/\alpha$ accords well with
that derived from the MCMC method. We show the \textsc{vpfit} results
compared to our MCMC algorithm in table \ref{tab:MCMC results}, and
give commentary on each of the absorbers and results below. 

\begin{table}[tbph]
\caption[MCMC results for 3 quasar absorbers]{Comparison of purported values of $\Delta\alpha/\alpha$ calculated by \textsc{vpfit}, and the results of our MCMC algorithm. Quoted uncertainties are $1\sigma$. $\Delta\alpha/\alpha$ is given in units of $10^{-5}$.}\medskip\label{tab:MCMC results}

\noindent \centering{}%
\begin{tabular}{cccc}
\hline 
Object & $z_{\mathrm{abs}}$ & $\Delta\alpha/\alpha$(\textsc{vpfit) } & $\Delta\alpha/\alpha$(MCMC)\tabularnewline
\hline 
LBQS 2206$-$1958 & 1.018 & $-0.51\pm1.07$ & $-0.51\pm0.88$\tabularnewline
LBQS 0013$-$0029 & 2.029 & $-0.86\pm0.94$ & $-0.83\pm0.78$\tabularnewline
Q 0051$-$366 & 1.748 & $-0.80\pm1.08$ & $-0.89\pm0.84$\tabularnewline
\hline 
\end{tabular}
\end{table}

\subsection{LBQS 2206$-$1958 (J220852-194359) $z=1.018$}

\begin{figure}[tbph]
\noindent \begin{centering}
\includegraphics[bb=34bp 58bp 554bp 727bp,clip,angle=-90,width=0.8\columnwidth]{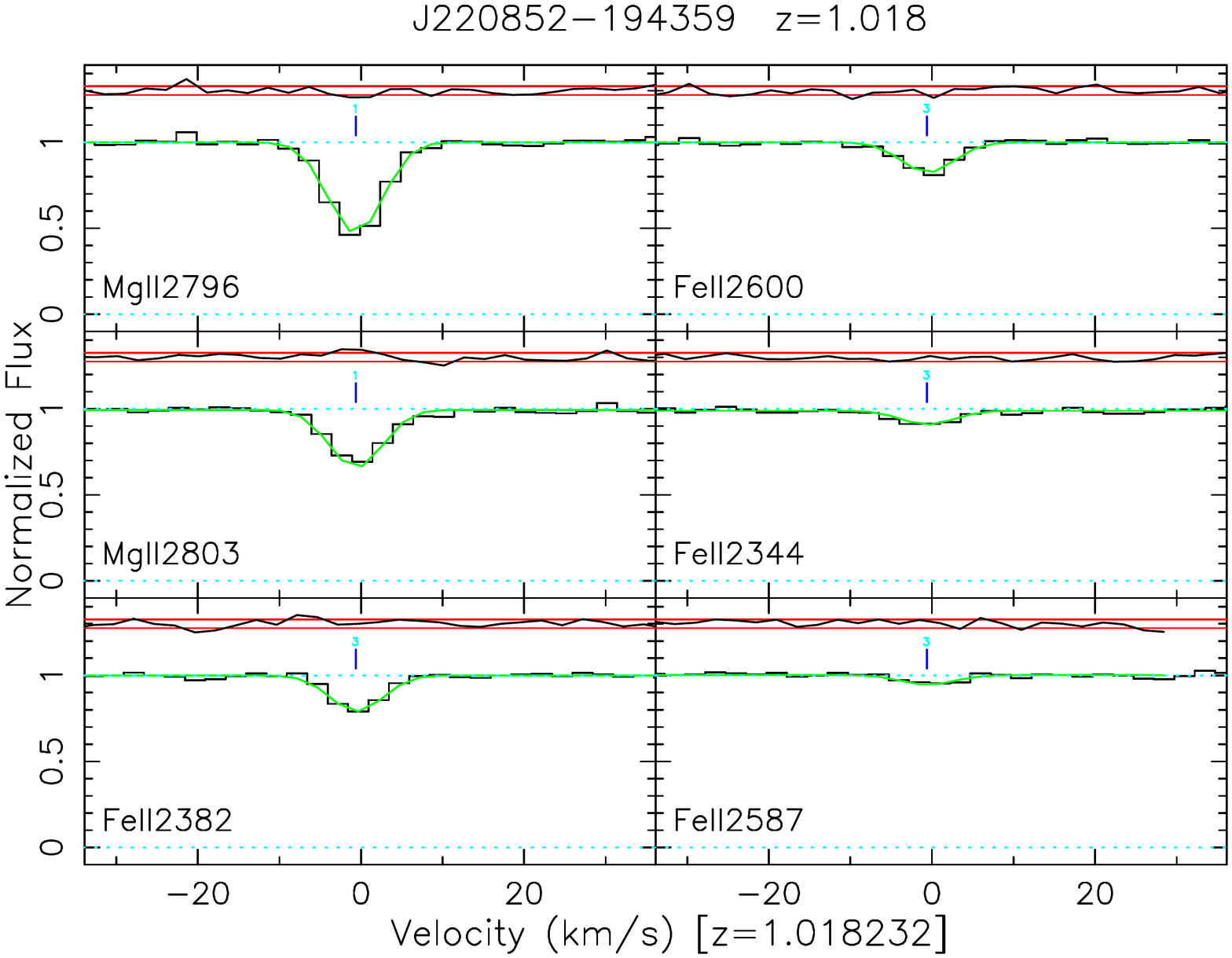}
\par\end{centering}

\caption[Fit for the $z=1.018$ absorber toward LBQS 2206$-$1958 (J220852$-$194359)]{Fit for the $z$=1.018 absorber toward LBQS 2206$-$1958 (J220852$-$194359) used for our MCMC analysis. \label{Flo:MCMC:LBQS2206-1958-fit}}
\end{figure}

This absorption system appears to be well fitted by a single Voigt
profile. We use the Mg~\iis\textsc{ $\lambda\lambda$}2796,2803\textsc{
}transitions, which are relatively insensitive to $\alpha$ variation,
and the Fe~ \iis $\lambda\lambda\lambda\lambda$ 2382,2600,2344,2587
transitions, which are strongly sensitive. We show the fit used for
this absorber in figure \ref{Flo:MCMC:LBQS2206-1958-fit} 

\begin{figure}[tbph]
\noindent \begin{centering}
\includegraphics[bb=0bp 0bp 438bp 658bp,clip,angle=-90,width=0.8\textwidth]{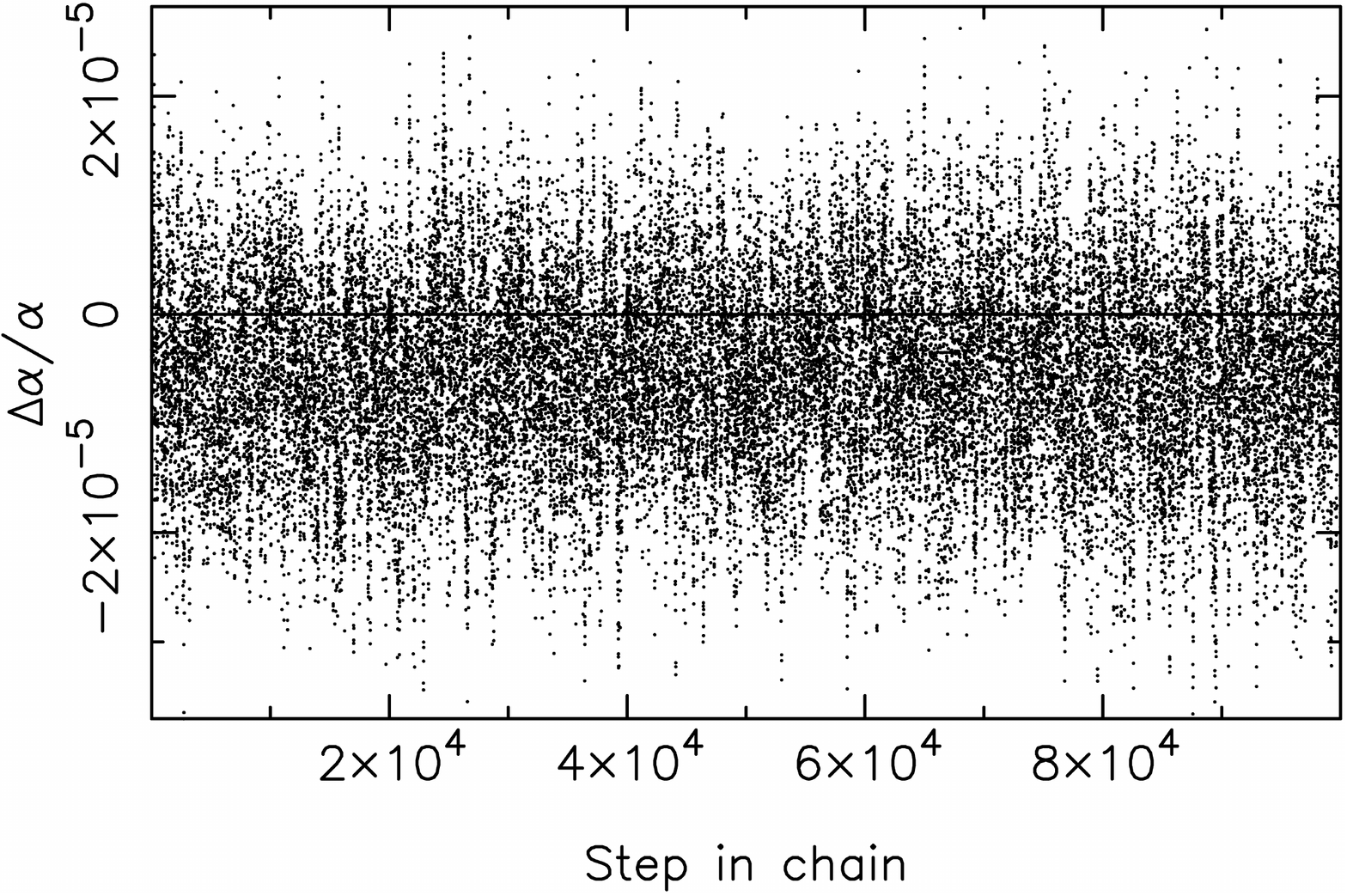}
\par\end{centering}

\caption[Chain values of $\Delta\alpha/\alpha$ for the $z=1.018$ absorber toward LBQS 2206$-$1958]{Chain of $\Delta\alpha/\alpha$ values for the $z=1.018$ absorber toward LBQS 2206$-$1958 (J220852$-$194359). There are no long range correlations visible by eye, suggesting the sampling is good. \label{Flo:MCMC:LBQS2206-1958-dachain}}
\end{figure}

\begin{figure}[tbph]
\noindent \begin{centering}
\includegraphics[bb=100bp 51bp 557bp 707bp,clip,angle=-90,width=0.7\textwidth]{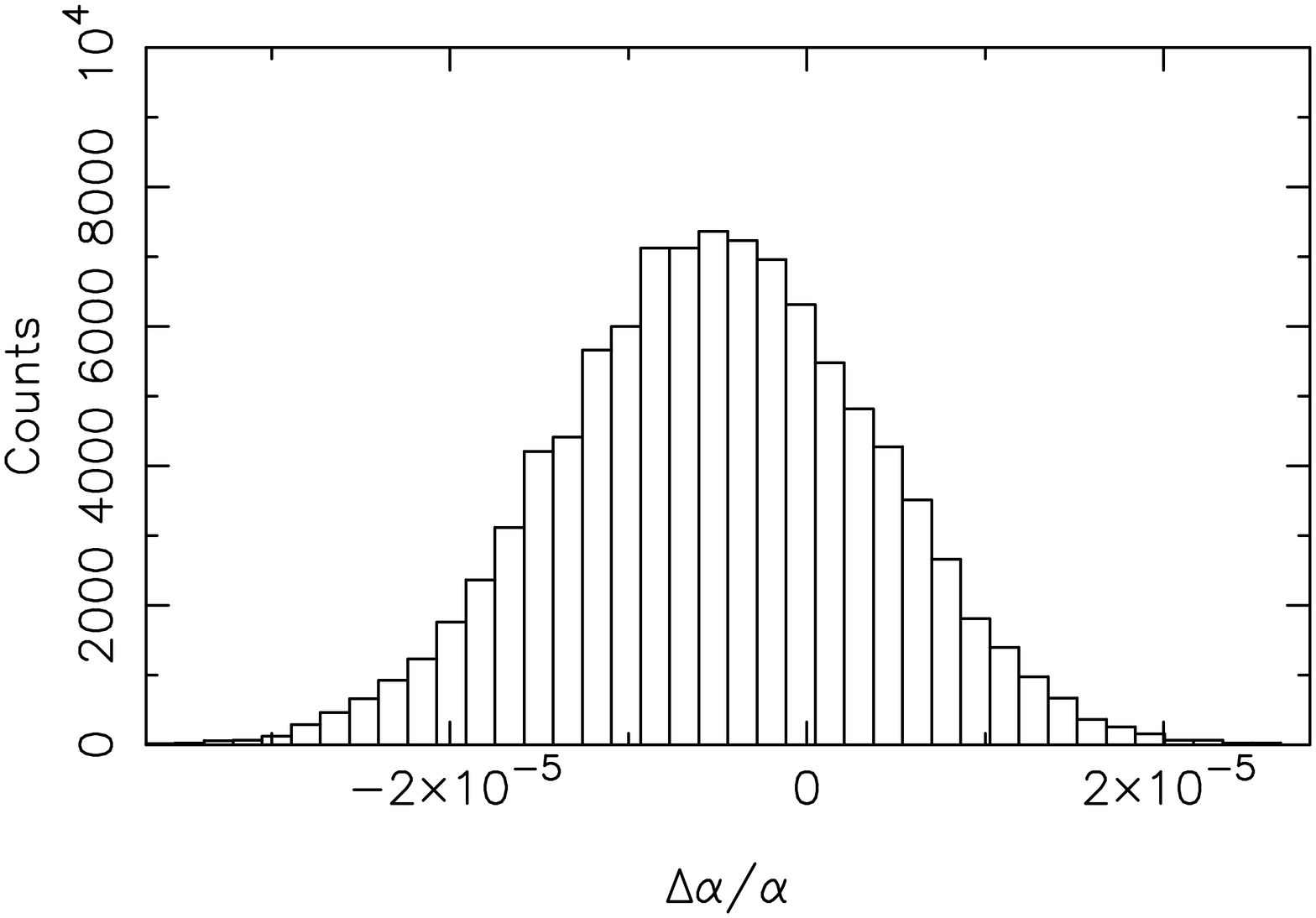}
\par\end{centering}

\caption[Histogram of $\Delta\alpha/\alpha$ for the $z=1.018$ absorber toward LBQS 2206$-$1958]{Histogram of $\Delta\alpha/\alpha$ chain values for the $z=1.018$ absorber toward LBQS 2206$-$1958 (J220852$-$194359). The resulting distribution appears to be well described by a Gaussian. \label{Flo:MCMC:LBQS2206-1958-dahistogram}}
\end{figure}

As only a single component is fitted, we naively expect that the chain
should be well-mixed even without re-estimation of the covariance
matrix. This is true, at least by eye. Nevertheless, we carried out
re-estimation of the covariance matrix in order to try to optimise
the efficiency of the final run. We show the chain of $\Delta\alpha/\alpha$
values in figure \ref{Flo:MCMC:LBQS2206-1958-dachain}. For the $\Delta\alpha/\alpha$
chain, $\eta=53$, which is much less than the chain length ($N=10^{5})$.
In figure \ref{Flo:MCMC:LBQS2206-1958-dahistogram} we show the histogram
of these chain values, which yields the distribution of $\Delta\alpha/\alpha$.
The distribution of $\Delta\alpha/\alpha$ is well approximated by
a Gaussian, and the mean value of the chain values corresponds extremely
well with that produced by \textsc{vpfit}. Interestingly, the standard
deviation of the chain values ($\sigma_{\mathrm{MCMC}}=0.88\times10^{-5}$)
is somewhat smaller the uncertainty estimate on $\Delta\alpha/\alpha$
returned by \textsc{vpfit} ($\sigma_{\mathrm{VPFIT}}=1.07\times10^{-5}$).

\begin{figure}[tbph]
\noindent \begin{centering}
\includegraphics[bb=0bp 0bp 436bp 658bp,clip,angle=-90,width=0.8\textwidth]{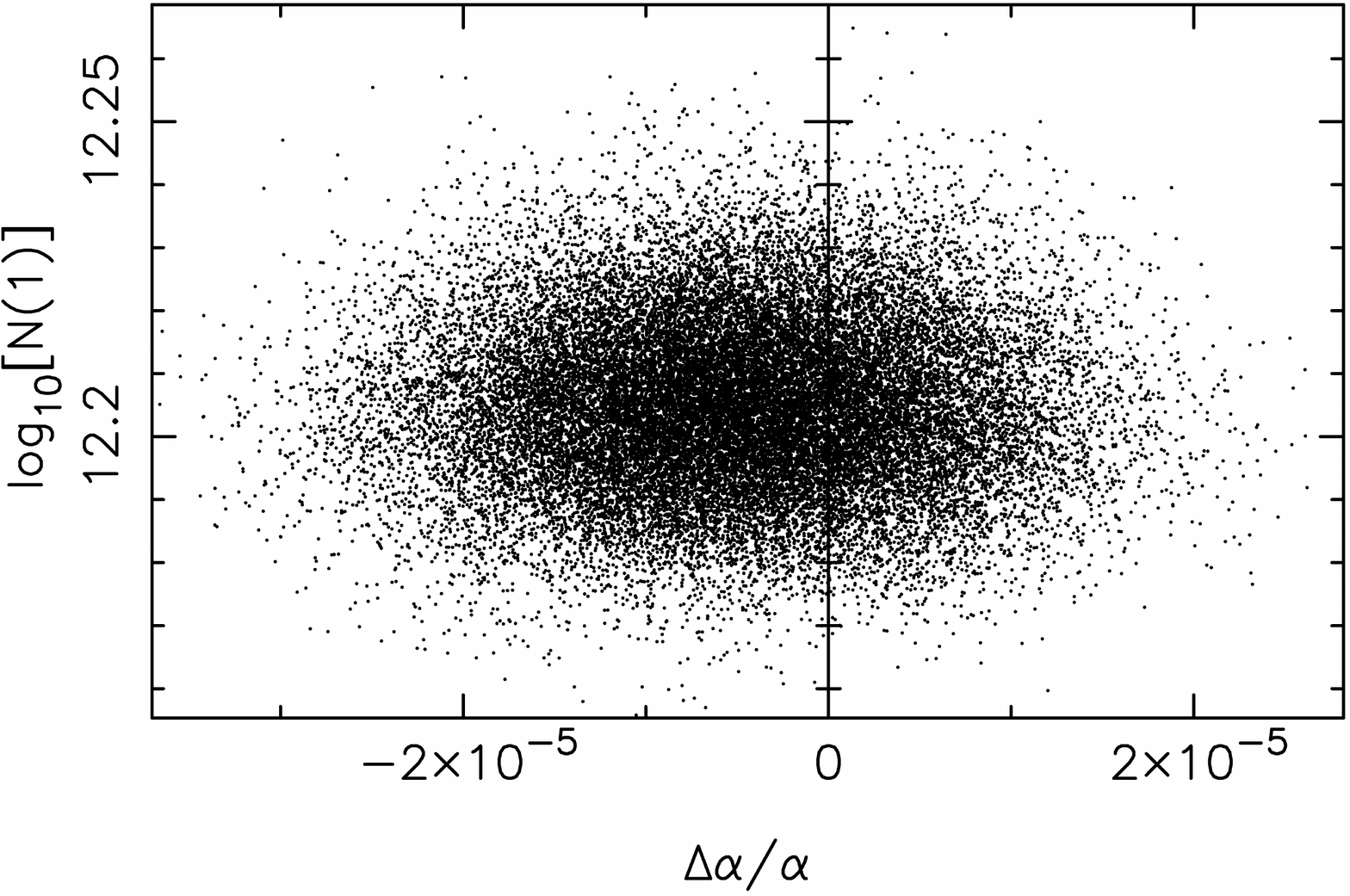}
\par\end{centering}

\caption[Chain values of $\Delta\alpha/\alpha$ vs N(1) for the $z=1.018$ absorber toward LBQS 2206$-$1958]{Chain values of $\Delta\alpha/\alpha$ vs the $\log_{10}[N(1)]$, where $N(1)$ is the column density of the Mg \textsc{ii} $\lambda 2796$ component in atoms/cm$^2$, for the $z=1.018$ absorber toward LBQS 2206$-$1958 (J220852$-$194359). The two parameters appear to be jointly Gaussian. \label{Flo:MCMC:LBQS2206-1958-davsN1}}
\end{figure}

The parameters are approximately jointly Gaussian. This is expected
for a single component fit --- the Voigt profile decomposition is
effectively unique with one component. We show the chain values of
$\Delta\alpha/\alpha$ plotted against the chain values of the column
density of the Mg~\iis $\lambda2796$ component in figure \ref{Flo:MCMC:LBQS2206-1958-davsN1}.
The probability density is larger where the density of points is greater.
The joint distribution here is elliptical, and the individual distributions
are Gaussian, suggesting that the joint distribution is well described
by a 2D Gaussian. 

\subsection{LBQS 0013$-$0029 (J001602$-$001225) $z=2.029$}

\begin{figure}[tbph]
\noindent \begin{centering}
\includegraphics[bb=34bp 58bp 554bp 727bp,clip,angle=-90,width=0.8\textwidth]{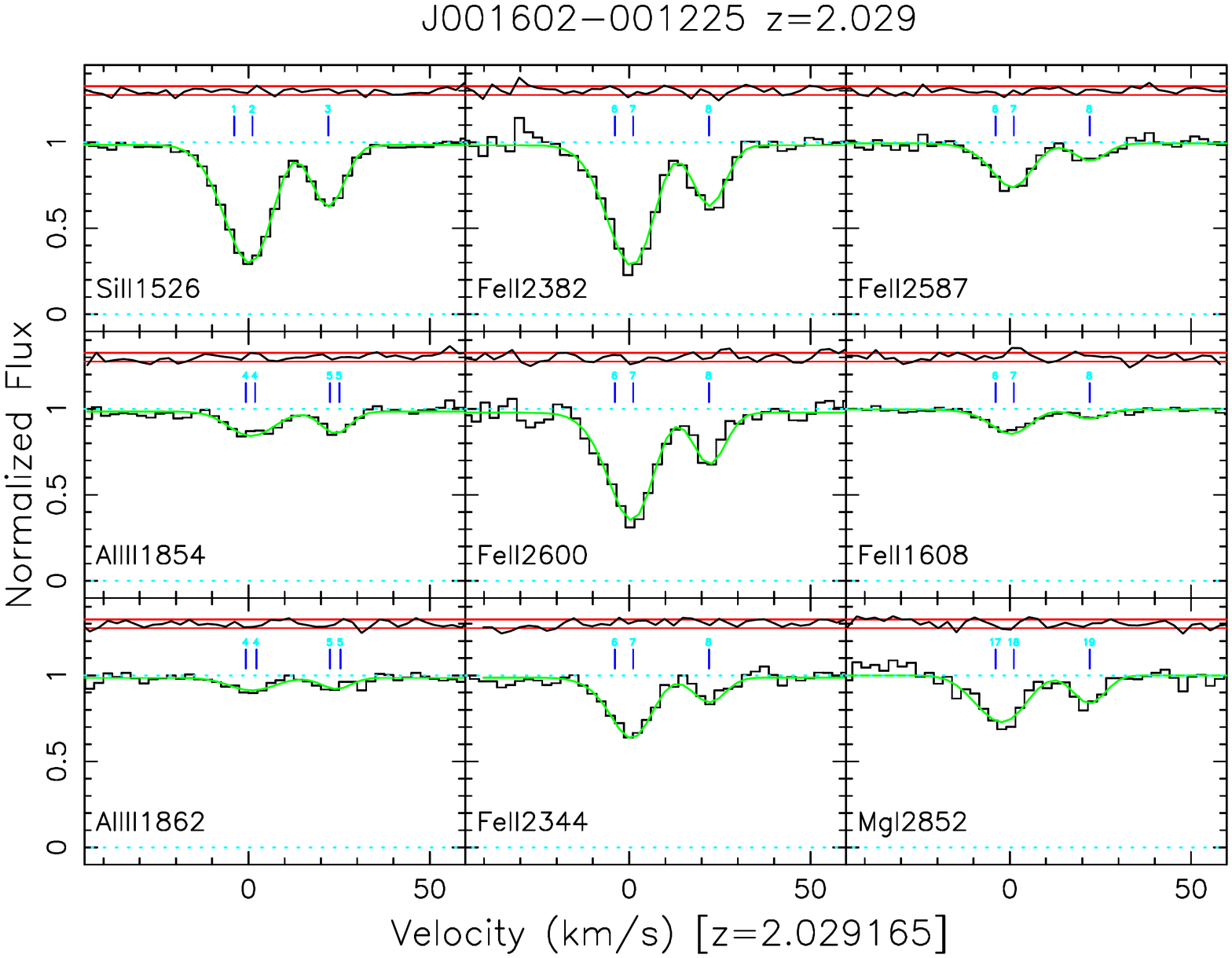}
\par\end{centering}

\caption[Fit for the $z=2.029$ absorber toward LBQS 0013$-$0029 (J001602$-$001225)]{Fit for the $z=2.029$ absorber toward LBQS 0013$-$0029  (J001602$-$001225) used for our MCMC analysis. The existence of two components in the blue feature is preferred over one on the basis of the AICC. However, because our parameter values for Al \textsc{iii} are not tied to those in the other components, and because the optical depth of Al~\textsc{iii} is relatively low, the fit will only support a single component in Al~\textsc{iii} in the blue feature. The presence of two tick marks per velocity component in the Al~\iiis fits is due to the fact that the hyperfine components are explicitly shown. \label{Flo:MCMC:LBQS0013-0029-fit}}
\end{figure}

This system appears with two obvious features. We find that the bluer
feature is better fitted by two components than one on the basis of
the AICC. Thus, we fit three components in total. Here, we use a wide
variety of transitions, namely: Si~\iis $\lambda1526$, Al~\iiis
$\lambda\lambda$1854,1862, Fe~\iis $\lambda\lambda\lambda\lambda\lambda$2382,2600,2344,2587,1608
and Mg~\iscs $\lambda2852$. We show the fit used in figure \ref{Flo:MCMC:LBQS0013-0029-fit}.

\begin{figure}[tbph]
\noindent \begin{centering}
\includegraphics[bb=0bp 0bp 436bp 658bp,clip,angle=-90,width=0.8\textwidth]{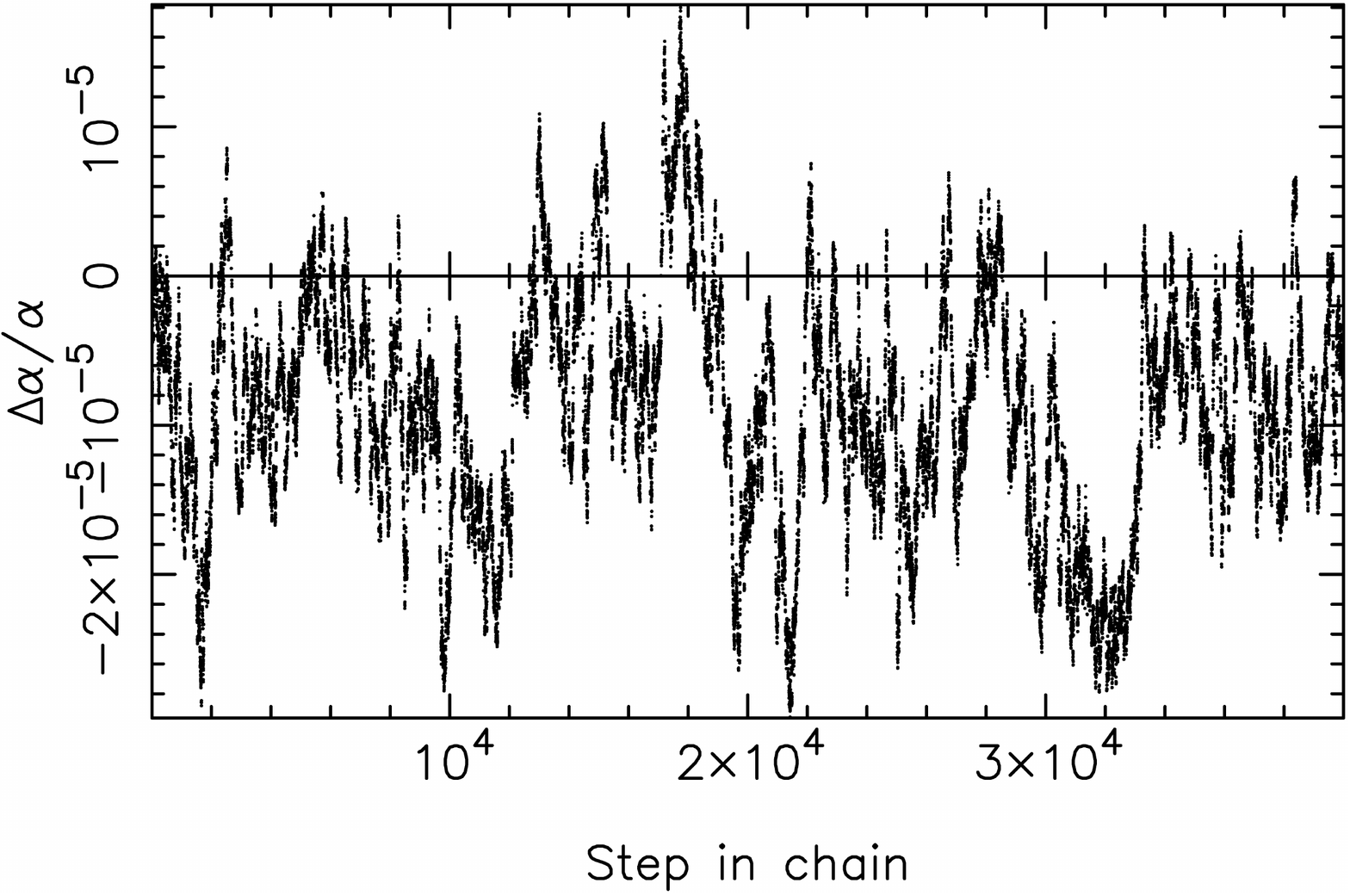}
\par\end{centering}

\caption[Example of a chain that is not well-mixed from the $z=2.029$ absorber toward LBQS 0013$-$0029]{Example of a chain that is not well-mixed for the $z=2.029$ absorber toward LBQS 0013$-$0029 (J001602$-$001225). This chain shows the chain values of $\Delta\alpha/\alpha$. Note that long-range correlations are easily visible by eye. Visually, the timescale taken to return to the central region is of order thousands of steps. This accords well with an explicit calculation of $\eta \sim 1100$. \label{Flo:MCMC:LBQS0013-0029-dawander}}
\end{figure}

In figure \ref{Flo:MCMC:LBQS0013-0029-dawander} we show an example
of a stage where the covariance matrix is poorly tuned to the target
distribution. One observes that the timescale required to retrace
the path to the $\sim$central value is on the order of thousands
of steps, implying that on needs thousands of samples to obtain the
equivalent of one independent sample. This intuition accords well
with an explicit calculation, which yields $\eta\sim1100$. Whilst
this chain will eventually adequately sample the parameter space,
the running time this would take is between tens and hundreds of times
longer than would be necessary if the covariance matrix was well tuned.
This demonstrates the utility of re-estimating the covariance matrix
multiple times.

\begin{figure}[tbph]
\noindent \begin{centering}
\includegraphics[bb=100bp 51bp 557bp 707bp,clip,angle=-90,width=0.8\textwidth]{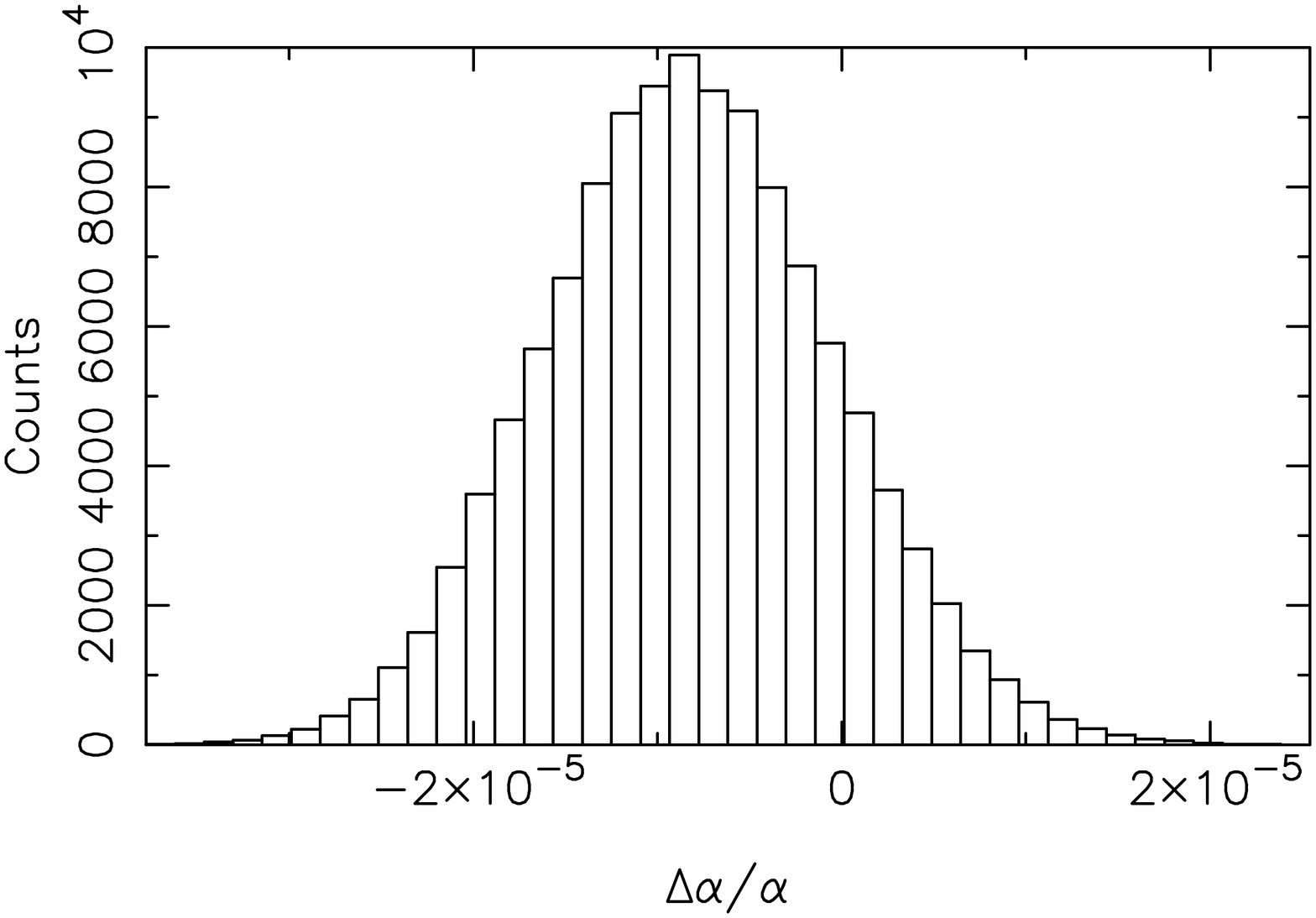}
\par\end{centering}

\caption[Histogram of chain values of $\Delta\alpha/\alpha$ for the $z=2.029$ absorber toward LBQS 0013$-$0029]{Histogram of chain values of $\Delta\alpha/\alpha$ for the $z=2.029$ absorber toward LBQS 0013$-$0029 (J001602$-$001225). The distribution appears to be well described by a Gaussian.\label{Flo:MCMC-LBQS0013-0029-dahistogram}}
\end{figure}

Our final MCMC run here consisted of 600,000 iterations, where the
chain was thinned by a factor of 5 to yield 120,000 samples. For the
chain of $\Delta\alpha/\alpha$ values, we find that the chain is
well-mixed, with $\eta\approx76$ after thinning. We show the histogram
of the chain of $\Delta\alpha/\alpha$ values in figure \ref{Flo:MCMC-LBQS0013-0029-dahistogram},
and note that it appears to be well described by a Gaussian.

\subsection{Q 0551$-$366 (J055246$-$363727) $z=1.748$\label{sub:MCMC:Q0551-366}}

\begin{figure}[tbph]
\noindent \begin{centering}
\includegraphics[bb=58bp 41bp 727bp 561bp,clip,width=0.8\textwidth]{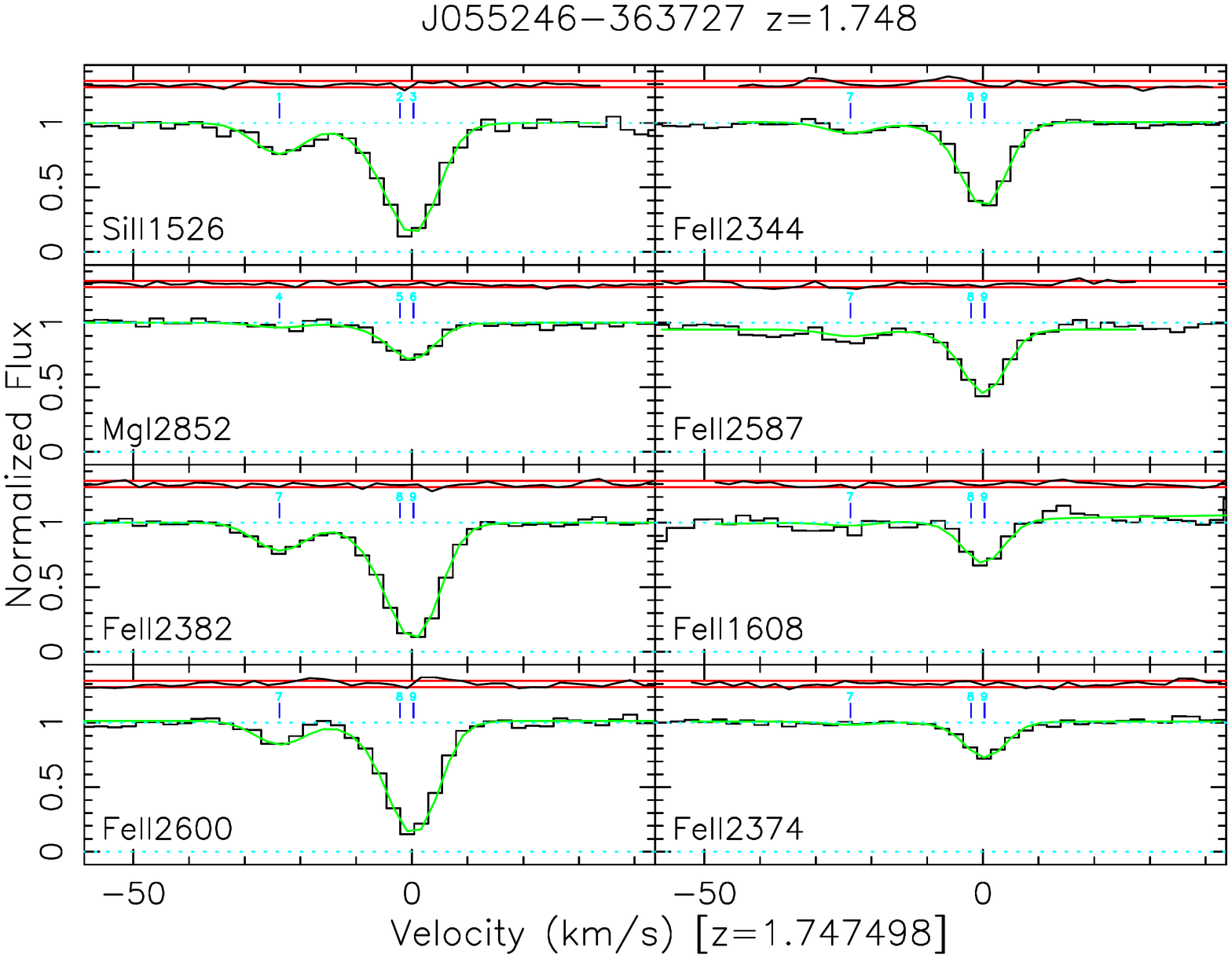}
\par\end{centering}

\caption[Fit for the $z=1.748$ absorber toward Q 0551$-$366 (J055246-363727)]{Fit for the $z$=1.748 absorber toward Q 0551$-$366 (J055246$-$363727) used for our MCMC analysis. The existence of two components in the central feature is preferred over one on the basis of the AICC. Note that because the two components are closely spaced, we expect that the parameters of these two components will be non-negligibly correlated.\label{Flo:MCMC:Q0551-366-fit}}
\end{figure}

This absorption feature appears as a single weak feature next to one
relatively strong feature, with some overlap. We find that the bluer
feature is well modelled by one component, however the higher wavelength
feature appears to require two closely spaced components to achieve
a statistical fit. We model the absorption with Si~\iis $\lambda1526$,
Mg~\iscs$\lambda2852$, and Fe~\iis $\lambda\lambda\lambda\lambda\lambda\lambda$
2382,2600,2344,2587,1608,2374. We show the fit used in figure \ref{Flo:MCMC:Q0551-366-fit}. 

\begin{figure}[tbph]
\noindent \begin{centering}
\includegraphics[bb=100bp 51bp 561bp 707bp,clip,angle=-90,width=0.7\textwidth]{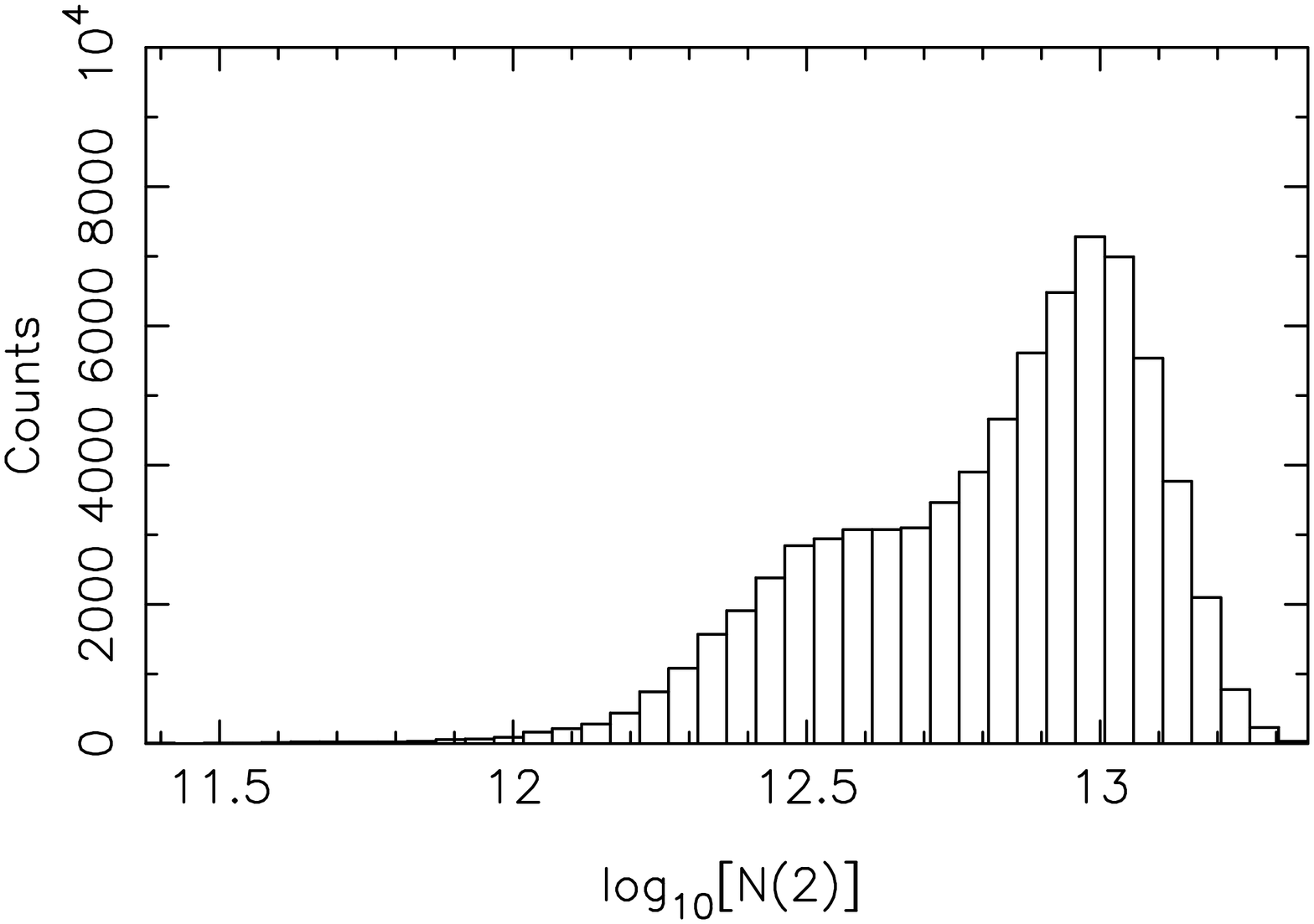}
\par\end{centering}

\caption[Histogram of chain values of N(2) for the $z=1.748$ absorber toward Q 0551$-$366]{Histogram of chain values of $\log_{10}[N(2)/\mathrm{cm}^2]$, where $N(2)$ is the column density of the central component to the fit in Si \textsc{ii} $\lambda 1526$ for the $z=1.748$ absorber toward Q 0551$-$366. The units of $N(2)$ are atoms/cm$^2$. \textsc{vpfit} correctly finds the maximum likelihood estimate of the column density as $\sim 10^{13}$ atoms/cm$^{2}$. However, there appears to be a probability shelf near $N=10^{12.6}$ atoms/cm$^2$. This implies that the \textsc{vpfit} uncertainty, which is based on the covariance matrix at the maximum likelihood solution, are not a full description of the probability space, being based on the assumption that the parameters are jointly Gaussian. For this parameter, \textsc{vpfit} gives $\log_{10} N = 13.0 \pm 0.2$. \label{Flo:MCMC-Q0551-N2histogram}}
\end{figure}

Our final MCMC run here consisted of 750,000 iterations. We note that
the parameters corresponding to the reddest two components do not
seem to be Gaussian. Figure \ref{Flo:MCMC-Q0551-N2histogram} shows
the histogram of the column density of the central component of the
Si~\iis $\lambda1526$ fit, which displays clear deviations from
Gaussianity. The column density returned by \textsc{vpfit} is, as
expected, that given by the mode of the distribution, near $N\sim10^{13}$
atoms/cm$^{2}$. The covariance matrix, which gives the parameter
uncertainty estimates, is determined at this point. As the distribution
of this column density is not Gaussian, the uncertainty returned by
\textsc{vpfit} is not a good description of the true probability density
for this component. In figure \ref{Flo:MCMC-Q0551-N2b2} we show the
chain values of the column density of this component in Si~\ii~$\lambda1526$
plotted against the velocity dispersion of this component. This shows
significant deviations from the expected elliptical shape, indicating
that a multivariate Gaussian is not a good description of the probability
density of the parameters.

\begin{figure}[h]
\noindent \begin{centering}
\includegraphics[bb=100bp 51bp 557bp 707bp,clip,angle=-90,width=0.8\textwidth]{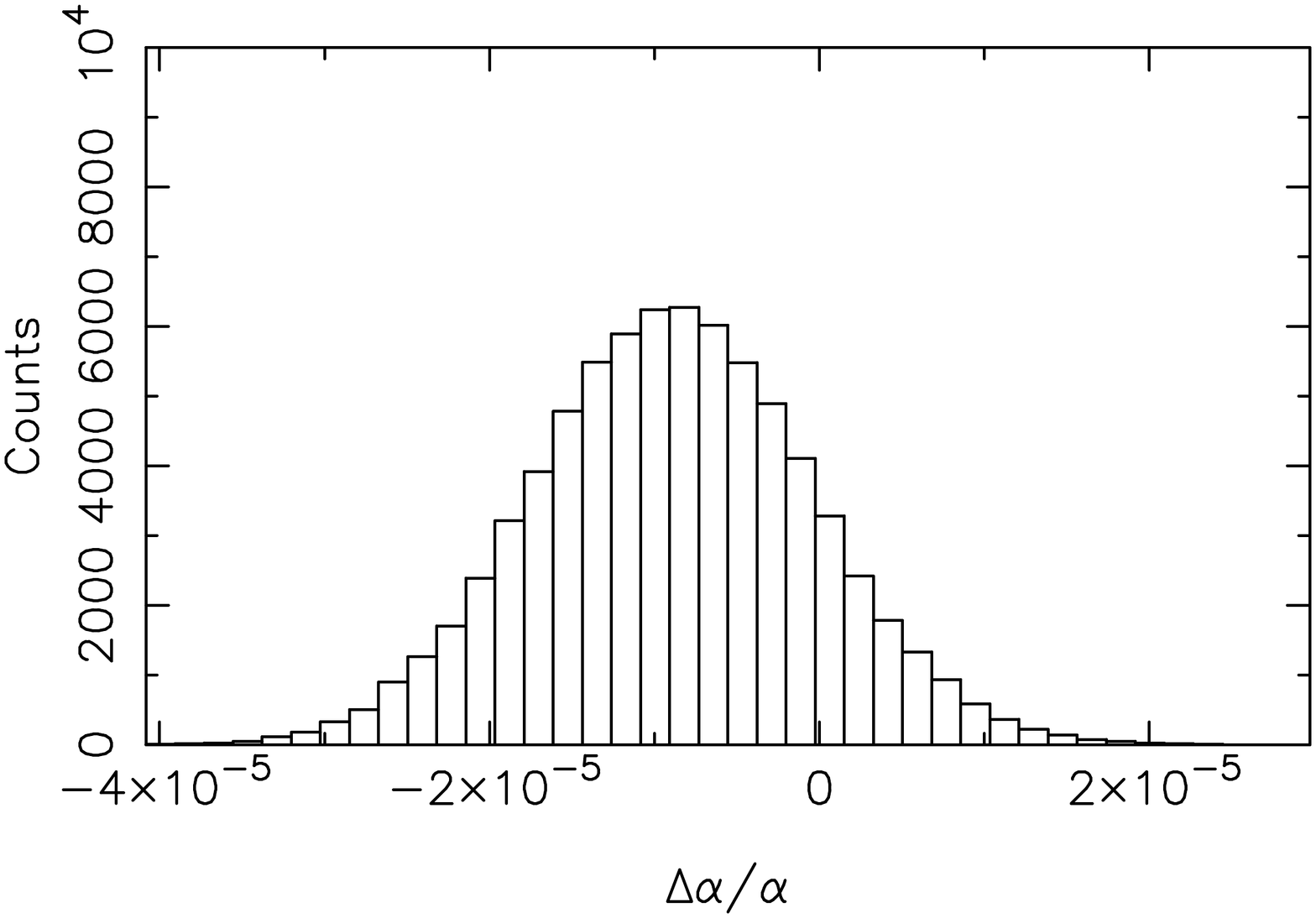}
\par\end{centering}

\caption[Histogram of chain values of $\Delta\alpha/\alpha$ for $z=1.748$ absorber toward Q 0551$-$366]{Histogram of chain values of $\Delta\alpha/\alpha$ for the $z=1.748$ absorber toward Q 0551$-$366. Note that $\Delta\alpha/\alpha$ appears to be Gaussian, despite the fact that other parameters are not. \label{Flo:MCMC-Q0551-dahistogram}}
\end{figure}
\begin{figure}[tbph]
\noindent \begin{centering}
\includegraphics[bb=0bp 0bp 441bp 658bp,clip,angle=-90,width=0.8\textwidth]{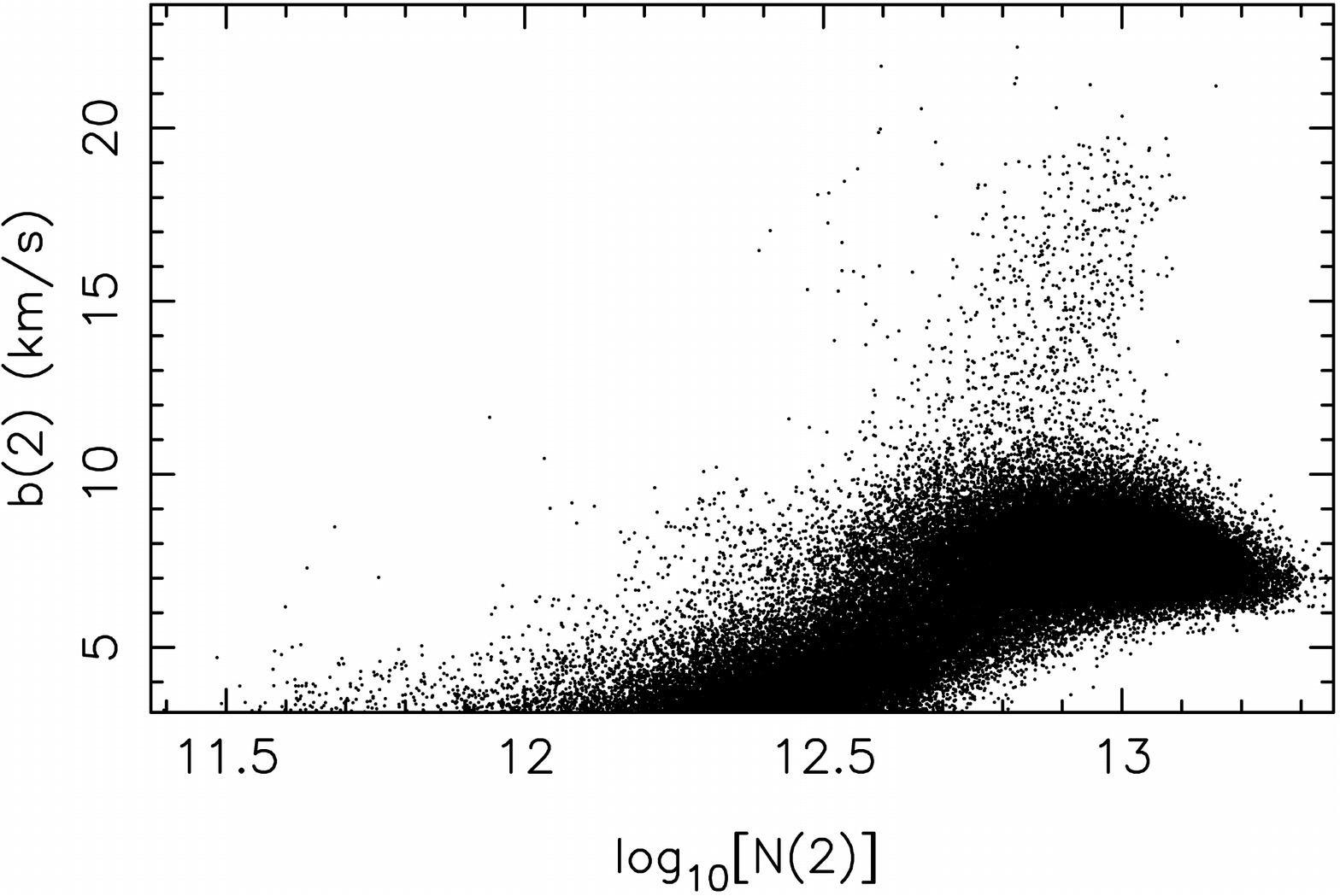}
\par\end{centering}

\caption[Chain values of N(2) vs b(2) for the $z=1.748$ absorber toward Q 0551$-$366]{Plot of the values of the chain for $\log_{10}[N(2)]$ vs $b(2)$, where $N(2)$ is the column density of the central component of the fit for Si~\iis $\lambda 1526$ in atoms/cm$^2$ and $b(2)$ is the velocity dispersion parameter of the same component in km/s, for the $z=1.748$ absorption system toward Q 0551$-$366. Note that the parameters are clearly not jointly Gaussian. The hard limit at the lower edge is caused by one of the other $b$ parameters hitting a hard limit, namely that the $b$ parameters must not decrease below 1 km/s. The rationale for this is given in section \ref{sub:MCMC hard limits}.\label{Flo:MCMC-Q0551-N2b2}}
\end{figure}

\begin{figure}[tbph]
\noindent \begin{centering}
\includegraphics[bb=0bp 0bp 436bp 658bp,clip,angle=-90,width=0.8\textwidth]{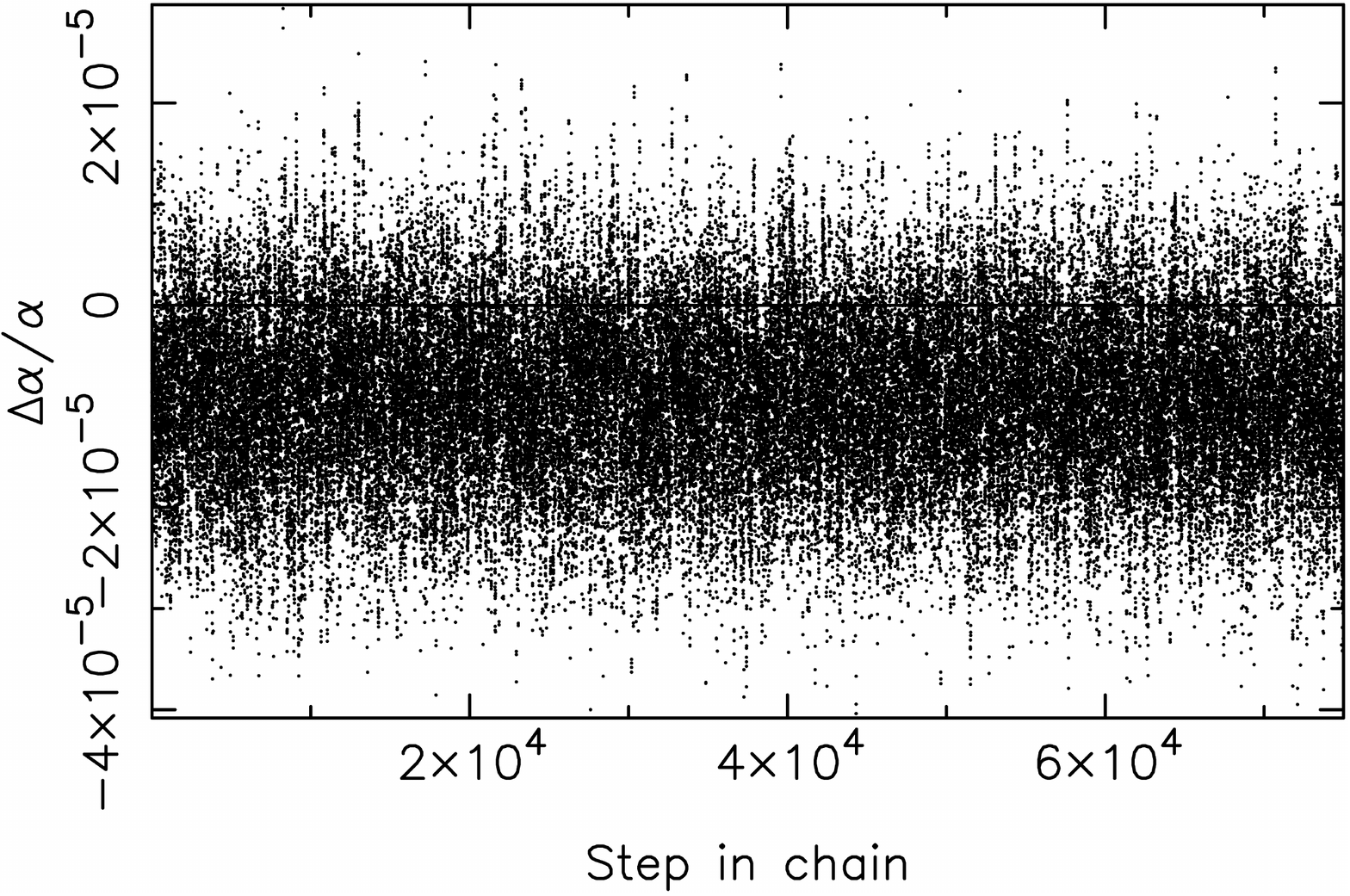}
\par\end{centering}

\caption[Chain values of $\Delta\alpha/\alpha$ for the $z=1.748$ absorber toward Q 0551$-$366]{Chain values of $\Delta\alpha/\alpha$ for the $z=1.748$ absorption system toward Q 0551$-$366. No large scale correlation is visible, implying that the chain is well-mixed. This run used 750,000 iterations, but the final chain has been thinned by a factor of 10. \label{Flo:MCMC-Q0551-dachain}}
\end{figure}
\begin{figure}[tbph]
\noindent \begin{centering}
\includegraphics[bb=120bp 51bp 557bp 735bp,clip,angle=-90,width=0.8\textwidth]{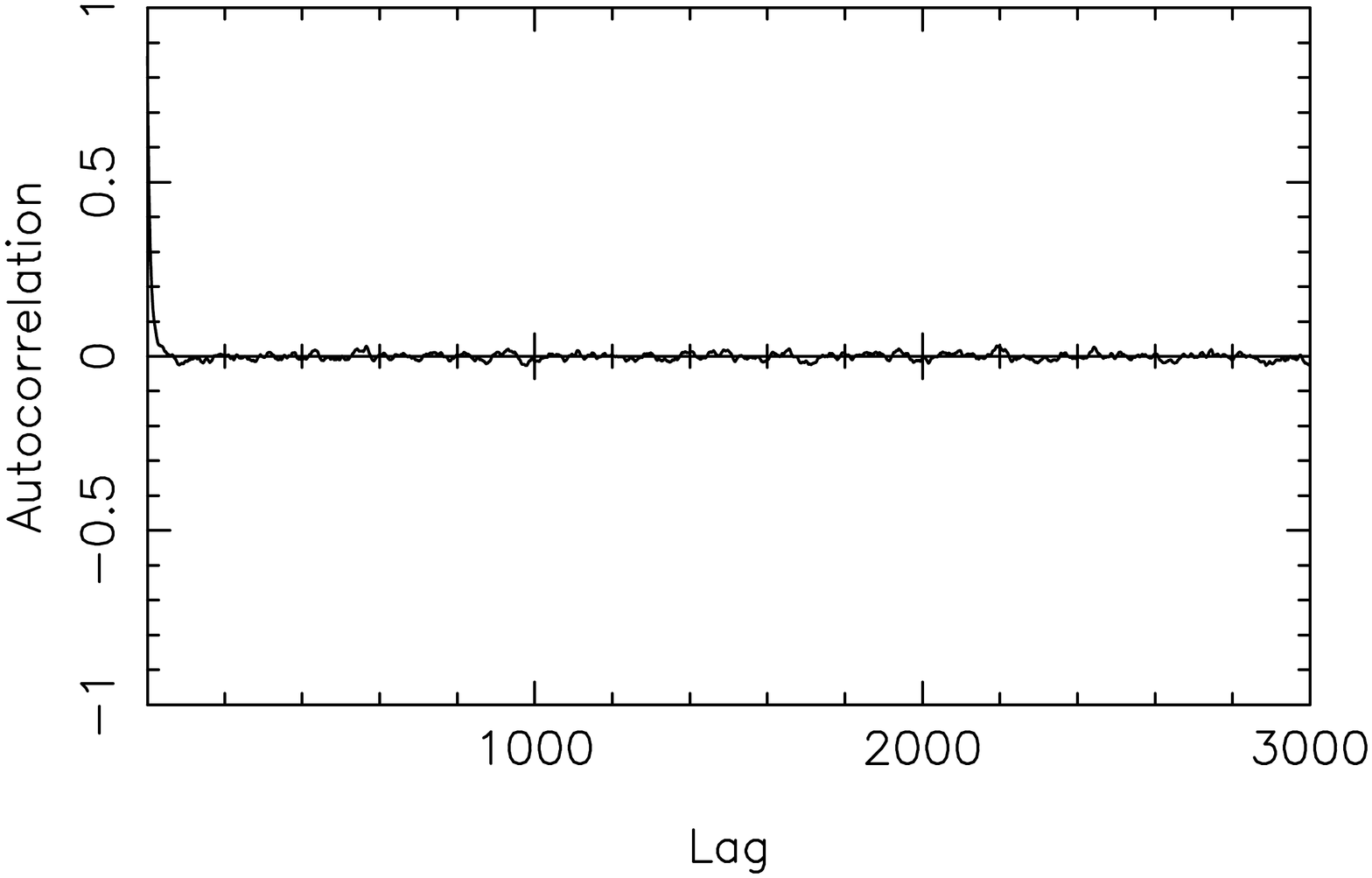}
\par\end{centering}

\caption[Autocorrelation function for the chain values of $\Delta\alpha/\alpha$ for the $z=1.748$ absorber toward Q 0551$-$366]{Autocorrelation function for the chain values of $\Delta\alpha/\alpha$ for the $z=1.748$ absorber toward Q 0551$-$366, where the chain has been thinned by a factor of 10. Note that the autocorelation function decays extremely rapidly compared to the chain length. As such, the chain as a whole provides many independent-equivalent samples of $\Delta\alpha/\alpha$.\label{Flo:MCMC-Q0551-dacorrel}}
\end{figure}
We show in figure \ref{Flo:MCMC-Q0551-dachain} the chain of samples
for $\Delta\alpha/\alpha$, which has been thinned by a factor of
10. For the thinned chain, we obtain $\eta\approx18$ (which implies
that $\eta\sim180$ for the unthinned chain), which means that the
chain is well-mixed. We show the autocorrelogram of the chain of $\Delta\alpha/\alpha$
values (thinned by a factor of 10) in figure \ref{Flo:MCMC-Q0551-dacorrel}.
The autocorrelogram demonstrates rapid decay of the autocorrelation
function compared to the chain length, which indicates that the chain
possesses many independent samples. Figure \ref{Flo:MCMC-Q0551-dahistogram}
shows the histogram of the chain values for $\Delta\alpha/\alpha$.

Importantly, even though several of the parameters are not Gaussian,
$\Delta\alpha/\alpha$ appears to be well described by a Gaussian
distribution. We naively expect this, as $\Delta\alpha/\alpha$ should
not be strongly correlated with the parameters and because there should
be a unique value of $\Delta\alpha/\alpha$ for each spectrum. The
MCMC results here confirm our \emph{a priori} beliefs about the distribution
of $\Delta\alpha/\alpha$. Additionally, the estimate of the best-fitting
value of $\Delta\alpha/\alpha$ returned by \textsc{vpfit} accords
well with the values obtained from the MCMC samples (see table \ref{tab:MCMC results}).
For our purposes, we are primarily interested in the value of $\Delta\alpha/\alpha$;
all other parameters are nuisance parameters. Therefore, it is extremely
reassuring that \textsc{vpfit} produces good parameter estimates and
uncertainties for $\Delta\alpha/\alpha$ even in the presence of non-Gaussianity
of some parameters.\afterpage{\clearpage}

\section{Discussion \& conclusion}

In this chapter, we have demonstrated successful application of MCMC
techniques to explicitly verifying the solution of \textsc{vpfit}
for simple metal absorbers. The MCMC technique is relatively robust,
and the application to more complicated systems is limited by the
computing power available. We note that the running time for Q 0551$-$366
(section \ref{sub:MCMC:Q0551-366}) is several days. Application of
this method to more complicated cases therefore requires either a
great deal of patience or the use of supercomputing facilities (or
both). More problematic is the fact that the running time to converge
is unknown \emph{a priori}. A more sophisticated version of our algorithm
would use a variable number of stages, with some termination criteria
based on autocorrelation, however we have not needed to implement
that for our cases. Ultimately, we would like to apply our algorithm
to substantially more complicated cases --- in particular, we would
like to verify the uncertainty on estimate of $\Delta\mu/\mu$ set
out earlier in this work. Unfortunately, these fits present with thousands
of parameters rather than tens. We nevertheless attempted to examine
whether this problem was remotely tractable with our MCMC algorithm,
and found that after one month of CPU time on a dual-core Pentium
D 3.6 GHz that convergence had not been achieved. Thus, we leave this
to future work.

Ultimately, the primary goal of this work was to verify that the uncertainties
produced by \textsc{vpfit} are reasonable, and we have demonstrated
that this is true for simple situations. Experience with \textsc{vpfit}
suggests that there does not appear to be any indication of failure
with moderately complicated circumstances, and so we argue both that
the optimisation algorithm used by \textsc{vpfit} is robust and that
the uncertainties produced are reasonable. An incidental consequence
of this work is demonstrating that the Gauss-Newton approximation
to the covariance matrix given by equation \ref{eq:Hessian approximation}
is good --- if it were not, the uncertainties produced by \textsc{vpfit}
would differ substantially from those given by the MCMC algorithm.

One intriguing possibility for future work is the use of nested sampling\index{nested sampling}
\citep{Skilling:04,Feroz:08}, which appears to cope well with both
multimodal distributions and high dimensionality. Not only does this
method produce samples from the posterior (as we obtain here with
MCMC), but importantly one obtains the Bayesian evidence, thereby
allowing the direct comparison of competing models. Nested sampling
transforms the multidimensional evidence integral (which is often
notoriously difficult to evaluate) into a one dimensional version
which can be approximated using the trapezium rule provided one can
sample from the prior, where the drawn sample must have $L(\mathbf{x})>L_{j}$
for some $L_{j}$. The work of \citet{Feroz:08} provides a method
to decompose likelihoods of arbitrary complexity into a series of
ellipsoidal approximations, to which the nested sampling algorithm
can be applied. Although in principle this method works well for high
dimensionality, the ellipses provide hard boundaries outside which
samples will not be drawn, and therefore the ellipse sizes must be
chosen carefully so as not to miss significant regions of the likelihood.
They proposed an enlargement factor, $(1+f)$, by which the ellipses
should be grown so as not to miss points. Unfortunately, this re-introduces
the curse of dimensionality unless $f\sim0$, as the regions of interest
will only constitute $\sim1/(1+f)^{D}$ of the sampled volume. However,
\citet{Feroz:08} noted that the time required is less than for MCMC
implementations, and therefore this technique shows significant promise.
Application of this method to the molecular hydrogen data may prove
fruitful, however to successfully tackle this challenge substantial
advances in computing speed will likely be required. It may simply
be, however, that with thousands of free parameters, full MCMC exploration
of a realistic molecular hydrogen fit may remain out of reach for
some time.

\chapter{Conclusions\label{cha:Conclusions}}

In this thesis, we have utilised the fact that high precision spectroscopy
allows precise redshift measurements of quasar absorption lines to
investigate the potential variation in the proton-to-electron mass
ratio, $\mu$, and fine-structure constant, $\alpha$. All the data
have been obtained with VLT/UVES, and all of the data are publicly
available through the ESO Science Archive, which helps to facilitate
verification of these results and rapid science generally. Below,
we summarise the main conclusions of this work
\begin{enumerate}
\item In chapter \ref{cha:mu}, we investigated possible variation of $\mu$
using molecular hydrogen absorbers in high quality spectra of the
quasars Q0405$-$443, Q0347$-$383 and Q0528$-$250. We attempted
to improve our analysis over that from previous works by modelling
the Lyman-$\alpha$ forest simultaneously with the H$_{2}$ transitions,
thereby accounting for a clear source of uncertainty in determining
the positions of the H$_{2}$ line centroids. The wavelength calibration
of our spectra utilises a more accurate ThAr calibration algorithm,
which should significantly reduce wavelength calibration errors compared
to previous analyses. We have also explicitly accounted for the under-estimation
of flux uncertainties in regions of low flux that occurs when the
spectra are reduced using the \textsc{midas} pipeline.
\item We found no statistically significant evidence for evolution in $\mu$
over cosmological timescales, with a weighted mean of the $\Delta\mu/\mu$
values from our direct $\chi^{2}$ minimisation method (DCMM) analysis
of $\Delta\mu/\mu=(2.6\pm3.0)\times10^{-6}$ (statistical). The individual
values of $\Delta\mu/\mu$ are themselves consistent with zero, being
$(10.1\pm6.6)\times10^{-6}$, $(8.2\pm7.5)\times10^{-6}$ and $(-1.4\pm3.9)\times10^{-6}$
for Q0405$-$443, Q0347$-$383 and Q0528$-$520 respectively. We are
therefore unable to reproduce the evidence for a change in $\mu$
found by \citet{Reinhold:06-1}.
\item We also analysed the absorbers toward Q0405$-$443 and Q0347$-$383
using the reduced redshift method (RRM), and found results consistent
with those derived from the DCMM. Importantly, our RRM results show
$\chi_{\nu}^{2}\approx1$ for the values of the reduced redshift,
$\zeta_{i}$, about the best linear fit. This is in contrast with
the results of \citet{Reinhold:06-1}, whose $\zeta_{i}$ values showed
excess scatter, with $\chi_{\nu}^{2}=2.1$. We can say, at least on
the basis of the observed scatter of our data, that there appears
to be no evidence for unmodelled systematics in our analysis. We attributed
this to a combination of better wavelength calibration and the fact
that we modelled the Lyman-$\alpha$ forest in the vicinity of the
H$_{2}$ transitions, which should lead to a more robust estimate
of uncertainties.
\item We noted explicit advantages of the DCMM over the RRM, in that the
DCMM allows analysis of systems with overlapping velocity components,
which is not possible within the RRM. By fitting the components in
the Q0528$-$250 absorber simultaneously, we were able to obtain an
extremely precise measurement of $\Delta\mu/\mu$.
\item We re-analysed the absorber toward Q0528$-$250 using new observations.
We investigated possible systematic errors, including: \emph{i) }systematic
distortions in the wavelength scale due to ThAr calibration uncertainties;
\emph{ii) }intra-order wavelength distortions; \emph{iii) }potential
velocity segregation between cold ($J\in[0,1]$) and warm ($J\in[2,4]$)
components, and; \emph{iv)} the effect of re-dispersion of the spectra.
We found that $\Delta\mu/\mu=(0.2\pm3.2_{\mathrm{stat}}\pm1.9_{\mathrm{sys}})\times10^{-6}$,
or $(0.2\pm3.7)\times10^{-6}$ if one aggregates the effect of statistical
and systematic errors. 
\item A weighted mean of all our values of $\Delta\mu/\mu$ yields $(1.7\pm2.4)\times10^{-6}$
--- an extremely stringent bound on any change in $\mu$. Including
the result obtained from J2123$-$0050 by \citet{Malec:10} yields
$(\Delta\mu/\mu)_{w}=(2.2\pm2.2)\times10^{-6}$. The results of chapter
\ref{cha:mu} are the best $z>1$ constraints on $\Delta\mu/\mu$
available.
\item In chapter \ref{cha:alpha}, we applied the many-multiplet method
to a large sample of quasar absorbers, the spectra for which have
been obtained over several years by many different observers on VLT/UVES.
Our aim was to produce a sample of comparable size to the Keck sample
\citep{Murphy:04:LNP}, with which we might be able to support or
contradict the $5\sigma$ evidence found from Keck/HIRES that $\Delta\alpha/\alpha<0$
at cosmological redshifts. In particular, with the result from \citet{Murphy:04:LNP}
that $\Delta\alpha/\alpha=(-0.57\pm0.11)$, we argued that a comparable
sample might be able to show inconsistency with Keck at the $\sim3.5\sigma$
level if $\Delta\alpha/\alpha=0$.
\item Our final VLT sample consists of 153%
\footnote{Excluding one absorber which was flagged as an outlier.%
} absorbers from 60 different sightlines. This is the largest statistical
sample of MM absorbers presented in any work so far. The data quality
is extremely good, representing the amalgamation of many exposures
taken over approximately 100 nights at the VLT. Even after accounting
for random errors, the data quality allow us to constrain $\Delta\alpha/\alpha$
at the few parts-per-million level.
\item For the VLT $\Delta\alpha/\alpha$ values taken by themselves, under
a weighted mean model we found that $\Delta\alpha/\alpha=(0.21\pm0.12)\times10^{-5}$.
This result is inconsistent with the Keck results at the $4.7\sigma$
level. However, we showed that both the VLT and Keck $\Delta\alpha/\alpha$
values can be made consistent if one assumes that spatial variation
in $\alpha$ exists. When we applied a simple model for angular variation
in $\alpha$, $\Delta\alpha/\alpha=A\cos\Theta+m$, to the VLT $\Delta\alpha/\alpha$
values we found a preference for a dipole+monopole model over a monopole-only
model at the $2.2\sigma$ level. Combining this with the Keck $\Delta\alpha/\alpha$
values, we found $4.1\sigma$ evidence for angular variation in $\alpha$
(in the sense that the dipole+monopole model is preferred over the
monopole-only model at the $4.1\sigma$ level), having amplitude $A=0.97_{-0.20}^{+0.22}\times10^{-5}$,
and pointing in the direction $\mathrm{RA}=(17.3\pm1.0)\,\mathrm{hr}$,
$\mathrm{dec=(-61\pm10)^{\circ}}$.
\item We showed that the VLT and Keck data demonstrate a remarkable internal
consistency, in that the dipole directions in a dipole+monopole model
fitted to the Keck and VLT $\Delta\alpha/\alpha$ values point in
a similar direction (with a chance probability of alignment of 6 percent),
and also that the dipole directions in a dipole+monopole model fitted
to low and high redshift cuts of the data (split at $z=1.6$) also
point in a similar direction (with a chance probability of alignment
of 2 percent). The joint probability of obtaining alignment as good
as is seen in both these cases is just 0.1 percent (equivalent to
$\approx3.3\sigma$). 
\item We noted the presence of a monopole at $z<1.6$, which is statistically
significant at the $3.6\sigma$ level. If real, this would represent
an angle-independent change in $\alpha$ relative to laboratory values.
Although the monopole is unusual, we showed that both the VLT and
Keck samples yield extremely consistent estimates of its value. This
means that, whatever the cause of the monopole, it can not be responsible
for the observed angular variation in $\alpha$. We discussed several
possible explanations for the presence of this monopole, and concluded
that evolution in the abundance of magnesium isotopes is the most
likely cause rather than universal temporal evolution in $\alpha$.
The lack of clear explanation of the monopole in the low-redshift
sample is a weakness of the results presented here, but on account
of the good consistency between the Keck and VLT results and the angular
component of low- and high-redshift samples we do not think that its
existence significantly affects the detection of angular variations
in $\alpha$. Future observations should be able to determine what
the cause of the monopole is.
\item We showed that the dipole effect demonstrated is not caused by small
numbers of outlying data points, by iteratively clipping away $\Delta\alpha/\alpha$
values about the model and demonstrating the effect this has on both
the significance of the dipole and the fitted direction. W also showed
that the effect is not being caused by a small number of aberrant
spectra, by exploring the influence of randomly removing spectra. 
\item On account of the fact that the observed angular variation in $\Delta\alpha/\alpha$
is larger at high redshifts, we explored simple distance-dependent
models, where the dipole amplitude scales as $z^{\beta}$ for some
$\beta$, and also where the amplitude scales with the lookback-time
distance to the absorbers, $r=ct$. We show that for the model $\Delta\alpha/\alpha=Br\cos\Theta+m$
that the statistical significance of the dipole effect increases to
$4.2\sigma$, and the dipole points in the direction $\mathrm{RA}=(17.5\pm1.0)\,\mathrm{hr}$,
$\mathrm{dec}=(-62\pm10)$, with amplitude $B=(1.1\pm0.2)\times10^{-6}\,\mathrm{GLyr^{-1}}$.
\item We concluded that the results set out in chapter \ref{cha:alpha}
present strong statistical evidence for spatial variations in $\alpha$.
\item In chapter \ref{cha:da systematic errors}, we considered the effect
of some possible systematic errors on $\Delta\alpha/\alpha$. We argued
there that the dipole effect seen is intrinsically difficult to emulate
through systematic effects, because the systematic effect must either
be well correlated with sky position (in both the Keck and VLT telescopes),
or there must be a conspiracy of systematic effects that by chance
produces angular variation in $\alpha$ in an extremely consistent
way between the two telescopes. One obvious consideration is the effect
of wavelength calibration at both the Keck and VLT telescopes; wavelength
scale distortions could easily lead to spurious values of $\Delta\alpha/\alpha$.
To empirically investigate possible wavelength distortions, we noted
that there are 7 quasars that appear in both the Keck and VLT samples.
We can use the fact that observations of absorption lines from both
telescopes should yield the same observed wavelengths to create the
$\Delta v$ test. In the $\Delta v$ test, one fits many absorption
lines at different wavelengths in each quasar spectral pair, but allows
for and estimates a velocity difference in corresponding spectral
regions between the two telescopes.
\item We utilised $\Delta v$ data for the 7 spectral pairs in the VLT and
Keck samples to investigate whether common wavelength distortions
exist. For all of the spectral pairs, there is no evidence for a common
wavelength distortion. Unfortunately, each spectral pair only provides
values of $\Delta v$ for a limited wavelength range. We combined
the $\Delta v$ data from six of the spectral pairs and modelled the
distortion with a simple linear function. Using both the LTS and SBLR
methods, we are unable to find statistically significant evidence
for a common linear wavelength distortion. Nevertheless, we modelled
the impact of the measured distortion on the dipole. This reduced
the statistical significance of the dipole model for the combined
Keck + VLT sample from $3.9\sigma$%
\footnote{Calculated from a reference set.%
} to $3.1\sigma$, thus not eliminating the dipole signal. Importantly,
this did not appreciably alter the location of the fitted dipole.
Therefore, despite the reduced statistical significance, the application
of this $\Delta v$ function does not destroy the good alignment between
the Keck and VLT dipole vectors, nor between dipole models fitted
to $z<1.6$ and $z>1.6$ sample cuts.
\item We also considered the 7th spectral pair, 2206$-$1958/J220852$-$194359.
This spectral pair displays significant relative wavelength distortions.
On account of the restricted wavelength range of the $\Delta v$ data,
we modelled the observed $\Delta v$ data with an arctangent approximation
and extrapolated to red wavelengths. We explored the impact of this
function on $\Delta\alpha/\alpha$ and found that a distortion of
this magnitude, if present in all spectra, would generate an extremely
strong signal for $\Delta\alpha/\alpha$ that is observed in neither
the Keck nor VLT data sets. 
\item We applied a Monte Carlo method to apply the (non-significant) common
linear $\Delta v$ function from 6 of the spectral pairs to 6/7 of
the VLT spectra chosen at random and the arctangent $\Delta v$ function
from the 2206$-$1958/J220852$-$194359 pair to the remaining 1/7
of the VLT spectra. We found that in almost all cases, this significantly
increases the AICC, allowing us to reject the presence of a distortion
of this type in most cases. We consider therefore consider it unlikely
that our data are affected by a combined wavelength distortion of
this type.
\item We considered the impact of the echelle intra-order distortions found
by \citet{Whitmore:10} on the combined results. Using a simple model
for the distortion in the VLT data, with a peak-to-peak amplitude
of $\approx300\,\mathrm{m\, s^{-1}}$, we found that the impact on
the location of the dipole was relatively small. The significance
of the VLT dipole is reduced from $2.2\sigma$ to $1.6\sigma$, but
this largely due to the randomising effect that a model of this type
has on the $\Delta\alpha/\alpha$ values. As a result, the significance
of the VLT+Keck dipole is reduced to $3.3\sigma$. Importantly, because
the intra-order distortions do not demonstrate any long-range component,
the effect of the distortions is to add random noise into the $\Delta\alpha/\alpha$
values; they can not manufacture a dipole or monopole. In fact, we
have already accounted for random effects like this by conservatively
increasing our error bars, and so we consider that the distortions
found by \citeauthor{Whitmore:10} are already accounted for adequately
in our VLT angular dipole significance estimate of $2.2\sigma$ (and,
thus, the VLT+Keck angular dipole significance estimate of $4.1\sigma$). 
\item Ultimately, the sample size we have for the $\Delta v$ test is small.
A strong priority for future work should be obtaining observations
of the same objects from both telescopes, so that wavelength dependent
systematics may be better constrained. Nevertheless, from the considerations
in chapter \ref{cha:da systematic errors} we argue that it is unlikely
that wavelength distortions are responsible for the observed angular
variation in $\alpha$.
\item We also considered the effect of variation in the heavy Mg isotope
fraction, $\Gamma$, as $\Delta\alpha/\alpha$ is sensitive to variation
in $\Gamma$ from the terrestrial value of $\Gamma_{t}=0.21$. We
investigated the extreme case $\Gamma=0$ by discarding the $^{25}$Mg
and $^{26}$Mg fraction (i.e.\ by fitting absorbers with only $^{24}$Mg).
We show that the effect of this is to push the $\Delta\alpha/\alpha$
values to be more negative, inducing a greater significance for any
monopole component of a model. Importantly, this experiment has no
effect of consequence on the fitted dipole locations, reinforcing
our earlier argument that any systematic which generates an angular
variation of $\alpha$ must be well correlated with sky position.
Despite the increased scatter introduced into the data as a result
of this investigation, the dipole model still remains significant
at the $3.5\sigma$ level. We also investigated the possibility that
the quasar clouds show an enriched heavy Mg fraction relative to terrestrial
values. By considering the effect this has on the $z<1.6$ monopole,
we show that the monopole could be made to disappear if $\Gamma=0.32\pm0.03$,
which is significantly higher than the terrestrial value of $\Gamma_{t}=0.21$.
Ultimately, more work is needed to resolve this issue, but differences
in the heavy Mg isotope fraction cannot be responsible for the observed
angular variation in $\alpha$.
\item Thus, we cannot find a systematic effect which is responsible for
the detected spatial variation in $\alpha$. We cannot rule out the
possibility that the detection of angular variations in $\alpha$
presented here is the result of a conspiracy of systematic effects,
or a systematic effect in both the VLT and Keck telescopes which is
well-correlated with sky position. We argued in chapter \ref{cha:da systematic errors}
that zenith-dependent systematics are unlikely. A systematic effect
which is correlated specifically with declination in the same way
in both telescopes would be very unusual, and we are not aware of
any process which could generate this. Future observations with a
third telescope will help to rule out telescope-dependent systematics. 
\item In chapter \ref{cha:Discussion}, we reviewed the consistency of the
$\Delta\mu/\mu$ and $\Delta\alpha/\alpha$ results. If both sets
of results are correct, they immediately imply that $|R|\lesssim3$
if $\Delta\mu/\mu=R(\Delta\alpha/\alpha)$, which contradicts the
quite general predictions of GUTs and string theory models that $|R|\sim35$.
Although the $\Delta\mu/\mu$ results including the two $z<1$ NH$_{3}$
constraints do not suggest spatial variation in $\mu$ that is consistent
with the $\alpha$ results, if we only fit the $\Delta\mu/\mu$ constraints
derived from the H$_{2}$ data we find that the fitted dipole points
in a similar direction to the $\Delta\alpha/\alpha$ dipole, with
the dipole vectors being separated by only $18^{\circ}$. The interpretation
of this is obviously hampered by a very limited sample size.
\item We also reviewed the $\Delta\alpha/\alpha$ results in the context
of other observations, both local and astrophysical, and concluded
that the results of chapter \ref{cha:alpha} are not in conflict with
any other existing constraints on the variation of fundamental constants. 
\item We discussed various claims for anisotropy in the universe, and noted
that there is an intriguing, but far from conclusive loose alignment
between different measurements of possible anisotropy in the universe.
\item In chapter \ref{cha:MCMC}, we investigated whether \textsc{vpfit}
produces correct parameter estimates and statistical uncertainties
by applying Markov Chain Monte Carlo methods. MCMC methods completely
dominate traditional Monte Carlo methods (random sampling of the likelihood
function) for high dimensions; degradation with increasing dimensionality
is only polynomial, whereas traditional Monte Carlo methods degrade
exponentially. We modified \textsc{vpfit} to allow for MCMC exploration
of the likelihood function of the Voigt profile fit using a modification
of the well-known Metropolis sampler, the Multiple Try Metropolis
method. We applied the resultant algorithm to several simple Voigt
profile fits. Despite the advantage of MCMC methods, reasonable exploration
of the parameter space nevertheless takes hours to a few days. We
verified what we set out to check: that the statistical estimates
of $\Delta\alpha/\alpha$ produced by \textsc{vpfit} are reasonable.
We also demonstrated that even where the joint likelihood function
is non Gaussian for individual line parameters, the likelihood for
$\Delta\alpha/\alpha$ is indeed Gaussian. This is expected, but reassuring,
and justifies the use of only a maximum likelihood estimate and standard
error when describing the estimate of $\Delta\alpha/\alpha$ for an
absorber; higher order moments (skewness, etc) can safely be neglected.
The results of this chapter give confidence to the results of \citet{Murphy:04:LNP}
and this work in investigating potential changes in $\mu$ and $\alpha$. 
\item We attempted to apply MCMC methods to $\Delta\mu/\mu$, but found
that even with ample computing resources the problem remains intractable
with a Metropolis-type sampler. We noted that this problem may become
directly amenable in the future as a result of better computing facilities,
but also noted that recent advances (e.g.\ nested sampling) may also
assist in directly investigating the likelihood function for complicated
molecular hydrogen fits.
\end{enumerate}

\section{Future work}

We have described in various places throughout this work how future
research may be able to shed more light on whether the fundamental
constants truly vary. The results of chapter \ref{cha:alpha} are
exciting, in that they yield evidence for variation in $\alpha$ that
is independent of and consistent with that obtained from Keck/HIRES.
The most obvious experimental path that is complimentary to the Keck/VLT
work is the use of a third telescope/spectrograph combination; Subaru/HDS
(High Dispersion Spectrograph) is currently the best choice. The coming
decade should see construction of one or more next-generation extremely
large optical telescopes, with primary mirror diameters of at least
$\sim20$ metres (and possibly as high as $\sim40$) depending on
the ultimate design. The spectrographs for these telescopes will be
built with extremely precision and stability in mind. 

ESPRESSO\index{ESPRESSO} (Echelle SPectrograph for Rocky Exoplanet--
and Stable Spectroscopic Observation) has recently been approved for
construction and installation at the VLT, with operation scheduled
to commence around 2014. Although the instrument will be able to operate
in 1-UT mode (using the light from a single VLT telescope), it will
also be able to operate in 4-UT mode, where the light from all four
VLT telescopes is collected at an incoherent focus, giving a collecting
area equivalent to a 16m telescope. The spectrograph is designed to
achieve $R=140,000$ and $10\,\mathrm{cm\, s^{-1}}$ precision for
radial-velocity planet searches, which would in principle allow the
detection of Earth-like planets. \citet{Molaro:07a} discussed the
science case for ESPRESSO in the context of variation of fundamental
constants, and suggests that $30\,\mathrm{m\, s^{-1}}$ precision
on narrow lines may be achievable with a few hours integration. A
shift of this magnitude corresponds to $\Delta\alpha/\alpha\approx1.4\times10^{-6}$
for the Fe~\iis $\lambda2382$ transition. This in principle enough
to accurately determine whether the results of chapter \ref{cha:alpha}
are correct or not unless systematic or random effects are significant.
The results of chapter \ref{cha:alpha} suggest that random effects
of order $\sim10^{-5}$ exist; these may reduce with higher quality
observations and instruments, but also may not. 

\citet{Liske:09a} discussed CODEX (COsmic Dynamics EXperiment)\index{CODEX},
the planned high-resolution optical spectrograph for the E-ELT (European
Extremely Large Telescope). The primary purpose of the spectrograph
is to directly observe the expansion of the universe by measuring
changes in the redshifts of absorption features. The intended target
is the Lyman-$\alpha$ forest, as it provides many lines over a large
redshift range. To achieve its science goals, CODEX will need to deliver
a radial velocity accuracy of $2\,\mathrm{cm\, s^{-1}}$ over a time-scale
of $\sim20$ years. If CODEX can deliver such precision then this
will potentially improve the current constraints on $\Delta\alpha/\alpha$
and $\Delta\mu/\mu$ by several orders of magnitude. However, it must
be said that it is not clear to what extent this precision will be
limited by uncontrollable systematics or random effects (for example,
the kinematics of the quasar absorbers).

Wavelength calibration uncertainties remain a significant problem
for optical spectroscopic measurements, as the results of \citet{Griest:09}
and \citet{Whitmore:10} demonstrate. Even if issues regarding the
quasar light path can be removed, the ThAr standard used is itself
problematic. The ThAr lines are unevenly distributed across the visual
spectrum, and the intensity of the lines differs greatly. The calibration
in some echelle orders is inevitably sub-optimal due to low numbers
of usable ThAr lines. Laser combs have recently held out promise of
vastly better wavelength calibration; laser combs can generate evenly
spaced transitions across the optical spectrum for which the absolute
calibration is known \emph{a priori}. \citet{Steinmetz:08a} discussed
the first implementation of laser comb calibration at an astronomical
observatory, achieving $9\,\mathrm{m\, s^{-1}}$ radial velocity precision
at $1.5\mu\mathrm{m}$, which they described as ``beyond state-of-the-art''.
\citet{Murphy:07c} discussed simulations of optical laser combs which
show that integration over a $4000\mathrm{\AA}$ range could produce
calibration uncertainties of as low as $1\,\mathrm{cm\, s^{-1}}$,
which has the potential to ``remove wavelength calibration uncertainties
from all practical spectroscopic experiments''. For these precisions
to be realised, combs will need to demonstrate increased pulse repetition
rates and more uniform intensity over the optical range compared to
what is available at present.

Near-term verification (or otherwise) of evolution of the fundamental
constants may occur more rapidly with radio measurements, as noted
earlier. New facilities are scheduled to commence operation shortly
which will offer extreme precision. For instance, the Square Kilometre
Array (SKA)\index{SKA} will be a radio telescope of unparalleled
sensitivity due to its enormous collecting area. Although observations
are not scheduled to start until 2017, the Australian and South African
pathfinder telescopes (ASKAP\index{ASKAP} and MEERKAT\index{MEERKAT}
respectively) will become operational before this. \citet{Curran:04a}
considered then-current results on the variation of fundamental constants
and existing biases in surveys for radio absorbers in the context
of the SKA. The Atacama Large Millimetre Array (ALMA\index{ALMA})
will also soon be operational. \citet{Combes:09a} reviews existing
constraints on fundamental constants with radio lines with some consideration
given to estimates of the increased number of sources detectable with
ALMA.

Ultimately, continuing improvements in atomic clocks and the new instrumentation
that will be available for astrophysical measurements over the next
decade means that the future for investigations into whether the fundamental
constants vary is bright.

\appendix

\chapter{Q0405$-$443 Voigt profile fits\label{cha:mu fits:Q0405}}

In this appendix, we provide the fits for the $z=2.595$ H$_{2}$
absorber toward Q0405$-$443. 

\begin{figure}[H]
\noindent \begin{centering}
\includegraphics[bb=86bp 180bp 544bp 801bp,clip,width=1\textwidth]{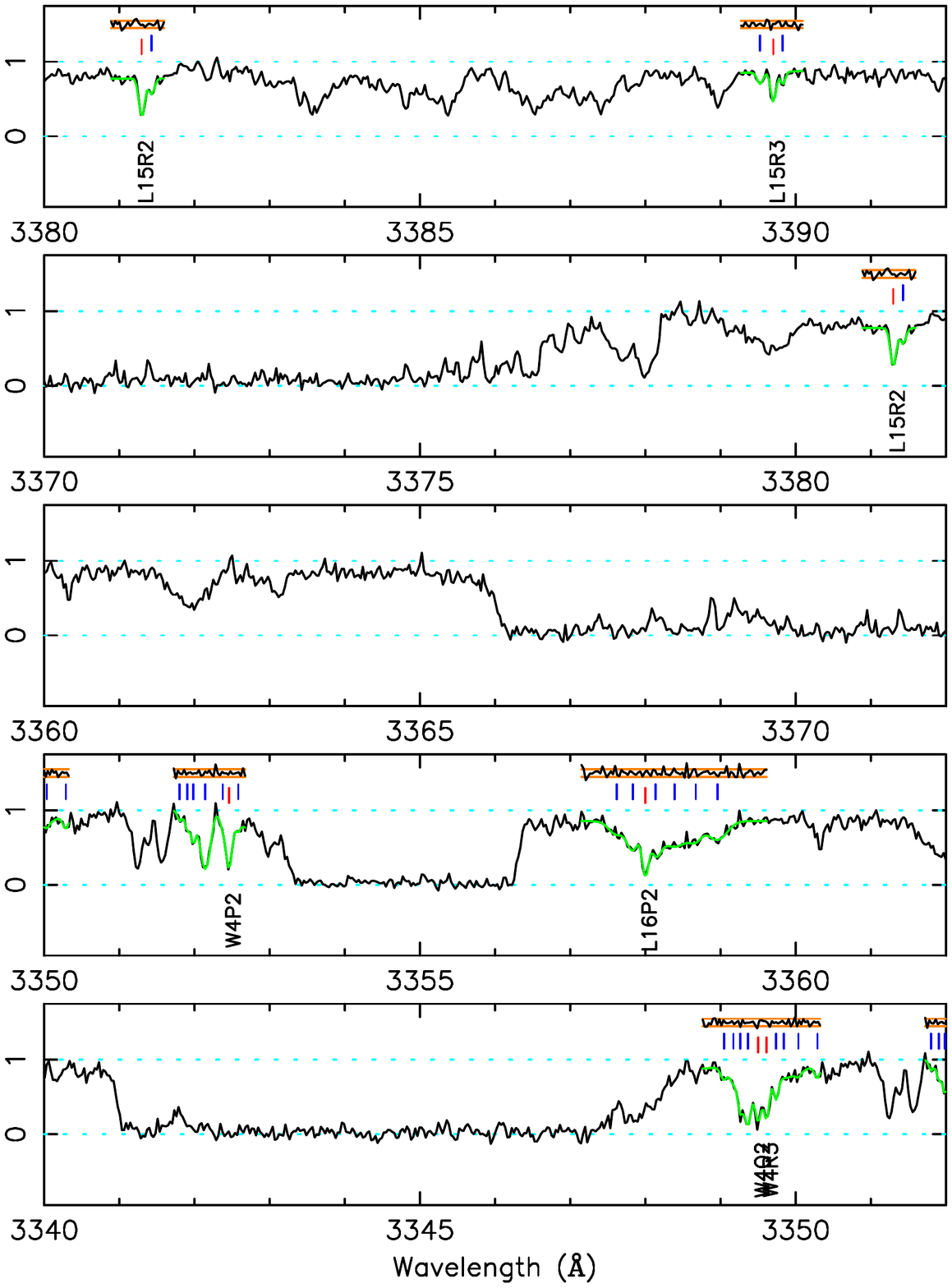}
\par\end{centering}

\caption[H$_2$ fit for the $z=2.595$ absorber toward Q0405$-$443 (1)]{H$_2$ fit for the $z=2.595$ absorber toward Q0405$-$443 (part 1). The vertical axis shows normalised flux. The model fitted to the spectra is shown in green. Red tick marks indicate the position of H$_2$ components, whilst blue tick marks indicate the position of blending transitions (presumed to be Lyman-$\alpha$). Normalised residuals (i.e. [data - model]/error) are plotted above the spectrum between the orange bands, which represent $\pm 1\sigma$. Labels for the H$_2$ transitions are plotted below the data.}
\end{figure}
\begin{figure}[H]
\noindent \begin{centering}
\includegraphics[bb=86bp 180bp 544bp 801bp,clip,width=1\textwidth]{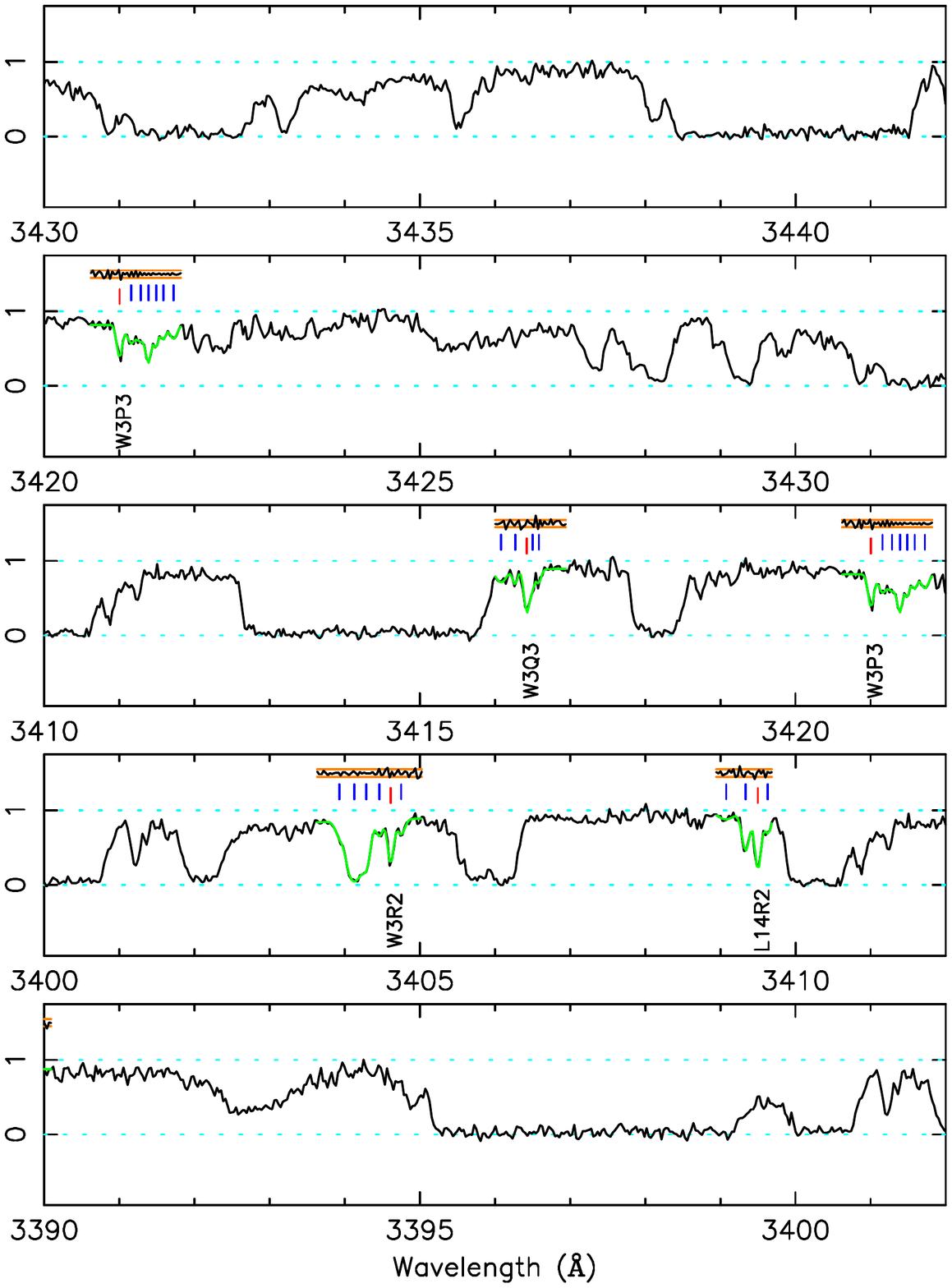}
\par\end{centering}

\caption[H$_2$ fit for the $z=2.595$ absorber toward Q0405$-$443 (2)]{H$_2$ fit for the $z=2.595$ absorber toward Q0405$-$443 (part 2). The vertical axis shows normalised flux. The model fitted to the spectra is shown in green. Red tick marks indicate the position of H$_2$ components, whilst blue tick marks indicate the position of blending transitions (presumed to be Lyman-$\alpha$). Normalised residuals (i.e. [data - model]/error) are plotted above the spectrum between the orange bands, which represent $\pm 1\sigma$. Labels for the H$_2$ transitions are plotted below the data.}
\end{figure}

\begin{figure}[H]
\noindent \begin{centering}
\includegraphics[bb=86bp 180bp 544bp 801bp,clip,width=1\textwidth]{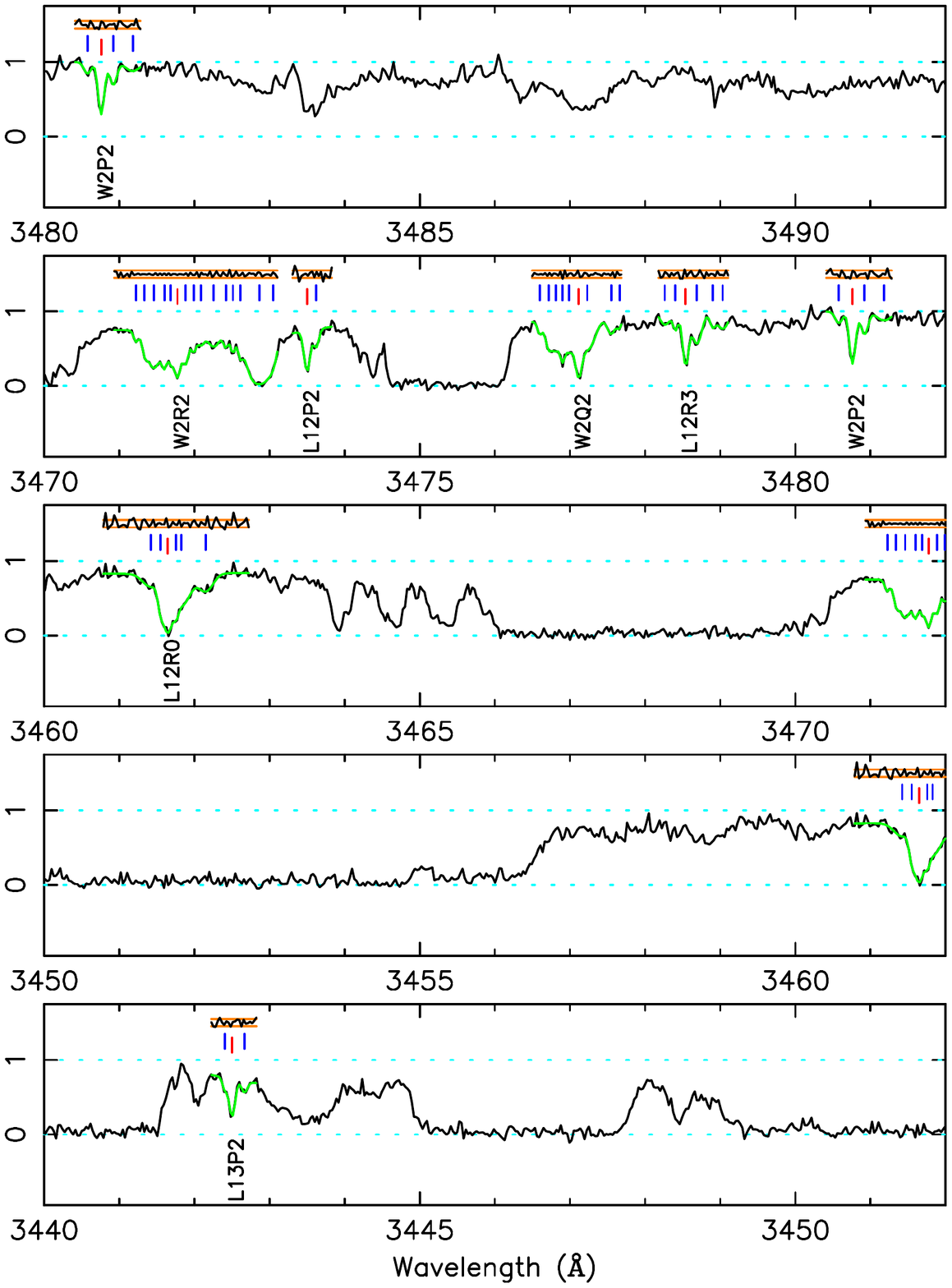}
\par\end{centering}

\caption[H$_2$ fit for the $z=2.595$ absorber toward Q0405$-$443 (3)]{H$_2$ fit for the $z=2.595$ absorber toward Q0405$-$443 (part 3). The vertical axis shows normalised flux. The model fitted to the spectra is shown in green. Red tick marks indicate the position of H$_2$ components, whilst blue tick marks indicate the position of blending transitions (presumed to be Lyman-$\alpha$). Normalised residuals (i.e. [data - model]/error) are plotted above the spectrum between the orange bands, which represent $\pm 1\sigma$. Labels for the H$_2$ transitions are plotted below the data.}
\end{figure}

\begin{figure}[H]
\noindent \begin{centering}
\includegraphics[bb=86bp 180bp 544bp 801bp,clip,width=1\textwidth]{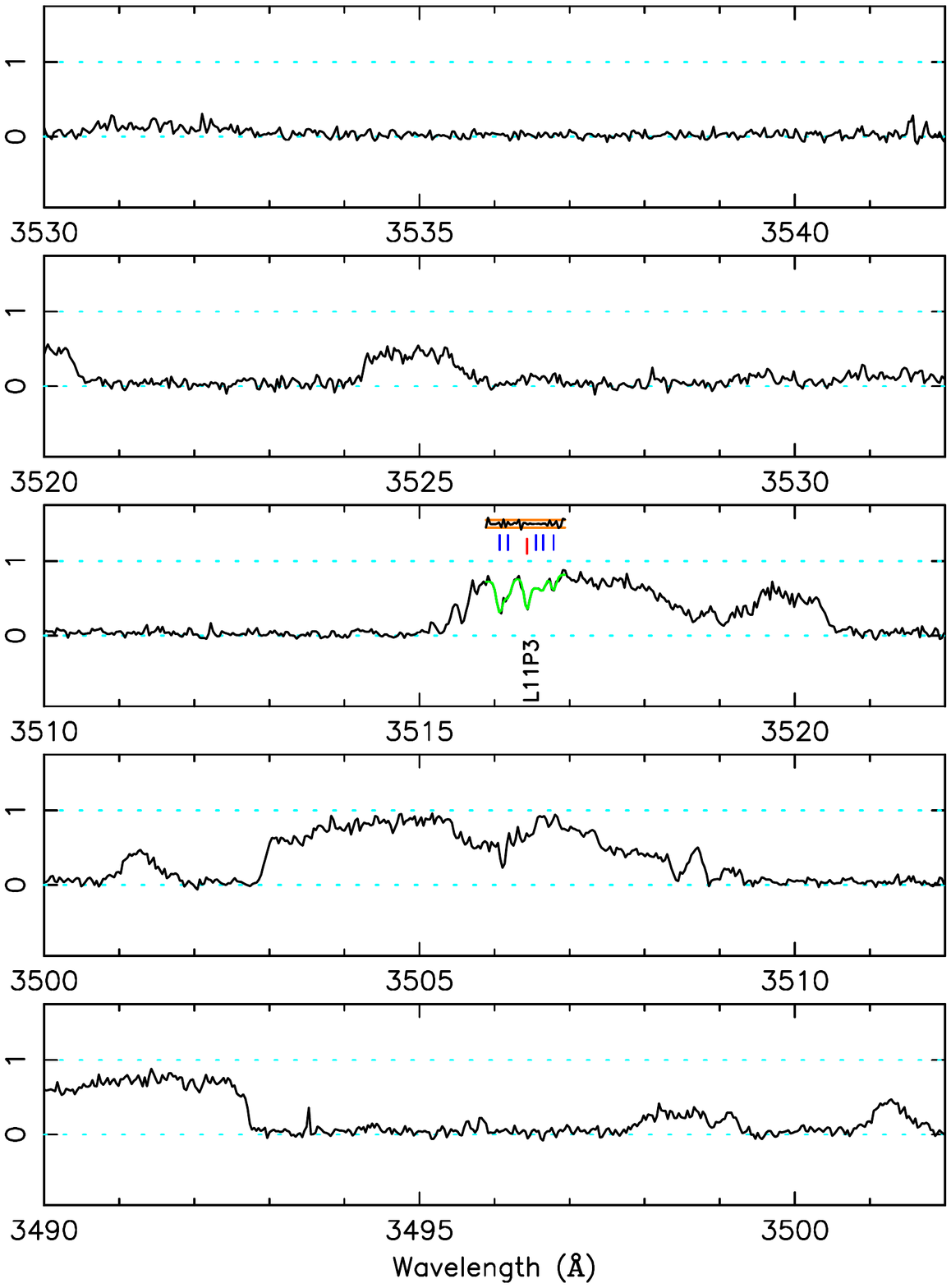}
\par\end{centering}

\caption[H$_2$ fit for the $z=2.595$ absorber toward Q0405$-$443 (4)]{H$_2$ fit for the $z=2.595$ absorber toward Q0405$-$443 (part 4). The vertical axis shows normalised flux. The model fitted to the spectra is shown in green. Red tick marks indicate the position of H$_2$ components, whilst blue tick marks indicate the position of blending transitions (presumed to be Lyman-$\alpha$). Normalised residuals (i.e. [data - model]/error) are plotted above the spectrum between the orange bands, which represent $\pm 1\sigma$. Labels for the H$_2$ transitions are plotted below the data.}
\end{figure}

\begin{figure}[H]
\noindent \begin{centering}
\includegraphics[bb=86bp 180bp 544bp 801bp,clip,width=1\textwidth]{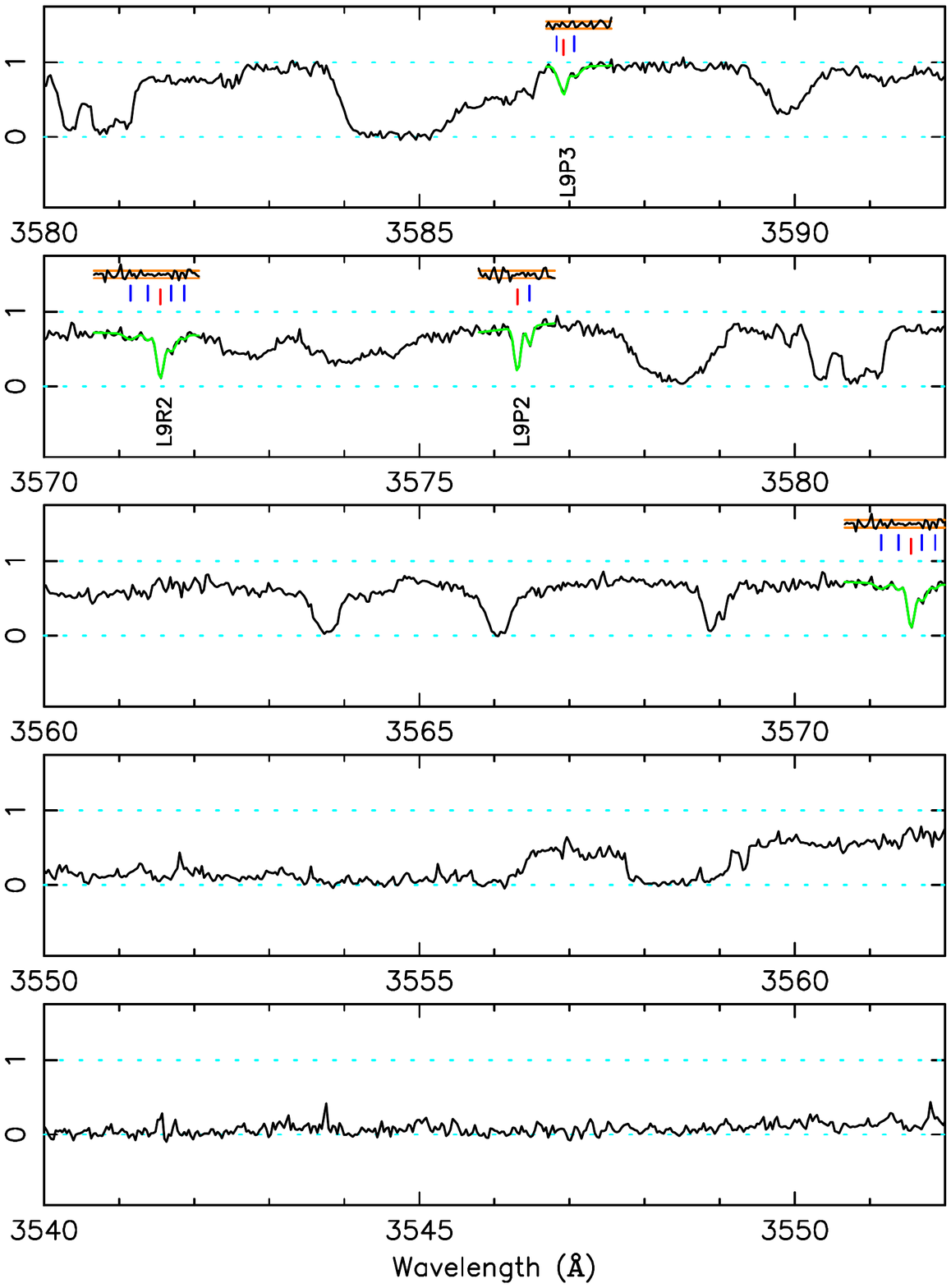}
\par\end{centering}

\caption[H$_2$ fit for the $z=2.595$ absorber toward Q0405$-$443 (5)]{H$_2$ fit for the $z=2.595$ absorber toward Q0405$-$443 (part 5). The vertical axis shows normalised flux. The model fitted to the spectra is shown in green. Red tick marks indicate the position of H$_2$ components, whilst blue tick marks indicate the position of blending transitions (presumed to be Lyman-$\alpha$). Normalised residuals (i.e. [data - model]/error) are plotted above the spectrum between the orange bands, which represent $\pm 1\sigma$. Labels for the H$_2$ transitions are plotted below the data.}
\end{figure}

\begin{figure}[H]
\noindent \begin{centering}
\includegraphics[bb=86bp 180bp 544bp 801bp,clip,width=1\textwidth]{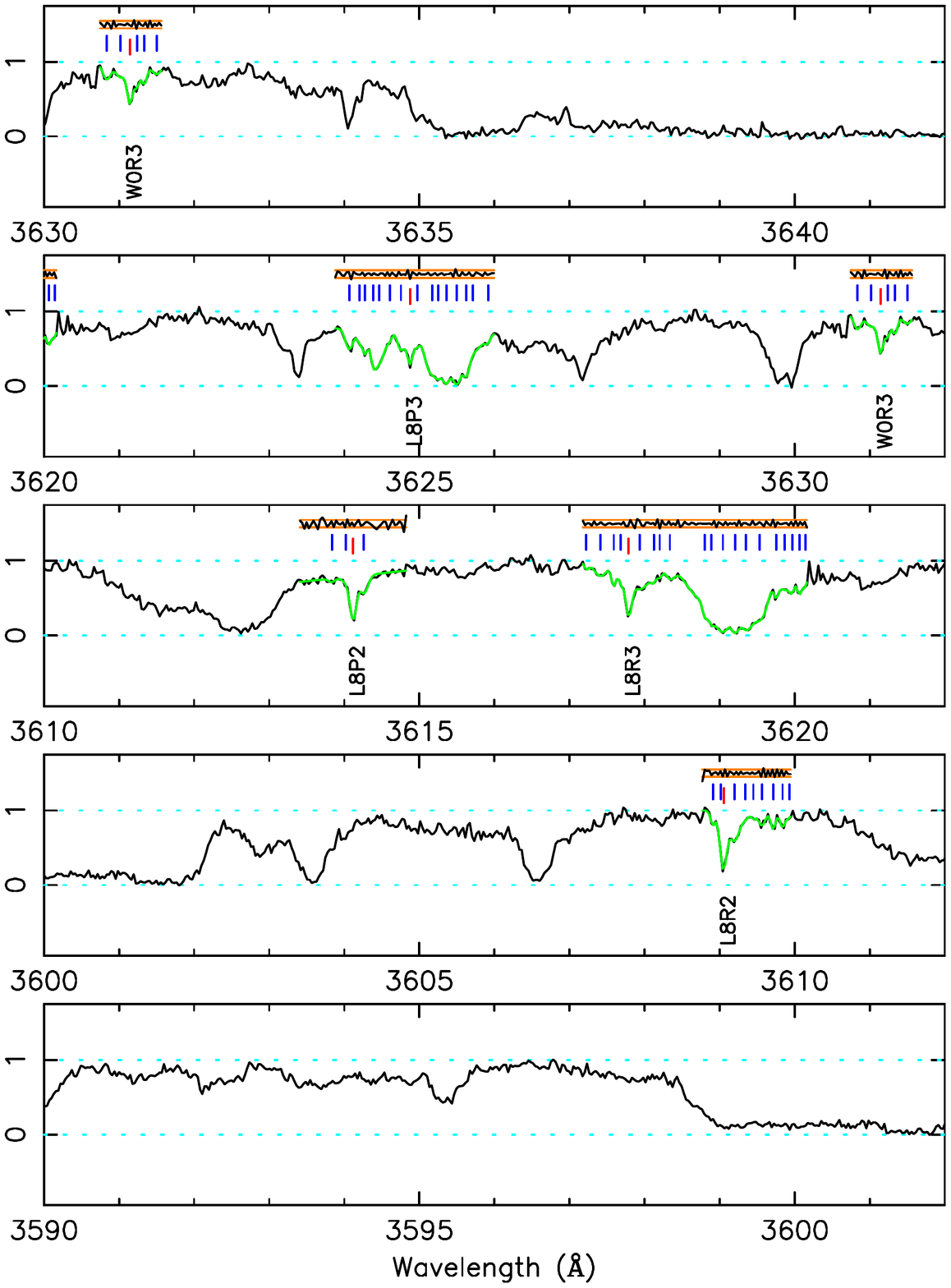}
\par\end{centering}

\caption[H$_2$ fit for the $z=2.595$ absorber toward Q0405$-$443 (6)]{H$_2$ fit for the $z=2.595$ absorber toward Q0405$-$443 (part 6). The vertical axis shows normalised flux. The model fitted to the spectra is shown in green. Red tick marks indicate the position of H$_2$ components, whilst blue tick marks indicate the position of blending transitions (presumed to be Lyman-$\alpha$). Normalised residuals (i.e. [data - model]/error) are plotted above the spectrum between the orange bands, which represent $\pm 1\sigma$. Labels for the H$_2$ transitions are plotted below the data.}
\end{figure}

\begin{figure}[H]
\noindent \begin{centering}
\includegraphics[bb=86bp 180bp 544bp 801bp,clip,width=1\textwidth]{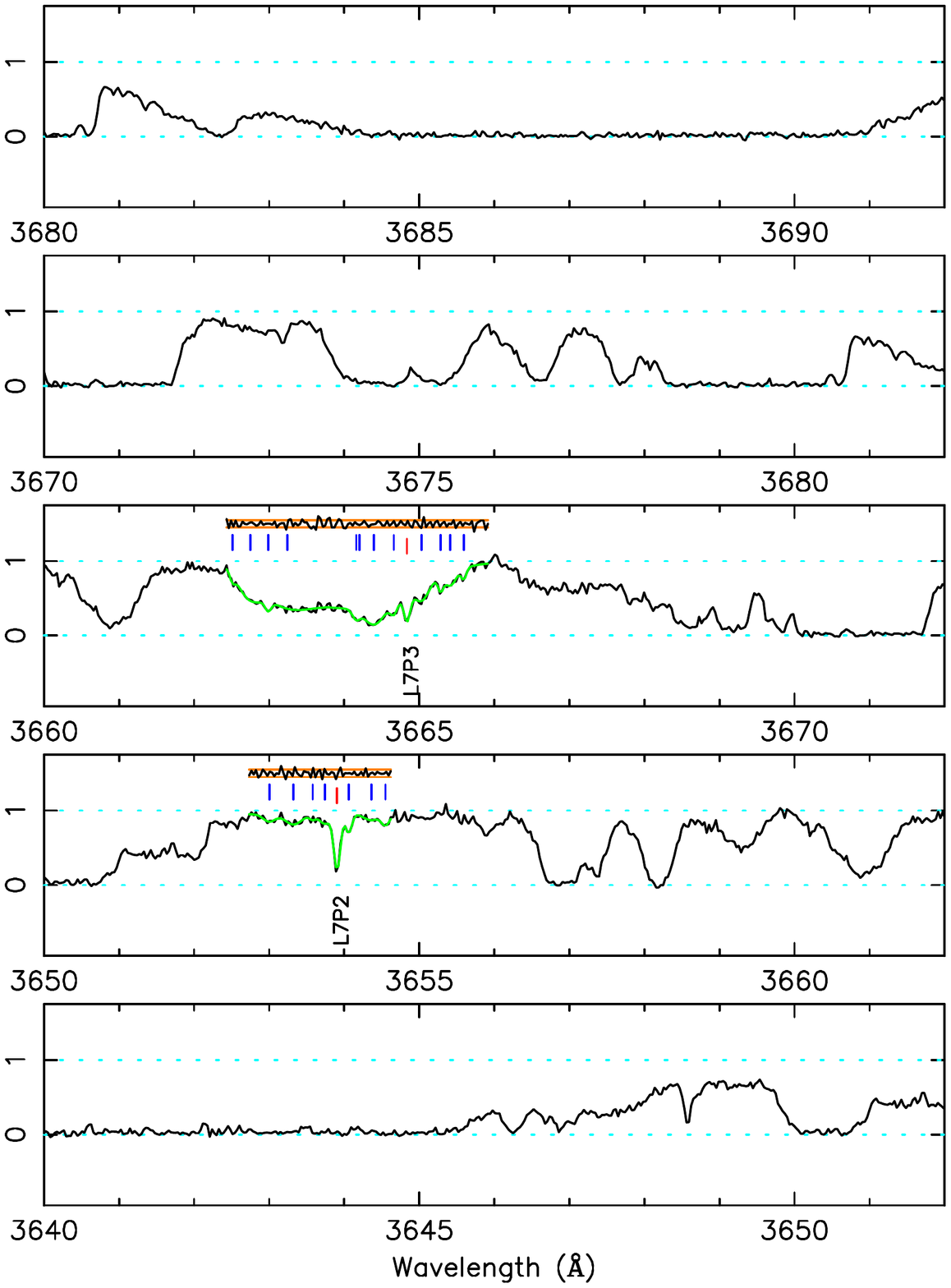}
\par\end{centering}

\caption[H$_2$ fit for the $z=2.595$ absorber toward Q0405$-$443 (7)]{H$_2$ fit for the $z=2.595$ absorber toward Q0405$-$443 (part 7). The vertical axis shows normalised flux. The model fitted to the spectra is shown in green. Red tick marks indicate the position of H$_2$ components, whilst blue tick marks indicate the position of blending transitions (presumed to be Lyman-$\alpha$). Normalised residuals (i.e. [data - model]/error) are plotted above the spectrum between the orange bands, which represent $\pm 1\sigma$. Labels for the H$_2$ transitions are plotted below the data.}
\end{figure}

\begin{figure}[H]
\noindent \begin{centering}
\includegraphics[bb=86bp 180bp 544bp 801bp,clip,width=1\textwidth]{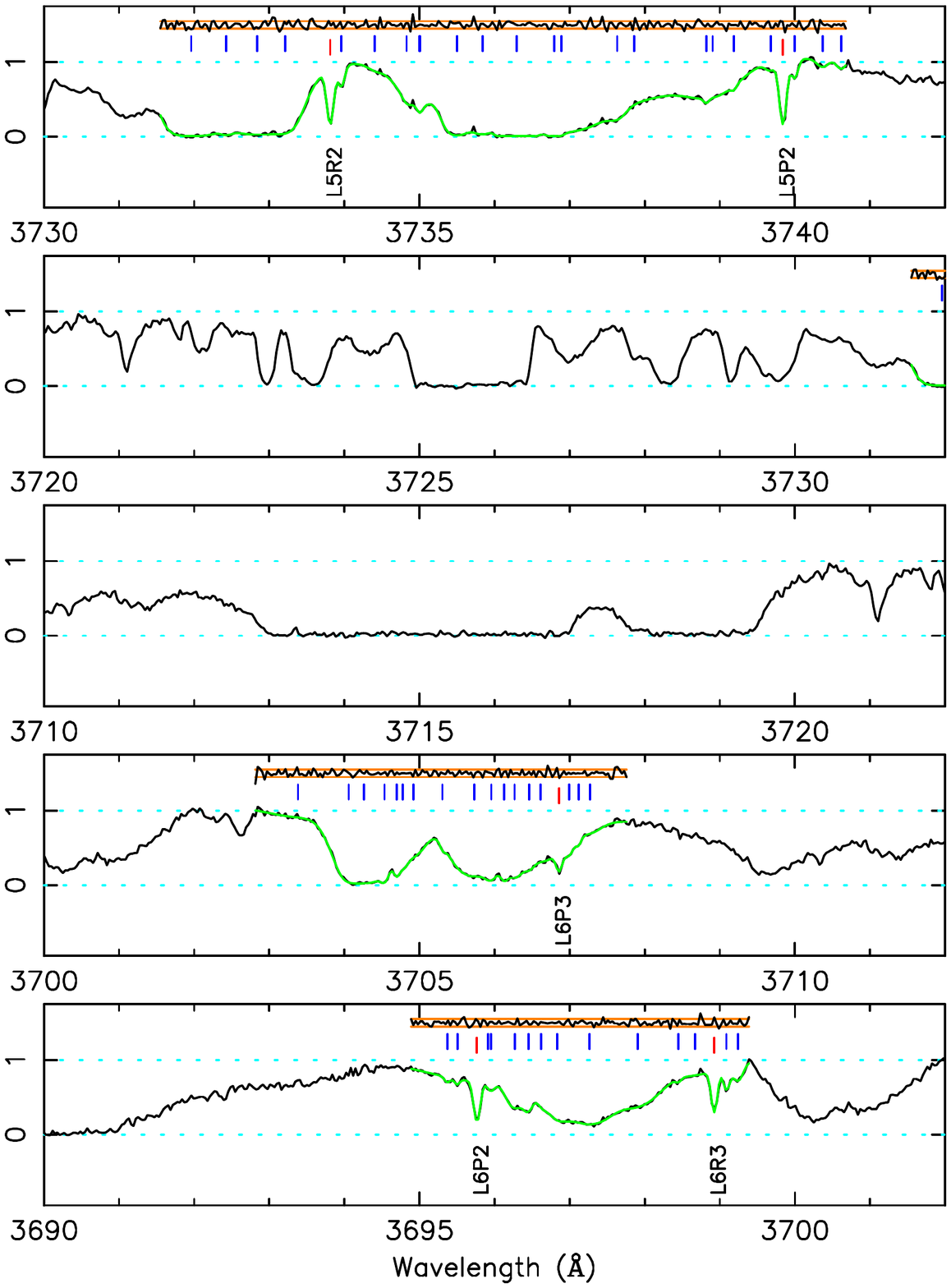}
\par\end{centering}

\caption[H$_2$ fit for the $z=2.595$ absorber toward Q0405$-$443 (8)]{H$_2$ fit for the $z=2.595$ absorber toward Q0405$-$443 (part 8). The vertical axis shows normalised flux. The model fitted to the spectra is shown in green. Red tick marks indicate the position of H$_2$ components, whilst blue tick marks indicate the position of blending transitions (presumed to be Lyman-$\alpha$). Normalised residuals (i.e. [data - model]/error) are plotted above the spectrum between the orange bands, which represent $\pm 1\sigma$. Labels for the H$_2$ transitions are plotted below the data.}
\end{figure}

\begin{figure}[H]
\noindent \begin{centering}
\includegraphics[bb=86bp 180bp 544bp 801bp,clip,width=1\textwidth]{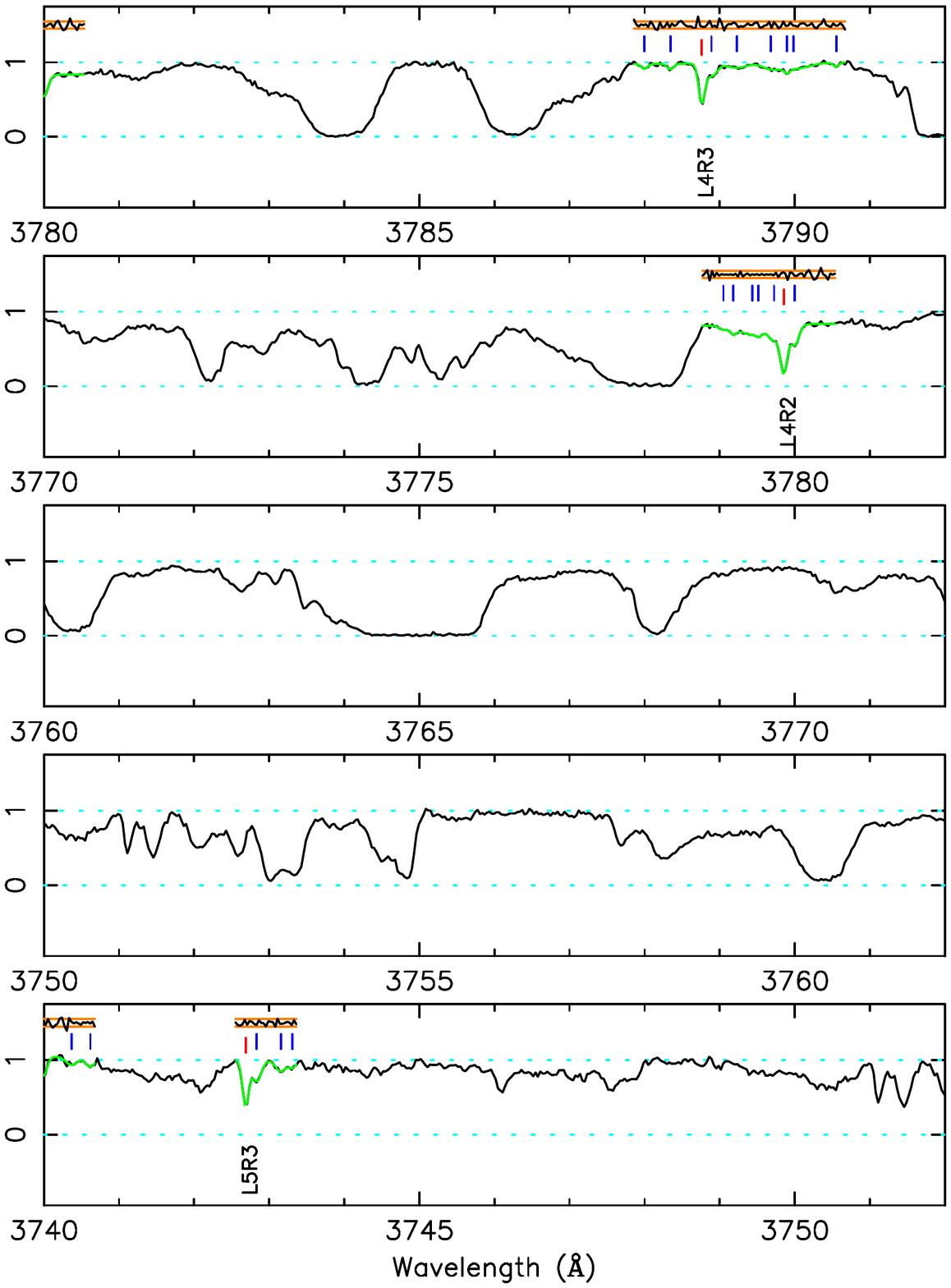}
\par\end{centering}

\caption[H$_2$ fit for the $z=2.595$ absorber toward Q0405$-$443 (9)]{H$_2$ fit for the $z=2.595$ absorber toward Q0405$-$443 (part 9). The vertical axis shows normalised flux. The model fitted to the spectra is shown in green. Red tick marks indicate the position of H$_2$ components, whilst blue tick marks indicate the position of blending transitions (presumed to be Lyman-$\alpha$). Normalised residuals (i.e. [data - model]/error) are plotted above the spectrum between the orange bands, which represent $\pm 1\sigma$. Labels for the H$_2$ transitions are plotted below the data.}
\end{figure}

\begin{figure}[H]
\noindent \begin{centering}
\includegraphics[bb=86bp 180bp 544bp 801bp,clip,width=1\textwidth]{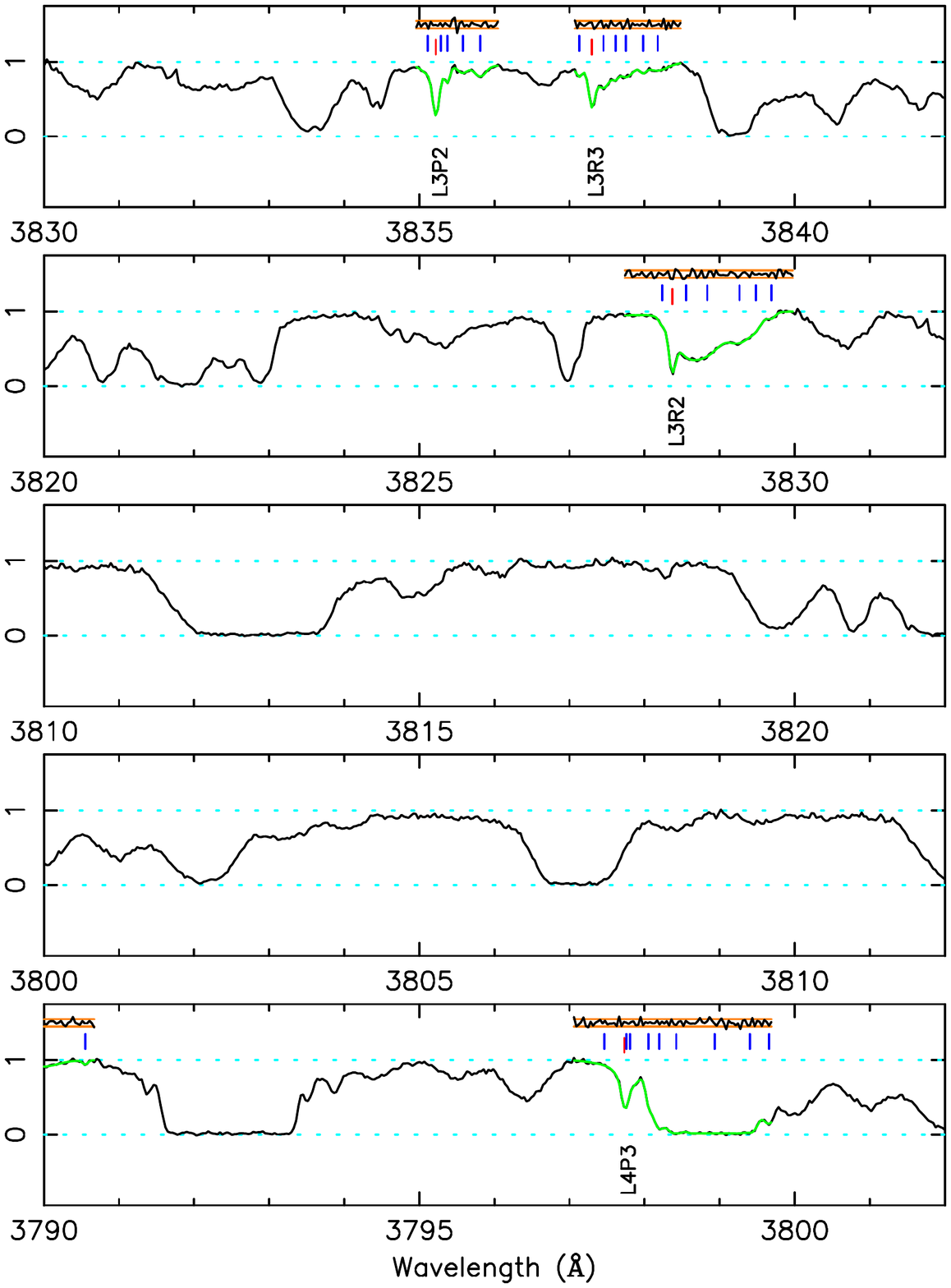}
\par\end{centering}

\caption[H$_2$ fit for the $z=2.595$ absorber toward Q0405$-$443 (10)]{H$_2$ fit for the $z=2.595$ absorber toward Q0405$-$443 (part 10). The vertical axis shows normalised flux. The model fitted to the spectra is shown in green. Red tick marks indicate the position of H$_2$ components, whilst blue tick marks indicate the position of blending transitions (presumed to be Lyman-$\alpha$). Normalised residuals (i.e. [data - model]/error) are plotted above the spectrum between the orange bands, which represent $\pm 1\sigma$. Labels for the H$_2$ transitions are plotted below the data.}
\end{figure}

\begin{figure}[H]
\noindent \begin{centering}
\includegraphics[bb=86bp 180bp 544bp 801bp,clip,width=1\textwidth]{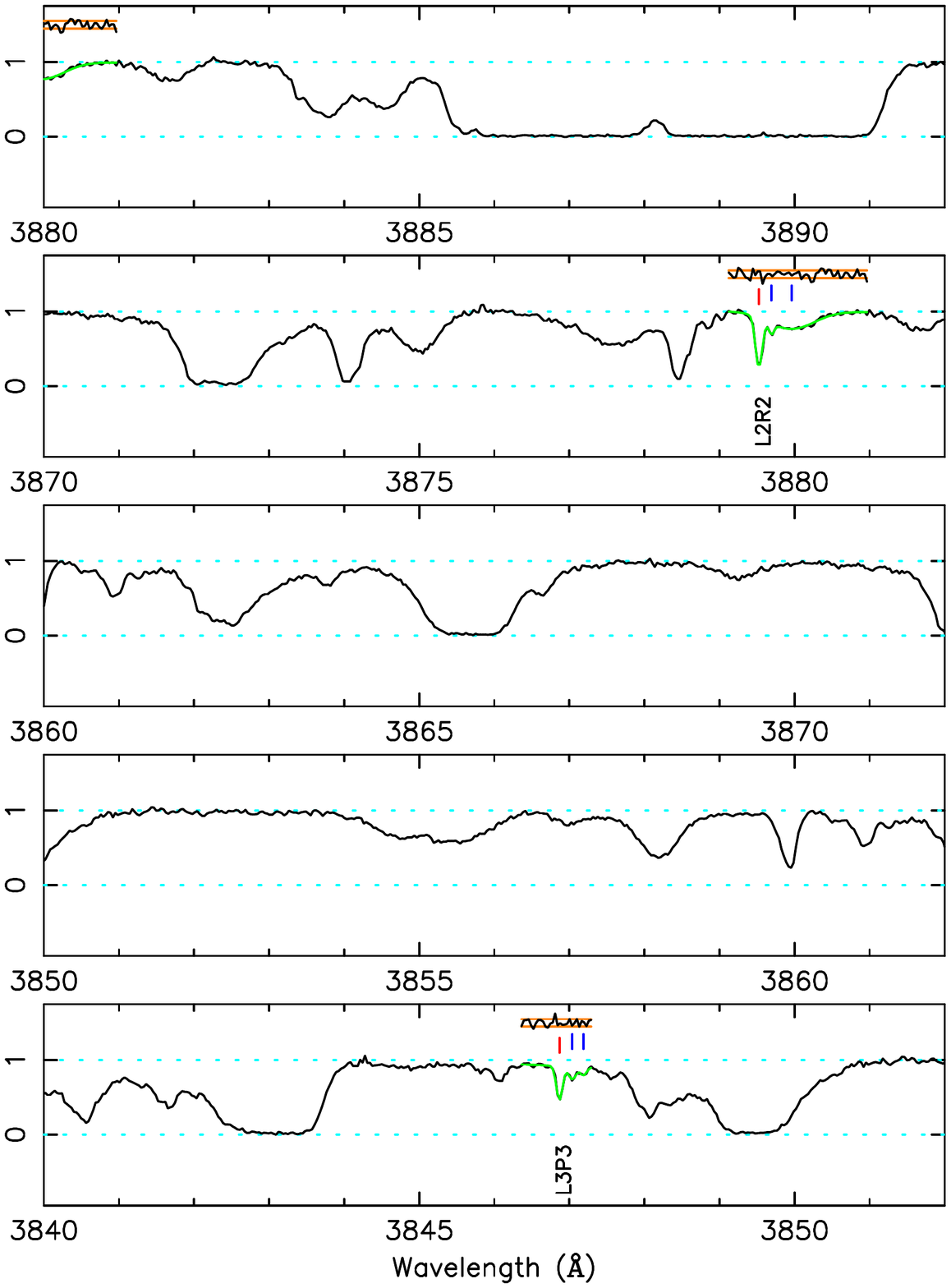}
\par\end{centering}

\caption[H$_2$ fit for the $z=2.595$ absorber toward Q0405$-$443 (11)]{H$_2$ fit for the $z=2.595$ absorber toward Q0405$-$443 (part 11). The vertical axis shows normalised flux. The model fitted to the spectra is shown in green. Red tick marks indicate the position of H$_2$ components, whilst blue tick marks indicate the position of blending transitions (presumed to be Lyman-$\alpha$). Normalised residuals (i.e. [data - model]/error) are plotted above the spectrum between the orange bands, which represent $\pm 1\sigma$. Labels for the H$_2$ transitions are plotted below the data.}
\end{figure}

\begin{figure}[H]
\noindent \begin{centering}
\includegraphics[bb=86bp 180bp 544bp 801bp,clip,width=1\textwidth]{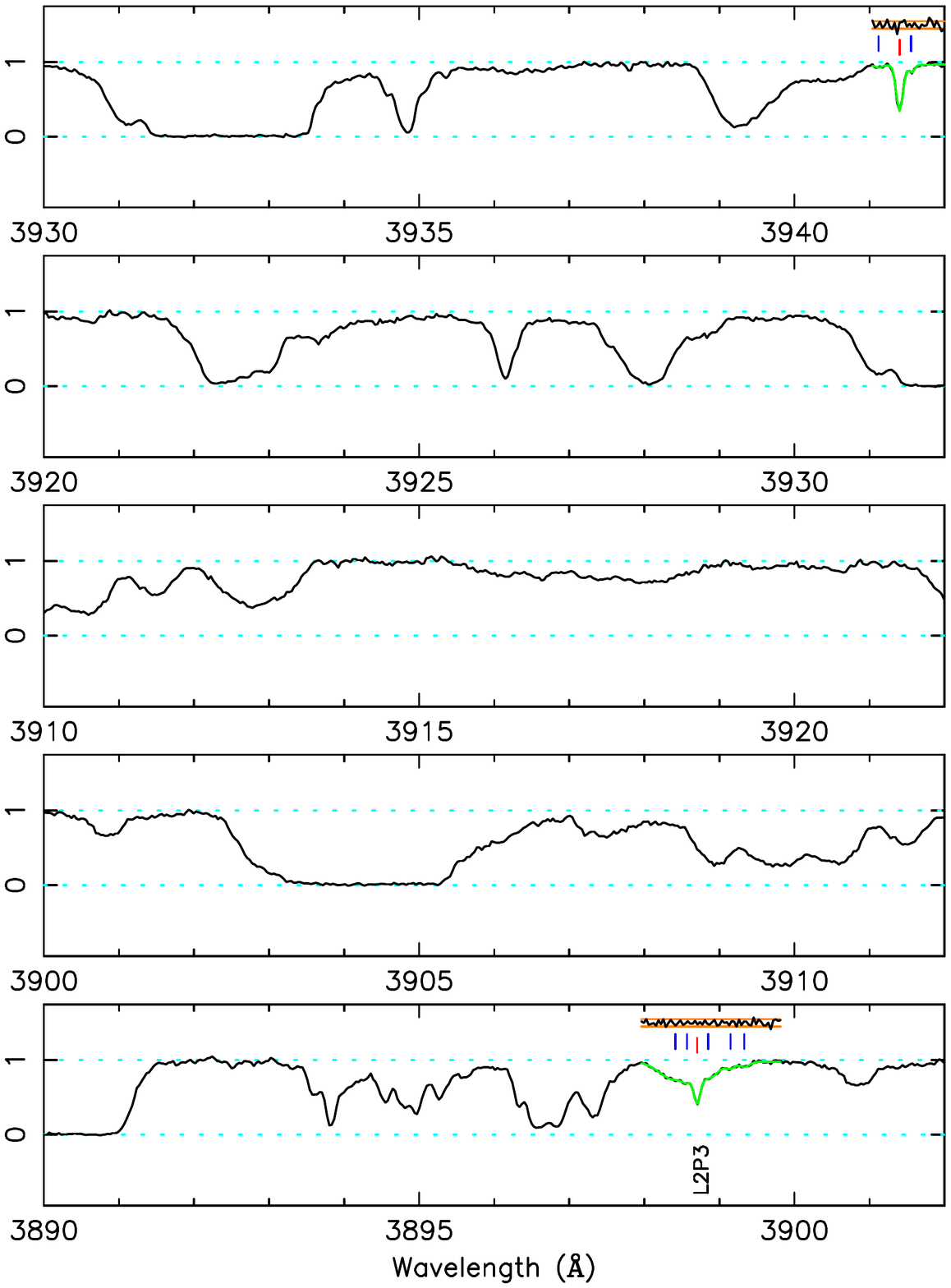}
\par\end{centering}

\caption[H$_2$ fit for the $z=2.595$ absorber toward Q0405$-$443 (12)]{H$_2$ fit for the $z=2.595$ absorber toward Q0405$-$443 (part 12). The vertical axis shows normalised flux. The model fitted to the spectra is shown in green. Red tick marks indicate the position of H$_2$ components, whilst blue tick marks indicate the position of blending transitions (presumed to be Lyman-$\alpha$). Normalised residuals (i.e. [data - model]/error) are plotted above the spectrum between the orange bands, which represent $\pm 1\sigma$. Labels for the H$_2$ transitions are plotted below the data.}
\end{figure}

\begin{figure}[H]
\noindent \begin{centering}
\includegraphics[bb=86bp 180bp 544bp 801bp,clip,width=1\textwidth]{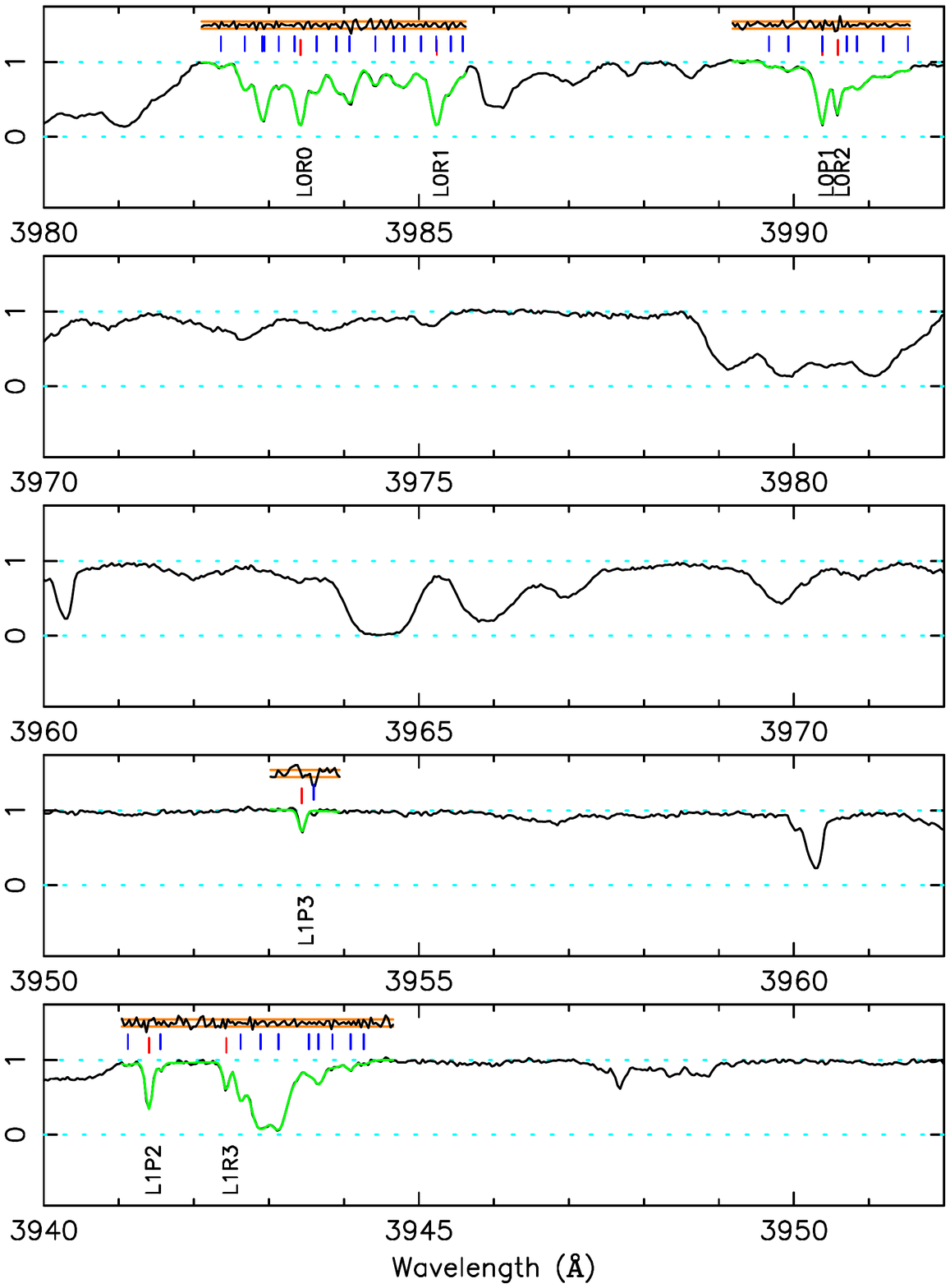}
\par\end{centering}

\caption[H$_2$ fit for the $z=2.595$ absorber toward Q0405$-$443 (13)]{H$_2$ fit for the $z=2.595$ absorber toward Q0405$-$443 (part 13). The vertical axis shows normalised flux. The model fitted to the spectra is shown in green. Red tick marks indicate the position of H$_2$ components, whilst blue tick marks indicate the position of blending transitions (presumed to be Lyman-$\alpha$). Normalised residuals (i.e. [data - model]/error) are plotted above the spectrum between the orange bands, which represent $\pm 1\sigma$. Labels for the H$_2$ transitions are plotted below the data.}
\end{figure}

\begin{figure}[H]
\noindent \begin{centering}
\includegraphics[bb=86bp 180bp 544bp 440bp,clip,width=1\textwidth]{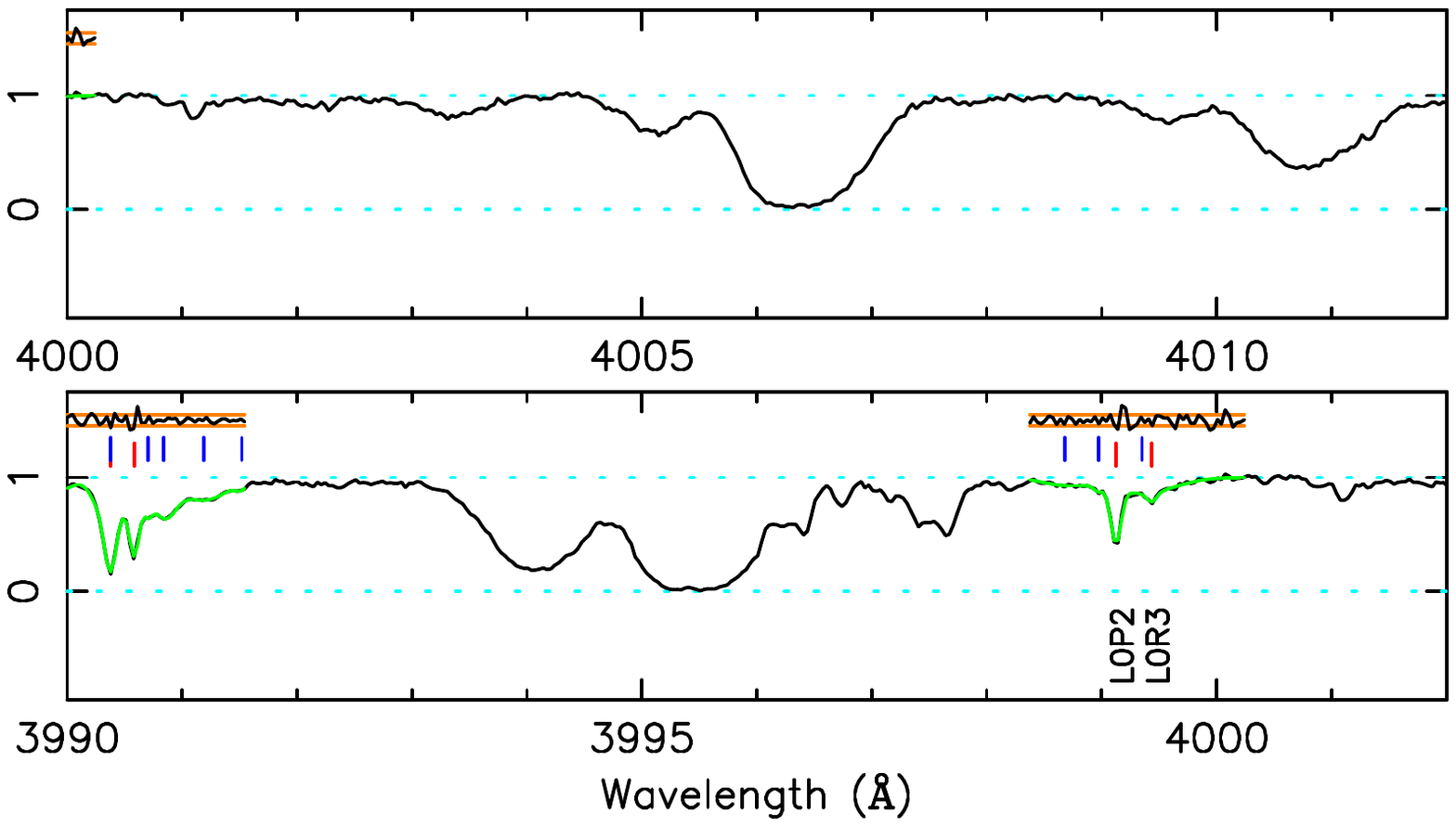}
\par\end{centering}

\caption[H$_2$ fit for the $z=2.595$ absorber toward Q0405$-$443 (14)]{H$_2$ fit for the $z=2.595$ absorber toward Q0405$-$443 (part 14). The vertical axis shows normalised flux. The model fitted to the spectra is shown in green. Red tick marks indicate the position of H$_2$ components, whilst blue tick marks indicate the position of blending transitions (presumed to be Lyman-$\alpha$). Normalised residuals (i.e. [data - model]/error) are plotted above the spectrum between the orange bands, which represent $\pm 1\sigma$. Labels for the H$_2$ transitions are plotted below the data.}
\end{figure}

\chapter{Q0347$-$383 Voigt profile fits\label{cha:mu fits:Q0347}}

In this appendix, we provide the fits for the $z=3.025$ H$_{2}$
absorber toward Q0347$-$383. 

\begin{figure}[H]
\noindent \begin{centering}
\includegraphics[bb=86bp 180bp 544bp 801bp,clip,width=1\textwidth]{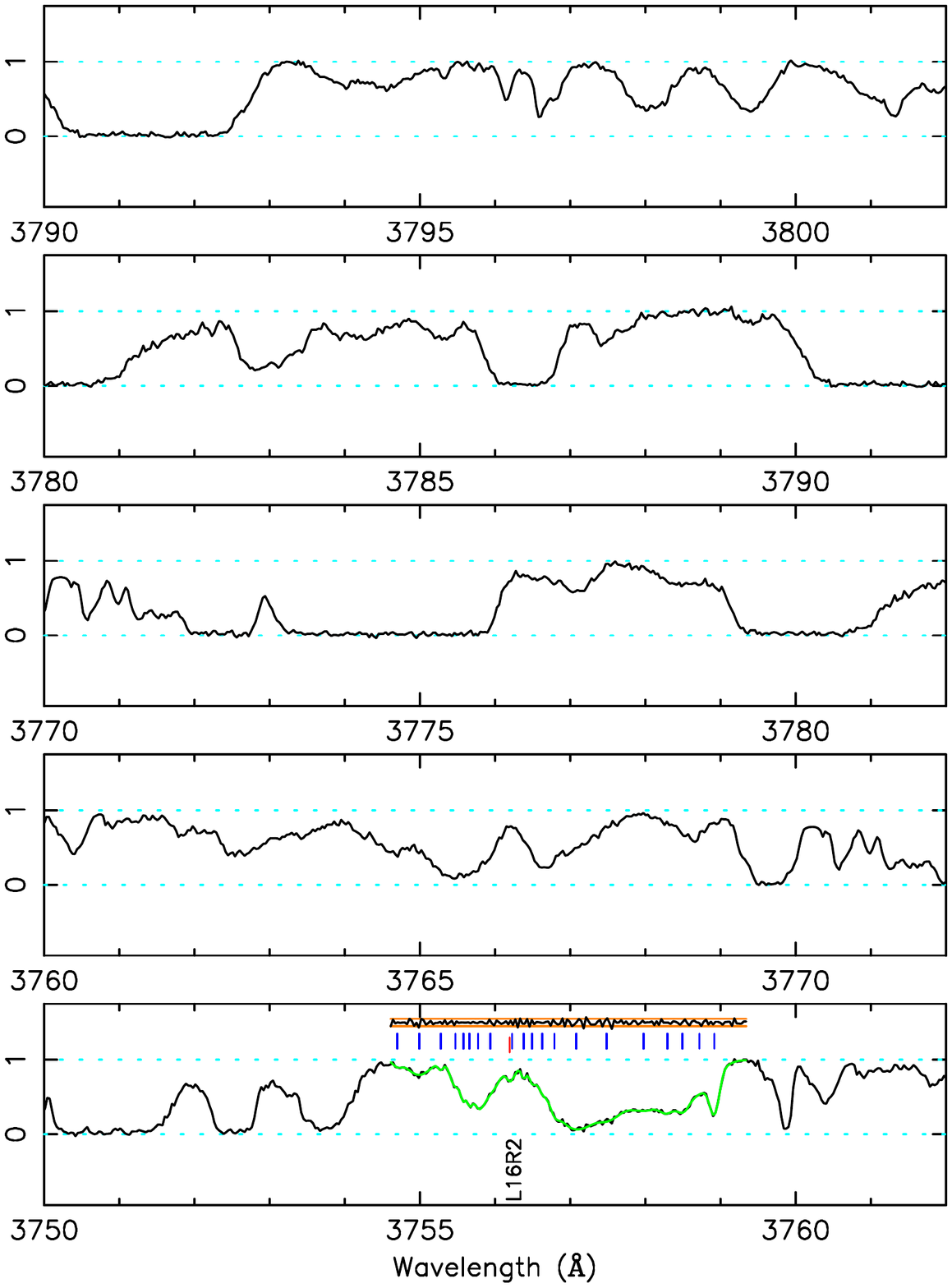}
\par\end{centering}

\caption[H$_2$ fit for the $z=3.025$ absorber toward Q0347$-$383 (1)]{H$_2$ fit for the $z=3.025$ absorber toward Q0347$-$383 (part 1). The vertical axis shows normalised flux. The model fitted to the spectra is shown in green. Red tick marks indicate the position of H$_2$ components, whilst blue tick marks indicate the position of blending transitions (presumed to be Lyman-$\alpha$). Normalised residuals (i.e. [data - model]/error) are plotted above the spectrum between the orange bands, which represent $\pm 1\sigma$. Labels for the H$_2$ transitions are plotted below the data.}
\end{figure}

\begin{figure}[H]
\noindent \begin{centering}
\includegraphics[bb=86bp 180bp 544bp 801bp,clip,width=1\textwidth]{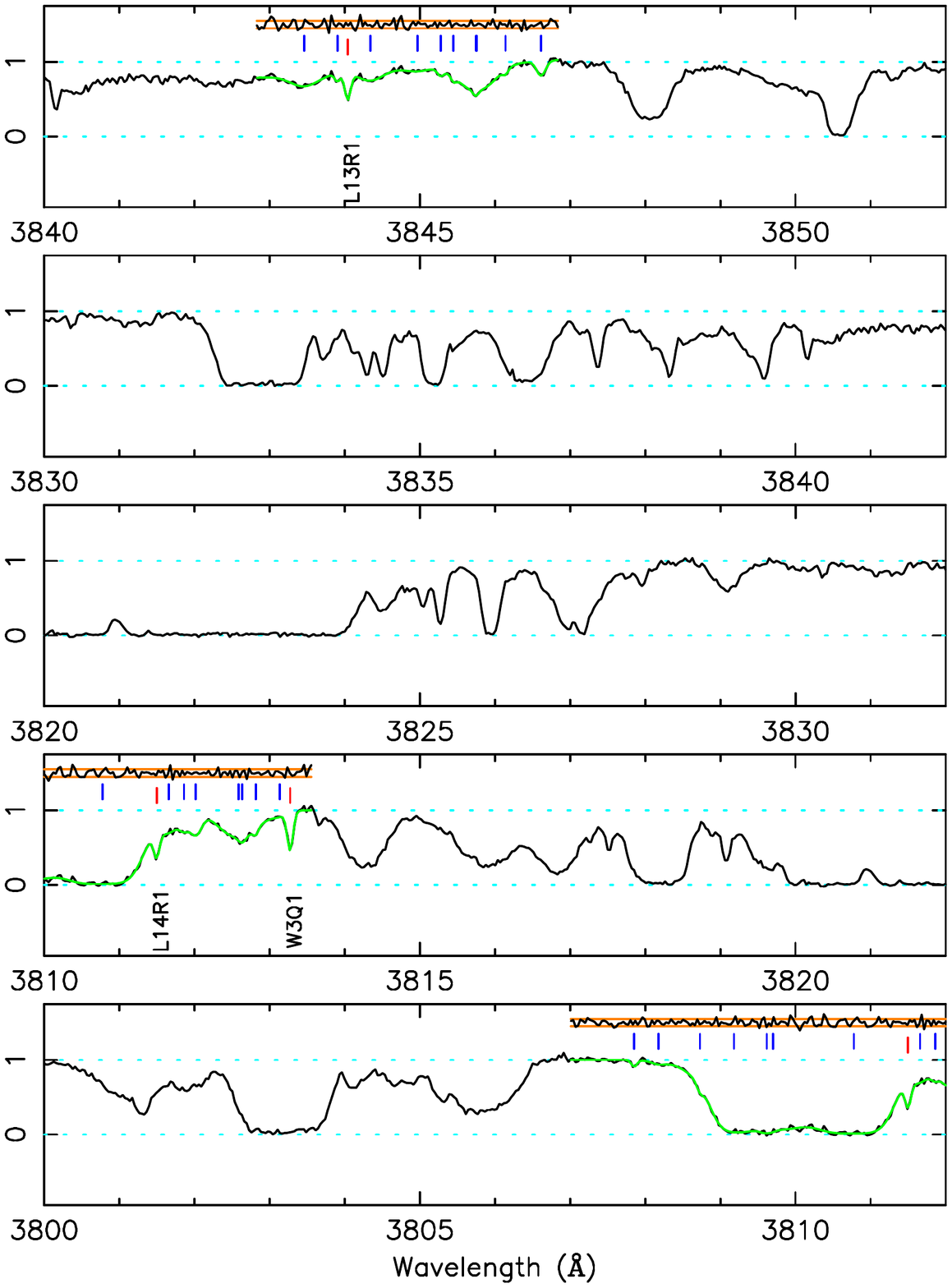}
\par\end{centering}

\caption[H$_2$ fit for the $z=3.025$ absorber toward Q0347$-$383 (2)]{H$_2$ fit for the $z=3.025$ absorber toward Q0347$-$383 (part 2). The vertical axis shows normalised flux. The model fitted to the spectra is shown in green. Red tick marks indicate the position of H$_2$ components, whilst blue tick marks indicate the position of blending transitions (presumed to be Lyman-$\alpha$). Normalised residuals (i.e. [data - model]/error) are plotted above the spectrum between the orange bands, which represent $\pm 1\sigma$. Labels for the H$_2$ transitions are plotted below the data.}
\end{figure}

\begin{figure}[H]
\noindent \begin{centering}
\includegraphics[bb=86bp 180bp 544bp 801bp,clip,width=1\textwidth]{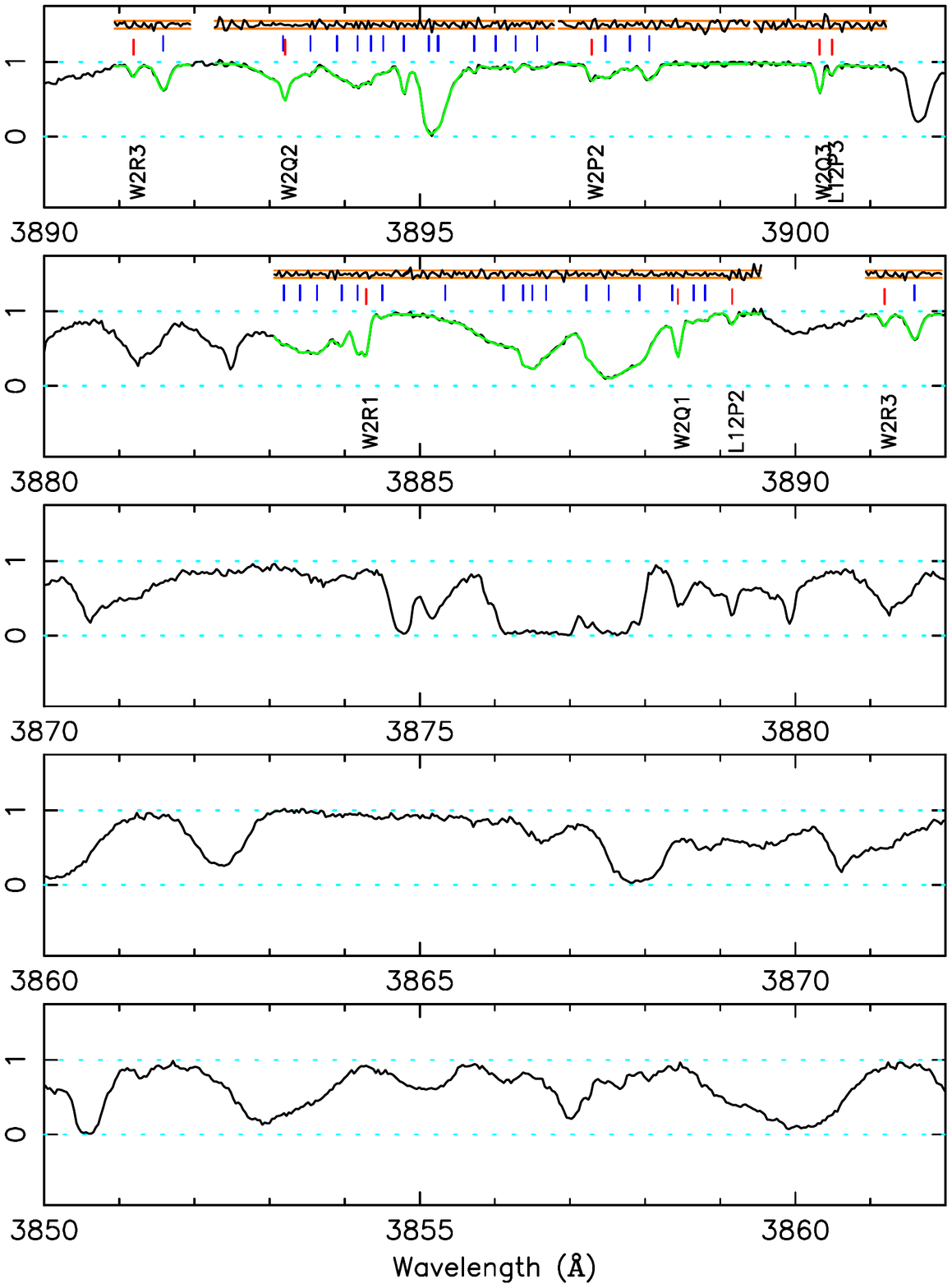}
\par\end{centering}

\caption[H$_2$ fit for the $z=3.025$ absorber toward Q0347$-$383 (3)]{H$_2$ fit for the $z=3.025$ absorber toward Q0347$-$383 (part 3). The vertical axis shows normalised flux. The model fitted to the spectra is shown in green. Red tick marks indicate the position of H$_2$ components, whilst blue tick marks indicate the position of blending transitions (presumed to be Lyman-$\alpha$). Normalised residuals (i.e. [data - model]/error) are plotted above the spectrum between the orange bands, which represent $\pm 1\sigma$. Labels for the H$_2$ transitions are plotted below the data.}
\end{figure}

\begin{figure}[H]
\noindent \begin{centering}
\includegraphics[bb=86bp 180bp 544bp 801bp,clip,width=1\textwidth]{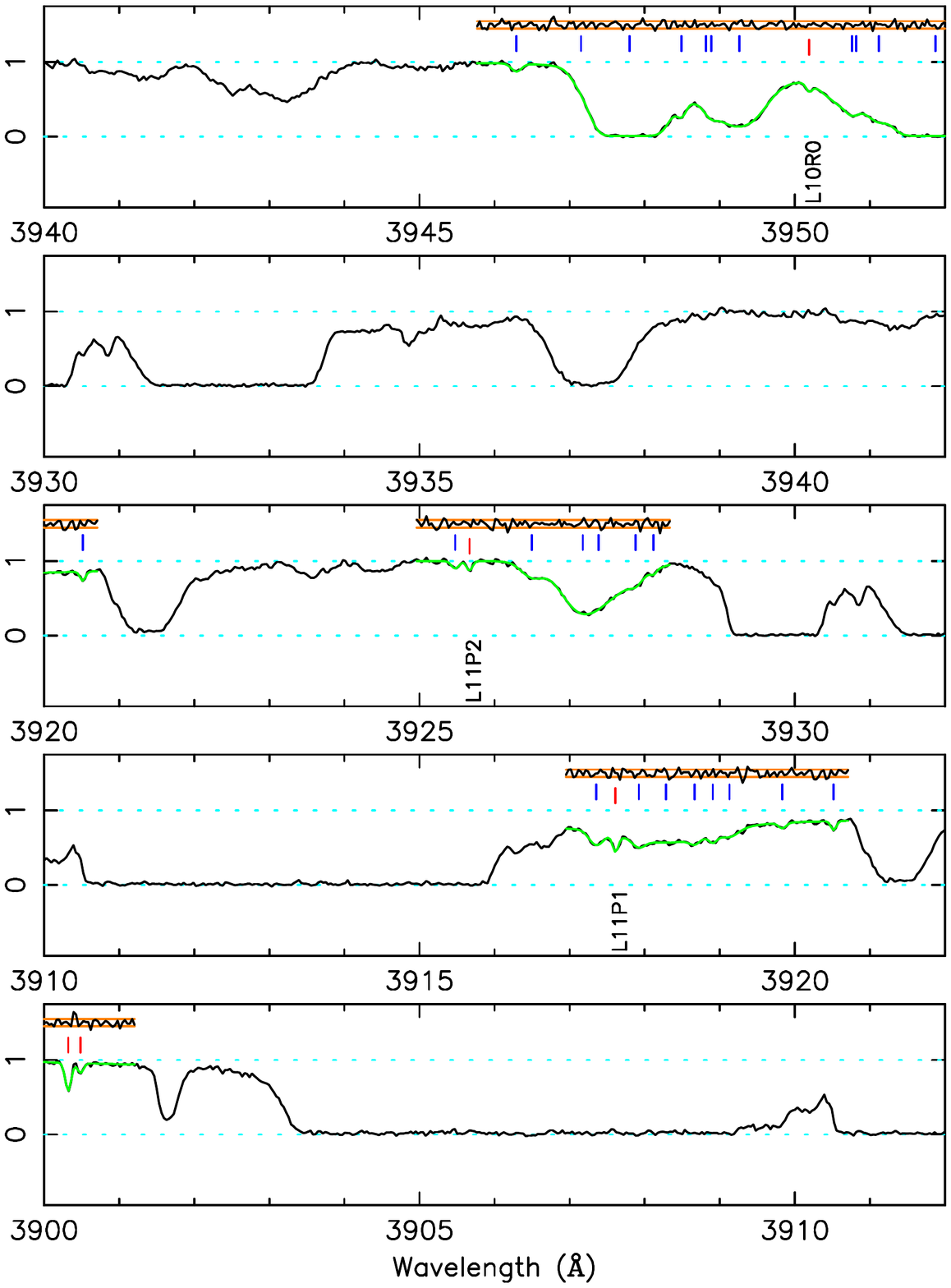}
\par\end{centering}

\caption[H$_2$ fit for the $z=3.025$ absorber toward Q0347$-$383 (4)]{H$_2$ fit for the $z=3.025$ absorber toward Q0347$-$383 (part 4). The vertical axis shows normalised flux. The model fitted to the spectra is shown in green. Red tick marks indicate the position of H$_2$ components, whilst blue tick marks indicate the position of blending transitions (presumed to be Lyman-$\alpha$). Normalised residuals (i.e. [data - model]/error) are plotted above the spectrum between the orange bands, which represent $\pm 1\sigma$. Labels for the H$_2$ transitions are plotted below the data.}
\end{figure}

\begin{figure}[H]
\noindent \begin{centering}
\includegraphics[bb=86bp 180bp 544bp 801bp,clip,width=1\textwidth]{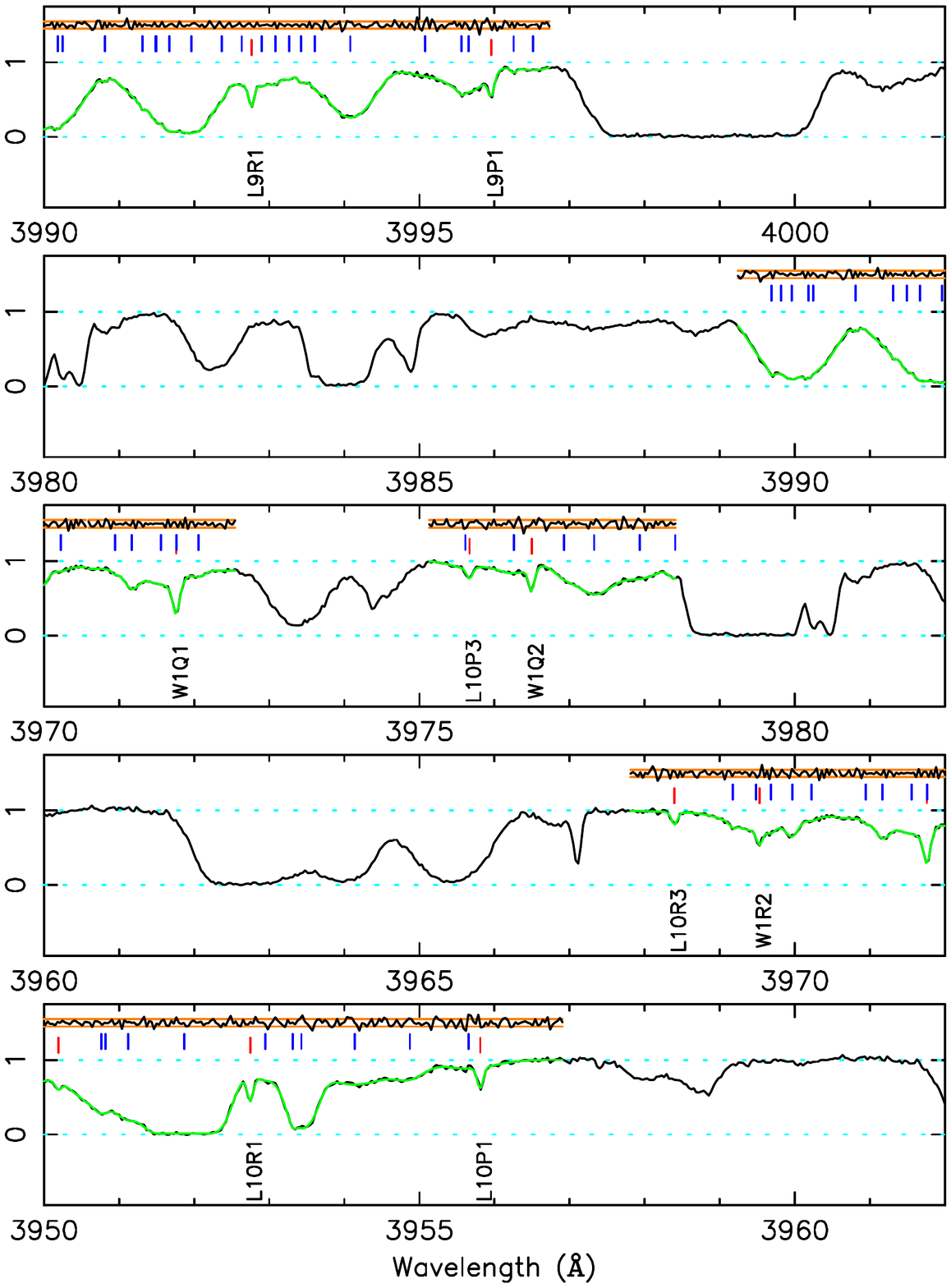}
\par\end{centering}

\caption[H$_2$ fit for the $z=3.025$ absorber toward Q0347$-$383 (5)]{H$_2$ fit for the $z=3.025$ absorber toward Q0347$-$383 (part 5). The vertical axis shows normalised flux. The model fitted to the spectra is shown in green. Red tick marks indicate the position of H$_2$ components, whilst blue tick marks indicate the position of blending transitions (presumed to be Lyman-$\alpha$). Normalised residuals (i.e. [data - model]/error) are plotted above the spectrum between the orange bands, which represent $\pm 1\sigma$. Labels for the H$_2$ transitions are plotted below the data.}
\end{figure}

\begin{figure}[H]
\noindent \begin{centering}
\includegraphics[bb=86bp 180bp 544bp 801bp,clip,width=1\textwidth]{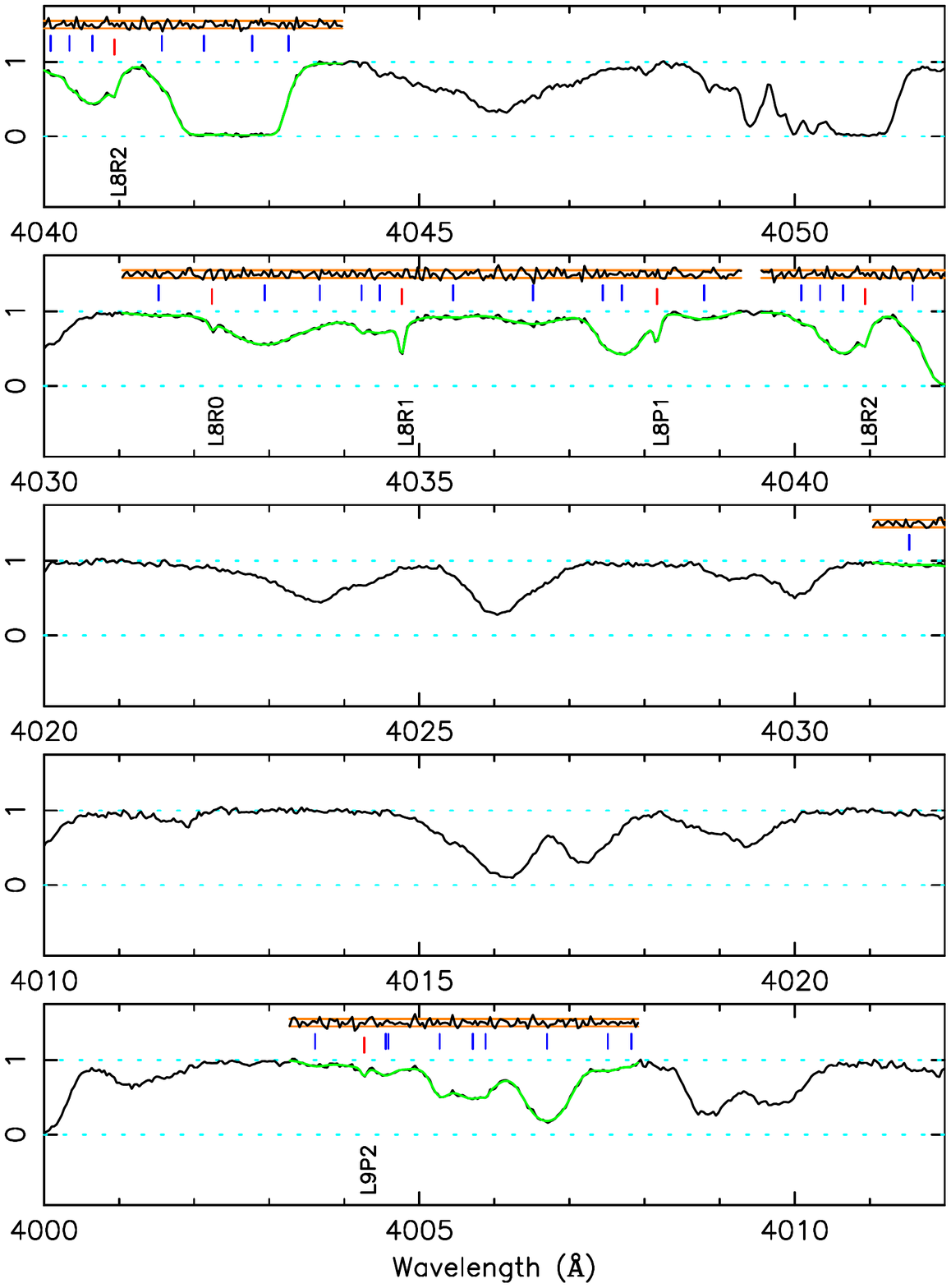}
\par\end{centering}

\caption[H$_2$ fit for the $z=3.025$ absorber toward Q0347$-$383 (6)]{H$_2$ fit for the $z=3.025$ absorber toward Q0347$-$383 (part 6). The vertical axis shows normalised flux. The model fitted to the spectra is shown in green. Red tick marks indicate the position of H$_2$ components, whilst blue tick marks indicate the position of blending transitions (presumed to be Lyman-$\alpha$). Normalised residuals (i.e. [data - model]/error) are plotted above the spectrum between the orange bands, which represent $\pm 1\sigma$. Labels for the H$_2$ transitions are plotted below the data.}
\end{figure}

\begin{figure}[H]
\noindent \begin{centering}
\includegraphics[bb=86bp 180bp 544bp 801bp,clip,width=1\textwidth]{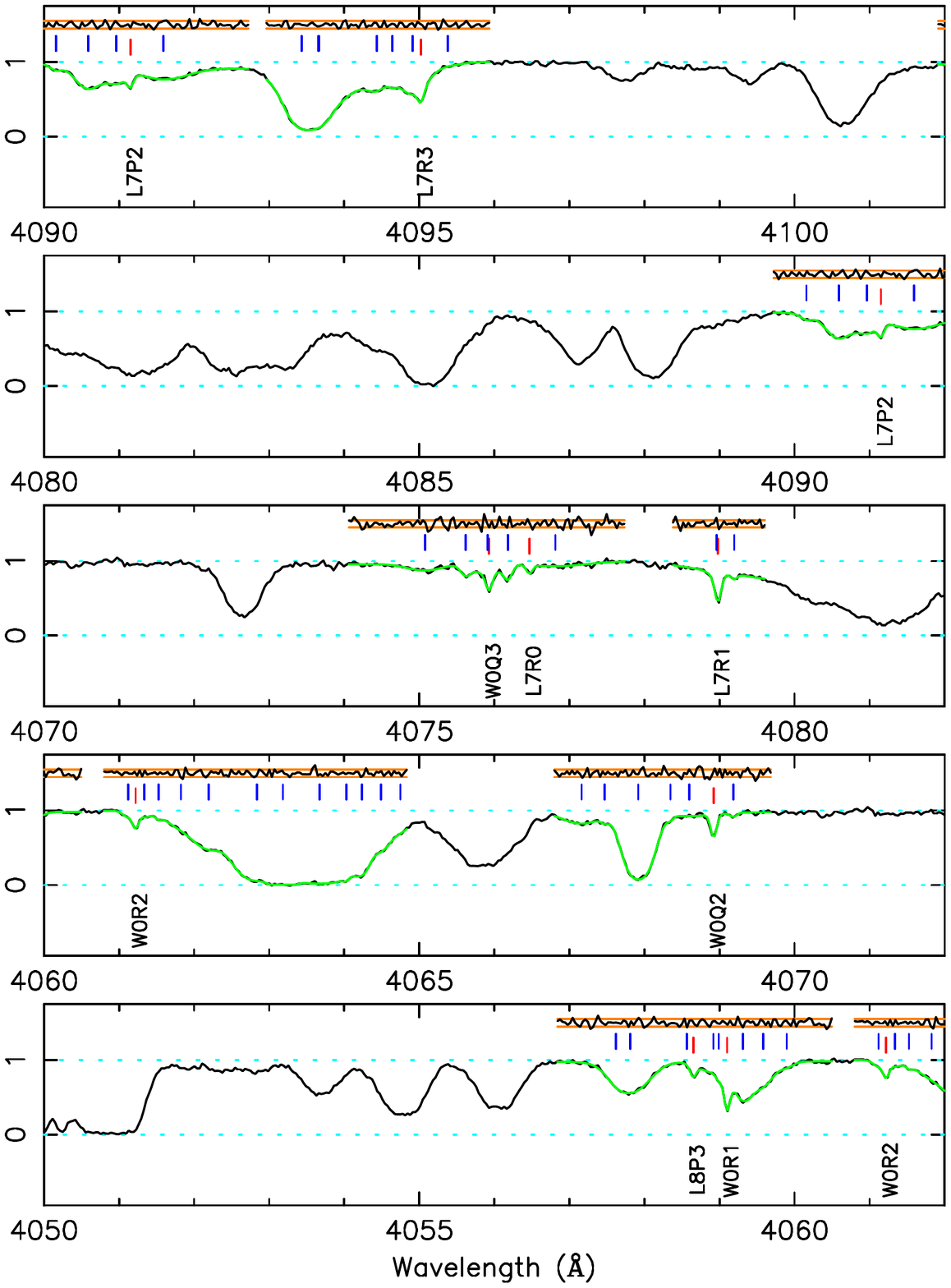}
\par\end{centering}

\caption[H$_2$ fit for the $z=3.025$ absorber toward Q0347$-$383 (7)]{H$_2$ fit for the $z=3.025$ absorber toward Q0347$-$383 (part 7). The vertical axis shows normalised flux. The model fitted to the spectra is shown in green. Red tick marks indicate the position of H$_2$ components, whilst blue tick marks indicate the position of blending transitions (presumed to be Lyman-$\alpha$). Normalised residuals (i.e. [data - model]/error) are plotted above the spectrum between the orange bands, which represent $\pm 1\sigma$. Labels for the H$_2$ transitions are plotted below the data.}
\end{figure}

\begin{figure}[H]
\noindent \begin{centering}
\includegraphics[bb=86bp 180bp 544bp 801bp,clip,width=1\textwidth]{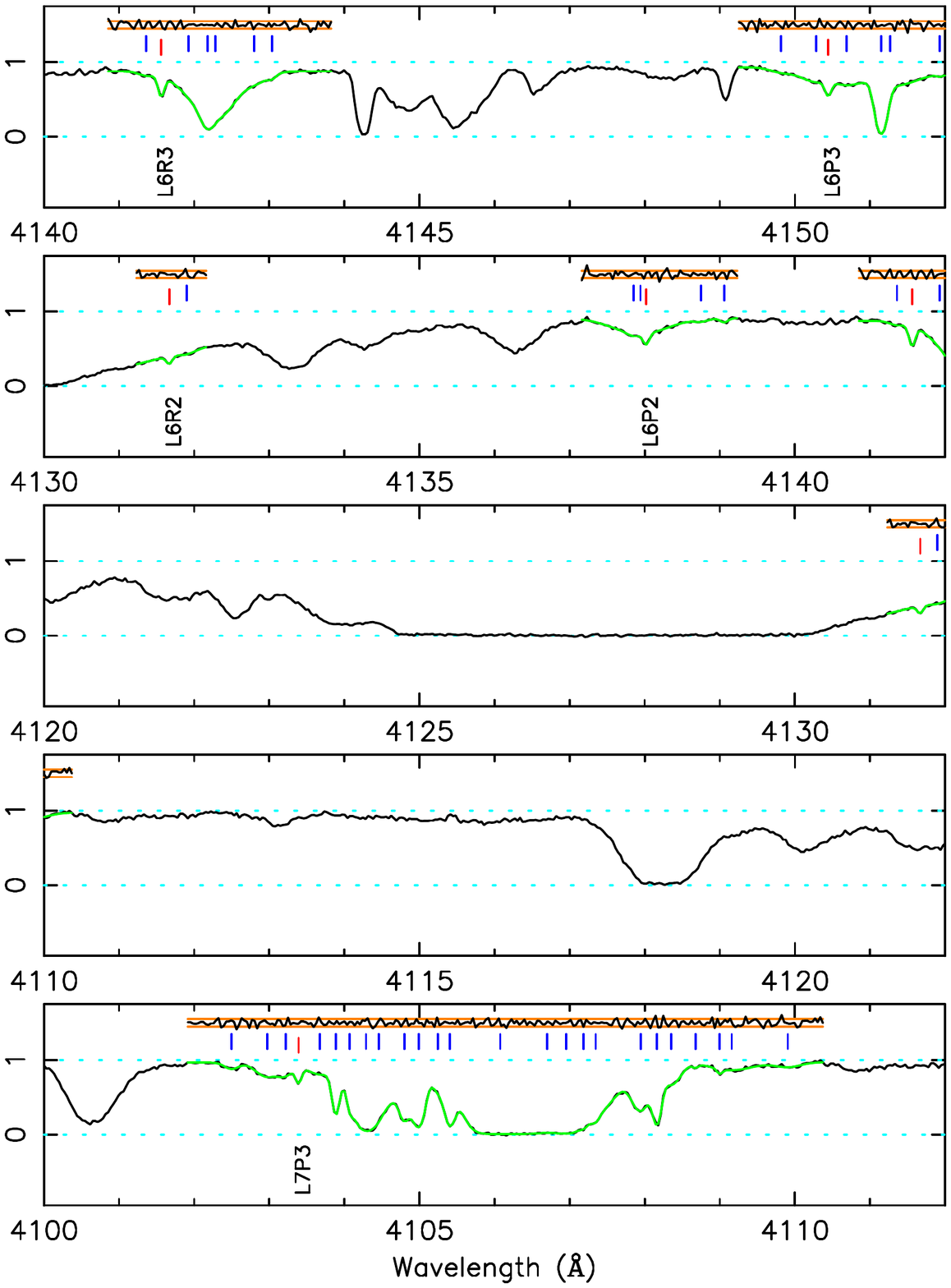}
\par\end{centering}

\caption[H$_2$ fit for the $z=3.025$ absorber toward Q0347$-$383 (8)]{H$_2$ fit for the $z=3.025$ absorber toward Q0347$-$383 (part 8). The vertical axis shows normalised flux. The model fitted to the spectra is shown in green. Red tick marks indicate the position of H$_2$ components, whilst blue tick marks indicate the position of blending transitions (presumed to be Lyman-$\alpha$). Normalised residuals (i.e. [data - model]/error) are plotted above the spectrum between the orange bands, which represent $\pm 1\sigma$. Labels for the H$_2$ transitions are plotted below the data.}
\end{figure}

\begin{figure}[H]
\noindent \begin{centering}
\includegraphics[bb=86bp 180bp 544bp 801bp,clip,width=1\textwidth]{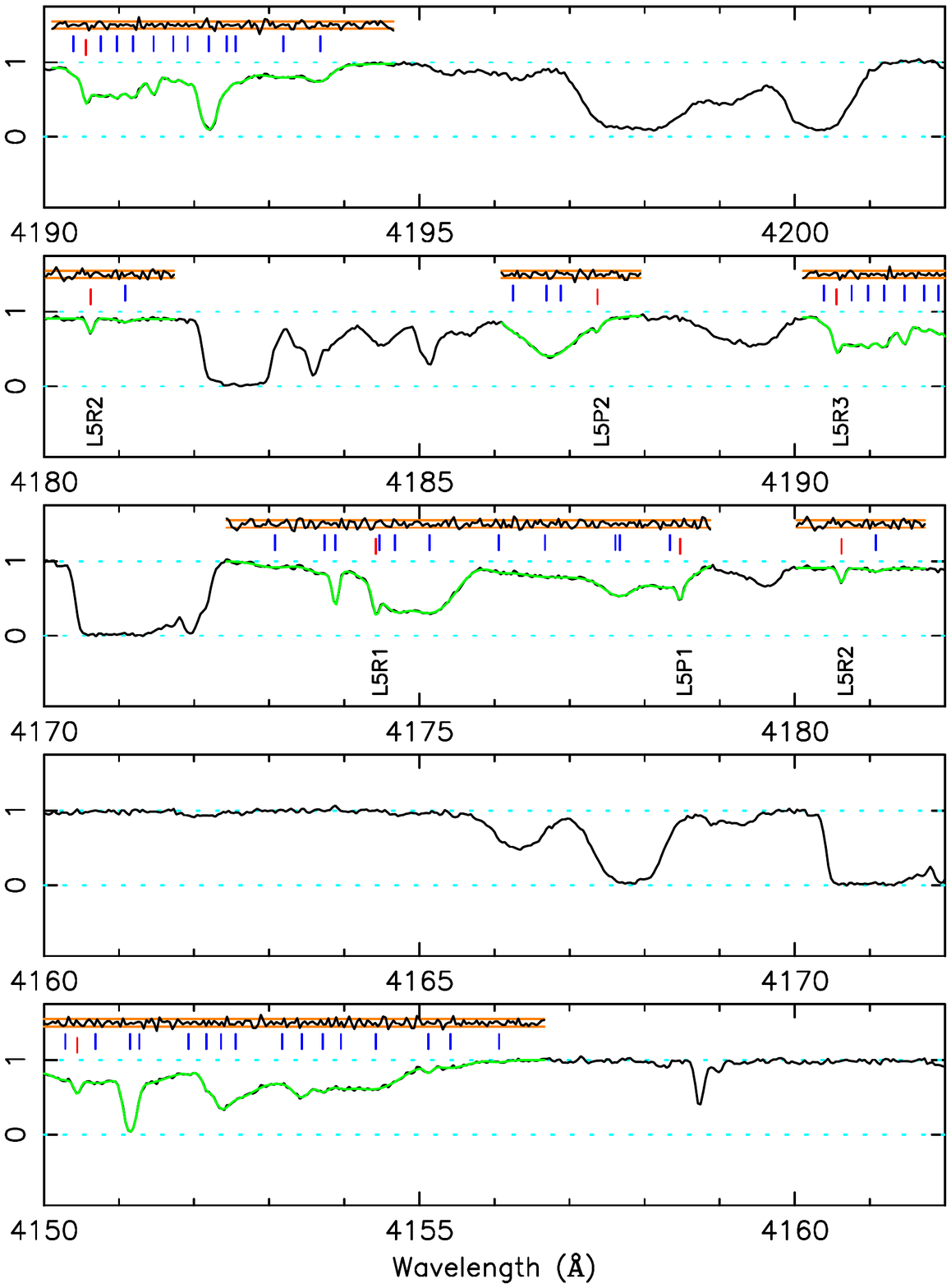}
\par\end{centering}

\caption[H$_2$ fit for the $z=3.025$ absorber toward Q0347$-$383 (9)]{H$_2$ fit for the $z=3.025$ absorber toward Q0347$-$383 (part 9). The vertical axis shows normalised flux. The model fitted to the spectra is shown in green. Red tick marks indicate the position of H$_2$ components, whilst blue tick marks indicate the position of blending transitions (presumed to be Lyman-$\alpha$). Normalised residuals (i.e. [data - model]/error) are plotted above the spectrum between the orange bands, which represent $\pm 1\sigma$. Labels for the H$_2$ transitions are plotted below the data.}
\end{figure}

\begin{figure}[H]
\noindent \begin{centering}
\includegraphics[bb=86bp 180bp 544bp 801bp,clip,width=1\textwidth]{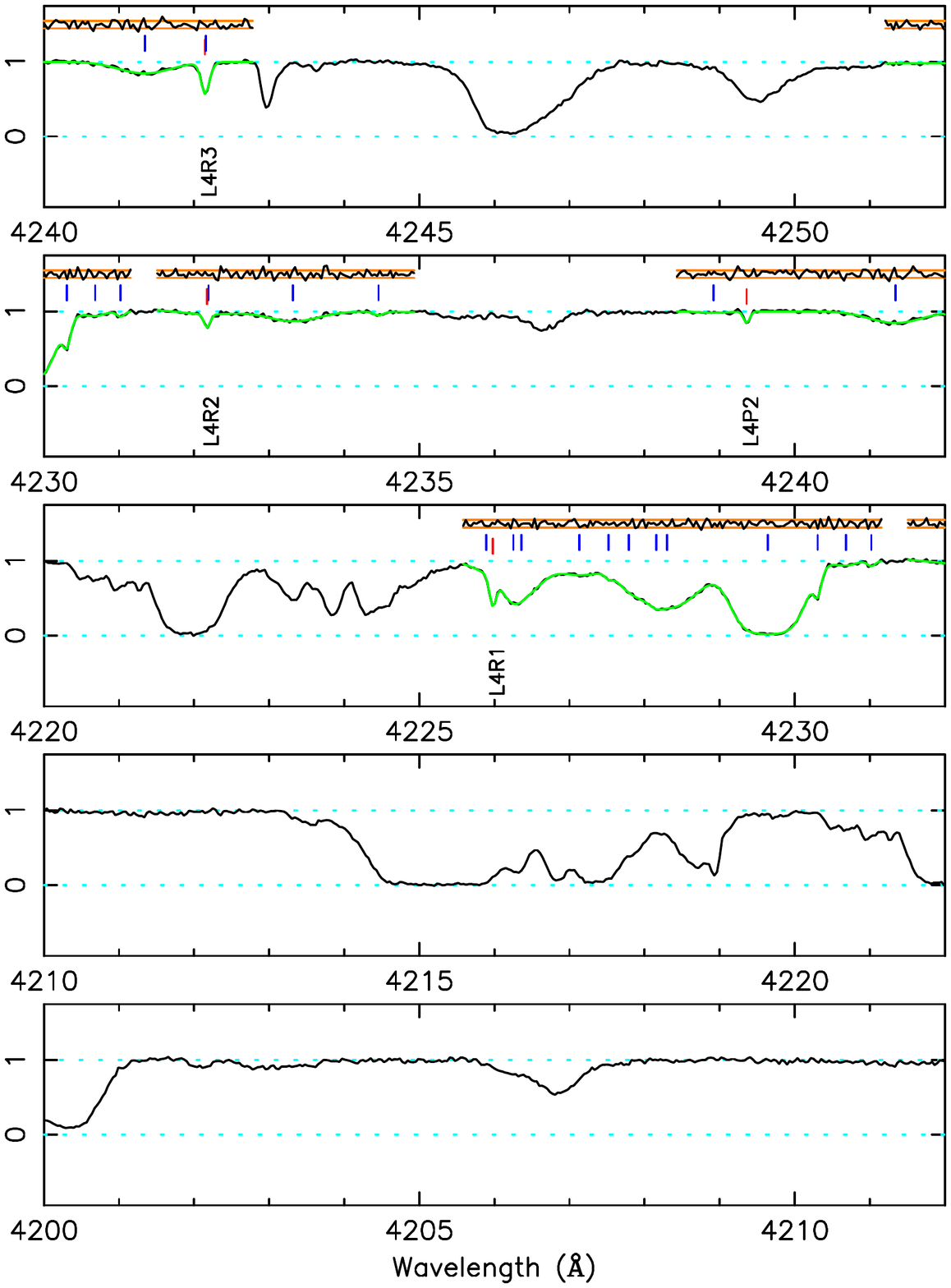}
\par\end{centering}

\caption[H$_2$ fit for the $z=3.025$ absorber toward Q0347$-$383 (10)]{H$_2$ fit for the $z=3.025$ absorber toward Q0347$-$383 (part 10). The vertical axis shows normalised flux. The model fitted to the spectra is shown in green. Red tick marks indicate the position of H$_2$ components, whilst blue tick marks indicate the position of blending transitions (presumed to be Lyman-$\alpha$). Normalised residuals (i.e. [data - model]/error) are plotted above the spectrum between the orange bands, which represent $\pm 1\sigma$. Labels for the H$_2$ transitions are plotted below the data.}
\end{figure}

\begin{figure}[H]
\noindent \begin{centering}
\includegraphics[bb=86bp 180bp 544bp 801bp,clip,width=1\textwidth]{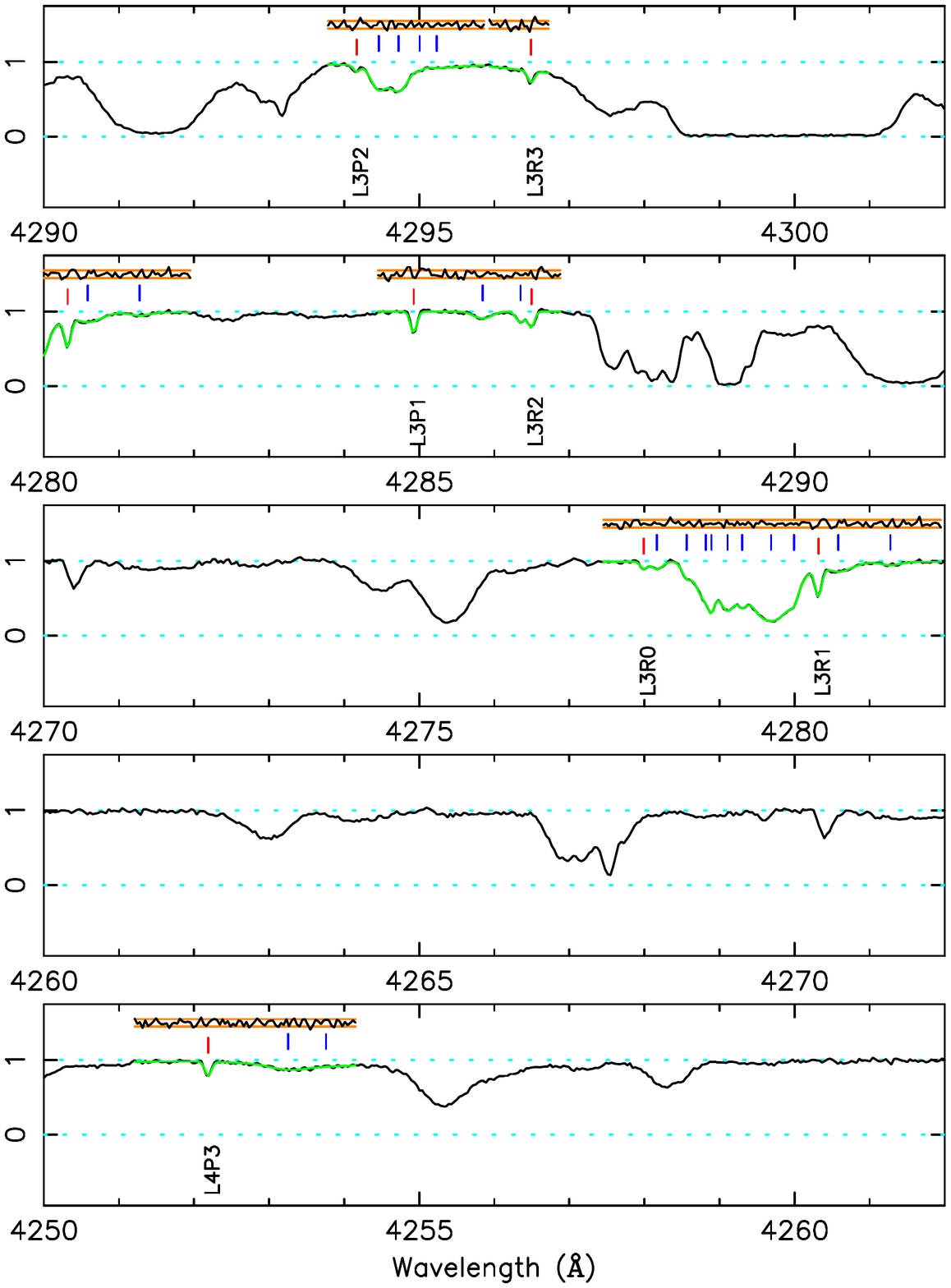}
\par\end{centering}

\caption[H$_2$ fit for the $z=3.025$ absorber toward Q0347$-$383 (11)]{H$_2$ fit for the $z=3.025$ absorber toward Q0347$-$383 (part 11). The vertical axis shows normalised flux. The model fitted to the spectra is shown in green. Red tick marks indicate the position of H$_2$ components, whilst blue tick marks indicate the position of blending transitions (presumed to be Lyman-$\alpha$). Normalised residuals (i.e. [data - model]/error) are plotted above the spectrum between the orange bands, which represent $\pm 1\sigma$. Labels for the H$_2$ transitions are plotted below the data.}
\end{figure}

\begin{figure}[H]
\noindent \begin{centering}
\includegraphics[bb=86bp 180bp 544bp 801bp,clip,width=1\textwidth]{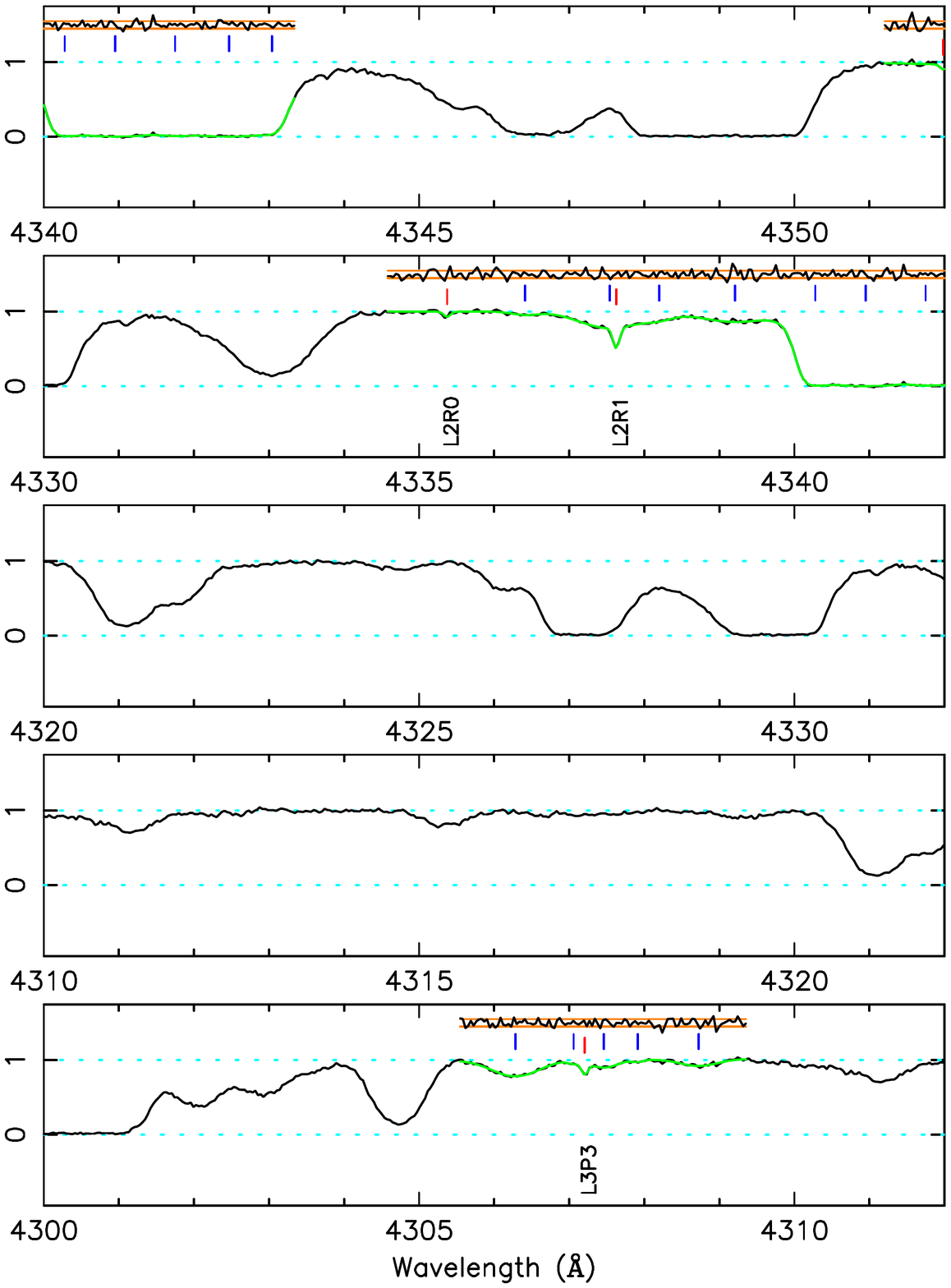}
\par\end{centering}

\caption[H$_2$ fit for the $z=3.025$ absorber toward Q0347$-$383 (12)]{H$_2$ fit for the $z=3.025$ absorber toward Q0347$-$383 (part 12). The vertical axis shows normalised flux. The model fitted to the spectra is shown in green. Red tick marks indicate the position of H$_2$ components, whilst blue tick marks indicate the position of blending transitions (presumed to be Lyman-$\alpha$). Normalised residuals (i.e. [data - model]/error) are plotted above the spectrum between the orange bands, which represent $\pm 1\sigma$. Labels for the H$_2$ transitions are plotted below the data.}
\end{figure}

\begin{figure}[H]
\noindent \begin{centering}
\includegraphics[bb=86bp 180bp 544bp 801bp,clip,width=1\textwidth]{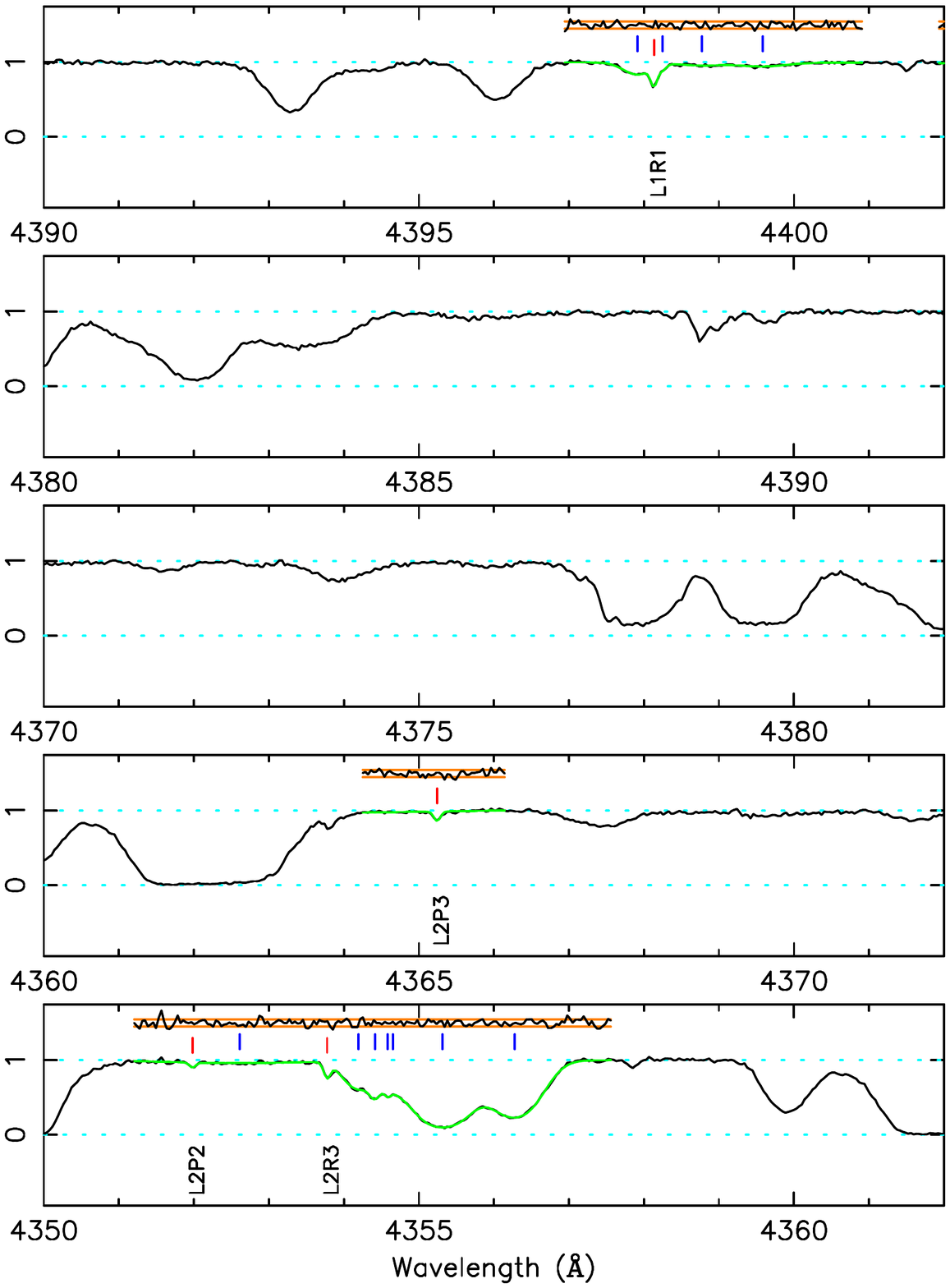}
\par\end{centering}

\caption[H$_2$ fit for the $z=3.025$ absorber toward Q0347$-$383 (13)]{H$_2$ fit for the $z=3.025$ absorber toward Q0347$-$383 (part 13). The vertical axis shows normalised flux. The model fitted to the spectra is shown in green. Red tick marks indicate the position of H$_2$ components, whilst blue tick marks indicate the position of blending transitions (presumed to be Lyman-$\alpha$). Normalised residuals (i.e. [data - model]/error) are plotted above the spectrum between the orange bands, which represent $\pm 1\sigma$. Labels for the H$_2$ transitions are plotted below the data.}
\end{figure}

\begin{figure}[H]
\noindent \begin{centering}
\includegraphics[bb=86bp 180bp 544bp 319bp,clip,width=1\textwidth]{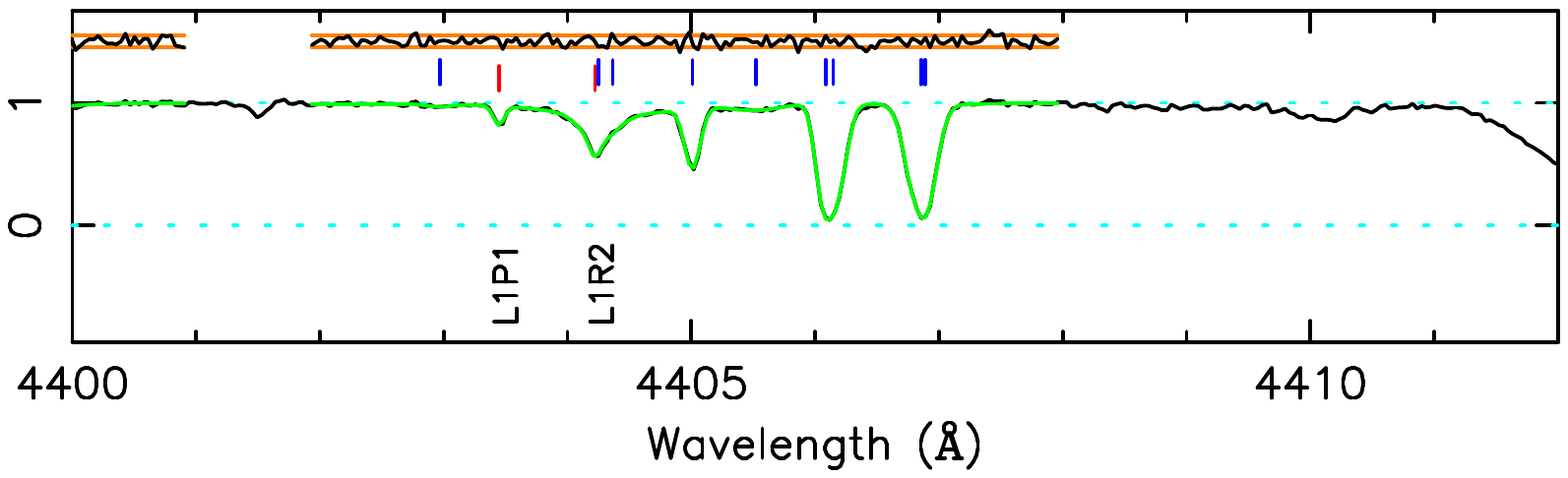}
\par\end{centering}

\caption[H$_2$ fit for the $z=3.025$ absorber toward Q0347$-$383 (14)]{H$_2$ fit for the $z=3.025$ absorber toward Q0347$-$383 (part 14). The vertical axis shows normalised flux. The model fitted to the spectra is shown in green. Red tick marks indicate the position of H$_2$ components, whilst blue tick marks indicate the position of blending transitions (presumed to be Lyman-$\alpha$). Normalised residuals (i.e. [data - model]/error) are plotted above the spectrum between the orange bands, which represent $\pm 1\sigma$. Labels for the H$_2$ transitions are plotted below the data.}
\end{figure}

\chapter{Q0528$-$250:A Voigt profile fits\label{cha:mu fits:Q0528}}

In this appendix, we provide the fits for the $z=2.811$ H$_{2}$
absorber toward Q0528$-$250. The fit relates to our first analysis
of this object, published in \citet{King:08}. The analysis for $\Delta\mu/\mu$
is set out in section \ref{sub:mu:first results}.

\begin{figure}[H]
\noindent \begin{centering}
\includegraphics[bb=86bp 180bp 544bp 801bp,clip,width=1\textwidth]{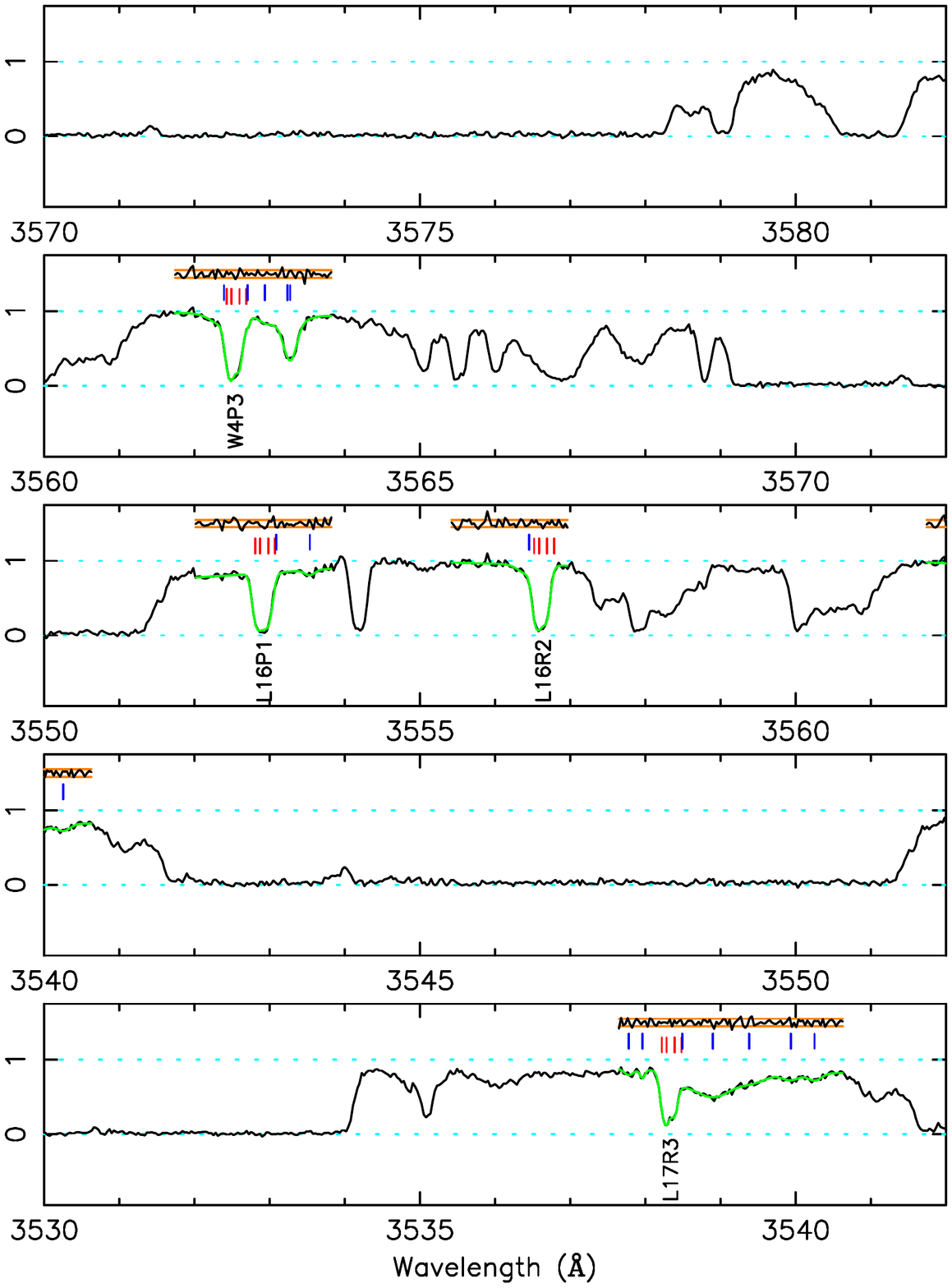}
\par\end{centering}

\caption[H$_2$ fit for the $z=2.811$ absorber toward Q0528$-$250:A (1)]{H$_2$ fit for the $z=2.811$ absorber toward Q0528$-$250:A (part 1). The vertical axis shows normalised flux. The model fitted to the spectra is shown in green. Red tick marks indicate the position of H$_2$ components, whilst blue tick marks indicate the position of blending transitions (presumed to be Lyman-$\alpha$). Normalised residuals (i.e. [data - model]/error) are plotted above the spectrum between the orange bands, which represent $\pm 1\sigma$. Labels for the H$_2$ transitions are plotted below the data.}
\end{figure}

\begin{figure}[H]
\noindent \begin{centering}
\includegraphics[bb=86bp 180bp 544bp 801bp,clip,width=1\textwidth]{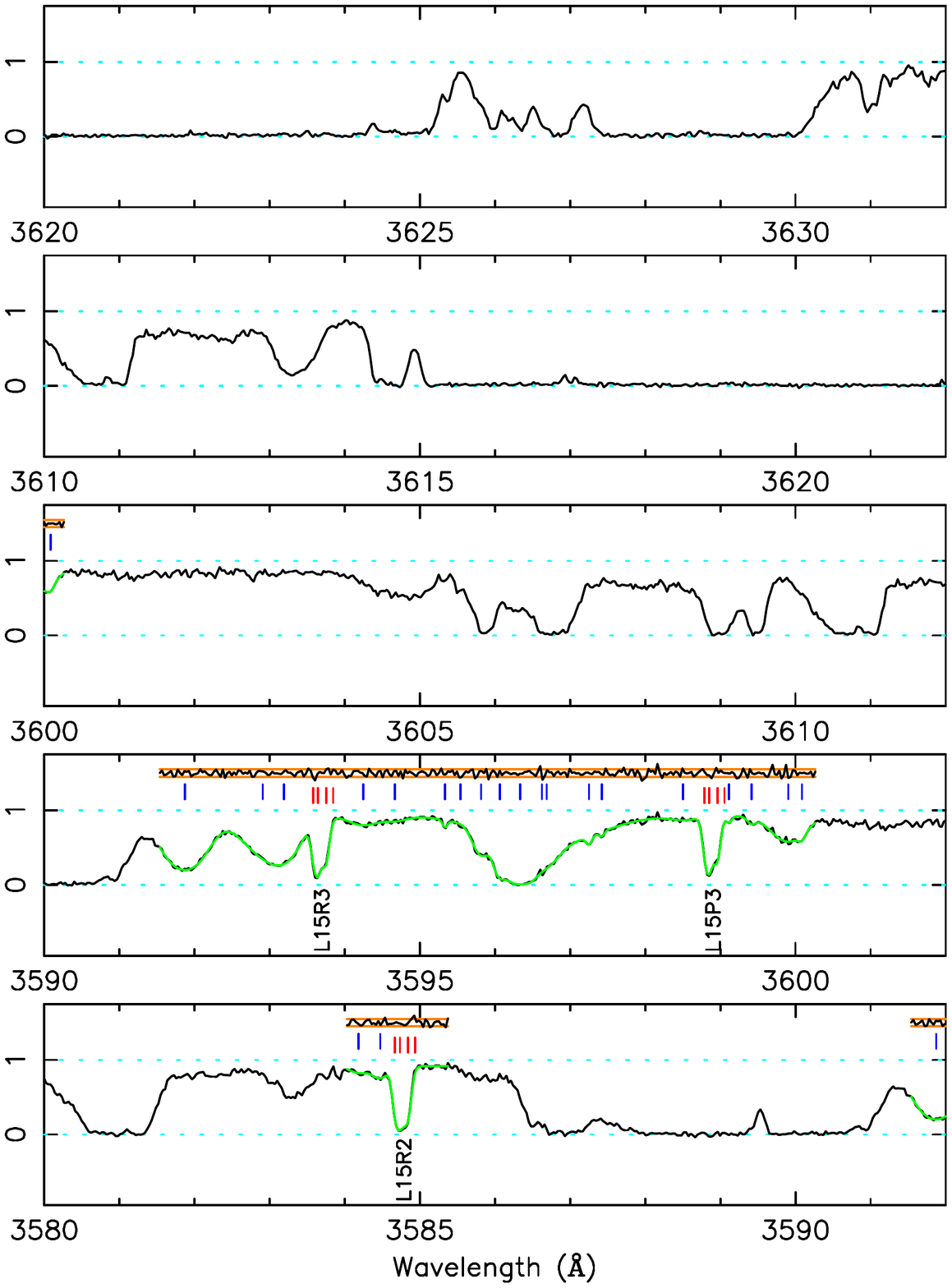}
\par\end{centering}

\caption[H$_2$ fit for the $z=2.811$ absorber toward Q0528$-$250:A (2)]{H$_2$ fit for the $z=2.811$ absorber toward Q0528$-$250:A (part 2). The vertical axis shows normalised flux. The model fitted to the spectra is shown in green. Red tick marks indicate the position of H$_2$ components, whilst blue tick marks indicate the position of blending transitions (presumed to be Lyman-$\alpha$). Normalised residuals (i.e. [data - model]/error) are plotted above the spectrum between the orange bands, which represent $\pm 1\sigma$. Labels for the H$_2$ transitions are plotted below the data.}
\end{figure}

\begin{figure}[H]
\noindent \begin{centering}
\includegraphics[bb=86bp 180bp 544bp 801bp,clip,width=1\textwidth]{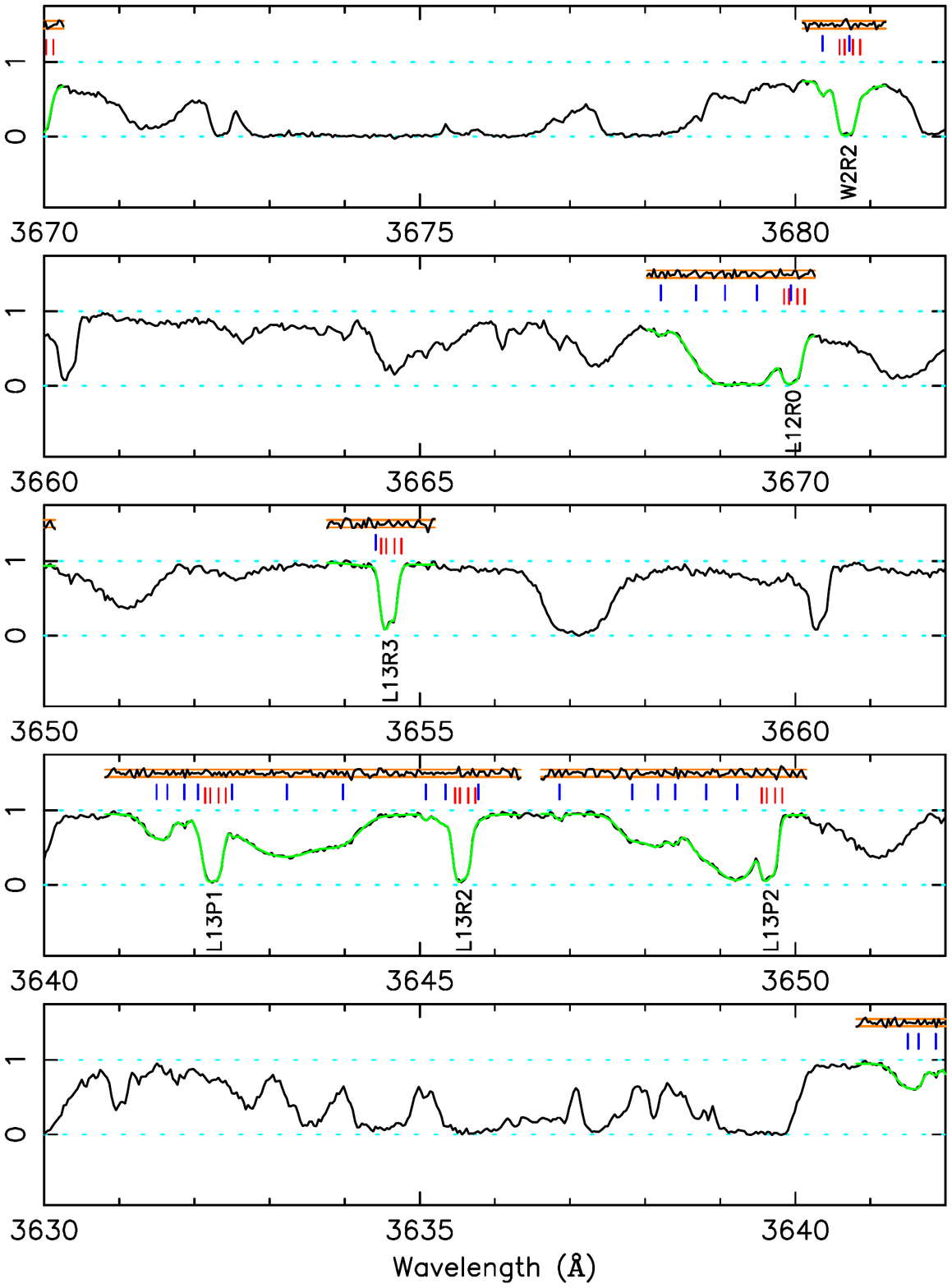}
\par\end{centering}

\caption[H$_2$ fit for the $z=2.811$ absorber toward Q0528$-$250:A (3)]{H$_2$ fit for the $z=2.811$ absorber toward Q0528$-$250:A (part 3). The vertical axis shows normalised flux. The model fitted to the spectra is shown in green. Red tick marks indicate the position of H$_2$ components, whilst blue tick marks indicate the position of blending transitions (presumed to be Lyman-$\alpha$). Normalised residuals (i.e. [data - model]/error) are plotted above the spectrum between the orange bands, which represent $\pm 1\sigma$. Labels for the H$_2$ transitions are plotted below the data.}
\end{figure}

\begin{figure}[H]
\noindent \begin{centering}
\includegraphics[bb=86bp 180bp 544bp 801bp,clip,width=1\textwidth]{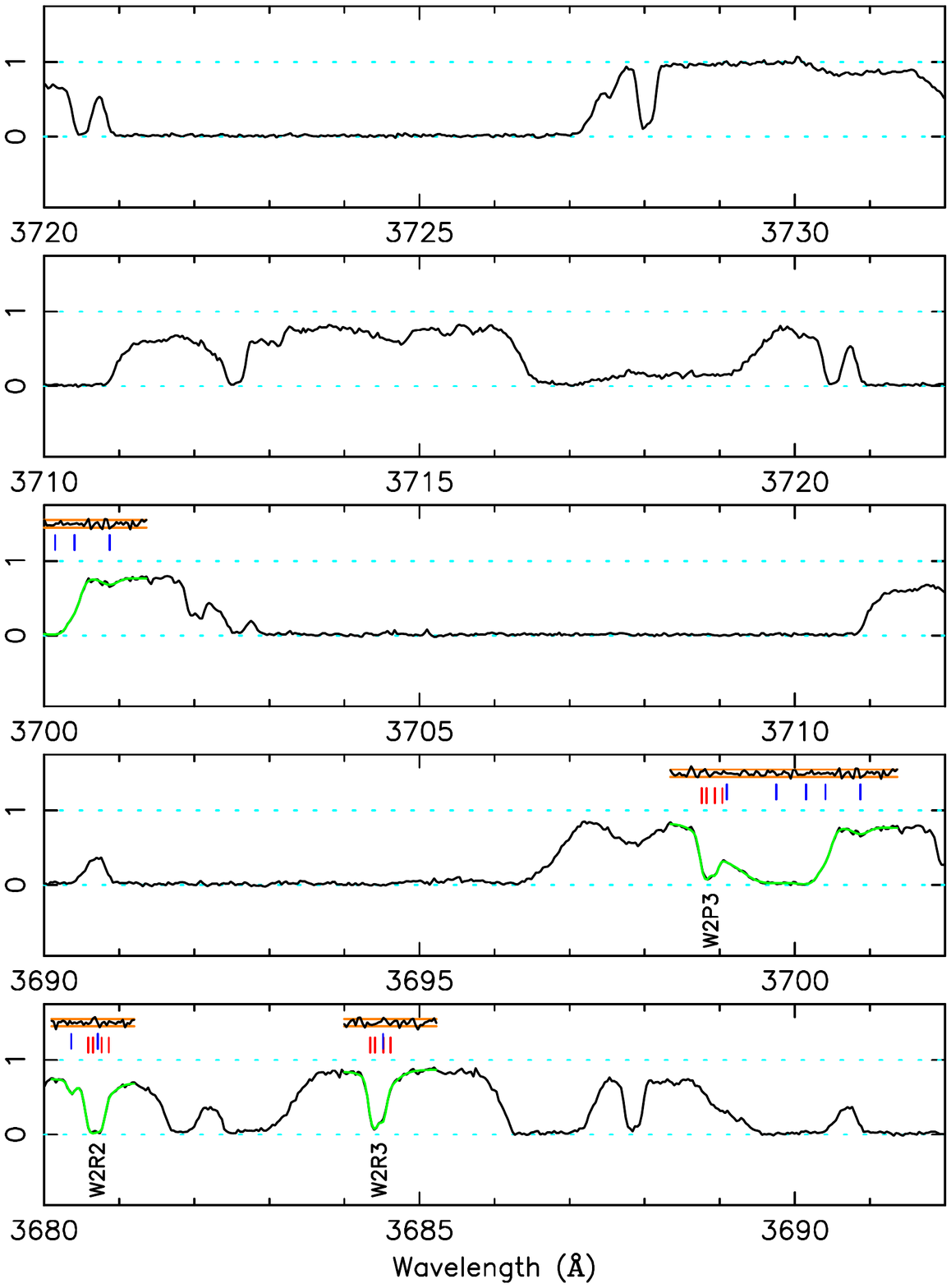}
\par\end{centering}

\caption[H$_2$ fit for the $z=2.811$ absorber toward Q0528$-$250:A (4)]{H$_2$ fit for the $z=2.811$ absorber toward Q0528$-$250:A (part 4). The vertical axis shows normalised flux. The model fitted to the spectra is shown in green. Red tick marks indicate the position of H$_2$ components, whilst blue tick marks indicate the position of blending transitions (presumed to be Lyman-$\alpha$). Normalised residuals (i.e. [data - model]/error) are plotted above the spectrum between the orange bands, which represent $\pm 1\sigma$. Labels for the H$_2$ transitions are plotted below the data.}
\end{figure}

\begin{figure}[H]
\noindent \begin{centering}
\includegraphics[bb=86bp 180bp 544bp 801bp,clip,width=1\textwidth]{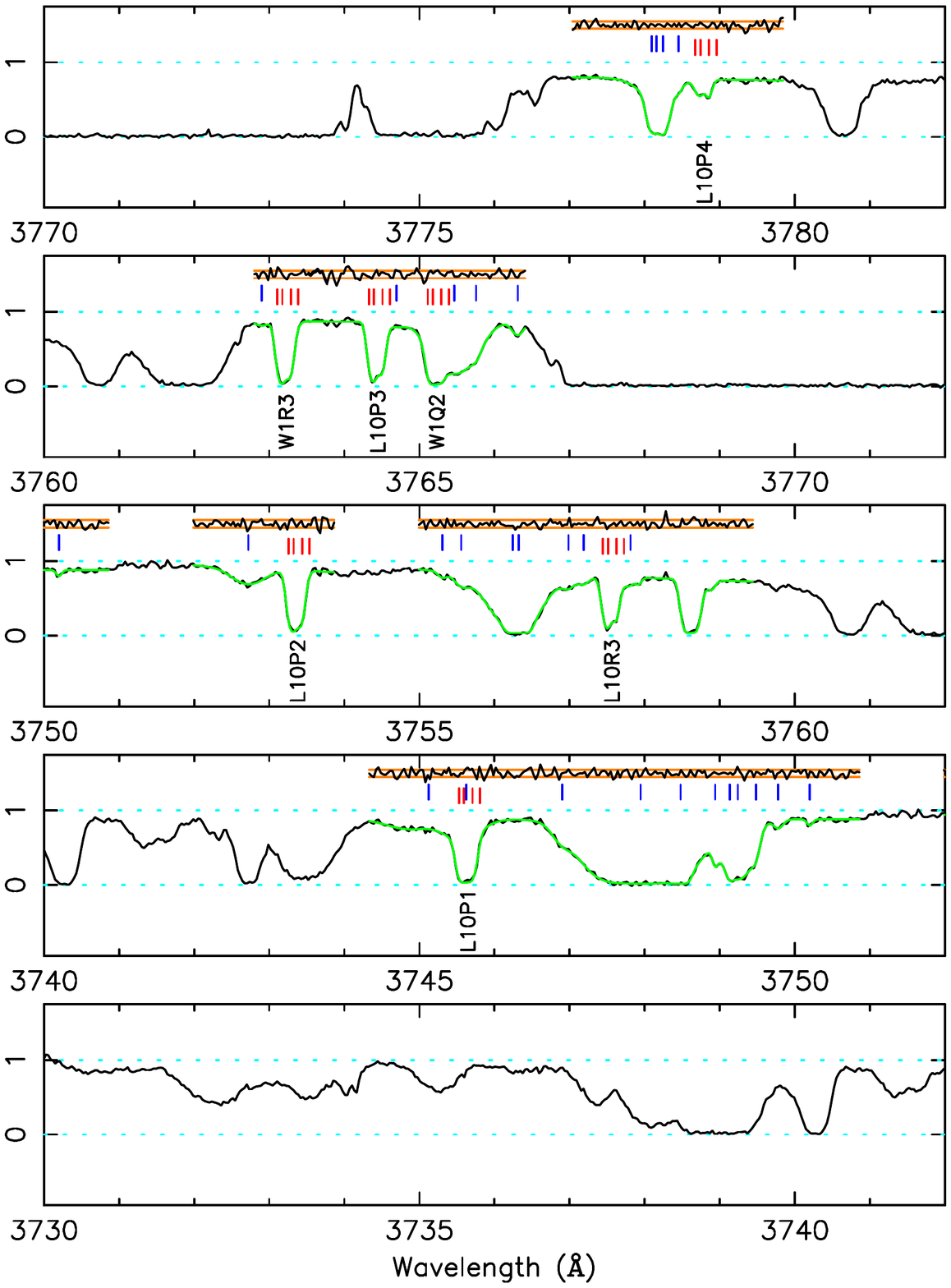}
\par\end{centering}

\caption[H$_2$ fit for the $z=2.811$ absorber toward Q0528$-$250:A (5)]{H$_2$ fit for the $z=2.811$ absorber toward Q0528$-$250:A (part 5). The vertical axis shows normalised flux. The model fitted to the spectra is shown in green. Red tick marks indicate the position of H$_2$ components, whilst blue tick marks indicate the position of blending transitions (presumed to be Lyman-$\alpha$). Normalised residuals (i.e. [data - model]/error) are plotted above the spectrum between the orange bands, which represent $\pm 1\sigma$. Labels for the H$_2$ transitions are plotted below the data.}
\end{figure}

\begin{figure}[H]
\noindent \begin{centering}
\includegraphics[bb=86bp 180bp 544bp 801bp,clip,width=1\textwidth]{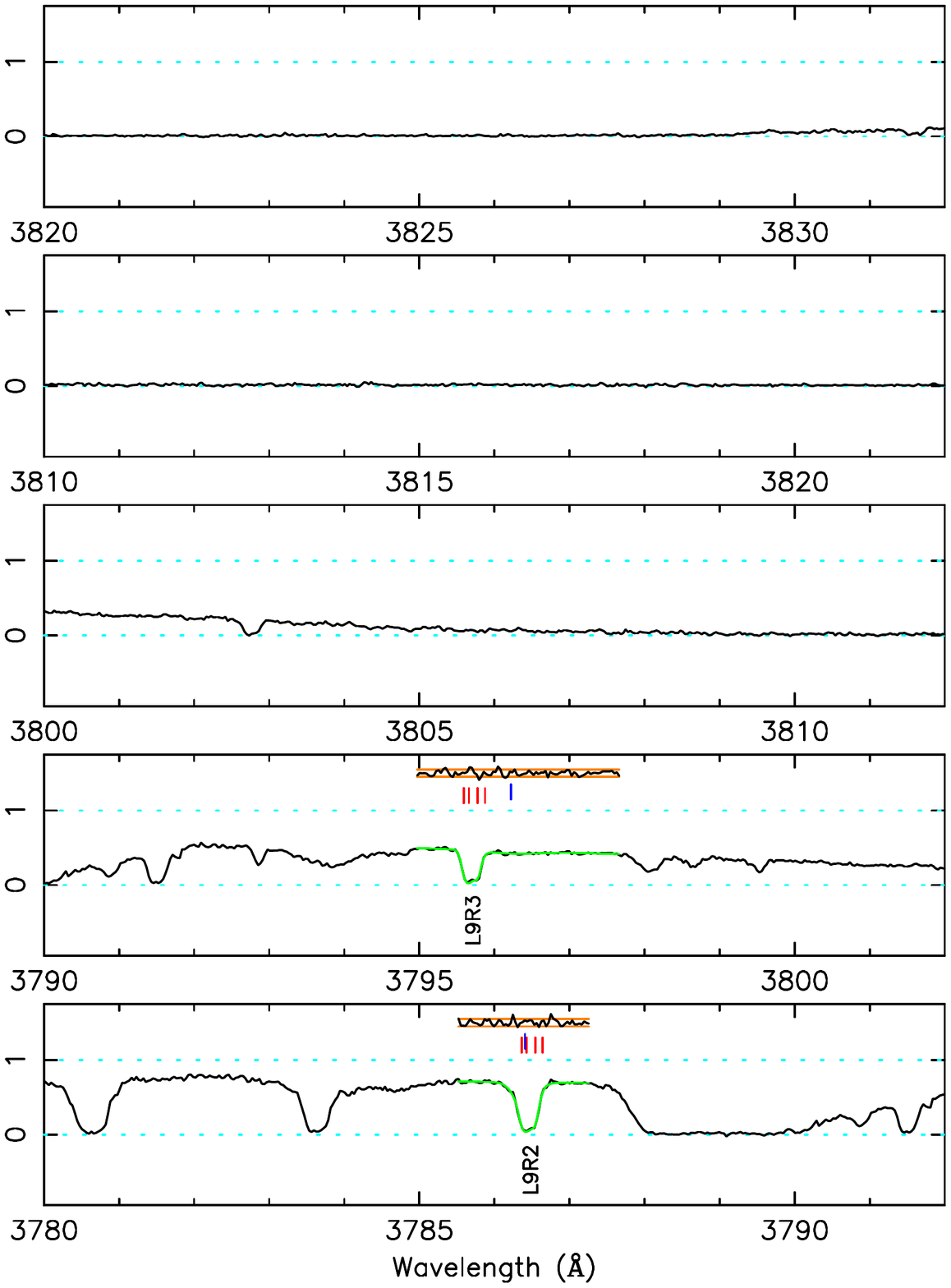}
\par\end{centering}

\caption[H$_2$ fit for the $z=2.811$ absorber toward Q0528$-$250:A (6)]{H$_2$ fit for the $z=2.811$ absorber toward Q0528$-$250:A (part 6). The vertical axis shows normalised flux. The model fitted to the spectra is shown in green. Red tick marks indicate the position of H$_2$ components, whilst blue tick marks indicate the position of blending transitions (presumed to be Lyman-$\alpha$). Normalised residuals (i.e. [data - model]/error) are plotted above the spectrum between the orange bands, which represent $\pm 1\sigma$. Labels for the H$_2$ transitions are plotted below the data.}
\end{figure}

\begin{figure}[H]
\noindent \begin{centering}
\includegraphics[bb=86bp 180bp 544bp 801bp,clip,width=1\textwidth]{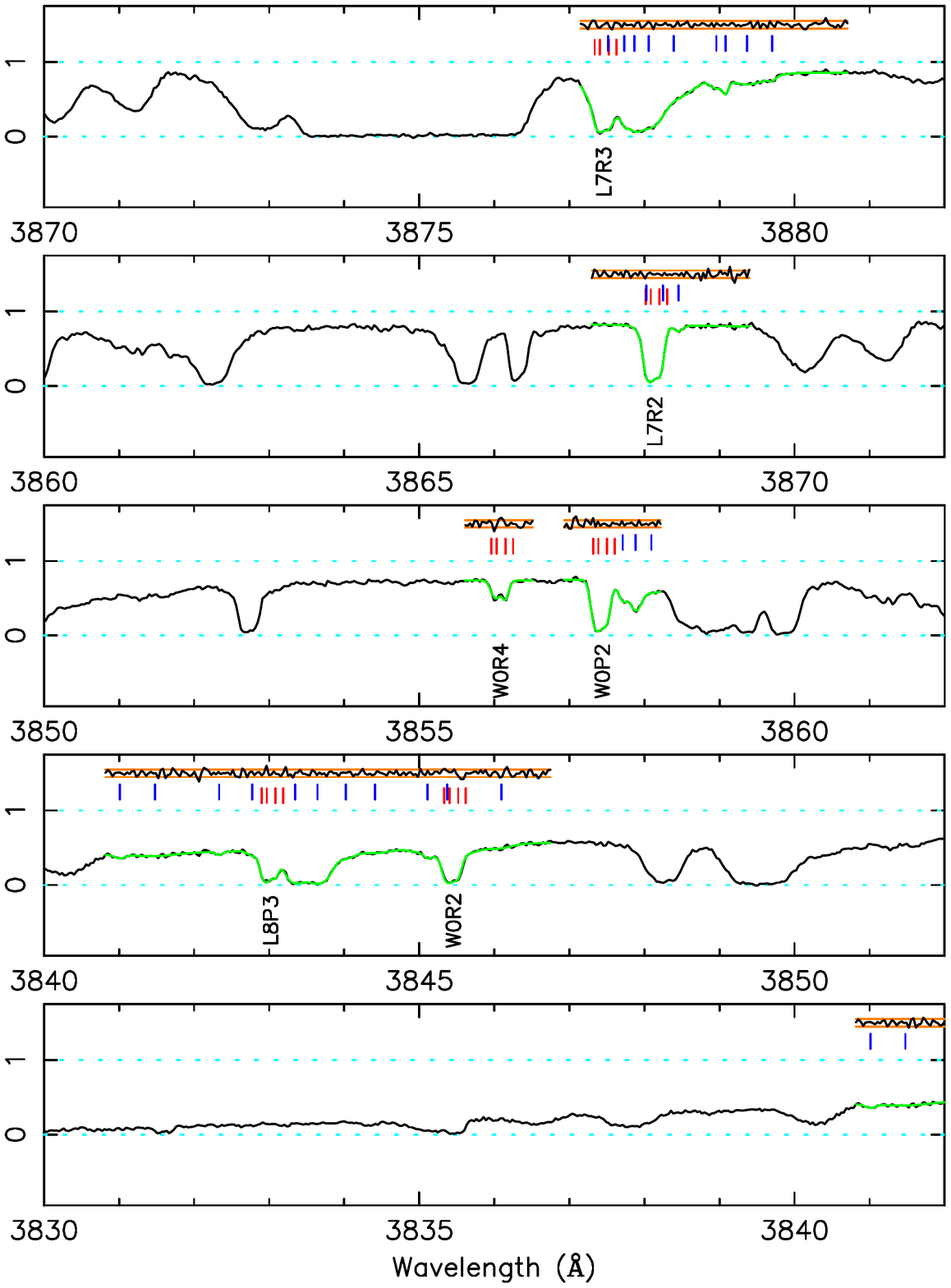}
\par\end{centering}

\caption[H$_2$ fit for the $z=2.811$ absorber toward Q0528$-$250:A (7)]{H$_2$ fit for the $z=2.811$ absorber toward Q0528$-$250:A (part 7). The vertical axis shows normalised flux. The model fitted to the spectra is shown in green. Red tick marks indicate the position of H$_2$ components, whilst blue tick marks indicate the position of blending transitions (presumed to be Lyman-$\alpha$). Normalised residuals (i.e. [data - model]/error) are plotted above the spectrum between the orange bands, which represent $\pm 1\sigma$. Labels for the H$_2$ transitions are plotted below the data.}
\end{figure}

\begin{figure}[H]
\noindent \begin{centering}
\includegraphics[bb=86bp 180bp 544bp 801bp,clip,width=1\textwidth]{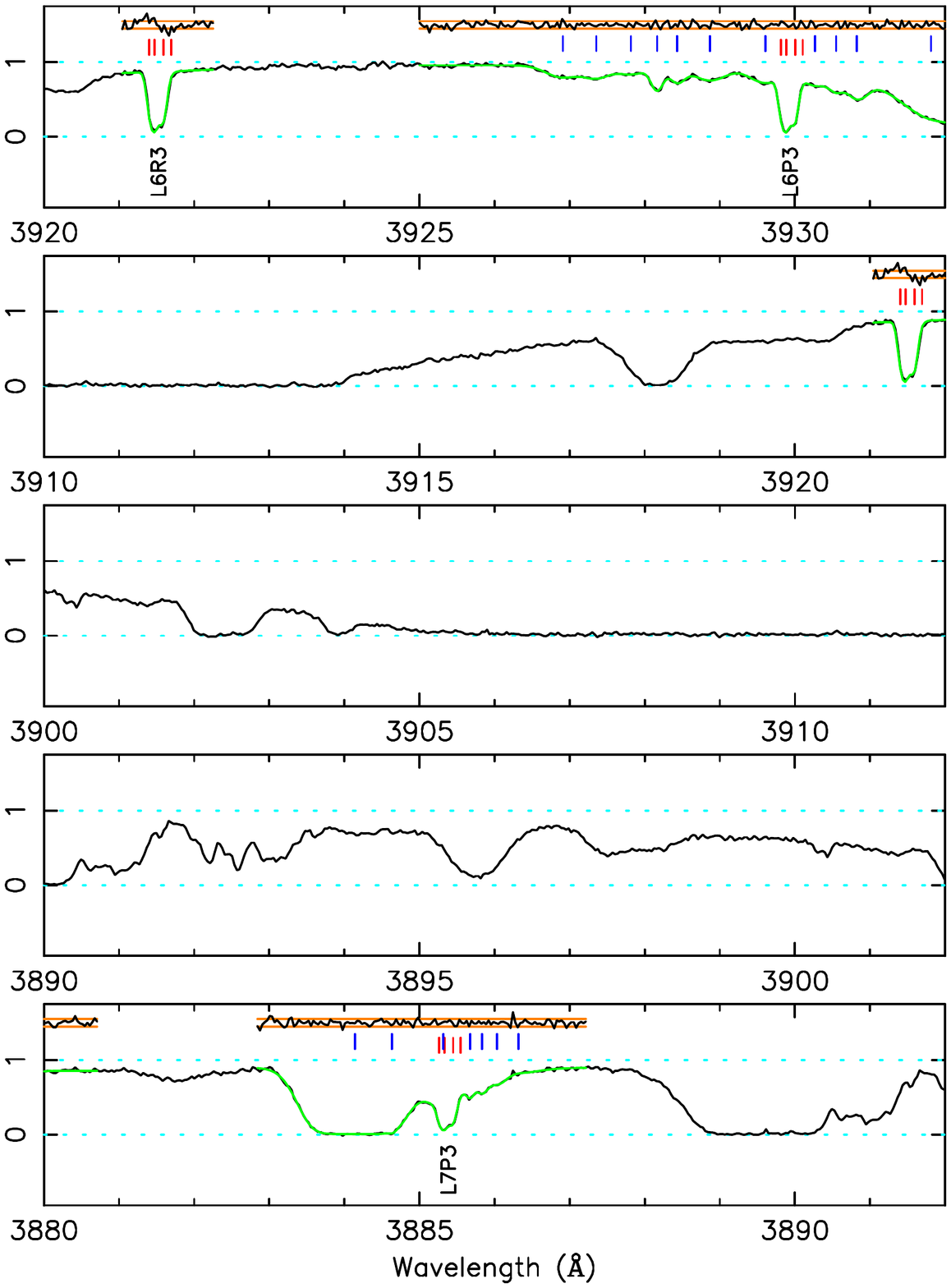}
\par\end{centering}

\caption[H$_2$ fit for the $z=2.811$ absorber toward Q0528$-$250:A (8)]{H$_2$ fit for the $z=2.811$ absorber toward Q0528$-$250:A (part 8). The vertical axis shows normalised flux. The model fitted to the spectra is shown in green. Red tick marks indicate the position of H$_2$ components, whilst blue tick marks indicate the position of blending transitions (presumed to be Lyman-$\alpha$). Normalised residuals (i.e. [data - model]/error) are plotted above the spectrum between the orange bands, which represent $\pm 1\sigma$. Labels for the H$_2$ transitions are plotted below the data.}
\end{figure}

\begin{figure}[H]
\noindent \begin{centering}
\includegraphics[bb=86bp 180bp 544bp 800bp,clip,width=1\textwidth]{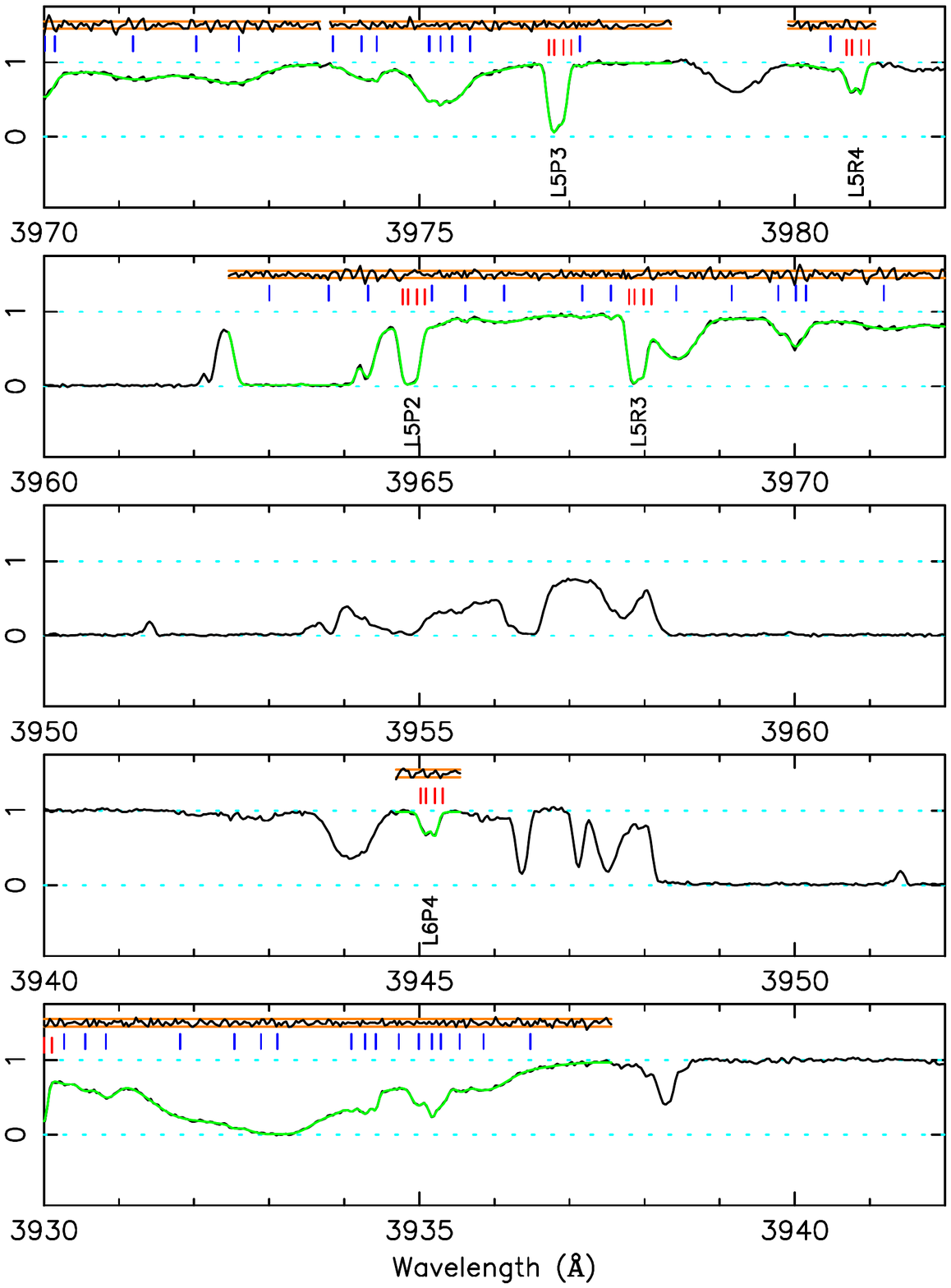}
\par\end{centering}

\caption[H$_2$ fit for the $z=2.811$ absorber toward Q0528$-$250:A (9)]{H$_2$ fit for the $z=2.811$ absorber toward Q0528$-$250:A (part 9). The vertical axis shows normalised flux. The model fitted to the spectra is shown in green. Red tick marks indicate the position of H$_2$ components, whilst blue tick marks indicate the position of blending transitions (presumed to be Lyman-$\alpha$). Normalised residuals (i.e. [data - model]/error) are plotted above the spectrum between the orange bands, which represent $\pm 1\sigma$. Labels for the H$_2$ transitions are plotted below the data.}
\end{figure}

\begin{figure}[H]
\noindent \begin{centering}
\includegraphics[bb=86bp 180bp 544bp 801bp,clip,width=1\textwidth]{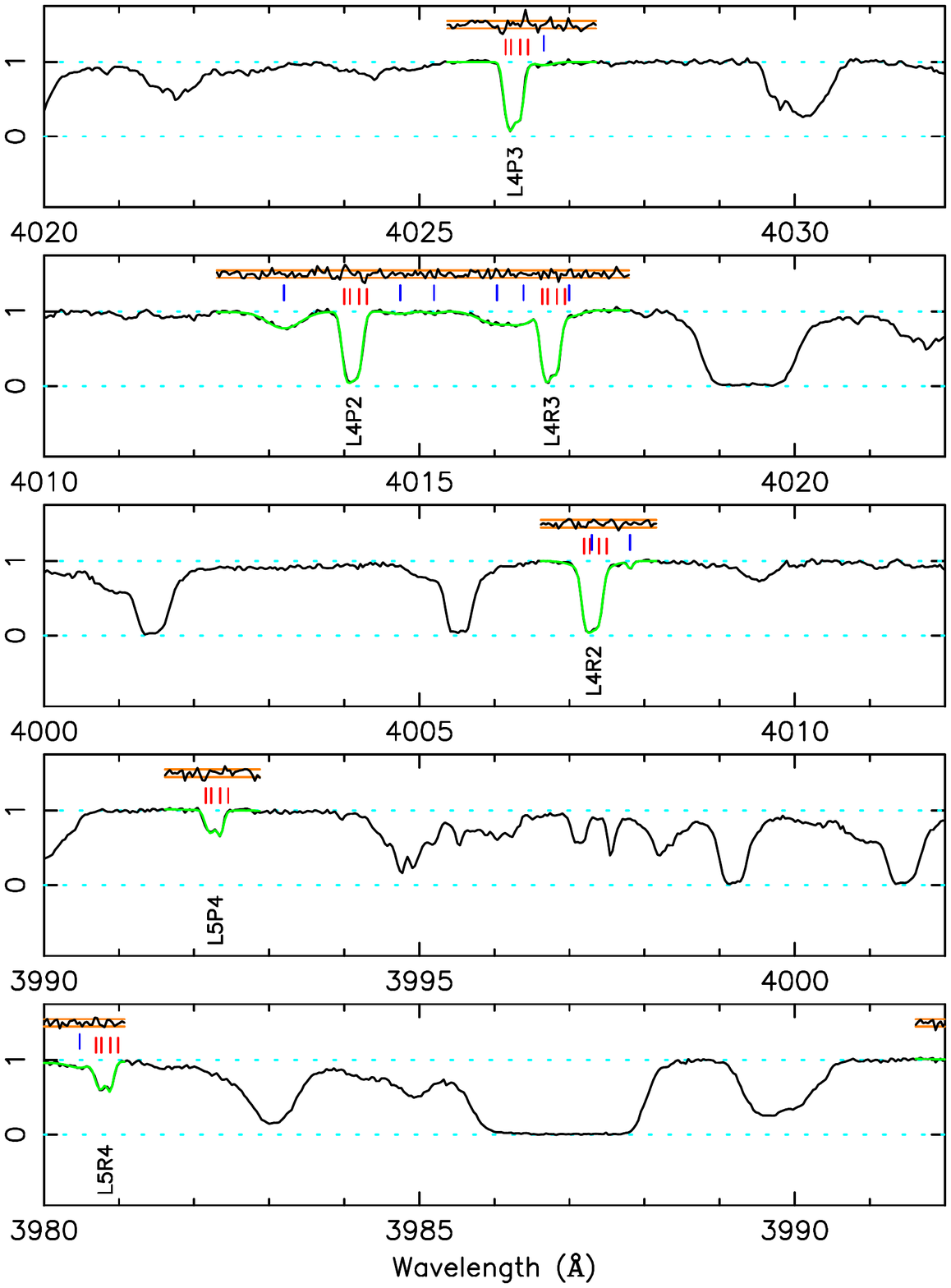}
\par\end{centering}

\caption[H$_2$ fit for the $z=2.811$ absorber toward Q0528$-$250:A (10)]{H$_2$ fit for the $z=2.811$ absorber toward Q0528$-$250:A (part 10). The vertical axis shows normalised flux. The model fitted to the spectra is shown in green. Red tick marks indicate the position of H$_2$ components, whilst blue tick marks indicate the position of blending transitions (presumed to be Lyman-$\alpha$). Normalised residuals (i.e. [data - model]/error) are plotted above the spectrum between the orange bands, which represent $\pm 1\sigma$. Labels for the H$_2$ transitions are plotted below the data.}
\end{figure}

\begin{figure}[H]
\noindent \begin{centering}
\includegraphics[bb=86bp 180bp 544bp 801bp,clip,width=1\textwidth]{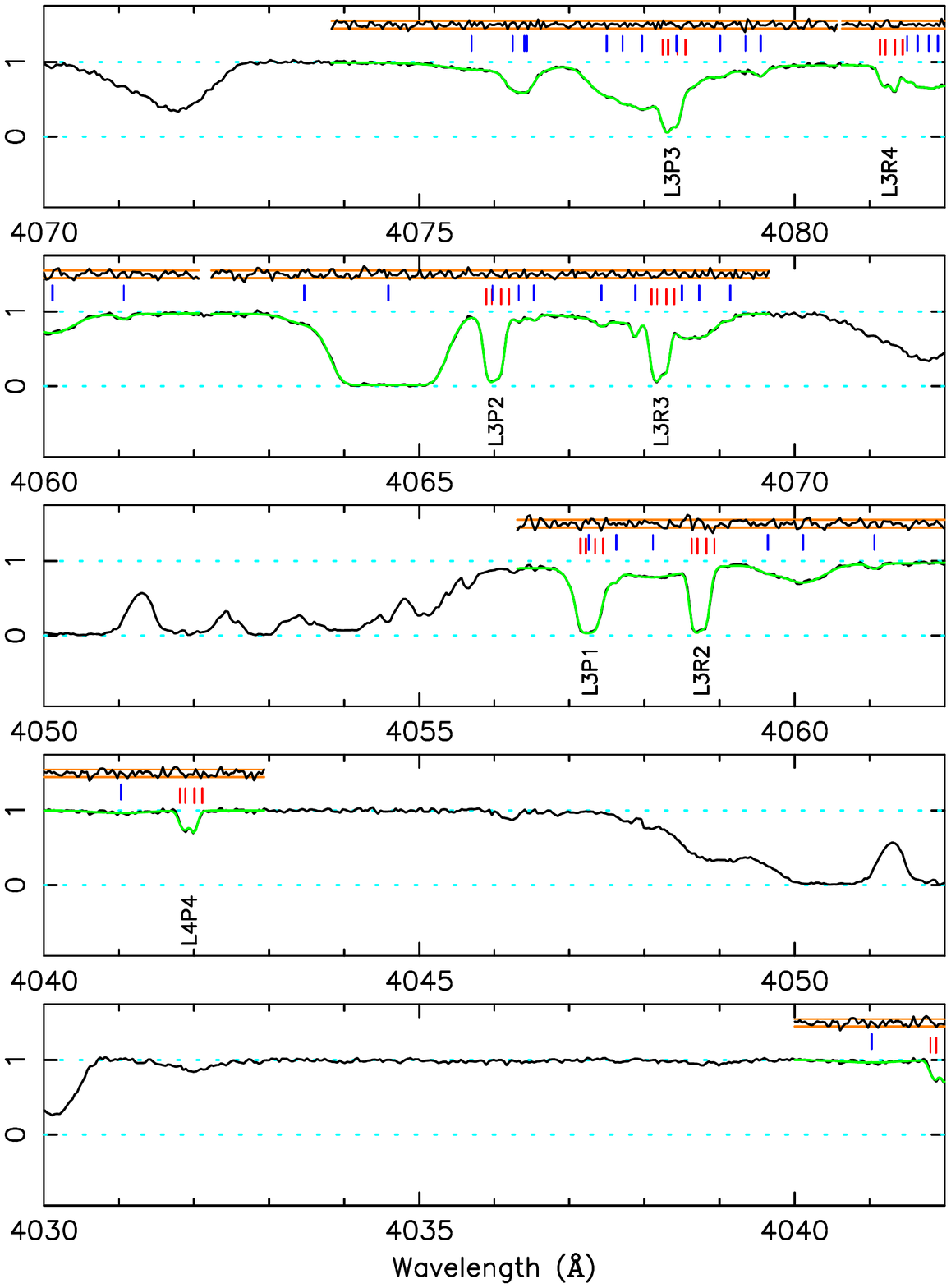}
\par\end{centering}

\caption[H$_2$ fit for the $z=2.811$ absorber toward Q0528$-$250:A (11)]{H$_2$ fit for the $z=2.811$ absorber toward Q0528$-$250:A (part 11). The vertical axis shows normalised flux. The model fitted to the spectra is shown in green. Red tick marks indicate the position of H$_2$ components, whilst blue tick marks indicate the position of blending transitions (presumed to be Lyman-$\alpha$). Normalised residuals (i.e. [data - model]/error) are plotted above the spectrum between the orange bands, which represent $\pm 1\sigma$. Labels for the H$_2$ transitions are plotted below the data.}
\end{figure}

\begin{figure}[H]
\noindent \begin{centering}
\includegraphics[bb=86bp 180bp 544bp 801bp,clip,width=1\textwidth]{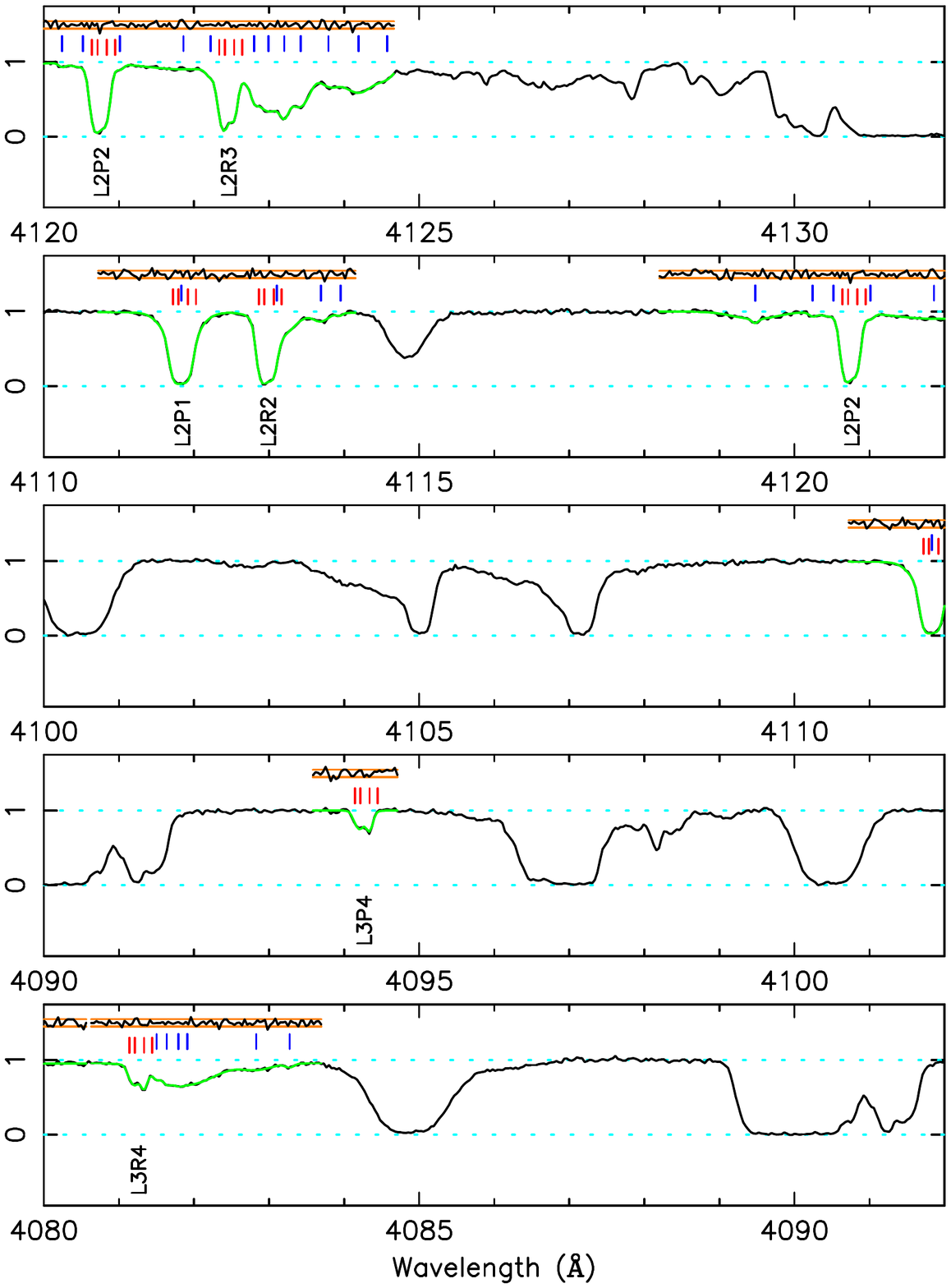}
\par\end{centering}

\caption[H$_2$ fit for the $z=2.811$ absorber toward Q0528$-$250:A (12)]{H$_2$ fit for the $z=2.811$ absorber toward Q0528$-$250:A (part 12). The vertical axis shows normalised flux. The model fitted to the spectra is shown in green. Red tick marks indicate the position of H$_2$ components, whilst blue tick marks indicate the position of blending transitions (presumed to be Lyman-$\alpha$). Normalised residuals (i.e. [data - model]/error) are plotted above the spectrum between the orange bands, which represent $\pm 1\sigma$. Labels for the H$_2$ transitions are plotted below the data.}
\end{figure}

\begin{figure}[H]
\noindent \begin{centering}
\includegraphics[bb=86bp 180bp 544bp 801bp,clip,width=1\textwidth]{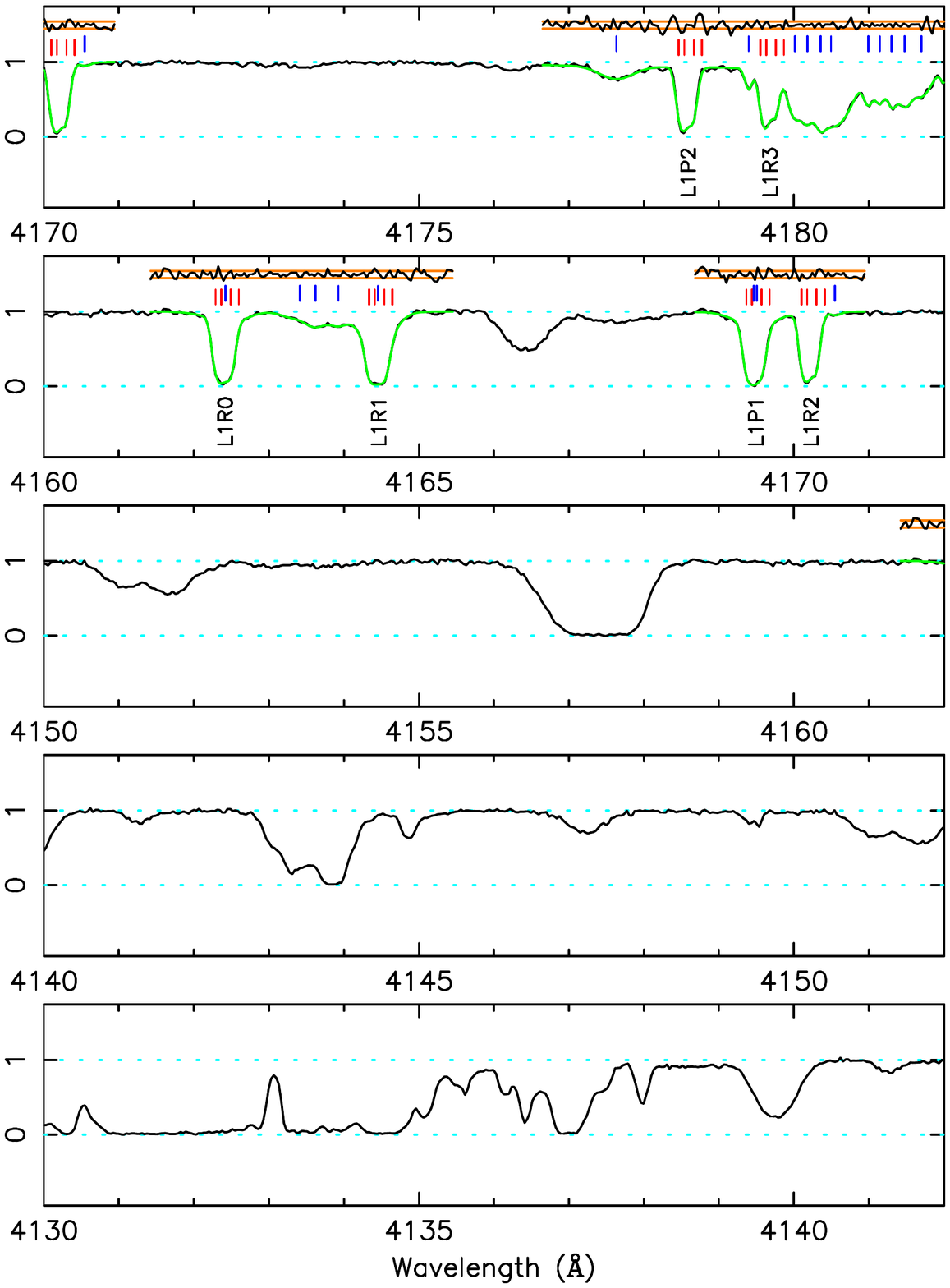}
\par\end{centering}

\caption[H$_2$ fit for the $z=2.811$ absorber toward Q0528$-$250:A (13)]{H$_2$ fit for the $z=2.811$ absorber toward Q0528$-$250:A (part 13). The vertical axis shows normalised flux. The model fitted to the spectra is shown in green. Red tick marks indicate the position of H$_2$ components, whilst blue tick marks indicate the position of blending transitions (presumed to be Lyman-$\alpha$). Normalised residuals (i.e. [data - model]/error) are plotted above the spectrum between the orange bands, which represent $\pm 1\sigma$. Labels for the H$_2$ transitions are plotted below the data.}
\end{figure}

\begin{figure}[H]
\noindent \begin{centering}
\includegraphics[bb=86bp 180bp 544bp 801bp,clip,width=1\textwidth]{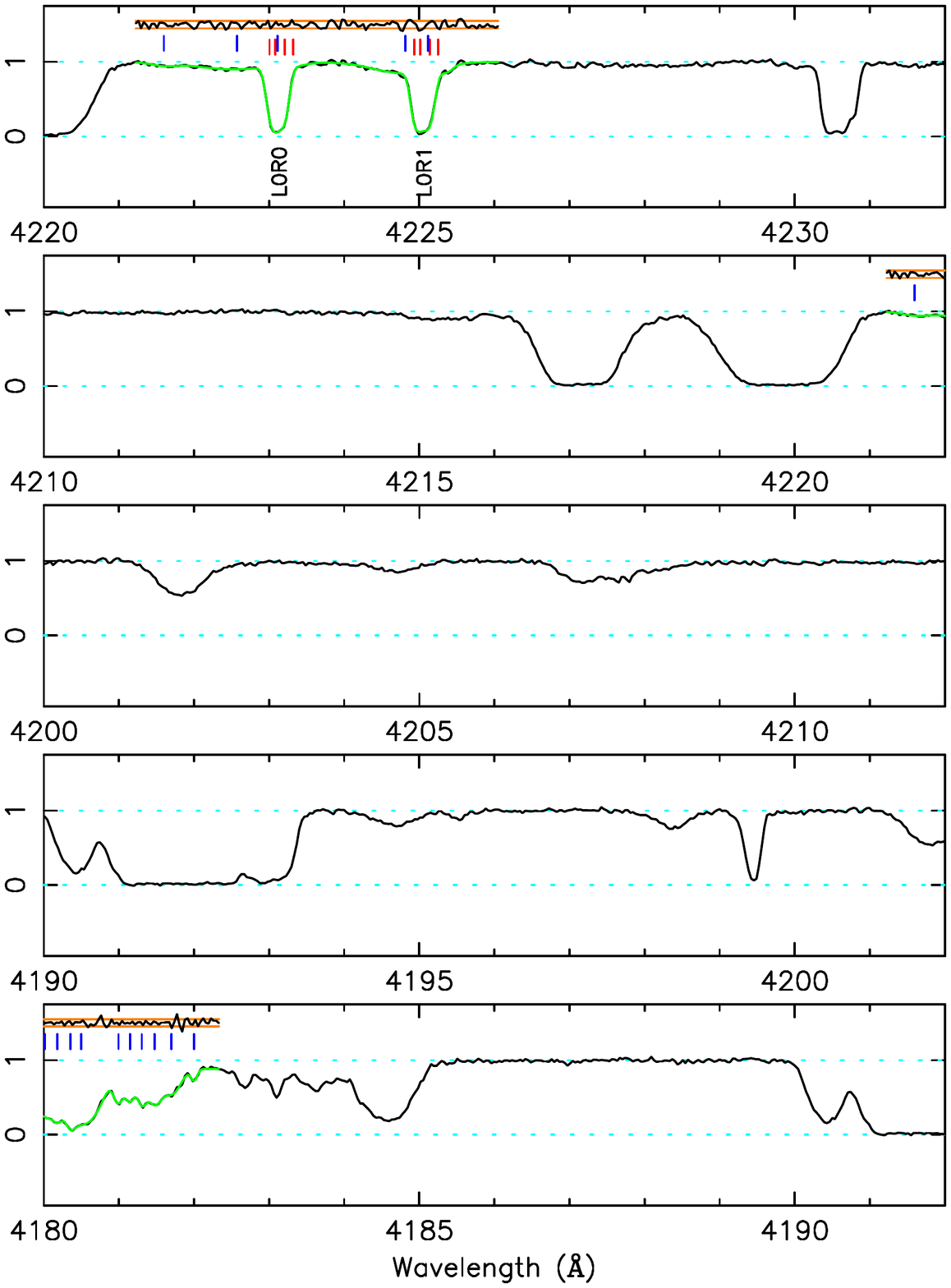}
\par\end{centering}

\caption[H$_2$ fit for the $z=2.811$ absorber toward Q0528$-$250:A (14)]{H$_2$ fit for the $z=2.811$ absorber toward Q0528$-$250:A (part 14). The vertical axis shows normalised flux. The model fitted to the spectra is shown in green. Red tick marks indicate the position of H$_2$ components, whilst blue tick marks indicate the position of blending transitions (presumed to be Lyman-$\alpha$). Normalised residuals (i.e. [data - model]/error) are plotted above the spectrum between the orange bands, which represent $\pm 1\sigma$. Labels for the H$_2$ transitions are plotted below the data.}
\end{figure}

\chapter{Q0528$-$250:B2 Voigt profile fits\label{cha:mu fits:Q0528:B2}}

We present here our reanalysis of the spectrum of the $z=2.811$ absorber
toward Q0528$-$250, using exposures obtained on VLT/UVES under program
ID 82.A-0087. We describe the results of this analysis in section
\ref{sub:Q0528-250 revisited}.

\begin{figure}[H]
\noindent \begin{centering}
\includegraphics[bb=86bp 180bp 544bp 801bp,clip,width=1\textwidth]{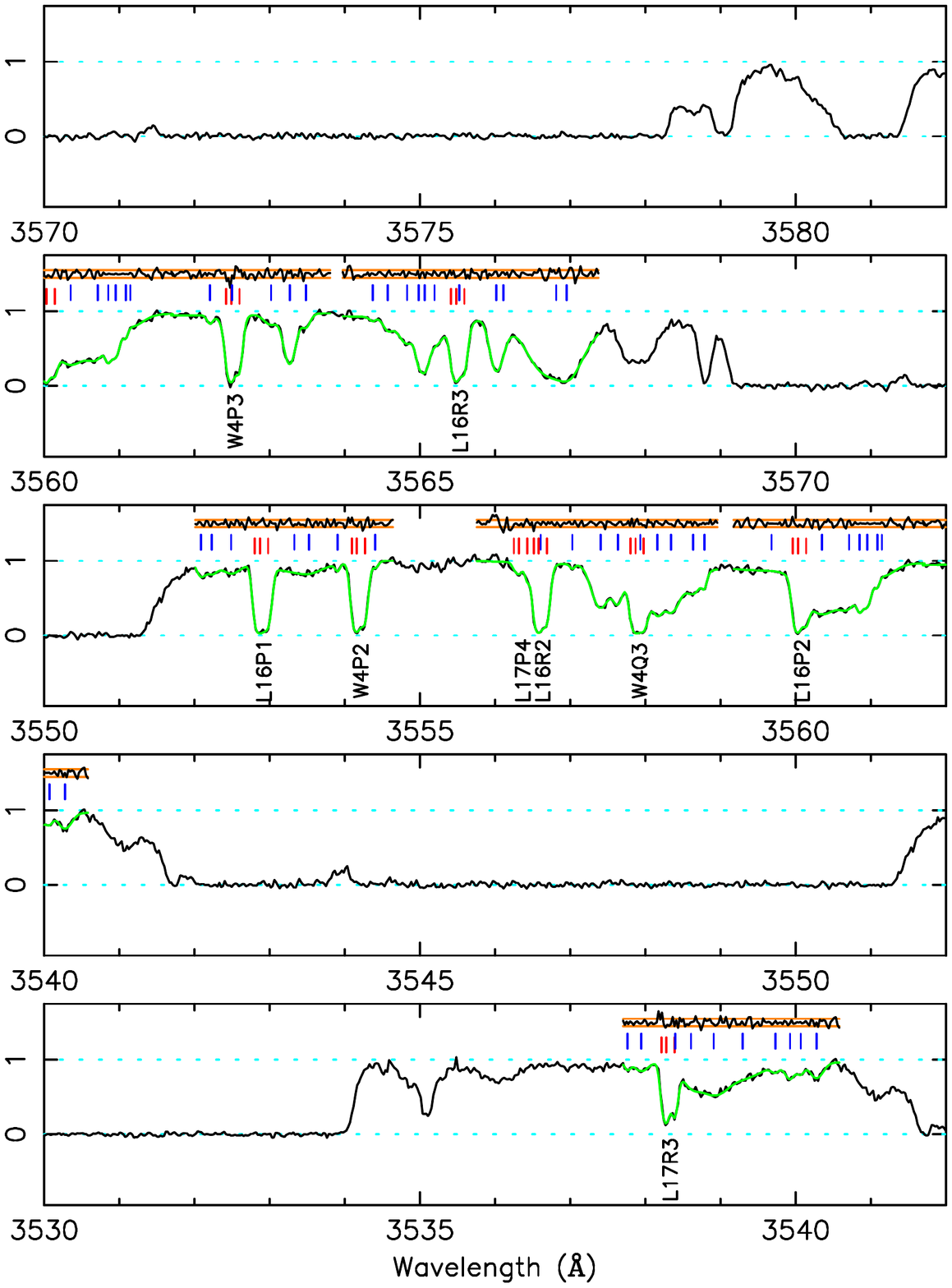}
\par\end{centering}

\caption[H$_2$ fit for the $z=2.811$ absorber toward Q0528$-$250:B2 (1)]{H$_2$ fit for the $z=2.811$ absorber toward Q0528$-$250:B2 (part 1). The vertical axis shows normalised flux. The model fitted to the spectra is shown in green. Red tick marks indicate the position of H$_2$ components, whilst blue tick marks indicate the position of blending transitions (presumed to be Lyman-$\alpha$). Normalised residuals (i.e. [data - model]/error) are plotted above the spectrum between the orange bands, which represent $\pm 1\sigma$. Labels for the H$_2$ transitions are plotted below the data.}
\end{figure}

\begin{figure}[H]
\noindent \begin{centering}
\includegraphics[bb=86bp 180bp 544bp 801bp,clip,width=1\textwidth]{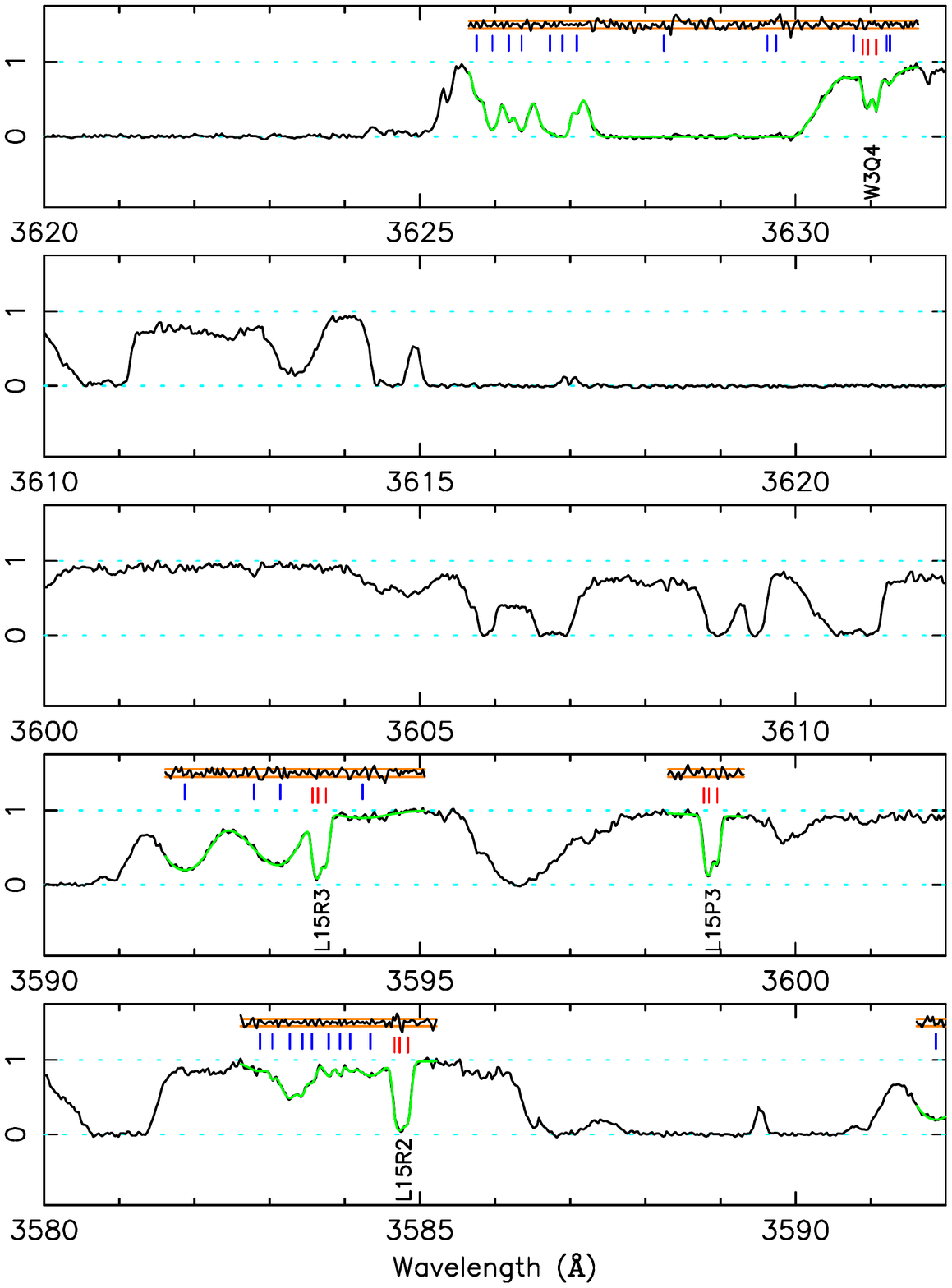}
\par\end{centering}

\caption[H$_2$ fit for the $z=2.811$ absorber toward Q0528$-$250:B2 (2)]{H$_2$ fit for the $z=2.811$ absorber toward Q0528$-$250:B2 (part 2). The vertical axis shows normalised flux. The model fitted to the spectra is shown in green. Red tick marks indicate the position of H$_2$ components, whilst blue tick marks indicate the position of blending transitions (presumed to be Lyman-$\alpha$). Normalised residuals (i.e. [data - model]/error) are plotted above the spectrum between the orange bands, which represent $\pm 1\sigma$. Labels for the H$_2$ transitions are plotted below the data.}
\end{figure}

\begin{figure}[H]
\noindent \begin{centering}
\includegraphics[bb=86bp 180bp 544bp 801bp,clip,width=1\textwidth]{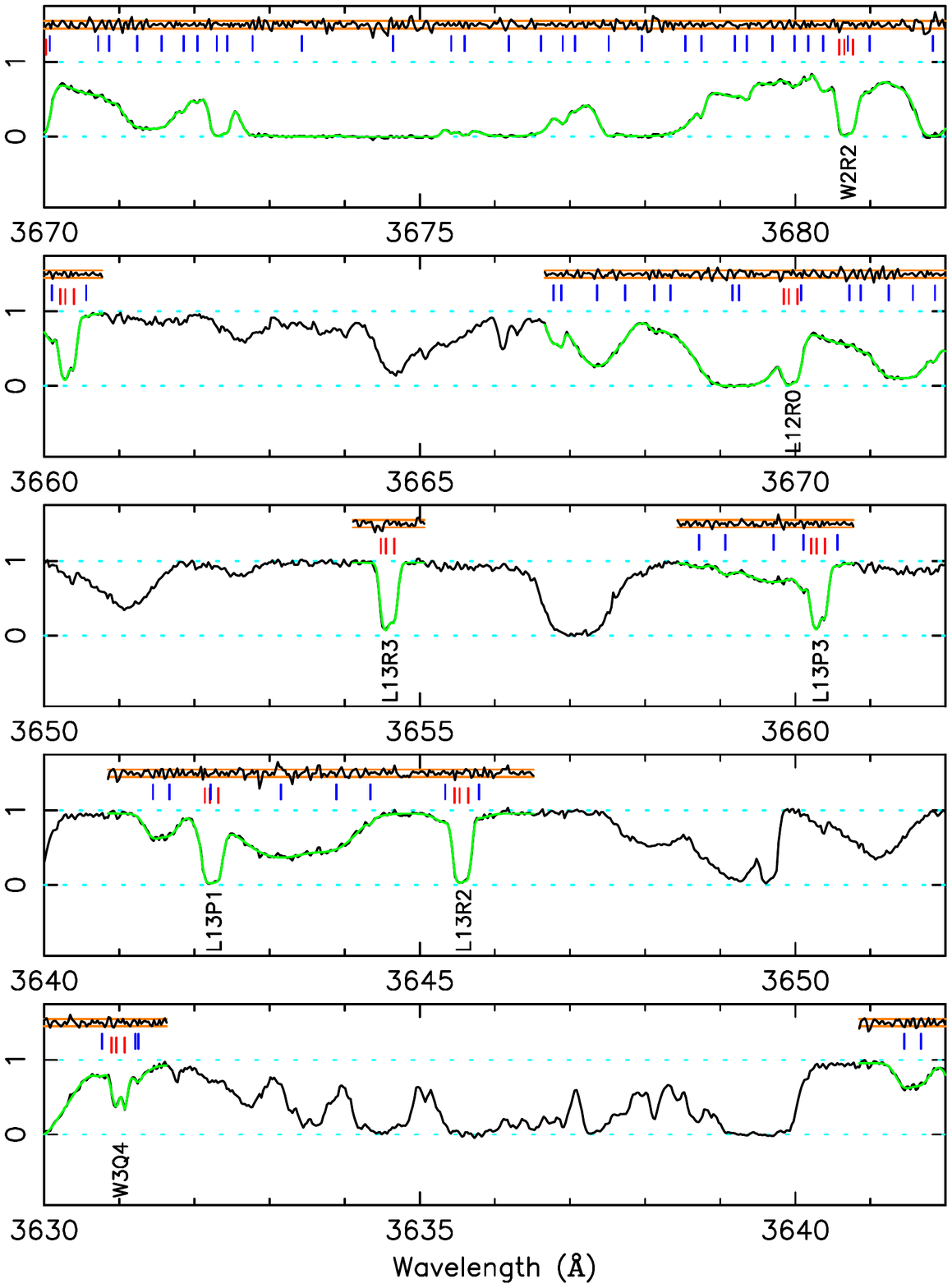}
\par\end{centering}

\caption[H$_2$ fit for the $z=2.811$ absorber toward Q0528$-$250:B2 (3)]{H$_2$ fit for the $z=2.811$ absorber toward Q0528$-$250:B2 (part 3). The vertical axis shows normalised flux. The model fitted to the spectra is shown in green. Red tick marks indicate the position of H$_2$ components, whilst blue tick marks indicate the position of blending transitions (presumed to be Lyman-$\alpha$). Normalised residuals (i.e. [data - model]/error) are plotted above the spectrum between the orange bands, which represent $\pm 1\sigma$. Labels for the H$_2$ transitions are plotted below the data.}
\end{figure}

\begin{figure}[H]
\noindent \begin{centering}
\includegraphics[bb=86bp 180bp 544bp 801bp,clip,width=1\textwidth]{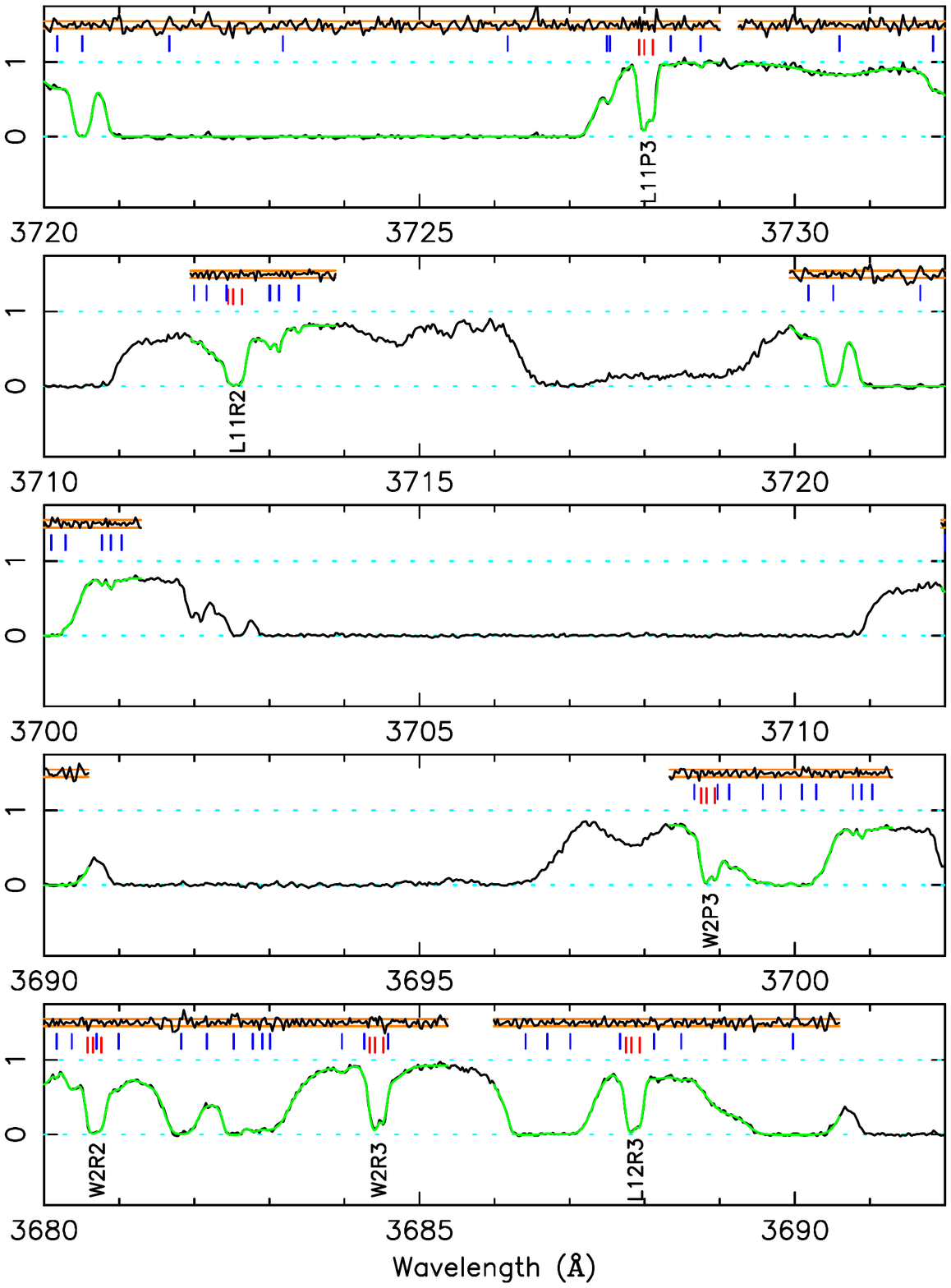}
\par\end{centering}

\caption[H$_2$ fit for the $z=2.811$ absorber toward Q0528$-$250:B2 (4)]{H$_2$ fit for the $z=2.811$ absorber toward Q0528$-$250:B2 (part 4). The vertical axis shows normalised flux. The model fitted to the spectra is shown in green. Red tick marks indicate the position of H$_2$ components, whilst blue tick marks indicate the position of blending transitions (presumed to be Lyman-$\alpha$). Normalised residuals (i.e. [data - model]/error) are plotted above the spectrum between the orange bands, which represent $\pm 1\sigma$. Labels for the H$_2$ transitions are plotted below the data.}
\end{figure}

\begin{figure}[H]
\noindent \begin{centering}
\includegraphics[bb=86bp 180bp 544bp 801bp,clip,width=1\textwidth]{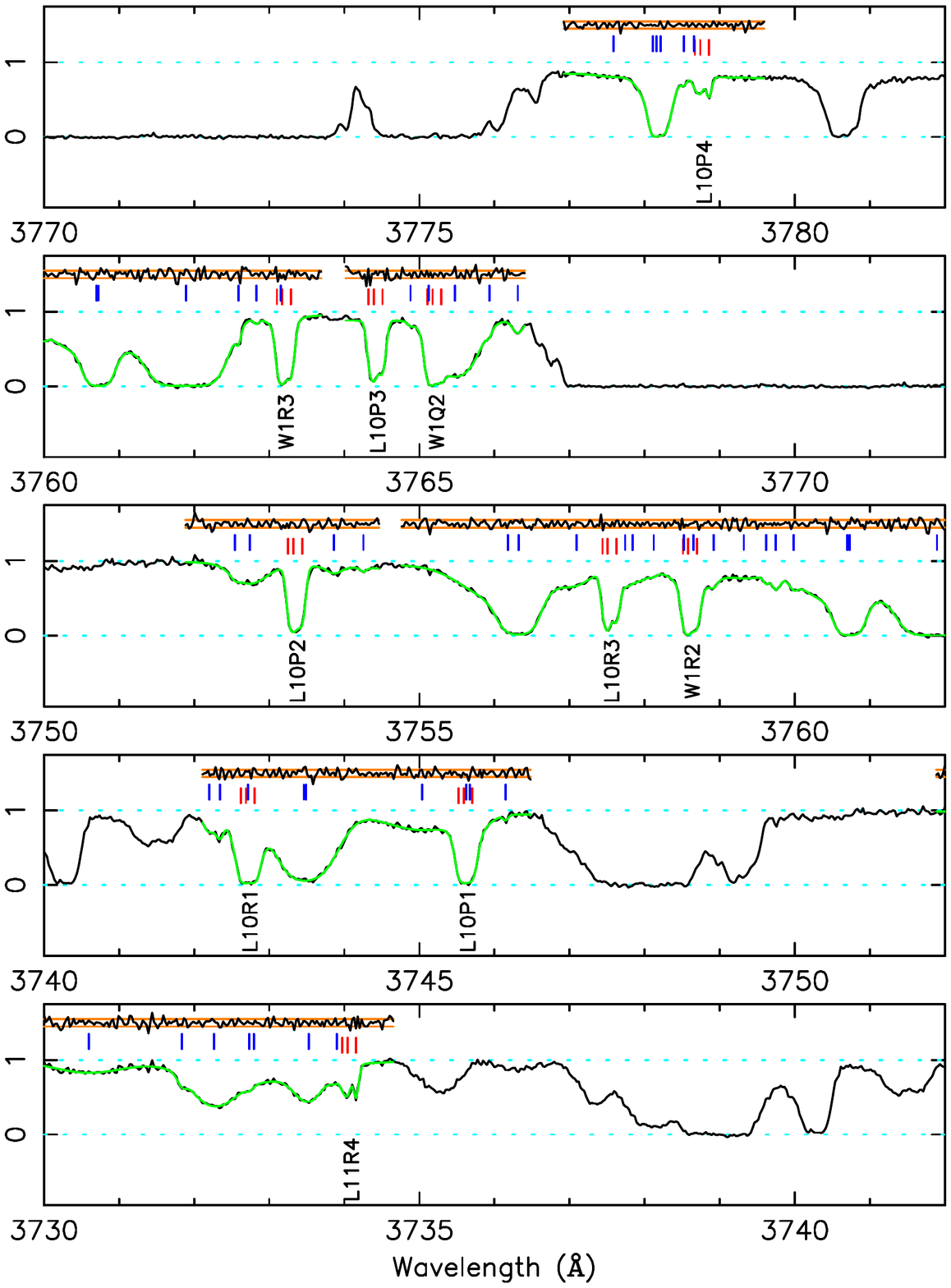}
\par\end{centering}

\caption[H$_2$ fit for the $z=2.811$ absorber toward Q0528$-$250:B2 (5)]{H$_2$ fit for the $z=2.811$ absorber toward Q0528$-$250:B2 (part 5). The vertical axis shows normalised flux. The model fitted to the spectra is shown in green. Red tick marks indicate the position of H$_2$ components, whilst blue tick marks indicate the position of blending transitions (presumed to be Lyman-$\alpha$). Normalised residuals (i.e. [data - model]/error) are plotted above the spectrum between the orange bands, which represent $\pm 1\sigma$. Labels for the H$_2$ transitions are plotted below the data.}
\end{figure}

\begin{figure}[H]
\noindent \begin{centering}
\includegraphics[bb=86bp 180bp 544bp 801bp,clip,width=1\textwidth]{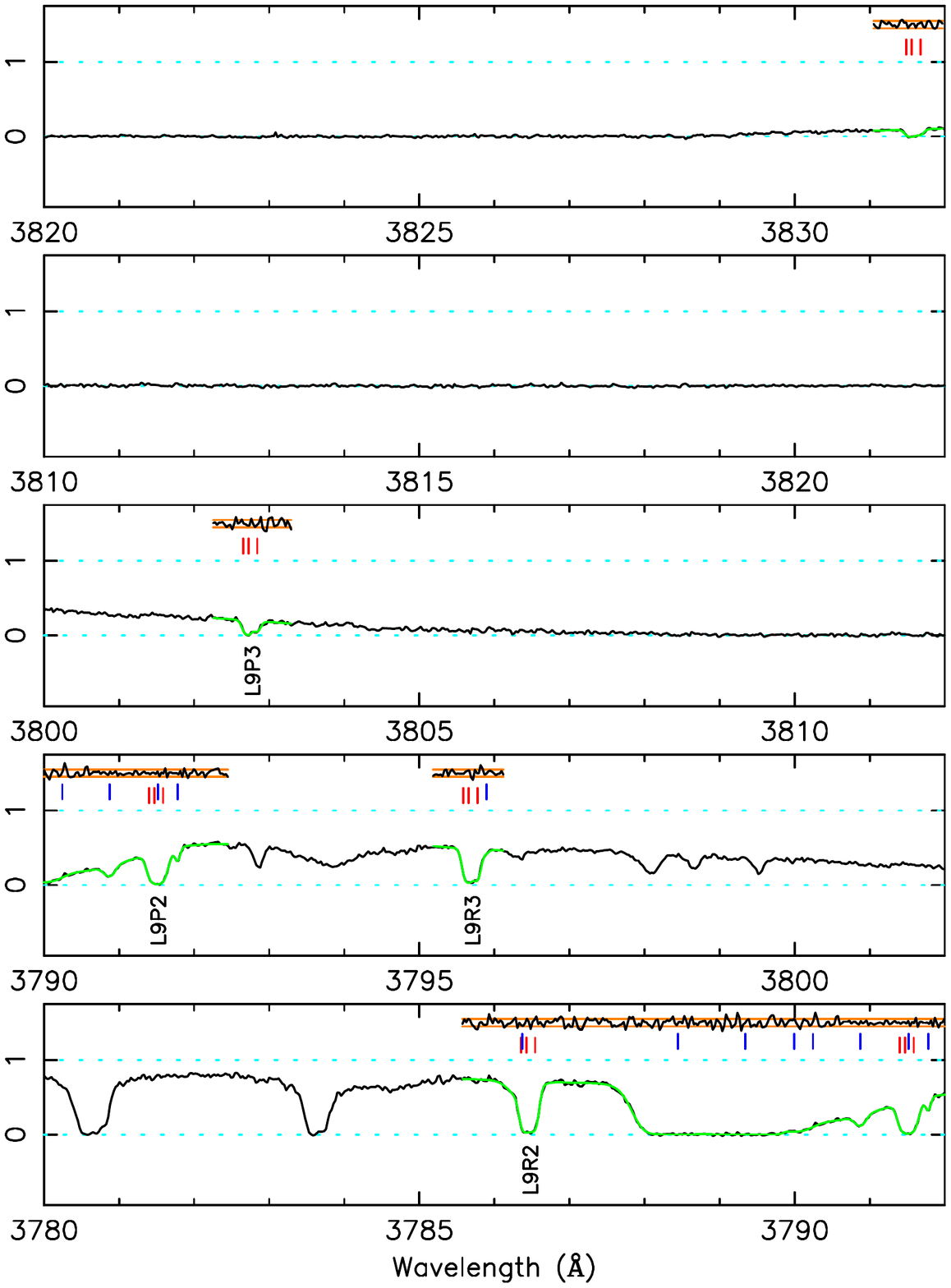}
\par\end{centering}

\caption[H$_2$ fit for the $z=2.811$ absorber toward Q0528$-$250:B2 (6)]{H$_2$ fit for the $z=2.811$ absorber toward Q0528$-$250:B2 (part 6). The vertical axis shows normalised flux. The model fitted to the spectra is shown in green. Red tick marks indicate the position of H$_2$ components, whilst blue tick marks indicate the position of blending transitions (presumed to be Lyman-$\alpha$). Normalised residuals (i.e. [data - model]/error) are plotted above the spectrum between the orange bands, which represent $\pm 1\sigma$. Labels for the H$_2$ transitions are plotted below the data.}
\end{figure}

\begin{figure}[H]
\noindent \begin{centering}
\includegraphics[bb=86bp 180bp 544bp 801bp,clip,width=1\textwidth]{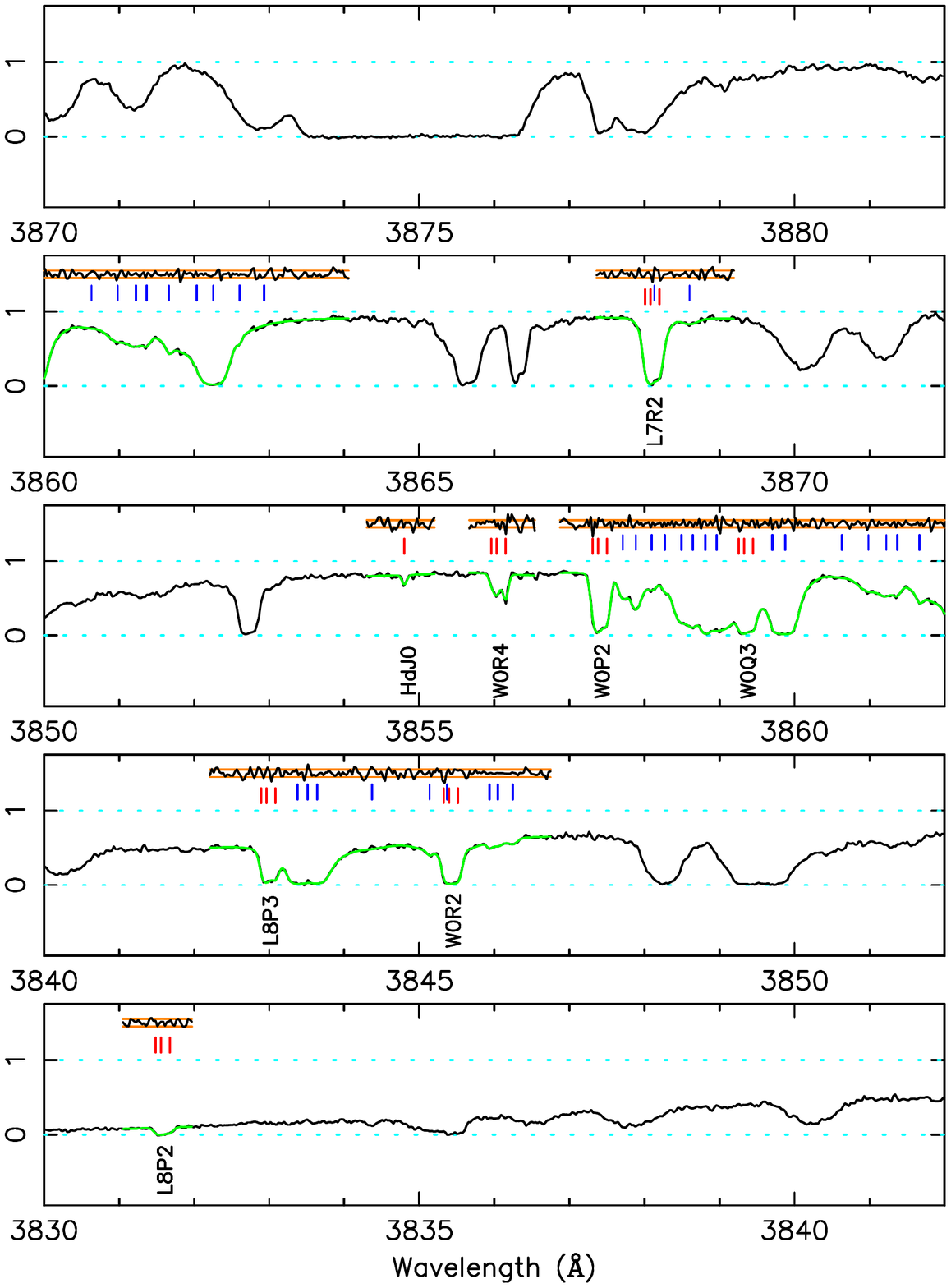}
\par\end{centering}

\caption[H$_2$ fit for the $z=2.811$ absorber toward Q0528$-$250:B2 (7)]{H$_2$ fit for the $z=2.811$ absorber toward Q0528$-$250:B2 (part 7). The vertical axis shows normalised flux. The model fitted to the spectra is shown in green. Red tick marks indicate the position of H$_2$ components, whilst blue tick marks indicate the position of blending transitions (presumed to be Lyman-$\alpha$). Normalised residuals (i.e. [data - model]/error) are plotted above the spectrum between the orange bands, which represent $\pm 1\sigma$. Labels for the H$_2$ transitions are plotted below the data.}
\end{figure}

\begin{figure}[H]
\noindent \begin{centering}
\includegraphics[bb=86bp 180bp 544bp 801bp,clip,width=1\textwidth]{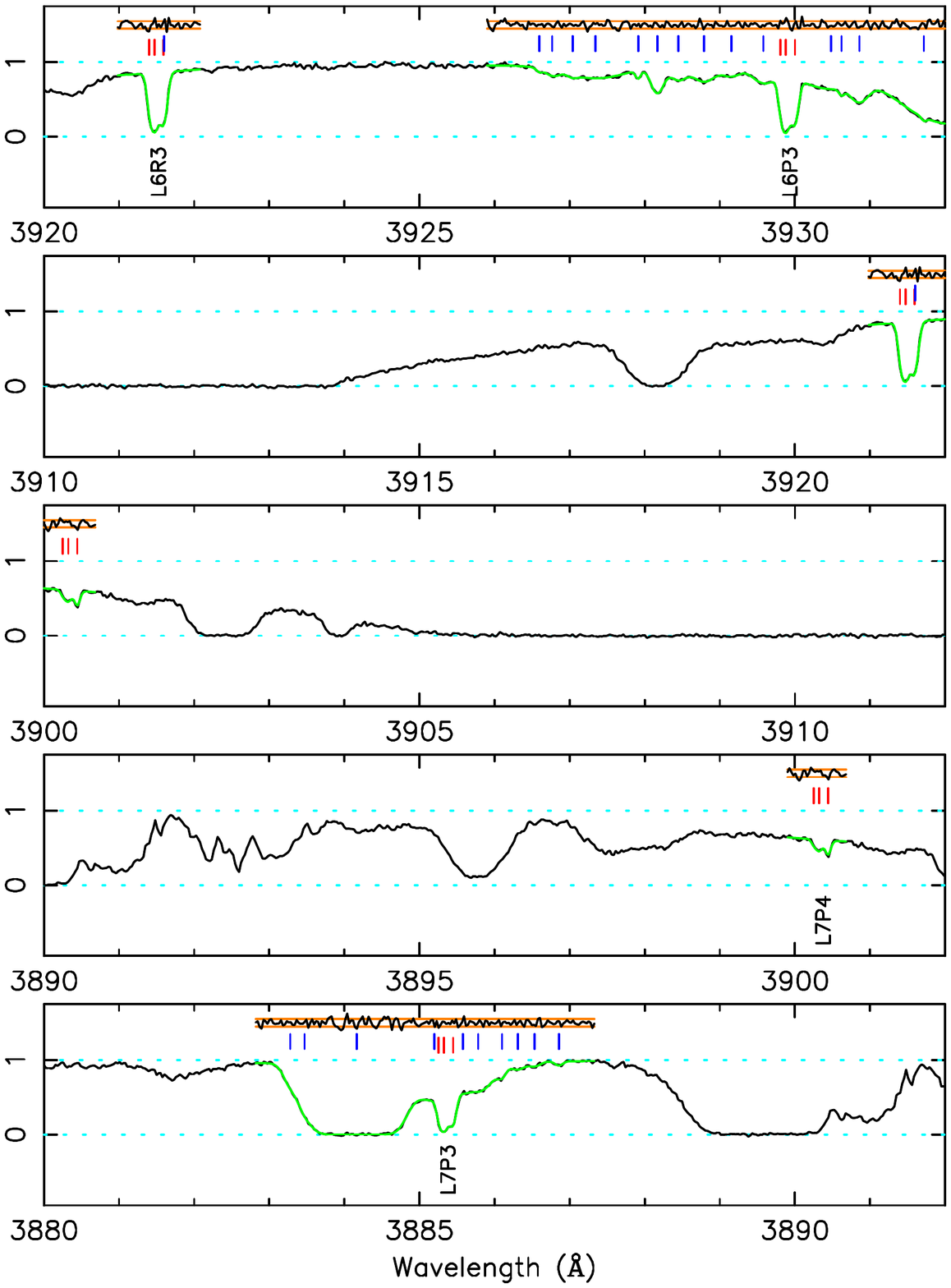}
\par\end{centering}

\caption[H$_2$ fit for the $z=2.811$ absorber toward Q0528$-$250:B2 (8)]{H$_2$ fit for the $z=2.811$ absorber toward Q0528$-$250:B2 (part 8). The vertical axis shows normalised flux. The model fitted to the spectra is shown in green. Red tick marks indicate the position of H$_2$ components, whilst blue tick marks indicate the position of blending transitions (presumed to be Lyman-$\alpha$). Normalised residuals (i.e. [data - model]/error) are plotted above the spectrum between the orange bands, which represent $\pm 1\sigma$. Labels for the H$_2$ transitions are plotted below the data.}
\end{figure}

\begin{figure}[H]
\noindent \begin{centering}
\includegraphics[bb=86bp 180bp 544bp 800bp,clip,width=1\textwidth]{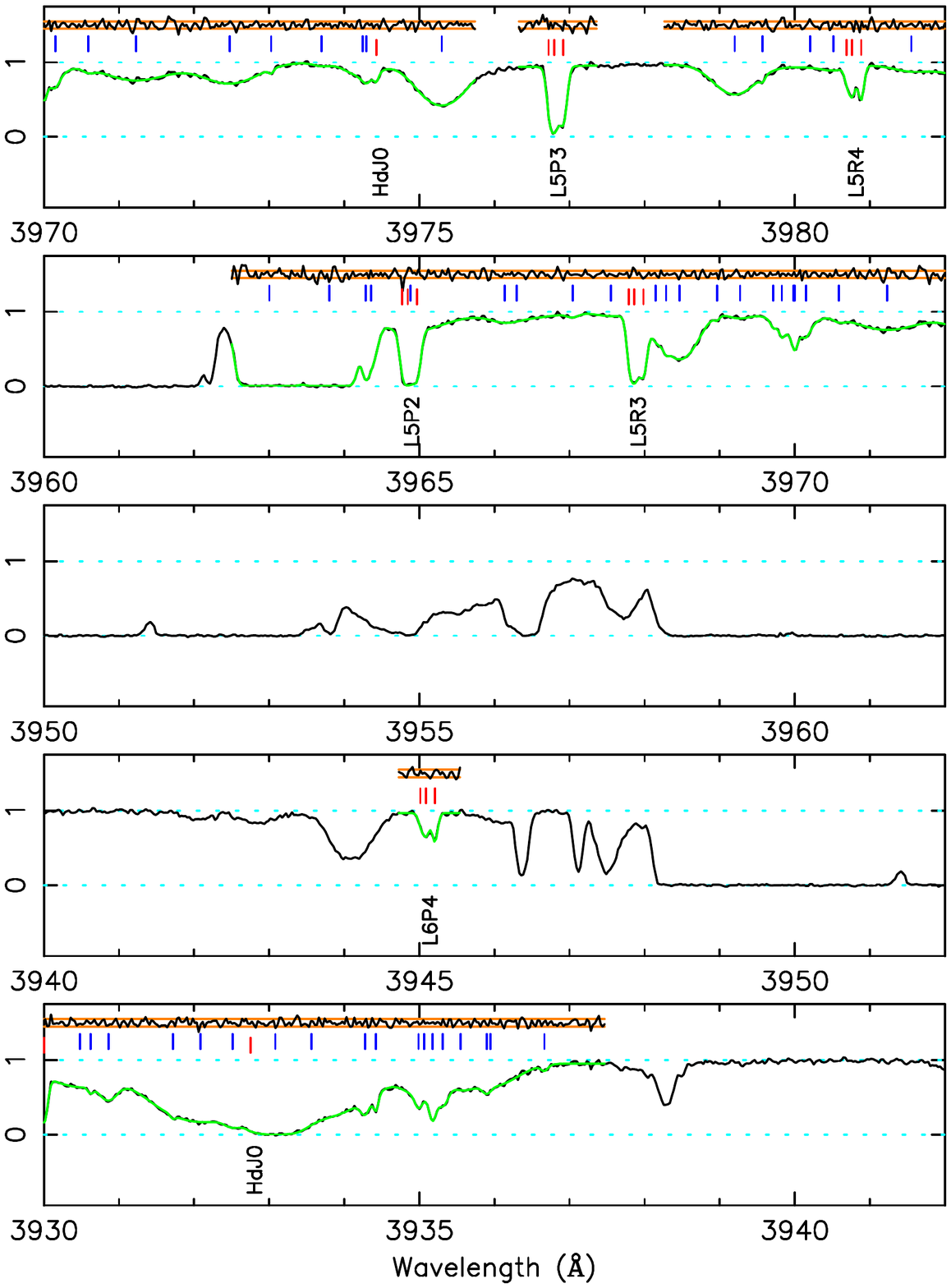}
\par\end{centering}

\caption[H$_2$ fit for the $z=2.811$ absorber toward Q0528$-$250:B2 (9)]{H$_2$ fit for the $z=2.811$ absorber toward Q0528$-$250:B2 (part 9). The vertical axis shows normalised flux. The model fitted to the spectra is shown in green. Red tick marks indicate the position of H$_2$ components, whilst blue tick marks indicate the position of blending transitions (presumed to be Lyman-$\alpha$). Normalised residuals (i.e. [data - model]/error) are plotted above the spectrum between the orange bands, which represent $\pm 1\sigma$. Labels for the H$_2$ transitions are plotted below the data.}
\end{figure}

\begin{figure}[H]
\noindent \begin{centering}
\includegraphics[bb=86bp 180bp 544bp 801bp,clip,width=1\textwidth]{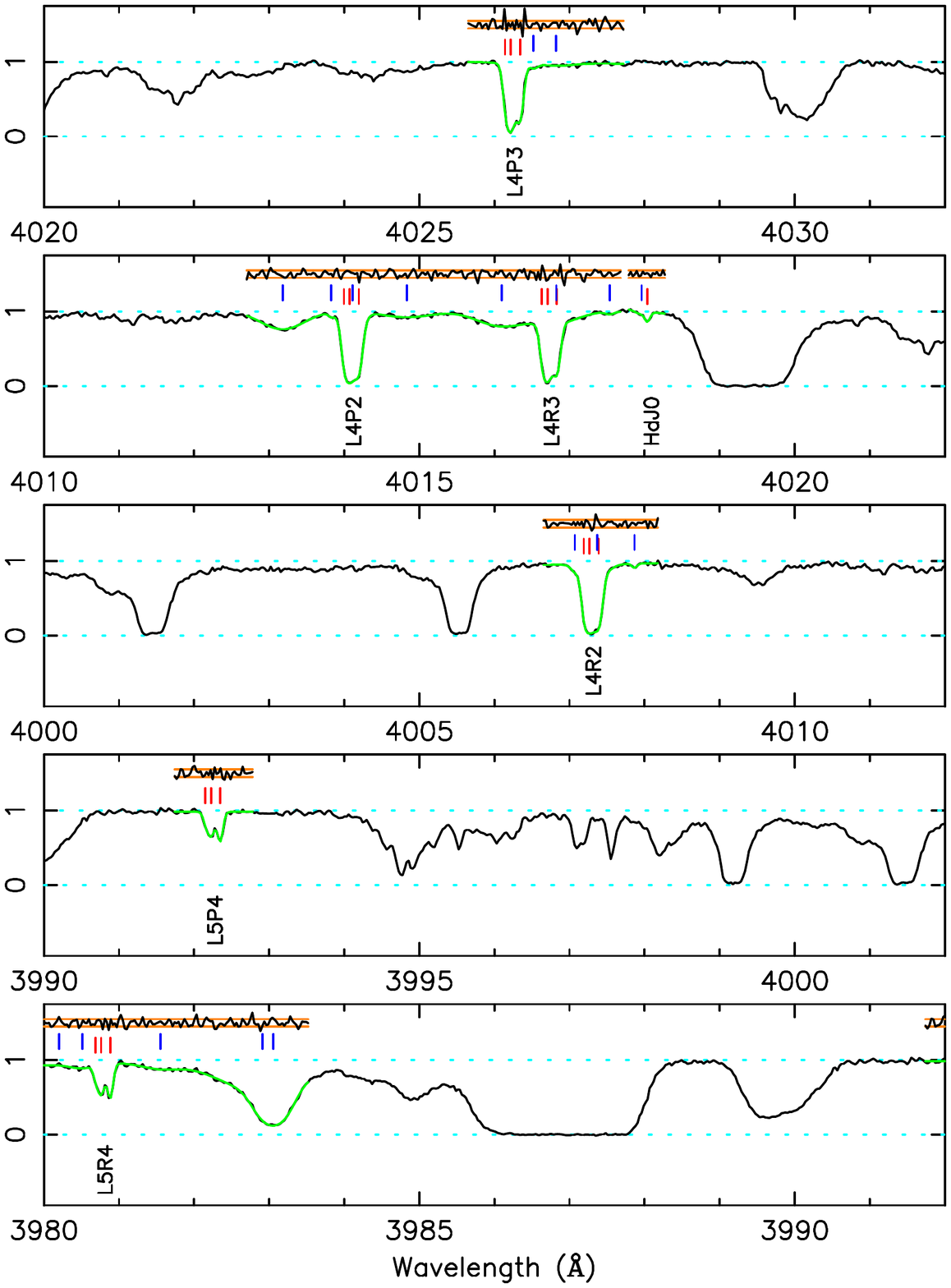}
\par\end{centering}

\caption[H$_2$ fit for the $z=2.811$ absorber toward Q0528$-$250:B2 (10)]{H$_2$ fit for the $z=2.811$ absorber toward Q0528$-$250:B2 (part 10). The vertical axis shows normalised flux. The model fitted to the spectra is shown in green. Red tick marks indicate the position of H$_2$ components, whilst blue tick marks indicate the position of blending transitions (presumed to be Lyman-$\alpha$). Normalised residuals (i.e. [data - model]/error) are plotted above the spectrum between the orange bands, which represent $\pm 1\sigma$. Labels for the H$_2$ transitions are plotted below the data.}
\end{figure}

\begin{figure}[H]
\noindent \begin{centering}
\includegraphics[bb=86bp 180bp 544bp 801bp,clip,width=1\textwidth]{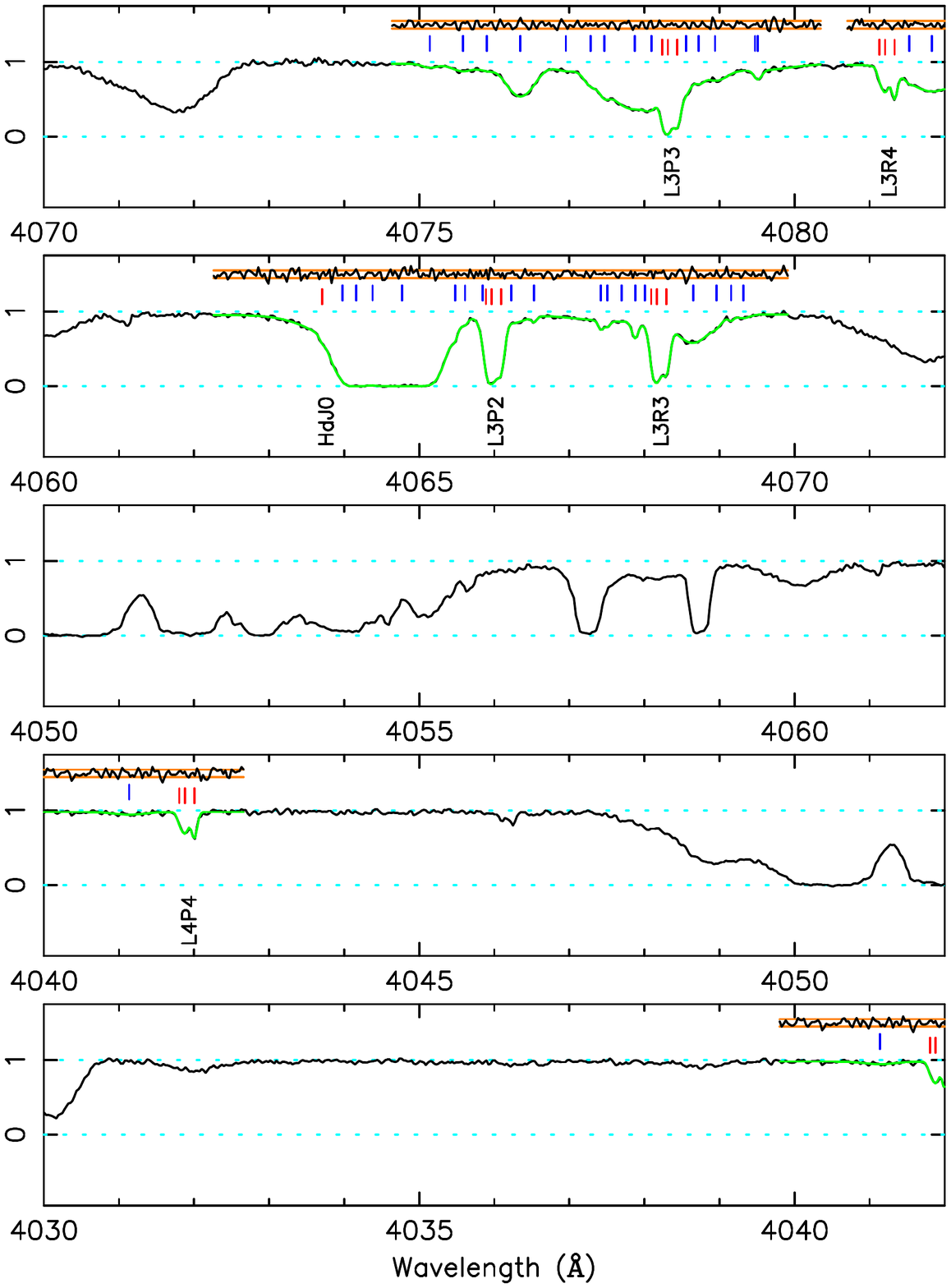}
\par\end{centering}

\caption[H$_2$ fit for the $z=2.811$ absorber toward Q0528$-$250:B2 (11)]{H$_2$ fit for the $z=2.811$ absorber toward Q0528$-$250:B2 (part 11). The vertical axis shows normalised flux. The model fitted to the spectra is shown in green. Red tick marks indicate the position of H$_2$ components, whilst blue tick marks indicate the position of blending transitions (presumed to be Lyman-$\alpha$). Normalised residuals (i.e. [data - model]/error) are plotted above the spectrum between the orange bands, which represent $\pm 1\sigma$. Labels for the H$_2$ transitions are plotted below the data.}
\end{figure}

\begin{figure}[H]
\noindent \begin{centering}
\includegraphics[bb=86bp 180bp 544bp 801bp,clip,width=1\textwidth]{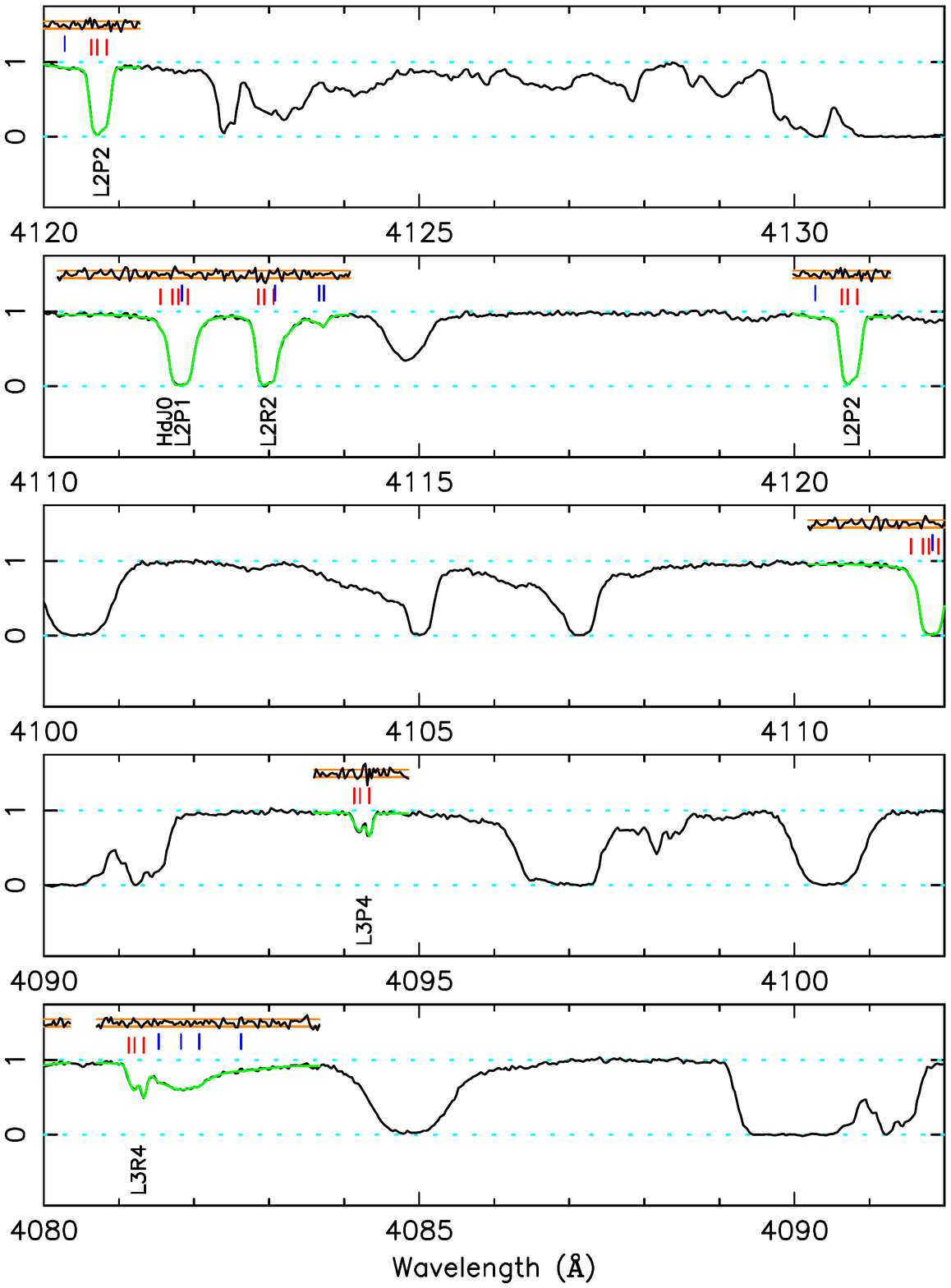}
\par\end{centering}

\caption[H$_2$ fit for the $z=2.811$ absorber toward Q0528$-$250:B2 (12)]{H$_2$ fit for the $z=2.811$ absorber toward Q0528$-$250:B2 (part 12). The vertical axis shows normalised flux. The model fitted to the spectra is shown in green. Red tick marks indicate the position of H$_2$ components, whilst blue tick marks indicate the position of blending transitions (presumed to be Lyman-$\alpha$). Normalised residuals (i.e. [data - model]/error) are plotted above the spectrum between the orange bands, which represent $\pm 1\sigma$. Labels for the H$_2$ transitions are plotted below the data.}
\end{figure}

\begin{figure}[H]
\noindent \begin{centering}
\includegraphics[bb=86bp 180bp 544bp 801bp,clip,width=1\textwidth]{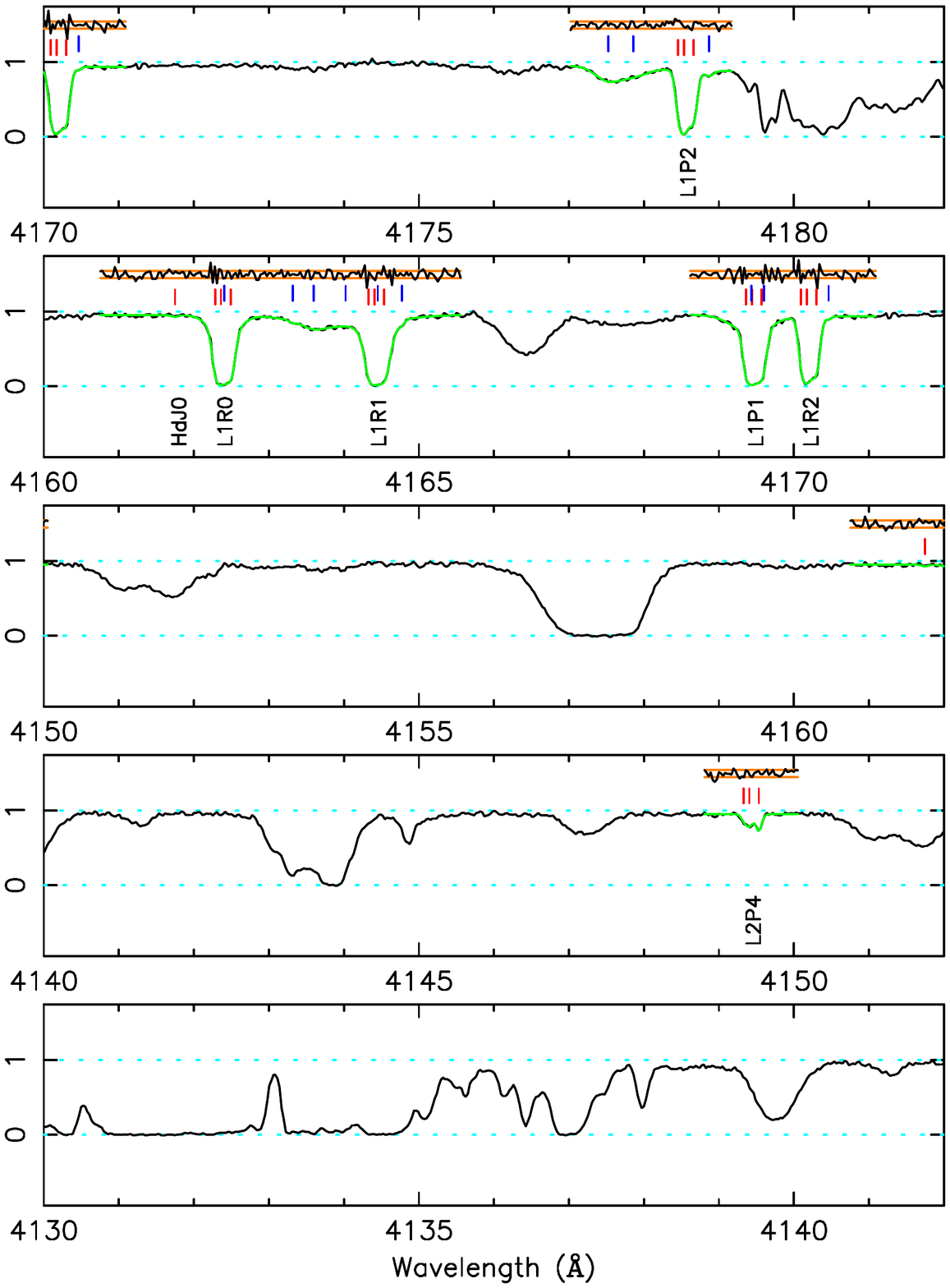}
\par\end{centering}

\caption[H$_2$ fit for the $z=2.811$ absorber toward Q0528$-$250:B2 (13)]{H$_2$ fit for the $z=2.811$ absorber toward Q0528$-$250:B2 (part 13). The vertical axis shows normalised flux. The model fitted to the spectra is shown in green. Red tick marks indicate the position of H$_2$ components, whilst blue tick marks indicate the position of blending transitions (presumed to be Lyman-$\alpha$). Normalised residuals (i.e. [data - model]/error) are plotted above the spectrum between the orange bands, which represent $\pm 1\sigma$. Labels for the H$_2$ transitions are plotted below the data.}
\end{figure}

\begin{figure}[H]
\noindent \begin{centering}
\includegraphics[bb=86bp 180bp 544bp 801bp,clip,width=1\textwidth]{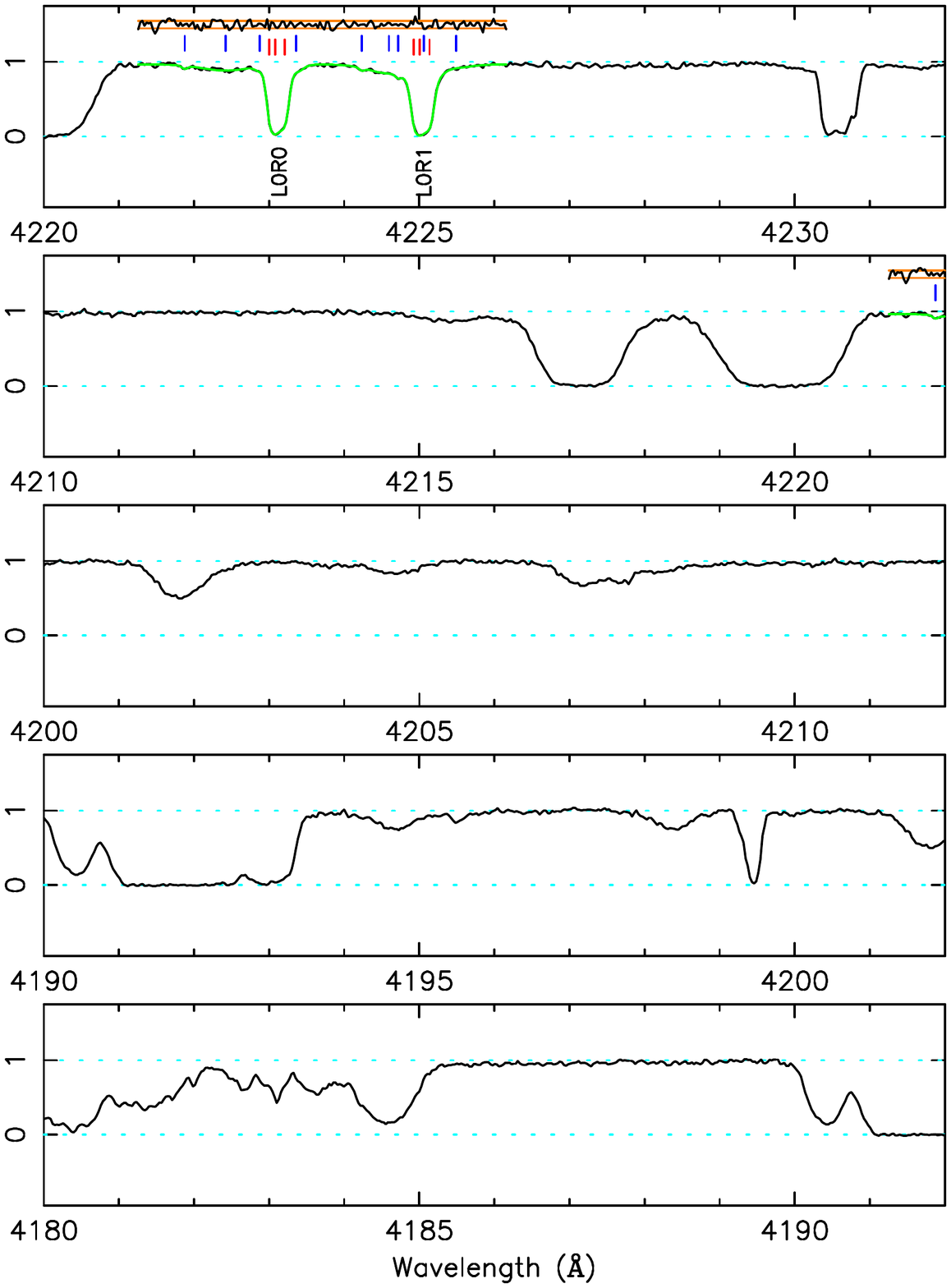}
\par\end{centering}

\caption[H$_2$ fit for the $z=2.811$ absorber toward Q0528$-$250:B2 (14)]{H$_2$ fit for the $z=2.811$ absorber toward Q0528$-$250:B2 (part 14). The vertical axis shows normalised flux. The model fitted to the spectra is shown in green. Red tick marks indicate the position of H$_2$ components, whilst blue tick marks indicate the position of blending transitions (presumed to be Lyman-$\alpha$). Normalised residuals (i.e. [data - model]/error) are plotted above the spectrum between the orange bands, which represent $\pm 1\sigma$. Labels for the H$_2$ transitions are plotted below the data.}
\end{figure}

\chapter{Many-multiplet Voigt profile fits\label{cha:MM fits}}

In this appendix, we provide the fits for the many-multiplet systems
considered in chapter \ref{cha:alpha}. Each absorber is plotted on
a velocity scale, such that corresponding components align vertically.
Velocities are given as differences from an arbitrary redshift, which
is usually chosen to be close to the maximum optical depth of the
absorber. The positions of fitted components are indicated by blue
tick marks. Plotted above each fit are the residuals of the fit, that
is {[}fit-data{]}/error, where the error is the $1\sigma$ uncertainty
associated with each flux pixel. The two red lines indicate $\pm1\sigma$,
within which the residuals are expected to occur about 68\% of the
time if the errors are Gaussian, the error array is correct and the
fitted model is a good representation of the data.

Each plot contains a maximum of 16 regions. In the event that there
are more fitting regions than this, the fit is split into several
parts. Each part may contain common transitions so as to provide a
common reference, and to illustrate the velocity structure more clearly.

\begin{figure}[H]
\noindent \begin{centering}
\includegraphics[bb=34bp 58bp 554bp 727bp,clip,width=1\textwidth]{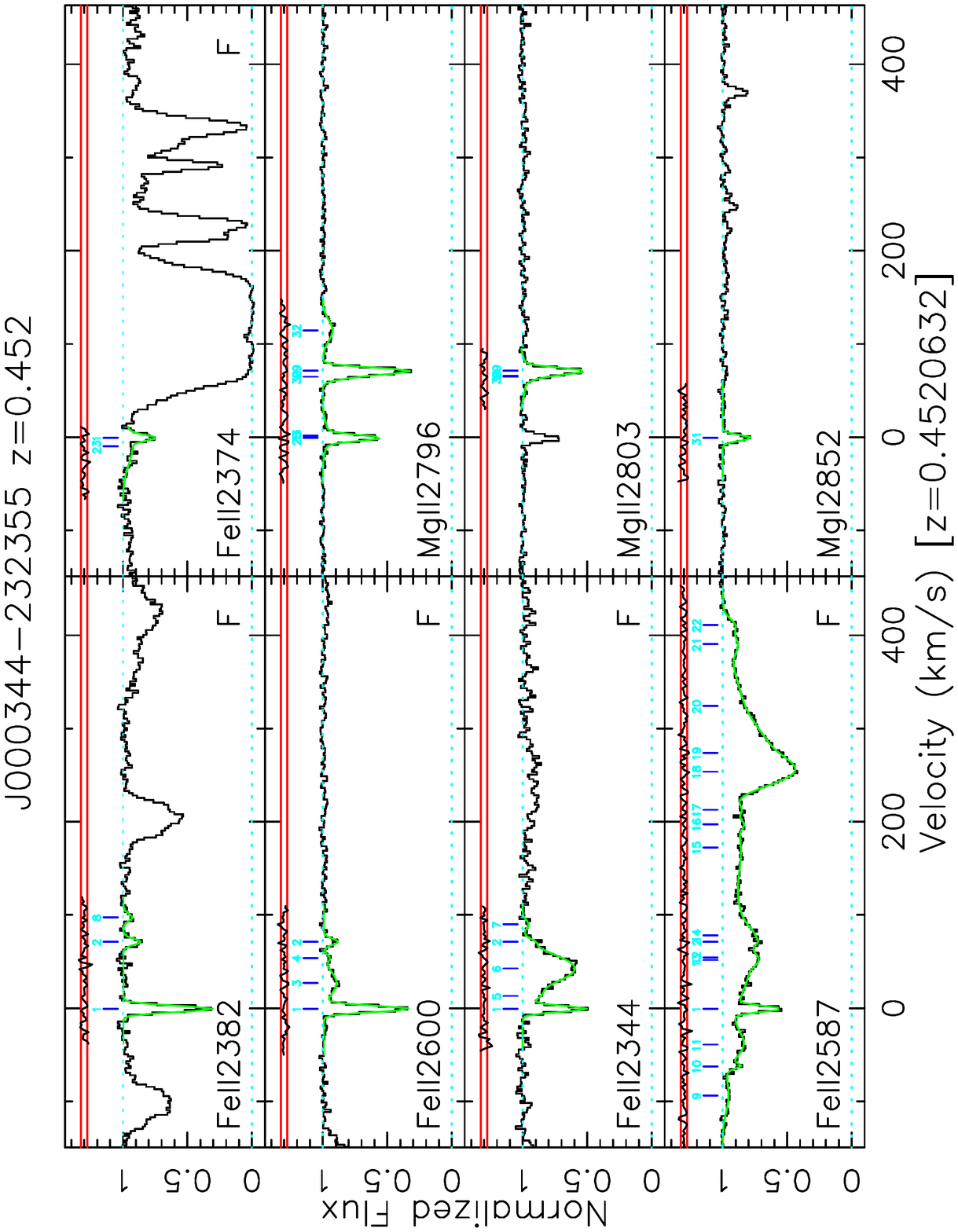}
\par\end{centering}

\caption[Fit for the $z=0.452$ absorber toward J000344$-$232355]{Many-multiplet fit for the $z=0.452$ absorber toward J000344$-$232355.}
\end{figure}

\begin{figure}[H]
\noindent \begin{centering}
\includegraphics[bb=34bp 58bp 554bp 727bp,clip,width=1\textwidth]{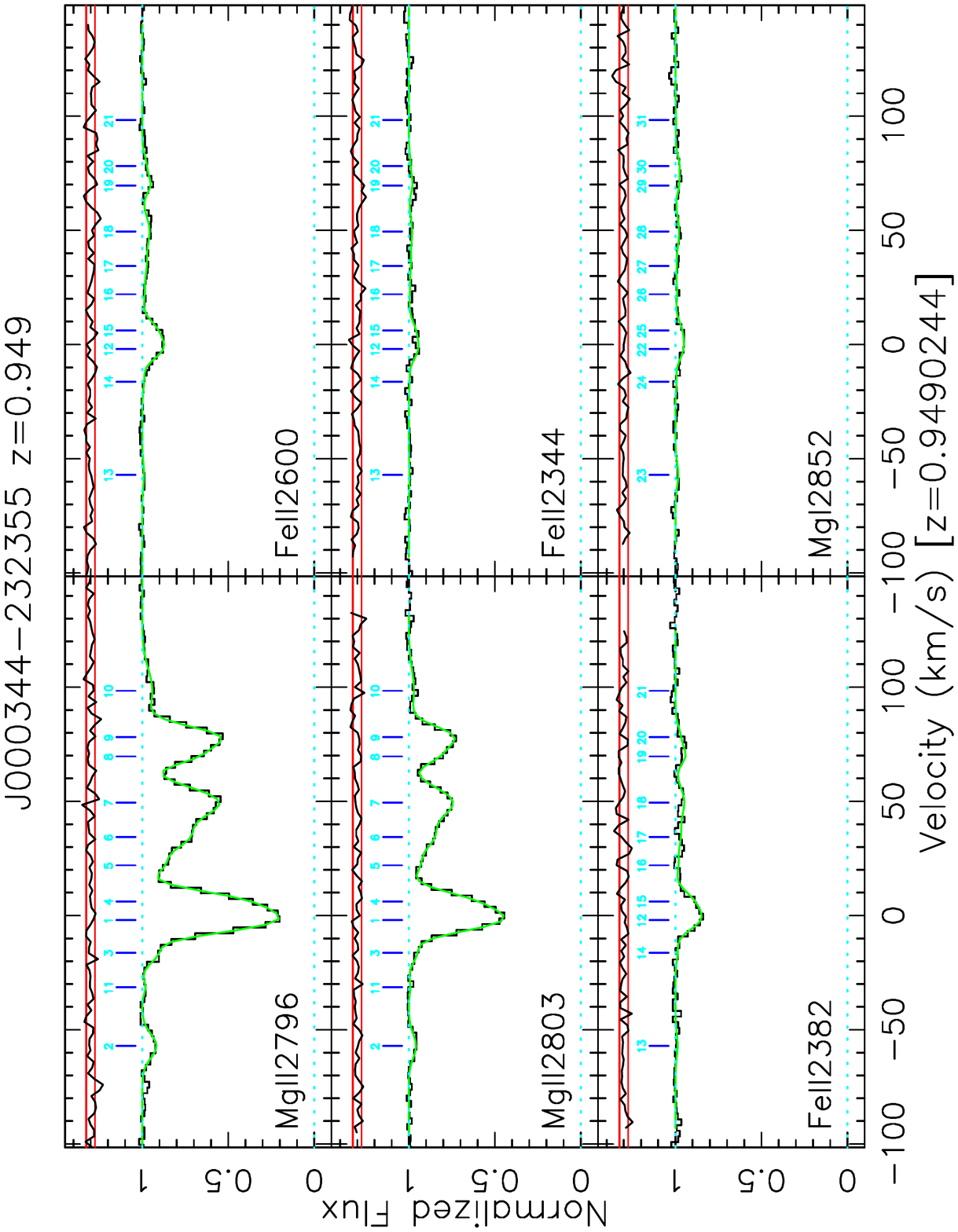}
\par\end{centering}

\caption[Fit for the $z=0.949$ absorber toward J000344$-$232355]{Many-multiplet fit for the $z=0.949$ absorber toward J000344$-$232355.}
\end{figure}

\begin{figure}[H]
\noindent \begin{centering}
\includegraphics[bb=34bp 58bp 554bp 738bp,clip,width=1\textwidth]{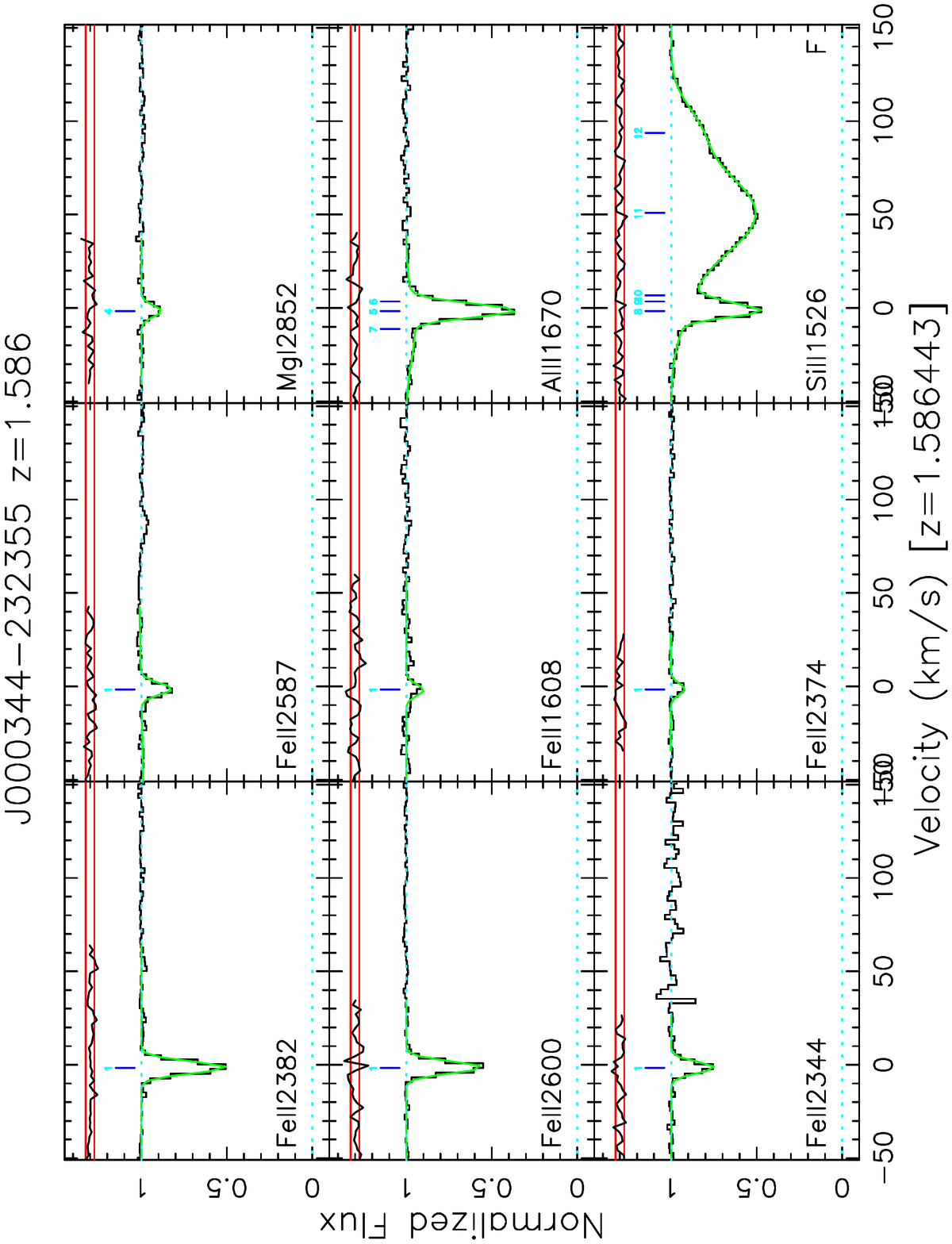}
\par\end{centering}

\caption[Fit for the $z=1.586$ absorber toward J000344$-$232355]{Many-multiplet fit for the $z=1.586$ absorber toward J000344$-$232355.}
\end{figure}
\begin{figure}[H]
\noindent \begin{centering}
\includegraphics[bb=34bp 58bp 554bp 738bp,clip,width=1\textwidth]{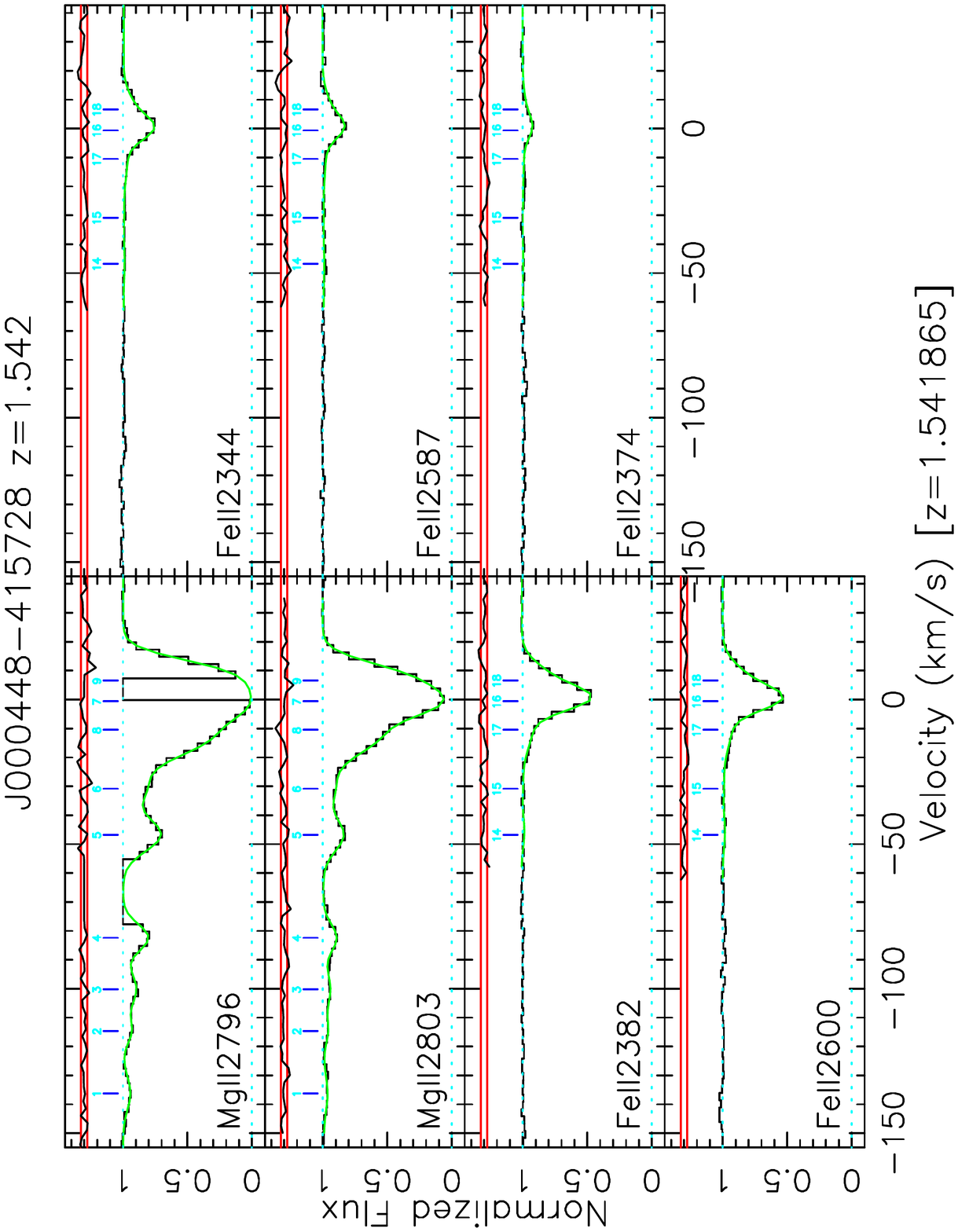}
\par\end{centering}

\caption[Fit for the $z=1.542$ absorber toward J000448$-$415728]{Many-multiplet fit for the $z=1.542$ absorber toward J000448$-$415728.}
\end{figure}
\begin{figure}[H]
\noindent \begin{centering}
\includegraphics[bb=34bp 58bp 554bp 738bp,clip,width=1\textwidth]{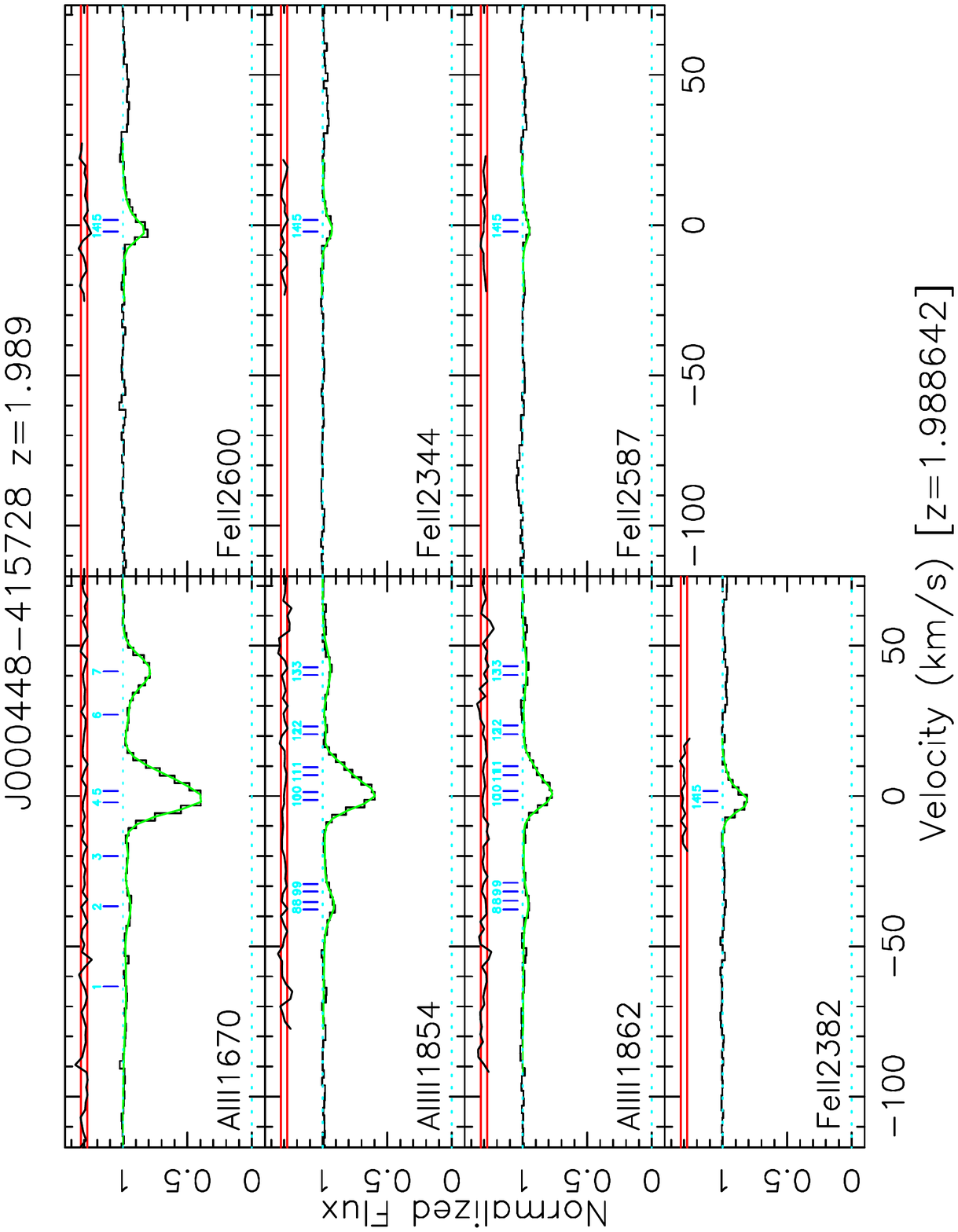}
\par\end{centering}

\caption[Fit for the $z=1.989$ absorber toward J000448$-$415728]{Many-multiplet fit for the $z=1.989$ absorber toward J000448$-$415728.}
\end{figure}
\begin{figure}[H]
\noindent \begin{centering}
\includegraphics[bb=34bp 58bp 554bp 738bp,clip,width=1\textwidth]{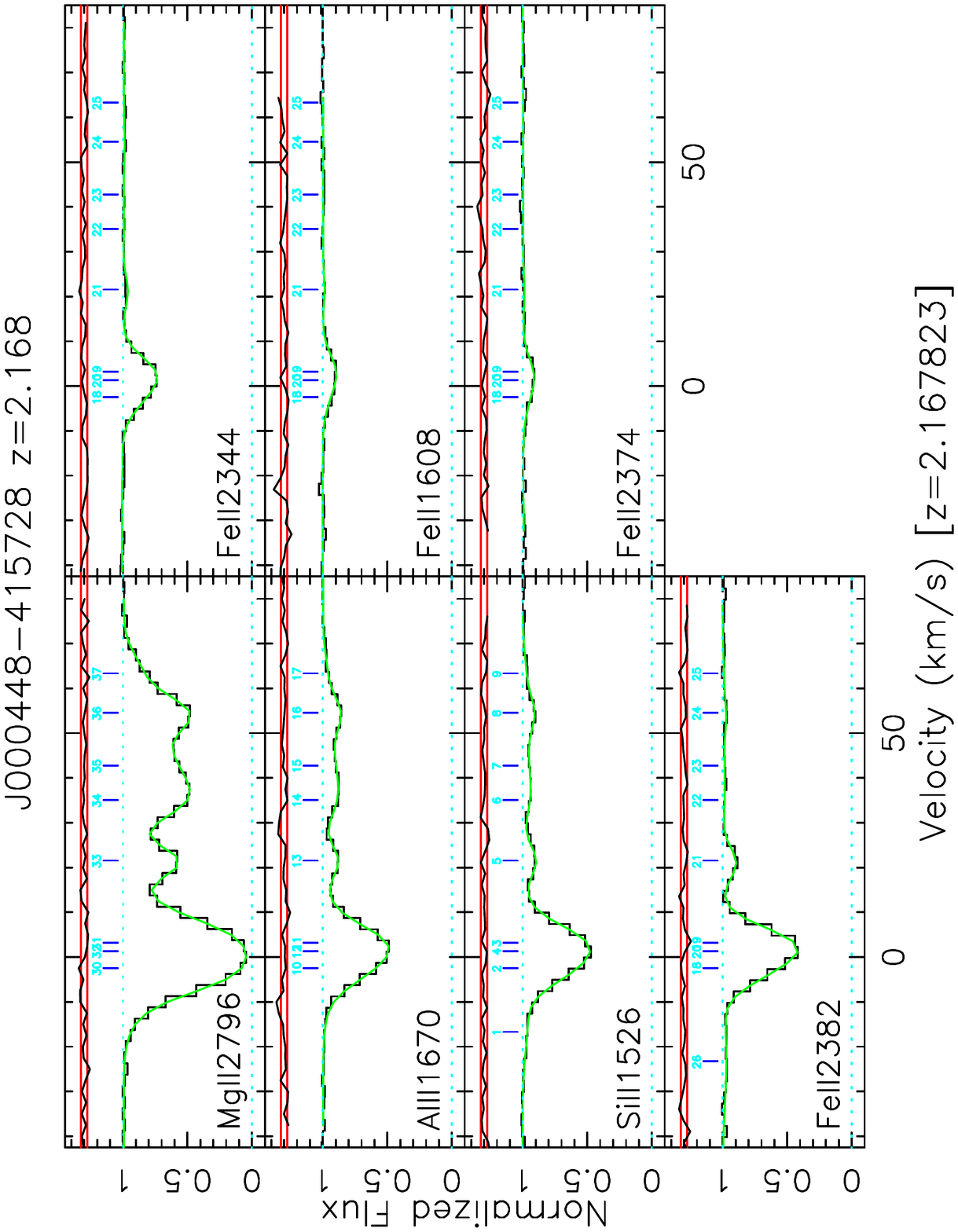}
\par\end{centering}

\caption[Fit for the $z=2.168$ absorber toward J000448$-$415728]{Many-multiplet fit for the $z=2.168$ absorber toward J000448$-$415728.}
\end{figure}
\begin{figure}[H]
\noindent \begin{centering}
\includegraphics[bb=34bp 58bp 554bp 738bp,clip,width=1\textwidth]{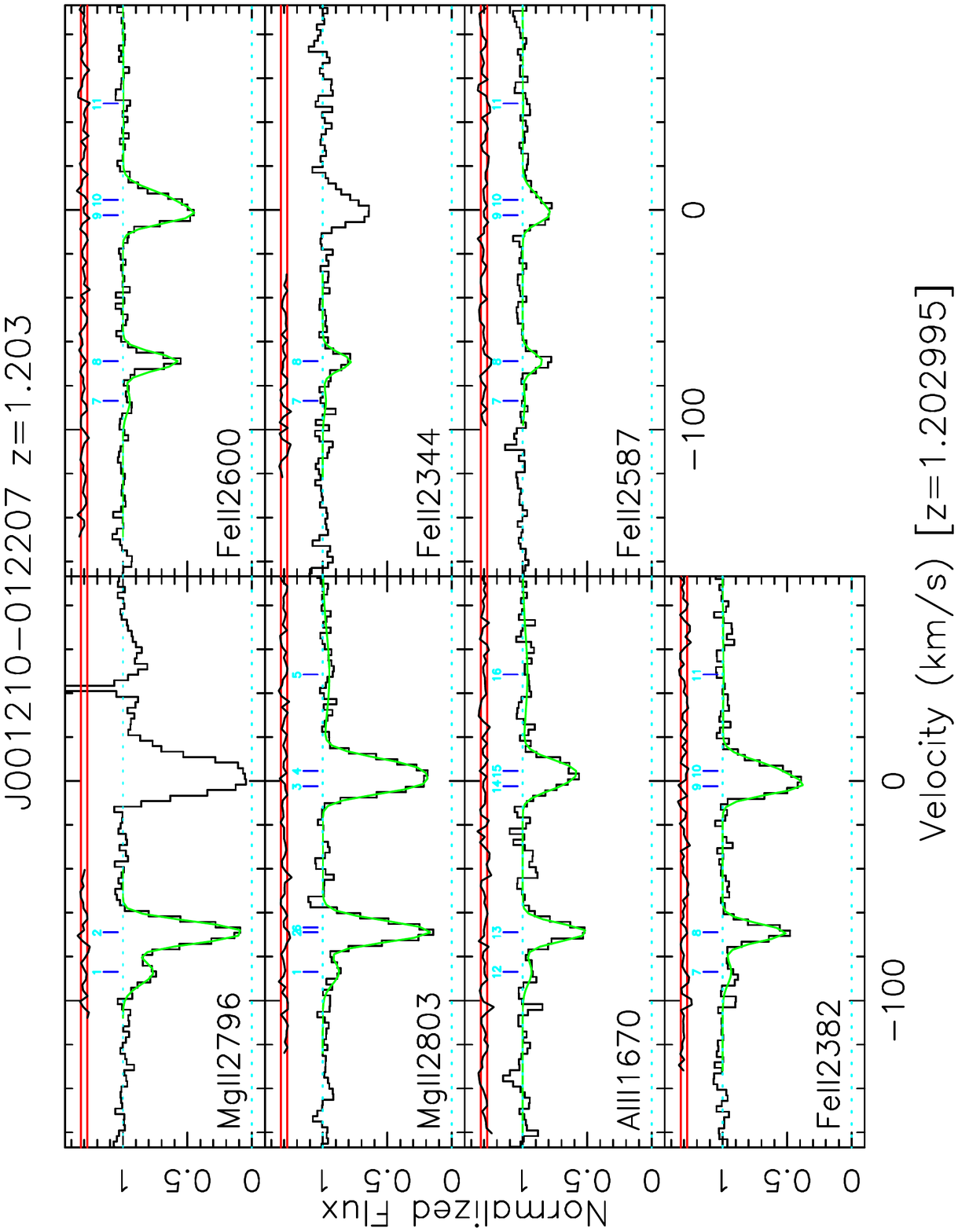}
\par\end{centering}

\caption[Fit for the $z=1.203$ absorber toward J001210$-$012207]{Many-multiplet fit for the $z=1.203$ absorber toward J001210$-$012207.}
\end{figure}

\begin{figure}[H]
\noindent \begin{centering}
\includegraphics[bb=34bp 58bp 554bp 738bp,clip,width=1\textwidth]{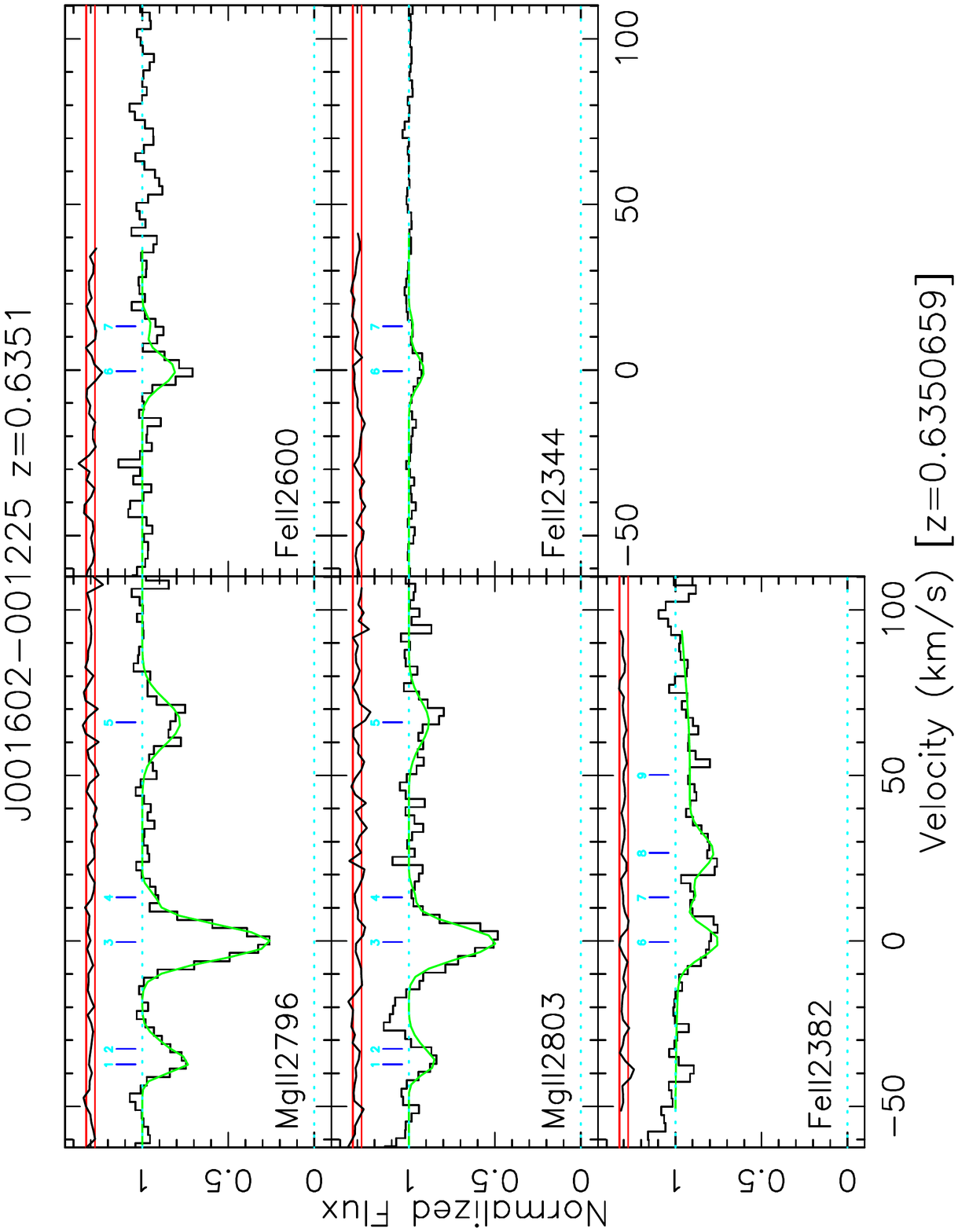}
\par\end{centering}

\caption[Fit for the $z=0.635$ absorber toward J001602$-$001225]{Many-multiplet fit for the $z=0.635$ absorber toward J001602$-$001225.}
\end{figure}
\begin{figure}[H]
\noindent \begin{centering}
\includegraphics[bb=34bp 58bp 554bp 738bp,clip,width=1\textwidth]{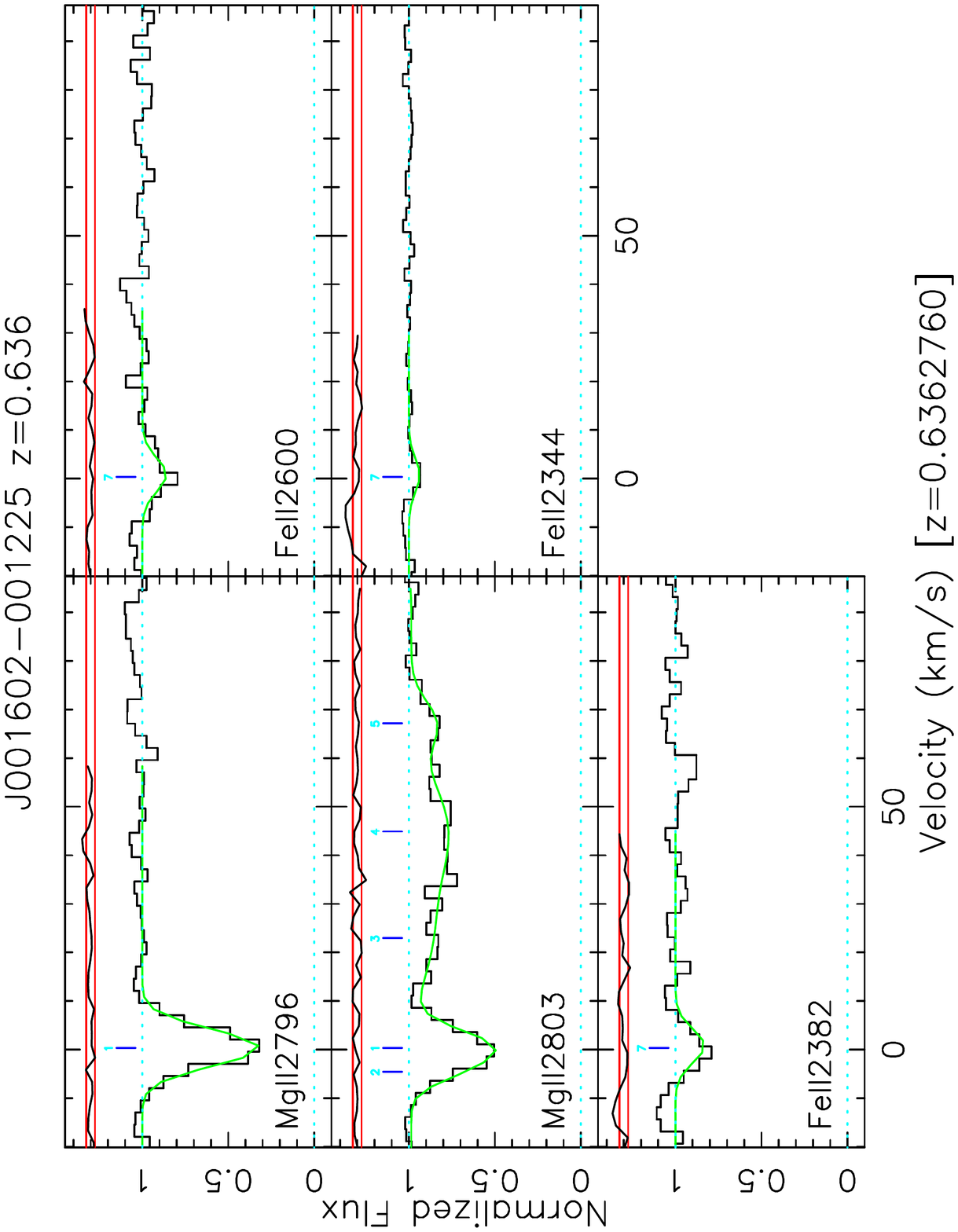}
\par\end{centering}

\caption[Fit for the $z=0.636$ absorber toward J001602$-$001225]{Many-multiplet fit for the $z=0.636$ absorber toward J001602$-$001225.}
\end{figure}
\begin{figure}[H]
\noindent \begin{centering}
\includegraphics[bb=34bp 58bp 554bp 738bp,clip,width=1\textwidth]{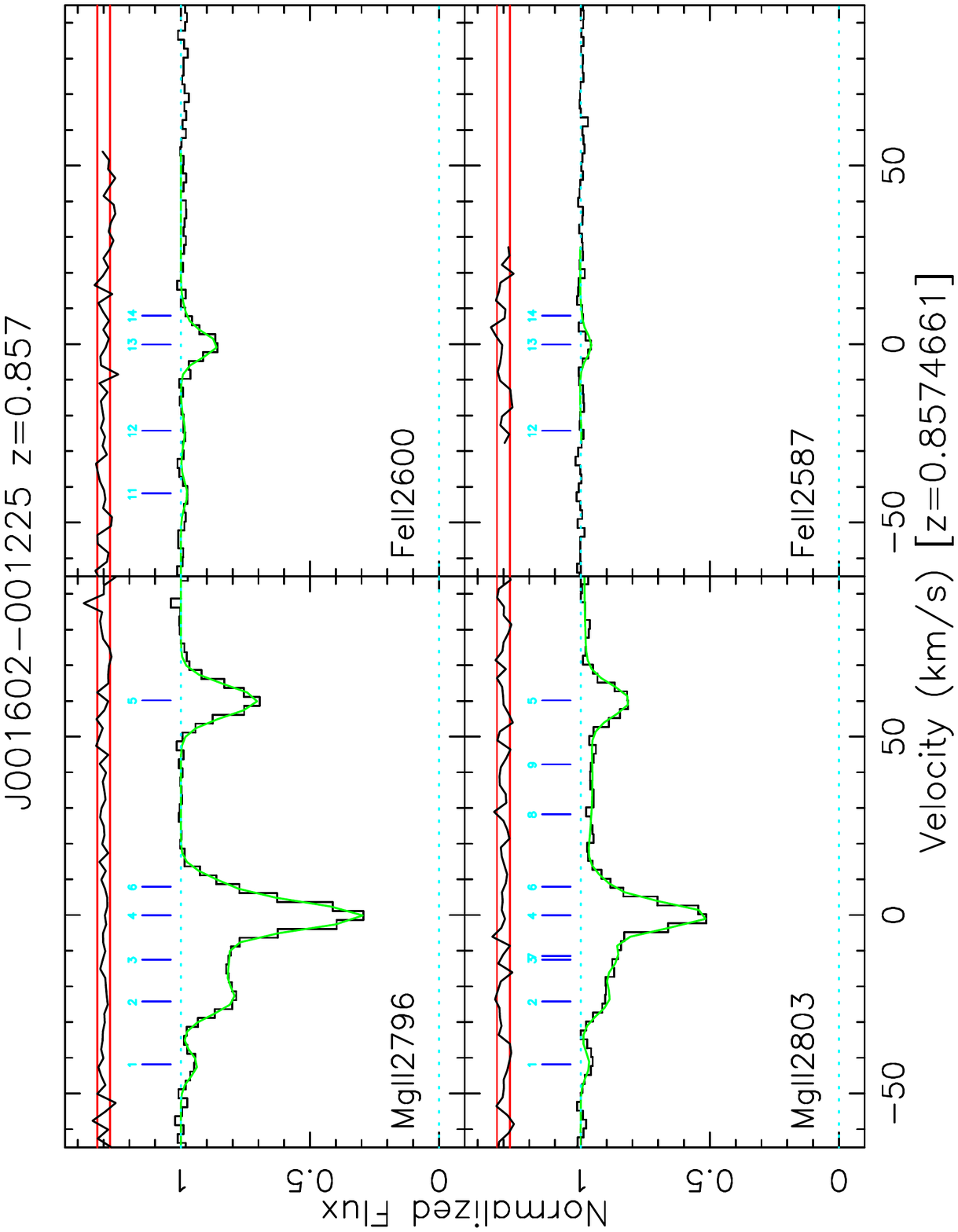}
\par\end{centering}

\caption[Fit for the $z=0.857$ absorber toward J001602$-$001225]{Many-multiplet fit for the $z=0.857$ absorber toward J001602$-$001225.}
\end{figure}
\begin{figure}[H]
\noindent \begin{centering}
\includegraphics[bb=34bp 58bp 554bp 738bp,clip,width=1\textwidth]{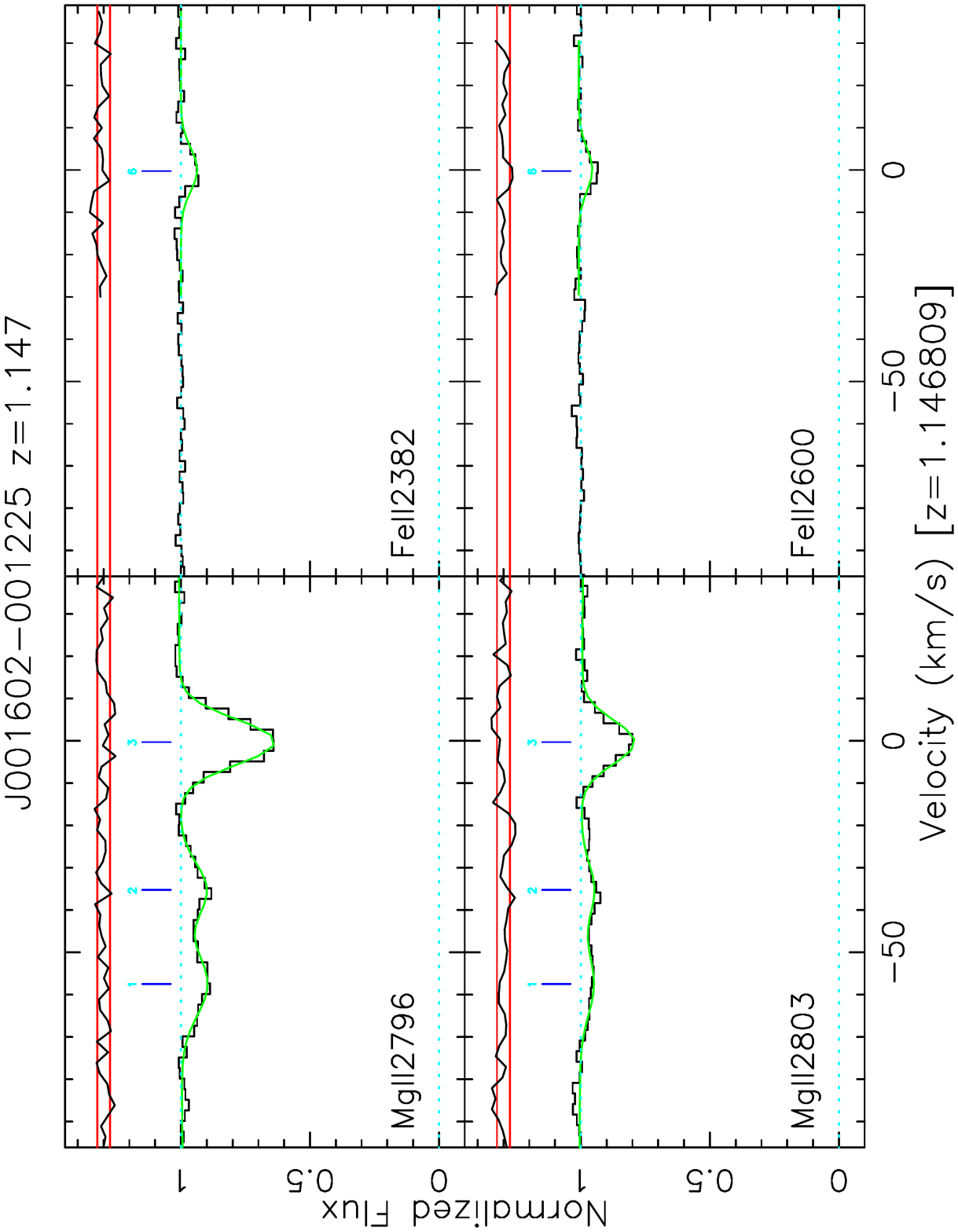}
\par\end{centering}

\caption[Fit for the $z=1.147$ absorber toward J001602$-$001225]{Many-multiplet fit for the $z=1.147$ absorber toward J001602$-$001225.}
\end{figure}
\begin{figure}[H]
\noindent \begin{centering}
\includegraphics[bb=34bp 58bp 554bp 738bp,clip,width=1\textwidth]{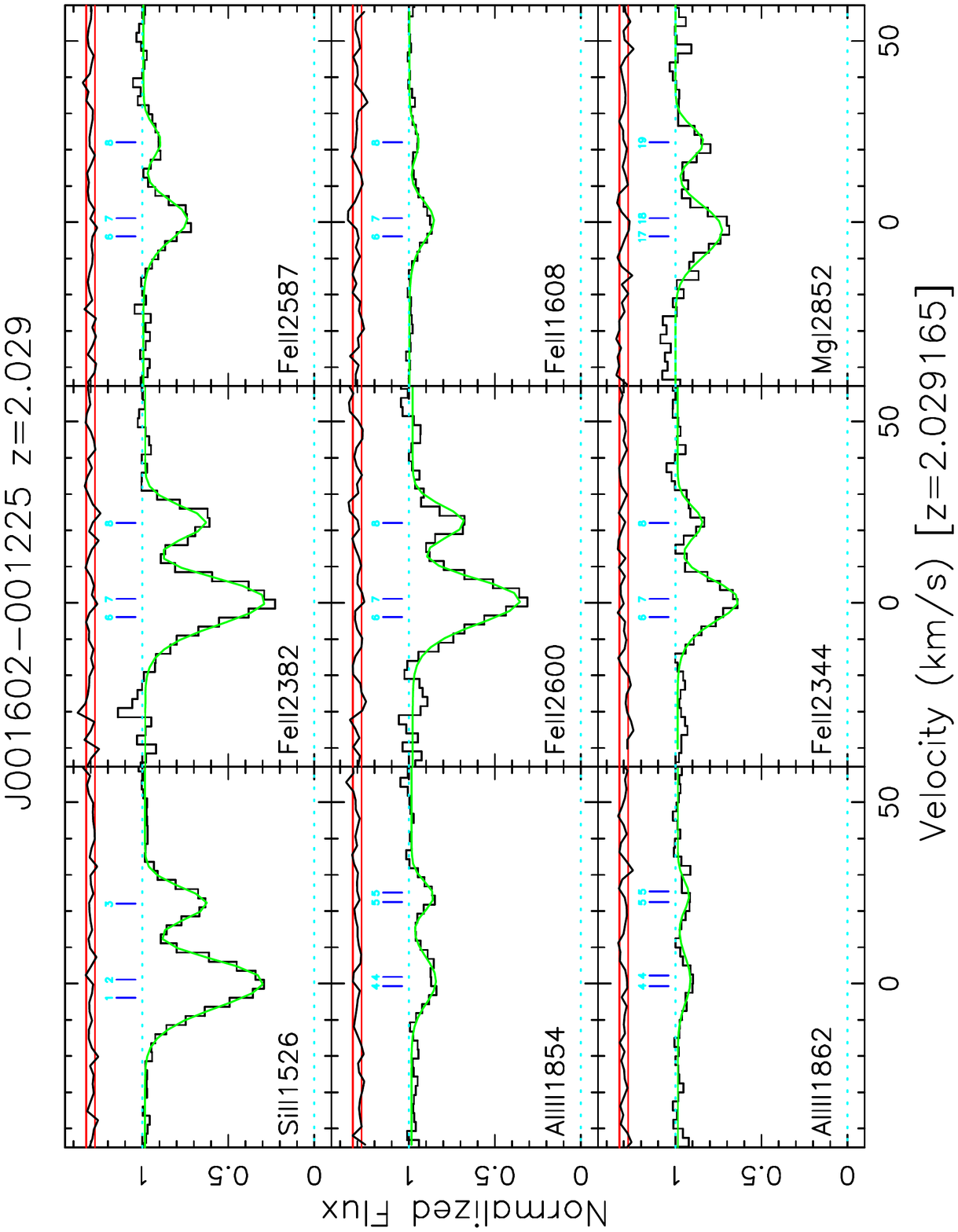}
\par\end{centering}

\caption[Fit for the $z=2.029$ absorber toward J001602$-$001225]{Many-multiplet fit for the $z=2.029$ absorber toward J001602$-$001225.}
\end{figure}
\begin{figure}[H]
\noindent \begin{centering}
\includegraphics[bb=34bp 58bp 554bp 738bp,clip,width=1\textwidth]{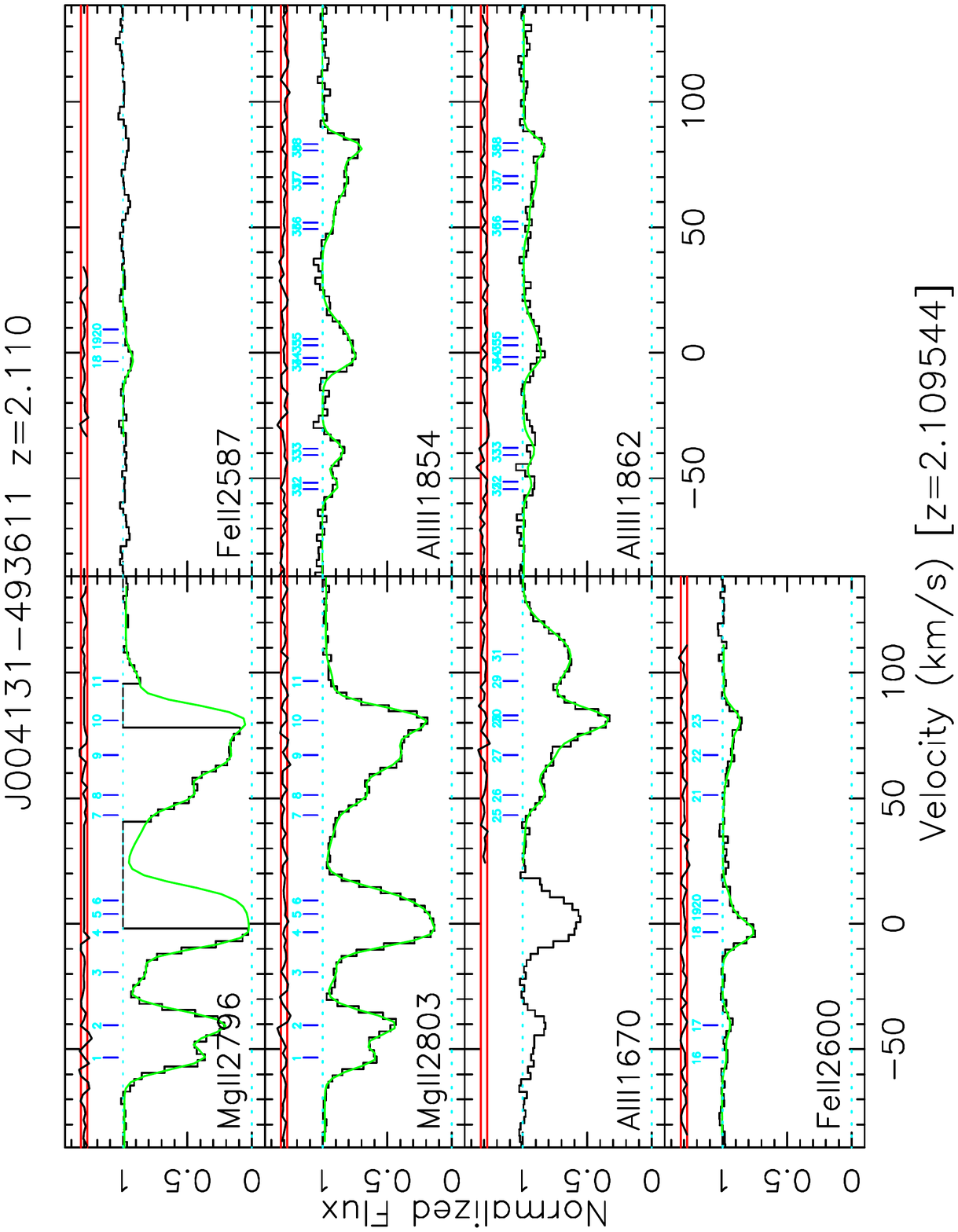}
\par\end{centering}

\caption[Fit for the $z=2.110$ absorber toward J004131$-$493611]{Many-multiplet fit for the $z=2.110$ absorber toward J004131$-$493611.}
\end{figure}
\begin{figure}[H]
\noindent \begin{centering}
\includegraphics[bb=34bp 58bp 554bp 738bp,clip,width=1\textwidth]{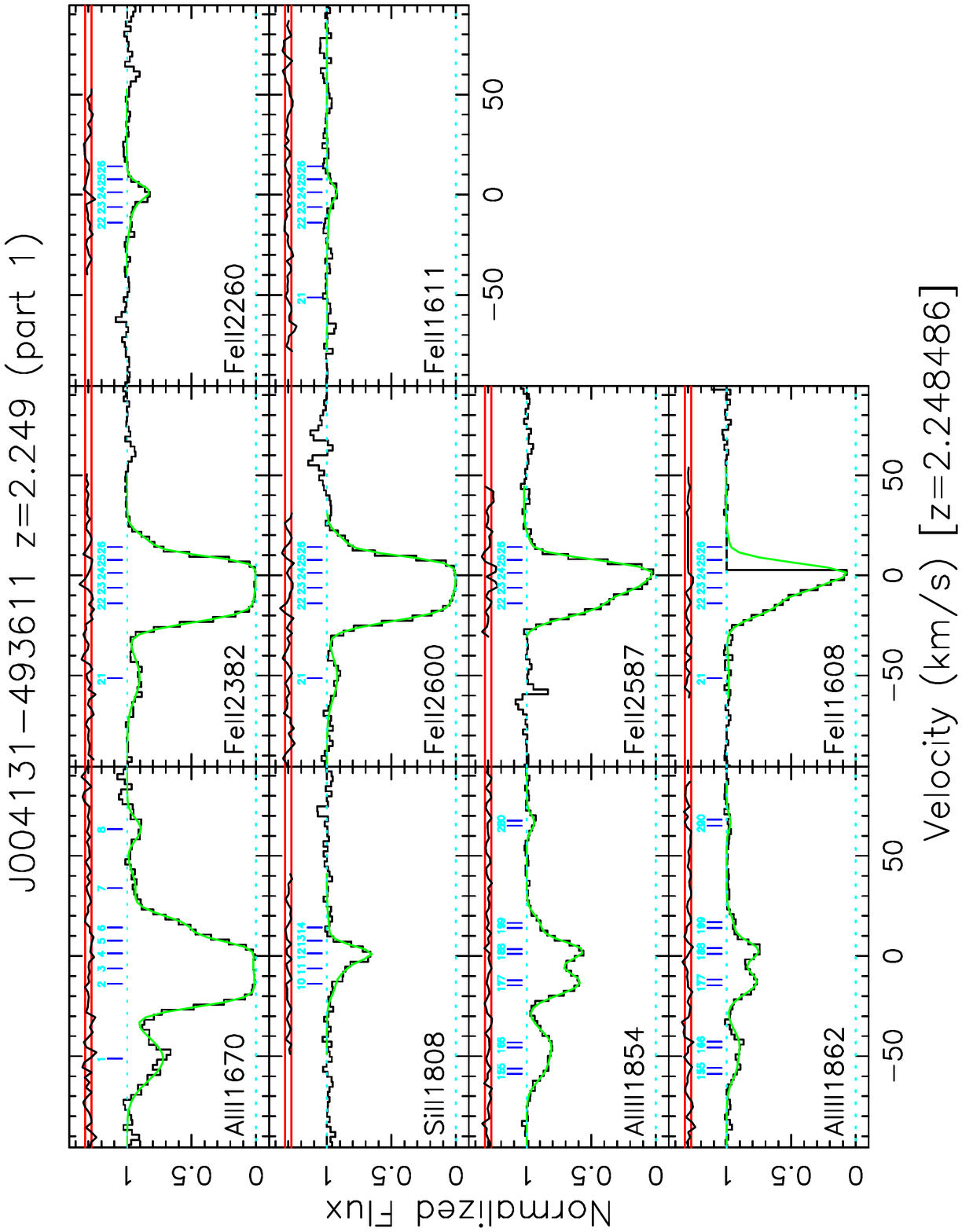}
\par\end{centering}

\caption[Fit for the $z=2.249$ absorber toward J004131$-$493611 (part 1)]{Many-multiplet fit for the $z=2.249$ absorber toward J004131$-$493611 (part 1).}
\end{figure}

\begin{figure}[H]
\noindent \begin{centering}
\includegraphics[bb=34bp 58bp 554bp 738bp,clip,width=1\textwidth]{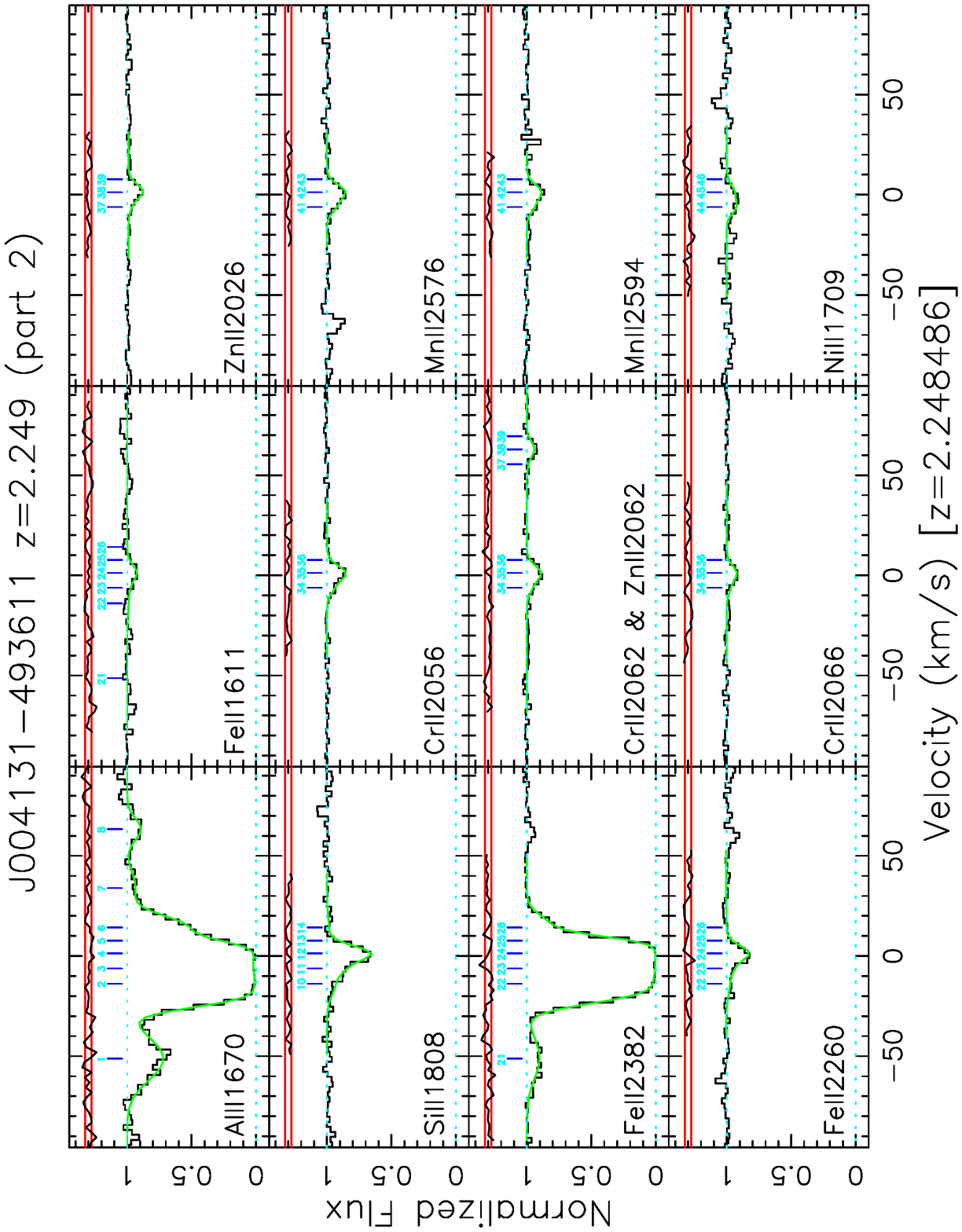}
\par\end{centering}

\caption[Fit for the $z=2.249$ absorber toward J004131$-$493611 (part 2)]{Many-multiplet fit for the $z=2.249$ absorber toward J004131$-$493611 (part 2).}
\end{figure}

\begin{figure}[H]
\noindent \begin{centering}
\includegraphics[bb=34bp 58bp 554bp 738bp,clip,width=1\textwidth]{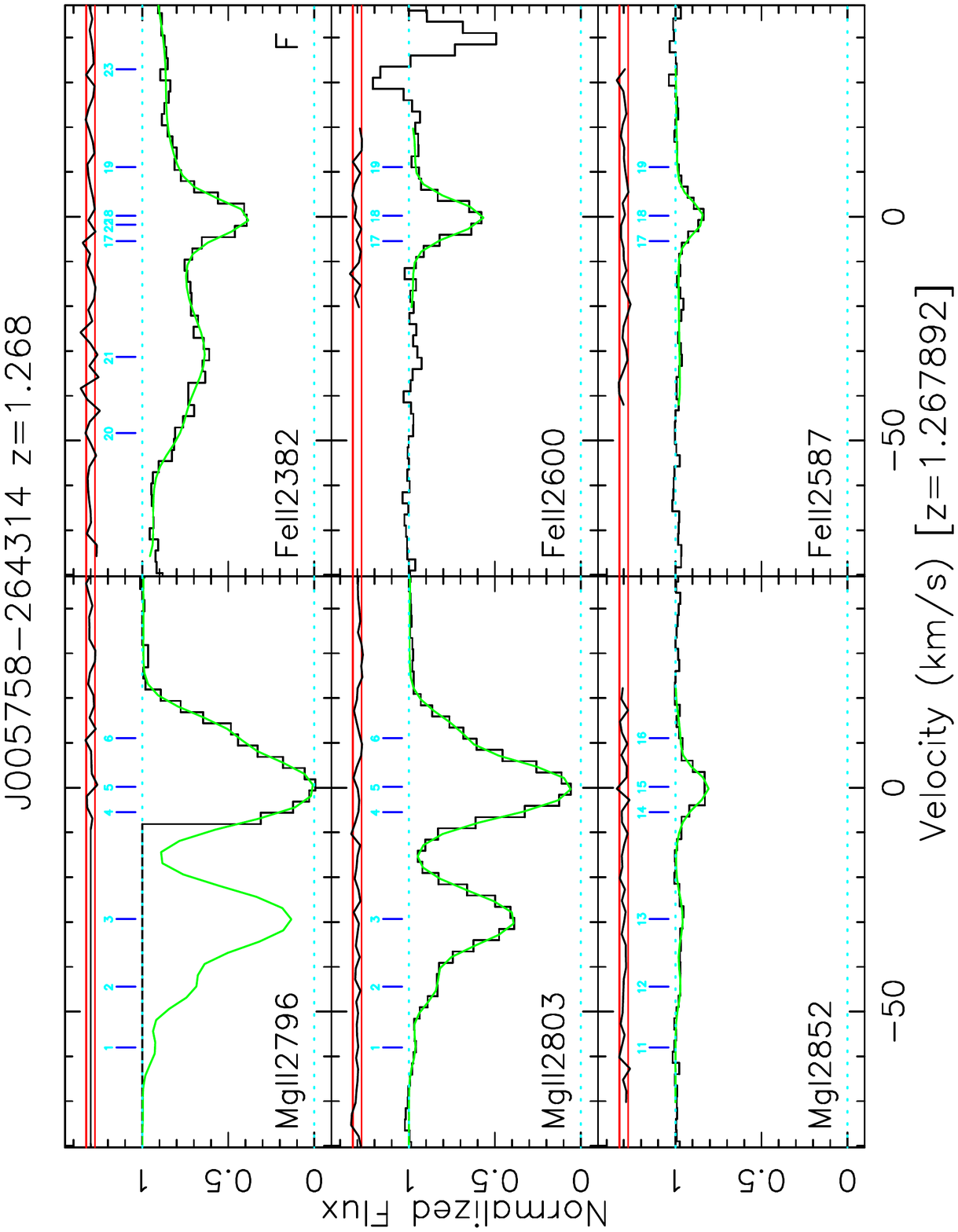}
\par\end{centering}

\caption[Fit for the $z=1.268$ absorber toward J005758$-$264314]{Many-multiplet fit for the $z=1.268$ absorber toward J005758$-$264314.\label{appfig:J005758-264314-z1.268}}
\end{figure}

\begin{figure}[H]
\noindent \begin{centering}
\includegraphics[bb=34bp 58bp 554bp 738bp,clip,width=1\textwidth]{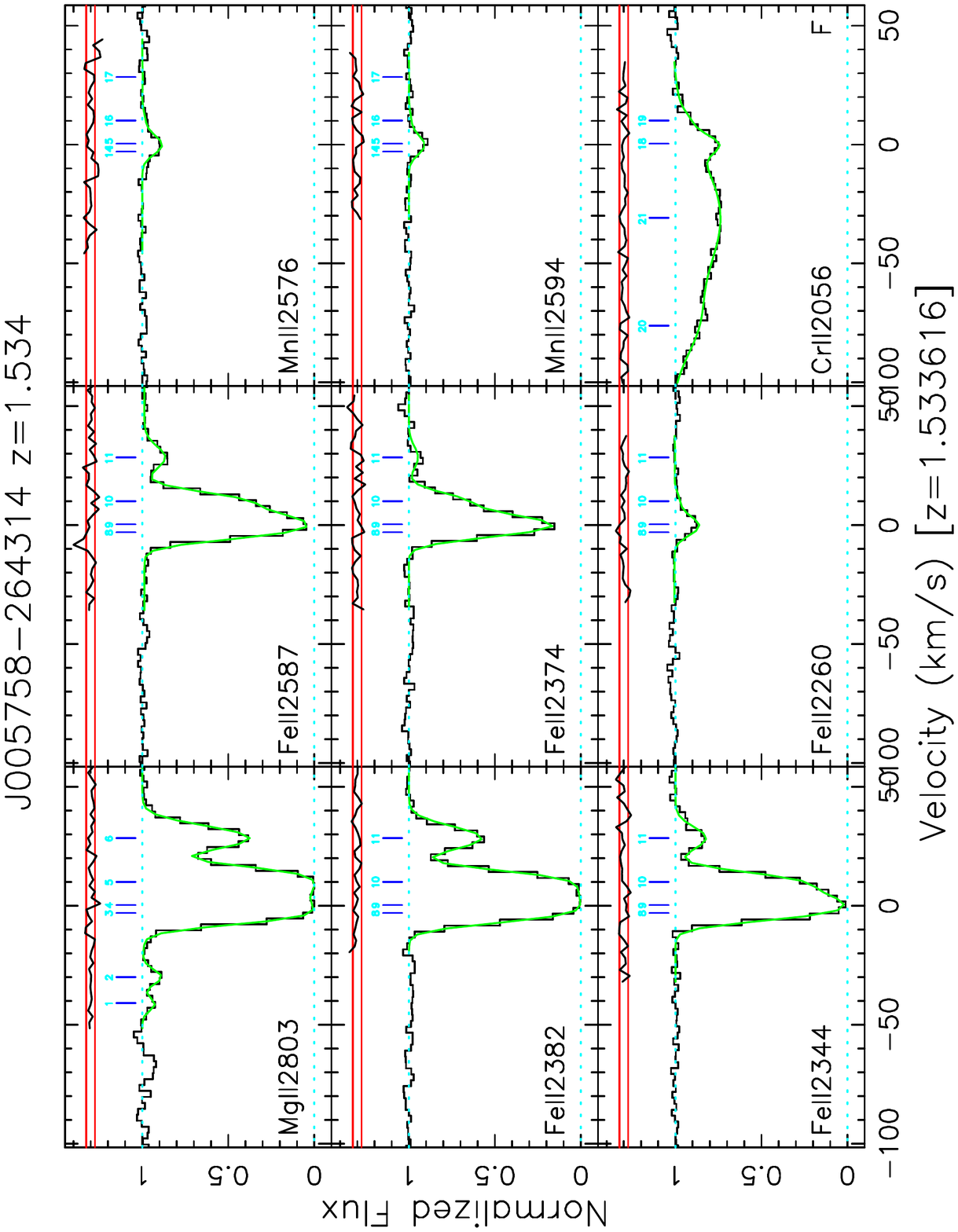}
\par\end{centering}

\caption[Fit for the $z=1.534$ absorber toward J005758$-$264314]{Many-multiplet fit for the $z=1.534$ absorber toward J005758$-$264314.}
\end{figure}
\begin{figure}[H]
\noindent \begin{centering}
\includegraphics[bb=34bp 58bp 554bp 738bp,clip,width=1\textwidth]{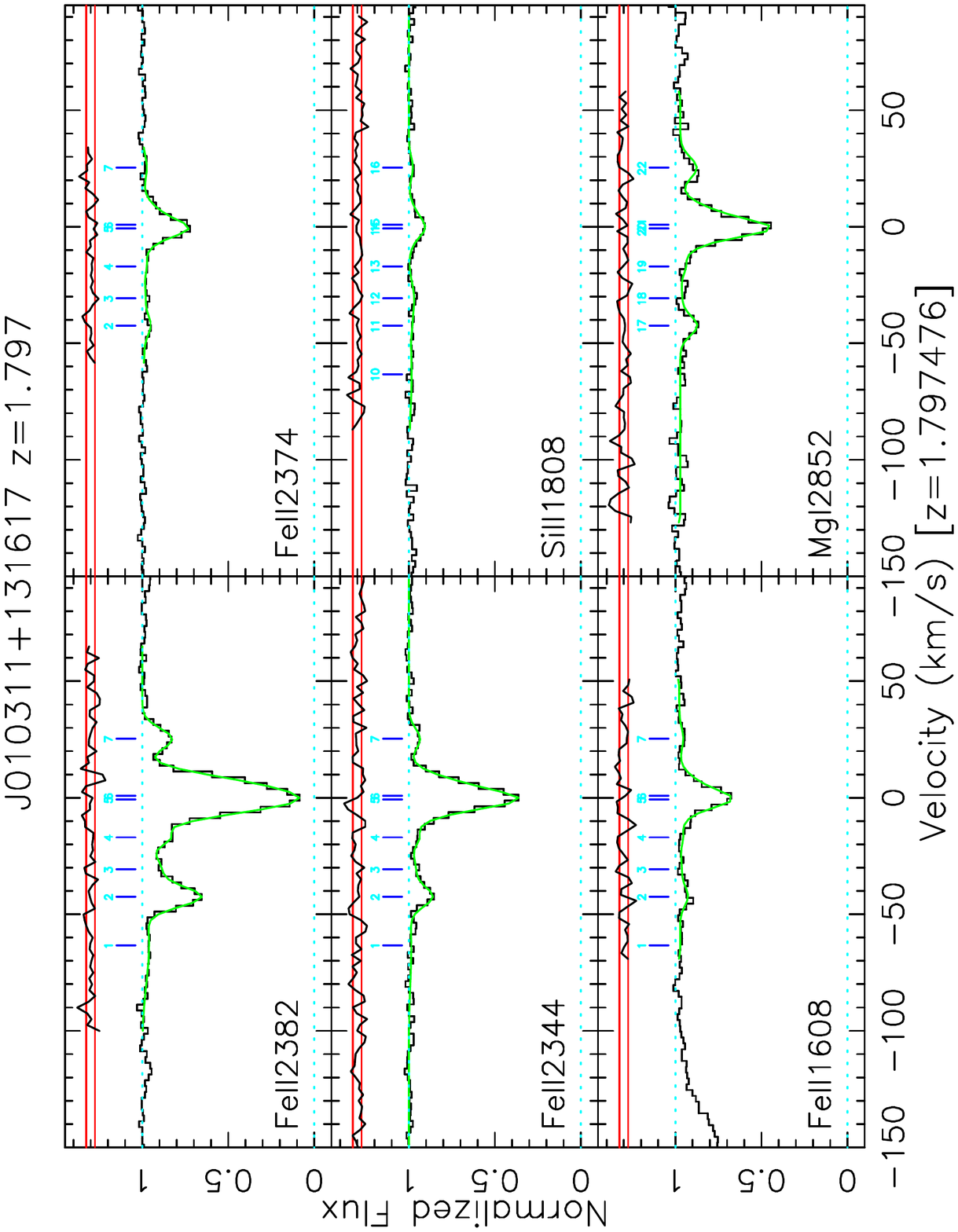}
\par\end{centering}

\caption[Fit for the $z=1.797$ absorber toward J010311+131617]{Many-multiplet fit for the $z=1.797$ absorber toward J010311+131617.}
\end{figure}
\begin{figure}[H]
\noindent \begin{centering}
\includegraphics[bb=34bp 58bp 554bp 738bp,clip,width=1\textwidth]{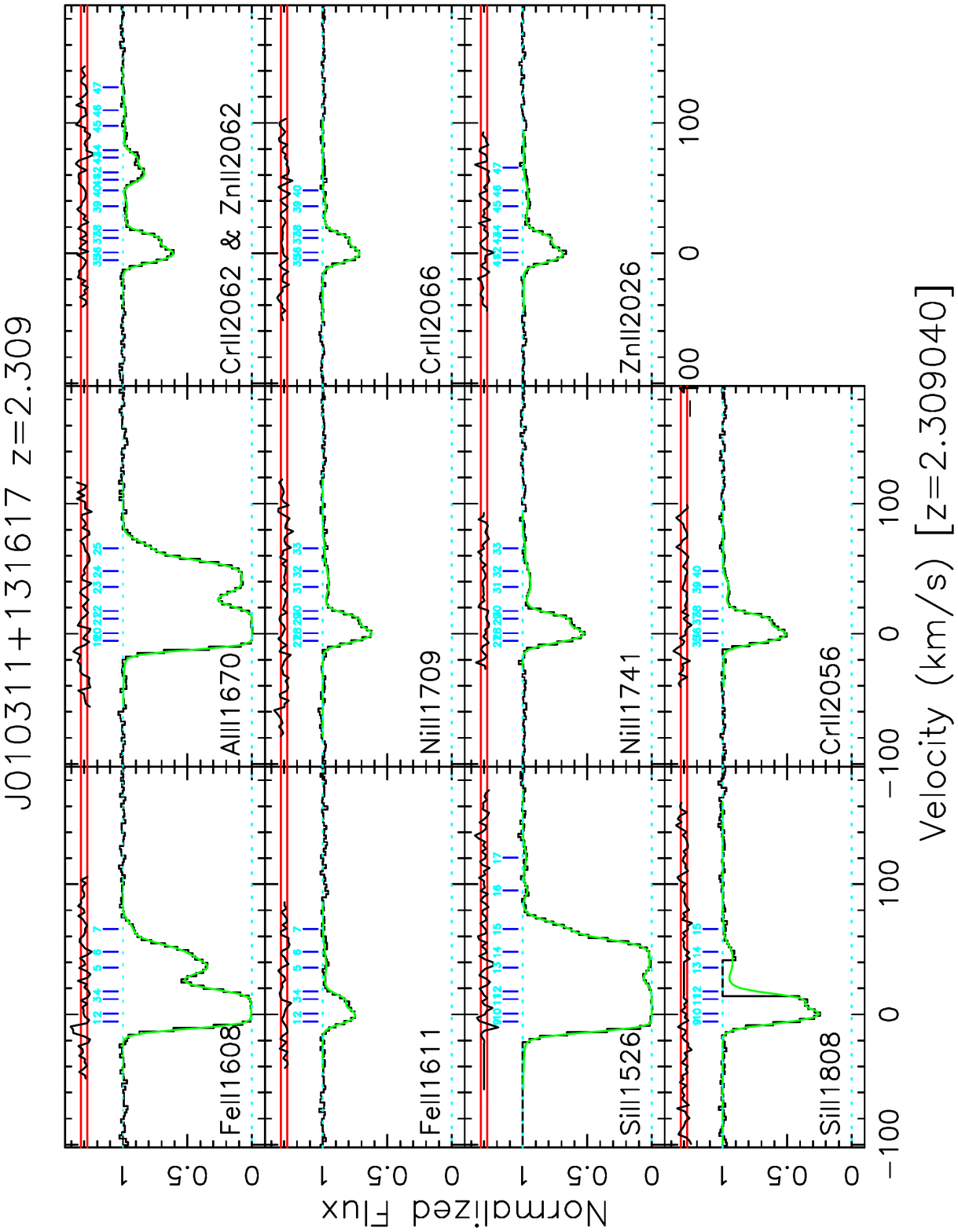}
\par\end{centering}

\caption[Fit for the $z=2.309$ absorber toward J010311+131617]{Many-multiplet fit for the $z=2.309$ absorber toward J010311+131617.}
\end{figure}
\begin{figure}[H]
\noindent \begin{centering}
\includegraphics[bb=34bp 58bp 554bp 738bp,clip,width=1\textwidth]{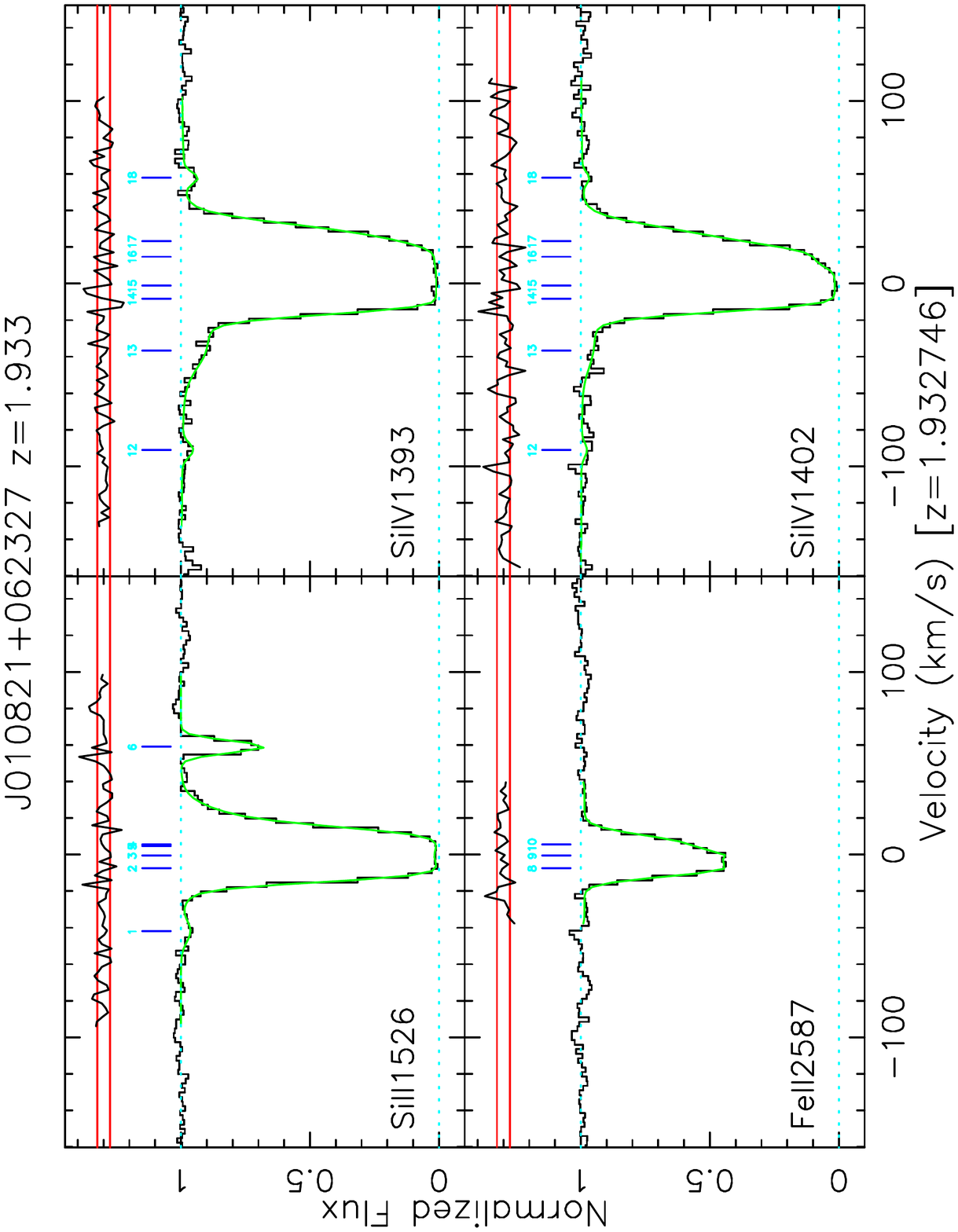}
\par\end{centering}

\caption[Fit for the $z=1.933$ absorber toward J010821+062327]{Many-multiplet fit for the $z=1.933$ absorber toward J010821+062327.}
\end{figure}
\begin{figure}[H]
\noindent \begin{centering}
\includegraphics[bb=34bp 58bp 554bp 738bp,clip,width=1\textwidth]{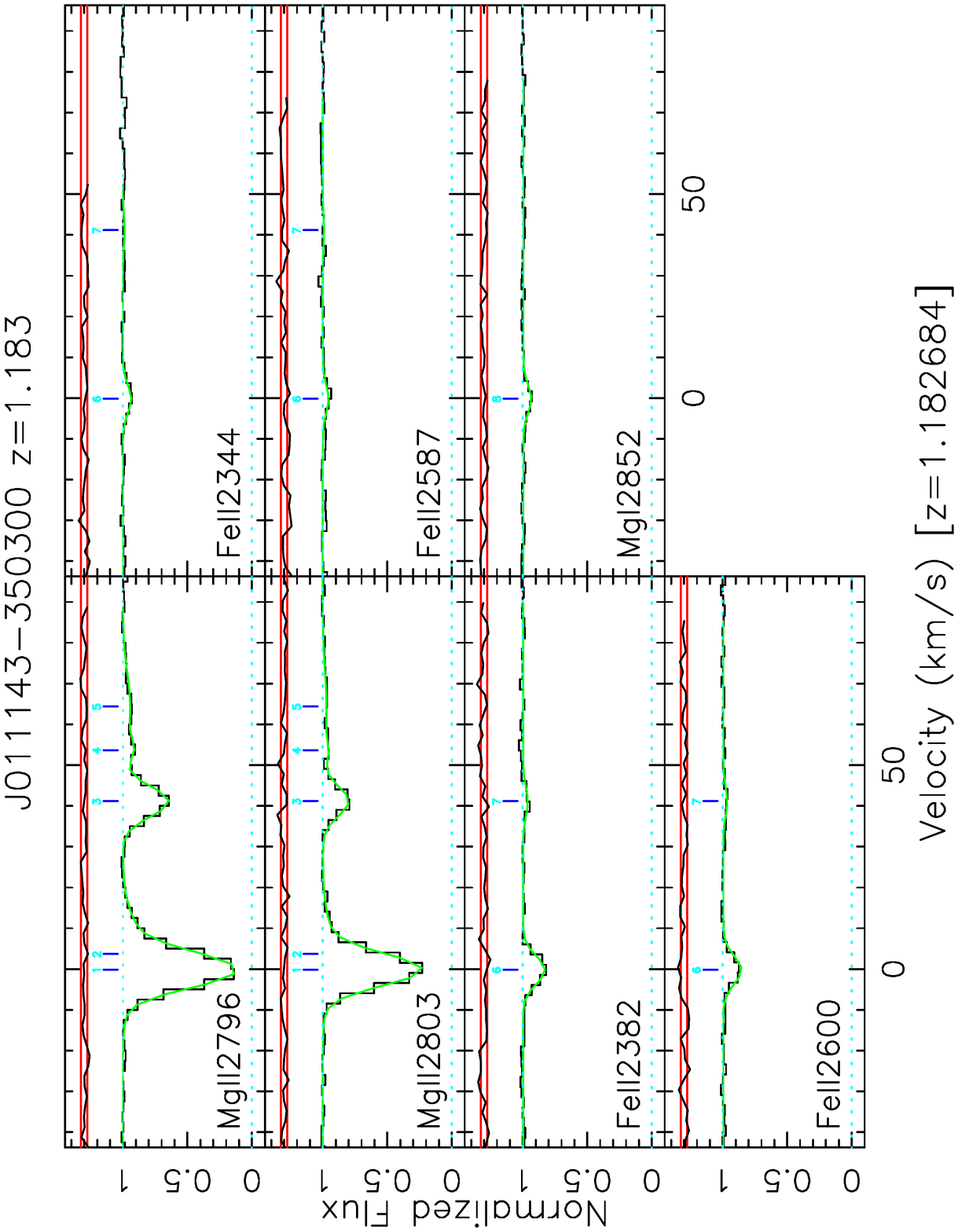}
\par\end{centering}

\caption[Fit for the $z=1.183$ absorber toward J011143$-$350300]{Many-multiplet fit for the $z=1.183$ absorber toward J011143$-$350300.}
\end{figure}
\begin{figure}[H]
\noindent \begin{centering}
\includegraphics[bb=34bp 58bp 554bp 738bp,clip,width=1\textwidth]{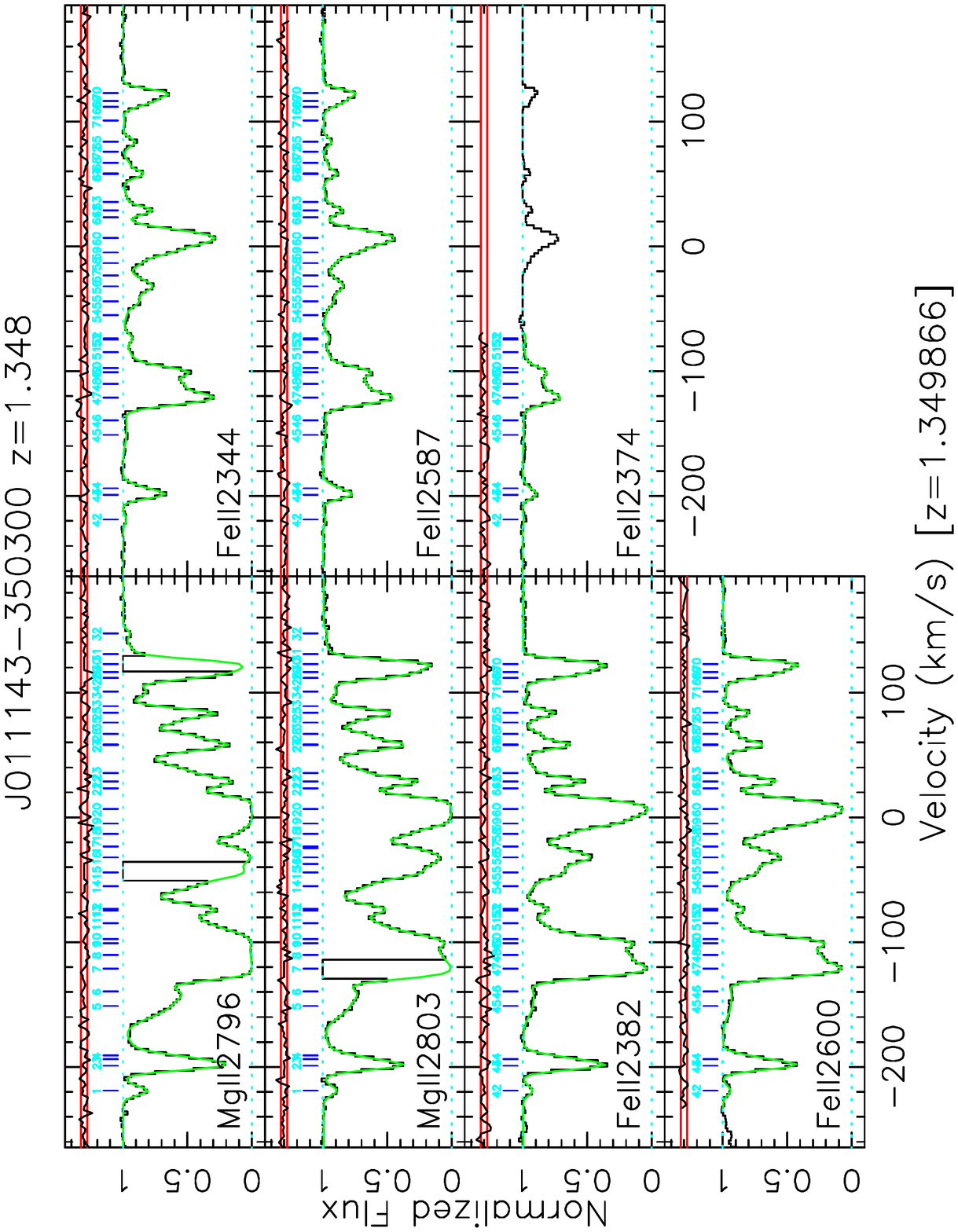}
\par\end{centering}

\caption[Fit for the $z=1.348$ absorber toward J011143$-$350300]{Many-multiplet fit for the $z=1.348$ absorber toward J011143$-$350300.}
\end{figure}
\begin{figure}[H]
\noindent \begin{centering}
\includegraphics[bb=34bp 58bp 554bp 738bp,clip,width=1\textwidth]{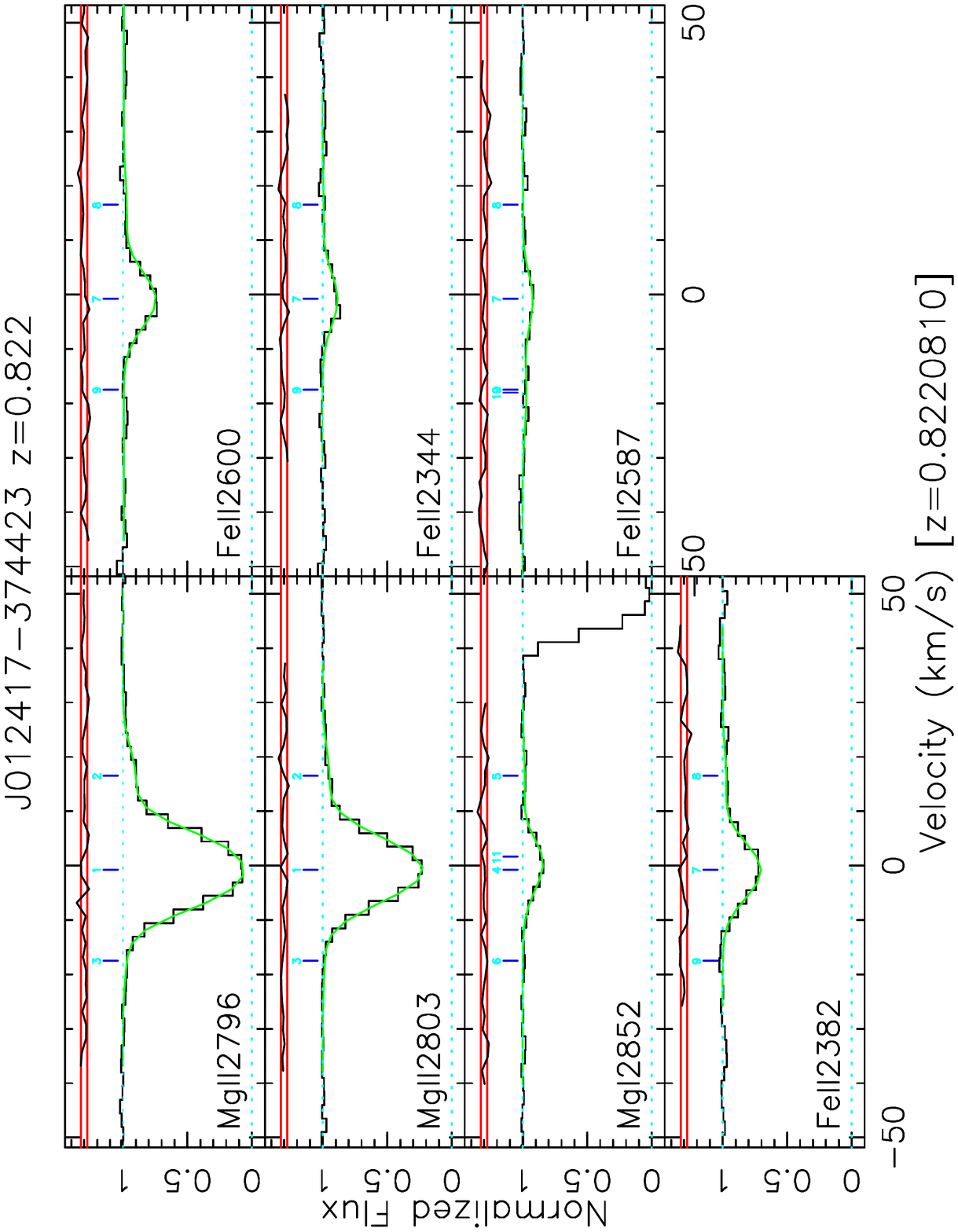}
\par\end{centering}

\caption[Fit for the $z=0.822$ absorber toward J012417$-$374423]{Many-multiplet fit for the $z=0.822$ absorber toward J012417$-$374423.}
\end{figure}
\begin{figure}[H]
\noindent \begin{centering}
\includegraphics[bb=34bp 58bp 554bp 738bp,clip,width=1\textwidth]{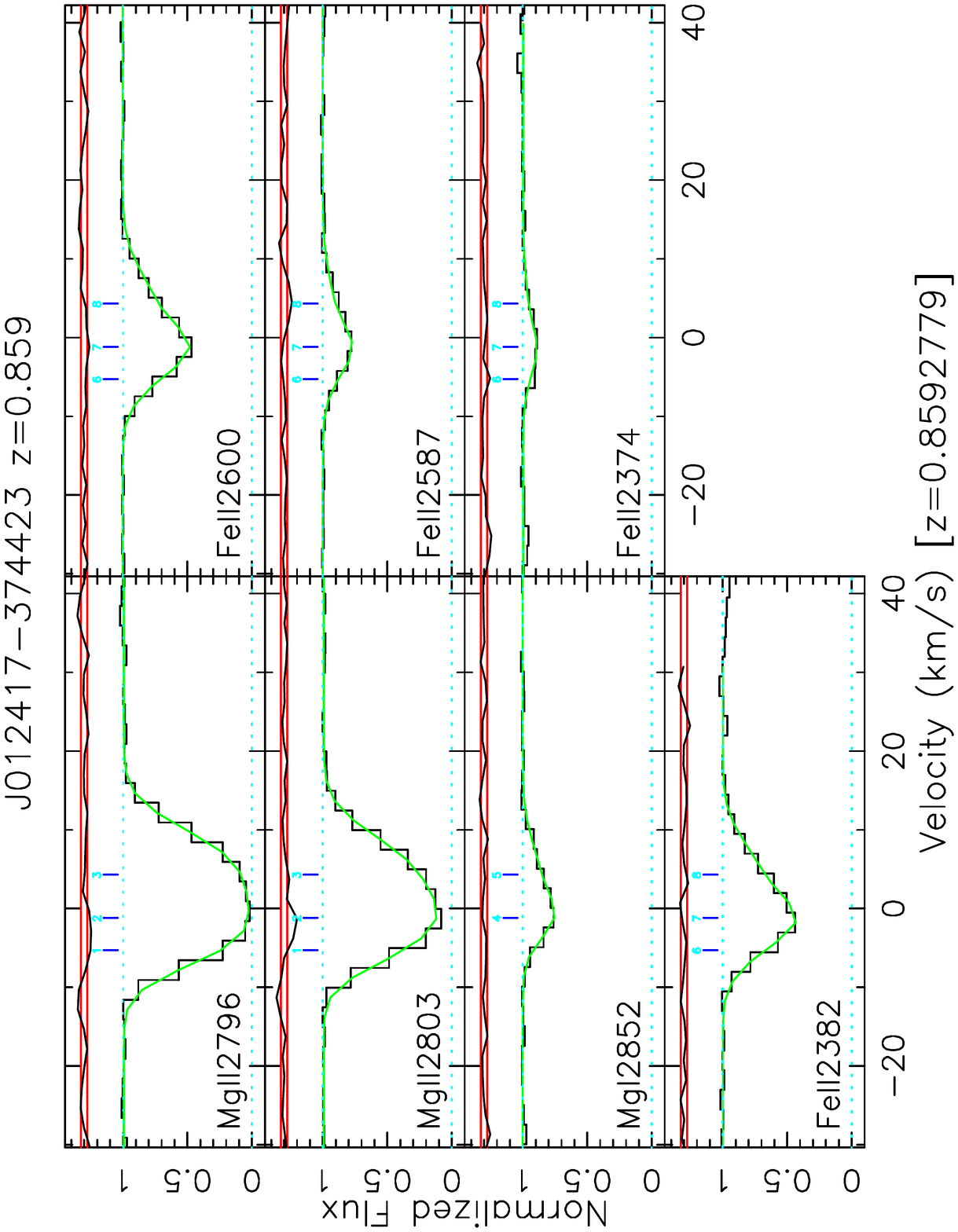}
\par\end{centering}

\caption[Fit for the $z=0.859$ absorber toward J012417$-$374423]{Many-multiplet fit for the $z=0.859$ absorber toward J012417$-$374423.}
\end{figure}
\begin{figure}[H]
\noindent \begin{centering}
\includegraphics[bb=34bp 58bp 554bp 738bp,clip,width=1\textwidth]{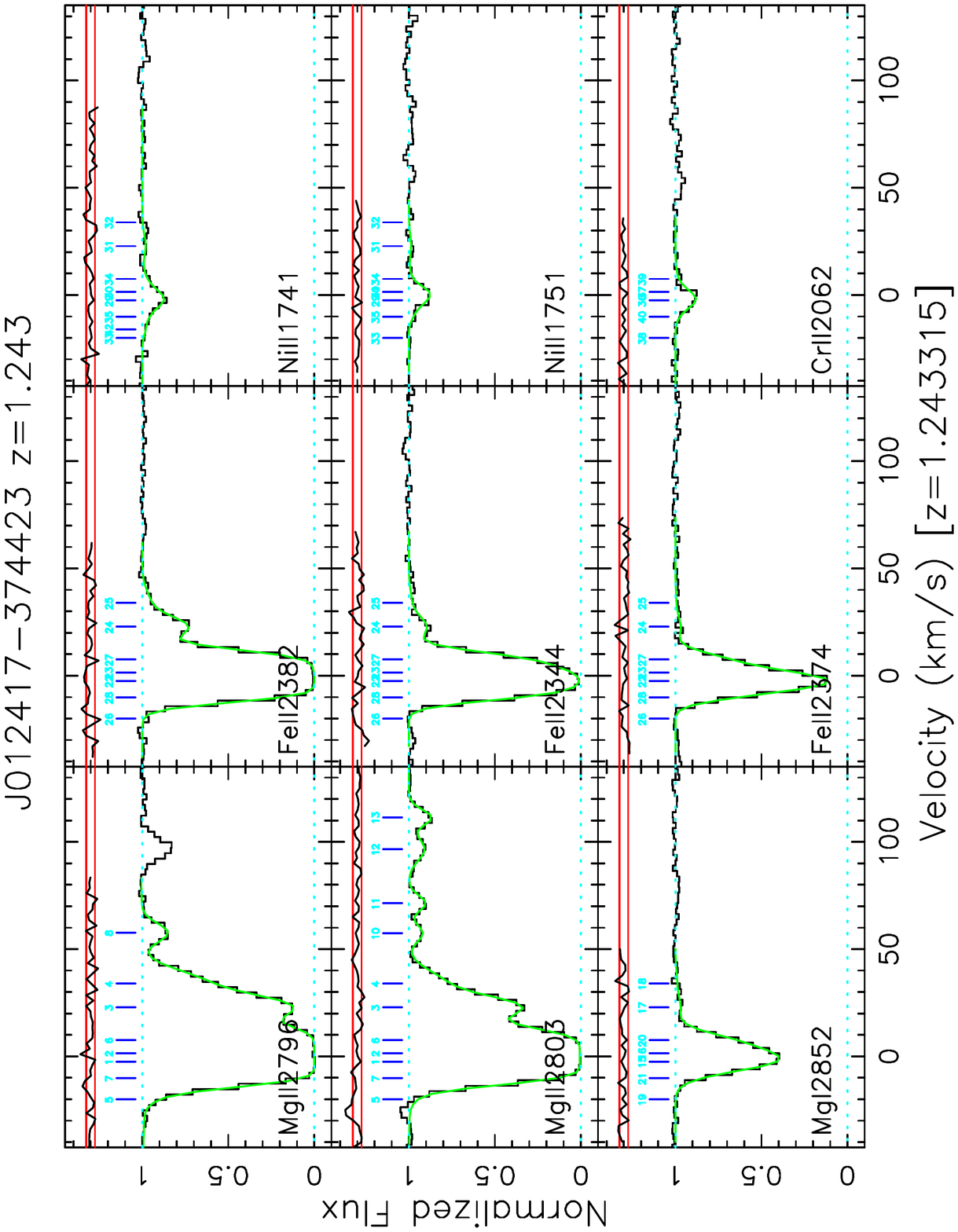}
\par\end{centering}

\caption[Fit for the $z=1.243$ absorber toward J012417$-$374423]{Many-multiplet fit for the $z=1.243$ absorber toward J012417$-$374423.}
\end{figure}
\begin{figure}[H]
\noindent \begin{centering}
\includegraphics[bb=34bp 58bp 554bp 738bp,clip,width=1\textwidth]{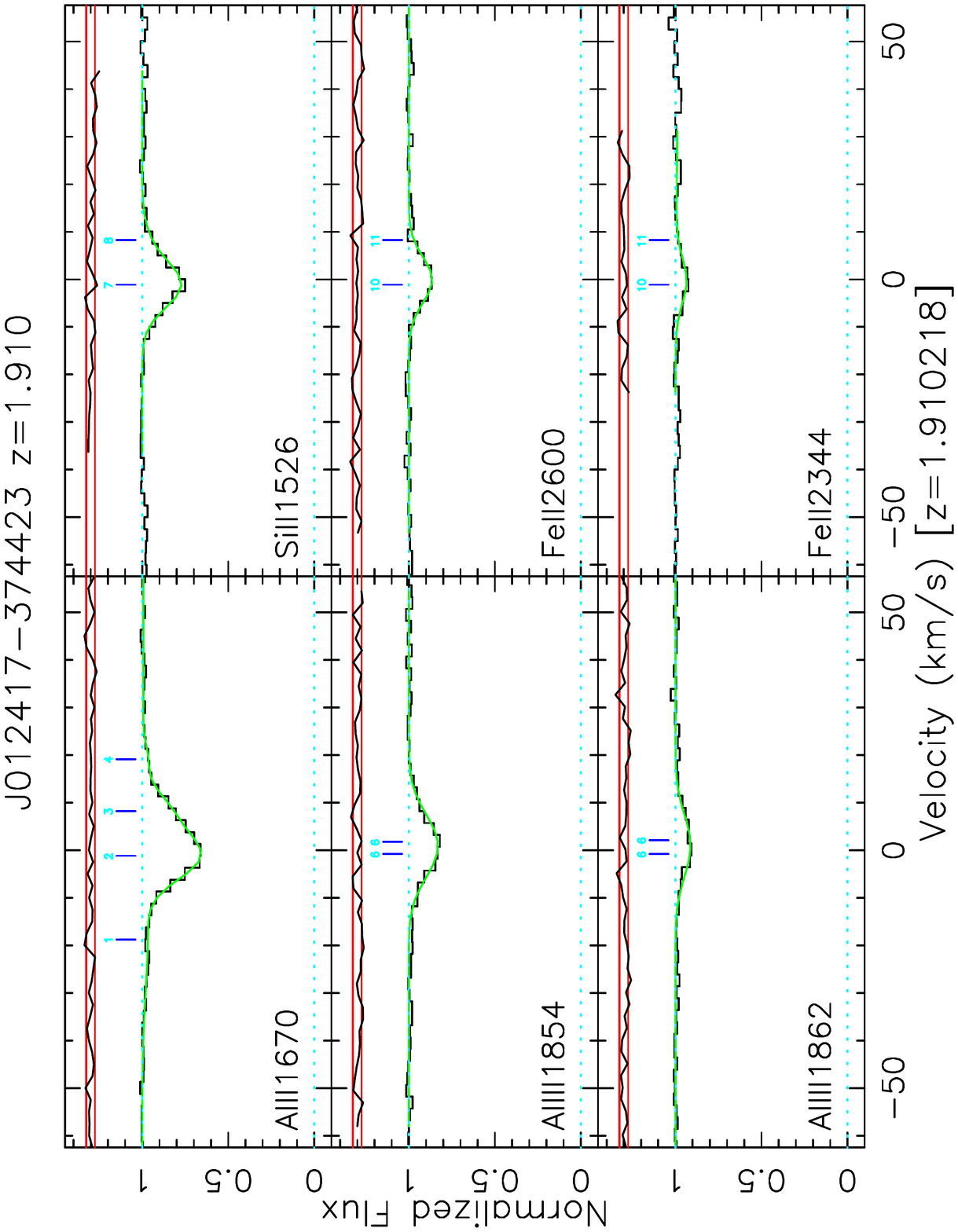}
\par\end{centering}

\caption[Fit for the $z=1.910$ absorber toward J012417$-$374423]{Many-multiplet fit for the $z=1.910$ absorber toward J012417$-$374423.}
\end{figure}
\begin{figure}[H]
\noindent \begin{centering}
\includegraphics[bb=34bp 58bp 554bp 738bp,clip,width=1\textwidth]{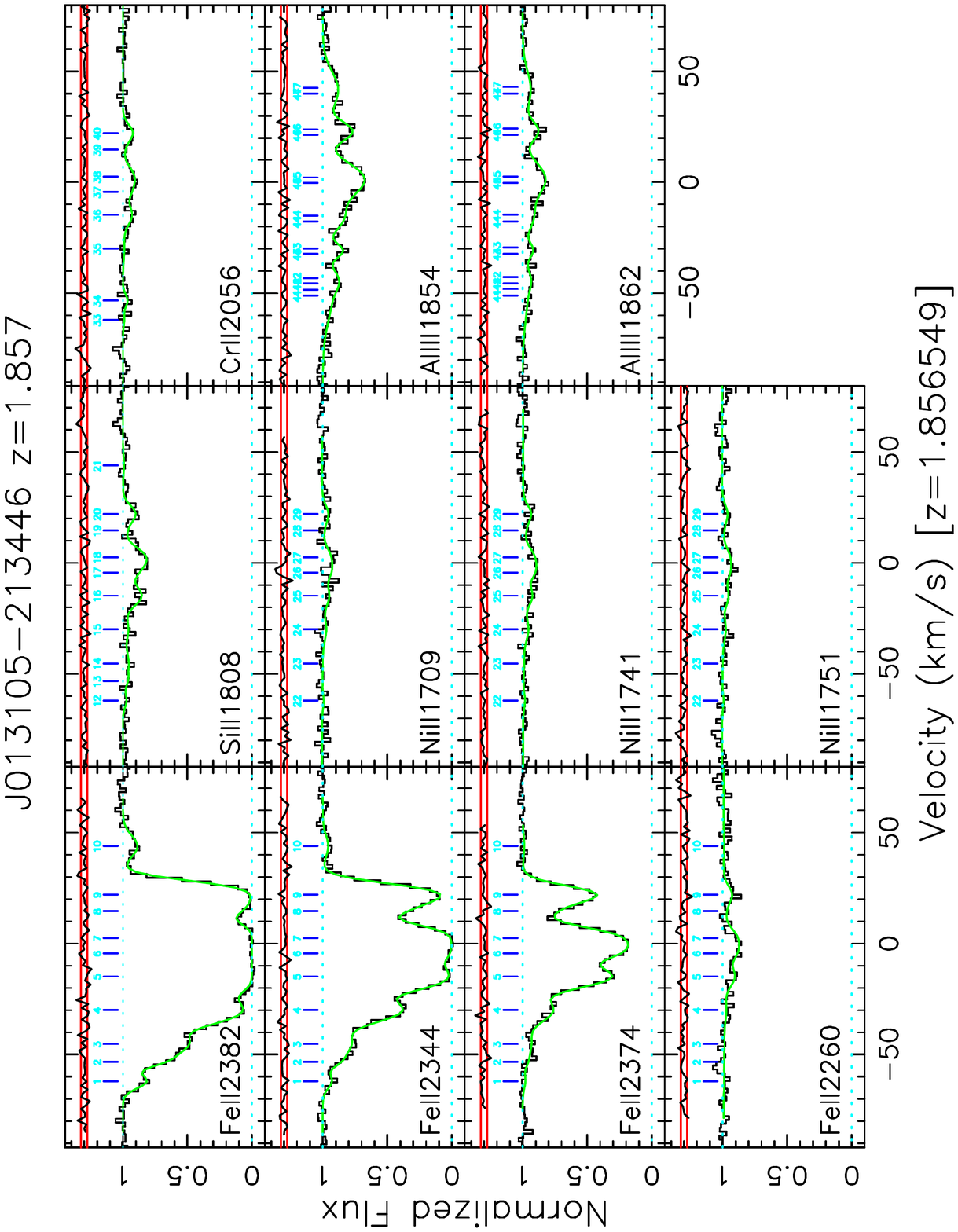}
\par\end{centering}

\caption[Fit for the $z=1.857$ absorber toward J013105$-$213446]{Many-multiplet fit for the $z=1.857$ absorber toward J013105$-$213446.}
\end{figure}
\begin{figure}[H]
\noindent \begin{centering}
\includegraphics[bb=34bp 58bp 554bp 738bp,clip,width=1\textwidth]{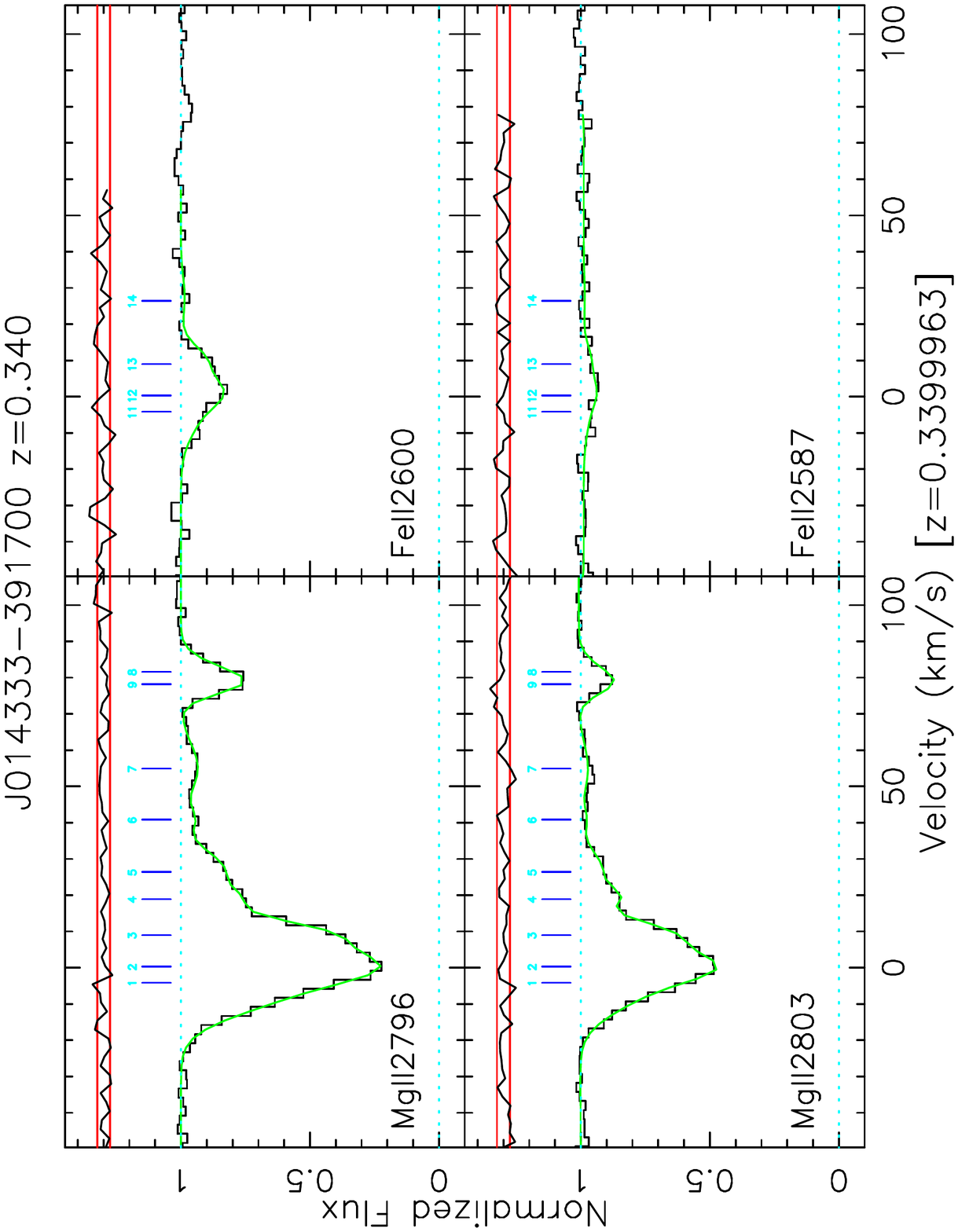}
\par\end{centering}

\caption[Fit for the $z=0.340$ absorber toward J014333$-$391700]{Many-multiplet fit for the $z=0.340$ absorber toward J014333$-$391700.}
\end{figure}
\begin{figure}[H]
\noindent \begin{centering}
\includegraphics[bb=34bp 58bp 554bp 738bp,clip,width=1\textwidth]{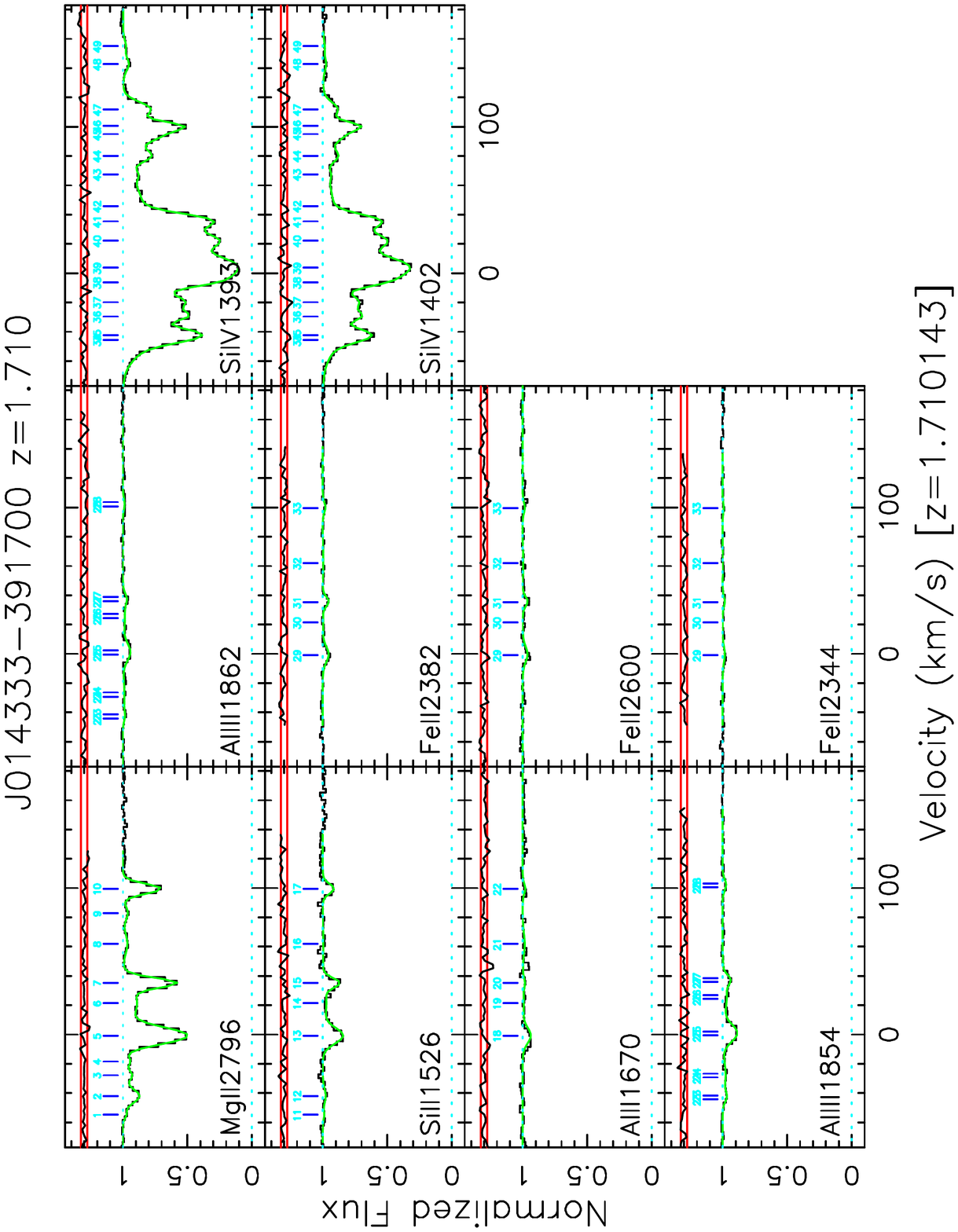}
\par\end{centering}

\caption[Fit for the $z=1.710$ absorber toward J014333$-$391700]{Many-multiplet fit for the $z=1.710$ absorber toward J014333$-$391700.}
\end{figure}
\begin{figure}[H]
\noindent \begin{centering}
\includegraphics[bb=34bp 58bp 554bp 738bp,clip,width=1\textwidth]{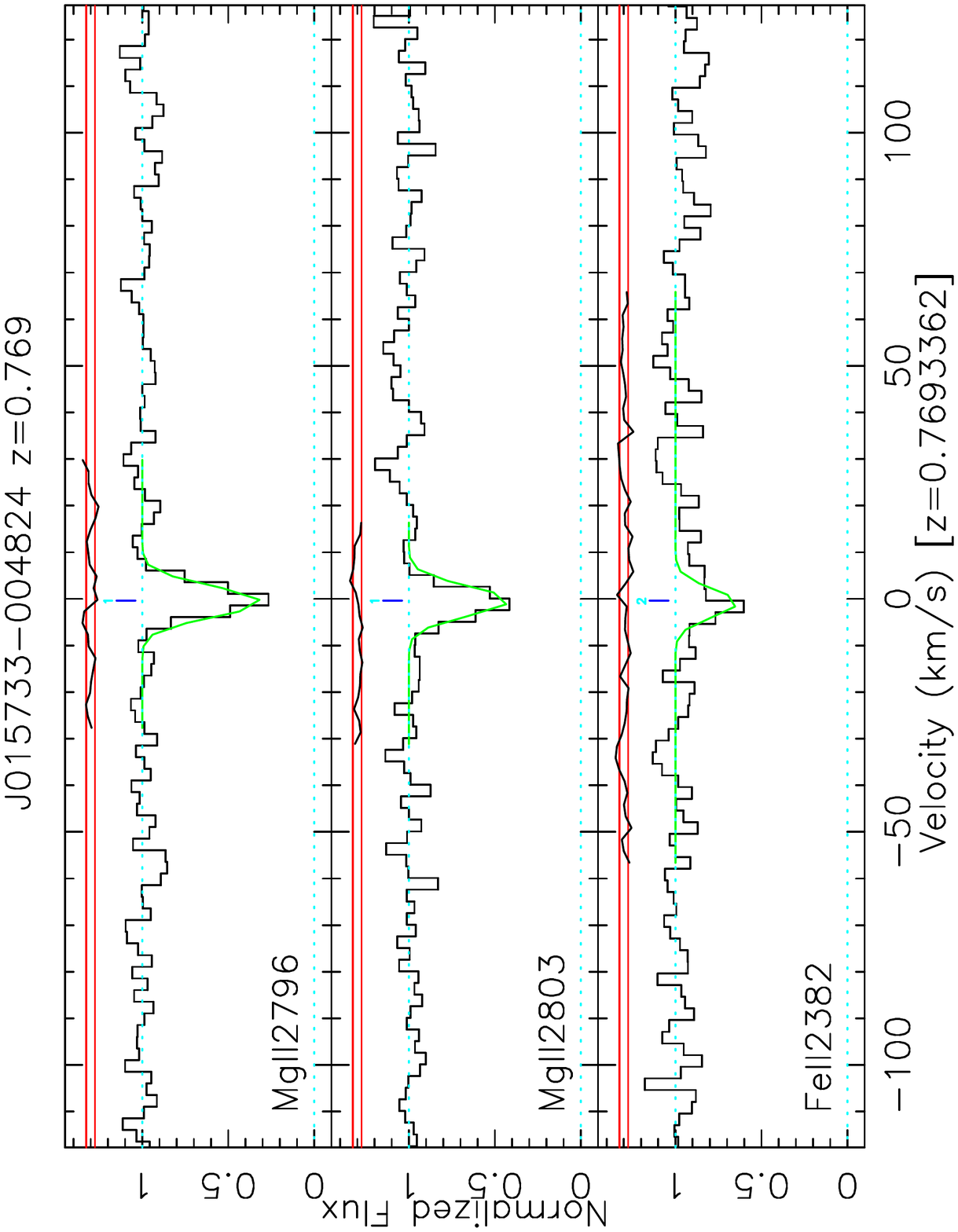}
\par\end{centering}

\caption[Fit for the $z=0.769$ absorber toward J015733$-$004824]{Many-multiplet fit for the $z=0.769$ absorber toward J015733$-$004824.}
\end{figure}
\begin{figure}[H]
\noindent \begin{centering}
\includegraphics[bb=34bp 58bp 554bp 738bp,clip,width=1\textwidth]{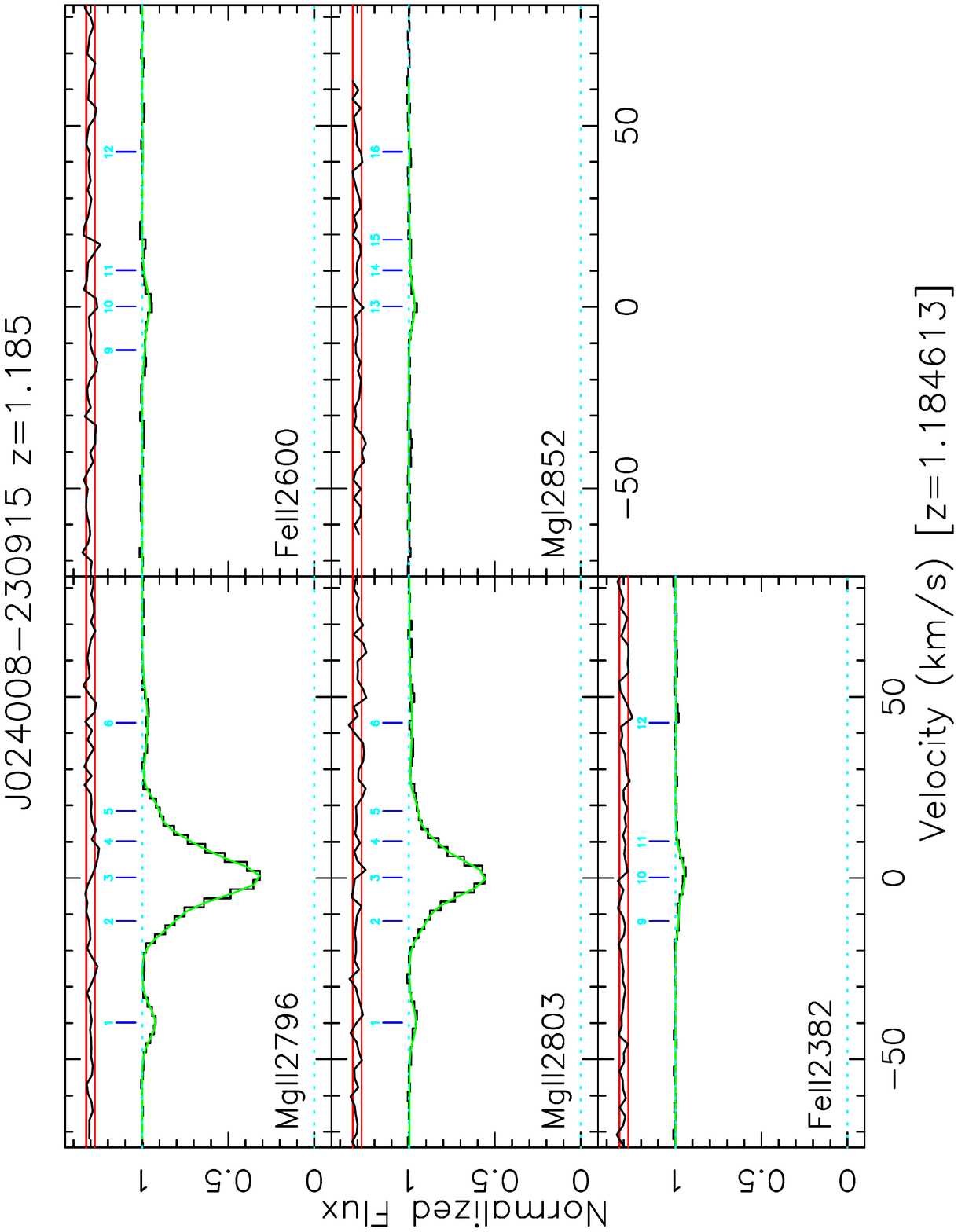}
\par\end{centering}

\caption[Fit for the $z=1.185$ absorber toward J024008$-$230915]{Many-multiplet fit for the $z=1.185$ absorber toward J024008$-$230915.}
\end{figure}
\begin{figure}[H]
\noindent \begin{centering}
\includegraphics[bb=34bp 58bp 554bp 738bp,clip,width=1\textwidth]{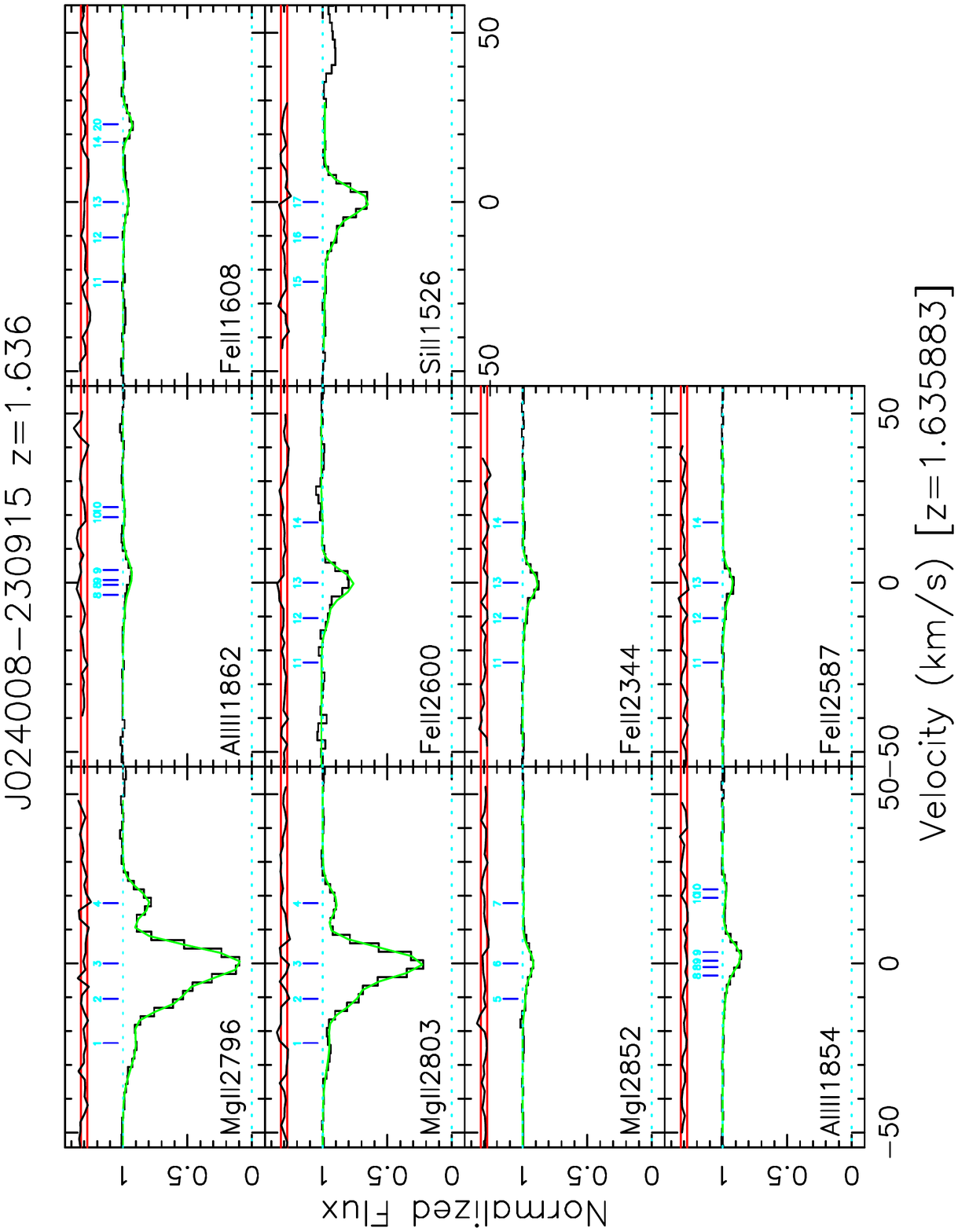}
\par\end{centering}

\caption[Fit for the $z=1.636$ absorber toward J024008$-$230915]{Many-multiplet fit for the $z=1.636$ absorber toward J024008$-$230915.}
\end{figure}
\begin{figure}[H]
\noindent \begin{centering}
\includegraphics[bb=34bp 58bp 554bp 738bp,clip,width=1\textwidth]{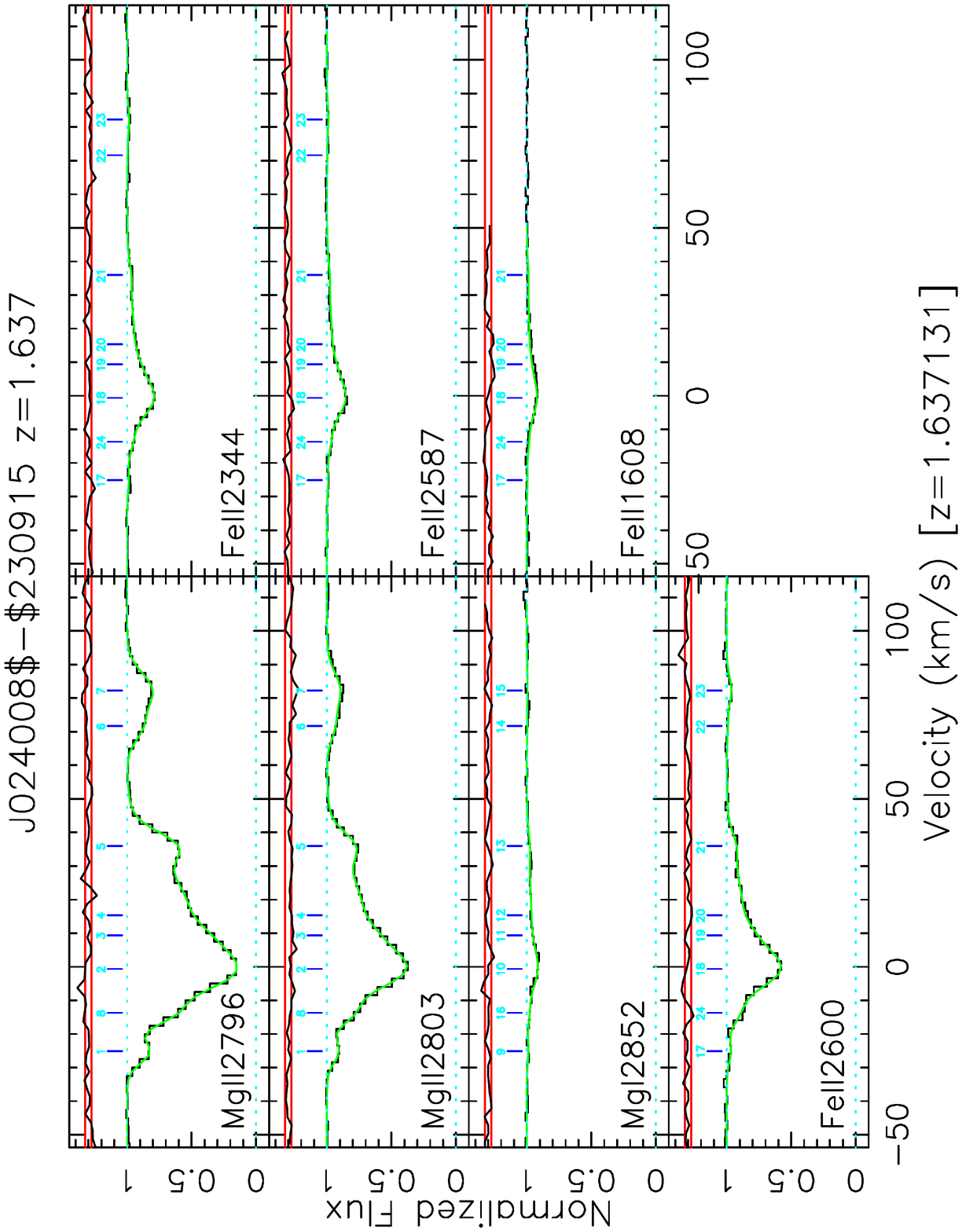}
\par\end{centering}

\caption[Fit for the $z=1.637$ absorber toward J024008$-$230915]{Many-multiplet fit for the $z=1.637$ absorber toward J024008$-$230915.}
\end{figure}
\begin{figure}[H]
\noindent \begin{centering}
\includegraphics[bb=34bp 58bp 554bp 738bp,clip,width=1\textwidth]{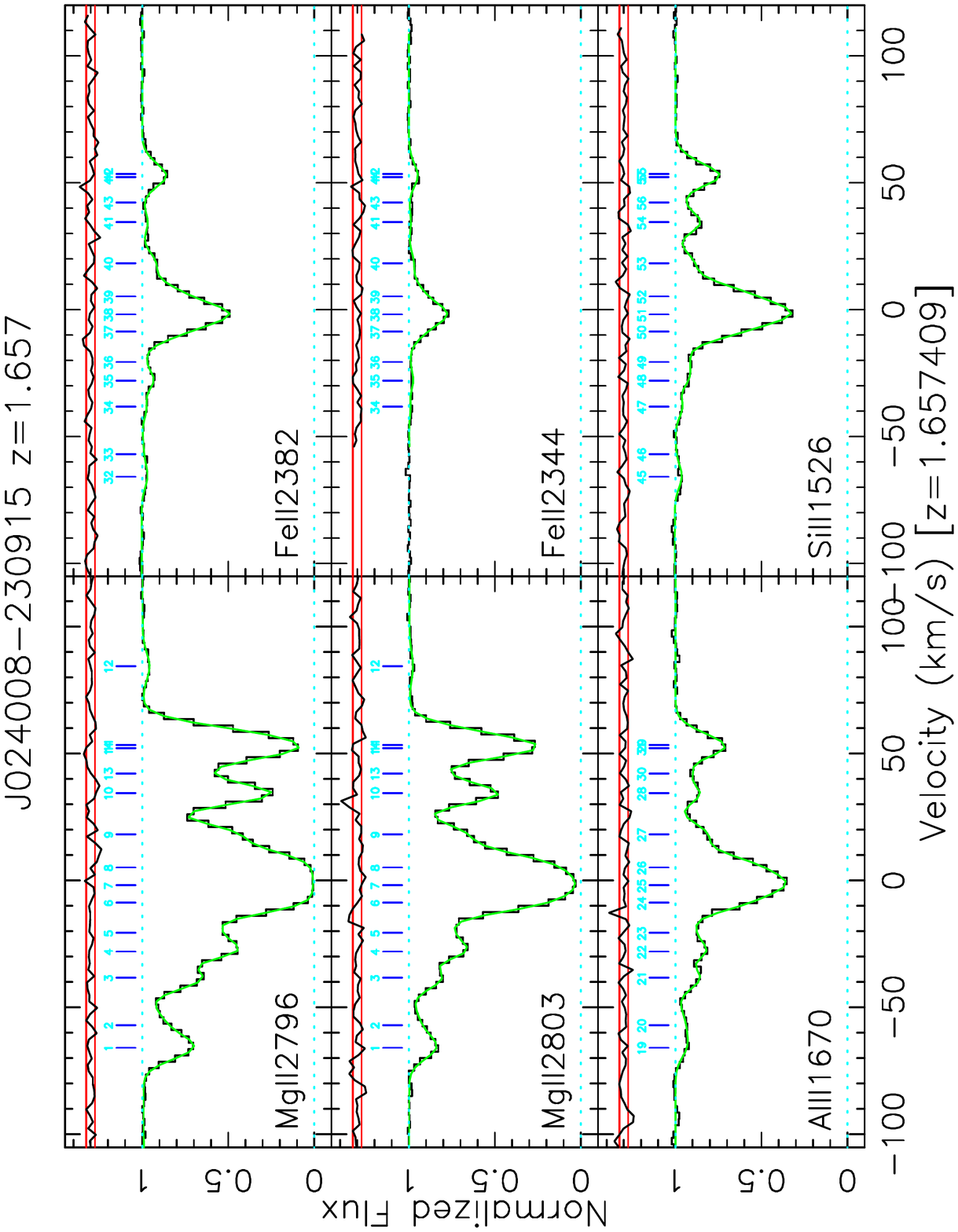}
\par\end{centering}

\caption[Fit for the $z=1.657$ absorber toward J024008$-$230915]{Many-multiplet fit for the $z=1.657$ absorber toward J024008$-$230915.}
\end{figure}
\begin{figure}[H]
\noindent \begin{centering}
\includegraphics[bb=34bp 58bp 554bp 738bp,clip,width=1\textwidth]{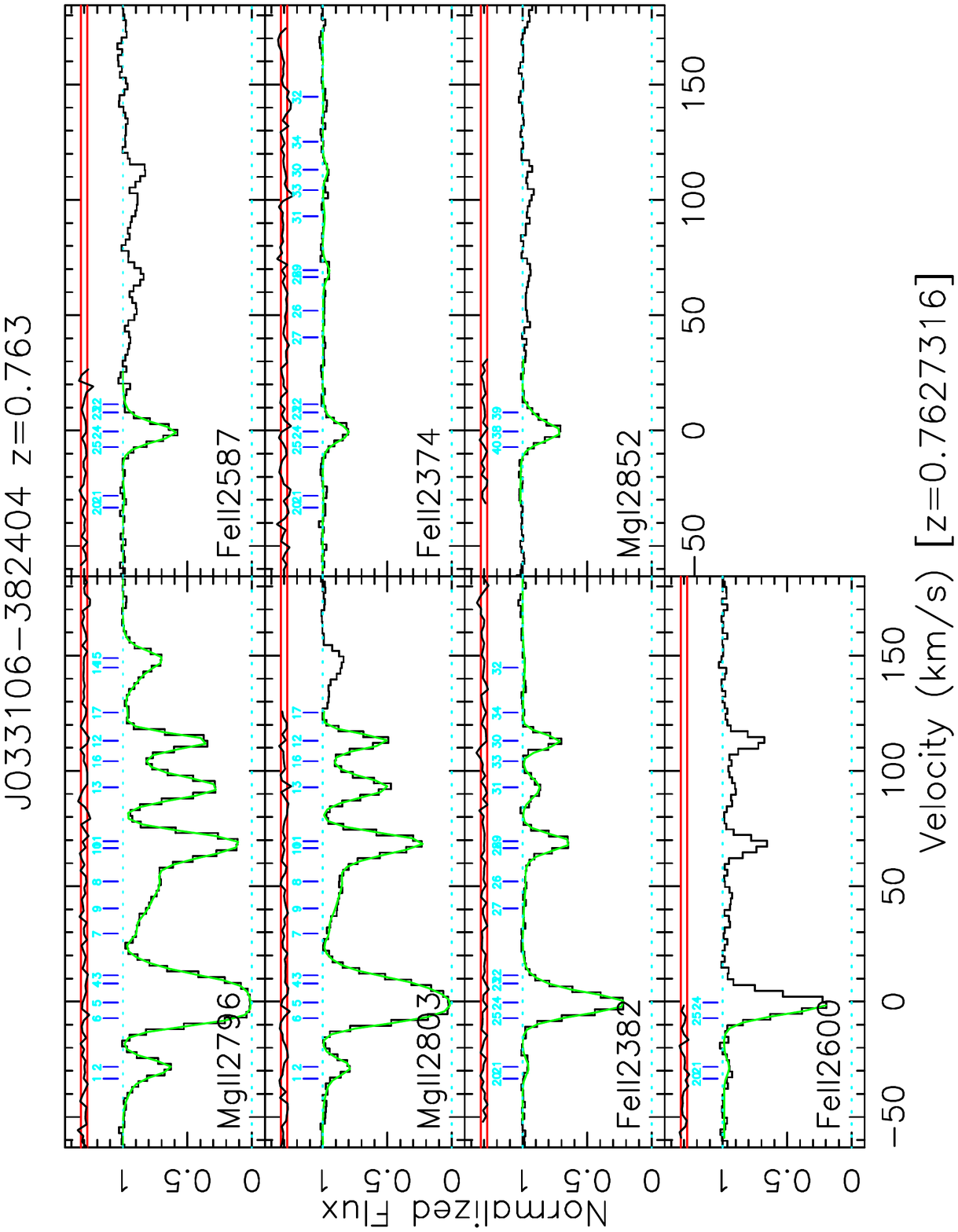}
\par\end{centering}

\caption[Fit for the $z=0.763$ absorber toward J033106$-$382404]{Many-multiplet fit for the $z=0.763$ absorber toward J033106$-$382404.}
\end{figure}
\begin{figure}[H]
\noindent \begin{centering}
\includegraphics[bb=34bp 58bp 554bp 738bp,clip,width=1\textwidth]{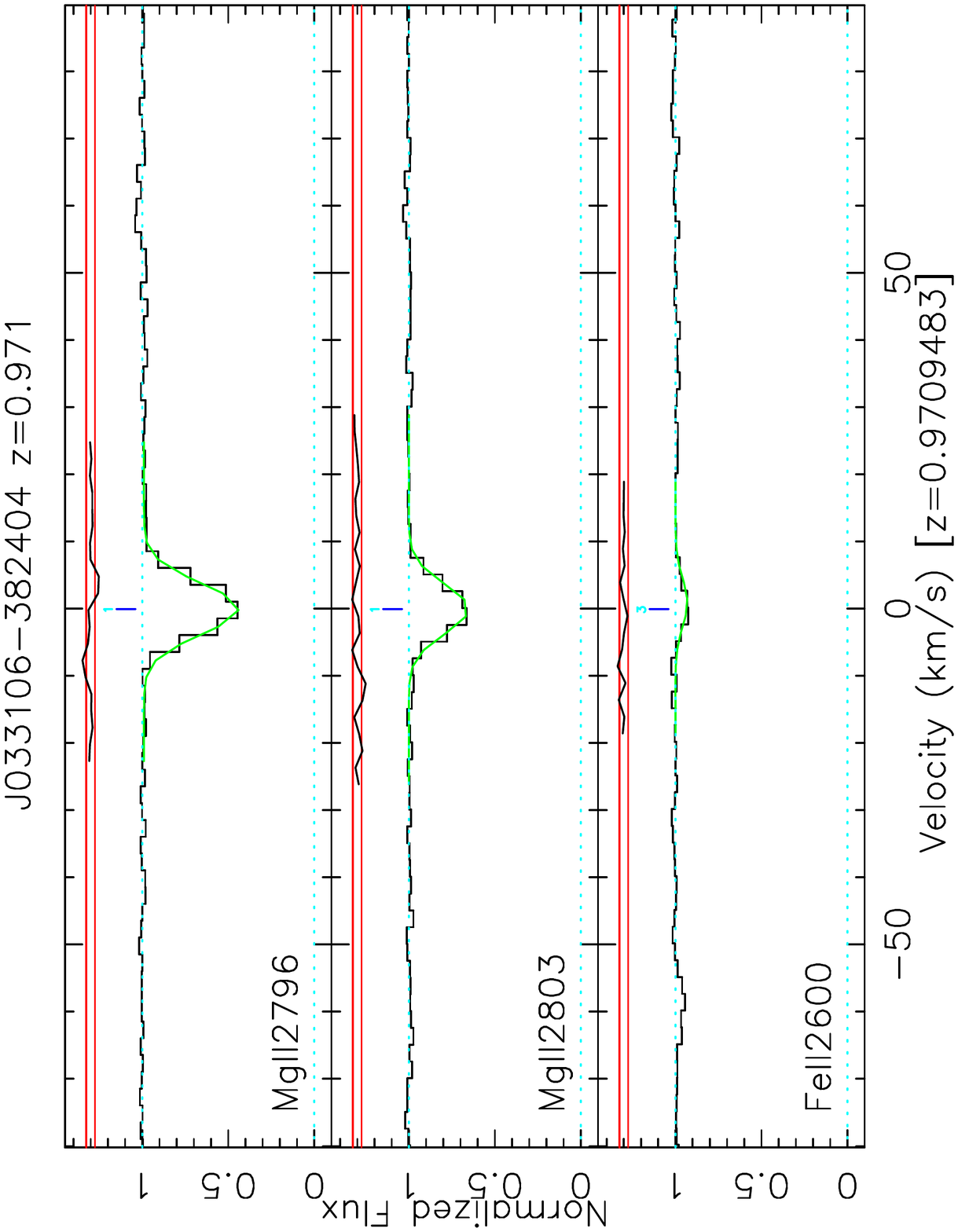}
\par\end{centering}

\caption[Fit for the $z=0.971$ absorber toward J033106$-$382404]{Many-multiplet fit for the $z=0.971$ absorber toward J033106$-$382404.}
\end{figure}
\begin{figure}[H]
\noindent \begin{centering}
\includegraphics[bb=34bp 58bp 554bp 738bp,clip,width=1\textwidth]{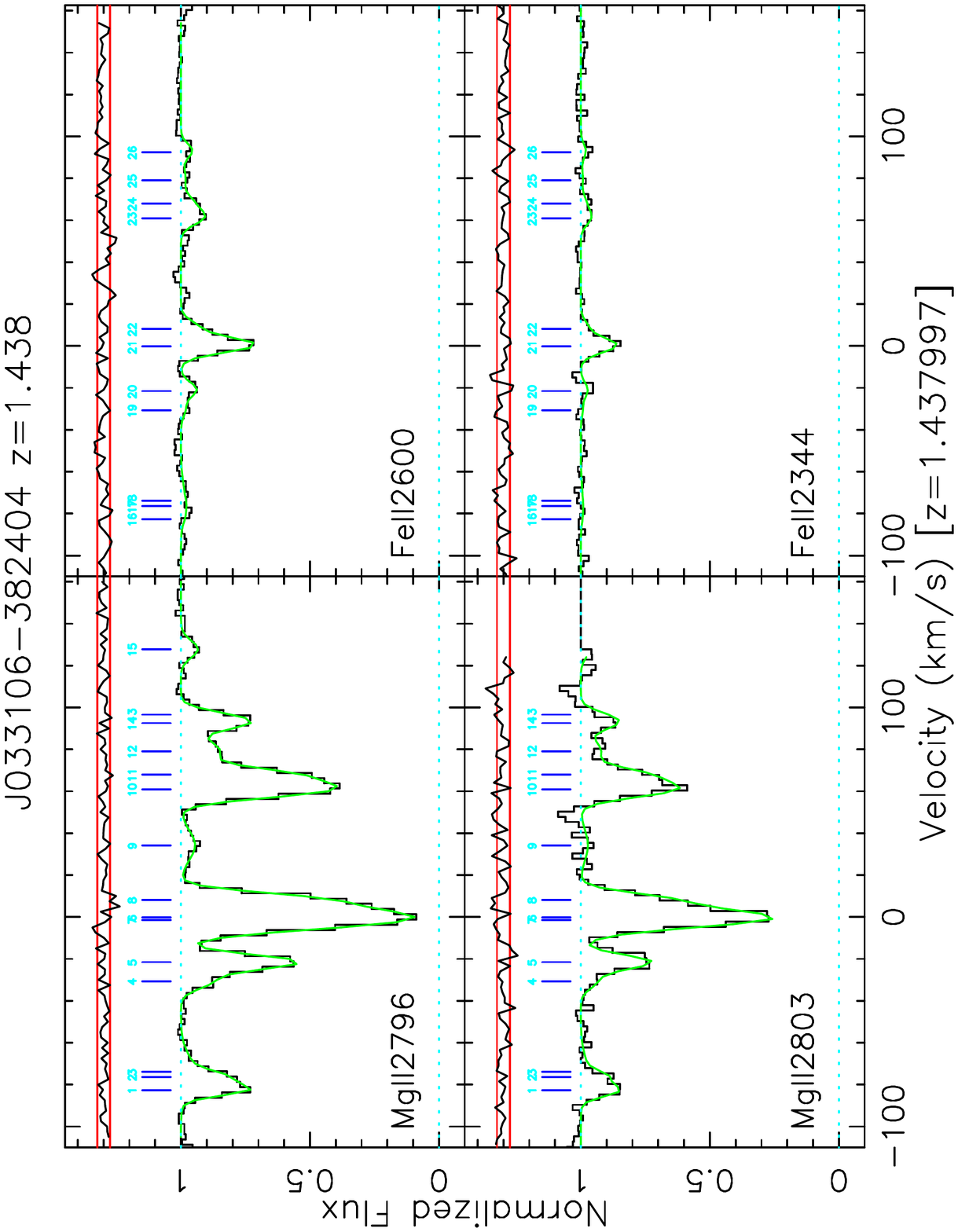}
\par\end{centering}

\caption[Fit for the $z=1.438$ absorber toward J033106$-$382404]{Many-multiplet fit for the $z=1.438$ absorber toward J033106$-$382404.}
\end{figure}
\begin{figure}[H]
\noindent \begin{centering}
\includegraphics[bb=34bp 58bp 554bp 738bp,clip,width=1\textwidth]{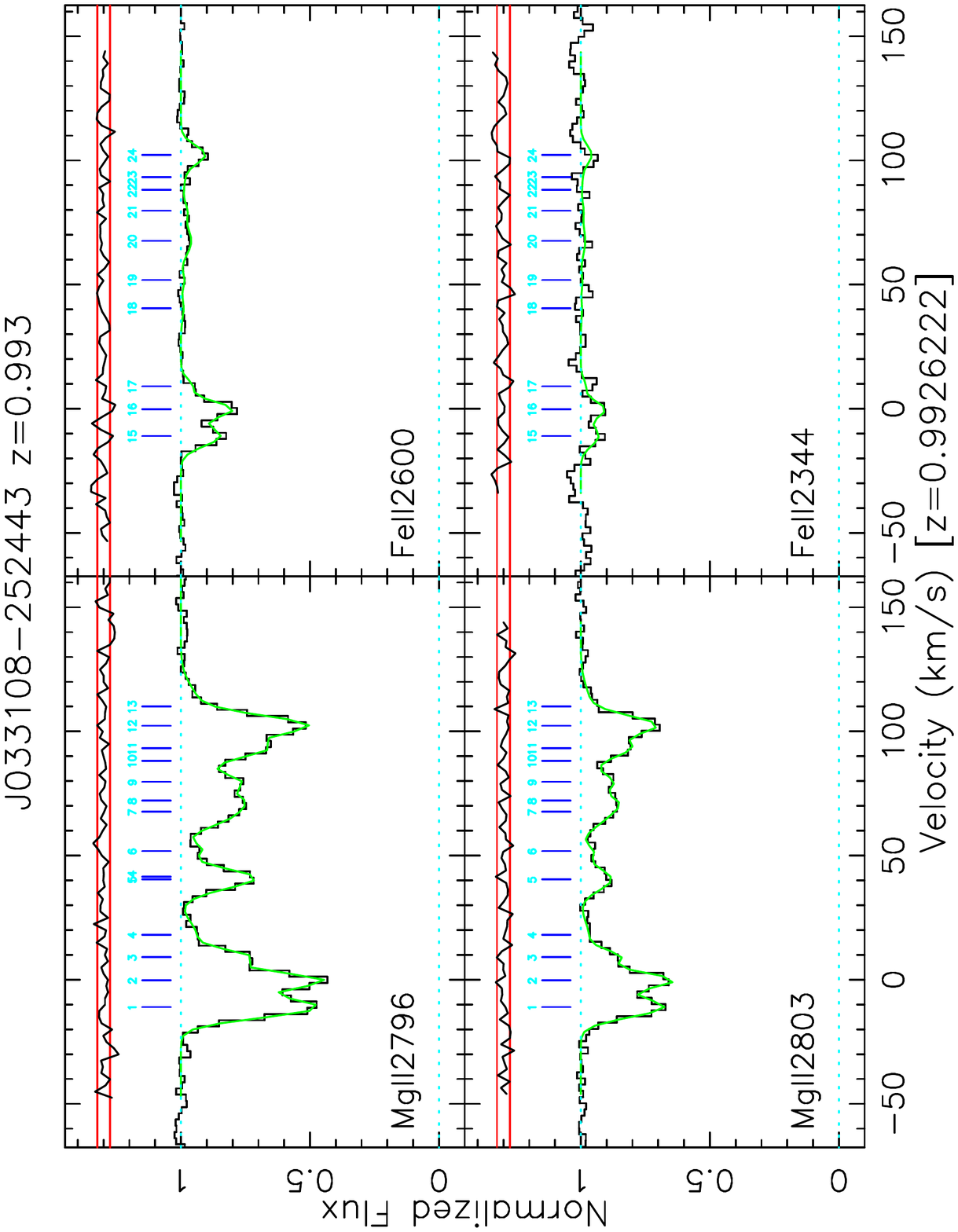}
\par\end{centering}

\caption[Fit for the $z=0.993$ absorber toward J033108$-$252443]{Many-multiplet fit for the $z=0.993$ absorber toward J033108$-$252443.}
\end{figure}
\begin{figure}[H]
\noindent \begin{centering}
\includegraphics[bb=34bp 58bp 554bp 738bp,clip,width=1\textwidth]{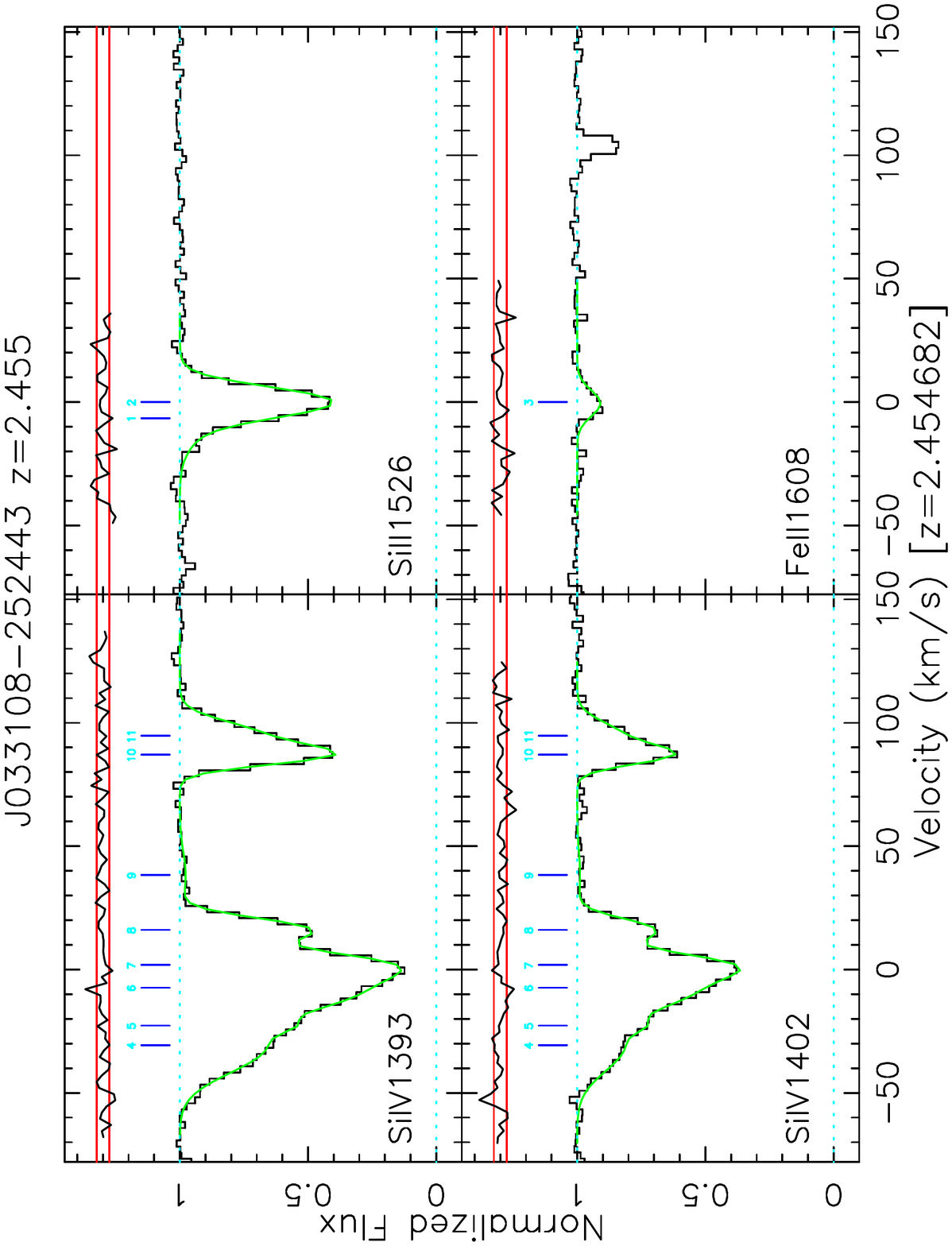}
\par\end{centering}

\caption[Fit for the $z=2.455$ absorber toward J033108$-$252443]{Many-multiplet fit for the $z=2.455$ absorber toward J033108$-$252443.}
\end{figure}
\begin{figure}[H]
\noindent \begin{centering}
\includegraphics[bb=34bp 58bp 554bp 738bp,clip,width=1\textwidth]{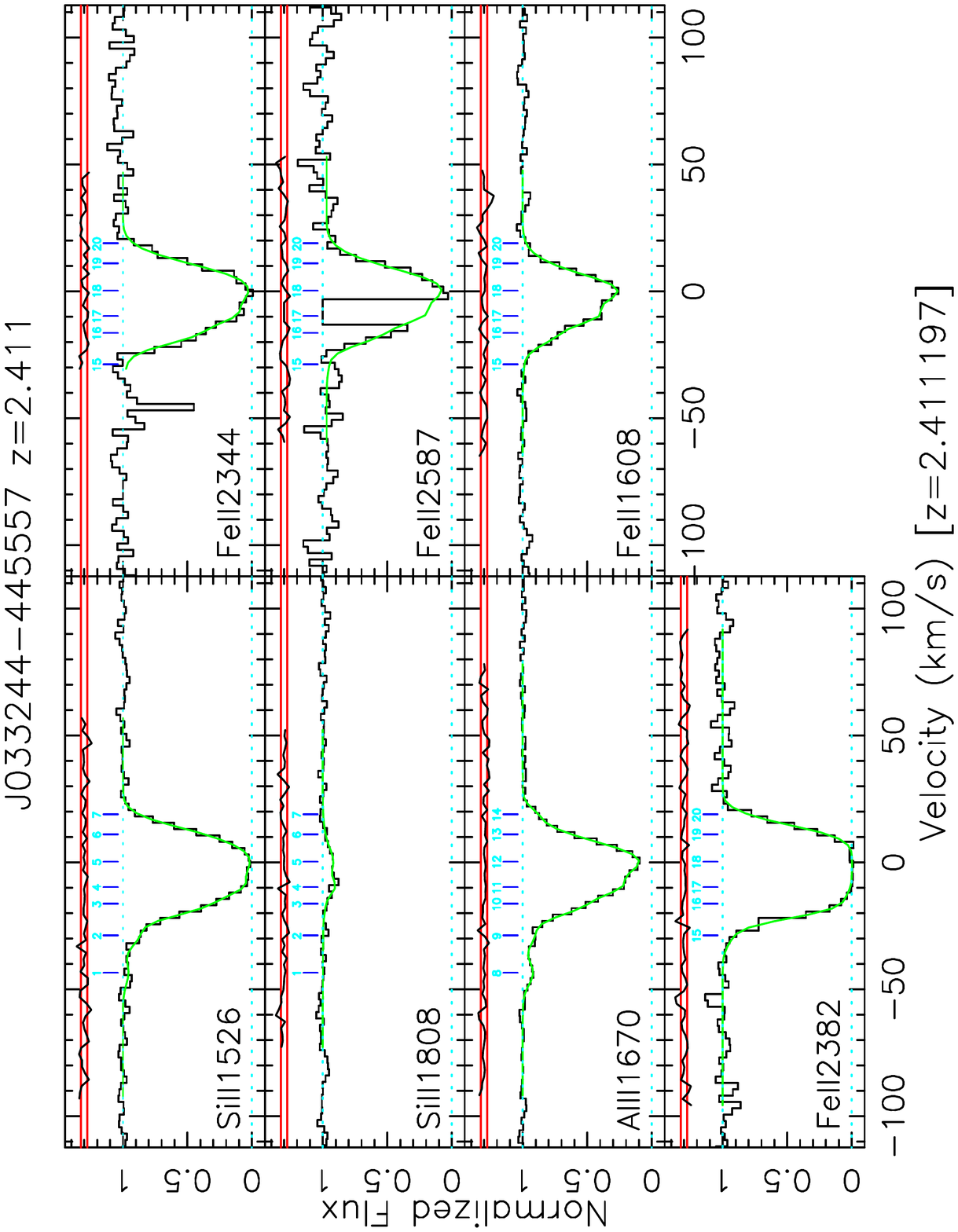}
\par\end{centering}

\caption[Fit for the $z=2.411$ absorber toward J033244$-$445557]{Many-multiplet fit for the $z=2.411$ absorber toward J033244$-$445557.}
\end{figure}
\begin{figure}[H]
\noindent \begin{centering}
\includegraphics[bb=34bp 58bp 554bp 738bp,clip,width=1\textwidth]{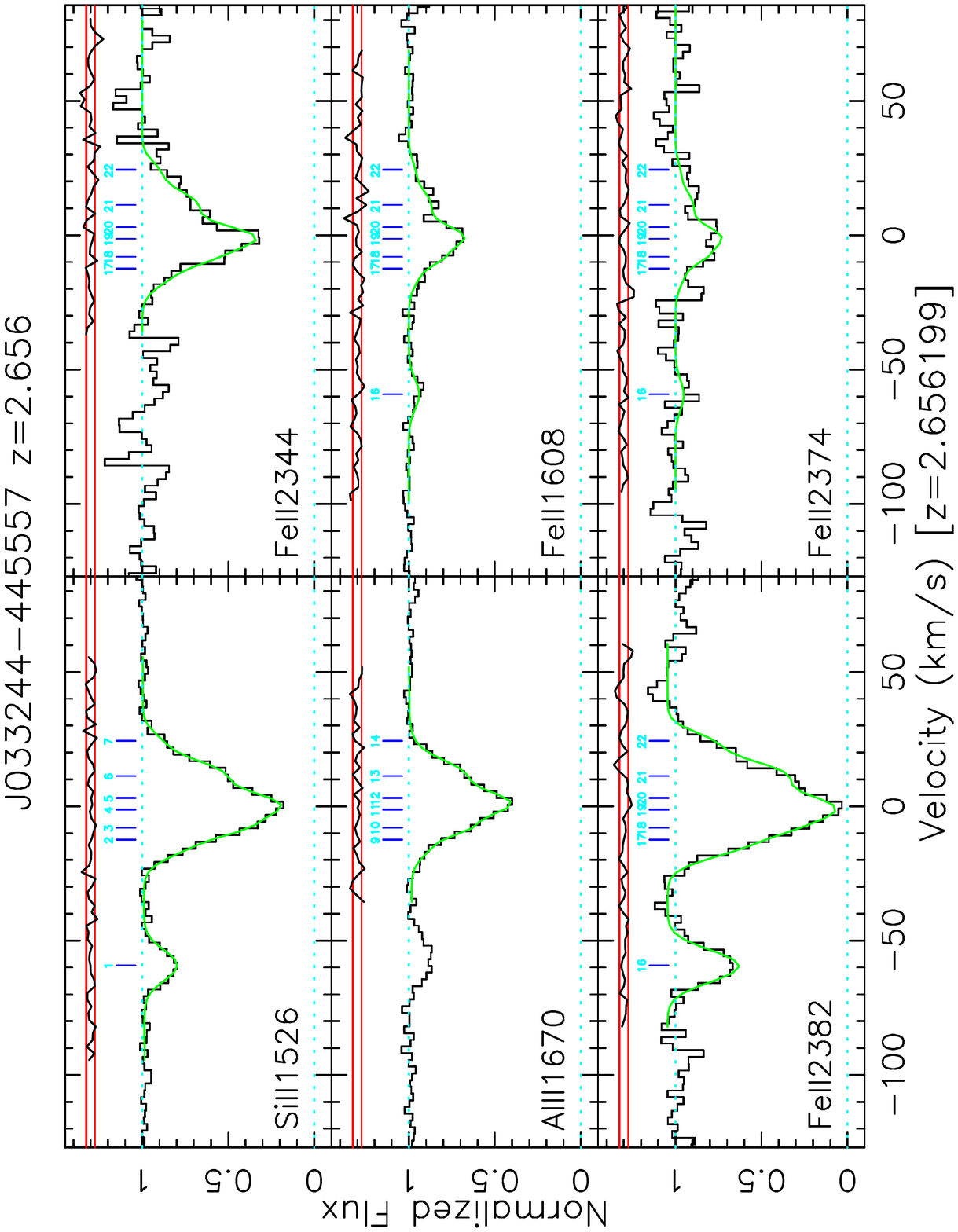}
\par\end{centering}

\caption[Fit for the $z=2.656$ absorber toward J033244$-$445557]{Many-multiplet fit for the $z=2.656$ absorber toward J033244$-$445557.}
\end{figure}
\begin{figure}[H]
\noindent \begin{centering}
\includegraphics[bb=34bp 58bp 554bp 738bp,clip,width=1\textwidth]{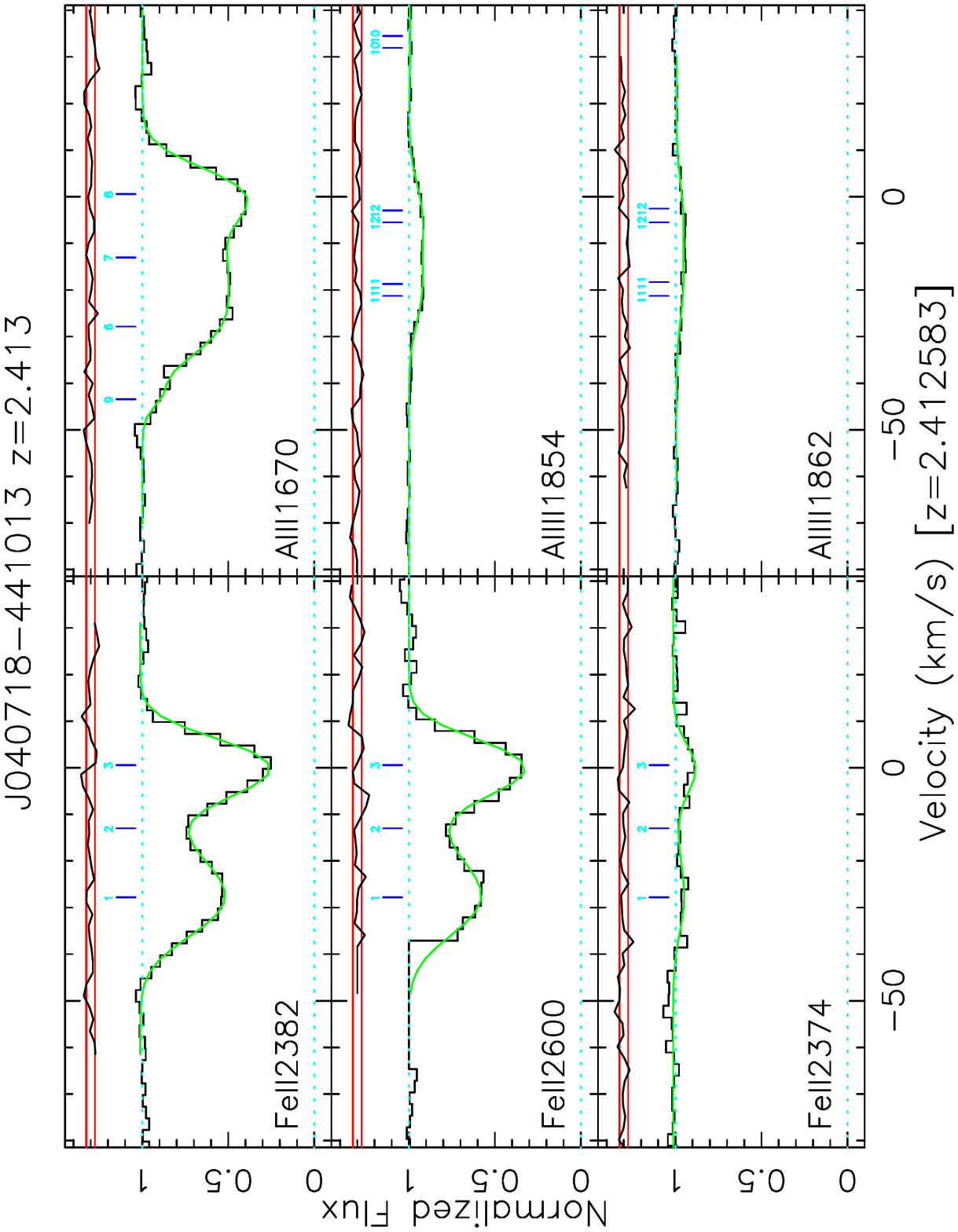}
\par\end{centering}

\caption[Fit for the $z=2.413$ absorber toward J040718$-$441013]{Many-multiplet fit for the $z=2.413$ absorber toward J040718$-$441013.}
\end{figure}
\begin{figure}[H]
\noindent \begin{centering}
\includegraphics[bb=34bp 58bp 554bp 738bp,clip,width=1\textwidth]{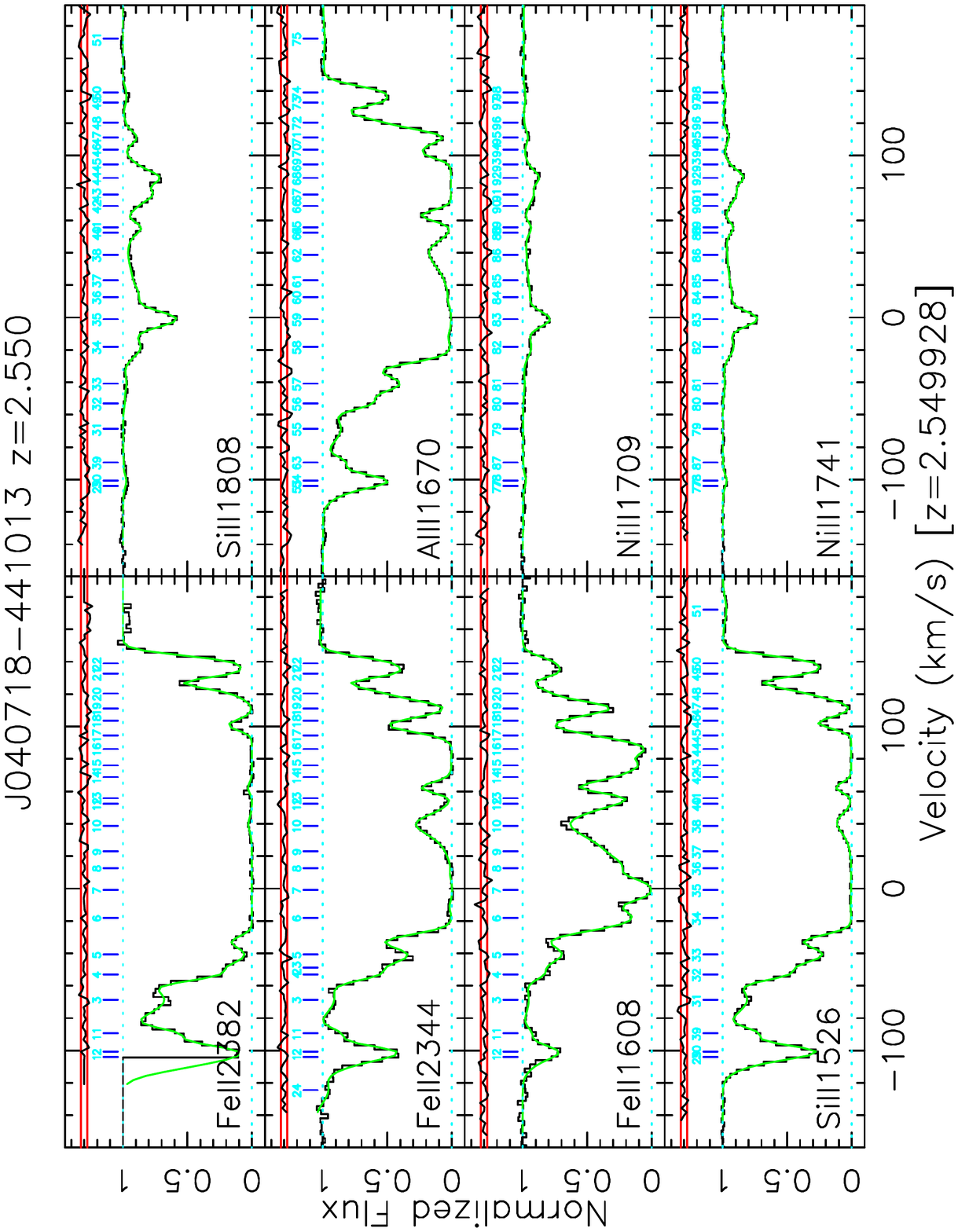}
\par\end{centering}

\caption[Fit for the $z=2.550$ absorber toward J040718$-$441013]{Many-multiplet fit for the $z=2.550$ absorber toward J040718$-$441013.}
\end{figure}
\begin{figure}[H]
\noindent \begin{centering}
\includegraphics[bb=34bp 58bp 554bp 738bp,clip,width=1\textwidth]{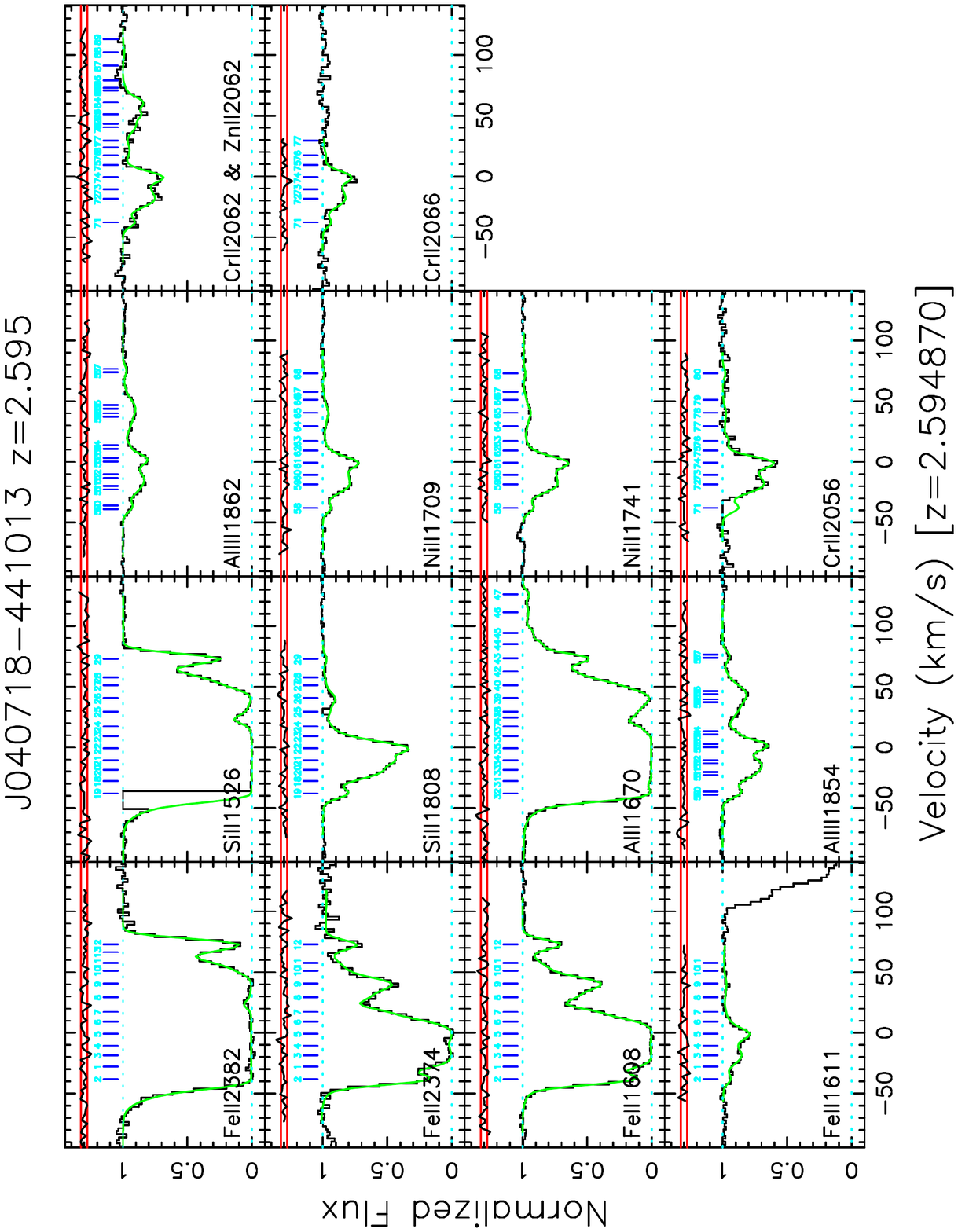}
\par\end{centering}

\caption[Fit for the $z=2.595$ absorber toward J040718$-$441013]{Many-multiplet fit for the $z=2.595$ absorber toward J040718$-$441013.}
\end{figure}
\begin{figure}[H]
\noindent \begin{centering}
\includegraphics[bb=34bp 58bp 554bp 738bp,clip,width=1\textwidth]{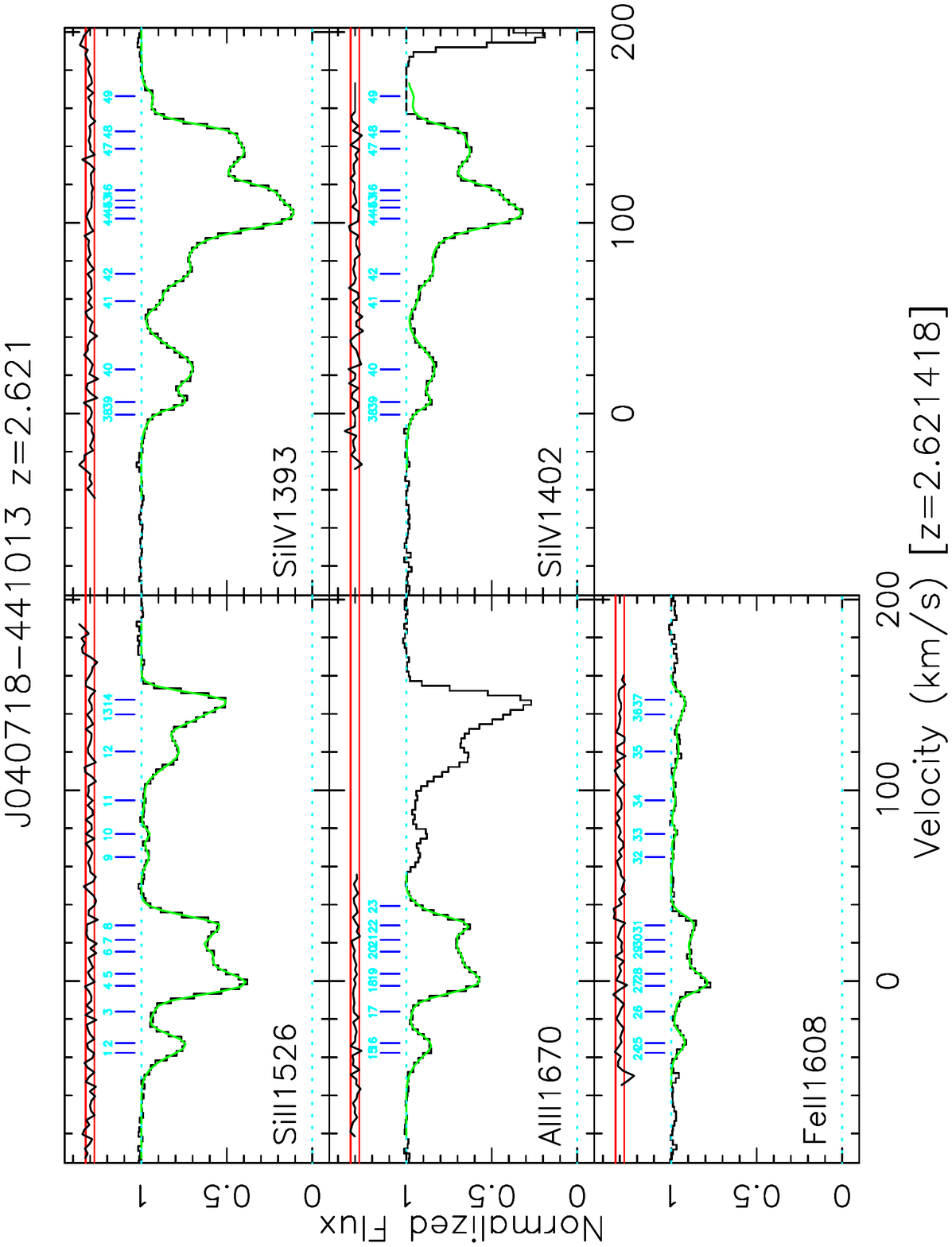}
\par\end{centering}

\caption[Fit for the $z=2.621$ absorber toward J040718$-$441013]{Many-multiplet fit for the $z=2.621$ absorber toward J040718$-$441013.}
\end{figure}
\begin{figure}[H]
\noindent \begin{centering}
\includegraphics[bb=34bp 58bp 554bp 738bp,clip,width=1\textwidth]{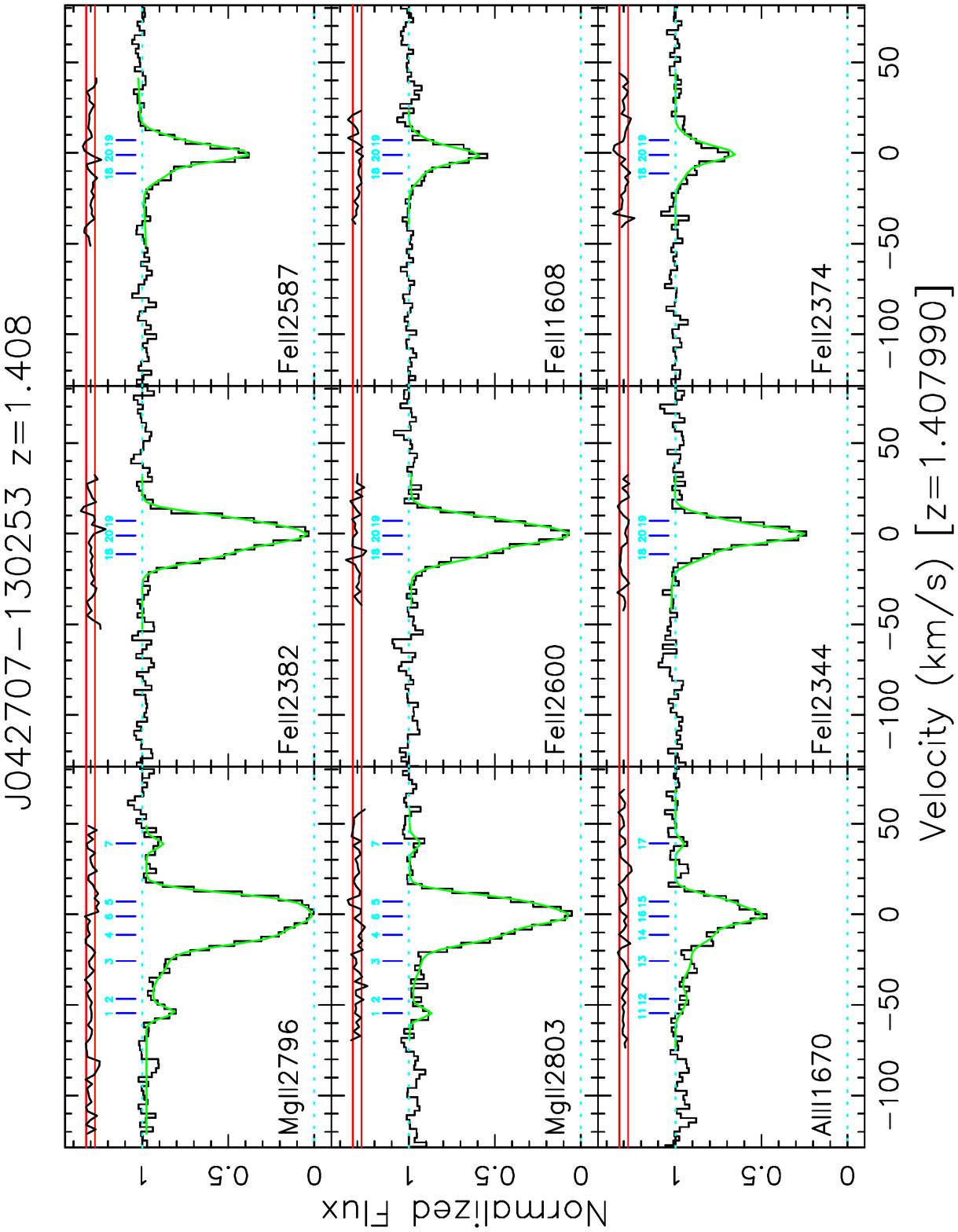}
\par\end{centering}

\caption[Fit for the $z=1.408$ absorber toward J042707$-$130253]{Many-multiplet fit for the $z=1.408$ absorber toward J042707$-$130253.}
\end{figure}
\begin{figure}[H]
\noindent \begin{centering}
\includegraphics[bb=34bp 58bp 554bp 738bp,clip,width=1\textwidth]{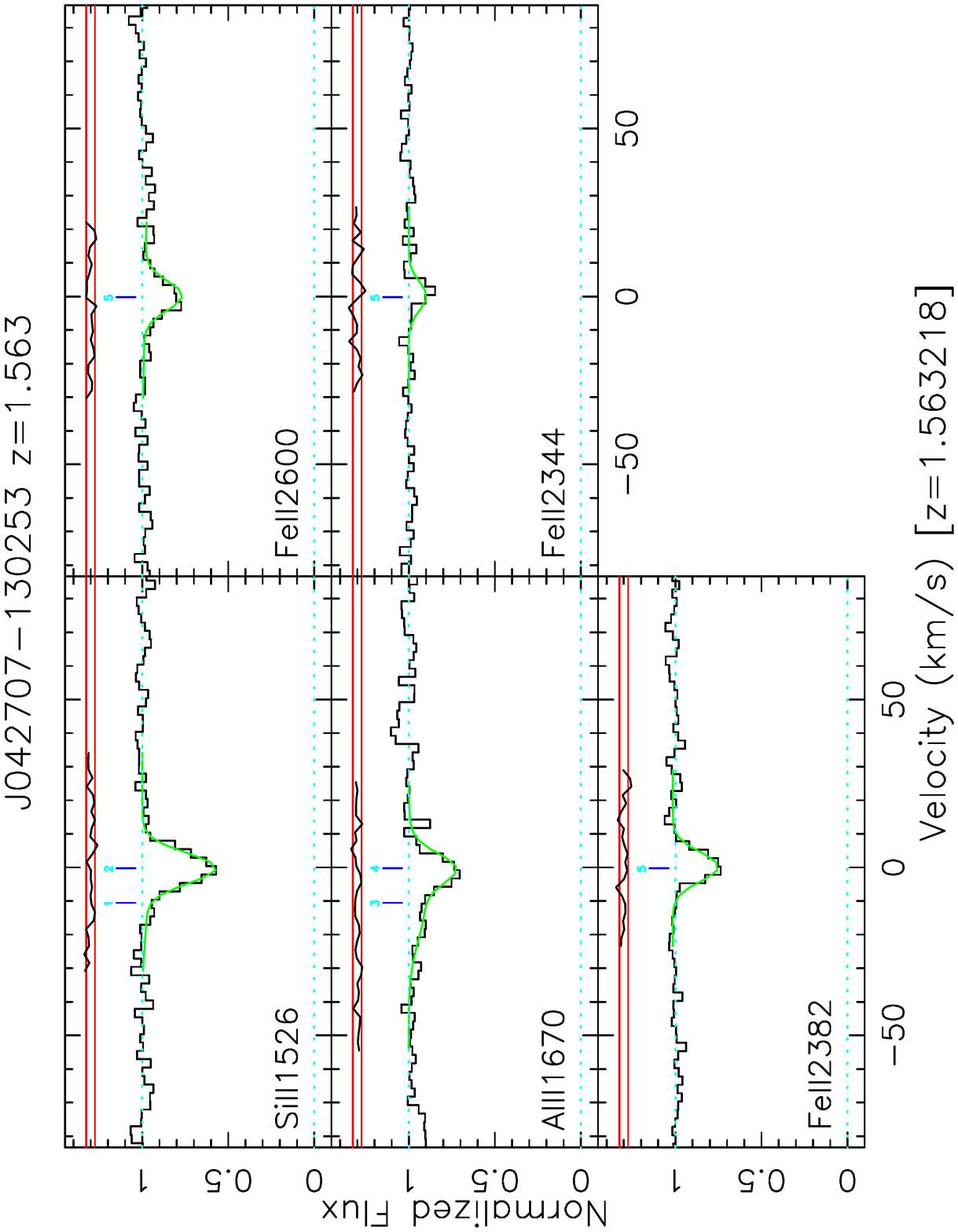}
\par\end{centering}

\caption[Fit for the $z=1.563$ absorber toward J042707$-$130253]{Many-multiplet fit for the $z=1.563$ absorber toward J042707$-$130253.}
\end{figure}
\begin{figure}[H]
\noindent \begin{centering}
\includegraphics[bb=34bp 58bp 554bp 738bp,clip,width=1\textwidth]{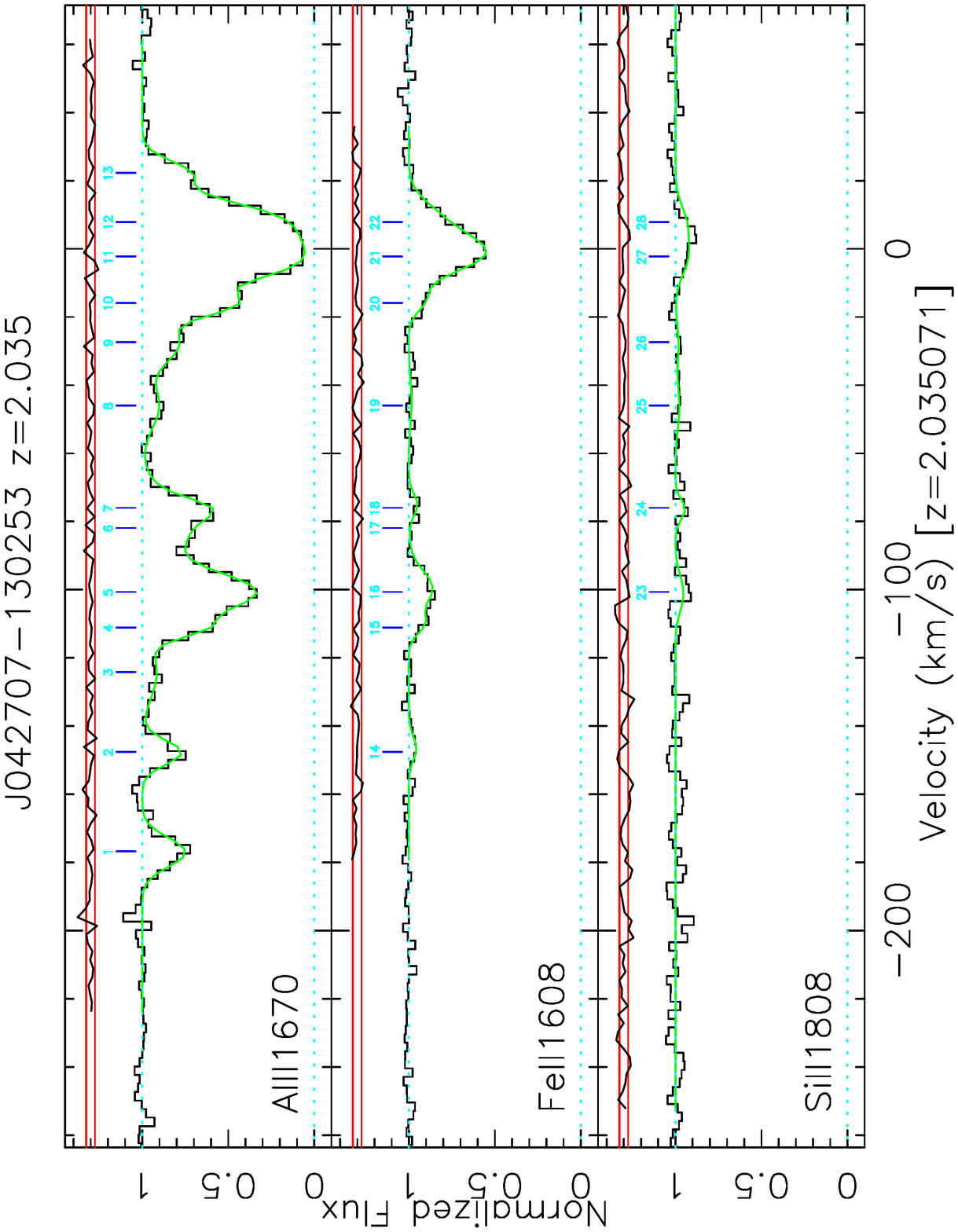}
\par\end{centering}

\caption[Fit for the $z=2.035$ absorber toward J042707$-$130253]{Many-multiplet fit for the $z=2.035$ absorber toward J042707$-$130253.}
\end{figure}
\begin{figure}[H]
\noindent \begin{centering}
\includegraphics[bb=34bp 58bp 554bp 738bp,clip,width=1\textwidth]{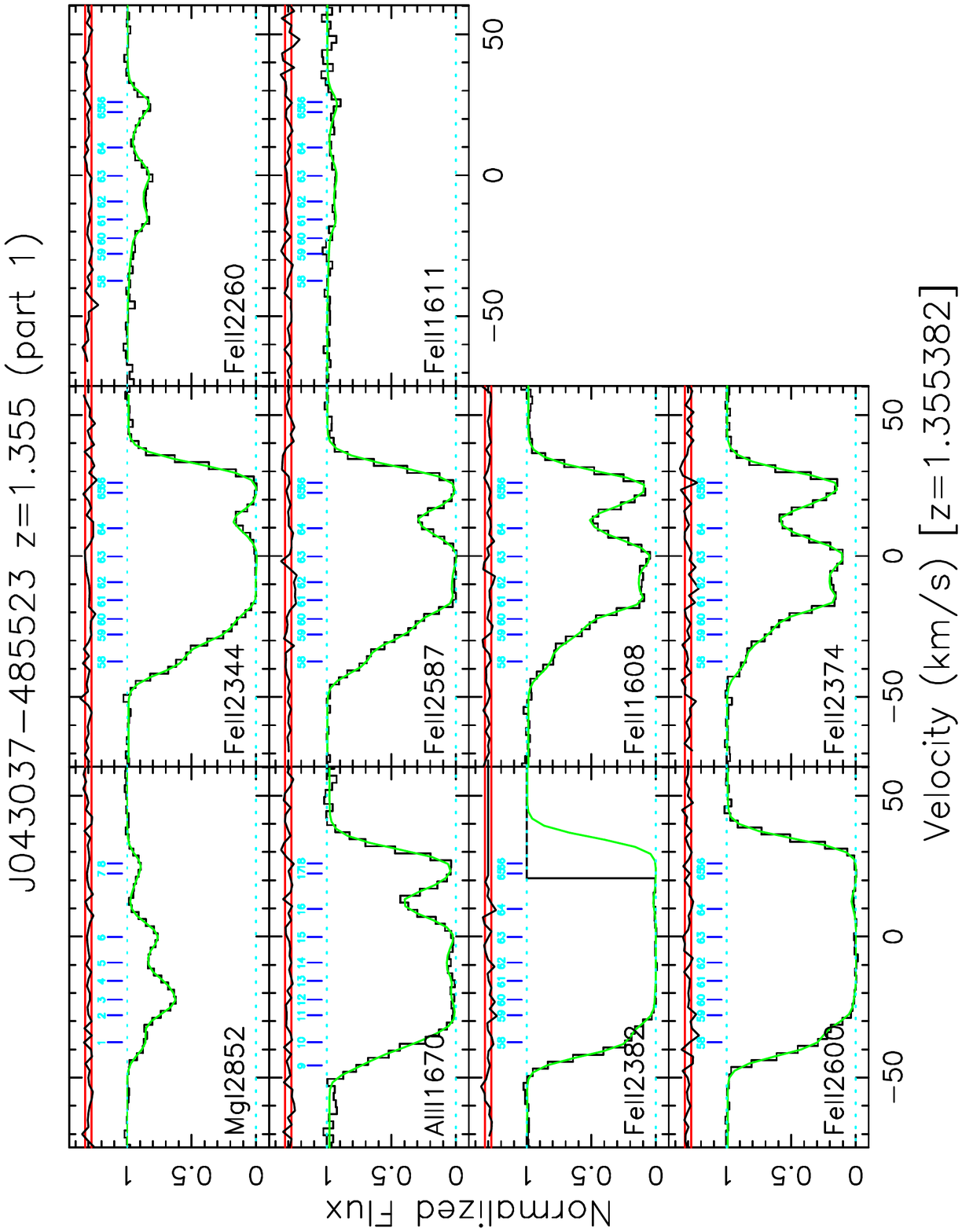}
\par\end{centering}

\caption[Fit for the $z=1.355$ absorber toward J043037$-$485523 (part 1)]{Many-multiplet fit for the $z=1.355$ absorber toward J043037$-$485523 (part 1).}
\end{figure}
\begin{figure}[H]
\noindent \begin{centering}
\includegraphics[bb=34bp 58bp 554bp 738bp,clip,width=1\textwidth]{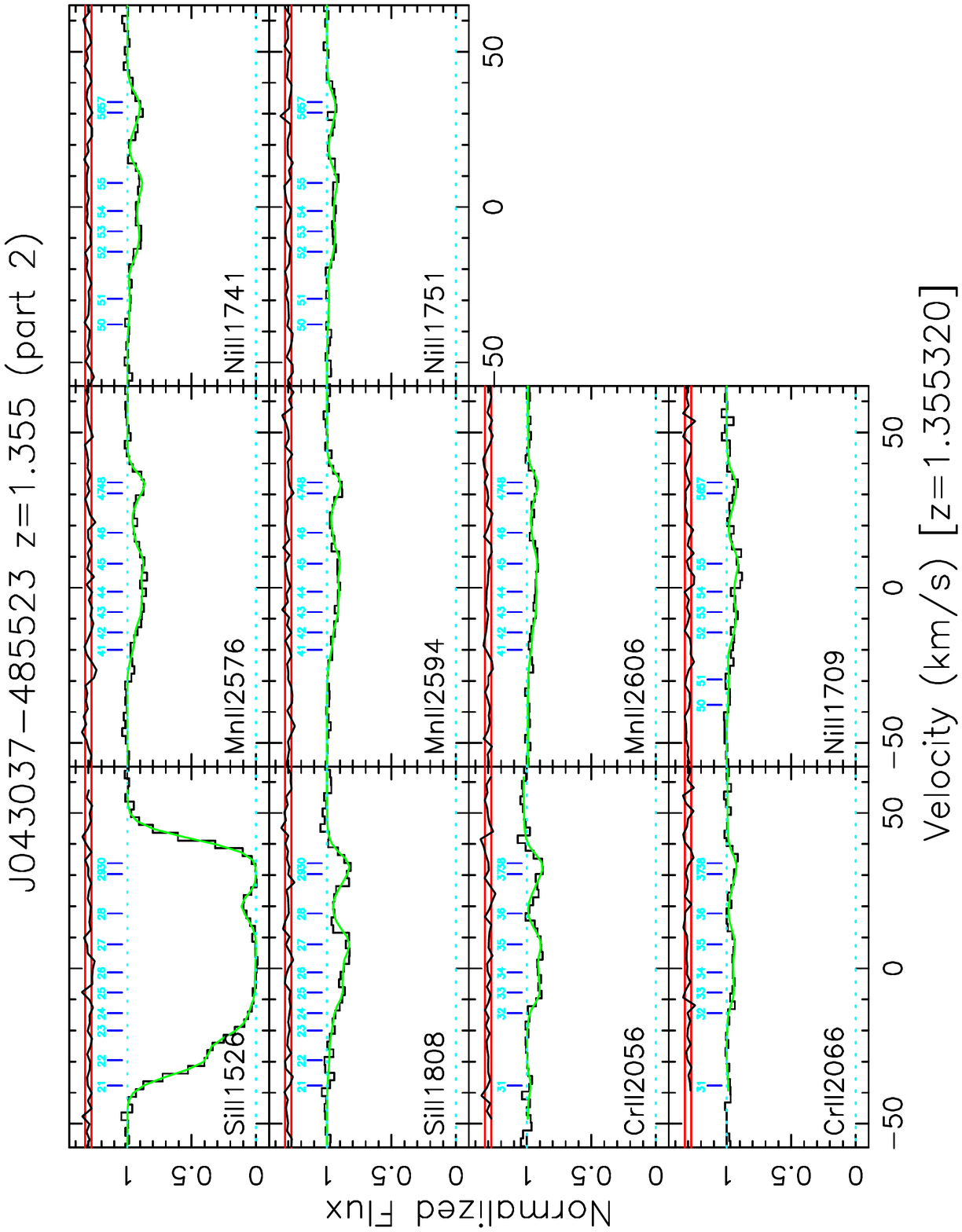}
\par\end{centering}

\caption[Fit for the $z=1.355$ absorber toward J043037$-$485523 (part 2)]{Many-multiplet fit for the $z=1.355$ absorber toward J043037$-$485523 (part 2).}
\end{figure}
\begin{figure}[H]
\noindent \begin{centering}
\includegraphics[bb=34bp 58bp 554bp 738bp,clip,width=1\textwidth]{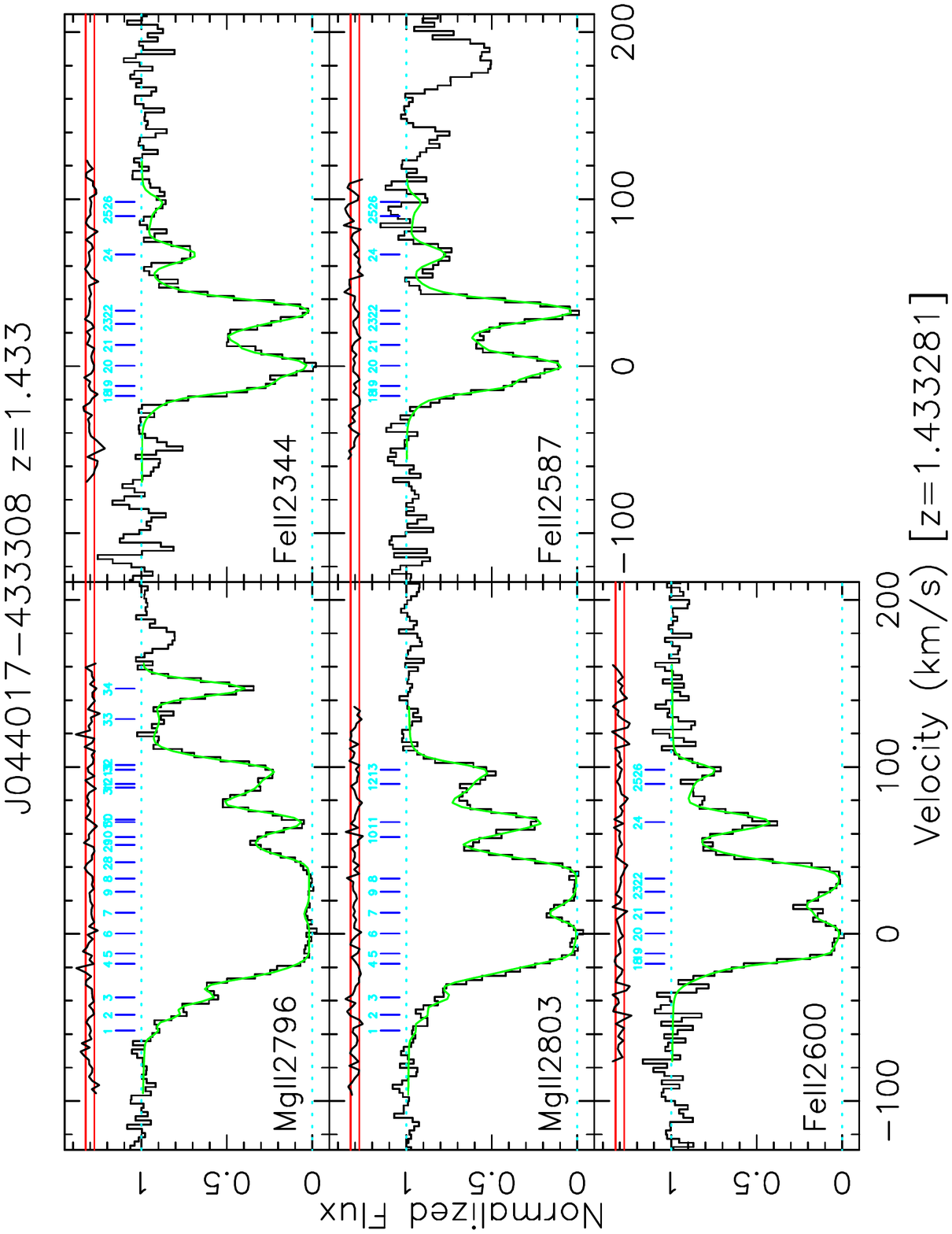}
\par\end{centering}

\caption[Fit for the $z=1.433$ absorber toward J044017$-$433308]{Many-multiplet fit for the $z=1.433$ absorber toward J044017$-$433308.}
\end{figure}
\begin{figure}[H]
\noindent \begin{centering}
\includegraphics[bb=34bp 58bp 554bp 738bp,clip,width=1\textwidth]{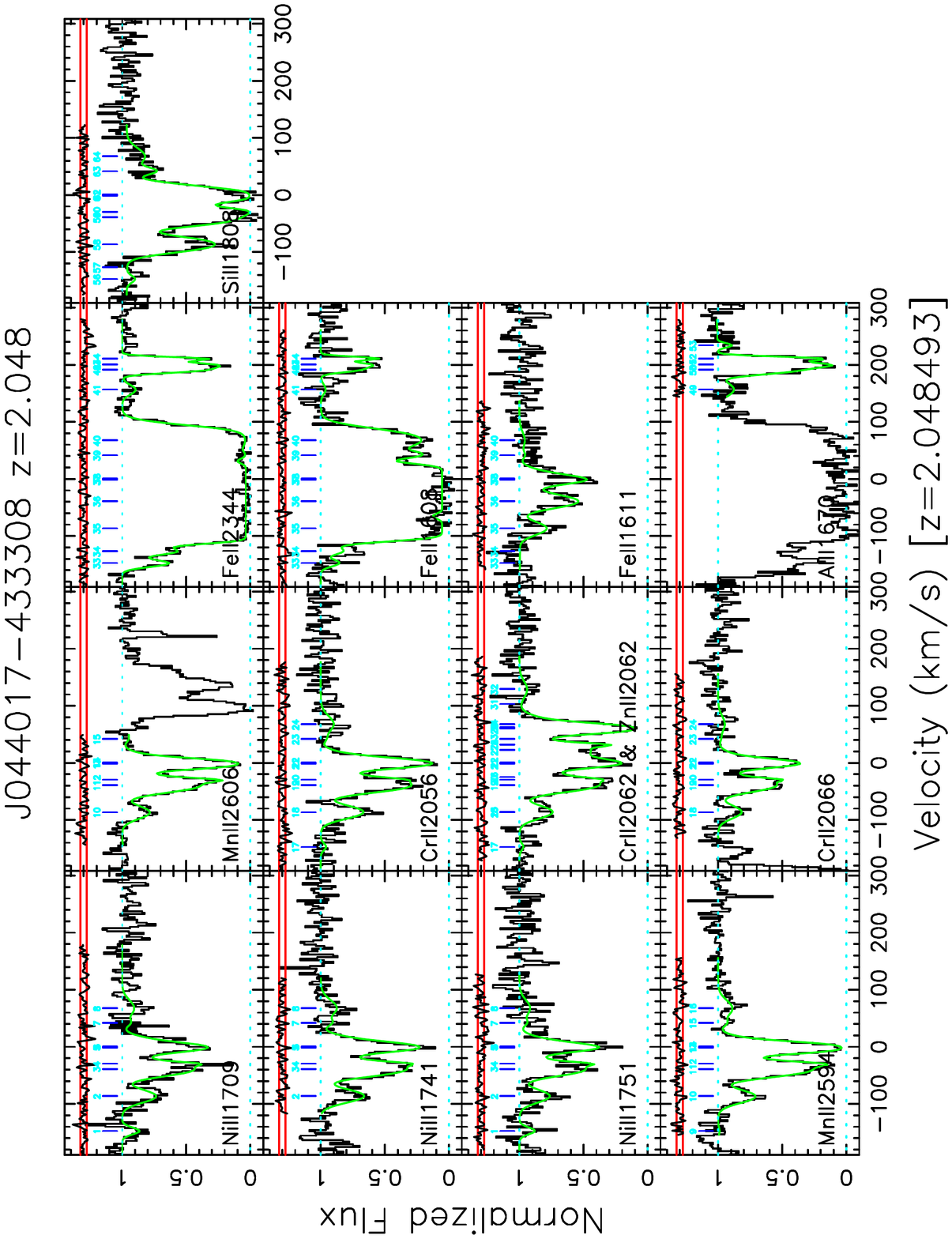}
\par\end{centering}

\caption[Fit for the $z=2.048$ absorber toward J044017$-$433308]{Many-multiplet fit for the $z=2.048$ absorber toward J044017$-$433308.}
\end{figure}
\begin{figure}[H]
\noindent \begin{centering}
\includegraphics[bb=34bp 58bp 554bp 738bp,clip,width=1\textwidth]{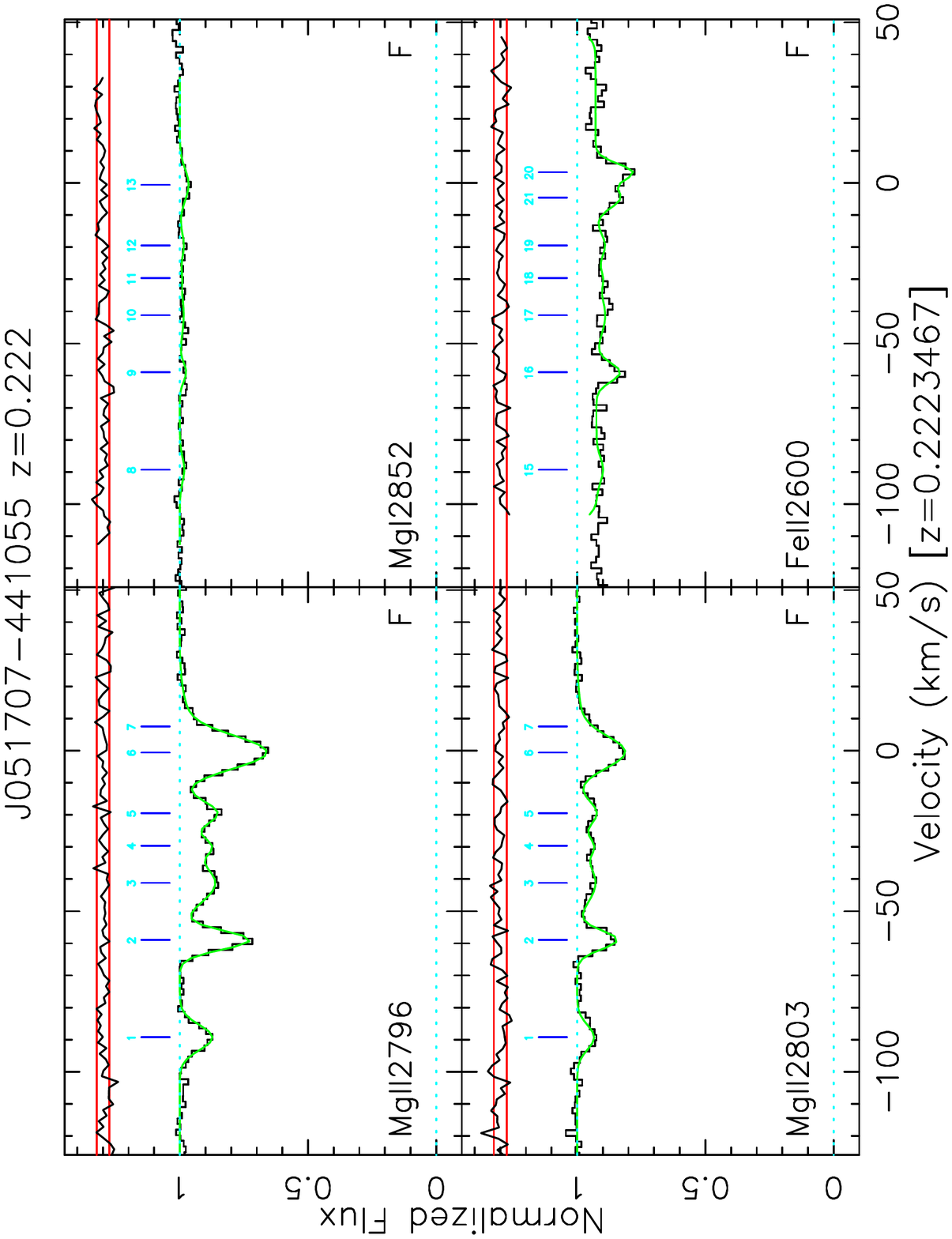}
\par\end{centering}

\caption[Fit for the $z=0.222$ absorber toward J051707$-$441055]{Many-multiplet fit for the $z=0.222$ absorber toward J051707$-$441055.}
\end{figure}
\begin{figure}[H]
\noindent \begin{centering}
\includegraphics[bb=34bp 58bp 554bp 738bp,clip,width=1\textwidth]{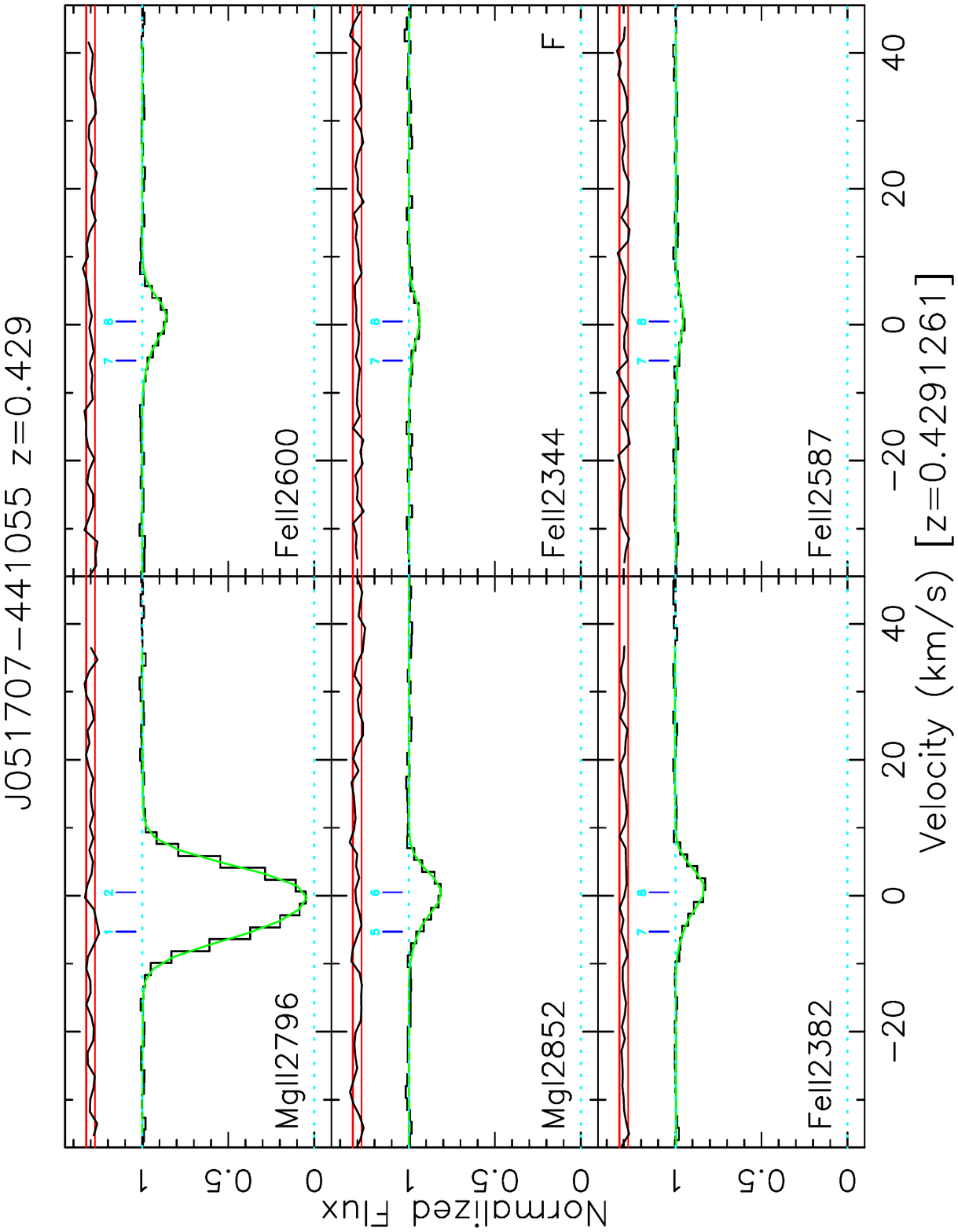}
\par\end{centering}

\caption[Fit for the $z=0.429$ absorber toward J051707$-$441055]{Many-multiplet fit for the $z=0.429$ absorber toward J051707$-$441055.}
\end{figure}
\begin{figure}[H]
\noindent \begin{centering}
\includegraphics[bb=34bp 58bp 554bp 738bp,clip,width=1\textwidth]{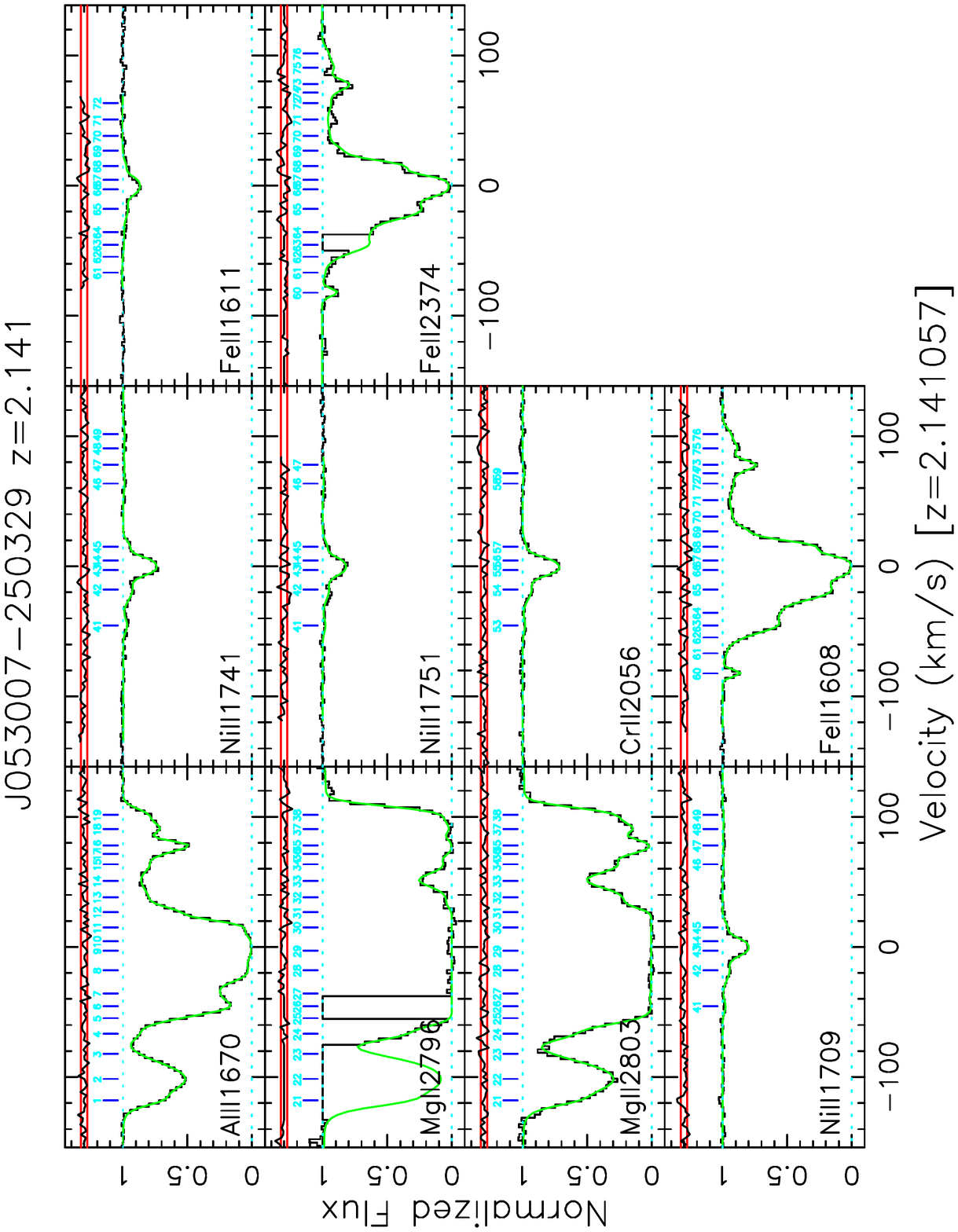}
\par\end{centering}

\caption[Fit for the $z=2.141$ absorber toward J053007$-$250329]{Many-multiplet fit for the $z=2.141$ absorber toward J053007$-$250329.}
\end{figure}
\begin{figure}[H]
\noindent \begin{centering}
\includegraphics[bb=34bp 58bp 554bp 738bp,clip,width=1\textwidth]{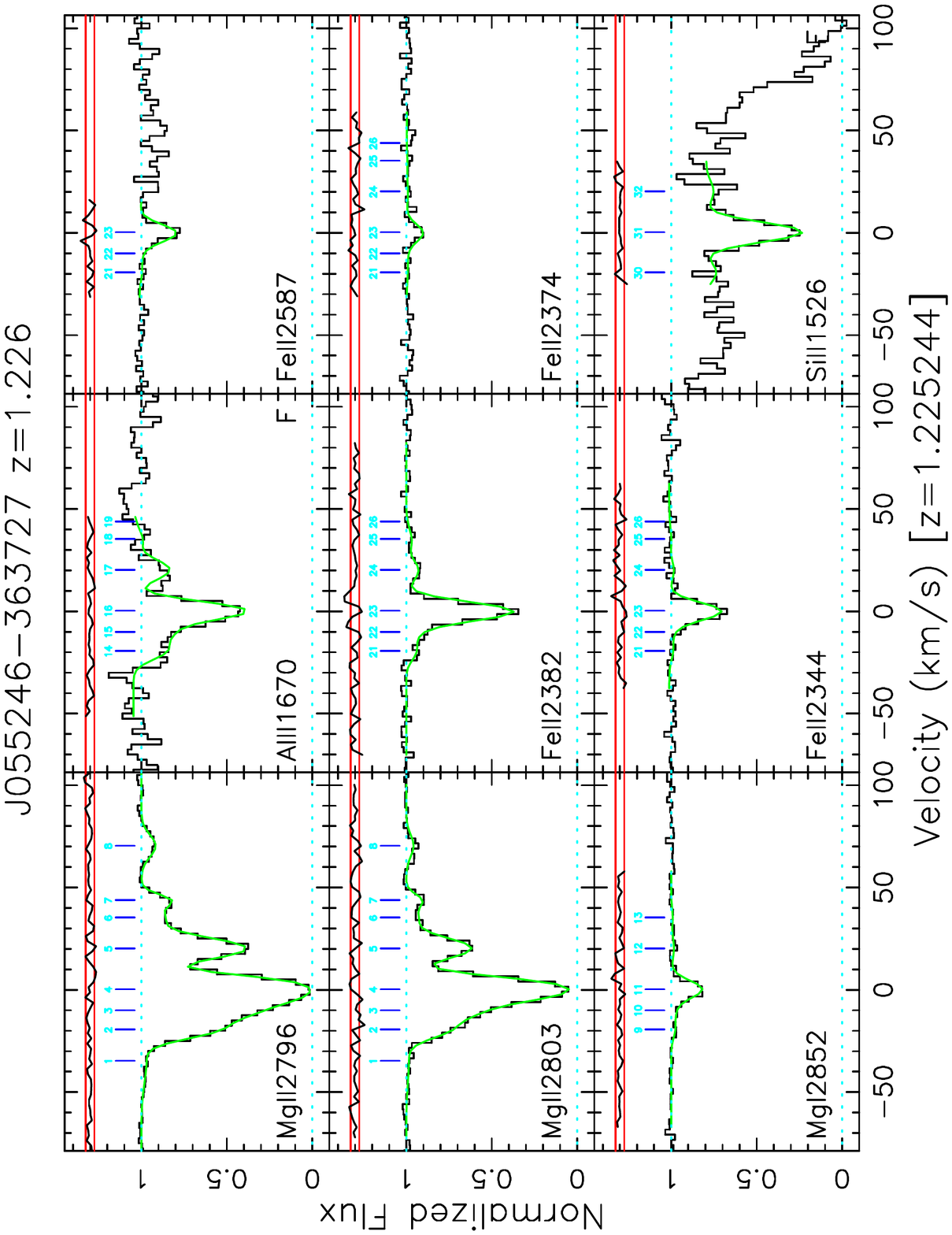}
\par\end{centering}

\caption[Fit for the $z=1.226$ absorber toward J055246$-$363727]{Many-multiplet fit for the $z=1.226$ absorber toward J055246$-$363727.}
\end{figure}
\begin{figure}[H]
\noindent \begin{centering}
\includegraphics[bb=34bp 58bp 554bp 738bp,clip,width=1\textwidth]{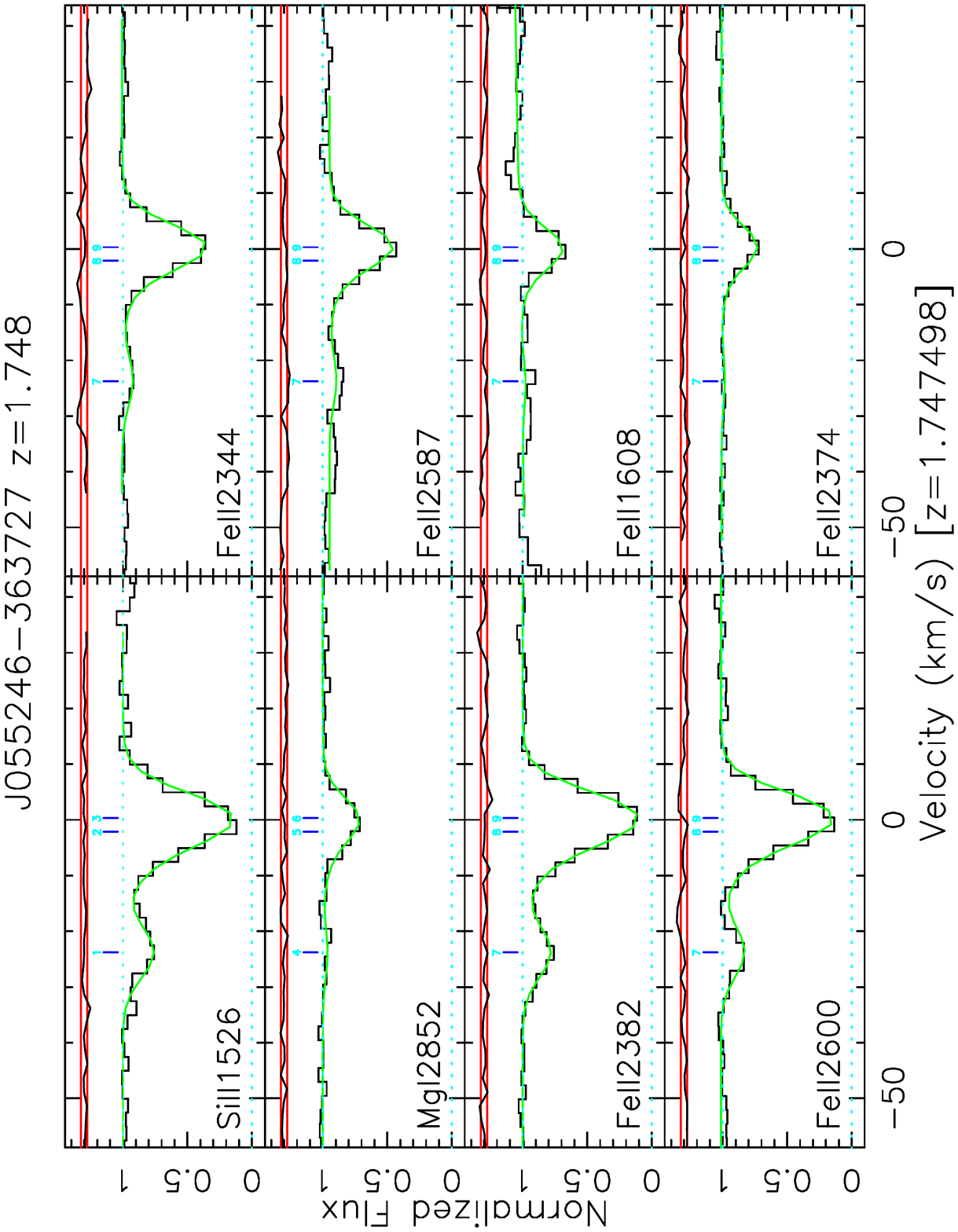}
\par\end{centering}

\caption[Fit for the $z=1.748$ absorber toward J055246$-$363727]{Many-multiplet fit for the $z=1.748$ absorber toward J055246$-$363727.}
\end{figure}
\begin{figure}[H]
\noindent \begin{centering}
\includegraphics[bb=34bp 58bp 554bp 738bp,clip,width=1\textwidth]{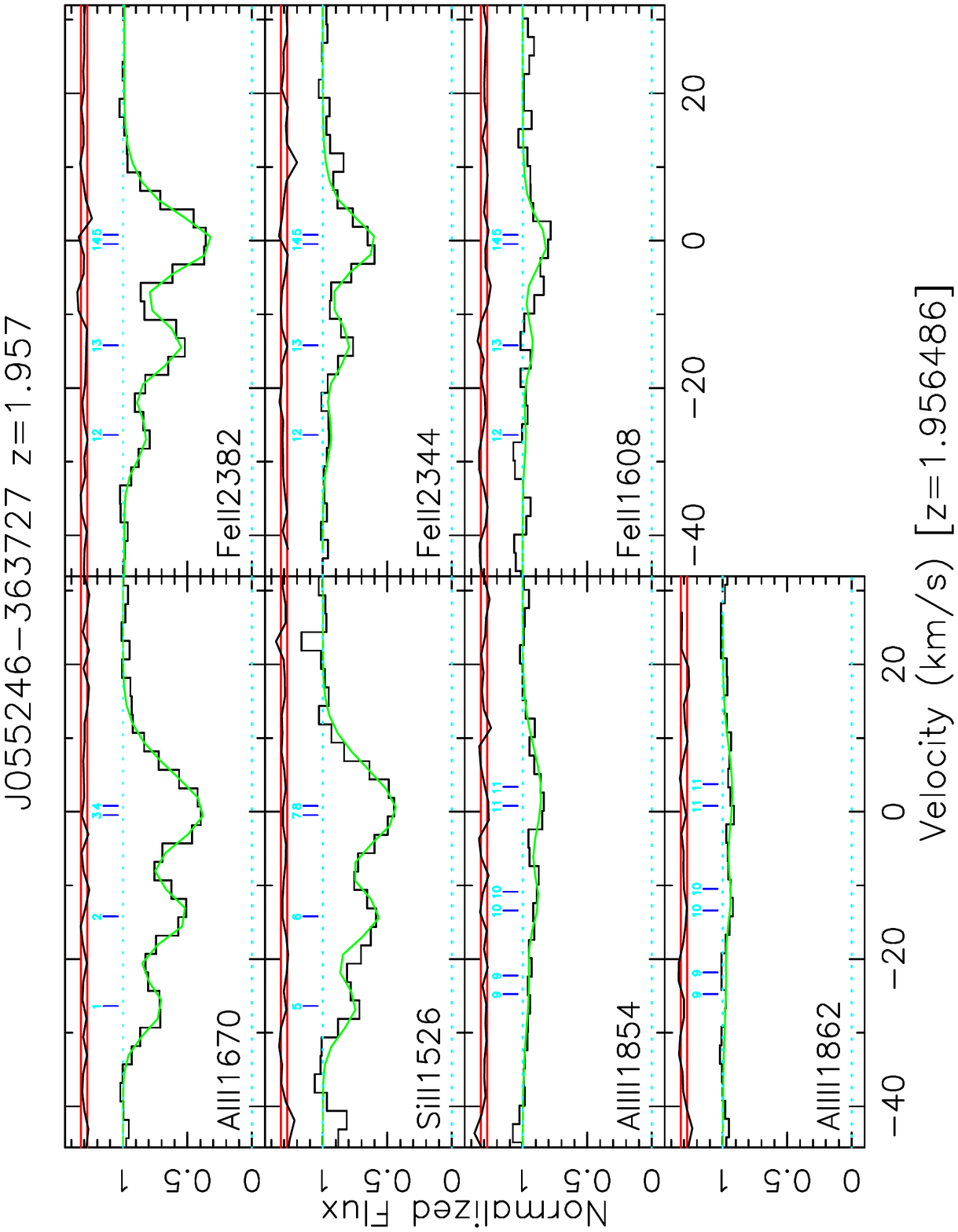}
\par\end{centering}

\caption[Fit for the $z=1.957$ absorber toward J055246$-$363727]{Many-multiplet fit for the $z=1.957$ absorber toward J055246$-$363727.}
\end{figure}
\begin{figure}[H]
\noindent \begin{centering}
\includegraphics[bb=34bp 58bp 554bp 738bp,clip,width=1\textwidth]{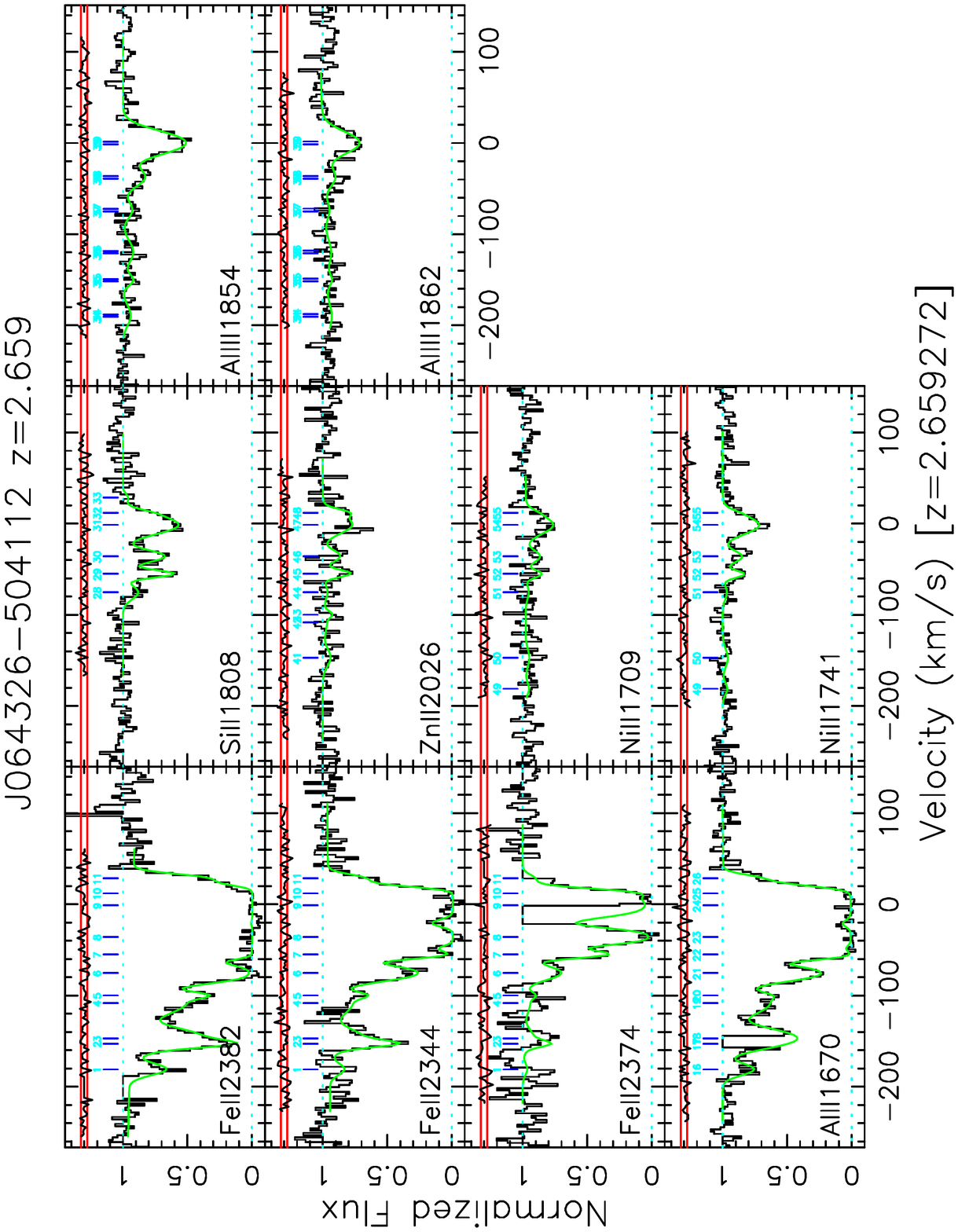}
\par\end{centering}

\caption[Fit for the $z=2.659$ absorber toward J064326$-$504112]{Many-multiplet fit for the $z=2.659$ absorber toward J064326$-$504112.}
\end{figure}
\begin{figure}[H]
\noindent \begin{centering}
\includegraphics[bb=34bp 58bp 554bp 738bp,clip,width=1\textwidth]{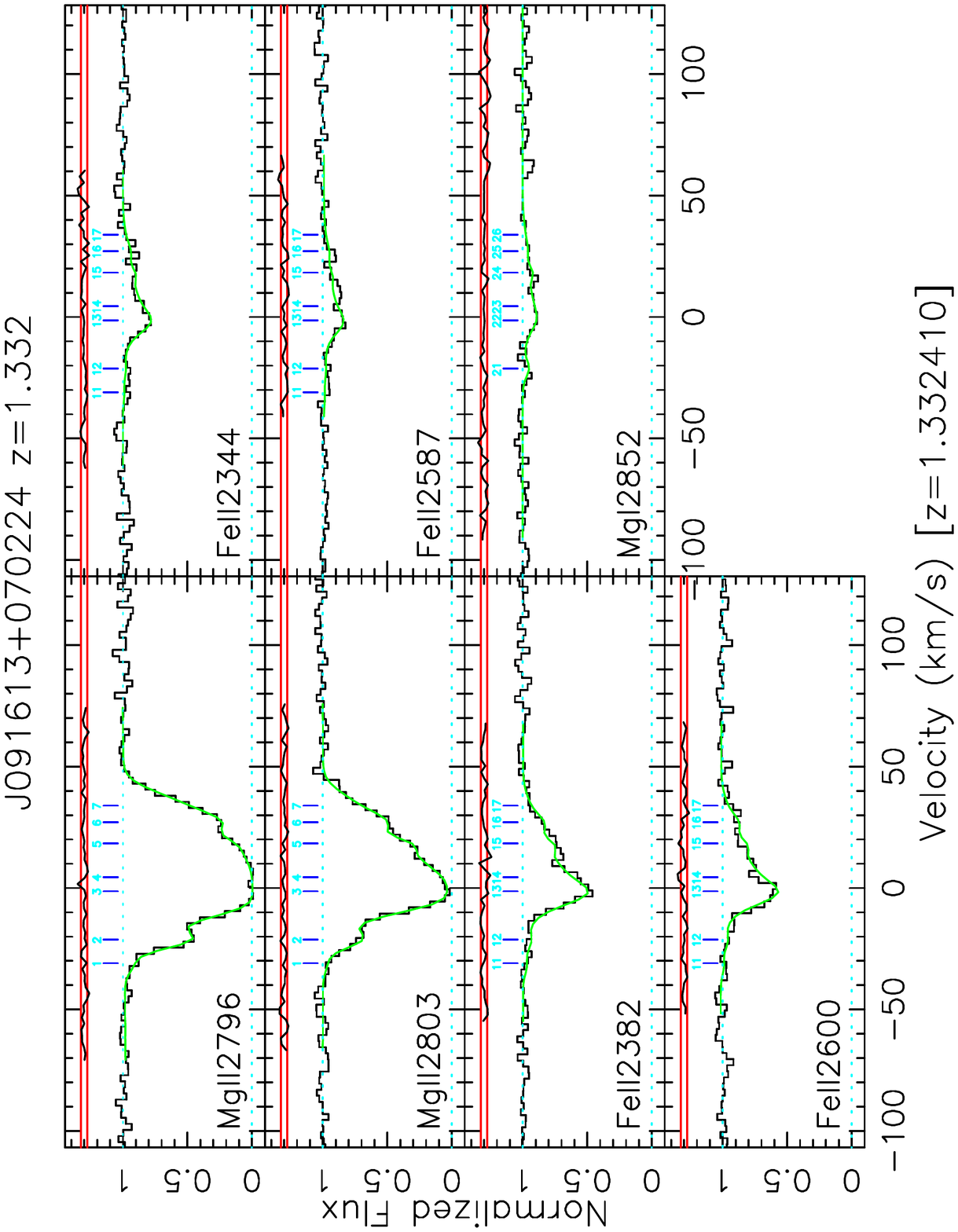}
\par\end{centering}

\caption[Fit for the $z=1.332$ absorber toward J091613+070224]{Many-multiplet fit for the $z=1.332$ absorber toward J091613+070224.}
\end{figure}
\begin{figure}[H]
\noindent \begin{centering}
\includegraphics[bb=34bp 58bp 554bp 738bp,clip,width=1\textwidth]{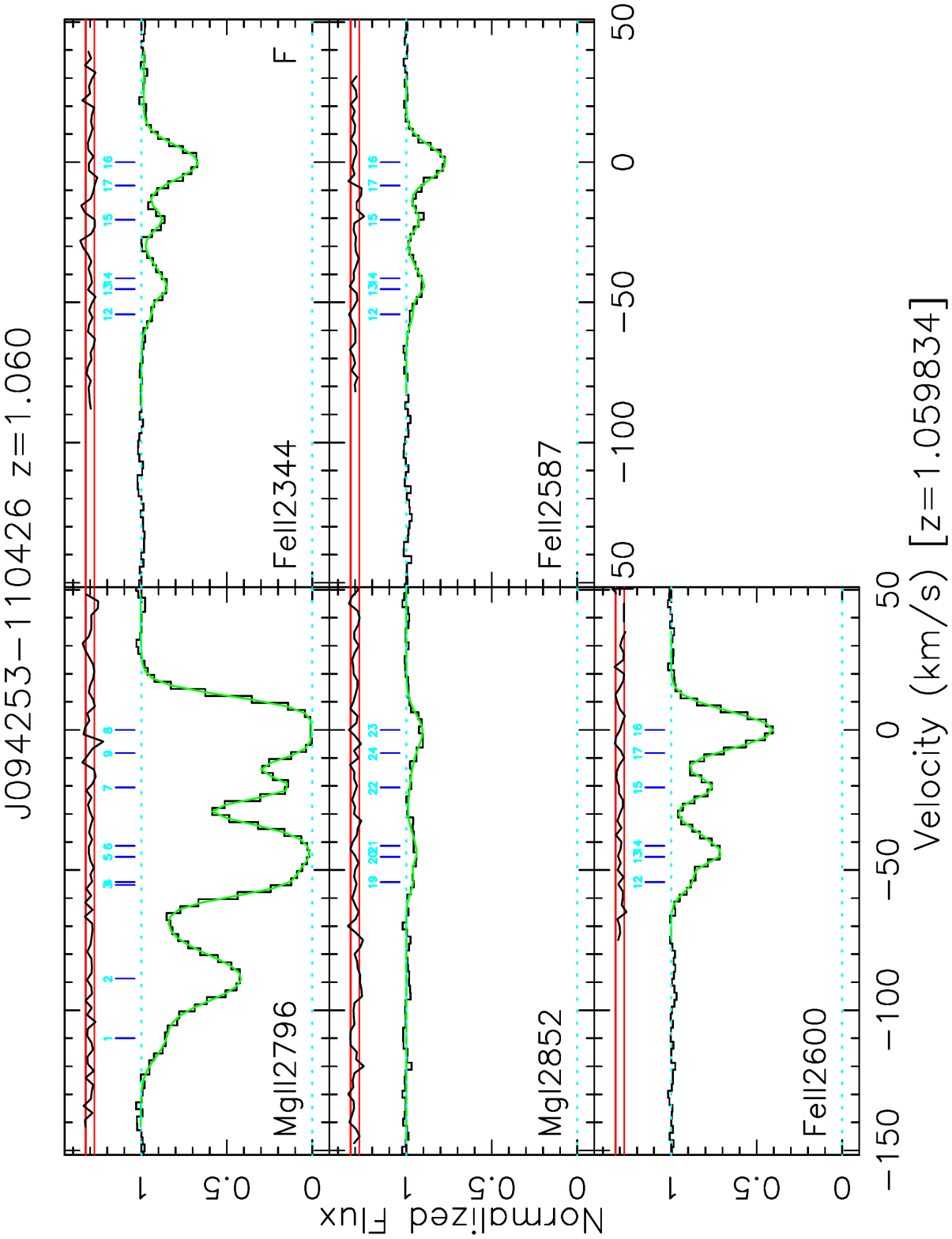}
\par\end{centering}

\caption[Fit for the $z=1.060$ absorber toward J094253$-$110426]{Many-multiplet fit for the $z=1.060$ absorber toward J094253$-$110426.}
\end{figure}
\begin{figure}[H]
\noindent \begin{centering}
\includegraphics[bb=34bp 58bp 554bp 738bp,clip,width=1\textwidth]{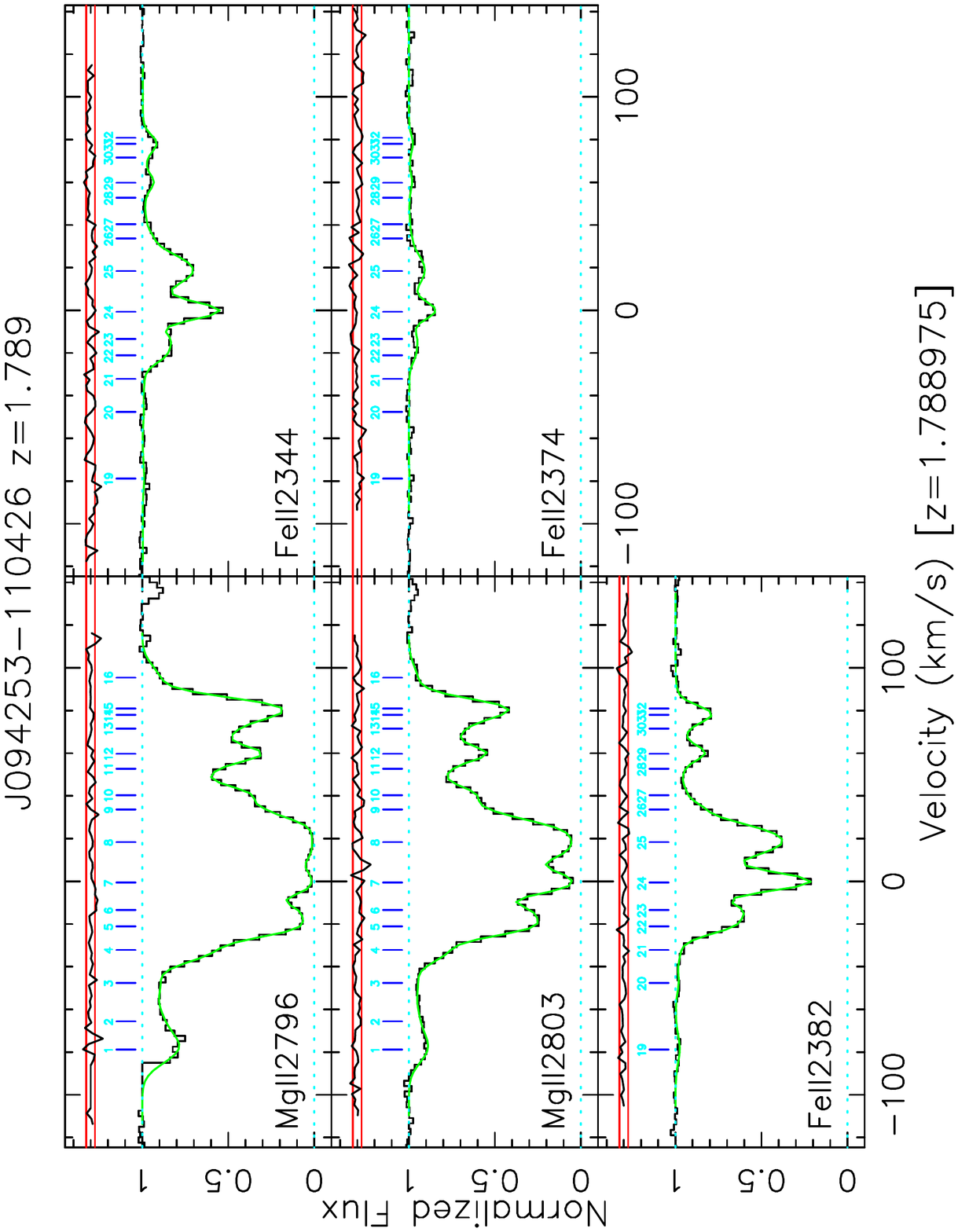}
\par\end{centering}

\caption[Fit for the $z=1.789$ absorber toward J094253$-$110426]{Many-multiplet fit for the $z=1.789$ absorber toward J094253$-$110426.}
\end{figure}
\begin{figure}[H]
\noindent \begin{centering}
\includegraphics[bb=34bp 58bp 554bp 738bp,clip,width=1\textwidth]{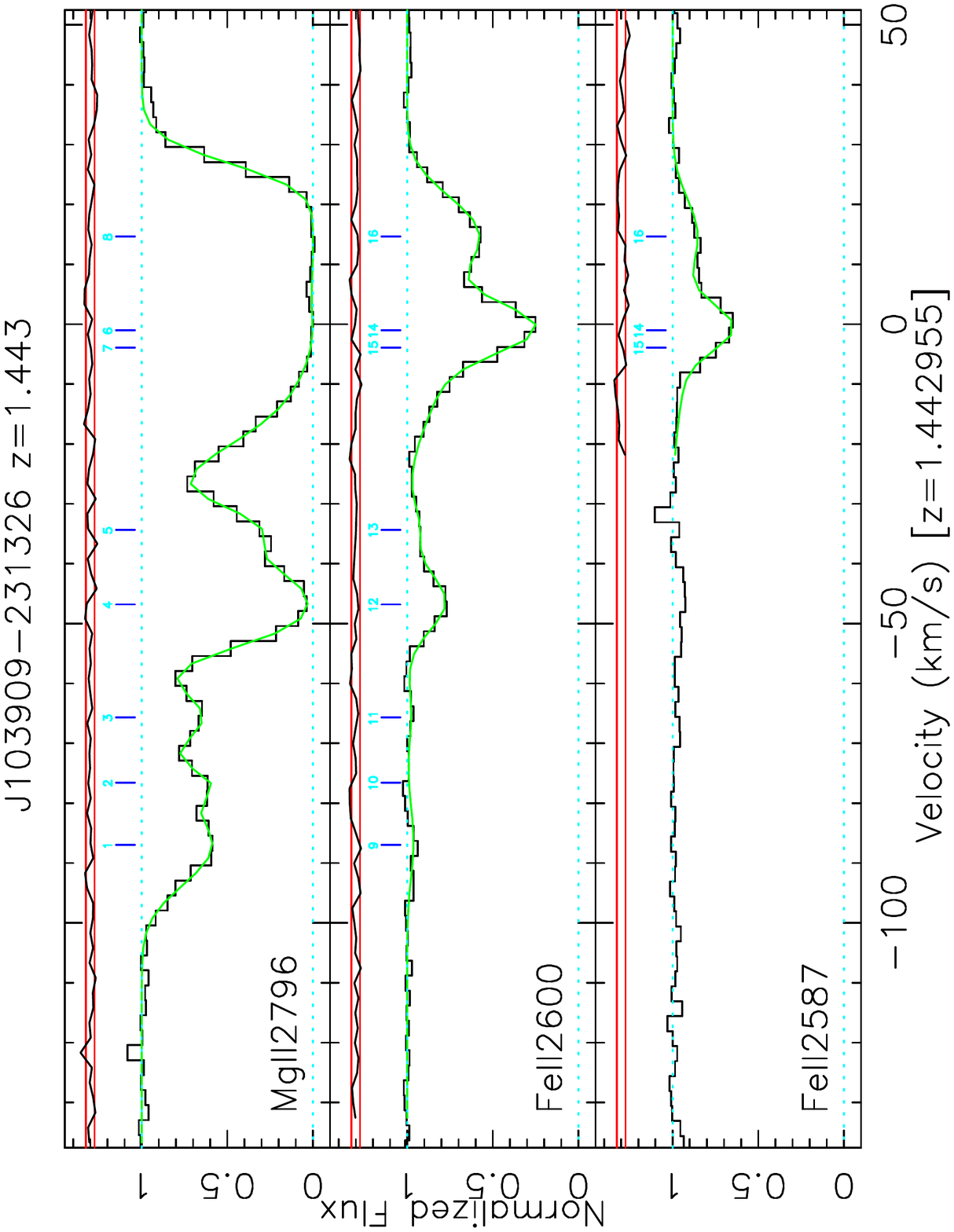}
\par\end{centering}

\caption[Fit for the $z=1.443$ absorber toward J103909$-$231326]{Many-multiplet fit for the $z=1.443$ absorber toward J103909$-$231326.}
\end{figure}
\begin{figure}[H]
\noindent \begin{centering}
\includegraphics[bb=34bp 58bp 554bp 738bp,clip,width=1\textwidth]{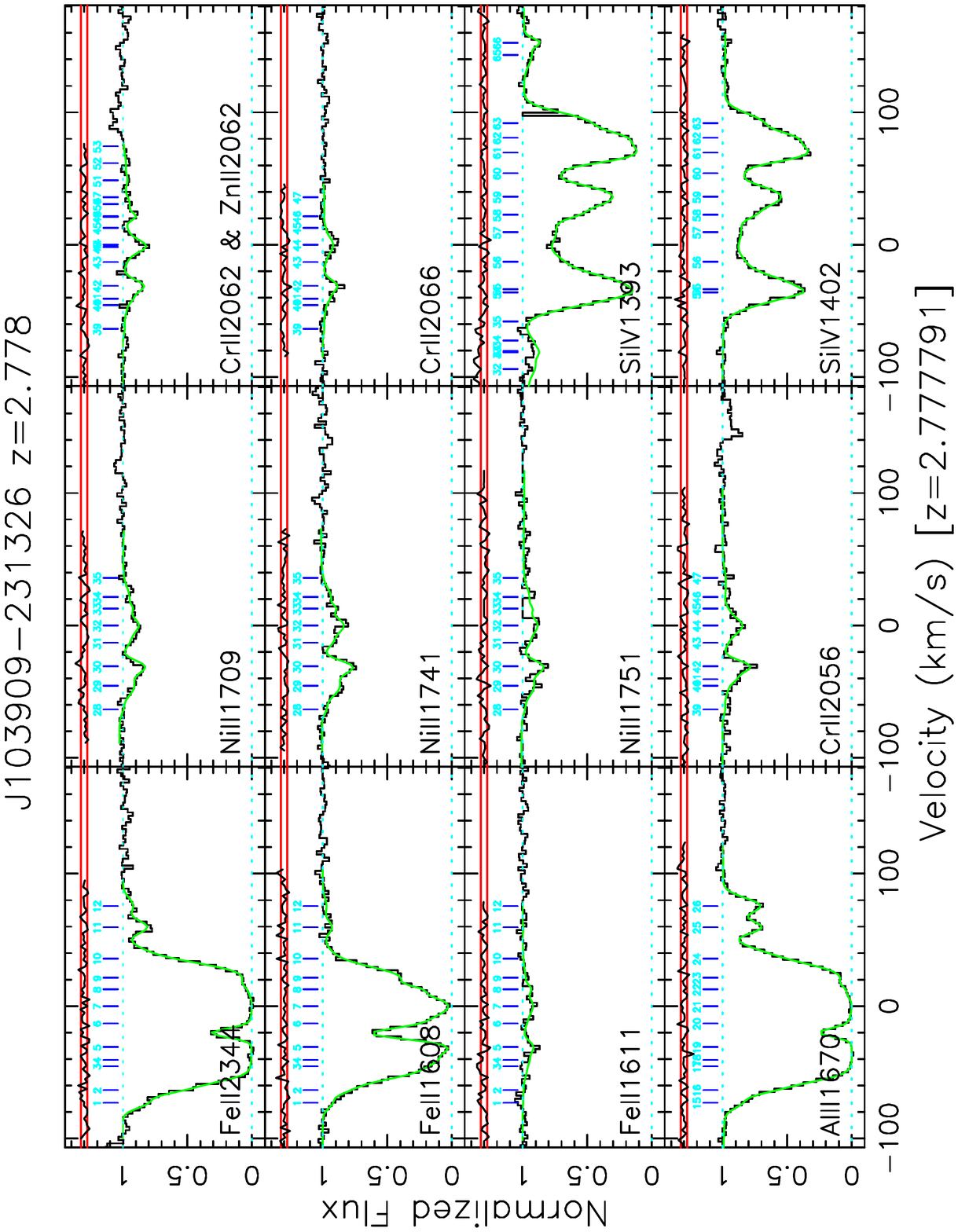}
\par\end{centering}

\caption[Fit for the $z=2.778$ absorber toward J103909$-$231326]{Many-multiplet fit for the $z=2.778$ absorber toward J103909$-$231326.}
\end{figure}
\begin{figure}[H]
\noindent \begin{centering}
\includegraphics[bb=34bp 58bp 554bp 738bp,clip,width=1\textwidth]{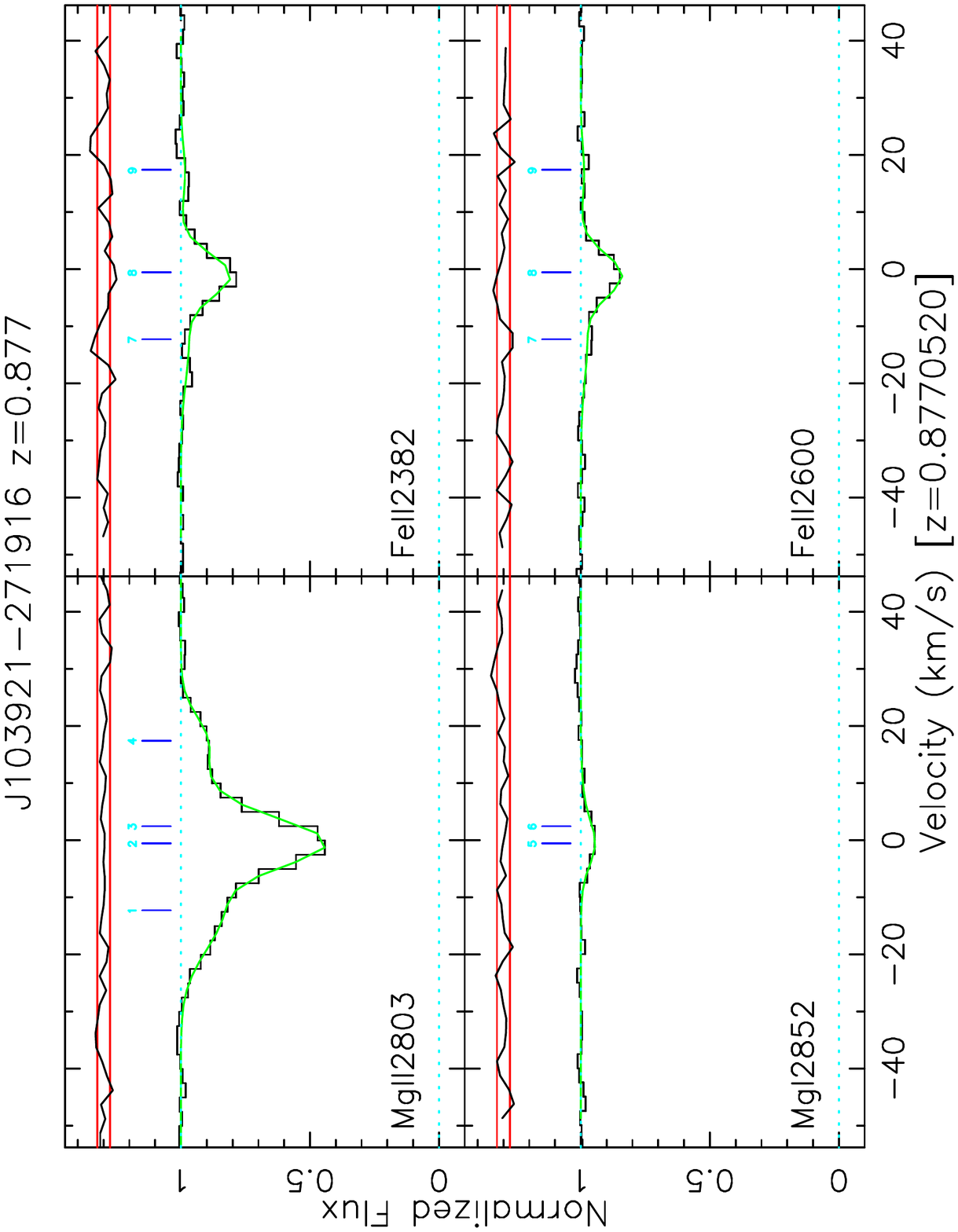}
\par\end{centering}

\caption[Fit for the $z=0.877$ absorber toward J103921$-$271916]{Many-multiplet fit for the $z=0.877$ absorber toward J103921$-$271916.}
\end{figure}
\begin{figure}[H]
\noindent \begin{centering}
\includegraphics[bb=34bp 58bp 554bp 738bp,clip,width=1\textwidth]{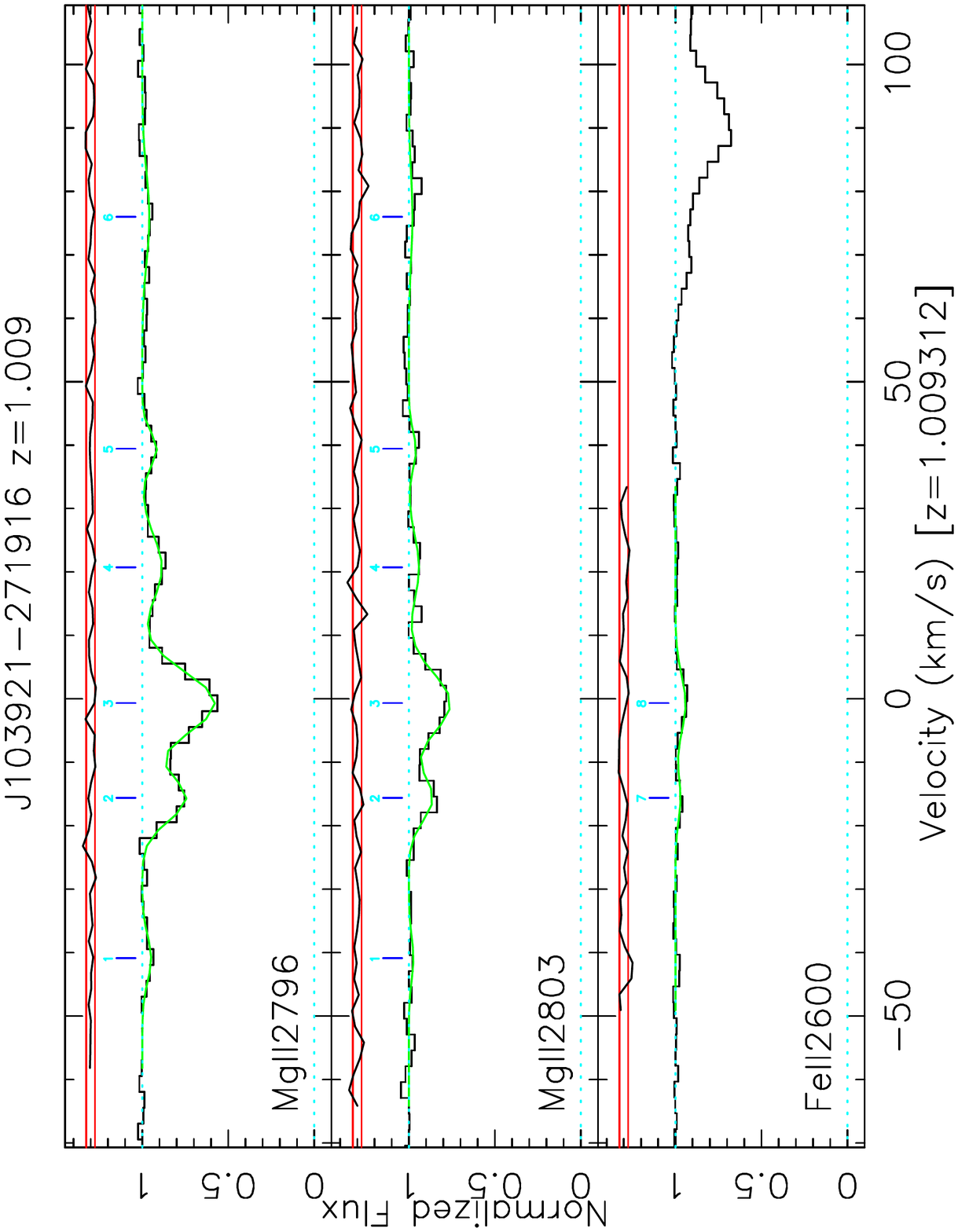}
\par\end{centering}

\caption[Fit for the $z=1.009$ absorber toward J103921$-$271916]{Many-multiplet fit for the $z=1.009$ absorber toward J103921$-$271916.}
\end{figure}
\begin{figure}[H]
\noindent \begin{centering}
\includegraphics[bb=34bp 58bp 554bp 738bp,clip,width=1\textwidth]{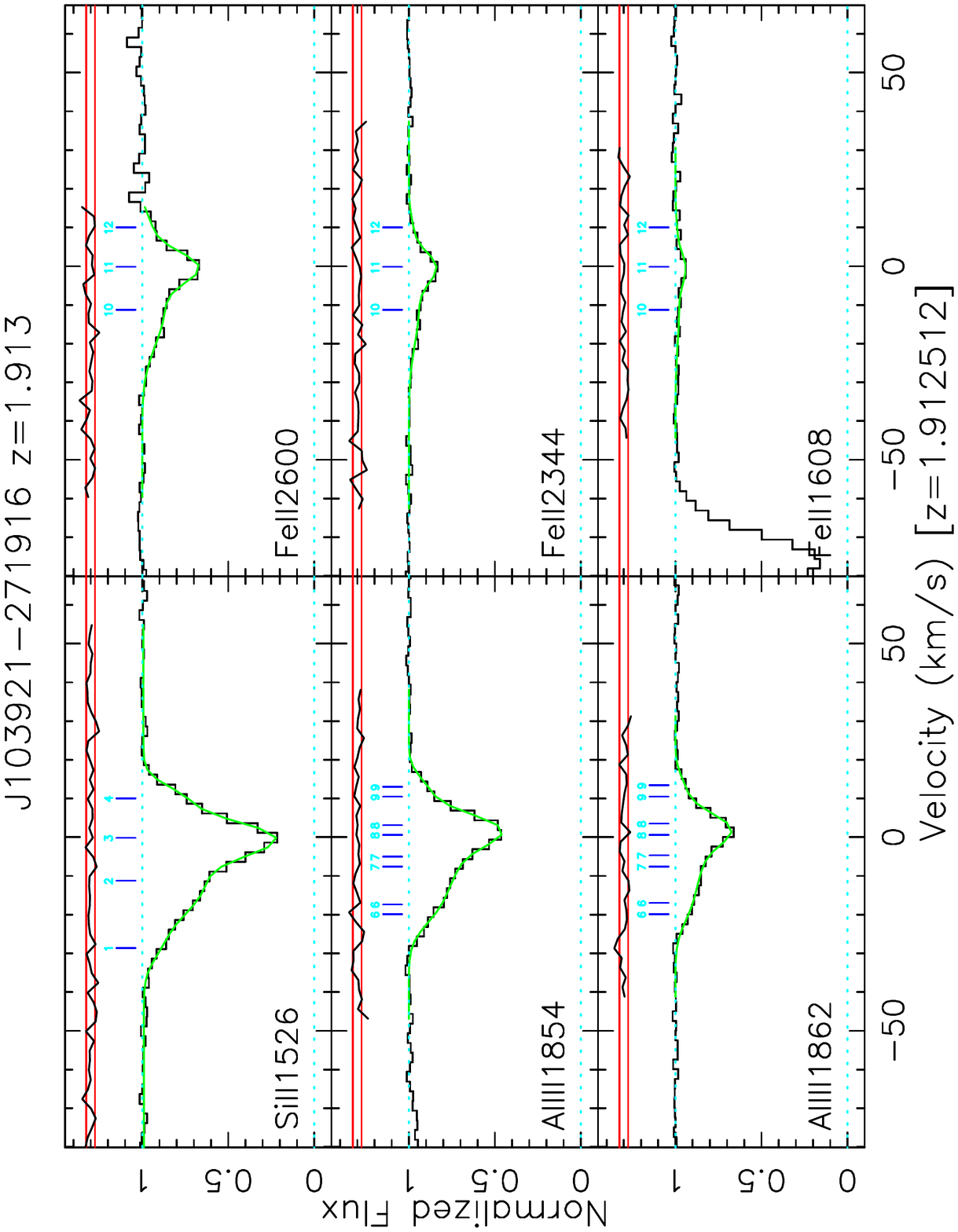}
\par\end{centering}

\caption[Fit for the $z=1.913$ absorber toward J103921$-$271916]{Many-multiplet fit for the $z=1.913$ absorber toward J103921$-$271916.}
\end{figure}
\begin{figure}[H]
\noindent \begin{centering}
\includegraphics[bb=34bp 58bp 554bp 738bp,clip,width=1\textwidth]{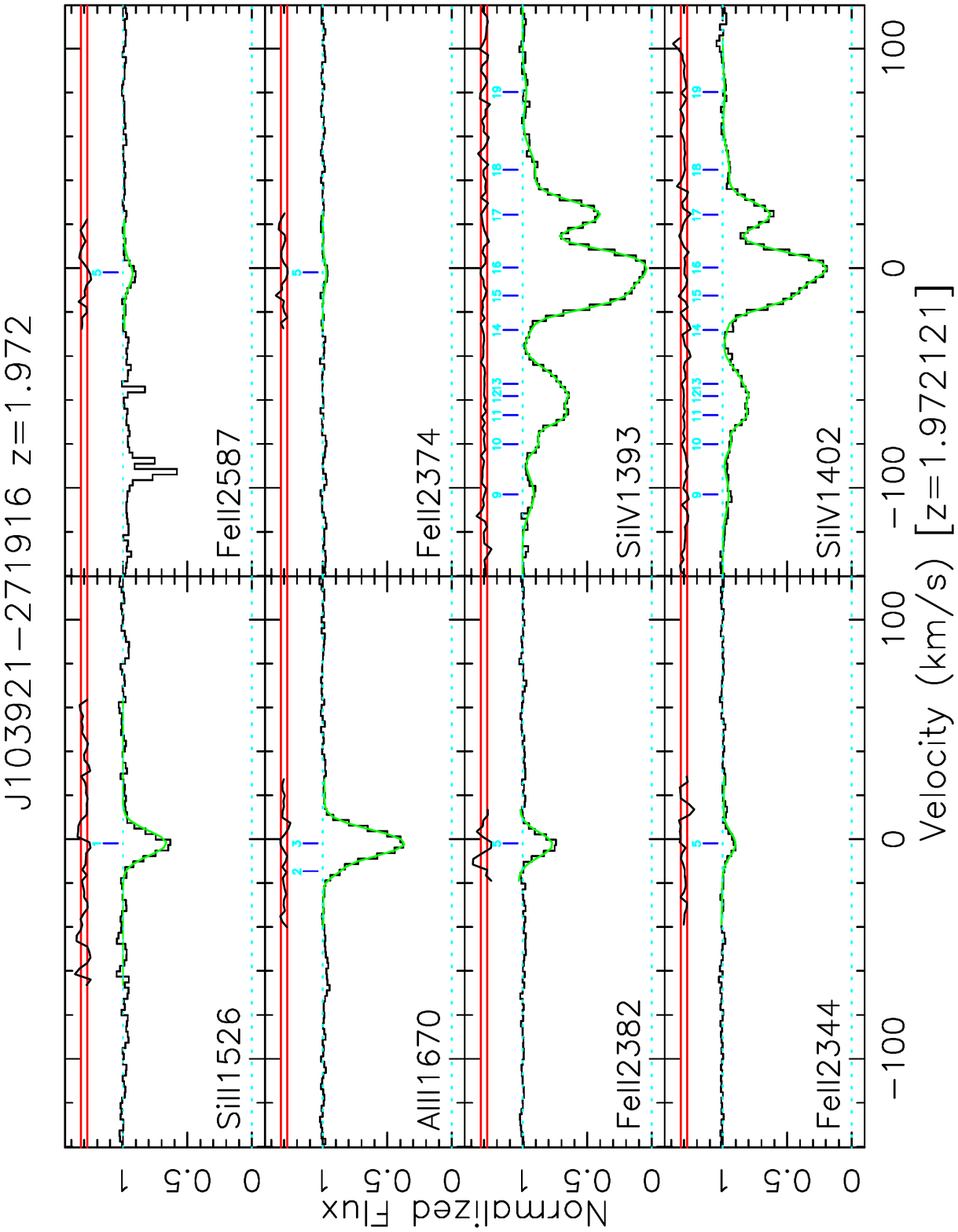}
\par\end{centering}

\caption[Fit for the $z=1.972$ absorber toward J103921$-$271916]{Many-multiplet fit for the $z=1.972$ absorber toward J103921$-$271916.}
\end{figure}
\begin{figure}[H]
\noindent \begin{centering}
\includegraphics[bb=34bp 58bp 554bp 738bp,clip,width=1\textwidth]{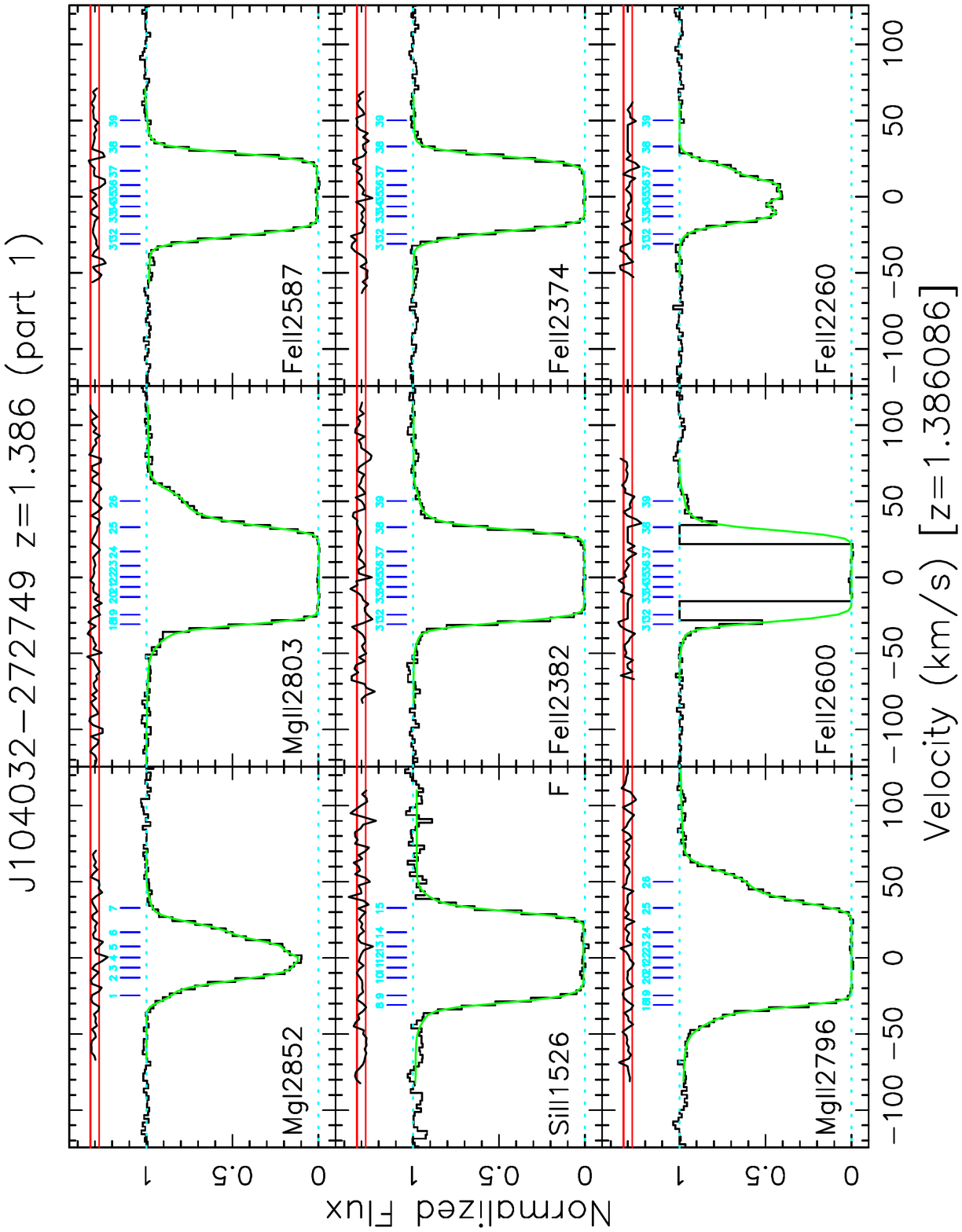}
\par\end{centering}

\caption[Fit for the $z=1.386$ absorber toward J104032$-$272749 (part 1)]{Many-multiplet fit for the $z=1.386$ absorber toward J104032$-$272749 (part 1).}
\end{figure}
\begin{figure}[H]
\noindent \begin{centering}
\includegraphics[bb=34bp 58bp 554bp 738bp,clip,width=1\textwidth]{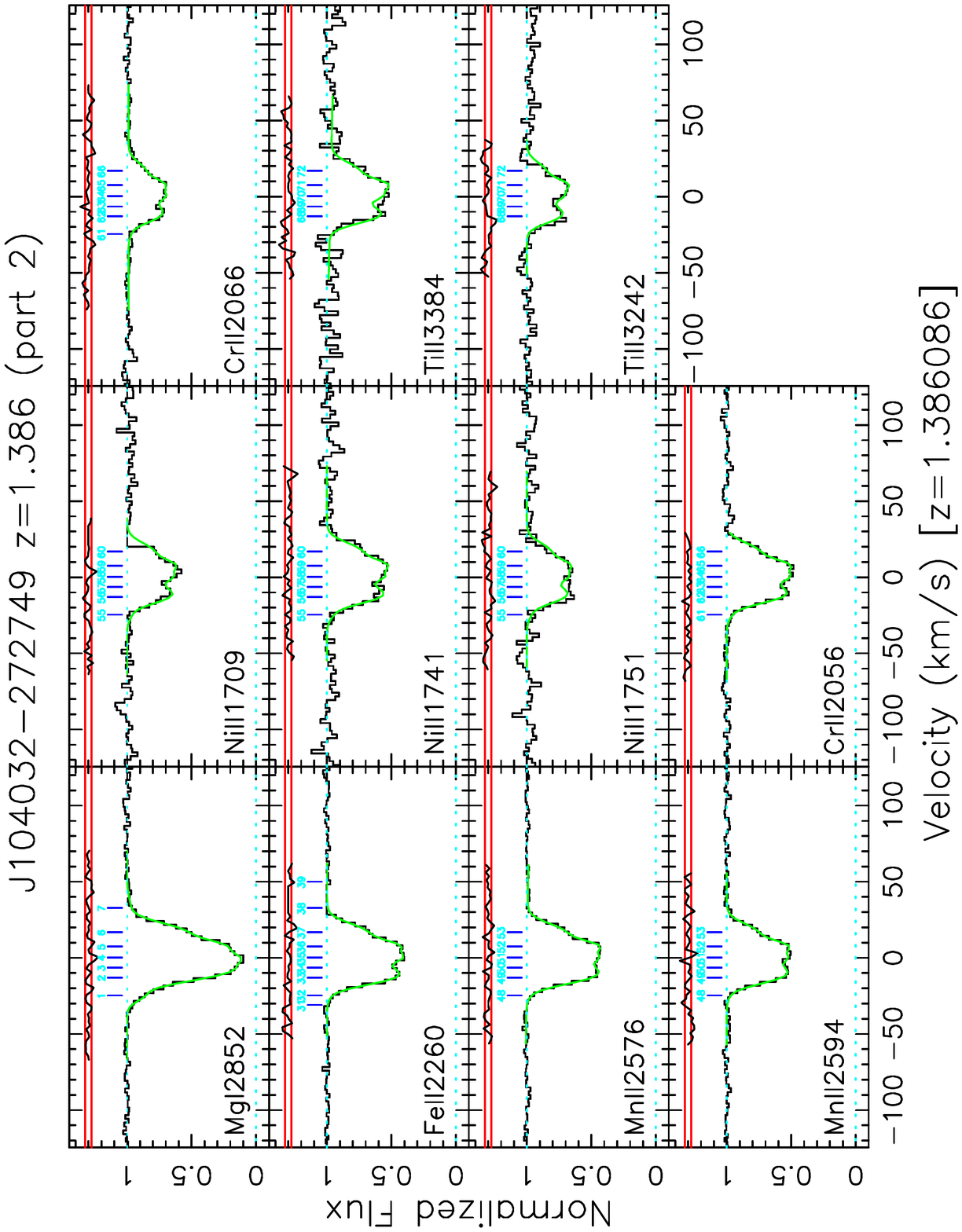}
\par\end{centering}

\caption[Fit for the $z=1.386$ absorber toward J104032$-$272749 (part 2)]{Many-multiplet fit for the $z=1.386$ absorber toward J104032$-$272749 (part 2).}
\end{figure}
\begin{figure}[H]
\noindent \begin{centering}
\includegraphics[bb=34bp 58bp 554bp 738bp,clip,width=1\textwidth]{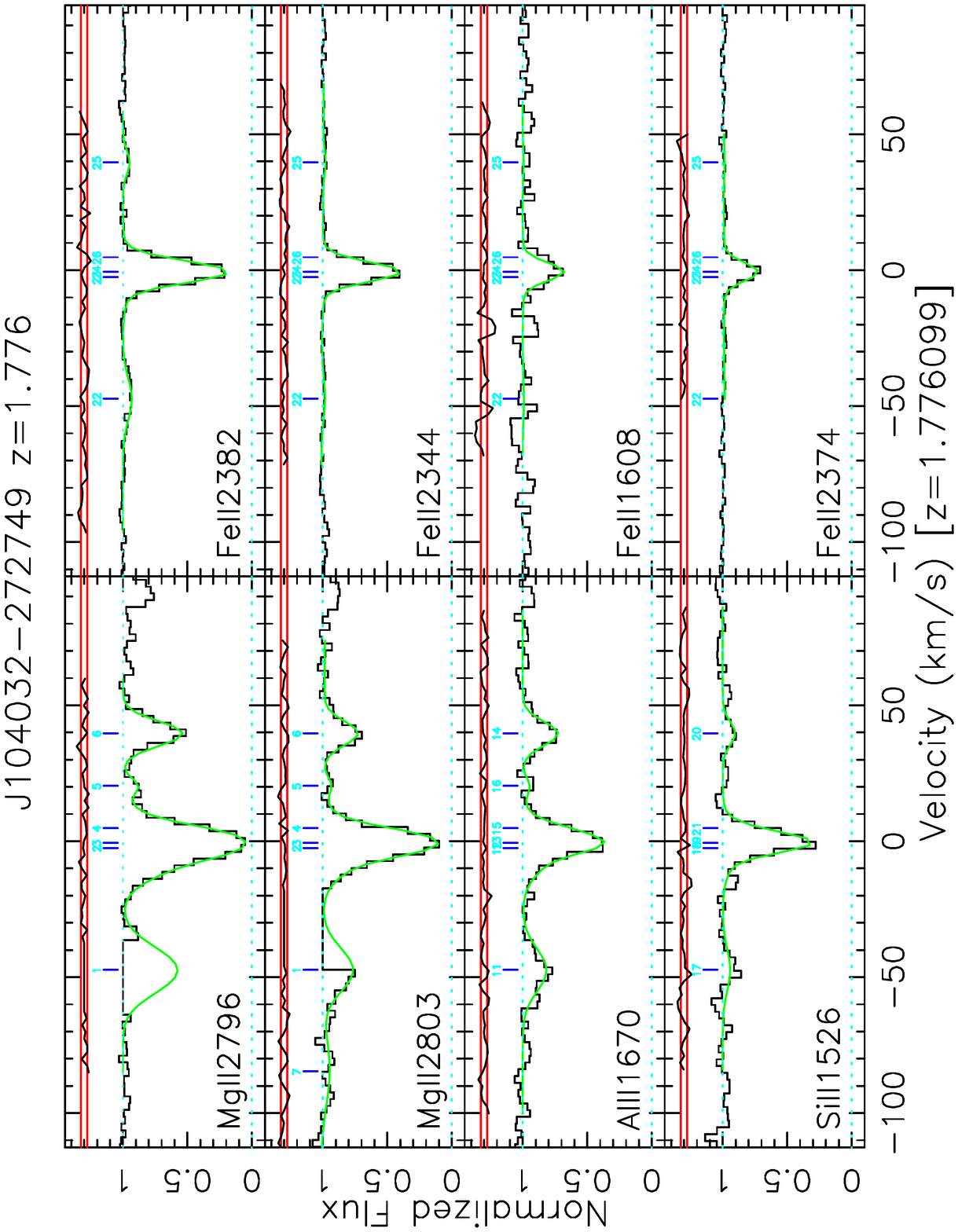}
\par\end{centering}

\caption[Fit for the $z=1.776$ absorber toward J104032$-$272749]{Many-multiplet fit for the $z=1.776$ absorber toward J104032$-$272749.}
\end{figure}
\begin{figure}[H]
\noindent \begin{centering}
\includegraphics[bb=34bp 58bp 554bp 738bp,clip,width=1\textwidth]{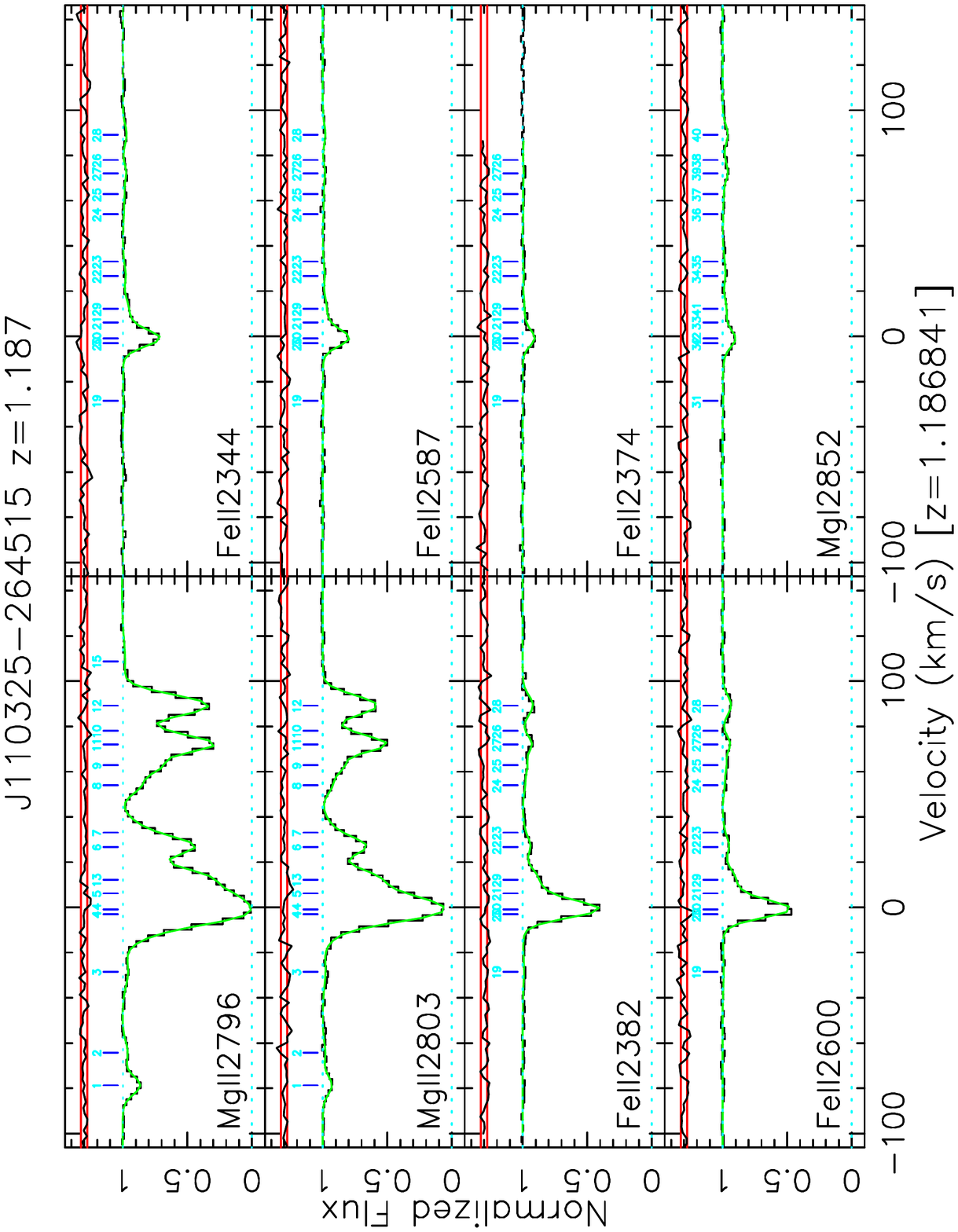}
\par\end{centering}

\caption[Fit for the $z=1.187$ absorber toward J110325$-$264515]{Many-multiplet fit for the $z=1.187$ absorber toward J110325$-$264515.}
\end{figure}
\begin{figure}[H]
\noindent \begin{centering}
\includegraphics[bb=34bp 58bp 554bp 738bp,clip,width=1\textwidth]{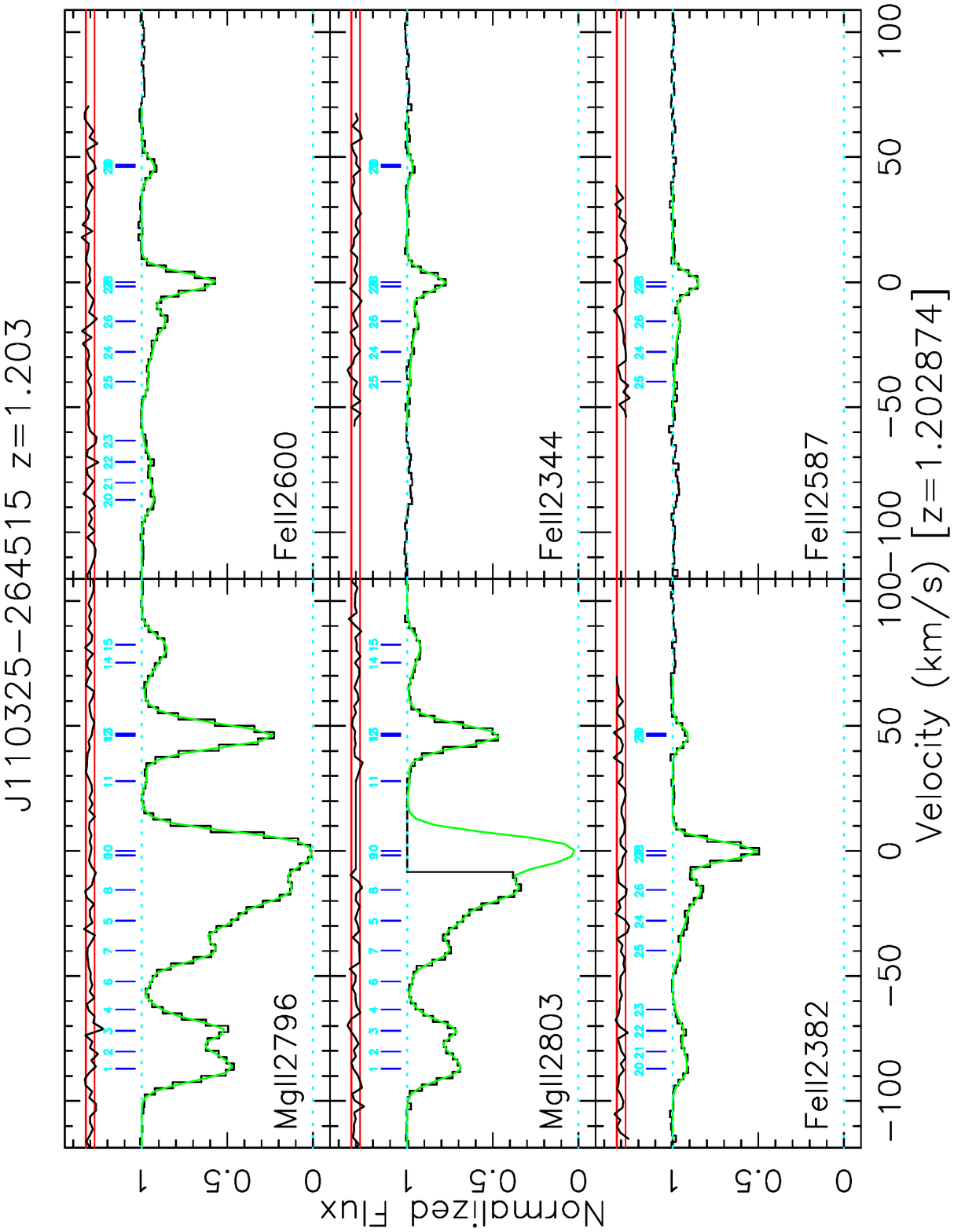}
\par\end{centering}

\caption[Fit for the $z=1.203$ absorber toward J110325$-$264515]{Many-multiplet fit for the $z=1.203$ absorber toward J110325$-$264515.}
\end{figure}
\begin{figure}[H]
\noindent \begin{centering}
\includegraphics[bb=34bp 58bp 554bp 738bp,clip,width=1\textwidth]{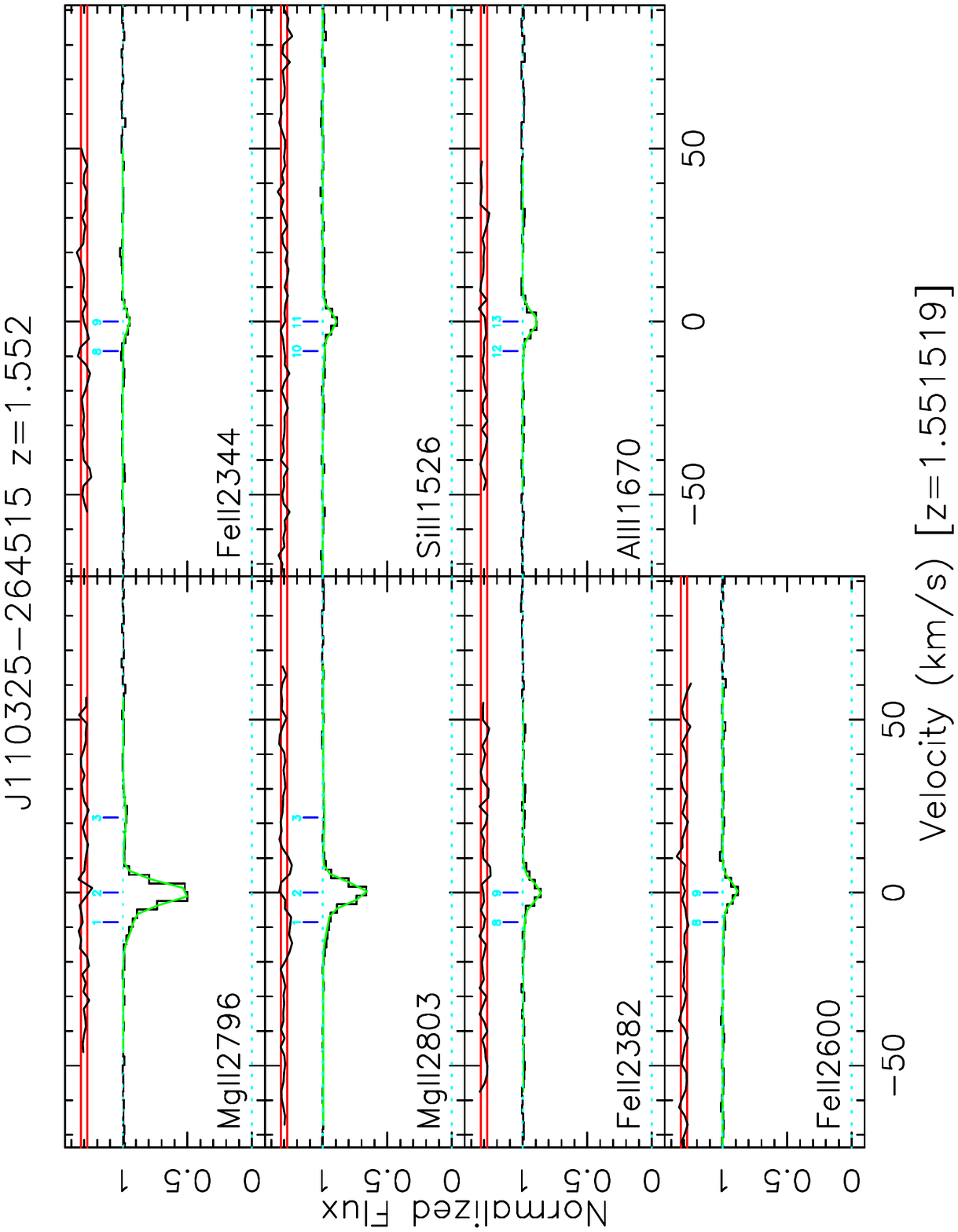}
\par\end{centering}

\caption[Fit for the $z=1.552$ absorber toward J110325$-$264515]{Many-multiplet fit for the $z=1.552$ absorber toward J110325$-$264515.}
\end{figure}
\begin{figure}[H]
\noindent \begin{centering}
\includegraphics[bb=34bp 58bp 554bp 738bp,clip,width=1\textwidth]{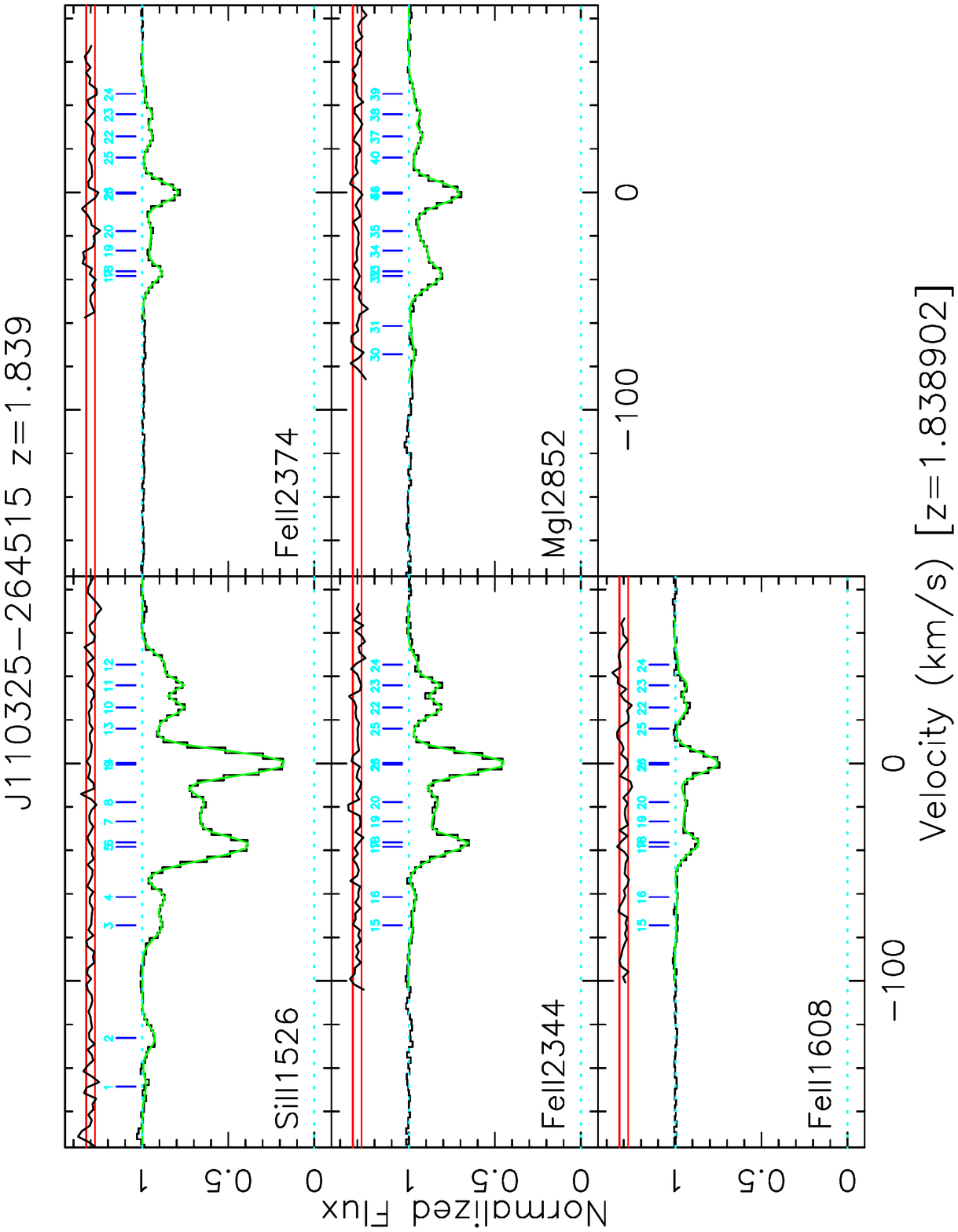}
\par\end{centering}

\caption[Fit for the $z=1.839$ absorber toward J110325$-$264515]{Many-multiplet fit for the $z=1.839$ absorber toward J110325$-$264515.}
\end{figure}
\begin{figure}[H]
\noindent \begin{centering}
\includegraphics[bb=34bp 58bp 554bp 738bp,clip,width=1\textwidth]{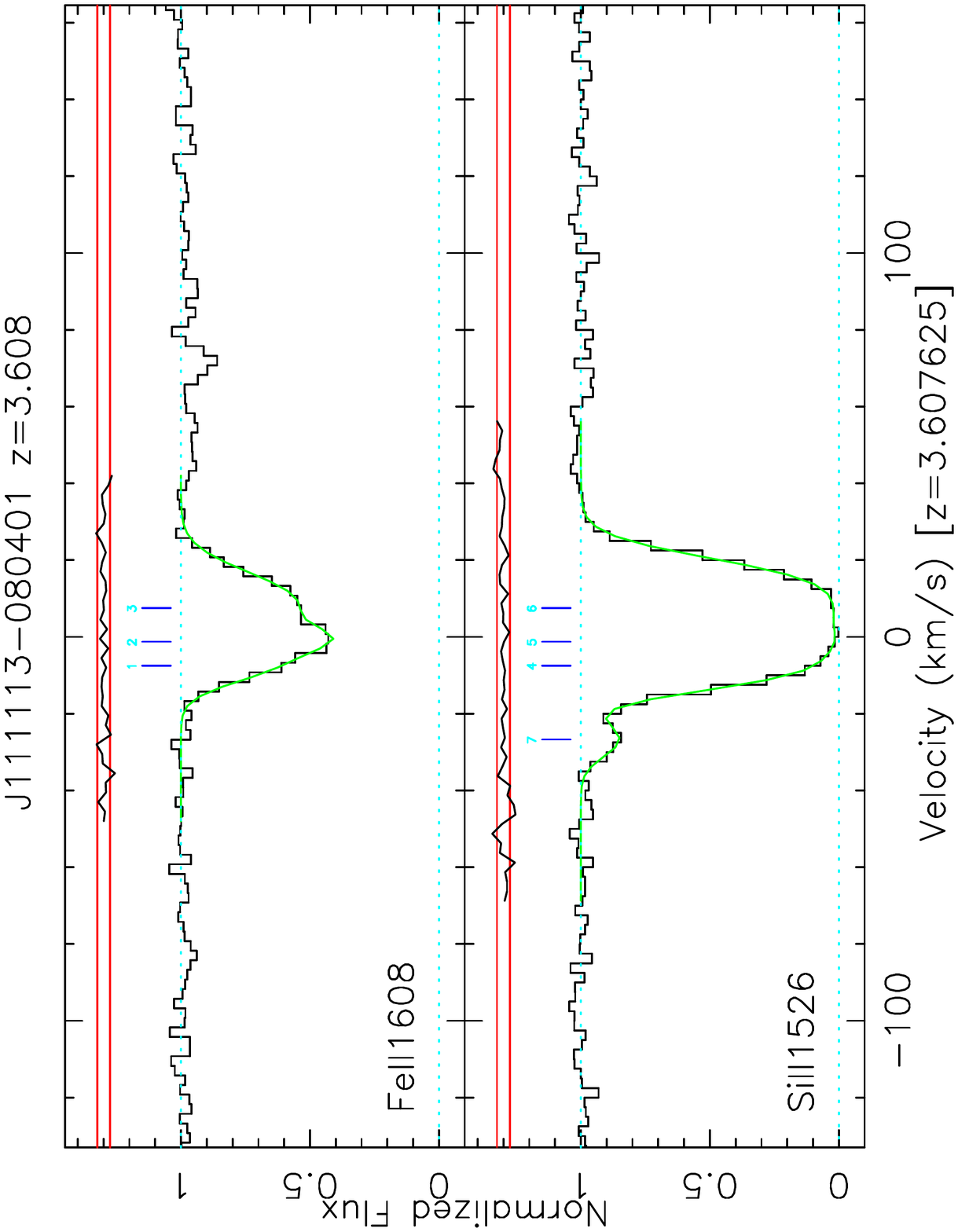}
\par\end{centering}

\caption[Fit for the $z=3.608$ absorber toward J111113$-$080401]{Many-multiplet fit for the $z=3.608$ absorber toward J111113$-$080401.}
\end{figure}
\begin{figure}[H]
\noindent \begin{centering}
\includegraphics[bb=34bp 58bp 554bp 738bp,clip,width=1\textwidth]{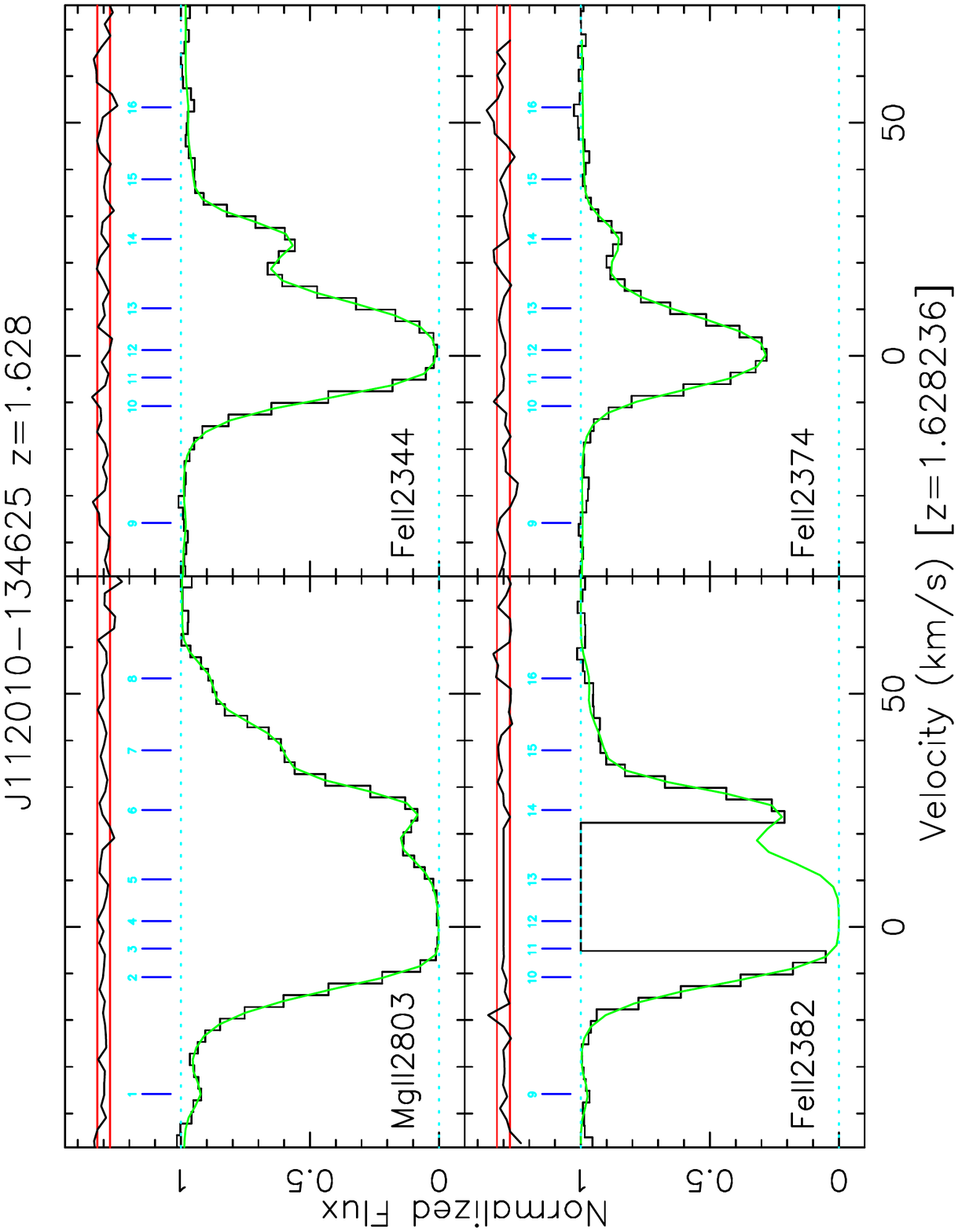}
\par\end{centering}

\caption[Fit for the $z=1.628$ absorber toward J112040$-$134625]{Many-multiplet fit for the $z=1.628$ absorber toward J112040$-$134625.}
\end{figure}
\begin{figure}[H]
\noindent \begin{centering}
\includegraphics[bb=34bp 58bp 554bp 738bp,clip,width=1\textwidth]{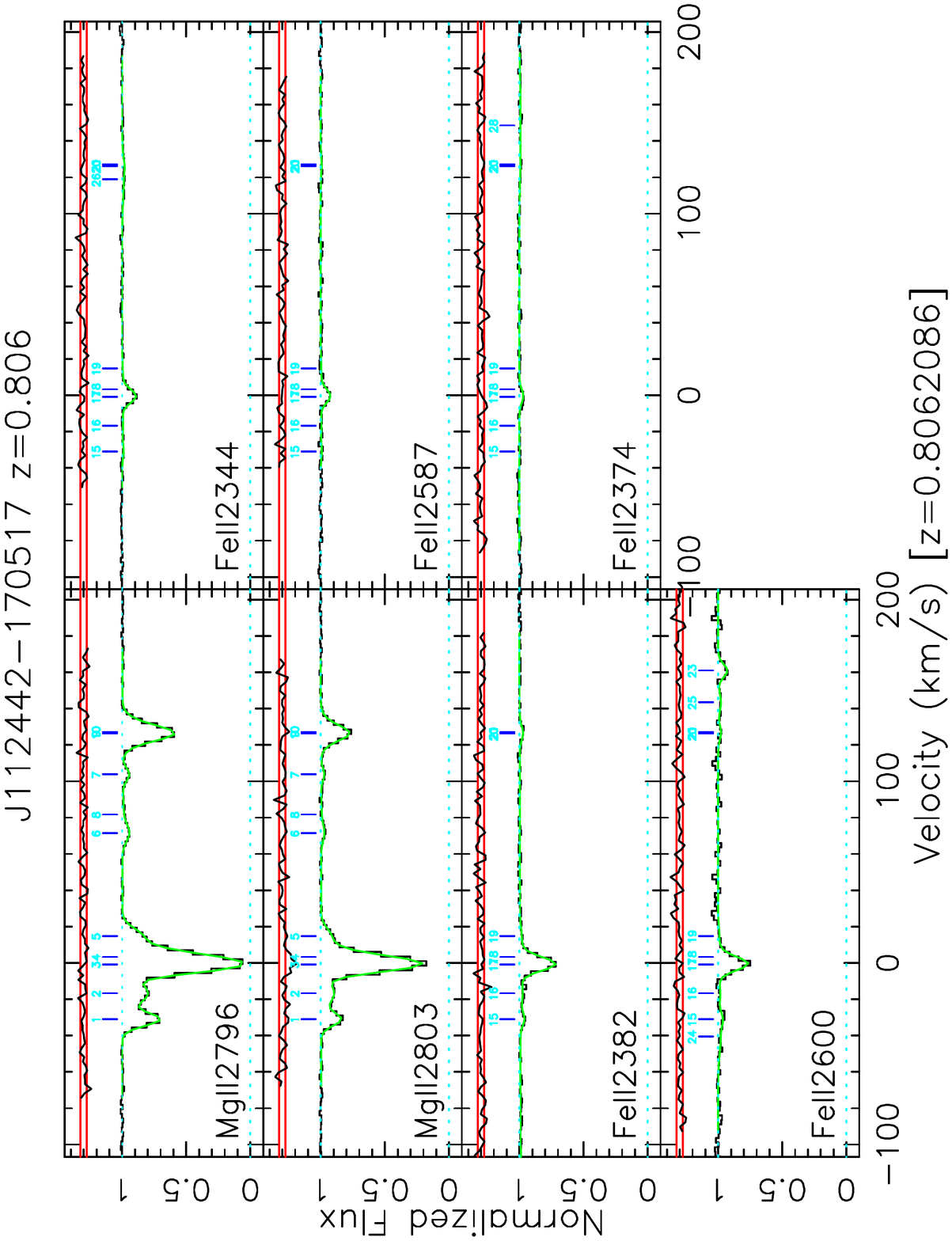}
\par\end{centering}

\caption[Fit for the $z=0.806$ absorber toward J112442$-$170517]{Many-multiplet fit for the $z=0.806$ absorber toward J112442$-$170517.}
\end{figure}
\begin{figure}[H]
\noindent \begin{centering}
\includegraphics[bb=34bp 58bp 554bp 738bp,clip,width=1\textwidth]{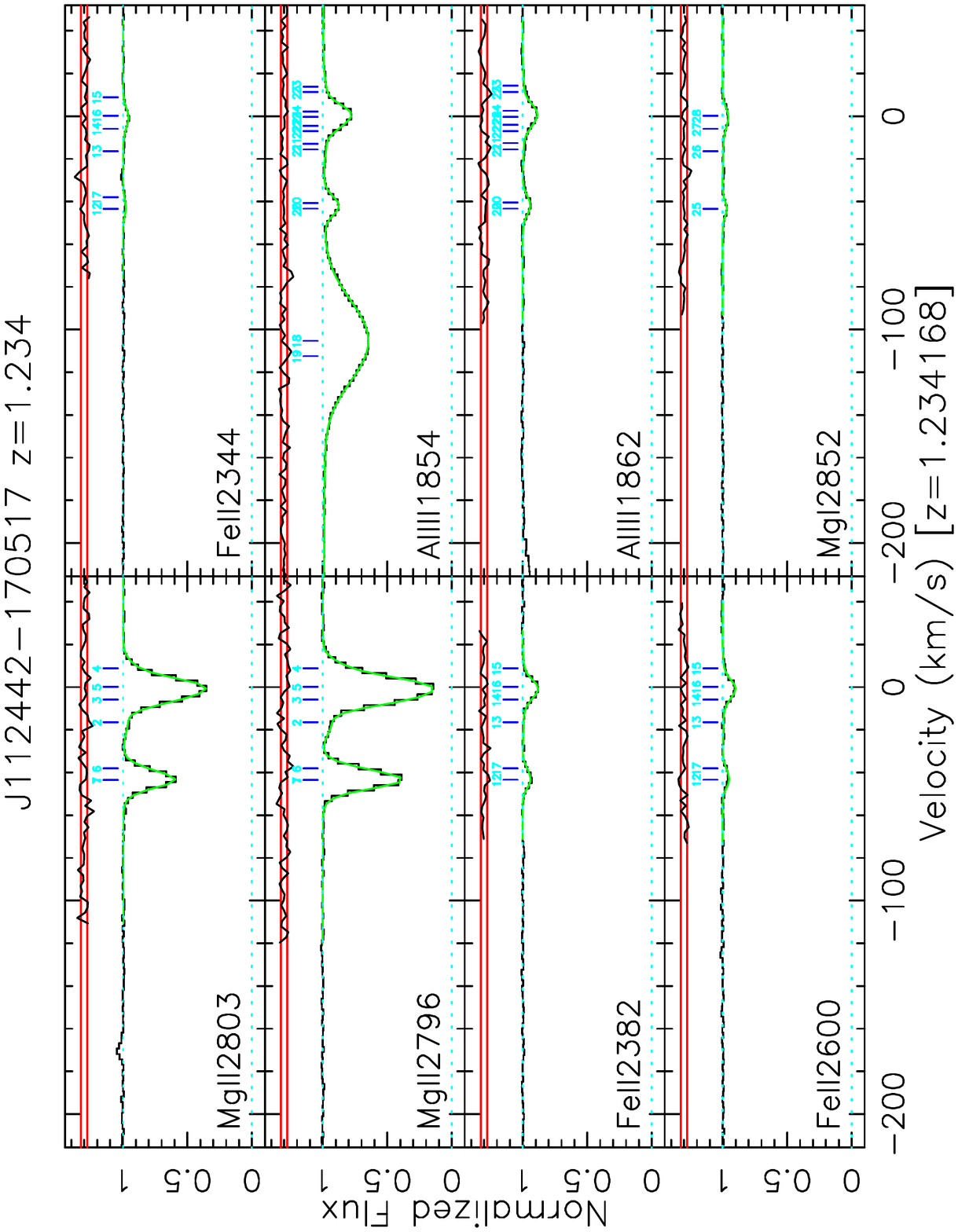}
\par\end{centering}

\caption[Fit for the $z=1.234$ absorber toward J112442$-$170517]{Many-multiplet fit for the $z=1.234$ absorber toward J112442$-$170517.}
\end{figure}

\begin{figure}[H]
\noindent \begin{centering}
\includegraphics[bb=34bp 58bp 554bp 738bp,clip,width=1\textwidth]{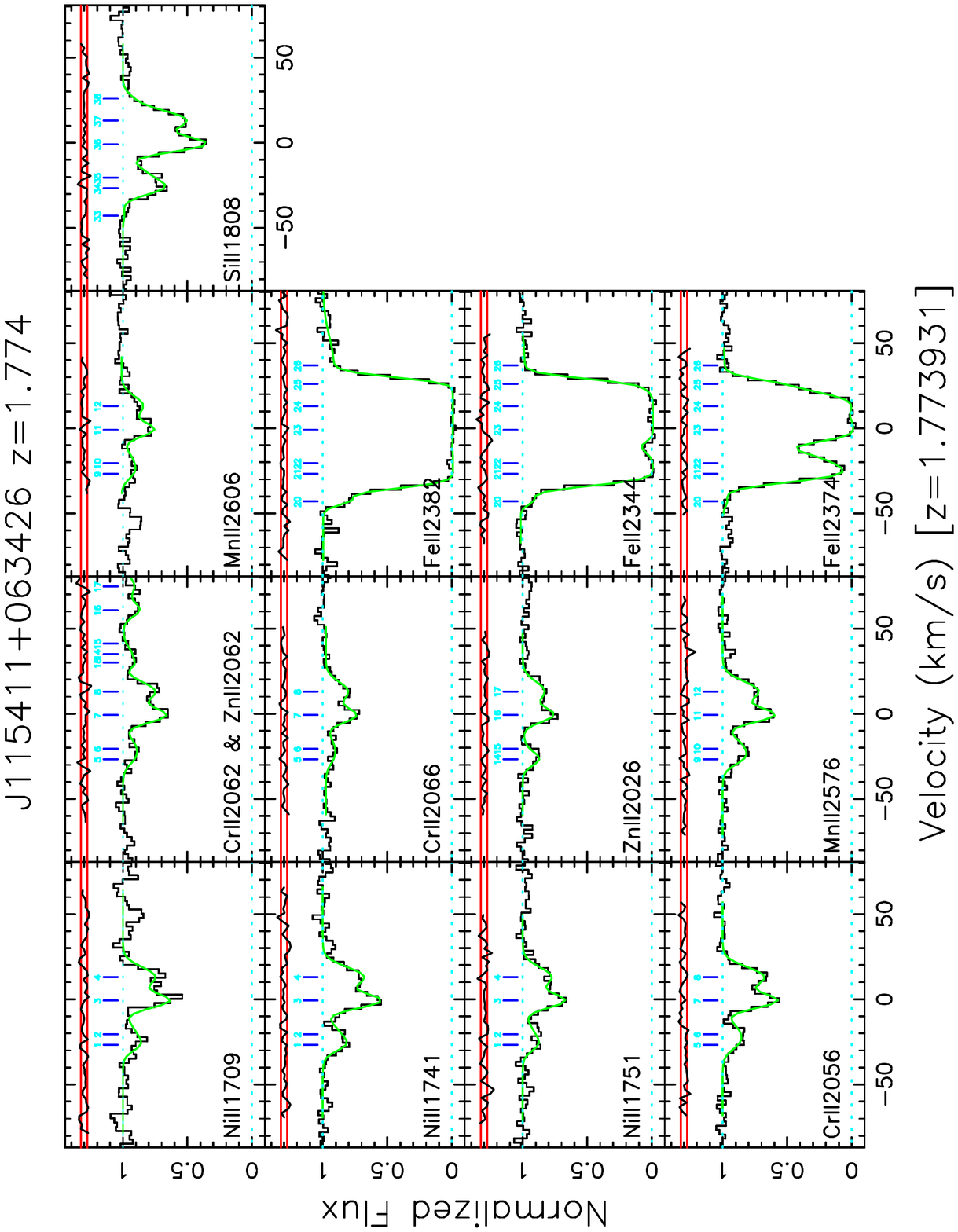}
\par\end{centering}

\caption[Fit for the $z=1.774$ absorber toward J115411+063426]{Many-multiplet fit for the $z=1.774$ absorber toward J115411+063426.}
\end{figure}
\begin{figure}[H]
\noindent \begin{centering}
\includegraphics[bb=34bp 58bp 554bp 738bp,clip,width=1\textwidth]{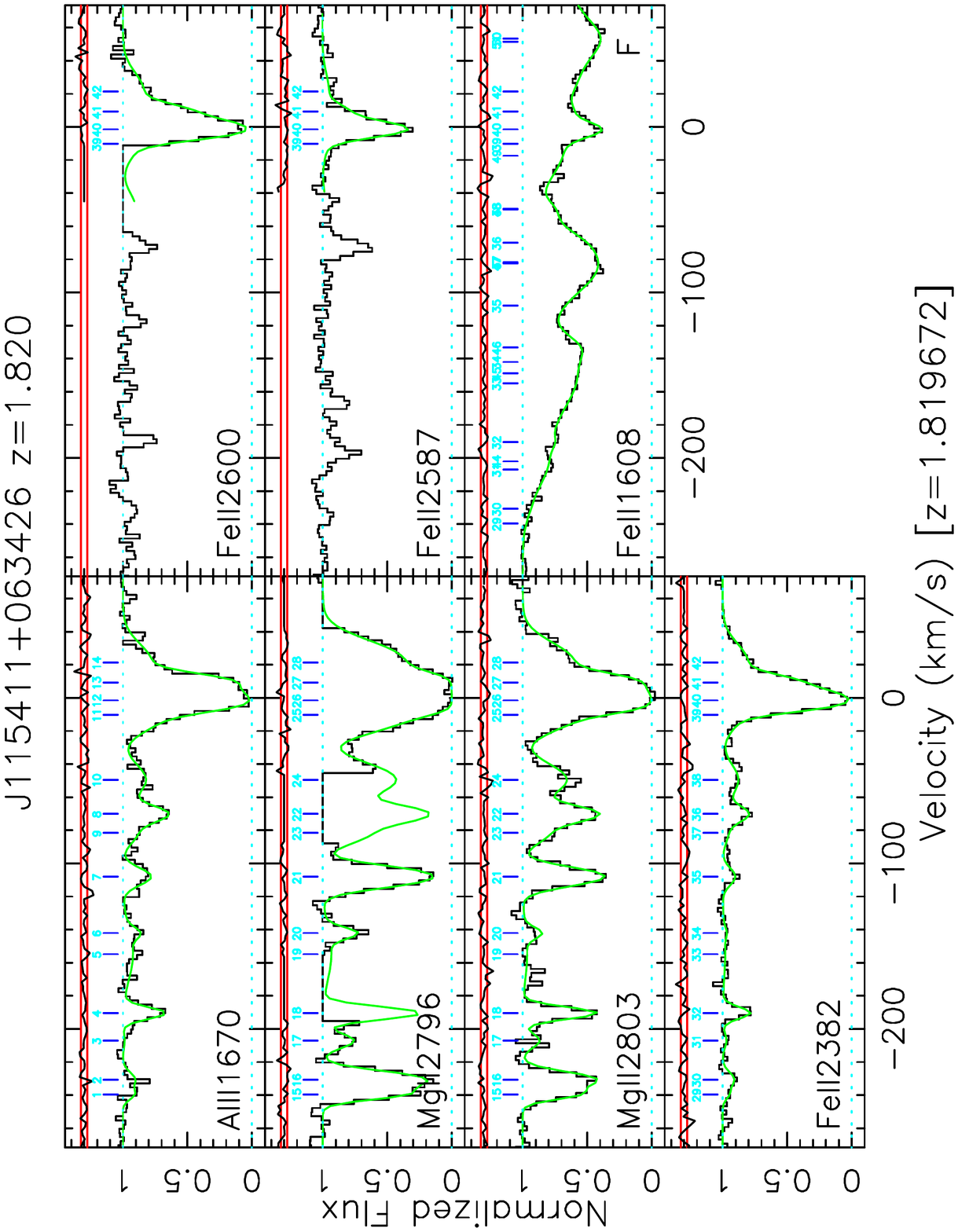}
\par\end{centering}

\caption[Fit for the $z=1.820$ absorber toward J115411+063426]{Many-multiplet fit for the $z=1.820$ absorber toward J115411+063426.}
\end{figure}
\begin{figure}[H]
\noindent \begin{centering}
\includegraphics[bb=34bp 58bp 554bp 738bp,clip,width=1\textwidth]{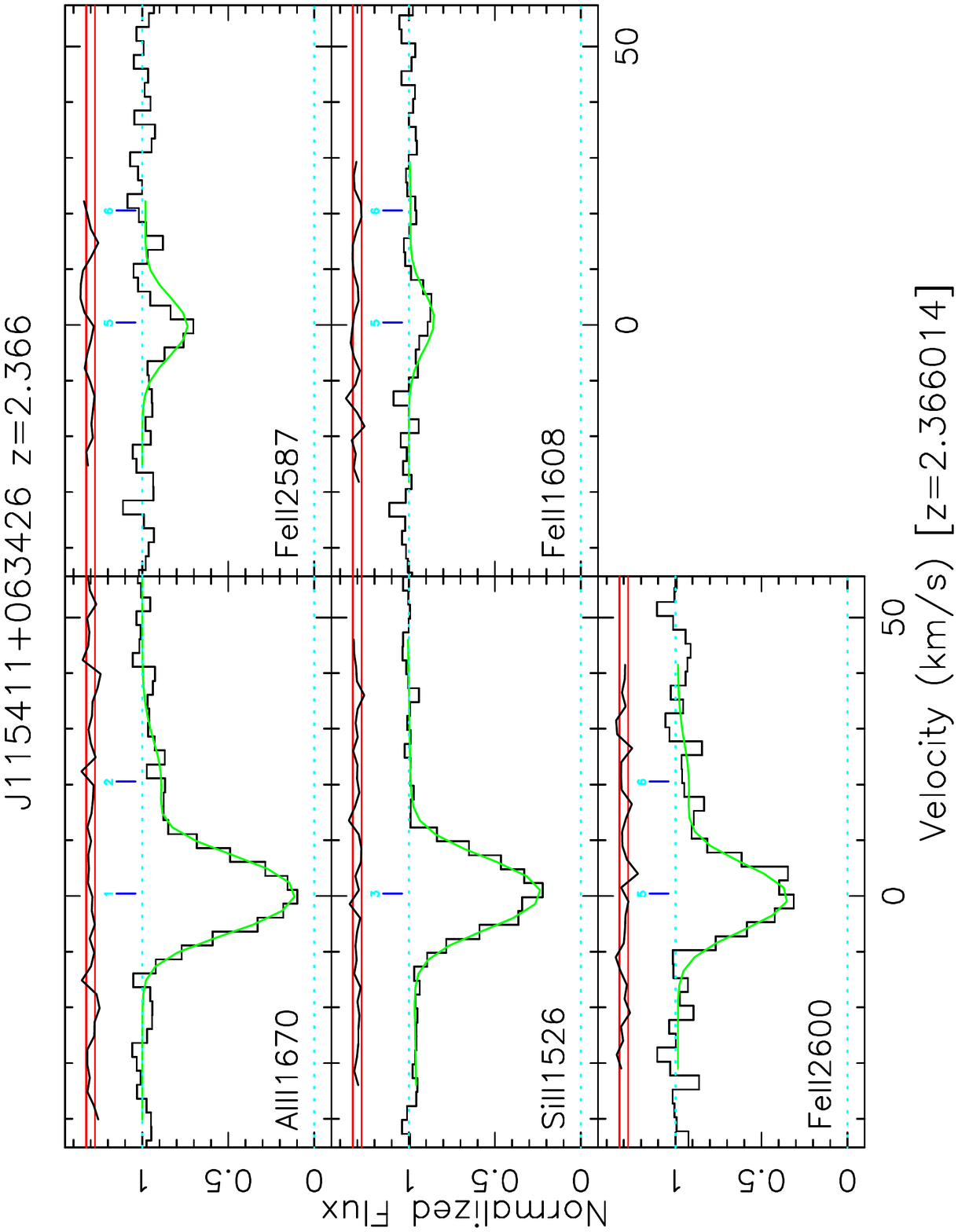}
\par\end{centering}

\caption[Fit for the $z=2.366$ absorber toward J115411+063426]{Many-multiplet fit for the $z=2.366$ absorber toward J115411+063426.}
\end{figure}
\begin{figure}[H]
\noindent \begin{centering}
\includegraphics[bb=34bp 58bp 554bp 738bp,clip,width=1\textwidth]{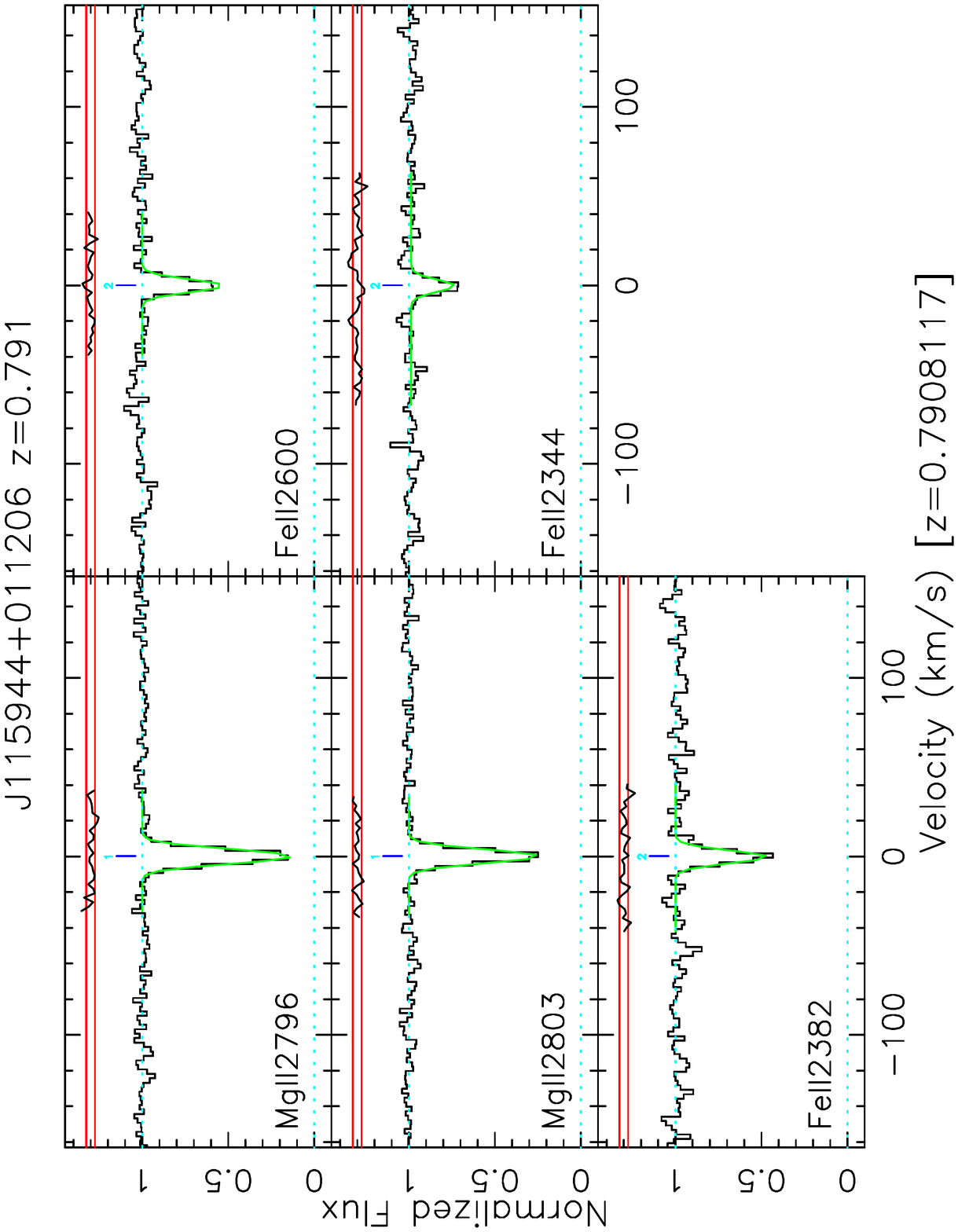}
\par\end{centering}

\caption[Fit for the $z=0.791$ absorber toward J115944+011206]{Many-multiplet fit for the $z=0.791$ absorber toward J115944+011206.}
\end{figure}
\begin{figure}[H]
\noindent \begin{centering}
\includegraphics[bb=34bp 58bp 554bp 738bp,clip,width=1\textwidth]{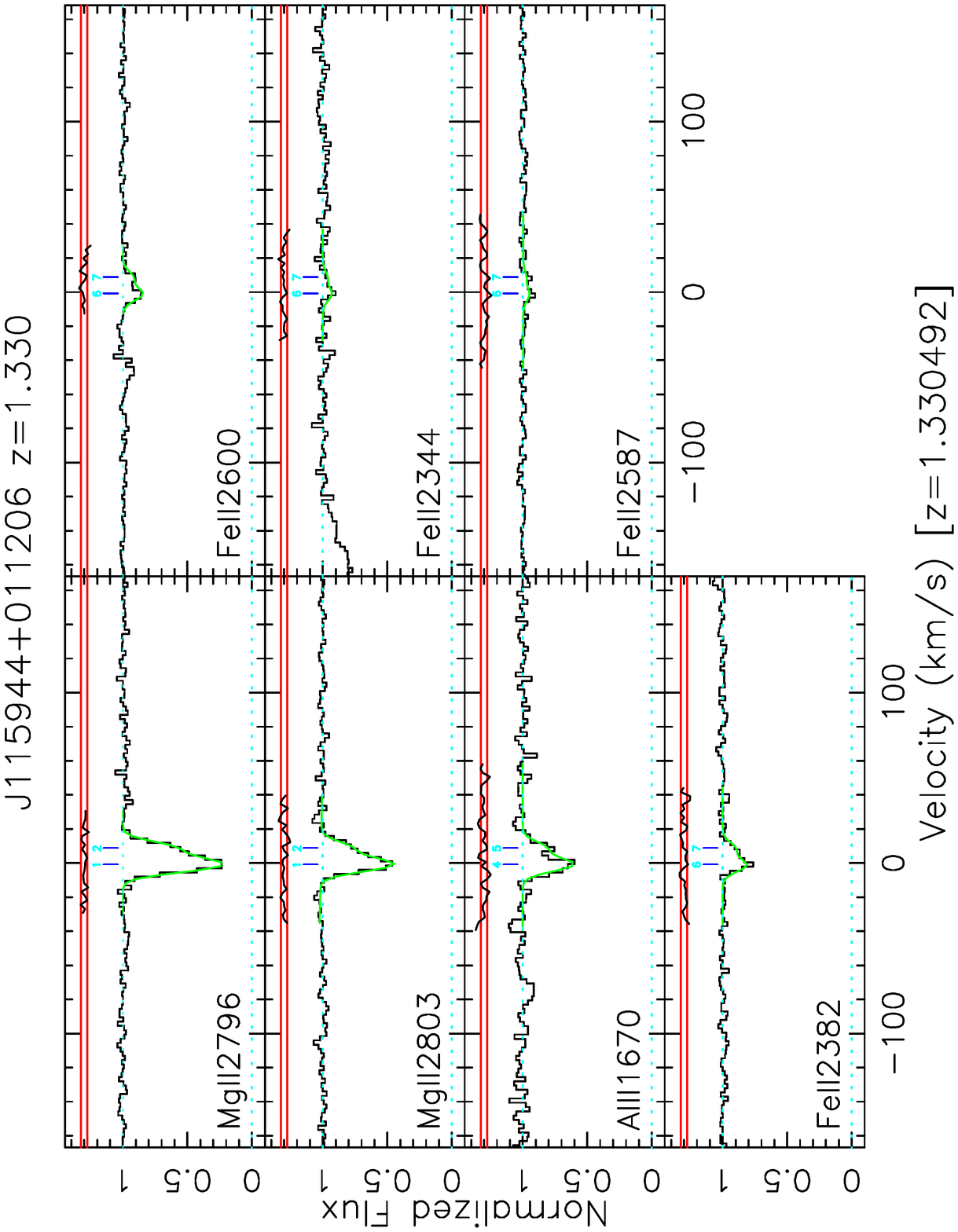}
\par\end{centering}

\caption[Fit for the $z=1.330$ absorber toward J115944+011206]{Many-multiplet fit for the $z=1.330$ absorber toward J115944+011206.}
\end{figure}
\begin{figure}[H]
\noindent \begin{centering}
\includegraphics[bb=34bp 58bp 554bp 738bp,clip,width=1\textwidth]{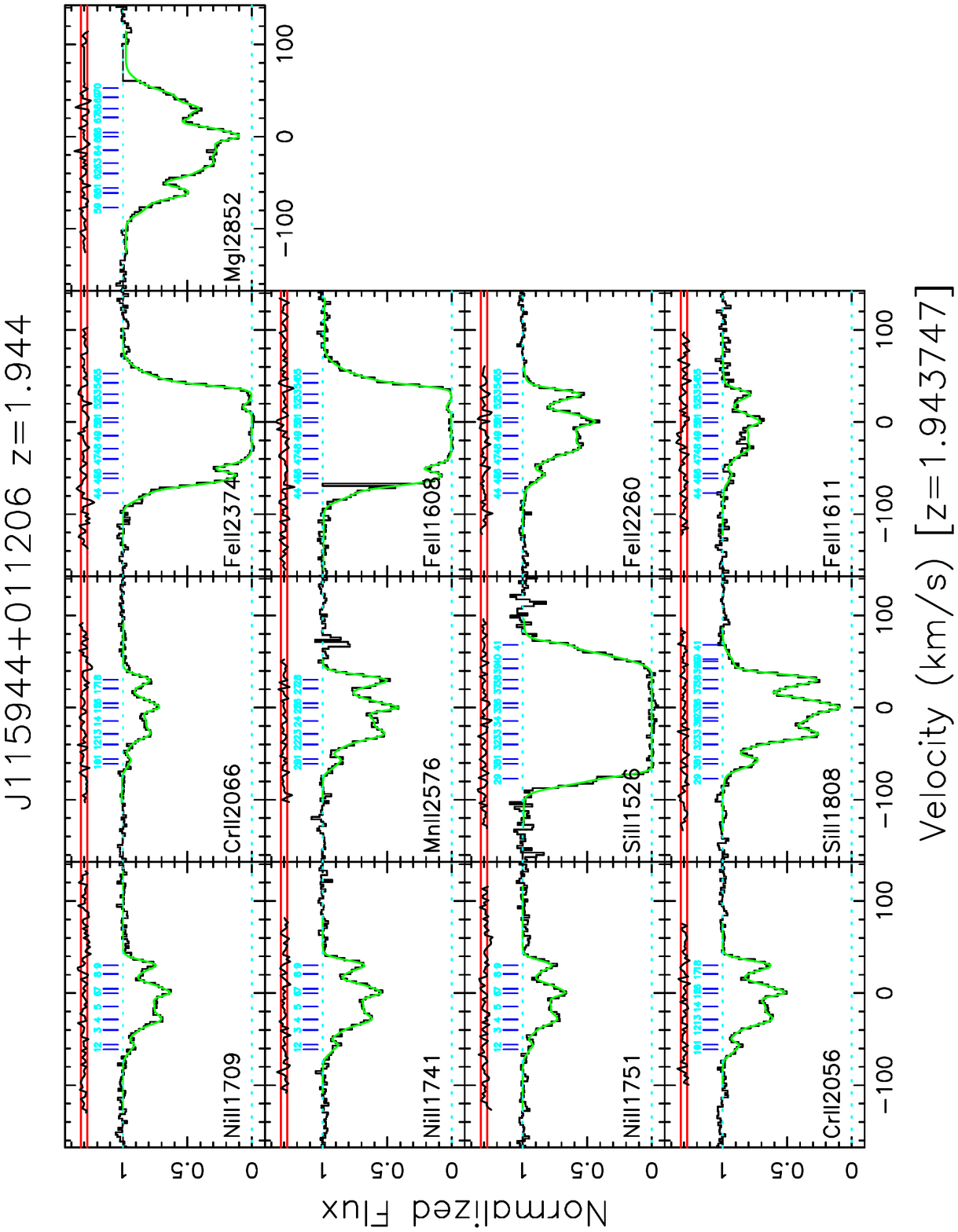}
\par\end{centering}

\caption[Fit for the $z=1.944$ absorber toward J115944+011206]{Many-multiplet fit for the $z=1.944$ absorber toward J115944+011206.}
\end{figure}
\begin{figure}[H]
\noindent \begin{centering}
\includegraphics[bb=34bp 58bp 554bp 738bp,clip,width=1\textwidth]{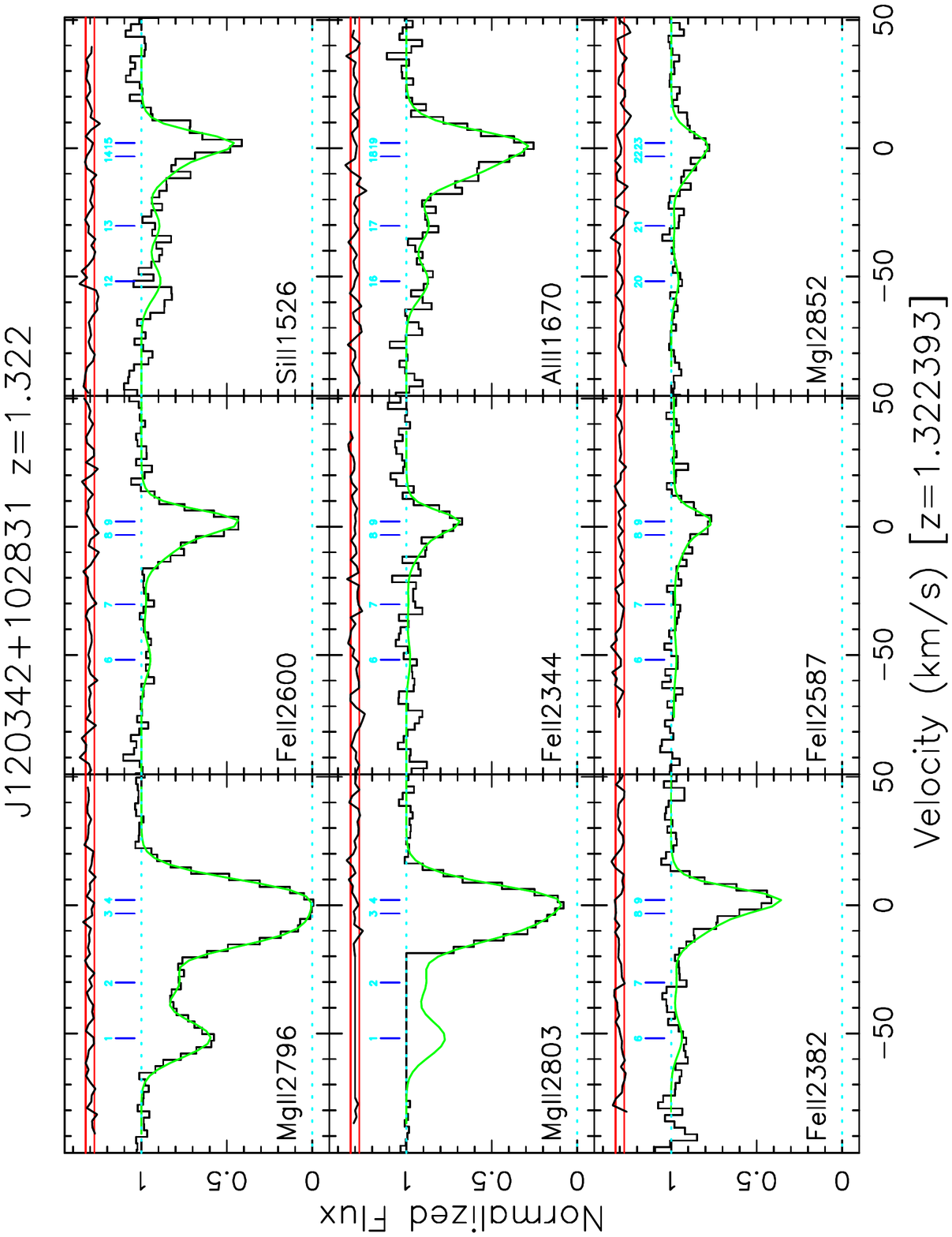}
\par\end{centering}

\caption[Fit for the $z=1.322$ absorber toward J120342+102831]{Many-multiplet fit for the $z=1.322$ absorber toward J120342+102831.}
\end{figure}
\begin{figure}[H]
\noindent \begin{centering}
\includegraphics[bb=34bp 58bp 554bp 738bp,clip,width=1\textwidth]{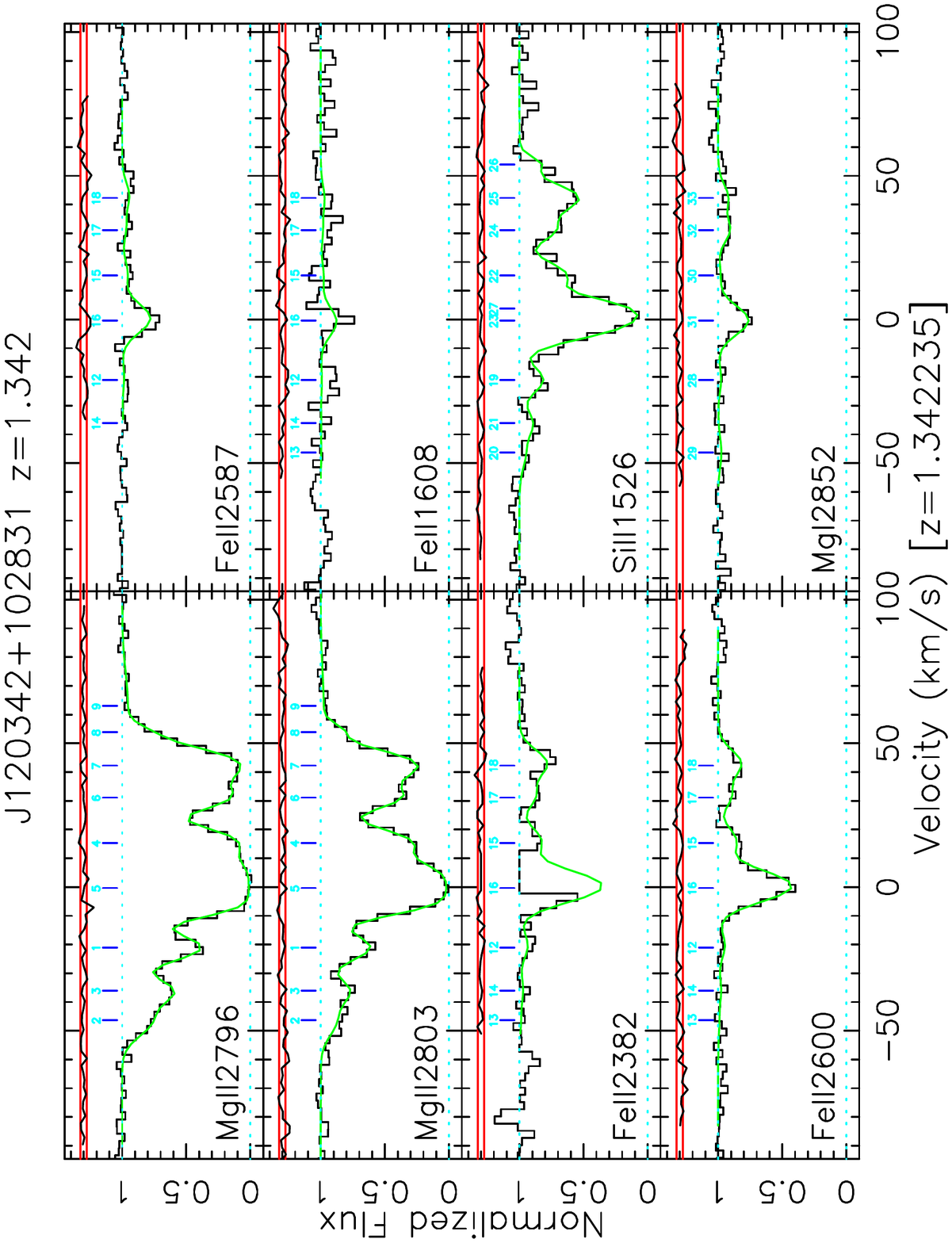}
\par\end{centering}

\caption[Fit for the $z=1.342$ absorber toward J120342+102831]{Many-multiplet fit for the $z=1.342$ absorber toward J120342+102831.}
\end{figure}
\begin{figure}[H]
\noindent \begin{centering}
\includegraphics[bb=34bp 58bp 554bp 738bp,clip,width=1\textwidth]{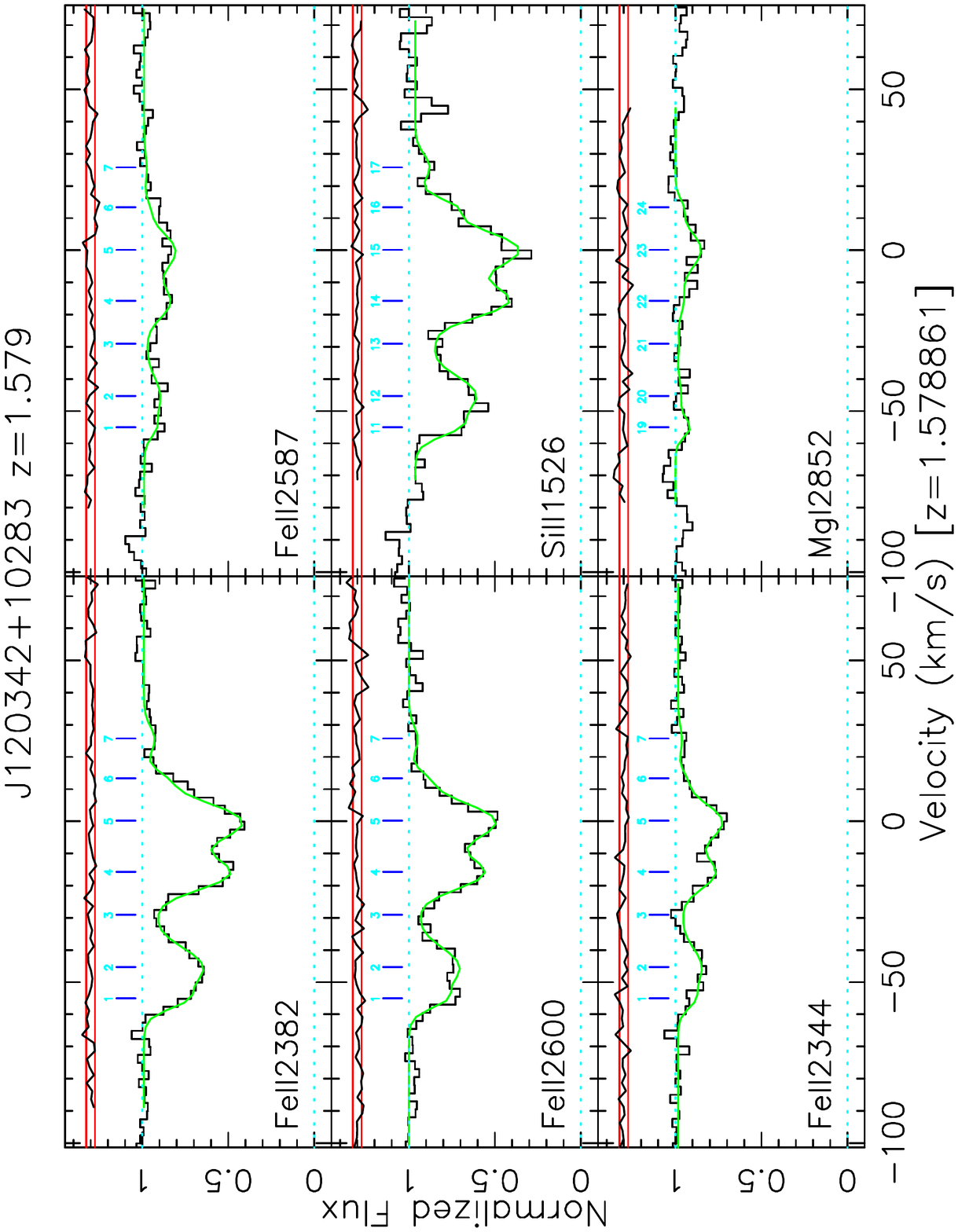}
\par\end{centering}

\caption[Fit for the $z=1.579$ absorber toward J120342+102831]{Many-multiplet fit for the $z=1.579$ absorber toward J120342+102831.}
\end{figure}
\begin{figure}[H]
\noindent \begin{centering}
\includegraphics[bb=34bp 58bp 554bp 738bp,clip,width=1\textwidth]{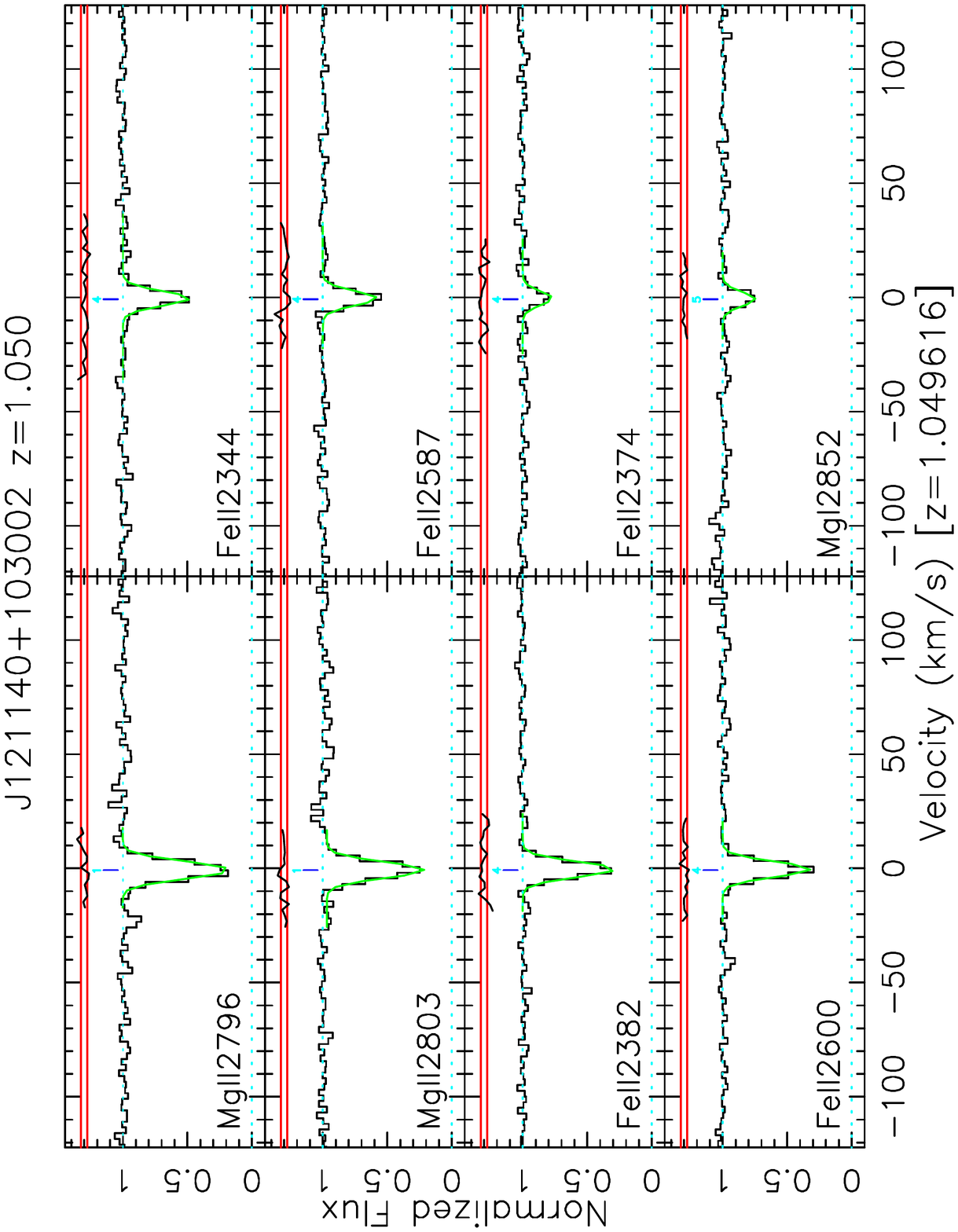}
\par\end{centering}

\caption[Fit for the $z=1.050$ absorber toward J121140+103002]{Many-multiplet fit for the $z=1.050$ absorber toward J121140+103002.}
\end{figure}
\begin{figure}[H]
\noindent \begin{centering}
\includegraphics[bb=34bp 58bp 554bp 738bp,clip,width=1\textwidth]{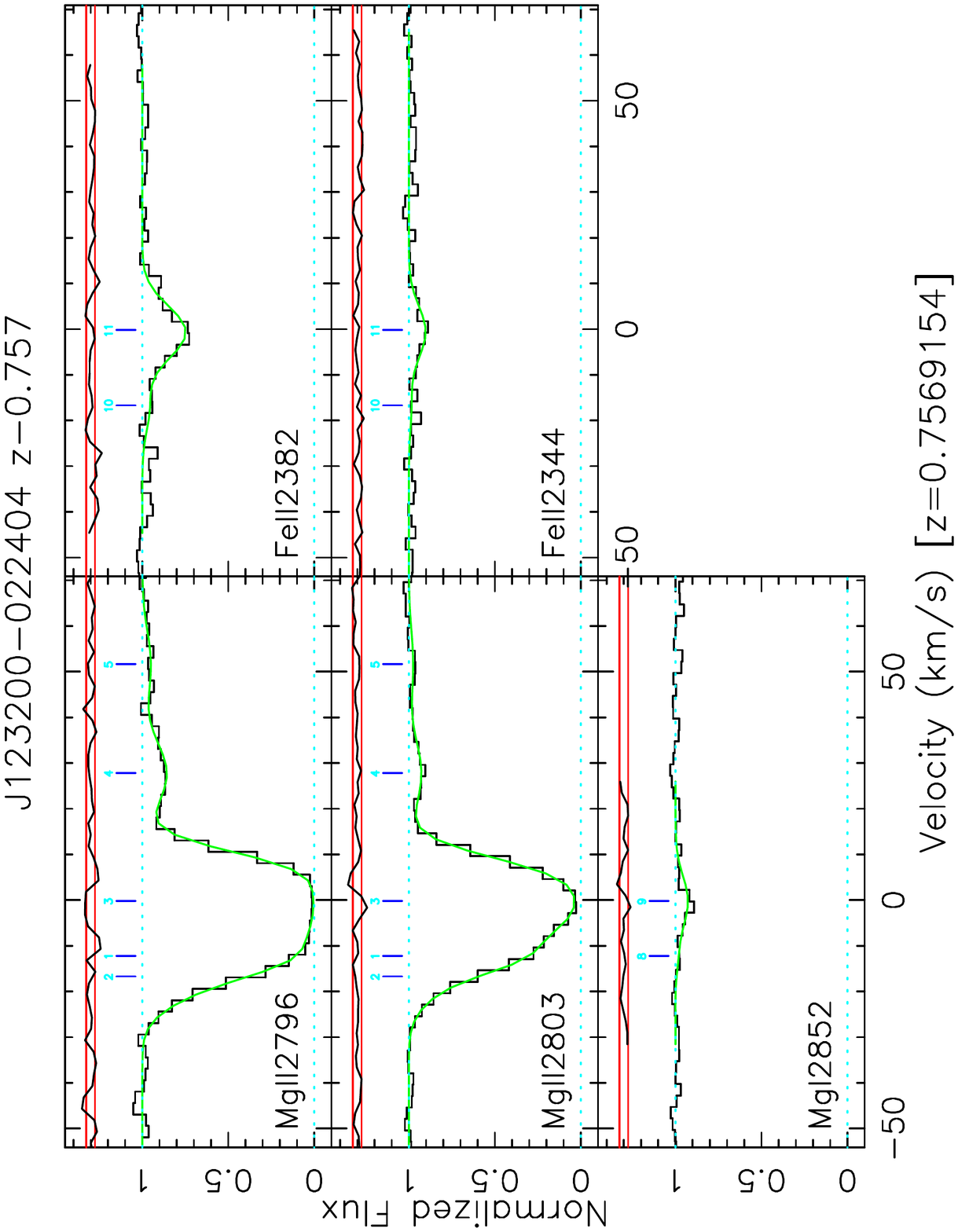}
\par\end{centering}

\caption[Fit for the $z=0.757$ absorber toward J123200$-$022404]{Many-multiplet fit for the $z=0.757$ absorber toward J123200$-$022404.}
\end{figure}
\begin{figure}[H]
\noindent \begin{centering}
\includegraphics[bb=34bp 58bp 554bp 738bp,clip,width=1\textwidth]{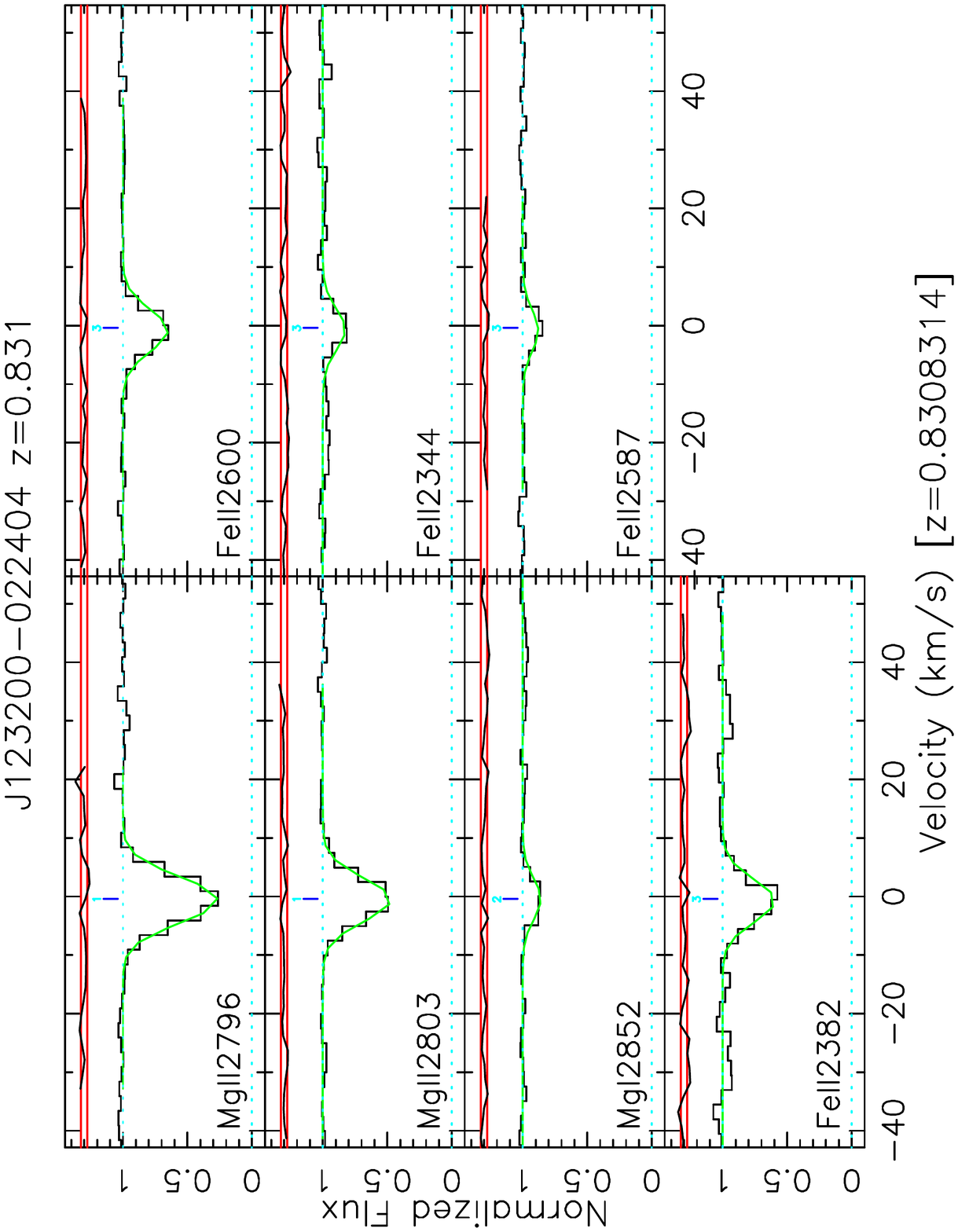}
\par\end{centering}

\caption[Fit for the $z=0.831$ absorber toward J123200$-$022404]{Many-multiplet fit for the $z=0.831$ absorber toward J123200$-$022404.}
\end{figure}
\begin{figure}[H]
\noindent \begin{centering}
\includegraphics[bb=34bp 58bp 554bp 738bp,clip,width=1\textwidth]{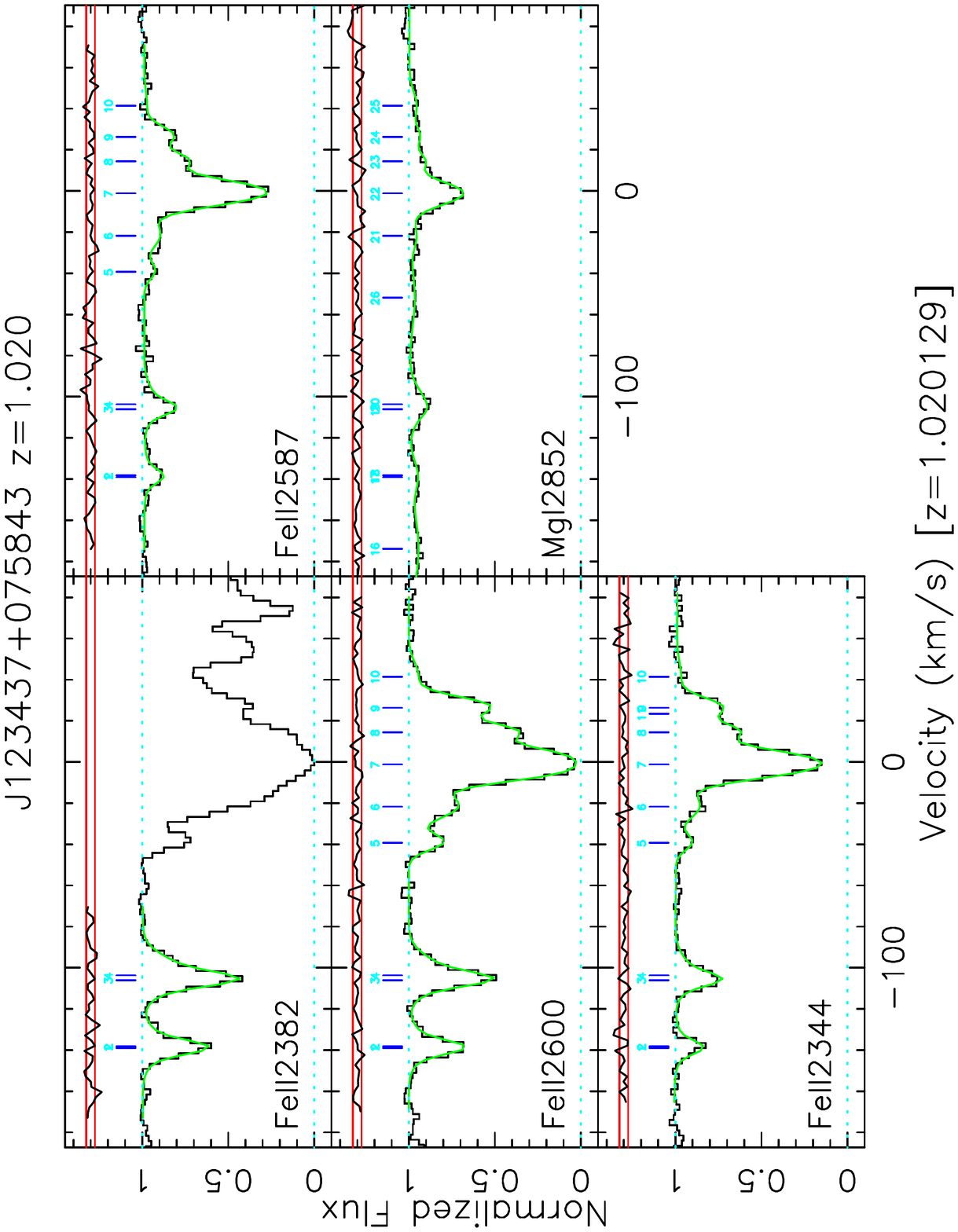}
\par\end{centering}

\caption[Fit for the $z=1.020$ absorber toward J123437+075843]{Many-multiplet fit for the $z=1.020$ absorber toward J123437+075843.}
\end{figure}
\begin{figure}[H]
\noindent \begin{centering}
\includegraphics[bb=34bp 58bp 554bp 738bp,clip,width=1\textwidth]{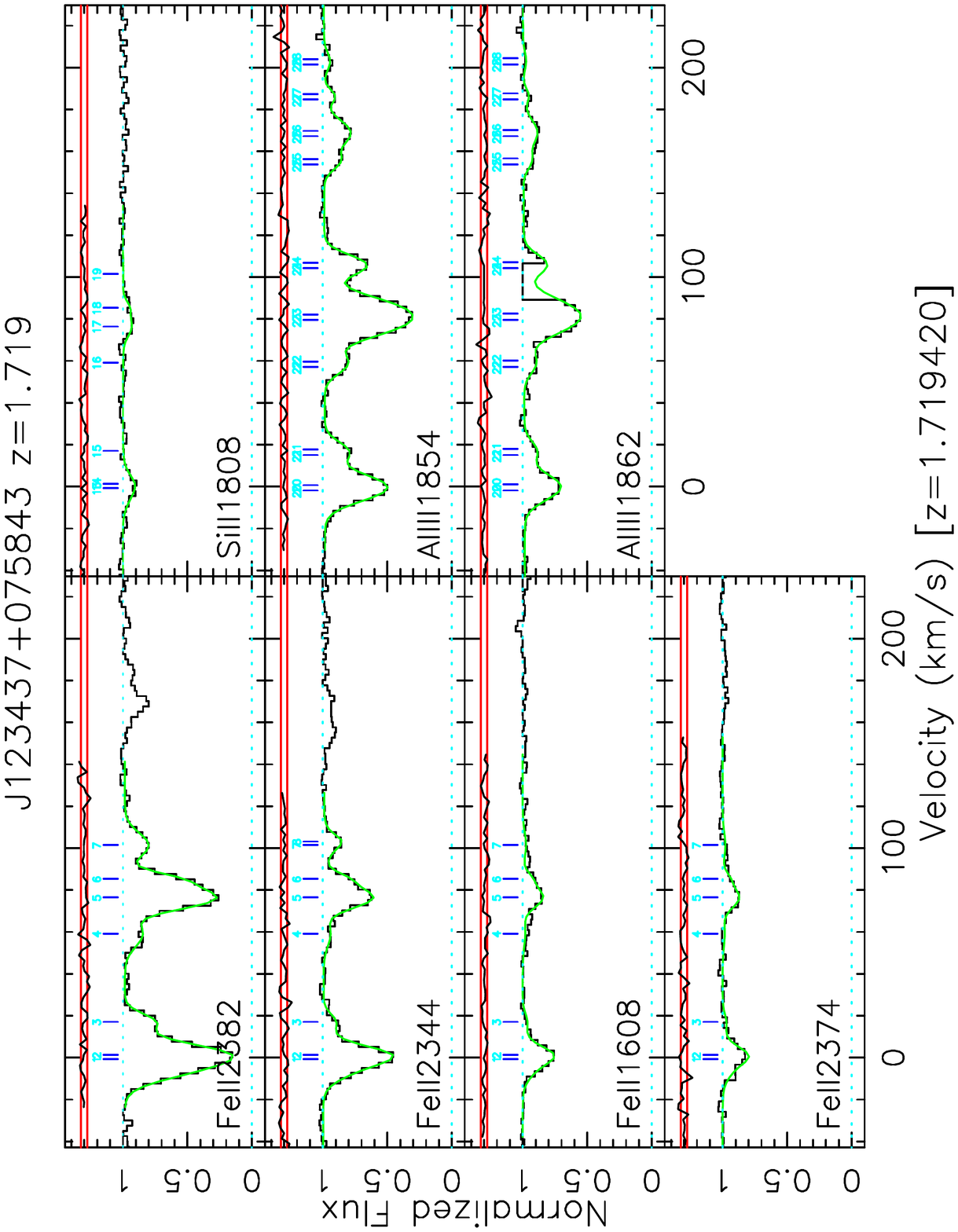}
\par\end{centering}

\caption[Fit for the $z=1.719$ absorber toward J123437+075843]{Many-multiplet fit for the $z=1.719$ absorber toward J123437+075843.}
\end{figure}
\begin{figure}[H]
\noindent \begin{centering}
\includegraphics[bb=34bp 58bp 554bp 738bp,clip,width=1\textwidth]{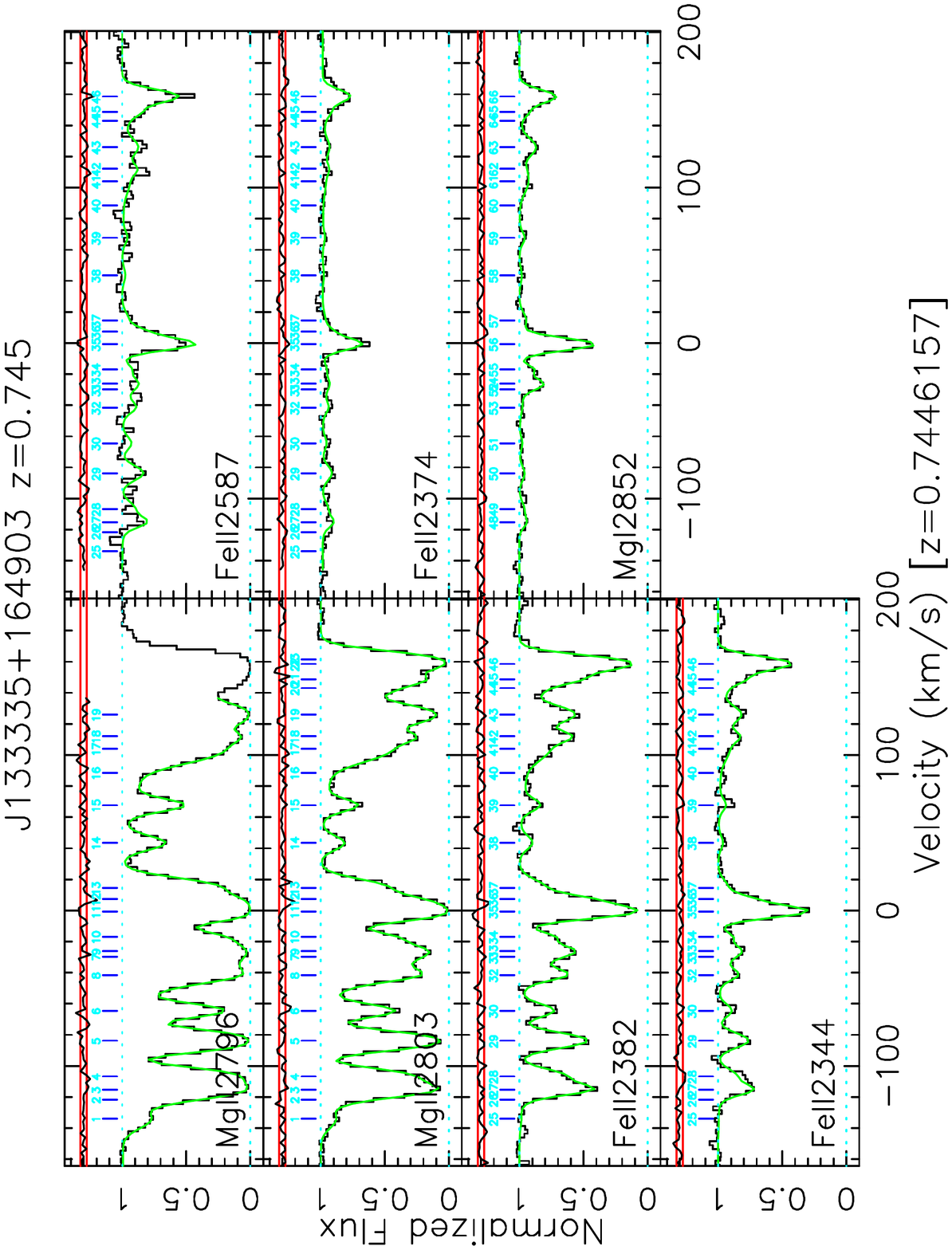}
\par\end{centering}

\caption[Fit for the $z=0.745$ absorber toward J133335+164903]{Many-multiplet fit for the $z=0.745$ absorber toward J133335+164903.}
\end{figure}
\begin{figure}[H]
\noindent \begin{centering}
\includegraphics[bb=34bp 58bp 554bp 738bp,clip,width=1\textwidth]{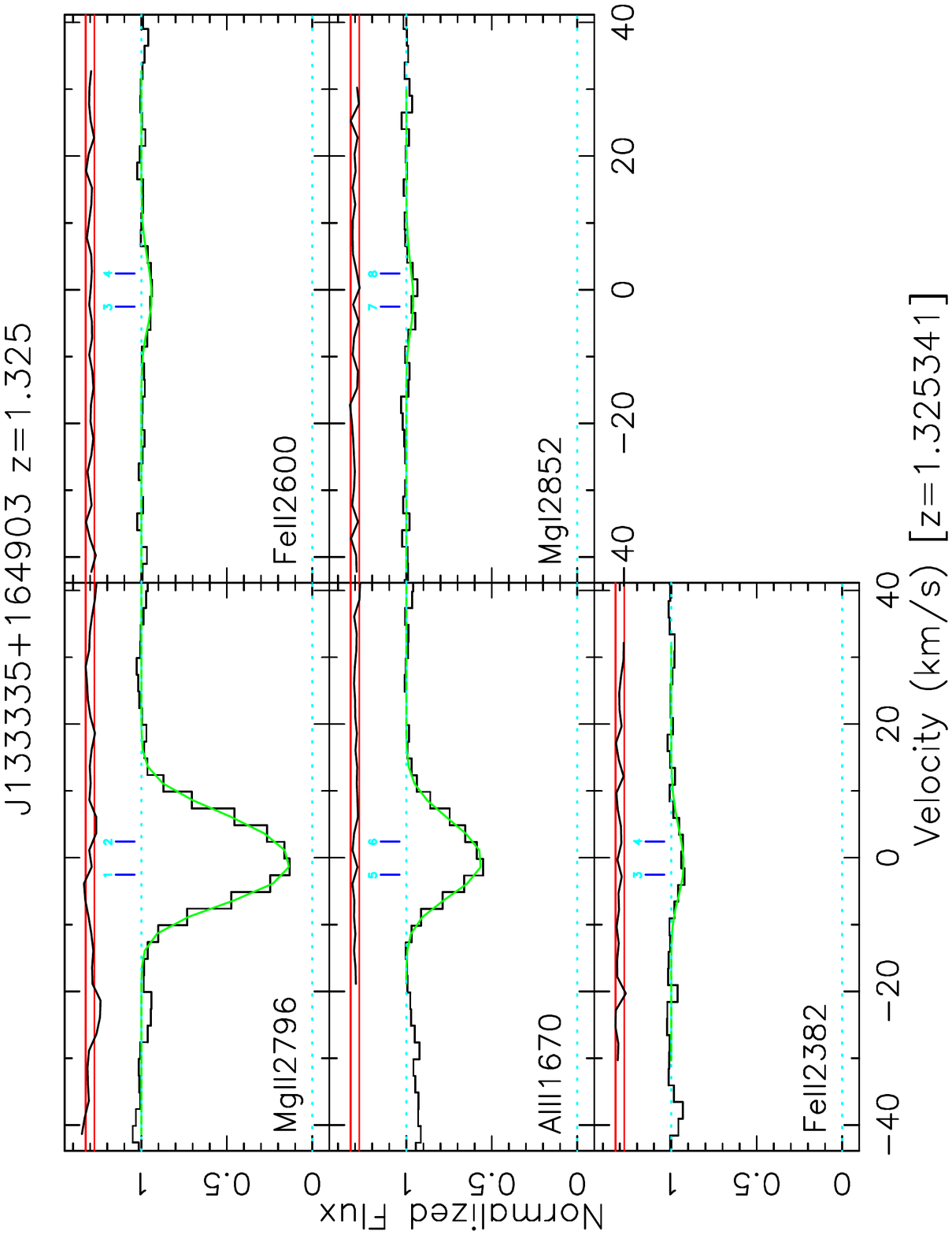}
\par\end{centering}

\caption[Fit for the $z=1.325$ absorber toward J133335+164903]{Many-multiplet fit for the $z=1.325$ absorber toward J133335+164903.}
\end{figure}
\begin{figure}[H]
\noindent \begin{centering}
\includegraphics[bb=34bp 58bp 554bp 738bp,clip,width=1\textwidth]{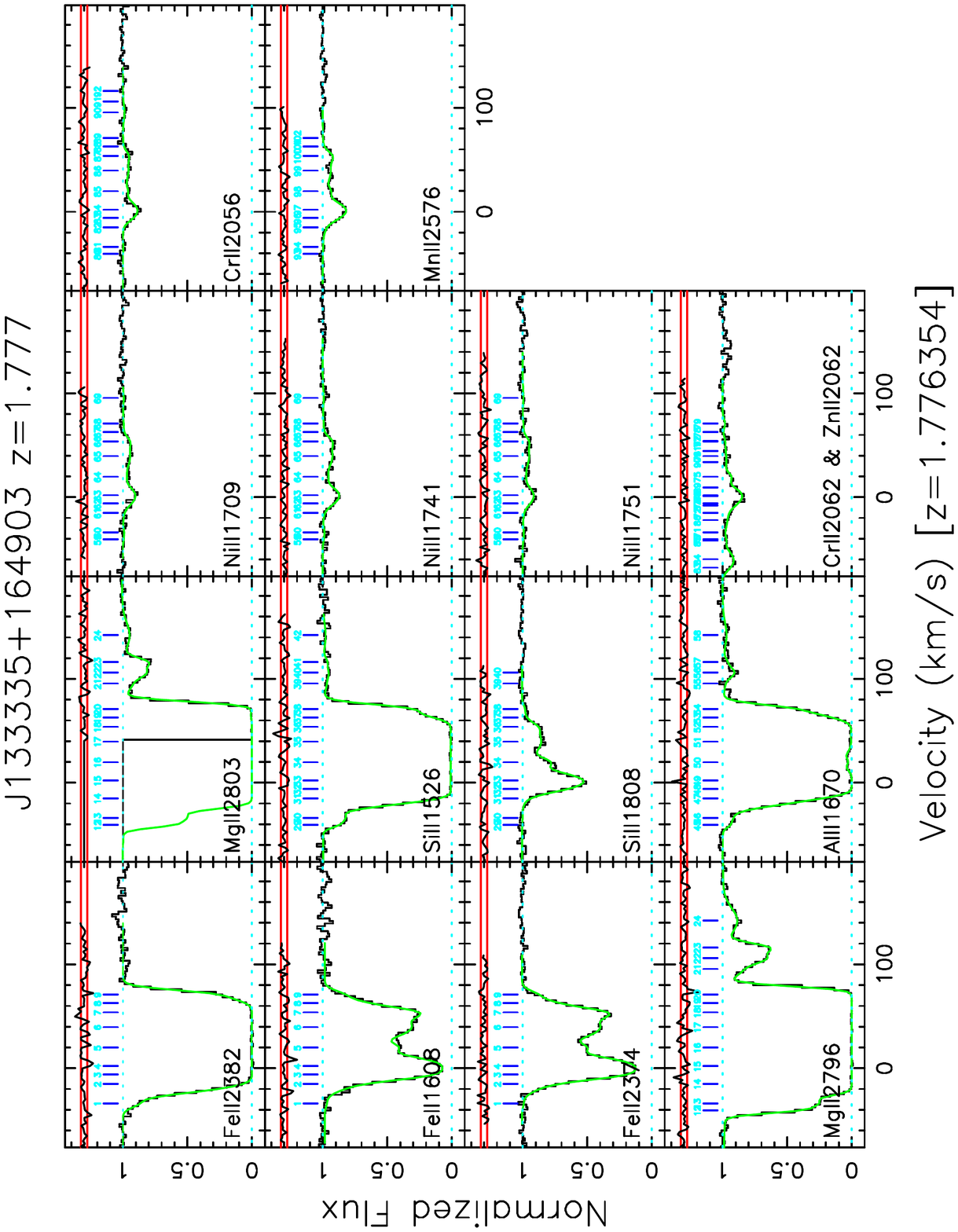}
\par\end{centering}

\caption[Fit for the $z=1.777$ absorber toward J133335+164903]{Many-multiplet fit for the $z=1.777$ absorber toward J133335+164903.}
\end{figure}
\begin{figure}[H]
\noindent \begin{centering}
\includegraphics[bb=34bp 58bp 554bp 738bp,clip,width=1\textwidth]{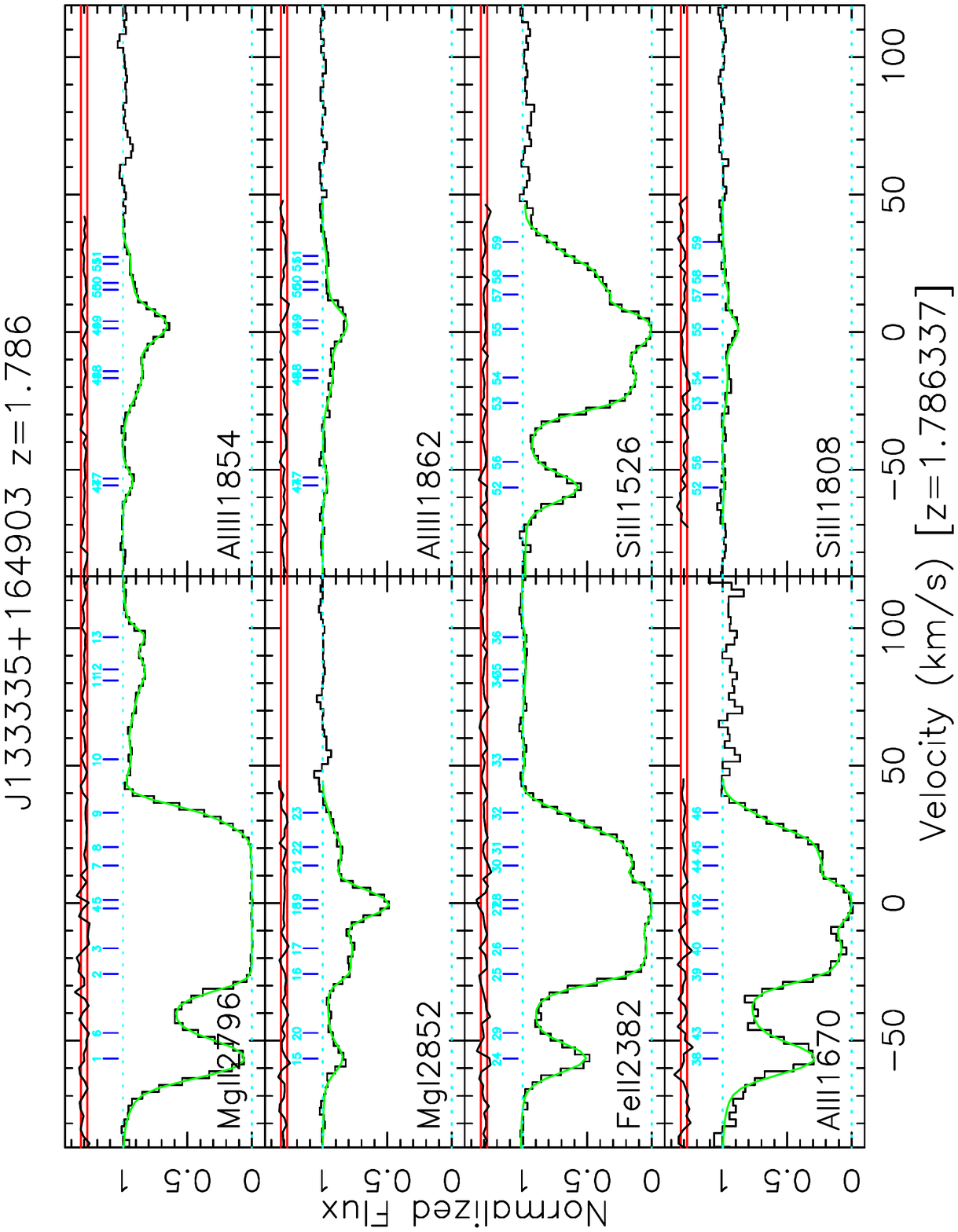}
\par\end{centering}

\caption[Fit for the $z=1.786$ absorber toward J133335+164903]{Many-multiplet fit for the $z=1.786$ absorber toward J133335+164903.}
\end{figure}
\begin{figure}[H]
\noindent \begin{centering}
\includegraphics[bb=34bp 58bp 554bp 738bp,clip,width=1\textwidth]{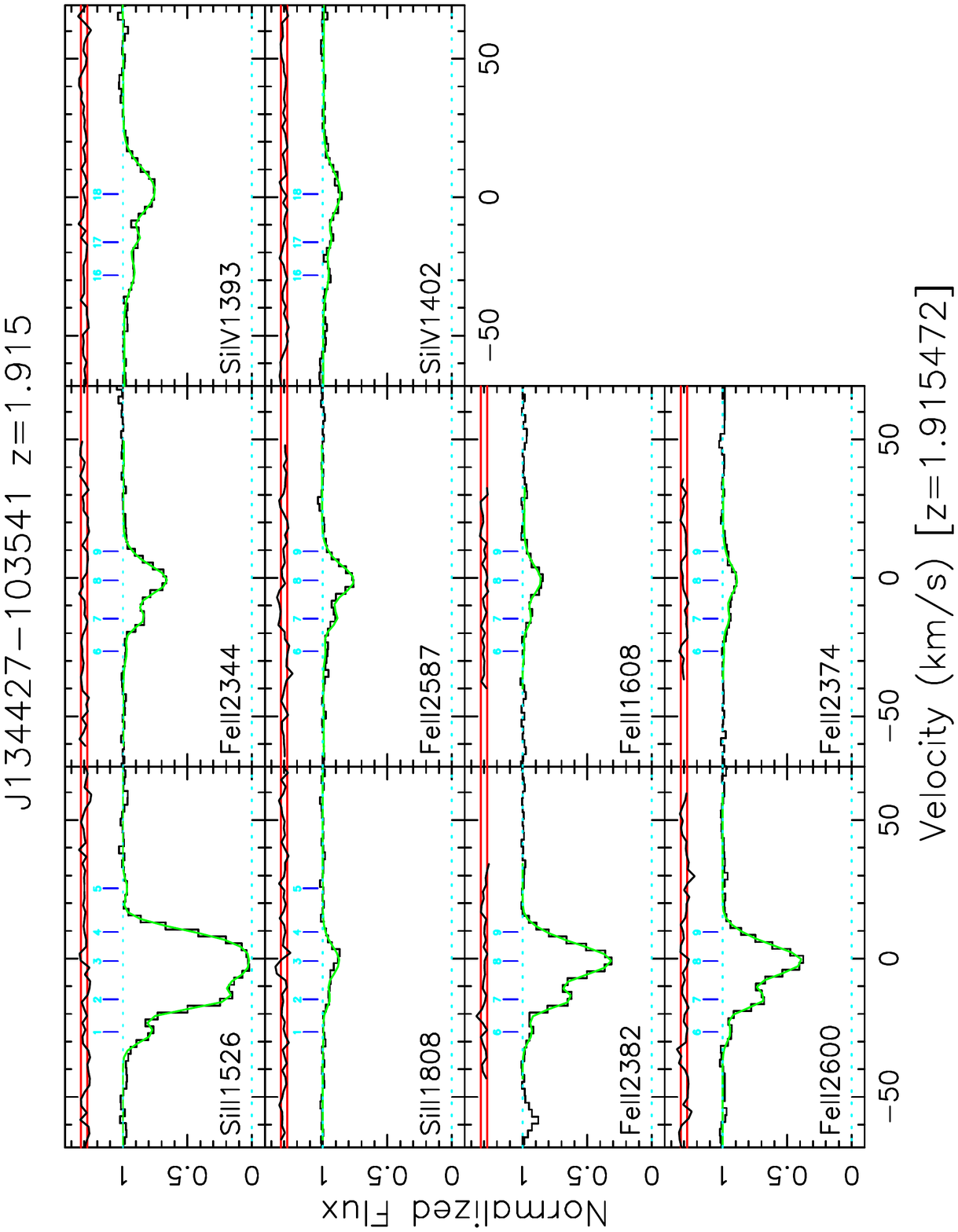}
\par\end{centering}

\caption[Fit for the $z=1.915$ absorber toward J134427$-$103541]{Many-multiplet fit for the $z=1.915$ absorber toward J133427$-$103541.}
\end{figure}
\begin{figure}[H]
\noindent \begin{centering}
\includegraphics[bb=34bp 58bp 554bp 738bp,clip,width=1\textwidth]{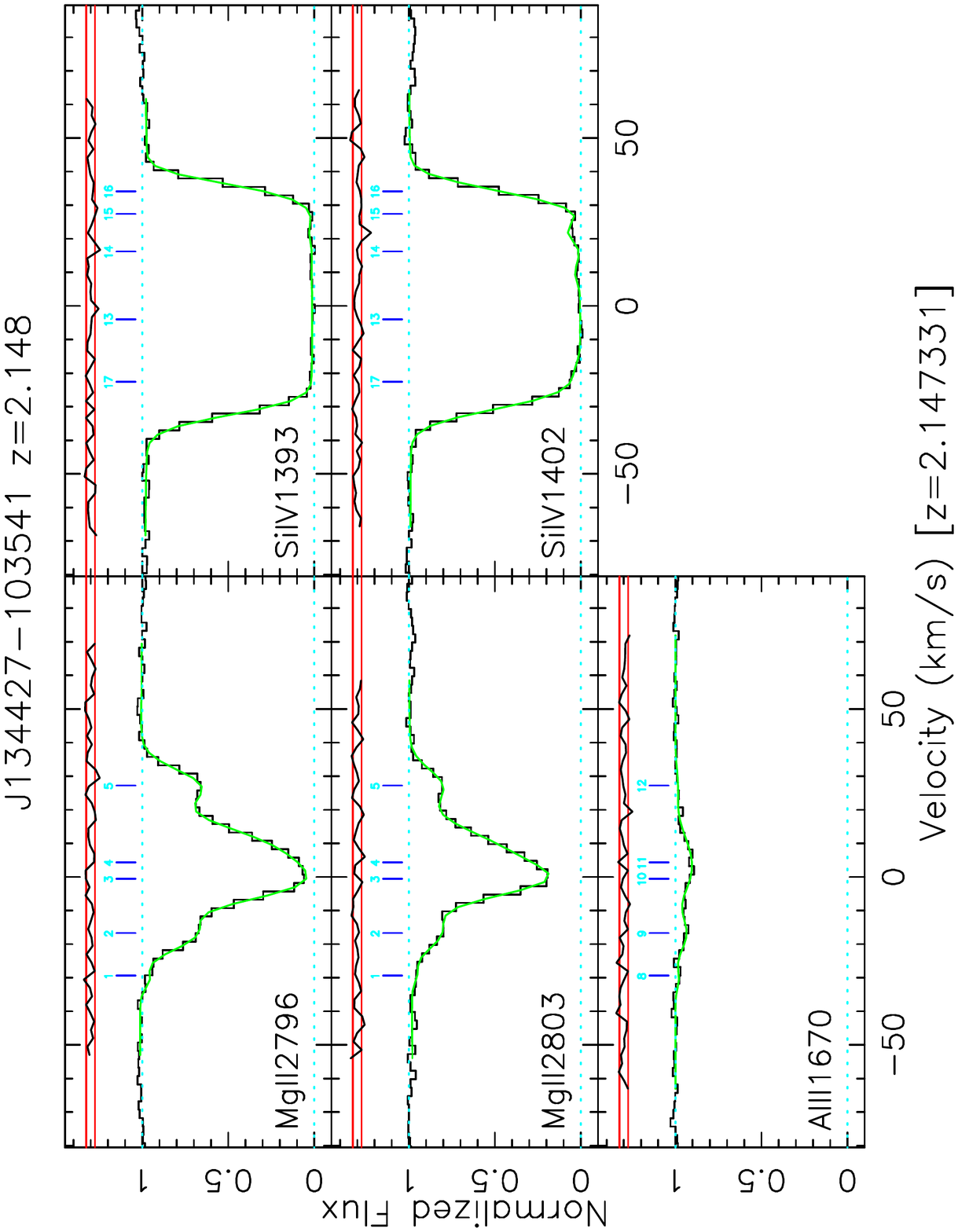}
\par\end{centering}

\caption[Fit for the $z=2.148$ absorber toward J134427$-$103541]{Many-multiplet fit for the $z=2.148$ absorber toward J133427$-$103541.}
\end{figure}
\begin{figure}[H]
\noindent \begin{centering}
\includegraphics[bb=34bp 58bp 554bp 738bp,clip,width=1\textwidth]{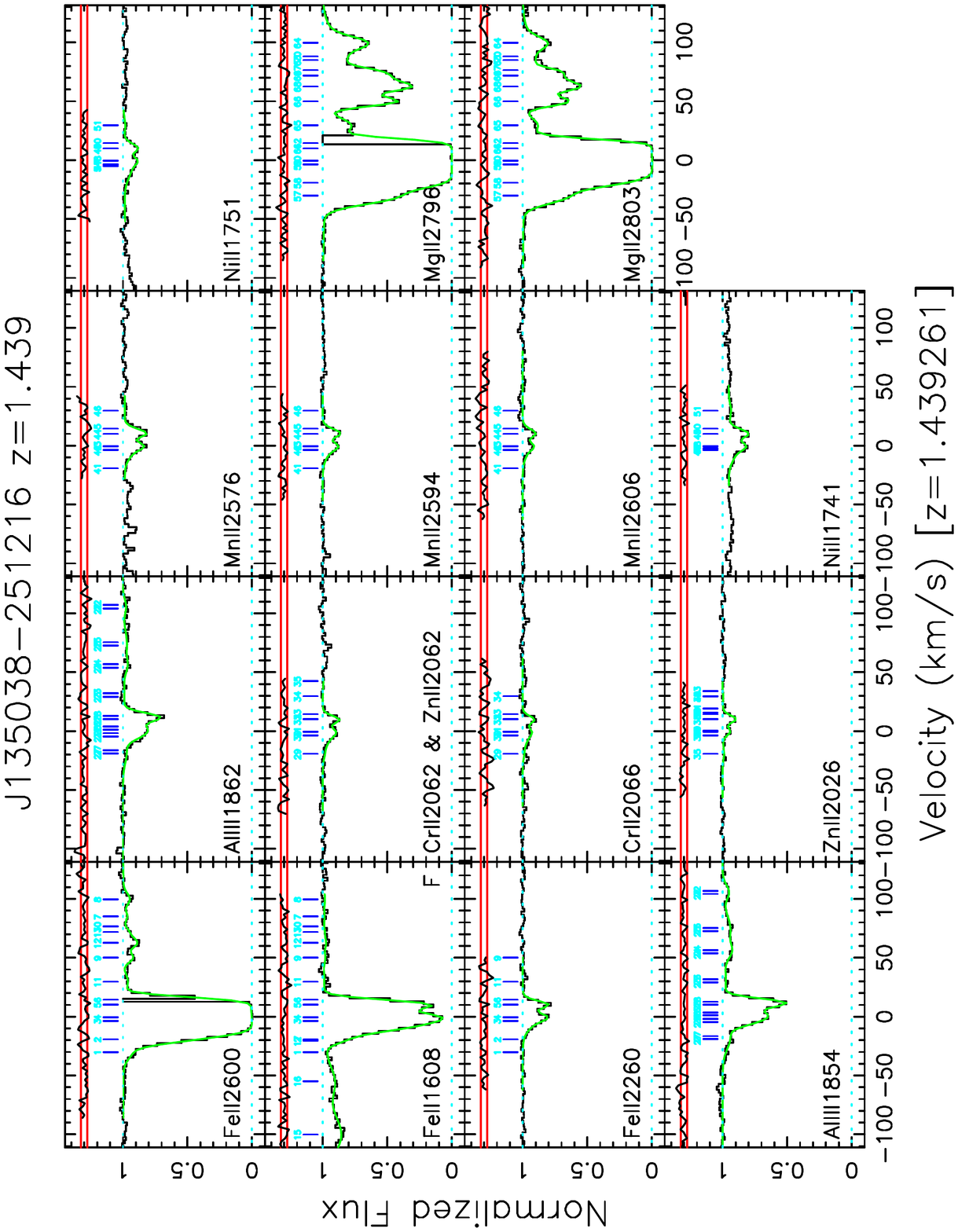}
\par\end{centering}

\caption[\ Fit for the $z=1.439$ absorber toward J135038$-$251216]{Many-multiplet fit for the $z=1.439$ absorber toward J135038$-$251216.}
\end{figure}
\begin{figure}[H]
\noindent \begin{centering}
\includegraphics[bb=34bp 58bp 554bp 738bp,clip,width=1\textwidth]{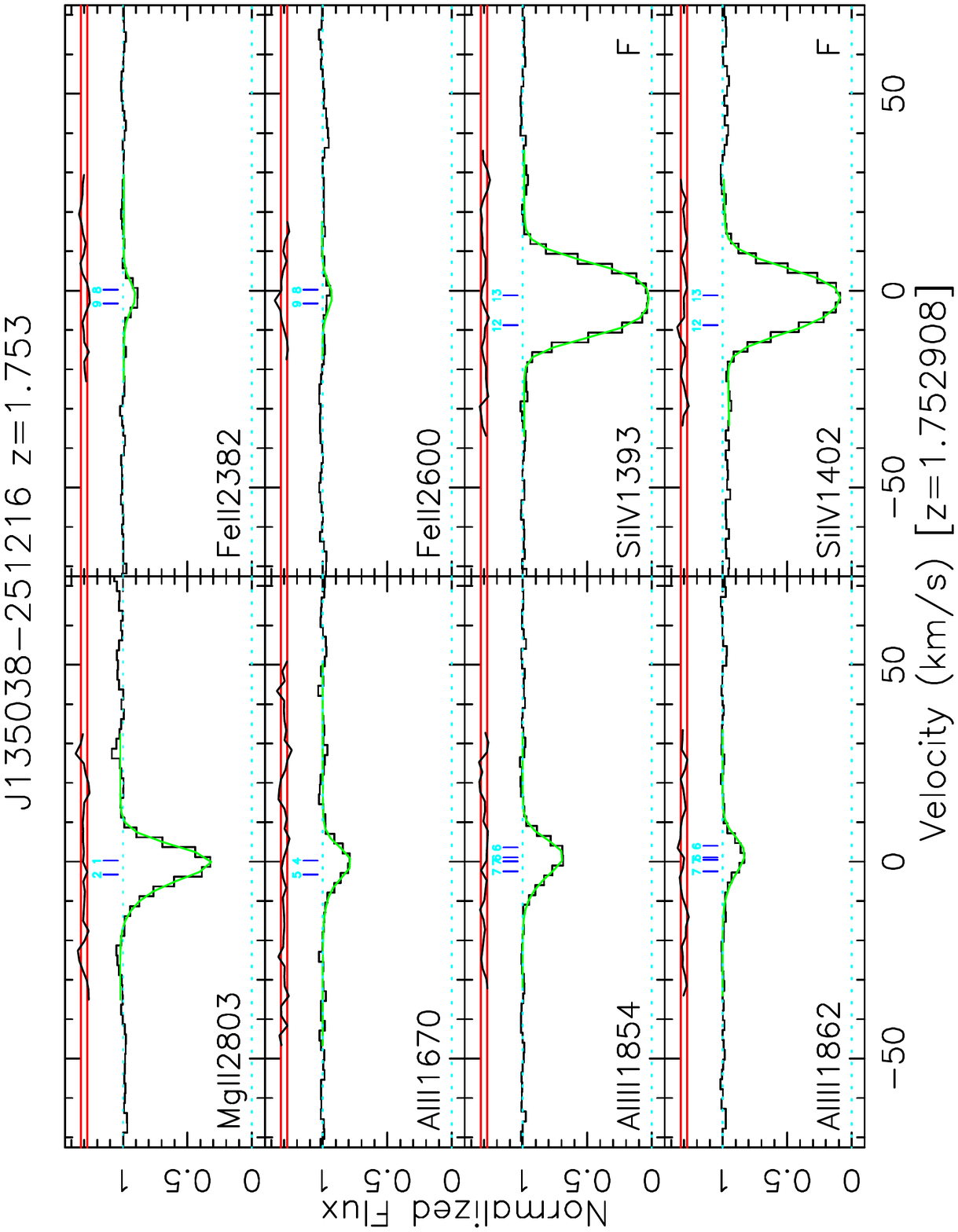}
\par\end{centering}

\caption[\ Fit for the $z=1.753$ absorber toward J135038$-$251216]{Many-multiplet fit for the $z=1.753$ absorber toward J135038$-$251216.}
\end{figure}
\begin{figure}[H]
\noindent \begin{centering}
\includegraphics[bb=34bp 58bp 554bp 738bp,clip,width=1\textwidth]{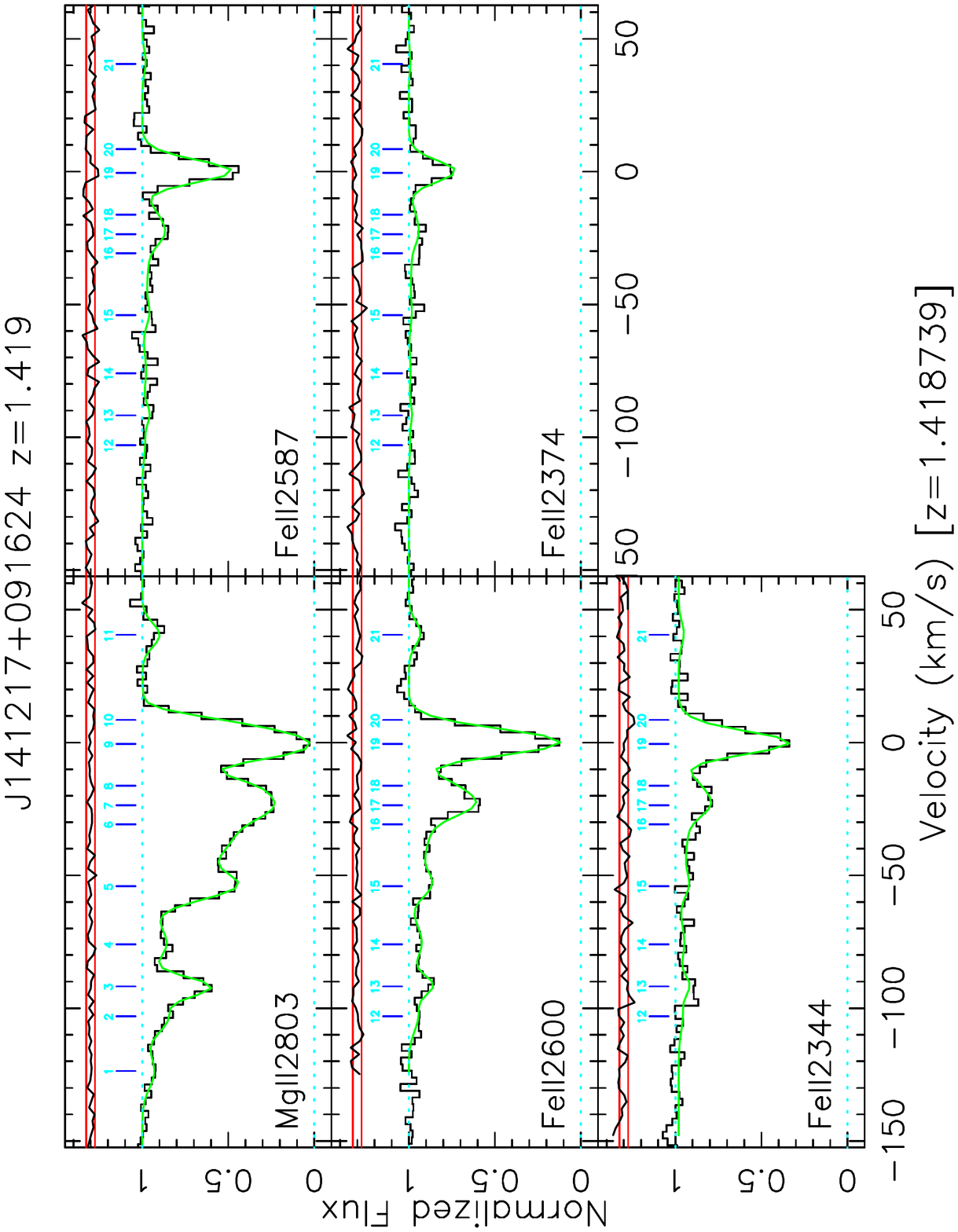}
\par\end{centering}

\caption[\ Fit for the $z=1.419$ absorber toward J141217+091624]{Many-multiplet fit for the $z=1.419$ absorber toward J141217+091624.}
\end{figure}
\begin{figure}[H]
\noindent \begin{centering}
\includegraphics[bb=34bp 58bp 554bp 738bp,clip,width=1\textwidth]{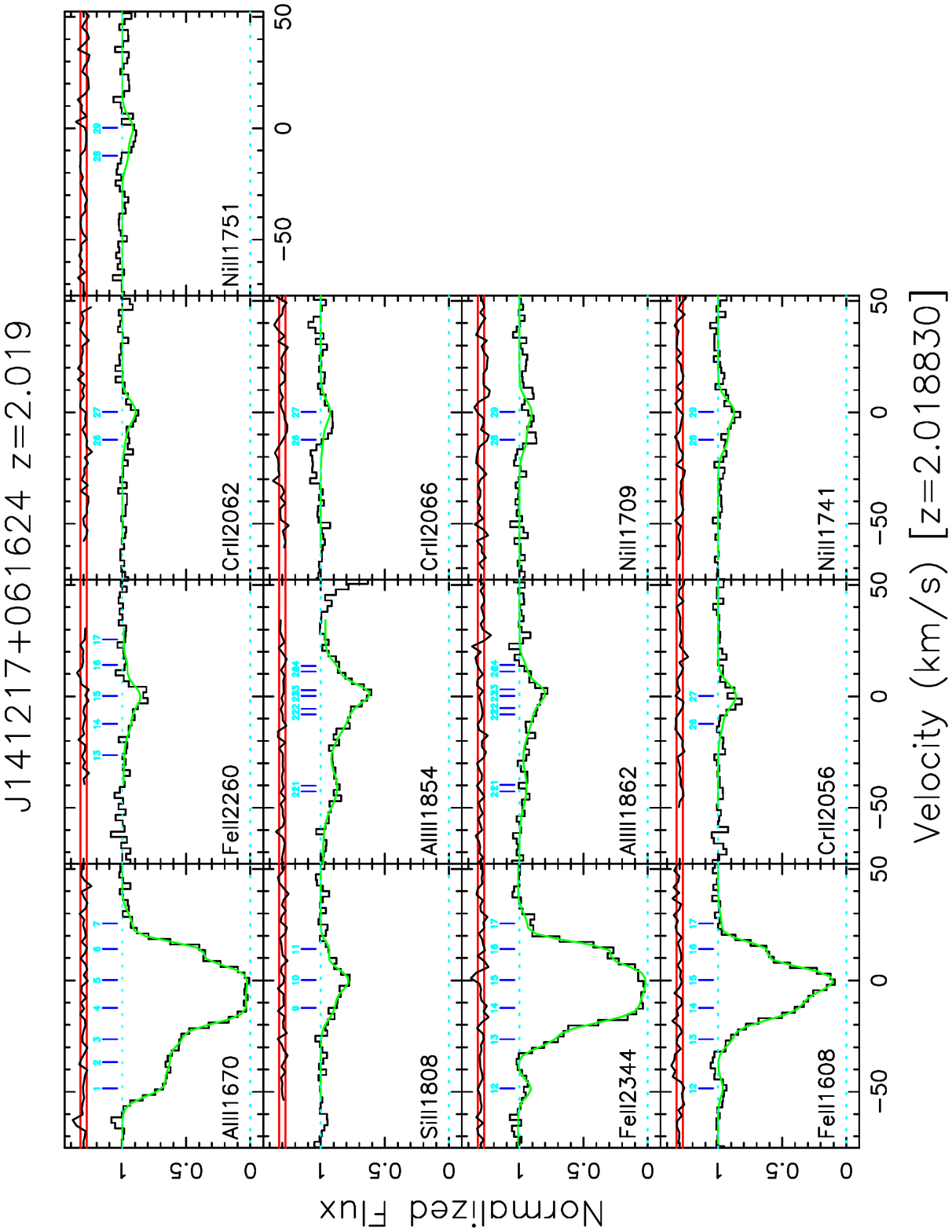}
\par\end{centering}

\caption[\ Fit for the $z=2.019$ absorber toward J141217+091624]{Many-multiplet fit for the $z=2.109$ absorber toward J141217+091624.}
\end{figure}
\begin{figure}[H]
\noindent \begin{centering}
\includegraphics[bb=34bp 58bp 554bp 738bp,clip,width=1\textwidth]{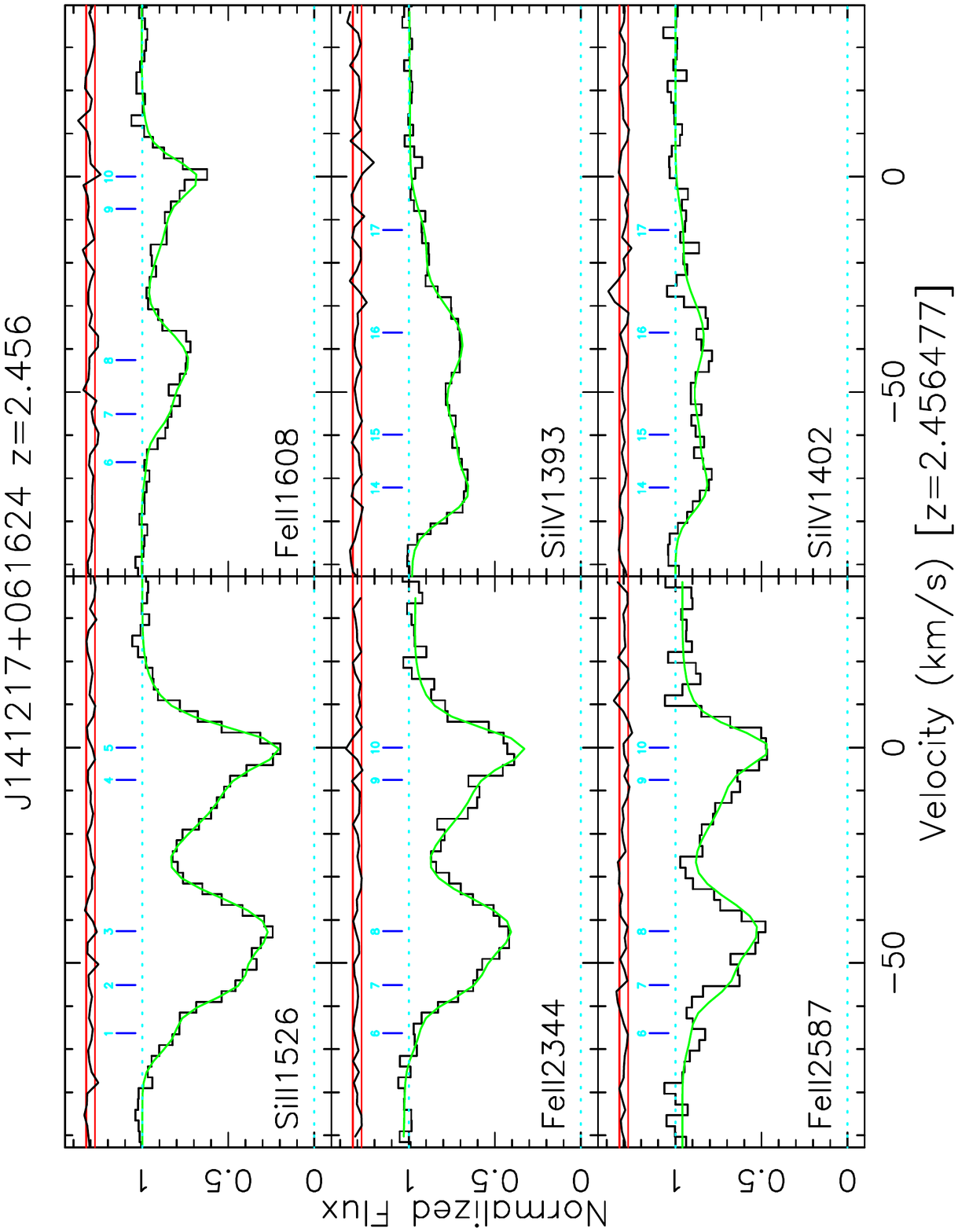}
\par\end{centering}

\caption[\ Fit for the $z=2.456$ absorber toward J141217+091624]{Many-multiplet fit for the $z=2.456$ absorber toward J141217+091624.}
\end{figure}
\begin{figure}[H]
\noindent \begin{centering}
\includegraphics[bb=34bp 58bp 554bp 738bp,clip,width=1\textwidth]{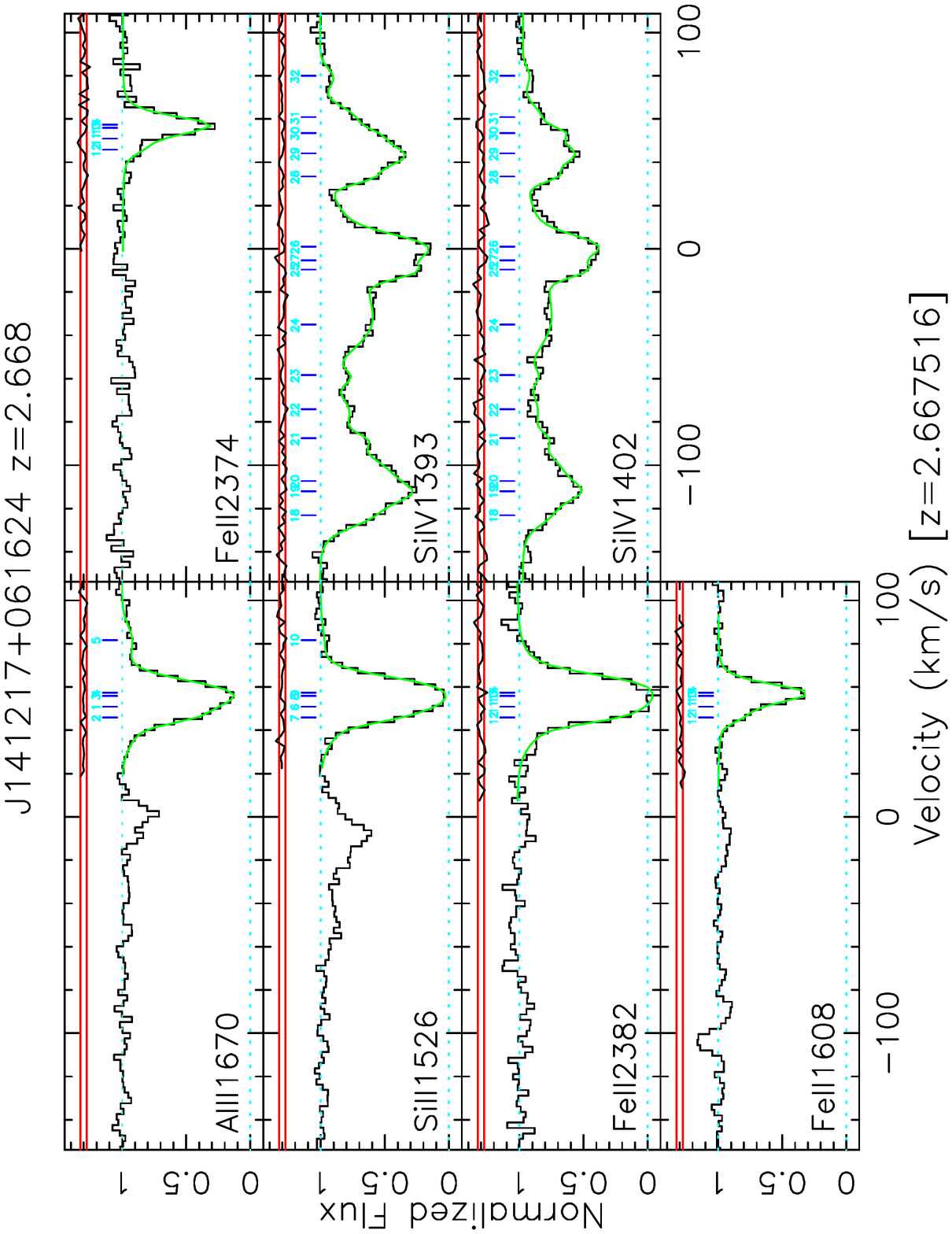}
\par\end{centering}

\caption[\ Fit for the $z=2.668$ absorber toward J141217+091624]{Many-multiplet fit for the $z=2.668$ absorber toward J141217+091624.}
\end{figure}
\begin{figure}[H]
\noindent \begin{centering}
\includegraphics[bb=34bp 58bp 554bp 738bp,clip,width=1\textwidth]{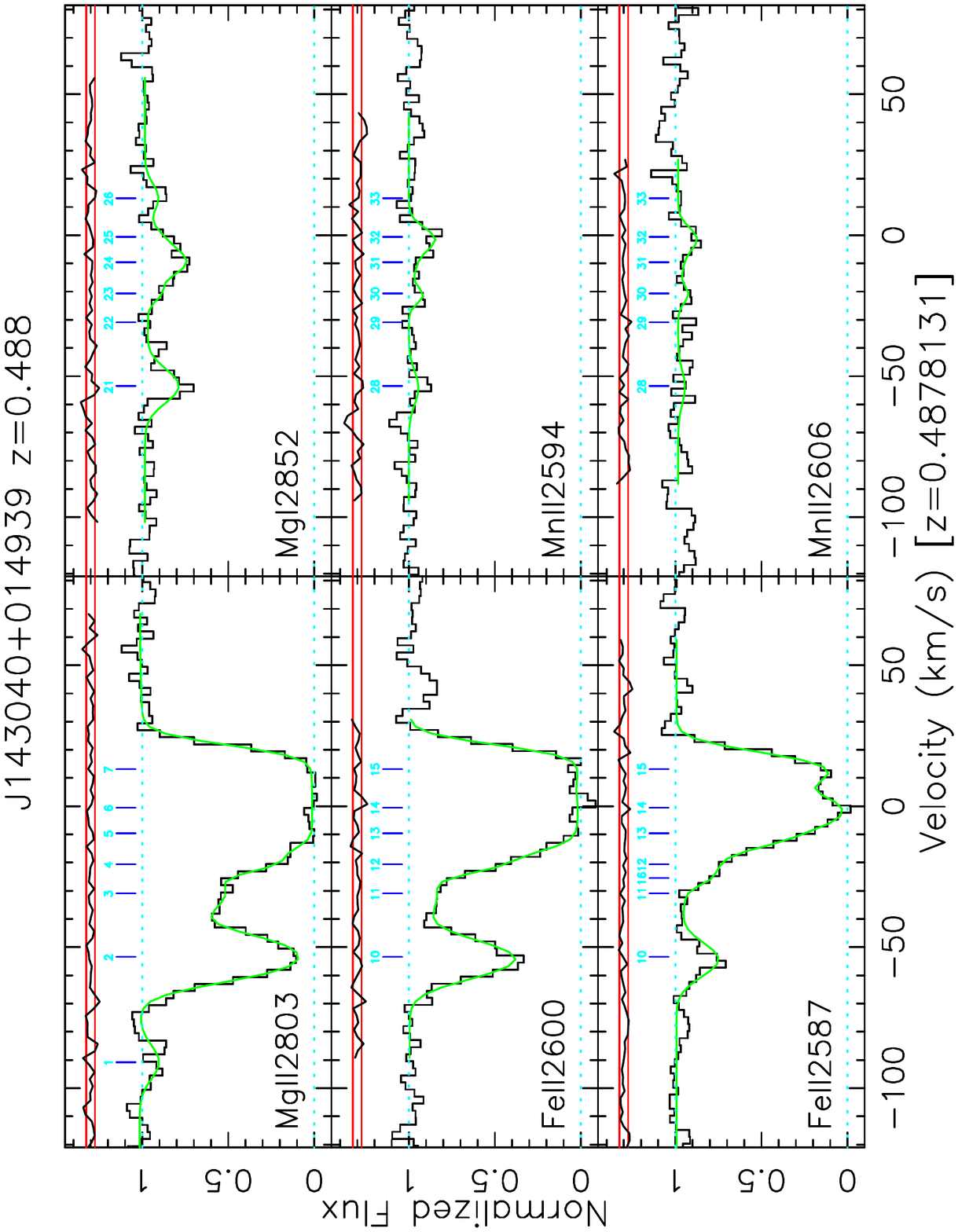}
\par\end{centering}

\caption[\ Fit for the $z=0.488$ absorber toward J143040+014939]{Many-multiplet fit for the $z=0.488$ absorber toward J143040+014939.}
\end{figure}
\begin{figure}[H]
\noindent \begin{centering}
\includegraphics[bb=34bp 58bp 554bp 738bp,clip,width=1\textwidth]{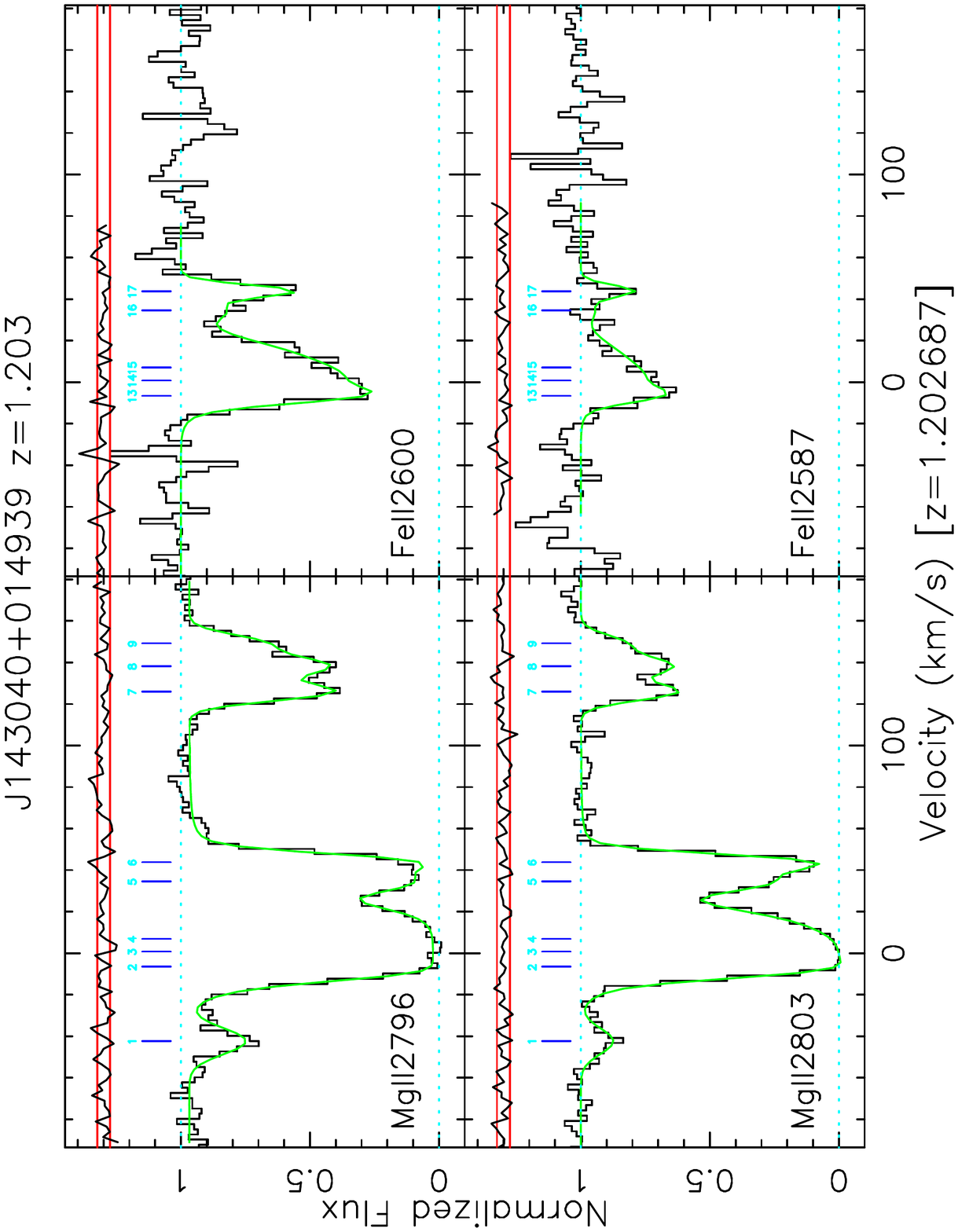}
\par\end{centering}

\caption[\ Fit for the $z=1.203$ absorber toward J143040+014939]{Many-multiplet fit for the $z=1.203$ absorber toward J143040+014939.}
\end{figure}
\begin{figure}[H]
\noindent \begin{centering}
\includegraphics[bb=34bp 58bp 554bp 738bp,clip,width=1\textwidth]{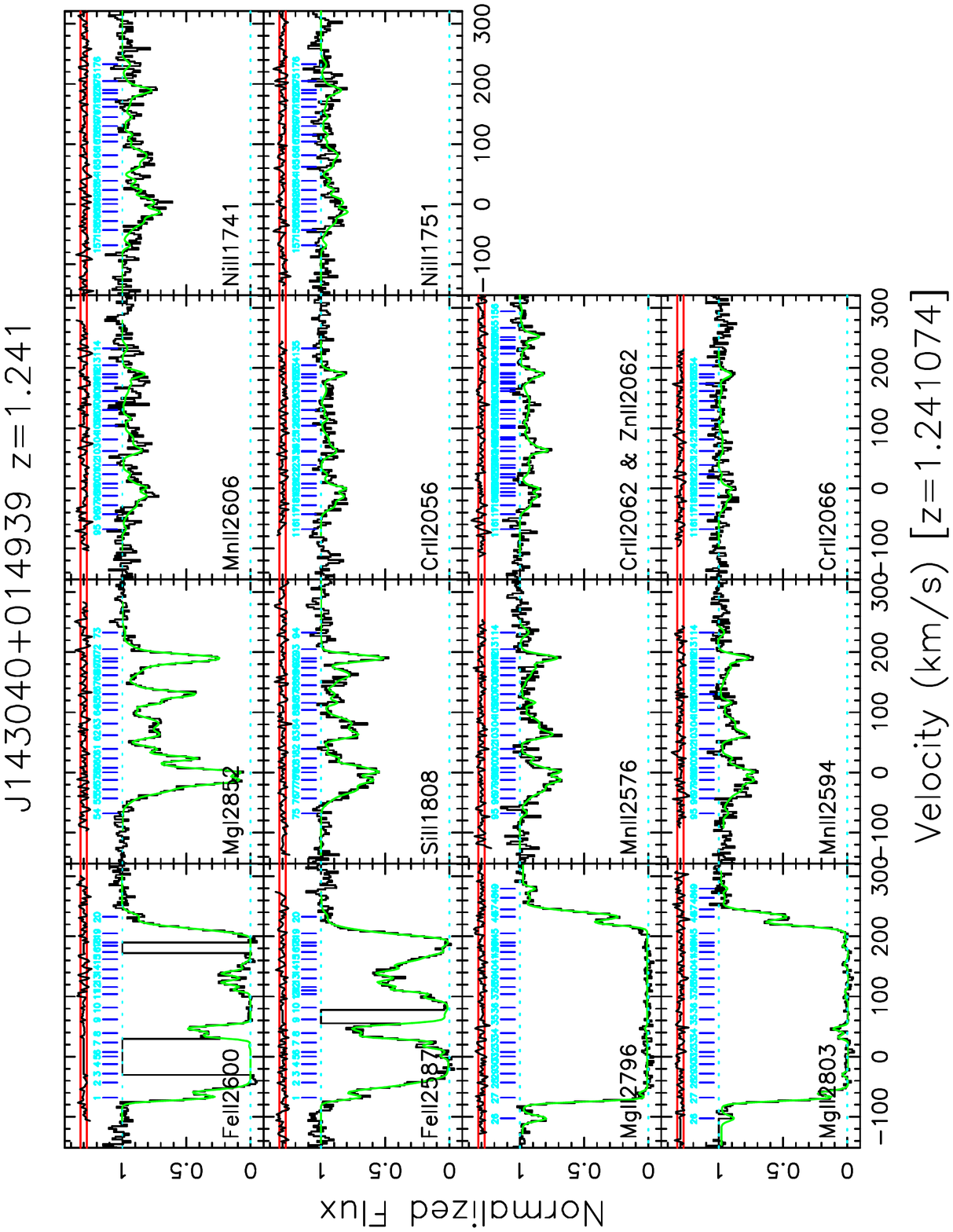}
\par\end{centering}

\caption[\ Fit for the $z=1.241$ absorber toward J143040+014939]{Many-multiplet fit for the $z=1.241$ absorber toward J143040+014939.}
\end{figure}
\begin{figure}[H]
\noindent \begin{centering}
\includegraphics[bb=34bp 58bp 554bp 738bp,clip,width=1\textwidth]{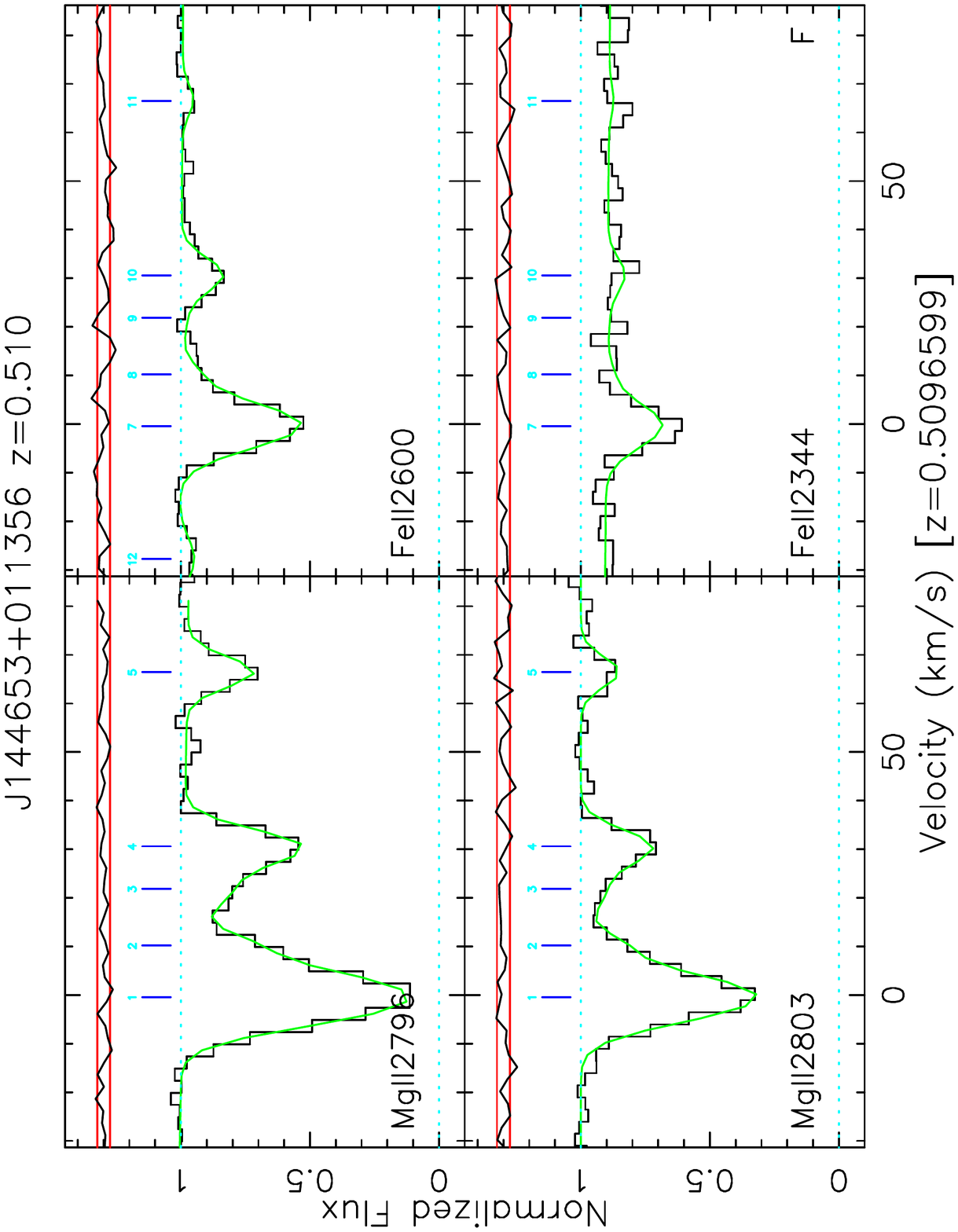}
\par\end{centering}

\caption[\ Fit for the $z=0.510$ absorber toward J144653+011356]{Many-multiplet fit for the $z=0.510$ absorber toward J144653+011356.}
\end{figure}
\begin{figure}[H]
\noindent \begin{centering}
\includegraphics[bb=34bp 58bp 554bp 738bp,clip,width=1\textwidth]{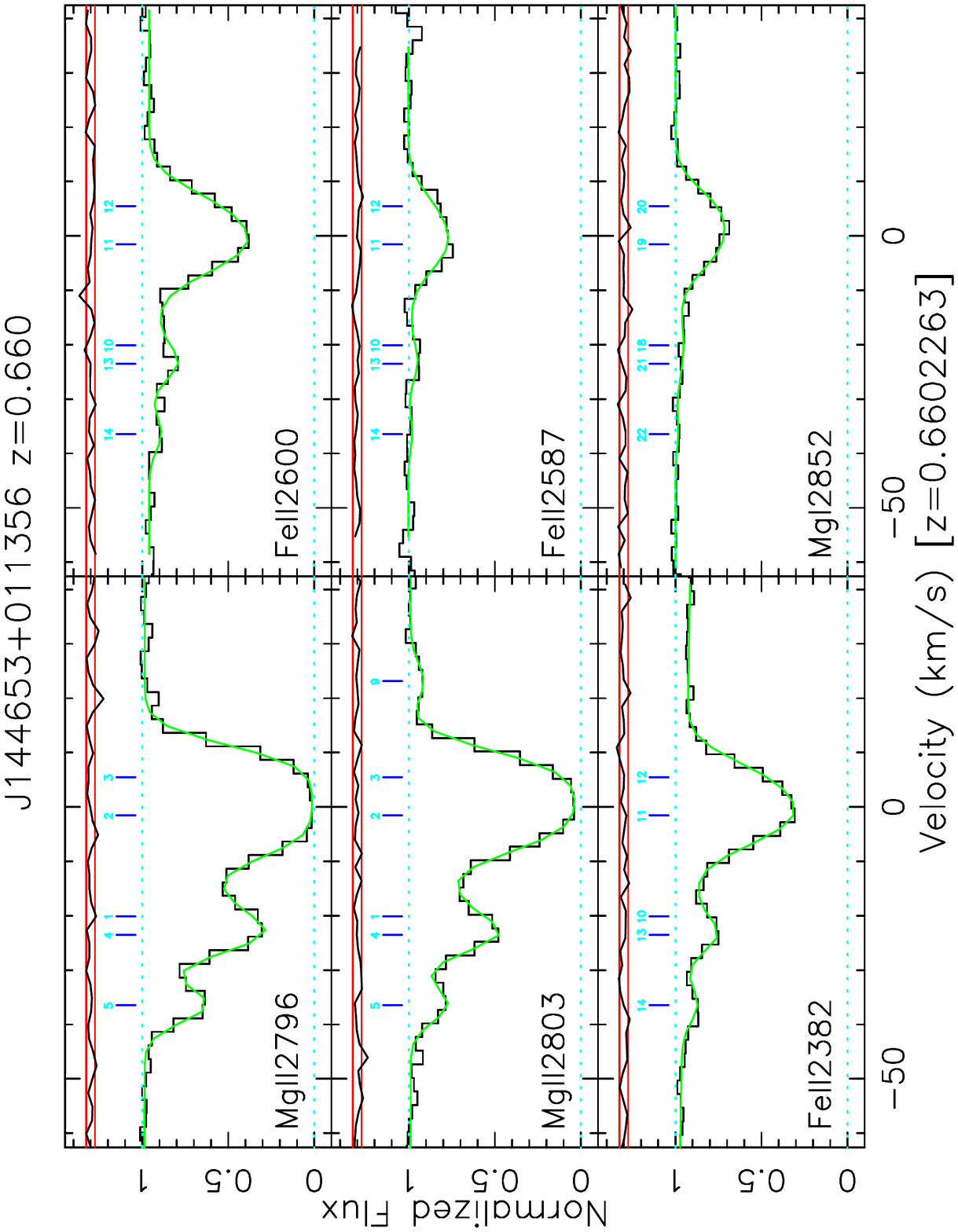}
\par\end{centering}

\caption[\ Fit for the $z=0.660$ absorber toward J144653+011356]{Many-multiplet fit for the $z=0.660$ absorber toward J144653+011356.}
\end{figure}
\begin{figure}[H]
\noindent \begin{centering}
\includegraphics[bb=34bp 58bp 554bp 738bp,clip,width=1\textwidth]{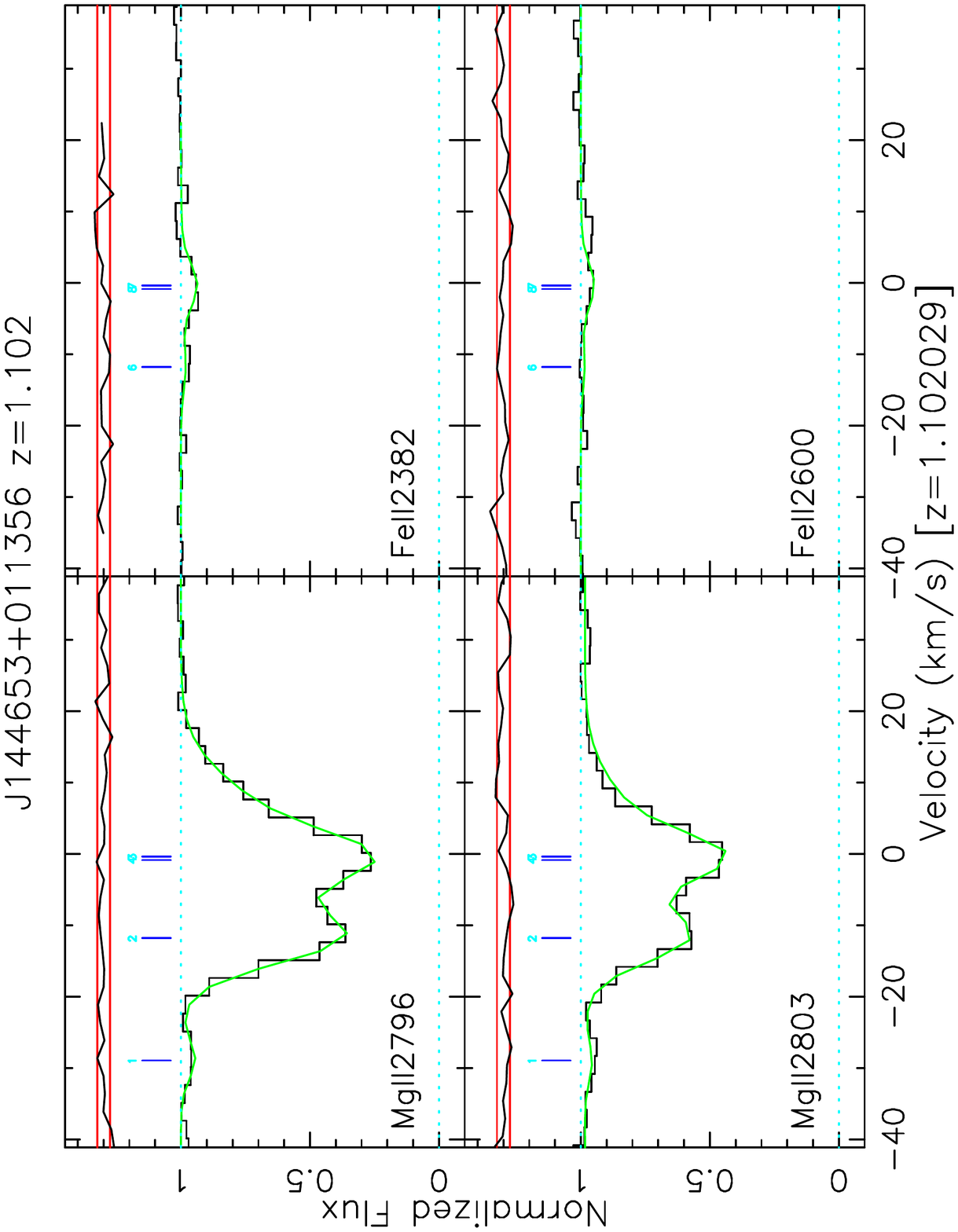}
\par\end{centering}

\caption[\ Fit for the $z=1.102$ absorber toward J144653+011356]{Many-multiplet fit for the $z=1.102$ absorber toward J144653+011356.}
\end{figure}
\begin{figure}[H]
\noindent \begin{centering}
\includegraphics[bb=34bp 58bp 554bp 738bp,clip,width=1\textwidth]{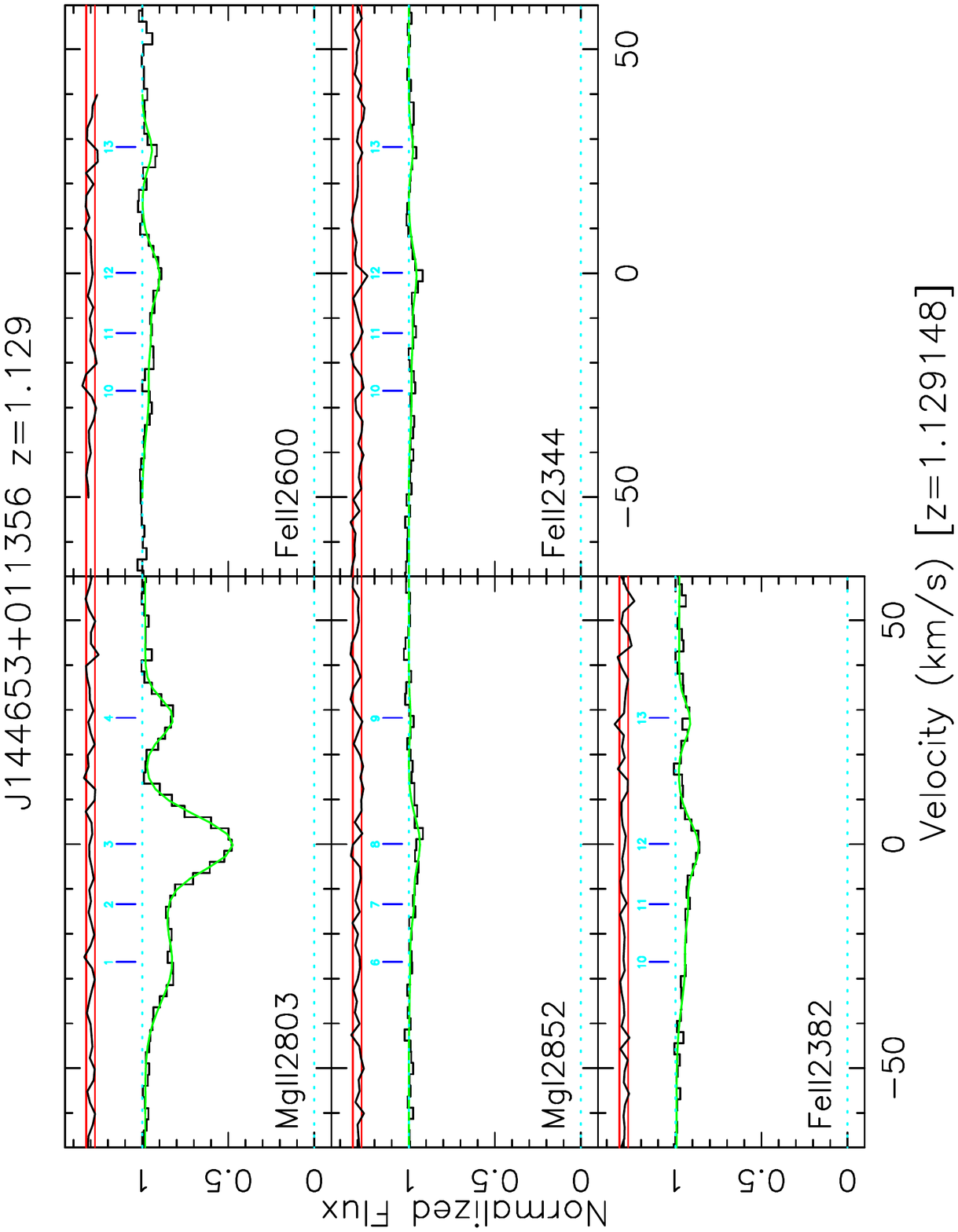}
\par\end{centering}

\caption[\ Fit for the $z=1.129$ absorber toward J144653+011356]{Many-multiplet fit for the $z=1.129$ absorber toward J144653+011356.}
\end{figure}
\begin{figure}[H]
\noindent \begin{centering}
\includegraphics[bb=34bp 58bp 554bp 738bp,clip,width=1\textwidth]{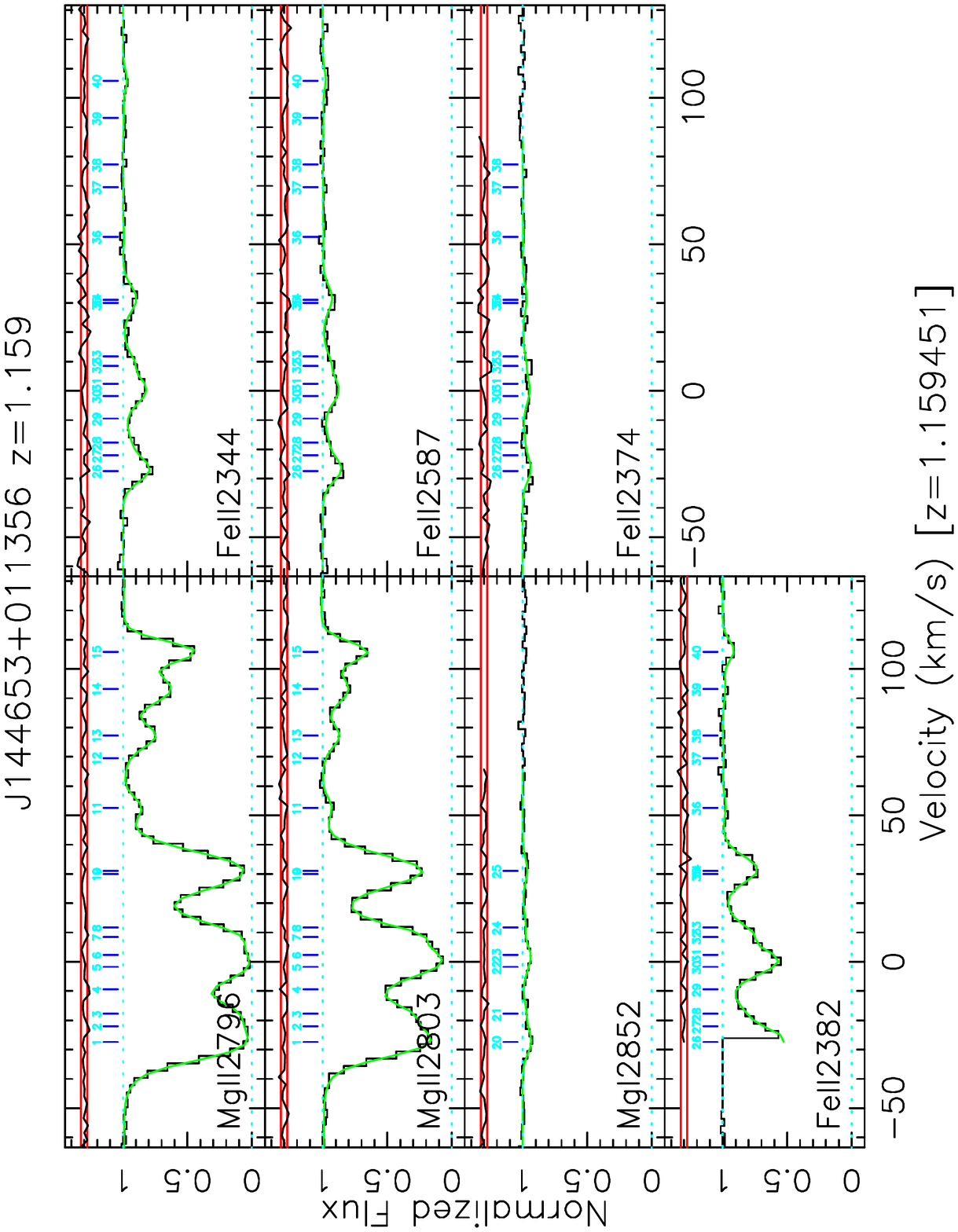}
\par\end{centering}

\caption[\ Fit for the $z=1.159$ absorber toward J144653+011356]{Many-multiplet fit for the $z=1.159$ absorber toward J144653+011356.}
\end{figure}
\begin{figure}[H]
\noindent \begin{centering}
\includegraphics[bb=34bp 58bp 554bp 738bp,clip,width=1\textwidth]{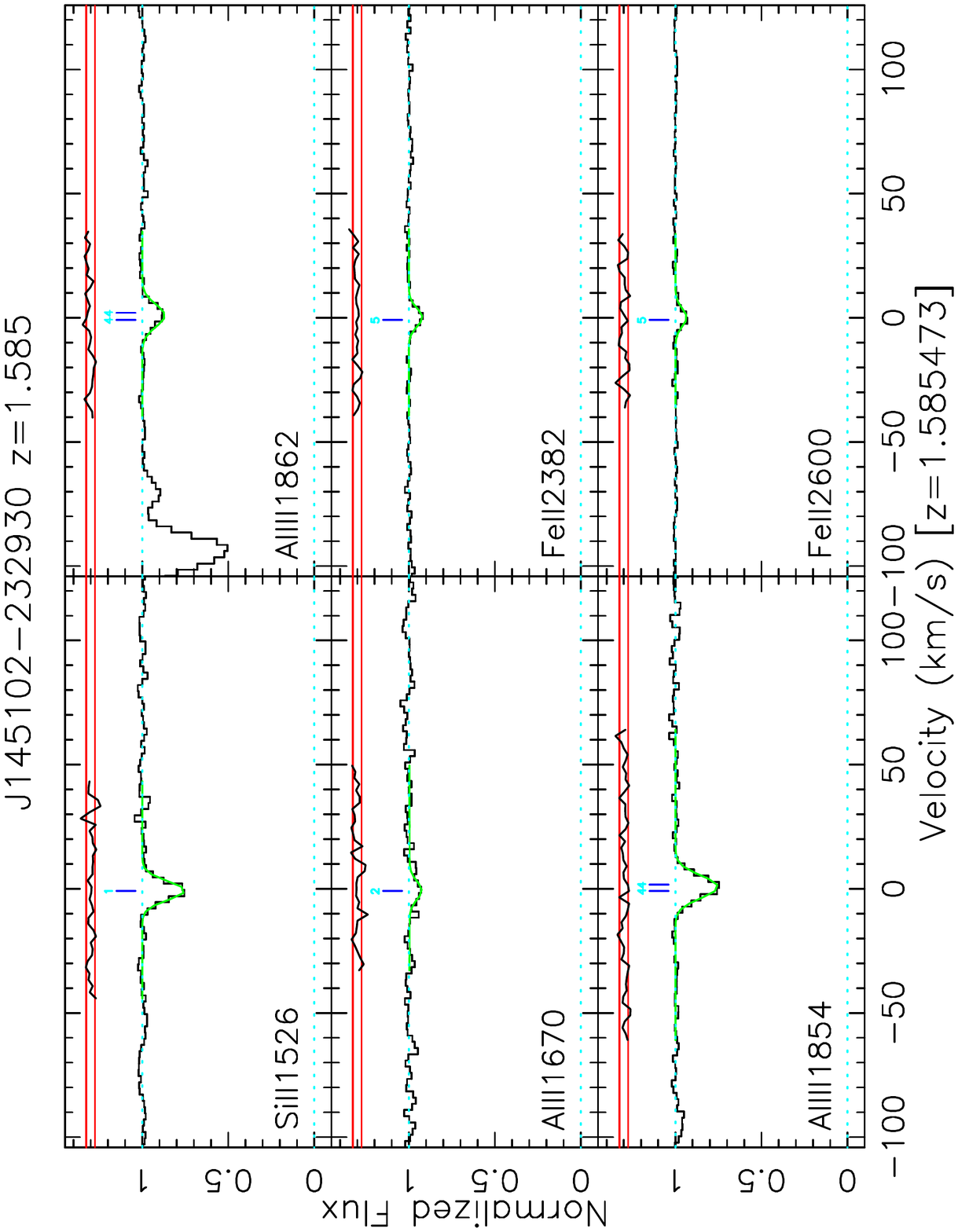}
\par\end{centering}

\caption[\ Fit for the $z=1.585$ absorber toward J145102$-$232930]{Many-multiplet fit for the $z=1.585$ absorber toward J145102$-$232930.}
\end{figure}
\begin{figure}[H]
\noindent \begin{centering}
\includegraphics[bb=34bp 58bp 554bp 738bp,clip,width=1\textwidth]{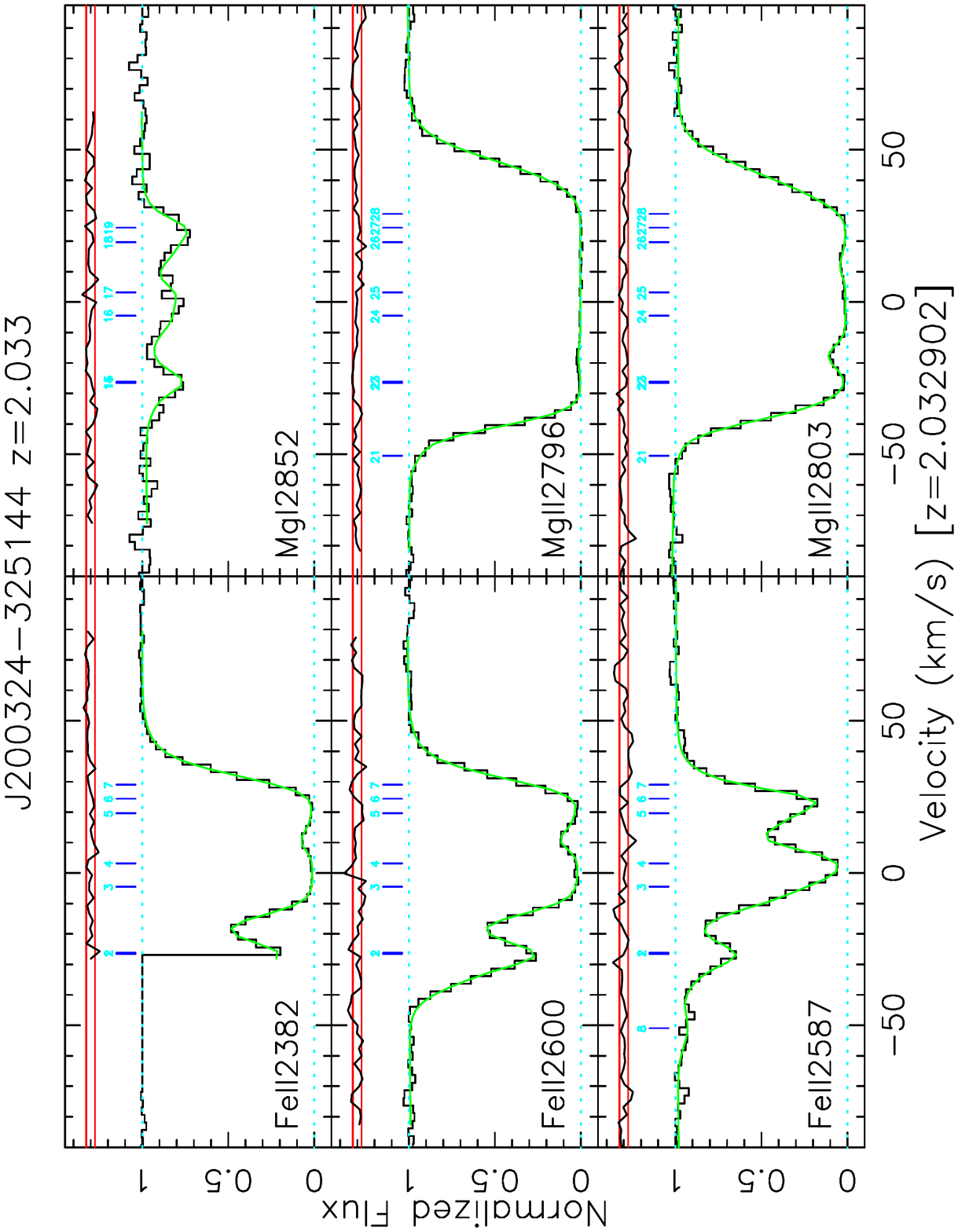}
\par\end{centering}

\caption[\ Fit for the $z=2.033$ absorber toward J200324$-$325144]{Many-multiplet fit for the $z=2.033$ absorber toward J200324$-$325144.}
\end{figure}
\begin{figure}[H]
\noindent \begin{centering}
\includegraphics[bb=34bp 58bp 554bp 738bp,clip,width=1\textwidth]{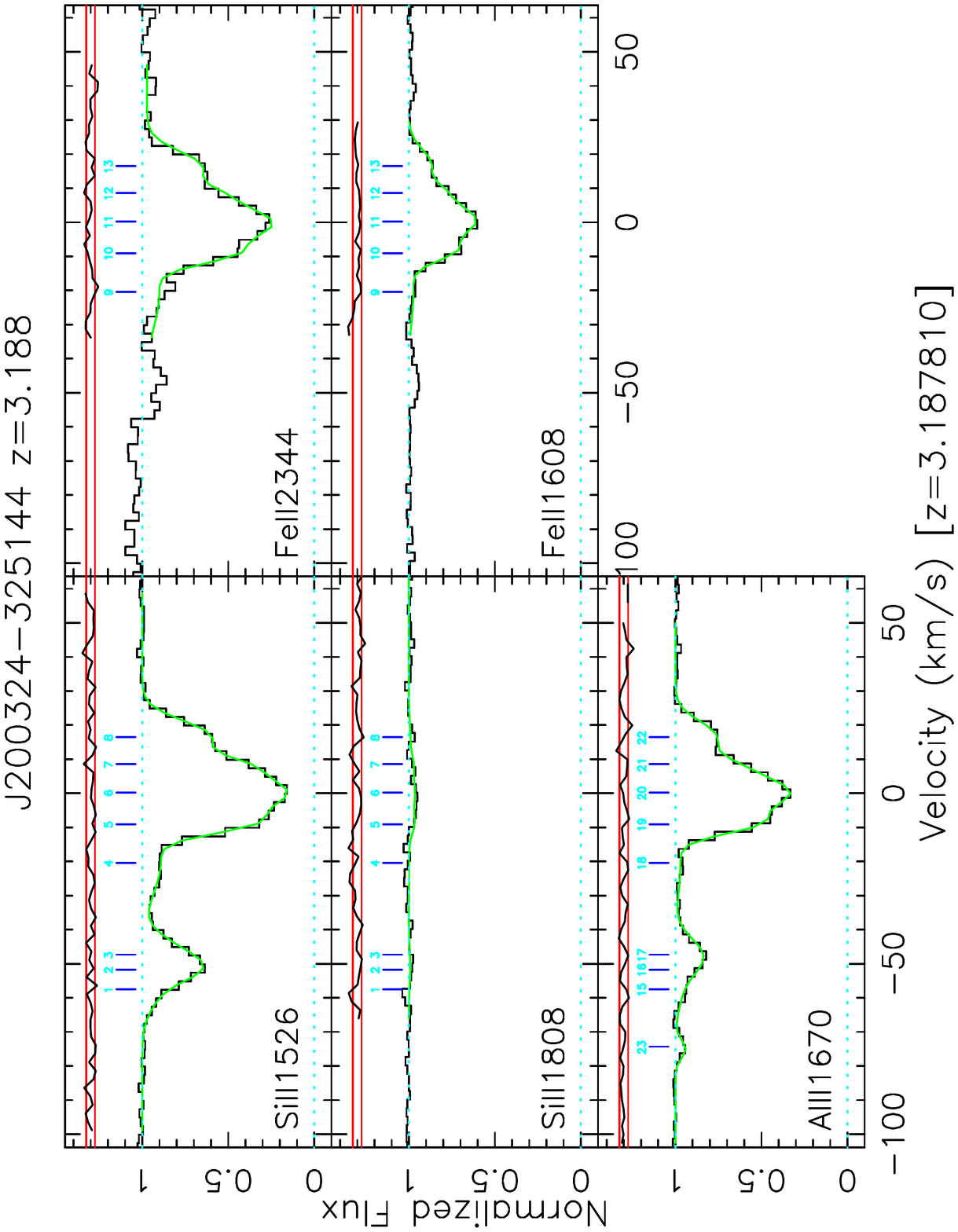}
\par\end{centering}

\caption[\ Fit for the $z=3.188$ absorber toward J200324$-$325144]{Many-multiplet fit for the $z=3.188$ absorber toward J200324$-$325144.}
\end{figure}
\begin{figure}[H]
\noindent \begin{centering}
\includegraphics[bb=34bp 58bp 554bp 738bp,clip,width=1\textwidth]{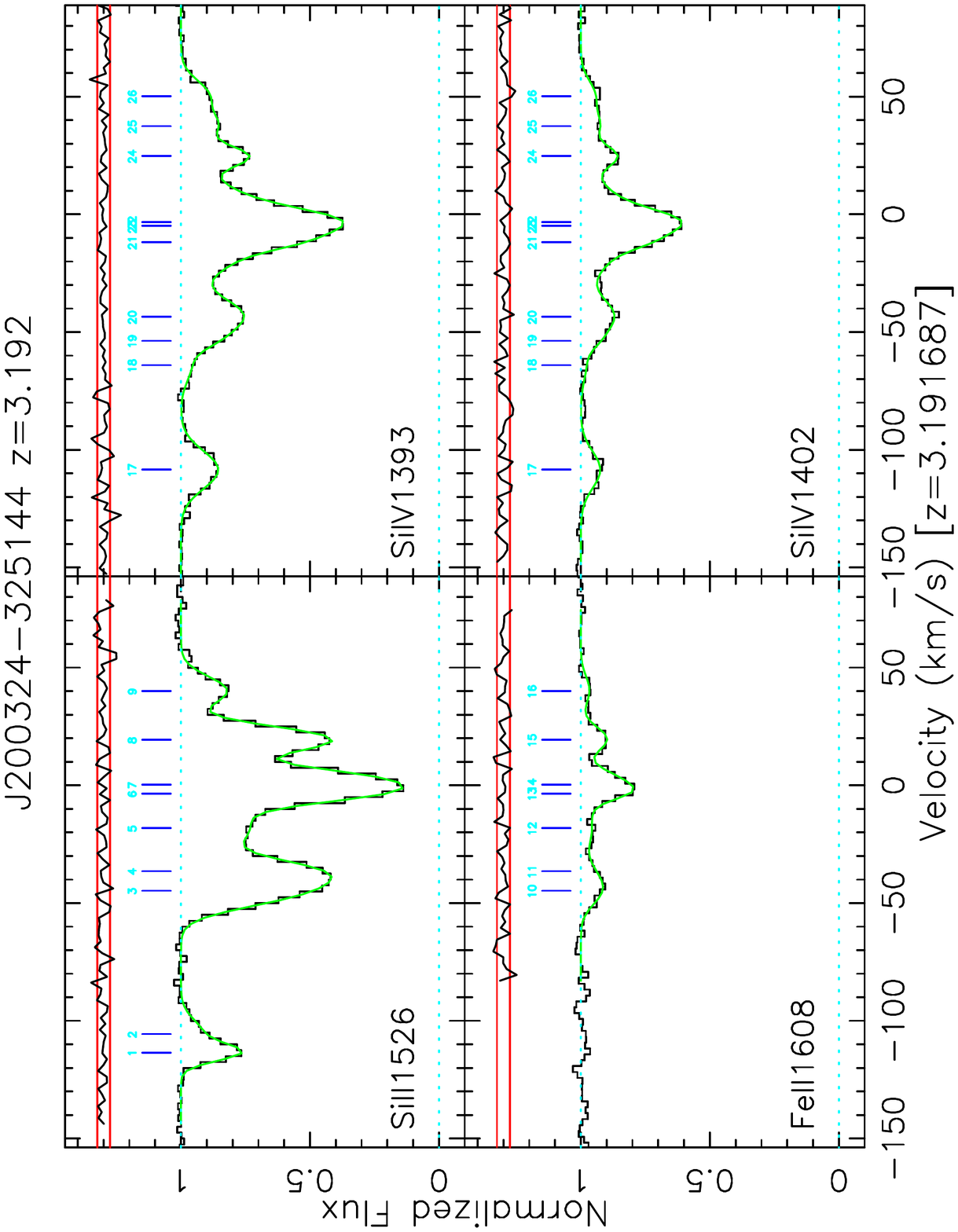}
\par\end{centering}

\caption[\ Fit for the $z=3.192$ absorber toward J200324$-$325144]{Many-multiplet fit for the $z=3.192$ absorber toward J200324$-$325144.}
\end{figure}
\begin{figure}[H]
\noindent \begin{centering}
\includegraphics[bb=34bp 58bp 554bp 738bp,clip,width=1\textwidth]{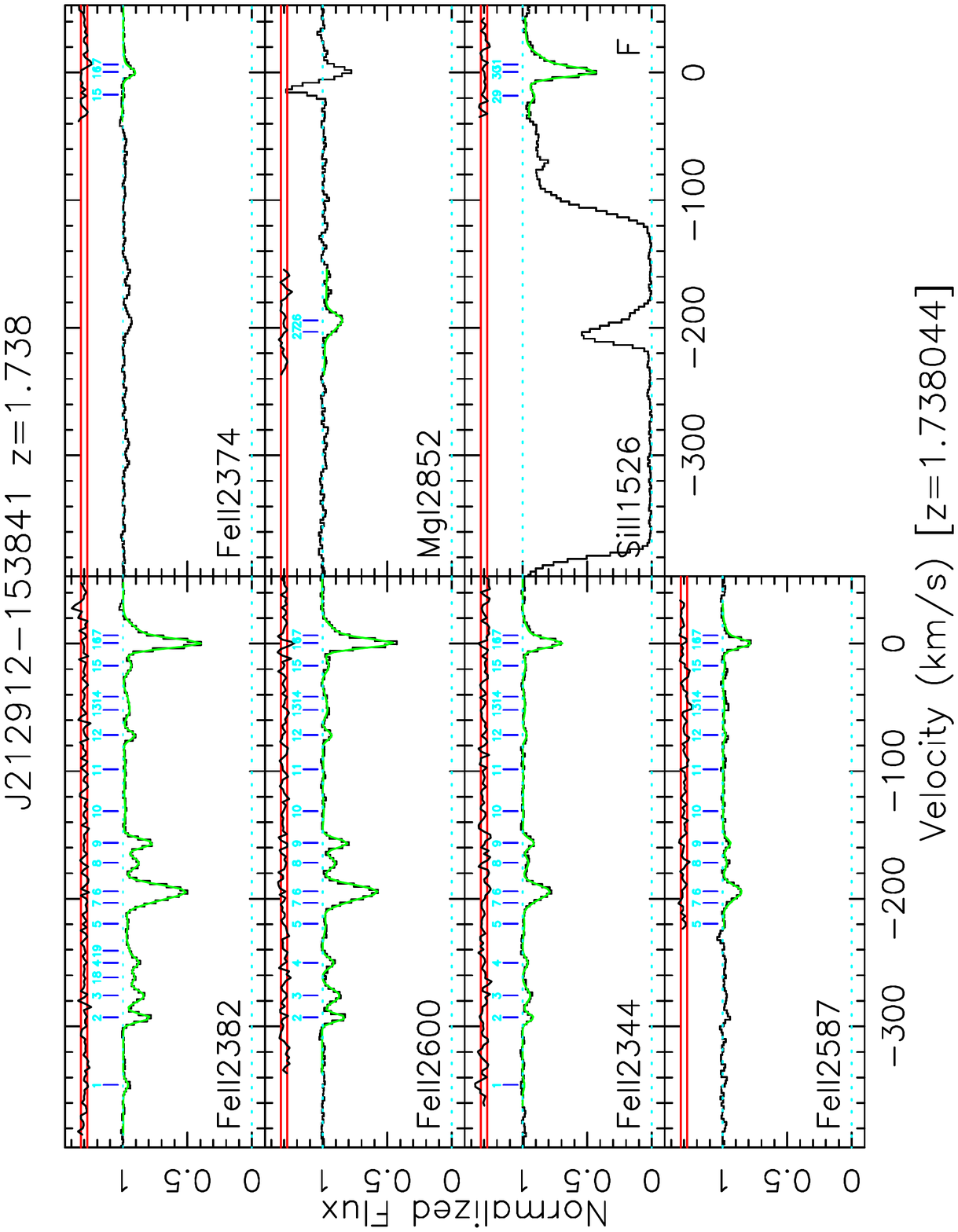}
\par\end{centering}

\caption[\ Fit for the $z=1.738$ absorber toward J212912$-$153841]{Many-multiplet fit for the $z=1.738$ absorber toward J212912$-$153841.}
\end{figure}
\begin{figure}[H]
\noindent \begin{centering}
\includegraphics[bb=34bp 58bp 554bp 738bp,clip,width=1\textwidth]{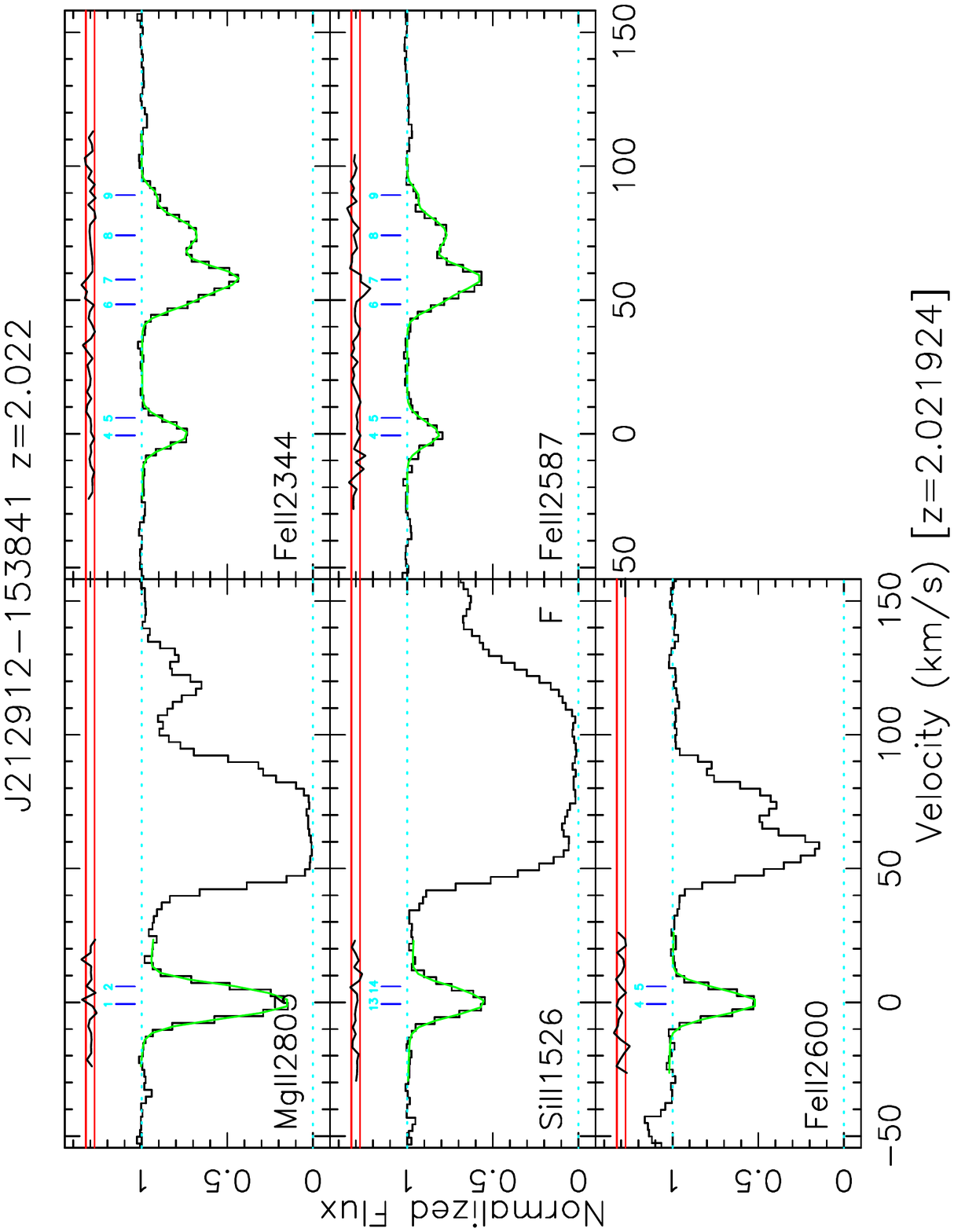}
\par\end{centering}

\caption[\ Fit for the $z=2.022$ absorber toward J212912$-$153841]{Many-multiplet fit for the $z=2.022$ absorber toward J212912$-$153841.}
\end{figure}
\begin{figure}[H]
\noindent \begin{centering}
\includegraphics[bb=34bp 58bp 554bp 738bp,clip,width=1\textwidth]{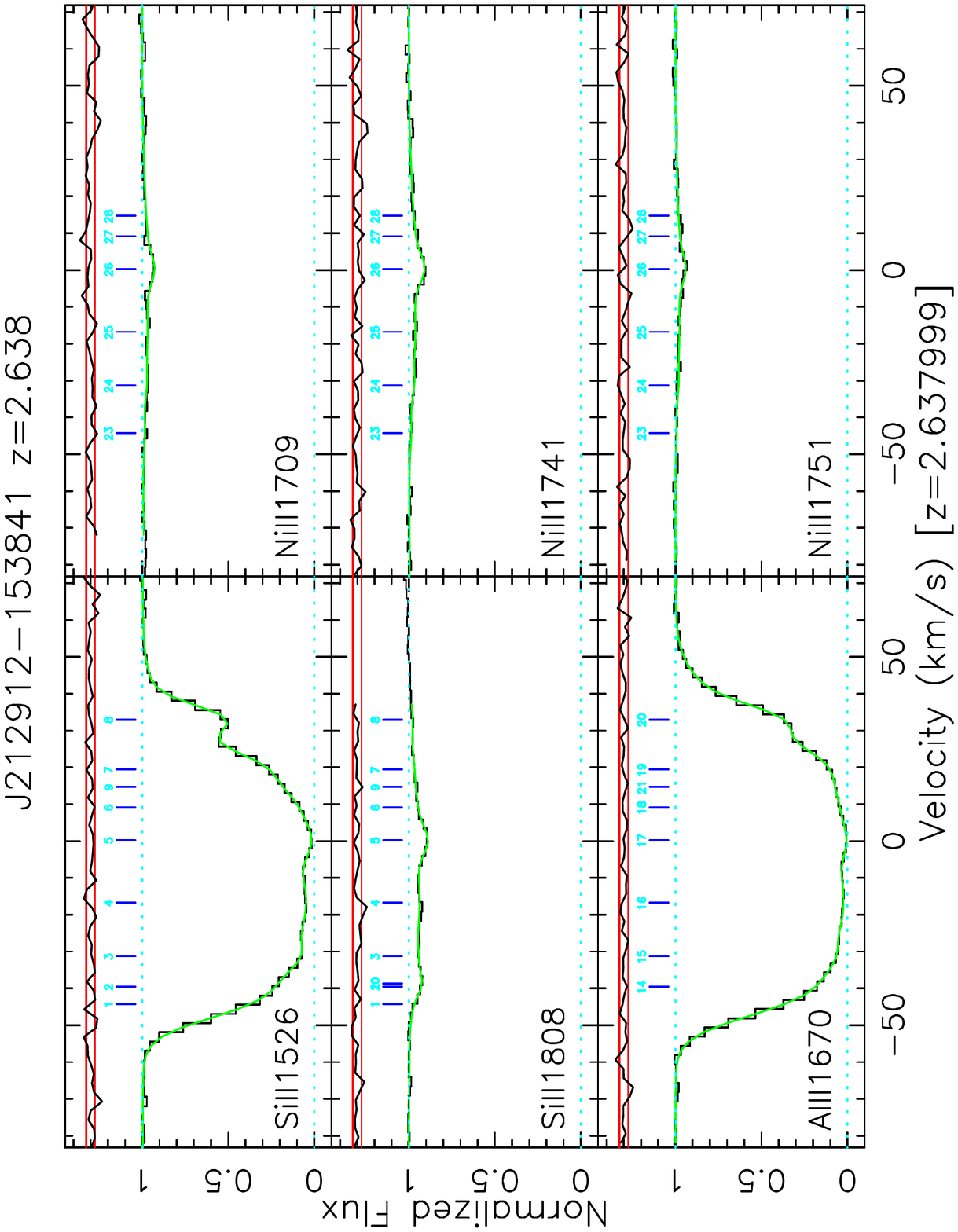}
\par\end{centering}

\caption[\ Fit for the $z=2.638$ absorber toward J212912$-$153841]{Many-multiplet fit for the $z=2.638$ absorber toward J212912$-$153841.\label{appfig:J212912-153841-z2.638}}
\end{figure}
\begin{figure}[H]
\noindent \begin{centering}
\includegraphics[bb=34bp 58bp 554bp 738bp,clip,width=1\textwidth]{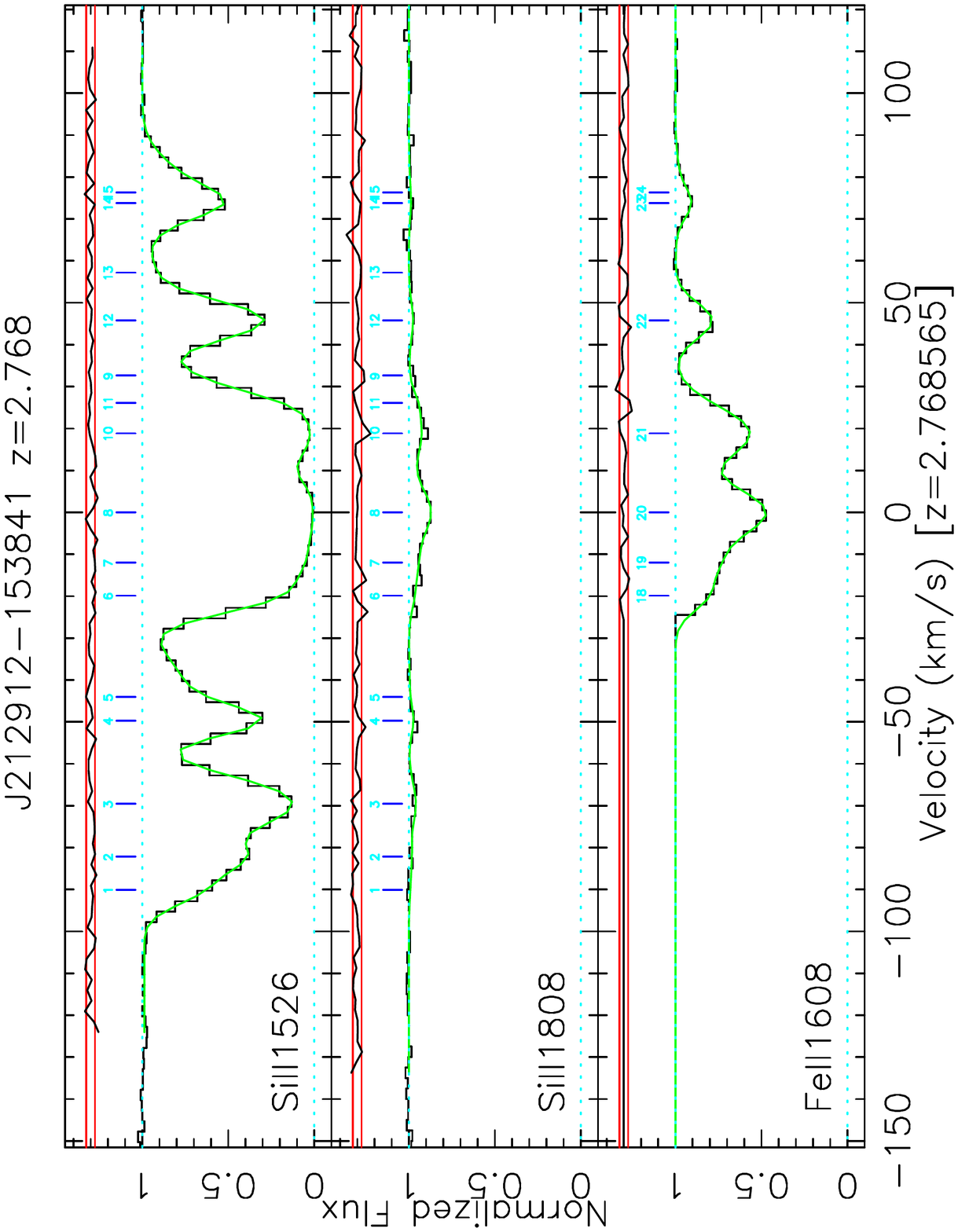}
\par\end{centering}

\caption[\ Fit for the $z=2.768$ absorber toward J212912$-$153841]{Many-multiplet fit for the $z=2.768$ absorber toward J212912$-$153841.}
\end{figure}
\begin{figure}[H]
\noindent \begin{centering}
\includegraphics[bb=34bp 58bp 554bp 738bp,clip,width=1\textwidth]{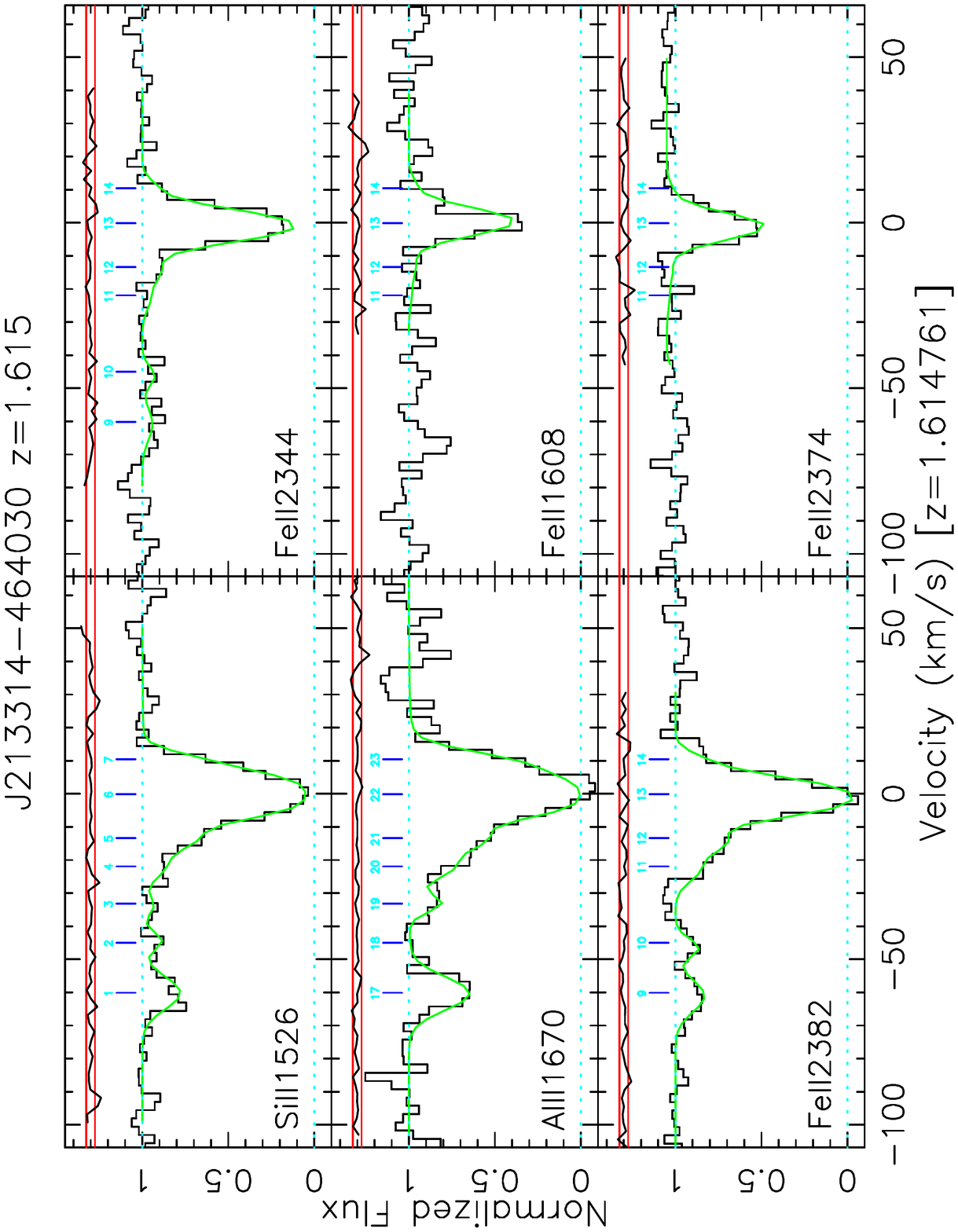}
\par\end{centering}

\caption[\ Fit for the $z=1.615$ absorber toward J213314$-$464030]{Many-multiplet fit for the $z=1.615$ absorber toward J213314$-$464030.}
\end{figure}
\begin{figure}[H]
\noindent \begin{centering}
\includegraphics[bb=34bp 58bp 554bp 738bp,clip,width=1\textwidth]{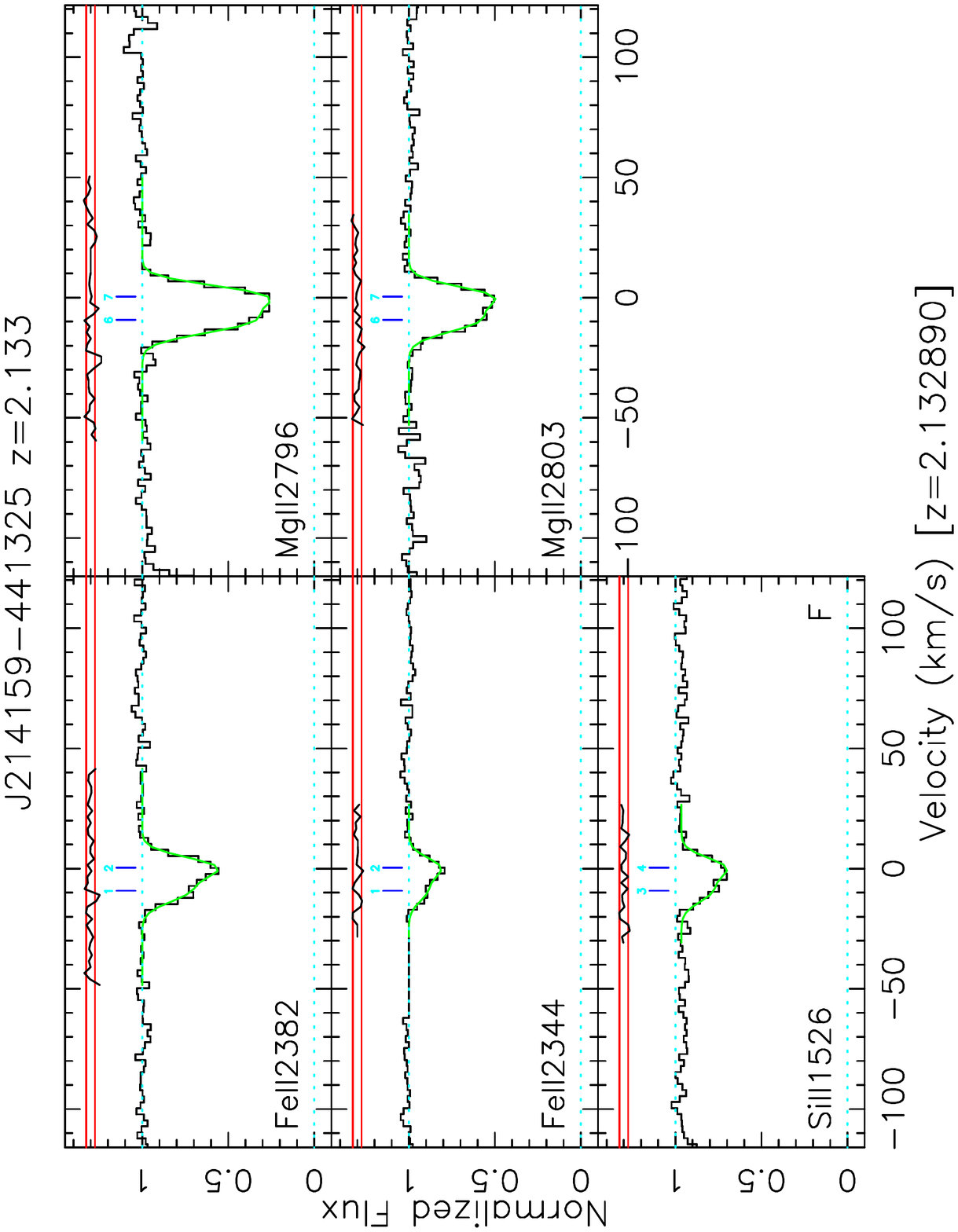}
\par\end{centering}

\caption[\ Fit for the $z=2.133$ absorber toward J214159$-$441325]{Many-multiplet fit for the $z=2.133$ absorber toward J214159$-$441325.}
\end{figure}
\begin{figure}[H]
\noindent \begin{centering}
\includegraphics[bb=34bp 58bp 554bp 738bp,clip,width=1\textwidth]{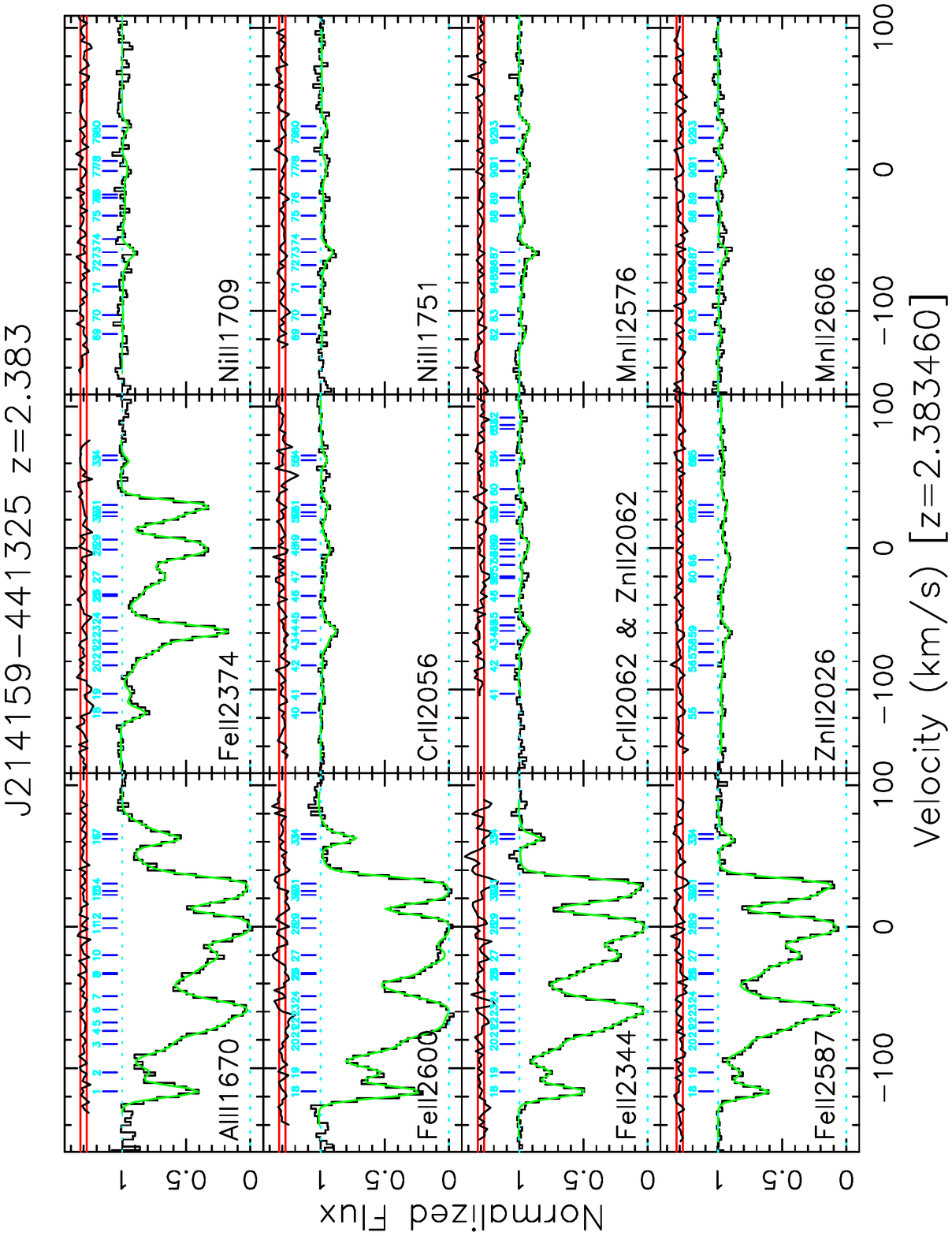}
\par\end{centering}

\caption[\ Fit for the $z=2.383$ absorber toward J214159$-$441325]{Many-multiplet fit for the $z=2.383$ absorber toward J214159$-$441325.}
\end{figure}
\begin{figure}[H]
\noindent \begin{centering}
\includegraphics[bb=34bp 58bp 554bp 738bp,clip,width=1\textwidth]{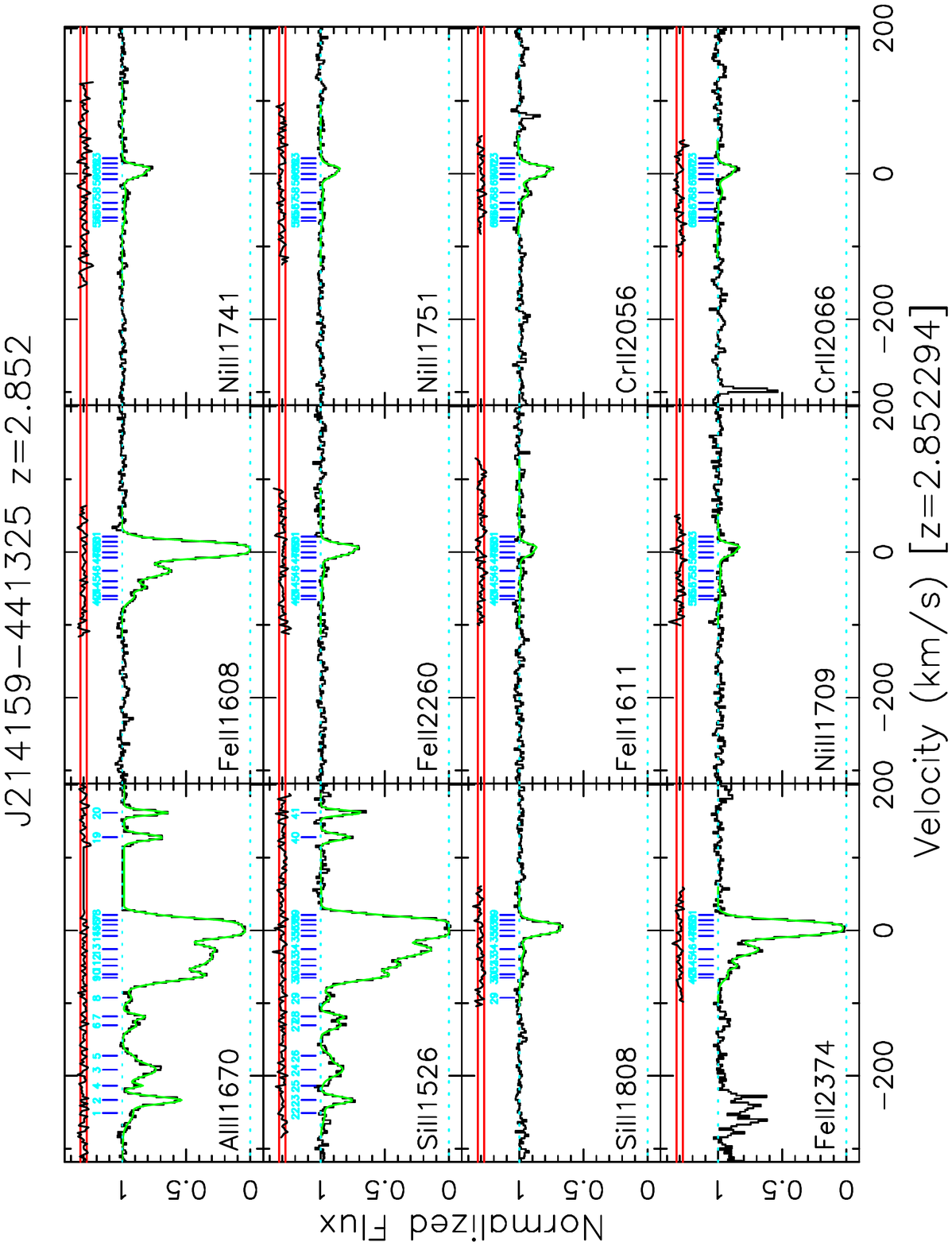}
\par\end{centering}

\caption[\ Fit for the $z=2.852$ absorber toward J214159$-$441325]{Many-multiplet fit for the $z=2.852$ absorber toward J214159$-$441325.}
\end{figure}
\begin{figure}[H]
\noindent \begin{centering}
\includegraphics[bb=34bp 58bp 554bp 738bp,clip,width=1\textwidth]{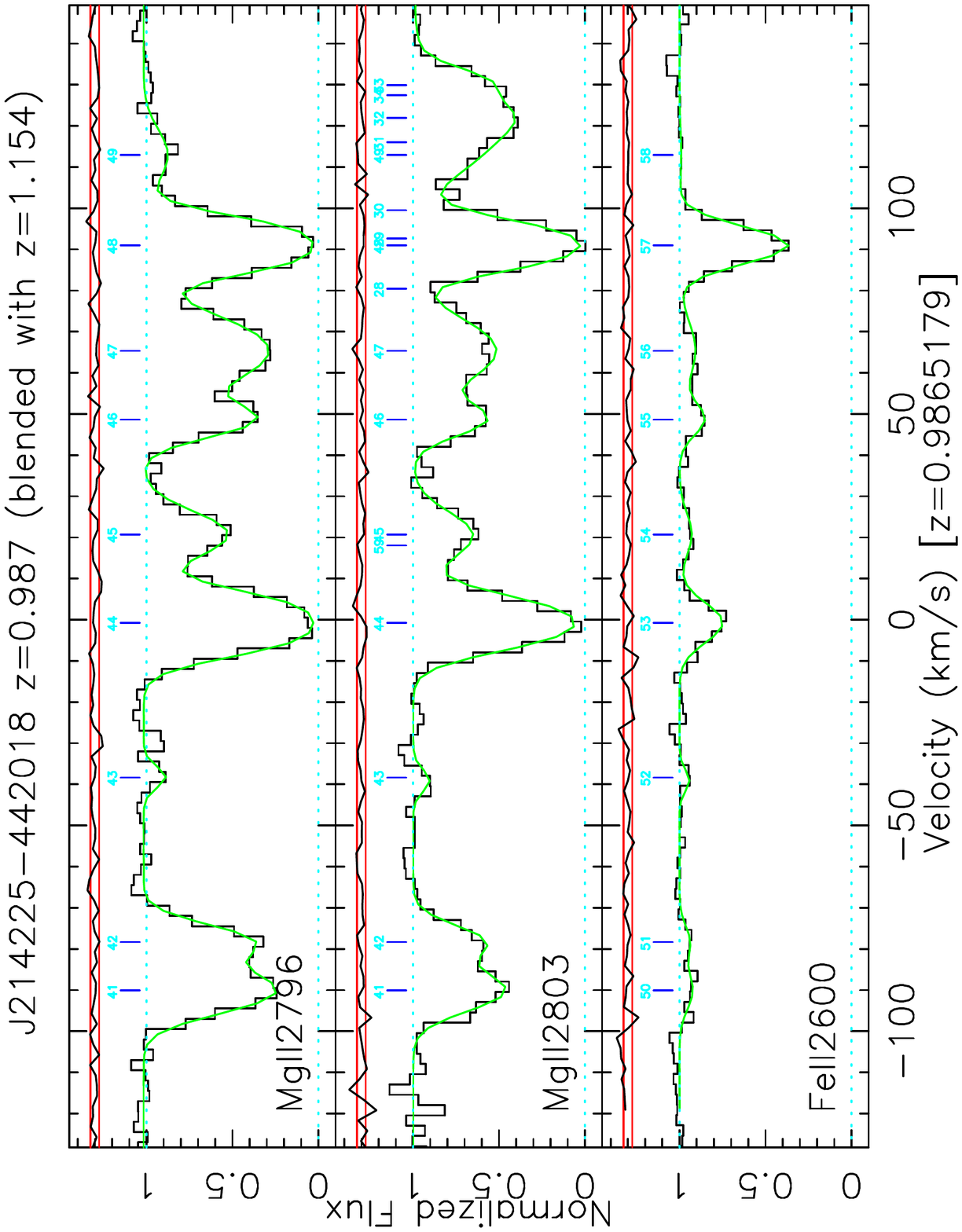}
\par\end{centering}

\caption[\ Fit for the $z=0.987$ absorber toward J214225$-$442018]{Many-multiplet fit for the $z=0.987$ absorber toward J214225$-$442018.}
\end{figure}
\begin{figure}[H]
\noindent \begin{centering}
\includegraphics[bb=34bp 58bp 554bp 738bp,clip,width=1\textwidth]{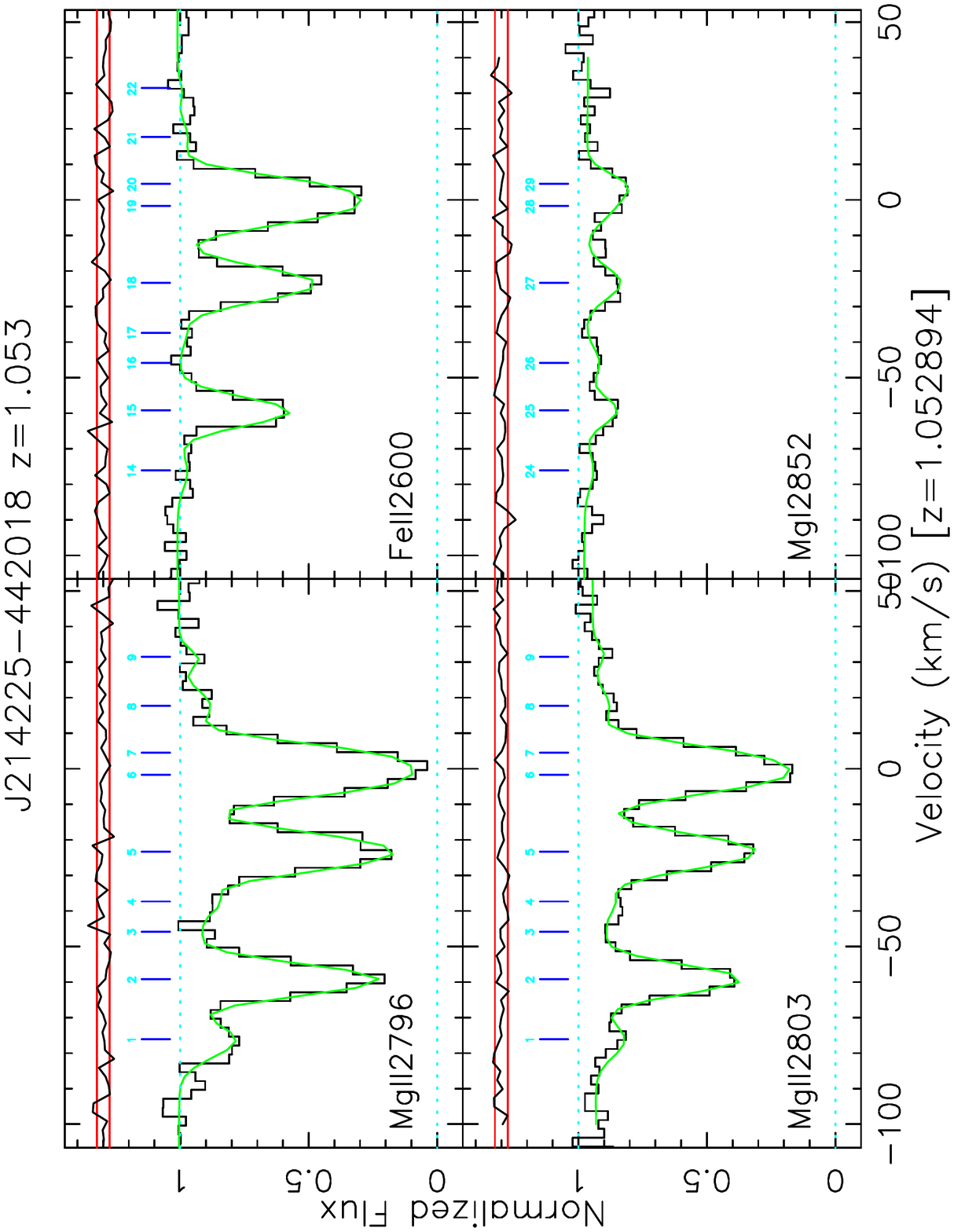}
\par\end{centering}

\caption[\ Fit for the $z=1.053$ absorber toward J214225$-$442018]{Many-multiplet fit for the $z=1.053$ absorber toward J214225$-$442018.}
\end{figure}
\begin{figure}[H]
\noindent \begin{centering}
\includegraphics[bb=34bp 58bp 554bp 738bp,clip,width=1\textwidth]{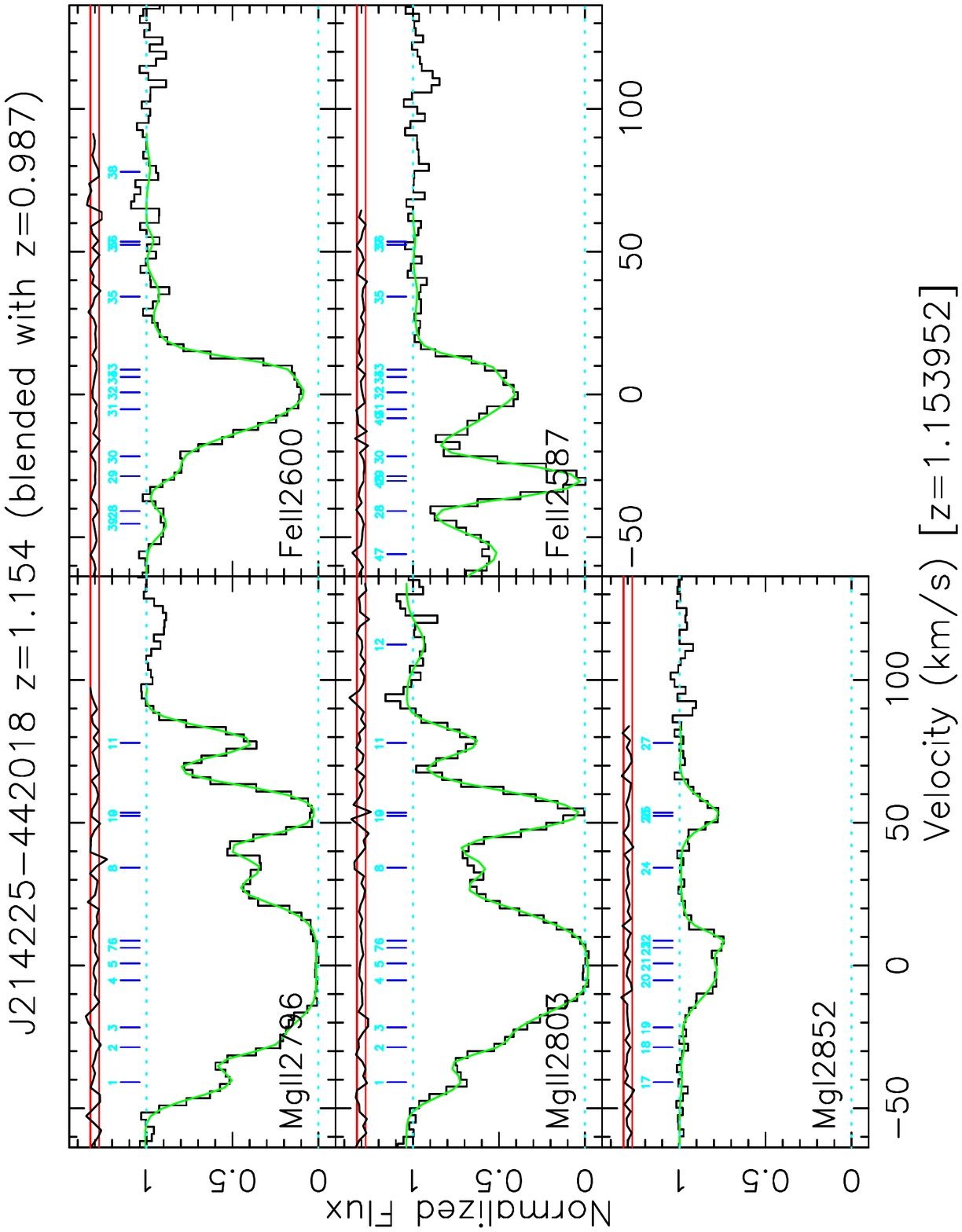}
\par\end{centering}

\caption[\ Fit for the $z=1.154$ absorber toward J214225$-$442018]{Many-multiplet fit for the $z=1.154$ absorber toward J214225$-$442018.}
\end{figure}
\begin{figure}[H]
\noindent \begin{centering}
\includegraphics[bb=34bp 58bp 554bp 738bp,clip,width=1\textwidth]{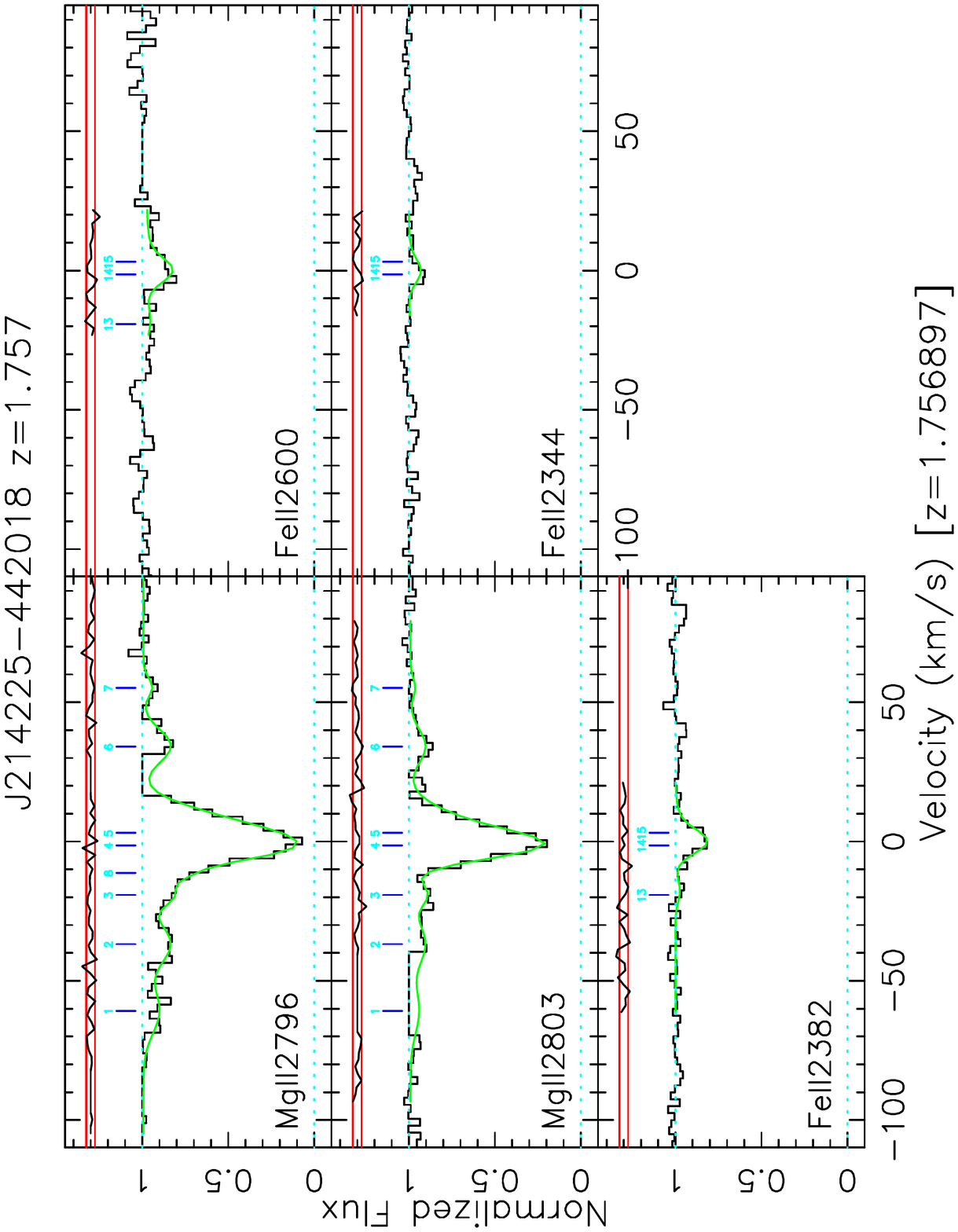}
\par\end{centering}

\caption[\ Fit for the $z=1.757$ absorber toward J214225$-$442018]{Many-multiplet fit for the $z=1.757$ absorber toward J214225$-$442018.}
\end{figure}
\begin{figure}[H]
\noindent \begin{centering}
\includegraphics[bb=34bp 58bp 554bp 738bp,clip,width=1\textwidth]{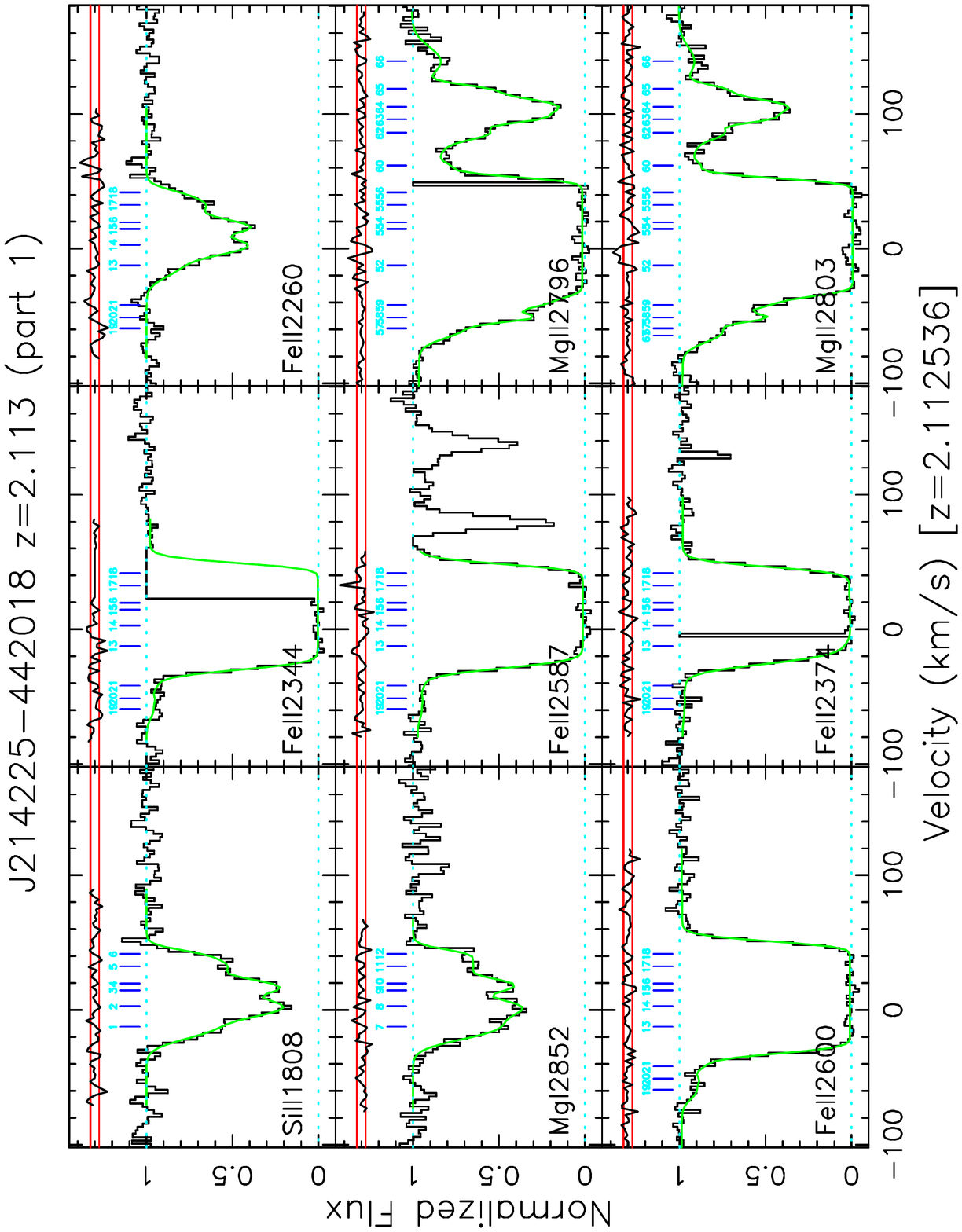}
\par\end{centering}

\caption[\ Fit for the $z=2.113$ absorber toward J214225$-$442018 (part 1)]{Many-multiplet fit for the $z=2.113$ absorber toward J214225$-$442018 (part 1).}
\end{figure}
\begin{figure}[H]
\noindent \begin{centering}
\includegraphics[bb=34bp 58bp 554bp 738bp,clip,width=1\textwidth]{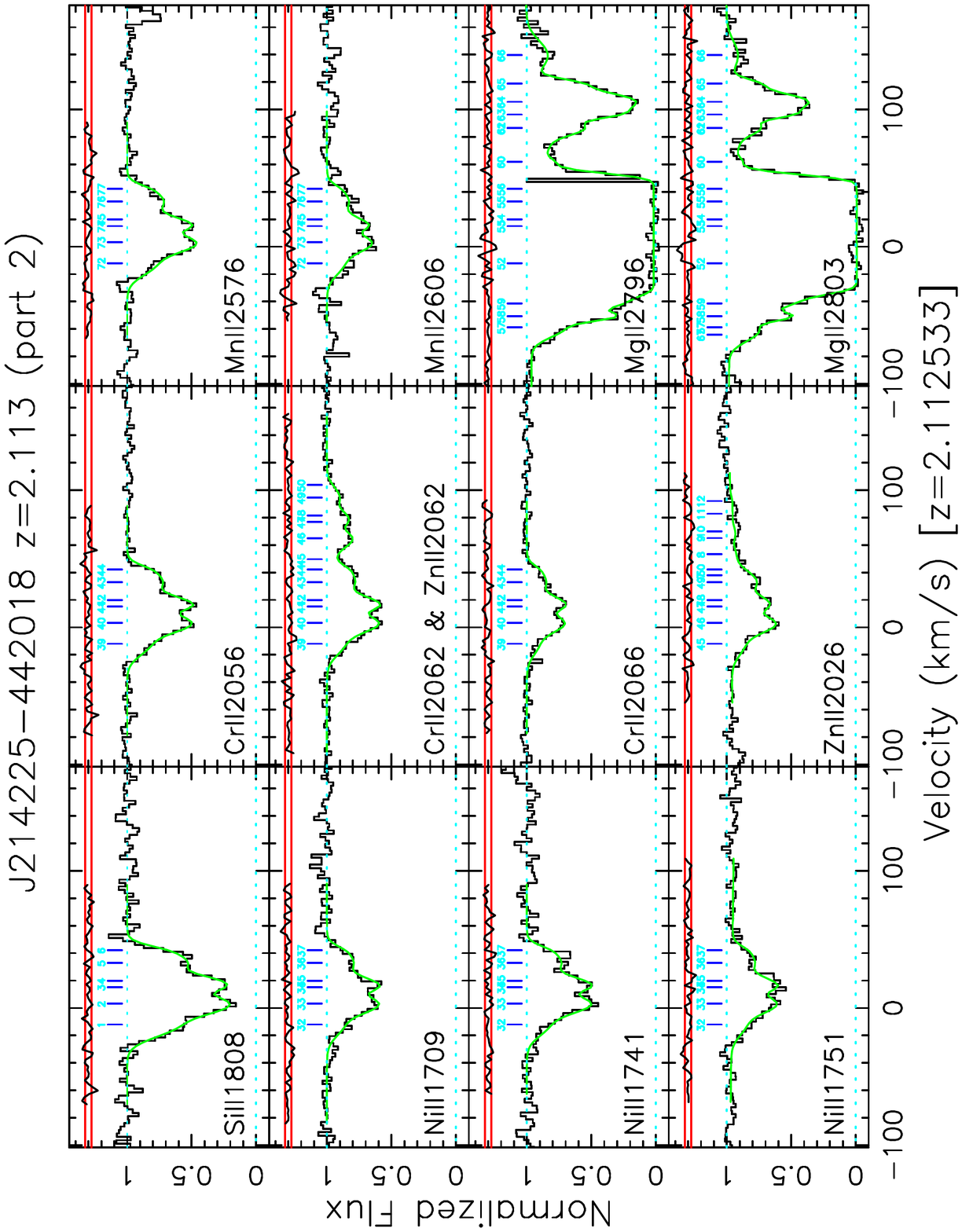}
\par\end{centering}

\caption[\ Fit for the $z=2.113$ absorber toward J214225$-$442018 (part 2)]{Many-multiplet fit for the $z=2.113$ absorber toward J214225$-$442018 (part 2).}
\end{figure}
\begin{figure}[H]
\noindent \begin{centering}
\includegraphics[bb=34bp 58bp 554bp 738bp,clip,width=1\textwidth]{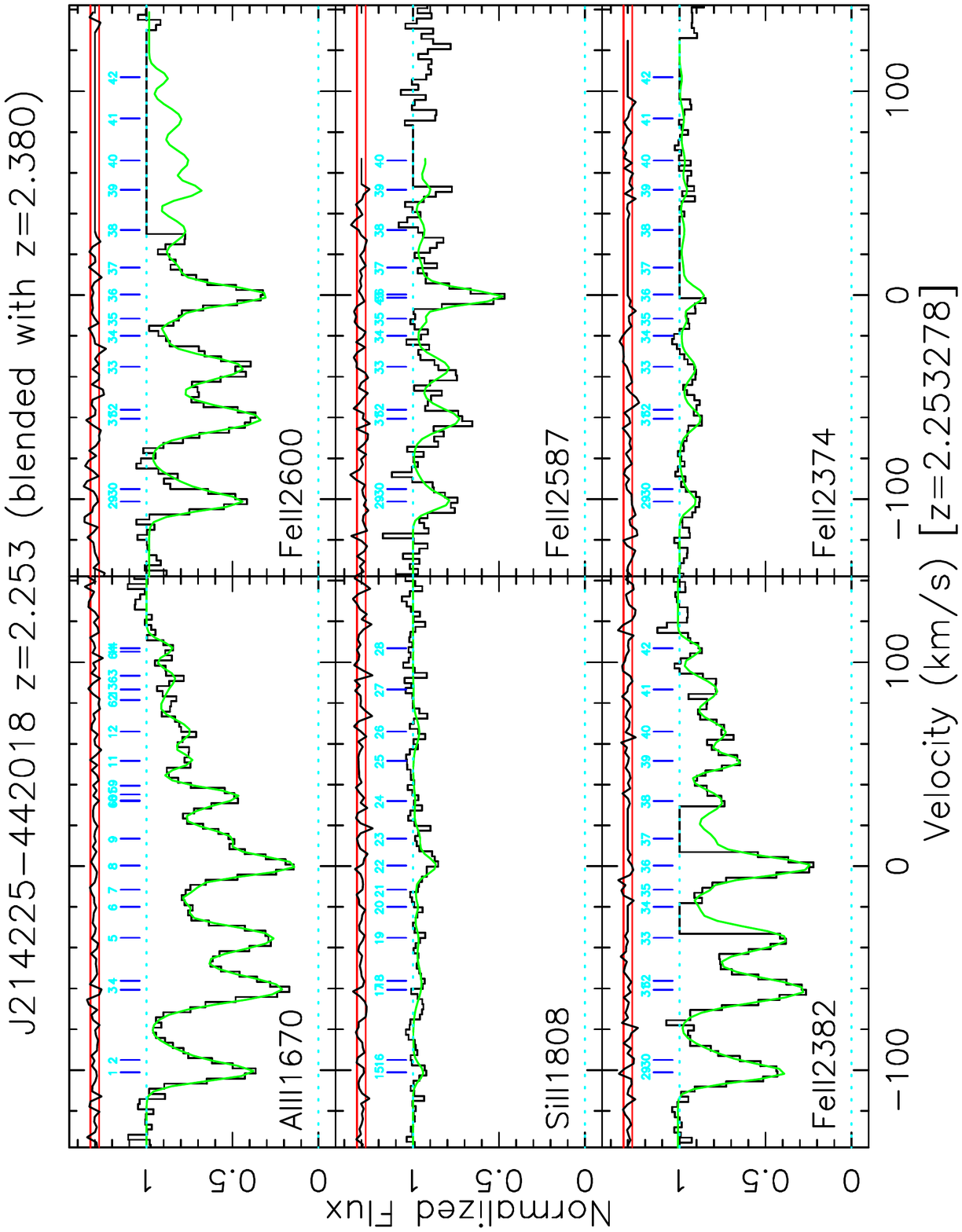}
\par\end{centering}

\caption[\ Fit for the $z=2.253$ absorber toward J214225$-$442018]{Many-multiplet fit for the $z=2.253$ absorber toward J214225$-$442018.}
\end{figure}
\begin{figure}[H]
\noindent \begin{centering}
\includegraphics[bb=34bp 58bp 554bp 738bp,clip,width=1\textwidth]{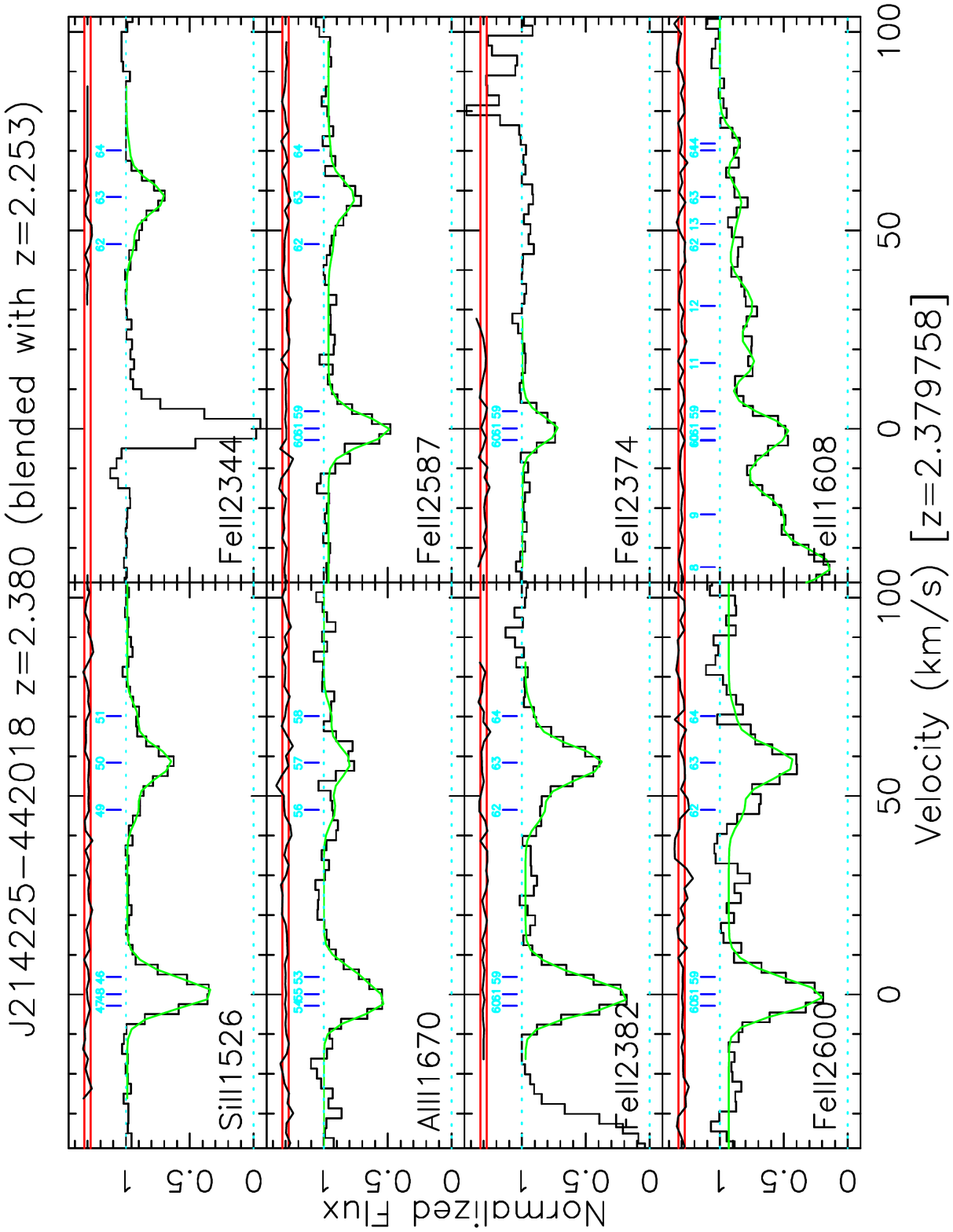}
\par\end{centering}

\caption[\ Fit for the $z=2.380$ absorber toward J214225$-$442018]{Many-multiplet fit for the $z=2.380$ absorber toward J214225$-$442018.}
\end{figure}
\begin{figure}[H]
\noindent \begin{centering}
\includegraphics[bb=34bp 58bp 554bp 738bp,clip,width=1\textwidth]{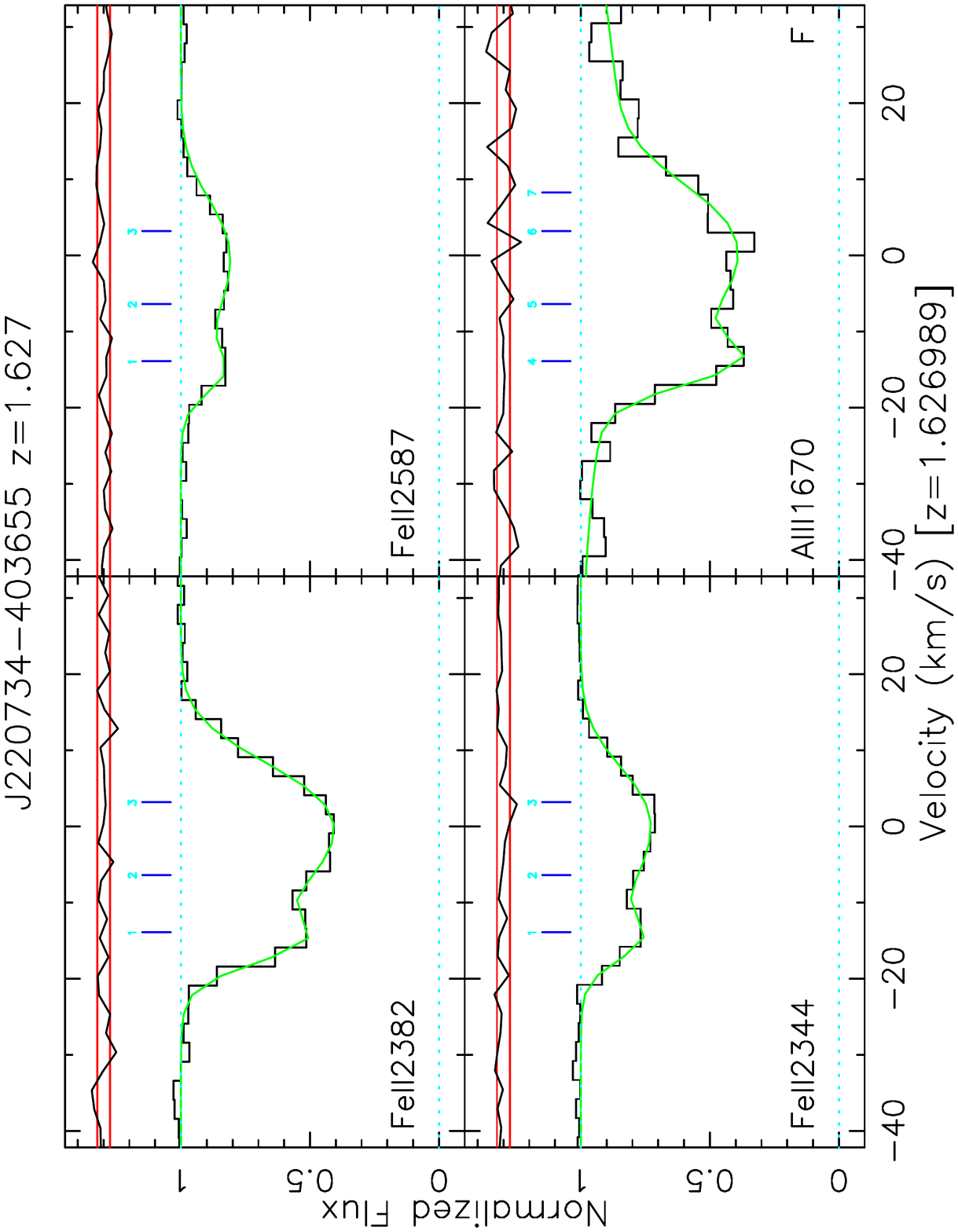}
\par\end{centering}

\caption[\ Fit for the $z=1.627$ absorber toward J220734$-$403655]{Many-multiplet fit for the $z=1.627$ absorber toward J220734$-$403655.}
\end{figure}
\begin{figure}[H]
\noindent \begin{centering}
\includegraphics[bb=34bp 58bp 554bp 738bp,clip,width=1\textwidth]{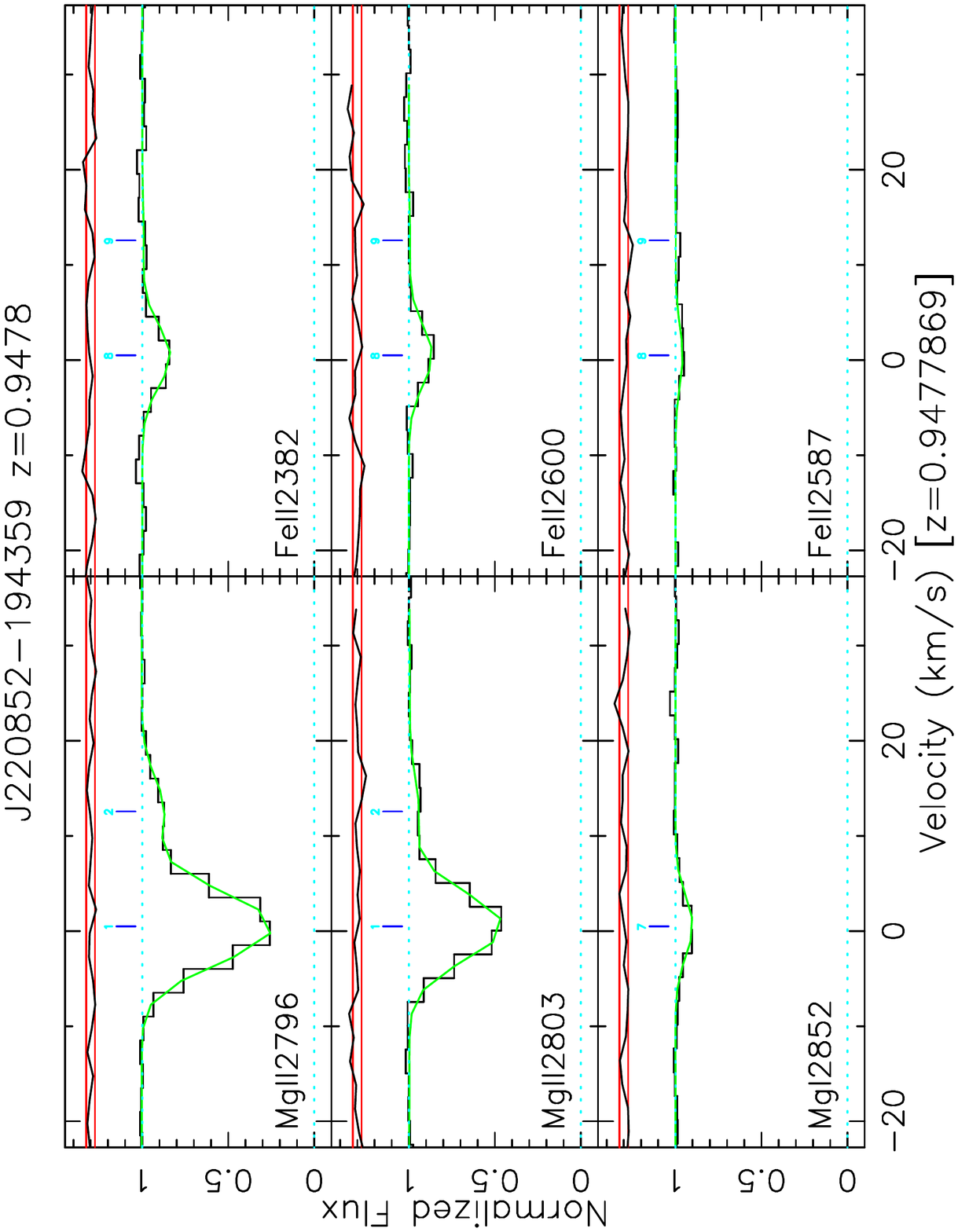}
\par\end{centering}

\caption[\ Fit for the $z=0.9478$ absorber toward J220852$-$194359]{Many-multiplet fit for the $z=0.9478$ absorber toward J220852$-$194359.}
\end{figure}
\begin{figure}[H]
\noindent \begin{centering}
\includegraphics[bb=34bp 58bp 554bp 738bp,clip,width=1\textwidth]{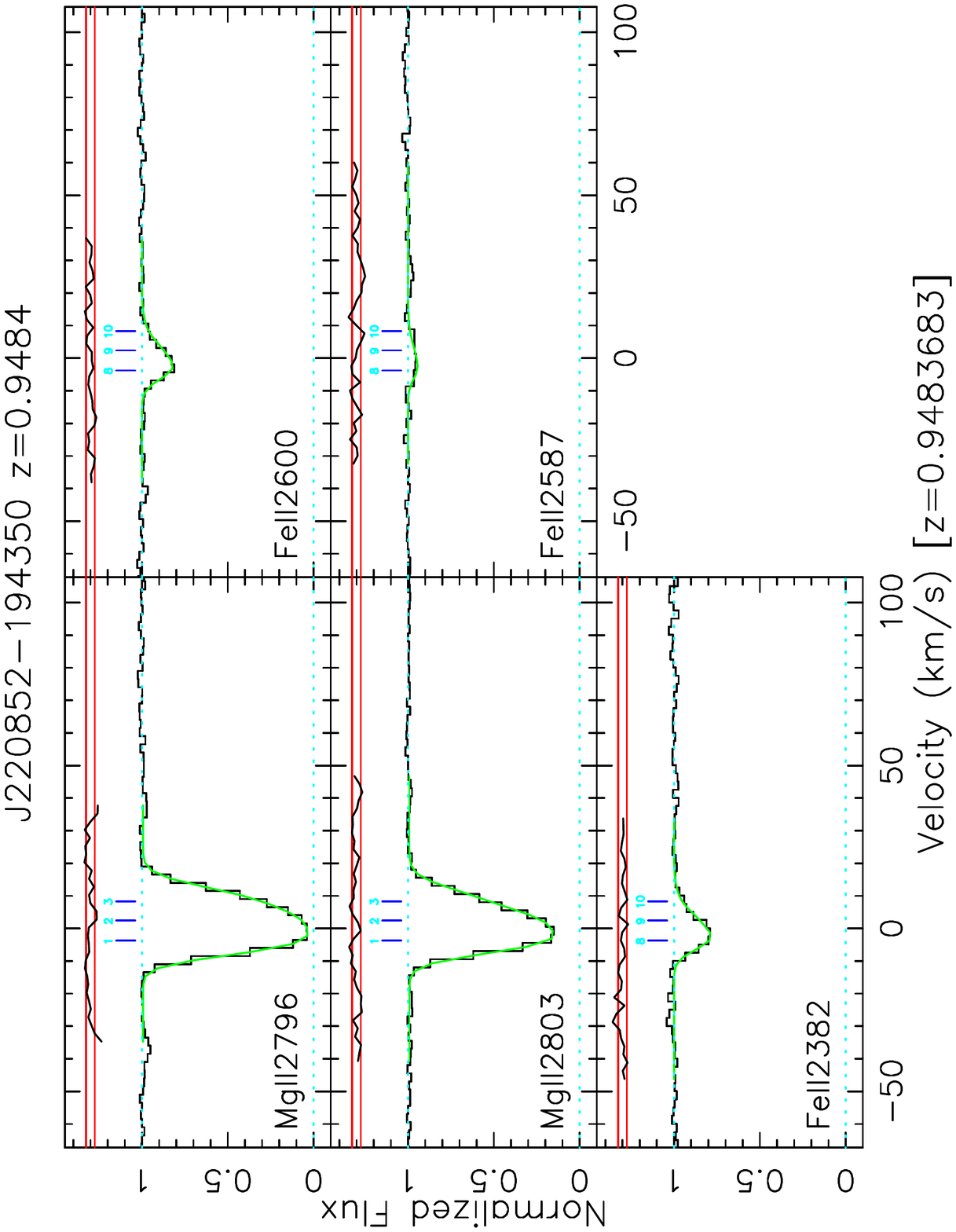}
\par\end{centering}

\caption[\ Fit for the $z=0.9484$ absorber toward J220852$-$194359]{Many-multiplet fit for the $z=0.9484$ absorber toward J220852$-$194359.}
\end{figure}
\begin{figure}[H]
\noindent \begin{centering}
\includegraphics[bb=34bp 58bp 554bp 738bp,clip,width=1\textwidth]{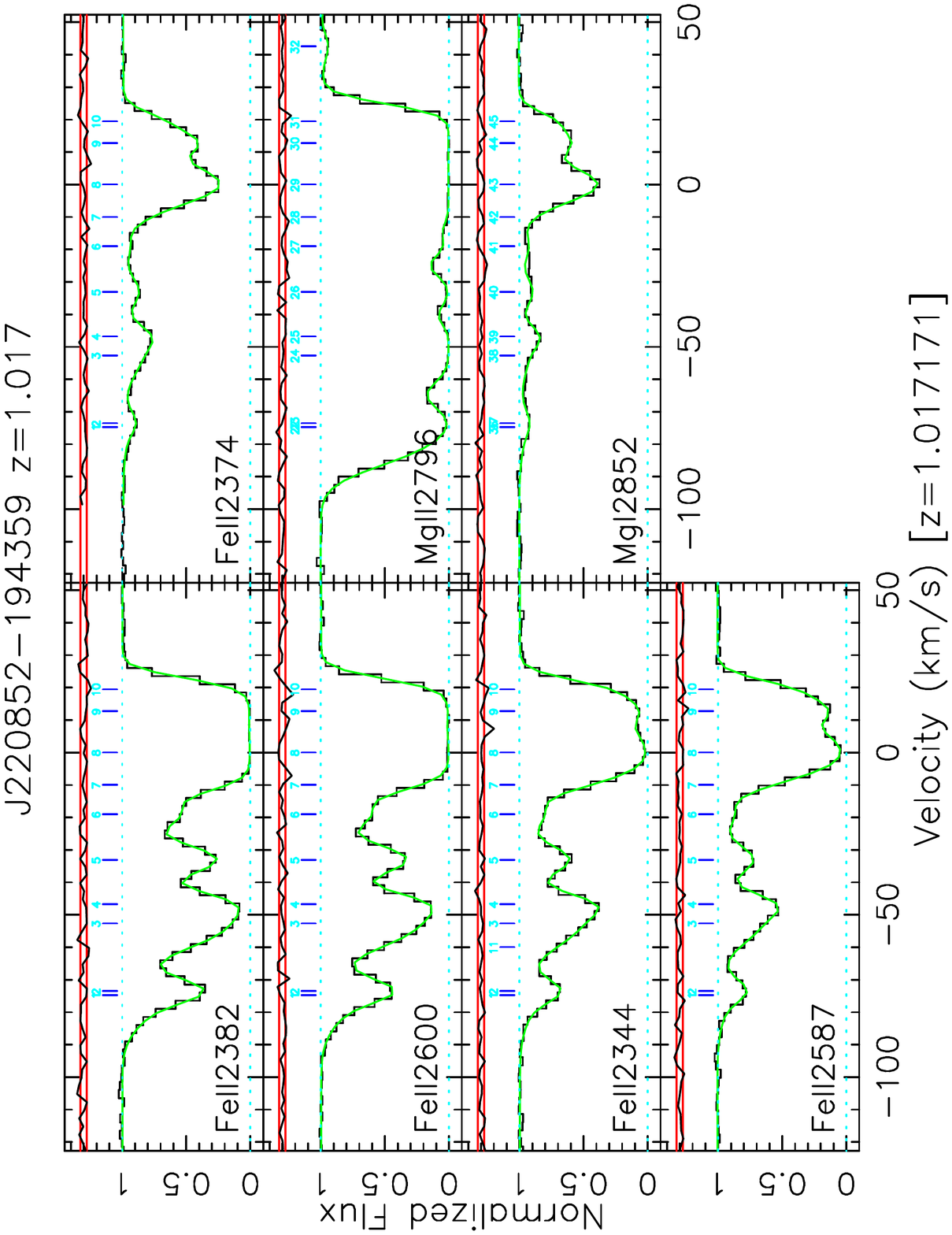}
\par\end{centering}

\caption[\ Fit for the $z=1.017$ absorber toward J220852$-$194359]{Many-multiplet fit for the $z=1.017$ absorber toward J220852$-$194359.}
\end{figure}
\begin{figure}[H]
\noindent \begin{centering}
\includegraphics[bb=34bp 58bp 554bp 738bp,clip,width=1\textwidth]{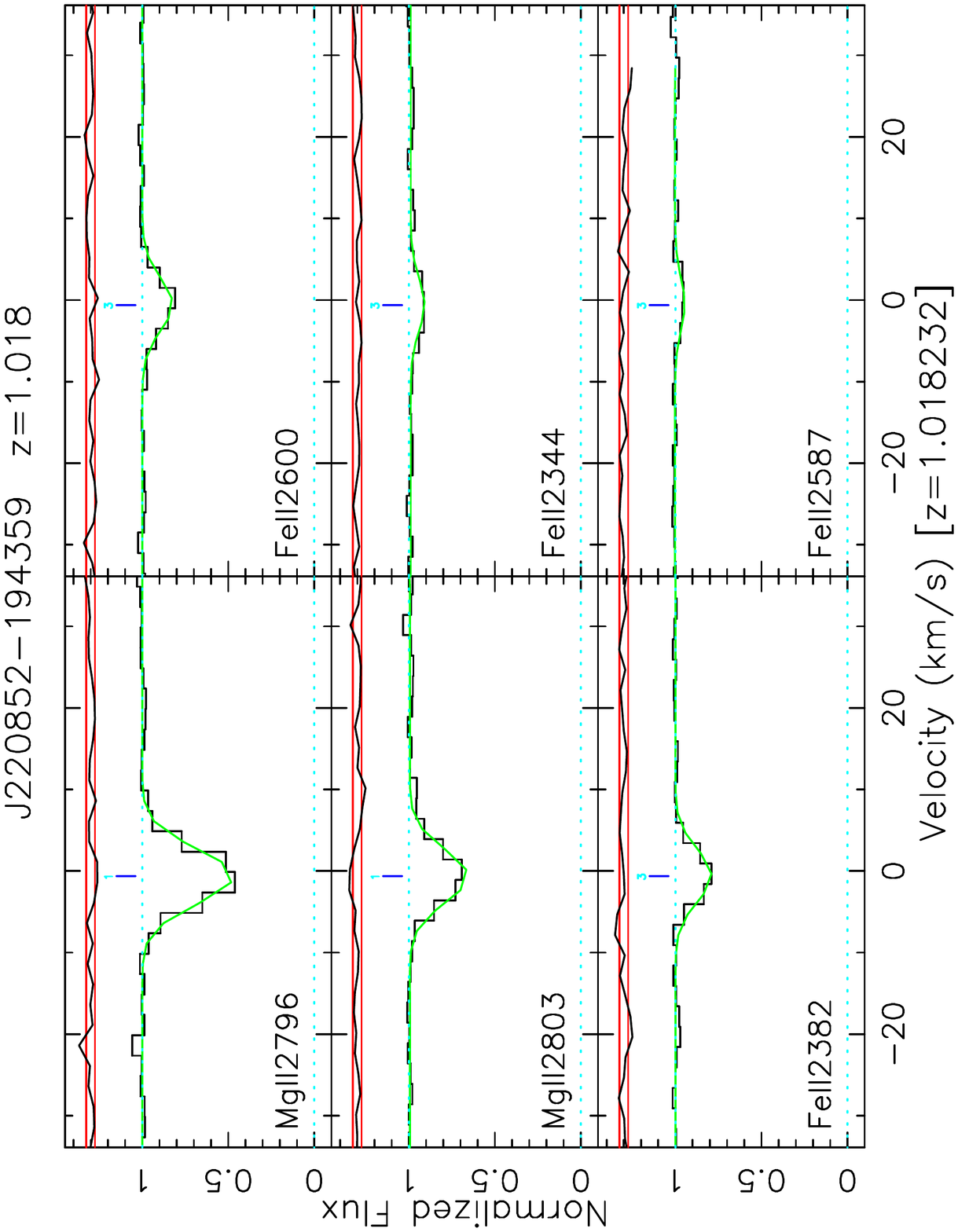}
\par\end{centering}

\caption[\ Fit for the $z=1.018$ absorber toward J220852$-$194359]{Many-multiplet fit for the $z=1.018$ absorber toward J220852$-$194359.}
\end{figure}
\begin{figure}[H]
\noindent \begin{centering}
\includegraphics[bb=34bp 58bp 554bp 738bp,clip,width=1\textwidth]{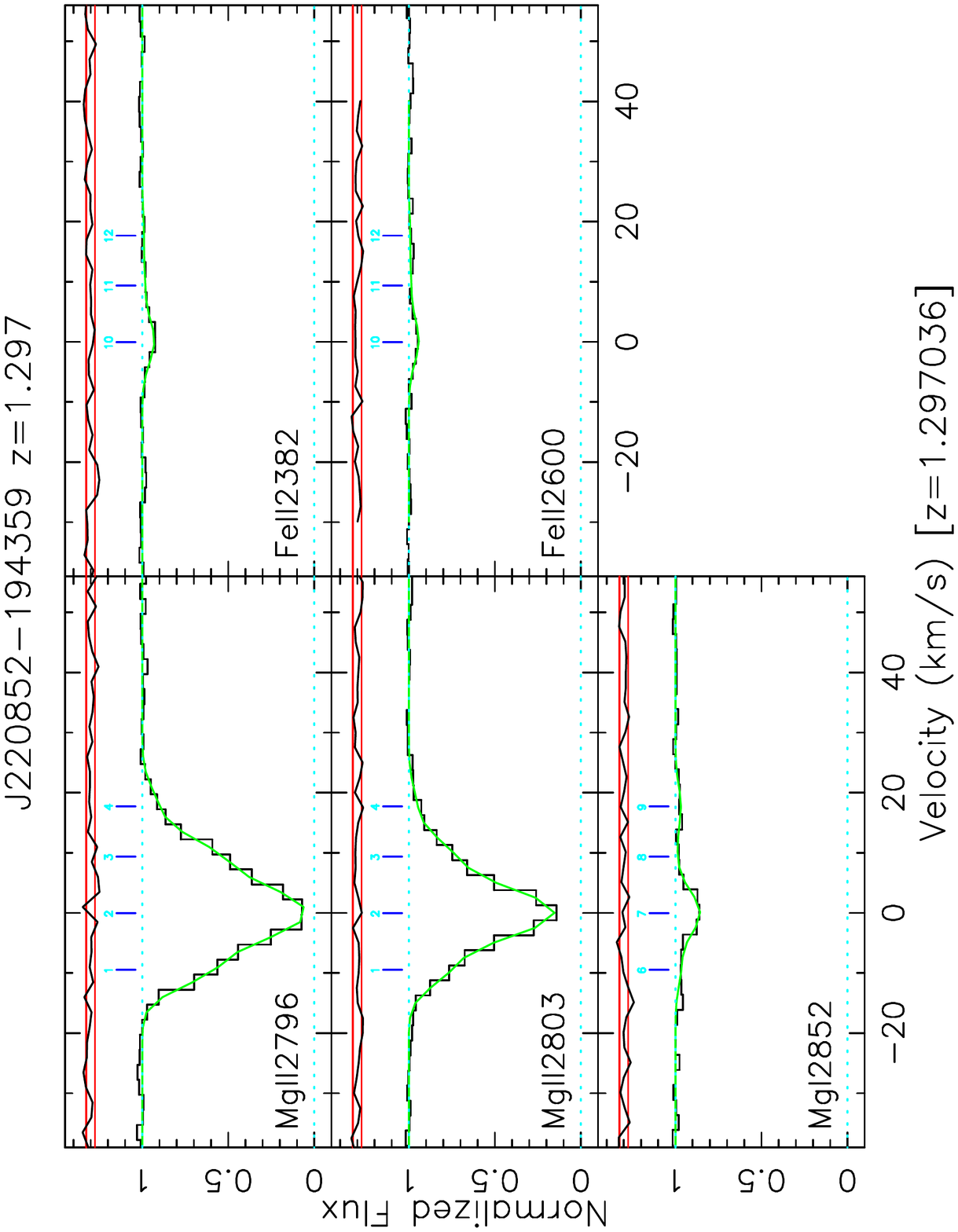}
\par\end{centering}

\caption[\ Fit for the $z=1.297$ absorber toward J220852$-$194359]{Many-multiplet fit for the $z=1.297$ absorber toward J220852$-$194359.}
\end{figure}
\begin{figure}[H]
\noindent \begin{centering}
\includegraphics[bb=34bp 58bp 554bp 738bp,clip,width=1\textwidth]{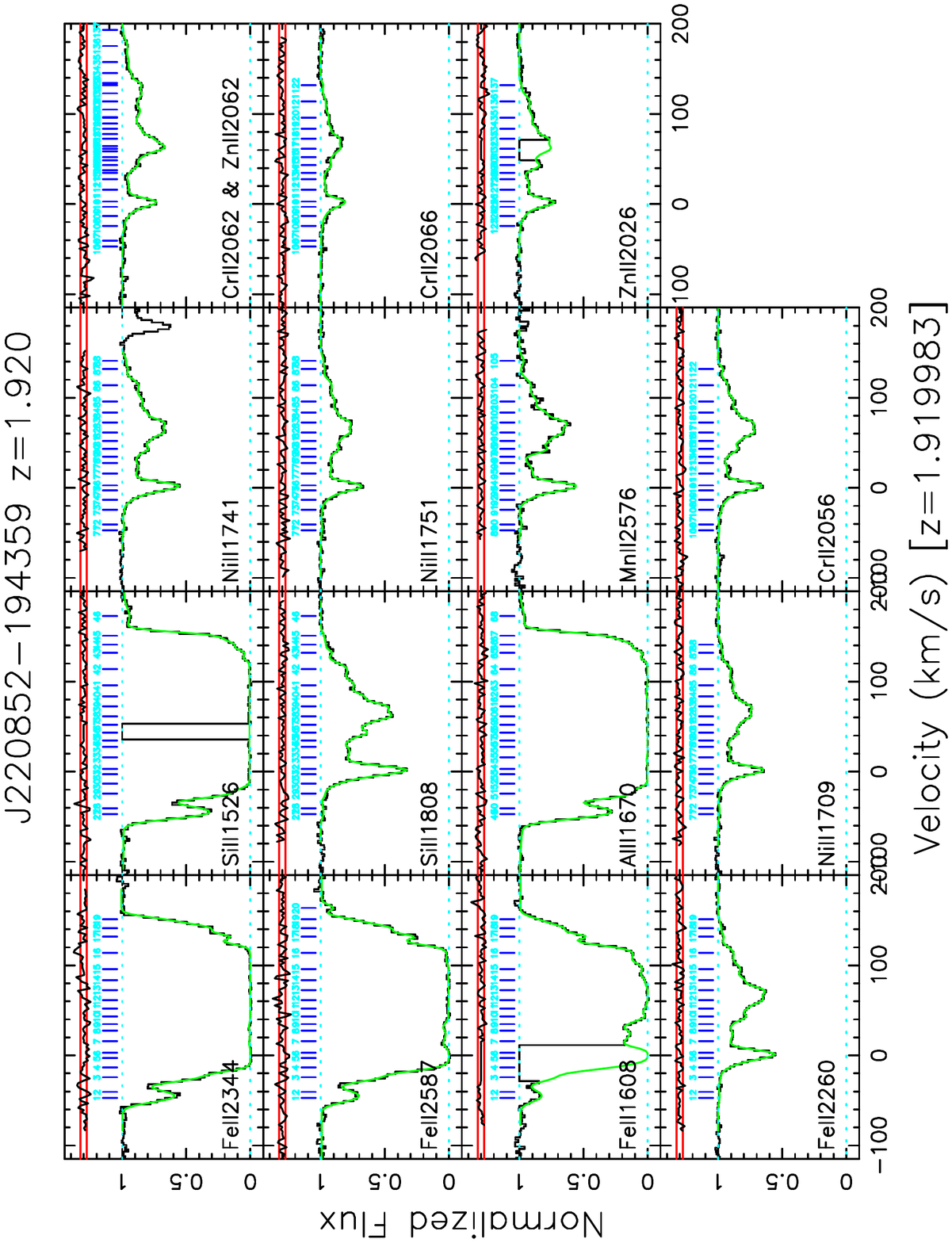}
\par\end{centering}

\caption[\ Fit for the $z=1.920$ absorber toward J220852$-$194359]{Many-multiplet fit for the $z=1.920$ absorber toward J220852$-$194359.}
\end{figure}
\begin{figure}[H]
\noindent \begin{centering}
\includegraphics[bb=34bp 58bp 554bp 738bp,clip,width=1\textwidth]{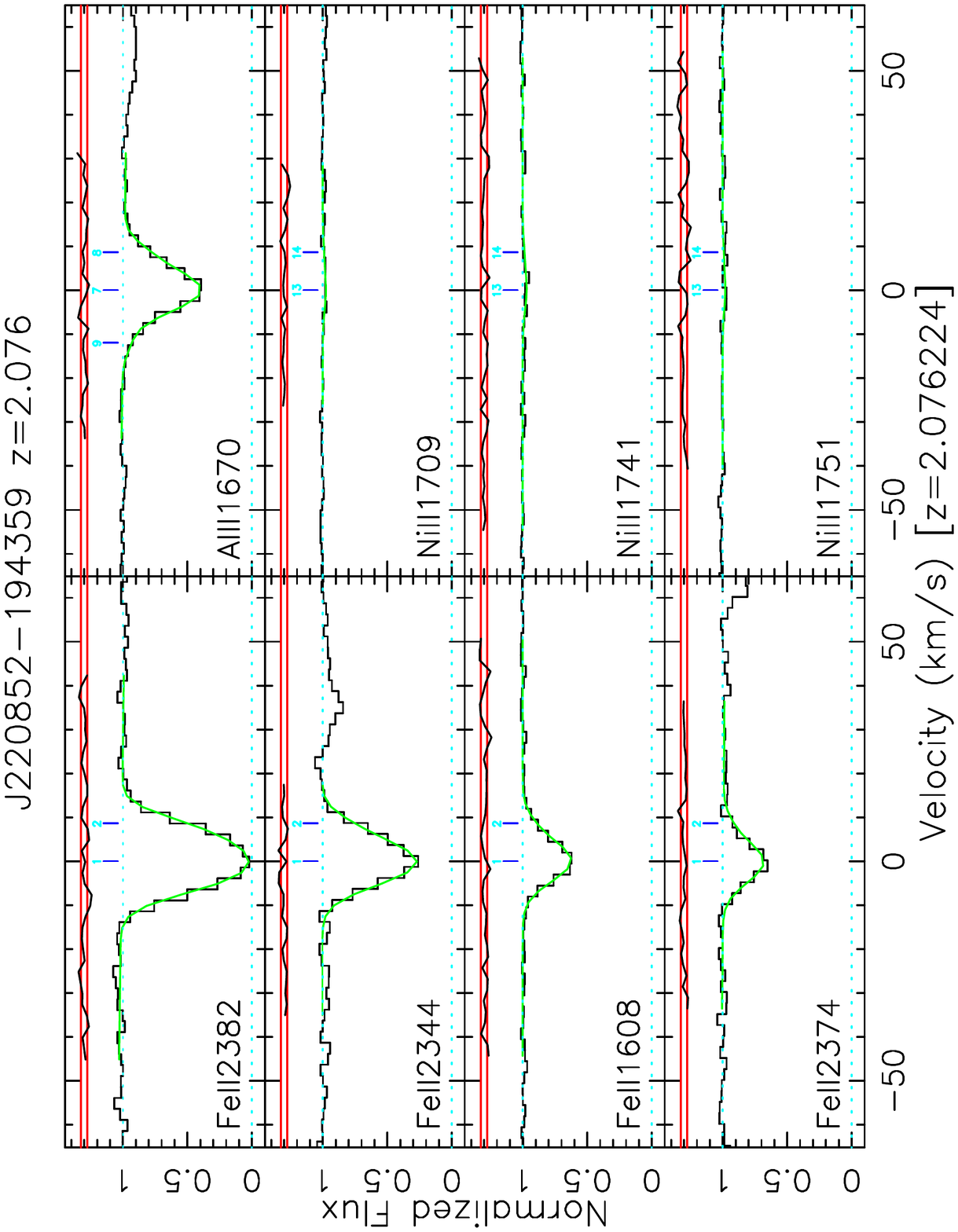}
\par\end{centering}

\caption[\ Fit for the $z=2.076$ absorber toward J220852$-$194359]{Many-multiplet fit for the $z=2.076$ absorber toward J220852$-$194359.}
\end{figure}
\begin{figure}[H]
\noindent \begin{centering}
\includegraphics[bb=34bp 58bp 554bp 738bp,clip,width=1\textwidth]{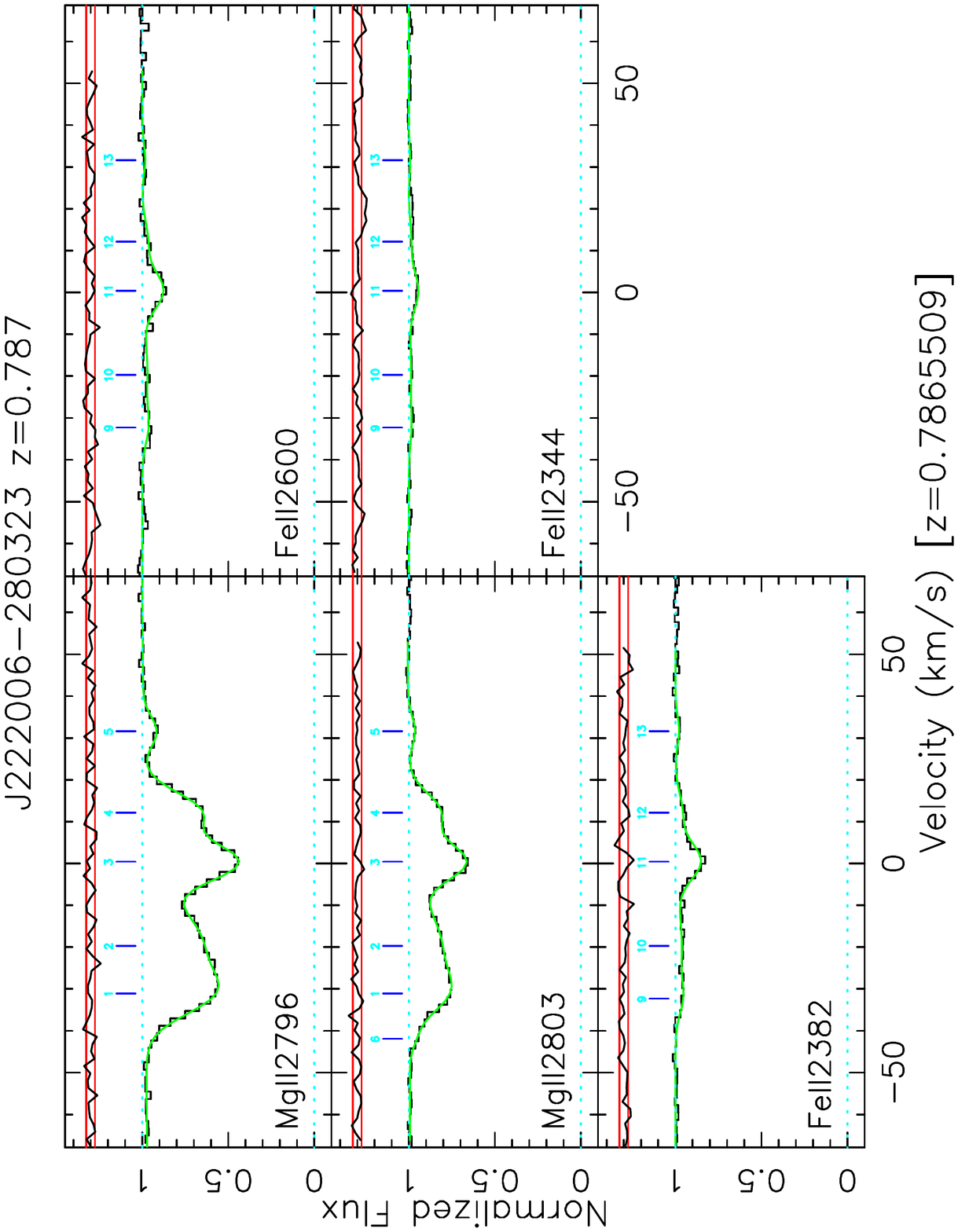}
\par\end{centering}

\caption[\ Fit for the $z=0.941$ absorber toward J222006$-$280323]{Many-multiplet fit for the $z=0.941$ absorber toward J222006$-$280323.}
\end{figure}
\begin{figure}[H]
\noindent \begin{centering}
\includegraphics[bb=34bp 58bp 554bp 738bp,clip,width=1\textwidth]{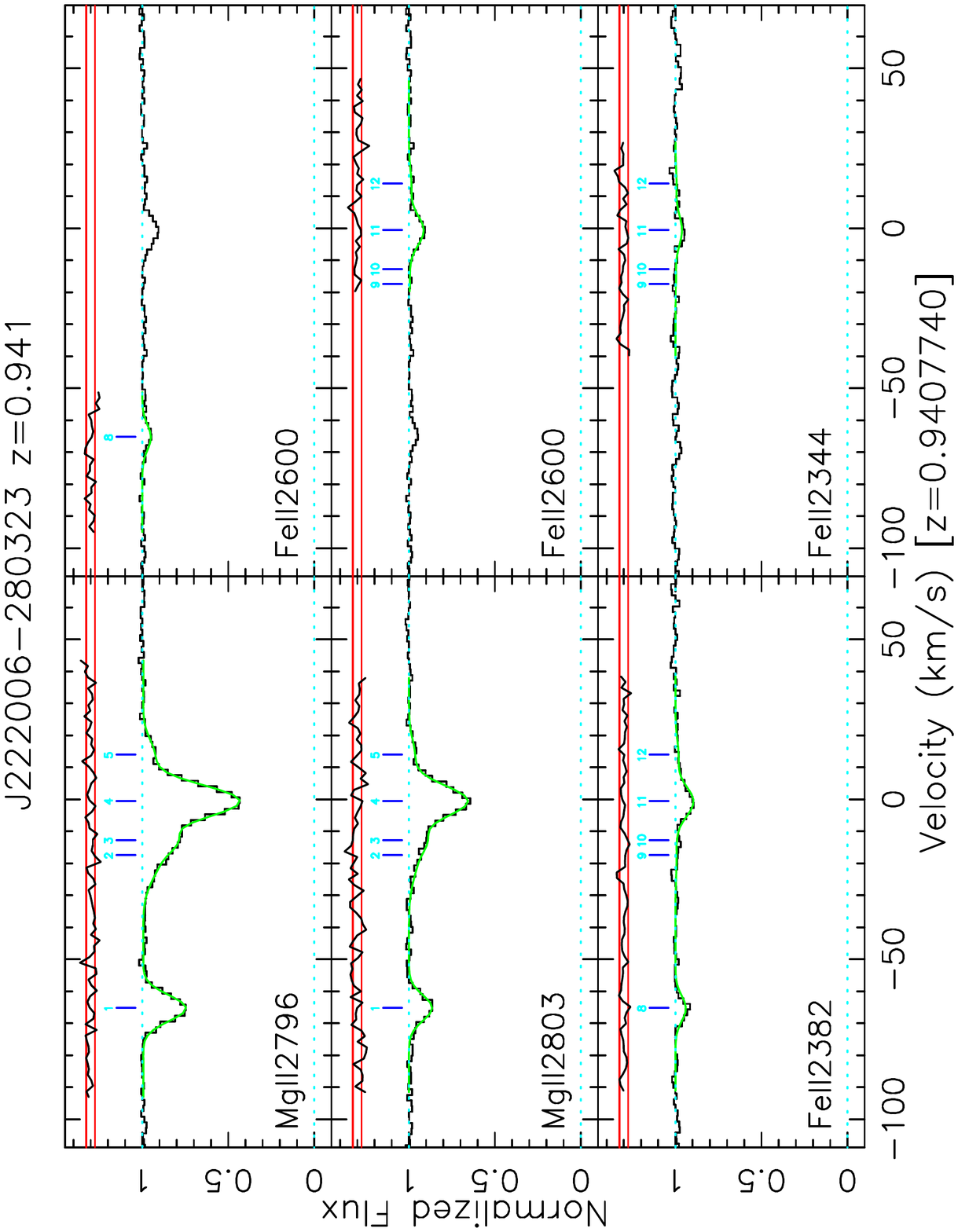}
\par\end{centering}

\caption[\ Fit for the $z=0.941$ absorber toward J222006$-$280323]{Many-multiplet fit for the $z=0.941$ absorber toward J222006$-$280323.}
\end{figure}
\begin{figure}[H]
\noindent \begin{centering}
\includegraphics[bb=34bp 58bp 554bp 738bp,clip,width=1\textwidth]{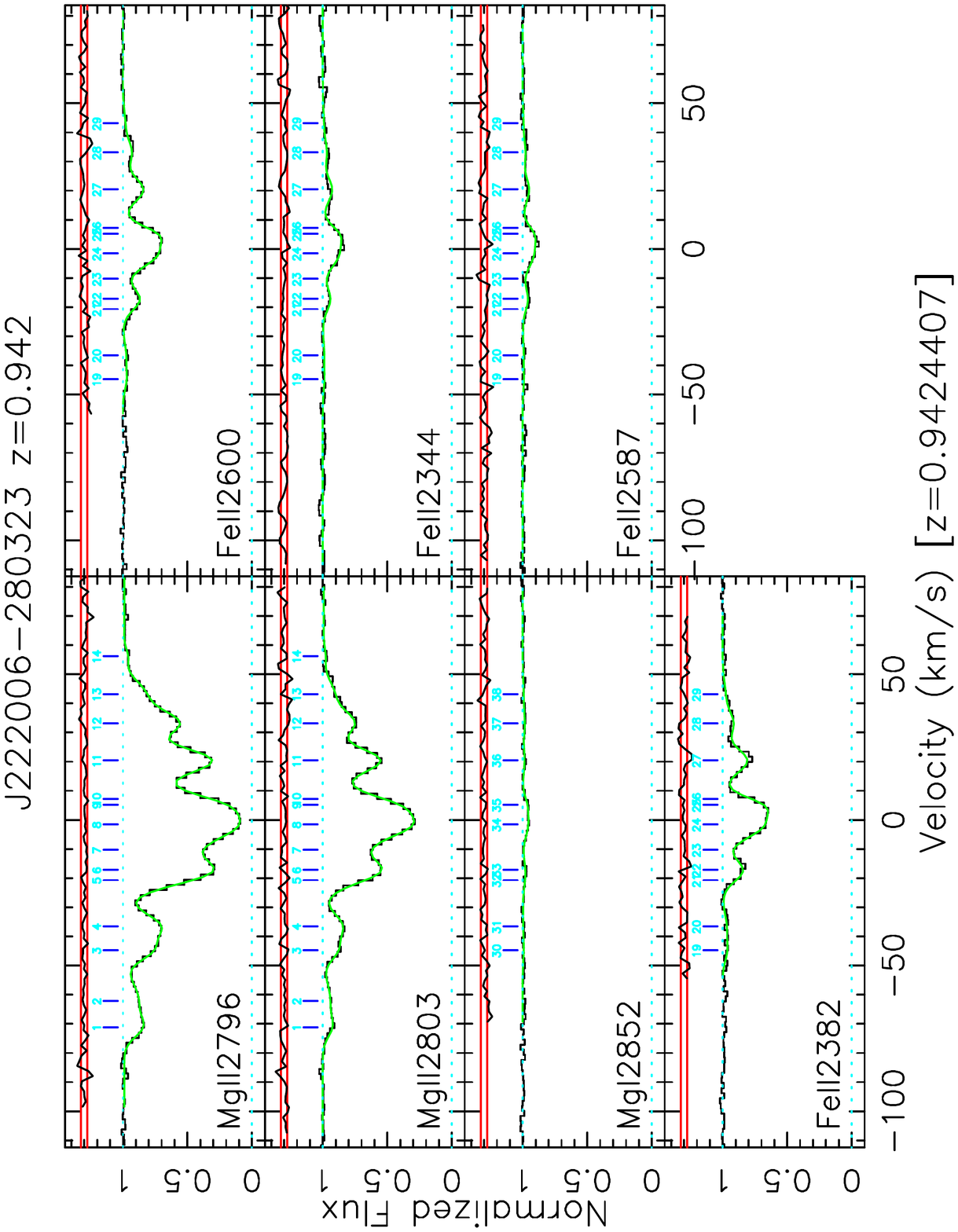}
\par\end{centering}

\caption[\ Fit for the $z=0.942$ absorber toward J222006$-$280323]{Many-multiplet fit for the $z=0.942$ absorber toward J222006$-$280323.}
\end{figure}
\begin{figure}[H]
\noindent \begin{centering}
\includegraphics[bb=34bp 58bp 554bp 738bp,clip,width=1\textwidth]{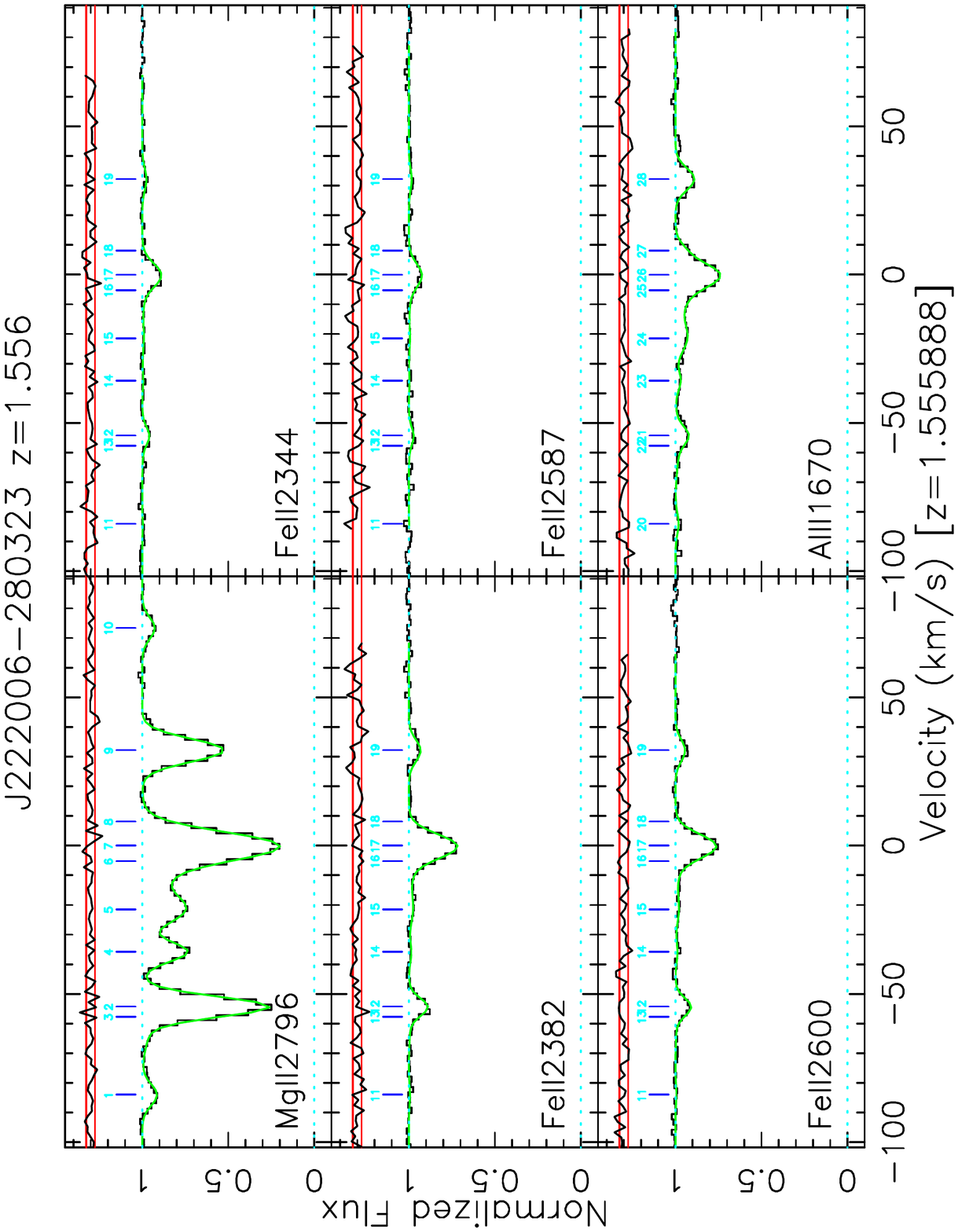}
\par\end{centering}

\caption[\ Fit for the $z=1.556$ absorber toward J222006$-$280323]{Many-multiplet fit for the $z=1.556$ absorber toward J222006$-$280323.}
\end{figure}
\begin{figure}[H]
\noindent \begin{centering}
\includegraphics[bb=34bp 58bp 554bp 738bp,clip,width=1\textwidth]{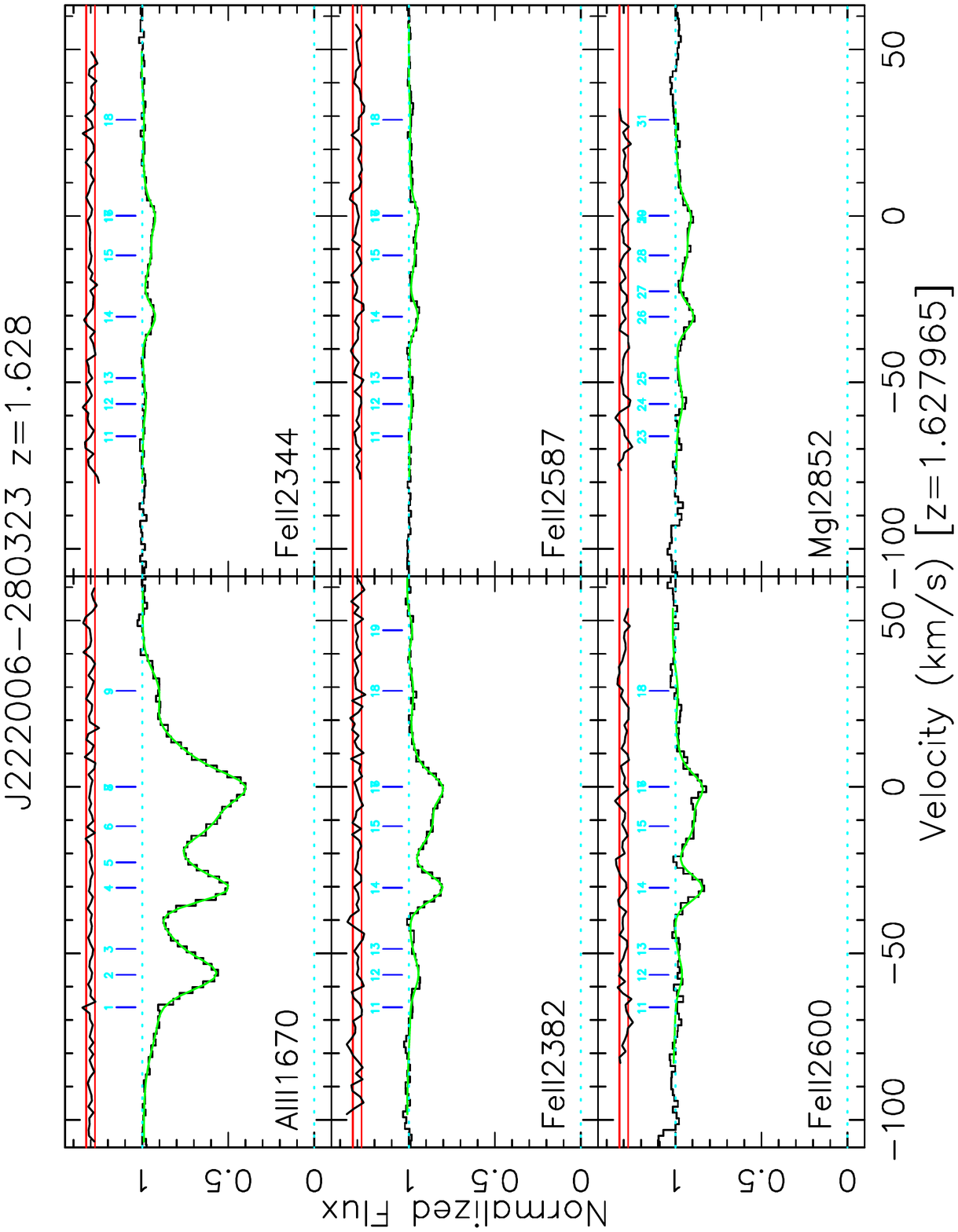}
\par\end{centering}

\caption[\ Fit for the $z=1.628$ absorber toward J222006$-$280323]{Many-multiplet fit for the $z=1.628$ absorber toward J222006$-$280323.}
\end{figure}
\begin{figure}[H]
\noindent \begin{centering}
\includegraphics[bb=34bp 58bp 554bp 738bp,clip,width=1\textwidth]{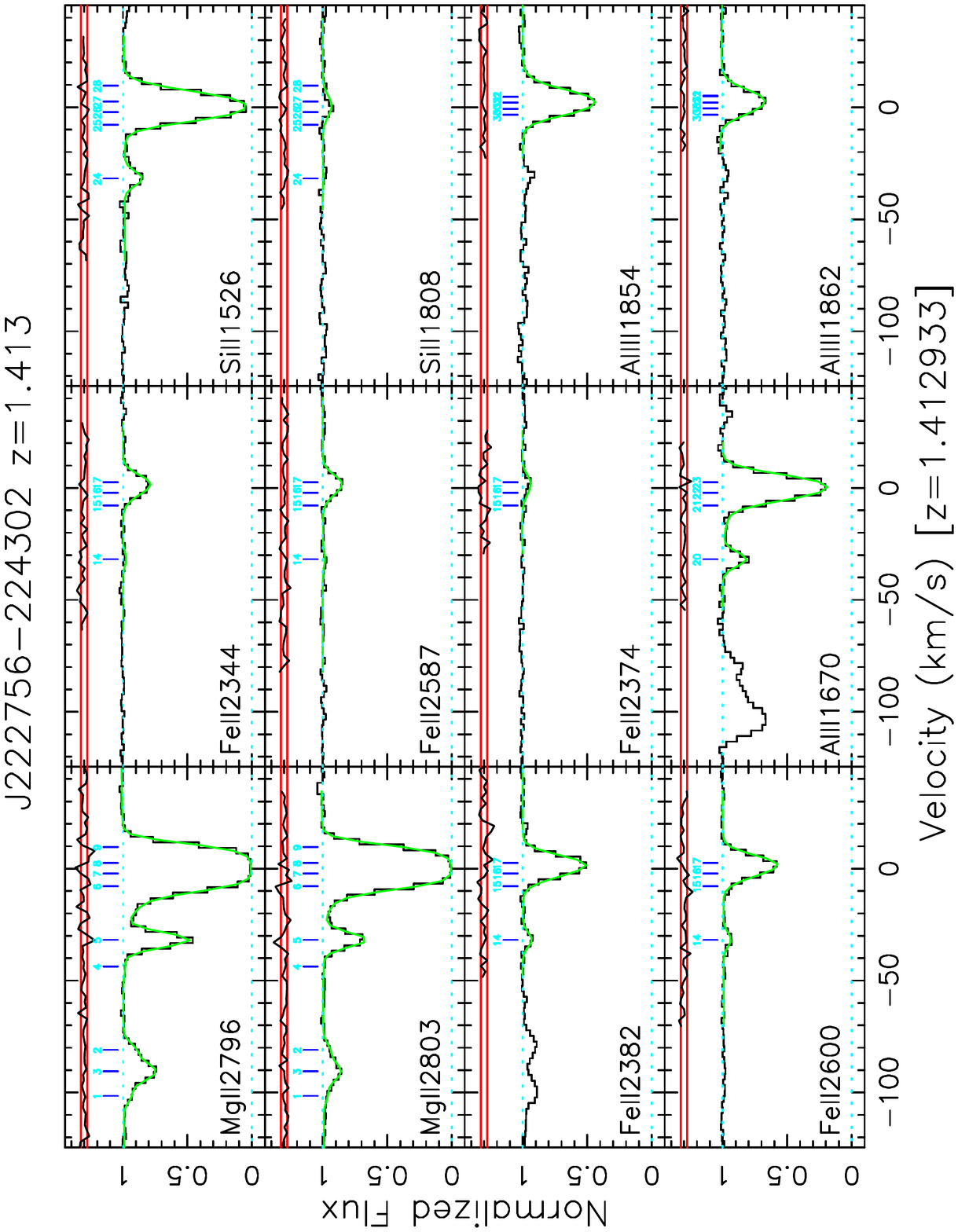}
\par\end{centering}

\caption[\ Fit for the $z=1.413$ absorber toward J222756$-$224302]{Many-multiplet fit for the $z=1.413$ absorber toward J222756$-$224302.}
\end{figure}
\begin{figure}[H]
\noindent \begin{centering}
\includegraphics[bb=34bp 58bp 554bp 738bp,clip,width=1\textwidth]{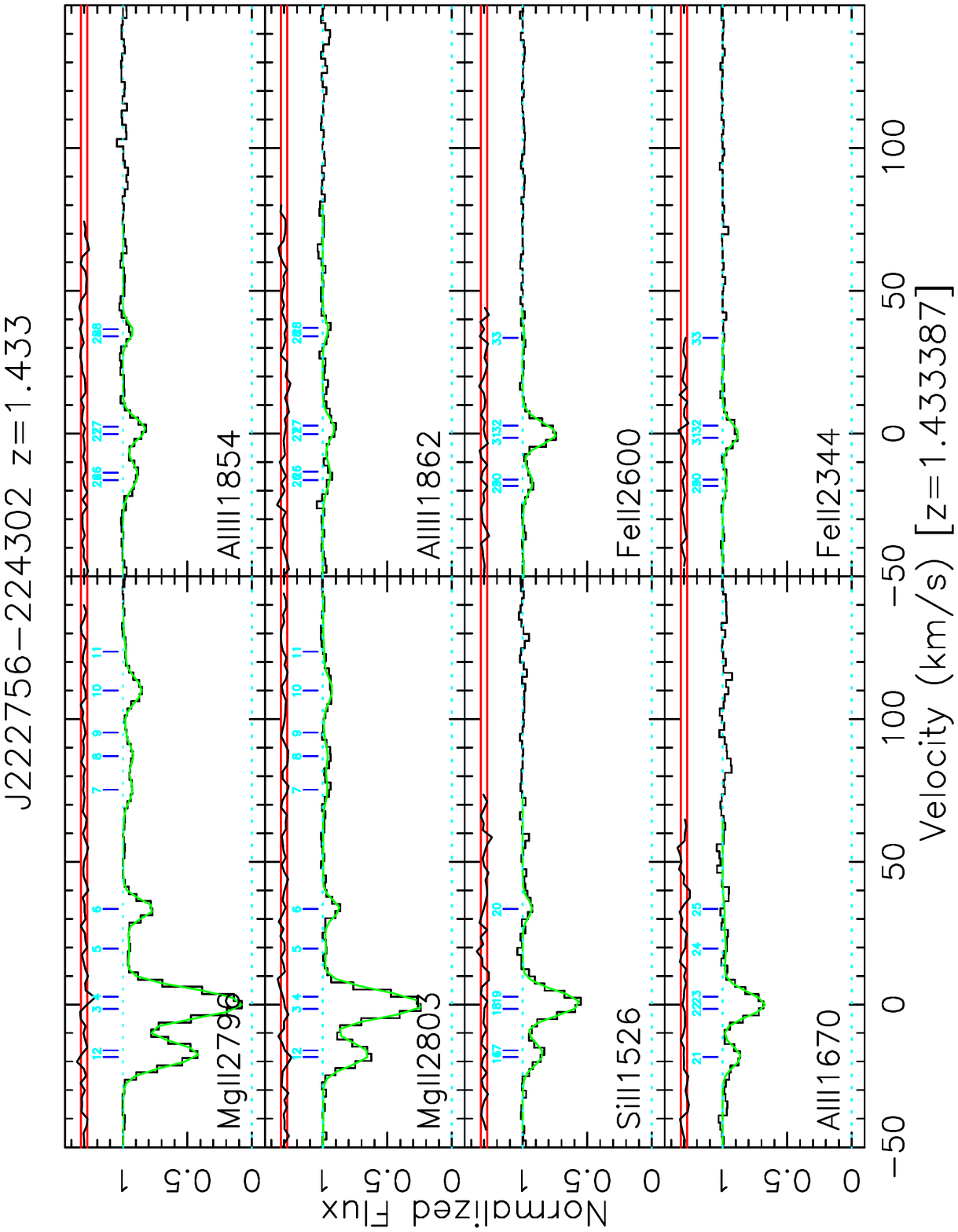}
\par\end{centering}

\caption[\ Fit for the $z=1.433$ absorber toward J222756$-$224302]{Many-multiplet fit for the $z=1.433$ absorber toward J222756$-$224302.}
\end{figure}
\begin{figure}[H]
\noindent \begin{centering}
\includegraphics[bb=34bp 58bp 554bp 738bp,clip,width=1\textwidth]{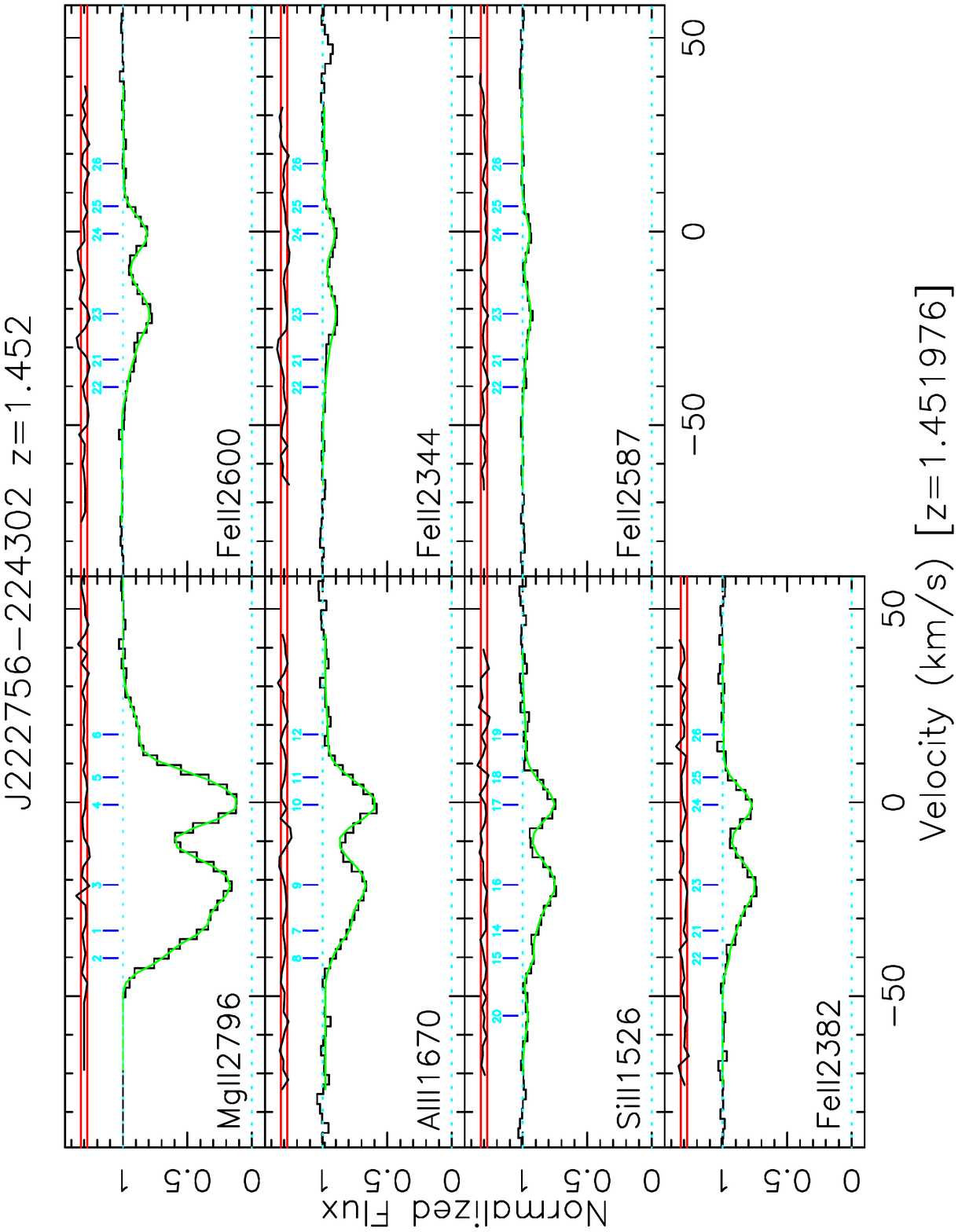}
\par\end{centering}

\caption[\ Fit for the $z=1.452$ absorber toward J222756$-$224302]{Many-multiplet fit for the $z=1.452$ absorber toward J222756$-$224302.}
\end{figure}
\begin{figure}[H]
\noindent \begin{centering}
\includegraphics[bb=34bp 58bp 554bp 738bp,clip,width=1\textwidth]{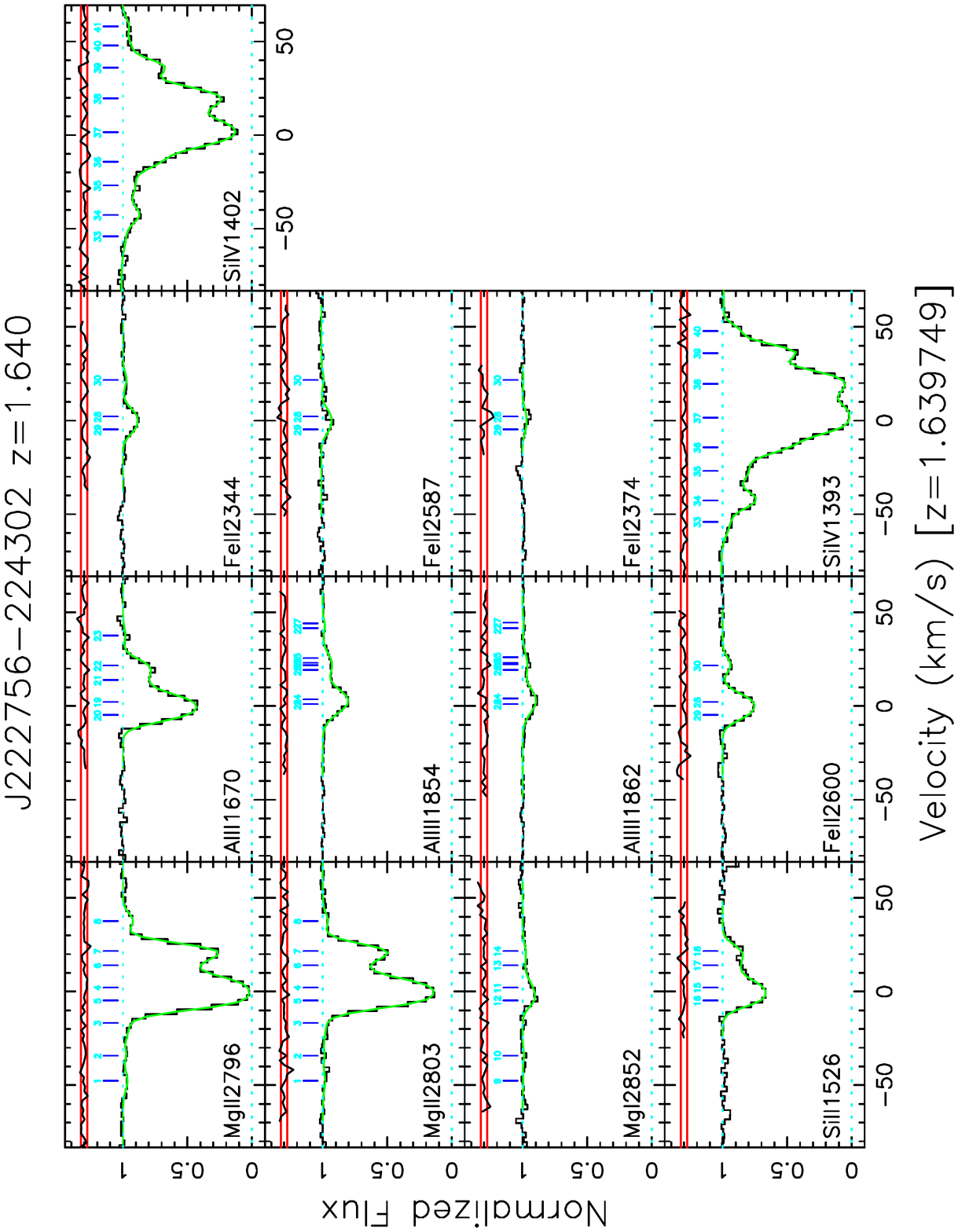}
\par\end{centering}

\caption[\ Fit for the $z=1.640$ absorber toward J222756$-$224302]{Many-multiplet fit for the $z=1.640$ absorber toward J222756$-$224302.}
\end{figure}
\begin{figure}[H]
\noindent \begin{centering}
\includegraphics[bb=34bp 58bp 554bp 738bp,clip,width=1\textwidth]{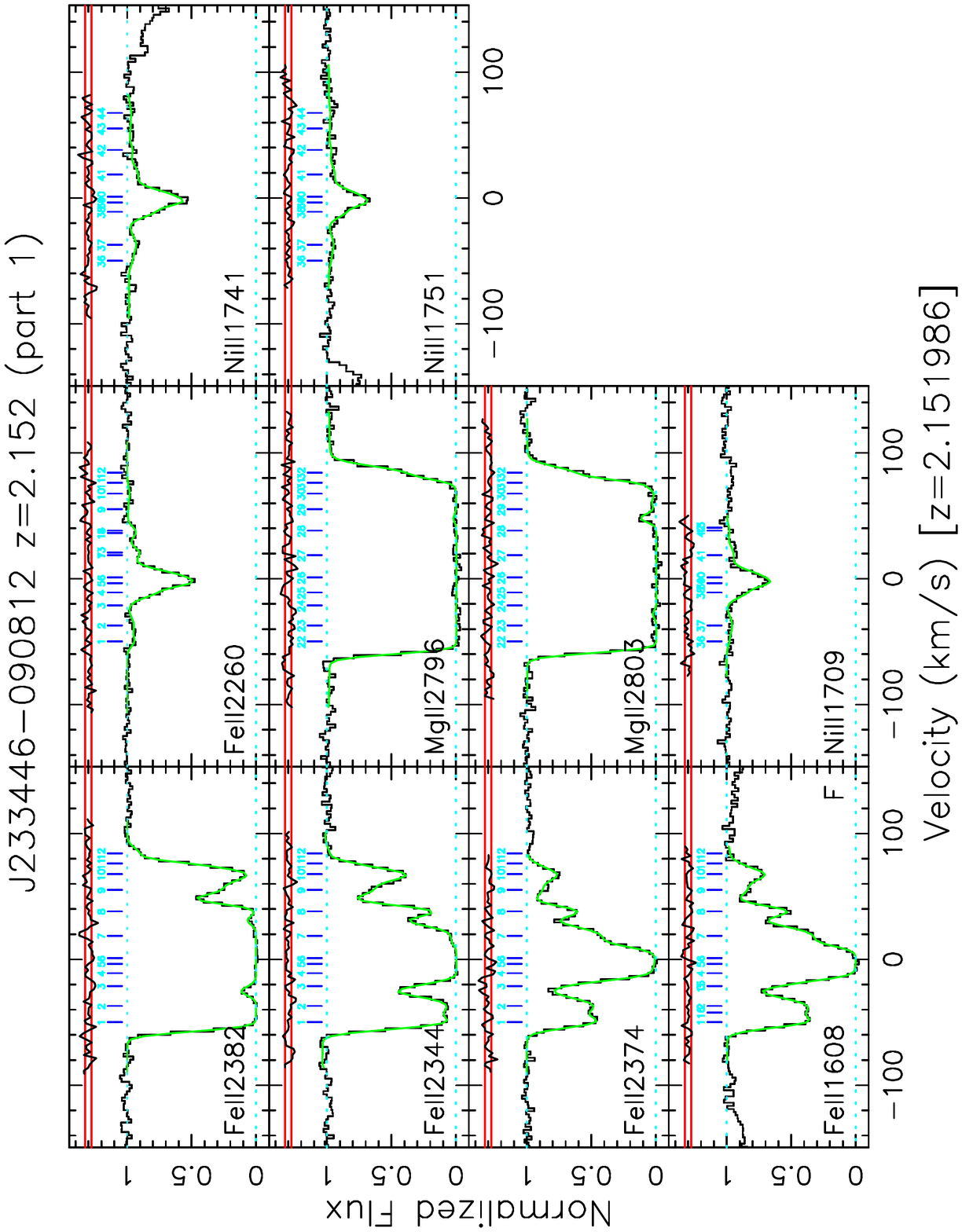}
\par\end{centering}

\caption[\ Fit for the $z=2.152$ absorber toward J233446$-$090812 (part 1)]{Many-multiplet fit for the $z=2.152$ absorber toward J233446$-$090812 (part 1).}
\end{figure}
\begin{figure}[H]
\noindent \begin{centering}
\includegraphics[bb=34bp 58bp 554bp 738bp,clip,width=1\textwidth]{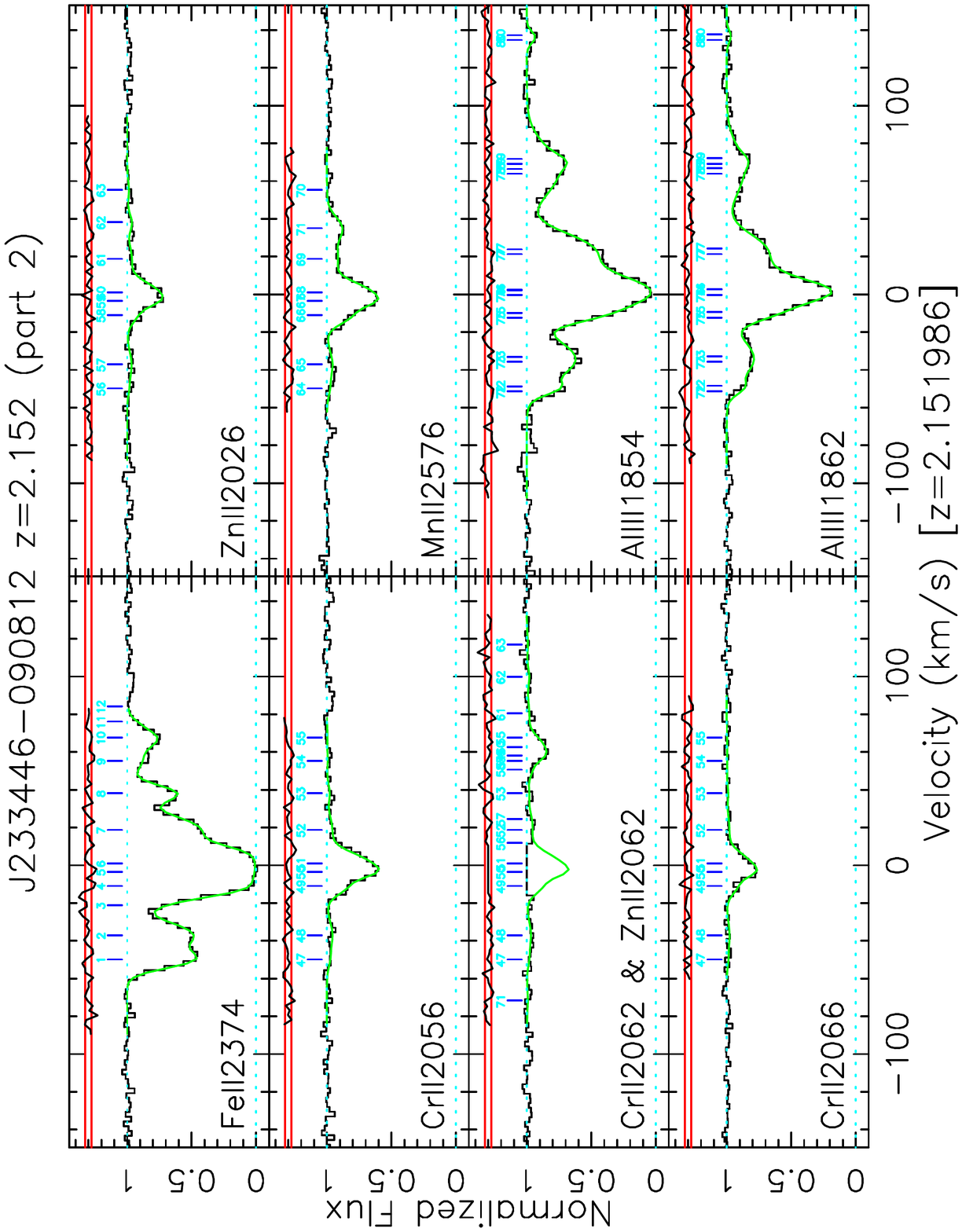}
\par\end{centering}

\caption[\ Fit for the $z=2.152$ absorber toward J233446$-$090812 (part 2)]{Many-multiplet fit for the $z=2.152$ absorber toward J233446$-$090812 (part 2).}
\end{figure}
\begin{figure}[H]
\noindent \begin{centering}
\includegraphics[bb=34bp 58bp 554bp 738bp,clip,width=1\textwidth]{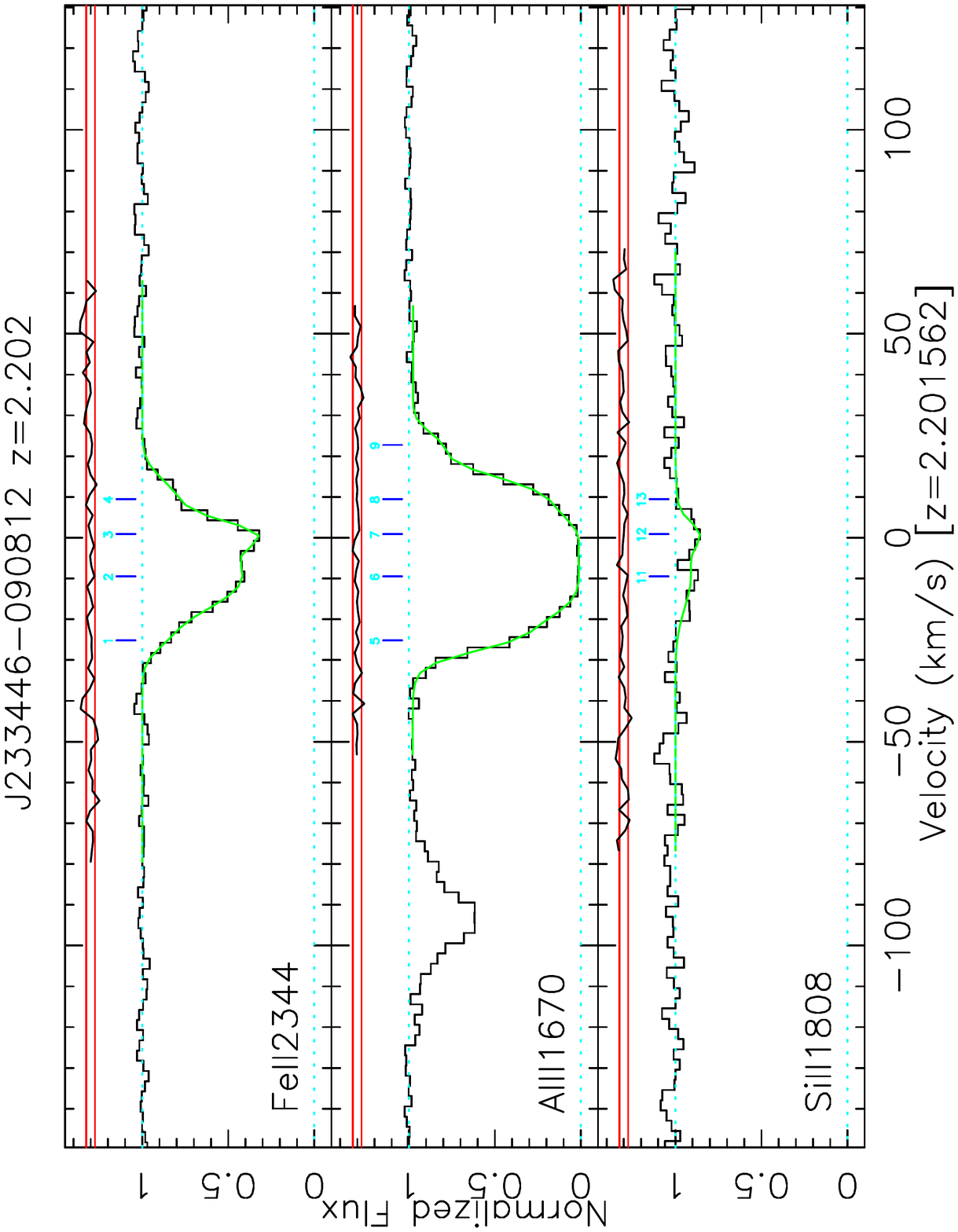}
\par\end{centering}

\caption[\ Fit for the $z=2.202$ absorber toward J233446$-$090812]{Many-multiplet fit for the $z=2.202$ absorber toward J233446$-$090812.}
\end{figure}
\begin{figure}[H]
\noindent \begin{centering}
\includegraphics[bb=34bp 58bp 554bp 738bp,clip,width=1\textwidth]{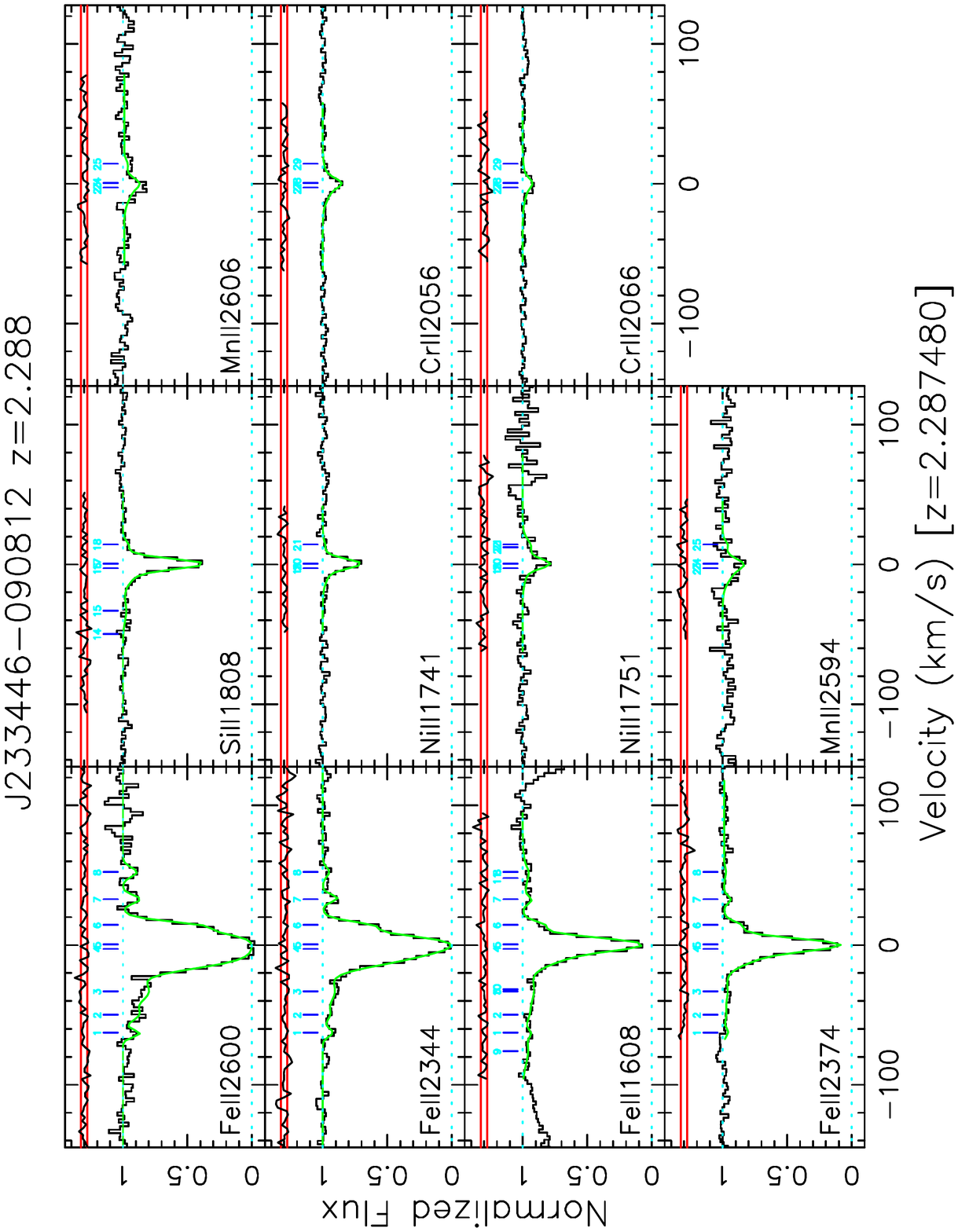}
\par\end{centering}

\caption[\ Fit for the $z=2.288$ absorber toward J233446$-$090812]{Many-multiplet fit for the $z=2.288$ absorber toward J233446$-$090812.}
\end{figure}
\begin{figure}[H]
\noindent \begin{centering}
\includegraphics[bb=34bp 58bp 554bp 738bp,clip,width=1\textwidth]{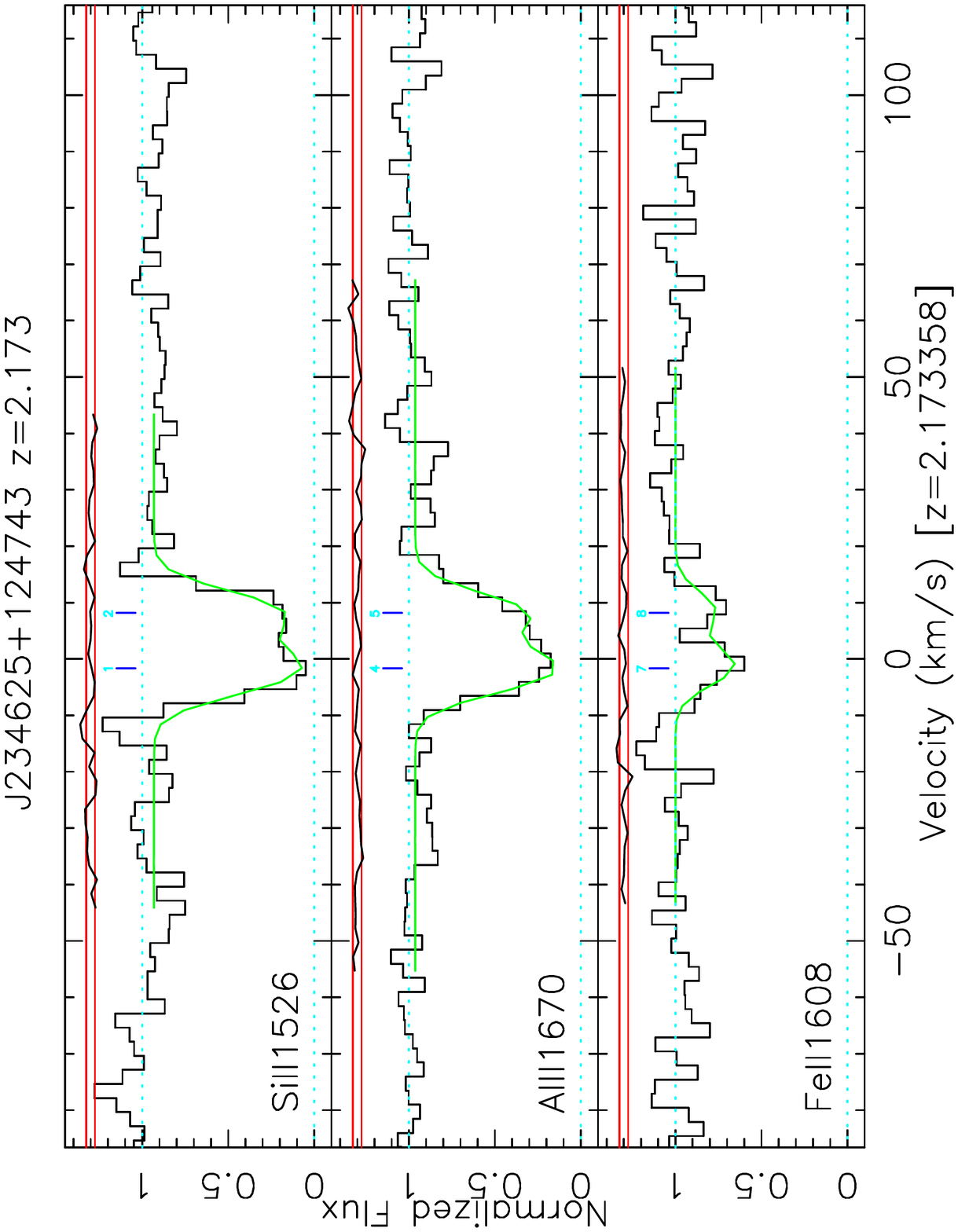}
\par\end{centering}

\caption[\ Fit for the $z=2.173$ absorber toward J234625+124743]{Many-multiplet fit for the $z=2.173$ absorber toward J234625+124743.}
\end{figure}
\begin{figure}[H]
\noindent \begin{centering}
\includegraphics[bb=34bp 58bp 554bp 738bp,clip,width=1\textwidth]{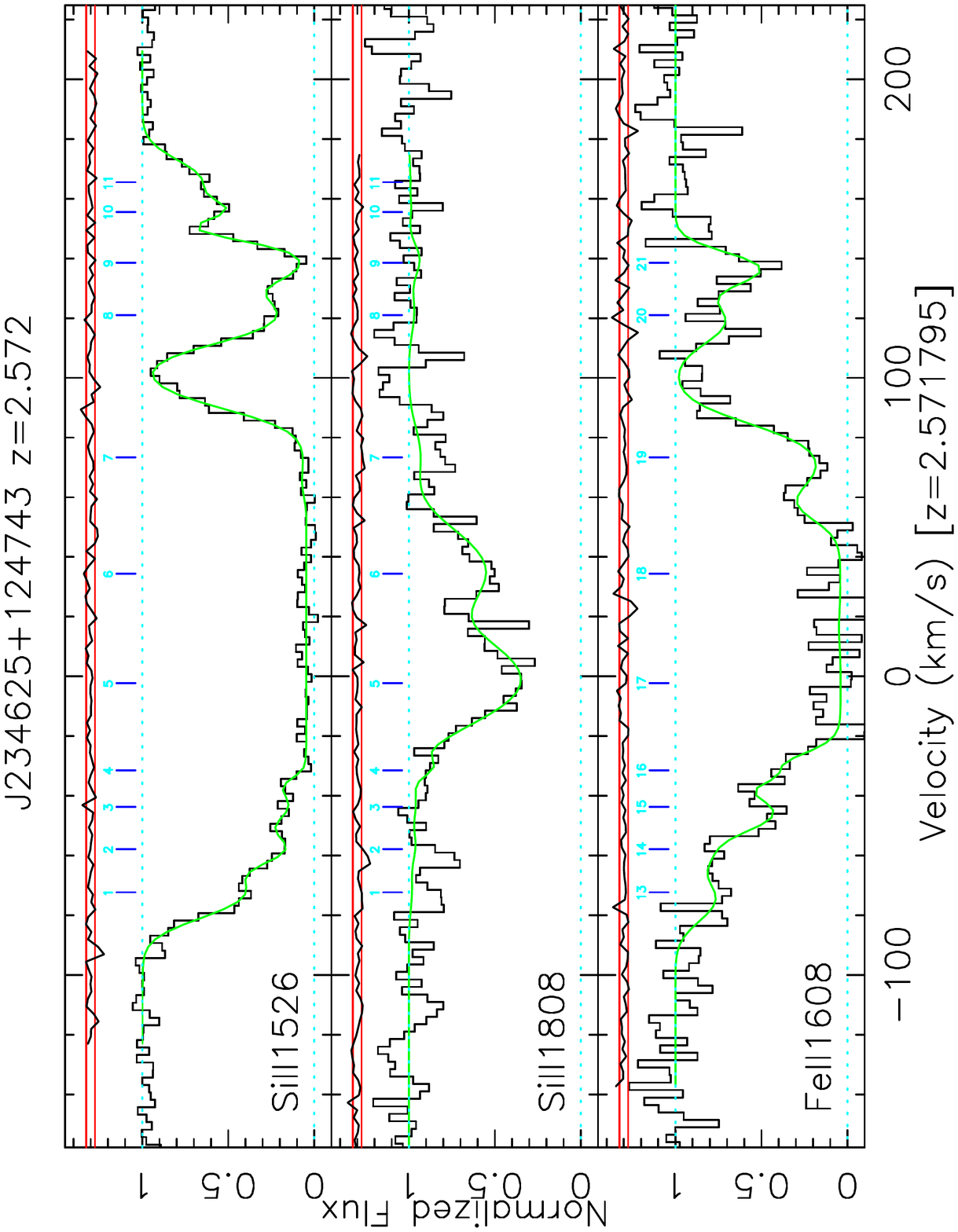}
\par\end{centering}

\caption[\ Fit for the $z=2.572$ absorber toward J234625+124743]{Many-multiplet fit for the $z=2.572$ absorber toward J234625+124743.}
\end{figure}
\begin{figure}[H]
\noindent \begin{centering}
\includegraphics[bb=34bp 58bp 554bp 738bp,clip,width=1\textwidth]{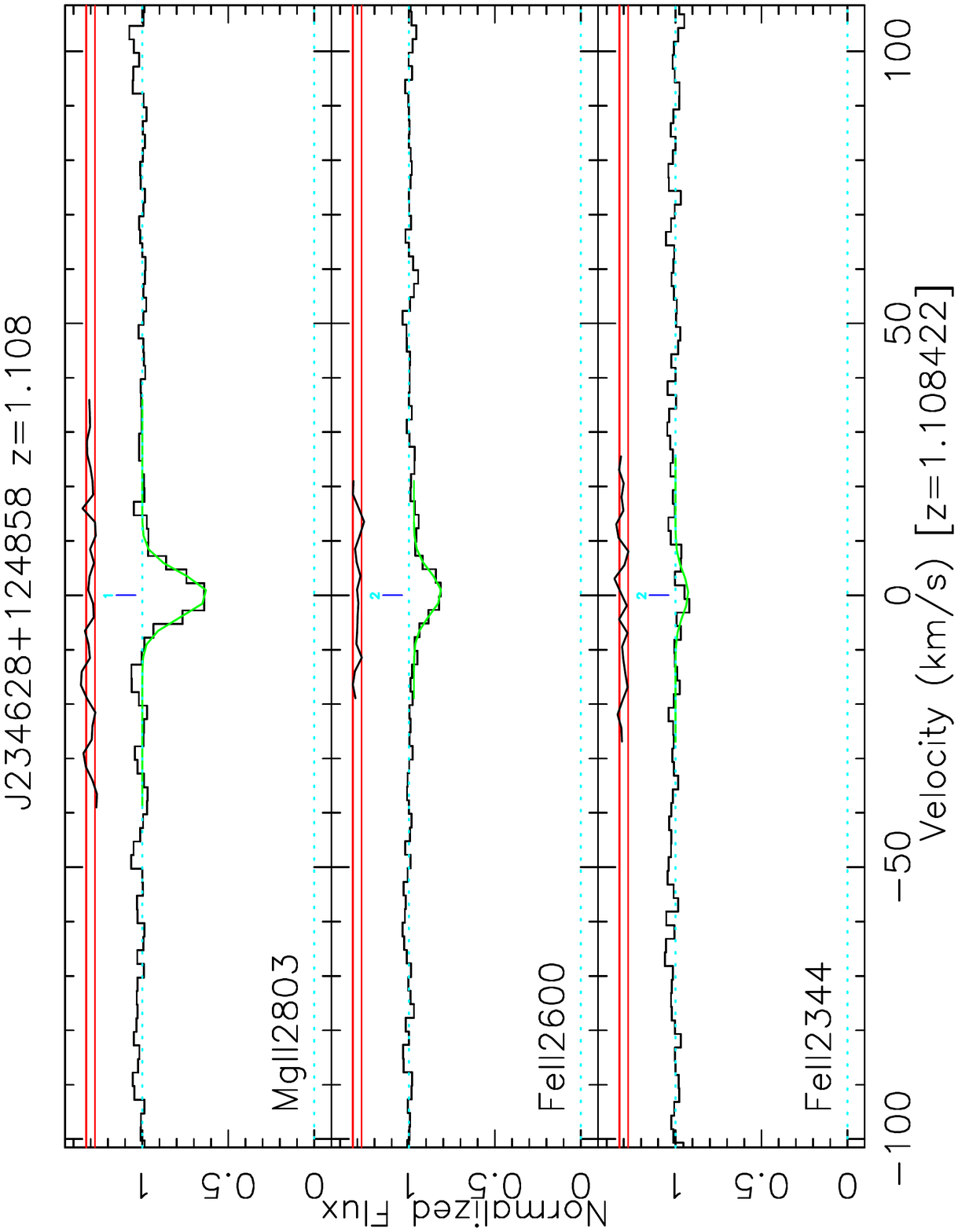}
\par\end{centering}

\caption[\ Fit for the $z=1.108$ absorber toward J234628+124858]{Many-multiplet fit for the $z=1.108$ absorber toward J234628+124858.}
\end{figure}
\begin{figure}[H]
\noindent \begin{centering}
\includegraphics[bb=34bp 58bp 554bp 738bp,clip,width=1\textwidth]{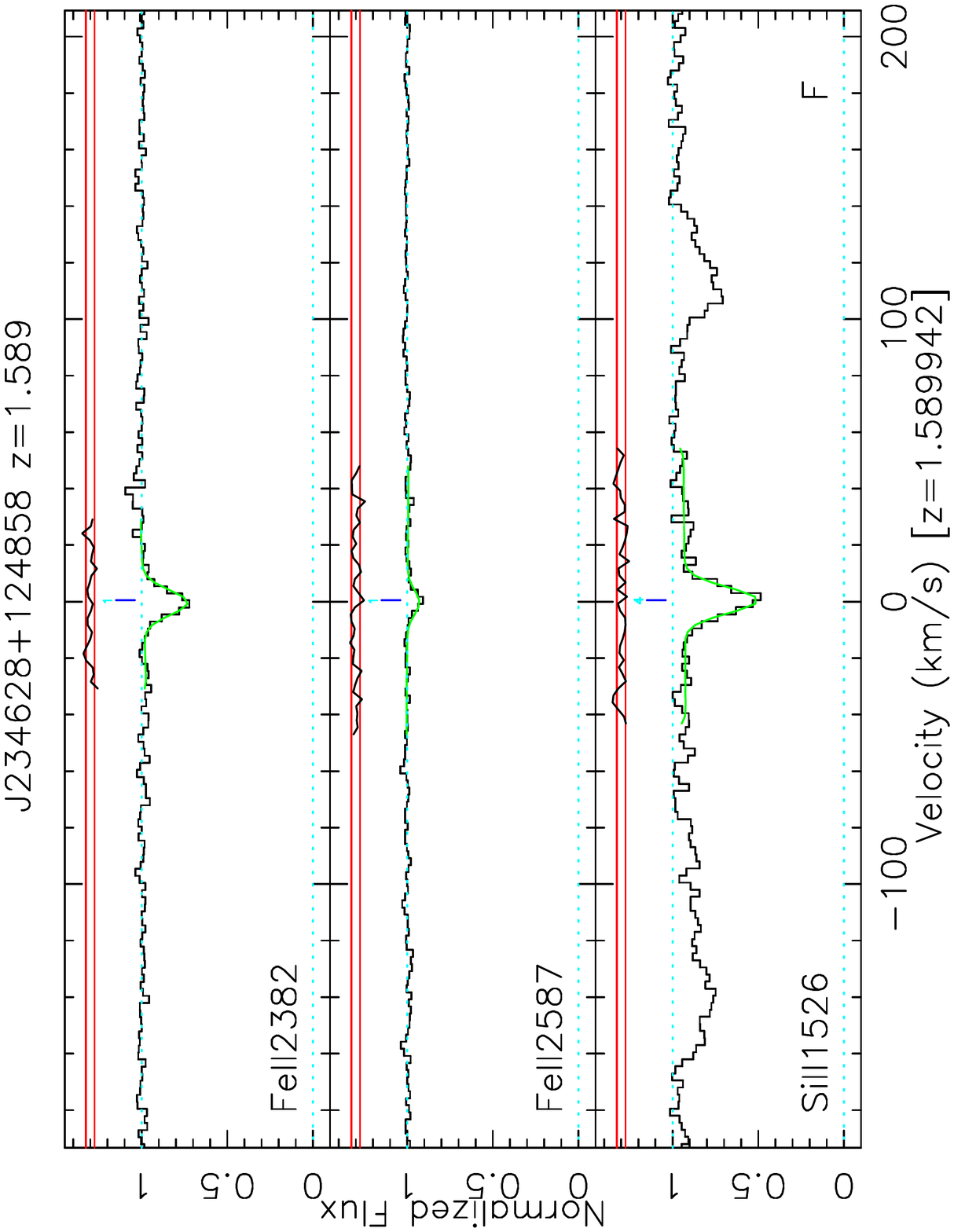}
\par\end{centering}

\caption[\ Fit for the $z=1.589$ absorber toward J234628+124858]{Many-multiplet fit for the $z=1.589$ absorber toward J234628+124858.}
\end{figure}
\begin{figure}[H]
\noindent \begin{centering}
\includegraphics[bb=34bp 58bp 554bp 738bp,clip,width=1\textwidth]{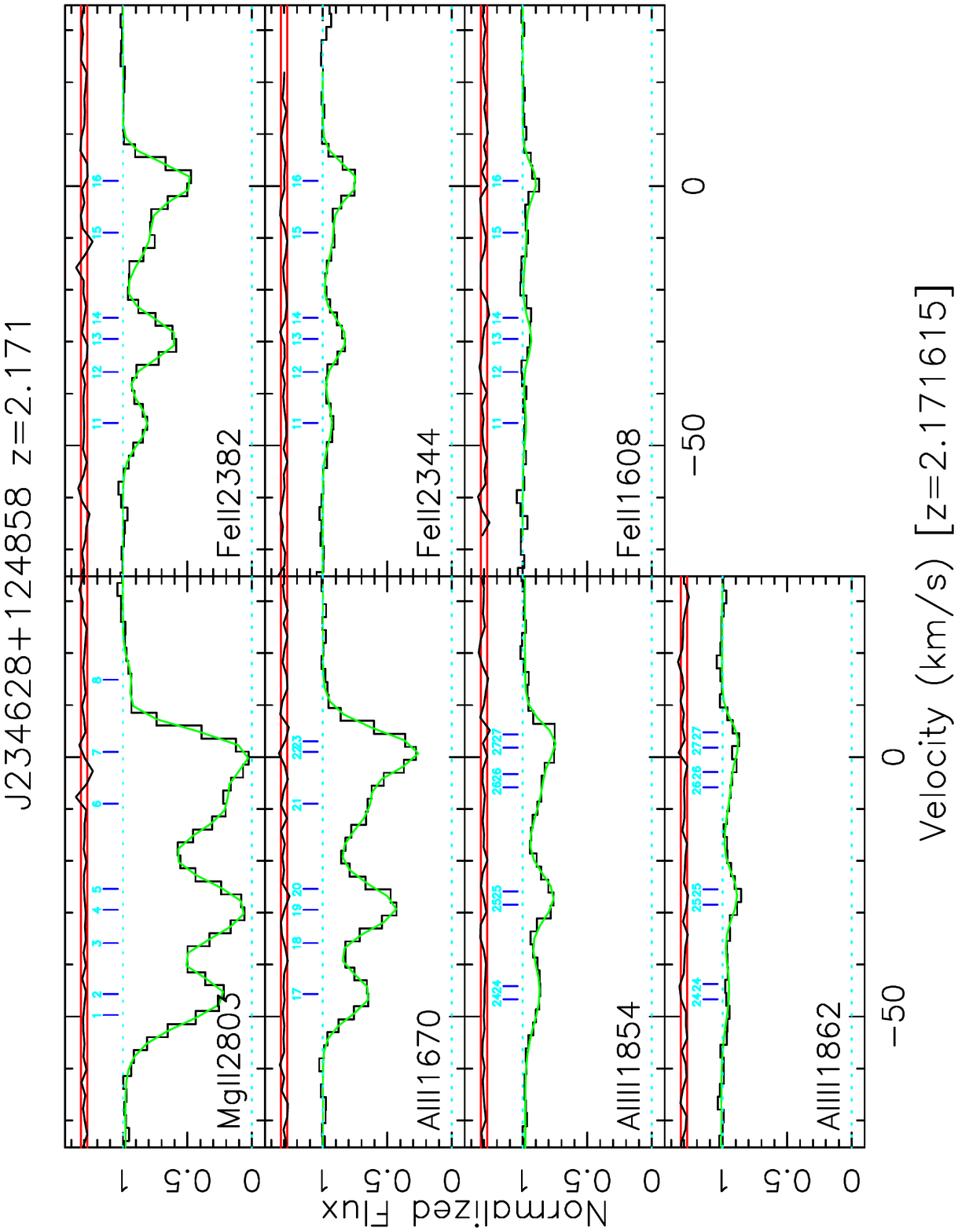}
\par\end{centering}

\caption[\ Fit for the $z=2.171$ absorber toward J234628+124858]{Many-multiplet fit for the $z=2.171$ absorber toward J234628+124858.}
\end{figure}
\begin{figure}[H]
\noindent \begin{centering}
\includegraphics[bb=34bp 58bp 554bp 738bp,clip,width=1\textwidth]{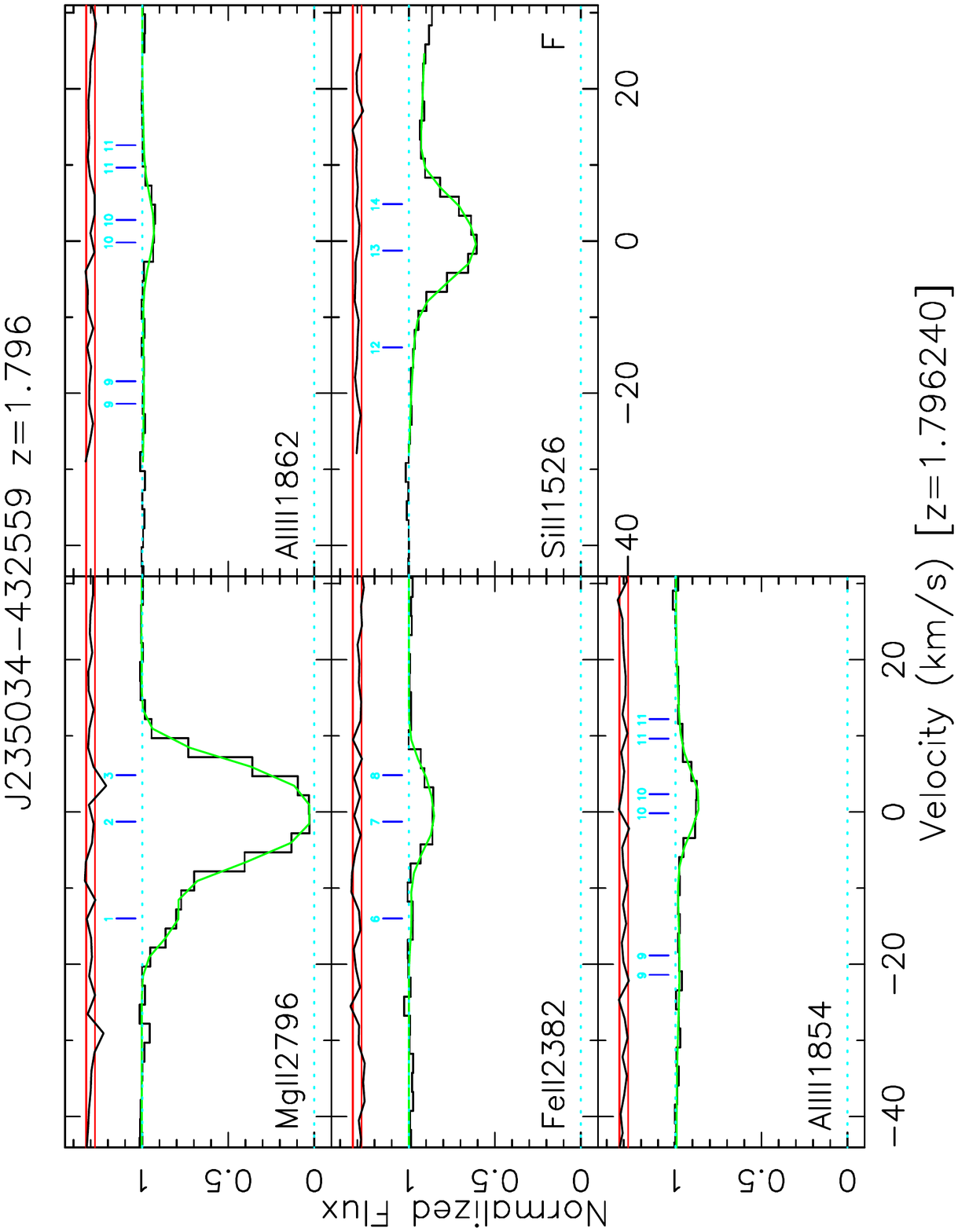}
\par\end{centering}

\caption[\ Fit for the $z=1.796$ absorber toward J235034+432559]{Many-multiplet fit for the $z=1.796$ absorber toward J235034+432559.}
\end{figure}

\cleardoublepage \phantomsection\bibliographystyle{perception}
\addcontentsline{toc}{chapter}{\bibname}\bibliography{thesis,MCMC,alpha,mu}

\cleardoublepage \phantomsection \addcontentsline{toc}{chapter}{Index}\printindex{}
\end{document}